\documentclass[final,3p,12pt]{elsarticle}
\usepackage{amsthm}
\usepackage{graphicx}
\usepackage{amsmath}
\usepackage{times,latexsym,euscript,bm}
\usepackage{amstext,amssymb,amsbsy,amsmath,amsfonts}
\usepackage{rsfs}
\usepackage{color}
\usepackage{bm}

\renewcommand{\theequation}{{\thesection.\arabic{equation}}}
\makeatletter
\@addtoreset{equation}{section}
\makeatother
\journal{Physics Report}

\begin{document}

\begin{frontmatter}

\begin{flushright}  
Preprint: CHIBA-EP-209 v3
\\
KEK Preprint 2014-23
\end{flushright}



\title{
Quark confinement: dual superconductor picture
based on a non-Abelian Stokes theorem and reformulations of  Yang-Mills theory 
}


\author[1]{Kei-Ichi~Kondo}
\ead{kondok@faculty.chiba-u.jp}
\address[1]{Department of Physics,  
Graduate School of Science, 
Chiba University, Chiba 263-8522, Japan}

\author[2]{Seikou Kato}
\ead{skato@fukui-nct.ac.jp}
\address[2]{Fukui National College of Technology, Sabae 916-8507, Japan
}

\author[3]{Akihiro Shibata}
\ead{Akihiro.Shibata@kek.jp}
\address[3]{Computing Research Center, High Energy Accelerator Research Organization (KEK), Tsukuba 305-0801, Japan
%and Graduate Univ. for Advanced Studies (Sokendai), Tsukuba 305-0801, Japan
}

\author[4]{Toru Shinohara}
\ead{sinohara@graduate.chiba-u.jp}
\address[4]{Department of Physics,  
Graduate School of Science, 
Chiba University, Chiba 263-8522, Japan}

\begin{abstract}

The purpose of this paper is to review the recent progress in understanding quark confinement.
The emphasis of this review is placed on how to obtain a manifestly gauge-independent picture for quark confinement supporting the dual superconductivity in the Yang-Mills theory, which should  be compared with the Abelian projection proposed by 't Hooft.  
The basic tools are novel reformulations of the Yang-Mills theory based on  change of variables extending the decomposition  of the $SU(N)$ Yang-Mills field due to Cho, Duan-Ge and Faddeev-Niemi, together with the combined use of extended versions of the Diakonov-Petrov version of the non-Abelian Stokes theorem for the $SU(N)$  Wilson loop operator. 

Moreover, we give the lattice gauge theoretical versions of the reformulation of the Yang-Mills theory which enables us to perform the numerical simulations on the lattice. 
In fact, we present some numerical evidences for supporting the dual superconductivity for quark confinement. 
The numerical simulations include the derivation of the linear potential for static interquark potential, i.e., non-vanishing string tension, in which the ``Abelian'' dominance and magnetic monopole dominance are established, confirmation of the dual Meissner effect by measuring the chromoelectric flux tube between quark-antiquark pair, the induced magnetic-monopole current, and the type of dual superconductivity, etc. 
In addition, we give a direct connection between the topological configuration of the Yang-Mills field such as instantons/merons and the magnetic monopole.

We show especially that magnetic monopoles in the Yang-Mills theory can be constructed in a manifestly gauge-invariant way starting from the gauge-invariant Wilson loop operator and thereby the contribution from the magnetic monopoles can be extracted from the Wilson loop in a gauge-invariant way through the non-Abelian Stokes theorem for the Wilson loop operator, which is a prerequisite for exhibiting magnetic monopole dominance for quark confinement. 
The Wilson loop average is calculated according to the new reformulation written in terms of new field variables obtained from the original Yang-Mills field based on change of variables. 
The Maximally Abelian gauge in the original Yang-Mills theory is also reproduced by taking a specific gauge fixing in the reformulated Yang-Mills theory. 
This observation justifies the preceding results obtained in the maximal Abelian gauge at least for gauge-invariant quantities for $SU(2)$ gauge group, which eliminates the  criticism of gauge artifact raised for the Abelian projection. 
The claim has been confirmed based on the numerical simulations. 

However, for $SU(N)$ ($N \ge 3$), such a gauge-invariant reformulation is not unique, although the extension along the line proposed by Cho, Faddeev and Niemi is possible. In fact, we have found that there are a number of possible options of the reformulations, which are discriminated by the maximal stability group $\tilde{H}$ of $G$, while there is a unique option of $\tilde{H}=U(1)$ for $G=SU(2)$. 
The maximal stability group depends on the representation of the gauge group, to that the quark source belongs.  For the fundamental quark for $SU(3)$, the maximal stability group is $U(2)$, which is different from the maximal torus group $U(1) \times U(1)$ suggested from the Abelian projection. 
Therefore, the chromomagnetic monopole inherent in the Wilson loop operator responsible for confinement of quarks in the fundamental representation for $SU(3)$ is the non-Abelian magnetic monopole, which is distinct from the Abelian magnetic monopole for the $SU(2)$ case. 
Therefore, we claim that the mechanism for quark confinement for $SU(N)$ ($N \ge 3$) is the non-Abelian dual superconductivity caused by condensation of non-Abelian magnetic monopoles. 
We give some theoretical considerations and numerical results supporting this picture. 
Finally, we discuss some issues to be investigated in future studies. 

\end{abstract}


\end{frontmatter}

%\newpage
%\thispagestyle{empty}
\tableofcontents
\newpage

\setcounter{page}{1}
\pagenumbering{arabic}


%\section{}
%\label{}
%%%%%%%%%%%%%%%%%%%%%%%%%%%%%%%%%%%%%%%%%%%%%%%%%%%%%%%%%%%%
%%%%%%%%%%%%%%%%%%%%%%%%%%%%%%%%%%%%%%%%%%%%%%%%%%%%%%%%%%%%
\section{Introduction}\label{sec:intro}
%%%%%%%%%%%%%%%%%%%%%%%%%%%%%%%%%%%%%%%%%%%%%%%%%%%%%%%%%%%%
%%%%%%%%%%%%%%%%%%%%%%%%%%%%%%%%%%%%%%%%%%%%%%%%%%%%%%%%%%%%

It is widely accepted that \textbf{quantum chromodynamics (QCD)} is the most fundamental quantum field theory for describing the \textbf{strong interactions} mediated by gluons whose dynamics is governed by the \textbf{Yang-Mills theory} \cite{YM54}.
In experiments, however, the fundamental degrees of freedom for QCD, i.e., \textbf{quarks} and \textbf{gluons}, have never been observed in the isolated form. 
Only  the \textbf{color singlet} combinations  known as  \textbf{hadrons (mesons and baryons)} and candidates of \textbf{glueballs}  are   observed.  
%Therefore, the physical scale suitable for describing quarks and gluons in terms of hadrons is around 1 fermi, 10$^{-15}$m or the energy around 1GeV. 
%(See Fig.~\ref{fig:glue-string}.)
This experimental fact can be explained from a hypothesis of \textbf{color confinement}: 
Colored objects cannot be observed and only color singlet objects (colorless composites)  can be observed in Nature. 
Indeed, quarks and gluons have their own color charges and hence they cannot be observed according to this hypothesis.  
%Thus we  raise a natural question why color confinement occurs. 
From this viewpoint, \textbf{quark confinement} is a special case of color confinement and seems to be more tractable than color confinement. 
For quark confinement, indeed, we have a well-known gauge-invariant criterion due to Wilson \cite{Wilson74}, i.e., \textbf{area law of the Wilson loop average}, which is equivalent to the \textbf{linear potential} for the static quark-antiquark potential. 
Whereas the gauge-invariant criterion for color confinement is still unknown to the best of the author knowledge.

We wish to clarify the basic mechanism for quark confinement, which enables us to explain how and why quark confinement occurs in Nature.
A promising scenario for quark confinement called the \textbf{dual superconductivity} picture has been proposed in 1970s by Nambu, 't Hooft,  Mandelstam \cite{dualsuper}.
% and others \cite{Polyakov77a,Polyakov77b}. 
The dual superconductivity is supposed to be realized as the \textbf{electric--magnetic duality} of the ordinary superconductivity:  
%The dual superconductivity which is believed as the promising mechanism for quark confinement is conjectured to occur 
The dual superconductivity could be realized as a consequence of condensation of \textbf{magnetic monopoles}, i.e., magnetically charged objects, just as the ordinary superconductivity is caused by condensation of the \textbf{Cooper pairs}, i.e., electrically charged objects. 
In the vacuum of dual superconductor, the \textbf{dual Meissner effect} squeezes the  \textbf{chromoelectric flux} between a quark and an antiquark into a tube like region to form the hadronic string. 
The key ingredients of the dual superconductor  picture for the Yang-Mills theory vacuum are the existence of \textbf{chromomagnetic monopole condensation} and the \textbf{dual Meissner effect}.
In order to establish the dual superconductivity, therefore, it is a first step to show the existence of  {magnetic monopole} in QCD, i.e., \textbf{chromomagnetic monopole}, which is to be condensed in the Yang-Mills theory \cite{YM54}, since chromomagnetic monopole is an indispensable ingredient for dual superconductivity in QCD.

At present, there is no analytical proof of the existence of the condensate of the magnetic monopole in gluodynamics and in QCD. 
%In some models simpler than the  Yang-Mills theory and QCD, however,  it has been assured that the dual superconductivity is true as a confinement mechanism:%
However, in all theories allowing for an analytical proof of confinement, confinement is due to the condensation of magnetic monopoles. 
4D compact electrodynamics (compact $U(1)$ lattice gauge theory) \cite{Polyakov75}, 3D Georgi-Glashow model \cite{Polyakov77} and a $\mathcal{N}=2$ supersymmetric Yang-Mills theory \cite{SW94}.

In view of this, we recall well-known examples of magnetic monopoles appeared so far in gauge field theories. 
In the Maxwell electromagnetism,  the Dirac magnetic monopole \cite{Dirac31,WY75} is realized by introducing singularities in the gauge potential.  Otherwise, the Bianchi identity leads to identically vanishing magnetic current. 
 In the non-Abelian gauge theory with (adjoint) matter fields such as the Georgi-Glashow model,  one can construct the 't Hooft-Polyakov magnetic monopole \cite{tHP74} without introducing the singularity in the Yang-Mills field thanks to the extra degrees of freedom coming from the matter fields. See e.g. \cite{GNO77,GO78} for reviews. 
\begin{itemize}
\item 
%\noindent $\bullet$  
\textbf{The Dirac magnetic monopole} \cite{Dirac31} 
%\footnote{
%P.A.M. Dirac, 
%Quantized Singularities in the Electromagnetic Field,
%Proc. Roy. Soc. London, A{\bf 133}, 60--72 (1931). 
%Tai Tsun Wu, Chen Ning Yang, 
%Concept of Nonintegrable Phase Factors and Global Formulation of Gauge Fields,
%Phys. Rev. D{\bf 12}, 3845--3857 (1975). 
%Dirac Monopole Without Strings: Monopole Harmonics, 
%Nucl. Phys. B107, 365--380  (1976).
%Dirac's Monopole Without Strings: Classical Lagrangian Theor,  
%For a review, see e.g.,
%P. Goddard and D. Olive, 
%New Developments in the Theory of Magnetic Monopoles,
%Rep. Prog. Phys. {\bf 41}, 1357--1437 (1978)
%}
in the $U(1)$ Maxwell theory is realized when and only when there are singularities in the gauge potential $A_\mu(x)$, leading to the violation of the Bianchi identity for $U(1)$ gauge field strength $F_{\mu\nu}(x)$. 
As far as  the gauge potential $A_\mu(x)$ is non-singular, indeed, the magnetic-monopole current $k^\mu$ vanishes identically:%
% due to the nilpotency $dd \equiv 0$%
\footnote{
Here 
$A=A_\mu dx^\mu$ and $k=k_\mu dx^\mu$ are 1-forms and $F=\frac12 F_{\mu\nu}dx^\mu \wedge dx^\nu$ is a 2-form.
And $d$ is the exterior derivative, $\delta$ is the codifferential and $ {}^{\displaystyle *}$ is the Hodge duality operation.
The exterior derivative $d$ has the nilpotency $dd \equiv 0$. 
} 
\begin{equation}
 k= \delta {}^{\displaystyle *}F = {}^{\displaystyle *}dF = {}^{\displaystyle *}ddA \equiv 0 
  .
\end{equation}

\item 
%\noindent $\bullet$ 
\textbf{The 't Hooft-Polyakov magnetic monopole} \cite{tHP74}  in the presence of the  Higgs fields (in the adjoint representation) is obtained  due to  the   \textbf{spontaneous symmetry breaking} of the original gauge symmetry $G$ to a subgroup $H$:
%\footnote{
%G. 't Hooft,
%Magnetic Monopoles in Unified Gauge Theories,
%Nucl. Phys. B{\bf 79}, 276--284 (1974).
%A. M. Polyakov,
%Particle Spectrum in the Quantum Field Theory,
%JETP Lett. {\bf 20}, 194--195 (1974).
%Pisma Zh. Eksp. Teor. Fiz. {\bf 20}, 430--433 (1974). 
%J. Arafune, P.G.O. Freund, and C.J. Goebel,
%Topology of Higgs Fields,
%J. Math. Phys. {\bf 16}, 433  (1975). 
%}
\begin{equation}
 G \to H 
  .
\end{equation}

\end{itemize}

Then we can raise a basic question: What kind of magnetic monopoles can be defined in QCD and Yang-Mills theory without the scalar fields?
Currently, there are at least two methods available to define the  chromomagnetic monopole  in the Yang-Mills theory, i.e., magnetic monopole in the non-Abelian gauge theory in the absence of matter fields:
 %In the non-Abelian gauge theory in the absence of matter fields, two methods are so far known in extracting magnetic monopole degrees of freedom from the Yang-Mills field: 

\begin{enumerate}
%\item
%Abelian projection \cite{tHooft81}.
%  and maximal Abelian gauge as a partial gauge fixing \cite{KLSW87} 
%\\{} \indent [Nucl. Phys. B190, 455 (1981)]

%\item
%Decomposition of the Yang-Mills field variable \cite{DG79,Cho80,FN98,Shabanov99},\cite{Cho80c,FN99a,BCK02} and change of variables \cite{KMS06,KMS05,Kondo06},\cite{KSM08,KKMSS05,KKMSSI06,IKKMSS06,SKKMSI07,KSSMKI08,SKS10}
  
%\\{} \indent [Phys. Rev. D21, 1080 (1980)]  [Phys. Rev. Lett. 82, 1624 (1999)]...
\item
  \textbf{Abelian projection} \cite{tHooft81}%
%\footnote{
%G. 't Hooft,
%Topology of the gauge condition and new confinement phases in non-Abelian gauge theories,
%Nucl.Phys. B{\bf 190} [FS3], 455--478 (1981).
%}
  %(by ' t Hooft \cite{tHooft81})

%partial gauge fixing of the gauge group $G$ to the maximal torus subgroup: $G \rightarrow U(1)^r$  (naive) %Abelian monopole

Abelian magnetic monopole is obtained   due to   \textbf{explicit breaking} (\textbf{partial gauge fixing}) of the original gauge group $G$ to the maximal torus subgroup $H=U(1)^r$ with $r$ being the rank of the gauge group $G$: $G=SU(N) \rightarrow H=U(1)^{N-1}$, e.g., $SU(2) \rightarrow U(1)$, $SU(3) \rightarrow U(1) \times U(1)$.
The magnetic  monopole is identified with the gauge-fixing defect.

\item
\textbf{Field decomposition} 
\cite{DG79,Cho80,FN98,Shabanov99},\cite{Cho80c,FN99a,BCK02} and change of variables \cite{KMS06,KMS05,Kondo06},\cite{KSM08,KKMSS05,KKMSSI06,IKKMSS06,SKKMSI07,KKS14,KSSMKI08,SKS10,KSSK11,SKKS13}

It is possible to realize magnetic monopoles in the pure non-Abelian gauge theory in the absence of the adjoint Higgs field,
without explicitly introducing singularities in the gauge connection and without breaking the original gauge symmetry and color symmetry.
This is done by performing a gauge-covariant decomposition of the gluon field by introducing the local color direction field.
% for separating the dominant mode for confinement. 
%(This method is however involved.)  
%[spin-charge separation]
\end{enumerate}

The first method, i.e., \textbf{Abelian projection} proposed by 't Hooft \cite{tHooft81} is well known and has been conventionally and extensively used to extract  magnetic monopoles from the Yang-Mills field.  
For instance, the Abelian projection for $SU(2)$ is regarded as a partial gauge fixing from the original gauge group $SU(2)$ to the maximal torus subgroup $U(1)$:
\begin{equation}
 G=SU(2) \rightarrow H=U(1) .
\end{equation}
This is an explicit breaking of the (local) gauge symmetry.%
\footnote{
The conventional Abelian gauge breaks the original local symmetry $G$ to leave the local $U(1)^{\rm local} (\subset G=SU(2)^{\rm local})$ and the global $U(1)^{\rm global}$ unbroken, but breaks the global $SU(2)_{\rm global}$.
The new reformulation based on the second method leaves the local $G'=SU(2)^{\rm local}$ and the global $SU(2)^{\rm global}$ unbroken (color rotation invariant).
}

Within the Abelian projection, the relevant data supporting the validity of the dual superconductor picture have been accumulated by numerical simulations on the lattice especially since 1990 and some of the theoretical predications have been confirmed by these investigations. 
In the \textbf{maximally Abelian gauge (MA gauge, MAG)} \cite{KLSW87} for $SU(2)$, as a realization of the Abelian projection, especially, the infrared Abelian dominance \cite{tHooft81,EI82} was confirmed by numerical simulations on the lattice \cite{SY90,BBMPS96,AS99}, and subsequently magnetic monopole dominance \cite{SS94,SNW94} was also confirmed in the string tension in the linear potential for the static quark-antiquark potential. 
The infrared Abelian dominance was also confirmed for the gluon propagator by exhibiting the rapid fall-off of the off-diagonal gluon propagator, leading to the non-vanishing off-diagonal gluon mass.
For the $SU(2)$ Yang-Mills theory, thus, the following  characteristic features supporting the dual superconductivity for confinement were established in the MA gauge for $SU(2)$:
\begin{itemize}
\item
infrared Abelian dominance in the string tension \cite{SY90,BBMPS96}: 

\item
magnetic monopole dominance in the string tension \cite{SS94,SNW94}:

\item
diagonal gluon propagator dominance, and 
rapid fall-off of the off-diagonal gluon propagator and non-vanishing off-diagonal gluon mass  \cite{AS99,BCGMP03}:

%\item
%center vortex dominance \cite{Greensite03}: 

\end{itemize}
However, they are confirmed only in specific gauges, i.e., \textbf{Abelian gauges}, such as the MA gauge, Laplacian Abelian gauge \cite{Sijs91,LAG} and maximal center   gauge \cite{Greensite03}.
All these gauge fixing break color symmetry explicitly.
See, e.g., \cite{Suzuki93,CP97,Haymaker99,Bali98} for reviews of Abelian projection and Abelian gauge and also \cite{Greensite03,Greensite11,Engelhardt05} for more topics. 
See e.g., \cite{Ripka03} for dual superconductor models.

The second method less familiar than the first one is based on the gauge-covariant decomposition of the Yang-Mills field, which has been proposed independently by Cho \cite{Cho80,Cho80c} and Duan and Ge \cite{DG79}  at nearly  the same time as the 't Hooft proposal of the Abelian projection, and later readdressed by Faddeev and Niemi \cite{FN98,FN99a} and Shabanov \cite{Shabanov99}.
Therefore, it is called the 
\textbf{Cho--Duan-Ge (CDG) decomposition} or \textbf{Cho--Duan-Ge--Faddeev-Niemi (CDGFN) decomposition}, or \textbf{Cho--Duan-Ge--Faddeev-Niemi--Shabanov (CDGFNS) decomposition}.
By using the second method, the magnetic monopole in the Yang-Mills theory is constructed in a gauge-independent way. 
The first method, i.e., Abelian projection is reproduced as a special limit of the second method, i.e., Field decomposition.
In other words, the first method can be regarded as a gauge-fixed version of the second method.  
The second method including the extension to $SU(N)$ explained later has been furthermore developed especially in the last decade by the Chiba University group \cite{KMS06,KMS05,Kondo06,KSM08} and Chiba/KEK/Fukui (Takamatsu) collaborations \cite{KKMSS05,KKMSSI06,IKKMSS06,SKKMSI07,KKS14,KSSMKI08,SKS10,KSSK11,SKKS13}.

The key ingredient of the decomposition is the introduction of a novel field $\bm{n}(x)$ which represents the (local) Abelian direction embedded in the non-Abelian gauge degrees of freedom at each space--time point. 
%The CDG decomposition enables us to extract the dominant degrees of freedom in the infrared region in a gauge-covariant manner and give a gauge-invariant definition of the chromomagnetic monopole in the Yang-Mills theory. 
The decomposition begins with the introduction of the unit field $\bm{n}(x)$, which we call the  \textbf{color direction field} or \textbf{color field} for short.
For $G=SU(2)$,  the  color field $\bm{n}(x)$ has three components and a unit length.  Therefore, it has two independent degrees of freedom.
In the Lie-algebra notation $\bm{n}$, the color field $\bm{n}(x)$ is expressed  as 
\begin{align}
 \bm{n}(x)=  n^A(x)\frac12 \sigma_A  , \quad \quad (A=1,2,3),
%T_A= \frac12 \sigma_A  
%\quad 
% \nonumber\\
% \Longleftrightarrow \mathbf{n}(x) = (n_1(x),n_2(x),n_3(x)) ,
%\nonumber\\
%{\rm i.e.,} {\rm tr}[\bm{n}(x)\bm{n}(x)]=1/2 \quad {\rm or} \quad  \mathbf{n}(x) \cdot \mathbf{n}(x) = n_A(x) n_A(x) = 1 , 
\end{align}
where $\sigma_A$ ($A=1,2,3$) denotes the Pauli matrices. 
The constraint of the unit length is represented   as $n^A(x) n^A(x) = 1$.
For $SU(2)$, therefore, the color field takes the value in the Lie algebra of $SU(2)/U(1)$, which is easily observed using the adjoint orbit representation without loss of generality: 
\begin{align}
 \bm{n}(x)=  U(x) \frac{\sigma_3}{2} U(x)^\dagger \in Lie[SU(2)/U(1)], \quad U(x) \in SU(2) .
\end{align}

According to the decomposition, the Yang-Mills gauge field $\mathscr{A}_\mu(x)$ can be separated into the \textbf{restricted field} $\mathscr{V}_\mu$ (corresponding to the {``Abelian'' part} in the Abelian projection) and the remaining field  $\mathscr{X}_\mu$:
\begin{align}
 \mathscr{A}_\mu(x) = \mathscr{V}_\mu(x) + \mathscr{X}_\mu(x) ,
\end{align}
in such a way  that 
\begin{enumerate}
\item
%$\bullet$ 
{[gauge-independent ``Abelian''  projection]}
The restricted part $\mathscr{V}_\mu$ is the dominant mode responsible for quark confinement.
The non-Abelian Wilson loop operator $W_C[A]$ defined in terms of the original field $\mathscr{A}_\mu$ is entirely rewritten in terms of the restricted field $\mathscr{V}_\mu$ alone, i.e., 
\begin{equation}
 W_C[\mathscr{A}]=W_C[\mathscr{V}] .
\end{equation}
Furthermore, it is rewritten into the surface integral of the $SU(2)$ invariant field strength $F^{[\mathscr{V}]}_{\mu\nu}(x)$ of the Abelian type over the surface $\Sigma$ bounded by the loop $C$: 
\begin{equation}
 W_C[\mathscr{V}]= \tilde W_\Sigma [F^{[\mathscr{V}]}] ,
\end{equation}
where $F^{[\mathscr{V}]}(x)$ is defined from the color direction field $\bm{n}(x)$ and the field strength $\mathscr{F}_{\mu\nu}^{[\mathscr{V}]}(x)$ of the restricted field $\mathscr{V}_\mu(x)$.  
\begin{equation}
 F^{[\mathscr{V}]}_{\mu\nu}(x) =  \bm{n}(x) \cdot \mathscr{F}_{\mu\nu}^{[\mathscr{V}]}(x) .
\end{equation}
This fact follows from the non-Abelian Stokes theorem for the Wilson loop operator. 
  (The restricted field corresponds to the Abelian part in the Abelian projection.)

\item
\noindent
%$\bullet$ 
{[infrared restricted field (``Abelian'')  dominance]}
The remaining part $\mathscr{X}_\mu$ decouples in the low-energy regime. 
The correlation function of the field $\mathscr{X}_\mu$ exhibits the rapid fall-off of the exponential type. 
This suggests that the field $\mathscr{X}_\mu$  acquires the gauge-invariant mass dynamically.
% through the non-vanishing vacuum condensation of mass dimension--two $\left< \mathscr{X}_\mu^2 \right> \ne 0$.  
%The remaining part $\mathscr{X}_\mu$ acquires the mass to decouple in the low-energy region, leading to infrared ``Abelian'' $\mathscr{V}_\mu$ dominance. 
This leads to the \textbf{infrared restricted field (``Abelian'')  dominance}.
The dynamical mass can originate from  the   gauge-invariant  \textbf{vacuum condensate  of mass dimension-two} $\left< \mathscr{X}_\mu^2 \right>\ne 0$. 
Note that $\mathscr{X}_\mu^2$ is a gauge-invariant operator of mass dimension two, which corresponds to the \textbf{BRST invariant operator of mass dimension two}, constructed from the gluon and ghost \cite{Kondo01}. 
The existence of dimension-two condensate $\left< \mathscr{X}_\mu^2 \right> \ne 0$ has been examined by an analytical method \cite{Kondo06}  and numerical one \cite{KKMSS05}. 

%Moreover, we have shown  \cite{Kato-lattice2009} that the string tension is also reproduced by a gauge-independent variable called the restricted field obtained by the decomposition, which we call the \textbf{``Abelian'' dominance} or \textbf{infrared restricted field dominance} to distinguish it from the Abelian dominance obtained under the gauge-dependent Abelian projection. 

\end{enumerate}
The restricted part $\mathscr{V}_\mu$ ({``Abelian'' part})  responsible for quark confinement can be extracted from the non-Abelian gauge field in the {gauge-independent way} without breaking color symmetry. 
Using the restricted field alone, we can give a definition of a gauge-invariant magnetic monopole current $k^\mu(x)$: \begin{equation}
 k^\mu(x) = \partial_\nu  {}^{\displaystyle *}F^{[\mathscr{V}] \mu\nu}(x) ,
\end{equation}
giving the magnetic charge $q_m:= \int d^3x k^0(x)$ subject to the Dirac quantization  condition
(which is confirmed by analytical and numerical methods \cite{IKKMSS06})
and magnetic monopole dominance in the string tension (confirmed by numerical method \cite{IKKMSS06,KKS14}). 

%In ref.\cite{KOSSM06}, we have discussed how the Faddeev model can be regarded as a low-energy effective theory of Yang-Mills theory to see the mass gap.

The second method is more involved than the first one, but it is now recognized to be superior in some aspects to the first one:  
\begin{itemize}
\item
%\noindent
%$\bullet$  
The gauge-invariance of the chromomagnetic monopole is guaranteed from the beginning by construction.

\item
%\noindent
%$\bullet$  
The direct relevance of the chromomagnetic monopole to the Wilson loop and ``Abelian'' dominance  in the operator level are manifest via  a non-Abelian Stokes theorem for the Wilson loop operator.

\end{itemize}

The CDG decomposition of the Yang-Mills field variable is translated to the change of variables. 
Then the Yang-Mills theory was reformulated in terms of new variables based on a new viewpoint of the CDG decomposition proposed in \cite{KMS06,KMS05}.  
%\begin{itemize}
%\item
%\noindent
%$\bullet$  
The path integral quantization has been completed  by giving the action and the integration measure in the closed form in terms of new field variables.

%\item
%\noindent
%$\bullet$  
%The relevant lattice gauge formulations become available for numerical simulations.

%\end{itemize}

The second method enables us to answer the following questions which are unclear in the first one.
\begin{enumerate}
\item 
How to extract the {``Abelian'' part} responsible for quark confinement in the {gauge-independent way}   from the non-Abelian gauge theory without losing characteristic features of non-Abelian gauge theory, e.g., asymptotic freedom \cite{KondoI,KK05}.

\item  
How to define the {magnetic monopole} to be condensed in  Yang-Mills theory in the {gauge-independent way} even in absence of any scalar field, in sharp contrast to the Georgi-Glashow model in which scalar field in the adjoint representation plays the important role in constructing the 't Hooft-Polyakov magnetic monopole.

\end{enumerate}

 The purpose of this paper is to give a review of recent developments for understanding quark confinement by paying special attention to the second method in comparison with the first one, which will give more insight into the mechanism for quark confinement.
In particular, we emphasize some aspects of the second method superior to the first one for establishing the dual superconductivity in Yang-Mills theory. 
In particular, the reformulation enables us to extract in a gauge-independent manner the dominant degrees of freedom that are relevant to quark confinement in the sense of the Wilson criterion in such a way that they reproduce the original string tension in the linear part of the static quark-antiquark potential.

Moreover, we have given a new reformulation for the lattice $SU(2)$ Yang-Mills theory, so that it reproduces the continuum $SU(2)$ Yang-Mills theory   in the naive continuum limit.
This  new reformulation of the Yang-Mills field  theory on a lattice enables us to perform numerical simulations on a lattice. 
In fact, we present the results of numerical simulations obtained based on this framework and give some numerical evidences supporting the  dual superconductivity for quark confinement. 
%For the $SU(2)$ Yang-Mills theory, the numerical simulations include the derivation of the linear potential for static interquark potential, i.e., non-vanishing string tension.
%in which the infrared ``Abelian'' dominance and magnetic monopole dominance are established, 
For the $SU(2)$ Yang-Mills theory, the second method has already reproduced all characteristic features supporting the dual superconductivity for confinement obtained so far by the first method:
 %i.e., Abelian projection by 't Hooft. 
\begin{itemize}
\item
infrared restricted field (or ``Abelian'') dominance in the string tension (calculated from the Wilson loop average) \cite{KKS14}: 
This is called the \textbf{``Abelian'' dominance} or \textbf{restricted field dominance} to distinguish it from the Abelian dominance obtained under the gauge-dependent Abelian projection. 

\item
magnetic monopole dominance in the string tension  (calculated from the Wilson loop average) \cite{IKKMSS06}:

\item
restricted field propagator dominance, or 
rapid fall-off of the remaining field propagator with a non-vanishing mass  \cite{SKKMSI07}:

\item
confirmation of the dual Meissner effect by measuring the chromoelectric flux tube between quark-antiquark pair, the induced magnetic-monopole current around the chromoelectric flux tube, and the type of dual superconductivity, etc. \cite{KKS14}:

\end{itemize}
In addition, we give a direct connection between the magnetic monopole and the other topological configurations of the Yang-Mills field such as instantons/merons. 

For the $SU(2)$ Yang-Mills theory, the gauge-independent results obtained in the second method guarantee the validity of those obtained by the preceding studies based on the first method, since the MA gauge is reproduced as a special gauge choice of the gauge-independent second method.  
This fact removes the criticism raised for the results obtained in the MA gauge for $SU(2)$. 
This is a great progress toward the goal of understanding the mechanism of quark confinement. 
But, it turns out that the cases of $SU(3)$ and $SU(N)$ ($N>3$) are not necessarily the straightforward extensions of the $SU(2)$.

Subsequently, we proceed to consider the general $SU(N)$ Yang-Mills theory ($N \ge 3$) including a realistic case of three colors $N=3$. 
Even in the $SU(N)$ Yang-Mills theory, we can extract magnetic monopoles using the Abelian projection, 
which breaks explicitly the original gauge group $G=SU(N)$ into the maximal torus subgroup $H=U(1)^{N-1}$:
\begin{equation}
 G=SU(N) \rightarrow H=U(1)^{N-1} .
\end{equation}
 Then, the Abelian projection yields $N-1$ kinds of Abelian magnetic monopoles with different magnetic charges, in agreement with the observation due to the non-trivial homotopy group: 
\begin{equation}
 \pi_{2}(SU(N)/U(1)^{N-1})=\pi_{1}(U(1)^{N-1})=\mathbb{Z}^{N-1}.
 \label{Homo1}
\end{equation}
 Therefore, it tends to assume that magnetic monopoles of $N-1$ kinds  are necessary to realize quark confinement in the $SU(N)$ Yang-Mills theory from the viewpoint of the dual superconductivity, while a single magnetic monopole is sufficient in the $SU(2)$ Yang-Mills theory. 
However, it is not yet established whether or not $N-1$ kinds of magnetic monopoles are necessary or indispensable to achieve quark confinement in $SU(N)$ Yang-Mills theory ($N \ge 3$). 
 
Keeping this question in mind, we proceed to extend the new formulation of $SU(2)$ Yang-Mills theory to the $SU(N)$ case. 
For the $SU(N)$ Yang-Mills field, the decomposition as an extension of the CDGFN decomposition was proposed by Cho \cite{Cho80c,BCK02} and Fadeev-Niemi \cite{FN99a}  
which we call the \textbf{Cho--Faddeev-Niemi (CFN) decomposition}.
They introduced $N-1$ color direction fields $\bm{n}_j(x)$ $(j=1, \dots, N-1)$ and decomposed the $SU(N)$ Yang-Mills field  by way of the color fields in the similar way to the $SU(2)$ case.
This is reasonable, since  the $SU(N)$ group has $N-1$ Abelian directions specified by the Cartan subalgebra $\{ H_1, \dots, H_{N-1} \}$ whose generators mutually commute, i.e., $[H_j, H_k]=0$ ($j,k \in \{ 1,\dots, N-1\}$).  
Then the   color field takes the value in the Lie algebra of $G/H$, which is represented by 
%$\bm{n}_j(x)=U(x)H_jU(x)^\dagger$.
\begin{align}
 \bm{n}_j(x)=U(x)H_jU(x)^\dagger \in Lie[SU(N)/U(1)^{N-1}], \quad U(x) \in SU(N) .
\end{align}
Consequently, the $SU(N)$ Yang-Mills theory is rewritten in terms of new variables constructed through the color fields just defined, just as done in the $SU(2)$ case. 
This decomposition and subsequent reformulation enables us to give a gauge-independent Abelian-projection.
The CFN scenario lead to the $N-1$ kinds of gauge-invariant  magnetic monopoles. 
However, whether or not this reformulation gives the best description for quark confinement is another problem to be examined. 
This is because the reformulation which is equivalent to the original Yang-Mills theory is not unique. Which reformulation gives the best description for confinement should be determined by imposing further requirements.

Rather, we claim that even in the $SU(N)$ Yang-Mills theory a single magnetic monopole is sufficient to achieve confinement, when quarks belong to the fundamental representation of the color gauge group. 
In fact, this scenario was originally proposed in \cite{KT00b,KT00,Kondo99Lattice99} based on a consideration from a non-Abelian Stokes theorem for the Wilson loop operator \cite{DP89,DP96,FITZ00,KondoIV}, which is more elaborated in \cite{Kondo08,Kondo08b,KS08,Kondo00} (See   \cite{HU99,Halpern79,Bralic80,Arefeva80,Simonov89,Lunev97,HM97} for other versions of non-Abelian Stokes theorem). 
In order to consider confinement of quarks in the specified representation based on the Wilson loop operator,
we adopt the color field $\bm{n}(x)$ taking the value in the Lie algebra of $G/\tilde H$, 
\begin{align}
 \bm{n}(x)  \in Lie[G/\tilde H], %\quad U(x) \in SU(N) .
\end{align}
where $\tilde{H}$ is a subgroup of $G$ called the \textbf{maximal  stability subgroup} which is determined once the highest-weight state of a given representation is chosen. 
This is the best theoretical choice deduced from the non-Abelian Stokes theorem for the Wilson loop operator. 
%This is  associated with the breaking of the original gauge group $G$ to the \textbf{maximal  stability subgroup} $\tilde{H}$ (defined in the text):
%\begin{equation}
%G=SU(N) \rightarrow \tilde{H}=U(N-1) \simeq SU(N-1) \times U(1) .
%\end{equation}
For the fundamental representation of $G=SU(N)$, we find that the maximal stability group is $\tilde H=U(N-1)$ due to the non-Abelian Stokes theorem for the Wilson loop operator. 
The fact that the target space of the color field is the coset $G/\tilde H$ suggests a single magnetic monopole  inherent in the Wilson loop operator is responsible for confining \textbf{quarks in the fundamental representation}, in agreement with the non-trivial Homotopy group:
\begin{equation}
 \pi_{2}(SU(N)/U(N-1))=\pi_{1}(SU(N-1) \times U(1))=\pi_{1}(U(1))=\mathbb{Z} ,
 \label{Homo2}
\end{equation}
which  is independent of $N$, in sharp contrast to the Abelian projection (\ref{Homo1}).

By adopting a single color field alone, we can reformulate the $SU(N)$ Yang-Mills theory in terms of new variables. 
This fact elucidates the crucial difference between the decomposition in the preceding approach and the change of variables in our approach, which we emphasize in this paper. 
In fact, this observation can be substantiated by making use of a novel reformulation of Yang-Mills theory using new variables completed in \cite{KSM08} combined with the non-Abelian Stokes theorem for the Wilson loop operator \cite{Kondo08}. 
The new formulation is an extension of the CFN proposals, since the CFN case is obtained as a special option of the new formulation. 
The reformulation can exhaust all the options discriminated by the possible choice of the maximal stability groups. 
Among them, the reformulation with a single color field is called the \textbf{minimal option}, while the conventional CFN option is called the \textbf{maximal option}.

For   $G=SU(3)$, in particular, there are only two possibilities:  \textbf{maximal option} and  \textbf{minimal option}. 
The minimal option with the {maximal  stability subgroup}: 
\begin{equation}
  \tilde{H}=U(2) \simeq SU(2)\times U(1) 
\end{equation}
 is a new option (overlooked so far) suited  for representing the Wilson loop in the fundamental representation, on which we focus in this review, 
while the maximal one with the maximal torus subgroup as a maximal stability group:
\begin{equation}
  \tilde{H}=H =U(1)\times U(1) 
\end{equation}
enables us to give a gauge-independent reformulation of the Abelian projection represented by the conventional MA gauge.
The maximal option is equivalent to the CFN case. 
For $G=SU(N)$ ($N \ge 4$), there are more options other than the maximal and minimal ones. 
For instance, see \cite{KSM08} for concrete examples of $G=SU(4)$ case. 
In this review we consider only maximal and minimal options and we do not discuss other options.

For $SU(3)$ in the minimal option, the resulting chromomagnetic monopole is not the Abelian magnetic monopole, rather a \textbf{non-Abelian magnetic monopole} with non-trivial internal structure $U(2)$, of which only the $U(1)$ degree of freedom is seen by a single magnetic charge in the long distance. 
Therefore, we propose the \textbf{non-Abelian dual superconductivity} as a mechanism for confining quarks in the fundamental representation in the $SU(3)$ Yang-Mills theory. 

In the static potential for a pair of  quark and antiquark in the fundamental representation, it has been demonstrated:
(i) the  infrared \textbf{restricted-field dominance} (or \textbf{``Abelian''  dominance})  (which is a gauge-independent (invariant) extension of the conventionally called infrared Abelian  dominance):  the string tension $\sigma_{\rm V}$ obtained from the   restricted  field $V$ reproduces the string tension $\sigma_{\rm full}$ of the original YM field,
$\sigma_{\rm V}/\sigma_{\rm full}=93\pm16\%$,
(ii)  the gauge-independent \textbf{non-Abelian magnetic monopole dominance}:  the string tension $\sigma_{\rm V}$ extracted from the restricted field is reproduced by only the (non-Abelian) magnetic monopole part $\sigma_{\rm mon}$,
$\sigma_{\rm mon}/\sigma_{\rm V}=94\pm9\%$. 
We have also confirmed the dual Meissner effect by measuring the chromoelectric flux tube between quark-antiquark pair, the induced magnetic-monopole current around the chromoelectric flux tube, and the type of dual superconductivity, etc.  the $SU(3)$ Yang-Mills theory in the minimal option.
These results are based on the lattice reformulation  of $SU(3)$ Yang-Mills theory constructed in \cite{KSSMKI08,SKS10} and the results of numerical simulations on the lattice performed in \cite{KSSK11,SKKS13}.

Finally, it is fair to mention that other mechanisms were proposed and investigated. 
In the $Z(N)$ \textbf{vortex (condensation) picture}, the the vacuum of the $SU(N)$ Yang-Mills theory is presumed to be dominated by long vortices carrying multiples of $Z(N)$ magnetic flux \cite{tHooft79,Mack80,Cornwall79,NO79}.  
In particular, another very popular one was the proposal based on \textbf{center vortices} \cite{center-vortex} as topological excitations that disorder the Wilson loop. The community was somewhat divided and arguing about one picture or the other, without reaching a consensus, until slowly people moved on to study other aspects of confinement.
This paper does not discuss the center vortex picture for confinement. 
See e.g., \cite{Greensite03} for the review on maximal center gauge and also \cite{DFGO97,DFGO97b,AGG00} for the relationship between the center vortex and magnetic monopole.

This paper is organized as follows. 
%This review is organized as follows.

In section 2, 
we give a short review on the original idea of the Abelian projection as a partial gauge fixing to demonstrate how  magnetic monopoles are obtained in the Yang-Mills theory in the absence of the scalar field. 
We give the basic knowledge on the MA gauge as a realization of the Abelian projection, and briefly summarize the remarkable results obtained  in the MA gauge  by the numerical simulations on the lattice in the last century.

In section 3, 
we give a review on the CDG decomposition of the $SU(2)$ Yang-Mills field, which is another method to define magnetic monopoles in the Yang-Mills theory in the absence of the scalar field. 
The decomposition is uniquely determined by solving the defining equations. 
We explain the relationship between the CDG decomposition and the Abelian projection.

In section 4, 
we give a reformulation of the $SU(2)$  Yang-Mills theory rewritten in terms of new variables.  The new variables are obtained from the original Yang-Mills field based on a (non-linear) change of variables.
The reformulated Yang-Mills theory is constructed based on the new viewpoint proposed by Kondo, Murakami and Shinohara \cite{KMS05,KMS06}, which resolves some questions raised for the original idea of the CDG decomposition and gives a clear guidance for further developments.
Indeed, it guarantees that the reformulated Yang-Mills theory written in terms of new variables is equipollent to the original Yang-Mills theory.  
This gives a definite prescription of defining the quantum Yang-Mills theory according to  the path-integral quantization so that the action integral and the integration measure are explicitly written in the closed form in terms of new variables.

In section 5, 
we show how the reformulation given for the $SU(2)$ Yang-Mills theory using new variables can be extended to the $SU(N)$ Yang-Mills theory ($N \ge 3$).
%The $SU(N)$ gauge group has the maximal torus subgroup $U(1)^{N-1}$ as the Cartan subgroup, i.e., $(N-1)$ Abelian directions in color space.
%The Abelian projection  as the partial gauge fixing $SU(N) \rightarrow U(1)^{N-1}$ breaks explicitly the original gauge symmetry and the color symmetry.
For the $SU(N)$ gauge group with rank $N-1$,  it possible to introduce $N-1$ color fields which will play the role of keeping the color symmetry (or recovering the color symmetry lost by adopting the Abelian projection) and defining a gauge-invariant magnetic monopole in $SU(N)$ case, which is the choice of Cho and Faddeev-Niemi. 
%From the viewpoint of reformulating the field theory, it is important to recognize what are the independent degrees of freedom to describe the theory in question. 
%In fact, this procedure  has been adopted so far to extend the $SU(2)$ CDGFN decomposition to the $SU(N)$ case \cite{Cho80c,FN99a}. 
%For the gauge group  $SU(3)$ with a rank of two, it is convenient to introduce two color fields ${\bf n}_3(x)$ and ${\bf n}_8(x)$. 
However, it is shown \cite{KSM08} that this procedure is not necessarily a unique starting point to reformulate the $SU(N)$ Yang-Mills theory and that only a single color field is sufficient to reformulate the $SU(N)$ Yang-Mills theory irrespective of the number of color $N$. 
%In order to obtain a new theory equipollent to the original theory, the number of independent degrees of freedom must be the same in the new theory and the original theory. 
Moreover, we find that the minimal case is indispensable  
in order to understand confinement of quarks in the fundamental representation, which will be shown in the next section based on the non-Abelian Stokes theorem for the Wilson loop operator.

In section 6, 
we give a (Diakonov-Petrov) version of non-Abelian Stokes theorem for the Wilson loop operator.
The presentation \cite{KondoIV,KT00b,KT00,Kondo08} based on the coherent state of the Lie group and the path-integral representation is more systematic or universal than the original derivation by Diakonov and Petrov \cite{DP89}.  
This form is quite useful to discuss quark confinement combined with the reformulation.  In fact, it turns out that the Wilson loop is a probe of the gauge-invariant magnetic monopole defined in this reformulation. 
We can define the gauge-invariant magnetic monopole from the Wilson loop operator which is gauge invariant by definition. 
The magnetic monopole is inherent in the Wilson loop operator. 
The non-Abelian Stokes theorem tells us why the restricted field dominance follows from the Wilson loop operator.

In section 7, we identify the restricted field $\mathscr{V}$ with the infrared dominant mode, and the remaining field $\mathscr{X}$ with the high-energy relevant mode (infrared decoupled mode) in the original Yang-Mills field. 
This is based on a very interesting interrelationship between the field decomposition and the non-Abelian Stokes theorem. 
We try to obtain a low-energy effective theory written in terms of the restricted field $\mathscr{V}$ alone, by integrating out the remaining field $\mathscr{X}$ in the Yang-Mills action. 
For this purpose, we discuss how to extract the independent field degrees of freedom.  
To demonstrate the validity of the identification and the resulting low-energy effective theory, we discuss the crossover between the confinement/deconfinement and chiral-symmetry-breaking/restoration phase transitions in two color QCD at finite temperature. 
This is regarded as the first-principle derivation of the Polyakov-loop extended Nambu-Jona-Lasinio (PNJL) model which has been extensively used to study the QCD phase diagram.

In section 8, 
we show in an analytical and numerical ways (without using the numerical simulations) that the color direction field can be determined by solving the differential equation for the reduction condition for a given Yang-Mills field, since the color field plays the key role  in the new reformulation of the Yang-Mills theory written in terms of the new variables.
Moreover, we examine when circular loops of magnetic monopole exist in the four-dimensional Euclidean $SU(2)$ Yang-Mills theory, since they are expected to be responsible for quark confinement in the dual superconductor picture. 
As  the given Yang-Mills fields, we examine  some known solutions of the classical Yang-Mills field equation with  non-trivial topology, i.e., one-meron, two-merons, one-instanton, and two-instantons.

In section 9, we give new lattice formulations which reduce in the continuum limit to the reformulation given  in the continuum form.  They enable us to study non-perturbative aspects by performing numerical simulations on the lattice.  
In fact, we give some numerical evidences for dual superconductivity in the Yang-Mills theory for $SU(2)$ and $SU(3)$ as enumerated in the above. 

In section 10, 
we refer to  some applications of the reformulations to other physical problems. 

The final section is devoted to conclusion and discussion. 
Several Appendices are added to give technical details which are removed from the text.

%\newpage
%%%%%%%%%%%%%%%%%%%%%%%%%%%%%%%%%%%%%%%%%%%%%%%%%%%%%%%%%%%%
\subsection{Notations}
\label{subsection:notations}
%\setcounter{equation}{0}
%%%%%%%%%%%%%%%%%%%%%%%%%%%%%%%%%%%%%%%%%%%%%%%%%%%%%%%%%%%%

In this review, we use the notation 
\begin{equation}
A:=B \quad \text{or} \quad B=: A
\end{equation}
 in the sense that $A$ is defined by $B$.
On the other hand, 
\begin{equation}
A \equiv B
\end{equation}
implies that $A$ is identically equal to $B$. 
Moreover, 
\begin{equation}
A \Longrightarrow B \quad (A \Longleftarrow B)
\end{equation}
implies that $B$ follows from $A$ or that $A$ leads to $B$ ($A$ follows from $B$ or that $B$ leads to $A$),
and  
\begin{equation}
A \Longleftrightarrow B
\end{equation}
implies that $A$ is equivalent to $B$.

We use the following two different notations to express field variables. 
\\
\noindent
\underline{The vector form}: 
\begin{equation}
\mathbf{X}(x) = (X^A(x))_{A=1}^{d}=(X^1(x),X^2(x), \ldots  , X^d(x)) ,
%   \mathbf{n}(x)=(n_A(x))_{A=1}^{d}=(n_1(x),n_2(x), \cdots, n_d(x)) ,
  \quad   d =  {\rm dim}G 
,
\end{equation}
or especially in the lattice gauge theory introduced later 
\begin{equation}
  \mathbb{X}(x) =  (X^A(x))_{A=1}^{d}=(X^1(x),X^2(x), \ldots ,  X^d(x)) ,
    \quad   d =  {\rm dim}G 
, 
\end{equation}
or
\begin{equation}
  \mathbb{X}_{x} =  (X^A_{x})_{A=1}^{d}=(X^1_{x},X^2_{x}, \ldots ,  X^d_{x}) ,
    \quad   d =  {\rm dim}G 
, 
\end{equation}
where $d$ denotes the dimension of a gauge group $G$, i.e., $d:={\rm dim}G$, e.g., ${\rm dim}G=N^2-1$ for $G=SU(N)$.
%\footnote{
%In this paper we do not discuss the gauge group other than $SU(N)$.  See \cite{} for other gauge groups. 
%}

%In order to lower and  raise the indices $A,B,C, \dots$, we need to introduce the Cartan metric $g_{AB}$ and its inverse $g^{AB}$ for the group $G$.  However, for $G=SU(N)$ group, this distinction can be avoided. This fact is explained in section 6. In this paper, therefore, we do not distinguish the lower and upper indices $A,B,C, \dots$ (except some parts in section 6).

\noindent
\underline{The Lie algebra form}: 
\begin{equation}
  \mathscr{X}(x)=\mathscr{X}^A(x) T_A  \ (A=1,2, \ldots  , d)
, \quad \mathscr{X}^A(x) \equiv X^A(x) ,
\end{equation}
where $T_A$ denote the generators  of the Lie algebra $\mathscr{G}$ for the Lie group $G$, e.g., $\mathscr{G}=su(N)$ for $G=SU(N)$.

The two notations give the equivalent description.
It should be understood that $T_A$ denotes the generator in the fundamental representation, unless otherwise stated.

We adopt the following normalization for the generators of the Lie algebra:
\begin{equation}
{\rm tr}(T_AT_B)=\frac12 \delta_{AB}   
\quad (A,B \in\{1,2, \ldots , d \}) 
 .
\end{equation}

In order to lower and  raise the indices $A,B,C, \dots$, we need to introduce the Cartan metric $g_{AB}$ and its inverse $g^{AB}$ for the group $G$.  However, for $G=SU(N)$ group, this distinction can be avoided.%
\footnote{ 
This fact is explained in  section 6.
} 
Therefore, we do not distinguish the lower and upper indices $A,B,C, \dots$ in this paper.

For $G=SU(N)$, we can define three types of products: $\cdot$,  $\times$ and $*$ in the vector form by 
\begin{subequations}
\begin{align}
\mathbf X \cdot \mathbf Y
 :=& X^A Y^A ,%= 2 {\rm tr}(\mathscr X \mathscr Y),
\\
(\mathbf X\times\mathbf Y)^C  
 :=& f_{ABC}X^AY^B  ,
%\quad 
% [\mathscr X , \mathscr Y ] :=  if_{ABC}X^AY^B T_C   ,
\\
(\mathbf X*\mathbf Y)^C   :=& d_{ABC}X^AY^B , 
%\quad 
%  \{ \mathscr X , \mathscr Y \} - \frac{1}{N} 2 {\rm tr}(\mathscr X \mathscr Y) \mathbf{1} = d_{ABC}X^AY^B T_C 
\end{align}
\end{subequations}
which correspond to three operations in the Lie algebra form: ${\rm tr}()$, $[, ]$ and $\{, \}$ as\begin{subequations}
\begin{align}
 2 {\rm tr}(\mathscr X \mathscr Y)  =& \mathscr  X^A \mathscr Y^A ,
\\
 [\mathscr X , \mathscr Y ] =&  if_{ABC} \mathscr X^A \mathscr Y^B T_C   ,
\\
  \{ \mathscr X , \mathscr Y \} - \frac{1}{N} 2 {\rm tr}(\mathscr X \mathscr Y) \mathbf{1} 
=& d_{ABC} \mathscr X^A \mathscr Y^B T_C 
 .
\end{align}
\end{subequations}
Here we define the structure constants $f_{ABC}$ of the Lie algebra   by 
\begin{equation}
 f_{ABC}=-2i{\rm tr}([T_A,T_B]T_C)  , 
 \quad (A,B,C\in\{1,2,\ldots N^2-1\})
\end{equation}
from the commutators among the generators:
\begin{equation}
[T_A,T_B]=if_{ABC}T_C
 ,
\end{equation}
while we use the anticommutator of the generators:
\begin{equation}
\{T_A,T_B\}
 =\frac1N \delta_{AB}\bm 1
  +d_{ABC}T_C   
 ,
 \label{C27-C27-anti-com}
\end{equation}
to define completely symmetric symbols:
\begin{equation}
d_{ABC}=2{\rm tr}(\{T_A,T_B\}T_C)
  .
  \label{C27-d-def}
\end{equation}
The product of two Lie algebra valued functions, $\mathscr X=\mathscr X^A T_A$ and $\mathscr Y=\mathscr Y^A T_A$, is rewritten as the sum of three types of products: 
%which is written in the vector form as 
\begin{align}
  \mathscr{X}  \mathscr Y =& \mathscr X^A \mathscr Y^B  T_A T_B 
=  \mathscr X^A \mathscr Y^B \left( \frac12 \{ T_A,T_B \} + \frac12 [T_A,T_B]  \right)
\nonumber\\
=&  \mathscr X^A \mathscr Y^B \frac12 \left( \frac{1}{N} \delta_{AB} \mathbf{1}
  +d_{ABC}T_C  + if_{ABC}T_C  \right)
\nonumber\\
=&  \frac{1}{2N} \mathscr X^A \mathscr Y^B  \delta_{AB} \mathbf{1}
  + \frac12d_{ABC}\mathscr X^A \mathscr Y^B  T_C  + \frac12 if_{ABC}\mathscr X^A \mathscr Y^B  T_C  
\nonumber\\
=&  \frac{1}{2N} (\mathbf X \cdot \mathbf Y) \mathbf{1}
+ \frac12  (\mathbf X * \mathbf Y)^C T_C  
+ \frac12 i  (\mathbf X \times \mathbf Y)^C T_C 
 .
 \label{C27-XY}
\end{align}
In other words, the  Lie algebra is closed under the three products: $\cdot, \times$ and $*$.

In what follows we use the same notation $ \cdot $ for two Lie-algebra valued functions $\mathscr{A}=\mathscr{A}^A T_A$ and $\mathscr{B}=\mathscr{B}^A T_A$ as that for the inner product in the vector notation in the sense that 
\begin{equation}
  \mathscr{A} \cdot \mathscr{B}  := 2 {\rm tr}(\mathscr{A} \mathscr{B}) = \mathscr{A}^A  \mathscr{B}^A   = \mathbf{A} \cdot \mathbf{B} ,
\end{equation}
and especially 
\begin{equation}
 \mathscr{A}^2:= \mathscr{A} \cdot \mathscr{A}  = \mathscr{A}^A  \mathscr{A}^A = \mathbf{A} \cdot \mathbf{A} .
\end{equation}
%But, we sometimes use the notation $\mathscr{A} \cdot \mathscr{B}$ also for the Lie algebra valued quantities instead of $(\mathscr{A} , \mathscr{B})$  for simplifying the notation. 
We use $(\mathscr{A} , \mathscr{B})$ for the $L^2$ inner product defined by 
\begin{equation}
 (\mathscr{A} , \mathscr{B}) := \int d^Dx \mathscr{A}(x)  \mathscr{B}(x) .
\end{equation}

The Lie algebra $\mathscr{G}$ of the group $G$ is sometimes denoted by 
\begin{equation}
 \mathscr{G} = Lie(G) .
\end{equation}
The degrees of freedom (d.o.f.) in the expression $[ \cdots ]$ is denoted by 
\begin{equation}
  \#[ \cdots ] :=  {\rm d.o.f.}[ \cdots ] .
\end{equation}

\newpage 
%%%%%%%%%%%%%%%%%%%%%%%%%%%%%%%%%%%%%%%%%%%%%%%%%%%%%%%%%%%%
%%%%%%%%%%%%%%%%%%%%%%%%%%%%%%%%%%%%%%%%%%%%%%%%%%%%%%%%%%%%
\section{Abelian projection and Maximally Abelian gauge}\label{sec:dual-super} 
\noindent
%%%%%%%%%%%%%%%%%%%%%%%%%%%%%%%%%%%%%%%%%%%%%%%%%%%%%%%%%%%%
%%%%%%%%%%%%%%%%%%%%%%%%%%%%%%%%%%%%%%%%%%%%%%%%%%%%%%%%%%%%

%%%%%%%%%%%%%%%%%%%%%%%%%%%%%%%%%%%%%%%%%%%%%%%%%%%%%%%%%%%%
\subsection{Abelian projection and magnetic monopole}\label{Abelian-projection}
%%%%%%%%%%%%%%%%%%%%%%%%%%%%%%%%%%%%%%%%%%%%%%%%%%%%%%%%%%%%

First of all, we review the idea of \textbf{Abelian projection} due to   't Hooft \cite{tHooft81}, a method which enables us to extract magnetic monopole degrees of freedom from the Yang-Mills theory  without the Higgs fields (i.e.,  scalar fields).

We consider the Yang-Mills theory with the gauge group $G=SU(N)$ on the $D$-dimensional space--time $\mathbb{R}^D$.  
We denote by $\mathscr{G}=su(N)$ the Lie algebra of the group $G=SU(N)$.  
%For $G=SU(N)$, $\mathscr{G}=su(N)$. 

\noindent
\begin{enumerate}
\item[
(1)] Let $\mathcal{X}(x)$ be a gauge-dependent (i.e., gauge non-invariant) local operator  as a functional of the Yang-Mills field $\mathscr{A}_\mu(x)$, which takes the value in the Lie algebra $\mathscr{G}$ of the gauge group $G$. 
Suppose that $\mathcal{X}(x) \in \mathscr{G}$ transforms according to the adjoint representation under the gauge transformation (gauge rotation):%
\begin{equation}
\mathcal{X}(x) \to \mathcal{X}^{\prime}(x) = U(x) \mathcal{X}(x) U^{\dagger}(x)   \in \mathscr{G} , \ U(x) \in G , \ x \in \mathbb{R}^D ,
  \label{gauge-rot}
\end{equation}
if the Yang-Mills field $\mathscr{A}_\mu(x)$ transforms \begin{equation}
\mathscr{A}_\mu(x) \to \mathscr{A}_\mu^\prime(x) = U(x)[\mathscr{A}_\mu(x)+ig^{-1} \partial_\mu] U^{\dagger}(x) .
  \label{gauge-transf}
\end{equation}

\item[(2)] 
 For the Hermitian operator $\mathcal{X}(x)$, i.e., $\mathcal{X}(x)^\dagger =\mathcal{X}(x)$, we choose the local unitary transformation $U(x)$ so that $\mathcal{X}(x)$ is diagonalized:
\begin{equation}
\mathcal{X}^{\prime}(x) = {\rm diag} ( \lambda_{1}(x), \lambda_{2}(x), \cdots , \lambda_{N}(x) ) , 
\end{equation}
where $\lambda_{a}(x)$ ($a=1, 2, \cdots , N$) are eigenvalues of $\mathcal{X}(x)$.

\end{enumerate}

After the Abelian projection, the original gauge group $G=SU(N)$ is explicitly broken to a subgroup $H \otimes {\rm Weyl} = U(1) {}^{N-1} \otimes  {\rm Weyl}$:
\begin{equation}
 G = SU(N) \to   H \otimes {\rm Weyl} =  U(1) ^{N-1} \otimes {\rm Weyl} ,
\end{equation}
where $H=U(1)^{N-1}$ is the so-called the \textbf{maximal torus subgroup} or \textbf{Cartan subgroup} of $SU(N)$, and the \textbf{Weyl group} is a discrete subgroup of $G$ which corresponds to the degrees of freedom for changing the ordering of the eigenvalues.%
\footnote{
See \ref{sec:Weyl} and the textbook of the group theory, e.g., \cite{Gilmore06}. 
%R. Gilmore,
%Lie Groups, Lie Algebras, and Some of Their Applications (Dover Books on Mathematics, 2006) 
}
If the eigenvalues are ordered such that $\lambda_{1} \geq \lambda_{2} \geq \cdots \geq \lambda_{N}$, the Weyl group is broken and only the maximal torus subgroup $H$ is left intact after the Abelian projection.

Therefore, the Abelian projection is regarded as \textbf{a partial gauge fixing}, since $\mathcal{X}(x)$ is a gauge-dependent quantity.  
This is indeed the gauge fixing because we cannot perform the original $SU(N)$ gauge transformation any more in order to keep the diagonal form of $\mathcal{X}$.   Here the partial comes from the fact that there are exceptional parts  as shown shortly.  
Consequently, the whole space--time  points are classified into two categories:
%Consequently, the field degrees of freedom are classified into two sets: 

\begin{enumerate}
\item[(a)]
 At the ``non-degenerate'' space--time point $x \in \mathbb{R}^D$ with  the non-degenerate eigenvalues $\lambda_{a}(x)$, i.e., all the eigenvalues $\lambda_{a}(x)$ at $x$ are different, the original gauge degrees of freedom are not completely fixed because any diagonal gauge rotation matrix:
\begin{align}
U(x) =& {\rm diag} ( e^{i \theta_{1}(x)} , e^{i \theta_{2}(x)} , \cdots , e^{i \theta_{N}(x)} ) \in  G  =  SU(N) , \nonumber\\
& \det  U(x) = 1  \Leftrightarrow  \sum_{a=1}^{N} \theta_{a}(x) =  0 \ , 
%\ ( U \in  G  =  SU(N) )
\end{align}
leaves $\mathcal{X}(x)$ invariant, i.e., $U(x) \mathcal{X}(x) U^{\dagger}(x) = \mathcal{X}(x)$ or $ [ U(x) , \mathcal{X}(x) ] = 0$.
Here $U(x)$ is an element of the maximal Abelian subgroup $H=U(1)^{N-1}$.
Therefore, the Abelian subgroup $H$ remains intact at the non-degenerate space--time point even after the Abelian projection.

\item[(b)]
 At the ``degenerate'' space--time point $x_0 \in \mathbb{R}^D$ with  degenerate eigenvalues.
For instance, two eigenvalues coincide with each other:
\begin{equation}
\lambda_{a}(x_0) = \lambda_{b}(x_0) , \ a \neq b , \ a,b  \in \{ 1, \cdots , N \} ,
\end{equation}
 the singularities leading to  magnetic monopoles appear in the diagonal part (the Cartan part) $a_{\mu}^j(x)$ of the gauge transformed non-Abelian gauge field $\mathscr{A}^{\prime}_{\mu}(x) = U(x) [ \mathscr{A}_{\mu} + i g^{-1} \partial_{\mu} ] U^{\dagger}(x)$, as will be shown below.

\end{enumerate}

In the Abelian projection, the magnetic monopole is identified with \textbf{the gauge fixing defect} (i.e., defect of the gauge fixing procedure): The magnetic monopole appears at the space--time point where the above gauge fixing procedure fails to work (and the original full gauge symmetry still remains intact at this point).  The remaining symmetry corresponding to the Abelian subgroup $H$ is called the \textbf{residual gauge symmetry}.

\par
By the Abelian projection, thus, the $G=SU(N)$ Yang-Mills theory reduces to the Abelian gauge theory which includes the $H=U(1)^{N-1}$ Abelian gauge field coupled with the matter fields plus magnetic monopole:
\begin{align}
 & \text{non-Abelian [$G=SU(N)$] Yang-Mills field ($\mathscr{A}_\mu^A$)}
\nonumber\\
  \rightarrow  &
 \text{Abelian [$H=U(1)^{N-1}$] gauge fields ($a_\mu^j$)}
\nonumber\\&
+ \text{magnetic monopoles ($k_\mu$)}
\nonumber\\&
+ \text{electrically charged matter fields ($A_\mu^a$)}
\end{align}
In view of this, the idea of Abelian projection is expected to be effective in discussing quark confinement in the light of dual superconductor picture.
The Abelian projection is nothing but a kind of gauge fixing, i.e., the \textbf{partial gauge fixing} from the original non-Abelian gauge group $G$ to the maximal torus subgroup $H$: $G \rightarrow H$.

As the choices of  the gauge-dependent local operator $\mathcal{X}(x)$ as a functional of the Yang-Mills field $\mathscr{A}_\mu(x)$ to be diagonalized,  
 't Hooft suggested  to adopt $\mathcal{X}(x)=\mathscr{F}_{12}(x)$, $\mathscr{F}_{\mu\nu}(x)\mathscr{F}_{\mu\nu}(x)$, and $\mathscr{F}_{\mu\nu}(x)\mathscr{D}^2\mathscr{F}_{\mu\nu}(x)$.
However, it is known that other choices are more effective in the actual numerical simulations, as will be discuss in the next section.

%Proof)
In what follows, we prove the statement (2.7) of the Abelian projection specified by the classification (a) and (b). 
First, we discuss the $G=SU(2)$ case.  
The $SU(N)$ case will be discussed later after preparing more technical tools.

The $su(2)$ Yang-Mills field defined by 
\begin{align}
  \mathscr{A}_\mu =  \mathscr{A}_\mu^A T^A 
=  A_\mu^3 T^3 +  \sum_{a=1,2}  A_\mu^{a} T^{a}    
%\\=& 
 = \frac12 
   \begin{pmatrix} 
    A_\mu^3  & A_\mu^1-iA_\mu^2    \\
    A_\mu^1+iA_\mu^2 & -A_\mu^3    \\
   \end{pmatrix}
   , \quad T^A := \frac{\sigma_A}{2}
\end{align}
has the Cartan decomposition:
\begin{align}
  \mathscr{A}_\mu 
%=&   \mathscr{A}_\mu^A T^A 
= A_\mu^3 H_1  +  W_\mu^*{} \tilde{E}_{+} + W_\mu \tilde{E}_{-} 
%\\=&
 =  \frac{1}{2}
 \begin{pmatrix} 
    A_\mu^3  & \sqrt{2}W_\mu{}^*   \\
    \sqrt{2}W_\mu    & -A_\mu^3  \\
   \end{pmatrix} 
   ,
\end{align}
where we have defined the complex field and the Cartan basis:
\begin{align}
  W_\mu  := \frac{1}{\sqrt{2}}(A_\mu^1+ i A_\mu^2),  \quad
      H_1 :=  T^3  , \quad
  \tilde{E}_{\pm} :=  \frac{1}{\sqrt{2}}(T^1 \pm  i T^2) . 
\end{align}

First we consider the case (a): For non-degenerate points,
$
U(x) = {\rm diag}(e^{i\theta_1(x)},  e^{i\theta_2(x)} )
$
with
$
  \theta_2(x)=-\theta_1(x) 
$
does not change the diagonal form. 
This is an element of the maximal torus group $U(1)$, i.e., the residual symmetry $U(1)$ ($\det U(x)=1  \Longleftrightarrow \sum_{j=1}^{2} \theta_j(x)=0$).
The gauge transformation by the Cartan subgroup
$
  U(x) = \exp \left( ig \theta(x) H_1 \right)
  = {\rm diag}(e^{\frac12 ig\theta},e^{-\frac12 ig\theta})
$
($\theta_2(x)=-\theta_1(x):=\theta(x)$)
reads
\begin{align}
  \mathscr{A}_\mu{}' =  U \left( \mathscr{A}_\mu + ig^{-1}  \partial_\mu \right) U^\dagger 
%=   U \left(  A_\mu^3 H_1  +   W_\mu^* \tilde{E}_{+} + W_\mu  \tilde{E}_{-}   + ig^{-1} \partial_\mu \right) U^\dagger 
%\nonumber\\
=  ( A_\mu^3  + \partial_\mu \theta ) H_1  
+ e^{ig \theta } W_\mu^*  \tilde{E}_{+} + e^{-ig \theta }  W_\mu  \tilde{E}_{-} .
\end{align}
This implies the gauge transformation law:
\begin{align}
  A_\mu^3{}' = A_\mu^3 + \partial_\mu \theta , \quad
  W_\mu{}' = e^{-ig \theta  }  W_\mu  , \quad 
  (W_\mu^* {}' = e^{ig \theta }  W_\mu^*  ). 
\end{align}
Thus, the diagonal vector fields $A_\mu^3 (x)$ transform as an Abelian gauge field $a_{\mu} (x)$ for the remaining subgroup $H=U(1)$, while the two off-diagonal vector fields $A_{\mu}^{a}(x)$($a=1,2$) transform as charged (i.e., complex) matter fields  $W_{\mu} (x)$.

Then, we observe that magnetic monopole indeed appears in the diagonal component of the non-Abelian gauge field at the  space--time point $x_0 \in \mathbb{R}^D$ with  degenerate eigenvalues.

Next, we consider the case (b): For a given Hermitian and traceless matrix:
\begin{equation}
\mathcal{X}(x)=\mathcal{X}_A(x) \frac{\sigma_A}{2} ,
\end{equation}
where $\sigma_A$ $(A=1,2,3)$ are the Pauli matrices, 
the two eigenvalues are given by 
 $\lambda_1(x)=-\sqrt{\mathcal{X}_A(x) \mathcal{X}_A(x)}/2$ and  
 $\lambda_2(x)=+\sqrt{\mathcal{X}_A(x) \mathcal{X}_A(x)}/2$.
Therefore, the degenerate point $x_0$, i.e., $\lambda_1(x_0)=\lambda_2(x_0)$, is determined by solving three equations simultaneously:
\begin{equation}
\mathcal{X}_1(x_0)=\mathcal{X}_2(x_0)=\mathcal{X}_3(x_0)=0 \Longleftrightarrow \mathcal{X}_A(x_0)=0 \quad (A=1,2,3) .
\label{C25-co-dim}
\end{equation}
Then the location of the magnetic monopole is given by \textbf{zeros} $x_0$ of $\mathcal{X}(x)$ as follows.
In the static case, $\mathcal{X}(x)$ is expanded around the zeros 
as
\begin{equation}
 \mathcal{X}(x) =\mathcal{X}_A(x) \frac{\sigma_A}{2} = (x-x_0)_j \partial_j \mathcal{X}_A(x_0) \frac{\sigma_A}{2} + \cdots .
 \label{C25-chi-expand}
\end{equation}

For $G=SU(2)$,  a $2 \times 2$ matrix $U$ is  expressed by three Euler angles $\alpha, \beta, \gamma$:
\begin{align}
 U(x) =& e^{i\gamma(x)\sigma_3/2 }  e^{i\beta(x)\sigma_2/2}  e^{i \alpha(x)\sigma_3/2}
 \nonumber\\
=& 
 \begin{pmatrix}
 e^{{i \over 2}[\alpha(x)+\gamma(x)]} \cos {\beta(x) \over 2} &
 e^{{i \over 2}[-\alpha(x)+\gamma(x)]} \sin {\beta(x) \over 2} \cr
 -e^{{i \over 2}[\alpha(x)-\gamma(x)]} \sin {\beta(x) \over 2} &
 e^{{i \over 2}[-\alpha(x)-\gamma(x)]} \cos {\beta(x) \over 2}   
 \end{pmatrix}
%\begin{pmatrix}
%e^{\frac{i}{2}(\alpha(x)+\gamma(x))}\cos\frac{\beta(x)}{2} & 
%e^{\frac{i}{2}(\alpha(x)-\gamma(x))}\sin\frac{\beta(x)}{2} \\
%e^{-\frac{i}{2}(\alpha(x)-\gamma(x))}\sin\frac{\beta(x)}{2} &
%e^{\frac{i}{2}(\alpha(x)+\gamma(x))}\cos\frac{\beta(x)}{2} \\
%\end{pmatrix} ,
\nonumber\\&
\alpha(x) \in [0, 2\pi )  , \  \beta(x) \in [0, \pi ] , \ \gamma(x) \in [0, 2\pi ) ,
\label{C25-U}
\end{align}
where the range of the angles corresponds to the single valued region which leads to the unit magnetic charge. 
%\begin{align}
%U(x) =& e^{i \chi(x) T_{1}} e^{i \theta(x) T_{2}} e^{i \varphi(x) T_{3}} \nonumber\\
%=& \left( \begin{array}{cc}
%e^{\frac{i}{2} [ \varphi(x) + \chi(x) ]} \cos \frac{\theta(x)}{2} &
%e^{- \frac{i}{2} [ \varphi(x) - \chi(x) ]} \sin \frac{\theta(x)}{2} \\
%- e^{\frac{i}{2} [ \varphi(x) - \chi(x) ]} \sin \frac{\theta(x)}{2} &
%e^{- \frac{i}{2} [ \varphi(x) + \chi(x) ]} \cos \frac{\theta(x)}{2} \\
%\end{array} \right) , \nonumber\\
%&\theta(x) \in [0, \pi ] , \ \varphi(x) \in [0, 2\pi ) , \ \chi(x) \in [0, 2\pi ) .
%\label{C25-U}
%\end{align}
It turns out that the gauge rotation matrix $U(x)$ diagonalizing the first term of the expansion of the right-hand side of $(\ref{C25-chi-expand})$ is given by choosing 
$\alpha=\varphi$, $\beta=\theta$ and $\gamma=\gamma(\varphi)$  
in $U(x)$ represented by the Euler angles
where $\theta, \varphi$ are two angles in the three-dimensional polar coordinates.

Even if the original Yang-Mills field $\mathscr{A}_{\mu}(x)$ is not singular, the  transformed field $\mathscr{A}_{\mu}^{\prime}(x)$ can involve the singularities in the inhomogeneous term:
\begin{equation}
\Omega_{\mu}(x) := i g^{-1} U(x) \partial_{\mu} U^{-1}(x)  =\Omega_\mu^A(x) \frac{\sigma_A}{2} 
= \Omega_{\mu}(x)^\dagger . 
\label{C25-UdU}
\end{equation}
In the Yang-Mills field
$\mathscr{A}_\mu^\prime(x) = U(x)[\mathscr{A}_\mu(x)+ig^{-1} \partial_\mu] U^\dagger(x)$
 after the gauge rotation,  the singularity appears in the inhomogeneous term $\Omega_\mu(x)$.
 In fact, by substituting (\ref{C25-U}) into (\ref{C25-UdU}), we have \cite{KondoII}
\begin{align}
\Omega_\mu 
%=& ig^{-1}U \partial_\mu U^\dagger 
%\nonumber\\
=&   g^{-1} \frac{1}{2}
 \begin{pmatrix}
 \cos \beta \partial_\mu \alpha + \partial_\mu \gamma & 
 [-i\partial_\mu \beta-\sin \beta \partial_\mu \alpha]e^{i\gamma} \\
  [i\partial_\mu \beta-\sin \beta \partial_\mu \alpha]e^{-i\gamma} &
- [\cos \beta \partial_\mu \alpha + \partial_\mu \gamma]  \\
 \end{pmatrix} ,
\nonumber\\
=& g^{-1} [ \cos \beta(x) \partial_{\mu} \alpha(x) + \partial_{\mu} \gamma(x) ] \frac{\sigma_{3}}{2}
\nonumber\\
 &+  g^{-1} [  \sin \gamma(x) \partial_{\mu} \beta(x) - \sin \beta(x) \cos \gamma(x) \partial_{\mu} \alpha(x) ] \frac{\sigma_{1}}{2}
\nonumber\\
 &+ g^{-1} [ \cos \gamma(x) \partial_{\mu} \beta(x) + \sin \beta(x) \sin \gamma(x) \partial_{\mu} \alpha(x)     ] \frac{\sigma_{2}}{2} 
%+ \cdots .
% = \mathscr{V}_\mu^A \sigma_A/2
\label{C25-Omega2}
\end{align}
%which generates the magnetic monopole of the Dirac type.
%\begin{equation}
%\Omega_{\mu}(x) =   g^{-1} [ \cos \theta(x) \partial_{\mu} \varphi(x) + \partial_{\mu} \chi(x) ] \frac{\sigma_{3}}{2} + \cdots .
%\end{equation}
The diagonal part $\Omega_\mu^3(x)$ of the inhomogeneous term $\Omega_\mu(x)$ contains the singular potential of the Dirac type leading to the \textbf{Dirac magnetic monopole}: 
\begin{equation}
 \Omega_\mu^3(x) = g^{-1}[\cos \beta(x) \partial_\mu \alpha(x) + \partial_\mu \gamma(x) ] .
\end{equation}
For  $D=3$, using the Cartesian coordinates ($x,y,z$) or the spherical coordinates ($r,\theta,\varphi$),  
the spatial component $\bm{\Omega}(x)$ of $\Omega^\mu(x)$ is written  by putting 
%it agrees with the well-known \textbf{Dirac magnetic potential} by putting 
$\alpha=\varphi$, $\beta=\theta$ and $\gamma=\gamma(\varphi)$ as   
\begin{align}
 \bm{\Omega}(x) = i g^{-1}U(x) \nabla U^\dagger(x)
 =&  \frac{\sigma_1}{2} \frac{g^{-1}}{r}
[ \sin \gamma  {\bf e}_\theta -  \cos \gamma  {\bf e}_\varphi]
%[ \sin\varphi  {\bf e}_\theta +  \cos \varphi\partial_\varphi \gamma  {\bf e}_\varphi]
 \nonumber\\
 +& \frac{\sigma_2}{2} \frac{g^{-1}}{r}
[ \cos \gamma  {\bf e}_\theta +  \sin \gamma  {\bf e}_\varphi]
%[- \cos\varphi {\bf e}_\theta +  \sin \varphi\partial_\varphi \gamma  {\bf e}_\varphi]
 \nonumber\\
 +& \frac{\sigma_3}{2} \frac{g^{-1}}{r} \frac{ \cos \theta+ \partial_\varphi \gamma}{\sin\theta}{\bf e}_\varphi .
\end{align}

%%%%%%%%%%%%%%%%%%%%%%%%%%%%%%%%%%%%%%%%%%%%%%%%%%%%%%%
\begin{figure}[tb] 
%\vspace{30mm}
\begin{center}
\includegraphics[scale=0.35]{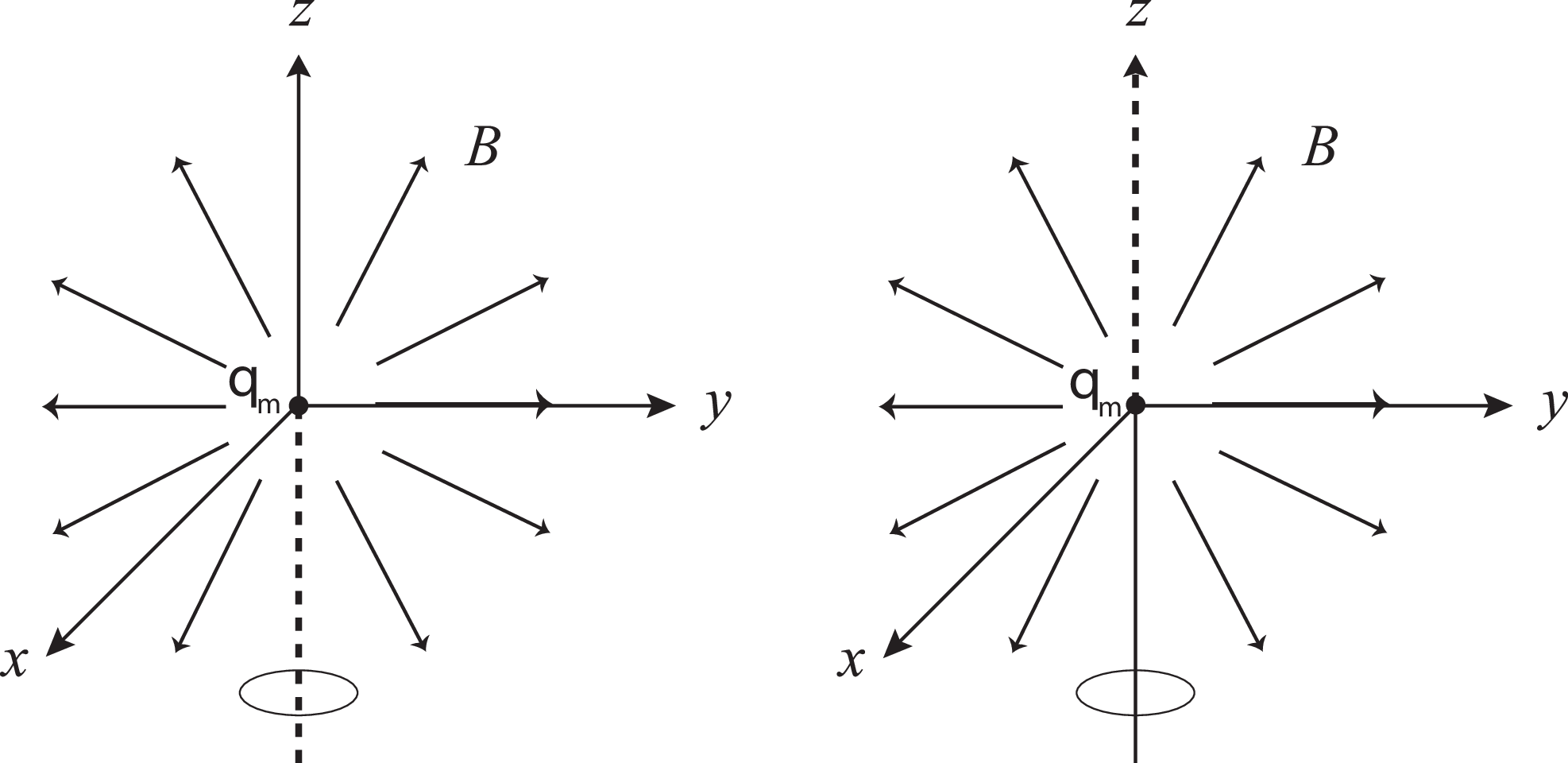}
\end{center}
%\capwidth90mm
\vskip -0.3cm
\caption{
The Dirac monopole defined from 
(left panel) the gauge potential $\bm{A}_{\rm I}$ regular except for singularities on the negative $z$ axis, 
(right panel) the gauge potential $\bm{A}_{\rm I\hspace{-.1em}I}$ regular except for singularities on the positive $z$ axis.
}
\label{C23-fig:Dirac-monopole}
\end{figure}
%%%%%%%%%%%%%%%%%%%%%%%%%%%%%%%%%%%%%%%%%%%%%%%%%%%%%%%

%it agrees with the well-known \textbf{Dirac magnetic potential} by putting 

The diagonal component $\bm{\Omega}^3(x)$ of  $\bm{\Omega}(x)=\bm{\Omega}^A(x)T_A$ agrees with  the magnetic potential giving the well-known  {Dirac magnetic monopole}, by choosing an appropriate $\gamma(x)$.
For instance, $\gamma(x) = - \varphi(x)$ reproduces the vector potential $A_{\rm I}(x)$  of the Dirac magnetic monopole, 
\begin{equation}
\bm{A}_{\rm I}(\bm{r}) = \frac{q_{m}}{4\pi} \left( \frac{-y}{r(r+z)},\frac{x}{r(r+z)},0 \right)
= \frac{q_{m}}{4\pi} \frac{1-\cos \theta}{r \sin \theta} \bm{e}_{\varphi} .
\label{C23-A1-sing}
\end{equation}
while $\gamma(x) = + \varphi(x)$ reproduces another vector potential $A_{\rm II}(x)$ given by
\begin{equation}
\bm{A}_{\rm I\hspace{-.1em}I} (\bm{r}) = \frac{q_{m}}{4\pi} \left( \frac{y}{r(r-z)}, \frac{-x}{r(r-z)},0 \right)
= \frac{q_{m}}{4\pi} \frac{-1-\cos \theta}{r \sin \theta}\bm{e}_{\varphi} ,
\end{equation}
%A way to describe the the magnetic field $\bm{B}$ for the magnetic monopole satisfying $\nabla \cdot \bm{B} \neq 0$ is to introduce the \textbf{singularity} in the vector potential $\bm{A}$.
%Such case is called the \textbf{Dirac magnetic monopole}.
It will turn out that the magnetic charge $q_{m}$ and the color charge $g$ 
satisfy the \textbf{Dirac quantization condition} $g q_{m} = \pm 4\pi$, 
which renders the Dirac string an unphysical object.
The magnetic monopole with the magnetic charge $q_{m}$ located at the origin would generate the magnetic field at the position $\bm{r}$:
\begin{equation}
\bm{B}_{\rm mono} ( \bm{r} ) = \frac{1}{4\pi} \frac{q_{m}}{r^{2}}\bm{e}_{r} , \ \bm{e}_{r}:=\frac{\bm{r}}{r} .
\label{C23-B-mag}
\end{equation}
The vector potential $\bm{A}_{\rm I}$ has the singularities at $z=-r$, i.e., on the negative $z$ axis, $z \in ( - \infty, 0]$ or $\theta=\pi$.
In the region $\Omega_{\rm I}$ except the singular points, see Fig.~\ref{C23-fig:Dirac-monopole}, indeed, the vector potential $\bm{A}_{\rm I}$ gives the magnetic field $\bm{B}_{\rm mono}$:
%[Exercise-2] \marginpar[uuu]{Ex-2} 
\begin{equation}
\nabla \times \bm{A}_{\rm I} (\bm{r}) = \frac{q_{m}}{4\pi r^{2}} \bm{e}_{r} 
= \bm{B}_{\rm mono} (\bm{r}) \quad {\rm for} \quad \ \bm{r} \in \Omega_{\rm I} .
\label{C23-A1}
\end{equation}
On the other hand, the vector potential   $\bm{A}_{\rm I\hspace{-.1em}I}$
has the singularities at $z=r$, i.e., on the positive $z$ axis, $z \in [ 0 , \infty )$, i.e., $\theta=0$, but gives the same magnetic field  $\bm{B}_{\rm mono}$ in the region   $\Omega_{\rm I\hspace{-.1em}I}$ except the singular points:
\begin{equation}
\nabla \times \bm{A}_{\rm I\hspace{-.1em}I} (\bm{r}) = \frac{q_{m}}{4\pi r^{2}}\bm{e}_{r}
= \bm{B}_{\rm mono} (\bm{r})  \quad {\rm for} \quad \bm{r} \in \Omega_{\rm I\hspace{-.1em}I} .
\end{equation}
Since the two vector potentials $\bm{A}_{\rm I}$ and $\bm{A}_{\rm I\hspace{-.1em}I}$ give the same magnetic field $\bm{B}_{\rm mono}$, they must be connected by a gauge transformation:
\begin{equation}
\bm{A}_{\rm I} (\bm{r}) = \bm{A}_{\rm I\hspace{-.1em}I} (\bm{r}) + \nabla \chi (\bm{r}) ,
\end{equation}
where $\chi$ is given by
\begin{equation}
\chi (\bm{r}) = \frac{q_{m}}{2\pi} \arctan \frac{y}{x} = \frac{q_{m}}{2\pi} \varphi  ,
\label{C23-gauge-1-2}
\end{equation}
with $\varphi$ being an angle around $z$ axis.
%Here, the magnetic charge $q_{m}$ and the color charge $g$ satisfy the \textbf{Dirac quantization condition} $g q_{m} = 4\pi$, which renders the Dirac string an unphysical object. 

The ``magnetic monopole'' appears differently depending on the dimension of the space--time $D$.
The magnetic monopole is a topological defect of \textbf{co-dimension} three, since three conditions (\ref{C25-co-dim}) are needed to specify the magnetic monopole. 
\begin{enumerate}
\item
[$D=3$]:  $0$-dimensional point defect $\rightarrow$ magnetic monopole

\item
[$D=4$]: $1$-dimensional line defect  $\rightarrow$ magnetic monopole loop. %(closed string):
Here  the loop follows from the conservation of the magnetic current $\partial_\mu k^\mu=0$. 

\end{enumerate}

The existence of the topological configuration for magnetic monopoles is suggested from the non-triviality of the \textbf{homotopy group} 
\begin{equation}
\Pi_{2}( SU(2)/U(1) ) = \Pi_{2} (S^2)  = \Pi_{1} ( U(1)) = \mathbb{Z}  
\end{equation}
of the map:
 \begin{equation}
 S^{2} \rightarrow SU(2)/U(1)  \simeq S^{2} .
\end{equation}
In the original formulation of the Yang-Mills theory (without any singularities), magnetic monopoles do not exist, since
\begin{equation}
\Pi_{2}(  SU(2) ) = 0 .
\end{equation}
In the Yang-Mills theory written in terms of the gauge fields alone without fundamental matter fields, therefore, there do  not exist magnetic monopoles  as a \textbf{topological soliton}%
\footnote{
In the pure Yang-Mills theory, the possible topological soliton is  restricted to \textbf{instantons} which exist only for $D=4$, corresponding to the map:
$
 S^3 \rightarrow SU(2) \cong S^3    , 
$
with non-trivial Homotopy group:
$
\Pi_3(SU(2)) =\Pi_3(S^3) = \mathbb{Z}
$,
and hence magnetic monopoles required do not exist as a usual topological soliton, as the \textbf{Derrick theorem} indicates. 
}.
Nevertheless, the magnetic monopoles appear  when the Abelian projection (partial gauge fixing) is performed.
This leads to the criticism that the magnetic monopole in the Yang-Mills theory may be a \textbf{gauge artifact} and the resulting magnetic monopole cannot be a gauge invariant physical quantity.
This is the most serious issue to be overcome in the scenario of  establishing  the dual superconductivity whose origin is the magnetic monopoles and their condensation, which will be discussed again later.

It should be remarked that in the above proof there is an implicit assumption that the first derivative at $x=x_0$ is non-vanishing in (\ref{C25-chi-expand}): 
\begin{equation}
\partial_j \mathcal{X}_A(x_0) \ne 0 ,
\end{equation}
namely, the zeros are first-order ones.
If this is not the case, i.e., $\partial_j \mathcal{X}_A(x_0) =0$, there are no magnetic monopoles.
For the second-order zeros, there exists another topological object called the \textbf{Hopfion} associated to the  \textbf{Hopf map}:%
\footnote{See section 8 for more details.}
\begin{equation}
 S^3 \rightarrow  SU(2)/U(1) \cong S^2   ,
\end{equation} 
with the non-trivial Homotopy group:
\begin{equation}
\Pi_{3}( SU(2)/U(1) ) = \Pi_{3} (S^2)   = \mathbb{Z}  
\end{equation}
while the magnetic monopole is associated with the non-trivial map  $S^2 \rightarrow S^2 \cong SU(2)/U(1)$.% 

The $G=SU(N)$ case is discussed as follows. 
We decompose the $\mathscr{G}=su(N)$ gauge field $\mathscr{A}_{\mu}(x) = \mathscr{A}_{\mu}^{A}(x) T^{A}$ into the \textbf{diagonal part} (Cartan part) $a_{\mu}^{j}(x)$ and \textbf{off-diagonal part} $A_{\mu}^{a}(x)$:
\begin{align}
\mathscr{A}_{\mu}(x) = \mathscr{A}_{\mu}^{A}(x) T^{A} =& a_{\mu}^{j}(x) H^{j} + A_{\mu}^{a}(x) T^{a} , 
\nonumber\\&
%\  \begin{array}{l}
j = 1, 2 , \cdots , N-1; \quad 
a = 1, 2, \cdots , N^{2}-N ,
%\end{array}
\end{align}
%\begin{align}
%  \mathscr{A}_\mu = \mathscr{A}_\mu^A T^A 
%=  \sum_{j=1,2,\cdots, N-1} a_\mu^j H_j   + \sum_{a=1,2,\cdots, N^2-N}  A_\mu^{a} T^{a}    
%\end{align}
where $a_{\mu}^{j}(x)$ ($j=1, \cdots , N-1$) are ($N-1$) diagonal vector fields and $A_{\mu}^{a}(x)$ ($a=1, \cdots , N^{2}-N$) are $N^{2}-1 -(N-1) = N^{2} - N$ off-diagonal vector fields.
It should be remarked that ($N-1$) diagonal vector fields $a_{\mu}^{j}(x)$ transform as Abelian gauge fields for the remaining subgroup $H=U(1) ^{N-1}$, while ($N^{2}-N$) off-diagonal vector fields $A_{\mu}^{a}(x)$ transform as charged matter fields.
This fact can be observed as follows.

On the other hand, the $\mathscr{G}=su(N)$ Yang-Mills field $\mathscr{A}_\mu$ has the \textbf{Cartan decomposition}:
\begin{align}
  \mathscr{A}_\mu 
=  \sum_{j=1}^{N-1} a_\mu^j H_j + \sum_{\alpha=1}^{(N^2-N)/2} (W_\mu^*{}^{\alpha} \tilde{E}_{\alpha} + W_\mu^{\alpha} \tilde{E}_{-\alpha})  ,
\end{align}
where we have redefined the generators in the \textbf{Cartan basis}:
\begin{align}
 \vec{H} =& (H_1, H_2, \cdots, H_{N-1}) = (T^3, T^8,  \cdots, T^{N^2-1}) ,
\nonumber\\
  \tilde{E}_{\pm 1} =& \frac{1}{\sqrt{2}}(T^1 \pm  i T^2) , \quad
  \tilde{E}_{\pm 2} = \frac{1}{\sqrt{2}}(T^4 \pm  i T^5) ,  
\quad
  \tilde{E}_{\pm 3} = \frac{1}{\sqrt{2}}(T^6 \pm  i T^7) ,   
\nonumber\\
  \cdots, & \quad 
  \tilde{E}_{\pm (N^2-N)/2} =  \frac{1}{\sqrt{2}}(T^{N^2-3} \pm i T^{N^2-2}) ,
\end{align}
and introduced the complex fields:
\begin{align}
  W_\mu^1 =& \frac{1}{\sqrt{2}}(A_\mu^1+ i A_\mu^2) , 
\quad
  W_\mu^2 = \frac{1}{\sqrt{2}}(A_\mu^4+ i A_\mu^5) , 
\quad 
  W_\mu^3 = \frac{1}{\sqrt{2}}(A_\mu^6+ i A_\mu^7) ,  
  \nonumber\\&
  \cdots, \quad 
  W_\mu^{(N^2-N)/2} = \frac{1}{\sqrt{2}}(A_\mu^{N^2-3}+ i A_\mu^{N^2-2})  
   .
\end{align}

The Cartan subgroup of $SU(N)$ represented by
\begin{align}
  U = \exp \left(  ig \theta^j H_j \right) \in U(1)^{N-1} 
\end{align}
causes the gauge transformation:
\begin{align}
  \mathscr{A}_\mu{}' =& U \left( \mathscr{A}_\mu + ig^{-1} \partial_\mu \right) U^\dagger 
  \nonumber\\
=&  U \left(  a_\mu^j H_j +   W_\mu^*{}^{\alpha} \tilde{E}_{\alpha} + W_\mu^{\alpha} \tilde{E}_{-\alpha}   +  ig^{-1}  \partial_\mu \right) U^\dagger 
\nonumber\\
=& ( a_\mu^j + \partial_\mu \theta^j) H_j  
+ e^{ ig \theta^j \alpha_j} W_\mu^*{}^{\alpha} \tilde{E}_{\alpha} + e^{-ig \theta^j \alpha_j}  W_\mu^{\alpha} \tilde{E}_{-\alpha} ,
\end{align}
where $\alpha_j$ is the root vectors and used the commutation relations for the Cartan basis:
\begin{align}
  [H_j, H_k]=0, \quad [H_j, \tilde{E}_{\pm\alpha} ]= \pm \alpha_j   \tilde{E}_{\pm\alpha} .
\end{align}
Therefore, the gauge transformation law under the Cartan subgroup (the residual gauge group) is obtained as
\begin{align}
  a_\mu^j{}' = a_\mu^j + \partial_\mu \theta^j, \quad
  W_\mu^{\alpha}{}' = e^{-ig \theta^j \alpha_j}  W_\mu^{\alpha} , \quad 
  W_\mu^*{}^{\alpha} {}' = e^{ ig \theta^j \alpha_j}  W_\mu^*{}^{\alpha} .
  \label{C25-transf-SUN}
\end{align}
For the maximal torus group $U(1)^r$ of  $SU(N)$ ($r=N-1$),  diagonal vector fields $a_\mu^j$ $(j=1,\cdots, r)$ transform as gauge fields of the residual gauge group  $U(1)^r$, while  $(N^2-N)/2$ pairs of off-diagonal vector fields $W_\mu^{\alpha}$ and $W_\mu^*{}^{\alpha}$ $(\alpha=1, \dots, (N^2-N)/2)$ behave as charged matter fields having $r$ kinds of charge $+\alpha_j$ and $- \alpha_j$, respectively.

%%%%%%%%%%%%%%%%%%%%%%%%%%%%%%%%%%%%%%%%%%%%%%%%%%%%%%%%%%%%
\subsection{Maximally Abelian gauge}\label{sec:MAgauge}
%%%%%%%%%%%%%%%%%%%%%%%%%%%%%%%%%%%%%%%%%%%%%%%%%%%%%%%%%%%%

Next, we consider the form of $\mathcal{X}(x)$.
The choice of $\mathcal{X}(x)$ is not unique.
Now we introduce the Abelian gauge called the \textbf{maximally Abelian gauge}  \textbf{(MAG or MA gauge)} \cite{KLSW87}.
%\footnote{ 
%A.S. Kronfeld, M.L. Laursen, G. Schierholz, and U.J. Wiese,
%Monopole Condensation and Color Confinement,
%Phys. Lett. B\textbf{198}, 516--520 (1987).
%}
The idea of MA gauge is to suppress the contribution from the off-diagonal components  as small as possible to ensure that the diagonal components become dominant in the low-energy regime.
The MA gauge  is a realization of the Abelian projection idea.
The  {Abelian projection} is a \textbf{partial gauge fixing} $G=SU(N) \rightarrow H=U(1)^{N-1}$. It is a kind of the \textbf{background field gauge} and is a \textbf{non-linear gauge}. 
The gauge freedom $\mathscr{A}_\mu(x) \rightarrow \mathscr{A}_\mu^\Omega(x):=\Omega(x)[\mathscr{A}_\mu(x)+ig^{-1}\partial_\mu ]\Omega^{-1}(x)$ is used to transform the non-Abelian gauge variable as close as possible to the Abelian components in  the maximal torus subgroup $H$ of the gauge group $G$. 
Consequently, the \textbf{magnetic monopole} of the Dirac type appears in the diagonal (Cartan) part $A_\mu^j$ of $\mathscr{A}_\mu(x)$  as defects of gauge fixing procedure.

A non-perturbative procedure of gauge fixing is given by minimizing (or maximizing)  a given functional $F[\mathscr{A}]$ of the gauge field $\mathscr{A}_\mu(x)$ under the (finite) gauge transformation $\Omega$:% 
\footnote{
On a lattice, $\mathscr{A}_\mu(x)$ is replaced by the gauge link variable $U_{x,\mu}$ and $F$ is replaced by the corresponding functional $F[U]$ of a set of the link variables $\{ U_{x,\mu} \}$. 
In fact, in order to impose MA gauge in numerical simulations on a lattice, the lattice version of (\ref{C25-reduction-cond-c}) is used, which will be given in section 9.%of lattice gauge theory.  
}
\begin{equation}
 \min_{\Omega} F [\mathscr{A}^\Omega] .
 \label{C25-reduction-cond}
\end{equation}

A nonperturbative definition of the MA gauge is given as follows.
We adopt the Euclidean formulation. 
The following functional $F_{\rm MAG}$ of the off-diagonal gluon field $A_{\mu}^{a}(x)$ is minimized, i.e.,  $\underset{\omega}\min F_{\rm MAG}[A^{\omega}]$ with respect to the gauge transformation $\omega$:
\begin{equation}
F_{\rm MAG} [A] = \int d^{D}x \frac{1}{2} A_{\mu}^{a}(x) A_{\mu}^{a}(x)  
=  \frac12 (A_\mu^a,A_\mu^a) \ge  0   ,
 \label{C25-reduction-cond-c}
\end{equation}
where $( \cdot, \cdot)$ is the $L_2$ inner product.

%In practice, MAG in the continuum formulation is given by minimizing the function $F_{\rm MAG}$ w.r.t. the gauge transformation.

For the continuum formulation, we need the local form of MA gauge obtained as follows. 
For the infinitesimal gauge transformation $\delta_{\omega}$, we have
\begin{align}
\delta_{\omega} F_{\rm MAG} [A]
=& \int d^{D} x A_{\mu}^{a}(x) \delta_{\omega} A_{\mu}^{a}(x) 
%\nonumber\\
=  - \int d^{D} x \left( \partial_{\mu} A_{\mu a}(x) + g f^{ajb} a_{\mu}^{j}(x) A_{\mu b}(x) \right) \omega^{a}(x) ,
\label{C25-MA-functional}
\end{align}
where we have used the fact that $f^{ABC}$ is completely antisymmetric and $f^{ajk}=0$ (following from  commutativity of two Cartan generators: $[H^{j} , H^{k}] = 0$).
 The \textbf{gauge fixing condition} is obtained from the first variation as the \textbf{stationary condition}.
%\begin{align}
% \delta_{\omega} F_{\rm MAG} 
%=& (\delta_{\omega}A_\mu^a,A_\mu^a)  
%= ((D_\mu[A]\omega)^a,A_\mu^a)  
%\nonumber\\
%=& ((\partial_\mu \omega^a + g\epsilon^{abc}A_\mu^b \omega^c + g\epsilon^{a3c} A_\mu^3 \omega^c, A_\mu^a) 
%\nonumber\\
%=& ((\partial_\mu \omega^a + g\epsilon^{a3c} A_\mu^3 \omega^c, A_\mu^a) 
%\nonumber\\
%=& - (\omega^a, \partial_\mu  A_\mu^a) 
%- (g\epsilon^{b3a} A_\mu^3 \omega^a, A_\mu^b) 
%= -(\omega^a, D_\mu^{ab}[A^3]A_\mu^b) 
%\end{align}
The stationary condition
$\delta_{\omega} F_{\rm MAG} [A] = 0$ for any $\omega^{a}$ 
leads to the differential form of MA gauge-fixing condition:%
\footnote{
The stationary means extrema including local maximum, local minimum, and saddle point.
} 
\begin{equation}
F^a := D_{\mu}[a] ^{ab} A_{\mu}^b(x)   := \partial_{\mu} A_{\mu}^{a}(x) + g f^{ajb} a_{\mu}^{ j}(x) A_{\mu}^{b}(x) = 0 .
\label{C25-MAG}
\end{equation}
This is indeed the \textbf{background field gauge} for the off-diagonal components $A_\mu^b(x)$ with the diagonal component $A_\mu^j(x)$ as the background field. 
Even after imposing MA gauge, therefore, the residual $U(1)^{N-1}$  gauge symmetry exists. 

In particular, 
for $G=SU(2)$, 
\begin{align}
 \delta_{\omega} F_{\rm MAG} [A]
=& (\delta_{\omega}A_\mu^a,A_\mu^a)  
= ((D_\mu[A]\omega)^a,A_\mu^a)  
\nonumber\\
=& (\partial_\mu \omega^a + g\epsilon^{abc}A_\mu^b \omega^c + g\epsilon^{a3c} A_\mu^3 \omega^c, A_\mu^a) 
\nonumber\\
=& (\partial_\mu \omega^a + g\epsilon^{a3c} A_\mu^3 \omega^c, A_\mu^a) 
\nonumber\\
=& - (\omega^a, \partial_\mu  A_\mu^a) 
- (g\epsilon^{b3a} A_\mu^3 \omega^a, A_\mu^b) 
= -(\omega^a, D_\mu^{ab}[A^3]A_\mu^b) .
\end{align}
The continuum form of MA gauge for the $SU(2)$ case reads 
\begin{equation}
 F^a := [\partial_\mu \delta^{ab} - g \epsilon^{ab3}A_\mu^3(x)] A_\mu^b(x) = 0 \quad (a,b=1,2) .
%\Longleftarrow 
% min_{\omega} \int d^4x A_\mu^\omega(x)A_\mu^\omega(x) < \infty ,
%\Longleftarrow 
% min_{\omega} \int d^4x A_\mu^\omega(x)A_\mu^\omega(x) < \infty ,
\end{equation}
This is indeed the background gauge for the off-diagonal components $A_\mu^b(x)$ in the diagonal part $A_\mu^3(x)$ as the background. 
Even after imposing MA gauge, therefore, the residual $U(1)$  gauge symmetry exists. 

It is instructive to compare the MA gauge with the well-known \textbf{Lorenz gauge} (Landau gauge) which is the complete gauge fixing $G=SU(N) \rightarrow H=\{ 0 \}$.
The gauge fixing functional of the \textbf{Landau gauge} for $G=SU(N)$  is given by
\begin{equation}
 F_{L}[\mathscr{A}] = \int d^Dx \frac12 \mathscr{A}_\mu^A(x) \mathscr{A}_\mu^A(x) = \frac12 (\mathscr{A}_\mu^A,\mathscr{A}_\mu^A)  . %\quad (A=1,2,3)  
\end{equation}
The first and second variations under the infinitesimal gauge transformation are 
\begin{align}
 \delta_{\omega} F_{L}  
=& (\delta_{\omega}\mathscr{A}_\mu^A,\mathscr{A}_\mu^A)  
= ((\mathscr{D}_\mu[\mathscr{A}]\omega)^A,\mathscr{A}_\mu^A)  
= -(\omega^A, (\mathscr{D}_\mu[\mathscr{A}]\mathscr{A}_\mu)^A) 
%\nonumber\\
=  -(\omega^A, \partial_\mu \mathscr{A}_\mu^A) ,
\nonumber\\
 \delta_{\omega}^2 F_{L} 
=& -(\omega^A, \partial_\mu \delta_{\omega} \mathscr{A}_\mu^A) 
=  (\omega^A, (-\partial_\mu \mathscr{D}_\mu[\mathscr{A}])^{AB} \omega^B) . 
%\quad \text{FP~determinant}
\end{align}
The first variation gives the Landau gauge condition:
\begin{equation}
  \partial_\mu \mathscr{A}_\mu^A(x) = 0 .
  \label{C25-Lc}
\end{equation}
The second variation gives the Faddeev-Popov operator $-\partial_\mu \mathscr{D}_\mu[\mathscr{A}]$.
The positivity (positive definiteness) of the FP operator gives the \textbf{local minima} of the functional:  
\begin{equation}
  (-\partial_\mu \mathscr{D}_\mu[\mathscr{A}])  > 0.
  \label{C25-FPpos}
\end{equation}
The two conditions (\ref{C25-Lc}) and (\ref{C25-FPpos}) restrict the space of the gauge field configurations to the first \textbf{Gribov region} $\Omega$ \cite{Gribov78}.
However, it is known that they are not sufficient to give the \textbf{absolute minimum} (global minimum) of the functional. 
This suggests the complete gauge fixing is quite a difficult problem in gauge field theories.
It is shown that the Gribov copies exist also in the MA gauge.
The Gribov problem in the MA gauge is more complicated and we do not discuss it. See  e.g., \cite{BHWT00}.

%%%%%%%%%%%%%%%%%%%%%%%%%%%%%%%%%%%%%%%%%%%%%%%%%%%%%%%%%%%%
\subsection{Modified MA gauge}\label{mMA-gauge}
%%%%%%%%%%%%%%%%%%%%%%%%%%%%%%%%%%%%%%%%%%%%%%%%%%%%%%%%%%%%

In order to adopt the MA gauge in quantizing Yang-Mills theory with the Lagrangian density $\mathscr{L}_{\rm YM}$, we add the \textbf{gauge fixing term} (GF) and \textbf{Faddeev-Popov (FP) ghost  term}  for the MA gauge:
According to the general prescription, the GF+FP term of the MA gauge (\ref{C25-MAG}) is given using the  Becchi-Rouet-Stora-Tyutin (BRST) transformation $\bm{\delta}$ by
\begin{equation}
\mathscr{L}_{\rm GF+FP}^{\rm MAG} = -i \bm{\delta} \biggl[ \bar{C}^{a} \left( ( D^{\mu}[a] A_{\mu} )^{a} + \frac{\alpha}{2} N^{a} \right) \biggr] .
\label{C25-GF+FP-MAG}
\end{equation}
Since the MA gauge is the partial gauge fixing $G \rightarrow H$,  we still have the Abelian gauge invariance under $H$ even after imposing the MA gauge.
In order to completely fix the gauge, we must add the GF+FP term for the maximal torus part $H$.
The most popular choice is the Lorenz type:
\begin{equation}
\mathscr{L}_{\rm GF+FP} = -i \bm{\delta} \biggl[ \bar{C}^{j} \left( \partial^{\mu} a_{\mu}^{j} + \frac{\beta}{2} N^{j} \right) \biggr] .
\end{equation}

For $G=SU(N)$,  the  GF+FP term for the MA gauge is given by
\begin{align}
\mathscr{L}_{\rm GF+FP}^{\rm MAG} 
%=& N^{a} F^{a} +  \frac{\alpha}{2} N^{a} N^{a} + i \bar{C}^{a} D_{\mu}^{ab} [a] D^{\mu bc} [a] C^{c} \nonumber\\&
%+ i \bar{C}^{a} D_{\mu}^{ab} [a] ( g f^{bck} A^{\mu c} C^{k} + g f^{bcd} A^{\mu c} C^{d} ) 
%\nonumber\\&
%+ i \bar{C}^{a} g f^{akb} ( \partial_{\mu} C^{k} + g f^{kcd} A_{\mu}^{c} C^{d} ) A^{\mu b} \nonumber\\
=&  N^{a} F^{a} + \frac{\alpha}{2} N^{a} N^{a} 
%\nonumber\\&
+ i \bar{C}^{a} D_{\mu}^{ab} [a] D^{\mu bc} C^{c} + i g^{2} f^{kba} f^{kcd} \bar{C}^{a} C^{d} A_{\mu}^{b} A^{\mu c} \nonumber\\
&+ i \bar{C}^{a} D_{\mu}^{ab} [a] ( g f^{bcd} A^{\mu c} C^{d} ) 
%\nonumber\\&
+ i \bar{C}^{a} D_{\mu}^{ab} [a] ( g f^{bck} A^{\mu c} C^{k} ) + i \bar{C}^{a} g f^{akb} \partial_{\mu} C^{k} A^{\mu b} .
\end{align}
For $G=SU(2)$, the GF+FP term of the MA gauge is relatively simple: using $D_{\mu}[a]^{ab}= \delta^{ab} \partial_{\mu}   -  g \epsilon^{abj} a_{\mu}^{j}$,
%[Exercise-7] \marginpar{Ex-7}
\begin{align}
\mathscr{L}_{\rm GF+FP}^{\rm MAG} =&
N^{a} D_{\mu}[a]^{ab} A^{\mu b} + \frac{\alpha}{2} N^{a} N^{a} 
%\nonumber\\&
+ i \bar{C}^{a} D_{\mu}[a]^{ac} D^{\mu}[a]^{cb} C^{b} 
\nonumber\\&
+  g^{2} \varepsilon^{ad} \varepsilon^{bc} i\bar{C}^{a} C^{b} A^{\mu c} A_{\mu}^{d} 
+ i \bar{C}^{a} g \varepsilon^{ab} ( D^{\mu}[a]^{bc} A_{\mu}^{c} ) C^{3} .
\label{C25-MAG-SU2}
\end{align}

However, the naive treatment for MA gauge given in the above has the following issues. 

%(1) 
In general, the GF+FP term must be introduced to maintain (or not to break) the renormalizability.
However, the Yang-Mills theory in the MA gauge   is not renormalizable if we adopt the naive GF+FP term (\ref{C25-GF+FP-MAG}).
This is easily understood by observing that the interaction term between the gluon and ghost $\bar{C}CAA$ in (\ref{C25-MAG-SU2}) inevitably generates the four-ghost interaction term of $\bar{C}C\bar{C}C$ type, which is not renormalizable, since this term does not exist in the original Lagrangian density of MA gauge.
%[Exercise-8] \marginpar{Ex-8}
%Verify that the interaction term  $\bar{C}CAA$ between the gluon and ghost  generates the effective  interaction term of the   $\bar{C}C\bar{C}C$ four-ghost type at one loop level, which has the ultraviolet divergence. 

In order to avoid this shortcoming, it is enough to include the extra FP ghost self-interaction term in the Lagrangian from the beginning which has the same form as the four-ghost interaction term to be generated in quantum correction \cite{MLP85}.
%\footnote{
%H. Min, T. Lee, and P.Y. Pac,
%Renormalization Of Yang-mills Theory In The Abelian Gauge,
%Phys.Rev. D\textbf{32}, 440--449 (1985).
%}
Since such an extra term is not unique, however, it is convenient to adopt a simple form with good symmetry.
For instance, we can take the \textbf{modified maximally Abelian gauge} or  \textbf{modified MA (mMA)  gauge } which is given using the BRST transformation $\bm{\delta}$ and the anti-BRST transformation $\bar{\bm{\delta}}$ by
\begin{equation}
\mathscr{L}_{\rm GF+FP}^{\rm mMAG} = i \bm{\delta} \bar{\bm{\delta}} \left( \frac{1}{2} A_{\mu}^{a} A^{\mu a} + \frac{\alpha}{2} i \bar{C}^{a} C^{a} \right) ,
\label{C25-mMAG1}
\end{equation}
which has the \textbf{orthosymplectic symmetry} $OSp(D|2)$.%
\footnote{
The mMA gauge   was proposed in the following paper in which remarkable non-perturbative properties of mMA gauge besides the perturbative renormalizability were shown: Parisi-Soulous \textbf{dimensional reduction} from $D$ to $D-2$ occurs \cite{KondoII}. 
%K. -I. Kondo,
%Yang-Mills theory as a deformation of topological field theory, dimensional reduction and quark confinement, 
%CHIBA-EP-103 
%hep-th/9801024, 
%Phys. Rev. D\textbf{58},  105019 (1998). 
}
This is a special case obtained by putting $\zeta = \alpha$ in a more general choice  $\mathscr{L}_{\rm GF+FP}^{{\rm mMAG} \prime}$  with the renormalizable four ghost self-interactions \cite{MLP85}, since the coupling constant $\zeta$ for the ghost self-interaction can be introduced independently:
\begin{align}
\mathscr{L}_{\rm GF+FP}^{{\rm mMAG} \prime} 
=&
-i \bm{\delta} \biggl[ \bar{C}^{a} \left( (D_{\mu}[a] A^{\mu})^{a} + \frac{\alpha}{2} N^{a} \right) 
%\nonumber\\&
- i \frac{\zeta}{2} g f^{abk} \bar{C}^{a} \bar{C}^{b} C^{k} - i \frac{\zeta}{4} g f^{abc} C^{a} \bar{C}^{b} \bar{C}^{c} \biggr] 
\nonumber\\
=& \mathscr{L}_{\rm GF+FP}^{{\rm  MAG} } 
 - \zeta g \epsilon^{ab} N^{a}  i \bar{C}^{b}C^{3} + \frac{\zeta}{4} g^{2} \epsilon^{ab} \epsilon^{cd} \bar{C}^{a} \bar{C}^{b} C^{c} C^{d}  
  .
\label{C25-mMAG2}
\end{align}
For $G=SU(2)$, after eliminating the NL field, the GF+FP term has the form:
%[Exercise-9] \marginpar{Ex-9}
\begin{align}
\mathscr{L}_{\rm GF+FP}^{{\rm mMAG} \prime} =&
- \frac{1}{2\alpha} (D_{\mu}^{ab} [a]A^{\mu b})^{2} %\nonumber\\&
+ i \bar{C}^{a} D_{\mu}^{ac}[a] D^{\mu cb}[a] C^{b} +  g^{2} \varepsilon^{ad} \varepsilon^{bc} i\bar{C}^{a} C^{b} A^{\mu c} A_{\mu}^{d} 
\nonumber\\&
+ \frac{\zeta}{4} g^{2} \varepsilon^{ab} \varepsilon^{cd} \bar{C}^{a} \bar{C}^{b} C^{c} C^{d} 
%\nonumber\\&
+ \left( 1 -  \zeta/\alpha  \right) i \bar{C}^{a} g \varepsilon^{ab} ( D^{\mu bc}[a] A_{\mu}^{c}) C^{3} ,
\label{C25-mMAG1a}
\end{align}
and
\begin{align}
\mathscr{L}_{\rm  GF+FP}^{{\rm mMAG} } =&
- \frac{1}{2\alpha} (D_{\mu}^{ab}[a] A^{\mu b})^{2} 
%\nonumber\\&
+ i \bar{C}^{a} D_{\mu}^{ac}[a] D^{\mu cb  }[a] C^{b} + g^{2} \varepsilon^{ad} \varepsilon^{bc} i\bar{C}^{a} C^{b} A^{\mu c} A_{\mu}^{d} 
\nonumber\\&
+ \frac{\alpha}{4} g^{2} \varepsilon^{ab} \varepsilon^{cd} \bar{C}^{a} \bar{C}^{b} C^{c} C^{d} .
\label{C25-mMAG2a}
\end{align}
In fact, the explicit form of the GF+FP term $\mathscr{L}_{\rm GF+FP}$ in the modified MA gauge reads
\begin{align}
\mathscr{L}_{\rm GF+FP}^{\rm mMAG} =& N^{a} F^{a} + \frac{\alpha}{2} N^{a} N^{a} + i \bar{C}^{a} D_{\mu}^{ab} [a] D^{\mu bc} [a] C^{c} \nonumber\\
&+ i \bar{C}^{a} D_{\mu}^{ab} [a] ( g \epsilon^{bc} A^{\mu c} C^{3} )
- i \bar{C}^{a} ( g \epsilon^{ab} ( \partial_{\mu} C^{3} + g \epsilon^{cd} A_{\mu}^{c} C^{d} ) A^{\mu b} ) \nonumber\\
&+ \alpha g \epsilon^{ab} i \bar{C}^{a} N^{b} C^{3} + \frac{\alpha}{4} g^{2}  \epsilon^{ab} \bar{C}^{a} \bar{C}^{b} \epsilon^{cd} C^{c} C^{d} \nonumber\\
%=& N^{a} F^{a} + \frac{\alpha}{2} N^{a} N^{a} + i \bar{C}^{a} D_{\mu}^{ab} [a] D^{\mu bc} [a] C^{c} 
%\nonumber\\&
%+ i \bar{C}^{a} \bigl[ \partial_{\mu} ( g \epsilon^{ac} A^{\mu c} C^{3} ) + g \epsilon^{ab} a_{\mu} ( g \epsilon^{bc} A^{\mu c} C^{3} ) \bigr] 
%\nonumber\\&
%- i  \bar{C}^{a} g \epsilon^{ab} \partial_{\mu} C^{3} A^{\mu b} - i g^{2} \epsilon^{ab} \epsilon^{cd} \bar{C}^{a} C^{d} A_{\mu}^{c} A^{\mu d} 
%\nonumber\\&
%+ \alpha g \epsilon^{ab} i \bar{C}^{a} N^{b} C^{3} + \frac{\alpha}{4} g^{2} \epsilon^{ab} \bar{C}^{a} \bar{C}^{b} \epsilon^{cd} C^{c} C^{d} \nonumber\\
=& N^{a} F^{a} + \frac{\alpha}{2} N^{a} N^{a}  
%\nonumber\\&
+ i \bar{C}^{a} D_{\mu}^{ab} [a] D^{\mu bc} [a] C^{c} 
+ g^{2} \epsilon^{ad} \epsilon^{bc} i \bar{C}^{a} C^{b} A_{\mu}^{c} A^{\mu d} \nonumber\\
&+ g  i \bar{C}^{a} \epsilon^{ab} ( D_{\mu}^{bc} [a] A^{\mu c} ) C^{3} 
%\nonumber\\&
+ \alpha g \epsilon^{ab} i \bar{C}^{a} N^{b} C^{3} + \frac{\alpha}{4} g^{2} \epsilon^{ab} \epsilon^{cd} \bar{C}^{a} \bar{C}^{b} C^{c} C^{d} .
\end{align}
For $\alpha \neq 0$, by completing the square: 
\begin{align}
&\frac{\alpha}{2} N^a N^a + F^a N^a - \alpha g \epsilon^{ab} i  \bar{C}^b C^3 N^a \notag \\
%&= \frac{\alpha}{2} \left( N^a + \alpha^{-1} F^a - g \epsilon^{ab} i  \bar{C}^b C^3 \right)^2 -\frac{1}{2\alpha} \left( F^a - \alpha g \epsilon^{ab} i  \bar{C}^b C^3 \right)^2 \notag \\
&=  \frac{\alpha}{2} \left( N^a + \alpha^{-1} F^a - g \epsilon^{ab} i  \bar{C}^b C^3 \right)^2  
\notag 
%\nonumber\\&
- \frac{1}{2\alpha} F^a F^a + F^a g \epsilon^{ab} i  \bar{C}^b C^3
-\frac{\alpha}{2} g^2 \left( \epsilon^{ab} i  \bar{C}^b C^3 \right)^2 ,
\end{align}
 the $N^a$ field can be integrated out to obtain 
\begin{align}
\mathscr{L}_{\rm GF+FP}^{\rm mMAG} 
 =& -\frac{1}{2\alpha} F^a F^a + i  \bar{C}^a D_\mu^{ab} [a] D^{\mu bc} [a] C^c 
%\notag \\                                       &
+ g^2 \epsilon^{ab} \epsilon^{cd} i  \bar{C}^a C^c A_\mu^b A^{\mu d} 
\notag \\ & 
+ \frac{\alpha}{4} g^2 \epsilon^{ab} \epsilon^{cd}  \bar{C}^a  \bar{C}^b C^c C^d \ , 
\ F^a = D_\mu^{ab} [a] A^{\mu b} ,
\end{align}
where the term $g \epsilon^{ab} i  \bar{C}^a \left( D_\mu^{bc} [a] A^{\mu c} \right) C^3 $ is canceled by $F^a g \epsilon^{ab} i  \bar{C}^b C^3$, and a term vanishes:  
$ \left( \epsilon^{ab} i  \bar{C}^b C^3 \right)^2 = \epsilon^{ab} i  \bar{C}^b C^3 \epsilon^{ac} i \Bar{C}^c C^3 = i^2  \bar{C}^a  \bar{C}^a ( C^3)^2 = 0$.

In the Lorenz gauge, we can introduce the four-ghost interaction by adopting the generalized Lorenz gauge, as already shown in \cite{KMSI02}.
In this case, however, there does not exist the term of $\bar{C}CAA$ type and hence the four-ghost interaction term proportional to $\alpha$ is never generated if the Landau gauge $\alpha = 0$ is chosen at the beginning.
In fact, $\alpha = 0$ is a fixed point of the renormalization group.
Whereas, $\alpha = 0$ is not a fixed point in MA gauge.%
\footnote{
See e.g., \cite{KS01}.
%K. -I. Kondo and T. Shinohara,
%Renormalizable Abelian projected effective gauge theory derived from quantum chromodynamics,
%CHIBA-EP-121, 
%hep-th/0005125,
%Prog. Theor. Phys. \textbf{105}, 649--665 (2001). 
For the most general MA gauge, see e.g., \cite{SIK03}.
%T. Shinohara, T. Imai, and K.-I. Kondo,  
%The Most general and renormalizable maximal Abelian gauge.
%CHIBA-EP-128 
%hep-th/0105268,
%Int.J.Mod.Phys. A\textbf{18}, 5733--5756 (2003). 
}
This is a crucial difference between the Lorenz gauge and the MA gauge. 

%(2)
 The MA gauge  extracts the diagonal part $H$ from the original gauge group $G$.
Hence the MA gauge  breaks explicitly the \textbf{global gauge symmetry}, i.e., \textbf{color symmetry}, in addition to the original \textbf{local gauge symmetry}.
In fact, MA gauge  condition $(D_{\mu}[a]A^{\mu})^{a} = 0$ is not covariant under the color rotation.
This fact is not desirable from a view point of considering \textbf{color confinement} as a generalization of quark confinement, or quark confinement as a special case of color confinement.
The important concept of \textbf{color singlet} in discussing color confinement loses its meaning if color symmetry is not maintained.
This should be compared with the \textbf{Lorenz gauge} $\partial_{\mu} \mathscr{A}^{\mu} = 0$, which maintains the color symmetry even after breaking the local gauge symmetry.

It is possible to maintain manifestly color symmetry by using a non-linear \textbf{change of variables}.
By using this method, moreover, we can give a gauge-invariant definition of magnetic monopole by keeping color symmetry intact, without using the Abelian projection.
\footnote{
The details will be given in the Sections 3, 4, and 5.% of Field decomposition and change of variables. 
}

%%%%%%%%%%%%%%%%%%%%%%%%%%%%%%%%%%%%%%%%%%%%%%%%%%%%%%%%%%%%
\subsection{Abelian dominance and magnetic monopole dominance}\label{subsec:dominance}
%%%%%%%%%%%%%%%%%%%%%%%%%%%%%%%%%%%%%%%%%%%%%%%%%%%%%%%%%%%%

For the Yang-Mills theory in the MA gauge, the following remarkable facts are obtained in the 1990s  according to numerical simulations on a lattice.

The Abelian (-projected) Wilson loop defined in terms of the diagonal component $A_\mu^3$ alone exhibits the area decay law, if the MA gauge is imposed on the $SU(2)$ Yang-Mills theory as a gauge fixing on a lattice:
\begin{align}
%  \text{Non-Abelian Wilson loop} \quad &
%\Big\langle \exp i g \oint_{C} dx^\mu \mathscr{A}_\mu(x) \Big\rangle \sim e^{-\%sigma_{NA} |S|} ,
%\\
W_{\rm Abel}(C) :=  \Big\langle \exp  \left\{ i g \oint_{C} dx^\mu A_\mu^3(x) \right\}    \Big\rangle_{\rm YM}^{\rm MAG} \sim e^{- \sigma_{A} |S|}  .
\end{align}
%where  we have used the SU(2) Cartan decomposition,
%$
%\mathscr{A}_\mu=A_\mu^a \frac{\sigma^a}{2} + A_\mu^3 \frac{\sigma^3}{2} \ (a=1,2) 
%$.
This result should be compared with a well known statement that quark confinement follows from the area law of the non-Abelian Wilson loop average \cite{Wilson74}%[Wilson,1974] 
\begin{align}
%  \text{Non-Abelian Wilson loop} \quad &
W(C) := \Big\langle {\rm tr} \left[ \mathscr{P} \exp \left\{ i g \oint_{C} dx^\mu \mathscr{A}_\mu(x) \right\} \right]  /{\rm tr}(\bm{1}) \Big\rangle_{\rm YM}^{\rm no~GF} \sim e^{-\sigma_{NA} |S|} . 
\end{align}
Here the most remarkable result obtained for the $SU(2)$ group  in the MA gauge is the fact:
 
\begin{enumerate}
\item[(i)]
\textbf{Abelian dominance} for $SU(2)$ \cite{SY90} 

The string tension $\sigma_{A}$ obtained from the Abelian Wilson loop average $W_{\rm Abel}(C)$  reproduces almost all the value (95$\%$) of the original string tension $\sigma_{NA} $ calculated from the non-Abelian Wilson loop average without gauge fixing:
\begin{equation}
\sigma_{A} \approx \sigma_{NA} .
\end{equation}
This is called the \textbf{infrared Abelian dominance} or \textbf{Abelian dominance} in the string tension.

The importance of Abelian dominance was stressed  by Ezawa and Iwazaki \cite{EI82} immediately after the proposal of the Abelian projection.  
%\footnote{
%Z.F. Ezawa and A. Iwazaki, 
%Abelian Dominance and Quark Confinement in Yang-Mills Theories,
%Phys. Rev. D\textbf{25}, 2681--2689  (1982).
%Abelian Dominance and Quark Confinement in Yang-Mills Theories, II. Oblique confinement and $\eta^\prime$ mass,
%Phys. Rev. D\textbf{26},  631--647 (1982).
%}
Later, Abelian dominance for the $SU(2)$ group  was confirmed by numerical simulations on a lattice  by Suzuki and Yotsuyanagi \cite{SY90}.%\footnote{
%T. Suzuki and I. Yotsuyanagi,
%A possible evidence for Abelian dominance in quark confinement,
%Phys. Rev. D\textbf{42}, 4257--4260 (1990).  
%} 

\item[(ii)]
\textbf{Magnetic monopole dominance}  for $SU(2)$ \cite{SNW94} 

 By separating the diagonal part $a_{\mu}(x)=A_\mu^3(x)$ into the magnetic monopole part $a_{\mu}^{\rm mono}(x)$ and the remaining part called the photon part $a_{\mu}^{ph}(x)$, $a_{\mu}(x) = a_{\mu}^{\rm mono}(x) + a_{\mu}^{ph}(x)$,
%(according to \cite{DT})
only the monopole part $a_{\mu}^{\rm mono}(x)$ exhibits the area law (decay):
\begin{align}
\Big\langle \exp  \left\{ i g \oint_{C} dx^\mu a_{\mu}^{\rm mono}(x) \right\}    \Big\rangle_{\rm YM}^{\rm MAG}   \sim e^{- \sigma_{\rm mono} |S|} .
\end{align}
and the monopole string tension $\sigma_{\rm mono}$ reproduces  almost all the value (95$\%$) of the Abelian string tension $\sigma_{A}$:
\begin{align}
&\sigma_{\rm mono} \approx \sigma_{A}   \quad ( \sigma_{ph} \approx 0 ) . 
\end{align}
This is called the \textbf{magnetic  monopole dominance} in the string tension. 
The magnetic monopole dominance for the $SU(2)$ group  was shown by numerical simulations on a lattice for $SU(2)$ \cite{SNW94}.
%\begin{equation}
% \Leftrightarrow \sigma_{\rm Abel}  \sim \sigma_{monopole}
%\end{equation}
%Moreover, if the diagonal gauge potential is decomposed into the so-called the photon and monopole parts: 
%$
%  A_\mu^3 = \text{Photon~part}+\text{Monopole part} 
%$ (according to \cite{DT}), 
%such that only the monopole part gives the non-vanishing magnetic charge, 
%then the monopole part reproduces almost all the value (95$\%$) of the string tension $\sigma_{Abel}$, which is called the monopole dominance in the string tension. 

\end{enumerate}

Another indication of  infrared Abelian dominance is the large \textit{non-zero mass of off-diagonal gluons} obtained in the MA gauge on a lattice \cite{AS99}. 

\begin{enumerate}
\item[(iii)] 
\textbf{Exponential fall-off of the off-diagonal gluon propagators}  for $SU(2)$  \cite{AS99}

The two-point correlation function exhibits the exponential fall-off in the large distance  for the off-diagonal gluon. 
For the $SU(2)$ group this means that   the off-diagonal gluon mass  $m_{A^1,A^2}$ is (much) larger than the possible diagonal gluon mass $m_{A^3}$:
\begin{align}
%   \text{off-diagonal gluon mass} \quad 
m_{A^1,A^2} (\cong 1.2 {\rm GeV}) \gg m_{A^3} 
%\ \text{diagonal gluon mass}
. 
\end{align}
It should be remarked that the precise statement for the diagonal gluon mass  $m_{A^3}=0$ or $m_{A^3} \ne 0$ is not sure at present. 

\end{enumerate}

Thus, non-perturbative studies performed so far in the MA gauge yield the following scenario for quark confinement.
\begin{enumerate}
\item[$\bullet$] The off-diagonal gluons decouple in the low-energy regime. 

This is because the the off-diagonal gluons  acquire their  mass dynamically.  
This leads to the infrared Abelian dominance.

\item[$\bullet$] The diagonal gluons are responsible for quark confinement.

The Wilson loop average for the diagonal gluon can reproduce the original string tension. The diagonal gluons generate magnetic monopoles with the magnetic charge subject to the Dirac quantization  condition.
%The magnetic monopole dominance in the string tension was confirmed by numerical method. 

\end{enumerate}

These facts enable  us to explain the \textbf{infrared Abelian dominance}:
In the low energy infrared region, the off-diagonal gluons are suppressed and the diagonal gluon dominates, since the massive off-diagonal gluons decouple and do not contribute to the low-energy physics.
This means that the off-diagonal gluons $A_{\mu}^{a}$ (and off-diagonal ghosts $C^{a}, \bar{C}^{a}$) are identified with the \textbf{high energy modes}, which are to be integrated out to obtain the \textbf{low-energy effective theory} valid below a certain energy (momentum scale) $M$ according to the idea of the \textbf{Wilsonian renormalization group}.
The effective theory $\mathscr{L}_{\rm eff} [a]$ written in terms of the diagonal gluon alone is suitable for discussing how the magnetic monopoles come into the play in the low-energy physics.

Indeed, by performing the functional integration of the off-diagonal gluons $A_{\mu}^{a}$ and off-diagonal ghosts $C^{a}, \bar{C}^{a}$, we can obtain a low-energy effective theory of the Abelian gauge type written in terms of the diagonal gluon $a_{\mu}$ alone, which is valid for the momenta $p$ in the region $p \leq M$:
\begin{equation}
\mathscr{L}_{\rm YM} + \mathscr{L}_{\rm GF+FP}^{\rm mMAG} \xrightarrow{\mathcal{D} A_{\mu}^{a} \mathcal{D} C^{a} \mathcal{D} \bar{C}^{a}} \mathscr{L}_{\rm eff}[a]   \quad ( p \leq M ) .
\end{equation}
Even in the low-energy region, the off-diagonal gluons and off-diagonal ghosts bring the renormalization effect and play the role of reflecting the characteristics of the original non-Abelian gauge theory.
In fact, it is shown \cite{KondoI} that the running gauge coupling in the resulting effective Abelian-like gauge theory $\mathscr{L}_{\rm eff}[a]$ obeys the beta function:
\begin{equation}
 \beta(g) :=  \mu \frac{\partial g(\mu)}{\partial \mu} = - \frac{\frac{11N}{3}}{(4\pi)^2} g^3(\mu) \quad (N=2),
\end{equation}
which reflects the \textbf{ultraviolet asymptotic freedom}  of the original non-Abelian gauge theory.
The resulting theory is called the \textbf{Abelian-projected low-energy effective theory} (APLEET).%
%\footnote{
%K.-I. Kondo,
%Abelian projected effective gauge theory of QCD with asymptotic freedom and quark confinement,
%CHIBA-EP-99, 
%e-Print: hep-th/9709109,
%Phys.Rev. D57, 7467--7487 (1998). 
%}

%The behavior of the diagonal gluon in the deep infrared region is  not yet established. The two-point correlation function does not show the simple  exponential fall-off behavior. How the diagonal gluon behaves in the deep infrared is a very important issue to be clarified, which is still under the debate in the current research from the viewpoint of the relationship between color confinement and the infrared behavior of the Green functions. This will be discussed in the chapter of Color confinement.

Therefore, the MA gauge is believed to be the most efficient way in performing the Abelian projection to demonstrate the dual superconductivity at least for $SU(2)$ Yang-Mills theory as a mechanism for quark confinement.
These results based on the Abelian projection, especially in the MA gauge are indeed remarkable progress toward the goal of understanding quark confinement based on the dual superconductivity in Yang-Mills theory. 
%See \cite{CP97} for a review. 
But, this is not the end of the story for quark confinement. 

(i) Partial gauge fixing: 
The MA gauge is a partial gauge fixing from the original non-Abelian gauge group $G$ to its subgroup $H$   in which the gauge degrees of  freedom of the coset $G/H$  are fixed.  Even after the MA gauge, there is a residual gauge group $H$ which is taken to be the maximal torus subgroup $H=U(1)^{N-1}$.  After the MA gauge, the magnetic monopole is expected to appear, since the Homotopy group $\pi_2(G/H)$ is non-trivial, i.e.,
\begin{equation}
 \pi_2(SU(N)/U(1)^{N-1}) = \pi_1(U(1)^{N-1}) = \mathbb{Z}^{N-1} .
\end{equation}
This implies that the breaking of gauge group $G \rightarrow H$ by partial gauge fixing leads to ($N-1$) species of magnetic monopoles.  
For the results of numerical simulations for  $SU(3)$ Yang-Mills theory in the MA gauge, see  e.g., \cite{SIMOY95,STW02,SAIIMT02,Langfeld04,Bornyakov04,GIS12,GS13,SS14}.

However, we do not necessarily need to consider the maximal breaking  $SU(N) \rightarrow U(1)^{N-1}$, although the maximal torus group is desirable as a gauge group of the low-energy effective {\it Abelian} gauge theory \cite{KondoI}.   Actually, even if we restrict $H$ to a continuous subgroup  of $G$, 
%\footnote{
%The possibility of a discrete subgroup has been extensively investigated recently from the viewpoint of the Abelian gauge, e.g., the center $Z_{N}$ for $SU(N)$ (see e.g. Ref.\cite{Greensite03}).  
%}
there are other possibilities for choosing $H$, e.g., we can choose a subgroup $\tilde H$ such that
\begin{equation}
  G \supset \tilde H \supset H = U(1)^{N-1} .
\end{equation}
The possible number of cases for choosing $\tilde H$ increases as $N$ increases.  
If we choose $\tilde H=U(N-1)$, then we find that a single kind of magnetic monopole:
% is sufficient for confining a fundamental quark, since 
\begin{equation}
 \pi_2(SU(N)/U(N-1)) = \pi_1(U(1)) = \mathbb{Z} .
\end{equation}
This kind of magnetic monopole might be sufficient for confining a fundamental quark. 
It is not clear which subgroup $\tilde H$  must be chosen for discussing quark confinement.

%We have found \cite{KT00} that the group $\tilde H$ may depend on the representation to which the quark belongs when $N \ge 3$  and that it suffices to take $\tilde H=U(N-1)$ for the fundamental quark to be confined in the sense of the area law of the Wilson loop under the partial gauge fixing.     
%Here $\tilde H$ is equal to the maximal stability group specified by the highest-weight state of the representation of the quark in the Wilson loop. This is a new feature which does not show up in the $SU(2)$ case. 
 
%Nevertheless, this does not mean that the choice of the maximal torus does not lead  to quark confinement.  In fact, even if we choose the maximal torus, the area law can be derived.  This is because the coset $G/\tilde H$ is contained in $G/H$, i.e., $G/\tilde H \subset G/H$, so that the Wilson loop does not feel the whole of $G/H$, but only feels the components of $G/\tilde H$ that are contained in $G/H$.  In other words, the variables belonging to $G/H - G/\tilde H$ are irrelevant for the expectation value of the Wilson loop, as can be seen from the non-Abelian Stokes theorem (NAST) that was presented in Ref.\cite{KT00} and is derived in this article.  
%Our results show that two partial gauge fixings  $SU(3) \rightarrow U(2)$ and $SU(3) \rightarrow U(1) \times U(1)$ lead  to the same  result for confinement as far as the fundamental quarks are concerned.
%\footnote{
%See Ref.\cite{Wensley99} for a result of the simulation on a lattice.
%}

(ii) Gauge invariance:
The partial gauge fixing is the explicit breaking of the gauge symmetry.  
Therefore, the results obtained in the partial gauge fixing cannot be guaranteed to be gauge-invariant physical results.

In the next section, we discuss how to surmount  these  drawbacks of the Abelian projection to obtain more reliable gauge-independent results. 
In view of these, we discuss a new approach called the CDGFN decomposition of the Yang-Mills field in the next section.

\newpage
%%%%%%%%%%%%%%%%%%%%%%%%%%%%%%%%%%%%%%%%%%%%%%%%%%%%%%%%%%%%
%Chapter :
% 
%%%%%%%%%%%%%%%%%%%%%%%%%%%%%%%%%%%%%%%%%%%%%%%%%%%%%%%%%%%%

%%%%%%%%%%%%%%%%%%%%%%%%%%%%%%%%%%%%%%%%%%%%%%%%%%%%%%%%%%%%
%%%%%%%%%%%%%%%%%%%%%%%%%%%%%%%%%%%%%%%%%%%%%%%%%%%%%%%%%%%%

\section{Cho-Duan-Ge-Faddeev-Niemi decomposition of Yang-Mills field}\label{sec: CDGFN-decomp} 

%%%%%%%%%%%%%%%%%%%%%%%%%%%%%%%%%%%%%%%%%%%%%%%%%%%%%%%%%%%%
%%%%%%%%%%%%%%%%%%%%%%%%%%%%%%%%%%%%%%%%%%%%%%%%%%%%%%%%%%%%

In this section, we review a decomposition of the Yang-Mills field called the \textbf{Cho--Duan-Ge (CDG) decomposition} or   \textbf{Cho--Duan-Ge--Faddeev-Niemi (CDGFN) decomposition}, or \textbf{Cho--Faddeev-Niemi (CFN) decomposition}, or \textbf{Cho--Duan-Ge--Faddeev-Niemi--Shabanov (CDGFNS) decomposition}, which was proposed by Cho \cite{Cho80}  and Duan \& Ge \cite{DG79} independently, and later readdressed by Faddeev and Niemi \cite{FN98} , and developed by Shabanov \cite{Shabanov99} and   Chiba University group \cite{KMS06,KMS05,Kondo06}.%
\footnote{
Cho's work \cite{Cho80} was done independently from that of Duan and De \cite{DG79} which was written in Chinese. 
The English version of \cite{DG79} is now available from Dr. Peng-ming Zhang.}

%%%%%%%%%%%%%%%%%%%%%%%%%%%%%%%%%%%%%%%%%%%%%%%%%%%%%%%%%%%%

\subsection{Abelian projection and the need for new approaches}\label{subsec:new-approach}

%%%%%%%%%%%%%%%%%%%%%%%%%%%%%%%%%%%%%%%%%%%%%%%%%%%%%%%%%%%%

The MA gauge is believed to be the most efficient choice for realizing the Abelian projection to demonstrate the dual superconductivity. 
%\footnote{
%  G. 't Hooft,
%Topology of the gauge condition and new confinement phases in non-Abelian gauge theories,
%Nucl.Phys. B{\bf 190} [FS3], 455--478 (1981).
%}
The results obtained for the $SU(2)$ Yang-Mills theory in the MA gauge are indeed remarkable progress toward the goal of understanding quark confinement based on the dual superconductivity in Yang-Mills theory. 
%See \cite{CP97} for a review. 

However, we are still unsatisfactory by the following reasons.

\begin{enumerate}
\item[$\bullet$]
 The Abelian projection  breaks  $SU(2)$ color symmetry  explicitly. 

This is inconvenient, because we wish to regard quark confinement as a special case of color confinement.  The color confinement is well-defined only if the color symmetry is preserved.  If the color symmetry is broken, we lose a chance of explaining color confinement as an extension of quark confinement.
% (quark confinement as a special case of color confinement). 

\item[$\bullet$] 
Infrared Abelian dominance has never been observed in  gauge fixings other than the Abelian gauges, such as the MA gauge,  Laplacian Abelian gauge and maximal center gauge \cite{Greensite03}. 
%\footnote{
%See e.g., \cite{Greensite03}. 
%\bibitem{Greensite03}
%J. Greensite,
%The Confinement problem in lattice gauge theory.
%[hep-lat/0301023],  
%Prog. Part. Nucl. Phys. {\bf 51}, 1 (2003).  
%}
This raises the dubious impression that the dual superconductivity obtained in the MA gauge might be a gauge artifact. 
This raises the dubious impression that the dual superconductivity obtained in the MA gauge might be a gauge artifact. 

\end{enumerate}

Therefore, we must answer whether the dual superconductivity can be a gauge-invariant concept or not.
In what follows, we wish to discuss how to cure these shortcomings encountered in MA gauge
and  obtain a gauge-invariant dual superconductivity picture.

To obtain \textit{gauge-invariant} dual superconductivity  in Yang-Mills theory, 
we ask once again: 
\begin{enumerate}
\item 
How to extract the \textit{``Abelian'' part} responsible for quark confinement  from the non-Abelian gauge theory in a \textit{gauge-invariant way}  without losing characteristic features of the non-Abelian gauge theory, e.g., asymptotic freedom.

\item  
How to define the \textbf{chromomagnetic monopole} to be condensed in  Yang-Mills theory even in absence of any  fundamental scalar field in the \textit{gauge-invariant way}.

\end{enumerate}

In this section,  we discuss how the Yang-Mills potential $\mathscr{A}_\mu(x)$ is decomposed into 
\begin{equation}
 \mathscr{A}_\mu(x) = \mathscr{V}_\mu(x) + \mathscr{X}_\mu(x) ,
\end{equation}
in such a way that
\begin{enumerate}
\item 
{[gauge-invariant ``Abelian''  projection]}

The restricted part $\mathscr{V}_\mu$ corresponding to the \textit{``Abelian'' part}  responsible for quark confinement can be extracted from the non-Abelian gauge theory
by using \textbf{a (nonlinear) change of variables} in the \textit{gauge-invariant way}  without breaking color symmetry. 

\item  
{[infrared ``Abelian''  dominance]} 

The remaining part $\mathscr{X}_\mu$ decouples (or is suppressed) in the low-energy region  leading to infrared ``Abelian'' $\mathscr{V}_\mu$ dominance. 
This is possible for the remaining part to acquire the mass. 
For instance, such a dynamical mass can originate from  the existence of dimension-2 vacuum condensate, e.g., $\left< \mathscr{X}_\mu^2 \right>\ne 0$. 
\end{enumerate}

In view of these, we discuss a new approach called the CDGFN decomposition of the Yang-Mills field in what follows. 

%\cite{DG79,Cho80,FN98,Shabanov99,Cho80c,FN99a,BCK02} and change of variables \cite{KMS06,KMS05,Kondo06,KSM08,KKMSS05,KKMSSI06,IKKMSS06,SKKMSI07,KSSMKI08,SKS10}

%For simplicity, we restrict the gauge group to $SU(2)$ in this section. The $SU(N)$ case will be treated later.

%%%%%%%%%%%%%%%%%%%%%%%%%%%%%%%%%%%%%%%%%%%%%%%%%%%%%%%%%%%%

\subsection{Decomposing the $SU(2)$ Yang-Mills field}\label{subsec:decomposition}

%%%%%%%%%%%%%%%%%%%%%%%%%%%%%%%%%%%%%%%%%%%%%%%%%%%%%%%%%%%%

 The key ingredient of the CDG   decomposition is the introduction of a  color direction field $\bm{n}(x)$ which represents the (local) Abelian direction embedded in the non-Abelian gauge degrees of freedom at each space--time point. 
The CDG decomposition enables us to extract the dominant degrees of freedom in the infrared region in a gauge-covariant manner and give a gauge-invariant definition of the chromomagnetic monopole in the Yang-Mills theory. 
 For simplicity, we restrict the gauge group to $SU(2)$ in this section. 
The $SU(N)$ case will be treated in the next section.
 
The decomposition begins with the introduction of the unit field $\bm{n}(x)$, which we call the  \textbf{color direction field} or \textbf{color field} for short.
For $G=SU(2)$,  the  ``color field'' $\bm{n}(x)$ has a unit length and  three components.  Therefore, it has two independent degrees of freedom.
In the Lie-algebra  notation  $\bm{n}$, the color field  is expressed  as 
\begin{align}
 \bm{n}(x)=  n^A(x)T_A, \quad T_A= \frac12 \sigma_A  \quad (A=1,2,3),
%\quad 
% \nonumber\\
% \Longleftrightarrow \mathbf{n}(x) = (n_1(x),n_2(x),n_3(x)) ,
%\nonumber\\
%{\rm i.e.,} {\rm tr}[\bm{n}(x)\bm{n}(x)]=1/2 \quad {\rm or} \quad  \mathbf{n}(x) \cdot \mathbf{n}(x) = n_A(x) n_A(x) = 1 , 
\end{align}
where $\sigma_A$ ($A=1,2,3$) denotes the Pauli matrices. 
In the vector  notation   $\mathbf{n}$, it is expressed  as
\begin{align}
 \mathbf{n}(x) = (n^1(x),n^2(x),n^3(x)) .
%\nonumber\\
%{\rm i.e.,} {\rm tr}[\bm{n}(x)\bm{n}(x)]=1/2 \quad {\rm or} \quad  \mathbf{n}(x) \cdot \mathbf{n}(x) = n_A(x) n_A(x) = 1 , 
\end{align}
The constraint of the unit length is represented respectively as
\begin{align}
\bm{n}(x) \cdot \bm{n}(x) = 2{\rm tr}[\bm{n}(x)\bm{n}(x)]=1 
\Longleftrightarrow
 \mathbf{n}(x) \cdot \mathbf{n}(x) = n_A(x) n_A(x) = 1 
 , 
\end{align}
where the summation over repeated indices $A$ should be understood. 
%\footnote{
%In what follows, we use two notations: the Lie algebra (matrix)  notation  and the vector  notation.
%The Lie algebra  notation  is represented by the italic letter, while the vector  notation  by the bold letter. 
%For the inner product, we use 
%\begin{align}
%\mathscr{A}(x) \cdot \mathscr{B}(x) := 2{\rm tr}[\mathscr{A}(x)\mathscr{B}(x)]=2\mathscr{A}^A(x)\mathscr{B}^B(x){\rm tr}[T^A T^B]=\mathscr{A}^A(x)\mathscr{B}^B(x)
% = \mathbf{A}(x) \cdot \mathbf{B}(x)   . 
%\end{align}
%} 

Why the color field has the unit length? From the physical point of view, the color field is introduced so as to allow the Abelian direction or the choice of the diagonal field to vary point to point in space--time according to the gauge principle, and hence its magnitude (length) is irrelevant for this purpose. 
From the technical point of view, the magnitude of the color field introduces an extra degree of freedom other than the original Yang-Mills field.  
This restriction enables one to obtain the new formulation which is  equivalent to the original Yang-Mills theory, see the reduction condition in the next section for details. 

The color  field plays the following roles, as will be explained in detail  later.

\begin{itemize}
\item 
Unbroken color symmetry [Recovery of color symmetry]: The color  field recovers the color symmetry which is lost in the convectional treatment of the Abelian projection or MA gauge, since the Abelian projection selects a special color direction over the whole space--time.
The Abelian projection is reproduced by choosing a uniform color field: 
\begin{align}
\bm{n}(x) = \bm{n}_0 :=T_3=\frac12 \sigma_3 \
( \text{or} \ n^A(x)=n^A_0(x):=\delta^{A3} ) 
%\nonumber\\
\Longleftrightarrow
\mathbf{n}(x) = \mathbf{n}_0:=(0,0,1)
 . 
\end{align}

\item
Carrier of topological defects: The color  field  carries topological defects responsible for quark confinement. 
The color field is identified with the fundamental field variable corresponding topological defects. 
In particular, it enables one to define a gauge invariant magnetic monopole. 
 
\end{itemize}

If  we can find a decomposition
of the $SU(2)$ gauge field: 
%$\mathbf{A}_\mu(x)=\mathbf{A}_\mu^A(x) \sigma^A/2$,
\begin{equation}
 \mathscr{A}_\mu(x)
 = \mathscr{V}_\mu(x) + \mathscr{X}_\mu(x) 
  ,
\end{equation}
such that  {the field strength $\mathscr{F}_{\mu\nu}[\mathscr{V}]$ of the restricted field $\mathscr{V}_\mu$  is proportional to a unit field $\bm{n}$ (i.e., $\bm{n} \cdot \bm{n}=1$)}:
\begin{equation}
\mathscr{F}_{\mu\nu}[\mathscr{V}](x)
 := \partial_\mu \mathscr{V}_\nu(x) - \partial_\nu \mathscr{V}_\mu(x) -ig [\mathscr{V}_\mu(x),  \mathscr{V}_\nu(x) ]
 = f_{\mu\nu}(x) \bm{n}(x) ,
%[\partial_\mu c_\nu - \partial_\nu c_\mu - g^{-1} \mathbf{n} \cdot (\partial_\mu \mathbf{n} \times \partial_\nu \mathbf{n})]  
%  \nonumber
\end{equation}
then we can introduce the gauge-invariant field strength $f_{\mu\nu}$ by
\begin{equation}
 f_{\mu\nu}(x) := \bm{n}(x) \cdot \mathscr{F}_{\mu\nu}[\mathscr{V}](x) \rightarrow  f_{\mu\nu}(x)  ,
%=  \partial_\mu c_\nu - \partial_\nu c_\mu - g^{-1} \bm{n} \cdot [ \partial_\mu \bm{n}, \partial_\nu \bm{n} ] .
\end{equation}
since  $\mathscr{F}_{\mu\nu}[\mathscr{V}]$ and $\bm{n}$ transform according to the adjoint representation under the gauge transformation:
\begin{equation}
 \mathscr{F}_{\mu\nu}[\mathscr{V}](x) \rightarrow U(x) \mathscr{F}_{\mu\nu}[\mathscr{V}](x) U^\dagger(x) , 
 \quad
 \bm{n}(x) \rightarrow U(x) \bm{n}(x) U^\dagger(x) .
%[\partial_\mu c_\nu - \partial_\nu c_\mu - g^{-1} \mathbf{n} \cdot (\partial_\mu \mathbf{n} \times \partial_\nu \mathbf{n})]  
\end{equation}
Therefore, we can define a gauge-invariant magnetic monopole current $k$ by 
\begin{equation}
 k_\mu(x) = \partial_\nu {}^*f_{\mu\nu}(x) .
%(=   (1/2) \epsilon_{\mu\nu\rho\sigma}\partial_{\nu}
%              f^{\rho\sigma}(x) \quad D=4) ,
\end{equation}
%The field strength $f_{\mu\nu}$ is gauge invariant:
%\begin{equation}
% f_{\mu\nu}(x) := \bm{n} \cdot \mathscr{F}_{\mu\nu}[\mathscr{V}] \rightarrow  f_{\mu\nu}(x)  .
%=  \partial_\mu c_\nu - \partial_\nu c_\mu - g^{-1} \bm{n} \cdot [ \partial_\mu \bm{n}, \partial_\nu \bm{n} ] .
%\end{equation}
Is such a decomposition ({spin--charge separation}) possible?
The answer is Yes! 
Indeed, the answer to this question was given by Cho (1980) and Duan \& Ge (1979) independently.
This is the CDG decomposition.

Suppose that the Lie algebra $su(2)$-valued Yang-Mills field: 
%$\mathscr{A}_\mu(x)=\mathscr{A}_\mu^A(x) T_A =\mathscr{A}_\mu^A(x) \sigma^A/2$  
\begin{equation}
 \mathscr{A}_\mu(x)=\mathscr{A}_\mu^A(x) T_A =\mathscr{A}_\mu^A(x)  \frac12 \sigma^A  \in su(2)
  ,
\end{equation}
is decomposed into two pieces $\mathscr{V}_\mu(x)$ and $\mathscr{X}_\mu(x)$: 
\begin{equation}
 \mathscr{A}_\mu(x)
 = \mathscr{V}_\mu(x) + \mathscr{X}_\mu(x) \in su(2), 
 \quad \mathscr{V}_\mu(x) , \mathscr{X}_\mu(x) \in su(2)
  ,
\end{equation}
such that the first piece $\mathscr{V}_\mu(x)$ transforms with  the inhomogeneous term under the gauge transformation $U(x) \in G$:
\begin{equation}
  \mathscr{V}_\mu(x) \rightarrow  \mathscr{V}^\prime_\mu(x) := U(x) \mathscr{V}_\mu(x) U(x)^\dagger + ig^{-1} U(x) \partial_\mu U(x)^\dagger  
  ,
\end{equation}
in the same way as the original Yang-Mills field:
\begin{equation}
  \mathscr{A}_\mu(x) \rightarrow \mathscr{A}^\prime_\mu(x)  := U(x) \mathscr{A}_\mu(x) U(x)^\dagger + ig^{-1} U(x) \partial_\mu U(x)^\dagger 
  .
\end{equation}
Consequently, the second piece $\mathscr{X}_\mu(x)$ transforms without the inhomogeneous term under the gauge transformation  as
\begin{equation}
  \mathscr{X}_\mu(x) \rightarrow  \mathscr{X}^\prime_\mu(x) := U(x) \mathscr{X}_\mu(x) U(x)^\dagger  
  .
\end{equation}

We find that the decomposition is uniquely determined by solving a set of   \textbf{defining equations} for $\mathscr{V}_\mu$ and  $\mathscr{X}_\mu$ for a given color field $\bm{n}$.
The form of the defining equation must be covariant under the gauge transformation (i.e., form-invariant) in agreement with the above requirement. 
For concreteness, we adopt the defining equations:
\begin{enumerate}
\item[(i)] covariant constantness (integrability) of color field $\bm{n}(x)$  in the background field $\mathscr{V}_\mu(x)$:
\begin{align}
%\text{(i) covariant constantness 
%(integrability) 
%of color field $\mathbf{n}$ in the background $\mathbf{V}_\mu$:}\quad
0 =& \mathscr{D}_\mu[\mathscr{V}]\bm{n}(x) 
\quad (\mathscr{D}_\mu[\mathscr{V}]\bm{n}(x):= \partial_\mu \bm{n}(x) - ig [\mathscr{V}_\mu(x), \bm{n}(x)] )
 \nonumber\\
\Longleftrightarrow 
 0 =&  D_\mu[\mathbf{V}]\mathbf{n}(x) 
\quad (D_\mu[\mathbf{V}]\mathbf{n}(x) := \partial_\mu \mathbf{n}(x) + g \mathbf{V}_\mu(x) \times \mathbf{n}(x) )
,
  \label{C26-def1}
\end{align}
with the covariant derivatives: 
\begin{align}
\mathscr{D}_\mu[\mathscr{A}] :=\partial_\mu -ig [ \mathscr{A}_\mu , \cdot ] ,
\quad 
%with the covariant derivative
 D_\mu[\mathbf{A}] :=\partial_\mu +g\mathbf{A}_\mu \times .
\end{align}

\item[(ii)] orthogonality of the color field $\bm{n}(x)$ to $\mathscr{X}_\mu(x)$:
\begin{equation}
%\text{(ii) orthogonality of  $\mathbf{X}_\mu$ to $\mathbf{n}$:}\quad\quad\quad\quad\quad\quad\quad\quad\quad\quad\quad\quad
 0 =  \bm{n}(x) \cdot \mathscr{X}_\mu(x)  = {\rm tr}[\bm{n}(x)   \mathscr{X}_\mu(x) ]
% 0 =  \bm{n}(x) \cdot  \mathscr{X}_\mu(x)  
\Longleftrightarrow 
 0 = \mathbf{n}(x) \cdot \mathbf{X}_\mu(x)   .
 \label{C26-def2}
\end{equation}
\end{enumerate}

The reason why imposing such defining equations can determine the decomposition uniquely will be discussed in detail later, since it is related to the issue of counting the number of independent degrees of freedom to be resolved finally. 

It is easy to check that the defining equations are form-invariant, namely, the same defining equations hold for the primed variables $\bm{n}^\prime(x), \mathscr{V}^\prime_\mu(x), \mathscr{X}^\prime_\mu(x)$ after the gauge transformation, provided that the color field transforms according to the adjoint representation under the gauge transformation:
\begin{equation}
  \bm{n}(x) \rightarrow  \bm{n}^\prime(x) := U(x) \bm{n}(x) U(x)^\dagger  
  .
\end{equation}
%[Exercise-1] \marginpar{Ex-1}
%Verify that the defining equations are form-invariant under the gauge transformation.

The color field $\bm{n}(x)$ is supposed to be given as a functional of $\mathscr{A}_\mu(x)$, i.e.,
\begin{align}
 \bm{n}(x) =  \bm{n}_{\mathscr{A}}(x) .
\end{align}
This issue  is discussed later in details and therefore we put this issue aside for a while.
Then, the defining equations are solved to determine the decomposition uniquely. 
In fact, the new variables are expressed in terms of $\mathscr{A}$ and $\bm{n}$ as follows, once the color field is given. 
%[Exercise-2] \marginpar{Ex-2}
%The decomposition (\ref{C26-CDG-decomp1}) or (\ref{C26-CDG-decomp2}) is uniquely obtained by solving the defining equations  (\ref{C26-def1}) and (\ref{C26-def2}) simultaneously. 

By using the first defining equation:
%\begin{align}
$  D_\mu[\mathbf{V}] \mathbf{n} 
%:= \partial_\mu \mathbf{n} + g\mathbf{V}_\mu \times \mathbf{n} 
 = 0   
%\label{C26-D[V]n=0}
%\end{align}
$,
we find the relation:
\begin{align}
  D_\mu[\mathbf{A}] \mathbf{n}
  := \partial_\mu \mathbf{n} +  g\mathbf{A}_\mu \times \mathbf{n} 
%  = \partial_\mu \bm{n} + g\mathbf{V}_\mu \times \bm{n} + g\mathbf{X}_\mu \times \bm{n} 
  = D_\mu[\mathbf{V}] \mathbf{n} + g\mathbf{X}_\mu \times \mathbf{n}
= g\mathbf{X}_\mu \times \mathbf{n} .
\label{C26-XnA}
\end{align}
Then, by taking the exterior product with the color field, we find
\begin{align}
 \mathbf{n} \times D_\mu[\mathbf{A}] \mathbf{n}
= \mathbf{n} \times( g\mathbf{X}_\mu \times \mathbf{n}) 
= (\mathbf{n} \cdot \mathbf{n}) g\mathbf{X}_\mu - (\mathbf{n} \cdot g\mathbf{X}_\mu) \mathbf{n} 
=  g\mathbf{X}_\mu 
 ,
 \label{C26-X-derivation}
\end{align}
where we have used the second defining equation $\mathbf{n} \cdot  \mathbf{X}_\mu=0$ and the constraint $\mathbf{n} \cdot \mathbf{n}=1$.

In the Lie-algebra notation:
\begin{subequations}
\begin{align}
  \mathscr{A}_\mu(x)= \mathscr{V}_\mu(x) +& \mathscr{X}_\mu(x),
  \nonumber
\\
   \mathscr{V}_\mu(x) =& c_\mu(x)\bm{n}(x) -i   g^{-1} [\partial_\mu \bm{n}(x) , \bm{n}(x) ] , 
%(\leftarrow \text{Cho connection})
\quad
   c_\mu(x) =   \mathscr{A}_\mu(x)  \cdot \bm{n}(x) ,
  \nonumber
\\
  \mathscr{X}_\mu(x) =&  -i g^{-1} [ \bm{n}(x) ,  \mathscr{D}_\mu[\mathscr{A}] \bm{n}(x) ] .
% \quad
% (\mathscr{D}_\mu[\mathscr{A}] :=\partial_\mu -ig [ \mathscr{A}_\mu , \cdot ] ) ,
\label{C26-CDG-decomp1}
\end{align}
%with the covariant derivative 
%$\mathscr{D}_\mu[\mathscr{A}] :=\partial_\mu -ig [ \mathscr{A}_\mu , \cdot ] $.

In the vector notation:
\begin{align}
  \mathbf{A}_\mu(x)= \mathbf{V}_\mu(x) +& \mathbf{X}_\mu(x),
  \nonumber
\\
   \mathbf{V}_\mu(x) =& c_\mu(x)\mathbf{n}(x) +   g^{-1} \partial_\mu \mathbf{n}(x) \times \mathbf{n}(x) , 
%(\leftarrow \text{Cho connection})
\quad
   c_\mu(x) =   \mathbf{A}_\mu(x)  \cdot \mathbf{n}(x) ,
  \nonumber
\\
  \mathbf{X}_\mu(x) =&  g^{-1} \mathbf{n}(x) \times D_\mu[\mathbf{A}] \mathbf{n}(x) .
% \quad
% (D_\mu[\mathbf{A}] :=\partial_\mu +g\mathbf{A}_\mu \times) .
\label{C26-CDG-decomp2}
\end{align}
\label{C26-decomp0}
\end{subequations}
%with the covariant derivative
%$D_\mu[\mathbf{A}] :=\partial_\mu +g\mathbf{A}_\mu \times$.

%($\Longleftarrow$) This recovers the original gauge field by the identification:
%\begin{equation}
%  \mathbf{A}_\mu(x)=\mathbf{V}_\mu(x)+\mathbf{X}_\mu(x),
%  \quad
%  \mathbf{V}_\mu(x)=\mathbf{n}(x)c_\mu(x) + \color{red}{g^{-1} \partial_\mu \mathbf{n}(x) \times \mathbf{n}(x)}
%\end{equation}

%We find that the decomposition is an identity, since the gluon field $\mathscr{A}_\mu(x)$ is cast into the equivalent form as 
%\begin{align}
% \mathbf{A}_\mu 
% =& (\mathbf{n} \cdot \mathbf{A}_\mu)\mathbf{n} + \mathbf{A}_\mu - (\mathbf{n} \cdot \mathbf{A}_\mu) \mathbf{n}
% \nonumber\\
% =& (\mathbf{n} \cdot \mathbf{A}_\mu)\mathbf{n} +   (\mathbf{n}\cdot\mathbf{n})  \mathbf{A}_\mu - (\mathbf{n} \cdot \mathbf{A}_\mu) \mathbf{n}
% \nonumber\\
% =& (\mathbf{n} \cdot \mathbf{A}_\mu)\mathbf{n} + \mathbf{n} \times (\mathbf{A}_\mu \times \mathbf{n})
% \nonumber\\
% =& (\mathbf{n} \cdot \mathbf{A}_\mu)\mathbf{n} - g^{-1}\mathbf{n} \times \partial_\mu \mathbf{n} + g^{-1}\mathbf{n} \times  \partial_\mu \mathbf{n}  + \mathbf{n} \times (\mathbf{A}_\mu \times \mathbf{n})
% \nonumber\\
% =& (\mathbf{n} \cdot \mathbf{A}_\mu)\mathbf{n} + g^{-1} \partial_\mu \mathbf{n} \times \mathbf{n} + g^{-1}\mathbf{n} \times D_\mu[\mathbf{A}]\mathbf{n} ,
%\label{C26-CFNform}
%\end{align}
%where we have used only a fact   $\mathbf{n}\cdot \mathbf{n}=1$  in the second equality. 

In fact, it is shown that the field strength   $\mathbf{F}_{\mu\nu}[\mathbf{V}]$ defined by
\begin{equation}
\mathbf{F}_{\mu\nu}[\mathbf{V}]
 := \partial_\mu \mathbf{V}_\nu(x) - \partial_\nu \mathbf{V}_\mu(x) + g \mathbf{V}_\mu(x) \times  \mathbf{V}_\nu(x) 
\end{equation}
is found to be proportional to $\mathbf{n}$: 
%[Ex-13] \marginpar{Ex-13}
\begin{equation}
 \mathbf{F}_{\mu\nu}[\mathbf{V}]
% := \partial_\mu \mathbb{V}_\nu - \partial_\nu \mathbb{V}_\mu + g \mathbb{V}_\mu \times \mathbb{V}_\nu 
 = \mathbf{n} [\partial_\mu c_\nu - \partial_\nu c_\mu - g^{-1} \mathbf{n} \cdot (\partial_\mu \mathbf{n} \times \partial_\nu \mathbf{n})]  .
\end{equation}
Then we find the explicit expression of a  {gauge-invariant field strength}:
\begin{equation}
 f_{\mu\nu} := \mathbf{n} \cdot \mathbf{F}_{\mu\nu}[\mathbf{V}]
  =  \partial_\mu c_\nu - \partial_\nu c_\mu - g^{-1} \mathbf{n} \cdot (\partial_\mu \mathbf{n} \times \partial_\nu \mathbf{n}) .
\end{equation}
%Remember that this is the same form as the 'tHooft tensor for the 't Hooft-Polyakov magnetic monopole, if  
%the color  field is identified with the normalized  {adjoint} scalar field in the Georgi-Glashow model:
%\begin{equation}
%\mathbf{n}^A(x) \leftrightarrow \hat{\phi}^A(x) := \phi^A(x)/||\phi(x)|| .
%\end{equation}

It should be remarked that $f_{\mu\nu}$ has the same form as the  tensor characterizing the 't Hooft-Polyakov magnetic monopole, 
%\footnote{
%G. 't Hooft,
%Nucl. Phys. B{\bf 79}, 276--284 (1974).
%A. M. Polyakov,
%JETP Lett. {\bf 20}, 194--195 (1974).
%Pisma Zh. Eksp. Teor. Fiz. {\bf 20}, 430--433  (1974). 
%}
if  the color unit field $\mathbf{n}^A(x)$ is identified with the normalized \textit{adjoint} scalar field $\hat{\phi}^A(x)$ in the Georgi-Glashow model:
\begin{equation}
\mathbf{n}^A(x) \leftrightarrow \hat{\phi}^A(x) := \phi^A(x)/||\phi(x)|| .
\end{equation}

%%%%%%%%%%%%%%%%%%%%%%%%%%%%%%%%%%%%%%%%%%%%%%%%%%%%%%%%%%%%

\subsection{Cartan decomposition and the Abelian projection}

%%%%%%%%%%%%%%%%%%%%%%%%%%%%%%%%%%%%%%%%%%%%%%%%%%%%%%%%%%%%

In particular, the Cartan decomposition used to define the conventional Abelian projection is reproduced by forcing the color field to be uniform as follows.
In the vector form,
\begin{subequations}
\begin{align}
 \mathbf{n}(x) = \mathbf{n}_0 \equiv (0,0,1) 
 \Longrightarrow \mathbf{A}_\mu(x)=& \mathbf{V}_\mu(x)+\mathbf{X}_\mu(x),
  \nonumber
\\
    \mathbf{V}_\mu(x)=& (0,0, c_\mu(x)) ,  
 \quad\   c_\mu(x) =   {A}_\mu^3(x)    ,
  \nonumber
\\
  \mathbf{X}_\mu(x) =&  ( {A}_\mu^1(x), {A}_\mu^2(x),0) ,
 \label{C26-n-abel1}
\end{align} 
or in the Lie algebra form,
\begin{align}
 \bm{n}(x) = \bm{n}_0 \equiv T_3 
 \Longrightarrow \mathscr{A}_\mu(x)=& \mathscr{V}_\mu(x)+\mathscr{X}_\mu(x),
  \nonumber
\\
   \mathscr{V}_\mu(x) =& \mathscr{A}_\mu^3(x) T_3 ,  
 \quad\   c_\mu(x) =   \mathscr{A}_\mu^3(x)    ,
  \nonumber
\\
   \mathscr{X}_\mu(x) =& \mathscr{A}_\mu^1(x) T_1+\mathscr{A}_\mu^2(x) T_2=\mathscr{A}_\mu^a(x) T_a
 .
 \label{C26-n-abel2}
\end{align}
\end{subequations}
Therefore, $\mathscr{V}_\mu$ corresponds to the diagonal part of $\mathscr{A}_\mu$, while $\mathscr{X}_\mu$ the off-diagonal part. 
We can easily check that they constitute a set of solutions of the defining equations for this choice of the uniform color field. 
%[Exercise-3] \marginpar{Ex-3}
%Verify that the solution of the defining equations is given by (\ref{C26-n-abel1}) or (\ref{C26-n-abel2}) if the color field is uniform: $\mathbf{n}(x) = \mathbf{n}_0 \equiv (0,0,1) $ or $\bm{n}(x) = \bm{n}_0 \equiv T_3$. 
For a uniform color field $\bm{n}(x) = {\rm const.}$, indeed, we find   $\partial_\mu \bm{n}(x)=0$.  Then the first defining equation means that $\mathbf{V}_\mu(x)$ is parallel to $\mathbf{n}(x)$: 
\begin{equation}
 0 =  \mathbf{V}_\mu(x) \times \mathbf{n}   \Longrightarrow  \mathbf{V}_\mu(x) \parallel \mathbf{n} \Longrightarrow \mathbf{V}_\mu(x) = c_\mu(x) \mathbf{n}  ,
  \label{C26-def1b}
\end{equation}
while $\mathbf{X}_\mu(x)$ is orthogonal to $\mathbf{n}(x)$ from the second defining equation: 
\begin{equation}
 0 =  \mathbf{X}_\mu(x) \cdot \mathbf{n}  \Longrightarrow  \mathbf{X}_\mu(x) \perp \mathbf{n}  .
\end{equation}
This reproduces the covariantly constant field strength. 
%\footnote{
%See the chapter of Chromomagnetic condensation. 
%}

%%%%%%%%%%%%%%%%%%%%%%%%%%%%%%%%%%%%%%%%%%%%%%%%%%%%%%%%%%%%

\subsection{Field strength for the decomposed $SU(2)$ Yang-Mills field}\label{subsec:field-strength}

%%%%%%%%%%%%%%%%%%%%%%%%%%%%%%%%%%%%%%%%%%%%%%%%%%%%%%%%%%%%

For the decomposition  
$\mathscr{A}  =  \mathscr{V} + \mathscr{X}$  of the gauge field (connection one-form) $\mathscr{A}$, 
the field strength (curvature two-form) is decomposed as
\begin{align}
 \mathscr{F}_{\mu\nu}[\mathscr{A}](x)
 :=&  \partial_\mu \mathscr{A}_\nu(x) - \partial_\nu \mathscr{A}_\mu(x) - ig [\mathscr{A}_\mu(x) , \mathscr{A}_\nu(x) ]
 \nonumber\\
 =&  \mathscr{F}_{\mu\nu}[\mathscr{V}](x) + \mathscr{D}_\mu[\mathscr{V}]\mathscr{X}_\nu(x) - \mathscr{D}_\nu[\mathscr{V}]\mathscr{X}_\mu(x) 
%\nonumber\\  &
-ig [\mathscr{X}_\mu(x) , \mathscr{X}_\nu(x) ]
  ,
\end{align}
where $\mathscr{F}[\mathscr{V}]$ is the field strength of $\mathscr{V}$ defined by
$
 \mathscr{F}_{\mu\nu}[\mathscr{V}](x)
 :=  \partial_\mu \mathscr{V}_\nu(x) - \partial_\nu \mathscr{V}_\mu(x) -ig [\mathscr{V}_\mu(x) , \mathscr{V}_\nu(x) ]  
$.

In the vector notation,   the decomposition:
$\mathbf{A}_{\mu}  =  \mathbf{V}_{\mu} + \mathbf{X}_{\mu} $ leads to  
\begin{align}
 \mathbf{F}_{\mu\nu}[\mathbf{A}](x) :=& \partial_\mu \mathbf{A}_\nu(x) - \partial_\nu \mathbf{A}_\mu(x) + g \mathbf{A}_\mu(x) \times \mathbf{A}_\nu(x)
\nonumber\\
=& \mathbf{F}_{\mu\nu}[\mathbf{V}](x) + D_\mu[\mathbf{V}] \mathbf{X}_\nu(x) - D_\nu[\mathbf{V}] \mathbf{X}_\mu(x) + g \mathbf{X}_\mu(x) \times \mathbf{X}_\nu(x) , 
\end{align}
where  
$
 \mathbf{F}_{\mu\nu}[\mathbf{V}](x)
 :=  \partial_\mu \mathbf{V}_\nu(x) - \partial_\nu \mathbf{V}_\mu(x) + g \mathbf{V}_\mu(x) \times \mathbf{V}_\nu(x)  
$.

The CDG decomposition has a remarkable property: 
The field strength $\mathscr{F}[\mathscr{V}](x)$ of $\mathscr{V}(x)$ is proportional to the color field $\bm{n}(x)$: 
\begin{subequations}
\begin{align}
 \mathscr{F}_{\mu\nu}[\mathscr{V}](x)
%:=& \partial_\mu \mathscr{V}_\nu(x) - \partial_\nu \mathscr{V}_\mu(x) -ig [\mathscr{V}_\mu(x) , \mathscr{V}_\nu(x) ]
%\nonumber\\
 =&  \bm{n}(x) \{ \partial_\mu c_\nu(x) - \partial_\nu c_\mu(x) + ig^{-1} \bm{n}(x) \cdot [\partial_\mu \bm{n}(x) , \partial_\nu \bm{n}(x)  ] \},
%[\partial_\mu c_\nu - \partial_\nu c_\mu - g^{-1} \mathbf{n} \cdot (\partial_\mu \mathbf{n} \times \partial_\nu \mathbf{n})]  
 \label{C26-F-decomp1}
\end{align}
or
\begin{align}
 \mathbf{F}_{\mu\nu}[\mathbf{V}](x)
% :=& \partial_\mu \mathbf{V}_\nu(x) - \partial_\nu \mathbf{V}_\mu(x) + g \mathbf{V}_\mu(x) \times \mathbf{V}_\nu(x) 
% \nonumber\\
 =  \mathbf{n}(x) \{ \partial_\mu c_\nu(x) - \partial_\nu c_\mu(x) - g^{-1} \mathbf{n}(x) \cdot (\partial_\mu \mathbf{n}(x) \times \partial_\nu \mathbf{n}(x)) \} .
 \label{C26-F-decomp2}
\end{align}
\end{subequations}
%This is shown by straightforward calculations. 
Thus, a piece $\mathbf{C}_\mu(x) := c_\mu(x)\mathbf{n}(x)$ of the gauge field $\mathbf{V}(x)$ parallel to $\mathbf{n}(x)$ is not sufficient to give a field strength $\mathbf{F}[\mathbf{V}](x)$ parallel to $\mathbf{n}(x)$, when the color field $\mathbf{n}(x)$ is not uniform. 
In order to obtain the field strength $\mathbf{F}_{\mu\nu}[\mathbf{V}](x)$ parallel to $\mathbf{n}(x)$, therefore, one needs the second piece perpendicular (orthogonal) to $\mathbf{n}(x)$:
\begin{equation}
\mathbf{B}_\mu(x) := g^{-1} \partial_\mu \mathbf{n}(x) \times \mathbf{n}(x) ,
\end{equation}
in addition to the first piece parallel to $\mathbf{n}(x)$: 
\begin{equation}
\mathbf{C}_\mu(x) := c_\mu(x)\mathbf{n}(x) , 
\quad 
c_\mu(x) :=   \mathbf{A}_\mu(x)  \cdot \mathbf{n}(x) ,
\end{equation}
so that the restricted gauge field is decomposed as
\begin{align}
   \mathbf{V}_\mu(x) :=&   \mathbf{B}_{\mu}(x) + \mathbf{C}_{\mu}(x) .
\end{align}
Such a restricted field $\mathbf{V}_\mu(x)$ (especially $\mathbf{B}_\mu(x)$) is called the \textbf{Cho connection}. 

%%%%%%%%%%%%%%%%%%%%%%%%%%%%%%%%%%%%%%%%%%%%%%%%%%%%%%%%%%%%
\begin{figure}[tbp]
\begin{center}
\includegraphics[scale=0.3]{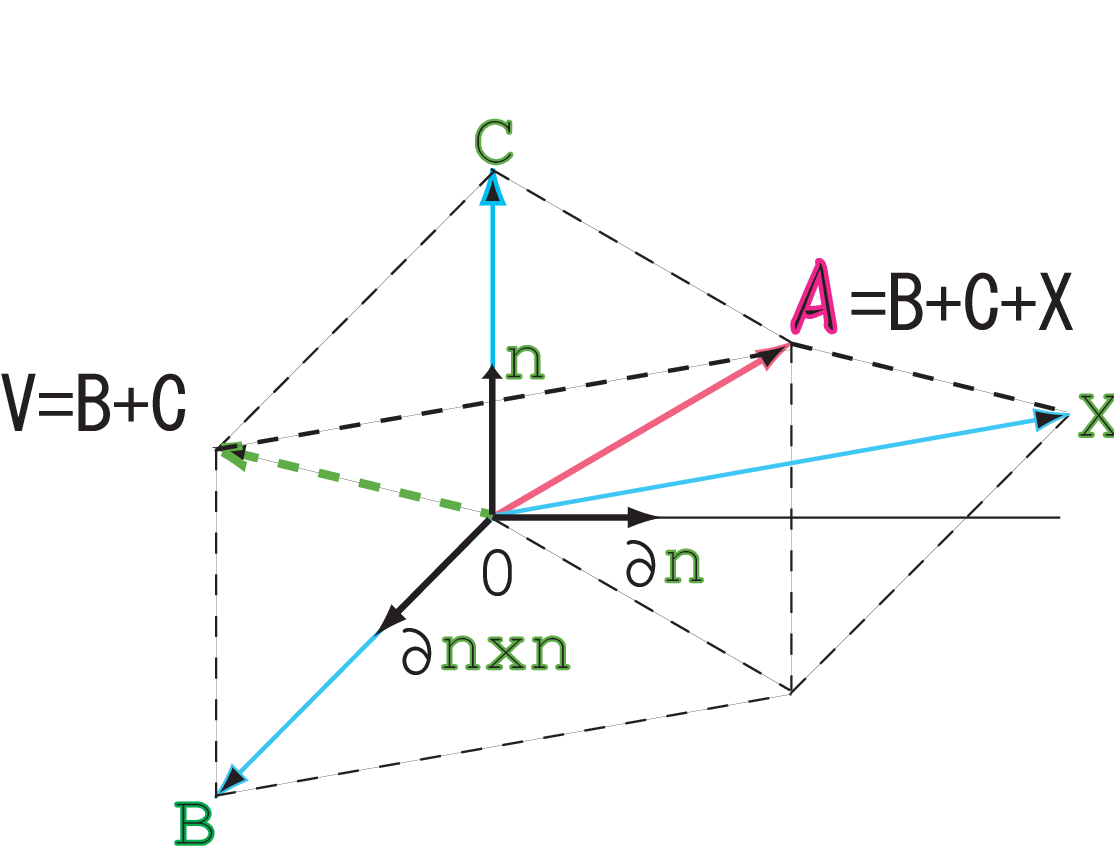}
\caption{\small 
The CDG decomposition of the $SU(2)$ gluon field $\mathscr{A}$ in color space.
}
\label{fig:CFNdecomp}
\end{center}
\end{figure}
%%%%%%%%%%%%%%%%%%%%%%%%%%%%%%%%%%%%%%%%%%%%%%%%%%%%%%%%%%%%

 Thus the original  CDG decomposition is written in the vector form with constraints:
\begin{subequations}
\begin{align}
 \mathbf{A}_\mu(x) 
%=  (\mathscr{A}_\mu^A(x)) 
%\nonumber\\
%=& \underbrace{A_\mu(x) \bm{n}(x)}_{\mathbf{A}_\mu} 
% +  \underbrace{\left(  g^{-1} + \sigma(x) \right) \partial_\mu \bm{n}(x) \times \bm{n}(x)  + \rho(x) \partial_\mu \bm{n}(x)}_{\mathbb{W}_\mu} 
%\nonumber\\
%=& 
:=& \overbrace{\underbrace{c_\mu(x) \mathbf{n}(x)}_{\mathbf{C}_\mu(x):\text{restricted~potential}}
 +    \underbrace{ g^{-1}  \partial_\mu \mathbf{n}(x)  \times \mathbf{n}(x) }_{\mathbf{B}_\mu(x):\text{magnetic~potential}}}^{\mathbf{V}_\mu(x)}  
 + \underbrace{ \mathbf{X}_\mu(x)}_{\text{covariant~potential}}
 ,
\\
 &  \mathbf{n}(x) \cdot \mathbf{n}(x) = 1   , 
\\
 & \mathbf{n}(x) \cdot \mathbf{X}_\mu(x) = 0 .
\end{align}
\end{subequations}

For the decomposition of the restricted gauge field   $\mathbf{V}_{\mu}$, 
$\mathbf{V}_{\mu}  =  \mathbf{B}_{\mu} + \mathbf{C}_{\mu} $, 
%\begin{align}
% \mathbf{A}_{\mu} :=& \mathbf{V}_{\mu} + \mathbf{X}_{\mu} 
%\nonumber\\
%& \mathbf{V}_{\mu} =  \mathbf{C}_{\mu} + \mathbf{B}_{\mu} ,
%\quad
%  \mathbf{C}_{\mu} := c_\mu \mathbf{n}, \quad   \mathbf{B}_{\mu} := g^{-1}  \partial_\mu \mathbf{n} \times  \mathbf{n} ,
%\end{align}
the field strength  $\mathbf{F}_{\mu\nu}[\mathbf{V}]$ is  further decomposed into the ``restricted electric field'' $\mathbf{E}_{\mu\nu}$ and the ``restricted magnetic field'' $\mathbf{H}_{\mu\nu}$:
%[Exercise-4] \marginpar{Ex-4}
%Verify that the field strength $\mathscr{F}[\mathscr{V}]$ of $\mathscr{V}$ is proportional to the color field $\bm{n}$ as given in (\ref{C26-F-decomp1}) or (\ref{C26-F-decomp2}).
\begin{align}
% \mathbf{F}_{\mu\nu}[\mathbf{A}] :=& \partial_\mu \mathbf{A}_\nu - \partial_\nu \mathbf{A}_\mu + g \mathbf{A}_\mu \times \mathbf{A}_\nu
%\nonumber\\
%=& \mathbf{F}_{\mu\nu}[\mathbf{V}] + D_\mu[\mathbf{V}] \mathbf{X}_\nu - D_\nu[\mathbf{V}] \mathbf{X}_\mu + g \mathbf{X}_\mu \times \mathbf{X}_\nu , 
%\\
   \mathbf{F}_{\mu\nu}[\mathbf{V}](x) =&    \mathbf{E}_{\mu\nu}(x) + \mathbf{H}_{\mu\nu}(x) 
= [E_{\mu\nu}(x)  + H_{\mu\nu}(x) ] \mathbf{n}(x),
\\
    \mathbf{E}_{\mu\nu}(x)   :=&   E_{\mu\nu}(x) \mathbf{n}(x), \quad
E_{\mu\nu}(x) := \partial_\mu c_\nu(x) - \partial_\nu c_\mu(x) ,
\nonumber\\
   \mathbf{H}_{\mu\nu}(x)  :=& \mathbf{F}_{\mu\nu}[\mathbf{B}](x)
 :=   \partial_\mu \mathbf{B}_\nu(x) - \partial_\nu \mathbf{B}_\mu(x) + g \mathbf{B}_\mu(x) \times \mathbf{B}_\nu(x) .
%\nonumber\\
%=&  - g \mathbf{B}_\mu(x) \times \mathbf{B}_\nu(x) 
%\nonumber\\
%=&  - g^{-1}   (\partial_\mu \mathbf{n}(x) \times %\partial_\nu \mathbf{n}(x))  
%\nonumber\\
%=:   H_{\mu\nu}(x) \mathbf{n}(x), 
%\quad  H_{\mu\nu} := - g^{-1} \mathbf{n} \cdot (\partial_\mu \mathbf{n} \times \partial_\nu \mathbf{n})  .
%=  \partial_\mu h_\nu - \partial_\nu h_\mu ,
\end{align}
Then we find that the restricted magnetic field $\mathbf{H}_{\mu\nu}(x)$ is proportional to $\mathbf{n}(x)$: 
\begin{align}
   \mathbf{H}_{\mu\nu}(x)  
%:=&  \mathbf{F}_{\mu\nu}[\mathbf{B}](x)
=&   - g \mathbf{B}_\mu(x) \times \mathbf{B}_\nu(x) 
\nonumber\\
=&  - g^{-1}   (\partial_\mu \mathbf{n}(x) \times \partial_\nu \mathbf{n}(x))  
%\nonumber\\
   =    H_{\mu\nu}(x) \mathbf{n}(x), 
\end{align}
where we have defined
\begin{align}
H_{\mu\nu}(x) := \mathbf{n}(x) \cdot \mathbf{H}_{\mu\nu}(x)  = - g^{-1} \mathbf{n}(x) \cdot (\partial_\mu \mathbf{n}(x) \times \partial_\nu \mathbf{n}(x))  
 .
 \label{C26-H-def}
\end{align}
%Here
%$\mathbf{E}_{\mu\nu}$, $\mathbf{H}_{\mu\nu}$, $g \mathbf{X}_\mu \times \mathbf{X}_\nu$, and  
%$\mathbf{E}_{\mu\nu} + \mathbf{H}_{\mu\nu} + g \mathbf{X}_\mu \times \mathbf{X}_\nu$ 
%are all parallel to 
%$\bm{n}$, while $D_\mu \mathbf{X}_\nu - D_\nu \mathbf{X}_\mu$
% is orthogonal to $\bm{n}$ (following from   $\bm{n} \cdot \mathbf{X}_\mu=0$). 
This is shown e.g., using 
\begin{align}
 \mathbf{F}_{\mu\nu}[\mathbf{V}](x) =&  \mathbf{F}_{\mu\nu}[\mathbf{B}](x) + D_\mu[\mathbf{B}] \mathbf{C}_\nu(x) - D_\nu[\mathbf{B}] \mathbf{C}_\mu(x) + g \mathbf{C}_\mu(x) \times \mathbf{C}_\nu(x) , 
\end{align}
and
\begin{align}
  \mathbf{C}_\mu(x) \times \mathbf{C}_\nu(x)  =&   c_\mu(x)\mathbf{n}(x) \times c_\nu(x)\mathbf{n}(x) \equiv   \bm{0}  ,
\\ 
  D_\mu[\mathbf{B}] \mathbf{C}_\nu(x) :=& \partial_\mu \mathbf{C}_\nu(x) + g \mathbf{B}_\mu(x) \times \mathbf{C}_\nu(x) 
\nonumber\\
=& \partial_\mu [c_\nu(x)\mathbf{n}(x)] +   [\partial_\mu \mathbf{n}(x) \times \mathbf{n}(x)] \times c_\nu(x)\mathbf{n}(x)  
\nonumber\\
=&  \partial_\mu c_\nu(x)\mathbf{n}(x)  + c_\nu(x) \partial_\mu \mathbf{n}(x)    
- c_\nu(x) \partial_\mu \mathbf{n}(x)  
%+  c_\nu(x) [\mathbf{n}(x) \cdot \partial_\mu \mathbf{n}(x) ] \mathbf{n}(x) 
\nonumber\\
=&  \partial_\mu c_\nu(x)\mathbf{n}(x) .
%+  c_\nu(x) [\mathbf{n}(x) \cdot \partial_\mu \mathbf{n}(x) ] \mathbf{n}(x) 
\end{align}
In the above calculations, we have used the property:
\begin{equation}
\mathbf{n}(x) \times \mathbf{n}(x) = \bm{0}, \quad
\mathbf{n}(x) \cdot   \mathbf{n}(x) = 1 \Longrightarrow  
\mathbf{n}(x) \cdot \partial_\mu \mathbf{n}(x) = \bm{0}, \
\end{equation}
and
\begin{align}
 \mathbf{A} \times ( \mathbf{B} \times \mathbf{C} ) = ( \mathbf{A} \cdot\mathbf{C}) \mathbf{B} - ( \mathbf{A} \cdot \mathbf{B}) \mathbf{C}  ,
 \nonumber\\
 (\mathbf{A} \times  \mathbf{B})  \times \mathbf{C}   = ( \mathbf{A} \cdot\mathbf{C}) \mathbf{B} - ( \mathbf{B} \cdot \mathbf{C}) \mathbf{A}  .
\end{align}

We find that 
$\mathbf{E}_{\mu\nu}(x)$, $\mathbf{H}_{\mu\nu}(x)$, 
$\mathbf{F}_{\mu\nu}[\mathbf{V}](x)$, and $g \mathbf{X}_\mu(x) \times \mathbf{X}_\nu(x)$ are all parallel to 
$\mathbf{n}(x)$ (This is also the case for  
$\mathbf{F}_{\mu\nu}[\mathbf{V}](x) + g \mathbf{X}_\mu(x) \times \mathbf{X}_\nu(x)$), while $D_\mu[\mathbf{V}] \mathbf{X}_\nu(x) - D_\nu[\mathbf{V}] \mathbf{X}_\mu(x)$
 is perpendicular to $\mathbf{n}(x)$, which follows from  the defining equations: 
% $\mathbf{n}(x) \cdot \mathbf{X}_\mu(x)=0$): 
% $\mathbf{n}(x) \cdot \mathbf{X}_\mu(x)=0$): 
%[Exercise-5] \marginpar{Ex-5}
%Verify that $\mathbf{F}_{\mu\nu}[\mathbf{V}](x)$, $g \mathbf{X}_\mu(x) \times \mathbf{X}_\nu(x)$, and  
%$\mathbf{F}_{\mu\nu}[\mathbf{V}](x) + g \mathbf{X}_\mu(x) \times \mathbf{X}_\nu(x)$ 
%are all parallel to 
%$\mathbf{n}(x)$, while $D_\mu[\mathbf{V}] \mathbf{X}_\nu(x) - D_\nu[\mathbf{V}] \mathbf{X}_\mu(x)$
% is perpendicular to $\mathbf{n}(x)$.
\begin{align}
& \mathbf{F}_{\mu\nu}[\mathbf{V}](x) \parallel \mathbf{n}(x), \quad
g \mathbf{X}_\mu(x) \times \mathbf{X}_\nu(x) \parallel \mathbf{n}(x), 
 %mathbb{F}_{\mu\nu}[\mathbf{V}](x) + g \mathbf{X}_\mu(x) \times \mathbf{X}_\nu(x) \parallel \bm{n}(x)
 \nonumber\\&
 D_\mu[\mathbf{V}] \mathbf{X}_\nu(x) ,  D_\nu[\mathbf{V}] \mathbf{X}_\mu(x) \perp \mathbf{n}(x) .
\end{align}
Then we obtain a \textit{gauge-invariant field strength} $f_{\mu\nu}$ as the magnitude of the field strength of the restricted field:%
\footnote{
We will see later that this $f_{\mu\nu}$ coincides exactly with  the two-form $f_{\mu\nu}$ appearing in the non-Abelian Stokes theorem for  the Wilson loop operator.
See section 6. 
}
\begin{subequations}
\begin{align}
 f_{\mu\nu}(x) :=& \bm{n}(x) \cdot \mathscr{F}_{\mu\nu}[\mathscr{V}](x) := 2 {\rm tr} \left\{ \bm{n}(x)   \mathscr{F}_{\mu\nu}[\mathscr{V}](x) \right\} 
 \nonumber\\
 =&  \partial_\mu c_\nu(x) - \partial_\nu c_\mu(x) + i g^{-1} \bm{n}(x) \cdot [\partial_\mu \bm{n}(x) , \partial_\nu \bm{n}(x) ] ,
\end{align}
or
\begin{align}
 f_{\mu\nu}(x) :=& \mathbf{n}(x) \cdot \mathbf{F}_{\mu\nu}[\mathbf{V}](x)
 \nonumber\\
 =&  \partial_\mu c_\nu(x) - \partial_\nu c_\mu(x) - g^{-1} \mathbf{n}(x) \cdot (\partial_\mu \mathbf{n}(x) \times \partial_\nu \mathbf{n}(x)) ,
\end{align}
\label{C26-field-inv}
\end{subequations}
since
$\mathscr{F}_{\mu\nu}[\mathscr{V}]$ and $\bm{n}$ transform according to the adjoint representation under the gauge transformation:
\begin{equation}
 \mathscr{F}_{\mu\nu}[\mathscr{V}](x) \rightarrow U(x) \mathscr{F}_{\mu\nu}[\mathscr{V}](x) U^{-1}(x) , 
 \quad
 \bm{n}(x) \rightarrow U(x) \bm{n}(x) U^{-1}(x) .
%[\partial_\mu c_\nu - \partial_\nu c_\mu - g^{-1} \mathbf{n} \cdot (\partial_\mu \mathbf{n} \times \partial_\nu \mathbf{n})]  
\end{equation}

There is another way to see that  the ``Abelian'' field strength  $f_{\mu\nu}$ (with no  group indices) in (\ref{C26-field-inv}) is $SU(2)$ gauge invariant, since it is cast into the manifestly $SU(2)$ invariant form, which is written in the Lie algebra notation and the vector notations, respectively:
%[Exercise-6] \marginpar{Ex-6}
%Show that the  ``Abelian'' field strength  $f_{\mu\nu}$ (\ref{C26-field-inv}) is cast into the manifestly $SU(2)$ invariant form (\ref{C26-f-inv-form}).
\begin{subequations}
\begin{align}
f_{\mu\nu}(x)
=&  \partial_\mu c_\nu(x) - \partial_\nu c_\mu(x) + i g^{-1} \bm{n}(x) \cdot [\partial_\mu \bm{n}(x)  , \partial_\nu \bm{n}(x) ] 
\nonumber\\
 =&  2 {\rm tr} \left\{  \bm{n}(x) \mathscr{F}_{\mu\nu}[\mathscr{A}](x)  
  + ig^{-1} \bm{n}(x)  [\mathscr{D}_\mu[\mathscr{A}] \bm{n}(x), \mathscr{D}_\nu[\mathscr{A}] \bm{n}(x) ] 
 \right\}  ,
\\
%\text{or} 
f_{\mu\nu}(x) =& \partial_\mu c_\nu(x) - \partial_\nu c_\mu(x) - g^{-1} \mathbf{n}(x) \cdot (\partial_\mu \mathbf{n}(x) \times \partial_\nu \mathbf{n}(x))
\nonumber\\
=&   \mathbf{n}(x) \cdot \mathbf{F}_{\mu\nu}[\mathbf{A}](x)  
  - g^{-1} \mathbf{n}(x) \cdot (D_\mu[\mathbf{A}] \mathbf{n}(x) \times D_\nu[\mathbf{A}] \mathbf{n}(x)) ,
\end{align}
  \label{C26-f-inv-form}
\end{subequations}
where the gauge transformations of the color field, the field strength and the covariant derivative are respectively given by
\begin{align}
  \bm{n}(x) & \rightarrow U(x) \bm{n}(x) U^{-1}(x) , 
 \nonumber\\
  \mathscr{F}_{\mu\nu}[\mathscr{A}](x) & \rightarrow U(x) \mathscr{F}_{\mu\nu}[\mathscr{A}](x) U^{-1}(x) ,
\nonumber\\
  \mathscr{D}_\mu[\mathscr{A}](x) & \rightarrow U(x) \mathscr{D}_\mu[\mathscr{A}](x) U^{-1}(x)  .
\end{align}
Note that $\mathscr{V}$ has the same transformation property as  $\mathscr{A}$, and  $\mathscr{V}$ satisfies  $\mathscr{D}_\mu[\mathscr{V}] \bm{n}(x)=0$.

%In fact, it is shown that the field strength   $\mathscr{F}_{\mu\nu}[\mathscr{V}]$ is found to be proportional to $\mathbf{n}$: [Exercise-13] \marginpar{Ex-13}
%\begin{equation}
% \mathscr{F}_{\mu\nu}[\mathscr{V}]
% := \partial_\mu \mathbf{V}_\nu - \partial_\nu \mathbf{V}_\mu + g \mathbf{V}_\mu \times \mathbf{V}_\nu 
% = \mathbf{n} [\partial_\mu c_\nu - \partial_\nu c_\mu - g^{-1} \mathbf{n} \cdot (\partial_\mu \mathbf{n} \times \partial_\nu \mathbf{n})]  
%\end{equation}
%Then we have a  {gauge-invariant field strength}:
%\begin{equation}
% f_{\mu\nu} := \mathbf{n} \cdot \mathscr{F}_{\mu\nu}[\mathscr{V}]
%  =  \partial_\mu c_\nu - \partial_\nu c_\mu - g^{-1} \mathbf{n} \cdot (\partial_\mu \mathbf{n} \times \partial_\nu \mathbf{n})
%\end{equation}

It should be remarked that $H_{\mu\nu}(x)$ is locally closed $(dH=0)$ and hence it can be exact $(H=dh)$ locally due to the Poincare lemma. Then it has the Abelian \textbf{magnetic potential} $h_\mu(x)$: 
\begin{align}
%\mathbf{H}_{\mu\nu} = H_{\mu\nu} \bm{n}, \quad
H_{\mu\nu}(x) = - g^{-1} \mathbf{n}(x) \cdot (\partial_\mu \mathbf{n}(x) \times \partial_\nu \mathbf{n}(x))  
=  \partial_\mu h_\nu(x) - \partial_\nu h_\mu(x) 
%\leftrightarrow E_{\mu\nu} = \partial_\mu C_\nu - \partial_\nu C_\mu
 .
 \label{C26-magnetic-potential}
\end{align}

%%%%%%%%%%%%%%%%%%%%%%%%%%%%%%%%%%%%%%%%%%%%%%%%%%%%%%%%%%
%%%%%%%%%%%%%%%%%%%%%%%%%%%%%%%%%%%%%%%%%%%%%%%%%%%%%%%%%%

%It should be remarked that $f_{\mu\nu}$ has the same form as the  tensor characterizing the 't Hooft-Polyakov magnetic monopole, if  the color unit field $\mathbf{n}^A(x)$ is identified with the normalized \textit{adjoint} scalar field $\hat{\phi}^A(x)$ in the Georgi--Glashow model:
%\begin{equation}
%\mathbf{n}^A(x) \leftrightarrow \hat{\phi}^A(x) := \phi^A(x)/||\phi(x)|| .
%\end{equation}

%%%%%%%%%%%%%%%%%%%%%%%%%%%%%%%%%%%%%%%%%%%%%%%%%%%%%%%%%%%%
\subsection{Magnetic monopole from the decomposed  $SU(2)$  Yang-Mills field}\label{subsec:magnetic-monopole}
%%%%%%%%%%%%%%%%%%%%%%%%%%%%%%%%%%%%%%%%%%%%%%%%%%%%%%%%%%%%

Taking into account the fact that   the  ``Abelian'' field strength  $f_{\mu\nu}$ (\ref{C26-field-inv}) is gauge invariant, therefore, we can introduce (a candidate of) a gauge-invariant \textbf{magnetic monopole current} by% 
\footnote{
This construction of magnetic monopole has a profound meaning in connection with the Wilson loop, as shown in  section 6. 
}
\begin{equation}
 k^\mu(x) = \partial_\nu  {}^{\displaystyle *}f^{\mu\nu}(x) , \quad  f^{\mu\nu}(x) = E^{\mu\nu}(x)  + H^{\mu\nu}(x) ,
%\equiv  \frac12 \epsilon_{\mu\nu\rho\sigma}\partial_{\nu}               f^{\rho\sigma}(x) ,
\end{equation}
where $ {}^{\displaystyle *}$ denotes the Hodge dual, e.g., for $D=4$, the dual tensor $ {}^{\displaystyle *}f^{\mu\nu}$ of $f^{\mu\nu}$ is defined by
\begin{equation}
  {}^{\displaystyle *}f^{\mu\nu}(x)  :=  \frac12 \epsilon^{\mu\nu\rho\sigma} f_{\rho\sigma}(x) .
\end{equation}

In fact, we show that the non-vanishing magnetic charge is obtained without introducing Dirac singularities in $c_\mu(x)$, and that even in the classical level, the magnetic charge obeys the quantization condition of Dirac type. 
The magnetic charge $g_m$ is defined by the integral of the charge density $k^0$ over the three-dimensional volume $V_3$:
\begin{align}
 g_m :=& \int_{V_3} d^3x k^0 
= \int_{V_3} d^3x \partial_\ell \left( \frac12 \epsilon^{\ell jk} f_{jk} \right)  .
\end{align}

When $c_\mu(x)$ has no singularities, the electric field $E_{\mu\nu}(x) := \partial_\mu c_\nu(x) - \partial_\nu c_\mu(x) 
$ does not contribute to the magnetic monopole current $k^\mu(x)$, i.e., using the exterior derivative $d$, the coderivative $\delta$, and the Hodge dual operation $ {}^{\displaystyle *}$, 
$$
k=\delta {}^{\displaystyle *}f= {}^{\displaystyle *}df= {}^{\displaystyle *}d E+ {}^{\displaystyle *}d H , \quad dE= ddc \equiv 0 ,
$$ 
which is the Bianchi identity. 
The non-trivial contribution comes from the magnetic field $H^{\mu\nu}(x)=- g^{-1} \mathbf{n}(x) \cdot (\partial_\mu \mathbf{n}(x) \times \partial_\nu \mathbf{n}(x))$. 
Using the Stokes theorem (Gauss theorem), the magnetic charge is cast into the surface integral:  
\begin{align}
 g_m 
%:=& \int d^3x k_0 
%= \int d^3x \partial_i \left( \frac12 \epsilon^{ijk} f_{jk} \right) 
=  \oint_{S^2_{\rm phy}=\partial V_3} d\sigma^{jk} 
  g^{-1} \mathbf{n} \cdot (\partial_j \mathbf{n}  \times \partial_k \mathbf{n}) , \quad
%d\sigma^{jk} := \frac12 d\sigma_\ell \epsilon^{\ell jk}  .
\end{align}
where $d\sigma^{jk} := \frac12 d\sigma_\ell \epsilon^{\ell jk}$ is the surface element. 
The (unit) color field $\mathbf{n}(x)$ with the target space $SU(2)/U(1) \cong S^2_{\rm int}$  can be parameterized using two angle variables $\alpha, \beta$ as 
\begin{align}
   \mathbf{n}(x) 
 =& \begin{pmatrix} n^1(x) \cr n^2(x) \cr n^3(x) \end{pmatrix}
= \begin{pmatrix} \sin \beta(x) \cos \alpha(x) \cr
 \sin \beta(x) \sin \alpha(x) \cr
\cos \beta(x)  
 \end{pmatrix} \in SU(2)/U(1) \cong S_{\rm int}^2 
 .
\end{align}
Then we have the representation:
\begin{align}
% \Longrightarrow &
  \mathbf{n} \cdot (\partial_\mu \mathbf{n}   \times \partial_\nu \mathbf{n}) 
  = \sin \beta (\partial_\mu \beta \partial_\nu \alpha
  - \partial_\mu \alpha \partial_\nu \beta) 
  = \sin \beta \frac{\partial(\beta, \alpha)}{\partial(x^\mu,x^\nu)} ,
\end{align}
where  
$\frac{\partial(\beta, \alpha)}{\partial(x^\mu,x^\nu)}$ 
is the Jacobian associated with the map:
$(x^\mu,x^\nu) \in S^2_{\rm phy}$ $\rightarrow$ $(\beta, \alpha) \in S^2_{\rm int} \simeq SU(2)/U(1)$.
Therefore, the magnetic charge $g_m$ is calculated as 
\begin{align}
 g_m 
%:=& \oint_{S^2_{phy}} d\sigma_{jk} 
%  g^{-1} \bm{n} \cdot (\partial_j \bm{n}   \times \partial_k \bm{n}) 
% \nonumber\\
=& g^{-1} \oint_{S^2_{\rm phy}} d\sigma_{jk} 
 \sin \beta \frac{\partial(\beta, \alpha)}{ \partial(x^j,x^k)}
% \nonumber\\
=    g^{-1}  \oint_{S^2_{int}} \sin \beta d\beta  d\alpha 
%=   4\pi g^{-1}  n  \quad (n=0, \pm1, \cdots) ,
 .
\end{align}
Taking into account that $\sin \beta d\beta  d\alpha$ is a surface element on $S^2$ and that a unit sphere has  area $4\pi$, thus, we find that 
 the magnetic charge obeys the \textbf{Dirac quantization condition}:
%\footnote{
%P.A.M. Dirac, 
%Quantized Singularities in the Electromagnetic Field,
%Proc. Roy. Soc. London, A{\bf 133}, 60--72 (1931)
%}
\begin{align}
 g_m =   \frac{4\pi}{g} n  \quad (n \in \mathbb{Z} = \{ 0, \pm1, \pm 2, \cdots \} ) .
\end{align}
The magnetic charge $g_m$ represents the number as to how many times  $S^2_{\rm int}$ is
wrapped by a mapping from  $S^2_{\rm phys} $ to 
$S^2_{\rm int}$. 
The non-trivial magnetic charge corresponds to the non-trivial \textbf{Homotopy group} of the map $\bm{n}: S^2 \rightarrow SU(2)/U(1)  \simeq S^2$:
\begin{equation}
 \pi_2(SU(2)/U(1)) = \pi_2(S^2) = \mathbb{Z} .
\end{equation}

%%%%%%%%%%%%%%%%%%%%%%%%%%%%%%%%%%%%%%%%%%%%%%%%%%%%%%%%%%%%

\subsection{On-shell decomposition for topological configurations}\label{subsec:on-shell-decomp}

%%%%%%%%%%%%%%%%%%%%%%%%%%%%%%%%%%%%%%%%%%%%%%%%%%%%%%%%%%%%

It should be remarked that the original CDG decomposition is an off-shell decomposition, i.e., decomposition for the gauge field off mass shell. 
Faddeev and Niemi have proposed the on-shell decomposition for the gauge field called the \textbf{Faddeev-Niemi   on-shell decomposition} satisfying the equation of motion \cite{FN99}:
%\footnote{  
%L.D. Faddeev  and Antti J. Niemi,
%Partially dual variables in SU(2) Yang-Mills theory, 
%Phys. Rev. Lett. {\bf 82}, 1624-1627  (1999). 
%}
\begin{equation}
%\text{On-shell decomposition:}  \  
  \mathbf{A}_\mu(x) 
%= \mathbf{V}_\mu(x) +  \mathbf{X}_\mu(x) 
=  c_\mu(x)\mathbf{n} + \partial_\mu \mathbf{n}(x) \times  \mathbf{n}(x) 
+ \rho(x)\mathbf{n}(x)   + \sigma(x)   \partial_\mu \mathbf{n}(x)  \times \mathbf{n}(x)  ,
\label{C26-FN-decomp}
\end{equation}
where $\rho(x)$ and $\sigma(x) $ are scalar functions. 
This gives the decomposition for some of known topological configurations.

%%%%%%%%%%%%%%%%%%%%%%%%%%%%%%%%%%%%%%%%%%%%%%%%%%%%%%%%%%%%

%\normalsize
%\setcounter{equation}{0}
%\noindent
%$\bullet$ Examples of CDGFN decomposition for topological configurations: 
%\begin{align}
% \text{On-shell}  \Longrightarrow  
%  \mathbf{X}_\mu(x) :=   \rho(x)  \partial_\mu \bm{n}(x)  + \sigma(x)   \partial_\mu \bm{n}(x)  \times \bm{n}(x)   .
%\end{align}

%\noindent
\underline{Example 1}.  
%{Wu-Yang magnetic monopole} configuration of hedgehog form
The \textbf{Wu-Yang magnetic monopole} configuration of hedgehog form \cite{WY75}:
%\footnote{  
%Tai Tsun Wu, Chen Ning Yang, 
%Phys. Rev. D{\bf 12}, 3845--3857 (1975)
%Tai Tsun Wu, Chen Ning Yang, 
%Nucl. Phys. B{\bf 107}, 365--380  (1976).
%Tai Tsun Wu, Chen Ning Yang,
%Phys.Rev. D{\bf 14}, 437--445  (1976).
%}
\begin{align}
  \mathscr{A}_j^A(x) = \epsilon_{Ajk} \frac{x_k}{r^2}, \quad
  \mathscr{A}_0^A(x) = 0, 
\end{align}
is reproduced from the color field and the resulting decomposition:
\begin{align}
  n^A(x) = \frac{x^A}{r} , \quad
c_\mu(x) = 0, \quad
\mathbf{X}_\mu(x) = 0 \quad (\rho(x) = 0, \quad  \sigma(x)=0) ,
\end{align}
which gives a spherical symmetric configuration for the magnetic field $H_{k}$ ($k=1,2,3$): 
\begin{align}
 H_{jk} = -g^{-1} \epsilon^{ABC} n^A \partial_j n^B \partial_k n^C =  - g^{-1} \frac{1}{r^2} \epsilon^{jk\ell} \frac{x_\ell}{r}  
\Longrightarrow 
 H_{k} = - g^{-1} \frac{1}{r^2} \frac{x_k}{r} . 
\end{align}

%\noindent
\underline{Example 2}. 
The \textbf{BPST one-instanton} solution: \cite{BPST75}
%\footnote{  
%A.A. Belavin, A.M. Polyakov, A.S. Schwartz, and Yu.S. Tyupkin,  
%Pseudoparticle Solutions of the Yang-Mills Equations,  
%Phys.Lett. B {\bf 59},  85--87  (1975).
%}
%\begin{picture}(0,0)%(0,-3000)
%\put(0,1500){\tframe[1000][200](3000,1500)}%
%\put(4000,-1300){%
%  \put(0,0){\includegraphics[height=15ex]{BPST.eps}}%
%}%
%\end{picture}%
\begin{align}
  \mathscr{A}_j^A(x) =   \epsilon_{Ajk} \frac{2x_k}{r^2+\lambda^2}, \quad
  \mathscr{A}_0^A(x) = 0, \quad
\end{align}
is reproduced by the decomposition
\begin{align}
 n^A(x) = \frac{x^A}{r} , \quad
c_\mu(x) = 0, \quad
\rho(x) = 0  , \quad
\sigma(x) = \frac{2r^2}{r^2+\lambda^2} -1 
%, \quad
%  \mathscr{A}_0^A(x) = 0 
 ,
%\nonumber\\
% \sigma \rightarrow -1 (r \rightarrow 0),  \quad +1 (r \rightarrow \infty)  . 
\end{align}
which exhibits the asymptotic behavior
$\sigma \rightarrow -1 (r \rightarrow 0),  \
+1 (r \rightarrow \infty)$.

%\noindent
\underline{Example 3}. 
The \textbf{Witten  multi-instanton} solution    obtained from the Ansatz for axially (cylindrically) symmetric self-dual solution to the $SU(2)$ Yang-Mills equation 
with arbitrary instanton numbers \cite{Witten77}:%
%\footnote{  
%E. Witten, 
%Phys. Rev. Lett. {\bf 38}, 121--124 (1977).
%}
\begin{align}
   \mathscr{A}_0^A(x) =& \frac{x^A}{r} a_0(r,t), 
\nonumber\\
  \mathscr{A}_j^A(x) =& \frac{\varphi_2(r,t)+1}{r^2} \epsilon_{jAk} x_k + {\varphi_1(r,t) \over r^2} (\delta_{jA} r^2 - x_j x_A) + a_1(r,t) \frac{x_j x_A}{r^2} ,
\end{align}
is reproduced by the decomposition:
\begin{align}
 n^A(x) = \frac{x^A}{r} , \quad
 c_0(x) =& a_0(r,t) ,  \quad
 c_j(x) = \frac{x_j}{r}a_1(r,t) , 
\nonumber\\
  \rho(x) =& \varphi_1(r,t) , \quad
 \sigma(x) = \varphi_2(r,t) ,
\end{align}
where $a_0(r,t),a_1(r,t),\varphi_1(r,t),\varphi_2(r,t)$ are functions of $r:=|\bm{x}|$ and $t$.
%\begin{picture}(0,0)%(0,-3000)
%\put(0,1500){\tframe[1000][200](3000,1500)}%
%\put(1500,-1500){%
%  \put(0,0){\includegraphics[height=20ex]{Witten_a.eps}}%
%  \put(3000,0){\includegraphics[height=20ex]{Witten_b.eps}}%
%}%
%\end{picture}%

%%%%%%%%%%%%%%%%%%%%%%%%%%%%%%%%%%%%%%%%%%%%%%%%%%%%%%%%%%%

%\subsection{Possible phases of QCD under the FN decomposition}
%\textcolor{red}{\Large\bf $\S$~~Possible phases of QCD under the CFN decomposition}
%\setcounter{equation}{0}

%%%%%%%%%%%%%%%%%%%%%%%%%%%%%%%%%%%%%%%%%%%%%%%%%%%%%%%%%%%%

By making use of the Faddeev-Niemi on-shell decomposition, the possible phases of QCD, i.e., confinement phase, Higgs phase and non-Abelian Coulomb phase, were discussed in \cite{KT01}.

%\noindent
%$\bullet$ Possible phases of QCD under the decomposition
%\\
\begin{enumerate}
%\noindent
\item[(i)]
 Confinement phase, $\langle \phi \rangle =0 \ (\phi=\rho+i\sigma)$:
\\  
$
\mathbf{X}_\mu(x) :=   \rho(x)  \partial_\mu \mathbf{n}(x)  + \sigma(x)   \partial_\mu \mathbf{n}(x)  \times \mathbf{n}(x)   .
$
%(dual Higgs phase); 
\\
monopole-like configuration: $\sigma(x) \rightarrow 0 (|x| \rightarrow \infty) \Longrightarrow |\phi(x)| \rightarrow 0$ or $\mathbf{X}_\mu(x) \rightarrow 0$
\\
If the $\mathbf{X}_\mu$ (or $\phi$) field fluctuates more strongly than the $\mathbf{n}$ field, 
the Faddeev-Skyrme or Faddeev-Niemi model is expected to be derived:
\begin{align}
  S_{\rm eff} = \int d^4x \left\{ \Lambda^2 (\partial_\mu \mathbf{n})^2 +  (\partial_\mu \mathbf{n} \times \partial_\nu \mathbf{n})^2  \right\} ,
\end{align}
with 
a) the unique action containing $\mathbf{n}$ field and allowing for Hamiltonian interpretation, 
b) with topological soliton (knot soliton).

%\noindent
\item[(ii)]
Higgs phase, $\langle \phi \rangle \not=0$: 
\\
instanton-like configuration: $\sigma(x) \rightarrow 1  \Longrightarrow |\phi(x)| \rightarrow 1$ or $\mathbf{X}_\mu(x) \rightarrow \text{nonzero}$.
\\
If the $\mathbf{n}$ field fluctuates more strongly than the other  fields, the $\mathbf{n}$ is integrated out,  and the effective theory reads
\begin{align}
  S_{\rm eff} = \int d^4x \left\{ F_{\mu\nu}^2 + ((\partial_\mu - iA_\mu) \phi)^* ((\partial_\mu - iA_\mu) \phi) + (|\phi|^2-1)^2   \right\} .
\end{align}
%where $D_\mu := \partial_\mu - iA_\mu$.
%\noindent
\item[(iii)]
non-Abelian Coulomb phase:

If none of the fields fluctuate strongly, then one would have the original Yang-Mills action at low energies. (expected to be realized for large flavors $N_f \gg 1$)
\end{enumerate}

%%%%%%%%%%%%%%%%%%%%%%%%%%%%%%%%%%%%%%%%%%%%%%%%%%%%%%%%%%%%

%\section{Interplay between monopole and instanton}

%%%%%%%%%%%%%%%%%%%%%%%%%%%%%%%%%%%%%%%%%%%%%%%%%%%%%%%%%%%%

%\noindent
%$\bullet$ 
Moreover, the interplay between monopole and instanton was discussed using the Faddeev-Niemi   on-shell decomposition \cite{TTF0}. 
%\footnote{
%T. Tsurumaru, I. Tsutsui, A. Fujii,
%Instantons, monopoles and the flux quantization in the Faddeev-Niemi decomposition 
%hep-th/0005064,
%Nucl.Phys. B{\bf 589},  659--668 (2000). 
%}
In the $SU(2)$ case, the relationship between the instanton number $\nu$ and the flux $\Phi$ was obtained:
\begin{align}
  \nu :=  \frac{1}{16\pi^2} \int {\rm tr}(\mathscr{F}  {}^{\displaystyle *}\mathscr{F} ) = Q_m \frac{\Phi}{2\pi} ,
\end{align}
where $\Phi$ is the flux associated with the $U(1)$ gauge field $A_\mu$ trapped by the monopole loop $C$:%
% added by sinohara
%\begin{picture}(0,0)%(0,-3000)
%\put(0,1500){\tframe[1000][200](3000,1500)}%
%\put(500,-1800){%
%  \put(0,0){\includegraphics[height=15ex]{flux.eps}}%
%}%
%\end{picture}%
\begin{align}
 \Phi := \int_{C} A = \int_{S: C=\partial S} F .
\end{align}
This result suggest a possible relationship among the topological charges for the following topological objects:
%$SU(2)/U(1) \cong S^2$
\begin{itemize}
\item
Instantons classified by the Pontryagin number,   suggested from the non-trivial Homotopy group $\pi_3(S^3)=\mathbb{Z}$

\item
Magnetic monopoles classified by the magnetic charge, suggested from the non-trivial Homotopy group  $\pi_2(S^2)=\mathbb{Z}$

\item
Knot solitons  classified by the Hopf number, suggested from the non-trivial Homotopy group  $\pi_3(S^2)=\mathbb{Z}$

\end{itemize}

However, it should be remarked that there are some debate whether or not the above Faddeev-Niemi on-shell decomposition (\ref{C26-FN-decomp}) is  equivalent to the Yang-Mills theory \cite{EG11,NW11}. 
%\footnote{
%J. Evslin and S. Giacomelli, 
%A Faddeev-Niemi Solution that Does Not Satisfy Gauss' Law,  
%arXiv:1010.1702 [hep-th],
%JHEP 1104, 022  (2011) .
%A.J. Niemi and A. Wereszczynski,
%On Solutions to the 'Faddeev-Niemi' Equations 
%arXiv:1011.6667 [hep-th],
%J. Math. Phys. {\bf 52},  072302  (2011).
%} 

In what follows, we  do not discuss the on-shell decomposition.

\newpage
%%%%%%%%%%%%%%%%%%%%%%%%%%%%%%%%%%%%%%%%%%%%%%%%%%%%%%%%%%%%
%%%%%%%%%%%%%%%%%%%%%%%%%%%%%%%%%%%%%%%%%%%%%%%%%%%%%%%%%%%%
\section{Reformulation of $SU(2)$ Yang-Mills theory}\label{sec:reform-SU2} 
%%%%%%%%%%%%%%%%%%%%%%%%%%%%%%%%%%%%%%%%%%%%%%%%%%%%%%%%%%%%
%%%%%%%%%%%%%%%%%%%%%%%%%%%%%%%%%%%%%%%%%%%%%%%%%%%%%%%%%%%%
 
%%%%%%%%%%%%%%%%%%%%%%%%%%%%%%%%%%%%%%%%%%%%%%%%%%%%%%%%%%%%
\subsection{From the decomposition to the change of variables}
\label{subsection:}
%\setcounter{equation}{0}
%%%%%%%%%%%%%%%%%%%%%%%%%%%%%%%%%%%%%%%%%%%%%%%%%%%%%%%%%%%%

In the preceding sections, the decomposition of the original Yang-Mills field $\mathscr{A}_\mu^A(x)$ was performed:
\begin{equation} 
 \mathscr{A}_\mu^A(x)  \rightarrow \bm{n}^A(x), c_\mu(x), \mathscr{X}_\mu^A(x) .
\end{equation}

 We wish to regard the above relationship obtained through the CDG decomposition between the original variables and the new variables as a (non-linear) \textbf{change of variables} (NLCV) rather than the decomposition, so that the original $SU(2)$ Yang-Mills theory (YM) written in terms of  the original Yang-Mills field $\mathscr{A}_\mu^A(x)$ is cast into an equivalent or equipollent theory (YM') written in terms of new variables $\bm{n}^A(x), c_\mu(x), \mathscr{X}_\mu^A(x)$, which we call the \textbf{CDG--Yang-Mills theory} or a \textbf{reformulated Yang-Mills theory}:%
\footnote{
Two sets A and B are said to be equipollent iff there is a one-to-one correspondence (i.e., a bijection) from A onto B.
}
\\
\\
\noindent
 $SU(2)$ Yang-Mills theory $\quad\quad\quad\quad$
 $\quad\quad\quad$ A reformulated Yang-Mills theory 
\\
 written in terms of 
 $\quad\quad\quad\quad$
 $\underbrace{\Longleftrightarrow}_{\text{NLCV}}$ 
 $\quad\quad\quad$
 written in terms of new variables:
\\
$\mathscr{A}_\mu^A(x)$  $(A=1,2,3)$
$\quad\quad\quad\quad$ 
%NLCV
$\quad\quad$$\quad\quad$$\quad$
$\bm{n}^A(x), c_\mu(x), \mathscr{X}_\mu^A(x)$  $(A=1,2,3)$
\\

For this purpose, therefore, the new field variables $\bm{n}^A(x), c_\mu(x), \mathscr{X}_\mu^A(x)$  should be given as functionals of the original gauge field variables $\mathscr{A}_\mu^A(x)$.

However, we immediately encounter an issue of the mismatch for  the independent degrees of freedom (d.o.f.) between two sets of variables, which does not allow the change of variables. 
In fact, we can count the independent degrees of freedom of the field variables in  $D$-dimensional $SU(2)$ Yang-Mills theory:
\begin{align}
& {\rm d.o.f}[\mathscr A_\mu^A]= 3D, 
\nonumber\\ 
& {\rm d.o.f}[n^A]= 3-1=2, \ 
 {\rm d.o.f}[\mathscr{X}_\mu^A]= 3D-D=2D, \ 
 {\rm d.o.f}[c_\mu]=  D ,
\end{align}
since $n^A$ obeys a condition $n^An^A-1=0$ and $\mathscr{X}_\mu^A$ obeys the $D$ conditions $n^A \mathscr{X}_\mu^A=0$. 
Therefore, the new field variables $\bm{n}^A(x), c_\mu(x), \mathscr{X}_\mu^A(x)$ have two extra degrees of freedom in total. 
This suggests one to impose two constraints 
\begin{align}
 \chi=0 ,
\end{align}
among the new field variables for consistency. 
%Comparing the field variables before and after the decomposition, we find 
%\\
%\\
%%$\bullet$ Counting the degrees of freedom for  $D$-dimensional $SU(2)$ Yang-Mills field:
%%($A=1,2,3$)
%%\tiny
%\small
%\begin{tabular}{c||c|c|c|c||c}
%   Before &  \multicolumn{4}{|l||}{$\mathscr A_\mu^A$:  3D }
%  & total  3D \\ \hline\hline
%   After &   $n^A $:     $3-1=2$
%&   $\mathscr{X}_\mu^A $: $3D-D=2D$
%&   $c_\mu $:  $D$
%&   $\bm{\chi}=0$:  $-2$ 
% &  total         3D
%\end{tabular}
%\normalsize
%\\
%where $\bm{\chi}=0$ is the constraint to be imposed in the identification. 

In summary, the following issues must be fixed to make the change of variable well-defined and to identify the color field with the degree of freedom which enables one to describe topological objects such as magnetic monopoles responsible for confinement.
\begin{enumerate}
\item[(A)]
How the \textbf{color field} $\bm{n}(x)$ is determined from the original Yang-Mills field $\mathscr{A}_\mu(x)$?

This was assumed so far.  We must give a procedure to achieve this.

\item[(B)]
How the mismatch for the independent degrees of freedom between the original field variables and the new field variables  is solved?

The new variables have two extra degrees of freedom which should be eliminated by imposing appropriate constraints.

\item[(C)]
How the gauge transformation properties of the new field variables are determined from the original gauge field to achieve the expected ones?

It is expected that  $f_{\mu\nu}(x) =  2 {\rm tr} \left\{ \bm{n}(x) \mathscr{F}_{\mu\nu}[\mathscr{V}](x) \right\}$ becomes gauge invariant, provided that $\bm{n}(x)$ and $\mathscr{F}_{\mu\nu}[\mathscr{V}](x)$ transform  in the adjoint representation under the gauge transformation. 

\end{enumerate}

All of these issues are simultaneously solved as follows. 
This is due to a   new viewpoint for the reformulated    Yang-Mills theory.

%%%%%%%%%%%%%%%%%%%%%%%%%%%%%%%%%%%%%%%%%%%%%%%%%%%%%%%%%%%%
\subsection{New viewpoint for the Yang-Mills theory: reduction from the master Yang-Mills theory}\label{subsec:new-viewpoint}
%%%%%%%%%%%%%%%%%%%%%%%%%%%%%%%%%%%%%%%%%%%%%%%%%%%%%%%%%%%%

 First, we answer the question (A): how to define or obtain the color   field $\mathbf{n}(x)$ from the original Yang-Mills theory written in terms of $\mathbf{A}_\mu(x)$ alone. 
A procedure has been given in the new reformulation of Yang-Mills theory in the continuum space--time.%
\footnote{ 
This is based on %Kondo, Murakami and Shinohara (2005,2006).
\cite{KMS05,KMS06}.
}
By introducing the color field $\mathbf{n}(x)$  in addition to the original Yang-Mills field $\mathbf{A}_\mu(x)$, the Yang-Mills theory is enlarged to  an extended gauge theory written in terms of $\mathbf{A}_\mu(x)$ and $\mathbf{n}(x)$, which is invariant under the \textit{enlarged} gauge transformations consisting of two infinitesimal transformations:
\begin{itemize}
\item
the  usual gauge transformation of the Yang-Mills field $\mathbf{A}_\mu(x)$ by $\bm\omega(x)$:
\begin{equation}
\delta_\omega \mathbf{A}_\mu(x)
  =D_\mu[\mathbf{A}]{\bm\omega}(x) .
\label{C26-gtA}
\end{equation}
The Yang-Mills  Lagrangian is invariant under this transformation. 
This symmetry is the local $G=SU(2)$ gauge symmetry and denoted by $SU(2)^{\rm local}_{\omega}$.

\item
the  local rotation of the color field $\mathbf{n}(x)$ by an angle ${\bm\theta}(x)$:%
\begin{equation}
\delta_\theta \mathbf{n}(x)
  =g\mathbf{n}(x) \times {\bm\theta}(x) 
  =g\mathbf{n}(x) \times {\bm\theta}_\perp(x) ,
\label{C26-gtn}
\end{equation}
where  ${\bm\theta}(x)$ is decomposed ${\bm\theta}(x)={\bm\theta}_\perp(x)+{\bm\theta}_\parallel(x)$ so that ${\bm\theta}_\parallel(x)$ is parallel to $\mathbf{n}(x)$ and  
${\bm\theta}_\perp(x)$ is  perpendicular to $\mathbf{n}(x)$, i.e., $\mathbf{n}(x)\cdot{\bm\theta}_\perp(x)=0$, with two independent components.

%The gauge transformation of $\mathbf{n}$ is nothing but the map from $S^2$ to $S^2$  at each space--time point, since 
The $SU(2)$ color field $\mathbf{n}$ is defined to be a three dimensional  vector field with a unit length: $\mathbf{n}(x) \cdot \mathbf{n}(x) =1$, which takes the value in $S^2 \simeq SU(2)/U(1)$. 
%The fact $\bm{n}(x)^2=1$ urges us to consider 
For the parallel component ${\bm\theta}_\parallel(x)=\theta_\parallel(x)\mathbf{n}(x)$, the vector field $\mathbf{n}(x)$ is invariant under this transformation (\ref{C26-gtn})  [a rotation around the axis of $\mathbf{n}(x)$].% 
\footnote{
It is a {\it redundant} symmetry, say U(1)$^{\theta}$ symmetry, of the CDG--Yang-Mills theory, since $c_\mu(x)$ and $\mathbf{X}_\mu(x)$ are also unchanged for a given $\mathbf{A}_\mu(x)$. 
%Note that $S^2 \simeq SU(2)/U(1)$. 
}
Therefore, this symmetry (\ref{C26-gtn}) is the local $SU(2)/U(1)$ symmetry and denoted by $[SU(2)/U(1)]^{\rm local}_{\theta}$.
\footnote{
Shabanov (1999) %\cite{Shabanov99b} 
argued that it is possible to consider more general transformation of the field $\bm{n}(x)$, even the nonlocal one, keeping the condition $\bm{n}(x)^2=1$.  However, it will be unrealistic to consider the explicit transformation other than the local rotation treated here. 
}

\end{itemize}

%%%%%%%%%%%%%%%%%%%%% figures %%%%%%%%%%%%%%%%%%%%%%%%%%%
\begin{figure}[tbp]
\begin{center}
\includegraphics[scale=0.5]{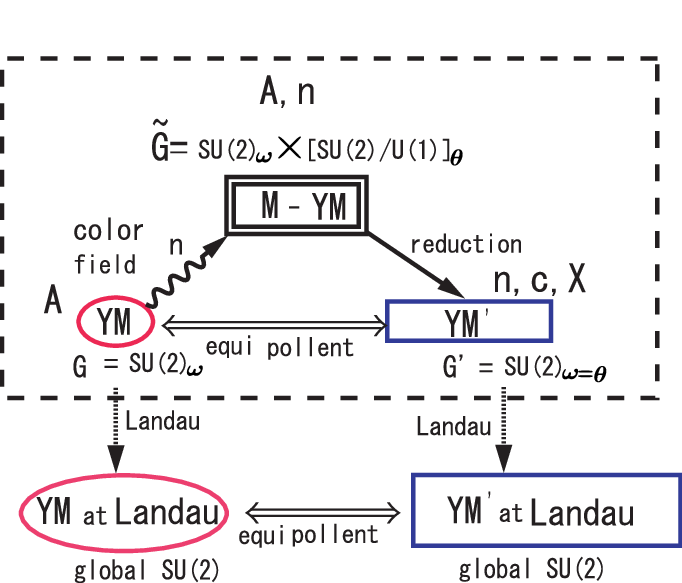}
\caption{\small  \cite{KMS06}
The relationship between the original Yang-Mills theory (YM) and the CDG--Yang-Mills theory (YM'). The master Yang-Mills theory (M-YM) obtained by the CDG decomposition has the larger  local and global gauge symmetries than the original Yang-Mills theory and becomes equipollent (equivalent) to the original Yang-Mills theory after the reduction condition is imposed.
}
\label{C26-fig:sym-cfn-ym} 
\end{center}
\end{figure}
%%%%%%%%%%%%%%%%%%%%% figures %%%%%%%%%%%%%%%%%%%%%%%%%%%

We call this extended gauge theory the \textbf{master Yang-Mills theory}.
See Fig.~\ref{C26-fig:sym-cfn-ym}. 
Note that 
${\bm\omega}(x)$ and ${\bm\theta}(x)$ are independent, since 
the original Yang-Mills Lagrangian is invariant irrespective of the choice of ${\bm\theta}(x)$.
For later convenience, we distinguish the  two transformations by $\delta_\theta$ and $\delta_\omega$, respectively:
\begin{subequations}
%\begin{itemize}
%\item 
%${\bm\theta}={\bm0}$, ${\bm\omega}\ne{\bm0}$:
%\begin{equation}
\begin{align}
 {\bm\theta}={\bm0} ,  {\bm\omega}\ne{\bm0}:
 \Longrightarrow & 
\delta_\omega \mathbf{n}(x)={\bm0},
\quad
\delta_\omega\mathbf{A}_\mu(x)
  =D_\mu[\mathbf{A}]{\bm\omega}(x) ,
%\end{equation}
\\
%\item 
%${\bm\theta}\ne{\bm0}$ ($\bm{n}\cdot{\bm\theta}=0$),
%${\bm\omega}={\bm0}$:
%\begin{equation}
{\bm\theta}\ne{\bm0},% (\bm{n}\cdot{\bm\theta}=0), 
{\bm\omega}={\bm0}:
 \Longrightarrow & 
\delta_\theta \mathbf{n}(x)
 =g \mathbf{n}(x)\times{\bm\theta}(x),
\quad
\delta_\theta\mathbf{A}_\mu(x)={\bm0}.
\end{align}
\end{subequations}
%\end{equation}
%\end{itemize}
The enlarged local gauge transformation in  the master Yang-Mills theory is obtained by combining 
 $\delta_\theta$ and $\delta_\omega$. 
Thus, the master Yang-Mills theory has the \textit{enlarged} local gauge symmetry: %$\tilde{G}:=SU(2)_{\omega} \times [SU(2)/U(1)]_{\theta}$, 
\begin{equation}
 \tilde{G}_{\omega,\theta}^{\rm local} := SU(2)^{\rm local}_{\omega} \times [SU(2)/U(1)]^{\rm local}_{\theta}  ,
\end{equation}
i.e., the direct product of $SU(2)^{\rm local}_{\omega}$ and $[SU(2)/U(1)]^{\rm local}_{\theta} $.

In order for the master Yang-Mills theory to reduce to a  gauge theory equipollent to  the original Yang-Mills theory, then, appropriate constraints must be imposed to break the enlarged gauge symmetry 
$\tilde{G}_{\omega,\theta} =SU(2)_{\omega} \times [SU(2)/U(1)]_{\theta}$ in the master Yang-Mills theory  down to a  subgroup: $G'=SU(2)_{\omega'}$ which corresponds to the original gauge symmetry.

Such a constraint  is not unique. 
%\footnote{
%} 
Our choice for the constraint is given by minimizing the functional: 
\begin{equation}
\int d^D x \frac12  \mathbf{X}_\mu^2 ,
\end{equation}
with respect to the  enlarged  gauge transformations in the master Yang-Mills theory:
\begin{align}
 \min_{\omega, \theta} \int d^D x \frac12  g^2  \mathbf{X}_\mu^2 
= \min_{\omega, \theta} \int d^D x  \frac12  (D_\mu[\mathbf{A}]\mathbf{n})^2
 ,
 \label{C26-MAGcond}
\end{align}
which is called the \textbf{reduction condition}.%
\footnote{
This condition was called the new Maximally Abelian gauge (nMAG) in %Kondo, Murakami and Shinohara (2006). 
\cite{KMS06}. 
However, this naming is misleading, since it turns out later that the reduction does not necessarily lead to the Abelian magnetic monopole in the case of $SU(N)$ ($N>2$) in general, see %Kondo, Shinohara and Murakami (2008).
 \cite{KSM08}. 
Therefore,  this terminology should not be used anymore. 
}
In the remainder of this section, we do not discriminate the upper and lower indexes in the summation convention. 

In order to show that this condition is indeed an appropriate choice, we calculate the infinitesimal variation:
\begin{align}
 0 = \delta_{\omega, \theta} \int d^D x \frac12 g^2 \mathbf{X}_\mu^2 
= \delta_{\omega, \theta} \int d^D x   \frac12 (D_\mu[\mathbf{A}]\mathbf{n})^2 ,
 \label{C26-MAGcond2}
\end{align}
since  the definition of $\mathbf{X}_\mu$ or (\ref{C26-X-derivation}) leads to 
\begin{align}
 g^2 \mathbf{X}_\mu \cdot \mathbf{X}_\mu  
= g^2 \mathbf{X}_\mu^2 
=  (\mathbf{n}\times D_\mu[\mathbf{A}]\mathbf{n})^2
=      %\left\{
     (D_\mu[\mathbf{A}]\mathbf{n})^2
     -(\mathbf{n}\cdot D_\mu[\mathbf{A}]\mathbf{n})^2
    %\right\}
=    (D_\mu[\mathbf{A}]\mathbf{n})^2 ,
\label{C26-X2}
\end{align}
where we have used
$
\mathbf{n}\cdot D_\mu[\mathbf{A}]\mathbf{n} = \mathbf{n}\cdot (g\mathbf{X}_\mu \times \mathbf{n}) = g\mathbf{X}_\mu \cdot ( \mathbf{n} \times \mathbf{n}) = 0 
$.
%The local gauge transformation II  does not change $\mathbf{X}^2$, since
Applying the infinitesimal enlarged gauge transformation to  the integrand leads to 
\begin{align}
\delta_{\omega, \theta}\frac12  g^2 \mathbf{X}_\mu^2
  = g D^\mu[\mathbf{A}](\bm\omega_\perp - \bm\theta_\perp)  \cdot \{  \mathbf{n} \times 
     (D_\mu[\mathbf{A}]\mathbf{n})  \}
 .
\label{C26-eq:dX^2}
\end{align}
%where we have used (\ref{C26-gtn}) and (\ref{C26-gtA}).
This is shown as follows.
The local enlarged gauge transformation of $\mathbf{X}^2$ is calculated as
\begin{align}
\delta_{\omega, \theta} \frac12 g^2  \mathbf{X}_\mu^2
%  &=\frac1{2g^2}\delta(\mathbf{n}\times D_\mu[\mathbf{A}]\mathbf{n})^2
%    \nonumber \\
%  &=\frac1{2g^2}\delta
%    \left\{
%     (D_\mu[\mathbf{A}]\mathbf{n})^2
%     -(\mathbf{n}\cdot D_\mu[\mathbf{A}]\mathbf{n})^2
%    \right\}
%    \nonumber \\
%  &=\frac1{2g^2}\delta(D_\mu[\mathbf{A}]\mathbf{n})^2
%    \nonumber \\
   =& 
     (D^\mu[\mathbf{A}]\mathbf{n}) \cdot \delta_{\omega, \theta} (D_\mu[\mathbf{A}]\mathbf{n})
    \nonumber \\
  =&  
     (D^\mu[\mathbf{A}]\mathbf{n}) \cdot 
     (D_\mu[\mathbf{A}]\delta_\theta \mathbf{n}
      +g\delta_\omega \mathbf{A}_\mu\times \mathbf{n})
    \nonumber \\
  =&  
     (D^\mu[\mathbf{A}]\mathbf{n}) \cdot 
     \{gD_\mu[\mathbf{A}](\mathbf{n} \times\bm\theta_\perp)
       +(gD_\mu[\mathbf{A}]\bm\omega)\times \mathbf{n} \}
    \nonumber \\
   =& 
     (D^\mu[\mathbf{A}]\mathbf{n}) \cdot 
     \{(gD_\mu[\mathbf{A}] \mathbf{n}) \times\bm\theta_\perp
      +\mathbf{n}\times (gD_\mu[\mathbf{A}]\bm\theta_\perp)
%\nonumber \\  &  
     +(gD_\mu[\mathbf{A}]\bm\omega)\times \mathbf{n} \}
    \nonumber \\
   =& g 
     (D^\mu[\mathbf{A}]\mathbf{n}) \cdot 
     \{D_\mu[\mathbf{A}](\bm\omega-\bm\theta_\perp) \times \mathbf{n} \}
    \nonumber \\
   =&  g 
     (D^\mu[\mathbf{A}]\mathbf{n}) \cdot 
     \{D_\mu[\mathbf{A}](\bm\omega_\perp - \bm\theta_\perp) \times \mathbf{n} \} ,
\label{C26-eq:dX^2b}
\end{align}
where we have used (\ref{C26-gtn}) and (\ref{C26-gtA}) in the third equality, the Leibniz rule for the covariant derivative in the fourth equality, 
and in the last equality   
we have decomposed $\bm\omega-\bm\theta_\perp$ into the parallel component $\bm\omega_\parallel= \omega_\parallel \mathbf{n}$ and perpendicular one $\bm\omega_\perp-\bm\theta_\perp$ and used the fact that the parallel part does not contribute, since 
$
%\begin{align}
 D_\mu[\mathbf{A}] \bm\omega_\parallel  \times \mathbf{n}
=   D_\mu[\mathbf{A}]( \omega_\parallel \mathbf{n})  \times \mathbf{n}
%\nonumber\\
=   \{ \mathbf{n} \partial_\mu  \omega_\parallel  +  \omega_\parallel  D_\mu[\mathbf{A}]\mathbf{n} \}  \times \mathbf{n} 
%\nonumber\\
=   \omega_\parallel   (D_\mu[\mathbf{A}]\mathbf{n} )  \times \mathbf{n}  
%\end{align}
$. 
%Therefore, the local gauge transformation II  does not change $\mathbf{X}^2$.
%[Exercise-7] \marginpar{Ex-7}

Then the minimization condition (the average over the space--time of  (\ref{C26-eq:dX^2b})) reads  
\begin{align}
0 = \delta_{\omega, \theta}\int d^D x\frac12 g^2 \mathbf{X}_\mu^2
  =- \int d^D x
     (\bm\omega_\perp-\bm\theta_\perp)\cdot
     D^\mu[\mathbf{V}]\mathbf{X}_\mu .
 \label{C26-minX2}
\end{align}
In fact,  this is derived as
\begin{align}
\delta_{\omega, \theta} \int d^Dx\frac12\mathbf{X}_\mu^2
  &=g^{-1} \int d^Dx
     (D^\mu[\mathbf{A}]\mathbf{n}) \cdot 
     \{ D_\mu[\mathbf{A}](\bm\omega_\perp-\bm\theta_\perp) \times \mathbf{n}\}
    \nonumber \\
  &= \int d^Dx
     \mathbf{X}_\mu\cdot
     D^\mu[\mathbf{A}](\bm\omega_\perp-\bm\theta_\perp)
    \nonumber \\
  &=- \int d^Dx
     (\bm\omega_\perp-\bm\theta_\perp)\cdot
     D^\mu[\mathbf{A}]\mathbf{X}_\mu
    \nonumber \\
  &=- \int d^Dx
     (\bm\omega_\perp-\bm\theta_\perp)\cdot
     D^\mu[\mathbf{V}]\mathbf{X}_\mu ,
 \label{C26-minX2b}
\end{align}
where we have used the definition of $\mathbf{X}_\mu$ or (\ref{C26-X-derivation}) in the second equality and integration by parts in the third equality. 
%[Exercise-8] \marginpar{Ex-8}
%where we have used 
%$
%\mathbf{X}_\mu(x) :=   g^{-1} \mathbf{n}(x) \times D_\mu[\mathbf{A}] \mathbf{n}(x)
%$
%and  integration by parts. 

For $\bm\omega_\perp = \bm\theta_\perp$, the minimizing condition imposes no constraint, while 
for $\bm\omega_\perp \not= \bm\theta_\perp$, 
the minimizing condition yields a condition in  the differential form:
\begin{equation}
   \bm{\chi} :=D^\mu[\mathbf{V}]\mathbf{X}_\mu  \equiv0 .
 \label{C26-dMAG}
\end{equation}
This is called the \textbf{differential reduction condition}. 
It should be noted that (\ref{C26-dMAG}) represents two conditions, since 
$\mathbf{n}(x) \cdot \bm{\chi}(x) =0$ which follows from 
$ \mathbf{n}(x) \cdot \mathbf{X}_\mu(x) = 0$. 
Thus, the issue (B) is resolved.% 
\footnote{
Note that (\ref{C26-MAGcond}) is more general than (\ref{C26-dMAG}), since (\ref{C26-dMAG}) is the differential form which is valid only in the absence of \textbf{Gribov copies}.  
The condition (\ref{C26-MAGcond}) is the most general gauge-fixing condition which can be used also in numerical simulations on a lattice and works even if the Gribov copies exist and leads to the true minimum, while (\ref{C26-dMAG}) leads only to the local minimum along the gauge orbit. 
}

Thus, if we impose the reduction condition (\ref{C26-MAGcond}) to the master Yang-Mills theory,   
 then the larger gauge symmetry  $ \tilde{G}:=SU(2)_{\omega} \times [SU(2)/U(1)]_{\theta}$ enlarged from the original gauge symmetry $G=SU(2)_{\omega}$ is broken down to the   subgroup $G'=SU(2)_{\omega^\prime}$: in the respective step in Fig.~\ref{C26-fig:sym-cfn-ym},
\begin{equation}
 G=SU(2)^{\rm local}_{\omega} \nearrow \tilde{G}:=SU(2)^{\rm local}_{\omega} \times [SU(2)/U(1)]^{\rm local}_{\theta} 
 \searrow G':=SU(2)^{\rm local}_{\omega^\prime} .
\end{equation}
Finally, we have an equipollent Yang-Mills theory with  \textbf{the residual local gauge symmetry} 
$G':=SU(2)^{\rm local}_{\omega^\prime}$ with the gauge transformation parameter: 
\begin{equation}
\bm\omega^\prime(x)
=(\bm\omega_\parallel(x),\bm\omega_\perp(x) )=(\bm\omega_\parallel(x), \bm\theta_\perp(x)) , 
\quad  \bm\omega_\perp(x)=\bm\theta_\perp(x) ,
\end{equation}
which is called a ``diagonal'' $SU(2)$ part $\tilde{G}_{\omega=\theta}^{\rm local}$ of   
$\tilde{G}_{\omega,\theta}^{\rm local}$.
This is the CDG--Yang-Mills theory.  
%The local gauge symmetry $G'=SU(2)^{\rm local}_{\omega=\theta}$ is the same as the gauge symmetry II. 
\footnote{
This theory  was called the Yang-Mills theory II  in \cite{KMS06}, which has the gauge symmetry II defined later.  
}
%\\
%which is a diagonal SU(2) part of   
%$\tilde{G}:=SU(2)_{\rm local}^{\omega} \times [SU(2)/U(1)]_{\rm local}^{\theta}$.  
%\\
%The local gauge symmetry $G'$  of the Yang-Mills theory II has  the gauge symmetry II. 

It is important to remark that the reduction condition  in the differential form (\ref{C26-dMAG}) has another expression written in terms of $\mathbf{A}$ and $\mathbf{n}$:
\begin{equation}
  \mathbf{n} \times (D_\mu[\mathbf{A}]D_\mu[\mathbf{A}]\mathbf{n})   = 0 
  ,
  \label{C26-rDE}
\end{equation}
since the above reduction condition is rewritten as
%[Exercise-9] \marginpar{Ex-9}
\begin{equation}
 gD_\mu[\mathbf{V}]\mathbf{X}_\mu 
 = gD_\mu[\mathbf{A}]\mathbf{X}_\mu
 =  D_\mu[\mathbf{A}]\{ \mathbf{n} \times (D_\mu[\mathbf{A}]\mathbf{n}) \} 
 =    \mathbf{n} \times (D_\mu[\mathbf{A}]D_\mu[\mathbf{A}]\mathbf{n})   
  \label{C26-rDE2}
  .
\end{equation}
Thus, \textit{the color field $\mathbf{n}(x)$ is determined by solving the differential reduction  equation (\ref{C26-rDE}) for a given Yang-Mills field $\mathbf{A}_\mu(x)$.}
This determines the color field $\bm{n}(x)$ as a functional of a given configuration of $\mathbf{A}_\mu(x)$. 
Thus, the issue (A) is resolved.% 
\footnote{
This is possible in principle. 
Some examples for the explicit solution for the reduction condition for the gauge field $\mathbf{A}_\mu(x)$ representing the topological configurations such as merons and instantons are given in the following papers. 
\cite{KFSS08,FKSS10,FKSS12,FKSS12b}
%K.-I. Kondo, N. Fukui, A. Shibata, and T. Shinohara,
%Magnetic Monopole Loops supported by a Meron pair as the Quark Confiner, 
%CHIBA-EP-172, KEK-2008-14 
%arXiv:0806.3913 [hep-th],
%Phys.Rev.D{\bf 78}, 065033 (2008). 
%N. Fukui, K.-I. Kondo, A. Shibata, and T. Shinohara, 
%CHIBA-EP-183-KEK-PREPRINT-2010-8 
%Jackiw-Nohl-Rebbi two-instanton as a source of magnetic monopole loop, 
%arXiv:1005.3157 [hep-th],
%Phys. Rev. D{\bf 82},  045015 (2010). 
%N. Fukui, K.-I. Kondo, A. Shibata, and T. Shinohara, 
%Magnetic monopole loops generated from two-instanton solutions: Jackiw-Nohl-Rebbi versus 't Hooft instanton, 
%CHIBA-EP-193-KEK-PREPRINT-2012-9 
%arXiv:1205.4972 [hep-th], 
%Phys.Rev. D{\bf 86},  065020 (2012). 
%However, there could exist the subtlety of the existence of zero modes in solving the reduction condition. 
}

On the other hand, the original Yang-Mills theory with the local gauge symmetry $G=SU(2)^{\rm local}_{\omega}$  is reproduced from the master Yang-Mills theory if  the color field variable $\mathbf{n}(x)$ is fixed to e.g., $\mathbf{n}(x) \equiv \mathbf{n}_{0}:=(0,0,1)$ on the whole space--time points. 

In the CDG--Yang-Mills theory, we can impose any further gauge-fixing condition (e.g., Landau gauge) for fixing the residual unbroken symmetry $G=SU(2)^{\rm local}_{\omega}$ even after the reduction condition is imposed.  
(In fact, we can furthermore impose the conventional MA gauge, if desired.)  
This is an advantage of the new viewpoint for the CDG--Yang-Mills theory. 
See  Fig.~\ref{C26-fig:sym-cfn-ym}. 
%This leads to a possibility of examining the gauge invariance even after the nMAG.   In the previous approach the MAG is one of the gauge fixing and there is no specific reason to take the MAG (except for the coincidence of the degrees of freedom).  But in our approach the MAG plays a different and distinguished role.  Even after imposing the nMAG, the CDG--Yang-Mills theory has the full $SU(2)$ symmetry.   

%The comparison between MA gauge and reduction condition is as follows. 
The reduction condition should be compared with the conventional MA gauge. 
The MA gauge breaking the local $SU(2)^{\rm local}$ to $U(1)^{\rm local}$ breaks also the global   $SU(2)^{\rm global}$ as well, while the MA gauge leaves the local $U(1)^{\rm local} (\subset  G=SU(2)^{\rm local})$ and global $U(1)^{\rm global}$ symmetries unbroken.
The reduction condition  leaves both the local $G'=SU(2)^{\rm local}$ symmetry and global $SU(2)^{\rm global}$ symmetry (color symmetry) unbroken.
In the CDG--Yang-Mills theory, therefore, the overall gauge fixing condition can be imposed without breaking color symmetry, e.g. using the Landau gauge. 
This is an advantage of the new reformulation.

%This determines the color field $\bm{n}(x)$ as a functional of a given configuration of $\mathbf{A}_\mu(x)$. 
%
%The local gauge transformation II  does not change $\mathbf{X}^2$, since
%\begin{align}
%\delta_{\omega, \theta}\frac12\mathbf{X}_\mu^2
%  = D_\mu[\mathbf{A}](\bm\omega_\perp - \bm\theta_\perp)  \cdot \{ g^{-1} \bm n \times  (D_\mu[\mathbf{A}]\bm{n})  \} .
%\label{C26-eq:dX^2}
%\end{align}
%where we have used (\ref{C26-gtn}) and (\ref{C26-gtA}).
%where we have used (\ref{C26-def:X})  and integration by parts. 
%\\
%which is a diagonal $SU(2)$ part of   
%$\tilde{G}:=SU(2)_{\rm local}^{\omega} \times [SU(2)/U(1)]_{\rm local}^{\theta}$.  
%\\
%The local gauge symmetry $G'$  of the Yang-Mills theory II has  the gauge symmetry II. 

The form of the reduction condition (\ref{C26-dMAG}) agrees exactly with the MA gauge fixing condition for the CDG variables.
However, the reduction condition  (\ref{C26-dMAG}) or  (\ref{C26-MAGcond}) is totally different from the conventional MA gauge which was used so far to fix the off-diagonal part $SU(2)/U(1)$ of the local gauge symmetry $SU(2)$ in the original Yang-Mills theory (based on the Cartan decomposition) keeping $U(1)$ part intact, 
while the reduction condition introduced above plays a role of eliminating the proliferated gauge symmetry generated by using the CDG field variables to  leave  the  desired $SU(2)$ local gauge symmetry.  
%Therefore, we call (\ref{C26-MAGcond}) (and (\ref{C26-dMAG})) the {\it new MAG} (the differential form).
%The reformulated Yang-Mills theory obtained by imposing the reduction condition on the master Yang-Mills theory is called the  CDG--Yang-Mills theory  hereafter. 
Among three gauge degrees $\bm{\omega}=(\bm{\omega}_\perp,\bm{\omega}_\parallel)$ and two degrees $\bm{\theta}_\perp$ in the master Yang-Mills theory, two extra gauge degrees of freedom  was eliminated by imposing the two conditions represented by the reduction condition $\bm{\chi}=0$ and then the remaining degrees in the CDG--Yang-Mills theory  are the same as the original Yang-Mills theory.

The  viewpoint given above for the CDG--Yang-Mills theory resolves in a natural way a crucial issue of the discrepancy in the independent degrees of freedom between both theories, i.e., the original Yang-Mills theory and the CDG--Yang-Mills theory.  
Moreover, it tells one the necessity of adopting the reduction condition in the master Yang-Mills theory, although the conventional MA gauge is merely one choice of the gauge fixings. 
%To the best of our knowledge, this point was not correctly understood so far.
% in the published literatures. 
\footnote{
A possible algorithm for the numerical simulation was proposed in  \cite{Shabanov02}. 
Moreover, the actual simulations were first attempted in \cite{DHW02}.  
However, from our point of view, the resulting theory can not be identified with the CDG--Yang-Mills theory.   
Here, we mention only a point that only the  field $\bm{n}$ was constructed in these works and the simulation results show the breaking of the global $SU(2)$ invariance even in the Landau gauge, which cannot be regarded as the correct implementation of the CDG decomposition on a lattice. It is the essence to preserve the color symmetry. 
In the work   
%done by Kato et al. (2002),  
\cite{KKMSS05}, 
  the Monte Carlo simulations of the CDG--Yang-Mills theory were performed for the first time by imposing the reduction condition.
In addition, we can impose the Lattice-Landau gauge (LLG) simultaneously.
Here, the LLG fixes the local gauge symmetry $G'=SU(2)^{\rm local}_{\omega'}$. 
In general, we can impose any gauge fixing condition (instead of LLG) after imposing the reduction condition in the numerical simulations. 
If desired, the conventional MA gauge can be further imposed instead of LLG, after the reduction condition is imposed. 
}

In other words, the CDG--Yang-Mills theory has been constructed on a vacuum selected in a gauge invariant way among possible vacua of the master Yang-Mills theory, since the reduction condition is satisfied for the CDG  field configurations realizing the minimum of the functional $\int d^Dx\frac12\mathbf{X}_\mu^2$ and the minimum 
$
 \min_{\bm\omega,\bm\theta} \int d^Dx\frac12\mathbf{X}_\mu^2 
$
 is gauge invariant in the sense that it does no longer change the value with respect to the  local gauge transformation. 
Therefore,  the reduction condition is a gauge-invariant criterion of choosing a vacuum on which the CDG--Yang-Mills theory is defined from the vacua of the master Yang-Mills theory, although the reduction condition is not necessarily a unique prescription of selecting out the gauge-invariant vacuum.  
This demonstrates a quite different role played by the reduction condition compared with the conventional MA gauge.

Moreover, in the  CDG--Yang-Mills theory defined in this way, the local operator $\mathbf{X}_\mu^2(x)$ itself is invariant under the residual $SU(2)$ local gauge transformation as the gauge transformation for the CDG--Yang-Mills theory.

%%%%%%%%%%%%%%%%%%%%%%%%%%%%%%%%%%%%%%%%%%%%%%%%%%%%%%%%%%%%
\subsection{Gauge transformation of new variables}\label{gt-new-variable}
%%%%%%%%%%%%%%%%%%%%%%%%%%%%%%%%%%%%%%%%%%%%%%%%%%%%%%%%%%%%

In the new formulation, two types of local gauge transformations are introduced:
% 
%by decomposing the original gauge transformation, 
%$
%  \delta_\omega \mathbf{A}_\mu(x)  =  D_\mu[\mathbf{A}] \bm{\omega}(x) .
%$

\underline{Local gauge transformation I}:  
\begin{subequations}
\begin{align}
  \delta_\omega \mathbf{n}(x)  =& 0  ,
\\
 \delta_\omega c_\mu(x) =& \mathbf{n}(x) \cdot  D_\mu[\mathbf{A}] \bm{\omega}(x) ,
\\
  \delta_\omega \mathbf{X}_\mu(x) =&     D_\mu[\mathbf{A}] \bm{\omega}(x) - \mathbf{n}(x) ( \mathbf{n}(x) \cdot  D_\mu[\mathbf{A}] \bm{\omega}(x)) ,
%\\ 
%  \Longrightarrow  & \delta_\omega \mathbf{B}_\mu  =   0 ,
%\quad
%  \delta_\omega \mathbf{V}_\mu 
%=   \bm{n}( \bm{n} \cdot  \mathscr{D}_\mu[\mathscr{A}] \bm{\omega})  . 
\end{align}
\end{subequations}

\underline{Local gauge transformation II}:  
\begin{subequations}
\begin{align}
  \delta_{\omega}^\prime \mathbf{n}(x)  =& g \mathbf{n}(x) \times \bm{\omega'}(x)  ,
\\
 \delta_{\omega}^\prime c_\mu(x) =&    \mathbf{n}(x) \cdot \partial_\mu \bm{\omega'}(x)   ,
\\
  \delta_{\omega}^\prime  \mathbf{X}_\mu(x) =&  g \mathbf{X}_\mu(x) \times \bm{\omega'}(x)  ,
%\\
%\Longrightarrow  & \delta_{\omega}' \mathbf{B}_\mu  
%=   D_\mu[\mathbf{B}] \bm{\omega'} - (\bm{n} \cdot \partial_\mu \bm{\omega'}) \bm{n}  ,
%\quad
%\delta_{\omega}' \mathbf{V}_\mu =   D_\mu[\mathbf{V}] \bm{\omega'}    . 
\end{align}
\end{subequations}
The gauge transformation I was called the passive or quantum gauge transformation, while II was called the active or background gauge transformation.  However, this classification is not necessarily independent, leading to sometimes confusing and misleading results.
%, since two gauge transformations I and II are not independent. 
%The local gauge transformation I defined in the previous paper \cite{Kondo04} is nothing but $\delta_\omega$.

In order to see how the gauge transformation II  is reproduced,  we apply the enlarged gauge transformation $\delta_{\omega,\theta}$ specified by
% (\ref{C26-gtA}) and (\ref{C26-gtn})
\begin{equation}
\delta\mathbf{A}_\mu(x)
  =D_\mu[\mathbf{A}]{\bm\omega}(x) , \quad
%\label{C26-gtA2}
\delta\mathbf{n}(x)
  =g\mathbf{n}(x) \times {\bm\theta}(x) 
  =g\mathbf{n}(x) \times {\bm\theta}_\perp(x) ,
%\label{C26-gtn2}
\end{equation}
 to
%   (\ref{C26-def:X}).  
\begin{align}
c_\mu(x)
 &=\mathbf{n}(x)\cdot\mathbf{A}_\mu(x), 
%\label{C26-def:c}
%\\
\quad 
\mathbf{X}_\mu(x)
  =g^{-1}\mathbf{n}(x)\times D_\mu[\mathbf{A}]\mathbf{n}(x) .
\label{C26-def:X2}
\end{align}
Then  the enlarged gauge transformation of the new variables $c_\mu(x)$ and $\mathbf{X}_\mu(x)$ are given by 
%[Exercise-10] \marginpar{Ex-10}
\footnote{
This transformation law was obtained by \cite{Shabanov99}.
In it, $\delta \bm{n}(x)$ is not specified and is left undermined on the right-hand side, based on the viewpoint that $\delta \bm{n}(x)$ should be determined by the choice of the constraint condition $\bm{\chi}(\bm{n},\mathscr{A}) \equiv \bm{\chi}(\bm{n},c,\mathbf{X})=0$ which reduces the degrees  to the original ones (by solving $\delta \bm{\chi}=0$ on the hypersurface $\bm{\chi}=0$). 
The reduction condition is consistent with the local rotation of $\bm{n}$, as a part of gauge transformation II.  
%Therefore, our result is in agreement with the claim \cite{Shabanov99b}, see (\ref{C26-dMAG}) and (\ref{C26-gtF}). 
}
\begin{align}
  \delta_{\omega,\theta} c_\mu(x) 
% =  \mathbf{n}(x) \cdot \partial_\mu \bm{\omega}(x)  + g(\mathbf{n}(x) \times \mathbf{A}_\mu(x)) \cdot (\bm{\omega}_\perp(x) - \bm{\theta}_\perp(x)) 
  =  \mathbf{n}(x) \cdot \partial_\mu \bm{\omega}(x)  + \mathbf{n}(x) \cdot g(\mathbf{A}_\mu(x) \times    (\bm{\omega}_\perp(x) - \bm{\theta}_\perp(x)))    ,
  \label{C26-gtc}
\\
  \delta_{\omega,\theta} \mathbf{X}_\mu(x) 
  = g \mathbf{X}_\mu(x) \times  (\bm{\omega}_\parallel(x)+\bm{\theta}_\perp(x)) + D_\mu[\mathbf{V}](\bm{\omega}_\perp(x)-\bm{\theta}_\perp(x))  ,
  \label{C26-gtX}
\end{align}
and
\begin{align}
  \delta_{\omega,\theta} \mathbf{V}_\mu(x) 
  = D_\mu[\mathbf{V}](\bm{\omega}_\parallel(x)+\bm{\theta}_\perp(x)) + g \mathbf{X}_\mu(x) \times  (\bm{\omega}_\parallel(x)-\bm{\theta}_\perp(x))  .
  \label{C26-gtX2}
\end{align}

We proceed to obtain the transformation of $\mathbf X_\mu$ for the $SU(2)$ case:
\begin{align}
  \delta_{\omega,\theta} \mathbf X_\mu
%   \nonumber\\
 &=\delta_{\omega,\theta} (g^{-1}\mathbf{n} \times  D_\mu[\mathbf A]\mathbf{n})
   \nonumber\\
 &=g^{-1}\delta_{\omega,\theta} \mathbf{n} \times  D_\mu[\mathbf A]\mathbf{n}
   +g^{-1}\mathbf{n} \times  D_\mu[\mathbf A]\delta_{\omega,\theta} \mathbf{n}
   +g^{-1}\mathbf{n} \times (g\delta_{\omega,\theta} \mathscr A_\mu \times \mathbf{n})
   \nonumber\\
 &=(\mathbf{n} \times \bm\theta_\perp) \times  D_\mu[\mathbf A]\mathbf{n}
   +\mathbf{n} \times  D_\mu[\mathbf A](\mathbf{n} \times \bm\theta_\perp)
   +\mathbf{n} \times (D_\mu[\mathbf A]\bm\omega \times \mathbf{n}) .
   \label{C26-dX}
\end{align}
Note that the Jacobi identities:
\begin{subequations}
\begin{align}
& (\mathbf A \times \mathbf B) \times \mathbf C
+(\mathbf B \times \mathbf C) \times \mathbf A
+(\mathbf C \times \mathbf A) \times \mathbf B
 =0 ,
\\
& \mathbf A \times (\mathbf B \times \mathbf C)
+\mathbf B \times (\mathbf C \times \mathbf A)
+\mathbf C \times (\mathbf A \times \mathbf B)
 =0 ,
\end{align}
\end{subequations}
yield
\begin{subequations}
\begin{align}
  (\mathbf A \times \mathbf B) \times \mathbf C
 =& \mathbf A \times (\mathbf B \times \mathbf C)
  -\mathbf B \times (\mathbf A \times \mathbf C) 
%\nonumber\\
 =  (\mathbf C \times \mathbf B)  \times  \mathbf A 
  + (\mathbf A \times \mathbf C)  \times  \mathbf B ,
\\
  \mathbf A \times (\mathbf B \times \mathbf C)
 =& \mathbf B \times (\mathbf A \times \mathbf C)
  -\mathbf C \times (\mathbf A \times \mathbf B) .
\end{align}
  \label{vector-3product}
\end{subequations}

By applying the Jacobi identity, therefore, the first term of (\ref{C26-dX}) reads
\begin{align}
(\mathbf{n} \times \bm\theta_\perp) \times  D_\mu[\mathbf A]\mathbf{n}
 &=  (D_\mu[\mathbf A]\mathbf{n} \times \bm\theta_\perp) \times \mathbf{n}
+ (\mathbf{n} \times  D_\mu[\mathbf A]\mathbf{n})  \times  \bm\theta_\perp
   \nonumber\\
 &= (D_\mu[\mathbf A]\mathbf{n} \times \bm\theta_\perp)  \times  \mathbf{n}  + g\mathbf X_\mu \times \bm\theta_\perp ,
\end{align}
while the second term of (\ref{C26-dX}) reads
\begin{align}
\mathbf{n} \times  D_\mu[\mathbf A](\mathbf{n} \times \bm\theta_\perp)
 &=\mathbf{n} \times (D_\mu[\mathbf A]\mathbf{n} \times \bm\theta_\perp)
   +\mathbf{n} \times (\mathbf{n} \times  D_\mu[\mathbf A]\bm\theta_\perp) ,
%   \nonumber\\
% &= -(D_\mu[\mathbf A]\mathbf{n} \times \bm\theta_\perp) \times \mathbf{n} +\mathbf{n} \times (\mathbf{n} \times  D_\mu[\mathbf A]\bm\theta_\perp) .
\end{align}
where we have used the Leibniz rule for the covariant derivative.
Thus we obtain 
\begin{align}
\delta_{\omega,\theta} \mathbf X_\mu
 &=g\mathbf X_\mu \times \bm\theta_\perp +  \mathbf{n} \times (\mathbf{n} \times  D_\mu[\mathbf A]\bm\theta_\perp) 
   +\mathbf{n} \times (D_\mu[\mathbf A]\bm\omega \times \mathbf{n}) 
   \nonumber\\
 &=g\mathbf X_\mu \times \bm\theta_\perp
  +\mathbf{n} \times \{D_\mu[\mathbf A](\bm\omega-\bm\theta_\perp) \times \mathbf{n}\} .
  \label{C26-dX-2}
\end{align}
The transformation is further simplified by reducing the number of the exterior product $ \times $.

The second term of (\ref{C26-dX-2}) is rewritten as 
\begin{align}
  \mathbf{n} \times \{D_\mu[\mathbf A](\bm\omega-\bm\theta_\perp) \times \mathbf{n}\}
%   \nonumber\\
 &=\mathbf{n} \times  D_\mu[\mathbf A]\{(\bm\omega-\bm\theta_\perp) \times \mathbf{n}\}
   -\mathbf{n} \times \{(\bm\omega-\bm\theta_\perp) \times  D_\mu[\mathbf A]\mathbf{n}\}
   \nonumber\\
 &=\mathbf{n} \times  D_\mu[\mathbf A]\{(\bm\omega-\bm\theta_\perp) \times \mathbf{n}\}
   +\{\mathbf{n}\cdot(\bm\omega-\bm\theta_\perp)\}D_\mu[\mathbf A]\mathbf{n}
   \nonumber\\
 &=\mathbf{n} \times  D_\mu[\mathbf A]\{(\bm\omega-\bm\theta_\perp) \times \mathbf{n}\}
   +(\mathbf{n}\cdot\bm\omega)D_\mu[\mathbf A]\mathbf{n}
   \nonumber\\
 &=\mathbf{n} \times  D_\mu[\mathbf A]\{(\bm\omega-\bm\theta_\perp) \times \mathbf{n}\}
   +g(\mathbf{n}\cdot\bm\omega)(\mathbf X_\mu \times \mathbf{n})
   \nonumber\\
 &=\mathbf{n} \times 
   D_\mu[\mathbf A]\{(\bm\omega_\perp-\bm\theta_\perp) \times \mathbf{n}\}
   +g\mathbf X_\mu \times \bm\omega_\parallel,
\end{align}
where we have used the Leibniz rule for the covariant derivative in the first equality, the identity 
\begin{align}
\mathbf A \times (\mathbf B \times \mathbf C) = 
(\mathbf A \cdot \mathbf C) \mathbf B   
- (\mathbf A \cdot  \mathbf B) \mathbf C ,
\label{vector-prod}
\end{align}
 and 
$\mathbf{n} \cdot D_\mu[\mathbf A]\mathbf{n}=0$ in the second equality, $D_\mu[\mathbf V]\mathbf{n}=0$ in the fourth equality, and $\bm\omega_\parallel=(\mathbf{n}\cdot\bm\omega)\mathbf{n}$ in the last equality. 
 
Thus, we obtain
\begin{align}
\delta\mathbf X_\mu
% &=g\mathbf X_\mu \times \bm\theta_\perp
%  +\mathbf{n} \times \{D_\mu[\mathbf A](\bm\omega-\bm\theta) \times \mathbf{n}\}
%   \nonumber\\
 &=g\mathbf X_\mu \times (\bm\theta_\perp+\bm\omega_\parallel)
   +\mathbf{n} \times 
    D_\mu[\mathbf A]\{(\bm\omega_\perp-\bm\theta_\perp) \times \mathbf{n}\} 
   \nonumber\\
 &=g\mathbf X_\mu \times (\bm\theta_\perp+\bm\omega_\parallel)
%\nonumber\\
%&\quad
   +D_\mu[\mathbf A]
    \bigl[\mathbf{n} \times \{(\bm\omega_\perp-\bm\theta_\perp) \times \mathbf{n}\}\bigr]
   -D_\mu[\mathbf A]\mathbf{n} \times 
    \{(\bm\omega_\perp-\bm\theta_\perp) \times \mathbf{n}\}
   \nonumber\\
 &=g\mathbf X_\mu \times (\bm\theta_\perp+\bm\omega_\parallel)
   +D_\mu[\mathbf A](\bm\omega_\perp-\bm\theta_\perp)
   +\{(\bm\omega_\perp-\bm\theta_\perp)\cdot D_\mu[\mathbf A]\mathbf{n}\}\mathbf{n}
   \nonumber\\
 &=g\mathbf X_\mu \times (\bm\theta_\perp+\bm\omega_\parallel)
   +D_\mu[\mathbf V](\bm\omega_\perp-\bm\theta_\perp)
%\nonumber\\
%&\quad
   +g\mathbf X_\mu \times (\bm\omega_\perp-\bm\theta_\perp)
   +g\{(\bm\omega_\perp-\bm\theta_\perp)\cdot(\mathbf X_\mu \times \mathbf{n})\}\mathbf{n}
   \nonumber\\
 &=g\mathbf X_\mu \times (\bm\theta_\perp+\bm\omega_\parallel)
   +D_\mu[\mathbf V](\bm\omega_\perp-\bm\theta_\perp)
%\nonumber\\
%&\quad
   +g\mathbf X_\mu \times (\bm\omega_\perp-\bm\theta_\perp)
   -g\{\mathbf{n}\cdot[\mathbf X_\mu \times (\bm\omega_\perp-\bm\theta_\perp)]\}\mathbf{n}
   \nonumber\\
 &=g\mathbf X_\mu \times (\bm\theta_\perp+\bm\omega_\parallel)
   +D_\mu[\mathbf V](\bm\omega_\perp-\bm\theta_\perp),
\end{align}
where we have used
the Leibniz rule for the covariant derivative in the second equality, the identity (\ref{vector-prod}), $\mathbf{n} \cdot \mathbf{n} =1$ and 
$\mathbf{n} \cdot D_\mu[\mathbf A]\mathbf{n}=0$ in the third equality, 
$D_\mu[\mathbf A]=D_\mu[\mathbf V]+ g\mathbf X_\mu \times$ and 
$D_\mu[\mathbf V]\mathbf{n}=0$ in the fourth equality, 
$\mathbf A \cdot (\mathbf B \times \mathbf C)=\mathbf C \cdot (\mathbf A \times \mathbf B)$ in the fifth equality,
and $\mathbf{n}\cdot \mathbf X_\mu=$ in the last equality.

If $\bm\omega_\perp(x)=\bm\theta_\perp(x)$ is satisfied,
the transformation (\ref{C26-gtc}) and (\ref{C26-gtX}) reduce to the gauge transformation II $\delta^\prime_\omega$ with the parameter $\bm\omega'(x)$.
%: 
%\begin{equation}
%\bm\omega'(x)=(\bm\omega_\parallel(x),\bm\omega_\perp(x)=\bm\theta_\perp(x)) .
%\end{equation} 
Therefore,  the gauge transformation II corresponds to a special case 
$\bm\omega_\perp(x)=\bm\theta_\perp(x)$.
Then it turns out that the new field variables in the CDG--Yang-Mills theory obey the gauge transformation II. 
Thus, the issue (C) is resolved.

Therefore, the minimization condition (\ref{C26-MAGcond}) works as a gauge fixing condition for the enlarged gauge symmetry 
except for the gauge symmetry II, i.e., $\bm\omega_\perp(x)=\bm\theta_\perp(x)$.
In fact, the condition (\ref{C26-dMAG}) does not transform covariantly $\delta \bm{\chi} \not= g \bm{\chi} \times   \bm{\omega}$ in general, and transforms covariantly
$\delta \bm{\chi} = g \bm{\chi} \times   \bm{\omega}$ 
only when $\bm\omega_\perp(x)=\bm\theta_\perp(x)$, since 
 the gauge transformation of the condition (\ref{C26-dMAG}) reads
%[Exercise-11] \marginpar{Ex-11}
\begin{align}
  \delta \bm{\chi}
  =& g \bm{\chi} \times  (\bm{\omega}_\parallel +\bm{\theta}_\perp ) 
  - g^2 \mathbf{X}_\mu \times [\mathbf{X}_\mu \times (\bm{\omega}_\perp -\bm{\theta}_\perp )]
%\nonumber\\&
  + D_\mu[\mathbf{V}]D_\mu[\mathbf{V}](\bm{\omega}_\perp -\bm{\theta}_\perp )  .
  \label{C26-gtF}
\end{align}
For  
$\bm\omega_\perp(x)=\bm\theta_\perp(x)$, the condition (\ref{C26-dMAG}) transforms covariantly 
$
 \delta \bm{\chi} = g \bm{\chi} \times   \bm{\omega}
$, 
since 
$
\delta \bm{\chi}
  = g \bm{\chi} \times  (\bm{\omega}_\parallel +\bm{\omega}_\perp ) 
  = g \bm{\chi} \times  \bm{\omega}.   
$
Here the local rotation of $\mathbf{n}$,  
$
\delta\mathbf{n}(x)  =g\mathbf{n}(x) \times {\bm\theta}_\perp(x)
$,
leads to  $\delta \bm{\chi}=0$ on $\bm{\chi}=0$. 
Moreover, the $U(1)_{\rm local}^{\omega}$ part in $G=SU(2)_{\rm local}^{\omega}$ is not affected by this condition. Hence, the gauge symmetry corresponding to $\bm\omega_\parallel(x)$ remains unbroken. 

The gauge transformation II yields 
\begin{subequations}
\begin{align}
%  \delta_{\omega}^\prime  \mathbf{n}  =& g \mathbf{n} \times \bm{\omega'}  ,
%\\
% \delta_{\omega}^\prime  c_\mu =& \mathbf{n} \cdot \partial_\mu \bm{\omega'}   ,
%\\
%  \delta_{\omega}^\prime   \mathbf{X}_\mu =&  g \mathbf{X}_\mu \times \bm{\omega'} ,
%\\
%\Longrightarrow  & 
 \delta_{\omega}^\prime  \mathbf{V}_\mu =   D_\mu[\mathbf{V}] \bm{\omega'}   
& \Longrightarrow
 \delta_{\omega}^\prime  \mathbf{A}_\mu =   D_\mu[\mathbf{A}] \bm{\omega'}   , 
\\
& \Longrightarrow
\delta_{\omega}^\prime  \mathbf{F}_{\mu\nu}[\mathbf{V}] 
=  g \mathbf{F}_{\mu\nu}[\mathbf{V}] \times \bm{\omega}' ,
%\delta_{\omega'}  \mathbf{B}_\mu  
%=   D_\mu[\mathbf{B}] \bm{\omega'} - (\mathbf{n} \cdot %\partial_\mu \bm{\omega'}) \mathbf{n}  ,
\end{align}
\end{subequations}
Hence, the inner product $f_{\mu\nu}=\mathbf{n} \cdot \mathbf{F}_{\mu\nu}[\mathbf{V}]$ is  $SU(2)^\prime$ invariant, i.e., invariant under the gauge transformation II:
\begin{align}
% \delta_{\omega'}( \bm{n} \cdot  \mathbb{G}_{\mu\nu}) 
% & \equiv 
\delta_{\omega}^\prime f_{\mu\nu} 
=    0 ,
\ 
 f_{\mu\nu} &=  \partial_\mu c_\nu - \partial_\nu c_\mu  -  g^{-1}  \mathbf{n} \cdot  (\partial_\mu \mathbf{n} \times \partial_\nu \mathbf{n}), \
c_\mu 
  =\mathbf{n} \cdot \mathbf{A}_\mu  .
\end{align}
Then $f_{\mu\nu}^2=\mathbf{F}_{\mu\nu}[\mathbf{V}]^2= \mathbf{F}_{\mu\nu}[\mathbf{V}] \cdot \mathbf{F}_{\mu\nu}[\mathbf{V}]$ is $SU(2)^\prime$ invariant:
\begin{align}
 \delta_{\omega}^\prime \mathbf{F}_{\mu\nu}[\mathbf{V}]^2 
= \delta_{\omega}^\prime f_{\mu\nu}^2 
=  0 .
\end{align}
Therefore, we can construct $SU(2)$ invariant ``Abelian" gauge theory.

As already pointed out, we can define the {\it gauge-invariant} \textbf{magnetic monopole current}  by
\begin{equation}
  k^\mu(x) := \partial_\nu {}^*f^{\mu\nu}(x) .
%=   (1/2) \epsilon^{\mu\nu\rho\sigma}\partial_{\nu} f_{\rho\sigma}(x) ,
%\nonumber\\
% G_{\mu\nu} :=& \bm{n} \cdot (\mathbf{E}_{\mu\nu}+\mathbf{H}_{\mu\nu})
%=  E_{\mu\nu}+H_{\mu\nu} 
%= \partial_\mu c_\nu - \partial_\nu c_\mu  -  g^{-1}  \bm{n} \cdot  (\partial_\mu \bm{n} \times \partial_\nu \bm{n}) .
%c_\mu   =\bm{n} \cdot \mathscr{A}_\mu  .
\label{C26-latCFN-monop}
\end{equation}
In addition, we find the invariance
\begin{align}
 \delta_{\omega}^\prime \mathbf{X}_\mu^2  =  0 .
\end{align}
This fact has an important implication for the gluon mass generation and the mass gap problem to be discussed later.

%\newpage
\noindent
%%%%%%%%%%%%%%%%%%%%%%%%%%%%%%%%%%%%%%%%%%%%%%%%%%%

\subsection{%$\bullet$ 
Quantization based on the functional integral (Path-integral) method
}  

%%%%%%%%%%%%%%%%%%%%%%%%%%%%%%%%%%%%%%%%%%%%%%%%%%%

Following the steps indicated in Fig.~\ref{C26-fig:sym-cfn-ym}, we rewrite the original Yang-Mills theory into the equipollent theory. 
In the Euclidean formulation, the original Yang-Mills theory has the partition function written in terms of   $\mathscr{A}_\mu(x)$:
\begin{align}
 Z_{{\rm YM}} = \int \mathcal{D}\mathscr{A}_\mu \exp (-S_{{\rm YM}}[\mathscr{A}]) .
\end{align}
To be precise, we need to add the factor $\delta(F[\mathscr{A}])$ for the gauge-fixing condition $F[\mathscr{A}]=0$ and the associated Faddeev-Popov determinant 
$\Delta_{\rm FP}^{F}:=\det  \left(\frac{\delta F[\mathscr{A}^\omega]}{\delta \omega}\right)
$ to the integration measure $\mathcal{D}\mathscr{A}_\mu$. 
But we omit to write them explicitly in the following to simplify the expression.

We introduce the  color field $\bm{n}(x)$ with a unit length as an auxiliary  field by inserting the unity:
\begin{align}
 1 = \int \mathcal{D}\bm{n}  \delta(\bm n\cdot\bm n-1) = 
 \prod_{x \in \mathbb{R}^D} \int  [d\mathbf{n}(x)]  \delta(\mathbf{n}(x) \cdot \mathbf{n}(x)-1) ,
\end{align}
into the original Yang-Mills theory to obtain the master Yang-Mills theory written in terms of $\bm{n}(x)$ and $\mathscr{A}_\mu(x)$:
\begin{align}
 Z_{{\rm mYM}}  = \int \mathcal{D}\mathscr{A}_\mu
\int \mathcal{D}\bm{n}  \delta(\bm n\cdot\bm n-1)
 \exp (-S_{{\rm YM}}[\mathscr{A}]) . 
 \label{C26-Z-mYM}
\end{align}

Here we point out that such a partition function  appears when calculating the expectation value of the Wilson loop operator, i.e., the \textbf{Wilson loop average}  
$W(C)$  defined by
\begin{equation}
 W(C):=\left< W_C[\mathscr{A}] \right>_{\rm YM}
= Z_{\rm YM}^{-1} \int \mathcal{D}\mathscr{A}_\mu  e^{ -S_{\rm YM}[\mathscr{A}] } W_C[\mathscr{A}] .
\end{equation}
We can use  a \textbf{non-Abelian Stokes theorem} for the Wilson loop operator:%
\footnote{
See section 6.%the chapter of non-Abelian Stokes theorem.
}
\begin{align}
W_C[\mathscr{A}]  =  \int \mathcal{D}\mu[\bm{n}] \tilde{W}_{\mathscr{A}}(S) , \  
 \tilde{W}_{\mathscr{A}}(S) 
:=  \exp \left\{ i g J  \int_{S:\partial S=C} d^2S^{\mu\nu} f_{\mu\nu} \right\}  ,
 \label{C26-reducedW}
\end{align} 
where the surface integral is over any surface $S$ bounding the loop $C$, $J$ is the index taking the half integer $J=\frac12, 1, \frac32, 2, \cdots$ which characterizes the representation of the Wilson loop, and 
$
  \mathcal{D}\mu[\bm{n}]
%:=   \mathcal{D}\bm{n}  \delta(\bm{n} \cdot \bm{n} -1)
%= \prod_{x \in \mathbb{R}^D}   [d\bm{n}(x)]  \delta(\bm{n}(x) \cdot \bm{n}(x)-1)  
$ is an invariant integration measure. 
Then we arrive at the expression for the Wilson loop average
$W(C)$:
\begin{align}
W(C)  
=& Z_{\rm mYM}^{-1}  \int \mathcal{D}\mathscr{A}_\mu  \int \mathcal{D}\mu[\bm{n}] 
  e^{ -S_{\rm YM}[\mathscr{A}] } \tilde{W}_{\mathscr{A}}(S) 
\nonumber\\
=& \frac{\int \mathcal{D}\mathscr{A}_\mu  \int \mathcal{D}\mu[\bm{n}]  e^{ -S_{\rm YM}[\mathscr{A}] } \tilde{W}_{\mathscr{A}}(S) }{ \int \mathcal{D}\mathscr{A}_\mu   \int \mathcal{D}\mu[\bm{n}] e^{ -S_{\rm YM}[\mathscr{A}] }} ,
\label{C26-W1}
\end{align} 
where we have inserted the unity,
$
1=\int \mathcal{D}\mu[\mathbf{n}]
$,
 into the functional integration of the denominator.  
Here the {\it reduced} Wilson loop operator $\tilde{W}_{\mathscr{A}}(S)$ is gauge invariant. 
In this way, the partition function of the form (\ref{C26-Z-mYM}) appears naturally in the calculation of the Wilson loop average through a non-Abelian Stokes theorem for the Wilson loop operator. 

In order to obtain the gauge theory which is equipollent to the original Yang-Mills theory, we must eliminate extra degrees of freedom which are carried by the new field  variables. 
For this purpose, we impose the constraint $\bm{\chi}[ \mathscr{A},\bm{n}]=0$ which we call the \textbf{reduction condition}.
For this purpose, we insert the unity to the functional integral:
\begin{align}
  1 = \int \mathcal{D} \bm{\chi}^\theta \delta(\bm{\chi}^\theta)
=   \int\! \mathcal{D}\bm\theta\delta(\bm\chi^\theta)
   \det\left(\frac{\delta\bm\chi^\theta}{\delta{\bm\theta}}\right) ,
\end{align}
where 
%$\bm{\chi}^\theta$ is the constraint written in terms of the gauge-transformed variable, i.e., 
$\bm{\chi}^\theta:=\bm{\chi}[ \mathscr{A},\bm{n}^\theta]$ 
is the reduction condition ($\bm{n}^\theta$ is the local rotation of $\bm{n}$ by $\bm\theta$)
and 
$
\det\left(\frac{\delta\bm\chi^\theta}{\delta{\bm\theta}}\right)
$
denotes  the Faddeev-Popov determinant associated with the reduction condition.
Note that $\bm\theta$ have the same degrees of freedom as $\bm\chi$, since $\bm\theta_\parallel$ does not affect the rotation. 
Then we obtain
\begin{align}
Z_{{\rm mYM}}   = \int \mathcal{D}\mathscr{A}_\mu 
\int \mathcal{D}\bm{n}  \delta(\bm n\cdot\bm n-1)  \int\! \mathcal{D}\bm\theta\delta(\bm\chi^\theta)
   \det\left(\frac{\delta\bm\chi^\theta}{\delta{\bm\theta}}\right) 
\exp (-S_{{\rm YM}}[\mathscr{A}]) .
\end{align}

Then  we cast the original Yang-Mills theory into an equipollent gauge theory which has the same gauge symmetry (degrees of freedom) as the original theory and is rewritten in terms of the new variables
$
(\bm n,c_\mu, \mathscr{X}_\mu)
$.
Therefore, using the notation $\tilde{Z}_{{\rm YM}}$ instead of $Z_{{\rm mYM}} $, we write
\begin{align}
 \tilde{Z}_{{\rm YM}}  
=&  \int \mathcal{D}c_\mu \int 
\mathcal{D}\mathscr{X}_\mu  \delta( \bm{n} \cdot \mathscr{X}_\mu)  
\int \mathcal{D}\bm{n}  \delta(\bm n\cdot\bm n-1) {J}  
%\nonumber\\
%& \times \int\! 
\mathcal{D}\bm\theta\delta(\bm\chi^\theta)
   \det\left(\frac{\delta\bm\chi^\theta}{\delta{\bm\theta}}\right)
 \exp (-\tilde S_{\rm YM}[\bm n,c,\mathscr{X}]) . 
\end{align}
where ${J}$ is the Jacobian associated with the change of variables from 
$
(\mathscr{A}_\mu^A, \bm n^B)
$ 
to 
$
(c_\mu, \mathscr{X}_\mu^B, \bm n^C)
$
and the action $\tilde S_{\rm YM}[\bm n,c, \mathscr{X}]$ is obtained by substituting the CDGFN decomposition of $\mathscr A_\mu$, i.e.,
\begin{equation}
\mathscr{A}_\mu (x)
 =\bm{n}(x)c_\mu(x) 
  +i g^{-1} [ \bm{n}(x)  , \partial_\mu \bm{n}(x) ]
  +\mathscr X_\mu(x) ,
\end{equation}
%\begin{align}
%\mathscr{A}_\mu^{CFN}(x)
%% =c_\mu(x) \bm{n}(x)
%%  +g^{-1}\partial_\mu \bm{n}(x)\times \bm{n}(x)
%%  +\mathbf{X}_\mu(x) 
%= \bm{n}(x) [\bm{n}(x)\cdot\mathscr{A}_\mu(x)]
%  +g^{-1}\partial_\mu \bm{n}(x)\times \bm{n}(x)
%+ g^{-1}\bm{n}(x)\times \mathscr{D}_\mu[\mathscr{A}]\bm{n}(x) .
%\end{align} 
 into the original Yang-Mills action $S_{{\rm YM}}[\mathscr{A}]$: 
\begin{align}
 S_{\rm YM}[\mathscr A]  = \tilde S_{\rm YM}[\bm n,c, \mathscr{X}] .
\end{align}
In order to define the Jacobian $J$, 
%for the change of variables from $
%(\mathscr{A}_\mu^A, \bm n^B)
%$ 
%to 
%$
%(c_\mu, \mathscr{X}_\mu^B, \bm n^C)
%$
the integration measures $\mathcal{D}\bm{n}$ and $\mathcal{D}\mathscr{X}_\mu$ are understood to be written in terms of independent degrees of freedom by taking into account the constraints $\bm n\cdot\bm n=1$ and $\bm n\cdot\mathscr{X}_\mu=0$.  
The details for calculating $J$ will be given in the next section.

%\newpage
%In order to fix the enlarged symmetry in the master-Yang-Mills theory and retain only the gauge symmetry II,

We perform the change of variables $\bm{n} \rightarrow \bm{n}^{\theta}$, i.e., the local rotation by the angle $\theta$ and the corresponding   gauge   transformations  II for other new variables $c_\mu$ and $\mathscr{X}_\mu$: 
$c_\mu, \mathscr{X}_\mu \rightarrow c_\mu^{\theta}, \mathscr{X}_\mu^{\theta}$. 
From the gauge invariance of the action $\tilde S_{\rm YM}[\bm n,c, \mathscr{X}]$ and the integration measure 
$
  \mathcal{D}c_\mu  \mathcal{D}\mathscr{X}_\mu 
 \mathcal{D}\bm{n}  \delta(\bm n\cdot\bm n-1)
$, 
we can rename the dummy integration variables $\bm{n}^{\theta}, c_\mu^{\theta}, \mathscr{X}_\mu^{\theta}$  as $\bm{n}, c_\mu, \mathscr{X}_\mu$ respectively.
Thus the integrand does not depend on $\theta$ and the gauge volume $\int\! \mathcal{D}\bm\theta$ can be factored out:%
%\footnote{
%Here we neglect the Gribov problem.
%See the chapter of Gribov problem. 
%}:
\begin{align}
 \tilde{Z}_{{\rm YM}}  
=& \int\! \mathcal{D}\bm\theta 
\int \mathcal{D}\bm{n}  \delta(\bm n\cdot\bm n-1)
 \int \mathcal{D}c_\mu \int 
\mathcal{D}\mathscr{X}_\mu  \delta( \bm{n} \cdot \mathscr{X}_\mu)  
 {J}
%\nonumber\\ & \times 
\delta(\bm\chi)
   \det\left(\frac{\delta\bm\chi}{\delta{\bm\theta}}\right)
 e^{-\tilde S_{\rm YM}[\bm n,c, \mathscr{X}]} .
\end{align}
Note that the Faddeev--Popov determinant
\footnote{
This is the same as the determinant called the Shabanov determinant:
$
\Delta_{S}[\mathscr{A}_\mu, \bm{n}]  
:= \det \left| \frac{\delta \bm{\chi}}{\delta \bm{n}} \right|_{\bm{\chi}=0} ,
$
which guarantees the equivalence between the original and reformulated Yang-Mills theories. 
Therefore, the Shabanov determinant is simply interpreted as  the Faddeev--Popov determinant associated with the reduction condition from the new viewpoint.  
}
 $\det\left(\frac{\delta\bm\chi}{\delta{\bm\theta}}\right)$ can be rewritten into another form:  
\begin{equation}
 \Delta_{\rm FP}^{\rm red}
:= \det\left(\frac{\delta\bm\chi}{\delta{\bm\theta}}\right)_{\bm{\chi}=0}
=   \det\left(\frac{\delta\bm\chi}{\delta\bm n^\theta}\right)_{\bm{\chi}=0}.
\end{equation}
Thus, we have arrived at the reformulated Yang-Mills theory  in which the independent variables are regarded as $\bm{n}(x)$, $c_\mu(x)$ and $\mathscr X_\mu(x)$ with the partition function:
\begin{align}
 Z_{{\rm YM}}^\prime  
=&  \int \mathcal{D}c_\mu \int 
\mathcal{D}\mathscr{X}_\mu  \delta( \bm{n} \cdot \mathscr{X}_\mu)     
\int \mathcal{D}\bm{n}  \delta(\bm n\cdot\bm n-1)
 {J} 
%\nonumber\\ & \times  
\delta(\tilde{\bm\chi})  
   \Delta_{\rm FP}^{\rm red}
 e^{-\tilde S_{\rm YM}[\bm n,c,\mathscr{X}]} , 
\end{align}
where the constraint is rewritten in terms of the new variables:
\begin{equation}
\tilde{\bm\chi} 
 :=\tilde{\bm\chi} [\bm n, c, \mathscr{X}]
 :=\mathscr{D}^\mu[\mathscr{V} ]\mathscr{X}_\mu .
% \quad \mathbf{V}_\mu \equiv c_\mu  \bm{n}   +g^{-1}\partial_\mu \bm{n} \times \bm{n} .
\end{equation}

Thus, the integration measure is obtained by replacing that of the original Yang-Mills theory by
%We can define the Jacobian $\tilde{J}$ between the original Yang-Mills theory and the reformulated one by
\begin{equation}
  \mathcal{D}\mathscr{A}_\mu 
\to   
\mathcal{D}\bm{n}   \delta(\bm n\cdot\bm n-1)
\mathcal{D}c_\mu  \mathcal{D}\mathscr{X}_\mu \delta( \bm{n} \cdot \mathscr{X}_\mu)      
\delta(\tilde{\bm\chi}) 
 \Delta_{\rm FP}^{\rm red}[\bm n,c, \mathscr X] {J} ,
\end{equation}
which is also written using only the independent field variables $(n^a,c_\mu,\mathscr{X}_\mu^b)$ obtained by solving the constraints $\bm n\cdot\bm n-1=0$ and $\bm{n} \cdot \mathscr{X}_\mu=0$  (which will be explained in the next subsection) as
\begin{equation}
  \mathcal{D}\mathscr{A}_\mu^A 
\to \mathcal{D}n^a   \mathcal{D}c_\mu  \mathcal{D}\mathscr{X}^b_\mu  \delta(\tilde{\bm\chi})  \Delta_{\rm FP}^{\rm red}  [\bm n,c, \mathscr X]  J .
\end{equation}
For  the Jacobian ${J}$ associated with the change of variables,  
%We have succeeded to separate the original variables  
%$
% (\mathscr{A}_\mu(x),\bm{n}(x)) \rightarrow (c_\mu(x), \mathscr{X}_\mu(x), \bm{n}(x))  
%$,
%i.e.,
%$\mathcal{D}\mathscr{A}_\mu^A \mathcal{D}n^B  =  {J} \mathcal{D}c_\mu \mathcal{D}X_\mu^b  \mathcal{D}n^c$.
the result is simple. As shown in the next section, the Jacobian is equal to one:
\begin{equation}
  {J} = 1 .
\end{equation}

For  the FP determinant $\Delta_{\rm FP}^{\rm red}$,  the precise form of $\Delta_{\rm FP}^{\rm red}$ is obtained using  (\ref{C26-gtF})  as
%[Exercise-12] \marginpar{Ex-12}
\footnote{
Although this result can be obtained by using the Faddeev-Popov trick, it was obtained originally using the BRST method which is applicable to more general setting. 
For the through treatment, see the following reference in which the explicit derivation of the FP ghost term has been worked out explicitly using the BRST method 
\cite{KMS05}.
%Kondo, Murakami and Shinohara (2005).
}
%\cite{KMS05}
\begin{equation}
  \Delta_{\rm FP}^{\rm red}[\bm n,c,\mathscr X] 
= \det [-\mathscr{D}_\mu[\mathscr V+\mathscr X]\mathscr{D}_\mu[\mathscr V-\mathscr  X]]  .
\label{C26-red-FP}
\end{equation}

The reformulated Yang-Mills theory obtained after imposing the reduction condition has still the original full gauge symmetry $G$.  
In order to obtain a completely gauge-fixed theory, we must start from the theory with the overall gauge fixing condition for $G$ symmetry, e.g., the Landau gauge $\partial^\mu \mathscr{A}_\mu(x)=0$, 
and take into account the Faddeev-Popov ghost associated with the overall gauge fixing simultaneously, in addition to the Faddeev-Popov ghost associated with the reduction condition. 
According to the clarification of the symmetry in the master-Yang-Mills theory explained above,  we can obtain the unique Faddeev-Popov ghost terms associated with the gauge fixing conditions adopted in quantization.  
This is  another advantage of the new viewpoint for the master-Yang-Mills theory. 
Thus, we have given a reformulation of Yang-Mills theory called the CDG--Yang-Mills theory  written in term of new variables obtained by using the non-linear change of variables.   

%%%%%%%%%%%%%%%%%%%%%%%%%%%%%%%%%%%%%%%%%%%%%%%%%%%%%%%%%%%%%%%%%%%%%

For the Wilson loop average  $W(C)$,  
by repeating the same steps as before in both the numerator and the denominator of (\ref{C26-W1}), we can obtain an expression of the Wilson loop average in the reformulated Yang-Mills theory:
\begin{align}
W(C)  
=& Z^\prime_{\rm YM}{}^{-1} 
\int \mathcal{D}n^a(x) \mathcal{D}c_\mu(x) \mathcal{D}\mathscr{X}^b_\mu(x) \delta(\tilde{\bm\chi}) \Delta_{\rm FP}^{\rm red} e^{- \tilde{S}_{\rm YM}[\bm{n},c,\mathscr{X}] } \tilde{W}_{\mathscr{A}}(S) 
\nonumber\\
=& \frac{ 
\int \mathcal{D}n^a(x)  \mathcal{D}c_\mu(x) \mathcal{D}\mathscr{X}^b_\mu(x) \delta(\tilde{\bm\chi}) \Delta_{\rm FP}^{\rm red} e^{- \tilde{S}_{\rm YM}[\bm{n},c,\mathscr{X}]} \tilde{W}_{\mathscr{A}}(S) 
}{
 \int \mathcal{D}n^a(x)  \mathcal{D}c_\mu(x) \mathcal{D}\mathscr{X}^b_\mu(x) \delta(\tilde{\bm\chi}) \Delta_{\rm FP}^{\rm red}  e^{-\tilde{S}_{\rm YM}[\bm{n},c,\mathscr{X}]} 
} .
\label{C26-W2}
\end{align}

%\newpage
%%%%%%%%%%%%%%%%%%%%%%%%%%%%%%%%%%%%%%%%%%%%%%%%%%

\subsection{Jacobian associated with the change of variables}
\label{subsec:path-integral-SU2}
%\setcounter{equation}{0}

%%%%%%%%%%%%%%%%%%%%%%%%%%%%%%%%%%%%%%%%%%%%%%%%%%

We consider the Jacobian $J$ associated with the change of variables. 
If the color field $\bm{n}$ is introduced into the  Yang-Mills theory, then we obtain the master (extended) Yang-Mills theory with extra degrees of freedom carried by the color field.  In order to obtain the gauge theory which is equipollent to the original Yang-Mills theory with the same gauge degrees of freedom, we must introduce the constraint $\bm{\chi}=0$ to eliminate the extra degrees of freedom, which is the reduction condition.

For $G=SU(2)$, the color field $\bm{n}$ has two independent degrees of freedom, since it is a unit vector with three components. %$\bm{n} \cdot \bm{n} := n^A n^A =1$. 
The reduction condition $\bm{\chi}=0$ gives two conditions to eliminate two extra degrees of freedom introduced by the color field $\bm{n}$. 
Therefore, we obtain the correspondence between the two measures: the integration measure in the respective step in Fig.~\ref{C26-fig:sym-cfn-ym} is given by
\begin{equation}
  [d\mathscr{A}_\mu^A]  \nearrow [d\mathscr{A}_\mu^A] [dn^B \delta(\bm{n} \cdot \bm{n} -1)] \searrow [d\mathscr{A}_\mu^A] [dn^B \delta(\bm{n} \cdot \bm{n} -1)] \delta^{(2)}(\bm\chi) .
\end{equation}

In the decomposition $\mathscr{A}=\mathscr{V}+\mathscr{X}$ for a given color field $\bm{n}$, the first defining equation $\mathscr{D}_\mu[\mathscr{V}]\bm{n}=0$ gives $2D$ conditions (per a space--time point) and eliminates $2D$ degrees of freedom in $3D$ degrees of freedom of $\mathscr{V}_\mu^E$: thereby $\mathscr{V}_\mu^E$ are reduced to $c_\mu$ with $D$ degrees of freedom.
Whereas the the second defining equation $\bm{n} \cdot \mathscr{X}_\mu=0$ gives $D$ conditions  and eliminates $D$ degrees of freedom in $3D$ degrees of freedom of $\mathscr{X}_\mu^F$: thereby, $\mathscr{X}_\mu^F$ are reduced to $X_\mu^f$ with $2D$ degrees of freedom ($f=1,2$).  Thus we have the total $3D$ degrees of freedom for $SU(2)$ Yang-Mills theory. 
Before gauge fixing, $\mathscr{A}_\mu^A(x)$ are  independent variables in the original Yang-Mills theory, while 
\begin{equation}
 c_\mu(x), X_\mu^b(x), n^a(x) \quad ( a,b=1,2 ,  \mu=1, \cdots, D  )
\end{equation}
are independent variables in the reformulated Yang-Mills theory. 
Thus, we find the correspondence among the integration measures:
\begin{align}
 [d\mathscr{A}_\mu^A]  \sim & [d\mathscr{A}_\mu^A] [dn^B \delta(\bm{n} \cdot \bm{n} -1)]   \delta^{(2)}(\bm{\chi}) 
 \nonumber\\
 \sim & {J} [d\mathscr{V}_\nu^E \delta^{(2D)}(\mathscr{D}_\nu[\mathscr{V}]\bm{n})] [d\mathscr{X}_\nu^F  \delta^{(D)}(\bm{n} \cdot \mathscr{X}_\nu)] [dn^C \delta(\bm{n} \cdot \bm{n} -1)]   \delta^{(2)}(\bm{\chi})
 \nonumber\\
  \sim & {J} [dc_\nu] [dX_\nu^f]  [dn^C \delta(\bm{n} \cdot \bm{n} -1)] \delta^{(2)}(\bm{\chi}) 
 ,
 \label{C26-JJ}
\end{align}
where 
$A,B,C,E,F=1,2,3$; $\mu,\nu=1, \cdots, D  
$, and $f=1,2$.

The reduction condition $\bm{\chi}=\bm{0}$ is transformed into the three equivalent forms:
\begin{align}
  \bm{\chi}_1  =& -ig^{-1}[\bm{n}, \mathscr{D}^\mu[\mathscr{A}] \mathscr{D}_\mu[\mathscr{A}] \bm{n}] ,
 \nonumber\\
 \bm{\chi}_2 =&  \mathscr{D}^\mu[\mathscr{V}]\mathscr{X}_\mu ,
 \nonumber\\
 \bm{\chi}_3 =&  \partial^\mu X_\mu^a -g(c^\mu+h^\mu) \epsilon^{ab}X_\mu^b
 ,
 \label{C26-chi}
\end{align}
where $h_\mu$ is the magnetic potential introduced in (\ref{C26-magnetic-potential}). 
Note that $\bm{\chi}_1=\bm{\chi}_2$ has been shown in (\ref{C26-rDE2}) and $\bm{\chi}_2=\bm{\chi}_3$ can be shown by explicit calculations.%in Appendix~2 \cite{Kondo06}. 
\footnote{
$\bm{\chi}_2=\bm{\chi}_3$ is shown in Appendix~2 of \cite{Kondo06}. %Kondo (2006). 
}

The Jacobian ${J}$ associated with the change of variables is calculated as follows.%
\footnote{
See e.g., \cite{Kondo06}. %Kondo (2006). 
}
Using the correspondence (\ref{C26-JJ}), we consider the change of $3D+2=D+2D+2$ variables:
\begin{equation}
 (\mathscr{A}_\mu^A,n^b) \rightarrow (c_\nu, X_\nu^b,n^c) \quad (b,c=1,2; A =1,2,3; \mu,\nu=1, \cdots, D ) , 
\end{equation}
to calculate the Jacobian which is the determinant for the $(3D+2)\times(D+2D+2)$ matrix:
\begin{align}
 d\mathscr{A}_\mu^A dn^B  = {J} dc_\nu dX_\nu^b  dn^C , 
\quad
 {J} = 
% \begin{vmatrix}
%  \frac{\partial \mathscr{A}_\mu^A}{\partial c_\nu} & \frac{\partial \mathscr{A}_\mu^A}{\partial X_\nu^a} & \frac{\partial \mathscr{A}_\mu^A}{\partial n^C}   \cr
%  \frac{\partial n^B}{\partial c_\nu} & \frac{\partial n^B}{\partial X_\nu^a} & \frac{\partial n^B}{\partial n^C} \cr
% \end{vmatrix} .
 \begin{vmatrix}
 \frac{\partial n^B}{\partial n^C} & \frac{\partial n^B}{\partial c_\nu} & \frac{\partial n^B}{\partial X_\nu^b} \cr
 \frac{\partial \mathscr{A}_\mu^A}{\partial n^C} & \frac{\partial \mathscr{A}_\mu^A}{\partial c_\nu} & \frac{\partial \mathscr{A}_\mu^A}{\partial X_\nu^b} \cr
  \end{vmatrix} 
   .
\end{align}
Here $n^B$ and $n^C$ should be understood as denoting two independent degrees of freedom obtained after solving the constraint $n^A n^A=1$. However, if we choose specific components (directions), the color symmetry is apparently broken. Therefore, we use this notation in the followings, keeping this convention in mind.

To  avoid  complications coming from  constraints, we rewrite the integration measure in terms of the independent variables and calculate the Jacobian associated to this change of variables. 
For this purpose, we introduce the ortho-normal basis $(\mathbf{n}_1(x), \mathbf{n}_2(x), \mathbf{n}_3(x))=(\bm{e}_1(x), \bm{e}_2(x), \mathbf{n}(x))$, i.e., 
$\mathbf{n}_j(x) \cdot \mathbf{n}_k(x) = \delta_{jk}$, 
$\mathbf{n}_j(x) \times \mathbf{n}_k(x) = \epsilon_{jk\ell}\mathbf{n}_\ell(x)$, $(j,k=1,2,3)$, or equivalently
\begin{align}
   \bm{e}_a(x) \cdot \bm{e}_b(x) =& \delta_{ab} , \ 
   \mathbf{n}(x) \cdot \bm{e}_a(x)  = 0, \ 
   \mathbf{n}(x) \cdot \mathbf{n}(x)  = 1 ,
 \nonumber\\
  \bm{e}_a(x) \times \bm{e}_b(x) =& \epsilon_{ab} \mathbf{n}(x) , \ 
 \mathbf{n}(x) \times \bm{e}_a(x) = \epsilon_{ab}  \bm{e}_b(x)  , 
 \ (a,b=1,2).
\label{C26-basis}
\end{align}
The local rotation (gauge transformation II) for the basis vector $\mathbf{n}_j(x)$ is given by
\begin{align}
& \delta_{\omega}' \mathbf{n}_j(x)  = g \mathbf{n}_j(x) \times \bm{\omega'}(x) 
 \nonumber\\
 \Leftrightarrow 
& \ \delta_{\omega}' \bm{e}_a(x)  = g \bm{e}_a(x) \times \bm{\omega'}(x) , \ 
 \delta_{\omega}' \mathbf{n}(x)  = g \mathbf{n}(x) \times \bm{\omega'}(x) .
\end{align}

Now we show that the Jacobian is equal to one for the transformation from the original variables $(n^B, \mathscr{A}_\mu^A)$ to the new variables 
$(n^C, c_\nu, X_\nu^b)$  in this basis $(\bm{e}_1(x), \bm{e}_2(x), \mathbf{n}(x))$.
Since $c_\nu, X_\nu^b, n^C$ are independent, we have
\begin{align}
 \frac{\partial n^B}{\partial n^C} = \delta^{BC} , \quad
 \frac{\partial n^B}{\partial c_\nu} = 0 , \quad
\frac{\partial n^B}{\partial X_\nu^b} = 0 .
\end{align}
Then the determinant is reduced to the $3D\times(D+2D)$ matrix:
\begin{align}
 {J} = 
% \begin{vmatrix}
%  \frac{\partial \mathscr{A}_\mu^A}{\partial c_\nu} & \frac{\partial \mathscr{A}_\mu^A}{\partial X_\nu^a} & \frac{\partial \mathscr{A}_\mu^A}{\partial n^C} \cr
% 0  & 0 & \delta^{BC} \cr
% \end{vmatrix} 
 \begin{vmatrix}
 \delta^{BC}  & 0  & 0 \cr
 \frac{\partial \mathscr{A}_\mu^A}{\partial n^C} &  \frac{\partial \mathscr{A}_\mu^A}{\partial c_\nu} & \frac{\partial \mathscr{A}_\mu^A}{\partial X_\nu^b}  \cr
 \end{vmatrix} 
=  \begin{vmatrix}
   \frac{\partial \mathscr{A}_\mu^A}{\partial c_\nu} & \frac{\partial \mathscr{A}_\mu^A}{\partial X_\nu^b} \cr
 \end{vmatrix} .
\end{align}
This implies that $J$ is independent of how $\bm{n}$ is related to the original field $\mathscr{A}$, that is to say, $J$ does not depend on the choice of the reduction condition, since the Jacobian does not depend on $\frac{\partial \mathscr{A}_\mu^A}{\partial n^C}$.

In order to calculate $\frac{\partial \mathscr{A}_\mu^A}{\partial X_\nu^b}$, 
%we introduce an orthonormal frame $(\bm{e}_1(x) , \bm{e}_2(x), \bm{n}(x))$ to 
we rewrite $\mathscr{A}_\mu^A$ in terms of independent degrees of freedom $c_\nu, X_\nu^b, n^c$. 
Then $\mathscr{X}_\mu(x)$ is rewritten in terms of  the orthonormal frame $\bm{e}_1(x)$ and $\bm{e}_2(x)$, since $\mathscr{X}_\mu(x)$ is orthogonal to $\mathbf{n}(x)$: 
\begin{align}
%   \bm{n}(x) =& n^a(x) \bm{e}_a(x)   \Longleftrightarrow  n^A(x)  = n^a(x) e_a^A(x)   , 
%\\
\mathscr{X}_\mu(x) =& 
%X_\mu^1(x) \bm{e}_1(x) + X_\mu^2(x) \bm{e}_2(x) = 
X_\mu^a(x) \bm{e}_a(x) 
%\nonumber\\
%   \mathbf{X}_\mu(x)  = X_\mu^a(x) \bm{e}_a(x)     
 \Longleftrightarrow
X^A_\mu(x) = X_\mu^a(x) e_a^A(x) 
\nonumber\\
  \text{or} \   
 X^a_\mu(x) =&  \mathbf{X}_\mu(x) \cdot \bm{e}_a(x) 
= X_\mu^A(x) e_a^A(x)   \  (A=1,2,3: a=1,2) . 
% X^A_\mu(x) = X_\mu^a(x) e_a^A(x) \quad (A=1,2,3: a=1,2) , 
\end{align}
This is also the case for $\mathscr{B}_\mu(x)$. 
By using the above bases, therefore, the field $\mathscr{A}_\mu^A$ is decomposed as
\begin{align}
   \mathscr{A}_\mu =& c_\mu \bm{n} + \mathscr{B}_\mu + \mathscr{X}_\mu 
= c_\mu \bm{n} + (B_\mu^a + X_\mu^a ) \bm{e}_a 
\nonumber\\
  \Longleftrightarrow
\mathscr{A}_\mu^A =& c_\mu n^A + (B_\mu^a + X_\mu^a ) e_a^A.
\label{C26-decomp}
\end{align}
Making use of (\ref{C26-decomp}), we have
\begin{align}
 \frac{\partial \mathscr{A}_\mu^A}{\partial c_\nu} = \delta_{\mu\nu} n^A , \quad
  \frac{\partial \mathscr{A}_\mu^A}{\partial X_\nu^b} = \delta_{\mu\nu} e_b^A .
\end{align}
Thus, we conclude that the Jacobian is equal to one:
%[Exercise-13] \marginpar{Ex-13}
\begin{align}
 {J} = 
  \begin{vmatrix}
    \delta_{\mu\nu} n^A  & \delta_{\mu\nu} e_b^A \cr
 \end{vmatrix}
=    \begin{vmatrix}
      n^A  &  e_b^A \cr
 \end{vmatrix}^D
= | \mathbf{n} \bm{e}_1 \bm{e}_2 |^D
= ( \mathbf{n} \cdot (\bm{e}_1 \times \bm{e}_2)  )^D
= 1  .
\label{C26-J=1}
\end{align}
The $SU(N)$ case will be discussed in the next chapter. 
%The SU(N) case can be proved in the similar way, see \cite{KSM08}.

%%%%%%%%%%%%%%%%%%%%%%%%%%%%%%%%%%%%%%%%%%%%%%%%%%

%\appendix
\subsection{The BRST symmetry and the ghost term}
\label{subsec:BRST-SU2}

%%%%%%%%%%%%%%%%%%%%%%%%%%%%%%%%%%%%%%%%%%%%%%%%%%

We introduce two kinds of ghost fields $\mathbf{C}_\omega(x)$ and $\mathbf{C}_\theta(x)$ which correspond to $\mathbb \omega(x)$ and $\mathbb \theta(x)$ in the enlarged gauge transformation, respectively. 
For the gauge field and the color field, then, the \textbf{Becchi-Rouet-Stora-Tyutin (BRST) transformation} is given by
\begin{align}
 \bm\delta \mathbf{A}_\mu(x) = D_\mu[\mathbf{A}]\mathbf{C}_\omega(x) ,
 \quad
 \bm\delta \mathbf{n}(x) = g\mathbf{n}(x) \times \mathbf{C}_\theta(x) .
\end{align}
The BRST transformation of $\mathbf{C}_\omega(x)$ is determined from the nilpotency of $\bm\delta  \bm\delta \mathbf{A}_\mu(x) \equiv 0$ in the usual way:
\begin{align}
 \bm\delta   \mathbf{C}_\omega(x) = -\frac12 g\mathbf{C}_\omega(x) \times \mathbf{C}_\omega(x) .
\end{align}
We require 
\begin{align}
    \bm n(x)  \cdot  \mathbf{C}_\theta(x) = 0 .
    \label{C26-nC}
\end{align}
Then the BRST transformation of $\mathbf{C}_\theta(x)$ is determined from the nilpotency of the BRST transformation $\bm\delta \bm\delta \mathbf{n}(x) \equiv 0$ and $\bm\delta [\bm n(x)  \cdot  \mathbf{C}_\theta(x)] \equiv 0$: 
\begin{align}
 \bm\delta   \mathbf{C}_\theta(x) = -g\mathbf{C}_\theta(x) \times \mathbf{C}_\theta(x) ,
\end{align}
We can introduce two kinds of  antighost and the Nakanishi-Lautrup field with the nilpotent BRST transformations:
\begin{align}
 \bm\delta \mathbf{\bar C}_\omega = i \mathbf{N}_\omega , 
 \quad \bm\delta \mathbf{N}_\omega = 0 ,
\end{align}
and
\begin{align}
 \bm\delta \mathbf{\bar C}_\theta = i \mathbf{N}_\theta , 
 \quad \bm\delta \mathbf{N}_\theta = 0 .
\end{align}
In harmony with (\ref{C26-nC}), we require also%
\footnote{
For more details on this section, see \cite{KMS05}.
%K.-I. Kondo, T. Murakami and T. Shinohara,
%BRST quantization of the Yang-Mills theory in the Cho--Faddeev--Niemi decomposition, 
%[hep-th/0504198],
%Eur. Phys. J. C {\bf 42}, 475--481 (2005).
}
\begin{align}
 \bm n(x) \cdot \mathbf{\bar C}_\theta (x) = 0 .%=  \bm n(x)  \cdot  \mathbf{C}_\theta(x) .
\end{align}

By identifying the reduction condition $\bm{\chi}=0$ with a gauge-fixing condition for the enlarged gauge symmetry, the gauge-fixing term and associated  ghost term is obtained from
\begin{align}
 {\cal L}_{\rm GF+FP}^\theta
 = -i \bm\delta [\mathbf{\bar C}_\theta \cdot  \bm{\chi}]
=  \mathbf{N}_\theta \cdot \bm{\chi}
+ i \mathbf{\bar C}_\theta \cdot \bm\delta \bm{\chi} , \quad 
\bm{\chi} = D_\mu[\mathbf{V}]\mathbf{X}_\mu .
\end{align}
Here it should be remarked that the antighost field $\mathbf{\bar C}_\theta$ must have the same degrees of freedom of the constraint $\bm{\chi}= D_\mu[\mathbf{V}]\mathbf{X}_\mu$. 
This is achieved by requiring 
$
 \mathbf{n}(x) \cdot  \mathbf{\bar C}_\theta(x)    = 0
$, 
since 
$
  \mathbf{n}(x) \cdot \bm{\chi}(x) =  0
$.
By using
\begin{align}
 \bm{\chi}  = D_\mu[\mathbf{A}]\mathbf{X}_\mu 
= D_\mu[\mathbf{A}] (g^{-1}\mathbf{n} \times D_\mu[\mathbf{A}]\mathbf{n})
= g^{-1} \mathbf{n}\times D_\mu[\mathbf{A}]D_\mu[\mathbf{A}]\mathbf{n}
,
\end{align}
the ghost term is calculated and is simplified by using  repeatedly the Leibniz rule for the covariant derivative and the formula 
(\ref{vector-3product})
%$
%(\mathbf A \times \mathbf B) \times \mathbf C = (\mathbf A \cdot \mathbf C) \mathbf B - \mathbf A (\mathbf B \cdot \mathbf C), 
%$
%$
%\mathbf A \times (\mathbf B \times \mathbf C) = (\mathbf A \cdot \mathbf C) \mathbf B - (\mathbf A \cdot \mathbf B) \mathbf C , 
%$
together with 
%$
% \mathbf{\bar C}_\theta \cdot \mathbf{n} =  \mathbf{n}  \cdot \mathbf{C}_\theta = 0
%$, 
\begin{align}
 \mathbf{n}(x) \cdot  \mathbf{\bar C}_\theta(x) =  0 =  \mathbf{n}(x)  \cdot \mathbf{C}_\theta(x) ,
\end{align}
and
$
  \mathbf{n}\cdot D_\mu[\mathbf{A}]\mathbf{n} =  0
$
following from 
$
\mathbf{n} \cdot \mathbf{n}=1
$.
The result is rewritten in terms of the new variables as
\begin{align}
i \mathbf{\bar C}_\theta \cdot \bm\delta \bm{\chi}
  =&
%\mathbb N_\theta^\perp\cdot(D_\mu[\mathbf{V}]\mathbf{X}_\mu)
%   \nonumber\\
% &\quad
  -i\mathbf{\bar C}_\theta\cdot
    D_\mu[\mathbf{V}-\mathbf{X}]D_\mu[\mathbf{V}+\mathbf{X}]
    (\mathbf{C}_\theta-\mathbf{C}_\omega) .
\end{align}
Thus we obtain
\begin{align}
 {\cal L}_{\rm GF+FP}^\theta
%= -i \bm\delta [\mathbf{\bar C}_\theta \cdot  \bm{\chi}]
=  \mathbf{N}_\theta \cdot D_\mu[\mathbf{V}]\mathbf{X}_\mu
-i\mathbf{\bar C}_\theta^\perp \cdot
    D_\mu[\mathbf{V}-\mathbf{X}]D_\mu[\mathbf{V}+\mathbf{X}]
    (\mathbf{C}_\theta^\perp-\mathbf{C}_\omega) ,
%\bm{\chi} := D_\mu[\mathbf{V}]\mathbf{X}_\mu .
\label{C26-FP1}
\end{align}
where the conditions imply that $\mathbf{\bar C}_\theta$ and $\mathbf{C}_\theta$ have only the perpendicular components $\mathbf{\bar C}_\theta^\perp$ and $\mathbf{C}_\theta^\perp$ respectively. 

Another way to derive the GF+FP term is as follows.
The BRST transformations of $\mathbf{X}_\mu$ and $\mathbf{V}_\mu$ are given by
\begin{align}
%  \delta_{\omega,\theta} c_\mu(x) 
% =  \mathbf{n}(x) \cdot \partial_\mu \bm{\omega}(x)  + g(\mathbf{n}(x) \times \mathbf{A}_\mu(x)) \cdot (\bm{\omega}_\perp(x) - \bm{\theta}_\perp(x)) 
%  =  \mathbf{n}(x) \cdot \partial_\mu \bm{\omega}(x)  + \mathbf{n}(x) \cdot g(\mathbf{A}_\mu(x) \times    (\bm{\omega}_\perp(x) - \bm{\theta}_\perp(x)))    ,
%\\
  \bm\delta \mathbf{X}_\mu(x) 
= g \mathbf{X}_\mu(x) \times  (\mathbf{C}_\omega^\parallel(x)+\mathbf{C}_\theta^\perp(x)) + D_\mu[\mathbf{V}](\mathbf{C}_\omega^\perp(x)-\mathbf{C}_\theta^\perp(x))  ,
  \nonumber\\
 \bm\delta \mathbf{V}_\mu(x) 
= D_\mu[\mathbf{V}](\mathbf{C}_\omega^\parallel(x)+\mathbf{C}_\theta^\perp(x)) + g \mathbf{X}_\mu(x) \times  (\mathbf{C}_\omega^\perp(x)-\mathbf{C}_\theta^\perp(x))     ,
 \end{align}
The ghost term is calculated  and  is rewritten in terms of the new variables as
\begin{align}
i \mathbf{\bar C}_\theta \cdot \bm\delta \bm{\chi}
=&  i\mathbf{\bar C}_\theta \cdot   \{  
%-g \mathbf{X}_\mu \times  D_\mu[\mathbf{V}+\mathbf{X}]\mathbf{C}_\omega    
%  \nonumber\\
% &\quad\quad
  g D_\mu[\mathbf{V}]  \mathbf{X}_\mu \times   \mathbf{C}_\omega   
%+  g \mathbf{X}_\mu \times  D_\mu[\mathbf{V}+\mathbf{X}]   \mathbf{C}_\omega 
  \nonumber\\
 &\quad\quad  
+   D_\mu[\mathbf{V}+\mathbf{X}] D_\mu[\mathbf{V}-\mathbf{X}] \mathbf{C}_\omega^\perp - D_\mu[\mathbf{V}+\mathbf{X}] D_\mu[\mathbf{V}-\mathbf{X}] \mathbf{C}_\theta^\perp   \} 
 ,
\end{align}
where we have  used the Leibniz rule for the covariant derivative.
Thus we obtain
\begin{align}
 {\cal L}_{\rm GF+FP}^\theta
%= -i \bm\delta [\mathbf{\bar C}_\theta \cdot  \bm{\chi}]
=&  \mathbf{N}_\theta \cdot D_\mu[\mathbf{V}]\mathbf{X}_\mu
+i\mathbf{\bar C}_\theta^\perp \cdot (g D_\mu[\mathbf{V}]  \mathbf{X}_\mu \times   \mathbf{C}_\omega  )
%\nonumber\\ &
-i\mathbf{\bar C}_\theta^\perp
D_\mu[\mathbf{V}+\mathbf{X}]D_\mu[\mathbf{V}-\mathbf{X}]
    (\mathbf{C}_\theta^\perp-\mathbf{C}_\omega^\perp) .
%\bm{\chi} := D_\mu[\mathbf{V}]\mathbf{X}_\mu .
\label{C26-FP2}
\end{align}
%It is directly shown that (\ref{C26-FP2}) agrees with (\ref{C26-FP1}). 
%[Exercise-14] \marginpar{Ex-14}

\newpage
%%%%%%%%%%%%%%%%%%%%%%%%%%%%%%%%%%%%%%%%%%%%%%%%%%%%%%%%%%%%
%%%%%%%%%%%%%%%%%%%%%%%%%%%%%%%%%%%%%%%%%%%%%%%%%%%%%%%%%%%%
\section{Reformulation of $SU(N)$ Yang-Mills theory}\label{sec:reform-SUN} 
%%%%%%%%%%%%%%%%%%%%%%%%%%%%%%%%%%%%%%%%%%%%%%%%%%%%%%%%%%%%
%%%%%%%%%%%%%%%%%%%%%%%%%%%%%%%%%%%%%%%%%%%%%%%%%%%%%%%%%%%%

In this section, we show how the reformulation given for $SU(2)$ Yang-Mills theory using new variables can be extended to the $SU(N)$ Yang-Mills theory ($N \ge 3$).
The $SU(N)$ gauge group has the maximal torus subgroup $U(1)^{N-1}$ as the Cartan subgroup, i.e., $(N-1)$ Abelian directions in color space.
The Abelian projection  as the partial gauge fixing $SU(N) \rightarrow U(1)^{N-1}$ breaks explicitly the original gauge symmetry and the color symmetry.
Therefore, it seems natural to introduce $N-1$ color fields which will play the role of recovering the color symmetry and defining a gauge-invariant magnetic monopole in $SU(N)$ case. 
%From the viewpoint of reformulating the field theory, it is important to recognize what are the independent degrees of freedom to describe the theory in question. 
In fact, this procedure  has been adopted so far to extend the $SU(2)$ CDGFN decomposition to the $SU(N)$ case \cite{Cho80c,FN99a}. 
For the gauge group  $SU(3)$ with a rank of two, it is convenient to introduce two unit vector fields ${\bf n}_3(x)$ and ${\bf n}_8(x)$. 

Recently, however, it has been pointed out that this procedure is not necessarily a unique starting point to reformulate the $SU(N)$ Yang-Mills theory.  
Indeed, it has been shown \cite{KSM08} that only a single color field is sufficient to reformulate the $SU(N)$ Yang-Mills theory irrespective of the number of color $N$. 
This reformulation with a single color field is called the minimal option, while the conventional option is called the maximal one. 
%In order to obtain a new theory equipollent to the original theory, the number of independent degrees of freedom must be the same in the new theory and the original theory. 
Moreover, we find that the minimal case is indispensable  
in order to understand confinement of quarks in the fundamental representation, which will be shown in the next section based on the non-Abelian Stokes theorem for the Wilson loop operator.

%%%%%%%%%%%%%%%%%%%%%%%%%%%%%%%%%%%%%%%%%%%%%%%%%%%%%%%%%%%%%
%%%%%%%%%%%%%%%%%%%%%%%%%%%%%%%%%%%%%%%%%%%%%%%%%%%%%%%%%%%%%
\subsection{Change of variables}\label{C27-appendix:CFNS}
%%%%%%%%%%%%%%%%%%%%%%%%%%%%%%%%%%%%%%%%%%%%%%%%%%%%%%%%%%%%%
%%%%%%%%%%%%%%%%%%%%%%%%%%%%%%%%%%%%%%%%%%%%%%%%%%%%%%%%%%%%%

\begin{figure}[tbp]
\begin{center}
\includegraphics[scale=0.65]{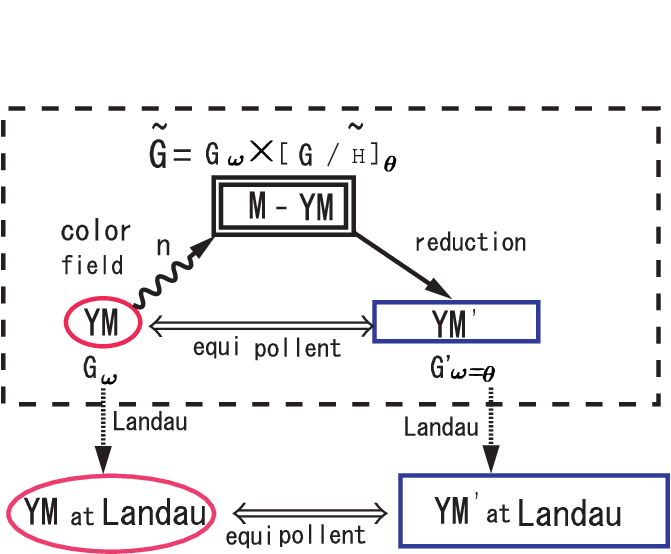}
\caption{\small \cite{KSM08}
The relationship between the original Yang-Mills (YM) theory and the reformulated Yang-Mills (YM') theory.  A single color field $\bm{n}$ is introduced to enlarge the original Yang-Mills theory with a gauge group $G$ into the master Yang-Mills (M-YM) theory  with the enlarged gauge symmetry $\tilde{G}=G \times G/\tilde{H}$.  The reduction conditions are imposed to reduce the master Yang-Mills theory to the reformulated Yang-Mills theory with the equipollent gauge symmetry $G^\prime$. 
In addition, we can impose any over-all gauge fixing condition, e.g., Landau gauge to both the original YM theory and the reformulated YM' theory. 
}
\label{R05-fig:enlarged-YM}
\end{center}
\end{figure}

Our strategy for reformulating the $SU(N)$ Yang-Mills theory is as follows. The schematic representation is given in Fig.~\ref{R05-fig:enlarged-YM}. 
Even for the gauge group $SU(N)$, we introduce a \textit{single} color field $\bm{n}$. 
In the preceding works, a set of $r$ color fields $\bm{n}^{(1)}, \dots, \bm{n}^{(r)}$ is introduced from the beginning for $SU(N)$ Yang-Mills theory where $r=N-1$ is the \textbf{rank} of the group $SU(N)$.%
\footnote{
It was pointed out that the single color field is enough for specifying the Wilson loop operator through the non-Abelian Stokes theorem \cite{KT00}. 
%K.-I. Kondo and Y. Taira,
%Non-Abelian Stokes Theorem and   Quark Confinement in SU(3) Yang-Mills gauge theory, 
%[hep-th/9906129],
%Mod. Phys. Lett. A{\bf 15}, 367--377  (2000). 
%K.-I. Kondo and Y. Taira,
%Non-Abelian Stokes Theorem and  Quark Confinement in SU(N) Yang-Mills gauge theory, 
%[hep-th/9911242],
%Prog. Theor. Phys. {\bf 104}, 1189--1265 (2000).
A  reformulation of the $SU(N)$ Yang-Mills theory based on this idea was completed in \cite{KSM08}.
% [KSM08].
}

Then the original  Yang-Mills theory written in terms of the variable $\mathscr{A}_\mu$ with a gauge group $G=SU(N)$ is extended to a gauge theory called the \textbf{master Yang-Mills theory} with the enlarged gauge symmetry $\tilde G =G_{\mathscr A}\times [G/\tilde{H}]_{\bm{n}}$, where the degrees of freedom $[G/\tilde{H}]_{\bm{n}}$ are possessed by the color field $\bm{n}$. 
Here $\tilde{H}$ is the \textbf{maximal stability subgroup}. 
By imposing a sufficient number of constraints, say, the \textbf{reduction conditions}, 
%\footnote{
%In preceding works, we called the reduction condition the new maximal Abelian gauge (new MAG).  However, this is misleading for $SU(N)$, $N \ge 3$, since the reduction condition does not necessarily reduce to the MA gauge.  Therefore, this terminology should be used no longer.
%}
to eliminate the extra degrees of freedom, 
the master Yang-Mills theory is reduced to the gauge theory  reformulated in terms of new variables  with the gauge symmetry $G^\prime=SU(N)$, say the \textbf{equipollent gauge symmetry},% 
\footnote{
In preceding works, the equipollent gauge transformation was called the gauge transformation II or the active or background gauge transformation. 
}
 which is respected by the new variables. 
 See Fig.~\ref{R05-fig:enlarged-YM}.
\begin{equation}
  G \nearrow (\text{enlargement}) \quad  \tilde G = G_{\mathscr A}\times [G/\tilde{H}]_{\bm{n}} \quad \searrow (\text{reduction}) \quad G^\prime
   .
\end{equation}
The reformulated Yang-Mills theory is written in terms of new variables, i.e., the color field variable $\bm{n}(x)$ and the other new field variables specified later.

%%%%%%%%%%%%%%%%%%%%%%%%%%%%%%%%%%%%%%%%%%%%%%%%%%%%%%%%%%%%

%%%%%%%%%%%%%%%%%%%%%%%%%%%%%%%%%%%%%%%%%%%%%%%%%%%%%%%%%%%%

In the reformulated Yang-Mills theory, the color field $\bm{n}(x)$   is supposed to be given as a functional of the $SU(N)$ Yang-Mills gauge field $\mathscr A_\mu(x)$:
\begin{equation}
 \bm{n}(x) = \bm{n}_{[\mathscr A]}(x) 
\label{C27-nAx}
 .
\end{equation}
In fact, the color field $\bm{n}_{[\mathscr A]}(x)$ is obtained by  solving  the reduction condition.%
\footnote{
This does not means that the solution is unique. 
}
  For the moment, therefore, we do not ask how this is achieved and we omit the subscript $\mathscr A$ of $\bm{n}_{[\mathscr A]}$ to simplify the notation.
Other new field variables are also obtained from the original $\mathscr A_\mu(x)$ by  change of variables thanks to the existence of the color field.

In this reformulation, we require that the original gauge field $\mathscr{A}_\mu(x)$ is decomposed into  two pieces, the restricted field  $\mathscr{V}_\mu(x)$ and the remaining field $\mathscr{X}_\mu(x)$:
\begin{equation}
 \mathscr{A}_\mu(x)
=\mathscr{V}_\mu(x)+\mathscr{X}_\mu(x)
%=\mathscr{C}_\mu(x)+\mathscr{B}_\mu(x)+\mathscr{X}_\mu(x) 
 ,
\end{equation}
such that 
the $\mathscr{V}_\mu(x)$ field transforms under the gauge transformation:  
\begin{align}
  \mathscr{V}_\mu(x) & \rightarrow \mathscr{V}_\mu^\prime(x) = \Omega(x) [\mathscr{V}_\mu(x) + ig^{-1} \partial_\mu] \Omega^{\dagger}(x) 
 ,
 \label{C27-V-ctransf}
\end{align}
 in the same way as the the original gauge field $\mathscr{A}_\mu(x)$:
\begin{align}
  \mathscr{A}_\mu(x) & \rightarrow \mathscr{A}_\mu^\prime(x) = \Omega(x) [\mathscr{A}_\mu(x) + ig^{-1} \partial_\mu] \Omega^{\dagger}(x) 
 . 
\end{align}
Consequently, the $\mathscr{X}_\mu(x)$ field transforms just like  an adjoint matter field:
%\begin{subequations}
\begin{align}
  \mathscr{X}_\mu(x) & \rightarrow \mathscr{X}_\mu^\prime(x) = \Omega(x)  \mathscr{X}_\mu(x) \Omega^{\dagger}(x) 
 \label{C27-X-ctransf}
 . 
\end{align}
Such a decomposition is achieved by way of a \textit{single}  color field $\bm{n}$, which transforms according to the adjoint representation: 
\begin{equation}
 \bm{n}(x)   \rightarrow \bm{n}^\prime(x) = \Omega(x)  \bm{n}(x) \Omega^{\dagger}(x)
 .
\label{C27-n-ctransf}
\end{equation}
We call the field $\bm{n}$ with this property  the \textbf{color field} or the \textbf{color direction field}. 
% \label{C27-gauge-transf}
%\end{subequations}
%The gauge transformation for the new variables is the  {\it equipollent gauge transformation}, which should be compared with the original gauge transformation for $\mathscr{A}_\mu(x)$:

In the following, we consider the decomposition of the Lie algebra valued function $\mathscr F$ into two parts, i.e., an $\tilde{H}$-commutative part $\mathscr F_{\tilde{H}}$  and the remaining part $\mathscr F_{G/\tilde{H}}$:
\begin{align}
  \mathscr F(x) 
:=& \mathscr F_{\tilde{H}}  + \mathscr F_{G/\tilde{H}} ,  
\quad
%\nonumber\\&
\mathscr F_{\tilde{H}}  \in Lie(\tilde{H}), \ 
\mathscr F_{G/\tilde{H}}  \in Lie(G/\tilde{H}), 
\end{align}
where the precise definition of the $\tilde{H}$-commutative part $\mathscr F_{\tilde{H}}$ is given later. 
%\begin{align}
%  [ \mathscr F_{\tilde{H}}(x) , \bm{n}(x)  ] = 0  .
%\end{align}
It is also possible to consider the decomposition using the the maximal torus subgroup $H=U(1)^{N-1}$:
\begin{align}
  \mathscr F 
= \mathscr F_{H}  + \mathscr F_{G/H} 
%:= \mathscr F_\parallel + \mathscr F_\perp 
 , \quad 
\mathscr F_{H} \in Lie(H), \ 
\mathscr F_{G/H} \in Lie(G/H)  
 .
\end{align}
In considering the decomposition, however, it should be remarked that the maximal stability subgroup $\tilde{H}$ of $G$  does not necessarily agree with the maximal torus subgroup $H$ of $G$  for $G=SU(N)$ ($N \ge 3$) except for $N=2$.

Only for $G=SU(2)$, the maximal stability subgroup agrees with the maximal torus subgroup,  $\tilde{H}=H=U(1)$, and the two parts are uniquely specified.
In the $SU(2)$ case, the decomposition is also written as
\begin{align}
  \mathscr F 
%= \mathscr F_{H}  + \mathscr F_{G/H} 
:= \mathscr F_\parallel + \mathscr F_\perp 
 ,  
\end{align}
which corresponds to the well-known decomposition of a three-dimensional vector $\bm{F}$ into the part $\bm{F}_\parallel$ parallel to $\mathbf{n}$  and the  part  $\bm{F}_\perp$ perpendicular to $\mathbf{n}$ as written in the vector form:
\begin{align}
  \bm{F} 
=  \bm{F}_\parallel +  \bm{F}_\perp 
= \mathbf{n} (\mathbf{n}  \cdot \bm{F}) + \mathbf{n}  \times (\bm{F} \times \mathbf{n} )
 ,
\end{align}
which follows from the simple identity,
$
 \mathbf{n}  \times (\mathbf{n}  \times \bm{F}) = \mathbf{n}  (\mathbf{n}  \cdot \bm{F}) - (\mathbf{n} \cdot \mathbf{n} ) \bm{F} 
$
in the vector analysis.

%%%%%%%%%%%%%%%%%%%%%%%%%%%%%%%%%%%%%%%%%%%%%%%%%%
\subsection{General consideration}
\label{subsection:general}
%\setcounter{equation}{0}
%%%%%%%%%%%%%%%%%%%%%%%%%%%%%%%%%%%%%%%%%%%%%%%%%%
%Suppose at each space--time point we have chosen a direction in color space for the gauge group $G$.  
In order to see that a single color field is sufficient for reformulating the $SU(N)$ Yang-Mills theory, we give general considerations. 

Suppose that the color direction at each point of space--time is specified by a field $\bm{n}(x)=\{ n_A(x) \}$ in $d={\rm dim}G$ dimensional color space for the gauge group $G$:
We prepare initially a color field $\bm{n}$:
\begin{equation}
 \bm{n}(x) = n^A(x) T_A \quad   (A=1, \cdots, d={\rm dim}G )   ,
\end{equation}
where its magnitude is fixed as
\begin{equation}
  2{\rm tr}[\bm{n}(x)\bm{n}(x)] = \mathbf{n}(x) \cdot \mathbf{n}(x) = n^A(x) n^A(x) = 1 \ (A=1, \cdots, d)   .
\end{equation}
Suppose that the initial color field $\bm{n}$ satisfies the covariantly-constant   equation:
\begin{equation}
 0 = \mathscr{D}_\mu[\mathscr{V}] \bm{n}(x) := \partial_\mu  \bm{n}(x) -ig [\mathscr{V}(x),\bm{n}(x)]  ,
\end{equation}
which has the the same form as the first defining equation in the $SU(2)$ case.

Then we can construct a new field by the multiplication.  By using the relation (\ref{C27-XY}), indeed, we find a new field $\mathbf{n} * \mathbf{n}$ apart from the unit matrix $\mathbf{1}$:
\begin{align}
  \bm{n} \bm{n} 
=   \frac{1}{2N} (\mathbf{n} \cdot \mathbf{n}) \mathbf{1}+ \frac12 i  (\mathbf{n} \times \mathbf{n})^C T_C + \frac12  (\mathbf{n} * \mathbf{n})^C T_C  
%\nonumber\\
=   \frac{1}{2N}  \mathbf{1}  + \frac12  (\mathbf{n} * \mathbf{n})^C T_C  
 ,
\end{align}
where we have used $\mathbf{n}  \cdot \mathbf{n} =1$ and $\mathbf{n} \times \mathbf{n}=0$.
The multiplication generates the unit matrix $\mathbf{1}$.

The unit matrix $\mathbf{1}$ is a trivial solution of the covariantly-constant equation.  
The resulting field $\bm{n}\bm{n}$ also satisfies the first defining equation, since the Leibniz rule holds for the covariant derivative:
\begin{equation}
 \mathscr{D}_\mu[\mathscr{V}] (\bm{n} \bm{n})= (\mathscr{D}_\mu[\mathscr{V}] \bm{n})  \bm{n} +  \bm{n}  (\mathscr{D}_\mu[\mathscr{V}] \bm{n}) = 0   .
\end{equation}
We can repeat this step and construct other vector fields satisfying the first defining equation by the successive multiplication.
Thus the set of color fields $\{ \mathbf{1}, \bm{n}, \bm{n}\bm{n}, \ldots \}$  satisfying the first defining equation is closed under the multiplications. 

This procedure must terminate at some step, since we eventually reach the situation in which the new field obtained by the multiplication is no longer independent of the fields already obtained, in the sense that the former is written as a linear combination of the latter. 
If we prepare a suitable set of independent bases $\{ \bm{n}^{(k)} \}$ ($k=0,1, \cdots, \#$) with unit length $\mathbf{n}^{(k)} \cdot \mathbf{n}^{(k)}=1$ in color space,  the initial color field $\bm{n}$ is expanded as $\bm{n}=\sum_{k=0}^{\#} c_k \bm{n}^{(k)}$ where we have introduced the normalization: $\bm{n}^{(0)} \propto \mathbf{1}$.  
Then the other fields can also be expanded by these bases.  Therefore, we can write the relationship by introducing a matrix $A$:
\begin{equation}
   (\mathbf{1}, \bm{n}, \bm{n}\bm{n}, \bm{n}\bm{n}\bm{n},\ldots, \overbrace{\bm{n} \cdots \bm{n}}^{\#} ) = (\bm{n}^{(0)}, \bm{n}^{(1)}, \bm{n}^{(2)}, \bm{n}^{(3)}, \dots, \bm{n}^{(\#)} ) A  ,
\end{equation}
By converting this equation, we can obtain a set of independent color fields $\{ \bm{n}^{(1)}, \bm{n}^{(2)}, \bm{n}^{(3)} \cdots \bm{n}^{(\#)} \}$ by starting from the initial color field $\bm{n}$ by multiplication:
\begin{equation}
  (\bm{n}^{(0)}, \bm{n}^{(1)}, \bm{n}^{(2)}, \bm{n}^{(3)}, \dots ,  \bm{n}^{(\#)}) =  (\mathbf{1}, \bm{n}, \bm{n}\bm{n}, \bm{n}\bm{n}\bm{n}, \ldots ,\overbrace{\bm{n} \cdots \bm{n}}^{\#}) A^{-1}   .
\end{equation}
Then all the basis vector fields $\bm{n}^{(k)}(x)$ satisfy the first defining equation by construction:
\begin{equation}
 0 = \mathscr{D}_\mu[\mathscr{V}] \bm{n}^{(k)}(x)   \quad (k=1, \cdots, \#) .
\end{equation}
Explicit examples will be given shortly. 

The color field is Hermitian:
\begin{equation}
 \bm{n}(x)^\dagger = n^A(x) T_A =   \bm{n}(x) ,
\end{equation}
since $n_A(x)$ are real fields and the generator is Hermitian $(T_A)^\dagger=T_A$.
Therefore, it can be diagonalized by using a suitable unitary matrix $U(x)$.
In other words,  a single color field $\bm{n}(x)$ is constructed by choosing an appropriate ($x$-independent) diagonal matrix $T$ and a unitary matrix $U(x)$ (for any representation) such that
\begin{equation}
  \bm{n}(x) = U^\dagger(x) T U(x)   \in G/\tilde{H}, \quad T = {\rm diag}(\Lambda_1, \Lambda_2, \cdots, \Lambda_M)
 ,
\end{equation}
where the matrix $T$ is chosen to  be as simple as possible. 
By using this expressions, we see that the product $\bm{n}  \bm{n} \cdots  \bm{n}$ of the color field $\bm{n}$ can be again a color field (up to the normalization):
\begin{equation}
  \bm{n}  \bm{n} \cdots  \bm{n} 
= (U^\dagger T U) (U^\dagger T U)  \cdots (U^\dagger T U)   
= U^\dagger  (T T \cdots T) U  
 ,
\end{equation}
since the product $T T \cdots T$ of the diagonal matrix $T$ is again a diagonal matrix.
But the diagonal matrix $T T \cdots T$ may have  more degenerate diagonal elements than $T$, as given shortly for explicit examples. 

In particular, we call $\#=N-1$ the \textbf{maximal option}.  In the preceding works on the CFN decomposition for $SU(N)$, ($N \ge 3$), only this case has been discussed.   
However, there are possible other options.%
\footnote{
This fact was discovered in \cite{KSM08}. 
This fact was overlooked in the preceding studies. 
}
 In particular, the \textbf{minimal option} is interesting from a viewpoint of quark confinement. 
The minimal case appears in the case of $\#=1$.  
In the minimal option of $G=SU(N)$, the following relation must be satisfied.%
\footnote{
By using a discrete symmetry, the signature of $T$ can be changed to yield the relation:
\begin{equation}
 \bm{n} \bm{n} =\frac{1}{2N}\mathbf{1} + \frac{N-2}{\sqrt{2N(N-1)}} \bm{n} .
 \nonumber
%\label{C27-eq:Hidentity0}%
\end{equation}
} 
\begin{equation}
 \bm{n} \bm{n} =\frac{1}{2N}\mathbf{1} - \frac{N-2}{\sqrt{2N(N-1)}} \bm{n} .
\label{C27-eq:Hidentity0}%
\end{equation}
This definition of the minimal option works for $SU(3)$. 
However, it is not valid for $N \ge 4$ for $SU(N)$, since $\#=1$ is not restricted to the minimal case.  
The precise definition of the minimal option for general $N$ will be given later. 
Even in the case of $\#=1$, the following case is not the minimal option, although the above procedure of creating new  fields terminates in the second step: 
\begin{align}
  \bm{n} \bm{n} 
=  {\rm const.}  \mathbf{1}   
 .
\end{align}
For example, in $SU(4)$, the maximal stability subgroup 
$\tilde{H}=[U(2) \times U(2)]/U(1)=U(2) \times SU(2)$ is realized for
\begin{equation}
 T =  \frac{1}{2\sqrt{2}}\left(
  \begin{array}{cccc}
   1 & 0 & 0 & 0 \\
   0 & 1 & 0 & 0 \\
   0 & 0 & -1 & 0 \\
   0 & 0 & 0 & -1 \\
 \end{array}
 \right)  
% = \frac{\sqrt{2}}{\sqrt{3}} \frac{\lambda_8}{2} + \frac{1}{\sqrt{3}} \frac{\lambda_{15}}{2}  
% , \quad
% \tilde{H}=\frac{U(2) \times U(2)}{U(1)} = U(2) \times SU(2)
\Longrightarrow  TT = \frac18 \mathbf{1}  
\Longrightarrow   \bm{n} \bm{n} 
= \frac18  \mathbf{1} 
  .
\end{equation}

The precise definition of the maximal and minimal options for $SU(N)$ are given from the viewpoint of degeneracy of the eigenvalues as follows. 
\begin{enumerate}
\item[(I)]
 \underline{Maximal option} in which all eigenvalues $\Lambda_1, \dots,\Lambda_N$ of $\bm{\Lambda}$ are distinct.  The maximal stability subgroup of $\bm{n}$ is $\tilde{H}=U(1)^{N-1}$, and 
  $\bm{n}$ covers the $(N^2-N)$-dimensional internal or target space $SU(N)/U(1)^{N-1}$, i.e., the \textbf{flag space} $F_{N-1}$:
\begin{equation}
\bm{n}  \in SU(N)/U(1)^{N-1}  = F_{N-1}
 .
\end{equation}

\item[(II)]
 \underline{Minimal option} in which $N-1$  out of $N$ eigenvalues  of $\bm{\Lambda}$ are equal. The maximal stability subgroup of $\bm{n}$ becomes $\tilde{H}=U(N-1)$, and  
 $\bm{n}$ covers only the $2(N-1)$-dimensional internal or target space $SU(N)/U(N-1)$, i.e., the \textbf{complex projective space} $P^{N-1}(\mathbb{C})$, which is a submanifold of $F_{N-1}$:
\begin{equation}
 \bm{n}   \in SU(N)/U(N-1)=P^{N-1}(\mathbb{C})  \subset  F_{N-1}
 .
\end{equation}

\end{enumerate}

For $SU(2)$, in particular, we find $\mathbf{n} * \mathbf{n}=0$ due to the absence of $d_{ABC}$ and hence we have
\begin{align}
  \bm{n} \bm{n} 
=  \frac{1}{4}  \mathbf{1}   
 .
\end{align}
Therefore, we obtain just a single color field $\bm{n}(x) = n_A(x) \sigma^A/2$ with three Pauli matrices $\sigma^A$.  There is no distinction between the maximal and the minimal options.

For the minimal option of $SU(N)$, the color field is written in the form: 
\begin{equation}
 \bm{n}_r 
=U^\dagger H_r U \in SU(N)/U(N-1) 
=P^{N-1}(\mathbb{C}) , 
\ {\rm dim}P^{N-1}(\mathbb{C})=2(N-1)
 ,
\end{equation}
using the last Cartan generator $H_r$ $(r=N-1)$.
By using a discrete symmetry as a subgroup of $SU(N)$,  the diagonal elements of the Cartan generator can be exchanged so that the Cartan generator becomes equal to $T_{N^2-1}$ in the \textbf{Gell-Mann basis}:
\begin{align}
 H_r =& \frac{1}{\sqrt{2N(N-1)}} {\rm diag}(\overbrace{1, 1, \cdots, 1}^{N-1}, -(N-1)) 
 .
\end{align}
The $SU(N-1)\times U(1)_{N^2-1}=U(N-1)$ rotations caused by $(N-1)^2$ generators in the Gell-Mann basis:  
\begin{equation}
 \{T_1,T_2,\cdots, T_{(N-1)^2-1},T_{N^2-1} \} ,
\end{equation}
do not change (the direction of) $\bm{n}_r$, as only the $(N-1)^2$  generators commute with  $H_r$. 
Consequently,  
$\bm{n}_r$ runs over only the $(2N-2)$-dimensional internal space $SU(N)/U(N-1)=CP^{N-1} =P^{N-1}(\mathbb{C})$.
Therefore, 
%$\bm{n}_1$ covers the largest space, while 
$\bm{n}_r$ covers the smallest target space. This is the minimal case. 

To obtain the largest internal or target space of the color field $\bm{n}$, i.e.,  the $(N^2-N)$-dimensional the flag space $F_{N-1}=SU(N)/U(1)^{N-1}$, 
we must determine the diagonal matrix $T$ in the expression: 
\begin{equation}
 \bm{n} = U^\dagger T U \in SU(N)/U(1)^{N-1} = F_{N-1},  
\quad  {\rm dim}F_{N-1}=N(N-1) 
 ,
\end{equation}
such that all the generators commuting with $T$ are exhausted  by $r=N-1$ diagonal generators  (Cartan generators)  $H_j$ including itself: 
\begin{equation}
[H_j,H_k]=0 
,
\end{equation}
and the $U(1)^{r}=U(1)^{N-1}$ rotations caused by the $r=N-1$ diagonal generators $H_j$ do not change the direction of $\bm{n}$  for the general unitary transformation $U=\exp(i\sum_{A=1}^{N^2-1} \theta_AT_A)$ specified by the $N^2-1$ generators $T_A$. 
The diagonal matrix $T$ can be constructed by a linear combination of all diagonal generators $H_j$ such that all diagonal elements in $T$ have distinct values. This is the maximal option.

The maximal option occurs when all  $N$ eigenvalues are distinct (no degeneracy), which corresponds to the smallest maximal stability subgroup $\tilde{H}=U(1)^{N-1}$. 
The minimal option occurs when $N-1$ eigenvalues are equal (maximally degenerate), which corresponds to the largest maximal stability subgroup $\tilde{H}=U(N-1)$.
If all the eigenvalues are the same, the color field reduces to the unit matrix, $\bm{n}(x) \propto U^\dagger(x) \mathbf{1} U(x) = \mathbf{1} \propto \bm{n}^{(0)} $.
But this contradicts with the fact ${\rm tr}(\bm{n})=0$. Therefore, this case does not exist. 
%In addition, there are intermediate cases in which some of the $N$ eigenvalues are degenerate.  
%To see the situation concretely, it is better to examine the $SU(4)$ case. See \cite{KSM08}.

For $N \ge 4$, there exist intermediate cases other than the maximal and minimal options. To see this, it is instructive to consider the $SU(4)$ case explicitly.%
\footnote{
See  \cite{KSM08} for details.
}
The $SU(4)$ has the maximal stability subgroups:
\begin{equation}
 \tilde H = U(1) \times U(1) \times U(1),\
 U(1) \times U(2), \
SU(2) \times U(2), \
U(3) . 
\end{equation}
 The above consideration also clarifies the meaning of this type of classification and helps to correct a misunderstanding in the preceding work  regarding the number of independent degrees of freedom.

%%%%%%%%%%%%%%%%%%%%%%%%%%%%%%%%%%%%%%%%%%%%%%%%%%%%%%%%%%%%%
\subsection{Constructing the color direction field for the $SU(3)$ group}
%%%%%%%%%%%%%%%%%%%%%%%%%%%%%%%%%%%%%%%%%%%%%%%%%%%%%%%%%%%%%

We now consider the $SU(3)$ case. 
We consider a real scalar field $\bm{\phi}(x)$ taking its value in the Lie algebra $su(3)$ of $SU(3)$:
\begin{equation}
 \bm{\phi}(x)=\phi_A(x) T_A  , \quad \phi_A(x) \in \mathbb{R}  . 
\ ( A=1, 2, \cdots, 8 )
\end{equation}
Thus, $\bm{\phi}$ is a traceless and Hermitian matrix for real $\phi_A$, i.e., ${\rm tr}(\bm{\phi})=0$ and $\bm{\phi}^\dagger=\bm{\phi}$, since we have chosen   the traceless generators $T_A$ to be Hermitian; ${\rm tr}(T_A)=0$ and $(T_A)^\dagger=T_A$.   The Hermitian matrix $\bm{\phi}$ can be cast into the diagonal form $\bm\Lambda$ using a suitable unitary matrix $U \in SU(3)( \subset U(3))$:
\begin{equation}
  U(x) \bm{\phi}(x) U^\dagger(x) = {\rm diag}(\Lambda_1(x),\Lambda_2(x),\Lambda_3(x)) := \bm{\Lambda}(x)
 ,
 \label{diag}
\end{equation}
with real elements $\Lambda_1,\Lambda_2,\Lambda_3$.
The traceless condition leads to   
\begin{equation}
  {\rm tr}(\bm{\phi}(x)) = {\rm tr}(\bm{\Lambda}(x)) = \Lambda_1(x)+\Lambda_2(x)+\Lambda_3(x)=0 
 .
\end{equation}
The diagonal matrix $\bm{\Lambda}$ is expressed as a linear combination of two diagonal  generators  $H_1$ and $H_2$ belonging to the Cartan subalgebra as
\begin{equation}
  U(x) \bm{\phi}(x) U^\dagger(x) 
%= {\rm diag}(\Lambda_1(x),\Lambda_2(x),\Lambda_3(x)) 
  = a(x) H_1 + b(x) H_2
  .
\end{equation}
From (\ref{diag}), we obtain the relation 
\begin{align}
    2 {\rm tr}(\bm{\phi}^2) 
%=& 2 \phi_A \phi_B {\rm tr}(T_A T_B)
%=& \phi_A \phi_B \delta^{AB} 
=& \phi_A \phi_A  
= \vec{\phi} \cdot \vec{\phi}
 \quad ( A =1, 2, \cdots, 8 ) 
  \nonumber\\
  =& 2  {\rm tr}(\bm{\Lambda}^2) 
  =2 (\Lambda_1^2+\Lambda_2^2+\Lambda_3^2) 
  = a^2+b^2 
 .
\end{align}

At this stage, the color vector field ${\bf n}(x)$ defined by
$\bm{n}(x)= \bm{\phi}(x)/|\bm{\phi}(x)|$
is an  eight-dimensional unit vector, i.e., 
\begin{equation}
 {\bf n}(x) \cdot {\bf n}(x) = n_A(x) n_A(x) =2 {\rm tr}(\bm{n}(x)^2)=  1
   \ ( A=1, 2, \cdots, 8 )
    .
\end{equation}
In other words,  ${\bf n}$ belongs to the 7-sphere, ${\bf n} \in S^7$, or the target space of the map ${\bf n}$ is $S^7$; 
$\bm{n}:\mathbb{R}^D \rightarrow S^7$.
Hence,  $\bm{n}(x)$ has 7 independent degrees of freedom at each $x$.
However, the group $G$ does not act \textbf{transitively}%
\footnote{
We say that the group $G$ acts transitively on the manifold $M$  if any two elements of $M$ are connected by a group transformation.  
} 
 on the manifold of the target space  $S^7$.
If the maximal stability subgroup $\tilde{H}$ of $G$ is nontrivial,%
\footnote{
We define the maximal stability subgroup or stationary subgroup as a subgroup of $G$ that consists of all the group elements $h$ that leave the reference state (or highest weight state) $|\Lambda  \rangle$ invariant up to a phase factor: 
$h |\Lambda \rangle =|\Lambda \rangle  e^{i\phi(h)}$.
}
 then the transitive target space $M$ is identified with the left coset space, $G/\tilde{H}$, where $G/\tilde{H} \subset M$. 
Thus, we have
\begin{equation}
x \in \mathbb{R}^D \rightarrow  \bm{n}(x) := \bm{\phi}(x)/|\bm{\phi}(x)| \in G/\tilde{H} \subset S^7 
  .
\end{equation} 
Then the unit color  field $\bm{n}(x) \in su(3)$
is expressed as 
\begin{align}
&  \bm{n}(x) 
  =  (\cos \vartheta(x))\bm{n}_3(x)  + (\sin \vartheta(x)) \bm{n}_8(x) ,
\quad
\vartheta(x)\in[0,2\pi),
\nonumber\\
& \bm{n}_3 (x)  =  U^\dagger(x)  \frac12\lambda_3 U(x) , \quad 
\bm{n}_8 (x)  = U^\dagger(x)  \frac12\lambda_8 U(x)   
\label{su3-n}
 ,
\end{align}
where  $a^2+b^2=1$ is used to rewrite $a$ and $b$ in terms of an angle $\vartheta$:
$
%\begin{align}
\cos \vartheta(x) = a(x)
%=& 2\Lambda_1(x)+  \Lambda_3(x) =-2\Lambda_2(x)-  \Lambda_3(x) 
$, 
$
%\nonumber\\
\sin \vartheta(x)  = b(x)
%=& - \sqrt{3}  \Lambda_3(x)
 .
%\end{align}
$
Note that $\bm{n}_3(x)$ and $\bm{n}_8(x)$  are  Hermitian and traceless unit fields: 
\begin{equation}
  \bm{n}_j^\dagger(x)=\bm{n}_j(x) , \quad 
  {\rm tr}(\bm{n}_j(x))=0 ,  \quad
 2 {\rm tr}(\bm{n}_j(x)^2)=1 \ (j \in \{3,8\}) 
 .
\end{equation}
This is also the case for the color field $\bm{n}(x)$:
\begin{equation}
  \bm{n}^\dagger(x)=\bm{n}(x) , \quad 
  {\rm tr}(\bm{n}(x))=0 ,  \quad
 2 {\rm tr}(\bm{n}(x)^2)=1 
 .
\end{equation}

In the vector form, there are following relations between $\mathbf{n}_3$ and $\mathbf{n}_8$.
\begin{subequations}
\begin{align}
\mathbf{n}_3\cdot \mathbf{n}_3=& 1,
\quad
\mathbf{n}_3\cdot \mathbf{n}_8= 0,\quad 
\mathbf{n}_8\cdot \mathbf{n}_8=1,
\quad
\label{C27-nSU3pro-a}
\\
\mathbf{n}_3\times \mathbf{n}_3=& 0,\quad 
\mathbf{n}_3\times \mathbf{n}_8= 0, \quad 
\mathbf{n}_8\times \mathbf{n}_8=0 
 ,
\label{C27-nSU3pro-b}
\\
\mathbf{n}_3*{\bf n}_3=& \frac1{\sqrt3} \mathbf{n}_8,
\quad
\mathbf{n}_3*{\bf n}_8= \frac1{\sqrt3} \mathbf{n}_3,
\quad
\mathbf{n}_8*{\bf n}_8=\frac{-1}{\sqrt3} \mathbf{n}_8
 ,
 \label{C27-nSU3pro-c}
\end{align}
 \label{C27-nSU3pro}
\end{subequations}
It should be remarked that $U$ used in the expression for $\bm{n}_3$ and $\bm{n}_8$ is regarded as an element of $SU(3)$ rather than $U(3)$, since a diagonal generator of a unit matrix in $U(3)$ commutes trivially with all the other generators of $U(3)$.
%It is easy to see that the $SU(2)$ case considered in the previous subsection is reproduced in a similar way to the $SU(3)$ case we have just considered. 

%%%%%%%%%%%%%%%%%%%%%%%%%%%%%%%%%%%%%%%%%%%%%%%%%%%%%%%%%%%%%
\subsubsection{Maximal and minimal cases defined by  degeneracies}
%%%%%%%%%%%%%%%%%%%%%%%%%%%%%%%%%%%%%%%%%%%%%%%%%%%%%%%%%%%%%

We now show that the unit color field $\bm{n}$ is classified into two categories, maximal and minimal, according to the degeneracies of the eigenvalues.
By taking into account the fact that the three eigenvalues obey two equations,
\begin{equation}
\Lambda_1(x)+\Lambda_2(x)+\Lambda_3(x)=0 ,
\quad
\Lambda_1(x)^2+\Lambda_2(x)^2+\Lambda_3(x)^2=\frac12 
 ,
\end{equation}
we find that only one degree of freedom is independent.
%, which corresponds to the choice of $\vartheta$.  
Therefore, the category to which  $\bm{n}$ belongs is specified by the value of $\vartheta$.

\begin{enumerate}

\item[(I)]
 Maximal case in which three eigenvalues $\Lambda_1,\Lambda_2,\Lambda_3$ of $\bm\Lambda$ are distinct.  In the maximal case, the maximal stability subgroup of $\bm{n}$ is 
\begin{equation}
 \tilde{H}=U(1) \times U(1)
 ,
\end{equation}
and 
  $\bm{n}$ covers the six-dimensional internal or target space $SU(3)/(U(1)\times U(1))$, i.e., the flag space $F_2$:
\begin{equation}
\bm{n}  \in G/\tilde{H} =SU(3)/(U(1)\times U(1)) = F_2
 .
\end{equation}
The maximal cases are realized for any angle $\vartheta \in [0,2\pi)$ except for 6 values of   $\vartheta$ in the minimal case, i.e., $\vartheta \notin \{ \frac{1}{6}\pi,\frac{1}{2}\pi,\frac{5}{6}\pi,\frac{7}{6}\pi,\frac{3}{2}\pi,\frac{11}{6}\pi \}$.

\item[(II)]
 Minimal case in which two of the three eigenvalues are  equal. In the minimal case, the maximal stability subgroup of $\bm{n}$ becomes 
\begin{equation}
 \tilde{H}=U(2)  
 ,
\end{equation}
and  
 $\bm{n}$ covers only the four dimensional internal or target space $SU(3)/U(2)$, i.e., the complex projective space $P^2(\mathbf{C})$:
\begin{equation}
 \bm{n}   \in G/\tilde{H} = SU(3)/U(2) =P^2(\mathbf{C}) 
 .
\end{equation}

The minimal cases are exhausted by 6 values of $\vartheta$, i.e., 
$\vartheta = \frac{(2n-1)}{6} \pi \ (n=1,2,\cdots,6) 
$ or
$\vartheta \in \{ \frac{1}{6}\pi,\frac{1}{2}\pi,\frac{5}{6}\pi,\frac{7}{6}\pi,\frac{3}{2}\pi,\frac{11}{6}\pi \}$:
\begin{align}
  \bm{n}(x) =& U^\dagger(x)  T U(x) , 
  \nonumber\\
  T  =&   \cos \vartheta  \frac12\lambda_3  +  \sin \vartheta  \frac12\lambda_8  
  \nonumber\\
=& \frac12 {\rm diag} \left(\cos \vartheta +\frac{1}{\sqrt{3}} \sin \vartheta , - \cos \vartheta +\frac{1}{\sqrt{3}} \sin \vartheta , -\frac{2}{\sqrt{3}} \sin \vartheta \right) 
  \nonumber\\
=& \frac{1}{2\sqrt{3}} (2,-1,-1), (1,1,-2), (-1,2,-1), (-2,1,1), (-1,-1,2),(1,-2,1) .
\label{su3-mini-n}
\end{align}

\end{enumerate}

These definitions for maximal and minimal cases based on the degeneracies of eigenvalues are equivalent to those given in the previous subsection based on multiplication properties, as shown in the following. 

This observation suggests that  in the maximal or minimal case  for $SU(3)$, one of the fields, ${\bf n}_3$ or ${\bf n}_8$, is sufficient as a representative of the color field $\bm{n}$  to rewrite   Yang-Mills theory based on change of variables, and that the choice of  ${\bf n}_3$ or ${\bf n}_8$ as a fundamental variable can be used  alternatively to obtain the equivalent reformulation of Yang-Mills theory.  
%This statement has been verified  in \cite{KSM08}.

It should be noted, however, that ${\bf n}_3$ carry 6 degrees of freedom, while ${\bf n}_8$ carry 4 degrees of freedom. Therefore, the other new field variables to be introduced for rewriting Yang-Mills theory must provide the remaining degrees of freedom in each case.   
It should be remarked that for the $SU(2)$ gauge group,  $\bm{n}$ carries always 2 degrees of freedom and there is no distinction between the maximal and minimal cases.

%The maximal case (I) (three eigenvalues are distinct) is realized for any angle $\vartheta$ except for  6 values of the angle $\vartheta$, i.e., $\vartheta \notin \{ \frac{1}{6}\pi,\frac{1}{2}\pi,\frac{5}{6}\pi,\frac{7}{6}\pi,\frac{3}{2}\pi,\frac{11}{6}\pi \}$.
%Whereas the minimal case (II) (two of the three eigenvalues get equal) is exhausted by $\vartheta \in \{ \frac{1}{6}\pi,\frac{1}{2}\pi,\frac{5}{6}\pi,\frac{7}{6}\pi,\frac{3}{2}\pi,\frac{11}{6}\pi \}$:

%%%%%%%%%%%%%%%%%%%%%%%%%%%%%%%%%%%%%%%%%%%%%%%%%%%%%%%%%%%%%
\subsubsection{Minimal case}
%%%%%%%%%%%%%%%%%%%%%%%%%%%%%%%%%%%%%%%%%%%%%%%%%%%%%%%%%%%%%

According to (I) in the above definition, the minimal cases are given by $\vartheta = \frac{(2n-1)}{6} \pi \ (n=1,2,\cdots,6) 
$.

\begin{itemize}
\item $\Lambda_2=\Lambda_3$: 
$\Lambda_2=\Lambda_3=-\frac{1}{2\sqrt{3}} < \Lambda_1= \frac{1}{\sqrt{3}} 
 \Longrightarrow 
 (a,b) =  \left( \frac{\sqrt{3}}{2},  \frac{1}{2}\right) \Longrightarrow  \vartheta=\frac{1}{6}\pi 
$

\item $\Lambda_3=\Lambda_1$: 
$\Lambda_3=\Lambda_1=-\frac{1}{2\sqrt{3}} < \Lambda_2= \frac{1}{\sqrt{3}} 
 \Longrightarrow 
 (a, b) =  \left(-\frac{\sqrt{3}}{2}, \frac{1}{2} \right) \Longrightarrow  \vartheta=\frac{5}{6}\pi
$

\item $\Lambda_1=\Lambda_2$: 
$\Lambda_1=\Lambda_2=-\frac{1}{2\sqrt{3}} < \Lambda_3=\frac{1}{\sqrt{3}} 
 \Longrightarrow 
 (a,b) = (0, -1) \Longrightarrow  \vartheta=\frac{3}{2}\pi
$
\end{itemize}
and their Weyl reflections (rotation by  angle $\pi$) in the weight diagram: 
\begin{itemize}
\item $\Lambda_2=\Lambda_3$: 
$\Lambda_2=\Lambda_3=\frac{1}{2\sqrt{3}} > \Lambda_1=-\frac{1}{\sqrt{3}} 
 \Longrightarrow 
 (a, b) = \left( -\frac{\sqrt{3}}{2},  -\frac{1}{2} \right)  \Longrightarrow  \vartheta=\frac{7}{6}\pi 
$

\item $\Lambda_3=\Lambda_1$: 
$\Lambda_3=\Lambda_1=\frac{1}{2\sqrt{3}} > \Lambda_2=-\frac{1}{\sqrt{3}} 
 \Longrightarrow 
 (a,b)=\left(   \frac{\sqrt{3}}{2},  -\frac{1}{2} \right) \Longrightarrow  \vartheta=\frac{11}{6}\pi
$

\item $\Lambda_1=\Lambda_2$: 
$\Lambda_1=\Lambda_2=\frac{1}{2\sqrt{3}} > \Lambda_3=-\frac{1}{\sqrt{3}} 
 \Longrightarrow 
 (a, b) = (0, 1) \Longrightarrow  \vartheta=\frac{\pi}{2} 
$ .

\end{itemize}
Note that each of three two-dimensional vectors $(a,b)$ is proportional to  a weight vector for the fundamental representations of $3$ and $3^*$ in the weight diagram, 
$(2/\sqrt{3})\Lambda_j$, see, for example \cite{Kondo08}.  The total set is invariant under the action of the Weyl group. 
%The eigenvalues are ordered such that $\Lambda_1 \ge \Lambda_2 \ge \Lambda_3$

In the minimal case, two of the three eigenvalues are  equal. 
In particular, for $(a,b)=(0,\pm1)$ or equivalently $\vartheta=\pi/2, 3\pi/2$, the color field $\bm{n}(x)$ can be written using only $\bm{n}_8(x)$, i.e., $\bm{n}(x)=\pm \bm{n}_8(x)$, and $\bm{n}_3(x)$ disappears from the color field. 
In this case, the matrix $T$ is a diagonal matrix with two distinct eigenvalues: 
\begin{equation}
 T_8 = \frac12 \lambda _{8}=\frac{1}{2\sqrt{3}}\left(
  \begin{array}{cccc}
   1 & 0 & 0   \\
   0 &  1 & 0  \\
   0 & 0 & -2   \\
 \end{array}
 \right)
 .  \label{C27-SU3-inter2-T}
\end{equation}
Therefore, the maximal stability subgroup of $\bm{n}_{8} =U^\dagger T_8 U$ becomes a non-Abelian subgroup:
\begin{equation}
\tilde{H}=\frac{U(2) \times U(1) }{U(1)}
 = U(2) \cong SU(2) \times U(1) .
\end{equation}
Here  $\bm{n} = \bm{n}_{8}$ covers only the four dimensional internal or target space $SU(3)/U(2)$, i.e., the complex projective space $P^2(\mathbf{C})$:
\begin{equation}
 \bm{n} = \bm{n}_{8} =U^\dagger T_8 U   \in G/\tilde{H} = SU(3)/U(2) =P^2(\mathbf{C}) 
 .
\end{equation}
This originates from the fact that the $SU(2)\times U(1) =U(2)$ rotations caused by four generators $\{T_1,T_2,T_3,T_8\}$  do not change (the direction of) $\bm{n}_8^\infty:=T_8$, since the only generators commuting with $T_8$ are  the four generators $\{T_1,T_2,T_3,T_8\}$ specified by the standard Gell-Mann matrices:
\begin{equation}
[T_8,T_A]=0  .
\quad (A=1,2,3,8) .
\end{equation}
Incidentally, in this case, the degeneracy of the matrix does not change by the multiplication:
\begin{align}
 T_8  T_8 
=& \frac14 \lambda _{8} \lambda _{8} =\frac{1}{12}\left(
  \begin{array}{cccc}
   1 & 0 & 0   \\
   0 &  1 & 0  \\
   0 & 0 & 4   \\
 \end{array}
 \right) , \quad
%\nonumber\\ 
 T_8  T_8 T_8 
=  \frac18 \lambda _{8} \lambda _{8} \lambda _{8} 
 =\frac{1}{24\sqrt3}\left(
  \begin{array}{cccc}
   1 & 0 & 0   \\
   0 &  1 & 0  \\
   0 & 0 & -8   \\
 \end{array}
 \right) , \cdots  .
\end{align}

In other minimal cases, $\bm{n}_3(x)$ reappears in the color field $\bm{n}(x)$, and the $\vartheta=\pi/2, 3\pi/2$ cases may appear to be special and distinct from other cases.   
As the structure of the degenerate matrix $\Lambda$ indicates, however, the color field $\bm{n}(x)$ has the same degrees of freedom in all the minimal cases. 
Therefore, the difference is apparent due to the special choice of the Gell-Mann matrices.

Thus, all choices of $\vartheta \in \{ \frac{1}{6}\pi,\frac{1}{2}\pi,\frac{5}{6}\pi,\frac{7}{6}\pi,\frac{3}{2}\pi,\frac{11}{6}\pi \}$ should be treated on  an equal footing, but the apparently simplest way to define the color field in the minimal case is to choose $\vartheta=\pi/2$, i.e., $\bm{n}(x)=\bm{n}_8(x)$. 
Then we can calculate the multiplication according to (\ref{C27-XY}) and (\ref{C27-nSU3pro}):
\begin{align}
  \bm{n}  \bm{n} 
=&  \frac{1}{6} (\mathbf{n}_{8} \cdot \mathbf{n}_{8}) \mathbf{1}+ \frac12 i (\mathbf{n}_{8} \times \mathbf{n}_{8})^C T_C + \frac12 (\mathbf{n}_{8} * \mathbf{n}_{8})^C T_C 
%\nonumber\\
=   \frac{1}{6} \mathbf{1} - \frac12 \frac1{\sqrt3}\bm{n}_8 
 .
\end{align}
Hence, $\bm{n}_{8}$ is just one independent basis.
This is the minimal option for   $SU(3)$.

%%%%%%%%%%%%%%%%%%%%%%%%%%%%%%%%%%%%%%%%%%%%%%%%%%%%%%%%%%%%%
\subsubsection{Maximal case}
%%%%%%%%%%%%%%%%%%%%%%%%%%%%%%%%%%%%%%%%%%%%%%%%%%%%%%%%%%%%%

In particular, for $(a,b)=(\pm 1,0)$ or equivalently $\vartheta=0, \pi$, the color field $\bm{n}(x)$ can be  written using only $\bm{n}_3(x)$, i.e., $\bm{n}(x)=\pm \bm{n}_3(x)$, and $\bm{n}_8(x)$ disappears from the expression of the color field $\bm{n}(x)$. 
In the other maximal cases with $\vartheta \ne 0, \pi$,  $\bm{n}(x)$ contains both $\bm{n}_3(x)$ and $\bm{n}_8(x)$. 
The appearances of the representation $\bm{n}(x)$ are considerably different from each other in the two cases.  However, both reveal  the same physical situation  
because ${\bf n}_8(x)$ is constructed from ${\bf n}_3(x)$ as ${\bf n}_8(x)={\sqrt3} {\bf n}_3(x)*{\bf n}_3(x)$. 
Therefore, it is sufficient for us to consider $\bm{n}_3(x)$ to define the color  field $\bm{n}(x)$. 
%In the maximal case,  three eigenvalues $\Lambda_1,\Lambda_2,\Lambda_3$ of $\bm\Lambda$ are distinct.  
In this case, the matrix $T$ is a diagonal matrix with three distinct eigenvalues: 
\begin{equation}
 T_3 = \frac12 \lambda_{3}=\frac{1}{2}\left(
  \begin{array}{cccc}
   1 & 0 & 0   \\
   0 & -1 & 0   \\
   0 & 0 & 0   \\
 \end{array}
 \right)
  .
  \label{C27-SU3-inter1-T}
\end{equation}
Therefore, the maximal stability subgroup of $\bm{n}_{3}=U^\dagger T_3 U$ is an Abelian group:
\begin{equation}
 \tilde{H}=\frac{U(1) \times U(1) \times  U(1)}{U(1)}
 \cong U(1) \times  U(1) 
 .
\end{equation}
Here 
  $\bm{n} = \bm{n}_{3}$ covers the six-dimensional internal or target space $SU(3)/(U(1)\times U(1))$, i.e., the flag space $F_2$:
%[Exercise-1] \marginpar{Ex-1}
\begin{equation}
\bm{n} = \bm{n}_{3}  =U^\dagger T_3 U \in G/\tilde{H} =SU(3)/(U(1)\times U(1)) = F_2
 .
\end{equation}
This is understood from the fact that the $U(1)\times U(1)$ rotations caused by the two diagonal generators $\{T_3,T_8\}$ do not change (the direction of) $\bm{n}_3^\infty:=T_3$, since  the only generators commuting with $T_3$ are the two generators $\{T_3,T_8\}$:
%among the eight generators $T_A$ for  the general unitary transformation $U=\exp(i\sum_{A=1}^{8} \theta_AT_A)$  
\begin{equation}
[T_3,T_3]=0,\quad
[T_3,T_8]=0 
 \Leftrightarrow [T_3,T_A]=0 \ (A=3,8).
\end{equation}
Thus the easiest way to define the color field in the maximal case is to choose %$\vartheta=0$, i.e., 
$\bm{n}(x)=\bm{n}_3(x)$. 
Of course, this does not prohibit the  introduction of  both $\bm{n}_3(x)$ and $\bm{n}_8(x)$ for convenience. 
\footnote{
In the original approach for decomposing the $SU(3)$ Yang-Mills gauge field \cite{Cho80c,FN99a}, two fields $\bm{n}_3$ and $\bm{n}_8$ are introduced from the beginning as they were fundamental variables.  
Therefore, this option is included in the maximal case. 
}
Incidentally, in this case, the degeneracy of the matrix  changes by the multiplication:
\begin{align}
 T_3 T_3 =& \frac14 \lambda _{3} \lambda _{3} 
=\frac{1}{4}\left(
  \begin{array}{cccc}
   1 & 0 & 0   \\
   0 & 1 & 0   \\
   0 & 0 & 0   \\
 \end{array}
 \right)
  , \
%\nonumber\\
 T_3 T_3  T_3  =  \frac18 \lambda _{3} \lambda _{3} \lambda _{3}
=\frac{1}{8}\left(
  \begin{array}{cccc}
   1 & 0 & 0   \\
   0 & -1 & 0   \\
   0 & 0 & 0   \\
 \end{array}
 \right) = \frac14 T_3 
 .
\end{align}
In the original approach for decomposing the $SU(3)$ Yang-Mills gauge field, two fields $\bm{n}_3$ and $\bm{n}_8$ are introduced from the beginning as they were fundamental variables.  
Therefore, this case is included in the maximal option. 
%\footnote{
%The maximal cases are realized for any angle $\vartheta \in [0,2\pi)$ except for 6 values of   $\vartheta$ in the minimal case, i.e., $\vartheta \notin \{ \frac{1}{6}\pi,\frac{1}{2}\pi,\frac{5}{6}\pi,\frac{7}{6}\pi,\frac{3}{2}\pi,\frac{11}{6}\pi \}$.
%}

%As in the previous approach, we can also introduce another color field $\bm{n}'(x)$ is given by 
%$\bm{n}'(x)=\bm n_8(x)={\sqrt3} \bm n_3(x)*\bm n_3(x)={\sqrt3} \bm n(x)*\bm n(x)$.

%See Appendix \ref{appendix:moreSU3} for more details. 

We can calculate the multiplication according to (\ref{C27-XY}) and (\ref{C27-nSU3pro}):
\begin{align}
  \bm{n}  \bm{n}  
=&  \frac{1}{6} (\mathbf{n}_{3} \cdot \mathbf{n}_{3}) \mathbf{1}+ \frac12 i  (\mathbf{n}_{3} \times \mathbf{n}_{3})^C T_C + \frac12  (\mathbf{n}_{3} * \mathbf{n}_{3})^C T_C  
\nonumber\\
=&  \frac{1}{6}  \mathbf{1}  + \frac12  (\mathbf{n}_{3} * \mathbf{n}_{3})^C T_C  
=  \frac{1}{6}  \mathbf{1}  + \frac12  \frac1{\sqrt3}\bm{ n}_8 
 ,
\end{align}
and
\begin{align}
  \bm{n}  \bm{n} \bm{n} 
=&  \frac{1}{6}  \bm{n}_{3}  + \frac12  \frac1{\sqrt3} \bm{ n}_8 \bm{n}_{3}
\nonumber\\
=& \frac{1}{6}  \bm{n}_{3}  + \frac12 \frac1{\sqrt3} \left[ 
\frac{1}{6} (\mathbf{n}_{3} \cdot \mathbf{n}_{8}) \mathbf{1}+ \frac12 i  (\mathbf{n}_{3} \times \mathbf{n}_{8})^C T_C + \frac12  (\mathbf{n}_{3} * \mathbf{n}_{8})^C T_C\right] 
\nonumber\\
=& \frac{1}{6}  \bm{n}_{3}  + \frac12  \frac1{\sqrt3} \left( 
   \frac12  \frac1{\sqrt3}\bm{n}_3 \right) 
   = \frac14 \bm{n}_3
 .
\end{align}
Thus we find that $\bm{n}_{3}$ and $\bm{n}_{8}$ are independent bases: 
\begin{equation}
   (\mathbf{1}, \bm{n} , \bm{n} \bm{n}  ) = (\mathbf{1}, \bm{n}_{3}, \bm{n}_{8} ) A  ,
 \quad  A =  \left(
 \begin{array}{cccc}
   1 & 0 &  \frac{1}{6}   \\
   0 & 1 & 0   \\
   0 & 0 & \frac12  \frac1{\sqrt3}   \\
 \end{array}
 \right) ,
\end{equation}
which is converted as
\begin{equation}
  (\mathbf{1}, \bm{n}_{3}, \bm{n}_{8} ) =  (\mathbf{1}, \bm{n} , \bm{n} \bm{n}  ) A^{-1}   ,
 \quad  A^{-1} =  \left(
 \begin{array}{cccc}
   1 & 0 & -1/\sqrt3   \\
   0 & 1 & 0   \\
   0 & 0 & 2\sqrt3   \\
 \end{array}
 \right) .
\end{equation}
This is the maximal option for $SU(3)$.

%%%%%%%%%%%%%%%%%%%%%%%%%%%%%%%%%%%%%%%%%%%%%%%%%%%%%%%%%%%%%
%%%%%%%%%%%%%%%%%%%%%%%%%%%%%%%%%%%%%%%%%%%%%%%%%%%%%%%%%%%%%
\subsection{Maximal option for $SU(N)$}
\label{subsection:maximal}%%%%%%%%%%%%%%%%%%%%%%%%%%%%%%%%%%%%%%%%%%%%%%%%%%%%%%%%%%%%%
%%%%%%%%%%%%%%%%%%%%%%%%%%%%%%%%%%%%%%%%%%%%%%%%%%%%%%%%%%%%%

Let $G$ be a gauge group and $\mathscr{G}$ be the Lie algebra of $G$. 
%We use $r$ to denote the rank of the group $G$, i.e., $r:=\text{rank} G$. 
The $G=SU(N)$ has the rank  $r=N-1$ and the Lie algebra $\mathscr{G}=su(N)$.

In the maximal option, we introduce a set of  the Lie algebra $\mathscr{G}$-valued fields:
\begin{align}
 \bm{n}_j(x) =n_j^A(x) T_A \in \mathscr{G} \quad  ( j=1, \cdots, r ) 
 ,
\end{align}
 so that they are orthonormal:
\begin{align}
  \bm{n}_j(x) \cdot \bm{n}_k(x)  &:= 2 {\rm tr}(\bm{n}_j(x)\bm{n}_k(x))
  = n_j^A(x) n_k^A(x) 
= \delta_{jk}, \nonumber\\&
 j,k \in \{ 1, 2, \cdots, r \} 
 ,
 \label{C27-orthonormal}
\end{align}
and they mutually commute: 
\begin{align}
  [ \bm{n}_j(x), \bm{n}_k(x)] = 0, \quad j,k \in \{ 1, 2, \cdots, r \} 
 .
\label{C27-com1}
\end{align}
We call $\bm{n}_j(x)$ the \textbf{color field} or \textbf{color direction field}, which constitutes the maximal set of color fields. 

Such color fields are constructed 
using the adjoint orbit representation:
\begin{align}
  \bm{n}_j(x) = U^{\dagger}(x) H_j U(x) , \quad j \in \{ 1, 2, \cdots, r \} 
 ,
\label{C27-ador}
\end{align}
where $H_j$ are the generators of the Cartan subalgebra of $\mathscr{G}$. 
In fact, the fields $\bm{n}_j(x)$ defined in this way satisfy the ortho-normality condition (\ref{C27-orthonormal}), since
\begin{align}
   \bm{n}_j(x) \cdot \bm{n}_k(x)  &= 2 {\rm tr}(\bm{n}_j(x)\bm{n}_k(x))
=  2 {\rm tr}(U^{\dagger}(x) H_j U(x) U^{\dagger}(x) H_k U(x)) 
\nonumber\\&
=  2 {\rm tr}( H_j  H_k ) 
=  H_j \cdot H_k  = \delta_{jk}
 .
\end{align}
Moreover, the commutativity (\ref{C27-com1}) is satisfied, 
since $H_j$ are the Cartan subalgebra obeying 
\begin{align}
  [ H_j ,H_k ] = 0, \quad j,k \in \{ 1, 2, \cdots, r \} 
 .
\end{align}

%%%%%%%%%%%%%%%%%%%%%%%%%%%%%%%%%%%%%%%%%%%%%%%%%%%%%%%%%%%%%
%%%%%%%%%%%%%%%%%%%%%%%%%%%%%%%%%%%%%%%%%%%%%%%%%%%%%%%%%%%%%
\subsubsection{Defining equation for the decomposition in the maximal option}
\label{subsubsection:decomposition-maximal}%%%%%%%%%%%%%%%%%%%%%%%%%%%%%%%%%%%%%%%%%%%%%%%%%%%%%%%%%%%%%
%%%%%%%%%%%%%%%%%%%%%%%%%%%%%%%%%%%%%%%%%%%%%%%%%%%%%%%%%%%%%

%We decompose the original gauge field  $\mathscr{A}_\mu(x)$ into two pieces $\mathscr{V}_\mu(x)$ and $\mathscr{X}_\mu(x)$:
%\begin{align}
%\mathscr A_\mu(x)  =\mathscr V_\mu(x)   +\mathscr X_\mu(x) ,
%\end{align}
Once a set of color  fields $\bm{n}_j(x)$ satisfying the above properties is given, 
  the respective pieces $\mathscr V_\mu(x)$ and $\mathscr X_\mu(x)$ of the decomposition $\mathscr A_\mu(x)=\mathscr V_\mu(x)+\mathscr X_\mu(x)$ are uniquely determined by imposing the following conditions, which we call the \textbf{defining equation}:

(I) all $\bm{n}_j(x)$ are covariantly constant in the background field $\mathscr{V}_\mu(x)$:
\begin{align}
  0 = \mathscr{D}_\mu[\mathscr{V}] \bm{n}_j(x) 
=\partial_\mu \bm{n}_j(x) -  ig [\mathscr{V}_\mu(x), \bm{n}_j(x)]
\quad (j=1,2, \cdots, r) 
 ,
\label{C27-defVL}
\end{align}

(II)  $\mathscr{X}_\mu(x)$  is orthogonal to all $\bm{n}_j(x)$:
\begin{align}
 0 =  \mathscr{X}_\mu(x) \cdot  \bm{n}_j(x) 
 =& 2{\rm tr}(\mathscr{X}_\mu(x) \bm{n}_j(x) )  = \mathscr{X}_\mu^A(x) n_j^A(x)  
%\nonumber\\&
\quad (j=1,2, \cdots, r) 
\label{C27-defXL}
 . 
\end{align}

The  defining equation (I) follows from 
\\
(I) The single color field $\bm{n}(x)$ is covariantly constant in the background field $\mathscr{V}_\mu(x)$:
\begin{align}
  0 = \mathscr{D}_\mu[\mathscr{V}] \bm{n}(x) 
=\partial_\mu \bm{n}(x) -  ig [\mathscr{V}_\mu(x), \bm{n}(x)] .
\label{C27-defVL-n}
\end{align}
The defining equation (II) is also given by 
\\
(II)  $\mathscr{X}^\mu(x)$  does not have the ${H}$-commutative part $\mathscr{X}^\mu(x)_{{H}}
%:= \left( {\bf 1} -     [\bm{n}_j(x) , [\bm{n}_j(x) ,  \cdot ]] \right)   \mathscr{X}^\mu(x)
:= \mathscr{X}^\mu(x) -  [\bm{n}_j(x) , [\bm{n}_j(x) ,    \mathscr{X}^\mu(x) ]]
$:
\begin{align}
 & 0 =  \mathscr{X}^\mu(x)_{{H}} %:= \left( {\bf 1} -     [\bm{n}_j(x) , [\bm{n}_j(x) ,  \cdot ]] \right) \mathscr{X}^\mu(x)   
%\nonumber\\
  \Longleftrightarrow \mathscr{X}^\mu(x)  =   [\bm{n}_j(x) , [\bm{n}_j(x) , \mathscr{X}^\mu(x) ]]
\label{C27-defXL2a}
 . 
\end{align}
The equivalence between (\ref{C27-defXL}) and (\ref{C27-defXL2a}) immediately follows from the Lemma given below. 

We find that the following identity holds:

\textbf{Lemma}:  
For a given set of the color fields $\bm{n}_j$ $(j=1,\dots,r)$, any $su(N)$   Lie algebra  valued function $\mathscr F$ can be  decomposed into the $H$-commutative part $\mathscr F_H$ and the remaining $G/H$ part $\mathscr F_{G/H}$ with $r$ being the rank of $G$, $r={\rm rank}SU(N)=N-1$: 
% \cite{Shabanov99,Shabanov02}:
%\footnote{
%In fact, the $SU(2)$ version of this identity is well-known in the vector form:
%\begin{align}
%  \bm{V} = \bm{n} (\bm{n} \cdot \bm{V}) + \bm{n} \times (\bm{V} \times \bm{n}) =  \bm{V}_\parallel +  \bm{V}_\perp  ,
%\end{align}
%which follows from a simple identity,
%$
% \bm{n} \times (\bm{n} \times \bm{V}) = \bm{n} (\bm{n} \cdot \bm{V}) - (\bm{n} \cdot \bm{n}) \bm{V}.
%$
%}
%The factor $N$ in (\ref{C27-idv}) is unnecessary. 
\footnote{
The derivation of this identity was given in Appendix~B  of \cite{KSM08}, which is presented in the \ref{section:formulae1}.
%[KSM,2008]\cite{KSM08}. %\ref{C27-appendix:idv}. 
In proving this identity,  we have used the identification (\ref{C27-ador}).
%\begin{align}
% \bm{n}_j (x)  :=  U^\dagger(x)  H_j U(x)  
% ,
%\end{align}
}
\begin{align}
&  \mathscr F 
= \mathscr F_{H} + \mathscr F_{G/H} ,
\quad 
\mathscr F_{H} := \sum_{j=1}^{r}  \bm{n}_j(\bm{n}_j \cdot  \mathscr F) , 
\quad 
\mathscr F_{G/H} :=  \sum_{j=1}^{r}  [\bm{n}_j, [\bm{n}_j, \mathscr F]]  ,
\label{C27-idv-max}
\end{align}
where 
$\mathscr F_{H}$ commutes with all $\bm{n}_j(x)$:
\begin{align}
    [ \bm{n}_k , \mathscr F_{H}  ] = 0 
    \quad (k=1,2, \cdots, r=N-1) 
 ,
\end{align}
while $\mathscr F_{G/H}$ is orthogonal to all $\bm{n}_j(x)$:
\begin{align}
   \bm{n}_k  \cdot  \mathscr F_{G/H}   = 2{\rm tr}[ \bm{n}_k \mathscr F_{G/H}  ] = 0 
    \quad (k=1,2, \cdots, r=N-1) 
 .
\end{align}

It turns out that the first and second defining equations are transformed respectively in a covariant and invariant way under the gauge transformation after imposing the reduction condition to obtain the gauge theory which is equipollent to the original Yang-Mills theory. 
In other words, the defining equations are form-invariant under the gauge transformations, (\ref{C27-V-ctransf}), (\ref{C27-X-ctransf}), and (\ref{C27-n-ctransf}), that is to say, the gauge transformed fields $\mathscr{V}^\prime_\mu(x)$, $\mathscr{X}^\prime_\mu(x)$ and $\bm{n}_j^\prime(x)$ satisfy the same defining equations. Therefore, the decomposition has the same form as the original one after the gauge transformation, which will be confirmed below by using the explicit form (solutions of the defining equations) for $\mathscr{V}_\mu(x)$ and $\mathscr{X}_\mu(x)$.

We proceed to show that the  respective pieces $\mathscr V_\mu(x)$ and $\mathscr X_\mu(x)$ are uniquely determined by solving the defining equation. 
First, we determine the $\mathscr{X}_\mu$ field by solving the defining equation. 
For this purpose, we apply the identity (\ref{C27-idv-max}) to $\mathscr{X}_\mu$:
\begin{align}
 \mathscr{X}_\mu = \sum_{j=1}^{r} (\mathscr{X}_\mu \cdot \bm{n}_j)\bm{n}_j +   \sum_{j=1}^{r}  [\bm{n}_j, [\bm{n}_j, \mathscr{X}_\mu]]
%=  \sum_{j=1}^{r}   [\bm{n}_j, [\bm{n}_j, \mathscr{X}_\mu]]
 .
\end{align}
We use the second defining equation (\ref{C27-defXL}) to obtain   
\begin{align}
 \mathscr{X}_\mu  =  \sum_{j=1}^{r}   [\bm{n}_j, [\bm{n}_j, \mathscr{X}_\mu]]
 .
\end{align}
Then we take into account the first defining equation (\ref{C27-defVL}): 
\begin{align}
 \mathscr{D}_\mu[\mathscr{A}]\bm{n}_j 
 =& \partial_\mu \bm{n}_j - ig[\mathscr{A}_\mu, \bm{n}_j]
%\nonumber\\
=  \mathscr{D}_\mu[\mathscr{V}]\bm{n}_j - ig [\mathscr{X}_\mu, \bm{n}_j] 
%\nonumber\\ 
%=  - ig [\mathscr{X}_\mu, \bm{n}_j] 
=   ig [ \bm{n}_j, \mathscr{X}_\mu] 
 .
\end{align}
Thus, the $\mathscr{X}_\mu(x)$ field is expressed in terms of $\mathscr{A}_\mu(x)$ and $\bm{n}_j(x)$ as
\begin{align}
 \mathscr{X}_\mu(x) = -ig^{-1} \sum_{j=1}^{r}  [\bm{n}_j(x), \mathscr{D}_\mu[\mathscr{A}]\bm{n}_j(x) ]
%= g^{-1}N  (\bm{n}_j \times \mathscr{D}_\mu[\mathscr{A}]\bm{n}_j)   .
 .
\end{align}

Next, the $\mathscr{V}_\mu$ field is expressed in terms of $\mathscr{A}_\mu(x)$ and $\bm{n}_j(x)$: 
\begin{align}
  \mathscr{V}_\mu(x) 
  =& \mathscr{A}_\mu(x)  - \mathscr{X}_\mu(x)  
  \nonumber\\
  =& \mathscr{A}_\mu(x)  + ig^{-1} \sum_{j=1}^{r}  [\bm{n}_j(x), \mathscr{D}_\mu[\mathscr{A}]\bm{n}_j(x) ] 
  \nonumber\\
  =& \mathscr{A}_\mu(x) - \sum_{j=1}^{r}  [\bm{n}_j(x), [ \bm{n}_j(x), \mathscr{A}_\mu(x) ] ]
+ ig^{-1} \sum_{j=1}^{r}    [\bm{n}_j(x) , \partial_\mu  \bm{n}_j(x) ]
 .
\label{C27-defV11}
\end{align}
We now apply the identity (\ref{C27-idv-max}) to $\mathscr{A}_\mu$ to obtain a simpler expression of (\ref{C27-defV11}):
\begin{align}
\mathscr{V}_\mu(x)
=\sum_{j=1}^{r}(\mathscr{A}_\mu(x) \cdot \bm{n}_j(x))
\bm{n}_j(x)+ig^{-1} \sum_{j=1}^{r} [\bm{n}_j(x) , \partial_\mu  \bm{n}_j(x)].
\label{C27-defV}
\end{align}
Thus, $\mathscr{V}_\mu(x)$ and $\mathscr{X}_\mu(x)$ are  written in terms of $\mathscr A_\mu(x)$, once $\bm{n}_j(x)$ are given as a functional of $\mathscr A_\mu(x)$.

We further decompose the background field$\mathscr V_\mu(x)$ into $\mathscr C_\mu(x)$ and $\mathscr B_\mu(x)$:
\begin{align}
 \mathscr V_\mu(x)
 =\mathscr C_\mu(x)
  +\mathscr B_\mu(x)
 ,
\end{align}
where $\mathscr{C}_\mu(x)$ is defined as a part which commutes with all $\bm{n}_j(x)$:
\begin{align}
    [ \mathscr{C}_\mu(x), \bm{n}_j(x) ] = 0 
    \quad (j=1,2, \cdots, r=N-1) 
\label{C27-specC}
 .
\end{align}
Such an $H$-commutative part $\mathscr{C}_\mu(x)$  in  $\mathscr{V}_\mu(x)$ is not determined uniquely from the first defining equation (\ref{C27-defVL}) alone. However, it is determined by imposing the second defining equation as shown above. 
Applying the identity (\ref{C27-idv-max}) to $\mathscr{C}_\mu(x)$ and   taking into account (\ref{C27-specC}), we obtain
\begin{align}
 \mathscr{C}_\mu(x)
 = \sum_{j=1}^{r}  (\mathscr{C}_\mu(x),\bm{n}_j(x)) \bm{n}_j(x) 
% = \sum_{j=1}^{r=N-1}  \bm{n}_j(x)(\bm{n}_j(x), \mathscr{A}_\mu(x)) 
% \\
%    c_\mu^j(x) = (\bm{n}_j(x), \mathscr{C}_\mu(x)) = (\bm{n}_j(x), \mathscr{A}_\mu(x)) 
 . 
\end{align}
If the remaining part $\mathscr{B}_\mu(x)$, which is not $H$-commutative, i.e., $[\mathscr{B}_\mu(x),\bm{n}_j(x)]  \ne 0$, is non-commutative to all $\bm{n}_j(x)$:
\begin{align}
 0 =  \mathscr{B}_\mu(x) \cdot \bm{n}_j(x) 
= 2{\rm tr}(\mathscr{B}_\mu(x) \bm{n}_j(x) )  
\quad (j=1,2, \cdots, r)
\label{C27-defBL}
 ,
\end{align}
then we have   
\begin{align}
   \mathscr{A}_\mu(x) \cdot \bm{n}_j(x) 
=   \mathscr{V}_\mu(x) \cdot \bm{n}_j(x)  
=  \mathscr{C}_\mu(x) \cdot \bm{n}_j(x) 
%:=   c_\mu^j(x) 
 . 
\end{align}
Consequently, the the $H$-commutative part $\mathscr{C}_\mu(x)$  of $\mathscr{V}_\mu(x)$ reads
\begin{align}
 \mathscr{C}_\mu(x)
% = \sum_{j=1}^{r=N-1}  \bm{n}_j(x)(\bm{n}_j(x), \mathscr{C}_\mu(x)) 
 = \sum_{j=1}^{r}  (\mathscr{A}_\mu(x) \cdot \bm{n}_j(x)) \bm{n}_j(x)
% \\
%    c_\mu^j(x) = (\bm{n}_j(x), \mathscr{C}_\mu(x)) = (\bm{n}_j(x), \mathscr{A}_\mu(x)) 
 , 
\end{align}
and the remaining part $\mathscr{B}_\mu(x)$ is determined as
%\footnote{
%The $SU(2)$ version in the vector form reads 
%$\mathbb{B}_\mu(x)=g^{-1}  \partial_\mu  \mathbf{n}(x) \times  \mathbf{n}(x)$.
%} 
\begin{align}
  \mathscr{B}_\mu(x) 
  =   ig^{-1}  \sum_{j=1}^{r} [ \bm{n}_j(x), \partial_\mu  \bm{n}_j(x) ]  
%   =   g^{-1}  (\partial_\mu  \bm{n}_j(x) \times \bm{n}_j(x) )
 .
\label{C27-B}
\end{align}
It is easy to check that this expression indeed satisfies (\ref{C27-defBL}) and that 
%[Exercise-2] \marginpar{Ex-2}
\begin{align}
 \mathscr{D}_\mu[\mathscr{B}] \bm{n}_j(x) 
=\partial_\mu \bm{n}_j(x) -  ig [\mathscr{B}_\mu(x), \bm{n}_j(x)] = 0
\quad (j=1,2, \cdots, r) 
 .
\label{C27-defBL2b}
\end{align}

There are other ways of deriving the same result.
%\footnote{
For example, the same expression for $\mathscr{V}_\mu$ is also obtained by solving the defining equations as follows. 
Taking into account the commutator of the first defining equation (\ref{C27-defVL}) with $\bm{n}_j$, we have 
\begin{align}
  ig^{-1}   [ \bm{n}_j(x), \partial_\mu  \bm{n}_j(x) ] 
=& ig^{-1}    [ \bm{n}_j(x), ig[\mathscr{V}_\mu(x), \bm{n}_j(x)] ] 
%\nonumber\\
=      [  \bm{n}_j(x), [\bm{n}_j(x),\mathscr{V}_\mu(x)]  ] 
%= - \mathscr{V}_\mu(x) +    \bm{n}_j(x)(\bm{n}_j(x), \mathscr{V}_\mu(x))
 .
%\label{C27-defVa}
\end{align}

Then we obtain by applying (\ref{C27-idv-max}) to $\mathscr{V}_\mu(x)$ the relation:
\begin{align}
  \mathscr{V}_\mu(x) 
=  \sum_{j=1}^{r}  (\mathscr{V}_\mu(x) \cdot \bm{n}_j(x)) \bm{n}_j(x) + ig^{-1} \sum_{j=1}^{r}    [ \bm{n}_j(x) , \partial_\mu  \bm{n}_j(x) ] 
 .
%\label{C27-defVaa}
\end{align}
The second defining equation (\ref{C27-defXL}) leads to 
\begin{equation}
 \mathscr{V}_\mu(x) \cdot \bm{n}_j(x)  =  \mathscr{A}_\mu(x) \cdot \bm{n}_j(x) 
 ,
\end{equation}
and hence we arrive at the same result as (\ref{C27-defV}).
%}

Thus, once a full set of color fields $\bm{n}_j(x)$ is given, the original gauge field  has the following decomposition in the Lie algebra form: 
\begin{subequations}
\begin{align}
\mathscr A_\mu(x)
 =& \mathscr V_\mu(x)
  +\mathscr X_\mu(x)
% =\mathscr C_\mu(x) + \mathscr B_\mu(x) + \mathscr X_\mu(x)
%\quad (\mu=0, 1, \cdots, D-1)  
,
\\
 & \mathscr{X}_\mu(x) = -ig^{-1}   \sum_{j=1}^{N-1}  [\bm{n}_j(x), \mathscr{D}_\mu[\mathscr{A}]\bm{n}_j(x) ] ,
\\
  & \mathscr{V}_\mu(x) =   \mathscr{C}_\mu(x) + \mathscr{B}_\mu(x) ,
\\
  & \mathscr{C}_\mu(x) =   \sum_{j=1}^{N-1}  \bm{n}_j(x)  (\bm{n}_j(x) \cdot \mathscr{A}_\mu(x)) 
= \sum_{j=1}^{N-1}  \bm{n}_j(x) c_\mu^j(x)  ,
\\
  & \mathscr{B}_\mu(x) =  
   ig^{-1} \sum_{j=1}^{N-1} [\bm{n}_j(x), \partial_\mu  \bm{n}_j(x)] 
 .
\end{align}
\label{C27-NLCV-maximal}
\end{subequations}
Equivalently, the decomposition (\ref{C27-NLCV-maximal})  is written in the vector form  as 
\begin{subequations}
\begin{align}
  \mathbf{A}_\mu(x) 
= \mathbf{V}_\mu(x)  + \mathbf{X}_\mu(x)  
= \mathbf{C}_\mu(x) + \mathbf{B}_\mu(x) + \mathbf{X}_\mu(x)  
 ,
\end{align}
where each part is expressed in terms of $\mathbf{A}_\mu(x)$ and ${\bf n}_j(x)$ as
\begin{align}
 \mathbf{X}_\mu(x)  =& g^{-1}  \sum_{j=1}^{N-1}  ({\bf n}_j(x) \times D_\mu[\mathbf{A}]{\bf n}_j(x))   
\\
  \mathbf{V}_\mu(x) =&   \mathbf{C}_\mu(x) + \mathbf{B}_\mu(x) ,
\\
  \mathbf{C}_\mu(x) =& \sum_{j=1}^{N-1} ( \mathbf{A}_\mu(x)  \cdot {\bf n}_j(x)) {\bf n}_j(x)  = \sum_{j=1}^{N-1} c_\mu^j(x) {\bf n}_j(x)  ,
\\
  \mathbf{B}_\mu(x) :=&   g^{-1} \sum_{j=1}^{N-1}  (\partial_\mu  {\bf n}_j(x) \times  {\bf n}_j(x)) 
%=   - g^{-1}  \sum_{j=1}^{r=N-1} (\bm{n}_j(x) \times \partial_\mu  \bm{n}_j(x)) 
 .
\label{C27-eqV}
\end{align}
\label{C27-NLCV-maximal2}
\end{subequations}
In what follows, the summation over the index $j$ should be understood when it is repeated, unless otherwise stated.

%There are other ways of deriving the same result \cite{KSM08}.

%%%%%%%%%%%%%%%%%%%%%%%%%%%%%%%%%%%%%%%%%%%%%%%%%%%%%%%%%%%%%
%%%%%%%%%%%%%%%%%%%%%%%%%%%%%%%%%%%%%%%%%%%%%%%%%%%%%%%%%%%%%
\subsubsection{Independent degrees of freedom in the maximal option}
\label{subsubsection:cov-maximal}%%%%%%%%%%%%%%%%%%%%%%%%%%%%%%%%%%%%%%%%%%%%%%%%%%%%%%%%%%%%%
%%%%%%%%%%%%%%%%%%%%%%%%%%%%%%%%%%%%%%%%%%%%%%%%%%%%%%%%%%%%%

Here we count the number of conditions imposed by  the respective defining equation (I) and (II), which is denoted by $\#[({\rm I})]$, or $\#[({\rm II})]$. 
In what follows, we use $\#[ \cdots ]$ to denote the degrees of freedom (d.o.f.) in the expression $[ \cdots ]$:
\begin{equation}
  \#[ \cdots ] :=  {\rm d.o.f.}[ \cdots ] .
\end{equation}\
At first glance, the first defining equation (\ref{C27-defVL}) has   $D \times r \times {\rm dim}(G/H)$ conditions.
However, this is not true except for $r=1$, i.e., $G=SU(2)$. 
This is because a single color field $\bm{n}$ is sufficient in the sense that all the other remaining color fields in the maximal case are constructed by the single color field $\bm{n}(x)$, as already demonstrated in the above.  
Therefore, a single color field $\bm{n}(x)$ is sufficient to specify the decomposition. 
%\footnote{ 
%See section \ref{C27-subsection:maximalSU3} for the explicit form. 
%}
Therefore, the first defining  equation (\ref{C27-defVL}) has $\#[({\rm I})]=D \times {\rm dim}(G/H)$ conditions ($\#[({\rm I})] \ne D \times r \times {\rm dim}(G/H)$), while the second defining equation (\ref{C27-defXL}) has $\#[({\rm II})]=D \times {\rm dim}H$ conditions, since ${\rm dim}H=r$.  
Therefore, the defining equations totally impose  $\#[({\rm I})+({\rm II})]=D \times {\rm dim}G$ conditions for $\#[\mathscr{V}_\mu^A]=D \times {\rm dim}G$ and $\#[\mathscr{X}_\mu^A]=D \times {\rm dim}G$. 
Therefore, the field decomposition $\mathscr{A}_\mu=\mathscr{V}_\mu+\mathscr{X}_\mu$ can be determined by imposing the defining equations, since 
\begin{equation}
\#[\mathscr{A}_\mu^A]=D \times {\rm dim}G 
= \#[\mathscr{V}_\mu^A]+\#[\mathscr{X}_\mu^A]-\#[({\rm I})+({\rm II})]
\end{equation}

We wish to regard the new variables $\bm{n}^a, c_\mu^j, \mathscr{X}_\mu^a$ as those obtained by  the change of variables from the original gauge field: 
\begin{align}
   & \mathscr{A}_\mu^A  \Longrightarrow  ( \bm{n}^a, c_\mu^j, \mathscr{X}_\mu^a )   ,
\nonumber\\ &
(A=1, \cdots, N^2-1; \mu=1, \cdots, D; j=1, \dots, N-1 )
\end{align}
In the maximal case,  the naive counting of independent degrees of freedom is as follows. 
\begin{itemize}
\item
$\mathscr A_\mu^A \in Lie(G)=su(N)  
%\\
\rightarrow \#[\mathscr A_\mu^A]=D \times {\rm dim}G=D(N^2-1)$ ,
\item
$c_\mu^j \in Lie(H)=u(1)+ \cdots +u(1) 
%\\
\rightarrow  \#[c_\mu^j]=D \times {\rm dim}H=D(N-1) 
= \#[\mathscr V_\mu^A]-\#[({\rm I})]$ ,

\item
$\mathscr X_\mu^a \in Lie(G/H)=su(N)-[u(1)+ \cdots +u(1)] 
\\
\rightarrow \#[\mathscr X_\mu^a]%=D(N^2-1)- D(N-1)
=D \times {\rm dim}(G/H)=D(N^2-N)   
= \#[\mathscr X_\mu^A]-\#[({\rm II})]$,
\item
$\bm{n}^a  \in Lie(G/H)=su(N)-[u(1)+ \cdots +u(1)] 
%\\
\rightarrow \#[\bm{n}^a]= {\rm dim}(G/H)%=N^2-1-(N-1)
=N^2-N$ .

\end{itemize}
In the decomposition just given, therefore, there is an issue of mismatch for the independent degrees of freedom. 
In fact, the new variables carry $N^2-N$  extra degrees of freedom after the decomposition.
Therefore, we must eliminate $N^2-N$ extra degrees of freedom. 
For this purpose,  we intend to impose $N^2-N$ constraints. %to eliminate  the extra degrees of freedom. 
We call such constraints the \textbf{reduction condition} in the maximal option.

%$\bm{n}_j^A$ have $N^2-1-(N-1)=N^2-N$ independent components, while $\mathbf{X}_\mu^A$ have $D(N^2-1)-D(N-1)-(N^2-N)=(D-1)(N^2-N)$independent components after imposing the $(N^2-N)$ constraints.
%Indeed$\#[n_j^A]+\#[c_\mu^j]+\#[X_\mu^A]=N^2-N+D(N-1)+(D-1)(N^2-N)=D(N^2-1) = \#[\mathscr A_\mu^A]$.

%%%%%%%%%%%%%%%%%%%%%%%%%%%%%%%%%%%%%%%%%%%%%%%%%%%%%%%%%%%%%
%%%%%%%%%%%%%%%%%%%%%%%%%%%%%%%%%%%%%%%%%%%%%%%%%%%%%%%%%%%%%
\subsubsection{Reduction condition in the maximal option}
\label{subsubsection:reduction-maximal}%%%%%%%%%%%%%%%%%%%%%%%%%%%%%%%%%%%%%%%%%%%%%%%%%%%%%%%%%%%%%
%%%%%%%%%%%%%%%%%%%%%%%%%%%%%%%%%%%%%%%%%%%%%%%%%%%%%%%%%%%%%

The transformation properties of the decomposed fields $\mathscr{B}_\mu, \mathscr{C}_\mu, \mathscr{X}_\mu$ are uniquely determined, once we specify those for $\mathscr{A}_\mu$ and $\bm{n}$,   as in the $SU(2)$ case discussed in the previous chapter.  
We consider the  infinitesimal version of the enlarged gauge transformation $\delta_{\omega,\theta}$,  which is obtained by combining the  local transformations for $\delta_\omega\mathscr{A}_\mu$ and $\delta_\theta \bm{n}$ 
in the Lie algebra form:
\begin{align}
\delta_\omega  \mathscr{A}_\mu(x)=& \mathscr{D}_\mu[\mathscr{A}]\mathscr{\omega}(x),\ \mathscr{\omega} \in Lie(G) ,
\nonumber\\
\delta_\theta \bm{n}(x) =& ig [ \theta(x) , \bm{n}(x)]
 , \  \theta \in Lie(G/H) .
\end{align}
The color field $\bm{n}$ can be taken to be a linear combinations of $\bm{n}_j$:
\begin{align}
  \bm{n} =  \sum_{j=1}^{r} c^j \bm{n}_j \in   Lie(G/H) .
\end{align} 
In particular, $\bm{n}$ can be chosen to be one of $\bm{n}_j$.
The Lie algebra valued $\omega$ and $\theta$ are decomposed as 
\begin{align}
  \omega =&  \omega_{G/H}  +   \omega_{H}  , \quad  \omega_{G/H}  \in  Lie(G/H), \   \omega_{H} = \sum_{j} \omega_\parallel^j \bm{n}_j \in   Lie(H) ,
\nonumber\\
  \theta =& \theta_{G/H}  +   \theta_{H}  , \quad  \theta_{G/H}  \in  Lie(G/H), \    \theta_{H} = \sum_{j} \theta_\parallel^j \bm{n}_j \in   Lie(H) .
\end{align} 
In what follows, we use the notation:
\begin{align}
  \omega_\perp :=&  \omega_{G/H}  \in  Lie(G/H), \quad     \omega_\parallel :=  \omega_{H}   \in   Lie(H) ,
\nonumber\\
  \theta_\perp :=&   \theta_{G/H}  \in  Lie(G/H), \quad
   \theta_\parallel :=  \theta_{H}  \in   Lie(H) .
\end{align} 
With this understanding, we use the vector notation: 
\begin{align}
\delta_\omega {\mathbf A}_\mu(x)=& D_\mu[\mathbf A]\bm\omega(x), %\ \bm\omega \in Lie(G)
%\nonumber\\
\quad
\delta_\theta {\bf n}(x) =  g{\bf n}(x) \times\bm\theta_\perp(x) 
 . %\ \bm\theta_\perp \in Lie(G/H) .
\label{C27-egt1}
\end{align}

%We propose a constraint,  which we call the reduction condition, as follows.
To find the reduction condition for $G=SU(N)$, we calculate the $SU(N)$ version of the squared  $\mathbf X^\mu$  as suggested from the $SU(2)$ case: 
%[Exercise-3] \marginpar{Ex-3}
\begin{align}
 g^2  \mathbf X^\mu\cdot\mathbf X_\mu
 &=\{ {\bf n}_j \times  D_\mu[\mathbf A]{\bf n}_j  \} \cdot \{ {\bf n}_k \times  D_\mu[\mathbf A]{\bf n}_k  \}
   \nonumber\\
 &= D_\mu[\mathbf A]{\bf n}_k \cdot [({\bf n}_j \times  D_\mu[\mathbf A]{\bf n}_j) \times {\bf n}_k]  
\nonumber\\
 &=  D_\mu[\mathbf A]{\bf n}_k \cdot [({\bf n}_k \times  D_\mu[\mathbf A]{\bf n}_j) \times {\bf n}_j]   
\nonumber\\
 &= D_\mu[\mathbf A]{\bf n}_k \cdot [({\bf n}_j \times  D_\mu[\mathbf A]{\bf n}_k) \times {\bf n}_j]
\nonumber\\
 &= D_\mu[\mathbf A]{\bf n}_k \cdot [D_\mu[\mathbf A]{\bf n}_k - ({\bf n}_j \cdot  D_\mu[\mathbf A]{\bf n}_k) {\bf n}_j]
\nonumber\\
 &= (D_\mu[\mathbf A]{\bf n}_j)^2
 ,
 \label{C27-X^2-max}
\end{align}
where 
%the summation is over $j,k$ and 
 we have used the Jacobi identity $({\bf n}_j \times  D_\mu[\mathbf A]{\bf n}_j) \times {\bf n}_k + (D_\mu[\mathbf A]{\bf n}_j \times {\bf n}_k  ) \times {\bf n}_j +(\mathbf{n}_k \times \mathbf{n}_j) \times D_\mu[\mathbf A]{\bf n}_j=0$ and the fact $\mathbf{n}_k \times \mathbf{n}_j=0$ due to $[\bm{n}_k , \bm{n}_j]=0$ in the third equality,  $0=D_\mu[\mathbf A](\mathbf{n}_k \times \mathbf{n}_j)=D_\mu[\mathbf A] \mathbf{n}_k   \times \mathbf{n}_j+ \mathbf{n}_k \times D_\mu[\mathbf A] \mathbf{n}_j$ in the fourth equality, $\mathbf{n}_j \cdot \mathbf{n}_j=1$ in the fifth equality and ${\bf n}_j \cdot D_\mu[\mathbf A]{\bf n}_k=0$ in the last step. 

We show that a reduction condition is obtained by minimizing the following functional under the enlarged gauge transformation.%
\footnote{
The summation over $j$ is supposed in the definition of $R$ according to the preceding works.
However, it turns out that the following arguments hold for the functional $R$ without summing over $j$, even for a single   $j$. 
Therefore, we can identify  such $\bm{n}_j$ with the color field $\bm{n}$: $\bm{n}(x) \equiv \bm{n}_j(x)$. 
For $SU(3)$ case, we choose $\bm{n}(x) \equiv \bm{n}_3(x)$. 
See \cite{KSM08}. 
}  
\begin{equation}
 R[\mathbf A, \{ \bm{n}_j \}]
 :=   \int d^Dx \frac12 (D_\mu[\mathbf A] {\bf n}_j)^2
  .
  \label{C27-F-max}
\end{equation}
In fact, the transformation of the integrand 
$
(D_\mu[\mathbf A]\bm{n}_j)^2
$
under the infinitesimal enlarged gauge transformation is calculated as
%[Exercise-4] \marginpar{Ex-4}
\begin{align}
  \delta_{\omega,\theta} \left\{ \frac12(D_\mu[\mathbf A]{\bf n}_j)^2 \right\} 
%\nonumber\\
 &=(D_\mu[\mathbf A]{\bf n}_j)\cdot\delta_{\omega,\theta} (D_\mu[\mathbf A]{\bf n}_j)
   \nonumber\\
 &=(D_\mu[\mathbf A]{\bf n}_j)\cdot
   \{D_\mu[\mathbf A]\delta_{\omega,\theta} {\bf n}_j
     +g\delta_{\omega,\theta} \mathbf A_\mu\times{\bf n}_j\}
   \nonumber\\
 &=(D_\mu[\mathbf  A]{\bf n}_j)\cdot
   \{D_\mu[\mathbf A](g{\bf n}_j\times\bm\theta_\perp)
     +gD_\mu[\mathbf A]\bm\omega\times{\bf n}_j\}
   \nonumber\\
 &=(D_\mu[\mathbf A]{\bf n}_j)\cdot
   \{ g(D_\mu[\mathbf A]{\bf n}_j) \times\bm\theta_\perp
   + g{\bf n}_j\times D_\mu[\mathbf A]\bm\theta_\perp
     +gD_\mu[\mathbf A]\bm\omega\times{\bf n}_j\}
   \nonumber\\
 &=g(D_\mu[\mathbf A]{\bf n}_j)\cdot
   \{D_\mu[\mathbf A](\bm\omega-\bm\theta_\perp)\times{\bf n}_j\}
   \nonumber\\
 &=g(D_\mu[\mathbf A]{\bf n}_j)\cdot
   \{D_\mu[\mathbf  A](\bm\omega_\perp-\bm\theta_\perp)\times{\bf n}_j\} 
   + g(D_\mu[\mathbf A]{\bf n}_j)\cdot
   (D_\mu[\mathbf  A] \bm\omega_\parallel \times{\bf n}_j) 
   \nonumber\\
 &=g({\bf n}_j\times D_\mu[\mathbf A]{\bf n}_j)\cdot
   D_\mu[\mathbf A](\bm\omega_\perp-\bm\theta_\perp) 
 ,
 \label{C27-Dn-2-max}
\end{align}
where we have used 
$D_\mu[\mathbf A]\bm\omega_\parallel
=D_\mu[\mathbf A](\omega_\parallel^k {\bf n}_k)
= \partial_\mu \omega_\parallel^k {\bf n}_k + \omega_\parallel^k (D_\mu[\mathbf A] {\bf n}_k)
$,
$\mathbf{n}_k \times \mathbf{n}_j=0$ and $D_\mu[\mathbf A] \mathbf{n}_k   \times \mathbf{n}_j= -  \mathbf{n}_k \times D_\mu[\mathbf A] \mathbf{n}_j$ following from $0=D_\mu[\mathbf A](\mathbf{n}_k \times \mathbf{n}_j)$
 to obtain the last equality.

Therefore, $(D_\mu[\mathbf A]{\bf n}_j)^2$ is invariant under the the subset $\bm\omega_\perp=\bm\theta_\perp$ of the enlarged gauge transformation (\ref{C27-egt1}).
The infinitesimal variation of the functional is
%[Exercise-5] \marginpar{Ex-5}
\begin{align}
\delta_{\omega,\theta} R[\mathbf A, \{ {\bf n}_j \}]
&=\int d^Dx  g({\bf n}_j\times D_\mu[\mathbf A]{\bf n}_j)\cdot
   D_\mu[\mathbf A](\bm\omega_\perp-\bm\theta_\perp) 
   \nonumber\\
 &=- \int d^Dx (\bm\omega_\perp-\bm\theta_\perp) \cdot D_\mu[\mathbf A] g({\bf n}_j\times D_\mu[\mathbf A]{\bf n}_j)
   \nonumber\\
&=- \int d^Dx (\bm\omega_\perp-\bm\theta_\perp) \cdot  g({\bf n}_j\times D_\mu[\mathbf A]D_\mu[\mathbf A]{\bf n}_j)
 .
 \label{C27-dR-max}
\end{align}
Thus, for $\bm\omega_\perp \not= \bm\theta_\perp$, we obtain the differential form of the reduction condition:
\begin{equation}
g \bm\chi[ \mathbf A,{\bf n} ]
 :={\bf n}_j\times D_\mu[\mathbf A]D_\mu[\mathbf A]{\bf n}_j
 \equiv0 
  .
\label{C27-eq:nMAG_minimal_diff}
\end{equation}
Using the Leibniz rule for the covariant derivative $D_\mu[\mathbf A]$, we find
\begin{equation}
g \bm\chi[ \mathbf A,{\bf n} ]
 ={\bf n}_j\times D_\mu[\mathbf A]D_\mu[\mathbf A]{\bf n}_j
 =  D_\mu[\mathbf A] \{ {\bf n}_j \times D_\mu[\mathbf A]{\bf n}_j \}
  .
\end{equation}
Hence, the differential reduction condition can  also be expressed in terms of  $\mathbf V_\mu$   and $\mathbf X_\mu$ in the vector form:
\begin{equation}
\bm\chi[{\bf n},\mathbf C,\mathbf X]
 :=D^\mu[\mathbf V]\mathbf X_\mu
 \equiv0
 ,
\end{equation}
or in the Lie algebra form:
\begin{align}
0= \bm{\chi}[\mathscr A ,\bm{n}] 
&:= \mathscr{D}_\mu[\mathscr{V}] \mathscr{X}_\mu(x) = 
 \partial_\mu \mathscr{X}_\mu(x) -  ig [\mathscr{V}_\mu(x), \mathscr{X}_\mu(x)]
\nonumber\\&
=\partial_\mu \mathscr{X}_\mu - ig c_\mu^j [\bm{n}_j, \mathscr{X}_\mu] 
-  [[ \partial_\mu \bm{n}_j, \bm{n}_j], \mathscr{X}_\mu]  
 .
\end{align}
Note that 
\begin{equation}
 \bm\chi \in Lie(SU(N)/U(1)^r),
\end{equation}
and the number of conditions for $\bm{\chi}=(\chi^A) =0$ $(A=1, \cdots, N^2-1)$ is  $N^2-1-(N-1)=N^2-N$ as expected, 
since $\bm\chi$ is subject to  $N-1$ orthogonality conditions: 
\begin{equation}
 \bm{n}_j(x) \cdot  \bm\chi(x) 
=  \bm{n}_j(x) \cdot \mathscr{D}_\mu[\mathscr{V}] \mathscr{X}_\mu(x) 
= 0 \ (j=1, \cdots, r=N-1)  
 . 
\end{equation}
This  follows from the defining equations (\ref{C27-defXL}) and (\ref{C27-defVL}) as 
\begin{align}
  \bm{n}_j \cdot \mathscr{D}_\mu[\mathscr{V}] \mathscr{X}_\mu   
&=   \bm{n}_j \cdot \partial_\mu \mathscr{X}_\mu   
-ig  \bm{n}_j  \cdot  [\mathscr{V}_\mu , \mathscr{X}_\mu ]  
\nonumber\\&
= \partial_\mu  \bm{n}_j  \cdot  \mathscr{X}_\mu  
-  \partial_\mu \bm{n}_j \cdot  \mathscr{X}_\mu   
-ig  \bm{n}_j  \cdot  [\mathscr{V}_\mu , \mathscr{X}_\mu ] 
\nonumber\\&
= -ig  [\mathscr{V}_\mu , \bm{n}_j]  \cdot  \mathscr{X}_\mu  
-ig  \bm{n}_j  \cdot  [\mathscr{V}_\mu , \mathscr{X}_\mu ]  
= 0 
 .
\end{align}
By solving the differential reduction condition for a given $\mathbf A_\mu(x)$, the color field $\mathbf{n}(x)$ is  in principle obtained, thereby, ${\bf n}(x)$ is obtained as a functional of the original gauge field $\mathbf A_\mu(x)$.

There may be other choices for the functional to be minimized. 
For $SU(3)$, we can adopt the  functional written in terms of a single color field:
\begin{equation}
  F[\mathbf A,{\bf n}_3] = \int d^Dx \frac12 (D_\mu[\mathbf A]{\bf n}_3(x))^2
   .
\end{equation}
This  is a new option overlooked so far.  
We can repeat the same calculations as those given for (\ref{C27-F-max}).  
Then we obtain the differential form of the reduction condition: 
$
 \bm{\chi}_{\rm rc}[\mathbf A,{\bf n}_3]
 = {\bf n}_3 \times D_\mu[\mathbf A] D_\mu[\mathbf A]{\bf n}_3 
$. 
However, this cannot be rewritten into such a simple form as $D^\mu[\mathbf V]\mathbf X_\mu = 0$.

%%%%%%%%%%%%%%%%%%%%%%%%%%%%%%%%%%%%%%%%%%%%%%%%%%%%%%%%%%%%%
%%%%%%%%%%%%%%%%%%%%%%%%%%%%%%%%%%%%%%%%%%%%%%%%%%%%%%%%%%%%%
\subsection{Minimal option for $SU(N)$}
\label{subsection:minimal}%%%%%%%%%%%%%%%%%%%%%%%%%%%%%%%%%%%%%%%%%%%%%%%%%%%%%%%%%%%%%
%%%%%%%%%%%%%%%%%%%%%%%%%%%%%%%%%%%%%%%%%%%%%%%%%%%%%%%%%%%%%

Now we consider the minimal option. 
 In this option,  $\mathscr A_\mu$ is decomposed into 
$\mathscr{V}_\mu(x)$ and $\mathscr{X}_\mu(x)$, i.e.,  
$
 \mathscr{A}_\mu(x) = \mathscr{V}_\mu(x) + \mathscr{X}_\mu(x) 
$, 
using  only a single color  field $\bm h(x)$ without other  fields $\bm{n}_j(x)$.

The color field $\bm h(x)$ has the value in the Lie algebra $Lie(\tilde{H})$ such that 
  it is normalized:
\begin{align}
   \bm h(x) \cdot \bm h(x)  &= 2 {\rm tr}(\bm h(x)\bm h(x))
  =   h^A(x)  h^A(x) 
= 1 , 
 \label{C27-orthonormal2}
\end{align}
and that it commutes with an arbitrary element $\tilde{{\bm h}} \in Lie(\tilde{H})$: 
\begin{align}
  [ \bm h(x), \tilde{{\bm h}}(x) ] = 0   
 .
\label{C27-com2}
\end{align}

%Here we require that $\mathscr{V}_\mu(x)$ and $\mathscr{X}_\mu(x)$ are expressed in terms of $\mathscr{A}_\mu(x)$ and $\bm h(x)$ so as to obey the expected transformation property 
%\begin{align}
% \delta_\omega \mathscr{V}_\mu(x) = D_\mu[\mathscr V] \bm \omega^\prime(x)  , \quad
% \delta_\omega \mathscr{X}_\mu(x) = -ig[\mathscr{X}_\mu(x) , \bm \omega^\prime(x)]
%,
%\end{align}
%for the given gauge transformations of $\mathscr{A}_\mu(x)$ and $\bm h(x)$.
%\begin{align}
% \delta_\omega \mathscr{A}_\mu(x) =  D_\mu[\mathscr A] \bm \omega^\prime(x) , \quad 
% \delta_\omega \bm h(x) = -ig[\bm h(x), \bm \omega^\prime(x)] 
%,
%\end{align}

%%%%%%%%%%%%%%%%%%%%%%%%%%%%%%%%%%%%%%%%%%%%%%%%%%%%%%%%%%%%%
%%%%%%%%%%%%%%%%%%%%%%%%%%%%%%%%%%%%%%%%%%%%%%%%%%%%%%%%%%%%%
\subsubsection{Defining equation for the decomposition in the minimal option}
\label{subsubsection:decomposition-minimal}%%%%%%%%%%%%%%%%%%%%%%%%%%%%%%%%%%%%%%%%%%%%%%%%%%%%%%%%%%%%%
%%%%%%%%%%%%%%%%%%%%%%%%%%%%%%%%%%%%%%%%%%%%%%%%%%%%%%%%%%%%%

The decomposed fields $\mathscr V_\mu(x)$ and $\mathscr X_\mu(x)$ are determined by solving the  defining equation:

\noindent
(I)  $\bm{h}(x)$ is a covariant constant in the background field $\mathscr{V}_\mu(x)$:
\begin{align}
  0 = \mathscr{D}_\mu[\mathscr{V}] \bm{h}(x) 
=\partial_\mu \bm{h}(x) -  ig [\mathscr{V}_\mu(x), \bm{h}(x)]
 ;
\label{C27-defVL2}
\end{align}
(II)  $\mathscr{X}^\mu(x)$  does not have the $\tilde{H}$-commutative part, i.e., $\mathscr{X}^\mu(x)_{\tilde{H}}=0$:
\begin{align}
 & 0 =  \mathscr{X}^\mu(x)_{\tilde{H}} := \left( {\bf 1} -   2\frac{N-1}{N}  [\bm{h}(x) , [\bm{h}(x) ,  \cdot ]]
\right) \mathscr{X}^\mu(x)   
%\nonumber\\
  \Longleftrightarrow \mathscr{X}^\mu(x)  = 2\frac{N-1}{N}  [\bm{h}(x) , [\bm{h}(x) , \mathscr{X}^\mu(x)  ]]
\label{C27-defXL2}
 . 
\end{align}
Two defining equations are transformed in a covariant way under the gauge transformation, after imposing the reduction condition to obtain the gauge theory which is equipollent to the original Yang-Mills theory. 
In other words, the gauge transformed fields $\mathscr{V}^\prime_\mu(x)$, $\mathscr{X}^\prime_\mu(x)$ and $\bm h^\prime(x)$ satisfy the same defining equations. Therefore, the decomposition has the same form as the original one after the gauge transformation, which will be explicitly confirmed below.

Note that  condition (II) is different from the orthogonality to $\bm{h}(x)$:
\\
(II')  $\mathscr{X}_\mu(x)$  is orthogonal to $\bm{h}(x)$:
\begin{align}
  0 =  \mathscr{X}_\mu(x) \cdot \bm{h}(x) 
 = 2{\rm tr}(\mathscr{X}_\mu(x) \bm{h}(x) ) 
 = \mathscr{X}_\mu^A(x) \bm{h}^A(x)  ,
\label{C27-defXL2'}
\end{align}
which is not sufficient for characterizing the $\tilde{H}$-commutative part, in contrast to the $H$-commutative part in the maximal option.%
\footnote{
Note that (\ref{C27-defXL2'}) follows from (\ref{C27-defXL2}). 
$
{\rm tr}( \bm{h}(x) \mathscr{X}^\mu(x))  
= 2\frac{N-1}{N}  {\rm tr}(\bm{h}(x)[\bm{h} , [\bm{h} , \mathscr{X}^\mu(x)  ]])
= 2\frac{N-1}{N}  {\rm tr}( [\bm{h}(x) ,\bm{h} ], [\bm{h} , \mathscr{X}^\mu(x)  ] )=0
$.
But the converse is not true. 
}
This is understood from an identity used in the minimal case:

\textbf{Lemma}:
For a given color field $\bm{h} \in su(N) -  u(N-1)$, any  $su(N)$  Lie algebra valued function $\mathscr{F}(x)$ is decomposed into the $\tilde{H}$-commutative part $\mathscr{F}_{\tilde{H}}$ and the remaining part $\mathscr{F}_{G/\tilde{H}}$ as 
%where $\bm{h} (\bm{h},\mathscr{F})=2{\rm tr}(\mathscr{F} \bm{h})\bm{h} $
\begin{align}
  \mathscr{F} 
%=  \sum_{A=1}^{N^2-1} \mathscr{F}^A T_A
=&   \mathscr{F}_{\tilde{H}} + \mathscr{F}_{G/\tilde{H}} \in su(N)
, 
\label{C27-idv2-min}
\end{align}
where  $\mathscr{F}_{\tilde{H}}$ and  $\mathscr{F}_{G/\tilde{H}}$ satisfy 
%$(\bm{h},\mathscr{F}_{G/\tilde{H}})=2{\rm tr}(\bm{h}\mathscr{F}_{G/\tilde{H}} )=0$ and   $\left[\bm{h},\mathscr{F}_{\tilde{H}} \right] =0$.
\begin{align}
  \left[\bm{h},\mathscr{F}_{\tilde{H}} \right] =  0 ,
\quad 
   \bm{h} \cdot \mathscr{F}_{G/\tilde{H}} :=  2{\rm tr}(\bm{h}\mathscr{F}_{G/\tilde{H}} )=0 .
\end{align}
in the sense that $\mathscr{F}_{G/\tilde{H}}$ part is specified for a given color field $\bm{h}$ by 
\begin{align}
 \mathscr{F}_{G/\tilde{H}} =  \frac{2(N-1)}{N}  [\bm{h} , [\bm{h} , \mathscr{F}]] \in su(N) -  u(N-1)
% := v_\parallel + v_\perp
 ,
\label{C27-idv2-min-2}
\end{align}
and $\mathscr{F}_{\tilde{H}}$ part is the remainder:
\begin{align}
  \mathscr{F}_{\tilde{H}} 
%=  \sum_{A=1}^{N^2-1} \mathscr{F}^A T_A
=&  \mathscr{F}  -  \mathscr{F}_{G/\tilde{H}} \in u(N-1)
.
\end{align}
This identity shows that the $\tilde{H}$-commutative part is not necessarily written in a form that is proportional to $\bm{h}$ and there is an additional contribution  $\tilde{\mathscr{F}}$ to the $\tilde{H}$-commutative part. 
This forces us to use (\ref{C27-defXL2}) for the $\tilde{H}$-commutative part.%
\footnote{ 
The derivation of this identity was given in Appendix~B  of \cite{KSM08}%[KSM,2008]
, which is presented later in this chapter.
}
In fact, we can extract a part $\tilde{\mathscr{F}}$ by
\begin{align}
 \tilde{\mathscr{F}}  :=&  \mathscr{F}_{\tilde{H}}   - \bm{h} (\bm{h} \cdot \mathscr{F})  \in su(N-1)  , 
\end{align}
so that 
%It is important to remark that
\begin{equation}
 [\bm{h}, \tilde{\mathscr{F}} ]=0 .
\end{equation}

By introducing a subset of generators   $\bm{u}^{k} \in su(N-1)$ commuting with $\bm{h}$:
%$\left[ \bm{h}, \bm{u}^{k}   \right]  =0$ ($k=1,...,(N-1)^{2}-1$) and
%$ [ \bm{h}, \tilde{\mathscr{F}} ]=0 .$
\begin{align}
\left[ \bm{h}, \bm{u}^{k}   \right]  =0 \quad (k=1,...,(N-1)^{2}-1) ,
%\quad 
%  [ \bm{h}, \bar{\mathscr{F}} ]=0 .
\end{align}
$\tilde{\mathscr{F}}$ is expressed as 
\begin{align}
\tilde{\mathscr{F}}= \sum_{k=1}^{(N-1)^{2}-1} \bm{u}^{k} \left( \bm{u}^{k} \cdot \mathscr{F}   \right) \in su(N-1) .
\end{align}
Consequently, we have an expression in the special basis:
\begin{align}
%\mathscr{F} %=  \sum_{A=1}^{N^2-1} \mathscr{F}^A T_A
%=&   \mathscr{F}_{\tilde{H}} + \mathscr{F}_{G/\tilde{H}} , 
%\nonumber\\
 &  \mathscr{F}_{\tilde{H}}  
=   \bm{h} (\bm{h} \cdot \mathscr{F}) +  \tilde{\mathscr{F}}  
= \bm{h} (\bm{h} \cdot \mathscr{F}) +  \sum_{k=1}^{(N-1)^{2}-1} \bm{u}^{k} \left( \bm{u}^{k} \cdot \mathscr{F}   \right) \in  u(N-1).
%\nonumber\\&
\end{align}

For concreteness, we construct the single color field $\bm h(x)$  explicitly using the  diagonal  matrix $H_r$ ($r={\rm rank}G$):
\begin{align}
 \bm h(x) =\bm{n}_r(x) :=U^\dagger(x) H_r U(x) , \quad U(x) \in G 
  ,
  \label{C27-color-field-min}
\end{align}
where $H_r$ is the last Cartan matrix  given by
\begin{align}
 H_r  
 =& \frac{1}{\sqrt{2N(N-1)}} {\rm diag}(1,\cdots,1,-N+1) 
%\nonumber\\
 =  \frac{1}{\sqrt{2N(N-1)}}  
\begin{pmatrix}
  \bm{1}_{_{N-1}} & \bm{0} \\
  {}^t\bm{0}  &  -N+1 
 \end{pmatrix} 
 ,
\end{align}
using the column vector $\bm{0}={}^t(0,\dots,0)$ with $t$ denoting the transpose. 
This is the ``last'' diagonal matrix $T_{N^2-1}=H_{N-1}$ in the Gell-Mann representation for the generators of $su(N)$. 
For the choice (\ref{C27-color-field-min}) for the color field, $\tilde{\mathscr{F}}$  is equal to  the  matrix  in which all the elements in both the last column and the last row are zero except for the last diagonal element:
\begin{equation}
\tilde{\mathscr{F}}
 = \begin{pmatrix}
  A & \bm{0} \\
  {}^t\bm{0}  & 1 
 \end{pmatrix} 
  .
\end{equation}

First, we apply the identity (\ref{C27-idv2-min}) to $\mathscr{X}^\mu(x)$ and use the second  defining equation  (\ref{C27-defXL2}) to obtain
\begin{align}
  \mathscr{X}_\mu(x) 
%=& \sum_{A=1}^{N^2-1} \mathscr{X}_\mu(x)^A T_A
%\nonumber\\
=& 
\mathscr{X}_\mu(x){}_{\tilde{H}}
% \tilde{\mathscr{X}}_\mu(x) + (\mathscr{X}_\mu(x),\bm{h})  \bm{h}
+   \frac{2(N-1)}{N}  [\bm{h}(x) , [\bm{h}(x) , \mathscr{X}_\mu(x)]]
%\nonumber\\
=    \frac{2(N-1)}{N}  [\bm{h}(x) , [\bm{h}(x) , \mathscr{X}_\mu(x)]]
% := v_\parallel + v_\perp
 .
\end{align}
By taking into account the first defining equation (\ref{C27-defVL2}), we find  
\begin{align}
 \mathscr{D}_\mu[\mathscr{A}]\bm{h}(x)  
% :=& \partial_\mu \bm{h}  - ig[\mathscr{A}_\mu, \bm{h} ]
%\nonumber\\
=  \mathscr{D}_\mu[\mathscr{V}]\bm{h}(x)  - ig [\mathscr{X}_\mu(x), \bm{h}(x) ] 
%\nonumber\\ 
=  
%- ig [\mathscr{X}_\mu, \bm{h} ] = 
 ig [ \bm{h}(x) , \mathscr{X}_\mu(x)] 
 .
\end{align}
Therefore, the $\mathscr{X}_\mu(x)$ field is expressed in terms of $\mathscr{A}_\mu(x)$ and $\bm{h}(x)$ as
\begin{align}
 \mathscr{X}_\mu(x) 
=   -ig^{-1}  \frac{2(N-1)}{N}  [\bm{h}(x), \mathscr{D}_\mu[\mathscr{A}]\bm{h}(x) ]
%= g^{-1}N  (\bm{n}_j \times \mathscr{D}_\mu[\mathscr{A}]\bm{n}_j)   .
 .
\label{C27-X-min}
\end{align}
Next, the $\mathscr{V}_\mu$ field is expressed in terms of $\mathscr{A}_\mu(x)$ and $\bm{h}(x)$: 
\begin{align}
  \mathscr{V}_\mu(x) 
  =& \mathscr{A}_\mu(x)  - \mathscr{X}_\mu(x)  
%\nonumber\\
  =  \mathscr{A}_\mu(x)  + ig^{-1}  \frac{2(N-1)}{N}  [\bm{h}(x), \mathscr{D}_\mu[\mathscr{A}]\bm{h}(x) ]
 .
\label{C27-defV12}
\end{align}
Thus, $\mathscr{V}_\mu(x)$ and $\mathscr{X}_\mu(x)$ are  written in terms of $\mathscr A_\mu(x)$, once $\bm h(x)$ is given as a functional of $\mathscr A_\mu(x)$.

We further decompose $\mathscr V_\mu(x)$ into the $\tilde{H}$-commutative part $\mathscr C_\mu(x)$ and the remaining part $\mathscr B_\mu(x)$:
\begin{align}
 \mathscr V_\mu(x)
 =\mathscr C_\mu(x)
  +\mathscr B_\mu(x)
 . 
\end{align}
We rewrite (\ref{C27-defV12}) as
\begin{align}
\mathscr{V}_\mu(x)
  =& \mathscr{A}_\mu(x)   - \frac{2(N-1)}{N}   [\bm{h}(x), [ \bm{h}(x), \mathscr{A}_\mu(x) ] ]
%\nonumber\\&
+ ig^{-1} \frac{2(N-1)}{N}   [\bm{h}(x) , \partial_\mu  \bm{h}(x) ]
%  \nonumber\\
%=& \tilde{\mathscr{A}}_\mu(x)   + (\mathscr{A}_\mu(x),\bm{h}(x)) \bm{h}(x)
%+ig^{-1} \frac{2(N-1)}{N}   [\bm{h}(x) , \partial_\mu  \bm{h}(x) ]
 .
\label{C27-defV2}
\end{align}
The first two terms on the right-hand side of (\ref{C27-defV2})  together  constitute the $\tilde{H}$-commutative part of $\mathscr{A}_\mu(x)$, i.e., $\mathscr{A}_\mu(x)_{\tilde H}$.  
Therefore, we obtain
\begin{subequations}
\begin{align}
  \mathscr{C}_\mu(x) :=& \mathscr{A}_\mu(x){}_{\tilde{H}}
=  \left( {\bf 1} -   2\frac{N-1}{N}  [\bm{h}(x) , [\bm{h} (x),  \cdot ]]
\right) \mathscr{A}_\mu(x)
%= \mathscr{V}_\mu - \mathscr{B}_\mu  
%= \mathscr{A}_\mu  - \mathscr{X}_\mu  - \mathscr{B}_\mu 
\nonumber\\
=& \mathscr{A}_\mu(x) - \frac{2(N-1)}{N}   [\bm{h}(x), [ \bm{h}(x), \mathscr{A}_\mu(x) ] ]
,
\label{C27-C-min}
\\
 \mathscr{B}_\mu(x)
=& ig^{-1} \frac{2(N-1)}{N}[\bm{h}(x) , \partial_\mu  \bm{h}(x) ]
 .
\label{C27-B-min}
\end{align}
\end{subequations}

It turns out that $\mathscr X_\mu$ constructed in this way belongs to the coset 
\begin{align}
  \mathscr X_\mu \in Lie(G/\tilde H) = \mathscr{G} - \tilde{\mathscr{H}} = su(N)-u(N-1) 
,
\end{align}
since for an arbitrary element $\tilde{{\bm h}} \in Lie(\tilde{H})$ (\ref{C27-X-min}) yields:%
\footnote{
We have used 
${\rm tr}(A[B,C])={\rm tr}([A,B]C)$.
}
\begin{align}
  \tilde{{\bm h}} \cdot \mathscr{X}_\mu 
=  2{\rm tr}(\tilde{{\bm h}} \mathscr{X}_\mu) 
%\nonumber\\
=   -i \frac{2(N-1)}{gN} 2{\rm tr}( \tilde{{\bm h}}  [ {\bm h}, \mathscr{D}_\mu{[\mathscr A]} {\bm h} ] )
%\nonumber\\
=   -i \frac{2(N-1)}{gN} 2{\rm tr}( [ \tilde{{\bm h}} , {\bm h}]  \mathscr{D}_\mu{[\mathscr A]} {\bm h} )
%\nonumber\\
=    0 
,
\end{align}
where we have  used 
$[ \tilde{{\bm h}} , {\bm h}]=0$ in the last step.
Similarly, it is shown by using (\ref{C27-B-min}) that
\begin{align}
  \mathscr B_\mu \in Lie(G/\tilde H) = \mathscr{G} - \tilde{\mathscr{H}} = su(N)-u(N-1) 
 .
\end{align}
%and $\mathscr{B}_\mu(x)$ is noncommutative, $[\mathscr{B}_\mu(x),\bm{h}(x)]  \ne 0$, and 
since it is orthogonal to  $\bm{h}(x)$:
\begin{align}
  \bm{h}(x)  \cdot \mathscr{B}_\mu(x) 
= 2{\rm tr}(\bm{h}(x) \mathscr{B}_\mu(x) )=0 
 .
\label{C27-defBL22}
\end{align}
Moreover, we can show that
\begin{align}
  \mathscr C_\mu \in  Lie(\tilde H) = \tilde{\mathscr{H}} =  u(N-1) 
 .
\end{align}
In fact, $\mathscr{C}_\mu(x)$ commutes with $\bm{h}(x)$:
\begin{align}
    [ \bm{h}(x)  , \mathscr{C}_\mu(x)  ] = 0 
 , 
\label{C27-specC2}
\end{align}
since (\ref{C27-C-min}) yields
\begin{align}
  [ {\bm h} , \mathscr{C}_\mu ]
%\nonumber\\
%=& [ {\bm h} , \mathscr{A}_\mu ]  + \frac{2(N-1)}{N} [ {\bm h} , [  {\bm h} ,  [\mathscr A_\mu  ,{\bm h}] ] ]
%\nonumber\\
=& [ {\bm h} , \mathscr{A}_\mu ]  - \frac{2(N-1)}{N} [ {\bm h} , [ [\mathscr A_\mu  ,{\bm h}],  {\bm h}  ] ]
\nonumber\\
=&  [ {\bm h} , \mathscr{A}_\mu ] - \frac{2(N-1)}{N} \Big( \frac{2}{N} [ {\bm h} ,  \mathscr{A}_\mu] + \frac{2(2-N)}{\sqrt{2N(N-1)}} [ {\bm h} ,  \{  {\bm h}  , \mathscr{A}_\mu \} ] 
%\nonumber\\& \quad\quad\quad\quad
 - [ {\bm h} , \{ {\bm h}, \{ {\bm h}, \mathscr{A}_\mu \} \} ]  \Big) 
\nonumber\\
=&  [ {\bm h} , \mathscr{A}_\mu ] - \frac{2(N-1)}{N} \Big( \frac{2}{N} [ {\bm h} ,  \mathscr{A}_\mu] + \frac{2(2-N)^2}{2N(N-1)}  [ {\bm h} , \mathscr{A}_\mu  ] 
%\nonumber\\& \quad\quad\quad\quad
- \frac{(2-N)^2}{2N(N-1)}  [ {\bm h} ,  \mathscr{A}_\mu   ]  \Big) 
\nonumber\\
=&  [ {\bm h} , \mathscr{A}_\mu ] - [ {\bm h} , \mathscr{A}_\mu ]
%\nonumber\\
=  0
,
\label{C27-Ch}
\end{align}
where we have used the following identities:%
\footnote{
We have used 
$[[A, B],C]=\{ A, \{ B, C \} \}- \{ B, \{ C, A \} \}$, and
$[B,\{ B, A \}]= \{ B, [B, A] \}=[BB,A]$.
}
%[Exercise-6] \marginpar{Ex-6}
\begin{align}
\bm{h} \bm{h} =& \frac{1}{2N}\mathbf{1}+\frac{2-N}{\sqrt{2N(N-1)}} \bm{h} ,
\label{C27-eq:Hidentity}%
\\
  [[\mathscr A  ,{\bm h}],  {\bm h}  ]  
=& \{ \mathscr A, \{ {\bm h}, {\bm h} \} \} - \{ {\bm h}, \{ {\bm h}, \mathscr A \} \} 
\nonumber\\
=&  \frac{2}{N} \mathscr A + \frac{2(2-N)}{\sqrt{2N(N-1)}} \{ {\bm h} ,\mathscr A    \} - \{ {\bm h}, \{ {\bm h}, \mathscr A \} \}   
,
\label{C27-id-h2}
\\
  [ {\bm h} ,  \{ {\bm h}, \mathscr{A}   \} ]
=&   [ {\bm h}{\bm h} ,    \mathscr{A}  ]
= \frac{(2-N)}{\sqrt{2N(N-1)}} [ {\bm h} , \mathscr A ]  
 ,
\label{C27-id-h3}
\end{align}
and
\begin{align}
  [ {\bm h} , \{ {\bm h}, \{ {\bm h}, \mathscr{A} \} \} ]
=&  \{ {\bm h}, [ {\bm h} ,  \{ {\bm h}, \mathscr{A}   \} ] \}
= \frac{(2-N)}{\sqrt{2N(N-1)}} \{ {\bm h}, [ {\bm h} , \mathscr A ] \}
\nonumber\\
=& \frac{(2-N)}{\sqrt{2N(N-1)}} [ {\bm h}  {\bm h} , \mathscr A ]  
= \frac{(2-N)^2}{2N(N-1)}  [ {\bm h} ,  \mathscr{A}   ]
 .
\label{C27-id-h4}
\end{align}
Thus,  new variables constructed in this way indeed satisfy the desired property: 
\begin{align}
 \mathscr{D}_\mu[\mathscr{V}] {\bm h}  
% = \partial_\mu {\bm h} -ig[ \mathscr{V}_\mu ,{\bm h}  ]
% = \partial_\mu {\bm h} -ig[ \mathscr{B}_\mu +\mathscr{C}_\mu ,{\bm h}  ]
 = \mathscr{D}_\mu[\mathscr{B}] {\bm h} -ig[ \mathscr{C}_\mu ,{\bm h}  ]
 = 0
 .
\end{align}

It is instructive to note that the above $\mathscr C_\mu$ is written in the form:
\begin{equation}
 \mathscr C_\mu = u_\mu^\alpha \bm{n}_\alpha, \quad u_\mu^\alpha =  \bm{n}_\alpha \cdot \mathscr A_\mu , \quad \bm{n}_\alpha = U^\dagger T_\alpha U \quad T_\alpha \in u(N-1) 
 ,
\end{equation}
where $\alpha$ runs over $\alpha=1, \cdots, (N-1)^2-1$ and $N^2-1$.
Note that these $\bm{n}_k$ for $k=1, \cdots, (N-1)^2$ are not uniquely defined.  This is because $\bm{n}_\alpha$ ($\alpha=1, \cdots, (N-1)^2-1$) can be changed using the rotation within $\tilde H=U(N-1)$ without changing $\bm{n}_r$, while $\bm{n}_r$ is invariant under the $\tilde H=U(N-1)$ rotation. 
%This feature is discussed for $SU(3)$ in more detail. 

Thus, once a single color field $\bm{h}(x)$ is given, we have the decomposition: 
\begin{subequations}
\begin{align}
\mathscr A_\mu(x)
 =& \mathscr V_\mu(x)
  +\mathscr X_\mu(x)
 =\mathscr C_\mu(x)
  +\mathscr B_\mu(x)
  +\mathscr X_\mu(x)
%\quad (\mu=0, 1, \cdots, D-1)  
,
\\
 \mathscr{X}_\mu(x) & 
= -ig^{-1}  \frac{2(N-1)}{N}  [\bm{h}(x), \mathscr{D}_\mu[\mathscr{A}]\bm{h}(x) ]
 \in Lie(G/\tilde H)
%= g^{-1}N  (\bm{n}_j \times \mathscr{D}_\mu[\mathscr{A}]\bm{n}_j)   .
 ,
\\
\mathscr{V}_\mu(x)
  =& \mathscr{A}_\mu(x)  + ig^{-1}  \frac{2(N-1)}{N}  [\bm{h}(x), \mathscr{D}_\mu[\mathscr{A}]\bm{h}(x) ] ,
\\
  \mathscr{C}_\mu(x) &
%= \mathscr{V}_\mu - \mathscr{B}_\mu  
%= \mathscr{A}_\mu  - \mathscr{X}_\mu  - \mathscr{B}_\mu 
=  \mathscr{A}_\mu(x) - \frac{2(N-1)}{N}   [\bm{h}(x), [ \bm{h}(x), \mathscr{A}_\mu(x) ] ]
\in Lie(\tilde H)
,
\\
 \mathscr{B}_\mu(x) &
= ig^{-1} \frac{2(N-1)}{N}[\bm{h}(x) , \partial_\mu  \bm{h}(x) ]
\in Lie(G/\tilde H)
 .
\end{align}
 \label{C27-NLCV-minimal}
\end{subequations}
Thus, all the new variables have been written in terms of $\bm h$ and $\mathscr A_\mu$.

%%%%%%%%%%%%%%%%%%%%%%%%%%%%%%%%%%%%%%%%%%%%%%%%%%%%%%%%%%%%%
%%%%%%%%%%%%%%%%%%%%%%%%%%%%%%%%%%%%%%%%%%%%%%%%%%%%%%%%%%%%%
\subsubsection{Independent degrees of freedom  in the minimal option}
\label{subsubsection:cov-minimal}%%%%%%%%%%%%%%%%%%%%%%%%%%%%%%%%%%%%%%%%%%%%%%%%%%%%%%%%%%%%%
%%%%%%%%%%%%%%%%%%%%%%%%%%%%%%%%%%%%%%%%%%%%%%%%%%%%%%%%%%%%%

We wish to regard  $\mathscr C_\mu$, $\mathscr X_\mu$ and $\bm h$ as  new field variables which are obtained by the (non-linear) change of variables from the original field variable $\mathscr A_\mu$: 
\begin{equation}
 \mathscr A_\mu  \Longrightarrow (\bm h, \mathscr C_\mu  ,  \mathscr X_\mu )  
 ,
\end{equation}
Here we do not include the variable $\mathscr B_\mu$ in this identification, since it is written  in terms of $\bm h$ alone. 
By introducing the color field $\bm h$ in the original $SU(N)$ Yang-Mills theory written in terms of the original gauge field $\mathscr {A}$, the Yang-Mills theory is extended to the gauge theory with the enlarged gauge symmetry $SU(N)_{\mathscr A}\times[SU(N)/U(N-1)]_{\bm h}$, which  is called the master Yang-Mills theory.

First, we need to count the independent degrees of freedom in this identification. 
Note that all components of new variables $(\bm h^A, \mathscr C_\mu^A  ,  \mathscr X_\mu^A )$ ($A=1, \cdots, N^2-1$) are not necessarily independent. 
For $G=SU(N)$, the maximal stability group in the minimal option is $\tilde H=U(N-1)$.  Therefore, the respective field variable has the following degrees of freedom at each space--time point:
\begin{itemize}
\item
$\mathscr A_\mu \in Lie(G)%=su(N)  
\rightarrow \#[\mathscr A_\mu^A]=D \times {\rm dim}G=D(N^2-1)$

$\downarrow$ 

\item
$\mathscr C_\mu \in Lie(\tilde{H})=u(N-1) \rightarrow  \#[\mathscr C_\mu^\alpha]=D \times {\rm dim}\tilde{H}=D(N-1)^2 
%= \#[\mathscr V_\mu^A]-\#[(I)]
$

\item
$\mathscr X_\mu \in Lie(G/\tilde{H})%=su(N)-u(N-1)
\rightarrow \#[\mathscr X_\mu^a]
=D \times {\rm dim}(G/\tilde{H})=2D(N-1)   
%= \#[\mathscr X_\mu^A]-\#[(II)]
$

\item
$\bm{h}  \in Lie(G/\tilde{H})%=su(N)-u(N-1)
\rightarrow \#[\bm{h}^a]= {\rm dim}(G/\tilde{H})%=N^2-1-(N-1)^2
=2(N-1)$ .

\end{itemize}

The change of variables is obtained by solving the defining equations. In other words, the defining equations impose constraints among the new field variables, which reduces the independent degrees of freedom. 
The number of constraints imposed by the defining equations I and II are 
\begin{itemize}
\item
$
\#[({\rm I})]   
= D \times \#[\bm{h}]
=D \times {\rm dim}(G/\tilde{H})
=2D(N-1)
$
\item
$ 
\#[({\rm II})]
=D \times {\rm dim}(\tilde{H})
=D(N-1)^2
$ 

\end{itemize}
This is consistent with the counting of the degrees of freedom for the new variables:
\begin{itemize}
\item
$
\#[\mathscr C_\mu^\alpha] 
= \#[\mathscr V_\mu^A]-\#[({\rm I})]   
=D \times {\rm dim}G- D \times {\rm dim}(G/\tilde{H})
= D \times {\rm dim}(\tilde{H})
%=D(N^2-1)-2D(N-1)
$
\item
$ 
\#[\mathscr X_\mu^a]
= \#[\mathscr X_\mu^A]-\#[({\rm II})]
=D \times {\rm dim}G- D \times {\rm dim}(\tilde{H})
= D \times {\rm dim}(G/\tilde{H})
%=D(N^2-1)-D(N-1)^2
$ 

\end{itemize}

 The new formulation written in terms of new variables $(\bm h^a, \mathscr C_\mu^\alpha  ,  \mathscr X_\mu^a )$ still has the $2(N-1)$ extra degrees of freedom.  
Therefore, we must give a procedure for eliminating the $2(N-1)$ extra degrees of freedom to obtain the new theory equipollent to the original one. For this purpose, we impose $2(N-1)$ constraints $\bm\chi=0$, which we call the \textbf{reduction condition} in the minimal option:
\begin{itemize}
\item
$
\bm\chi \in Lie(G/\tilde{H}) \rightarrow \#[\bm\chi^a]= {\rm dim}(G/\tilde{H}) 
=2(N-1)=\#[\bm{h}^a]
$.
\end{itemize}

By imposing  the minimal version of the reduction condition, the master Yang-Mills theory is reduced to the gauge theory with the gauge symmetry $SU(N)$, which is equipollent to original Yang-Mills theory.

%%%%%%%%%%%%%%%%%%%%%%%%%%%%%%%%%%%%%%%%%%%%%%%%%%%%%%%%%%%%%
%%%%%%%%%%%%%%%%%%%%%%%%%%%%%%%%%%%%%%%%%%%%%%%%%%%%%%%%%%%%%
\subsubsection{Reduction condition in the minimal option}
\label{subsubsection:reduction-minimal}%%%%%%%%%%%%%%%%%%%%%%%%%%%%%%%%%%%%%%%%%%%%%%%%%%%%%%%%%%%%%
%%%%%%%%%%%%%%%%%%%%%%%%%%%%%%%%%%%%%%%%%%%%%%%%%%%%%%%%%%%%%

An explicit form of the reduction condition in the minimal option is given by minimizing the functional:%
\footnote{
It is possible to choose other forms of the reductional functionals.  This is just a choice of them. 
In the limit $N=2$, this reduction functional reduces to that for the $SU(2)$ Yang-Mills theory. 
}
\begin{align}
  \int d^Dx  \frac12 g^2  \mathbf X_\mu\cdot\mathbf X^\mu
  =&  \frac{2(N-1)^2}{N^2}\int d^Dx(\mathbf{h}\times D_\mu[\mathbf A]\mathbf{h})^2
 \nonumber\\
  =&  \frac{N-1}{N}\int d^Dx(D_\mu[\mathbf A]\mathbf{h})^2,
  \nonumber
\end{align}
with respect to the enlarged gauge transformation: 
\begin{align}
\delta\mathbf A_\mu =& D_\mu[\mathbf A]\bm\omega , \quad \bm\omega \in Lie(G) ,
\nonumber\\
\delta\mathbf{h}
 =& g\mathbf{h}\times\bm\theta  =g\mathbf{h}\times\bm\theta_\perp ,
 \quad \bm\theta_\perp \in Lie(G/\tilde{H})  .
\end{align}
In fact, the enlarged gauge transformation of the functional $R[\mathbf A, \mathbf{h}]$:
\begin{align}
R[\mathbf A, \mathbf{h}]
 :=   \int d^Dx \frac12 (D_\mu[\mathbf A]\mathbf{h})^2,
\end{align}
 is given by
%[Exercise-7] \marginpar{Ex-7}
\begin{align}
\delta R[\mathbf A, \mathbf{h}]
=g
   \int d^Dx
   (\bm\theta_\perp-\bm\omega_\perp)
   \cdot\left(\mathbf{h}\times D_\mu[\mathbf A]D_\mu[\mathbf A]\mathbf{h}\right)
 ,
 \label{C27-dR-min}
\end{align}
where $\bm\omega_\perp$ denotes the component of $\bm\omega$ in the direction $\mathscr{L}(G/\tilde{H})$. 

The minimization of the reduction functional, i.e., $\delta R[\mathbf A, \mathbf{h}]=0$ imposes no condition for $\bm\omega_\perp = \bm\theta_\perp$ (``diagonal'' part of $G \times G/\tilde{H}$), while it implies the constraint $\bm\chi =\mathbf{h}\times D^\mu[\mathbf A]D_\mu[\mathbf A]\mathbf{h}=0$ for $\bm\omega_\perp   \not=  \bm\theta_\perp$ (off-diagonal part of $G \times G/\tilde{H}$).
The number of constraint is 
$\#[\bm\chi]= {\rm dim}(G \times G/\tilde{H})- {\rm dim}G= {\rm dim}(G/\tilde{H})$ as desired. 

Thus, we obtain the differential form of the reduction condition:
\begin{equation}
\bm\chi[\mathbf A,\mathbf{h}]
  =\mathbf{h}\times D^\mu[\mathbf A]D_\mu[\mathbf A]\mathbf{h}
 \equiv0 
  .
\label{C27-eq:nMAG_minimal_differential}
\end{equation}
This is also expressed in terms of $\mathbf X_\mu$ and $\mathbf V_\mu$ ($\mathbf{h}$ and $\mathbf C_\mu$):
\begin{equation}
\bm\chi[\mathbf{h},\mathbf C,\mathbf X]
 :=D^\mu[\mathbf V]\mathbf X_\mu
 \equiv0
 .
\end{equation}
%This should be compared with the Maximally Abelian gauge.
%\begin{equation}
%G \stackrel{\bm{h}}{\rightarrow} G \times G/\tilde{H} \stackrel{\bm{\chi}}{\rightarrow} G 
% .
%\end{equation}

To determine which part of the symmetry is left after imposing the minimal version of the reduction condition (\ref{C27-eq:nMAG_minimal_differential}), we perform the gauge transformation on the enlarged gauge-fixing functional $\bm\chi$ to obtain
%[Exercise-8] \marginpar{Ex-8}
\begin{align}
\delta\bm\chi
 &=\delta\mathbf{h}\times D^\mu[\mathbf A]D_\mu[\mathbf A]\mathbf{h}
%   \nonumber\\
% &\quad
   +\mathbf{h}\times D^\mu[\mathbf A]D_\mu[\mathbf A]\delta\mathbf{h}
   \nonumber\\
 &\quad
   +\mathbf{h}\times(g\delta\mathbf A_\mu\times D_\mu[\mathbf A]\mathbf{h})
%   \nonumber\\
% &\quad
   +\mathbf{h}\times D^\mu[\mathbf A](g\delta\mathbf A_\mu\times\mathbf{h})
   \nonumber\\
 &=g(\mathbf{h}\times D^\mu[\mathbf A]D_\mu[\mathbf A]\mathbf{h})
   \times(\bm\theta_\perp+\bm\omega_\parallel)
   \nonumber\\
 &\quad
   +2g\mathbf{h}
    \times\{D^\mu[\mathbf A]\mathbf{h}
            \times D_\mu[\mathbf A](\bm\theta_\perp-\bm\omega_\perp)\}
   \nonumber\\
 &\quad   +g\mathbf{h}
    \times\{\mathbf{h}\times D^\mu[\mathbf A]D_\mu[\mathbf A]
                            (\bm\theta_\perp-\bm\omega_\perp)\} ,
\label{C27-eq:dchi}
\end{align}
where $\bm\omega_\parallel$ and $\bm\omega_\perp$ denote the components  of $\bm\omega$ in $Lie(\tilde{H})$ and $Lie(G/\tilde{H})$, respectively.
($\bm\omega=\bm\omega_\parallel+\bm\omega_\perp$).
Here we have used the relation following from $\bm\omega_\parallel\times\mathbf{h}=0$:
\begin{align}
0&=D^\mu[\mathbf A]D_\mu[\mathbf A](\bm\omega_\parallel\times\mathbf{h})
   \nonumber\\
 &=D^\mu[\mathbf A]D_\mu[\mathbf A]\bm\omega_\parallel\times\mathbf{h}
   +2D^\mu[\mathbf A]\bm\omega_\parallel\times D_\mu[\mathbf A]\mathbf{h}
   +\bm\omega_\parallel\times D^\mu[\mathbf A]D_\mu[\mathbf A]\mathbf{h}
 .
\end{align}
This result shows that the minimal version of the reduction  condition $\bm\chi\equiv0$ leaves 
$\bm\theta_\perp=\bm\omega_\perp$ intact. When $\bm\theta_\perp=\bm\omega_\perp$, we find from  (\ref{C27-eq:dchi})
\begin{equation}
 \delta\bm\chi=g\bm\chi\times\bm\alpha 
 , \quad 
\bm\alpha=(\bm\alpha_\parallel,\bm\alpha_\perp)
=(\bm\omega_\parallel,\bm\omega_\perp=\bm\theta_\perp) 
.
\end{equation}

%Even after imposing the reduction condition, the gauge symmetry, i.e., the equipollent gauge symmetry $G$ identified with the diagonal part of $G \times G/\tilde{H}$ remains intact. Hence, we obtain the new theory with the original gauge symmetry $G$: 
%\begin{equation}
%G \stackrel{\bm{h}}{\rightarrow} G \times G/\tilde{H} \stackrel{\bm{\chi}}{\rightarrow} G 
% .
%\end{equation}
%The reduction condition (\ref{eq:nMAG_minimal_differential}) is also expressed in terms of $\mathbf X_\mu$ and $\mathbf V_\mu$ ($\mathbf{h}$ and $\mathbf C_\mu$):
%\begin{equation}
%\bm\chi[\mathbf{h},\mathbf C,\mathbf X]
% :=D^\mu[\mathbf V]\mathbf X_\mu
% \equiv0
% .
%\end{equation}
%This should be compared with the Maximally Abelian gauge.

%%%%%%%%%%%%%%%%%%%%%%%%%%%%%%%%%%%%%%%%%%%%%%%%%%%%%%%%%%%%%
%\subsubsection{Residual Symmetry}
%%%%%%%%%%%%%%%%%%%%%%%%%%%%%%%%%%%%%%%%%%%%%%%%%%%%%%%%%%%%%

%%%%%%%%%%%%%%%%%%%%%%%%%%%%%%%%%%%%%%%%%%%%%%%%%%%%%%%%%%%%%
\subsection{Path-integral quantization and Jacobian}\label{subsec:path-integral-SUN}
%%%%%%%%%%%%%%%%%%%%%%%%%%%%%%%%%%%%%%%%%%%%%%%%%%%%%%%%%%%%%

We now reformulate a quantum Yang-Mills theory in terms of new variables  based on the functional integral quantization. 
The following steps can be clearly understood by comparing them with the corresponding ones in Fig.~\ref{R05-fig:enlarged-YM}.

First, the original Yang-Mills theory in the Euclidean formulation is defined by the partition function (see the middle left  part in Fig.~\ref{R05-fig:enlarged-YM}):
\begin{align}
 Z_{{\rm YM}} = \int \mathcal{D}\mathscr{A}_\mu^A \exp (-S_{{\rm YM}}[\mathscr{A}]) .
\end{align}
In the continuum, a well-defined Yang-Mills theory is obtained by imposing a gauge fixing condition to completely fix the  original gauge symmetry, which is called the overall gauge fixing condition  (see the down arrows in Fig.~\ref{R05-fig:enlarged-YM}). 
If we specify the gauge-fixing condition $F[\mathscr{A}]=0$, we must consider the associated Faddeev-Popov determinant 
$\Delta_{\rm FP}^{F}[\mathscr{A}]:=\det  \left(\frac{\delta F[\mathscr{A}^\omega]}{\delta \omega}\right)
$. 
Consequently, the integration measure is modified to
\begin{align}
 \mathcal{D}\mathscr{A}_\mu^A \rightarrow \mathcal{D}\mathscr{A}_\mu^A \delta(F[\mathscr{A}]) \Delta_{\rm FP}^{F}[\mathscr{A}] .
% \\
%Z_{{\rm YM}} = \int \mathcal{D}\mathscr{A}_\mu^A \delta(F[\mathscr{A}]) \Delta_{\rm FP}^{F}[\mathscr{A}]  \exp (-S_{{\rm YM}}[\mathscr{A}]) .
\end{align}
For instance, one chooses the Landau gauge%
\footnote{Here we neglect the Gribov problem associated with the existence of Gribov copies.
} 
 $\partial^\mu \mathscr{A}_\mu^A(x)=0$.  For a while, we disregard this procedure, as it can be carried out following the standard method. 
To simplify the argument,  we omit the procedure of the overall gauge fixing for the original $SU(N)$ gauge symmetry, which must be done in a consistent way with the following procedures using the BRST method.

Second, the  color field  $\bm{n}(x)$ is introduced as an auxiliary field to extend the original Yang-Mills theory (see the left up arrow in Fig.~\ref{R05-fig:enlarged-YM})   to the enlarged Yang-Mills theory called the master Yang-Mills   theory, which is defined by a partition function written in terms of both $n^\alpha(x)$ and $\mathscr{A}_\mu^A(x)$ (see the top part in Fig.~\ref{R05-fig:enlarged-YM}),
\begin{align}
  {Z}_{{\rm mYM}}  = \int \mathcal{D}n^\alpha
%\bm{n}  \delta(\bm{n}\cdot\bm{n}-1)
\int \mathcal{D}\mathscr{A}_\mu^A \exp (-S_{{\rm YM}}[\mathscr{A}]) 
 . 
\label{C27-Z}
\end{align}
by inserting  
$
 1 
%= \int \mathcal{D}\bm{n}  \delta(\bm{n}\cdot\bm{n}-1) 
:= \int \mathcal{D}n^\alpha   
$.
%where $n^\alpha$  denotes  independent degrees of freedom,
The details of this integration measure is discussed later in the calculation of the Jacobian.
%after solving the constraint $n^A n^A=1$.

In the functional integral quantization, we must specify the action and the integration measure. 
We can immediately rewrite the Yang-Mills action $S_{\rm YM}[\mathscr{A}]$ in terms of the new variables  by substituting the transformation law, e.g., (\ref{C27-NLCV-maximal}) for the maximal option or (\ref{C27-NLCV-minimal}) for the minimal option, into the original gauge field $\mathscr{A}$ in the Yang-Mills Lagrangian $\mathscr{L}_{\rm YM}[\mathscr{A}]$. 
To write the integration measure explicitly in terms of the new variables, we need to know the Jacobian $\tilde J$  associated with the change of variables from the original Yang-Mills gauge field to the new variables in the reformulated Yang-Mills theory.

Third, we regard (\ref{C27-Z}) as 
\begin{align}
 \tilde{Z}_{{\rm YM}}  
= \int \mathcal{D}n^\beta
%  \delta(\bm{n}\cdot\bm{n}-1)
 \int \mathcal{D}C_\nu^k \int 
\mathcal{D}X_\nu^b 
%\delta(\bm{n}\cdot\mathbf X_\mu)  
\tilde{J}
 \exp (-\tilde S_{\rm YM}[\bm{n}, \mathscr{C},\mathscr{X}]) 
 , 
 \label{C27-Z2}
\end{align}
where 
%$\tilde{J}$ is the Jacobian associated with the change of variables, and 
the action $\tilde S_{\rm YM}[\bm{n}, \mathscr{C},\mathscr{X}]$ is obtained by substituting the transformation law  of $\mathscr A_\mu^A$
 into the original action $S_{{\rm YM}}[\mathscr{A}]$: 
\begin{align}
\tilde S_{\rm YM}[\bm{n}, \mathscr{C},\mathscr{X}]
 =S_{\rm YM}[\mathscr A] .
\end{align}
 
Starting with this form, we wish to obtain a new Yang-Mills theory written in terms of new variables 
$
(n^\beta, C_\nu^k, X_\nu^b)
$
(see the middle right part in Fig.~\ref{R05-fig:enlarged-YM}), which is equipollent to the original Yang-Mills theory.
Here, the replacement of field variables: 
\begin{align}
 (n^\alpha(x), \mathscr{A}_\mu^A(x)) \rightarrow (n^\beta(x), C_\nu^k(x), X_\nu^b(x))
\end{align}
should  be considered as a change of variables so  that $n^\beta(x), C_\nu^k(x), X_\nu^b(x)$ become {\it independent} field variables in the new Yang-Mills  theory.

However, the master Yang-Mills has the enlarged gauge symmetry, which is larger than the original gauge symmetry $G=SU(N)$. Therefore,  to obtain a new Yang-Mills theory that is equipollent to the original Yang-Mills theory, the extra gauge degrees of freedom must be eliminated. 
To fix the enlarged gauge symmetry in the master Yang-Mills theory and retain only the same gauge symmetry as that in the original Yang-Mills, therefore, 
we impose the constraint $\bm{\chi}[ \mathscr{A},\bm{n}]=0$ called the reduction condition, 
which is also written in terms of the new variables as 
$
\tilde{\bm\chi} 
 :=\tilde{\bm\chi} [\bm{n}, \mathscr{C},\mathscr{X}]=0
$.
(see the right down arrow in Fig.~\ref{R05-fig:enlarged-YM}).

The constraint $\bm{\chi}[ \mathscr{A},\bm{n}]=0$ is introduced in the functional integral as follows.
  We write  unity in the form 
\begin{align}
  1 = \int \mathcal{D} \bm{\chi}^\theta \delta(\bm{\chi}^\theta)
=   \int\! \mathcal{D}\bm\theta\delta(\bm\chi^\theta)
   {\rm Det} \left(\frac{\delta\bm\chi^\theta}{\delta{\bm\theta}}\right) ,
\end{align}
where $\bm{\chi}^\theta$ is the constraint written in terms of the gauge-transformed variables, i.e., 
$\bm{\chi}^\theta:=\bm{\chi}[ \mathscr{A},\bm{n}^\theta]$.
As an intermediate step for moving from the master Yang-Mills theory to the new Yang-Mills theory, we insert  this into the functional integral   (\ref{C27-Z2}), 
%\begin{align}
% Z_{{\rm YM}} =  \int \mathcal{D}n^\alpha
%%\bm{n}  \delta(\bm{n}\cdot\bm{n}-1)
%\int \mathcal{D}\mathscr{A}_\mu^A 
%\int\! \mathcal{D}\bm\theta\delta(\bm\chi^\theta)
%   {\rm Det}  \left(\frac{\delta\bm\chi^\theta}{\delta{\bm\theta}}\right) 
%\exp (-S_{{\rm YM}}[\mathscr{A}]) .
%\end{align}
and cast the partition function of the master Yang-Mills theory into an intermediate form 
(see the middle right part in Fig.~\ref{R05-fig:enlarged-YM})
\begin{align}
 \tilde{Z}_{{\rm YM}}  
=& \int \mathcal{D}n^\beta
%  \delta(\bm{n}\cdot\bm{n}-1)
 \int \mathcal{D}C_\nu^k \int 
\mathcal{D}X_\nu^b 
%\delta(\bm{n}\cdot\mathbf X_\mu)  
\tilde{J}
%\nonumber\\& \times \int\! 
\mathcal{D}\bm\theta\delta(\bm\chi^\theta)
   {\rm Det}  \left(\frac{\delta\bm\chi^\theta}{\delta{\bm\theta}}\right)
 \exp (-\tilde S_{\rm YM}[\bm{n}, \mathscr{C},\mathscr{X}])
 . 
\end{align}

We next perform the change of variables $\bm{n} \rightarrow \bm{n}^{\theta}$ obtained through a local rotation by   angle $\theta$ and the corresponding gauge   transformations for the other new variables $\mathscr{C}_\mu$ and $\mathscr X_\mu$: 
$\mathscr{C}_\mu, \mathscr{X}_\mu \rightarrow \mathscr{C}_\mu^{\theta}, \mathscr{X}_\mu^{\theta}$. 
From the gauge invariance of the action $\tilde S_{\rm YM}[\bm{n}, \mathscr{C},\mathscr{X}]$ and the integration measure 
$
 \mathcal{D}n^\beta  \mathcal{D}C_\nu^k \mathcal{D}X_\nu^b 
$, 
we can rename the dummy integration variables $\bm{n}^{\theta}, \mathscr{C}_\mu^{\theta}, \mathscr{X}_\mu^{\theta}$  as $\bm{n}, \mathscr{C}_\mu, \mathscr{X}_\mu$, respectively.
Thus, the integrand does not depend on $\theta$, and the gauge volume $\int\! \mathcal{D}\bm\theta$ can be factored out: 
\begin{align}
 \tilde{Z}_{{\rm YM}}  
=& \int\! \mathcal{D}\bm\theta 
\int \mathcal{D}n^\beta
%  \delta(\bm{n}\cdot\bm{n}-1)
 \int \mathcal{D}C_\nu^k \int 
\mathcal{D}X_\nu^b 
%\delta(\bm{n}\cdot\mathbf X_\mu)  
\tilde{J} 
%\nonumber\\& \times 
\delta(\bm\chi)
   {\rm Det} \left(\frac{\delta\bm\chi}{\delta{\bm\theta}}\right)
 \exp (-\tilde S_{\rm YM}[\bm{n}, \mathscr{C},\mathscr{X}]) .
\end{align}
Thus, we have arrived at the reformulated Yang-Mills theory with the partition function:
\begin{align}
 Z_{{\rm YM}}^\prime  
=  \int \mathcal{D}n^\beta
%  \delta(\bm{n}\cdot\bm{n}-1)
   \mathcal{D}C_\nu^k   
\mathcal{D}X_\nu^b 
%\delta(\bm{n}\cdot\mathbf X_\mu)  
\tilde{J} 
%\nonumber\\ & \times  
\delta(\tilde{\bm\chi})  
   \Delta_{\rm FP}^{\rm red}
 \exp (-\tilde S_{\rm YM}[\bm{n}, \mathscr{C},\mathscr{X}]) , 
\end{align}
where the reduction condition is written in terms of the new  variables:
\begin{equation}
\tilde{\bm\chi} 
 :=\tilde{\bm\chi} [\bm{n}, \mathscr{C},\mathscr{X}]
 :=\mathscr{D}^\mu[\mathscr{V}]\mathscr{X}_\mu 
 , 
%\quad
%\mathscr{V}_\mu \equiv c_\mu  \bm{n}  +g^{-1}\partial_\mu \bm{n} \times \bm{n} .
\end{equation}
and $\Delta_{\rm FP}^{\rm red}$ is  the Faddeev-Popov determinant associated with the reduction condition:
\begin{equation}
 \Delta_{\rm FP}^{\rm red}
:= {\rm Det}  \left(\frac{\delta\bm\chi}{\delta{\bm\theta}}\right)_{\bm{\chi}=0}
=   {\rm Det}  \left(\frac{\delta\bm\chi}{\delta\bm{n}^\theta}\right)_{\bm{\chi}=0} .
\end{equation}
Consequently, the integration measure of the original Yang-Mills theory is transformed to 
\begin{equation}
  \mathcal{D}\mathscr{A}_\mu^A
\to  \mathcal{D}n^\beta   \mathcal{D}C_\nu^k   
\mathcal{D}X_\nu^b    
\delta(\tilde{\bm\chi}) 
  \Delta_{\rm FP}^{\rm red}[\bm{n},\mathscr{C}, \mathscr X] \tilde{J}   .
 \end{equation}
It is important to note that the independent variables are regarded as $n^\beta(x)$, $C_\nu^k(x)$ and $X_\nu^b(x)$ in the reformulated Yang-Mills theory, which simplifies the Jacobian $\tilde{J}$. 

In the following, we show that the Jacobian can be simplified by choosing appropriate bases in the group space without changing the physical content. 

For $G=SU(N)$ ($N \ge 3$), we introduce a complete set of orthonormal bases $\{  \bm{e}_a , \bm{u}_j  \}$ of $su(N)$: 
\begin{align}
  \bm{e}_a \in Lie(G/\tilde{H}), \
\bm{u}_j \in Lie(\tilde{H})   ,
\
( a=1, \cdots, {\rm dim}(G/\tilde{H});   
  j=1, \cdots, {\rm dim} \tilde{H} )
\end{align}
 satisfy  the relations in  vector form using   $\mathbf{e}_a(x)=(e_a^A(x))$ , $\mathbf{u}_j(x) =(\mathbf{u}_j^A(x))$:
\begin{align}
  \mathbf{e}_a(x) \cdot \mathbf{e}_b(x) = \delta_{ab}  ,   
    \quad
   \mathbf{e}_a(x) \cdot \mathbf{u}_j(x)  = 0,  
    \quad
  \mathbf{u}_j(x) \cdot \mathbf{u}_k(x) = \delta_{jk}  ,
\nonumber\\ 
\ ( a,b=1, \cdots, {\rm dim}(G/\tilde{H}); j,k=1, \dots, {\rm dim} \tilde{H} )
 ,
\end{align}
which are written in the Lie algebra form using $\bm{e}_a(x)= e_a^A(x) T_A$ , $\bm{u}_j(x) = \bm{u}_j^A(x) T_A$:
\begin{align}
   {\rm tr}[\bm{e}_a(x) \bm{e}_b(x)] =& \frac12 \delta_{ab}  , \
   {\rm tr}[\bm{e}_a(x) \bm{u}_j(x)]     = 0, \
   {\rm tr}[\bm{u}_j(x) \bm{u}_k(x)]  = \frac12 \delta_{jk} 
    .
\end{align} 
It is instructive to note that $\bm{u}_j$ reduces to a single color field $\bm{n}$ in $SU(2)$. 

The fields $\mathscr{B}_\mu(x)$ and $\mathscr{X}_\mu(x)$ are orthogonal to (or non-commutative with) all $\bm{u}_j(x)$:  
\begin{equation}
    \mathscr{B}_\mu(x) \cdot \bm{u}_j(x)  = 0, \quad
    \mathscr{X}_\mu(x) \cdot \bm{u}_j(x)  = 0
    \ (j=1, \cdots, {\rm dim} \tilde{H} )
 .
\end{equation}
Hence they can be written using the bases $\{ \bm{e}_a(x) \}$: 
\begin{align}
    \mathscr{B}_\mu(x) =& B_\mu^a(x) \bm{e}_a(x) \in Lie(G/\tilde{H}) ,
 \nonumber\\
    \mathscr{X}_\mu(x) =& X_\mu^a(x) \bm{e}_a(x) \in Lie(G/\tilde{H})  
    \ (a=1, \cdots, {\rm dim}(G/\tilde{H}))
%\quad [ \bm{e}_a(x) , \bm{n}(x)] \ne 0 
  .
  \label{C27-fdec2}
\end{align}
By definition, 
$\mathscr{C}_\mu(x)$ can be written as
\begin{equation}
    \mathscr{C}_\mu(x) = C_\mu^j(x) \bm{u}_j(x) \in Lie(\tilde{H}) 
    \quad  (j=1, \cdots, {\rm dim} \tilde{H} )
% [ \bm{u}_k(x) , \bm{n}(x)] = 0 
 %   \mathscr{C}_\mu^A(x) = C_\mu^k(x) \bm{u}_k^A(x)  
  .
    \label{C27-fdec1}
\end{equation}
In the minimal case, thus, $\tilde{H} =U(N-1)$, and the decomposition is 
\begin{align}
&   \mathscr{A}_\mu^A(x) =   C_\mu^j(x) \bm{u}_j^A(x)   
+ B_\mu^a(x) \bm{e}_a^A(x)    
+ X_\mu^a(x) \bm{e}_a^A(x)  
\nonumber\\
 &   \ (j=1, \cdots, {\rm dim} \tilde{H} =(N-1)^2; 
\ a=1, \cdots, {\rm dim}(G/\tilde{H})=2N-2)
  .
  \label{C27-fdec3}
\end{align}
In the maximal case, on the other hand,  $\tilde{H}=H=U(1)^{N-1}$ and  
$\bm{u}_j(x)$ reduce to $\bm{n}_j(x)$ ($j=1, \cdots, {\rm dim} \tilde{H} =N-1$), which mutually commute, i.e., 
$
  [ \bm{n}_j(x), \bm{n}_k(x)] = 0
$.
Therefore, we obtain
\begin{align}
 &  \mathscr{A}_\mu^A(x) =   C_\mu^j(x) \bm{n}_j^A(x)   
+ B_\mu^a(x) \bm{e}_a^A(x)    
+ X_\mu^a(x) \bm{e}_a^A(x)  
\nonumber\\
 &   \ (j=1, \cdots, {\rm dim} \tilde{H} =N-1; 
\ a=1, \cdots, {\rm dim}(G/\tilde{H})=N^2-N)
   .
  \label{C27-fdec4}
\end{align}

We now consider the change of variables from 
$\{ n^\alpha, \mathscr{A}_\mu^A \}$ to the new variables 
$\{ n^\beta, C_\nu^k, X_\nu^b \}$ defined in the $su(N)$ bases $\{  \bm{e}_b , \bm{u}_j  \}$. 
Here $n^\alpha$ and $n^\beta$ denote  independent degrees of freedom of the color field $\bm{n}$ after solving the constraint $n^A n^A=1$. 
Then the integration measure is transformed as
\begin{equation}
 \mathcal{D}n^\alpha \mathcal{D}\mathscr{A}_\mu^A  
 =   \mathcal{D}n^\beta \mathcal{D}C_\nu^k \mathcal{D}X_\nu^b J  
 .
\end{equation}
The Jacobian  $\tilde{J}$ is given by
%[Exercise-9] \marginpar{Ex-9}
\begin{align}
 \tilde{J}
:=& \begin{vmatrix}
 \frac{\partial n^\alpha}{\partial n^\beta} & \frac{\partial n^\alpha}{\partial C_\nu^k} & \frac{\partial n^\alpha}{\partial X_\nu^b} \cr
 \frac{\partial \mathscr{A}_\mu^A}{\partial n^\beta} & \frac{\partial \mathscr{A}_\mu^A}{\partial C_\nu^k} & \frac{\partial \mathscr{A}_\mu^A}{\partial X_\nu^b} \cr
  \end{vmatrix}
=  \begin{vmatrix}
 \delta^\alpha_\beta & 0 & 0 \cr
 \frac{\partial \mathscr{A}_\mu^A}{\partial n^\beta} & \delta_{\mu\nu} u_k^A & \delta_{\mu\nu} e_b^A \cr
  \end{vmatrix}
%\nonumber\\
=  \begin{vmatrix}
   \delta_{\mu\nu} u_k^A & \delta_{\mu\nu} e_b^A \cr
  \end{vmatrix} 
=  \begin{vmatrix}
    u_k^A &  e_b^A \cr
  \end{vmatrix}^D
= 1
 ,
 \label{C27-Jacobian-min}
\end{align}
where we have used the fact that $\{ n^\beta, C_\nu^k, X_\nu^b \}$ are independent variables in the second equality,  that $\mu$ and $\nu$ run from $1$ to $D$ in the fourth equality, and that $\{  e_b^A, u_k^A  \}$ constitute  the orthonormal base in the last equality. 
Thus, the Jacobian  $\tilde{J}$  is  very simple, irrespective of the choice of the reduction condition:
\begin{equation}
  \tilde{J} = 1 .
\end{equation}
For the Jacobian $J$ to be simplified to $ \tilde{J}=1$, the fact that $C_\nu^k, X_\nu^b$ are independent of $n^\beta$ is important.

The precise form of $\Delta_{\rm FP}^{\rm red}$ is obtained  using the BRST method as \cite{KMS05}
%[Exercise-10] \marginpar{Ex-10}%
%\footnote{
%  K.-I. Kondo, T. Murakami and T. Shinohara,
%BRST quantization of the Yang-Mills theory in the Cho--Faddeev--Niemi decomposition, 
%[hep-th/0504198],
%Eur. Phys. J. C{\bf 42}, 475--481 (2005).
%K.-I. Kondo, T. Murakami and T. Shinohara,
%Yang-Mills theory constructed from Cho--Faddeev--Niemi decomposition, 
%Preprint CHIBA-EP-151, 
%[hep-th/0504107], 
%Prog. Theor. Phys. {\bf 115}, 201--216  (2006). 
%} 
\begin{equation}
  \Delta_{\rm FP}^{\rm red}[\bm{n},\mathscr{C},\mathscr X] 
= {\rm Det}  \{-\mathscr{D}_\mu[\mathscr V+\mathscr X]\mathscr{D}_\mu[\mathscr V-\mathscr  X]\}  .
\label{C27-FP-red}
\end{equation}
At one-loop level this reduces to the simple form:
\begin{equation}
  \Delta_{\rm FP}^{\rm red}[\bm{n},\mathscr{C},\mathscr X] 
\simeq {\rm Det}  \{-\mathscr{D}_\mu[\mathscr V]\mathscr{D}_\mu[\mathscr V]\} ,
\end{equation}
which leads to the partition function suited for the calculations at one-loop level:
\begin{align}
 Z_{{\rm YM}}^\prime  
\simeq  \int \mathcal{D}n^\beta
%  \delta(\bm{n}\cdot\bm{n}-1)
   \mathcal{D}C_\nu^k   
\mathcal{D}X_\nu^b 
%\delta(\bm{n}\cdot\mathbf X_\mu)  
 %\nonumber\\ & \times  
\delta(\tilde{\bm\chi})  
   {\rm Det}  \{-\mathscr{D}_\mu[\mathscr V]\mathscr{D}_\mu[\mathscr V]\} 
 \exp (-\tilde S_{\rm YM}[\bm{n}, \mathscr{C},\mathscr{X}]) . 
\end{align}

%In order to obtain a well-defined Yang-Mills theory by completely fixing the gauge, we must impose a gauge fixing condition for the $SU(N)$ gauge symmetry in addition to the reduction condition, for example, the Landau gauge $\partial^\mu \mathscr{A}_\mu^A(x)=0$, which is called the overall gauge fixing condition  (see the down arrows in Fig.~\ref{C27-fig:enlarged-YM}). 

The reformulated Yang-Mills theory obtained after imposing the reduction condition has still the original full gauge symmetry $G$.  
As already mentioned above, in order to obtain a well-defined quantum Yang-Mills theory by completely fixing the gauge, we must impose the  overall gauge fixing condition  for the $SU(N)$ gauge symmetry (for example, the Landau gauge $\partial^\mu \mathscr{A}_\mu^A(x)=0$, see the down arrows in Fig.~\ref{R05-fig:enlarged-YM}) in addition to the reduction condition. 
When we discuss the Faddeev-Popov ghost associated to the reduction condition, we must take into account the Faddeev-Popov ghost associated to the overall gauge fixing simultaneously.% 
\footnote{
For the thorough treatment  for $G=SU(2)$, see \cite{KMS05}
 in which the explicit derivation of the FP ghost term has been worked out explicitly. 
} 
According to the clarification of the symmetry in the master Yang-Mills theory explained above,  we can obtain the unique Faddeev-Popov ghost terms associated to the gauge fixing conditions adopted in quantization.  
This is  another advantage of the new viewpoint for the master-Yang-Mills theory.

However, we have omitted to write explicitly this procedure for  simplifying the presentation, since the overall gauge fixing can be performed according to the standard procedures. 
Moreover, a systematic treatment of  gauge fixing and the associated Faddeev-Popov determinant or the introduction of ghost fields can be carried out using the BRST symmetry of the new theory. 
Although the BRST treatment can also be performed for the $SU(N)$ case following the method already performed for $SU(2)$,
such a treatment is rather involved. 
  Therefore, we have given a heuristic explanation based on the Faddeev-Popov trick for $SU(N)$ in the above, even though it is possible to develop the BRST approach.
See the final subsection in this section.

%%%%%%%%%%%%%%%%%%%%%%%%%%%%%%%%%%%%%%%%%%%%%%%%%%%%%%%%%%
\begin{table}[t] 
\begin{center} 
 \begin{tabular}{l|cl}
    &  original theory & $\Longrightarrow$ reformulated  theory \\ \hline
  field variables & $\mathscr A_\mu^A \in \mathscr{L}(G)$ & $\Longrightarrow$ $\bm n^\beta, \mathscr C_\nu^k  ,  \mathscr X_\nu^b$ \\ 
  action & $S_{\rm YM}[\mathscr A]$ & $\Longrightarrow$ $\tilde S_{\rm YM}[\bm n, \mathscr{C},\mathscr{X}]$ \\
  integration measure & $\mathcal{D}\mathscr{A}_\mu^A$ 
& $\Longrightarrow$  $\mathcal{D}n^\beta \mathcal{D}\mathscr{C}_\nu^k   \mathcal{D}\mathscr{X}_\nu^b
\tilde{J}  \delta( {\tilde{\bm\chi}}) 
  \Delta_{\rm FP}^{\rm red}[\bm n, \mathscr{C}, \mathscr X]$ \\
%  Wilson loop operator & $ W_C[\mathscr{A}]$ 
%& $\Longrightarrow$ $\int  [d\mu(g)]_{\mathbb{R}^D} \exp \left\{  ig_{{}_{\rm YM}} \sqrt{\frac{2(N-1)}{N}} [(k, \omega_{\Sigma}) + ...
%ig_{{}_{\rm YM}} (j, N_{\Sigma}) ] \right\}$ \\
 \end{tabular}
\label{table:reformulation2}
\end{center}
\end{table} 
%%%%%%%%%%%%%%%%%%%%%%%%%%%%%%%%%%%%%%%%%%%%%%%%%%%%%%%%%%%

 %%%%%%%%%%%%%%%%%%%%%%%%%%%%%%%%%%%%%%%%%%%%%%%%%%%%%%%%%%%%%
%%%%%%%%%%%%%%%%%%%%%%%%%%%%%%%%%%%%%%%%%%%%%%%%%%%%%%%%%%%%%
\subsection{Field strength}
\label{subsection:field-strength}%%%%%%%%%%%%%%%%%%%%%%%%%%%%%%%%%%%%%%%%%%%%%%%%%%%%%%%%%%%%%
%%%%%%%%%%%%%%%%%%%%%%%%%%%%%%%%%%%%%%%%%%%%%%%%%%%%%%%%%%%%%

For the decomposition of the Yang-Mills field:
$
 \mathscr{A}_\mu =  \mathscr{V}_\mu + \mathscr{X}_\mu 
  ,
$
the field strength $\mathscr{F}[\mathscr{A}]$ of $\mathscr{A}$ is decomposed as
%in terms of new variables as 
\begin{align}
   \mathscr{F}_{\mu\nu}[\mathscr{A}] 
%\nonumber\\
:=& \partial_\mu \mathscr{A}_\nu - \partial_\nu \mathscr{A}_\mu - ig [\mathscr{A}_\mu , \mathscr{A}_\nu ]
\nonumber\\
=& \mathscr{F}_{\mu\nu}[\mathscr{V}]
+ \partial_\mu  \mathscr{X}_\nu  - \partial_\nu  \mathscr{X}_\mu 
- ig [ \mathscr{V}_\mu ,   \mathscr{X}_\nu ]
- ig [  \mathscr{X}_\mu ,  \mathscr{V}_\nu  ]
%\nonumber\\&
  - ig [  \mathscr{X}_\mu ,   \mathscr{X}_\nu ]
\nonumber\\
=& \mathscr{F}_{\mu\nu}[\mathscr{V}]
+ \mathscr{D}_\mu[\mathscr{V}]  \mathscr{X}_\nu  - \mathscr{D}_\nu[\mathscr{V}]  \mathscr{X}_\mu  - ig [  \mathscr{X}_\mu ,   \mathscr{X}_\nu ]
 ,
\end{align}
where the covariant derivative $\mathscr{D}_\mu[\mathscr{V}]$ in the background gauge field $\mathscr{V}_\mu$ is defined by
\begin{equation}
\mathscr{D}_\mu[\mathscr{V}] := \partial_\mu \mathbf{1} -ig [ \mathscr{V}_\mu , \cdot ]   \quad
%\end{equation}
%or, equivalently, 
%\begin{equation}
  ( \mathscr{D}_\mu[\mathscr{V}]^{AC} := \partial_\mu \delta^{AC} + g f^{ABC} \mathscr{V}_\mu^B ) .
\end{equation}

Moreover, 
for the decomposition of the restricted field:
$
 \mathscr{V}_\mu =\mathscr{B}_\mu +\mathscr{C}_\mu  
  ,
$
the field strength $\mathscr{F}[\mathscr{V}]$ of the restricted field $\mathscr{V}$ is further decomposed   as 
\begin{align}
   \mathscr{F}_{\mu\nu}[\mathscr{V}]
 :=& \partial_\mu \mathscr{V}_\nu - \partial_\nu \mathscr{V}_\mu - ig [\mathscr{V}_\mu , \mathscr{V}_\nu ]
\nonumber\\
=& \mathscr{F}_{\mu\nu}[\mathscr{B}] 
+ \partial_\mu  \mathscr{C}_\nu  - \partial_\nu  \mathscr{C}_\mu  
- ig [ \mathscr{B}_\mu  ,  \mathscr{C}_\nu  ]
- ig [ \mathscr{C}_\mu  ,  \mathscr{B}_\nu  ]
- ig [ \mathscr{C}_\mu  ,  \mathscr{C}_\nu  ]
\nonumber\\
=& \mathscr{F}_{\mu\nu}[\mathscr{B}]
+ \mathscr{D}_\mu[\mathscr{B}]  \mathscr{C}_\nu  - \mathscr{D}_\nu[\mathscr{B}]  \mathscr{C}_\mu  - ig [  \mathscr{C}_\mu ,   \mathscr{C}_\nu ]
 ,
\nonumber\\  
=& \mathscr{H}_{\mu\nu} + \mathscr{E}_{\mu\nu} 
 ,
\end{align}
where we have defined
\begin{align}
  \mathscr{H}_{\mu\nu} :=& \mathscr{F}_{\mu\nu}[\mathscr{B}]
 =  \partial_\mu \mathscr{B}_\nu - \partial_\nu \mathscr{B}_\mu
  -i g [\mathscr{B}_\mu , \mathscr{B}_\nu] ,
%= -i \sum_{j}  \bm{n}_j(\bm{n}_j,[\mathscr{B}_\mu, \mathscr{B}_\nu])
%\\
%=  - g^{-1}   (\partial_\mu \bm{n} \times \partial_\nu \bm{n}) ?  .
\\
  \mathscr{E}_{\mu\nu} :=& \mathscr{D}_\mu[\mathscr{B}]  \mathscr{C}_\nu  - \mathscr{D}_\nu[\mathscr{B}]  \mathscr{C}_\mu  - ig [  \mathscr{C}_\mu ,   \mathscr{C}_\nu ]  .
\end{align}
Thus   we obtain the decomposition of the field strength:
\begin{align}
  \mathscr{F}_{\mu\nu}[\mathscr{A}]  
=  \mathscr{E}_{\mu\nu} + \mathscr{H}_{\mu\nu} + \mathscr{D}_\mu[\mathscr{V}] \mathscr{X}_\nu - \mathscr{D}_\nu[\mathscr{V}] \mathscr{X}_\mu - ig [\mathscr{X}_\mu , \mathscr{X}_\nu ] .
\end{align}

%%%%%%%%%%%%%%%%%%%%%%%%%%%%%%%%%%%%%%%%%%%%%%%%%%%%%%%%%%%%%
%%%%%%%%%%%%%%%%%%%%%%%%%%%%%%%%%%%%%%%%%%%%%%%%%%%%%%%%%%%%%
\subsubsection{Field strength in the maximal option}
\label{subsubsection:field-strength-maximal}
%%%%%%%%%%%%%%%%%%%%%%%%%%%%%%%%%%%%%%%%%%%%%%%%%%%%%%%%%%%%%
%%%%%%%%%%%%%%%%%%%%%%%%%%%%%%%%%%%%%%%%%%%%%%%%%%%%%%%%%%%%%

In the maximal option,% 
\footnote{
In what follows,  the summation symbol over $j,k$ should be understood.
} 
$ \mathscr{E}_{\mu\nu}$ is simplified as follows.
Using 
\begin{align}
\mathscr{C}_\mu(x) = c_\mu^j(x) \bm{n}_j(x) ,
\end{align}
we find 
%[Exercise-11] \marginpar{Ex-11} 
%$\mathscr{F}_{\mu\nu}[\mathscr{V}]$ is further decomposed as (omitting the summation symbol over $j$):
\begin{align}
    \mathscr{E}_{\mu\nu} 
 =&  
  \partial_\mu  \mathscr{C}_\nu  - \partial_\nu  \mathscr{C}_\mu  
- ig [ \mathscr{B}_\mu  ,  \mathscr{C}_\nu  ]
+ ig [ \mathscr{B}_\nu ,  \mathscr{C}_\mu   ]
- ig [ \mathscr{C}_\mu  ,  \mathscr{C}_\nu  ]
\nonumber\\  
=&   \bm{n}_j \partial_\mu  c_\nu^j  - \bm{n}_j \partial_\nu  c_\mu^j  +  c_\nu^j \partial_\mu  \bm{n}_j  - c_\mu^j \partial_\nu  \bm{n}_j
%\nonumber\\  &
 - ig [ \mathscr{B}_\mu  ,  c_\nu^j \bm{n}_j ]
+ ig [\mathscr{B}_\nu , c_\mu^j \bm{n}_j  ]
- ig [ c_\mu^j \bm{n}_j  ,  c_\nu^k \bm{n}_k  ]
\nonumber\\  
=&   \bm{n}_j \partial_\mu  c_\nu^j  - \bm{n}_j \partial_\nu  c_\mu^j  
%\nonumber\\ &
+ c_\nu^j \mathscr{D}_\mu[\mathscr{B}]  \bm{n}_j  - c_\mu^j \mathscr{D}_\nu[\mathscr{B}]  \bm{n}_j - ig c_\mu^j c_\nu^k [  \bm{n}_j  ,   \bm{n}_k  ] 
\nonumber\\  
=&   \bm{n}_j \partial_\mu  c_\nu^j  - \bm{n}_j \partial_\nu  c_\mu^j  
,
\label{C27-E-max}
\end{align}
where  we have used  (\ref{C27-defBL2b}) and (\ref{C27-com1}) in the last step.
Therefore, we obtain 
\begin{align}
  \mathscr{E}_{\mu\nu}(x) =& %\sum_{j=1}^{N-1} 
\bm{n}_j(x) E_{\mu\nu}^j(x) , \
E_{\mu\nu}^j(x) := \partial_\mu c_\nu^j(x) - \partial_\nu c_\mu^j(x)  .
\end{align}

On the other hand,  $\mathscr{H}_{\mu\nu}$ is simplified as follows.
Using
\begin{align}
\mathscr{B}_\mu(x) =    
   ig^{-1} %\sum_{j=1}^{N-1}
 [\bm{n}_j(x), \partial_\mu  \bm{n}_j(x)] ,
\end{align}
we find  
%[Exercise-12] \marginpar{Ex-12} 
\begin{subequations}
\begin{align}
  \partial_\mu \mathscr{B}_\nu - \partial_\nu \mathscr{B}_\mu
%\nonumber\\
=&  ig^{-1}  [ \partial_\mu \bm{n}_j , \partial_\nu \bm{n}_j] - (\mu \leftrightarrow \nu) 
\label{C27-a}
\\
=&  ig^{-1}  (ig)^2 [ [\mathscr{B}_\mu ,\bm{n}_j], [\mathscr{B}_\nu ,\bm{n}_j]] - (\mu \leftrightarrow \nu) 
\label{C27-b}
\\
=&  ig    [[ \bm{n}_j, [\mathscr{B}_\nu ,\bm{n}_j]], \mathscr{B}_\mu]
+ ig [[[ \mathscr{B}_\nu ,\bm{n}_j], \mathscr{B}_\mu], \bm{n}_j]  
- (\mu \leftrightarrow \nu)  
\label{C27-c}
\\
=& 
   ig  [ \mathscr{B}_\mu, [\bm{n}_j, [ \bm{n}_j, \mathscr{B}_\nu ]]]  
+   ig [\bm{n}_j, [[ \bm{n}_j, \mathscr{B}_\nu ], \mathscr{B}_\mu]] - (\mu \leftrightarrow \nu) 
\label{C27-d}
\\
=& 
   ig  [ \mathscr{B}_\mu, [\bm{n}_j, [ \bm{n}_j, \mathscr{B}_\nu ]]] - (\mu \leftrightarrow \nu) 
- ig    [\bm{n}_j, [\bm{n}_j, [\mathscr{B}_\mu  , \mathscr{B}_\nu]]] 
\label{C27-e}
\\
=& 
   ig   ([ \mathscr{B}_\mu,  \mathscr{B}_\nu ]  - [ \mathscr{B}_\nu,  \mathscr{B}_\mu ])
- ig    [\bm{n}_j, [\bm{n}_j, [\mathscr{B}_\mu  , \mathscr{B}_\nu]]] 
\label{C27-f}
\\
=& 
  2ig   [ \mathscr{B}_\mu,  \mathscr{B}_\nu ]  
- ig  [\mathscr{B}_\mu  , \mathscr{B}_\nu]
+   ig \bm{n}_j (\bm{n}_j \cdot [\mathscr{B}_\mu  , \mathscr{B}_\nu]) 
\label{C27-g}
\\
=&  i g [\mathscr{B}_\mu , \mathscr{B}_\nu] 
+ig   \bm{n}_j(\bm{n}_j \cdot [\mathscr{B}_\mu, \mathscr{B}_\nu])
,
\label{C27-h}
\end{align}
\label{C27-dB-min}
\end{subequations}
where we have used 
(\ref{C27-B}) in (\ref{C27-a}), 
(\ref{C27-defBL22}) in (\ref{C27-b}), 
the Jacobi identity for $\mathscr{B}_\mu$, $\bm{n}_j$ and $[\mathscr{B}_\nu ,\bm{n}_j]$ in (\ref{C27-c}), 
interchanged the ordering of the commutators in (\ref{C27-d}), 
the Jacobi identity in (\ref{C27-e}), 
the algebraic identity (\ref{C27-idv}) with (\ref{C27-defBL}) to obtain the first term of (\ref{C27-f})  
and again 
the algebraic identity (\ref{C27-idv}) for $[\mathscr{B}_\mu , \mathscr{B}_\nu]$ in (\ref{C27-g}).
Therefore we obtain 
\begin{align}
 \mathscr{H}_{\mu\nu}(x)  
%= \partial_\mu \mathscr{B}_\nu - \partial_\nu \mathscr{B}_\mu
%  -i g [\mathscr{B}_\mu , \mathscr{B}_\nu] 
=  ig   \bm{n}_j(x)(\bm{n}_j(x) \cdot [\mathscr{B}_\mu(x), \mathscr{B}_\nu(x)])
 .
\end{align}
Thus we find $\mathscr{H}_{\mu\nu}$ is written as the linear combination of all the color fields just like $\mathscr{E}_{\mu\nu}$:
\begin{align}
 \mathscr{H}_{\mu\nu}(x)   = \mathscr{F}_{\mu\nu}[\mathscr{B}](x)  = %\sum_{j=1}^{N-1} 
\bm{n}_j(x) H^j_{\mu\nu}(x)
%=  ig \bm{n}_j(\bm{n}_j,[\mathscr{B}_\mu, \mathscr{B}_\nu])
, 
\ 
H^j_{\mu\nu}(x) = ig  (\bm{n}_j(x) \cdot [\mathscr{B}_\mu(x), \mathscr{B}_\nu(x)]) .
\end{align}

Moreover, we find 
\begin{align}
 \mathscr{H}_{\mu\nu}  
=&  ig   \bm{n}_j(\bm{n}_j \cdot [\mathscr{B}_\mu, \mathscr{B}_\nu])
%\nonumber\\
%=&  ig   \bm{n}_j( [\bm{n}_j, \mathscr{B}_\mu ] \mathscr{B}_\nu )
%=  - \bm{n}_j( \partial_\mu \bm{n}_j \mathscr{B}_\nu )
%\nonumber\\
=   ig   \bm{n}_j(\mathscr{B}_\mu \cdot [ \mathscr{B}_\nu ,\bm{n}_j ] )
%\nonumber\\
=    \bm{n}_j(\mathscr{B}_\mu \cdot \partial_\nu \bm{n}_j   )
%\nonumber\\
=   ig^{-1} \bm{n}_j(   [\bm{n}_k , \partial_\mu  \bm{n}_k ] \cdot \partial_\nu \bm{n}_j   )
\nonumber\\
=&  ig^{-1}   \bm{n}_j(  \bm{n}_k \cdot [\partial_\mu  \bm{n}_k    , \partial_\nu \bm{n}_j]   )
.
\end{align}

%%%%%%%%%%%%%%%%%%%%%%%%%%%%%%%%%%%%%%%%%%%%%%%%%%%%%%%%%%%%%
%%%%%%%%%%%%%%%%%%%%%%%%%%%%%%%%%%%%%%%%%%%%%%%%%%%%%%%%%%%%%
\subsubsection{Field strength in the minimal option}
\label{subsubsection:field-strength-minimal-1}%%%%%%%%%%%%%%%%%%%%%%%%%%%%%%%%%%%%%%%%%%%%%%%%%%%%%%%%%%%%%
%%%%%%%%%%%%%%%%%%%%%%%%%%%%%%%%%%%%%%%%%%%%%%%%%%%%%%%%%%%%%

First, we apply the formula (\ref{C27-idv2-min}) to $\mathscr{E}_{\mu\nu}$ to obtain
\begin{align}
  \mathscr{E}_{\mu\nu}  =  \tilde{\mathscr{E}}_{\mu\nu} +  (\mathscr{E}_{\mu\nu} \cdot \bm{h})\bm{h} 
+  2\frac{N-1}{N}  [\bm{h} , [\bm{h} , \mathscr{E}_{\mu\nu}]]
% := v_\parallel + v_\perp
 ,
\end{align}
where
\begin{equation}
 \tilde{\mathscr{E}}_{\mu\nu} 
=  \mathscr{E}_{\mu\nu}^k \bm{u}_k
=   (\mathscr{E}_{\mu\nu} \cdot \bm{u}_k) \bm{u}_k 
=   2{\rm tr}(\mathscr{E}_{\mu\nu} \bm{u}_k)\bm{u}_k
 . 
\end{equation}
The second term is 
\begin{align}
%2{\rm tr}(\mathscr{H}_{\mu\nu} \bm{h}) 
%=& 
  \bm{h} \cdot \mathscr{E}_{\mu\nu} 
%\nonumber\\
=& \bm{h} \cdot (\partial_\mu \mathscr{C}_\nu-\partial_\nu \mathscr{C}_\mu)  
-i g  (\bm{h} \cdot [ \mathscr{B}_\mu  ,  \mathscr{C}_\nu  ])
-i g  (\bm{h} \cdot [ \mathscr{C}_\mu  ,  \mathscr{B}_\nu  ])
%\nonumber\\&
-i g  (\bm{h} \cdot [\mathscr{C}_\mu , \mathscr{C}_\nu]) 
\nonumber\\
=& \bm{h} \cdot (\partial_\mu \mathscr{C}_\nu-\partial_\nu \mathscr{C}_\mu)  
-i g  ( \mathscr{B}_\mu \cdot [\mathscr{C}_\nu  , \bm{h}] )
-i g  ([\bm{h},   \mathscr{C}_\mu ] \cdot \mathscr{B}_\nu )
%\nonumber\\ &
-i g  ([\bm{h},  \mathscr{C}_\mu ] \cdot \mathscr{C}_\nu) 
\nonumber\\
=& \bm{h} \cdot (\partial_\mu \mathscr{C}_\nu-\partial_\nu \mathscr{C}_\mu)  
 ,
\end{align}
where we have used 
${\rm tr}(AB)={\rm tr}(BA)$ and 
${\rm tr}([A,B]C)={\rm tr}(A,[B,C])$ 
and
$[\bm{h},  \mathscr{C}_\mu ]=0$.
The last term vanishes, since  
\begin{align}
    [\bm{h} , [\bm{h} , \mathscr{E}_{\mu\nu}]]
=& [\bm{h} , [\bm{h} , \mathscr{D}_\mu[\mathscr{B}]  \mathscr{C}_\nu ]] - [\bm{h} , [\bm{h} , \mathscr{D}_\nu[\mathscr{B}]  \mathscr{C}_\mu ]] - ig [\bm{h} , [\bm{h} , [\mathscr{C}_\mu , \mathscr{C}_\nu] ]]
\nonumber\\
=& [\bm{h} , [-\mathscr{D}_\mu[\mathscr{B}] \bm{h} ,  \mathscr{C}_\nu ]] - [\bm{h} , [-\mathscr{D}_\nu[\mathscr{B}]\bm{h} ,   \mathscr{C}_\mu ]] - ig [\bm{h} , [\bm{h} , [\mathscr{C}_\mu , \mathscr{C}_\nu] ]]
\nonumber\\
=&    ig [\bm{h} , [ \mathscr{C}_\mu , [ \mathscr{C}_\nu, \bm{h} ] ]] 
+ ig [\bm{h} , [ \mathscr{C}_\nu, [\bm{h}  , \mathscr{C}_\mu] ]] = 0 
   ,
\end{align}
where we have used 
$
0=\mathscr{D}_\mu[\mathscr{B}] [\bm{h} , \mathscr{C}_\nu ]=[ \bm{h}, \mathscr{D}_\mu[\mathscr{B}] \mathscr{C}_\nu ]+[ \mathscr{D}_\mu[\mathscr{B}]\bm{h} , \mathscr{C}_\nu ]
$ 
following from 
$[\bm{h},  \mathscr{C}_\nu  ]=0$ 
in the second equality, 
$
 0= \mathscr{D}_\mu[\mathscr{B}] \bm{h}= \partial_\mu  \bm{h} - ig [\mathscr{B}_\mu ,\bm{h}]
$
and the Jacobi identity 
$
[A,[B,C]]+[B,[C,A]]+[C,[A,B]]=0
$
in the third equality, 
and
$[ \mathscr{C}_\mu ,\bm{h}]=0$ again 
in the last equality. 
Therefore, we obtain
\begin{align}
  \mathscr{E}_{\mu\nu}  =    \bm{h} (\bm{h} \cdot (\partial_\mu \mathscr{C}_\nu-\partial_\nu \mathscr{C}_\mu)) + \bm{u}_k (\bm{u}_k \cdot \mathscr{E}_{\mu\nu}) 
  .
\end{align}

Second, we apply the formula (\ref{C27-idv2-min}) to $\mathscr{H}_{\mu\nu}=\mathscr{F}_{\mu\nu}[\mathscr{B}]$ to obtain
\begin{align}
  \mathscr{H}_{\mu\nu}  =  \tilde{\mathscr{H}}_{\mu\nu} +  (\mathscr{H}_{\mu\nu} \cdot \bm{h})\bm{h} 
+  2\frac{N-1}{N}  [\bm{h} , [\bm{h} , \mathscr{H}_{\mu\nu}]]
% := v_\parallel + v_\perp
 ,
\end{align}
where
\begin{equation}
 \tilde{\mathscr{H}}_{\mu\nu} =  \mathscr{H}_{\mu\nu}^k \bm{u}_k
=   (\mathscr{H}_{\mu\nu} \cdot \bm{u}_k) \bm{u}_k 
=   2{\rm tr}(\mathscr{H}_{\mu\nu} \bm{u}_k)\bm{u}_k
 . 
\end{equation}
The second term is 
\begin{align}
%2{\rm tr}(\mathscr{H}_{\mu\nu} \bm{h}) 
%=& 
  \bm{h} \cdot \mathscr{H}_{\mu\nu}  
%\nonumber\\
=&   \bm{h} \cdot \partial_\mu \mathscr{B}_\nu  -  \bm{h} \cdot \partial_\nu \mathscr{B}_\mu  -i g \bm{h} \cdot [\mathscr{B}_\mu , \mathscr{B}_\nu]  
\nonumber\\
=& - \partial_\mu  \bm{h} \cdot \mathscr{B}_\nu +  \partial_\nu \bm{h} \cdot \mathscr{B}_\mu -i g \bm{h} \cdot [\mathscr{B}_\mu , \mathscr{B}_\nu]  
\nonumber\\
=& - ig [\mathscr{B}_\mu ,\bm{h}] \cdot \mathscr{B}_\nu  + ig [\mathscr{B}_\nu ,\bm{h}] \cdot \mathscr{B}_\mu -i g \bm{h} \cdot [\mathscr{B}_\mu , \mathscr{B}_\nu]  
\nonumber\\
=& - ig [\mathscr{B}_\nu , \mathscr{B}_\mu ] \cdot \bm{h}    + ig [\mathscr{B}_\mu , \mathscr{B}_\nu ] \cdot \bm{h}  -i g \bm{h} \cdot [\mathscr{B}_\mu , \mathscr{B}_\nu]  
\nonumber\\
=&   ig \bm{h} \cdot [\mathscr{B}_\mu , \mathscr{B}_\nu ]   
 ,
\end{align}
where we have used 
$
0 = \partial_\mu (\bm{h}(x) \cdot \mathscr{B}_\nu(x) )
= \bm{h}(x) \cdot \partial_\mu \mathscr{B}_\nu(x) +  \partial_\mu \bm{h}(x) \cdot \mathscr{B}_\nu(x) 
$
which follows from 
$
  \bm{h}(x) \cdot \mathscr{B}_\mu(x) 
= 2{\rm tr}( \bm{h}(x) \mathscr{B}_\mu(x))=0 
$
in the third equality, 
$
 0= \mathscr{D}_\mu[\mathscr{B}] \bm{h}= \partial_\mu  \bm{h} - ig [\mathscr{B}_\mu ,\bm{h}]
$
in the fourth equality, 
and 
${\rm tr}(AB)={\rm tr}(BA)$ and 
${\rm tr}([A,B]C)={\rm tr}(A,[B,C])$ in the fifth equality.
The last term vanishes, since  
\begin{align}
    [\bm{h} , [\bm{h} , \mathscr{H}_{\mu\nu}]]
=& [\bm{h} , [\bm{h} , \partial_\mu \mathscr{B}_\nu ]] - [\bm{h} , [\bm{h} , \partial_\nu \mathscr{B}_\mu ]] - ig [\bm{h} , [\bm{h} , [\mathscr{B}_\mu , \mathscr{B}_\nu] ]]
\nonumber\\
=& [\bm{h} , [\bm{h} , \partial_\mu \mathscr{B}_\nu ]] - [\bm{h} , [\bm{h} , \partial_\nu \mathscr{B}_\mu ]] 
%\nonumber\\&
+ ig [\bm{h} , [\mathscr{B}_\nu , [\bm{h}  , \mathscr{B}_\mu] ]]
+ ig [\bm{h} , [\mathscr{B}_\mu , [\mathscr{B}_\nu , \bm{h}  ] ]]
\nonumber\\
=& - [\bm{h} , [\partial_\mu \mathscr{B}_\nu , \bm{h}  ]] + [\bm{h} , [ \partial_\nu \mathscr{B}_\mu , \bm{h} ]] 
%\nonumber\\ &
-   [\bm{h} , [\mathscr{B}_\nu ,  \partial_\mu \bm{h} ]]
+   [\bm{h} , [\mathscr{B}_\mu , \partial_\nu   \bm{h}   ]]
\nonumber\\
=& - [\bm{h} , \partial_\mu [ \mathscr{B}_\nu , \bm{h}  ]] + [\bm{h} , \partial_\nu [  \mathscr{B}_\mu , \bm{h} ]] 
\nonumber\\
=& ig^{-1} [\bm{h} , \partial_\mu \partial_\nu \bm{h}  ] -ig^{-1} [\bm{h} , \partial_\nu \partial_\mu \bm{h}  ] 
= 0
   ,
\end{align}
where we have used the Jacobi identity 
$
[A,[B,C]]+[B,[C,A]]+[C,[A,B]]=0
$
in the second equality and 
$
 0= \mathscr{D}_\mu[\mathscr{B}] \bm{h}= \partial_\mu  \bm{h} - ig [\mathscr{B}_\mu ,\bm{h}]
$
 in the third and fifth equalities. 
By using the explicit form:
\begin{align}
 \mathscr{B}_\mu 
 = ig^{-1} \frac{2(N-1)}{N} [ \bm{h}, \partial_\mu \bm{h}] 
 ,
 \label{C27-B-form}
\end{align}
 we can rewrite 
\begin{align}
   ig   \bm{h} (\bm{h} \cdot [\mathscr{B}_\mu, \mathscr{B}_\nu])
%\nonumber\\
%=&  ig   \bm{n}_j( [\bm{n}_j, \mathscr{B}_\mu ] \mathscr{B}_\nu )
%=  - \bm{n}_j( \partial_\mu \bm{n}_j \mathscr{B}_\nu )
%\nonumber\\
=&  ig   \bm{h}(\mathscr{B}_\mu \cdot [ \mathscr{B}_\nu ,\bm{h}] )
\nonumber\\
=&   \bm{h}(\mathscr{B}_\mu \cdot \partial_\nu \bm{h}   )
\nonumber\\
=& \frac{2(N-1)}{N} ig^{-1} \bm{h}(   [\bm{h} , \partial_\mu  \bm{h} ] \cdot \partial_\nu \bm{h}   )
\nonumber\\
=& \frac{2(N-1)}{N} ig^{-1}   \bm{h}(  \bm{h} \cdot [\partial_\mu \bm{h}    , \partial_\nu \bm{h}]   )
.
\end{align}
Therefore, we obtain
\begin{align}
  \mathscr{H}_{\mu\nu}  
=  ig \bm{h}(\bm{h} \cdot [\mathscr{B}_\mu , \mathscr{B}_\nu ] ) +  \bm{u}_k (\bm{u}_k \cdot \mathscr{H}_{\mu\nu})  
=  \frac{2(N-1)}{N} ig^{-1}   \bm{h}(  \bm{h} \cdot [\partial_\mu \bm{h}    , \partial_\nu \bm{h}]   ) +  \bm{u}_k (\bm{u}_k \cdot \mathscr{H}_{\mu\nu}) 
% := v_\parallel + v_\perp
 .
\end{align}

Thus, we obtain
\begin{align}
   \mathscr{F}_{\mu\nu}[\mathscr{V}]
%:=& \mathscr{H}_{\mu\nu} + \mathscr{E}_{\mu\nu} 
=&  \bm{h} (\bm{h} \cdot (\partial_\mu \mathscr{C}_\nu-\partial_\nu \mathscr{C}_\mu)) + ig \bm{h}(\bm{h} \cdot [\mathscr{B}_\mu , \mathscr{B}_\nu ] ) +  \bm{u}_k (\bm{u}_k \cdot \mathscr{F}_{\mu\nu}[\mathscr{V}])  
\nonumber\\
=&  \bm{h} (\bm{h} \cdot (\partial_\mu \mathscr{C}_\nu-\partial_\nu \mathscr{C}_\mu)) 
+ \frac{2(N-1)}{N} ig^{-1}   \bm{h}(  \bm{h} \cdot [\partial_\mu \bm{h}    , \partial_\nu \bm{h}]   )
+  \bm{u}_k (\bm{u}_k \cdot \mathscr{F}_{\mu\nu}[\mathscr{V}])  
 .
\end{align}

In what follows, we focus on the ``parallel'' part of the field strength $\mathscr{F}_{\mu\nu}[\mathscr{V}]$ defined by
%[Exercise-13] \marginpar{Ex-13} 
%\begin{subequations}
\begin{align}
  {\rm tr}( \bm{h}\mathscr{F}_{\mu\nu}[\mathscr{V}] )
%\nonumber\\
=& {\rm tr}( \bm{h}\mathscr{F}_{\mu\nu}[\mathscr{B}] )
+ {\rm tr}( \bm{h} \mathscr{D}_\mu[\mathscr{B}]  \mathscr{C}_\nu)
- {\rm tr}( \bm{h} \mathscr{D}_\nu[\mathscr{B}]  \mathscr{C}_\mu)
- ig {\rm tr}( \bm{h} [  \mathscr{C}_\mu ,   \mathscr{C}_\nu ])
\nonumber\\
=& {\rm tr}( \bm{h}\mathscr{F}_{\mu\nu}[\mathscr{B}] )
- {\rm tr}( (\mathscr{D}_\mu[\mathscr{B}] \bm{h})  \mathscr{C}_\nu)
+ \partial_\mu {\rm tr}(  \bm{h} \mathscr{C}_\nu)
\nonumber\\&
+ {\rm tr}( (\mathscr{D}_\nu[\mathscr{B}] \bm{h})   \mathscr{C}_\mu)
-  \partial_\nu  {\rm tr} ( \bm{h} \mathscr{C}_\mu)
-  ig {\rm tr}( \mathscr{C}_\nu  [ \bm{h},  \mathscr{C}_\mu ])
\nonumber\\
=& {\rm tr}( \bm{h}\mathscr{F}_{\mu\nu}[\mathscr{B}] )
+ \partial_\mu {\rm tr}(  \bm{h} \mathscr{C}_\nu)
-  \partial_\nu  {\rm tr} ( \bm{h} \mathscr{C}_\mu)
 ,
 \label{C27-hF-min}
\end{align}
%\end{subequations}
where we have used the defining equations for $\mathscr{B}_\mu$ and $\mathscr{C}_\mu$:
$\mathscr{D}_\mu[\mathscr{B}] \bm{h}=0$ and $[ \bm{h},  \mathscr{C}_\mu ]=0$.
Moreover, the ``parallel'' part of the field strength $\mathscr{F}_{\mu\nu}[\mathscr{B}]=\mathscr{H}_{\mu\nu}$ reads
%[Exercise-14] \marginpar{Ex-14} 
\begin{subequations}
\begin{align}
 {\rm tr}( \bm{h} \mathscr{H}_{\mu\nu})
 =& {\rm tr}( \bm{h}\mathscr{F}_{\mu\nu}[\mathscr{B}] )
 \nonumber\\
=& 
 {\rm tr}( \bm{h} \partial_\mu \mathscr{B}_\nu - \bm{h} \partial_\nu \mathscr{B}_\mu
  -i g \bm{h} [\mathscr{B}_\mu , \mathscr{B}_\nu])
\\
=& 
 {\rm tr}( \bm{h} \partial_\mu \mathscr{B}_\nu - \bm{h} \partial_\nu \mathscr{B}_\mu
  +i g \mathscr{B}_\nu  [\mathscr{B}_\mu , \bm{h}])
\\
=& 
 {\rm tr}( \bm{h} \partial_\mu \mathscr{B}_\nu - \bm{h} \partial_\nu \mathscr{B}_\mu
  + \mathscr{B}_\nu \partial_\mu \bm{h} )
\\
=& 
 \partial_\mu {\rm tr}(\bm{h} \mathscr{B}_\nu) - {\rm tr}(\bm{h} \partial_\nu \mathscr{B}_\mu )
\\
=& 
  {\rm tr}(\mathscr{B}_\mu  \partial_\nu  \bm{h}) - \partial_\nu {\rm tr}(\bm{h} \mathscr{B}_\mu) 
\\
=& 
  {\rm tr}(\mathscr{B}_\mu  \partial_\nu  \bm{h}) 
 ,
\end{align}
 \label{C27-hH-min}
\end{subequations}
where we have used ${\rm tr}(\bm{h} \mathscr{B}_\mu)=0$ twice and an identity:
$
 {\rm tr} (A[B,C]) =  {\rm tr} ( [A,B]C) =  {\rm tr} ([C,A]B)  
$.
Using the explicit form of $\mathscr{B}_\mu$ (\ref{C27-B-form}),
%\begin{align}
% \mathscr{B}_\mu 
% = ig^{-1} \frac{2(N-1)}{N} [ \bm{h}, \partial_\mu \bm{h}] 
% ,
%\end{align}
we have
\begin{align}
 {\rm tr}( \bm{h} \mathscr{H}_{\mu\nu})
 =&   {\rm tr}( \bm{h}\mathscr{F}_{\mu\nu}[\mathscr{B}] )
%\nonumber\\
=    \frac{2(N-1)}{N} {\rm tr}( ig^{-1}[ \bm{h}, \partial_\mu \bm{h}] \partial_\nu \bm{h} )
%\nonumber\\
=    \frac{2(N-1)}{N} 
 {\rm tr}(ig^{-1} \bm{h} [\partial_\mu \bm{h}, \partial_\nu \bm{h}])
 .
\end{align}
It should be remarked that  the equality does not hold   without the trace:
$\bm{h}\mathscr{F}_{\mu\nu}[\mathscr{B}] \ne \frac{2(N-1)}{N} 
 ig^{-1} \bm{h} [\partial_\mu \bm{h}, \partial_\nu \bm{h}] $.
 Thus, we obtain
\begin{align}
  {\rm tr}( \bm{h}\mathscr{F}_{\mu\nu}[\mathscr{V}] )
%\nonumber\\
 =  \frac{2(N-1)}{N} ig^{-1}
 {\rm tr}( \bm{h} [\partial_\mu \bm{h}, \partial_\nu \bm{h}])
+ \partial_\mu {\rm tr}(  \bm{h} \mathscr{C}_\nu)
-  \partial_\nu  {\rm tr} ( \bm{h} \mathscr{C}_\mu)
 .
 \label{C27-hF-min2}
\end{align}

The same result is obtained by calculating explicitly the field strength
$
\mathscr{F}_{\mu\nu}[\mathscr{B}]  
  :=  \partial_\mu \mathscr{B}_\nu - \partial_\nu \mathscr{B}_\mu  -i g [\mathscr{B}_\mu , \mathscr{B}_\nu] 
$
using the property of $\bm{h}$ (\ref{C27-eq:Hidentity})  
%\begin{align}
% \bm{h}\bm{h} = \frac{1}{2N} {\bf 1} + \frac{2-N}{\sqrt{2N(N-1)}} \bm{h} 
% ,
%\end{align}
which follows from the relation for the last Cartan generator $H_{N-1}$:
\begin{align}
 H_{N-1} H_{N-1}  = \frac{1}{2N} {\bf 1} + \frac{2-N}{\sqrt{2N(N-1)}} H_{N-1}
 .
\end{align}
%Alternatively, the same result is obtained without using the explicit form of $\mathscr{B}_\mu$ as  
%[Exercise-13] \marginpar{Ex-13} 
%\begin{align}
%[\bm{h}, [ \bm{h}, \mathscr{Z} ]]
%=& \{ \bm{h}^2, \mathscr{Z} \} - 2\bm{h} \mathscr{Z} \bm{h} 
%\nonumber\\
%=& \frac{1}{N} \mathscr{Z} + \frac{2-N}{\sqrt{2N(N-1)}} ( \bm{h}   %\mathscr{Z} +    \mathscr{Z} \bm{h}) - 2\bm{h} \mathscr{Z} \bm{h}  
% \label{C27-hhX-min}
%\end{align}

%%%%%%%%%%%%%%%%%%%%%%%%%%%%%%%%%%%%%%%%%%%%%%%%%%%%%%%%%%%%%
\subsubsection{Yang-Mills action in the minimal option}
\label{subsubsect:action-minimal}%%%%%%%%%%%%%%%%%%%%%%%%%%%%%%%%%%%%%%%%%%%%%%%%%%%%%%%%%%%%%
%%%%%%%%%%%%%%%%%%%%%%%%%%%%%%%%%%%%%%%%%%%%%%%%%%%%%%%%%%%%%

It is shown (see \ref{section:field-strength-minimal-2}) that the field strength $\mathscr{F}_{\mu\nu}[\mathscr{A}]$ in the minimal case is decomposed into the $\tilde{H}$-commutative part $\mathscr{F}_{\mu\nu}^{\tilde{H}}$ and the remaining $\tilde{H}$-non-commutative part $\mathscr{F}_{\mu\nu}^{G/\tilde{H}}$, which are orthogonal to each other:
\begin{align}
  \mathscr{F}_{\mu\nu}[\mathscr{A}]  
=& \mathscr{F}_{\mu\nu}^{\tilde{H}} + \mathscr{F}_{\mu\nu}^{G/\tilde{H}}  , 
\nonumber\\
 & \mathscr{F}_{\mu\nu}^{\tilde{H}} =  \mathscr{F}_{\mu\nu}[\mathscr{V}] -i g [ \mathscr{X}_\mu , \mathscr{X}_\nu ]  
 \in  \tilde{\mathscr{H}},  
\nonumber\\
 & \mathscr{F}_{\mu\nu}^{G/\tilde{H}} =  \mathscr{D}_\mu[\mathscr{V}] \mathscr{X}_\nu - \mathscr{D}_\nu[\mathscr{V}] \mathscr{X}_\mu 
 \in \mathscr{G} - \tilde{\mathscr{H}} .
\end{align}

For the $SU(N)$ Yang-Mills theory in the minimal option, therefore,  
the  Lagrangian density is decomposed as
\begin{align}
   \mathscr{L}_{\rm YM} =& 
   -\frac{1}{2} {\rm tr}(\mathscr{F}_{\mu\nu}[\mathscr{A}]  \mathscr{F}^{\mu\nu}[\mathscr{A}] )
\nonumber\\
=&  -\frac{1}{2} {\rm tr}(\mathscr{F}_{\mu\nu}[\mathscr{A}]^{\tilde{H}} \mathscr{F}^{\mu\nu}[\mathscr{A}]^{\tilde{H}})
-\frac{1}{2} {\rm tr}(\mathscr{F}_{\mu\nu}[\mathscr{A}]^{G/\tilde{H}} \mathscr{F}^{\mu\nu}[\mathscr{A}]^{G/\tilde{H}})
 ,
\end{align}
where the cross term vanishes identically: 
\begin{align}
   {\rm tr}(\mathscr{F}_{\mu\nu}[\mathscr{A}]^{\tilde{H}} \mathscr{F}^{\mu\nu}[\mathscr{A}]^{G/\tilde{H}})
%\nonumber\\
=& {\rm tr}\{ (\mathscr{F}_{\mu\nu}[\mathscr{V}] -i g [ \mathscr{X}_\mu , \mathscr{X}_\nu ] ) (\mathscr{D}_\mu[\mathscr{V}] \mathscr{X}_\nu - \mathscr{D}_\nu[\mathscr{V}] \mathscr{X}_\mu) \} 
\nonumber\\
=& {\rm tr}\{ \mathscr{F}_{\mu\nu}[\mathscr{V}] (\mathscr{D}_\mu[\mathscr{V}] \mathscr{X}_\nu - \mathscr{D}_\nu[\mathscr{V}] \mathscr{X}_\mu) \}
\nonumber\\&
 + {\rm tr}\{ -i g [ \mathscr{X}_\mu , \mathscr{X}_\nu ] (\mathscr{D}_\mu[\mathscr{V}] \mathscr{X}_\nu - \mathscr{D}_\nu[\mathscr{V}] \mathscr{X}_\mu) \} .
\end{align}
Consequently, the Lagrangian density does not contain the linear and trilinear terms in $\mathscr{X}$. 
Thus the  Lagrangian density of the $SU(N)$ Yang-Mills theory in the minimal option is decomposed as
\begin{align}
   \mathscr{L}_{\rm YM} 
%=& -\frac{1}{2} {\rm tr}(\mathscr{F}_{\mu\nu}[\mathscr{A}] \mathscr{F}^{\mu\nu}[\mathscr{A}])
%\\
=& -\frac{1}{4} \mathscr{F}_{\mu\nu}[\mathscr{A}] \cdot \mathscr{F}^{\mu\nu}[\mathscr{A}]
\\
=& -\frac{1}{4} \mathscr{F}_{\mu\nu}[\mathscr{V}]^2
%\nonumber\\&
%- \frac{1}{2} \mathscr{F}^{\mu\nu}[\mathscr{V}] \cdot (\mathscr{D}_\mu[\mathscr{V}] \mathscr{X}_\nu - \mathscr{D}_\nu[\mathscr{V}] \mathscr{X}_\mu)
%\nonumber\\&
+ \frac{1}{2}  \mathscr{F}_{\mu\nu}[\mathscr{V}] \cdot ig[ \mathscr{X}^\mu , \mathscr{X}^\nu ]
%\nonumber\\&
%+ \frac{1}{2}  (\mathscr{D}_\mu[\mathscr{V}] \mathscr{X}_\nu - \mathscr{D}_\nu[\mathscr{V}] \mathscr{X}_\mu) \cdot ig[ \mathscr{X}^\mu , \mathscr{X}^\nu ]
%\nonumber\\&
- \frac14 (i g [ \mathscr{X}_\mu , \mathscr{X}_\nu ])^2
%+ \mathcal{O}(\mathscr{X}^3) 
\nonumber\\&
- \frac{1}{4} (\mathscr{D}_\mu[\mathscr{V}] \mathscr{X}_\nu - \mathscr{D}_\nu[\mathscr{V}] \mathscr{X}_\mu)^2 
 .
\label{C27-YM-Lagrangian1}
\end{align}
%where the cross terms,  
%$
%- \frac{1}{2} \mathscr{F}^{\mu\nu}[\mathscr{V}] \cdot (\mathscr{D}_\mu[\mathscr{V}] \mathscr{X}_\nu - \mathscr{D}_\nu[\mathscr{V}] \mathscr{X}_\mu)
%$
%and
%$
%\frac{1}{2} ig[ \mathscr{X}^\mu , \mathscr{X}^\nu ]  \cdot (\mathscr{D}_\mu[\mathscr{V}] \mathscr{X}_\nu - \mathscr{D}_\nu[\mathscr{V}] \mathscr{X}_\mu) 
%$ 
%vanish identically. 

Then the last term in the Lagrangian density is rewritten using integration by parts (or up to   total derivatives) as 
\begin{align}
   \frac{1}{4} (\mathscr{D}_\mu[\mathscr{V}] \mathscr{X}_\nu - \mathscr{D}_\nu[\mathscr{V}] \mathscr{X}_\mu)^2
%\nonumber\\
=&   \frac{1}{2} (- \mathscr{X}_\mu \cdot \mathscr{D}_\nu[\mathscr{V}]\mathscr{D}^\nu[\mathscr{V}] \mathscr{X}^\mu + \mathscr{X}_\mu  \cdot \mathscr{D}_\nu[\mathscr{V}]\mathscr{D}^\mu[\mathscr{V}]   \mathscr{X}^\nu )  
\nonumber\\
=& \frac{1}{2} \mathscr{X}^\mu  \cdot  \{ - \mathscr{D}_\rho[\mathscr{V}]\mathscr{D}^\rho[\mathscr{V}] g_{\mu\nu} +  \mathscr{D}_\nu[\mathscr{V}]\mathscr{D}_\mu[\mathscr{V}] \} \mathscr{X}^\nu
\nonumber\\
=& \frac{1}{2} \mathscr{X}^{\mu A}  \{ - (\mathscr{D}_\rho[\mathscr{V}]\mathscr{D}_\rho[\mathscr{V}])^{AB} g_{\mu\nu} 
- [ \mathscr{D}_\mu[\mathscr{V}], \mathscr{D}_\nu[\mathscr{V}]]^{AB}
\nonumber\\&
+  (\mathscr{D}_\mu[\mathscr{V}]  \mathscr{D}_\nu[\mathscr{V}])^{AB} \} \mathscr{X}^{\nu B} 
\nonumber\\
=& \frac{1}{2} \mathscr{X}^{\mu A}  \{ - (\mathscr{D}_\rho[\mathscr{V}]\mathscr{D}_\rho[\mathscr{V}])^{AB} g_{\mu\nu} 
+ gf^{ABC} \mathscr{F}_{\mu\nu}^{C}[\mathscr{V}] 
\nonumber\\&
+  \mathscr{D}_\mu[\mathscr{V}]^{AC} \mathscr{D}_\nu[\mathscr{V}]^{CB} \} \mathscr{X}^{\nu B}  ,
\end{align}
where we have used the relation:
\begin{align}
[ \mathscr{D}_\mu[\mathscr{V}], \mathscr{D}_\nu[\mathscr{V}]]^{AB}
=&  [ \mathscr{D}_\mu[\mathscr{V}]^{AC}, \mathscr{D}_\nu[\mathscr{V}]^{CB}] 
%\nonumber\\ 
=  - gf^{ABC} \mathscr{F}_{\mu\nu}^{C}[\mathscr{V}]  .
\end{align}
Thus we obtain
\begin{align}
   \mathscr{L}_{\rm YM} 
=& - \frac{1}{4} \mathscr{F}_{\mu\nu}[\mathscr{V40}]^2
%\nonumber\\&
%- \frac{1}{2} \mathscr{F}^{\mu\nu}[\mathscr{V}] \cdot (\mathscr{D}_\mu[\mathscr{V}] \mathscr{X}_\nu - \mathscr{D}_\nu[\mathscr{V}] \mathscr{X}_\mu)
%\nonumber\\&
-     \frac{1}{2} \mathscr{X}^{\mu A}  W_{\mu\nu}^{AB} \mathscr{X}^{\nu B} 
%+ \mathcal{O}(\mathscr{X}^3)  
%=& \frac{1}{4} \mathscr{F}_{\mu\nu}[\mathscr{V40}]^2
%- \frac{1}{2} ( \mathscr{D}_\mu[\mathscr{V}] \mathscr{F}_{\mu\nu}[\mathscr{V}] \cdot  \mathscr{X}_\nu - \mathscr{D}_\nu[\mathscr{V}] \mathscr{F}_{\mu\nu}[\mathscr{V}] \cdot  \mathscr{X}_\mu)
%+     \frac{1}{2} \mathscr{X}_\mu^A  W_{\mu\nu}^{AB} \mathscr{X}_\nu^B 
%+ \mathcal{O}(\mathscr{X}^3) .
%\nonumber\\&
%+ \frac{1}{2}  (\mathscr{D}_\mu[\mathscr{V}] \mathscr{X}_\nu - \mathscr{D}_\nu[\mathscr{V}] \mathscr{X}_\mu) \cdot ig[ \mathscr{X}^\mu , \mathscr{X}^\nu ]
%\nonumber\\&
- \frac14 (i g [ \mathscr{X}_\mu , \mathscr{X}_\nu ])^2 
 ,
\label{C27-YM-Lagrangian2}
\end{align}
where we have defined
\begin{align}
W_{\mu\nu}^{AB}  :=  - (\mathscr{D}_\rho[\mathscr{V}]\mathscr{D}^\rho[\mathscr{V}])^{AB} g_{\mu\nu} 
+ 2gf^{ABC} \mathscr{F}_{\mu\nu}^{C}[\mathscr{V}] 
%\nonumber\\&
+  \mathscr{D}_\mu[\mathscr{V}]^{AC} \mathscr{D}_\nu[\mathscr{V}]^{CB} . 
\label{W2}
\end{align}

In order for the reformulated theory written in terms of the new variables to be equivalent to the original Yang-Mills theory, we must impose the reduction condition:
\begin{equation}
  \mathscr{D}_\mu[\mathscr{V}] \mathscr{X}^\mu = 0 .
  \label{reduction-cond}
\end{equation}
This eliminate the last term of $W_{\mu\nu}^{AB}$ in (\ref{W2}). 
Finally, the Yang-Mills Lagrangian density reads
\begin{align}
   \mathscr{L}_{\rm YM}  
=& -\frac{1}{4} \mathscr{F}_{\mu\nu}^A[\mathscr{V}]^2
%+ \frac{1}{2} \mathscr{F}_{\mu\nu}[\mathscr{V}] \cdot (\mathscr{D}_\mu[\mathscr{V}] \mathscr{X}_\nu - \mathscr{D}_\nu[\mathscr{V}] \mathscr{X}_\mu)
-  \frac{1}{2} \mathscr{X}^{\mu A}  Q_{\mu\nu}^{AB} \mathscr{X}^{\nu B} 
%+ \mathcal{O}(\mathscr{X}^3)  ,
%R\nonumber\\&
%+ \frac{1}{2}  (\mathscr{D}_\mu[\mathscr{V}] \mathscr{X}_\nu - \mathscr{D}_\nu[\mathscr{V}] \mathscr{X}_\mu) \cdot ig[ \mathscr{X}^\mu , \mathscr{X}^\nu ]
%\nonumber\\&
- \frac14 (i g [ \mathscr{X}_\mu , \mathscr{X}_\nu ])^2
 ,
\label{C27-YM-Lagrangian3}
\end{align}
where we have defined 
\begin{equation}
Q_{\mu\nu}^{AB}  := - (\mathscr{D}_\rho[\mathscr{V}]\mathscr{D}^\rho[\mathscr{V}])^{AB} g_{\mu\nu} 
+ 2gf^{ABC} \mathscr{F}_{\mu\nu}^{C}[\mathscr{V}] .
\end{equation}
This should be compared with the background field method.
In order to obtain the final form for the total Lagrangian, we need the Faddeev-Popov ghost term corresponding to the reduction condition. 
This is obtained by using the BRST method.

%%%%%%%%%%%%%%%%%%%%%%%%%%%%%%%%%%%%%%%%%%%%%%%%%%%%%%%%%%%%%
%%%%%%%%%%%%%%%%%%%%%%%%%%%%%%%%%%%%%%%%%%%%%%%%%%%%%%%%%%%%%
\subsection{BRST symmetry and the ghost term}
\label{subsec:BRST-SUN} %%%%%%%%%%%%%%%%%%%%%%%%%%%%%%%%%%%%%%%%%%%%%%%%%%%%%%%%%%%%%
%%%%%%%%%%%%%%%%%%%%%%%%%%%%%%%%%%%%%%%%%%%%%%%%%%%%%%%%%%%%%

We introduce two kinds of ghost fields $\mathbf{C}_\omega(x)$ and $\mathbf{C}_\theta(x)$ which correspond to $\mathbb \omega(x)$ and $\mathbb \theta(x)$ in the enlarged gauge transformation, respectively. 
Then the BRST transformations are given by
\begin{align}
 \bm\delta \mathbf{A}_\mu(x) = D_\mu[\mathbf{A}]\mathbf{C}_\omega(x) := \partial_\mu \mathbf{C}_\omega(x) + g \mathbf{A}_\mu(x) \times \mathbf{C}_\omega(x) .
% \quad
%\bm\delta \mathbf{n}(x) = g\mathbf{n}(x) \times \mathbb C_\theta(x) .
\end{align}
The BRST transformation of $\mathbf{C}_\omega(x)$ is determined from the nilpotency of $\bm\delta  \bm\delta \mathbf{A}_\mu(x) \equiv 0$ in the usual way:
\begin{align}
 \bm\delta   \mathbf{C}_\omega(x) = -\frac12 g\mathbf{C}_\omega(x) \times \mathbf{C}_\omega(x) .
\end{align}

The BRST transformation of the color field $\mathbf{n}(x)$ is given by
\begin{align}
 \bm\delta \mathbf{n}(x) = g\mathbf{n}(x) \times \mathbf{C}_\theta(x)
 .
 \label{C27-BRST-n}
\end{align}
Then the nilpotency 
\begin{align}
0 =  \bm\delta  \bm\delta \mathbf{n}(x) 
=& g\bm\delta \mathbf{n}(x) \times \mathbf{C}_\theta(x) +  g\mathbf{n}(x) \times \bm\delta \mathbf{C}_\theta(x)  
\nonumber\\
=& g^2(\mathbf{n}(x) \times \mathbf{C}_\theta(x)) \times \mathbf{C}_\theta(x) +  g\mathbf{n}(x) \times \bm\delta \mathbf{C}_\theta(x)  
 .
\end{align}
leads to a relation:
\begin{align}
  g (\mathbf{n}(x) \times \mathbf{C}_\theta(x)) \times \mathbf{C}_\theta(x) = - \mathbf{n}(x) \times \bm\delta \mathbf{C}_\theta(x)  
 .
 \label{C27-nc-rel}
\end{align}

For $SU(2)$, $\mathbf{C}_\theta(x)$ is orthogonal to $\mathbf{n}(x)$ and hence it does not have the parallel part to $\mathbf{n}(x)$. 
For $SU(N)$ in the minimal option, we impose that $\mathbf{C}_\theta(x)$ has the vanishing  $\tilde H$ commutative part: 
\begin{align}
 0 =  \mathbf{C}_\theta(x)^{{\tilde H}} :=  \mathbf{C}_\theta(x) + \frac{2(N-1)}{N}  \mathbf{n}(x) \times (\mathbf{n}(x) \times \mathbf{C}_\theta(x))  
 ,
 \label{C27-C-cond-0}
\end{align}
which is equivalent to 
\begin{align}
  \mathbf{C}_\theta(x) 
=  - \frac{2(N-1)}{N}  \mathbf{n}(x) \times (\mathbf{n}(x) \times \mathbf{C}_\theta(x))  
 . 
 \label{C27-C-cond}
\end{align}
Hence we find 
\begin{align}
    \bm n(x)  \cdot  \mathbf{C}_\theta(x) = 0 .
    \label{C27-nC}
\end{align}

Then the BRST transformation of $\mathbf{C}_\theta$ is 
\begin{align}
  \bm\delta \mathbf{C}_\theta  
=&  - \frac{2(N-1)}{N} [
 \bm\delta \mathbf{n} \times (\mathbf{n} \times \mathbf{C}_\theta )  
+ \mathbf{n} \times (\bm\delta \mathbf{n} \times \mathbf{C}_\theta )  
+ \mathbf{n} \times (\mathbf{n} \times \bm\delta \mathbf{C}_\theta )  
] 
\nonumber\\
=& - \frac{2(N-1)}{N} [
 \bm\delta \mathbf{n} \times (\mathbf{n} \times \mathbf{C}_\theta )  
+  \mathbf{n} \times  (g(\mathbf{n} \times \mathbf{C}_\theta ) \times \mathbf{C}_\theta )  
+ \mathbf{n} \times (\mathbf{n} \times \bm\delta \mathbf{C}_\theta )  
] 
\nonumber\\
=& - \frac{2(N-1)}{N} [
 \bm\delta \mathbf{n} \times (\mathbf{n} \times \mathbf{C}_\theta )  
- \mathbf{n} \times  (\mathbf{n} \times \bm\delta \mathbf{C}_\theta )  
+ \mathbf{n} \times (\mathbf{n} \times \bm\delta \mathbf{C}_\theta )  
] 
 ,
\end{align}
where we have used (\ref{C27-BRST-n}) and (\ref{C27-nc-rel}). 
Thus we obtain the BRST transformation of the ghost field $\mathbf{C}_\theta$:
\begin{align}
  \bm\delta \mathbf{C}_\theta  
=   - \frac{2(N-1)}{N} [
 \bm\delta \mathbf{n} \times (\mathbf{n} \times \mathbf{C}_\theta )  
 ] 
 = - \frac{2(N-1)}{N} g [
 (\mathbf{n} \times \mathbf{C}_\theta ) \times (\mathbf{n} \times \mathbf{C}_\theta )  
 ] 
 .
\end{align}
It is easier to use
\begin{align}
  \mathbf{C}_\theta(x) 
=  - \frac{2(N-1)}{N} g^{-1} \mathbf{n}(x) \times \bm\delta \mathbf{n}(x)   
 ,
\end{align}
to derive 
\begin{align}
  \bm\delta \mathbf{C}_\theta  
=   - \frac{2(N-1)}{N} g^{-1} (
 \bm\delta \mathbf{n} \times \bm\delta \mathbf{n}  
) 
 .
\end{align}
This leads immediately to the nilpotency:
\begin{align}
  \bm\delta \bm\delta \mathbf{C}_\theta  
=   - \frac{2(N-1)}{N} g^{-1} (
 \bm\delta \bm\delta \mathbf{n} \times \bm\delta \mathbf{n}  
) 
  + \frac{2(N-1)}{N} g^{-1} (
 \bm\delta \mathbf{n} \times \bm\delta \bm\delta \mathbf{n}  
 ) 
 = 0 
 .
\end{align}

By using the Jacobi identity and (\ref{C27-C-cond}), we find 
\begin{align}
  \bm\delta \mathbf{C}_\theta  
 =&  \frac{2(N-1)}{N}  g [
 -(\mathbf{n} \times \mathbf{C}_\theta ) \times (\mathbf{n} \times \mathbf{C}_\theta )  
 ] 
 \nonumber\\
 =&  \frac{2(N-1)}{N}  g [
   \mathbf{C}_\theta \times ((\mathbf{n} \times \mathbf{C}_\theta ) \times \mathbf{n})  
 + \mathbf{n} \times (\mathbf{C}_\theta \times (\mathbf{n} \times \mathbf{C}_\theta ) )  
 ] 
 \nonumber\\
 =&   g [
   -\mathbf{C}_\theta \times \mathbf{C}_\theta  
 + \frac{2(N-1)}{N} \mathbf{n} \times (\mathbf{C}_\theta \times (\mathbf{n} \times \mathbf{C}_\theta ) )  
 ] 
 .
\end{align}
For $N=2$, this reproduces the $SU(2)$ result:
\begin{align}
  \bm\delta \mathbf{C}_\theta  
  =    - g  
  \mathbf{C}_\theta \times \mathbf{C}_\theta  
  ,
\end{align}
since 
$
\mathbf{C}_\theta \times (\mathbf{n} \times \mathbf{C}_\theta )
= (\mathbf{C}_\theta\cdot  \mathbf{C}_\theta)  \mathbf{n} - (\mathbf{C}_\theta \cdot \mathbf{n} ) \mathbf{C}_\theta =0
$.

We can introduce two kinds of  antighost and the Nakanishi-Lautrup field with the nilpotent BRST transformations:
\begin{align}
 \bm\delta \mathbf{\bar C}_\omega = i \mathbf{N}_\omega , 
 \quad \bm\delta \mathbf{N}_\omega = 0 ,
\end{align}
and
\begin{align}
 \bm\delta \mathbf{\bar C}_\theta = i \mathbf{N}_\theta , 
 \quad \bm\delta \mathbf{N}_\theta = 0 .
\end{align}

In harmony with (\ref{C27-C-cond-0}), we require that $\mathbf{\bar C}_\theta$ satisfies 
\begin{align}
 0 =  \mathbf{\bar C}_\theta(x)^{{\tilde H}} :=  \mathbf{\bar C}_\theta(x) + \frac{2(N-1)}{N}  \mathbf{n}(x) \times (\mathbf{n}(x) \times \mathbf{\bar C}_\theta(x))  
 ,
 \label{C27-C-cond-1}
\end{align}
which is 
\begin{align}
  \mathbf{\bar C}_\theta(x) 
=  - \frac{2(N-1)}{N}  \mathbf{n}(x) \times (\mathbf{n}(x) \times \mathbf{\bar C}_\theta(x))  
 . 
 \label{C27-C-cond-2}
\end{align}

By identifying the reduction condition $\bm{\chi}=0$ with a gauge-fixing condition for the enlarged gauge symmetry, the gauge-fixing term and the associated  ghost term is obtained from
\begin{align}
 {\cal L}_{\rm GF+FP}^\theta
 = -i \bm\delta [\mathbf{\bar C}_\theta \cdot  \bm{\chi}]
=  \mathbf{N}_\theta \cdot \bm{\chi}
+ i \mathbf{\bar C}_\theta \cdot \bm\delta \bm{\chi} , \quad 
\bm{\chi} := D_\mu[\mathbf{V}]\mathbf{X}_\mu .
\end{align}
Here, it should be remarked that the antighost field $\mathbf{\bar C}_\theta$ must have the same degrees of freedom of the constraint $\bm{\chi}:= D_\mu[\mathbf{V}]\mathbf{X}_\mu$. 
This is achieved by requiring 
$
 \mathbf{n}(x) \cdot  \mathbf{\bar C}_\theta(x)    = 0
$, 
since 
$
  \mathbf{n}(x) \cdot \bm{\chi}(x) =  0
$.
We can show (see \ref{sec:FP-ghost-SUN}) that
\begin{align}
i \mathbf{\bar C}_\theta \cdot \bm\delta \bm{\chi}
=    \frac{N}{2(N-1)} 
i\mathbf{\bar C}_\theta \cdot 
    D_\mu[\mathbf{V}-\mathbf{X}] D_\mu[\mathbf{A}](\mathbf{C}_\omega  - \mathbf{C}_\theta )
.  
\label{C27-FP-ghost-SUN}
\end{align}
Thus, we have obtained the gauge-fixing and the Faddeev-Popov ghost term for the reduction condition $D_\mu[\mathbf{V}]\mathbf{X}_\mu=0$ for the $SU(N)$ Yang-Mills theory in the minimal option:
\begin{align}
 {\cal L}_{\rm GF+FP}^\theta
%= -i \bm\delta [\mathbf{\bar C}_\theta \cdot  \bm{\chi}]
 =&   \mathbf{N}_\theta \cdot D_\mu[\mathbf{V}]\mathbf{X}_\mu 
  + \frac{N}{2(N-1)} 
i\mathbf{\bar C}_\theta \cdot 
    D_\mu[\mathbf{V}-\mathbf{X}] D_\mu[\mathbf{A}](\mathbf{C}_\omega  - \mathbf{C}_\theta )
.  
\end{align}
Indeed, this result reproduces the $SU(2)$ case by taking $N=2$, i.e., (\ref{C26-FP1}) in the previous section.

\newpage 
%%%%%%%%%%%%%%%%%%%%%%%%%%%%%%%%%%%%%%%%%%%%%%%%%%%%%%%%%%%%
%%%%%%%%%%%%%%%%%%%%%%%%%%%%%%%%%%%%%%%%%%%%%%%%%%%%%%%%%%%%

\section{Wilson loop operator via a non-Abelian Stokes theorem}\label{sec:NAST} 

%%%%%%%%%%%%%%%%%%%%%%%%%%%%%%%%%%%%%%%%%%%%%%%%%%%%%%%%%%%%
%%%%%%%%%%%%%%%%%%%%%%%%%%%%%%%%%%%%%%%%%%%%%%%%%%%%%%%%%%%%

The non-Abelian Wilson loop operator is defined as a path-ordered product of an exponential with an argument of the line integral along a loop, i.e., a closed path $C$.  
Using the ordinary Stokes theorem, the Abelian Wilson loop operator is rewritten into a surface integral on the surface $S$ whose boundary is given by $C$.  
Such an expression is  possible even for the non-Abelian Wilson loop operator, although it is more complicated than the Abelian version. 
In contrast to the ordinary Stokes theorem, there exist many versions for the \textbf{non-Abelian Stokes theorem (NAST)} \cite{DP89,DP96,Halpern79,Bralic80,Arefeva80,Simonov89,Lunev97,HM97}.
%\footnote{
%See e.g., 
%Halpern79,
%Bralic80,
%Arefeva80,
%Simonov89,
%DP89,
%DP96,
%Lunev97,
%HM97
%\cite{DP89,DP96,Halpern79,Bralic80,Arefeva80,Simonov89,Lunev97,HM97}.  
%}
In this section, we treat a version of the NAST derived by Diakonov and Petrov \cite{DP89,DP96}.
%\footnote{
%DP89,
%DP96
% \cite{DP89,DP96}. 
%}
 This version of NAST is able to remove the path ordering from the expression of the non-Abelian Wilson loop operator.
Instead, we must perform the functional integration.
The NAST is manifestly gauge invariant, since the Wilson loop operator is defined in the gauge invariant way. 
The NAST is very useful to extract  magnetic monopoles inherent in the the Yang-Mills theory through the Wilson loop operator and to understand quark confinement based on the dual superconductivity picture. 

We use the coherent state to derive the NAST in the unified way according to the path integral formalism.  This enables us to derive the general NAST for any compact Lie group $G$, which generalizes the result originally derived for $SU(2)$ by Diakonov and Petrov. 
For $SU(2)$, this reproduces the NAST for the gauge group $G=SU(2)$ using the spin coherent state representation, which clarifies the relationship between the inherent magnetic monopole and the \textbf{Berry phase} (\textbf{geometric phase}).

%%%%%%%%%%%%%%%%%%%%%%%%%%%%%%%%%%%%%%%%%%%%%%%%%%
%%%%%%%%%%%%%%%%%%%%%%%%%%%%%%%%%%%%%%%%%%%%%%%%%%
\subsection{Coherent states and maximal stability subgroup}
%%%%%%%%%%%%%%%%%%%%%%%%%%%%%%%%%%%%%%%%%%%%%%%%%%
%%%%%%%%%%%%%%%%%%%%%%%%%%%%%%%%%%%%%%%%%%%%%%%%%%

We consider a \textbf{Lie group} $G$  characterized by  continuous parameters $\theta_1, \cdots, \theta_{{\rm dim}G}$ where a total number ${\rm dim}G$ of the parameters is called the \textbf{dimension of the group} $G$. 
Let $\mathscr{G}$ be the \textbf{Lie algebra} of a Lie group $G$.  The generators of  $\mathscr{G}$ are denoted by $T_A$ $(A= 1, \cdots , {\rm dim}G)$.

 The gauge group 
 $G$ has the Lie algebra
$\mathscr{G}$ with the generators $\{ T_A  \}$, which obey the commutation relations 
\begin{equation}
 [T_A, T_B ] = i f_{AB}{}^C T_C , \quad ( A,B,C \in \{ 1, \cdots , {\rm dim}G \} )
\end{equation} 
where the numerical coefficients $f_{AB}{}^C$ are called the \textbf{structure constants} of the Lie algebra.%
\footnote{
The superscript and the subscript in a structure constant can be distinguish using the Cartan metric later. 
} 
From the definition, we find 
\begin{equation}
 f_{AB}{}^C = - f_{BA}{}^C. 
\end{equation}
We define the  \textbf{Cartan metric}  by
\begin{equation}
 g_{AB} :=  f_{AE}{}^C f_{BC}{}^E  .
\end{equation}
Then the Cartan metric is symmetric:
\begin{equation}
 g_{AB}  = g_{BA} .
\end{equation}
The inverse of the Cartan metric is defined by
\begin{equation}
 g^{AB}   g_{BC} = \delta^{A}_{C}  .
\end{equation}
%Then the inner product between $U=u^A T_A \in \mathscr{G}$ and $V=v^A T_A \in \mathscr{G}$  is defined by the \textbf{Killing form}:
%\begin{equation}
%  (U,V) = (u,v) := g_{AB} u^{A} v^{B}    .
%\end{equation}
Using the Cartan metric, we can define the structure constant $f_{ABC}$ by
\begin{equation}
 f_{ABC} := g_{CE} f_{AB}{}^E   
\end{equation}
The structure constant $f_{ABC}$ is completely antisymmetric in the indices $A,B,C$, which is shown using the relation 
\begin{equation}
 f_{AE}{}^F  f_{BC}{}^E  + f_{BE}{}^F  f_{CA}{}^E  + f_{CE}{}^F  f_{AB}{}^E   = 0 ,
\end{equation}
derived from the Jacobi identity:
\begin{equation}
 [T_A, [T_B, T_C]] + [T_B, [T_C, T_A]] + [T_C, [T_A, T_B]] = 0   .
\end{equation}

The \textbf{Cartan subgroup} $H$ of a group  $G$ is defined to be  the maximal commutative \textbf{semi-simple} subgroup of $G$.  Let $\mathscr{H}$ be the \textbf{Cartan subalgebra} of $\mathscr{G}$, i.e., the Lie algebra of the Cartan  subgroup $H$.
The generators of the Cartan subalgebra $\mathscr{H}$ are denoted by $H_j$ ($j=1,, \cdots, r$) where  $r$ is called the \textbf{rank} of the group $G$, i.e., $r:={\rm rank}G$.

If the Lie algebra is \textbf{semi-simple},%  
\footnote{
Note that any compact \textbf{semi-simple} Lie group is a \textbf{direct product} of
compact \textbf{simple} Lie group.  Therefore, it is sufficient to consider
the case of a compact simple Lie group.  In the following we assume that $G$ is a compact  {simple} Lie group, i.e., a compact Lie group with no
closed connected invariant subgroup.
}
it is more convenient to rewrite the Lie algebra in terms of the \textbf{Cartan basis} 
$\{ H_j, E_\alpha, E_{-\alpha} \}$ instead of $T_A$.
There are two types of basic operators in the Cartan basis, $H_j$ and $E_{\pm\alpha}$.  
The operators $H_j$ in the Cartan subalgebra may be taken as diagonal operators (matrices), while $E_{\pm\alpha}$ are the off-diagonal \textbf{shift operators} (matrices).  They obey the commutation relations 
\begin{subequations}
\begin{align}
 [H_j, H_k] =& 0, \quad (j,k \in \{ 1, 2, \cdots, r \} )
\\
[ H_j, E_\alpha ] =& \alpha_j E_\alpha, \quad (j=1, \cdots, r  )
\\
 [ E_\alpha, E_\beta ] 
=& \begin{cases}
N_{\alpha;\beta} E_{\alpha+\beta}  &(\alpha+\beta \in \mathcal{R}) \\ 
 0  &(\alpha+\beta \not\in \mathcal{R}, \alpha+\beta \not= 0) 
   \end{cases}
    ,
\\
 [ E_\alpha, E_{-\alpha} ] =& \alpha^j H_j,
\end{align}
\end{subequations}
where $\mathcal{R}$ is the root system, i.e., a set of \textbf{root vectors}
$\{ \alpha_1, \cdots, \alpha_r \}$, with $r$ being the  {rank} of $G$.

We consider a \textbf{representation} $R$ of the group $G$. 
Let $d_R$  be the \textbf{dimension of the representation} $R$ of the group $G$.  
Let the Hilbert space $V^\Lambda$ be a carrier, i.e., the
\textbf{representation space} of the \textbf{unitary irreducible representation}
$\Gamma^\Lambda$ of $G$.   
We use a reference state 
$|\Lambda \rangle$ within the Hilbert space $V^\Lambda$.

As a reference state $\left| {\Lambda} \right>$, we can use the \textbf{highest-weight state} which is defined as follows. 
First, the  {highest-weight state}  $\left| {\Lambda} \right> $ is the (normalized) common eigenvector of the generators $H_1, H_2, \cdots, H_r$ in the {Cartan subalgebra}:
\begin{equation}
H_j \left|   {\Lambda} \right> 
=\Lambda_j \left|   {\Lambda} \right> 
 \quad ( j=1, \cdots, r ) 
   ,
   \label{C28-eigen}
\end{equation}
  with the eigenvalues $\Lambda_1, \Lambda_2, \cdots, \Lambda_r$.
 In other words, $|\Lambda \rangle$  is
mapped into itself by all diagonal operators $H_j$. 
%$H_j$ ($j=1,, \cdots, r=N-1$) are the generators of the Cartan subalgebra and 
Second, the highest-weight state satisfies
\begin{enumerate}
%\item[(i)] 
%$|\Lambda \rangle$  is mapped into itself by all diagonal operators $H_j$,
%$
% H_{j} |\Lambda \rangle = \Lambda_j |\Lambda \rangle ;
%$

\item[(i)]
 $|\Lambda \rangle$  is
annihilated by all the (off-diagonal) \textbf{shift-up operators} $E_{\alpha}$ with
$\alpha \in \mathcal{R}_+$: 
%$
% E_{\alpha} |\Lambda \rangle = 0 \ (\alpha \in R_+)  
%$,
\begin{equation}
% H_j | \Lambda \rangle = \Lambda_j | \Lambda \rangle
% , \quad
 E_\alpha | \Lambda \rangle = 0 \quad (\alpha \in \mathcal{R}_+)
 , 
\end{equation}

\item[(ii)] 
$|\Lambda \rangle$  is
annihilated by some \textbf{shift-down operators} $E_{\alpha}$ with
$\alpha \in \mathcal{R}_-$, not by other $E_{\beta}$ with $\beta \in \mathcal{R}_-$:
\begin{equation}
 E_{\alpha} |\Lambda \rangle = 0 \ ({\rm some~} \alpha \in \mathcal{R}_-) ;
 \quad
 E_{\beta} |\Lambda \rangle = |\Lambda+\beta \rangle 
 \ ({\rm some~} \beta \in \mathcal{R}_-) .
 \label{C28-ii}
\end{equation}
where
$\mathcal{R}_+$ $(\mathcal{R}_-)$ is a subsystem of \textbf{positive (negative) roots}.%
\footnote{
The root vector is defined to be the weight vector of the adjoint representation. 
A weight $\vec{\nu}_j$ is called positive if its {\it last} non-zero component is positive. With this definition, the weights satisfy 
$
 \nu^1 > \nu^2 > \cdots > \nu^N  
%\label{order}
$.
}

\end{enumerate}

Therefore, the highest-weight state $\left|   {\Lambda} \right> $ is labeled by $r$ eigenvalues $\Lambda_1, \cdots, \Lambda_r$.
Here the $r$-dimensional vector $\Lambda =( \Lambda_1, \dots,  \Lambda_r)$  is called  the \textbf{highest--weight vector} of the representation.
% (in which the Wilson loop is considered). 
Similarly, we can define  the \textbf{lowest-weight state}  to characterize the representation instead of the highest-weight state.

We denote a reference state $\left|   {\Lambda} \right> $
and its conjugate $\left<  {\Lambda} \right|$ by
\begin{equation}
\left|  {\Lambda} \right>
= \begin{pmatrix}
  \lambda_1 \cr
  \lambda_2 \cr
  \vdots \cr
  \lambda_{d_R} 
 \end{pmatrix}
= (\lambda_1, \cdots, \lambda_{d_R})^T
 ,
\quad
\left<  {\Lambda} \right| = (\lambda_1^*, \lambda_2^*, \cdots, \lambda_{d_R}^*) ,
\end{equation}
where the  reference state is normalized to unity:
\begin{equation}
  \left<   \Lambda |   \Lambda \right>
= \sum_{a=1}^{d_R} \lambda_a^*  \lambda_a 
=1 
 .
 \label{C28-form1}
\end{equation}
%We restrict the following argument to $G=SU(N)$ for concreteness. 
The rank $r$ of $G=SU(N)$ is given by $r={\rm rank}G=N-1$. 
For  $G=SU(N)$,  the representation has the dimension:  $d_R=N$ for the \textbf{fundamental representation},  $d_R=N^2-1$ for the \textbf{adjoint representation} and so on.

For the fundamental representation, 
the \textbf{highest-weight state} is represented by
\begin{equation}
\left|  \bm{\Lambda} \right>
= (1, \cdots,0,  0)^T
= \begin{pmatrix}
  1 \cr
  0 \cr
  \vdots \cr
  0 
 \end{pmatrix}
 \quad \text{or} \quad 
\left< \bm{\Lambda} \right| = (1, \ldots, 0, 0)  
 ,
\end{equation}
while the \textbf{lowest-weight state} is represented by
\begin{equation}
\left|  \bm{\Lambda} \right>
= (0, \cdots,0,  1)^T
= \begin{pmatrix}
  0 \cr
  \vdots \cr
  0 \cr
  1 
 \end{pmatrix}
 \quad \text{or} \quad 
\left< \bm{\Lambda} \right| = (0, \ldots, 0, 1)  
 .
\end{equation}

The \textbf{coherent state}
$
  |\xi, \Lambda \rangle 
$
corresponding to the coset representatives $\xi \in G/\tilde H$ is constructed as follows.% 
\footnote{
We follow the method developed by
Feng, Gilmore and Zhang and others \cite{FGZ90,Perelomov87,PP85} 
%\bibitem{FGZ90}
%D.H. Feng, R. Gilmore and W.-M. Zhang,
%Coherent states: theory and some applications,
%Rev. Mod. Phys. {\bf 62}, 867 (1990).
%
% 
%\bibitem{Perelomov87}
%A.M. Perelomov,
%Chiral models: geometrical aspects,
%Phys. Report {\bf 146}, 135 (1987).
%\\
%\bibitem{PP85}
%A.M. Perelomov and M.C. Prati,
%Exact Gell-Mann--Low function of supersymmetric K\"ahler $\sigma$-models (II),
%Nucl. Phys. B {\bf 258}, 647--660 (1985).
}

\par
We define the \textbf{maximal stability  subgroup} (or \textbf{isotropy subgroup}) $\tilde H$ as a subgroup of $G$ that consists of all the group elements $h$ that leave the reference state $|\Lambda \rangle$ invariant up to a phase factor:
\begin{equation}
  h |\Lambda \rangle = |\Lambda \rangle e^{i\phi(h)}, \quad
 h \in \tilde H .
\end{equation}
%Let $H$ be the  {Cartan subgroup} of $G$, i.e., the maximal commutative semi-simple subgroup in $G$,  and Let $\mathscr{H}$ be the  {Cartan subalgebra} in $\mathscr{G}$, i.e., the Lie algebra for the group $H$.
The maximal stability subgroup $\tilde H$ includes the Cartan subgroup $H=U(1)^r$, i.e., 
\begin{equation}
  H=U(1)^r \subset \tilde H .
\end{equation}

\par
For every element $g\in G$, there is a unique decomposition of $g$ into a product of two group elements $\xi$ and $h$: 
\begin{equation}
   g = \xi h  \in G, \quad \xi \in G/\tilde H, \quad h \in \tilde H .
\end{equation}
We can obtain a unique  \textbf{coset space} $G/\tilde H$ for a given $|\Lambda \rangle$.
The action of arbitrary group element $g\in G$  on $|\Lambda \rangle$ is given by
\begin{equation}
  |g,\Lambda \rangle :=g |\Lambda \rangle = \xi h  |\Lambda \rangle
  = \xi  |\Lambda \rangle  e^{i\phi(h)}
= |\xi , \Lambda \rangle  e^{i\phi(h)}
 ,
\end{equation}
or
\begin{equation}
  \langle g,\Lambda | := \langle \Lambda |g^\dagger = \langle \Lambda | h^\dagger \xi^\dagger 
  = e^{-i\phi(h)} \langle \Lambda |  \xi^\dagger 
=  e^{-i\phi(h)} \langle \xi,\Lambda |
 .
\end{equation}
Here we have defined the \textbf{coherent state} by
\begin{equation}
 |\xi, \Lambda \rangle := \xi |\Lambda \rangle  
 .
\end{equation}
The coherent states 
$|\xi, \Lambda \rangle$
are in one-to-one correspondence with the coset representatives $\xi \in G/\tilde H$:
\begin{equation}
 |\xi, \Lambda \rangle \leftrightarrow G/\tilde H ,
\end{equation}
%This definition of the coherent state is in one-to-one correspondence with the  {coset space} $G/\tilde H$ 
and the coherent states preserve all the algebraic and topological properties of the coset space $G/\tilde H$.
In other  words, $|\xi, \Lambda \rangle$
and $\xi \in G/\tilde H$ are topologically equivalent.

The phase factor is unimportant in the following discussion because we consider the expectation value of operators $\mathscr{O}$ in the coherent state:
\begin{equation}
\left< g, \Lambda |\mathscr{O} |g, \Lambda \right>
=e^{-i\phi(h)} \left< \xi, \Lambda |\mathscr{O} |\xi, \Lambda \right> e^{i\phi(h)}
=\left< \xi, \Lambda |\mathscr{O} |\xi, \Lambda \right>
 ,
\end{equation}
and the \textbf{projection operator} defined by
\begin{equation}
g | \Lambda \rangle  \langle  \Lambda | g^\dagger 
= |g,\Lambda \rangle  \langle g,\Lambda |
= |\xi , \Lambda \rangle  e^{i\phi(h)} e^{-i\phi(h)} \langle \xi,\Lambda |
= |\xi , \Lambda \rangle   \langle \xi,\Lambda |
= \xi |  \Lambda \rangle   \langle  \Lambda | \xi^\dagger 
 .
\end{equation}

\par
We recall that every group element $g \in G$ can be written as  the exponential of a complex linear combination of diagonal operators $H_j$ and off-diagonal shift operators $E_{\pm\alpha}$.  
If the irreducible representation $\Gamma^\Lambda(\mathscr{G})$ is unitary, then $H_j^\dagger=H_j$ and $E_{\alpha}^\dagger = E_{-\alpha}$.  
Then the coherent state is given by 
\begin{align}
  |\xi, \Lambda \rangle 
  = \xi |\Lambda \rangle
  = \exp \left[ 
  \sum_{\beta\in \mathcal{R}_{-}} \left( \eta_\beta E_\beta - 
  \bar{\eta}_\beta E_{\beta}^{\dagger}\right) \right] |\Lambda \rangle,
\quad \eta_\beta \in \mathbb{C},
\label{C28-cohrentdef}
\end{align}
where $\mathcal{R}_+$ $(\mathcal{R}_-)$ is a subsystem of  {positive (negative) roots} 
and the sum $\sum_{\beta}$ is restricted to those shift operators
$E_{\beta}$ obeying  (\ref{C28-ii}) in (ii).

The coherent states are normalized to unity: 
\begin{equation}
 \langle \xi, \Lambda | \xi, \Lambda \rangle = 1 ,
\end{equation}
since the normalization condition of the coherent state is trivial:
\begin{equation}
 1 
= \lambda_a^*  \lambda_a 
= \left<   \Lambda |   \Lambda \right>
= \left<   \Lambda | g^\dagger g| \Lambda \right>
= \left< g , \Lambda | g , \Lambda \right>
=\left< \xi , \Lambda | \xi , \Lambda \right>
%= \left<   \Lambda | \xi^\dagger \xi | \Lambda \right>
 .
 \label{C28-form1b}
\end{equation}
The coherent state spans the entire space $V^\Lambda$.
However, the coherent states are non-orthogonal: 
\begin{equation}
 \langle \xi' , \Lambda | \xi, \Lambda \rangle \not= 0 \quad \text{for any} \ \xi, \xi' \in G/H .
\end{equation}
By making use of the the \textbf{group-invariant measure} $d\mu(\xi)$ of $G$ which is appropriately normalized, we obtain the \textbf{resolution of identity}:
\begin{align}
  \int |\xi, \Lambda \rangle d\mu(\xi) 
  \langle \xi,  \Lambda |
  = I 
 .
 \label{C28-resolution}
\end{align}
This shows that the coherent states are complete, but in fact \textbf{over-complete}. 
This is because the Hilbert space of the state is spanned by a countable basis, 
while the coherent state is labeled by a continuous index. 
The resolution of identity (\ref{C28-resolution}) becomes very important later to obtain a non-Abelian Stokes theorem for the Wilson loop  operator, which is given as a \textbf{path integral representation of the Wilson loop operator}.%in the next chapter.

Here it is quite important to remark that for $G=SU(2)$ the maximal stability subgroup agrees with the \textbf{maximal torus group}, i.e., 
\begin{equation}
 G=SU(2) \Longrightarrow H=U(1)=\tilde H    .
\end{equation}
For a given group $G=SU(N)$  ($N \ge 3$), however, the maximal stability group is not unique and hence it is not restricted to the maximal stability group:  
\begin{equation}
   \tilde H   \supseteq  H =U(1)^{r} \quad (r=N-1 > 1)  .
\end{equation} 
For instance, the possible maximal stability groups are as follows. 
\begin{align}
% & G=SU(2) \Longrightarrow  \tilde H  = H = U(1) ,
% \nonumber\\
 & G=SU(3) \Longrightarrow  \tilde H = U(2), \quad U(1)^2  , 
 \nonumber\\
 & G=SU(4) \Longrightarrow \tilde H =  U(3),  \quad SU(2) \times U(2),  \quad U(1) \times U(2),  \quad U(1)^3    
  \nonumber\\
 & G=SU(N) \Longrightarrow  \tilde H = U(N-1),  \cdots,  U(1)^{N-1}.   
\end{align}
This fact is an important point to be kept in mind for studying $G=SU(3)$ case.

%%%%%%%%%%%%%%%%%%%%%%%%%%%%%%%%%%%%%%%%%%%%%%%%%%
%%%%%%%%%%%%%%%%%%%%%%%%%%%%%%%%%%%%%%%%%%%%%%%%%%
\subsection{Highest-weight of the representation and the stability group}\label{subsec:stability-group}
%%%%%%%%%%%%%%%%%%%%%%%%%%%%%%%%%%%%%%%%%%%%%%%%%%
%%%%%%%%%%%%%%%%%%%%%%%%%%%%%%%%%%%%%%%%%%%%%%%%%%

%In order to see the physical meaning of the requirements (I) and(II),  
We prepare some formulae for the coherent state which are needed for later applications.
We restrict the following argument to $G=SU(N)$ for concreteness.

For the {reference state} $\left|   {\Lambda} \right> $ of a representation $R$ of a group $G$, we define a matrix $\rho$ with the matrix element $\rho_{ab}$ given by
\begin{equation}
 \rho :=  \left|  \Lambda \right> \left< \Lambda \right| 
, \quad 
 \rho_{ab} :=  \left|  \Lambda \right>_a \left< \Lambda \right|_b = \lambda_a \lambda_b^*
 .
\end{equation}
Then the trace of $\rho$ has a unity:
\begin{equation}
 {\rm tr}(\rho) = \rho_{aa} =  \left|  \Lambda \right>_a \left< \Lambda \right|_a = \lambda_a \lambda_a^* = 1 
 ,
 \label{C28-form11}
\end{equation}
since  the reference state is normalized to unity, i.e., 
$\left<   \Lambda |   \Lambda \right>=\lambda_a^* \lambda_a=1$. 
Moreover, the matrix element $\left<  \Lambda \right|   \mathscr{O}  \left|  \Lambda \right>$ of an arbitrary matrix $\mathcal{O}$ is written in the trace form:
\begin{equation}
  \left<  \Lambda \right|   \mathscr{O}  \left|  \Lambda \right> 
 =  {\rm tr}(\rho  \mathscr{O}  )
 ,
 \label{C28-form3}
\end{equation}
since 
\begin{align}
  \left<  \Lambda \right|   \mathscr{O}  \left|  \Lambda \right>
  = \lambda_b^*  \mathscr{O}_{ba} \lambda_a 
  = \rho_{ab}  \mathscr{O}_{ba}   
 = {\rm tr}(\rho  \mathscr{O}  )
 .
\end{align}

For any representation of $SU(N)$,  we can construct the traceless matrix $\mathcal{H}$ through $\rho$ according to  
\begin{equation}
  \mathcal{H}  := \frac12 \left( \rho - \frac{1}{{\rm tr}(\bm{1})} \bm{1}  \right)  \quad ({\rm tr}(\mathcal{H} )=0)  \Longrightarrow 
  \rho  = \frac{1}{{\rm tr}(\bm{1})} \bm{1} + 2 \mathcal{H}  
   .
   \label{C28-H-def1}
\end{equation}
Then, it is shown that $\mathcal{H}$ agrees with %
\footnote{
For the derivation, see e.g.,  
\cite{DP89,Kondo08}.
%\bibitem{Kondo08}
%K.-I. Kondo,
%Wilson loop and magnetic monopole through a non-Abelian Stokes theorem, 
% Chiba Univ. Preprint, CHIBA-EP-169, 
 % January 2008.
%arXiv:0801.1274 [hep-th],
%Phys. Rev. D {\bf 77}, 085029 (2008).
 In this reference,  a special form for $\rho=\bm{e}_{dd}$ or $\left|  \bm{\Lambda} \right> $ is chosen to study the fundamental representation where a matrix $\bm{e}_{dd}$ has only one  non-vanishing  diagonal element: $(\bm{e}_{dd})_{bc}=\delta_{db}\delta_{dc}$ (no sum over $d$). 
However, the above derivation shows that the fundamental relationships for integrations that are necessary to obtain the path-integral representation for the Wilson loop operator 
%(\ref{C28-rel1}), (\ref{C28-rel2}) 
hold without specifying the explicit form of $\rho$ or $\left|  \bm{\Lambda} \right> $. 
Therefore, the results of section II of this reference hold for the holonomy operator and the Wilson loop operator in the arbitrary representation. 
} 
\begin{equation}
 \mathcal{H} = 
\vec{\Lambda} \cdot \vec{H}  = \sum_{j=1}^{r} \Lambda_j H_j \quad (j=1,, \cdots, r) ,
 \label{C28-H-def2}
\end{equation}
where  $H_j$ ($j=1,, \cdots, r$) are the generators in the Cartan subalgebra  and 
$r$-dimensional vector $\Lambda_j$ ($j=1,, \cdots, r$) is the weight of the representation.
% ($r=N-1$ is the rank of the gauge group $G=SU(N)$).
This is explicitly checked for $SU(2)$ and $SU(3)$ in what follows. 
%[Exercise-1] \marginpar{Ex-1}
% (in which the Wilson loop is considered). 

%For an arbitrary group element $g\in G$, we define 
%\begin{equation}
%  |g,\Lambda \rangle :=g |\Lambda \rangle   , \quad
%  \langle g,\Lambda | := \langle \Lambda |g^\dagger  
%  .
%\end{equation}
In the arbitrary representation, the  matrix element of any Lie algebra valued operator $ \mathscr{O}$  in the coherent states is cast into the form of the trace:
\begin{subequations}
\begin{align}
 \langle g,\Lambda| \mathscr{O}    | \tilde g, \Lambda \rangle 
%\nonumber\\
=   \langle \Lambda| g^\dagger  \mathscr{O}  \tilde g |  \Lambda \rangle 
%\nonumber\\
=& {\rm tr}[ g^\dagger \mathscr{O}  \tilde g \rho    ]
%\\
=  {\rm tr}[ \mathscr{O}  \tilde g \rho  g^\dagger   ]
 .
\end{align}
\end{subequations}
Hence the  matrix element of any Lie algebra valued operator $ \mathscr{O}$  in the coherent state  reads
\begin{subequations}
\begin{align}
  \langle g,\Lambda| \mathscr{O}    | \tilde g, \Lambda \rangle 
= {\rm tr}[ \mathscr{O}  \tilde g  \rho g^\dagger   ]
%\nonumber\\
%=& {\rm tr} \left[ \mathscr{O}  \tilde\xi \left( \frac{1}{{\rm tr}(\bm{1})} \bm{1} + 2 \mathcal{H}   \right) \xi^\dagger   \right]
=& {\rm tr} \left[ g^\dagger \mathscr{O} \tilde g  \left( \frac{1}{{\rm tr}(\bm{1})} \bm{1}   + 2 \mathcal{H} \right)   \right]
\label{C28-xOx}
\\
=& {\rm tr} \left[   \left( \frac{1}{{\rm tr}(\bm{1})} \tilde g g^\dagger  + 2 \tilde g \mathcal{H} g^\dagger   \right)   \mathscr{O}  \right]
%\nonumber\\
%=& {\rm tr} \left[ \mathscr{O}   \left( \frac{1}{{\rm tr}(\bm{1})} \bm{1} + 2 \tilde\xi \mathcal{H} \tilde\xi^\dagger   \right)  \tilde\xi \xi^\dagger  \right]
%\nonumber\\
%=& {\rm tr} \left[ \mathscr{O}  \tilde\xi \xi^\dagger \left( \frac{1}{{\rm tr}(\bm{1})} \bm{1} + 2 \xi \mathcal{H} \xi^\dagger   \right)  \  \right]
 .
 \label{C28-matrix-element}
\end{align}
\end{subequations}

We define a color field $\bm{m}(x)$ having its value in the Lie algebra $\mathscr{G}=su(N)$ by 
\begin{equation}
  \bm{m}(x)  
 := g(x) \mathcal{H} g(x)^\dagger \in  \mathscr{G}=su(N) , \quad g(x)  \in SU(N)  
%= \xi(x) \mathcal{H} \xi(x)^\dagger 
%=  \sum_{j=1}^{r}  \Lambda_j  \xi(x)H_j  \xi(x)^\dagger 
%=  \sum_{j=1}^{r}  \Lambda_j \bm{n}_j(x) 
   ,
   \label{m-def1}
\end{equation}
which  is traceless
\begin{equation}
  {\rm tr} [\bm{m}(x)]  = {\rm tr} [\mathcal{H} ]  = 0 .
\end{equation}
Then the (local) diagonal matrix element is cast into 
%\cite{FGZ90,Perelomov87} 
\begin{equation} 
  \langle g(x),\Lambda|\mathscr{O}(x)  | g(x), \Lambda \rangle 
=   {\rm tr} \left\{   \left[ \frac{1}{{\rm tr}(\bm{1})}\bm{1} + 2 \bm{m}(x) \right]  \mathscr{O}(x) \right\} 
 ,
\end{equation}
which has the special case:
\begin{equation}
  \langle  \Lambda| \mathscr{O}(x)    |   \Lambda \rangle 
=  {\rm tr} \left[   \left(    \frac{1}{{\rm tr}(\bm{1})}\bm{1} + 2   \mathcal{H}     \right)  \mathscr{O}(x)  \right]
%\nonumber\\
%=& {\rm tr} \left[ \mathscr{O}   \left( \frac{1}{{\rm tr}(\bm{1})} \bm{1} + 2 \tilde\xi \mathcal{H} \tilde\xi^\dagger   \right)  \tilde\xi \xi^\dagger  \right]
%\nonumber\\
%=& {\rm tr} \left[ \mathscr{O}  \tilde\xi \xi^\dagger \left( \frac{1}{{\rm tr}(\bm{1})} \bm{1} + 2 \xi \mathcal{H} \xi^\dagger   \right)  \  \right]
 .
\end{equation}
In particular, the traceless operator $\mathscr{O}(x)$ obeys    simpler relations:
\begin{align}  
\langle \xi(x),\Lambda|\mathscr{O}(x)  | \xi(x), \Lambda \rangle 
=&   2{\rm tr} \left\{ \bm{m}(x) \mathscr{O}(x) \right\}
 ,
\nonumber\\
\langle  \Lambda|\mathscr{O}(x)  |   \Lambda \rangle 
=&   2{\rm tr} \left\{ \mathcal{H} \mathscr{O}(x) \right\}
 .
\label{C28-aaa}
\end{align}

It should be remarked  that  $\bm{m}(x)$  does not depend on $\tilde H$:
\begin{align}
   \bm{m}(x)  =& \frac12 [
% g(x) \rho g(x)^\dagger - {\rm tr}(g(x) \rho g(x)^\dagger)/{\rm tr}(\mathbf{1}) \mathbf{1}  =
  g(x) \rho g(x)^\dagger -  \mathbf{1}/{\rm tr}(\mathbf{1})    ]
% \nonumber\\
 =  \frac12 [
%\xi(x) \rho \xi(x)^\dagger - {\rm tr}(\xi(x) \rho \xi(x)^\dagger)/{\rm tr}(\mathbf{1}) \mathbf{1}  =
 \xi(x) \rho \xi(x)^\dagger -  \mathbf{1}/{\rm tr}(\mathbf{1}) ] ,
\end{align}
which follows from
\begin{equation}
   h(x) \rho h(x)^\dagger  = \rho  \Longleftrightarrow  
   h(x) \mathcal{H} h(x)^\dagger  = \mathcal{H} . 
\end{equation}
By using (\ref{C28-H-def2}), we find that  the (unnormalized) color field $\bm{m}(x)$   defined by (\ref{m-def1}) is rewritten into
\begin{equation}
  \bm{m}(x)  
 = \xi(x) h(x) \mathcal{H} h(x)^\dagger \xi(x)^\dagger 
 = \xi(x) \mathcal{H} \xi(x)^\dagger 
%=  \sum_{j=1}^{r}  \Lambda_j  \xi(x)H_j  \xi(x)^\dagger 
=  \sum_{j=1}^{r}  \Lambda_j \bm{n}_j(x) 
    ,
\end{equation}
which is a linear combination of the $r=N-1$ new fields $\bm{n}_j(x)$ defined by
\begin{equation}
 \bm{n}_j(x)   :=    \xi(x)H_j  \xi(x)^\dagger \quad
 (j=1, \cdots, N-1)
    .
\end{equation}
For later convenience, we can introduce $\tilde{\bm{n}}(x)$ and $\bm{n}(x)$ defined by
\begin{equation}
 \tilde{\bm{n}}(x) := 2 \bm{m}(x)   
= \sqrt{\frac{2(N-1)}{N}} \bm{n}(x) 
= 2 g(x) \mathcal{H} g(x)^\dagger  .
\end{equation}
Moreover, we can introduce the normalized color field $\bm{n}(x)$ by
\begin{equation}
\bm{n}(x) 
= \sqrt{\frac{N}{2(N-1)}} \tilde{\bm{n}}(x)
= \sqrt{\frac{2N}{N-1}}   \bm{m}(x) 
= \sqrt{\frac{2N}{N-1}}   g(x) \mathcal{H} g(x)^\dagger 
 ,
\label{C28-unit-n}
\end{equation}
since the  field $\bm{m}(x)$  can be normalized by multiplying a factor $\sqrt{\frac{2N}{N-1}}$, 
\begin{equation}
2 {\rm tr} [\bm{m}(x) \bm{m}(x)]
= 2 {\rm tr}(  \mathcal{H} \mathcal{H}  )
= 2 \Lambda_j  \Lambda_k {\rm tr}( H_j H_k )
=   \Lambda_j^2 
= \frac{N-1}{2N} 
 .
\end{equation}
In this way we can introduce a vector field $\bm{n}(x)$ of unit length.

%In section IV, the resulting expression is further rewritten into another form in terms of a vector field $\mathbf{m}$ or a unit vector field $\mathbf{n}$, which are called color fields. 

%%%%%%%%%%%%%%%%%%%%%%%%%%%%%%%%%%%%%%%%%%%%%%%%%%
%%%%%%%%%%%%%%%%%%%%%%%%%%%%%%%%%%%%%%%%%%%%%%%%%%
\subsection{Explicit construction for $SU(2)$}
%%%%%%%%%%%%%%%%%%%%%%%%%%%%%%%%%%%%%%%%%%%%%%%%%%
%%%%%%%%%%%%%%%%%%%%%%%%%%%%%%%%%%%%%%%%%%%%%%%%%%

For $G=SU(2)$,  $\mathrm{dim}G=2^2-1=3$ implies that there are three generators $T_A$ ($A=1,2,3$), and   
 $\mathrm{rank} G=1$ implies that one of them is diagonal, making the Cartan subalgebra.
For the \textbf{fundamental representation}, the generators $T_A$ ($A=1,2,3$) of  $\mathscr{G}=su(2)$ are constructed from the Pauli matrices $\sigma_A$:
\begin{align}
& T_A=\frac{1}{2}\sigma_A\quad(A=1,2,3), \quad
%\nonumber\\
& \sigma_1=\left(
  \begin{array}{cc}
   0 & 1 \\
   1 & 0 \\
  \end{array}
 \right),\quad\sigma_2=\left(
  \begin{array}{cc}
   0 & -i \\
   i & 0 \\
  \end{array}
 \right),\quad\sigma_3=\left(
  \begin{array}{cc}
   1 & 0 \\
   0 & -1 \\
  \end{array}
 \right) ,
\end{align}
where $\sigma_3$ is chosen to be diagonal.
The generators satisfy the relations:
\begin{equation}
 \left[ \frac{\sigma _A}{2},\frac{\sigma _B}{2} \right] =   i\varepsilon _{ABC}\frac{\sigma _C}{2}, \quad 
\left\{ \frac{\sigma _A}{2},\frac{\sigma _B}{2} \right\}=\frac{1}{2}\delta _{AB} {\bf 1} .
\end{equation}
They show that the structure constant of the group $SU(2)$ is  equal to the Levi-Civita symbol $\varepsilon _{ABC}$,  a completely antisymmetric tensor of rank 3 with $\varepsilon_{123}:=1$, i.e., $f_{AB}{}^{C}=\varepsilon _{ABC}$.
Then the Cartan metric is diagonal given by 
\begin{equation}
 g_{AB}  = g_{BA} =  -f_{AE}{}^C f_{BC}{}^E  
= - \varepsilon _{AEC} \varepsilon _{BCE} 
= 2\delta_{AB} ,
\end{equation}
and the structure constant $f_{ABC}$ differ from $f_{AB}{}^C$ by the factor 2:
\begin{equation}
 f_{ABC} := g_{CE} f_{AB}{}^E = 2  f_{AB}{}^C .
\end{equation}
The inverse Cartan metric is 
\begin{equation}
 g^{AB}   = \frac12 \delta_{AB} .
\end{equation}

\begin{figure}[ptb]
\begin{center}
\includegraphics[height=1.5in]{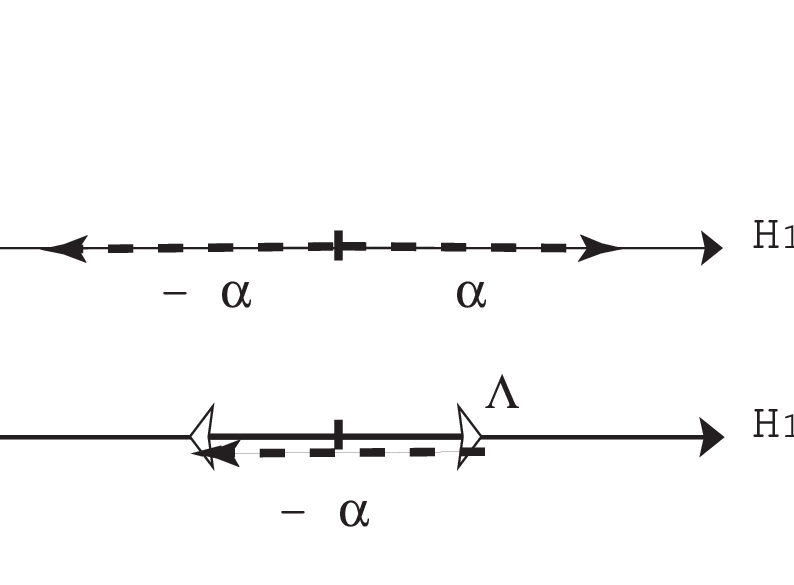}
\end{center}
\vskip -0.5cm
\caption{Root diagram and weight diagram of the  fundamental representation of $SU(2)$ where  $\Lambda$ is the highest weight of the fundamental representation.}
 \label{C28-fig:root_SU2}
\end{figure}

As the \textbf{Cartan subalgebra} we have chosen 
\begin{equation}
H_1=T_3 .
\end{equation}
Then the representation matrix of the \textbf{adjoint representation} is given by 
\begin{align}
 \{ {\rm Ad}(H_1) \}^{A}{}_{B}  = -i \epsilon_{3AB}
 = \begin{pmatrix}
    0 & -i & 0 \\
    i & 0  & 0 \\
    0 & 0  & 0 
   \end{pmatrix} .
\end{align}
Let  $v_\alpha=(v_\alpha^A)$ be the eigenvector associated with the eigenvalue $\alpha$ of the eigenvalue equation given by
\begin{align}
  \{ {\rm Ad}(H_1) \}^{A}{}_{B} v_\alpha^B = \alpha v_\alpha^A \ (A,B=1,2,3) . 
\end{align}
By solving this equation, we can obtain  two eigenvalues (other than zero): 
\begin{align}
 \alpha=\pm1 ,
\end{align}
and the corresponding three-dimensional (normalized) eigenvectors:%
\footnote{
By using the Cartan metric, the normalization of the vector $v_\alpha$ yields 
$1=g_{AB}v_\alpha^A{}^* v_\alpha^B=2v_\alpha^A{}^* v_\alpha^A=2|v_\alpha^A|^2$.
}
%[Exercise-2] \marginpar{Ex-2}
\begin{align}
  v_\alpha^A = v_{\pm}^A := \frac{1}{2} \begin{pmatrix} 1 \\ \pm i \\ 0 \end{pmatrix}  .
  \label{C28-eigen-SU2}
\end{align}
These eigenvalues $\alpha=\pm1$ are called the \textbf{roots}  (See Fig.~\ref{C28-fig:root_SU2}).
The root space of $SU(2)$ is one dimension.
The generator  of the Lie algebra corresponding to the respective root is given by
\begin{align}
 E_{+} = v_{+}^A T_A = \frac{1}{2}  (T_1 + iT_2) , \quad
 E_{-} = v_{-}^A T_A = \frac{1}{2}  (T_1-iT_2) = (E_{+})^\dagger .
\end{align}
We find that $\{ H_1, E_{+}, E_{-} \}$ constitutes the \textbf{Cartan standard form}:
%[Exercise-3] \marginpar{Ex-3}
\begin{align}
 [H_1, E_{\pm}] = \pm E_{\pm}, \quad 
  [E_{+}, E_{-}] =  \frac12  H_1 .
  \label{C28-Csf-SU2}
\end{align}

%%%%%%%%%%%%%%%%%%%%%%%%%%%%%%%%%%%%%%%%%%%%%%%%%%%%%%%%%%%
%%%%%%%%%%%%%%%%%%%%%%%%%%%%%%%%%%%%%%%%%%%%%%%%%%%%%%%%%%%
%%%%%%%%%%%%%%%%%%%%%%%%%%%%%%%%%%%%%%%%%%%%%%%%%%%%%%%%%%%
%%%%%%%%%%%%%%%%%%%%%%%%%%%%%%%%%%%%%%%%%%%%%%%%%%%%%%%%%%%

For $SU(2)$, thus, the \textbf{Cartan basis} $\{ H_1, E_{+}, E_{-} \}$ is obtained from the Pauli basis $\sigma_A$ $(A=1,2,3)$ as
\begin{align}
  H_1=\frac{\sigma_3}{2} = \frac12
\begin{pmatrix}
  1 & 0 \cr
  0 & -1 
 \end{pmatrix}
  ,
\
  \sqrt{2} E_{+} =& \frac{1}{\sqrt{2}} \frac{\sigma_1 + i   \sigma_2}{2}  
= \frac{1}{\sqrt{2}}   \begin{pmatrix}
  0 & 1 \cr
  0 & 0 
 \end{pmatrix}
 ,  
\nonumber\\ 
  \sqrt{2} E_{-} =& \frac{1}{\sqrt{2}} \frac{\sigma_1 - i   \sigma_2}{2} 
=  \frac{1}{\sqrt{2}}  \begin{pmatrix}
  0 & 0 \cr
  1 & 0 
 \end{pmatrix}
 .
\end{align}
The \textbf{highest-weight state} 
$
| \Lambda \rangle 
$ 
for the fundamental representation of $SU(2)$ is given by 
\begin{equation}
| \Lambda \rangle 
=  
\begin{pmatrix}
 1 \\
 0 
\end{pmatrix} 
 ,
\end{equation}
which is indeed the eigenvector of the Cartan generator $H_1$ with the eigenvalue $1/2$:
\begin{equation}
 H_1 
 \begin{pmatrix}
  1 \cr
  0 
 \end{pmatrix}
=   \frac12  
\begin{pmatrix}
  1 \cr
  0 
 \end{pmatrix} 
 \Longrightarrow \Lambda_1 = \frac12  
% \left|  \bm{\Lambda} \right> \text{highest}
 .
\end{equation}
The \textbf{shift-up operator} $E_{+}$ and \textbf{shift-down operator} $E_{-}$ act on the highest-weight state $| \Lambda \rangle$ as
\begin{equation}
E_{+}
 \begin{pmatrix}
  1 \cr
  0 
 \end{pmatrix}
=  \frac{1}{2}  \begin{pmatrix}
  0 \cr
  0 
 \end{pmatrix}
 =  \bm{0}
 , \quad
E_{-}
 \begin{pmatrix}
  1 \cr
  0 
 \end{pmatrix}
=   \frac{1}{2} \begin{pmatrix}
  0 \cr
  1 
 \end{pmatrix} 
 .
\end{equation}
This leads to the \textbf{lowest-weight state}:
\begin{equation}
| \Lambda^\prime \rangle 
=  
\begin{pmatrix}
 0 \\
 1 
\end{pmatrix},
\quad 
 H_1 
 \begin{pmatrix}
  0 \cr
  1 
 \end{pmatrix}
=  - \frac12  
\begin{pmatrix}
  0 \cr
  1 
 \end{pmatrix} 
% \left|  \bm{\Lambda} \right> \text{highest}
  \Longrightarrow \Lambda_1 = -\frac12 .
\end{equation}
The shift-up and shift-down operators act on the lowest-weight state as
\begin{equation}
E_{+}
 \begin{pmatrix}
  0 \cr
  1 
 \end{pmatrix}
=   \frac{1}{2}  \begin{pmatrix}
  1 \cr
  0 
 \end{pmatrix}
 , \quad
E_{-}
 \begin{pmatrix}
  0 \cr
  1 
 \end{pmatrix}
=   \frac{1}{2}  \begin{pmatrix}
  0 \cr
  0 
 \end{pmatrix} 
 = \bf{0} .
\end{equation}

Then the \textbf{projection operator} is obtained as
\begin{align}
\rho :=&  | \Lambda \rangle \langle \Lambda |
= 
\begin{pmatrix}
 1 \\
 0 
\end{pmatrix}
 (1,0)
= 
\begin{pmatrix}
 1 & 0 \\
 0 & 0 
\end{pmatrix} 
\Longrightarrow 
 \rho - \frac12 \mathbf{1} = \frac{\sigma_3}{2} .
\end{align}

Every representation of $SU(2)$ is specified by a  half integer $J$:
\begin{equation}
 \Lambda_1 =J = \frac12 , 1, \frac32, 2, \cdots 
    .
\end{equation}  
The fundamental representation $J=\frac12$ of $SU(2)$ leads to 
\begin{equation}
 \mathcal{H}  
= \frac12 \left( \rho - \frac{1}{2} \bm{1}  \right) 
 = \frac{1}{2} {\rm diag} \left( \frac{1}{2}, \frac{-1}{2} \right) 
= \frac12 \frac{\sigma_3}{2}  = \Lambda_1  H_1 
  .
\end{equation}  
Thus, the two expressions (\ref{C28-H-def1}) and (\ref{C28-H-def2}) for $\mathcal{H}$ are equivalent.

For $SU(2)$, we recall the relationship:
\begin{equation}
\bm{n}(x) 
=  \tilde{\bm{n}}(x)
=    2  \bm{m}(x) 
=   2 g(x) \mathcal{H} g(x)^\dagger 
.
\end{equation}
Then the color field is unique (up to the signature): 
\begin{equation}
  \bm{n}(x)=\bm{n}_1(x) = \xi(x) H_1  \xi(x)^\dagger 
  = \xi(x) \frac{\sigma_3}{2}   \xi(x)^\dagger 
    ,
\end{equation}
and
\begin{equation}
  \bm{m}(x)   
=   \Lambda_1 \bm{n}(x) 
=    J \bm{n}(x) 
= \frac12 \xi(x) \frac{\sigma_3}{2}  \xi(x)^\dagger 
%  , \quad J = \frac12, 1, \frac32, 2, \cdots, 
    .
    \label{C28-m-SU2}
\end{equation}

%%%%%%%%%%%%%%%%%%%%%%%%%%%%%%%%%%%%%%%%%%%%%%%%%%
%%%%%%%%%%%%%%%%%%%%%%%%%%%%%%%%%%%%%%%%%%%%%%%%%%
\subsection{Explicit construction for $SU(3)$}
%%%%%%%%%%%%%%%%%%%%%%%%%%%%%%%%%%%%%%%%%%%%%%%%%%
%%%%%%%%%%%%%%%%%%%%%%%%%%%%%%%%%%%%%%%%%%%%%%%%%%

For $G=SU(3)$, 
  $\mathrm{dim}G=3^2-1=8$  implies that there are eight  generators $T_A$, and $\mathrm{rank} G=2$ implies that two  of them are diagonal, making the Cartan subalgebra.

For the fundamental representation, we can adopt the \textbf{Gell-Mann matrices} $\lambda_A$ to denote the eight generators $T_A=\frac12 \lambda_A$ ($A=1,\cdots,8$) of $su(3)$:
\begin{align}
 \lambda _1=& \left(
  \begin{array}{ccc}
   0 & 1 & 0 \\
   1 & 0 & 0 \\
   0 & 0 & 0 \\
  \end{array}
 \right),\quad
 \lambda_2=\left(
  \begin{array}{ccc}
   0 & -i & 0 \\
   i & 0 & 0 \\
   0 & 0 & 0 \\
  \end{array}
 \right),\quad
 \lambda _3=\left(
  \begin{array}{ccc}
   1 & 0 & 0 \\
   0 & -1 & 0 \\
   0 & 0 & 0 \\
  \end{array}
 \right) ,
 \nonumber\\
%\end{equation*}
%\begin{equation*}
 \lambda _4=& \left(
  \begin{array}{ccc}
   0 & 0 & 1 \\
   0 & 0 & 0 \\
   1 & 0 & 0 \\
  \end{array}
 \right),\quad
 \lambda_5=\left(
  \begin{array}{ccc}
   0 & 0 & -i \\
   0 & 0 & 0 \\
   i & 0 & 0 \\
  \end{array}
 \right),\quad
 \nonumber\\
%\end{equation*}
%\begin{equation}
 \lambda _6=& \left(
  \begin{array}{ccc}
   0 & 0 & 0 \\
   0 & 0 & 1 \\
   0 & 1 & 0 \\
  \end{array}
 \right),\quad
 \lambda _7=\left(
  \begin{array}{ccc}
   0 & 0 & 0 \\
   0 & 0 & -i \\
   0 & i & 0 \\
  \end{array}
 \right),\quad
 \lambda _8=\frac{1}{\sqrt{3}}\left(
  \begin{array}{ccc}
   1 & 0 & 0 \\
   0 & 1 & 0 \\
   0 & 0 & -2 \\
  \end{array}
 \right)
 .
\end{align}

  For $G=SU(3)$, the commutation relations for the Gell-Mann matrices are given by
\begin{equation}
 \left[\frac{\lambda _A}{2},\frac{\lambda _B}{2}\right]= i f_{AB}{}^{C}\frac{\lambda _C}{2} ,
\end{equation}
The structure constant $f_{AB}{}^{C}$ is completely antisymmetric in the indices $A,B,C$. 
and the non-vanishing structure constants take the values:
\begin{align}
  f_{12}{}^{3}=1  ,
\quad
  f_{45}{}^{8}=f_{67}{}^{8}=\frac{\sqrt{3}}{2} ,
\nonumber\\
\quad
  f_{14}{}^{7}=f_{24}{}^{6}=f_{25}{}^{7}= -f_{15}{}^{6}=f_{45}{}^{3}= -f_{67}{}^{3} = \frac{1}{2}  .
\end{align}
  For $G=SU(3)$,  the three blocks correspond to the off-diagonal indices, 
$(a,b) \in \{ (1,2)$, $(4,5)$, $(6,7) \}$: 
\begin{align}
 f_{12}{}^{3}=1,   \quad 
f_{45}{}^{3}=\frac{1}{2},  
\quad   f_{67}{}^{3}=\frac{-1}{2}, 
\quad f_{45}{}^{8}=\frac{\sqrt{3}}{2},   \quad  f_{67}{}^{8}=\frac{\sqrt{3}}{2}  .
\end{align}
Other non-zero values are
\begin{align}
 f_{14}{}^{7}=\frac{1}{2},   \quad 
f_{15}{}^{6}=\frac{-1}{2}, \quad f_{24}{}^{6}=\frac{1}{2},   \quad 
f_{25}{}^{7}=\frac{1}{2} .
\end{align}
Then the Cartan metric is diagonal: 
\begin{equation}
 g_{AB}  = g_{BA} =  -f_{AE}{}^C f_{BC}{}^E  
= 3\delta_{AB} ,
\end{equation}
and the structure constant $f_{ABC}$ differ from $f_{AB}{}^C$ by the factor 3:
\begin{equation}
 f_{ABC} = g_{CE} f_{AB}{}^E = 3  f_{AB}{}^C .
\end{equation}
Then the inverse Cartan metric is 
\begin{equation}
 g^{AB}   = \frac13 \delta_{AB} .
\end{equation}
In the following applications, we do not distinguish $f_{AB}{}^C$ and $f_{ABC}$, since the Cartan metric is proportional to the unit matrix and hence the difference can be absorbed into the redefinition of the base, i.e., the rescaling of the overall factor of the basis vector.

In view of ${\rm rank} SU(3)=2$, as the \textbf{Cartan subalgebra} of $su(3)$, we choose 
\begin{equation}
H_1=T_3, \quad H_2=T_8
\end{equation}
Then the \textbf{adjoint representation} has the representation matrices:
\begin{align}
 \{ {\rm Ad}(H_1) \}^{A}{}_{B} =-i f_{AB}{}^{3} %= -if_{AB3}   
 = \begin{pmatrix}
    0 & -i & 0 & 0 & 0 & 0 & 0 & 0 \\
    +i & 0 & 0 & 0 & 0 & 0 & 0 & 0 \\
%    \hline
    0 & 0 & 0 & 0 & 0 & 0 & 0 & 0 \\
%    \hline
    0 & 0 & 0 & 0 & -i\frac{1}{2}  & 0 & 0 & 0 \\
    0 & 0 & 0 & +i\frac{1}{2} & 0 & 0 & 0 & 0 \\
%    \hline
    0 & 0 & 0 & 0 & 0 & 0 & +i\frac{1}{2} & 0 \\
    0 & 0 & 0 & 0 & 0 & -i\frac{1}{2} & 0 & 0 \\
%    \hline
    0 & 0 & 0 & 0 & 0 & 0 & 0 & 0  
   \end{pmatrix} ,
  \nonumber 
\end{align}
\begin{align}
 \{ {\rm Ad}(H_2) \}^{A}{}_{B} = -i f_{AB}{}^{8} %= -if_{AB8}   
 = \begin{pmatrix} 
    0 & 0 & 0 & 0 & 0 & 0 & 0 & 0 \\
    0 & 0 & 0 & 0 & 0 & 0 & 0 & 0 \\
%    \hline
    0 & 0 & 0 & 0 & 0 & 0 & 0 & 0 \\
%    \hline
    0 & 0 & 0 & 0 & -i\frac{\sqrt{3}}{2} & 0 & 0 & 0 \\
    0 & 0 & 0 & +i\frac{\sqrt{3}}{2} & 0 & 0 & 0 & 0 \\
%    \hline
    0 & 0 & 0 & 0 & 0 & 0 & -i\frac{\sqrt{3}}{2} & 0 \\
    0 & 0 & 0 & 0 & 0 & +i\frac{\sqrt{3}}{2} & 0 & 0 \\
%    \hline
   0 & 0 & 0 & 0 & 0 & 0 & 0 & 0  
   \end{pmatrix} .
\end{align}
Let  $v_{\alpha_j}$ be the eigenvectors associated with the eigenvalues $\alpha_j$ of the  two eigenvalue equations given by 
\begin{align}
  \{ {\rm Ad}(H_j) \}^{A}{}_{B} v_{\alpha_j}^B = \alpha_j v_{\alpha_j}^A \ (j=1,2; A, B =1,2, \cdots, 8) .
\end{align}
By solving these equations, we can obtain six pairs of eigenvalues  $\vec{\alpha} =(\alpha_1, \alpha_2)$ (other than two zeros, i.e., $2(0,0)$) for each $j=1,2$, and the six simultaneous eight-dimensional eigenvectors $v_{\alpha_j}=(v_{\alpha_j}^A)$ ($A=1,2, \cdots, 8$):
%[Exercise-4] \marginpar{Ex-4}
\begin{align}
 \vec{\alpha} =& \pm \vec{\alpha}^{(1)} := \pm ( 1,  0 )  \rightarrow 
  v_{\vec{\alpha}}  = v_{\pm \vec{\alpha}^{(1)}}  := \frac{1}{\sqrt{6}} (1, \pm i, 0, 0, 0, 0, 0, 0)^T ,
  \nonumber\\
 \vec{\alpha} =&  \pm \vec{\alpha}^{(2)} := \pm (  1/2,   \sqrt{3}/2 )  \rightarrow 
  v_{\vec{\alpha}} = v_{\pm \vec{\alpha}^{(2)}}  :=  \frac{1}{\sqrt{6}} (0, 0, 0, 1, \pm i, 0, 0, 0, 0)^T ,
  \nonumber\\
 \vec{\alpha} =& \pm \vec{\alpha}^{(3)} := \pm (- 1/2,    \sqrt{3}/2 )  \rightarrow 
  v_{\vec{\alpha}}  = v_{\pm \vec{\alpha}^{(3)}}  := \frac{1}{\sqrt{6}} (0, 0, 0, 0, 0, 1, \pm i, 0)^T ,
  \label{C28-eigen-SU3}
\end{align}
where $T$ is the transpose for writing a vector  $v_{\alpha_j}$ to a column vector.
Here $\vec{\alpha} =(\alpha_1, \alpha_2)$ are the \textbf{root vectors}.  The \textbf{root space} of $SU(3)$ is two dimension.
The generators of the Lie algebra corresponding  to the respective root are given by  
\begin{align}
 E_{\pm \alpha^{(1)}} = v_{\pm \vec{\alpha}^{(1)}}^A T_A = \frac{1}{\sqrt{6}} (T_1 \pm iT_2) , \quad
%\nonumber\\
 E_{\pm \alpha^{(2)}} = v_{\pm \vec{\alpha}^{(2)}}^A T_A = \frac{1}{\sqrt{6}} (T_4 \pm iT_5)  ,
  \nonumber\\
 E_{\pm \alpha^{(3)}} = v_{\pm \vec{\alpha}^{(3)}}^A T_A = \frac{1}{\sqrt{6}} (T_6 \pm iT_7)  .
 \label{C28-shift-ope-su3}
\end{align}

%%%%%%%%%%%%%%%%%%%%%%%%%%%%%%%%%%%%%%%%%%%%%%%%%%%%%%%%%
\begin{figure}
\begin{center}
% \leavevmode
% \epsfxsize=60mm
% \epsfysize=60mm
% \epsfbox{root_SU3rev.eps}
\includegraphics[scale=0.3]{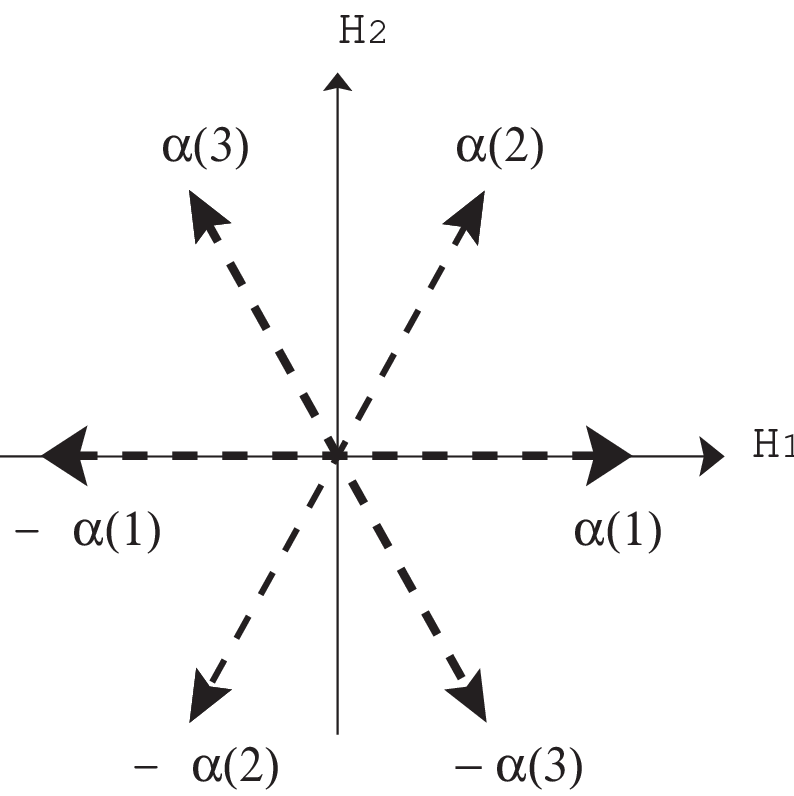}
\end{center} 
\vskip -0.5cm
\caption[]{
The root diagram of $SU(3)$, where positive root vectors are given by 
$\vec{\alpha}^{(1)}=(1,0)$,
$\vec{\alpha}^{(2)}=(\frac{1}{2},\frac{\sqrt{3}}{2})$, and
$\vec{\alpha}^{(3)}=(\frac{-1}{2},\frac{\sqrt{3}}{2})$.
%Here we have used the same weight ordering as in the $SU(N)$ case (see (\ref{C28-order}))  in defining the simple roots. Then 
The two simple roots are given by $\alpha^1:=
  \vec{\alpha}^{(1)} 
$
and
$ \alpha^2:= \vec{\alpha}^{(3)}$.
}
 \label{C28-fig:root}
\end{figure}%%%%%%%%%%%%%%%%%%%%%%%%%%%%%%%%%%%%%%%%%%%%%%%%%%%%%%%%%

We find that 
$\{ H_1 , H_2 , E_{+ \alpha^{(1)}}, E_{-  \alpha^{(1)}}, E_{+ \alpha^{(2)}}, E_{-  \alpha^{(2)}} , E_{+ \alpha^{(3)}}, E_{-  \alpha^{(3)}} \}$ forms the \textbf{Cartan standard form}:% 
%[Exercise-5] \marginpar{Ex-5}%
\footnote{
This form for the commutation relations is sometimes  useful to treat all the root vectors on equal footing so that $\alpha+\beta+\gamma=0$.
%The last commutation relation is concretely written as
%\begin{align}
%[E_{\alpha_3}, E_{\alpha_1}] =&  - \frac{1}{\sqrt{6}}   E_{\alpha_2} = - \frac{1}{\sqrt{6}}   E_{-\alpha_2}^\dagger  ,
%\quad 
%[E_{\alpha_2}, E_{-\alpha_3}] =   \frac{1}{\sqrt{6}}   E_{\alpha_1} =   \frac{1}{\sqrt{6}}   E_{-\alpha_1}^\dagger ,
%\nonumber\\
%[E_{\alpha_1}, E_{-\alpha_2}] =&  - \frac{1}{\sqrt{6}}   E_{-\alpha_3} = - \frac{1}{\sqrt{6}}   E_{\alpha_3}^\dagger  .
%\end{align}
}
\begin{align}
[H_j, E_{\alpha}] =& \alpha_j  E_{\alpha} , \quad
  ([H_j, E_{\alpha}^\dagger] =  -\alpha_j E_{\alpha}^\dagger ),
  \nonumber\\
[E_{\alpha}, E_{\beta}^\dagger] =&   \alpha^j  H_j \delta_{\alpha\beta} ,
  \nonumber\\
[E_{\alpha}, E_{\beta}] =& - \frac{1}{\sqrt{6}} \epsilon^{\alpha\beta\gamma}  E_{\gamma}^\dagger \quad (\alpha+\beta+\gamma=0) .
\label{C28-Csf-SU3}
\end{align}

Each block is characterized by 
 the \textbf{positive roots} given by  
(See Fig.~\ref{C28-fig:root})
\begin{align}
  \vec{\alpha}^{(1)} = (1,0), \quad 
  \vec{\alpha}^{(2)} = \left(\frac{1}{2},\frac{\sqrt{3}}{2} \right), \quad 
  \vec{\alpha}^{(3)} = \left(\frac{-1}{2},\frac{\sqrt{3}}{2} \right), \quad 
\end{align}
where the \textbf{simple roots} are given by
\begin{align}
 \alpha_1 := \vec{\alpha}^{(1)} , \quad \alpha_2 := \vec{\alpha}^{(3)} . 
\end{align}
%\footnote{
%Note that the \textbf{Cartan metric} is given by 
%$\hat{g}^{AA}=1/3$ and $\alpha^k=g^{kk}\alpha_k= \alpha_k/3$.
%We can set up the generators in the Cartan basis which obey  the commutation relations:
%\begin{align}
%[\vec{H}, E_{\alpha}] = \vec{\epsilon}_{\alpha} E_{\alpha} , \quad
%  [\vec{H}, E_{-\alpha}] = -\vec{\epsilon}_{\alpha} E_{-\alpha} ,
%\end{align}
%where the three roots vectors $\alpha^{(\alpha)}$ are given by 
%\begin{align}
% \vec{\epsilon}_{1} = \vec{\alpha}^{(1)} = (1,0), \  
% \vec{\epsilon}_{2} = - \vec{\alpha}^{(2)} =  \left(\frac{-1}{2},\frac{-\sqrt{3}}{2} \right), \  
%  \vec{\epsilon}_{3} = \vec{\alpha}^{(3)} = \left(\frac{-1}{2},\frac{\sqrt{3}}{2} \right) ,
%\end{align}
%so that 
%$\vec{\epsilon}_{1}+\vec{\epsilon}_{2}+\vec{\epsilon}_{3}$=0.
%}

  For $G=SU(3)$, the \textbf{Cartan basis} $\{ H_1, H_2, E_{\pm \alpha^{(1)}}, E_{\pm \alpha^{(2)}}, E_{\pm \alpha^{(3)}} \}$ is defined from the Gell-Mann basis $\lambda_A$ ($A=1,...,8$) by
\begin{align}
 &  H_1 = T_3, \  H_2 :=  T_8  , \
   E_{\pm \alpha^{(1)}} :=  \frac{1}{\sqrt{2}}(T_1 \pm  i T_2) , \ 
%\nonumber\\&
  E_{\pm \alpha^{(2)}} := \frac{1}{\sqrt{2}}(T_4 \pm  i T_5) , \ 
  E_{\pm \alpha^{(3)}} := \frac{1}{\sqrt{2}}(T_6 \pm i T_7) ,
\end{align}
or the matrix form:%
%\footnote{
%It is sometimes useful to define three kinds of shift-up and shift-down operators $I_\pm,V_\pm,U_\pm$ and the associated diagonal operators $I_3,V_3,U_3$ as 
%\begin{align}
% & I_\pm = \sqrt{2} E_{\pm \alpha^{(1)}} , \quad
%I_3 = [ E_{+ \alpha^{(1)}} , E_{-\alpha^{(1)}} ] 
%= {\rm diag }(1,-1,0) ,
%= \begin{pmatrix}
%  1 & 0 & 0 \cr
%  0 & -1 & 0 \cr
%  0 & 0 & 0 
% \end{pmatrix} ,
%\nonumber\\ 
% & V_\pm = \sqrt{2} E_{\pm \alpha^{(2)}} , \quad
%V_3 = [ E_{+ \alpha^{(2)}} , E_{-\alpha^{(2)}} ] 
%= {\rm diag }(1,0,-1) ,
%= \begin{pmatrix}
%  1 & 0 & 0 \cr
%  0 & 0 & 0 \cr
%  0 & 0 & -1 
% \end{pmatrix} ,
%\nonumber\\ 
% & U_\pm = \sqrt{2} E_{\pm \alpha^{(3)}} , \quad
%U_3 = [ E_{+ \alpha^{(3)}} , E_{-\alpha^{(3)}} ] 
%= {\rm diag }(0,1,-1) .
%= \begin{pmatrix}
%  0 & 0 & 0 \cr
%  0 &  1 & 0 \cr
%  0 & 0 & -1 
% \end{pmatrix} .
%\end{align}
%}
%G=SU(3): 
\begin{align}
& H_1=  \frac12 \lambda_3
= \frac12
\small
\begin{pmatrix}
  1 & 0 & 0 \cr
  0 & -1 & 0 \cr
  0 & 0 & 0 
 \end{pmatrix}
, 
\quad
H_2=\frac12 \lambda_8
= \frac12 \frac{1}{\sqrt{3}}
\small
\begin{pmatrix}
  1 & 0 & 0 \cr
  0 & 1 & 0 \cr
  0 & 0 & -2 
 \end{pmatrix}
 ,
\nonumber\\
& \sqrt{2} E_{\pm \alpha^{(1)}} :=  \frac12 (\lambda_1 \pm  i \lambda_2) 
=  \small \begin{pmatrix}
  0 & 1 & 0 \cr
  0 & 0 & 0 \cr
  0 & 0 & 0
 \end{pmatrix}
 , 
   \begin{pmatrix}
  0 & 0 & 0 \cr
  1 & 0 & 0 \cr
  0 & 0 & 0
 \end{pmatrix}
 , 
\nonumber\\
& \sqrt{2} E_{\pm \alpha^{(2)}} :=  \frac12 (\lambda_4 \pm  i \lambda_5) 
=  \small \begin{pmatrix}
  0 & 0 & 1 \cr
  0 & 0 & 0 \cr
  0 & 0 & 0
 \end{pmatrix}
 , 
   \begin{pmatrix}
  0 & 0 & 0 \cr
  0 & 0 & 0 \cr
  1 & 0 & 0
 \end{pmatrix}
 , 
\nonumber\\
& \sqrt{2} E_{\pm \alpha^{(3)}} :=  \frac12 (\lambda_6 \pm  i \lambda_7) 
= \small  \begin{pmatrix}
  0 & 0 & 0 \cr
  0 & 0 & 1 \cr
  0 & 0 & 0
 \end{pmatrix}
 , 
   \begin{pmatrix}
  0 & 0 & 0 \cr
  0 & 0 & 0 \cr
  0 & 1 & 0
 \end{pmatrix}
 . 
\end{align}

%%%%%%%%%%%%%%%%%%%%%%%%%%%%%%%%%%%%%%%%%%%%%%%%%%%%%%%%%
\begin{figure}[ptb]
\begin{center}
\includegraphics[scale=0.4]{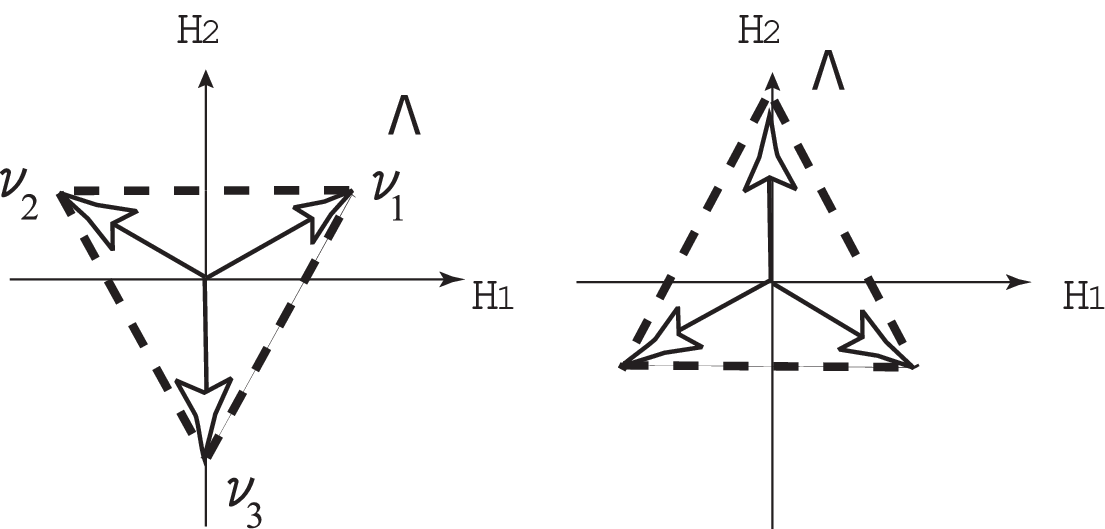}
\end{center}
\vskip -0.5cm
\caption{
The weight diagram and root vectors required to define the coherent state in the fundamental representations $[1,0]={\bf 3}$, $[0,1]={\bf 3}^*$  of $SU(3)$ where  $\vec{\Lambda} = \vec h_1=\vec \nu_1:=(\frac{1}{2},\frac{1}{ 2\sqrt{3}})$ is the highest weight of the fundamental representation, and the other weights are 
 $\vec\nu_2:=(-\frac{1}{2},\frac{1}{2\sqrt{3}})$ and
 $\vec\nu_3:=(0,-\frac{1}{\sqrt{3}})$.
}
 \label{C28-fig:fundamental-weight}
\end{figure}
%%%%%%%%%%%%%%%%%%%%%%%%%%%%%%%%%%%%%%%%%%%%%%%%%%%%%%%%%

For the fundamental representation of $SU(3)$,  
  the \textbf{highest-weight state} is given by
\begin{equation}
| \Lambda \rangle 
=  \small
\begin{pmatrix}
 1 \\
 0 \\
 0 
\end{pmatrix} ,
\end{equation}
 since this state is the simultaneous eigenstates of two Cartan generators $H_1, H_2$ with eigenvalues represented by the weight vector $\vec{\Lambda}$:
\begin{equation}
[1,0]: 
%\frac12 \lambda_3 \left|  \bm{\Lambda} \right> 
  H_1 
\small \begin{pmatrix}
  1 \cr
  0 \cr
  0 
 \end{pmatrix}
= \frac12 
\small \begin{pmatrix}
  1 \cr
  0 \cr
  0 
 \end{pmatrix}  , 
\quad
%\frac12 \lambda_8 \left|  \bm{\Lambda} \right> 
  H_2 
\small  \begin{pmatrix}
  1 \cr
  0 \cr
  0 
 \end{pmatrix}
= \frac{1}{2\sqrt{3}}   
\small  \begin{pmatrix}
  1 \cr
  0 \cr
  0 
 \end{pmatrix}  
\Longrightarrow 
 \vec{ \Lambda} = \vec{\nu}_1 :=  ( \frac12, \frac{1}{2\sqrt{3}})  ,
\end{equation}
where $[1,0]$ is the Dynkin index explained later. 
The \textbf{shift-up operators} act on the highest-weight state to yield the trivial result:
\begin{align}
 E_{+ \alpha^{(1)}} 
\small \begin{pmatrix}
 1 \\
 0 \\
 0 
\end{pmatrix} 
= \small \begin{pmatrix}
 0 \\
 0 \\
 0 
\end{pmatrix} = \bf{0} 
 , 
\ 
  E_{+ \alpha^{(2)}} 
\small \begin{pmatrix}
 1 \\
 0 \\
 0 
\end{pmatrix} 
%= \small \begin{pmatrix}
% 0 \\
% 0 \\
% 0 
%\end{pmatrix} 
= \bf{0}
 , 
  \ 
 E_{+ \alpha^{(3)}} 
\small \begin{pmatrix}
 1 \\
 0 \\
 0 
\end{pmatrix} 
%= \small \begin{pmatrix}
% 0 \\
% 0 \\
% 0 
%\end{pmatrix} 
= \bf{0}
 , 
\end{align}
while the \textbf{shift-down operators} leads to the non-trivial result:
\begin{align}
\sqrt{2} E_{- \alpha^{(1)}} 
\small \begin{pmatrix}
 1 \\
 0 \\
 0 
\end{pmatrix} 
= \small \begin{pmatrix}
 0 \\
 1 \\
 0 
\end{pmatrix}
,
\ 
 \sqrt{2}  E_{- \alpha^{(2)}} 
\small \begin{pmatrix}
 1 \\
 0 \\
 0 
\end{pmatrix} 
= 
\small \begin{pmatrix}
 0 \\
 0 \\
 1 
\end{pmatrix}
,
  \ 
\sqrt{2}  E_{- \alpha^{(3)}} 
\small \begin{pmatrix}
 1 \\
 0 \\
 0 
\end{pmatrix} 
%= \small \begin{pmatrix}
% 0 \\
% 0 \\
% 0 
%\end{pmatrix}  
= \bf{0}
 .
\end{align}
For other states in the fundamental representation, we find
\begin{align}
& [-1,1]: % [0,-1]: 
%\frac12 \lambda_3 \left|  \bm{\Lambda} \right> 
  H_1
 \begin{pmatrix}
  0 \cr
  1 \cr
  0 
 \end{pmatrix}
= \frac{-1}{2} 
 \begin{pmatrix}
  0 \cr
  1 \cr
  0 
 \end{pmatrix}  , 
\ 
%\frac12 \lambda_8 \left|  \bm{\Lambda} \right> 
  H_2 
 \begin{pmatrix}
  0 \cr
  1 \cr
  0 
 \end{pmatrix}
= \frac{1}{2\sqrt{3}}   
 \begin{pmatrix}
  0 \cr
  1 \cr
  0 
 \end{pmatrix}  
 ,
 \
 \vec{\nu}_2 :=   (  \frac{-1}{2}, \frac{1}{2\sqrt{3}}) ,
\nonumber\\
& [0,-1]:%[-1,1]: 
%\frac12 \lambda_3 \left|  \bm{\Lambda} \right> 
  H_1 
 \begin{pmatrix}
  0 \cr
  0 \cr
  1
 \end{pmatrix}
= 0 
 \begin{pmatrix}
  0 \cr
  0 \cr
  1 
 \end{pmatrix}  , 
\ 
%\frac12 \lambda_8 \left|  \bm{\Lambda} \right> 
  H_2 
 \begin{pmatrix}
  0 \cr
  0 \cr
  1 
 \end{pmatrix}
= \frac{-1}{\sqrt{3}}   
 \begin{pmatrix}
  0 \cr
  0 \cr
  1 
 \end{pmatrix}  
 ,
 \ 
 \vec{\nu}_3 :=  ( 0, \frac{-1}{\sqrt{3}})   .
\end{align}

%%%%%%%%%%%%%%%%%%%%%%%%%%%%%%%%%%%%%%%%%%%%%%%%%%%%%%%%%
\begin{figure}[ptb]
\begin{center}
\includegraphics[scale=0.4]{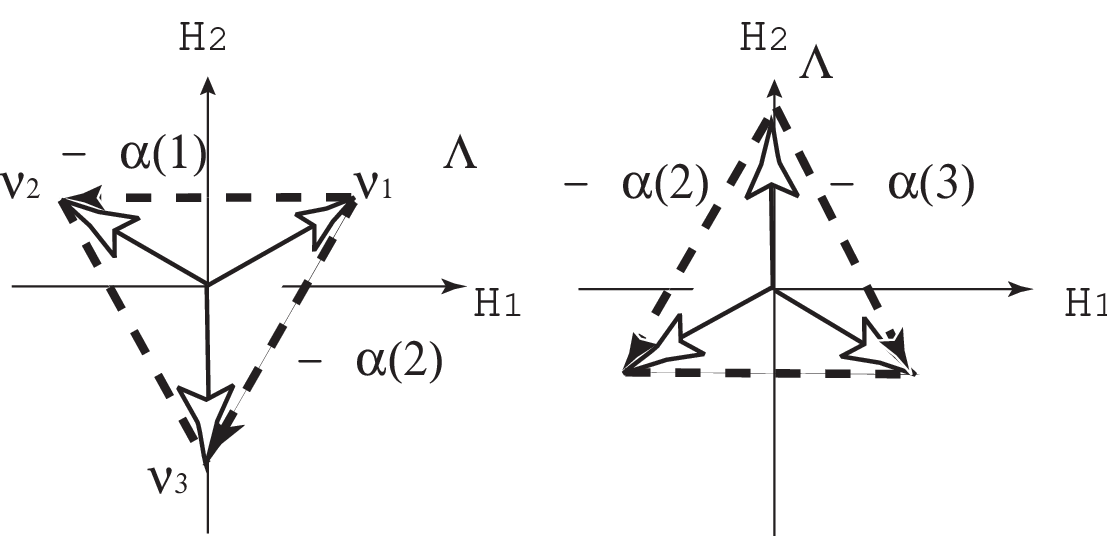}
\end{center}
\vskip -0.5cm
\caption{
The weight diagram and root vectors required to define the coherent state in the fundamental representations $[1,0]={\bf 3}$, $[0,1]={\bf 3}^*$  of $SU(3)$ where  $\vec{\Lambda} = \vec h_1=\vec \nu^1:=(\frac{1}{2},\frac{1}{2\sqrt{3}})$ is the highest weight of the fundamental representation ${\bf 3}$.
}
 \label{C28-fig:fundamental-weight2}
\end{figure}
%%%%%%%%%%%%%%%%%%%%%%%%%%%%%%%%%%%%%%%%%%%%%%%%%%%%%%%%%

The shift-up and shift-down operators act on the highest-weight state $\left|\bm{\Lambda} \right>=\left| \nu_1 \right>$ to yield the result:
See Fig.~\ref{C28-fig:fundamental-weight2}.
\begin{align}
 & E_{+ \alpha^{(1)}} \left|  \bm{\Lambda} \right> = \bm{0} , \quad 
  E_{- \alpha^{(1)}} \left|  \bm{\Lambda} \right>=  \left| \nu_2 \right> ,
\nonumber\\
  & E_{+ \alpha^{(2)}} \left|  \bm{\Lambda} \right> = \bm{0} , \quad  
 E_{- \alpha^{(2)}}  \left|  \bm{\Lambda} \right>  =  \left| \nu_3 \right> ,
\nonumber\\
 & E_{+ \alpha^{(3)}} \left|  \bm{\Lambda} \right> =  \bm{0}   , \quad 
  E_{- \alpha^{(3)}} \left|  \bm{\Lambda} \right>  
 =   \bm{0} .
\end{align}
%\begin{align}
% E_{\pm \alpha^{(1)}} \left|  \bm{\Lambda} \right> 
% = E_{\pm \alpha^{(1)}}   \nu_3  = \bm{0}, \bm{0} ,
%\nonumber\\
% \quad
% E_{\pm \alpha^{(2)}} \left|  \bm{\Lambda} \right> 
% = E_{\pm \alpha^{(2)}}   \nu_3  =  \nu_1 , \bm{0},
%\nonumber\\
% \quad
% E_{\pm \alpha^{(3)}} \left|  \bm{\Lambda} \right> 
% = E_{\pm \alpha^{(3)}}   \nu_3  =  \nu_2 , \bm{0},
%\end{align}

For $G=SU(3)$, the rank is two $r=2$ and every representation is specified by the \textbf{Dynkin indices} $[m,n]$.
%\footnote{
%The \textbf{Dynkin index} $[m^1, \cdots, m^r]$ for the representation with the weight vector $\vec{\Lambda}=(\Lambda_1, \cdots, \Lambda_r)$ is obtained from the \textbf{simple roots} $\alpha^{(1)}, \cdots, \alpha^{(r)}$:
%\begin{equation}
% m^k := 2 \frac{\vec{\alpha}^{(k)} \cdot \vec{\Lambda}}{\vec{\alpha}^{(k)} \cdot \vec{\alpha}^{(k)}} .
%\end{equation}
%For $SU(3)$, $\alpha_1=\vec{\alpha}^{(1)}= (1,0)$ and $\alpha_2=\vec{\alpha}^{(3)} =\left(\frac{1}{2} \cdot \frac{\sqrt{3}}{2} \right)$. 
%For $\vec{\Lambda}=\vec{\nu}_1$, $\vec{\alpha}^{(1)} \cdot \vec{\nu}_1 =1$ and $\vec{\alpha}^{(3)} \cdot \vec{\nu}_1 =0$. 
%For $\vec{\Lambda}=\vec{\nu}_2$, $\vec{\alpha}^{(1)} \cdot \vec{\nu}_2 =-1$ and $\vec{\alpha}^{(3)} \cdot \vec{\nu}_2 =1$. 
%For $\vec{\Lambda}=\vec{\nu}_3$, $\vec{\alpha}^{(1)} \cdot \vec{\nu}_3 =0$ and $\vec{\alpha}^{(3)} \cdot \vec{\nu}_3 =-1$. 
%In general, for 
%$
%\vec{\Lambda} = m \vec{h}_1  + n \vec{h}_2 =  m \vec{\nu}_1 - n \vec{\nu}_3
%$, 
%we find 
%$
%(\vec{\alpha}^{(1)} \cdot \vec{\Lambda})=m (\vec{\alpha}^{(1)} \cdot \vec{\nu}_1 )-n (\vec{\alpha}^{(1)} \cdot \vec{\nu}_3 )=\frac12 m
%$ 
%and 
%$
%(\vec{\alpha}^{(3)} \cdot \vec{\Lambda})=m (\vec{\alpha}^{(3)} \cdot \vec{\nu}_1 )-n (\vec{\alpha}^{(3)} \cdot \vec{\nu}_3 )=\frac12 n
%$. 
%}
  The two-dimensional highest weight vector of the representation $[m,n]$ is given by 
\begin{equation}
 \vec{\Lambda} = m \vec{h}_1  + n \vec{h}_2  
 ,
\end{equation}
using the highest weight $\vec{h}_1$ of a \textbf{fundamental representation} $[1,0]={\bf 3}$ and $\vec{h}_2$ of another  fundamental representation,  i.e., the \textbf{conjugate  representation} $[0,1]={\bf 3^*}$. 
The weight diagrams for fundamental representations ${\bf 3}$ and ${\bf 3^*}$ are given in Fig.~\ref{C28-fig:fundamental-weight}. 
As the highest weight of ${\bf 3}$ we adopt the standard one: 
\begin{equation}
  \vec{h}_1  
= \left( \frac{1}{2}, \frac{1}{2\sqrt{3}} \right)
=  \vec{\nu}_1  
 .
\end{equation}
As the highest weight of the conjugate representation ${\bf 3^*}$, on the other hand, we choose%
%\footnote{
%This is the choice of %Kondo and Taira (2000).
%\cite{KT00}. 
%The following results hold also for other choices of $\vec{h}_2$, e.g., 
%\begin{equation}
%  \vec{h}_1 
%= \left( \frac{1}{2}, \frac{1}{2\sqrt{3}} \right)
%=  \vec{\nu}_1  
% , \quad
%  \vec{h}_2  
%= \left( \frac{1}{2}, \frac{-1}{2\sqrt{3}} \right) 
%= - \vec{\nu}_2 
% ,   \quad
%  \vec{\Lambda} 
%= \left( \frac{m+n}{2}, \frac{m-n}{2\sqrt{3}} \right) 
% .
%\end{equation} 
%}
\begin{equation}
  \vec{h}_2 
= \left( 0, \frac{1}{\sqrt{3}} \right) 
= - \vec{\nu}_3
 .
\end{equation}
Then the highest weight vector of $[m.n]$ is given by 
\begin{equation}
 \vec{\Lambda} 
 = (\Lambda_1,\Lambda_2) 
=  \left( \frac{m}{2}, \frac{m+2n}{2\sqrt{3}} \right) 
 .
\end{equation}
%\footnote{
%This choice of $h_2$ is different from that in Ref.\citen{KT99a} 
%It is adopted so as to obtain the $SU(3)$ case when considering the $N=3$ case of $SU(N)$ case studied in the next subsecton.}
The generators  of $SU(3)$ in the Cartan basis are written as
\begin{equation}
\{ H_1, H_2, E_{\alpha}, E_{\beta}, E_{\alpha+\beta},
   E_{-\alpha}, E_{-\beta}, E_{-\alpha-\beta} \} .
\end{equation}
%where the two simple roots are 
%\begin{equation}
%\alpha_1=\alpha=\vec{\alpha}^{(1)} , \quad \alpha_2=\beta=\vec{\alpha}^{(3)} .
%\end{equation}
See Fig.~\ref{C28-fig:root} for the explicit choice.

%%%%%%%%%%%%%%%%%%%%%%%%%%%%%%%%%%%%%%%%%%%%%%%%%%%%%%%%%%%
%%%%%%%%%%%%%%%%%%%%%%%%%%%%%%%%%%%%%%%%%%%%%%%%%%%%%%%%%%%
\begin{figure}[ptb]
\begin{center}
\includegraphics[scale=0.3]{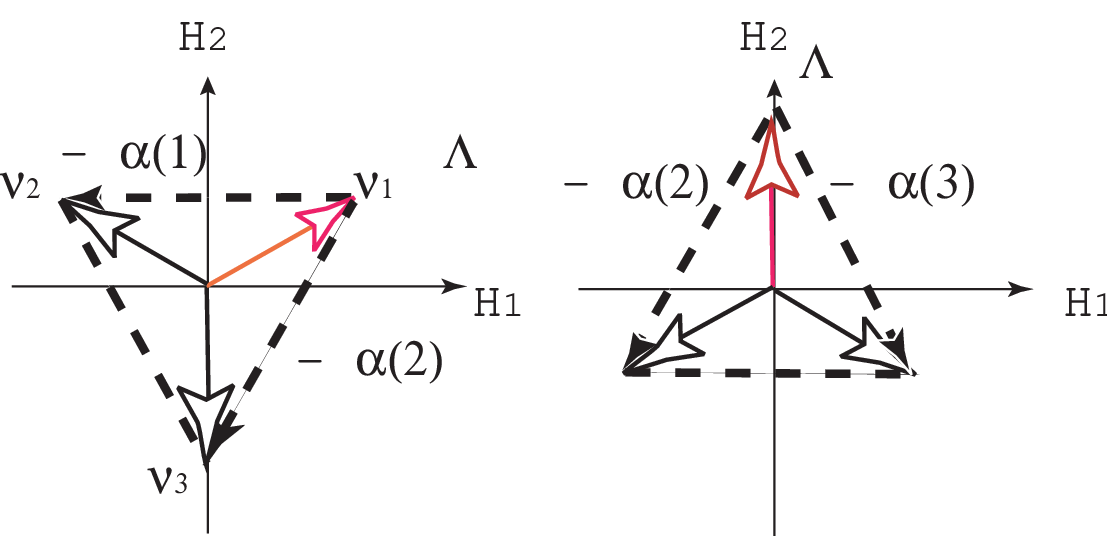}
\includegraphics[scale=0.3]{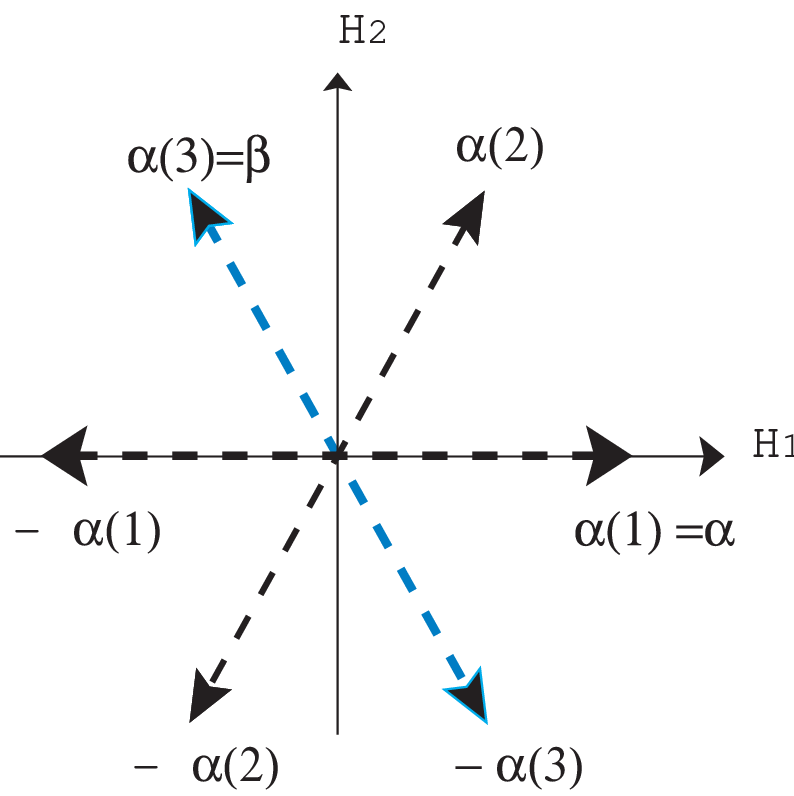}
\end{center}
\vskip -0.5cm
\caption{
The relationships among the weight vectors $\vec{\nu}_1, \vec{\nu}_2, \vec{\nu}_3$ in the fundamental representations ${\bf 3}$     and the root vectors $\vec{\alpha}^{(1)}, \vec{\alpha}^{(2)}, \vec{\alpha}^{(3)}$  in $SU(3)$.
We find
$
\vec{\nu}_1  \perp  \vec{\alpha}^{(3)}, - \vec{\alpha}^{(3)}
$.
Here   $\vec\Lambda  =\vec{\nu}_1:=(\frac{1}{2},\frac{1}{2\sqrt{3}})$ is the highest weight of the fundamental representation ${\bf 3}$.
}
 \label{C28-fig:fundamental-weight3}
\end{figure}
%%%%%%%%%%%%%%%%%%%%%%%%%%%%%%%%%%%%%%%%%%%%%%%%%%%%%%%%%%%
%%%%%%%%%%%%%%%%%%%%%%%%%%%%%%%%%%%%%%%%%%%%%%%%%%%%%%%%%%%

\begin{figure}
\begin{center}
% \leavevmode
% \epsfxsize=60mm
% \epsfysize=60mm
% \epsfbox{weight-adjointrev.eps}
\includegraphics[scale=0.27]{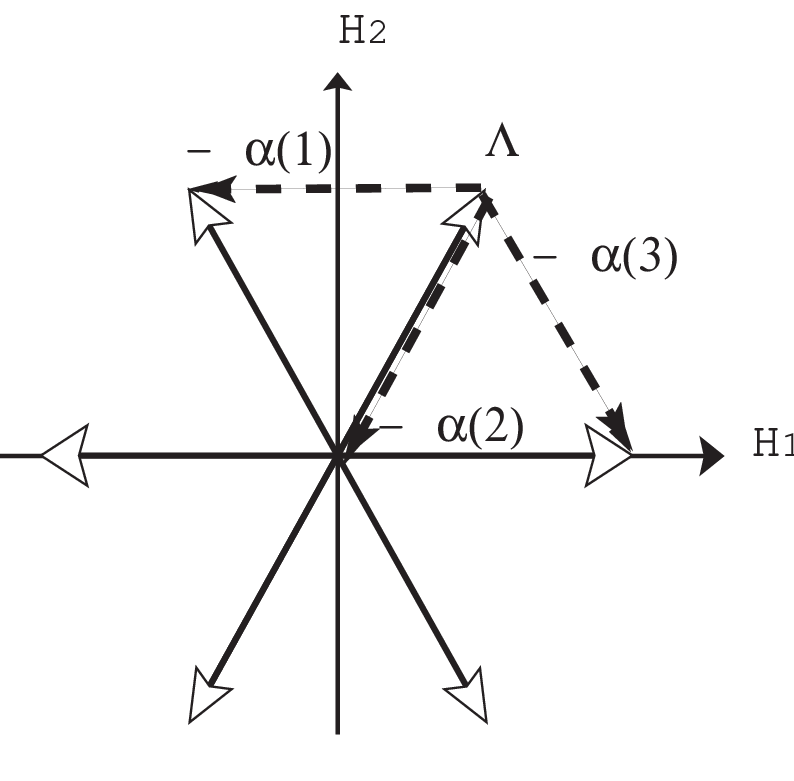}
\end{center} 
\vskip -0.5cm
\caption[]{
The weight vectors and root vectors required to define the coherent state in the adjoint representation $[1,1]={\bf 8}$ of $SU(3)$, where  $\vec{\Lambda}=(\frac{1}{2},\frac{\sqrt{3}}{2})$ is the highest weight of the adjoint
representation.
}
 \label{C28-fig:adjoint-weight}
\end{figure}

Every representation $R$ of $SU(3)$ specified by the Dynkin index $[m,n]$ belongs to (I) or (II):
\begin{enumerate}
\item[(I)]
[Minimal case] 
If $mn=0$, ($m=0$ or $n=0$), the maximal stability group
$\tilde H$ is given by 
\begin{equation}
 \tilde H=U(2) ,
\end{equation}
with generators 
$\{ H_1, H_2, E_\beta, E_{-\beta} \}$.
In the {minimal} case,  ${\rm dim}(G/\tilde{H})$ is minimal. 
Such a degenerate case occurs when the highest-weight vector $\vec{\Lambda}$ is orthogonal to some root vectors.%
%\footnote{
%The orthogonality of the highest-weight vector $\vec{\Lambda}$  to the respective root vector is realized as
%\begin{align}
% 0 =& \vec{\alpha}^{(1)} \cdot \vec{\Lambda} = \frac{m}{2} 
% \Longrightarrow m=0 \Longrightarrow [0,n] ,
% \nonumber\\
% 0 =& \vec{\alpha}^{(2)} \cdot \vec{\Lambda} = \frac{m+n}{2} 
% \Longrightarrow m=-n \Longrightarrow [m,-m] ,
% \nonumber\\
% 0 =& \vec{\alpha}^{(3)} \cdot \vec{\Lambda} = \frac{n}{2} 
% \Longrightarrow n=0 \Longrightarrow [m,0] ,
%\end{align}
%}   
\\
For example, the fundamental reprentation $[1,0]$ has the maximal stability subgroup $U(2)$ with the generators  
$\{ H_1, H_2, E_{\alpha^{(3)}}, E_{-\alpha^{(3)}} \} \in u(2)$,
where% $\Lambda \perp \beta, -\beta$. 
\begin{equation}
 \vec{\Lambda}=\vec{\nu}_1  \perp  \vec{\alpha}^{(3)}, - \vec{\alpha}^{(3)}  . 
\end{equation}
See Fig.~\ref{C28-fig:fundamental-weight3}.

\item[(II)]
[Maximal case]
If $mn \not=0$ 
($m\not=0$ and $n\not=0$), $H$ is the maximal torus group 
\begin{equation}
\tilde H=H=U(1) \times U(1) ,
\end{equation}
with generators 
$\{ H_1, H_2 \}$.
In the {maximal} case II,  ${\rm dim}(G/\tilde{H})$ is maximal. 
This is a non-degenerate case.
\\
For example, the adjoint representation $[1,1]$ has the maximal stability subgroup $U(1) \times U(1)$ with the generators $\{ H_1, H_2  \} \in u(1)+u(1)$.
See Fig.\ref{C28-fig:adjoint-weight}.
\end{enumerate}

In the minimal case, the coset $G/\tilde H$ is given by
\begin{equation}
  SU(3)/U(2)=SU(3)/(SU(2)\times U(1))=CP^2,
\end{equation}
whereas in the maximal case, 
\begin{equation}
  SU(3)/(U(1)\times U(1))=F_2 .
\end{equation}
Here, $CP^{n}$ is the \textbf{complex projective space} and $F_n$ is the \textbf{flag space}. 
Therefore, \textit{the   fundamental representations ($\bm{3}$ and $\bm{3^*}$) of $SU(3)$ belong to the minimal case (I), and hence the maximal stability group is $U(2)$, rather than the maximal torus group $U(1) \times U(1)$}.

%%%%%%%%%%%%%%%%%%%%%%%%%%%%%%%%%%%%%%%%%%%%%%%%%%%%%%%%%
%%%%%%%%%%%%%%%%%%%%%%%%%%%%%%%%%%%%%%%%%%%%%%%%%%%%%%%%%

For $SU(3)$, the explicit form of $\mathcal{H}$  for the fundamental representations is calculated  using the diagonal set of the Gell-Mann matrices $\lambda_3$ and $\lambda_8$:
\begin{align}
 \mathcal{H} 
=& \frac{1}{2} (\Lambda_1 \lambda_3 + \Lambda_2 \lambda_8)
%\nonumber\\
=  \frac{1}{2} \left( \frac{m}{2} \lambda_3 + \frac{m+2n}{2\sqrt{3}}   \lambda_8 \right)
\nonumber\\
=& %\frac{1}{2}
 \begin{pmatrix} 
  \frac{2m+n}{6} & 0 & 0 \cr
  0 & \frac{-m+n}{6} & 0 \cr
  0 & 0 & \frac{-m-2n}{6}  \cr
 \end{pmatrix} 
 =  \frac{1}{6} {\rm diag} \left(  2m+n ,  -m+n ,   -m-2n  \right)
 .
\end{align}
We enumerate all fundamental representations $\textbf{3}$:%
\footnote{For another choice of $h_2=-\vec{\nu}_2$, the same results are obtained if the following replacement is performed.
[-1,1] $\rightarrow$ [0,-1], 
[0,-1] $\rightarrow$ [-1,1],
[0,1] $\rightarrow$ [1,-1],
[1,-1] $\rightarrow$ [0,1].
$
 \mathcal{H} 
=   \frac{1}{2} (\Lambda_1 \lambda_3 + \Lambda_2 \lambda_8)
%\nonumber\\
%=& \frac{1}{2}
% \begin{pmatrix} 
%  \frac{2m+n}{3} & 0 & 0 \cr
%  0 & \frac{-m-2n}{3} & 0 \cr
%  0 & 0 & \frac{-m+n}{3}  \cr
% \end{pmatrix} 
= {\rm diag} \left( \frac{2m+n}{6}, \frac{-m-2n}{6} , \frac{-m+n}{6} \right)
 .
$

}
\begin{subequations}
\begin{align}
%\begin{enumerate}
%\item[[1,0]]:
%\begin{equation}
[1,0]: & \
 \vec{\Lambda}
= \left( \frac{1}{2}, \frac{1}{2\sqrt{3}} \right) = \vec{\nu}_1 
 , \quad
  \mathcal{H} = -\frac{1}{6} {\rm diag} \left( -2,  1,  1 \right) 
   ,
   \label{C28-fr1}
%\end{equation}
\\
%\item[[-1,1]]:
%\begin{equation}
[-1,1]: & \
 \vec{\Lambda}
= \left( \frac{-1}{2}, \frac{1}{2\sqrt{3}} \right)  = \vec{\nu}_2 
 , \quad
  \mathcal{H} = -\frac{1}{6} {\rm diag} \left(  1,-2,  1 \right)
   ,
   \label{C28-fr2}
%\end{equation}
\\
%\item[[0,-1]]:
%\begin{equation}
[0,-1]: & \
 \vec{\Lambda} 
= \left( 0, \frac{-1}{\sqrt{3}} \right)  = \vec{\nu}_3 
 , \quad
  \mathcal{H} = -\frac{1}{6} {\rm diag} \left(  1,  1, -2 \right)
  = \frac{-1}{\sqrt{3}} \frac{\lambda_8}{2} 
   ,
   \label{C28-fr3}
%\end{equation}
%\end{enumerate}
\end{align}
\end{subequations}
and their conjugates $\textbf{3$^*$}$
\begin{subequations}
\begin{align}
%\begin{enumerate}
%\item[[0,1]]:
%\begin{equation}
[0,1]: & \
 \vec{\Lambda} 
= \left( 0, \frac{1}{\sqrt{3}} \right) = -\vec{\nu}_3 
 , \ 
  \mathcal{H} = - \frac{1}{6} {\rm diag} \left( 1,  1 ,-2  \right)
  = \frac{-1}{\sqrt{3}} \frac{\lambda_8}{2} 
   ,
%\end{equation}
\\
%\item[[1,-1]]:
%\begin{equation}
[1,-1]: & \
\vec{\Lambda}
= \left( \frac{1}{2}, \frac{-1}{2\sqrt{3}} \right) = -\vec{\nu}_2 
 , \ 
  \mathcal{H} = - \frac{1}{6} {\rm diag} \left( 1, -2, 1 \right)
   ,
%\end{equation}
\\
%\item[[-1,0]]:
%\begin{equation}
[-1,0]: &
 \vec{\Lambda}
= \left( \frac{-1}{2}, \frac{-1}{2\sqrt{3}} \right) = -\vec{\nu}_1 
 , \ 
  \mathcal{H} = - \frac{1}{6} {\rm diag} \left(  -2, 1, 1 \right)
   .
%\end{equation}
%\end{enumerate}
\end{align}
\end{subequations}

For three fundamental representations (\ref{C28-fr1}), (\ref{C28-fr2}) and (\ref{C28-fr3}), the eigenvectors (\ref{C28-eigen}) are found to be
\begin{align}
\left|  \bm{\Lambda} \right> 
=(1,0,0)^T , 
\quad
\left|  \bm{\Lambda} \right> 
=(0,1,0)^T ,
\quad \& \quad 
\left|  \bm{\Lambda} \right> 
=(0,0,1)^T
 ,
\end{align}
respectively.

For $SU(3)$, the  $\bm{m}$ field is a linear combination of two color fields:
\begin{equation}
 \bm{n}_1(x)
% :=  U^{-1}(x)  H_1  U(x) 
  =  g(x)  \frac{\lambda_3}{2}  g^{\dagger}(x) 
  , \quad 
  \bm{n}_2(x)
% :=  U^{-1}(x)  H_2  U(x) 
  =  g(x)  \frac{\lambda_8}{2}  g^{\dagger}(x) 
   .
\end{equation}
The   $\bm{m}$ field reads
 for $[1,0]$ and $[-1,0]$, 
\begin{subequations}
\begin{equation}
  \bm{m}(x)  
  =  \pm  \left[  \frac{1}{2}  \bm{n}_1(x)  +   \frac{1}{2\sqrt{3}} \bm{n}_2(x) \right]
   =  \pm   \frac{1}{\sqrt{3}} \left[  \frac{\sqrt{3}}{2}  \bm{n}_1(x)  +   \frac{1}{2} \bm{n}_2(x) \right]
 , 
 \label{C28-m1}
\end{equation}
for  $[-1, 1]$ and $[1,-1]$, 
\begin{equation}
  \bm{m}(x)  
  =  \pm \left[- \frac{1}{2}  \bm{n}_1(x)  +   \frac{1}{2\sqrt{3}} \bm{n}_2(x) \right] 
   =  \pm   \frac{1}{\sqrt{3}} \left[  -\frac{\sqrt{3}}{2}  \bm{n}_1(x)  +   \frac{1}{2} \bm{n}_2(x) \right]
  .
  \label{C28-m2}
\end{equation}
In particular, for $[0,-1]$ and $[0, 1]$, $\Lambda_1=0$ and hence the    $\bm{m}$ field   is written using only $\bm{n}_2(x)$:
\begin{equation}
  \bm{m}(x)  
%=   \pm  \frac{1}{2} \left[  \frac{-2}{\sqrt{3}} \bm{n}_2(x) \right]
   =   \pm   \frac{-1}{\sqrt{3}}    \bm{n}_2(x)  
   \label{C28-m3}
  .
\end{equation}
\end{subequations}

For the fundamental representation of $SU(3)$, the highest-weight state $| \Lambda \rangle$ yields the projection operator and its traceless version:
\begin{align}
& \rho  =  | \Lambda \rangle \langle \Lambda |
%= 
%\begin{pmatrix}
% 1 \\
% 0 \\
% 0 
%\end{pmatrix}
% (1,0,0)
= 
\begin{pmatrix}
 1 & 0 & 0 \\
 0 & 0 & 0 \\
 0 & 0 & 0 \\
\end{pmatrix} 
\Longrightarrow 
\mathcal{H} =  \frac12 \left(  \rho - \frac13 \mathbf{1} \right) 
= \frac{1}{6} 
\begin{pmatrix}
 2 & 0 & 0 \\
 0 & -1 & 0 \\
 0 & 0 & -1 \\
\end{pmatrix} 
.
\end{align}
Then the two expressions  (\ref{C28-H-def1}) and (\ref{C28-H-def2}) of $\mathcal{H}$ are equivalent: 
\begin{align}
 \mathcal{H} 
=  \frac{1}{2}  \left( \rho - \frac13 \mathbf{1}  \right) 
=  \Lambda_1 H_1  + \Lambda_2 H_2  
 .
\end{align}
The  color field is constructed as
\begin{equation}
\bm{n}(x) 
= \frac{\sqrt{3}}{2}  \tilde{\bm{n}}(x)
= \sqrt{3} \bm{m}(x) 
= \sqrt{3} g(x) \mathcal{H} g(x)^\dagger \  (g(x) \in SU(3))
.
\end{equation}
\begin{align}
  \bm{n}(x) 
%=  \frac{\sqrt{3}}{2} g_{x} \left(  \rho - \frac13 \mathbf{1} \right) g_{x}^\dagger 
=   {g(x)}
\frac{-1}{2\sqrt{3}} 
\begin{pmatrix}
 -2 & 0 & 0 \\
 0 &  {1} & 0 \\
 0 & 0 &  {1} \\
\end{pmatrix} 
   {g(x)^\dagger} 
\in SU(3)/U(2) \simeq \mathbb{C}P^2 .%\simeq CP^2 .
\end{align}
%with the Pauli matrix $\sigma_3:={\rm diag.}(1,-1)$.
The matrix ${\rm diag.}(-2,1,1)$ is degenerate. Using the Weyl symmetry (discrete global symmetry as a subgroup of color symmetry), it is changed into $\lambda_8$. 
This color field describes a \textbf{non-Abelian magnetic monopole}, which corresponds to the spontaneous symmetry breaking $SU(3) \to U(2)$ in the gauge-Higgs model.

Note that  
\begin{equation}
  \bm{m}(x) = m^A(x) T_A , \
  m^A(x) = 2 {\rm tr}(\bm{m}(x) T_A)
 = 2 {\rm tr}(g(x) \mathcal{H}  g^{\dagger}(x)T_A)
   ,
\end{equation}
for the normalization 
${\rm tr}(T_A T_B)=\frac12 \delta_{AB}$. 
For three fundamental representations (\ref{C28-fr1}), (\ref{C28-fr2}) and (\ref{C28-fr3}) of $SU(3)$, $m^A(x)$ is equal to the first, second and third diagonal elements of $g(x)T_A g^{\dagger}(x)$ respectively: 
\begin{equation}
  m^A(x) =  (g^{\dagger}(x)T_A g(x))_{ff} \ (f=1,2,3;  \ \text{no sum over $f$})
   .
\end{equation}
This is checked easily, e.g., for $[1,0]$,
\begin{align}
m^A(x) =& 2 {\rm tr}(\mathcal{H}  g^{\dagger}(x)T_A g(x) )
\nonumber\\
=& \frac23 (g^{\dagger}(x)T_A g(x))_{11} - \frac13 (g^{\dagger}(x)T_A g(x))_{22} - \frac13 (g^{\dagger}(x)T_A g(x))_{33} 
\nonumber\\
=&  (g^{\dagger}(x)T_A g(x))_{11}
 ,
\end{align}
where we have used a fact that $g^{\dagger}(x)T_A g(x)$ is traceless. 
Therefore, we have 
\begin{equation}
  m^A(x)  
 =  \left< \bm{\Lambda} |  g^{\dagger}(x)T_A g(x) |\bm{\Lambda} \right>
 =  \left< g(x), \bm{\Lambda} |  T_A  |g(x), \bm{\Lambda} \right>
  =  \left< \xi(x), \bm{\Lambda} |  T_A  | \xi(x), \bm{\Lambda} \right>
  ,
\end{equation}
using   the highest--weight state $\left|  \Lambda \right>$ of the respective fundamental representation.

The state  $\left| \xi(x), \bm{\Lambda} \right>$ is regarded as the coherent state describing the subspace corresponding to the subgroup $G/\tilde{H}=SU(3)/U(2) \simeq CP^2$, the two-dimensional complex projective space.
This is also assured in the following way.
The component $m^A$ of $\bm{m}$ is rewritten as 
\begin{equation}
  m^A(x)  = \Phi^*(x) T_A \Phi(x)
% =  \left< \Lambda |  U(x)T_A U^{-1}(x) |\Lambda \right>
 = \phi_a^*(x) (T_A)_{ab} \phi_b(x)  , \quad \phi_a(x) \in \mathbb{C} , \ \ (a,b=1,2,3)
   ,  
\end{equation}
by introducing the $CP^2$ variable $\phi^a(x)$:
\begin{equation}
   \phi_a(x)  := (g(x) \left|\bm{\Lambda}  \right>)_a 
   .
\end{equation}
The complex field $\phi_a(x)$ is indeed the $CP^2$ variable, since there are only two independent complex degrees of freedom among three complex variables $\phi^a(x)$($a=1,2,3$).
This is because the variables are subject to a constraint:
\begin{equation}
  \phi^\dagger (x) \phi(x)
:= \sum_{a=1}^{3}  \phi_a^* (x) \phi_a(x)  
=  \left< \bm{\Lambda} \right| g^{\dagger}(x)  g(x) \left|\bm{\Lambda}  \right> 
 =  \left< \bm{\Lambda}   |\bm{\Lambda}  \right>
 = 1
  ,
\end{equation}
and the invariance under the $U(1)$ phase transformation $e^{i\theta}$ eliminates  one degree of freedom. 
This result suggests that the Wilson loop operator in fundamental representations of $SU(N)$ can be studied by the $CP^{N-1}$ valued field effectively, rather than $F_{N-1}$. 
%\footnote{
%For more details, see section 6.%the chapter of the non-Abelian Stokes theorem.
%Kondo and Taira (2000).
%\cite{KT99}.
%}
This parameterization becomes useful in the large $N$ expansion \cite{KT00}.

%The  $SU(3)$ Yang-Mills field in the Cartan basis is written as
%\begin{align}
%  \mathscr{A}_\mu = \mathscr{A}_\mu^A T_A 
%=& a_\mu T_3 + a'_\mu T_8 + \sum_{a=1,2,4,5,6,7}  A_\mu^{a} T_{a}   
%\nonumber\\
%=& a_\mu H_1 + a'_\mu H_2 + \sum_{a=1}^{3} (W_\mu^*{}^{a} E_{a} + W_\mu^{a} E_{-a})  ,
%\end{align}
%where
%$
% A \in \{1,2,3,4,5,6,7,8\};  a \in \{ 1,2,4,5,6,7 \},
%$
%\begin{align}
%  W_\mu^1 = \frac{1}{\sqrt{2}}(A_\mu^1+ i A_\mu^2) , %\quad
%  W_\mu^2 = \frac{1}{\sqrt{2}}(A_\mu^4- i A_\mu^5) , %\quad
%  W_\mu^3 = \frac{1}{\sqrt{2}}(A_\mu^6+ i A_\mu^7) .
%\end{align}

 For $SU(N)$, we can introduce $N-1$ color fields $\bm{n}_j(x)$ ($j=1, \cdots, N-1$) corresponding to the degrees of freedom of the maximal torus group $U(1)^{N-1}$ of $SU(N)$.  This is just the way adopted in the conventional approach.  
However, this is not necessarily effective to see the physics extractable  from the Wilson loop.
This is because only the specific combination $\bm{m}(x)$ of the color fields $\bm{n}_j(x)$ has a physical meaning as shown in the next chapter and this nice property of $\bm{m}(x)$ will be lost once $\bm{m}(x)$ is separated into the respective color field, except for the $SU(2)$ case in which  $\bm{m}(x)$ agrees with the unique color field $\bm{n}(x)$ of $SU(2)$. 
In view of this, only the last color field $\bm{n}_{N-1}(x)$ is enough for investigating quark confinement through the Wilson loop in fundamental representations of $SU(N)$.%
\footnote{
This is called the minimal option proposed in Kondo, Shinohara and Murakami (2008).
%\cite{KSM08}. 
Indeed, the above combinations (\ref{C28-m1}), (\ref{C28-m2}) and (\ref{C28-m3}) correspond to 6 minimal cases discussed.
A unit vector $\bm{n}(x)$ introduced in (\ref{C28-unit-n})   is related to $\bm{m}$ as
$\sqrt{3} \bm{m}(x)=\bm{n}(x)=(\cos \vartheta(x)) \bm{n}_1(x)+(\sin \vartheta(x)) \bm{n}_2(x)$
where $\vartheta(x)$ denotes the angle of a weight vector in the weight diagram measured anticlockwise from the $H_1$ axis. 
Here 
(\ref{C28-m1}), (\ref{C28-m2}) and (\ref{C28-m3}) correspond to $\vartheta(x)=\frac{1}{6}\pi (\frac{7}{6}\pi)$ $\frac{5}{6}\pi (\frac{11}{6}\pi)$, and $\frac{3}{2}\pi (\frac{1}{2}\pi)$ respectively. 
}

%%%%%%%%%%%%%%%%%%%%%%%%%%%%%%%%%%%%%%%%%%%%%%%%%%%%%%%%%%%%%
%%%%%%%%%%%%%%%%%%%%%%%%%%%%%%%%%%%%%%%%%%%%%%%%%%%%%%%%%%%%%
\subsection{Non-Abelian Stokes theorem as a path-integral representation}
\label{subsection:derivation-NAST}%%%%%%%%%%%%%%%%%%%%%%%%%%%%%%%%%%%%%%%%%%%%%%%%%%%%%%%%%%%%%
%%%%%%%%%%%%%%%%%%%%%%%%%%%%%%%%%%%%%%%%%%%%%%%%%%%%%%%%%%%%%

Let $\mathscr{A}$ be the Lie algebra valued \textbf{connection one-form}: 
\begin{equation}
 \mathscr{A}(x) :=  \mathscr{A}_\mu(x)  dx^\mu  = \mathscr{A}_\mu^A(x) T_A dx^\mu
  .
\end{equation}
For a given loop, i.e.,  a closed path $C$,   the  \textbf{Wilson loop operator} $W_{\rm C}[\mathscr{A}]$ in the representation $R$ is defined by
\begin{equation}
 W_C[\mathscr{A}] := \mathcal{N}^{-1} {\rm tr}_{R} \left\{ \mathscr{P} \exp \left[ -ig_{{}_{\rm YM}} \oint_C \mathscr{A} \right] \right\} ,
 \quad  \mathscr{N}:=d_R = {\rm tr}_R({\bf 1}) ,
\end{equation}
where $\mathscr{P}$ denotes the \textbf{path ordering} (defined precisely later) and  the normalization factor $\mathcal{N}$ is equal to the dimension $d_R$ of the representation $R$, to which the probe of the Wilson loop belongs,   
ensuring $W_C[0]=1$.
We introduce the Yang-Mills coupling constant $g_{{}_{\rm YM}}$  for later convenience, although this can be removed by scaling the field $\mathscr{A}^\prime=g_{{}_{\rm YM}}\mathscr{A}$.

We show that  the non-Abelian Wilson loop operator $W_C[\mathscr{A}]$ defined by the line integral along a closed path $C$ can be rewritten into a surface integral form.  
This fact is called a \textbf{non-Abelian Stokes theorem}. 
It is derived through the \textbf{path-integral  representation} of the Wilson loop operator.

Let  $L$ be a  curve (path) starting at $x_0$ and ending at $x$ which is parameterized by a parameter $s$: $x=x(s)$ ($x_0=x(s_0)$).  Then we define the \textbf{parallel transporter} $W_L[\mathscr{A}](s,s_0)$ by 
\begin{equation}
 W_L[\mathscr{A}] (s,s_0) 
:=   \mathscr{P} \exp \left[ -ig_{{}_{\rm YM}} \int_{L:x_0 \rightarrow x}  \mathscr{A}  \right]  
=    \mathscr{P} \exp \left[ -ig_{{}_{\rm YM}} \int_{s_0}^{s} d\tau \mathscr{A}(\tau) \right]  ,
\end{equation}
where we have defined the tangent component $\mathscr{A}(\tau)$ of $\mathscr{A}(x)$ along the curve:
\begin{equation}
 \mathscr{A}(\tau) =  \mathscr{A}_\mu(x) dx^\mu/d\tau =  \mathscr{A}_\mu^A(x) T_A dx^\mu/d\tau .
\end{equation}

The Wilson loop operator $W_C[\mathscr{A}]$ for a closed loop $C$ is obtained by taking the trace of $W_L[\mathscr{A}]$ for a closed path $L=C$:
\begin{equation}
  W_C[\mathscr{A}] = \mathcal{N}^{-1} {\rm tr} (W_C[\mathscr{A}](s,s_0)) 
  = \mathcal{N}^{-1} \sum_{a=1}^{\mathcal{N}} W_C[\mathscr{A}]_{aa}(s,s_0) ,
\end{equation}
where $W_{ab}$ is a matrix element  of $W$   specified by two indices $a$,$b$.

The parallel transporter $W_L[\mathscr{A}]$ satisfies the differential equation which has the same form as  the \textbf{Schr\"odinger equation}:
\begin{equation}
 i \frac{d}{ds}W_L[\mathscr{A}](s,s_0) =   g_{{}_{\rm YM}}  \mathscr{A}(s) \ W_L[\mathscr{A}](s,s_0) .
\end{equation}
Therefore, $W_L[\mathscr{A}]$ represents  the time-evolution operator of a quantum mechanical system with the Hamiltonian:
\begin{equation}
\mathcal{H}(s) =  g_{{}_{\rm YM}} \mathscr{A}(s)
\end{equation}
on the one-dimensional space parameterized by a parameter $s$, if $s$ is identified with the time. 
This suggests that it is possible to write the path-integral representation  of the parallel transporter $W_L$ and  the Wilson loop operator $W_C$.

%%%%%%%%%%%%%%%%%%%%% figures %%%%%%%%%%%%%%%%%%%%%%%%%%%
\begin{figure}[tbp]
\begin{center}
\includegraphics[height=2.5cm]{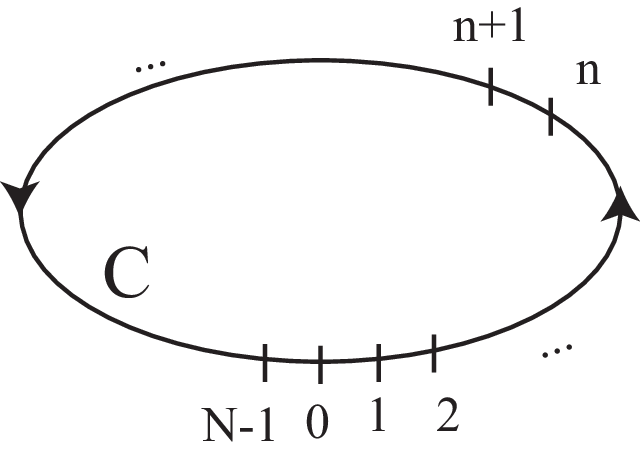}\end{center} 
\vskip -0.5cm
\caption[]{
A loop $C$ for defining the Wilson loop operator is divided into $N$ infinitesimal segments. 
}
\label{fig:W_loop-path-ordering}
\end{figure}
%%%%%%%%%%%%%%%%%%%%% figures %%%%%%%%%%%%%%%%%%%%%%%%%%%

The \textbf{path ordering} is defined as follows. 
Divide the path $L$ into $N$ infinitesimal segments 
so that   the parallel transporter is divided into that of the each infinitesimal segment (See Fig.~\ref{fig:W_loop-path-ordering}):  
\begin{align}
  \mathscr{P} \exp \left[ -ig_{{}_{\rm YM}}  \int_{s_0}^{s} d\tau \mathscr{A}(\tau) \right] 
%\nonumber\\
:=&  \mathscr{P} \prod_{n=0}^{N-1}  U(s_{n+1},s_n),  
\nonumber\\
=&    \exp \left[-ig_{{}_{\rm YM}}  \int_{s_{N-1}}^{s_{N}} d\tau \mathscr{A}(\tau)  \right] 
 \exp \left[-ig_{{}_{\rm YM}}  \int_{s_{N-2}}^{s_{N-1}} d\tau \mathscr{A}(\tau)  \right] \times \dots
\nonumber\\&
  \dots \times \exp \left[-ig_{{}_{\rm YM}}  \int_{s_0}^{s_{1}} d\tau \mathscr{A}(\tau)  \right] 
%\nonumber\\
%=&  \mathscr{P} \prod_{n=0}^{N-1}  \exp [-i   g_{{}_{\rm YM}} \epsilon \mathscr{A}(s_n) ] 
%=    \mathscr{P} \prod_{n=0}^{N-1} [1-ig_{{}_{\rm YM}}  \epsilon \mathscr{A}(s_n)] 
   ,
\end{align}
where $\epsilon :=(s-s_0)/N$ and $s_n :=s_0+n\epsilon$, and the infinitesimal parallel transporter $U(s_{n+1},s_n)$ is defined by
\begin{equation}
U(s_{n+1},s_n) :=   \exp \left[-ig_{{}_{\rm YM}}  \int_{s_n}^{s_{n+1}} d\tau \mathscr{A}(\tau)  \right] .
\end{equation}
We set $s=s_N$ and $x=x_N =x(s_N)$.
Then the non-Abelian Wilson loop operator should be understood as
\begin{equation}
 W_C[\mathscr{A}]  = \lim_{N \rightarrow \infty, \epsilon \rightarrow 0}  {\rm tr}_{R} \left[ \mathscr{P} \prod_{n=0}^{N-1}  U(s_{n+1},s_n)   \right]/{\rm tr}_R({\bf 1}) .
\end{equation}

We follow the standard procedures of deriving the path-integral formula:
\begin{enumerate}
\item
Replace the trace of the operator $\mathscr{O}$ by the integral:
\begin{equation}
 {\rm tr}_{\rm R}(\mathscr{O})/{\rm tr}_R({\bf 1}) 
 = \int d\mu({g}_0) \left< {g}_0, \Lambda \right| \mathscr{O}\left| {g}_0, \Lambda \right> ,
 \label{C29-trace-integral}
\end{equation}
where 
$d\mu({g})$ is an invariant measure on $G$
and  the state is normalized 
\begin{equation}
\left< {g}_n, \Lambda | {g}_n, \Lambda \right>=1 
 .
\end{equation}
The proof of the relation (\ref{C29-trace-integral}) is given in \ref{section:integration-formula}.

\item
Insert a \textbf{complete set} of states at each partition point $x_n:=x(s_n)$ $(n=1, \cdots, N-1)$ on $L$: 
\begin{equation}
 {\bf 1} = \int  d\mu({g}_n)  \left| {g}_n, \Lambda \right> \left< {g}_n, \Lambda \right| , 
\  {g}_n := {g}(x_n) \  (n=1, \cdots, N-1) ,
\end{equation}

\item
Take the limit $N \rightarrow \infty$ and $\epsilon \rightarrow 0$ appropriately such that $N\epsilon=s-s_0$ is fixed.

\end{enumerate}

As the complete set to be inserted, we adopt the \textbf{coherent state}. 
%\footnote{
%For details, see the chapter of coherent state and the stability group.
%}
The coherent state $\left| {g} , \Lambda \right>$ is constructed by operating a group element ${g} \in G$ to a reference state $\left| \Lambda \right>$:
\begin{equation}
 \left| {g} , \Lambda \right> = {g} \left| \Lambda \right> , \quad g \in G .
\end{equation}
Note that the coherent states are non-orthogonal:
\begin{equation}
\langle {g}^\prime ,\Lambda| {g}, \Lambda \rangle
\ne 0 .
\end{equation}

Putting aside the issue of what type of complete set is chosen, we obtain
\begin{align}
% &    {\rm tr}_{\rm R} \left\{ \mathscr{P} \exp \left[  -ig_{{}_{\rm YM}} \int_{s_0}^s d\tau \mathscr{A}(\tau)\right] \right\}/{\rm tr}_R({\bf 1})
%\nonumber\\
 W_C[\mathscr{A}]  =& \lim_{N \rightarrow \infty, \epsilon \rightarrow 0} 
   \int d\mu({g}_{0}) \langle {g}_0,\Lambda | U(s_{N},s_{N-1})  | {g}_{N-1}, \Lambda \rangle
\nonumber\\
&  \times
\int d\mu({g}_{N-1}) \langle {g}_{N-1},\Lambda | U(s_{N-1},s_{N-2})  | {g}_{N-2}, \Lambda \rangle
\nonumber\\
& 
\cdots 
\int d\mu({g}_{1})  \langle {g}_{1},\Lambda| U(s_{1},s_0) | {g}_{0}, \Lambda \rangle 
%  =& \lim_{N \rightarrow \infty, \epsilon \rightarrow 0} 
%\int \cdots \int d\mu({g}_N)  \langle {g}_N,\Lambda|[1+i \epsilon g\mathscr{A}(s_{N-1})] | {g}_{N-1}, \Lambda \rangle d\mu({g}_{N-1})
%\nonumber\\
%&  \times
%\langle {g}_{N-1},\Lambda |[1+i \epsilon g\mathscr{A}(s_{N-2})] | {g}_{N-2}, \Lambda \rangle d\mu({g}_{N-2}) 
%\nonumber\\
%& 
%\cdots d\mu({g}_{1}) \langle {g}_1,\Lambda |[1+i \epsilon g\mathscr{A}(s_0)] | {g}_N, \Lambda \rangle
\nonumber\\
 =& \lim_{N \rightarrow \infty, \epsilon \rightarrow 0}
\prod_{n=0}^{N-1} \int d\mu({g}_n) \prod_{n=0}^{N-1}
\langle {g}_{n+1}, \Lambda | U(s_{n+1},s_n) | {g}_{n}, \Lambda \rangle  
%\nonumber\\
 %=& \lim_{N \rightarrow \infty, \epsilon \rightarrow 0}
%\prod_{n=0}^{N-1} \int d\mu({g}_n) 
%\prod_{n=0}^{N-1} [\langle {g}_{n}, \Lambda | {g}_{n+1}, \Lambda \rangle +i \epsilon g \langle {g}_{n}, \Lambda | \mathscr{A}(s_{n})] | {g}_{n+1}, \Lambda \rangle ]
 ,
\end{align}
where we have used ${g}_0={g}_N$.

%Apart from $O(\epsilon^2)$ terms, we find 
%\begin{align}
%% \epsilon \bar A(s_n) :=
%&  \epsilon \langle {g}_{n},\Lambda| \mathscr{A}(s_n)  | {g}_{n+1}, \Lambda \rangle  
%\nonumber\\
%=&  \epsilon \langle {g}_{n},\Lambda| \mathscr{A}(s_n)  | {g}_{n}, \Lambda \rangle + O(\epsilon^2) 
%\nonumber\\
%=&  \epsilon \langle \Lambda| {g}(s_{n})^\dagger \mathscr{A}(s_n) {g}(s_{n})   | \Lambda \rangle + O(\epsilon^2)  ,
%\end{align}
%and
%\begin{align}
% \langle {g}_{n},\Lambda | {g}_{n+1}, \Lambda \rangle 
% =& \langle {g}(s_{n}),\Lambda | {g}(s_{n}), \Lambda \rangle 
%+ \epsilon \langle {g}(s_{n}),\Lambda | \dot {g}(s_{n}), \Lambda \rangle  +   O(\epsilon^2) 
%\nonumber\\
% =& 1 + \epsilon \langle {g}(s_{n}),\Lambda | \dot {g}(s_{n}), \Lambda \rangle  +   O(\epsilon^2)
%\nonumber\\
% =& \exp [ \epsilon \langle  {g}(s_{n}),\Lambda 
%| \dot{g}(s_{n}), \Lambda \rangle  +   O(\epsilon^2) ] 
%\nonumber\\
% =& \exp [ - i \epsilon \langle \Lambda |
% i  {g}(s_{n})^\dagger \dot{g}(s_{n}) | \Lambda \rangle  +  
%O(\epsilon^2) ] 
%\nonumber\\
% =& \exp [ i \epsilon \langle \Lambda |
% i {g}(s_{n})^\dagger \dot {g}(s_{n}) | \Lambda \rangle  +  O(\epsilon^2) ]  ,
%\end{align}
%where we have used the normalization condition,
%$\langle {g}(s_{n}),\Lambda | {g}(s_{n}), \Lambda \rangle=1$ and the dot denotes the differentiation with respect to $s$.
We define the  transformed variable $U^{g}(s_{n+1},s_n)$ of $U(s_{n+1},s_n)$ by the action of a group element $g$:
\begin{align}
  \langle {g}_{n+1}, \Lambda |  U(s_{n+1},s_n) | {g}_{n}, \Lambda \rangle
%\nonumber\\
= & \langle  \Lambda |   {g}(x_{n+1})^\dagger  U(s_{n+1},s_n)  g(x_{n}) | \Lambda \rangle
%\nonumber\\
=   \langle  \Lambda | U^{g}(s_{n+1},s_n)  | \Lambda \rangle
  ,
\end{align}
where  
\begin{equation}
U^{g}(s_{n+1},s_n) :=    {g}(x_{n+1})^\dagger  U(s_{n+1},s_n)  g(x_{n}) 
=   \exp \left[-ig_{{}_{\rm YM}}  \int_{s_n}^{s_{n+1}} d\tau \mathscr{A}^{g}(\tau)  \right] 
%=  \exp [-ig_{{}_{\rm YM}}  \epsilon \mathscr{A}^g(s_{n}) ]  
.
 \label{C29-def-Ag0}
\end{equation} 
For taking the limit $\epsilon \rightarrow 0$ in the final step, it is sufficient to retain the $O(\epsilon)$ terms. 
Therefore, the transition amplitude can be rewritten in the exponential form by introducing the gauge transformation $\mathscr{A}^{g}$ of the gauge variable $\mathscr{A}$ by a group element $g$ as
\begin{align}
  \langle   \Lambda |  U^{g}(s_{n+1},s_n) |  \Lambda \rangle
%\nonumber\\
%= & \langle  \Lambda |   {g}(x_{n+1})^\dagger  U(s_{n+1},s_n)  g(x_{n}) | \Lambda \rangle
%\nonumber\\
= & \langle  \Lambda |    \exp \left[-ig_{{}_{\rm YM}}  \int_{s_n}^{s_{n+1}} d\tau \mathscr{A}^{g}(\tau)  \right]   | \Lambda \rangle
\nonumber\\
= &  \langle   \Lambda |   \left[ 1-ig_{{}_{\rm YM}}  \int_{s_n}^{s_{n+1}} d\tau \mathscr{A}^{g}(\tau)   +   O(\epsilon^2) \right] |  \Lambda \rangle
\nonumber\\
= & \langle  \Lambda |  \Lambda \rangle  -ig_{{}_{\rm YM}}  \int_{s_n}^{s_{n+1}} d\tau   \langle  \Lambda |\mathscr{A}^{g}(\tau) |  \Lambda \rangle
 +   O(\epsilon^2) 
%\nonumber\\
%= & \langle  \Lambda |  \Lambda \rangle  -i \epsilon g_{{}_{\rm YM}}  \langle  \Lambda | \mathscr{A}^g(s_{n})  |  \Lambda \rangle
% +   O(\epsilon^2) 
\nonumber\\
=& \exp \left[ -i \epsilon g_{{}_{\rm YM}}  \int_{s_n}^{s_{n+1}} d\tau   \langle  \Lambda |\mathscr{A}^{g}(\tau) |  \Lambda \rangle  \right] +   O(\epsilon^2) 
  ,
\end{align}
where we have used the normalization condition,
$\langle \Lambda |   \Lambda \rangle=1$.
Here $\mathscr{A}^{g}(x)$ agrees with the gauge transformation of $\mathscr{A}(x)$ by the group element $g$: 
\begin{align}
\mathscr{A}^{g}(x)  :=  {g}(x)^\dagger \mathscr{A}(x) {g}(x)
+ ig_{{}_{\rm YM}}^{-1} {g}(x)^\dagger d {g}(x) .
\end{align}
Defining the one-form $A^{g}$ from the  Lie algebra valued one-form $\mathscr{A}^g$ by 
\begin{align}
A^g  :=& \langle \Lambda | \mathscr{A}^{g}      |\Lambda \rangle ,
\quad
%\mathscr{A}^{g}(x)  :=  {g}(x)^\dagger \mathscr{A}(x) {g}(x)
%+ ig_{{}_{\rm YM}}^{-1} {g}(x)^\dagger d {g}(x)  ,
%\nonumber\\
  A^{g} =  A^{g}_\mu(x) dx^\mu  =  \langle \Lambda | \mathscr{A}_\mu^{g} (x)    |\Lambda \rangle  dx^\mu ,
\quad 
\mathscr{A}^{g} =  \mathscr{A}^{g}_\mu(x) dx^\mu ,  
%\mathscr{A}^{g}_\mu(x)  :=  {g}(x)^\dagger \mathscr{A}_\mu(x) {g}(x)
%+ ig_{{}_{\rm YM}}^{-1} {g}(x)^\dagger \partial_\mu {g}(x)  ,
%\\
%  \mathscr{A}_\mu^{g}(x) :=&  {g}(x)^\dagger \mathscr{A}_\mu(x) {g}(x) 
%+ ig_{{}_{\rm YM}}^{-1} {g}(x)^\dagger \partial_\mu {g}(x)  .
\label{C29-pre-NAST0}
\end{align}
 we arrive at a path-integral representation of the Wilson loop operator:
%[Exercise-2] \marginpar{Ex-2}
%Verify that $\mathscr{A}^{g}$ defined by (\ref{C29-def-Ag0}) is written in the form (\ref{C29-Ag2}). 
\begin{align}
 W_C[\mathscr{A}] 
 =&  \int [d\mu({g})]_C \exp \left( 
-ig_{{}_{\rm YM}}  \oint_C  A^g \right) , 
%\quad 
%[d\mu({g})]_C =  \prod_{x \in C}  d\mu({g}(x)) 
%:= \lim_{N \rightarrow \infty, \epsilon \rightarrow 0} \prod_{n=0}^{N-1}  d\mu({g}_n) ,
%\nonumber\\
% A^g  :=& \langle \Lambda | \mathscr{A}^{g}      |\Lambda \rangle ,
%\nonumber\\
% \mathscr{A}^{g}(x) :=&  \mathscr{A}^{g}_\mu(x) dx^\mu , \ \mathscr{A}^{g}_\mu(x)  :=  {g}(x)^\dagger \mathscr{A}_\mu(x) {g}(x)
%+ ig_{{}_{\rm YM}}^{-1} {g}(x)^\dagger \partial_\mu {g}(x)  ,
\label{C29-Ag2}
\end{align}
where 
$[d\mu({g})]_C$ is the product of the integration measure $d\mu({g}_n)$ at each partition point along the loop $C$:
\begin{align}
  [d\mu({g})]_C =  \prod_{x \in C}  d\mu({g}(x)) := 
\lim_{N \rightarrow \infty, \epsilon \rightarrow 0}
\prod_{n=0}^{N-1}  d\mu({g}_n) ,
\end{align}
and $d$ denotes the exterior derivative:
\begin{equation}
 d = ds \frac{d}{ds} = ds \frac{dx^\mu}{ds} \frac{\partial}{\partial x^\mu} =    dx^\mu \frac{\partial}{\partial x^\mu} = dx^\mu \partial_\mu  .
\end{equation}
%Finally, we arrive at a path-integral representation of the Wilson loop operator:
%\begin{align}
% W_C[\mathscr{A}] 
% =&   \int [d\mu({g})]_C \exp \left( 
%-ig_{{}_{\rm YM}}  \oint_C  A^{g}    \right) , 
%\nonumber\\&  
%\quad 
%[d\mu({g})]_C :=  \prod_{x \in C}  d\mu({g}(x)) , 
%\end{align}
%where  we have introduced the one-form $A^{g}$ by
%\begin{align}
%  A^{g}(x)  =&  A^{g}_\mu(x) dx^\mu :=  \langle \Lambda | \mathscr{A}_\mu^{g} (x)    |\Lambda \rangle  dx^\mu .
%\\
%  \mathscr{A}_\mu^{g}(x) :=&  {g}(x)^\dagger \mathscr{A}_\mu(x) {g}(x) 
%+ ig_{{}_{\rm YM}}^{-1} {g}(x)^\dagger \partial_\mu {g}(x)  .
%\label{C29-pre-NAST0}
%\end{align}
It should be remarked that the path-ordering has disappeared in the resulting path-integral representation. 
We call this result the pre-non-Abelian Stokes theorem (pre-NAST).

\begin{figure}[tbp]
\begin{center}
\includegraphics[height=3.5cm]{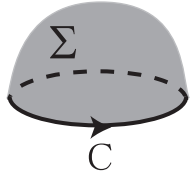}\end{center} 
\vskip -0.5cm
\caption[]{
A closed loop $C$ for defining the Wilson loop operator and the surface $\Sigma$ whose boundary is given by the loop $C$. 
}
\label{C29-fig:W_loop-Sigma}
\end{figure}

Now the argument of the exponential is an Abelian quantity, since $A^{g}_\mu$ is no longer a matrix, just a number. 
Therefore, we can apply  the (usual) \textbf{Stokes theorem}:
\begin{equation}
 \oint_{C=\partial S} \omega = \int_S d \omega   ,
\end{equation}
to replace the line integral   to the surface integral.
Thus we obtain a  \textbf{non-Abelian Stokes theorem} (NAST)
(See Fig.~\ref{C29-fig:W_loop-Sigma}):
\begin{equation}
 W_C[\mathscr{A}]  =\int [d\mu(g)]_{\Sigma}
\exp \left[ -ig_{{}_{\rm YM}} \int_{\Sigma: \partial \Sigma=C} F^g  \right] , %\quad
%[d\mu(g)]_{\Sigma} :=\prod_{x \in \Sigma: \partial \Sigma=C}   d\mu(g(x))  ,
\label{C29-NAST1}
\end{equation}
where  the $F$ is the two-form  defined by  
\begin{equation}
 F^g  =dA^g  = \frac12 F^g_{\mu\nu}(x) dx^\mu \wedge dx^\nu \end{equation}
Here we have replaced the integration measure on the loop $C$ by the integration measure on the surface $\Sigma$:
\begin{equation}
 [d\mu(g)]_{\Sigma} :=\prod_{x \in \Sigma: \partial \Sigma=C}   d\mu(g(x))  ,
\end{equation}
by inserting additional integral measures,  $1=\int d\mu(g(x))$ for $x \in \Sigma - C$.
The explicit expression for $F^g$ will be obtained in the next section.%
\footnote{
This version of the non-Abelian Stokes theorem was derived for the first time for the $SU(2)$ Wilson loop operator based on a different method in \cite{DP89}.
%[Diakonov and Petrov, Phys.~Lett. B224, 131 (1989)].
It is also derived based on the $SU(2)$ coherent state in \cite{KondoIV}.
%[Kondo, hep-th/9805153, Phys.~Rev. D58, 105016 (1998)]. 
In the similar way, it has been extended into the gauge group $SU(3)$ in \cite{KT00b},
%[Kondo \& Taira, hep-th/9906129, Mod.~Phys.~Lett. A15, 367 (2000)].
and gauge group $SU(N)$ in \cite{KT00} and \cite{Kondo00}.
%[Kondo \& Taira, hep-th/9911242, Prog.~Theor.~Phys. 104, 1189 (2000)]
%and 
%[Kondo, hep-th/0009152]. 
The presentation for $SU(N)$  of this chapter is based on the paper  \cite{Kondo08}.
%[Kondo, arXiv:0801.1274, Phys.~Rev.D77, 085029 (2008)]. 
The Diakonov-Petrov version of the non-Abelian Stokes theorem can be derived in a unified way using the coherent state, as given in this chapter. 
There exist other versions of the non-Abelian Stokes theorem, see e.g., \cite{HU99,Halpern79,Bralic80,Arefeva80,Simonov89,Lunev97,HM97}. 
}

%%%%%%%%%%%%%%%%%%%%% figures %%%%%%%%%%%%%%%%%%%%%%%%%%%
%%%%%%%%%%%%%%%%%%%%% figures %%%%%%%%%%%%%%%%%%%%%%%%%%%

%%%%%%%%%%%%%%%%%%%%%%%%%%%%%%%%%%%%%%%%%%%%%%%%%%
\subsection{Restricted variable  in the Wilson  loop operator}
%%%%%%%%%%%%%%%%%%%%%%%%%%%%%%%%%%%%%%%%%%%%%%%%%%

We proceed to obtain the explicit expression for $F^g$ in the non-Abelian Stokes theorem for the Wilson loop operator.
 
%For the \textbf{highest-weight state} $\left|   {\Lambda} \right> $ of a representation $R$ of a group $G$, we define a matrix $\rho$ with the matrix element $\rho_{ab}$ by
%\begin{equation}
% \rho :=  \left|  \Lambda \right> \left< \Lambda \right| 
%, \quad 
% \rho_{ab} :=  \left|  \Lambda \right>_a \left< \Lambda \right|_b = \lambda_a \lambda_b^* .
%\end{equation}
%Then the trace of $\rho$ has a unity:
%\begin{equation}
% {\rm tr}(\rho) = \rho_{aa} =  \left|  \Lambda \right>_a \left< \Lambda \right|_a = \lambda_a \lambda_a^* = 1  ,
% \label{C29-form11}
%\end{equation}
%since  the highest-weight state is normalized
%$\left<   \Lambda |   \Lambda \right>=1$. 
%Moreover, the matrix element $\left<  \Lambda \right|   \mathscr{O}  \left|  \Lambda \right>$ of an arbitrary matrix $\mathcal{O}$ is written in the trace form:
%\begin{equation}
%  \left<  \Lambda \right|   \mathscr{O}  \left|  \Lambda \right> 
% =  {\rm tr}(\rho  \mathscr{O}  ) ,
% \label{C29-form3}
%\end{equation}
%since 
%\begin{align}
%  \left<  \Lambda \right|   \mathscr{O}  \left|  \Lambda \right>
%  = \lambda_b^*  \mathscr{O}_{ba} \lambda_a 
%  = \rho_{ab}  \mathscr{O}_{ba}   
% = {\rm tr}(\rho  \mathscr{O}  ) .
%\end{align}
%For more details on the coherent state, see the chapter of Maximal stability group and coherent state. 

By using the operator 
$\rho :=  \left|  \Lambda \right> \left< \Lambda \right|$ introduced in section 6.2, the ``Abelian'' field $A^g$ in (\ref{C29-pre-NAST0}) is written in the trace form of a matrix: 
\begin{align}
   A^g_\mu(x) :=& 
 \langle \Lambda | \mathscr{A}_\mu^g (x)    |\Lambda \rangle  
 \nonumber\\
=& {\rm tr}\{\rho \mathscr{A}^g_\mu(x)   \} 
%\nonumber\\
=  {\rm tr}\{ g(x) \rho  g^\dagger(x)  \mathscr{A}_\mu(x) \} + ig_{{}_{\rm YM}}^{-1} {\rm tr}\{ \rho  g^\dagger(x)  \partial_\mu   g(x)  \} 
 .
 \label{C29-def-Ag}
\end{align}
By introducing the traceless field $\tilde{\bm{n}}(x)$ defined by [which we call the \textbf{color (direction) field} after the normalization]
%and the normalized and traceless field $\bm{n}(x)$
\begin{equation}
 \tilde{\bm{n}}(x) :=  g(x) \left[ \rho - \frac{\bm{1}}{{\rm tr}(\bm{1})} \right]  g^\dagger(x) 
= g(x)  \rho g^\dagger(x) - \frac{\bm{1}}{{\rm tr}(\bm{1})}    ,
%\quad
% \bm{n}(x) := - \sqrt{\frac{N}{2(N-1)}} \tilde{\bm{n}}(x)
  \label{C29-tilde-n}
\end{equation}
the ``Abelian'' field $A^g$   is rewritten as
\begin{align}
   A^g_\mu(x) =& {\rm tr}\{ \tilde{\bm{n}}(x)  \mathscr{A}_\mu(x) \} + ig_{{}_{\rm YM}}^{-1} {\rm tr}\{ \rho  g^\dagger(x)  \partial_\mu   g(x)  \} 
 ,
 \label{C29-def-Ag2}
\end{align}
where we have used
$
g(x)  \rho g^\dagger(x) = \tilde{\bm{n}}(x) + \frac{\bm{1}}{{\rm tr}(\bm{1})} 
$
and $\mathscr{A}_\mu$ is traceless: 
 ${\rm tr}(\mathscr{A}_\mu)=0$, i.e., $\mathscr{A}_\mu \in su(N)$.
 We can introduce also the normalized and traceless field $\bm{n}(x)$ defined by
\begin{equation}
 \bm{n}(x) :=   \sqrt{\frac{N}{2(N-1)}} \tilde{\bm{n}}(x)
 = \sqrt{\frac{N}{2(N-1)}} g(x) \left[ \rho - \frac{\bm{1}}{{\rm tr}(\bm{1})} \right]  g^\dagger(x)  
 ,
 \label{C29-n-def}
\end{equation}
which we call the \textbf{color (direction) field}.

We proceed to decompose the original gauge field $\mathscr{A}_\mu(x)$  into two pieces $\mathscr{V}_\mu(x)$ and $\mathscr{X}_\mu(x)$: 
%Here we consider the decomposition of the original gauge field $\mathscr{A}_\mu(x)$  into two pieces $\mathscr{V}_\mu(x)$ and $\mathscr{X}_\mu(x)$:
\begin{equation}
 \mathscr{A}_\mu(x)
=\mathscr{V}_\mu(x)+\mathscr{X}_\mu(x)
%=\mathscr{C}_\mu(x)+\mathscr{B}_\mu(x)+\mathscr{X}_\mu(x) 
 .
\end{equation}
 with different characters from the viewpoint of the contribution to the Wilson loop operator.
We simply require that $\mathscr{X}_\mu(x)$ satisfies the condition:[\textbf{defining equation}]
%is perpendicular to $\bm{n}(x)$ in the sense that
\begin{align}
\text{(ii)}  \quad  \bm{n}(x) \cdot \mathscr{X}_\mu(x) = 2{\rm tr}\{ \bm{n}(x)  \mathscr{X}_\mu(x) \} = 0.
 \label{defeq-X-1}
\end{align}
Then $\mathscr{X}_\mu(x)$ disappears from the Wilson loop operator, since $A^g_\mu(x)$ is written without $\mathscr{X}_\mu(x)$:
\begin{align}
   A^g_\mu(x) =& {\rm tr}\{ \tilde{\bm{n}}(x)  \mathscr{V}_\mu(x) \} + ig_{{}_{\rm YM}}^{-1} {\rm tr}\{ \rho  g^\dagger(x)  \partial_\mu   g(x)  \} 
 .
%\label{C29-def-Ag2}
\end{align}
Consequently, the Wilson loop operator $W_C[\mathscr{A}]$ defined in terms of the original Yang-Mills field $\mathscr{A}_\mu(x)$ can be reproduced by the \textbf{restricted field} variable   $\mathscr{V}_\mu(x)$ alone and the remaining field variable  $\mathscr{X}_\mu(x)$ is redundant for the Wilson loop operator: 
For arbitrary loop $C$ and any representation $R$, the Wilson loop operators satisfies
\begin{equation}
\text{(a)}  \quad  W_C[\mathscr{A}]= W_C[\mathscr{V}]
 ,
 \label{C29-W-dominant0}
\end{equation} 
This is the \textbf{restricted field dominance} in the Wilson loop operator.
This does not necessarily imply the restricted field dominance for the Wilson loop average: 
\begin{equation}
\langle W_C[\mathscr{A}] \rangle_{\rm YM}
= \langle W_C[\mathscr{V}] \rangle_{\rm YM} ,
\end{equation} 
which holds only when the cross term between $\mathscr{V}$ and $\mathscr{X}$ in the action can be neglected. 

%This does not necessarily imply
%$
%\langle W_C[\mathscr{A}] \rangle_{\rm YM}
%= \langle W_C[\mathscr{V}] \rangle_{\rm YM} 
%$,
%which holds only when the cross term between $\mathscr{V}$ and $\mathscr{X}$ are neglected. 

We look for the \textbf{gauge covariant decomposition}, which  means that the decomposition holds after the gauge transformation:
\begin{equation}
 \mathscr{A}_\mu^\prime(x)
=\mathscr{V}_\mu^\prime(x)+\mathscr{X}_\mu^\prime(x)
%=\mathscr{C}_\mu(x)+\mathscr{B}_\mu(x)+\mathscr{X}_\mu(x) 
 .
\end{equation}
For the condition (ii) [eq.(\ref{defeq-X-1})] to be gauge covariant (or gauge invariant), 
the transformation of the color field $\bm{n}$ given by  (\ref{C29-n-def})   
\begin{equation}
 g(x) \to U(x) g(x) \Longrightarrow  \bm{n}(x) \to \bm{n}^\prime(x) = U(x)  \bm{n}(x) U^{\dagger}(x) .
\end{equation}
requires that $\mathscr{X}_\mu(x)$ transforms as an adjoint (matter) field:
\begin{align}
  \mathscr{X}_\mu(x) & \rightarrow \mathscr{X}_\mu^\prime(x) = U(x)  \mathscr{X}_\mu(x) U^{\dagger}(x) .
 \label{C29-X-ctransf}
\end{align}
This immediately means  that $\mathscr{V}_\mu(x)$ must transform under the gauge transformation just like the original gauge field $\mathscr{A}_\mu(x)$:
\begin{align}
  \mathscr{V}_\mu(x) & \rightarrow \mathscr{V}_\mu^\prime(x) = U(x)  \mathscr{V}_\mu(x) U^{\dagger}(x) + ig_{{}_{\rm YM}}^{-1} U(x)  \partial_\mu U^{\dagger}(x) 
 ,
 \label{C29-V-ctransf}
%  \\
%  \mathscr{X}_\mu(x) & \rightarrow \mathscr{X}_\mu^\prime(x) = U(x)  \mathscr{X}_\mu(x) U^{\dagger}(x) .
% \label{C29-X-ctransf}
\end{align}
since the original gauge field $\mathscr{A}_\mu(x)$ must transform  in the usual way:
\begin{align}
  \mathscr{A}_\mu(x) & \rightarrow \mathscr{A}_\mu^\prime(x) = U(x)  \mathscr{A}_\mu(x) U^{\dagger}(x) + ig_{{}_{\rm YM}}^{-1} U(x)  \partial_\mu U^{\dagger}(x) 
 .
 \label{C29-A-ctransf}
\end{align}

These transformation properties impose restrictions on the requirement to be imposed on the restricted field $\mathscr{V}_\mu(x)$. 
Such a candidate is [covariant constantness of the color field] which we call the [first \textbf{defining equation}]:
\begin{align}
 \text{(I)}  \quad  \mathscr{D}_\mu[\mathscr{V}] \bm{n} = 0 \quad
(\mathscr{D}_\mu[\mathscr{V}]  :=   \partial_\mu   -ig_{{}_{\rm YM}} [ \mathscr{V}_\mu,  \cdot ]  ) ,
 \label{defeq-V-1}
\end{align}
since the covariant derivative transforms in the adjoint way:
$\mathscr{D}_\mu[\mathscr{V}(x)] \to U(x)(\mathscr{D}_\mu[\mathscr{V}](x))U^\dagger(x)$. 

For $G=SU(2)$, it is indeed shown in section 3 that the two conditions (\ref{defeq-X-1}) and (\ref{defeq-V-1}) which we call the defining equations for the decomposition determine the decomposition uniquely. 
By solving the defining equations, we obtain
\begin{align}
  \mathscr{A}_\mu(x)= \mathscr{V}_\mu(x) +& \mathscr{X}_\mu(x),
  \nonumber
\\
   \mathscr{V}_\mu(x) =& c_\mu(x)\bm{n}(x) -ig_{{}_{\rm YM}}^{-1} [\partial_\mu \bm{n}(x) , \bm{n}(x) ] , 
%(\leftarrow \text{Cho connection})
\quad
   c_\mu(x) =   \mathscr{A}_\mu(x)  \cdot \bm{n}(x) ,
  \nonumber
\\
  \mathscr{X}_\mu(x) =&  -ig_{{}_{\rm YM}}^{-1} [ \bm{n}(x) ,  \mathscr{D}_\mu[\mathscr{A}] \bm{n}(x) ] .
%\label{C26-CDG-decomp1}
\end{align}
This is the same as the \textbf{Cho--Duan-Ge--Faddeev-Niemi (CDGFN) decomposition}.

For $G=SU(N)$ ($N \ge 3$), the conditions (I) and (ii) are not sufficient to uniquely determine the decomposition.  The condition (ii) [eq.(\ref{defeq-X-1})] must be modified:
\\  
(II)  $\mathscr{X}^\mu(x)$  does not have the $\tilde{H}$-commutative part, i.e., $\mathscr{X}^\mu(x)_{\tilde{H}}=0$:
\begin{align}
\text{(II)}  \quad   & 0 =  \mathscr{X}^\mu(x)_{\tilde{H}} := \mathscr{X}^\mu(x)  -   2\frac{N-1}{N}  [\bm{n}(x) , [\bm{n}(x) , \mathscr{X}^\mu(x) ]] 
\nonumber\\
& \Longleftrightarrow \mathscr{X}^\mu(x)  = 2\frac{N-1}{N}  [\bm{n}(x) , [\bm{n}(x) , \mathscr{X}^\mu(x)  ]]
\label{C29-defXL2b}
 . 
\end{align}
This condition is also gauge covariant. 
Note that the condition (ii)[eq.(\ref{defeq-X-1})] follows from (II)[eq.(\ref{C29-defXL2b})]. 
See section 5 for details. 
For $G=SU(2)$, i.e., $N=2$, the condition (II)[eq.(\ref{C29-defXL2b})] reduces to (ii)[eq.(\ref{defeq-X-1})].
By solving (I)[eq.(\ref{defeq-V-1})] and (II)[eq.(\ref{C29-defXL2b})], as shown in section 5, 
  $\mathscr{X}_\mu(x)$ is determined as
\begin{align}
 \mathscr{X}_\mu(x) & 
= -ig_{{}_{\rm YM}}^{-1}  \frac{2(N-1)}{N}  [\bm{n}(x), \mathscr{D}_\mu[\mathscr{A}]\bm{n}(x) ]
 \in Lie(G/\tilde H)
%= g_{{}_{\rm YM}}^{-1}N  (\bm{n}_j \times \mathscr{D}_\mu[\mathscr{A}]\bm{n}_j)   .
 .
\label{C29-X-def1}
\end{align}

%In order to connect the Wilson loop operator with the new reformulation, we identify $\bm{n}(x)$ with the color field $\bm{n}(x)$ to obtain the new expression of the Wilson loop operator written in terms of new variables used in the new reformulation.  

%Suppose that a given {single} color field $\bm{n}$    transforms as 
%\begin{equation}
%\bm{n}(x)   \rightarrow \bm{n}^\prime(x) = U(x)  \bm{n}(x) U^{\dagger}(x)  ,
%\end{equation}  
%and  that 
%the original gauge field $\mathscr{A}_\mu(x)$ is decomposed into two pieces $\mathscr{V}_\mu(x)$ and $\mathscr{X}_\mu(x)$:
%\begin{equation}
% \mathscr{A}_\mu(x) =\mathscr{V}_\mu(x)+\mathscr{X}_\mu(x)
%=\mathscr{C}_\mu(x)+\mathscr{B}_\mu(x)+\mathscr{X}_\mu(x)  ,
%\end{equation}
%such that  
%$\mathscr{V}_\mu(x)$ transforms under the gauge transformation just like the original gauge field $\mathscr{A}_\mu(x)$, while $\mathscr{X}_\mu(x)$ transforms like an adjoint matter field:
%\begin{align}
%  \mathscr{V}_\mu(x) & \rightarrow \mathscr{V}_\mu^\prime(x) = U(x)  \mathscr{V}_\mu(x) U^{\dagger}(x) + ig_{{}_{\rm YM}}^{-1} U(x)  \partial_\mu U^{\dagger}(x)  ,
%\label{C29-V-ctransf}
%  \\
%  \mathscr{X}_\mu(x) & \rightarrow \mathscr{X}_\mu^\prime(x) = U(x)  \mathscr{X}_\mu(x) U^{\dagger}(x) 
%\label{C29-X-ctransf}
%\end{align}

%We require that $\mathscr{X}_\mu(x)$ is perpendicular to $\bm{n}(x)$ in the sense that
%\begin{align}
% {\rm tr}\{ \bm{n}(x)  \mathscr{X}_\mu(x) \} = 0.
%\end{align}
Moreover, the field $\mathscr{V}_\mu(x)$ is  decomposed as
\begin{equation}
 \mathscr{V}_\mu(x)
=\mathscr{C}_\mu(x)+\mathscr{B}_\mu(x)
 ,
\end{equation}
such that $\mathscr{B}_\mu(x)$ is $\tilde{H}$-noncommutative and $\mathscr{C}_\mu(x)$ is $\tilde{H}$-commutative in the sense that
%such that $\mathscr{B}_\mu(x)$ is perpendicular to $\bm{n}(x)$ and $\mathscr{C}_\mu(x)$ is parallel to $\bm{n}(x)$ in the sense that
\begin{align}
 {\rm tr}\{ \bm{n}(x) \mathscr{B}_\mu(x) \} =  0, 
\quad
 [\bm{n}(x) , \mathscr{C}_\mu(x) ]=0 .
%{\rm tr}\{ \bm{n}(x)  \mathscr{A}_\mu(x) \}
%=& {\rm tr}\{ \bm{n}(x)  \mathscr{V}_\mu(x) \} 
%=  {\rm tr}\{ \bm{n}(x)  \mathscr{C}_\mu(x) \}.
\end{align}
The decomposed fields $\mathscr{C}_\mu(x)$ and $\mathscr{B}_\mu(x)$ are explicitly written in terms of $\mathscr{A}_\mu(x)$ and $\bm{n}(x)$:%
%\footnote{ 
%For the details, see the chapter of Change of variables for  $SU(N)$ Yang-Mills theory.
%They are regarded as those obtained by the (non-linear) change of variables from the original gauge field. 
%For the $SU(2)$ case, this is well-known as the Cho-Duan-Ge-Faddeev-Niemi-Shabanov (CDGFNS) decomposition.
% \cite{Cho80,DG79,FN98,Shabanov99,Cho00}.
%The extension to the  $SU(N)$ case was done. 
%See the chapter of .
%See \cite{KSM08}.
%}
\begin{subequations}
\begin{align}
%\mathscr A_\mu(x)
% =& \mathscr V_\mu(x)  +\mathscr X_\mu(x)
%=\mathscr C_\mu(x)  +\mathscr B_\mu(x)  +\mathscr X_\mu(x)
%\quad (\mu=0, 1, \cdots, D-1)  
% ,
%\\
  \mathscr{C}_\mu(x) &
%= \mathscr{V}_\mu - \mathscr{B}_\mu  
%= \mathscr{A}_\mu  - \mathscr{X}_\mu  - \mathscr{B}_\mu 
=  \mathscr{A}_\mu(x) - \frac{2(N-1)}{N}   [\bm{n}(x), [ \bm{n}(x), \mathscr{A}_\mu(x) ] ]
\in Lie(\tilde H)
,
\label{C29-C-def1}
\\
 \mathscr{B}_\mu(x) &
= ig_{{}_{\rm YM}}^{-1} \frac{2(N-1)}{N}[\bm{n}(x) , \partial_\mu  \bm{n}(x) ]
\in Lie(G/\tilde H)
 .
\label{C29-B-def1}
%\\
% \mathscr{X}_\mu(x) & 
%= -ig_{{}_{\rm YM}}^{-1}  \frac{2(N-1)}{N}  [\bm{n}(x), \mathscr{D}_\mu[\mathscr{A}]\bm{n}(x) ] \in Lie(G/\tilde H)
%= g_{{}_{\rm YM}}^{-1}N  (\bm{n}_j \times \mathscr{D}_\mu[\mathscr{A}]\bm{n}_j)   .
% .
%\label{C29-X-def1}
\end{align}
 \label{C29-NLCV-minimal}
\end{subequations}

The field strength $F^g$ in (\ref{C29-NAST1}) is calculated to be 
%[Exercise-3] \marginpar{Ex-3}
\begin{align}
 F^g_{\mu\nu}(x) :=& \partial_\mu A^g_\nu(x) - \partial_\nu A^g_\mu(x)   
\nonumber\\ 
%=&  \partial_\mu {\rm tr}\{ \tilde{\bm{n}}(x) \mathscr{A}_\nu(x) \}
%-  \partial_\nu {\rm tr}\{ \tilde{\bm{n}}(x)   \mathscr{A}_\mu(x) \}
%\nonumber\\&
%   +  ig_{{}_{\rm YM}}^{-1} {\rm tr}\{ \rho  [ \partial_\mu g^\dagger(x)  \partial_\nu  g(x) - \partial_\nu g^\dagger(x)  \partial_\mu   g(x) ] \} 
%\nonumber\\&
% + ig_{{}_{\rm YM}}^{-1} {\rm tr}\{ \rho  g^\dagger(x) [\partial_\mu , \partial_\nu] g(x)  \} 
%\nonumber\\ 
% =&  \partial_\mu {\rm tr}\{ \tilde{\bm{n}}(x) \mathscr{A}_\nu(x) \}
%-  \partial_\nu {\rm tr}\{ \tilde{\bm{n}}(x)   \mathscr{A}_\mu(x) \}
%\nonumber\\&
%   +  ig_{{}_{\rm YM}}^{-1} {\rm tr}\{ g(x) \rho  g^\dagger(x)  [ g(x) \partial_\mu g^\dagger(x)  \partial_\nu   g(x) g^\dagger(x)  
%   -     g(x) \partial_\nu g^\dagger(x)  \partial_\mu   g(x) g^\dagger(x) ] \} 
%\nonumber\\&
% + ig_{{}_{\rm YM}}^{-1} {\rm tr}\{ \rho  g^\dagger(x) [\partial_\mu , \partial_\nu] g(x)  \} 
%\nonumber\\ 
 =&  \partial_\mu {\rm tr}\{ \tilde{\bm{n}}(x) \mathscr{A}_\nu(x) \}
-  \partial_\nu {\rm tr}\{ \tilde{\bm{n}}(x)   \mathscr{A}_\mu(x) \}
\nonumber\\&
   + ig_{{}_{\rm YM}} {\rm tr}\{ g(x) \rho  g^\dagger(x) [ig_{{}_{\rm YM}}^{-1} g(x)  \partial_\mu g^\dagger(x), ig_{{}_{\rm YM}}^{-1}  g(x)  \partial_\nu g^\dagger(x)]   \}
\nonumber\\&
 + ig_{{}_{\rm YM}}^{-1} {\rm tr}\{ \rho  g^\dagger(x) [\partial_\mu , \partial_\nu] g(x)  \} 
 ,
 \label{C29-Fg}
\end{align}
where we have used $g^\dagger(x) g(x) =1$ and $\partial_\mu   g(x) g^\dagger(x)=- g(x) \partial_\mu  g^\dagger(x)$ following from $g(x) g^\dagger(x)=1$. 
Therefore, the field strength $F^g$ is rewritten   as
\begin{align}
 F^g_{\mu\nu}(x) 
%:= \partial_\mu A_\nu - \partial_\nu A_\mu   
=&  \partial_\mu {\rm tr}\{ \tilde{\bm{n}}(x)  \mathscr{A}_\nu(x) \}
-  \partial_\nu {\rm tr}\{ \tilde{\bm{n}}(x)  \mathscr{A}_\mu(x) \}
%\nonumber\\&
   + ig_{{}_{\rm YM}}  {\rm tr}\{ \tilde{\bm{n}}(x) [ \Omega_\mu(x) , \Omega_\nu(x) ]   \}
  \nonumber\\& 
+ ig_{{}_{\rm YM}}^{-1} {\rm tr}\{ \rho  g^\dagger(x) [\partial_\mu , \partial_\nu] g(x)   \} 
 ,
\end{align}
where we have defined
\begin{equation}
 \Omega_\mu(x) :=  ig_{{}_{\rm YM}}^{-1} g(x) \partial_\mu g^\dagger(x) 
  .
\end{equation}
%Here we have used
%$
%g(x)  \rho g^\dagger(x) = \tilde{\bm{n}}(x) + \frac{\bm{1}}{{\rm tr}(\bm{1})} 
%$
%and $\mathscr{A}_\mu$ is traceless: 
%${\rm tr}(\mathscr{A}_\mu)=0$, i.e., $\mathscr{A}_\mu \in su(N)$.
The field strength $F^g$ is modified using the normalized color field as
\begin{align}
 F^g_{\mu\nu}(x) 
%:= \partial_\mu A_\nu - \partial_\nu A_\mu   
=&  \sqrt{\frac{2(N-1)}{N}}  (    \partial_\mu {\rm tr}\{  {\bm{n}(x)}  \mathscr{A}_\nu(x) \}
-  \partial_\nu {\rm tr}\{ {\bm{n}(x)}  \mathscr{A}_\mu(x) \} 
%\nonumber\\&
   + ig_{{}_{\rm YM}}  {\rm tr}\{  {\bm{n}(x)} [ \Omega_\mu(x) , \Omega_\nu(x) ]   \}  )  
\nonumber\\& 
+ ig_{{}_{\rm YM}}^{-1} {\rm tr}\{ \rho  g^\dagger(x) [\partial_\mu , \partial_\nu] g(x)   \} 
 .
 \label{C29-Fg1}
\end{align}

In what follows, we show that the field strength $F_{\mu\nu}^g$ in (\ref{C29-Fg1}) except for  $ig_{{}_{\rm YM}}^{-1} {\rm tr}\{ \rho  g^\dagger(x) [\partial_\mu , \partial_\nu] g(x)   \}$  is written in terms of the field strength $\mathscr{F}_{\mu\nu} [\mathscr{V}]$ of the restricted field $\mathscr{V}_\mu$  alone:
\begin{align}
  F_{\mu\nu}^g - ig_{{}_{\rm YM}}^{-1} {\rm tr}\{ \rho  g^\dagger(x) [\partial_\mu , \partial_\nu] g(x)   \}
=&    \partial_\mu {\rm tr}\{  {\bm{n}}  \mathscr{A}_\nu \}
-  \partial_\nu {\rm tr}\{ {\bm{n}}  \mathscr{A}_\mu \} 
   + ig_{{}_{\rm YM}}  {\rm tr}\{  {\bm{n}} [ \Omega_\mu , \Omega_\nu ]   \}  
\nonumber\\ 
=&   {\rm tr}\{ \bm{n} \mathscr{F}_{\mu\nu} [\mathscr{V}] \} 
 .
\end{align}
%and
%\begin{align}
% F^g_{\mu\nu}  
% =   \sqrt{\frac{2(N-1)}{N}} 
%  {\rm tr}\{ \bm{n}  \mathscr{F}_{\mu\nu} [\mathscr{V}]  \} 
% + ig_{{}_{\rm YM}}^{-1} {\rm tr}\{ \rho  g^\dagger [\partial_\mu , \partial_\nu] g   \} 
% .
%\end{align}

%First, we find
%\begin{align} 
%{\rm tr}\{ \bm{n}(x)  \mathscr{A}_\mu(x) \}
%= {\rm tr}\{ \bm{n}(x)  \mathscr{C}_\nu(x) \} ,
%\end{align}
%which follows from
%\begin{equation}
% {\rm tr}\{ \bm{n}(x) \mathscr{B}_\mu(x) \} = 0, 
%\quad 
% {\rm tr}\{ \bm{n}(x)  \mathscr{X}_\mu(x) \} = 0.
%\end{equation}

%First, we show that 
%This is shown as follows. 
  As the first step, we show that $\Omega_\mu$ differs from $\mathscr{B}_\mu$ by a $\tilde{H}$-commutative part $a_\mu$: 
\begin{equation}
 \Omega_\mu(x) =    \mathscr{B}_\mu(x) + a_\mu(x), \quad
 [a_\mu(x), \bm{n}(x)] = 0 .
 \label{C29-a-n}
\end{equation}
On the one hand, we find
\begin{align}
\partial_\mu \bm{n} = ig_{{}_{\rm YM}} [ \mathscr{B}_\mu, \bm{n} ] ,
\label{del-n1}
\end{align}
which follows from the defining equation (\ref{defeq-V-1}):  
\begin{align}
  0 = \mathscr{D}_\mu[\mathscr{V}] \bm{n} 
=&  \partial_\mu \bm{n} -ig_{{}_{\rm YM}} [ \mathscr{V}_\mu, \bm{n} ] 
 =   \partial_\mu \bm{n} -ig_{{}_{\rm YM}} [ \mathscr{B}_\mu, \bm{n} ] ,
%\nonumber\\&
%\Longrightarrow 
%\partial_\mu \bm{n} = ig_{{}_{\rm YM}} [ \mathscr{B}_\mu, \bm{n} ] .
\end{align}
by taking into account 
\begin{equation}
[ \mathscr{C}_\mu, \bm{n} ]=0 .
\end{equation}
On the other hand, we find  
\begin{align}
\partial_\mu \bm{n} = ig_{{}_{\rm YM}} [ \Omega_\mu , \bm{n} ] ,
\label{del-n2}
\end{align}
which follows from the relation for the derivative of $\tilde{\bm{n}}$ defined by (\ref{C29-tilde-n}):
\begin{align}
   \partial_\mu \tilde{\bm{n}} 
 =  \partial_\mu (g \rho g^\dagger) 
& = \partial_\mu  g g^\dagger g \rho g^\dagger  +   g  \rho g^\dagger g \partial_\mu g^\dagger 
\quad (    g^\dagger g =1)
\nonumber\\ &
= -   g \partial_\mu g^\dagger g \rho g^\dagger  +   g  \rho g^\dagger g \partial_\mu g^\dagger 
\quad (  \partial_\mu  g g^\dagger =-g \partial_\mu g^\dagger)
\nonumber\\ &
=  -[ g \partial_\mu g^\dagger  , g \rho g^\dagger ] 
%\nonumber\\ &
%= -[ g \partial_\mu g^\dagger  , \tilde{\bm{n}} ] 
\nonumber\\ &
=  ig_{{}_{\rm YM}}[ \Omega_\mu ,\tilde{\bm{n}} ] ,
%\nonumber\\&
%\quad \Longrightarrow \quad 
%\partial_\mu \bm{n} = ig_{{}_{\rm YM}} [ \Omega_\mu , \bm{n} ] ,
\end{align}
where we have used $g^\dagger g=\bm{1}=g g^\dagger$ in the second equality and $\partial_\mu  g g^\dagger =-g \partial_\mu g^\dagger$ following from $\partial_\mu(g g^\dagger )=0$ in the third equality.

Combining (\ref{del-n1}) and (\ref{del-n2}), we have
\begin{equation}
[ \Omega_\mu , \bm{n} ]=[ \mathscr{B}_\mu, \bm{n} ]  \Longrightarrow
    [\Omega_\mu -  \mathscr{B}_\mu, \bm{n} ] = 0,
\end{equation}
which means that $\Omega_\mu -  \mathscr{B}_\mu$ is  $\tilde{H}$-commutative, as claimed above (\ref{C29-a-n}). 

 As the second step, we show that this difference between $\Omega_\mu$ and $\mathscr{B}_\mu$ does not show up in the   quantity ${\rm tr}\{ \bm{n} [ \Omega_\mu , \Omega_\nu ] \}$:
\begin{align} 
{\rm tr}\{ \bm{n}(x) [ \Omega_\mu(x), \Omega_\nu(x) ]   \}  
= {\rm tr}\{ \bm{n}(x) [ \mathscr{B}_\mu(x) , \mathscr{B}_\nu(x) ]   \} .
\end{align}
This is shown using
$\Omega = \mathscr{B} + a$ repeatedly, as follows. 
\begin{align}
     {\rm tr}\{ \bm{n} [ \Omega_\mu , \Omega_\nu ] \}
=&  {\rm tr}\{ \bm{n} [ \mathscr{B}_\mu, \Omega_\nu ] \} +  {\rm tr}\{ \bm{n} [ a_\mu , \Omega_\nu ] \}
\nonumber\\ 
=&  {\rm tr}\{ \bm{n} [ \mathscr{B}_\mu, \mathscr{B}_\nu ] \} 
+ {\rm tr}\{ \bm{n} [ \mathscr{B}_\mu, a_\nu ] \}  
+  {\rm tr}\{ \bm{n} [ a_\mu , \Omega_\nu ] \}
\nonumber\\ 
=& {\rm tr}\{ \bm{n} [ \mathscr{B}_\mu , \mathscr{B}_\nu ]   \}
+ {\rm tr}\{   \mathscr{B}_\mu  [ a_\nu , \bm{n} ] \}  
+  {\rm tr}\{ [ \bm{n} , a_\mu ] \Omega_\nu  \}
\nonumber\\ 
=& {\rm tr}\{ \bm{n} [ \mathscr{B}_\mu , \mathscr{B}_\nu ]   \}
 ,
\end{align}
where we have used in the third equality the relation ${\rm tr}\{ A [ B, C ]\}={\rm tr}\{ B [ C , A ]\}={\rm tr}\{ C [ A , B ]\}$ and ${\rm tr}\{ A [ B, C ]\}={\rm tr}\{ [A , B]  C  \}$ due to the cyclicity of the trace, and $[ \bm{n} , a_\mu ]=0$ in the last equality.

Therefore, we have
\begin{align}
 F_{\mu\nu}^g -  ig_{{}_{\rm YM}}^{-1} {\rm tr}\{ \rho  g^\dagger [\partial_\mu , \partial_\nu] g \} 
=  \partial_\mu {\rm tr}\{ \tilde{\bm{n}} \mathscr{C}_\nu \}
-  \partial_\nu {\rm tr}\{ \tilde{\bm{n}} \mathscr{C}_\mu \}
   + ig_{{}_{\rm YM}} {\rm tr}\{ \tilde{\bm{n}}[ \mathscr{B}_\mu , \mathscr{B}_\nu ]   \}
 .
\end{align}

  As the third step, we show that ${\rm tr}\{ \bm{n} \mathscr{F}_{\mu\nu} [\mathscr{V}]\}$ is decomposed as
\begin{align}
 {\rm tr}\{ \bm{n} \mathscr{F}_{\mu\nu} [\mathscr{V}]\} 
 =   \partial_\mu {\rm tr}\{ \bm{n}  \mathscr{C}_\nu \}
-  \partial_\nu {\rm tr}\{ \bm{n}  \mathscr{C}_\mu \}
   + {\rm tr}\{ \bm{n} \mathscr{F}_{\mu\nu} [\mathscr{B}]\}
 .
 \label{C29-nF1}
\end{align}
In fact, the field strength $\mathscr{F}_{\mu\nu}[\mathscr{V}]$ is written in terms of $\mathscr{B}$ and $\mathscr{C}$  as 
\begin{align}
  \mathscr{F}_{\mu\nu}[\mathscr{V}] 
 =& \partial_\mu \mathscr{V}_\nu - \partial_\nu \mathscr{V}_\mu - ig_{{}_{\rm YM}} [\mathscr{V}_\mu , \mathscr{V}_\nu ]
\nonumber\\
=& \mathscr{F}_{\mu\nu}[\mathscr{B}]
+ \partial_\mu  \mathscr{C}_\nu  - \partial_\nu  \mathscr{C}_\mu 
- ig_{{}_{\rm YM}} [ \mathscr{B}_\mu ,   \mathscr{C}_\nu ]
- ig_{{}_{\rm YM}} [  \mathscr{C}_\mu ,  \mathscr{B}_\nu  ]
%\nonumber\\&
 - ig_{{}_{\rm YM}} [  \mathscr{C}_\mu ,   \mathscr{C}_\nu ]
\nonumber\\
=& \mathscr{F}_{\mu\nu}[\mathscr{B}]
+ \mathscr{D}_\mu[\mathscr{B}]  \mathscr{C}_\nu  - \mathscr{D}_\nu[\mathscr{B}]  \mathscr{C}_\mu  - ig_{{}_{\rm YM}} [  \mathscr{C}_\mu ,   \mathscr{C}_\nu ]
 ,
\end{align}
and hence% 
\footnote{This was shown as eq.(E9) in Appendix~E of  \cite{Kondo08}.
  Typos in eq.(E9) are corrected here.
} 
\begin{align}
  {\rm tr}( \bm{n}\mathscr{F}_{\mu\nu}[\mathscr{V}] )
=& {\rm tr}( \bm{n}\mathscr{F}_{\mu\nu}[\mathscr{B}] )
+ {\rm tr}( \bm{n} \mathscr{D}_\mu[\mathscr{B}]  \mathscr{C}_\nu)
- {\rm tr}( \bm{n} \mathscr{D}_\nu[\mathscr{B}]  \mathscr{C}_\mu)
%\nonumber\\& 
- ig_{{}_{\rm YM}} {\rm tr}( \bm{n} [  \mathscr{C}_\mu ,   \mathscr{C}_\nu ])
\nonumber\\
=& {\rm tr}( \bm{n}\mathscr{F}_{\mu\nu}[\mathscr{B}] )
- {\rm tr}( (\mathscr{D}_\mu[\mathscr{B}] \bm{n})  \mathscr{C}_\nu)
+ \partial_\mu {\rm tr}(  \bm{n} \mathscr{C}_\nu)
\nonumber\\&
+ {\rm tr}( (\mathscr{D}_\nu[\mathscr{B}] \bm{n})   \mathscr{C}_\mu)
-  \partial_\nu  {\rm tr} ( \bm{n} \mathscr{C}_\mu)
-  ig_{{}_{\rm YM}} {\rm tr}(   [ \bm{n},  \mathscr{C}_\mu ] \mathscr{C}_\nu)
\nonumber\\
=& {\rm tr}( \bm{n}\mathscr{F}_{\mu\nu}[\mathscr{B}] )
+ \partial_\mu {\rm tr}(  \bm{n} \mathscr{C}_\nu)
-  \partial_\nu  {\rm tr} ( \bm{n} \mathscr{C}_\mu)
 ,
\end{align}
where we have used the property of $\mathscr{B}_\mu$ and $\mathscr{C}_\mu$:
$[ \bm{n},  \mathscr{C}_\mu ]=0$ 
and
$\mathscr{D}_\mu[\mathscr{B}] \bm{n}=0$
following from (\ref{del-n1}).
%$
%0=\mathscr{D}_\mu[\mathscr{V}] \bm{n}
%= \partial_\mu \bm{n} - ig_{{}_{\rm YM}} [ \mathscr{V}_\mu, \bm{n} ] 
%= \partial_\mu \bm{n} - ig_{{}_{\rm YM}} [ \mathscr{B}_\mu, \bm{n} ] 
%=\mathscr{D}_\mu[\mathscr{B}] \bm{n}
%$.

As the fourth step, we show  that $ {\rm tr}\{ \bm{n} \mathscr{F}_{\mu\nu} [\mathscr{B}]\}$ is proportional to ${\rm tr}\{ \bm{n} [ \partial_\mu \bm{n} , \partial_\nu \bm{n} ] \}$:
\begin{align}
 {\rm tr}\{ \bm{n} \mathscr{F}_{\mu\nu} [\mathscr{B}]\} 
 =   ig_{{}_{\rm YM}}^{-1} \frac{2(N-1)}{N} {\rm tr}\{ \bm{n} [ \partial_\mu \bm{n} , \partial_\nu \bm{n} ] \} .
\end{align}
In fact,  we find the following relation:%
\footnote{
This is  eq.(E10)  in Appendix~E of \cite{Kondo08}.
}
\begin{subequations}
\begin{align}
 {\rm tr}( \bm{n}\mathscr{F}_{\mu\nu}[\mathscr{B}] )
=& 
 {\rm tr}( \bm{n} \partial_\mu \mathscr{B}_\nu - \bm{n} \partial_\nu \mathscr{B}_\mu
  -i g_{{}_{\rm YM}} \bm{n} [\mathscr{B}_\mu , \mathscr{B}_\nu])
\\
=& 
 {\rm tr}( \bm{n} \partial_\mu \mathscr{B}_\nu - \bm{n} \partial_\nu \mathscr{B}_\mu
  +i g_{{}_{\rm YM}} \mathscr{B}_\nu  [\mathscr{B}_\mu , \bm{n}])
\\
=& 
 {\rm tr}( \bm{n} \partial_\mu \mathscr{B}_\nu - \bm{n} \partial_\nu \mathscr{B}_\mu
  + \mathscr{B}_\nu \partial_\mu \bm{n} )
\\
=& 
 \partial_\mu {\rm tr}(\bm{n} \mathscr{B}_\nu) - {\rm tr}(\bm{n} \partial_\nu \mathscr{B}_\mu )
\\
=& 
  {\rm tr}(\mathscr{B}_\mu  \partial_\nu  \bm{n}) - \partial_\nu {\rm tr}(\bm{n} \mathscr{B}_\mu) 
  \quad ({\rm tr}(\bm{n} \mathscr{B}_\nu)=0)
\\
=& 
  {\rm tr}(\mathscr{B}_\mu  \partial_\nu  \bm{n}) 
  \quad ({\rm tr}(\bm{n} \mathscr{B}_\mu)=0)
 ,
\end{align}
 \label{C29-FB}
\end{subequations}
where we have used 
$
\partial_\mu \bm{n} = ig_{{}_{\rm YM}} [ \mathscr{B}_\mu, \bm{n} ]
$
in the third equality 
and
${\rm tr}(\bm{n} \mathscr{B}_\mu)=0$ twice in the fifth and the last equalities. 
Substituting the explicit form (\ref{C29-B-def1})   
%\begin{align}
% \mathscr{B}_\mu 
% = ig_{{}_{\rm YM}}^{-1} \frac{2(N-1)}{N} [ \bm{n}, \partial_\mu \bm{n}]  ,
%\end{align}
into  (\ref{C29-FB}), we obtain the desired result:%
\footnote{
This is eq.(E13) in Appendix~E of \cite{Kondo08}.%Kondo~(2008).
}\begin{align}
  {\rm tr}( \bm{n}\mathscr{F}_{\mu\nu}[\mathscr{B}] )
&=   ig_{{}_{\rm YM}}^{-1} \frac{2(N-1)}{N} {\rm tr}( [ \bm{n}, \partial_\mu \bm{n}] \partial_\nu \bm{n} )
%\nonumber\\&
= ig_{{}_{\rm YM}}^{-1} \frac{2(N-1)}{N} 
 {\rm tr} ( \bm{n} [\partial_\mu \bm{n}, \partial_\nu \bm{n}])
 .
\end{align}

Therefore, we find that ${\rm tr}\{ \bm{n} \mathscr{F}_{\mu\nu} [\mathscr{V}]\}$ of (\ref{C29-nF1}) is written as 
\begin{align}
 {\rm tr}\{ \bm{n} \mathscr{F}_{\mu\nu} [\mathscr{V}]\} 
  =&   \partial_\mu {\rm tr}\{ \bm{n}  \mathscr{C}_\nu \}
-  \partial_\nu {\rm tr}\{ \bm{n}  \mathscr{C}_\mu \}
   +  ig_{{}_{\rm YM}}^{-1} \frac{2(N-1)}{N}  {\rm tr}\{ \bm{n} [ \partial_\mu \bm{n} , \partial_\nu \bm{n} ] \} 
%   \nonumber\\
% =&   \partial_\mu {\rm tr}\{ \bm{n}  \mathscr{C}_\nu \}
%-  \partial_\nu {\rm tr}\{ \bm{n}  \mathscr{C}_\mu \}
%   + ig_{{}_{\rm YM}} {\rm tr}\{ \bm{n} [ \mathscr{B}_\mu , \mathscr{B}_\nu ]   \}
 .
\end{align}

As the final step, we find that $ {\rm tr}\{ \bm{n} [ \mathscr{B}_\mu , \mathscr{B}_\nu ] \}$ is related to ${\rm tr}\{ \bm{n} [ \partial_\mu \bm{n} , \partial_\nu \bm{n} ] \}$:
\begin{align}
   ig_{{}_{\rm YM}}    {\rm tr}\{ \bm{n} [ \mathscr{B}_\mu , \mathscr{B}_\nu ] \}
  =& - {\rm tr}\{ ig_{{}_{\rm YM}}[ \mathscr{B}_\mu , \bm{n}  ] \mathscr{B}_\nu   \}
\nonumber\\
  =& - {\rm tr}\{ \partial_\mu \bm{n}   \mathscr{B}_\nu  \}
\nonumber\\
  =& - {\rm tr}\{ \partial_\mu \bm{n}  ig_{{}_{\rm YM}}^{-1} \frac{2(N-1)}{N} [\bm{n} , \partial_\nu \bm{n}   ] \}
\nonumber\\
  =& - ig_{{}_{\rm YM}}^{-1} \frac{2(N-1)}{N} {\rm tr}\{ [\bm{n} , \partial_\nu \bm{n}   ] \partial_\mu \bm{n}   \}
\nonumber\\
 =&    ig_{{}_{\rm YM}}^{-1} \frac{2(N-1)}{N} {\rm tr}\{ \bm{n} [ \partial_\mu \bm{n} , \partial_\nu \bm{n} ] \} .
\end{align}
Then we obtain the relationship among $ {\rm tr}\{ \bm{n} \mathscr{F}_{\mu\nu} [\mathscr{B}]\}$,  ${\rm tr}\{ \bm{n} [ \partial_\mu \bm{n} , \partial_\nu \bm{n} ] \}$ and ${\rm tr}\{ \bm{n} [ \mathscr{B}_\mu , \mathscr{B}_\nu ] \}$:
\begin{align}
 {\rm tr}\{ \bm{n} \mathscr{F}_{\mu\nu} [\mathscr{B}]\} 
=   ig_{{}_{\rm YM}}^{-1} \frac{2(N-1)}{N} {\rm tr}\{ \bm{n} [ \partial_\mu \bm{n} , \partial_\nu \bm{n} ] \}   
 =   ig_{{}_{\rm YM}}  {\rm tr}\{ \bm{n} [ \mathscr{B}_\mu , \mathscr{B}_\nu ] \} .
\end{align}
Therefore, we find that ${\rm tr}\{ \bm{n} \mathscr{F}_{\mu\nu} [\mathscr{V}]\}$   is written as 
\begin{align}
 {\rm tr}\{ \bm{n} \mathscr{F}_{\mu\nu} [\mathscr{V}]\} 
%=&   \partial_\mu {\rm tr}\{ \bm{n}  \mathscr{C}_\nu \}
%-  \partial_\nu {\rm tr}\{ \bm{n}  \mathscr{C}_\mu \}
%   +  ig_{{}_{\rm YM}}^{-1} \frac{2(N-1)}{N}  {\rm tr}\{ \bm{n} [ \partial_\mu \bm{n} , \partial_\nu \bm{n} ] \} 
%   \nonumber\\
 =    \partial_\mu {\rm tr}\{ \bm{n}  \mathscr{C}_\nu \}
-  \partial_\nu {\rm tr}\{ \bm{n}  \mathscr{C}_\mu \}
   + ig_{{}_{\rm YM}} {\rm tr}\{ \bm{n} [ \mathscr{B}_\mu , \mathscr{B}_\nu ]   \}
 .
\end{align}

Finally, the field strength $F_{\mu\nu}^g$   is cast into the form:
\begin{align}
 F_{\mu\nu}^g(x) = \sqrt{\frac{2(N-1)}{N}} {\rm tr}\{ \bm{n}(x) \mathscr{F}_{\mu\nu} [\mathscr{V}](x) \} 
 + ig_{{}_{\rm YM}}^{-1} {\rm tr}\{ \rho  g^\dagger(x)[\partial_\mu , \partial_\nu] g(x)   \} 
 .
\end{align}
Incidentally, the last part $ig_{{}_{\rm YM}}^{-1} {\rm tr} \{ \rho g(x)^\dagger [\partial_\mu, \partial_\nu] g(x) \}$ in $F_{\mu\nu}^g(x)$ corresponds  to the \textbf{Dirac string}.%
\footnote{
This is explained later. See section \ref{C29-section:SU2-monopole}.
See %Kondo~(1997) and Kondo~(1998) 
 \cite{KondoI,KondoII}. 
}
This term is not gauge invariant and does not contribute to the Wilson loop operator in the end, since it disappears after the group integration $d\mu(g)$ is performed. 
This is reasonable, since the Wilson loop operator is gauge invariant by definition.

Thus the Wilson loop operator can be rewritten   in terms of new variables:
\begin{align}
 W_C[\mathscr{A}] =& \int [d\mu(g)]_{\Sigma}
 \exp \Big[  -ig_{{}_{\rm YM}} \frac12  \sqrt{\frac{2(N-1)}{N}}
 \int_{\Sigma: \partial \Sigma=C} 2{\rm tr} \{ \bm{n}  \mathscr{F}[\mathscr{V}]  \} \Big] ,
\end{align}
where 
\begin{align}
{\rm tr} \{ \bm{n}  \mathscr{F}[\mathscr{V}] \}  = \frac12 {\rm tr}\{ \bm{n} \mathscr{F}_{\mu\nu} [\mathscr{V}]\}  dx^\mu \wedge dx^\nu ,
\end{align}
with
\begin{align}
 {\rm tr}\{ \bm{n} \mathscr{F}_{\mu\nu} [\mathscr{V}]\} 
  =   \partial_\mu {\rm tr}\{ \bm{n}  \mathscr{C}_\nu \}
-  \partial_\nu {\rm tr}\{ \bm{n}  \mathscr{C}_\mu \}
   + \frac{2(N-1)}{N} ig_{{}_{\rm YM}}^{-1}  {\rm tr}\{ \bm{n} [ \partial_\mu \bm{n} , \partial_\nu \bm{n} ] \} 
 .
\end{align}

For the gauge group $SU(3)$, 
\begin{align}
  \tilde{\bm{n}}(x) =& \frac{2}{\sqrt{3}} \bm{n}(x) 
  , \quad \bm{n}(x) = \frac{\sqrt{3}}{2} \tilde{\bm{n}}(x)  .
\\
  \tilde{F}_{\mu\nu}^g =& \frac{1}{\sqrt{3}}
  2{\rm tr}( \bm{n}(x) \mathscr{F}_{\mu\nu}[\mathscr{V}](x) )
  \nonumber\\
  =& \frac{1}{\sqrt{3}} [\partial_\mu 2{\rm tr}(\bm{n}(x) \mathscr{C}_\nu(x)) - 
  \partial_\nu 2{\rm tr}(\bm{n}(x) \mathscr{C}_\mu(x)) 
  + 2{\rm tr}(\frac{4}{3}ig_{{}_{\rm YM}}^{-1} \bm{n} [\partial_\mu \bm{n}, \partial_\nu \bm{n}]) ]
   .
\end{align}
This should be compared with the gauge group $SU(2)$ case: 
\begin{align}
 \tilde{\bm{n}}(x) =& \bm{n}(x) 
  ,
\\
\tilde{F}_{\mu\nu}^g =& \frac{1}{2}
  {\rm tr}( 2\bm{n}(x) \mathscr{F}_{\mu\nu}[\mathscr{V}](x) )
  \nonumber\\
  =&  \frac{1}{2} [\partial_\mu 2{\rm tr}(\bm{n}(x) \mathscr{C}_\nu(x)) - 
  \partial_\nu 2{\rm tr}(\bm{n}(x) \mathscr{C}_\mu(x)) 
  + 2{\rm tr}(ig_{{}_{\rm YM}}^{-1} \bm{n} [\partial_\mu \bm{n}, \partial_\nu \bm{n}]) ]
   .
\end{align}

For the gauge group $SU(2)$, in particular, arbitrary representation is characterized by  a single index  $J=\frac12, 1, \frac32, 2, \frac52, \cdots$. The $SU(2)$  Wilson loop operator in the representation $J$ obeys the non-Abelian Stokes theorem \cite{DP89,KondoIV}:
\begin{align}
  W_C[\mathscr{A}]  
 =&  \int  [d \mu(g)]_\Sigma \exp \left\{ -i  g_{{}_{\rm YM}} J \int_{\Sigma: \partial \Sigma=C} dS^{\mu\nu} f_{\mu\nu}^g \right\} , 
% \ \text{no path-ordering}
  \nonumber\\
  f_{\mu\nu}^g(x) =& \partial_\mu [ {n}^A(x)   \mathscr{A}^A_\nu(x) ] -  \partial_\nu [{n}^A(x)  \mathscr{A}^A_\mu(x)  ]   
%\nonumber\\ &
  - g_{{}_{\rm YM}}^{-1} \epsilon^{ABC}  {n}^A(x)  \partial_\mu  {n}^B(x)  \partial_\nu  {n}^C(x) 
%  f_{\mu\nu}(x) =& \partial_\mu [\bm{n}(x) \cdot \mathscr{A}_\nu^A(x)] -  \partial_\nu [\bm{n}(x) \cdot \mathscr{A}_\mu(x)]   
%  - g^{-1} \bm{n}(x) \cdot [\partial_\mu \bm{n}(x) \times \partial_\nu \bm{n}(x) ]
 ,
  \nonumber\\
%  \bm{n}(x) &:= 
 n^A(x) \sigma^A   =& g(x)  \sigma^3 g^\dagger(x) , \ g(x) \in SU(2) \ (A,B,C \in \{ 1,2,3 \}) ,
\end{align}
and $ [d \mu(g)]_\Sigma$ is the product measure of an invariant measure on $SU(2)/U(1)$ over $\Sigma$: 
\begin{align}
 [d \mu(g)]_\Sigma :=\prod_{x \in \Sigma}d\mu(\bm{n}(x)) ,
  \ 
  d\mu(\bm{n}(x)) = \frac{2J+1}{4\pi} \delta(\bm{n}^A(x)   \bm{n}^A(x)-1) d^3 \bm{n}(x) .  
\end{align}
This is discussed later in more detail.

%%%%%%%%%%%%%%%%%%%%%%%%%%%%%%%%%%%%%%%%%%%%%%%%%%
%%%%%%%%%%%%%%%%%%%%%%%%%%%%%%%%%%%%%%%%%%%%%%%%%%
\subsection{Magnetic monopole inherent in the Wilson loop operator}
%%%%%%%%%%%%%%%%%%%%%%%%%%%%%%%%%%%%%%%%%%%%%%%%%%
%%%%%%%%%%%%%%%%%%%%%%%%%%%%%%%%%%%%%%%%%%%%%%%%%%

%$\bullet$ Wilson loop and magnetic monopole
\par
Let $\sigma=(\sigma^1,\sigma^2)$ be the world sheet coordinates on the two-dimensional surface $\Sigma_C$ which is bounded by the Wilson loop $C$,    
and $x(\sigma)$  be the target space coordinate of the surface $\Sigma_C$ in $\mathbb{R}^D$.

First of all, we rewrite the surface integral of a two-form $G$ over the surface $\Sigma_C$,  i.e., 
$\int_{\Sigma_C}  G = \int_{\Sigma_C} dS^{\mu\nu} G_{\mu\nu}$  
 into the volume integral:
\begin{align}
\int_{\Sigma_C}  G = \int_{\Sigma_C} dS^{\mu\nu}(x(\sigma)) G_{\mu\nu}(x(\sigma)) 
% =& \frac{1}{2} \int_{\Sigma_C} d^2 \sigma \epsilon^{ab}{\partial x^\mu \over \partial \sigma^a}{\partial x^\nu \over \partial \sigma^b} G_{\mu\nu}(x(\sigma))
%\nonumber\\
% =& {1 \over 2} \int_{\Sigma_C} d^2 \sigma J^{\mu\nu}(\sigma) G_{\mu\nu}(x(\sigma))  
%\nonumber\\
% =& {1 \over 2} \int _{\Sigma_C}d^2 \sigma J^{\mu\nu}(\sigma)  \int d^Dx G_{\mu\nu}(x) \delta^D(x-x(\sigma))
%\nonumber\\
 =& \int d^Dx G_{\mu\nu}(x) \Theta^{\mu\nu}_{\Sigma_C} (x)  ,
\end{align}
where we have introduced   the surface element $dS^{\mu\nu}$ of $\Sigma_C$ and an antisymmetric tensor of rank two:
\begin{align}
 \Theta^{\mu\nu}_{\Sigma_C} (x) 
:=   \int_{\Sigma_C: \partial \Sigma_C=C}  d^2 S^{\mu\nu}(x(\sigma)) \delta^D(x-x(\sigma)) 
%=  {1 \over 2} \int_{\Sigma_C} d^2 \sigma J^{\mu\nu}(\sigma) \delta^D(x-x(\sigma)) 
 .
%=  - \Theta_{\nu\mu}(x) .
\end{align}
We call $\Theta^{\mu\nu}_{\Sigma_C}(x)$ the \textbf{vorticity tensor} which has   the support only on the surface $\Sigma_C$ spanned by the  loop $C$.

Let $J$ be the Jacobian  from $x^\mu, x^\nu$ of the surface element $dS^{\mu\nu}$  to $\sigma^1,\sigma^2$.
Then the surface element $dS^{\mu\nu}$ is written as 
\begin{align}
dS^{\mu\nu}(x(\sigma)) 
=& \frac{1}{2} J^{\mu\nu}(\sigma) d^2 \sigma , \quad d^2 \sigma  = d\sigma^1 d\sigma^2 ,
\nonumber\\
 J^{\mu\nu}(\sigma) :=&  {\partial(x^\mu,x^\nu) \over \partial(\sigma^1,\sigma^2)} 
= \frac{\partial x^\mu}{\partial \sigma^1} \frac{\partial x^\nu }{\partial \sigma^2}  - \frac{\partial x^\mu}{\partial \sigma^2} \frac{\partial x^\nu }{\partial \sigma^1} 
= \epsilon^{ab} \frac{\partial x^\mu}{\partial \sigma^a} \frac{\partial x^\nu }{\partial \sigma^b} 
  .
\end{align}

Second, the integral is rewritten in terms of the  $(D-3)$-form  $k$ and the one-form $j$ in $D=d+1$ dimensions:
\begin{align}
 j:=& \delta G , \quad
k:=  \delta  {}^{\displaystyle *}G =  {}^{\displaystyle *}dG , \quad
G  := 2{\rm tr}\{ \bm{n} \mathscr{F} [\mathscr{V}]\} .
\end{align}
We call  $k$  the ``magnetic-monopole current'' and   $j$ the ``electric current''.
Both currents $k$ and $j$ are gauge invariant and conserved, $\delta k=0=\delta j$ due to nilpotency $\delta^2 \equiv 0$.
%Note that $k$ is $(D-3)$-form and $j$ is one-form in $D=d+1$ dimensions.   
The inner product between two two-forms $\Theta$ and $G$ is decomposed as (\textbf{Hodge decomposition})
%[Exercise-4] \marginpar{Ex-4}
\begin{align}
%   \int_{\Sigma_C} d^2S^{\mu\nu}(x(\sigma)) G_{\mu\nu}(x(\sigma))  
%=&
 \int d^Dx \Theta^{\mu\nu}(x) G_{\mu\nu}(x)
%\\
=& (\Theta,G) = ( {}^{\displaystyle *} \Theta, {}^{\displaystyle *}G) 
\nonumber\\
%=& (*\Theta,\Delta^{-1}(d\delta+\delta d)*G)
=& ( \Theta,\Delta^{-1}(d\delta+\delta d) G)
\nonumber\\
%=& (*\Theta,\Delta^{-1}d\delta *G) + (*\Theta,\Delta^{-1}\delta d*G)
%=& (\Delta^{-1} \Theta,d\delta G) + ( \Delta^{-1} \Theta, \delta d G)
%\nonumber\\
%=& (\delta \Delta^{-1} *\Theta, \delta *G) + (\Theta,*\Delta^{-1}\delta * \delta G) 
=& (\delta \Delta^{-1}  \Theta, \delta  G) + (d \Delta^{-1} \Theta, d     G)  
\nonumber\\
%=& (\delta \Delta^{-1} *\Theta, \delta *G) + (\Theta,\Delta^{-1}d  \delta G)  
=& (\delta \Delta^{-1}  \Theta, \delta  G) + ( {}^{\displaystyle *}d \Delta^{-1} \Theta,  {}^{\displaystyle *}d     G)  
\nonumber\\
=& (\delta \Delta^{-1} \Theta, j) + (\delta  \Delta^{-1} {}^{\displaystyle *} \Theta, k)  ,
\label{C29-Hodge}
\end{align}
where
 $\Delta$ is the Laplacian in the $D$-dimensional Euclidean space (d'Alembertian in the Minkowski space--time) $\Delta:=d\delta+\delta d$.
Here we have used some properties:
$(A,d B)=(\delta A,B)$, $(A,\delta B)=(dA,B)$, $\delta= {}^{\displaystyle *}d {}^{\displaystyle *}$, $ {}^{\displaystyle *} {}^{\displaystyle *}=1$, etc.%
%\footnote{
%For a given $p$-form,
%\begin{align}
% \omega_p = \frac{1}{p!} \omega_{\mu_1 \dots \mu_p}(x) dx^{\mu_1}  \wedge \dots  \wedge dx^{\mu_p} ,
%\end{align}  
%the exterior derivative $d\omega$ is the $(p+1)$-form given by
%\begin{align}
% d\omega_p = \frac{1}{p!} \partial_\nu \omega_{\mu_1 \dots \mu_p}(x) dx^\nu  \wedge dx^{\mu_1} \wedge \dots  \wedge  dx^{\mu_p} ,
%\end{align} 
%and the dual form $ {}^{\displaystyle *}\omega$ is the $(D-p)$-form given by
%\begin{align}
%  {}^{\displaystyle *}\omega_p 
%=& \frac{1}{(D-p)!} \frac{1}{p!} \omega_{\mu_1 \dots \mu_p}(x)   \epsilon^{\mu_1 \dots \mu_p}{}_{\mu_{p+1} \dots \mu_D} dx^{\mu_{p+1}}  \wedge \dots  \wedge dx^{\mu_D} 
%\nonumber\\
%=& \frac{1}{(D-p)!} \frac{1}{p!} \omega^{\mu_1 \dots \mu_p}(x)   \epsilon_{\mu_1 \dots \mu_p \mu_{p+1} \dots \mu_D} dx^{\mu_{p+1}}  \wedge  \dots  \wedge  dx^{\mu_D}  .
%\end{align}  
%}
Here we have defined the inner product for two $p$-forms   by
\begin{align}
 & ( F,G) 
:= \frac{1}{p!} \int d^Dx \ F^{\mu_1 \cdots \mu_{p}}(x) G^{\mu_1 \cdots \mu_{p}}(x) .
\end{align}

In this way we obtain another expression of the NAST for the Wilson loop operator in the \textbf{fundamental representation} for $SU(N)$:
\begin{equation}
 W_C[\mathscr{A}] 
= \int [d\mu({g})] \exp \left\{ -ig_{{}_{\rm YM}}  \frac12  \sqrt{\frac{2(N-1)}{N}} [ (  \omega_{\Sigma_C}, k) +    (  N_{\Sigma_C},j  ) ] \right\} ,
\end{equation}
where 
%we have defined the $(D-3)$-form $k$ and one-form $j$ by
%\begin{align}
%k:=  \delta *G  , \quad  j:=& \delta G  , \quad
%G_{\mu\nu} := {\rm tr}\{ \bm{n} \mathscr{F}_{\mu\nu} [\mathscr{V}]\} ,
%\end{align}
%and
we have defined the $(D-3)$-form $\omega_{\Sigma_C}$ and one-form $N_{\Sigma_C}$  by
\begin{equation}
 \omega_{\Sigma_C} :=  {}^{\displaystyle *} d \Delta^{-1} \Theta_{\Sigma_C}  = \delta  \Delta^{-1}  {}^{\displaystyle *}\Theta_{\Sigma_C}  , \quad
 N_{\Sigma_C} := \delta  \Delta^{-1}  \Theta_{\Sigma_C} ,
\end{equation}
with the inner product for two forms defined by
\begin{align}
 & ( \omega_{\Sigma_C},k) 
= \frac{1}{(D-3)!} \int d^Dx k^{\mu_1 \cdots \mu_{D-3}}(x) \omega^{\mu_1 \cdots \mu_{D-3}}_{\Sigma_C}(x) ,
  \nonumber\\
 &
  (N_{\Sigma_C},j) =  \int d^Dx j^{\mu}(x) N^{\mu}_{\Sigma_C}(x) .
\end{align}
Here we have replaced the measure $[d\mu({g})]_{\Sigma}$ by $[d\mu({g})]:=[d\mu({g})]_{\mathbb{R}^D}=\prod_{x \in \mathbb{R}^D} d\mu({g}(x))$ over all the space--time points.

Thus, the Wilson loop operator can  be expressed by the electric current $j$ and the magnetic-monopole current $k$.
We show that the first factor 
\begin{equation}
W_C^m:=\exp \left[-ig_{{}_{\rm YM}}  \frac12 \sqrt{\frac{2(N-1)}{N}} (  \omega_{\Sigma_C},k ) \right]
\end{equation}
originates from the contributions of the magnetic monopole and  has geometrical and topological meanings, as will be explained shortly. 
If the regular potential one-form $A$ exists so that $G=dA$, then the magnetic current $k$ is identically vanishing: $k= {}^{\displaystyle *}dG= {}^{\displaystyle *}ddA \equiv 0$, which is the Bianchi identity $dG=0$. Therefore, the existence of non-vanishing magnetic monopole $k$ means the violation of the \textbf{Bianchi identity} for the two-form $F^g$.

%%%%%%%%%%%%%%%%%%%%%%%%%%%%%%%%%%%%%%%%%%%%%%%%%%%%%%%%%%%%%
%%%%%%%%%%%%%%%%%%%%%%%%%%%%%%%%%%%%%%%%%%%%%%%%%%%%%%%%%%%%%
%\section{$SU(2)$ chromomagnetic monopole}
%%%%%%%%%%%%%%%%%%%%%%%%%%%%%%%%%%%%%%%%%%%%%%%%%%%%%%%%%%%%%
%%%%%%%%%%%%%%%%%%%%%%%%%%%%%%%%%%%%%%%%%%%%%%%%%%%%%%%%%%%%%

In particular, the $SU(2)$ Wilson loop operator in any  representation characterized by a half-integer $J=\frac12, 1, \frac32, 2, \dots$ obeys the representation:
\begin{align}
& W_C[\mathscr{A}] 
= \int [d\mu(g)]_{\Sigma_C} \exp \left[ -ig_{{}_{\rm YM}}  J  (  \omega_{\Sigma_C}, k) - ig_{{}_{\rm YM}}  J  ( N_{\Sigma_C} , j) \right] 
%  \nonumber\\
%  &   {k:= \delta {}^*F ={}^*dF, \quad
%  \omega_\Sigma := \delta  \triangle^{-1}  }  {}^*\Theta_\Sigma\leftarrow \quad \text{(D-3)-forms}
%  \nonumber\\
%  & j:= \delta F, \quad
%  N_\Sigma := \delta  \triangle^{-1}   \Theta_\Sigma \leftarrow \quad \text{1-forms (D-indep.)}
 .
\end{align}

Note that $N_{\Sigma_C}$ is one-form in any dimension $D$
having the component:
\begin{align}
 N_{\Sigma}^\mu(x)  =& \partial_\nu^x \int d^Dy \Theta^{\mu\nu}(y) \Delta_{(D)}^{-1}(x-y )
%\nonumber\\
 =    \partial_\nu^x \int_{\Sigma_C} d^2 S^{\mu\nu}(x(\sigma)) \Delta_{(D)}^{-1}(x-x(\sigma) ) .
\end{align}
Whereas, $\omega_{\Sigma_C}$ is $(D-3)$-form for $D=d+1$ dimensional case. The explicit form is obtained by using the   Laplacian (d'Alembertian) $\Delta_{(D)}$ in the $D$-dimensional space--time   as follows. 

\noindent
The magnetic monopole described by the current $k$ is a topological object of \textbf{co-dimension} 3:
\begin{itemize}
\item
$D=3$:  $0$-dimensional point defect $\rightarrow$ point-like magnetic monopole (cf. Wu-Yang type)

\item
$D=4$: $1$-dimensional line defect  $\rightarrow$ {  magnetic monopole loop (closed loop due to the topological conservation law $\delta k=0$)}   
\end{itemize}

Here, it will be pedagogic to compare the above result with the Abelian Wilson loop.
For the Abelian $U(1)$ gauge field $A_\mu(x)$, the  $U(1)$ Wilson loop operator $W_C[A]$ is defined without the need of introducing the trace and the path ordering by the line integral  along a closed loop $C$:
\begin{align}
 W_C[A] = \exp \left[ i e \oint_{C} dx^\mu A_\mu \right] 
= \exp \left[ i e \oint_{C}  A  \right] \in U(1) ,
\end{align}
which is a complex number of modulus one, i.e., an element of the $U(1)$ group. 
By using the Stokes theorem, the $U(1)$  Wilson loop operator is rewritten into the surface integral over the area $\Sigma$ bounded by $C$, i.e., $\partial {\Sigma_C}=C$:
\begin{align}
 W_C[A] = \exp \left[ i e \int_{{\Sigma_C}} dS^{\mu\nu} F_{\mu\nu}  \right]
= \exp \left[ i e \int_{{\Sigma_C}} F  \right] \in U(1) ,
\quad F=dA,
\end{align}
By introducing the vorticity tensor $\Theta_{\Sigma}$, i.e., antisymmetric tensor field of rank 2 with the support only on $\Sigma$, it is cast into the volume integral over the whole space--time:
\begin{align}
& W_C[A] 
=  \exp \left[   ie   ( N_{\Sigma_C} , j) + ie  (  \omega_{\Sigma_C}, k) \right] ,
\quad j:= \delta F, \quad k:=  \delta  {}^{\displaystyle *}F 
 .
\end{align}
Here the electric current $j$ is non-vanishing:
\begin{align}
 j= \delta F \ne 0 ,
\end{align}
while the magnetic current $k$ is vanishing due to the Bianchi identity:
\begin{align}
 k=  \delta  {}^{\displaystyle *}F 
=  {}^{\displaystyle *}dF 
=  {}^{\displaystyle *}ddA  = 0 ,
\end{align}
and there is no magnetic contribution to the Wilson loop, 
as far as there are no singularities in the gauge field $A$.
Note that there is a difference of factor 1/2 between the $U(1)$ and $SU(N)$ in the argument of the exponential. 

If we introduce the Dirac magnetic monopole in the Abelian gauge theory, then the magnetic current $k$ can be non-vanishing. The Dirac magnetic monopole can be calculable  once  the specific singularities are introduced into the gauge field. However, there is no general prescription which enables us to make the systematic estimate on how the Dirac magnetic monopole contributes to the Wilson loop average.
The Dirac monopole is not defined as the solution of a field equation, in sharp contrast to the topological solitons, e.g., instanton defined as the solution of the self-dual equation for the Yang-Mills field, and the 't Hooft-Polyakov magnetic monopole defined as the solution of the classical field equation for the Georgi-Glashow model as a gauge-Higgs model. 
The topological solitons are characterized by a finite number of collective coordinates (parameters of the moduli space) and then the integration measure can be defined in terms of them. This fact enables one to perform the systematic calculations in principle.

\subsubsection{$D=3$ magnetic monopole}

In the three-dimensional case $D=3$, $\omega$ is the zero-form with the component:
\begin{align}
 \omega_{\Sigma}(x)  =& {1 \over 2} \epsilon^{jk\ell} \partial_\ell^x \int d^3y \Theta_{jk}(y) \Delta_{(3)}^{-1}(x-y )
\nonumber\\
=& {1 \over 2} \epsilon^{jk\ell} \partial_\ell^x \int_{\Sigma_C} d^2 S_{jk}(x(\sigma)) \Delta_{(3)}^{-1}(x-x(\sigma) )  
\nonumber\\
=& {1 \over 2} \epsilon^{jk\ell} \partial_\ell^x \int_{\Sigma_C} d^2 S_{jk}(x(\sigma)) \frac{-1}{4\pi|x-x(\sigma)|} ,
\end{align}
and $k$ is also zero-form, i.e., the \textbf{magnetic charge density} function:
\begin{equation}
 k(x)  = {1 \over 2} \epsilon^{jk\ell}  \partial_\ell G_{jk}(x)  := \rho_m(x) .
\end{equation}

\begin{figure}%[thpb]
\begin{center}
%\begin{picture}(0,0)%(0,-3000)
%\put(8700,-800){\includegraphics[height=2.5in]{Fig-PR/Wilson-solid-angle.eps}}
\includegraphics[scale=0.4]{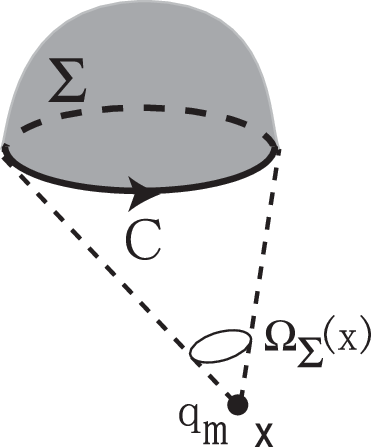}
%\end{picture}
\end{center}
%\vskip -0.5cm
\caption[]{
The magnetic monopole $q_m$ at $x$ and the solid angle $\Omega_{\Sigma}(x)$ at $x$ subtended by the surface $\Sigma$ bounding the Wilson loop $C$.
}
\label{C29-fig:Wilson-solid-angle}
\end{figure}

It is known that the \textbf{solid angle} $\Omega(x)$ 
  at the point $x$ subtended by the surface $\Sigma$ bounding the Wilson loop $C$
%under which the surface $\Sigma$ shows up to an observer at the point $x$ 
is written as
\begin{align}
 \Omega_{\Sigma}(x) 
:=& \frac{\partial}{\partial x^\ell} \int_{\Sigma_C} d^2S^\ell(y) \frac{1}{|x-y|}   
= \frac{1}{2} \epsilon^{jk\ell}  \frac{\partial}{\partial x^\ell} \int_{\Sigma_C} d^2S_{jk}(y) \frac{1}{|x-y|} 
\nonumber\\
 =& 4\pi \omega_{\Sigma}(x) , 
\end{align}
where we have used the surface element:
\begin{align}
 d^2S^\ell := \frac{1}{2} \epsilon^{jk\ell} d^2S_{jk}  .
\end{align}
See Fig.~\ref{C29-fig:Wilson-solid-angle}. 
In the case when $\Sigma$ is a closed surface surrounding the point $x$, one has the total solid angle $\Omega=4\pi$, since due to the Gauss law:
\begin{align}
 \Omega_{\Sigma}(x) 
=& \oint_{\Sigma} d^2S^\ell(y)  \frac{\partial}{\partial x^\ell}  \frac{1}{|x-y|} 
\nonumber\\
=& \int_{V:\partial V=\Sigma} d^3y  \frac{\partial}{\partial y^\ell}  \frac{\partial}{\partial x^\ell}  \frac{1}{|x-y|} 
\nonumber\\
=& - \int_{V:\partial V=\Sigma} d^3y  \frac{\partial}{\partial x^\ell}  \frac{\partial}{\partial x^\ell}  \frac{1}{|x-y|} 
=  \int_{V} d^3y  4\pi \delta^3(x-y)  
= 4\pi
  .  
\end{align}
This is a standard result for the total solid angle in three dimensions. 

%%%%%%%%%%%%%%%%%%%%% figures %%%%%%%%%%%%%%%%%%%%%%%%%%%
%%%%%%%%%%%%%%%%%%%%% figures %%%%%%%%%%%%%%%%%%%%%%%%%%%

For $D=3$, thus, $\omega_{\Sigma}$  is the (normalized) solid angle $\Omega_{\Sigma}$ ($\Omega_{\Sigma}$ divided by the total solid angle $4\pi$):
\begin{equation}
\omega_{\Sigma}(x)=\Omega_{\Sigma}(x)/(4\pi) .
\end{equation}

For $SU(2)$,  the magnetic contribution in the Wilson loop operator reads 
\begin{equation}
W_C^m = \exp \left[ -i g_{{}_{\rm YM}}  J \int d^3x k(x) \omega_{\Sigma}(x) \right]
= \exp \left[ -i g_{{}_{\rm YM}}  J \int d^3x \rho_m(x) \frac{\Omega_{\Sigma}(x)}{4\pi} \right] .
\end{equation}

For $SU(2)$, the \textbf{Wu-Yang configuration} 
\begin{align}
n^A= \frac{x^A}{r}   \quad ( A=1,2,3 ) 
\end{align}
leads to the magnetic-monopole density 
%[Exercise-5] \marginpar{Ex-5}
\begin{align}
 k(x) = \rho_m(x) =& \frac12 \epsilon^{jk\ell} \partial_\ell  F_{jk}(x)
\nonumber\\
=& \frac12 \epsilon^{jk\ell} \partial_\ell 
\left[ -g_{{}_{\rm YM}} ^{-1} \epsilon_{jkm} \frac{x^m}{r^3} \right] 
\nonumber\\
=&     -g_{{}_{\rm YM}} ^{-1}  \partial_\ell 
\left[   \frac{x^\ell}{r^3} \right] 
\nonumber\\
=&     - 4\pi g_{{}_{\rm YM}} ^{-1} \delta^3(x) ,
\label{magnetic-monopole-density}
\end{align}
where we have used 
\begin{align}
 F_{\mu\nu} = \partial_\mu (n^A \mathscr{A}_\nu^A) -  \partial_\nu (n^A \mathscr{A}_\mu^A) + H_{\mu\nu}, \  H_{\mu\nu} := - g_{_{\rm YM}}^{-1} \epsilon_{ABC} n^A \partial_\mu n^B \partial_\nu n^C .
\end{align}
This corresponds to a magnetic monopole with a unit magnetic charge $q_m=4\pi g_{{}_{\rm YM}} ^{-1}$ located at the origin. 
Hence, we obtain the contribution of a magnetic monopole to the Wilson loop operator:
\begin{align}
W_C^m
%= \exp \left[ i g \frac12 \int d^3x k(x) \omega_{\Sigma}(x) \right]
=  \exp \left[ -i g_{{}_{\rm YM}}  J \int d^3x  4\pi g_{{}_{\rm YM}} ^{-1} \delta^3(x) \frac{\Omega_{\Sigma}(x)}{4\pi} \right]  
 =  \exp \left[ -i J  \Omega_{\Sigma}(0)  \right] .
\end{align}
Therefore,   $\exp [-i J  g_{{}_{\rm YM}}(k, \omega_{\Sigma})]$ gives a non-trivial contribution: 
\begin{equation}
 \exp [ \pm iJ (2 \pi) ]=(e^{\pm i \pi})^{2J}=(-1)^{2J} ,
\end{equation}
 for a half-integer $J$ from a magnetic monopole with a unit magnetic charge $q_m=4\pi g_{{}_{\rm YM}} ^{-1}$ at the origin, since 
$\Omega_{\Sigma}(0)=\pm 2\pi$ for the upper or lower hemisphere $\Sigma$. 
This is also the case for a magnetic monopole at an arbitrary point on the minimal surface $S$ spanned by the Wilson loop $C$. 
This result does not depend on which surface bounding $C$ is chosen in the non-Abelian Stokes theorem.

%%%%%%%%%%%%%%%%%%%%% figures %%%%%%%%%%%%%%%%%%%%%%%%%%%
\begin{figure}%[htbp]
\begin{center}
\includegraphics[scale=0.35,clip]{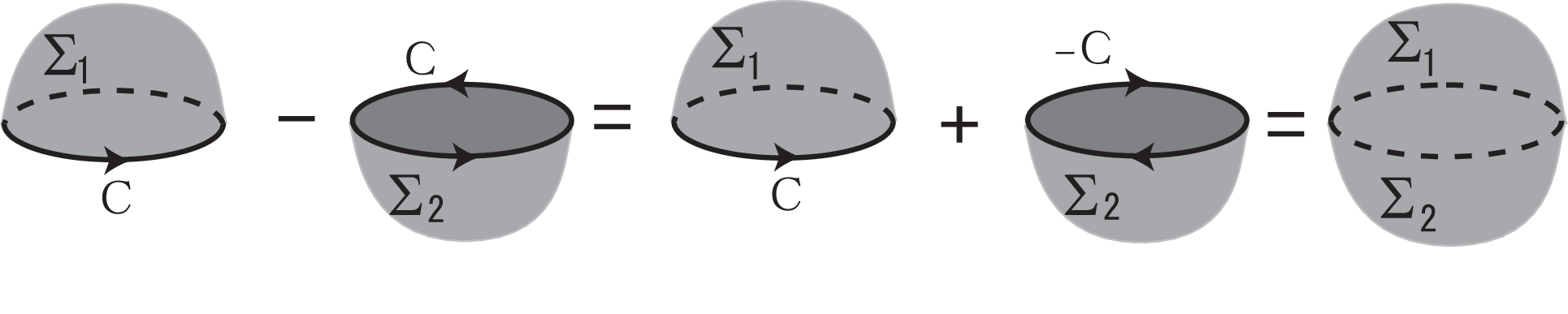}
\end{center} 
%\vskip -0.5cm
\caption[]{
Quantization of the magnetic charge.  The difference between  the surface integrals of the two-form $F$ over two surfaces $\Sigma_{1}$ and $\Sigma_{2}$ with the same boundary $C$ is equal to the surface integral $\int_{\Sigma} F$ of $F$ over one closed surface $\Sigma=\Sigma_{1}+\Sigma_{2}$. Here the direction of the normal vector to the surface $\Sigma_{2}$ must be consistent with the direction of the line integral over $C$.  The magnetic charge $Q_m$ is non-zero if the magnetic monopole exists inside $\Sigma$, otherwise it is zero. 
}
\label{C29-fig:Wilson-solid-angle2}
\end{figure}
%%%%%%%%%%%%%%%%%%%%% figures %%%%%%%%%%%%%%%%%%%%%%%%%%%

In general, we can show that the magnetic charge $q_m$ obeys the \textbf{Dirac quantization condition}:   
\begin{equation}
 q_m :=  \int d^3x \rho_m(x) =4\pi g_{{}_{\rm YM}} ^{-1} n \quad (n \in \mathbb{Z} = \{ \cdots, -2,  - 1, 0, +1, +2, \cdots \} )
 .
 \label{C29-Dirac-qc}
\end{equation}
In fact, the {Dirac quantization condition} $q_m=4\pi g_{{}_{\rm YM}}^{-1} n$ for the magnetic charge $q_m$ is obtained   from the condition:   the non-Abelian Stokes theorem does not depend on the surface which is chosen for spanning the surface bounded by the loop $C$, remembering that the original Wilson loop is defined for a given  closed loop $C$. 
The difference of the argument must be an integer multiple of $2\pi$:
\begin{align}
2\pi n 
=& 
 \frac12 g_{{}_{\rm YM}}  \int d^3x \rho_m(x) \frac{\Omega_{\Sigma_{1}}(x)}{4\pi} 
-    \frac12 g_{{}_{\rm YM}} \int d^3x \rho_m(x) \frac{\Omega_{\Sigma_{2}}(x)}{4\pi} 
\nonumber\\
=&   \frac12 g_{{}_{\rm YM}}  \int d^3x \rho_m(x) \frac{\Omega_{\Sigma_{1}}(x)-\Omega_{\Sigma_{2}}(x)}{4\pi} 
\nonumber\\
=&   \frac12 g_{{}_{\rm YM}}  \int d^3x \rho_m(x)  
=   \frac12 g_{{}_{\rm YM}}  q_m 
 ,
\end{align}
where we have used
$\Omega_{\Sigma_{1}}(x)-\Omega_{\Sigma_{2}}(x)=4\pi$.
Thus, the Wilson loop operator can probe  gauge-invariant magnetic monopole with the magnetic charge subject to the Dirac quantization condition. 
See Fig.~\ref{C29-fig:Wilson-solid-angle2}. 
%\footnote{
%for SU(N) irrespective of $N\ge 2$: 
%\begin{equation}
%  q_m=4\pi g_{{}_{\rm YM}}^{-1} n , \quad n \in \mathbb{Z}   .
%\end{equation}
%Therefore, one need not to introduce $N-1$ kinds of magnetic monopoles, which are usually supposed to be deduced from the maximal torus group $U(1)^{N-1}$ of SU(N).
%}

For an ensemble of point-like magnetic charges located at $x=z_a$ ($a=1, \cdots, n$) represented by
\begin{equation}
 k(x) = \rho_m(x) = \sum_{a=1}^{n} q_m^a \delta^{(3)}(x-z_a) 
  ,
  \quad 
  q_m^a=4\pi g_{{}_{\rm YM}} ^{-1} \ell_a , \quad \ell_a \in \mathbb{Z} 
   ,
\end{equation}
we have a geometric representation:
\begin{equation}
W_C^m  
= \exp \left\{ -i  g_{{}_{\rm YM}} J  \sum_{a=1}^{n} \frac{q_m^a}{4\pi} \Omega_{\Sigma}(z_a)  \right\}
= \exp \left\{ i J  \sum_{a=1}^{n} \ell_a \Omega_{\Sigma}(z_a)  \right\}
 , \quad \ell_a \in \mathbb{Z} 
  .
  \label{C29-mono-Wcont}
\end{equation}
The magnetic monopoles in the neighborhood of the Wilson surface $\Sigma$ ($\Omega_\Sigma(z_a)= \pm 2\pi$) contribute to the Wilson loop
\footnote{
This helps us to understand the  \textbf{N-ality} dependence of the asymptotic string tension.
See \cite{Kondo08b}.
%K.-I. K., 
%arXiv:0802.3829, 
%J.Phys.G35, 085001 (2008).
% 
}
\begin{equation}
W_{C}^m  
= \prod_{a=1}^{n} \exp [ \pm iJ (2 \pi) \ell_a ]
= \begin{cases}
 \prod_{a=1}^{n} (-1)^{\ell_a} &(J=1/2,3/2, \cdots) \cr
 = 1 &(J=1,2,\cdots) 
 \end{cases} .
\end{equation}
Here,   $\exp [i  g_{{}_{\rm YM}}J ( \omega_{\Sigma},k)]$ gives a non-trivial contribution, i.e., a center element: 
\begin{equation}
 \exp [ \pm iJ (2 \pi)    ]=(e^{\pm i \pi})^{2J}=(-1)^{2J} = \{ 1, -1 \} \in \mathbb{Z}(2) = \text{Center}(SU(2)) ,
\end{equation}
This result does not depend on which surface bounding $C$ is chosen in the non-Abelian Stokes theorem. 
[This helps us to understand the  \textbf{N-ality} dependence of the asymptotic string tension.]
%See \cite{Kondo08b}.]
%K.-I. K., 
%arXiv:0802.3829, 
%J.Phys.G35, 085001 (2008).
% 
%}
Here the magnetic flux from a magnetic monopole is assumed to be distributed isotropically in the space. 
However, the contribution of a magnetic monopole to the Wilson loop average depends on the location of the monopole relative to the Wilson loop.  
%The vortex condensation picture gives an easy way to understand the area law. 

\subsubsection{$D=4$ magnetic monopole}

In the four-dimensional case $D=4$, $\omega$ is one-form with the component:
\begin{align}
 \omega^\mu(x)  =& {1 \over 2} \epsilon^{\mu\nu\rho\sigma} \partial_\nu^x \int d^4y \Theta_{\rho\sigma}(y) \Delta_{(4)}^{-1}(x-y)
\nonumber\\
=& {1 \over 2} \epsilon^{\mu\nu\rho\sigma} \partial_\nu^x \int_S d^2 S_{\rho\sigma}(x(\sigma)) \Delta_{(4)}^{-1}(x-x(\sigma)) 
\nonumber\\
=&  {1 \over 2} \epsilon^{\mu\nu\rho\sigma} \partial_\nu^x \int_S d^2 S_{\rho\sigma}(x(\sigma)) \frac{1}{4\pi^2|x-x(\sigma)|^2} ,
\end{align}
and $k$ is also one-form, i.e., magnetic current:
\begin{equation}
 k^\mu(x) = {1 \over 2} \epsilon^{\mu\nu\rho\sigma} \partial_\nu G_{\rho\sigma}(x) .
\end{equation}
The magnetic current $k$ is conserved, $\partial_\mu k^\mu=0$. 

Note that $\omega$ agrees with the four-dimensional solid angle  given by
\begin{equation}
 \Omega^\mu_{\Sigma}(x) 
= \frac{1}{8\pi^2} \epsilon_{\mu\nu\rho\sigma} \frac{\partial}{\partial x^\nu} \int_{\Sigma} d^2S_{\rho\sigma}(y)    \frac{1}{|x-y|^2} 
= \omega^\mu_{\Sigma}(x) .
\end{equation}
Consequently, we have for $D=4$,
\begin{equation}
 W_C^m =  \exp \left[ -i g_{{}_{\rm YM}}  J \int d^4x k_\mu(x) \omega^\mu_{\Sigma}(x) \right]
=  \exp \left[ -i g_{{}_{\rm YM}}  J \int d^4x k_\mu(x) \Omega_{\Sigma}^\mu(x) \right].
\end{equation}

\begin{figure}%[htbp]
\begin{center}
\includegraphics[scale=0.35,clip]{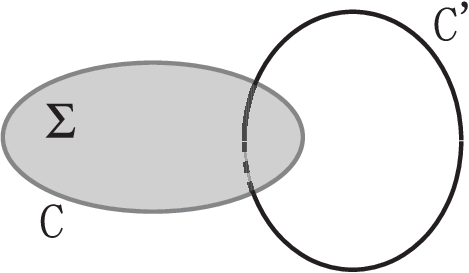}
\end{center} 
\vskip -0.5cm
\caption[]{
The linking  $L(\Sigma,C')$  between the closed loop  $C^\prime$ and the surface $\Sigma$ bounding the closed loop $C$.
}
\label{C29-fig:4d-linking}
\end{figure}

For an ensemble of magnetic monopole loops $C_a^\prime$ ($a=1,\cdots,n$):
\begin{equation}
 k^\mu(x) = \sum_{a=1}^{n} q_m^a \oint_{C_a^\prime} dy^\mu_a \delta^{(4)}(x-y_a) , 
 \quad 
 q_m^a = 4\pi g_{{}_{\rm YM}} ^{-1} n_a 
  ,
\  n_a \in \mathbb{Z} 
  ,
\end{equation}
we obtain
\begin{equation}
 W_C^m    
= \exp \left\{ -i J g_{{}_{\rm YM}}  \sum_{a=1}^{n} q_m^a L(\Sigma,C_a^\prime)  \right\}
= \exp \left\{ -i 4\pi J \sum_{a=1}^{n} n^a L( \Sigma,C_a^\prime)  \right\}
 , 
\end{equation}
where
$L(\Sigma,C_a^\prime)$ is the \textbf{linking number} between the surface $\Sigma$ and the curve $C^\prime$: 
\begin{equation}
  L(\Sigma, C^\prime)  = L(C^\prime, \Sigma)   
  := \oint_{C^\prime} dy^\mu(\tau) \omega^\mu_{\Sigma}(y(\tau)) \in \mathbb{Z}  .
\end{equation} 
Here the curve $C^\prime$ is identified with the trajectory of a magnetic monopole and the surface $\Sigma$ with the world sheet of a hadron (meson) string for a quark-antiquark pair. 
See Fig.~\ref{C29-fig:4d-linking}.

%%%%%%%%%%%%%%%%%%%%% figures %%%%%%%%%%%%%%%%%%%%%%%%%%%
%%%%%%%%%%%%%%%%%%%%% figures %%%%%%%%%%%%%%%%%%%%%%%%%%%

Thus we have shown that the Wilson loop operator is a probe of the magnetic monopole defined in the gauge invariant way in  the pure Yang-Mills theory even in the absence of any scalar field.
Therefore, calculating the Wilson loop average reduces to the summation over the contributions coming from the distribution of magnetic monopole charges ($D=3$)  or currents ($D=4$)  with the geometric factor, the solid angle ($D=3$)  or linking number  ($D=4$). 
Therefore, we do not need to use the Abelian projection  to define magnetic monopoles in Yang-Mills theory!

%%%%%%%%%%%%%%%%%%%%%%%%%%%%%%%%%%%%%%%%%%%%%%%%%%%%%%%%%%%%%
\subsection{Magnetic monopole and  the associated magnetic vortex}
%%%%%%%%%%%%%%%%%%%%%%%%%%%%%%%%%%%%%%%%%%%%%%%%%%%%%%%%%%%%%

For $D=4$, however, we fall in a trouble about the magnetic monopole just defined. 
In fact, such magnetic loops carrying the magnetic charge obeying the Dirac quantization condition (\ref{C29-Dirac-qc}) do not give non-trivial contributions to the Wilson loop, since $n_a$ and $L$ are integers. 
If the quantization condition (\ref{C29-Dirac-qc}) is naively taken at its face value, the magnetic monopole cannot be the topological defects responsible for quark confinement. 
In the following, we discuss how this dilemma is resolved according to \cite{Kondo08b}.

To understand the area law of the Wilson loop average in $D$-dimensional space--time, 
we want to find a $(D-1)$-dimensional geometric object $V_{D-1}$ such that 
\\
i) $V_{D-1}$ has an intersection with the Wilson loop $C$ as the one-dimensional object, 
\\
ii) $V_{D-1}$ has  relevance to the magnetic-monopole current as the $(D-3)$-form $k$.
%Suppose that $k$ has the support on the closed $(D-3)$-dimensional subspace $C_{D-3}\prime$: 
%The object must be $(D-1)$-dimensional one $V_{D-1}$, since the loop $C$ is one-dimensional object which has an intersection only with the $(D-1)$-dimensional object in $D$-dimensional space--time. 
%On the other hand, $k$ is a $(D-3)$-form with the support only on the closed $(D-3)$-dimensional subspace $C_{D-3}$. 

In what follows, we consider how the magnetic flux emanating from the magnetic monopole behaves in the vacuum.  
We first consider the $D=4$ case. 
Suppose that the magnetic current $k^\mu$ has the support on a loop $C_1^\prime$ in $D=4$ dimensions:%
\footnote{
In what follows, we extensively use the techniques developed  in constructing a continuum analogue of the maximal center gauge and center projection  by 
 Engelhardt and Reinhardt \cite{ER00}.
%, but from a different angle in this paper. 
}
\begin{align}
 k^\mu(x) = k^\mu(x;C_1^\prime) := \Phi \oint_{C_1^\prime} dy^\mu \delta^4(x-y) 
  ,
\end{align}
where  
$\Phi$ is a real number representing the magnetic flux carried by the magnetic current $k$.
The magnetic charge $q_m$ is defined by 
\begin{align}
 q_m=\int d^3 \tilde{\sigma}_\mu k^\mu 
  ,
\end{align}
where $\bar{x}^\mu$ denotes a parameterization of the 3-dimensional volume $V_3$ and 
 $d^3 \tilde{\sigma}_{\mu} $ is the dual of the 3-dimensional volume element $d^3 \sigma^{\gamma_1\gamma_2\gamma_3}$:
\begin{align}
 d^3 \tilde{\sigma}_{\mu} 
:=  \frac{1}{3!}   \epsilon_{\mu\gamma_1\gamma_2\gamma_3} d^3 \sigma^{\gamma_1\gamma_2\gamma_3}
 , \quad
% \nonumber\\
 d^3 \sigma^{\gamma_1\gamma_2\gamma_3}
 :=  \epsilon_{\beta_1\beta_2\beta_3} 
 \frac{\partial \bar{x}^{\gamma_1}}{\partial \sigma_{\beta_1}} \frac{\partial \bar{x}^{\gamma_2}}{\partial \sigma_{\beta_2}}  \frac{\partial \bar{x}^{\gamma_3}}{\partial \sigma_{\beta_3}}  d\sigma_{1} d\sigma_{2} d\sigma_{3} 
  .
\end{align}

First of all, we look for the dual field strength $ {}^{\displaystyle *}f^{\mu\nu}(x;S_2)$ representing the magnetic flux, which has the support on a two-dimensional surface $S_2$ bounding the loop $C_1^\prime$, $\partial S_2=C_1^\prime$ as the support of the magnetic current $k^\mu$, so that%  (\ref{C29-def-k-j}) holds:
\begin{align}
  \partial_\nu  {}^{\displaystyle *}f^{\mu\nu}(x;S_2)  = k^\mu(x;C_1^\prime) 
% , 
%\nonumber\\
% {}^*f^{\mu\nu}(x;C) 
% =& Q \int_{S} d^2 \sigma^{\mu\nu} \delta^4(x-\bar{x}(\sigma)) 
  .
\end{align}
Such a solution is given by
\begin{align}
  {}^{\displaystyle *}f^{\mu\nu}(x;S_2) 
 = \Phi \int_{S_2:\partial S_2=C_1^\prime} d^2 \sigma^{\mu\nu} \delta^4(x-\bar{x}(\sigma)) 
%  ,
%  \quad
%  f^{\mu\nu}(x;C) 
% = \Phi \int_{S} d^2 \tilde{\sigma}^{\mu\nu} \delta^4(x-\bar{x}(\sigma)) 
  .
  \label{C29-*f-sol}
\end{align}
In fact, it satisfies the equation:
\begin{align}
 \partial_\nu   {}^{\displaystyle *}f^{\mu\nu}(x;S_2) 
 =& \Phi \int_{S_2} d^2 \sigma^{\mu\nu} \partial_\nu^{x}   
\delta^4(x-\bar{x}(\sigma)) 
\nonumber\\
 =&  \Phi \int_{S_2} d^2 \sigma^{\mu\nu} \partial_\nu^{\bar{x}}   \delta^4(x-\bar{x}(\sigma)) 
\nonumber\\
 =&  \Phi \oint_{\partial S_2=C_1^\prime} d \bar{x}^{\mu}   \delta^4(x-\bar{x}(\sigma)) 
 = k^\mu(x;C_1^\prime)
  ,
\end{align}
where we have used the Stokes theorem in the last step.

Next, we proceed to obtain the gauge potential $v_\mu(x;V_3)$ giving the field strength $f^{\mu\nu}(x;S_2)$ whose dual satisfies (\ref{C29-*f-sol}):
\begin{align}
 f_{\mu\nu}(x;S_2) =& \partial_\mu v_\nu(x;V_3) - \partial_\nu v_\mu(x;V_3) 
 , 
%5\nonumber\\
%   v_\mu(x;V) =& \Phi   \int_{V:\partial V=S} d^3 \tilde{\sigma}_{\mu}  \delta^4(x-\bar{x}(\sigma))
%    ,
\end{align}
such that $v_\mu(x;V_3)$ has the support only on the open set $V_3$, the three-dimensional volume%
\footnote{
The precise position of the open set $V_3$ is irrelevant for the value of the Wilson loop as shown below. 
In fact, the open set $V_3$ can be deformed arbitrarily by singular gauge transformations in such a way that its boundary $\partial V_3$ representing the position of the magnetic flux of the vortex is fixed. 
}
whose boundary is $S_2$: $\partial V_3=S_2$.
The gauge field $v_\mu(x;V_3)$ is called the \textbf{ideal vortex field}. 
Note that $ {}^{\displaystyle *}f^{\mu\nu}(x;S_2) $ is cast into  
\begin{align}
  {}^{\displaystyle *}f^{\mu\nu}(x;S_2) 
% =& \Phi \int_{S} d^2 \sigma^{\mu\nu} \delta^4(x-\bar{x}(\sigma)) 
%\nonumber\\
 =& \Phi \int_{V_3:\partial V_3=S_2} d^3 \sigma^{\mu\nu\kappa}  \partial_\kappa^{\bar{x}}  \delta^4(x-\bar{x}(\sigma)) 
\nonumber\\
 =& \Phi \epsilon^{\mu\nu\alpha\beta} \frac{1}{3!} \epsilon_{\beta\gamma_1\gamma_2\gamma_3} \int_{V_3:\partial V_3=S_2} d^3 \sigma^{\gamma_1\gamma_2\gamma_3}  \partial_\alpha^{x}  \delta^4(x-\bar{x}(\sigma)) 
\nonumber\\
 =&  \epsilon^{\mu\nu\alpha\beta} \partial_\alpha^{x}   \left[\Phi  \frac{1}{3!} \epsilon_{\beta\gamma_1\gamma_2\gamma_3} \int_{V_3:\partial V_3=S_2} d^3 \sigma^{\gamma_1\gamma_2\gamma_3}  \delta^4(x-\bar{x}(\sigma)) \right]
\nonumber\\
 =& \frac12 \epsilon^{\mu\nu\alpha\beta} \left\{ \partial_\alpha^{x}   \left[\Phi   \int_{V_3:\partial V_3=S_2} d^3 \tilde{\sigma}_{\beta}  \delta^4(x-\bar{x}(\sigma)) \right] - (\alpha \leftrightarrow \beta) \right\}
  ,
\end{align}
where we have used the Gauss (Stokes) theorem in the first equality.
 Therefore, the gauge potential $v_\mu(x;V_3)$ is determined up to a gauge transformation:
\begin{align}
% f_{\mu\nu}(x;S) =& \partial_\mu v_\nu(x;V) - \partial_\nu v_\mu(x;V) 
% , 
%\nonumber\\
   v_\mu(x;V_3) =  \Phi   \int_{V_3:\partial V_3=S_2} d^3 \tilde{\sigma}_{\mu}  \delta^4(x-\bar{x}(\sigma))
    .
\end{align}
%This corresponds to an explicit (singular) gauge field representation of an {\it ideal vortex configuration} in 4 space--time dimensions \cite{ER00}.

\subsubsection{$D$-dimensional case}

%%%%%%%%%%%%%%%%%%%%%%%%%%%%%%%%%%%%%%%%%%%%%%%%%%%%%%%%%%%%%
\begin{figure}[tb]
\begin{center}
\includegraphics[scale=0.35]{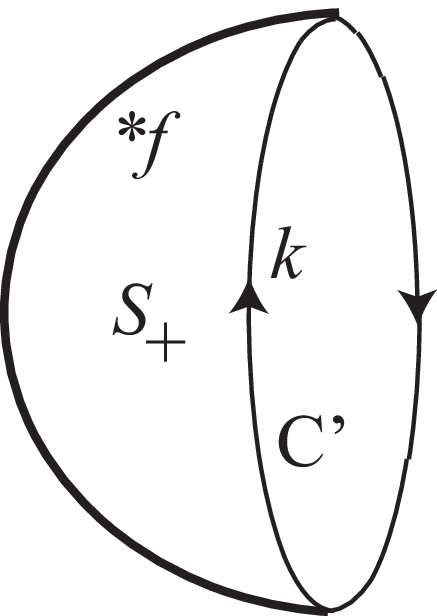}
\quad\quad
\includegraphics[scale=0.35]{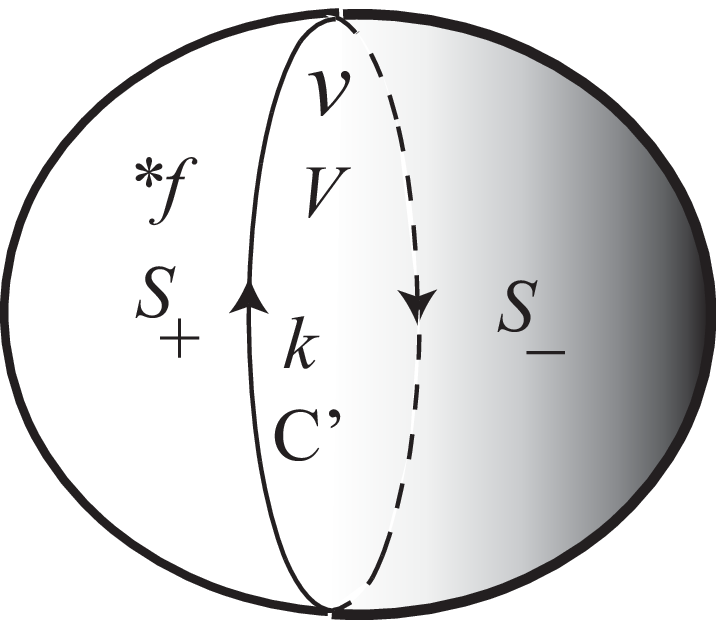}
\end{center} 
\vskip -0.5cm
\caption[]{ %\cite{Kondo08b}
%\small
The supports of the vortex and the magnetic monopole. 
(Left) The dual field strength $({}^{\displaystyle *}f_2)_{D-2}(x,S_{D-2})$ on $S_{D-2}$ and the magnetic-monopole current $k_{D-3}(x,C_{D-3}^\prime)$ on the boundary $\partial S_{D-2}=C_{D-3}^\prime$,  
(Right) The ideal vortex $v_1$ on $V_{D-1}$ and its boundary $\partial V_{D-1}=S_{D-2}$.
}
 \label{C29-fig:C-S-V-rel}
\end{figure}
%%%%%%%%%%%%%%%%%%%%%%%%%%%%%%%%%%%%%%%%%%%%%%%%%%%%%%%%%%%%%

We now proceed to consider the general $D$-dimensional case. 
In $D$ dimensions, the magnetic current $k$ is a $(D-3)$-form with the support on a closed $(D-3)$-dimensional subspace $C^\prime_{D-3}$:
\begin{align}
  k_{\mu_1 \dots \mu_{D-3}}(x;C^\prime_{D-3}) := \Phi \oint_{C^\prime_{D-3}} d^{D-3} \sigma_{\mu_1 \dots \mu_{D-3}} \delta^D(x-\bar{x}(\sigma)) .
%  (D \ge 4).
\end{align}
First, we consider the \textbf{dual field strength} $ ({}^{\displaystyle *}f_{2})^{\mu_1 \dots \mu_{D-2}}(x;S_{D-2})$ representing the magnetic flux, which has the support on an open $(D-2)$-dimensional surface $S_{D-2}$ bounding a closed  $(D-3)$-dimensional subspace $C_{D-3}^\prime$, $\partial S_{D-2}=C_{D-3}^\prime$, as the support of the magnetic current $k^{\mu_1 \dots \mu_{D-3}}$ so that%  (\ref{C29-def-k-j}) holds:
\begin{align}
  \partial_{\mu_{D-2}} ({}^{\displaystyle *}f_{2})^{\mu_1 \dots \mu_{D-3}\mu_{D-2}}(x;S_{D-2})  = k^{\mu_1 \dots \mu_{D-3}}(x;C^\prime_{D-3}) 
% , 
%\nonumber\\
% {}^*f^{\mu\nu}(x;C) 
% =& Q \int_{S} d^2 \sigma^{\mu\nu} \delta^4(x-\bar{x}(\sigma)) 
  .
\end{align}
See Fig.~\ref{C29-fig:C-S-V-rel}.
In the similar way to the above, we find that a solution is given by
\begin{align}
  ({}^{\displaystyle *}f_{2})^{\mu_1 \dots \mu_{D-2}}(x;S_{D-2}) 
 = \Phi \int_{S_{D-2}:\partial S_{D-2}=C_{D-3}^\prime} d^{D-2} \sigma_{\mu_1 \dots \mu_{D-2}}   \delta^D(x-\bar{x}(\sigma)) 
%  ,
%  \quad
%  f^{\mu\nu}(x;C) 
% = \Phi \int_{S} d^2 \tilde{\sigma}^{\mu\nu} \delta^4(x-\bar{x}(\sigma)) 
  .
  \label{C29-*f-sol-D}
\end{align}
Next, we consider the \textbf{ideal vortex field} $v^\mu(x;V_{D-1})$, giving the field strength field $f^{\mu\nu}(x;S_{D-2})$:
\begin{align}
 f_{\mu\nu}(x;S_{D-2}) =& \partial_\mu v_\nu(x;V_{D-1}) - \partial_\nu v_\mu(x;V_{D-1}) 
 , 
% \nonumber\\
%   v_\mu(x;V) =& \Phi   \int_{V:\partial V=S} d^3 \tilde{\sigma}_{\mu}  \delta^4(x-\bar{x}(\sigma))
%    ,
\end{align}
such that $v^\mu(x;V_{D-1})$ has the support only on the $(D-1)$-dimensional volume $V_{D-1}$ whose boundary is $S_{D-2}$: $\partial V_{D-1}=S_{D-2}$.
%Then  an explicit (singular) gauge field representation of an ideal vortex configuration in $D$ space--time dimensions is given by
We find such an explicit (singular) gauge field representation   in $D$ space--time dimensions:
\begin{align}
 v_\mu(x;V_{D-1}) =& \Phi   \int_{V_{D-1}:\partial V_{D-1}=S_{D-2}} d^{D-1} \tilde{\sigma}_{\mu} \ \delta^D(x-\bar{x}(\sigma)) \quad (D \ge 4)
  ,
\end{align}
where $V_{D-1}$ is the $(D-1)$-dimensional hypersurface (the string, surface, or volume in $D=2,3$ and $4$, respectively) parameterized by $\bar{x}^\mu(\sigma)=\bar{x}^\mu(\sigma_1, \sigma_2, \dots, \sigma_{D-1})$,
and $d^{D-1} \tilde{\sigma}_{\mu}$  is the dual of the surface element $d^{D-1} {\sigma}_{\alpha_{1}\dots\alpha_{D-1}}$:
\begin{align}
d^{D-1} \tilde{\sigma}_{\mu}
:=&   \frac{1}{(D-1)!} \epsilon_{\mu\alpha_{1}\dots\alpha_{D-1}}
d^{D-1} {\sigma}_{\alpha_{1}\dots\alpha_{D-1}}
\nonumber\\
d^{D-1} {\sigma}_{\alpha_{1}\dots\alpha_{D-1}}
 =& \epsilon_{\beta_1 \dots \beta_{D-1}} 
 \frac{\partial \bar{x}^{\alpha_1}}{\partial \sigma_{\beta_1}} \dots  \frac{\partial \bar{x}^{\alpha_{D-1}}}{\partial \sigma_{\beta_{D-1}}}  d\sigma_{1} \dots d\sigma_{D-1}
  .
\end{align}
The volume element $d^{D-1} {\sigma}_{\alpha_{1}\dots\alpha_{D-1}}$  defines an orientation of the vortex. In general, two different orientations of vortices can be distinguished. 
Note that the ideal vortex field $v$ must be singular. Otherwise, the magnetic current becomes trivial $k=0$, since $k=\delta {}^{\displaystyle *}f= \delta {}^{\displaystyle *}dv={}^{\displaystyle *}ddv=0$.

It will be shown later that when the hypersurface $V_{D-1}$ of the ideal vortex configuration intersects with a Wilson loop $C$, the ideal vortex  contribute a center group element to the Wilson loop operator. 
The boundary $S_{D-2}=\partial V_{D-1}$ gives the location of the thin vortex $s_\mu(x;S_{D-2})$ to which the ideal vortex configuration $v_\mu(x;V_{D-1})$ is gauge equivalent.

The ideal vortex field $v^\mu(x;V_{D-1})$  is not unique as mentioned above. 
Actually, the ideal vortex field $v^\mu(x;V_{D-1})$ can be gauge transformed to a \textbf{thin vortex field} $s^\mu(x;S_{D-2})$ which has the support only on the boundary $S_{D-2}=\partial V_{D-1}$ of $V_{D-1}$:  
\begin{equation}
  s_\mu(x;S_{D-2}) =  v_\mu(x;V_{D-1}) + iU(x;V_{D-1}) \partial_\mu U^\dagger(x;V_{D-1}) 
  , \quad (S_{D-2}=\partial V_{D-1})
  \label{C29-gt}
\end{equation}
so that 
$s^\mu(x;S_{D-2})$ carries the same magnetic flux located on $S_{D-2}=\partial V_{D-1}$ as that carried by $v^\mu(x;V)$.
In fact, for $v=s-d\Omega$, $dv$ agrees with $ds$, since 
$f=dv=d(s-d\Omega)=ds-d^2\Omega=ds$.
In other words, $s^\mu$ and $v^\mu$ are gauge equivalent.

It is shown that such a gauge transformation is given by 
\begin{equation}
  U(x;V_{D-1}) = \exp \left[ i\Phi \Omega(x;V_{D-1}) \right] 
   ,
\end{equation}
where $\Omega(x;V_{D-1})$ is the solid angle taken up by the volume $V_{D-1}$ when viewed from $x$:
\begin{align}
   \Omega(x;V_{D-1}) 
:= \frac{-1}{A_{D-1}} \int_{V_{D-1}} d^{D-1} \tilde{\sigma}_\mu \ \frac{x^\mu-\bar{x}^\mu(\sigma)}{[(x^\mu-\bar{x}(\sigma))^2]^{D/2}} ,
   \quad 
  A_{D-1} := \frac{2\pi^{D/2}}{\Gamma(D/2)} 
   ,
\end{align}
with $A_{D-1}$ being the area of the unit sphere $S^{D-1}$ in $D$ dimensions, e.g., $A_{2}=4\pi$, $A_{3}=2\pi^2$.  
The solid angle $\Omega(x;V_{D-1})$ defined in this way is normalized to unity for a point $x$ inside the volume $V$.  
Note that the solid angle is defined with a sign depending on the orientation of $V_{D-1}$ as rays emanating from $x$ pierce $V_{D-1}$. 
A deformation of $V_{D-1}$ keeping its boundary $S_{D-2}$ fixed leaves the solid angle invariant unless $x$ crosses $V_{D-1}$.  When $x$ crosses $V_{D-1}$, the solid angle changes by an integer.% 
\footnote{
Whether the flux of a vortex is electric or magnetic depends on the position of the $D-2$ dimensional vortex surface $S$ in $D$-dimensional space--time. For example, in $D=4$ the vortex defined by the boundary of a purely spatial 3-dimensional volume $V$ carries only electric flux, which is directed normal to the vortex surface $S=\partial V$.  On the other hand, a vortex $S=\partial V$ defined by a volume $V$ evolving in time represents at a fixed time a closed loop and carries the magnetic flux, which is tangential to the vortex loop \cite{ER00}. 
}

It is easy to check that the solid angle is rewritten as
\begin{equation}
   \Omega(x;V_{D-1}) := \int_{V_{D-1}} d^{D-1} \tilde{\sigma}_\mu \partial_\mu^{x} G(x-\bar{x}(\sigma)) 
   ,
   \label{C29-solid-angle-D2}
\end{equation}
by using   the Green function $G(x-\bar{x}(\sigma))$ of the $D$-dimensional Laplacian defined by
\begin{align}
  - \partial_\mu^{x} \partial_\mu^{x} G(x-\bar{x}(\sigma)) = \delta^D(x-\bar{x}(\sigma))
  ,
\end{align}
which has the explicit form:
\begin{align}
 G(x-y)
 = \frac{\Gamma(D/2-1)}{4\pi^{D/2}[(x-y)^2]^{(D-2)/2}} 
  .
\end{align}

Now we show that the thin vortex field $s_\mu(x;S)$ manifestly depends only on the boundary $S=\partial V$ where the magnetic flux associated with the vortex located: 
\begin{align}
 s_\mu(x;S_{D-2}) 
=& \Phi   \int_{S_{D-2}=\partial V_{D-1}} d^{D-2} \tilde{\sigma}_{\mu\lambda} \ \partial_\lambda G(x-\bar{x}(\sigma))
\nonumber\\
=& \Phi   \int_{S_{D-2}=\partial V_{D-1}} d^{D-2}  {\sigma}_{\alpha_{1}\dots\alpha_{D-2}} \  \frac{1}{(D-2)!} \epsilon_{\mu\lambda\alpha_{1}\dots\alpha_{D-2}} \partial_\lambda G(x-\bar{x}(\sigma))
 ,
  \label{C29-tv-field}
\end{align}
where 
$
d^{D-2} \tilde{\sigma}_{\mu\lambda}
=   \frac{1}{(D-2)!} \epsilon_{\mu\lambda\alpha_{1}\dots\alpha_{D-2}}
d^{D-2} {\sigma}_{\alpha_{1}\dots\alpha_{D-2}}
$
 is the dual of the  volume element:
\begin{align}
d^{D-2} {\sigma}_{\alpha_{1}\dots\alpha_{D-2}}
 = \epsilon_{\beta_1 \dots \beta_{D-2}} 
 \frac{\partial \bar{x}^{\alpha_1}}{\partial \sigma_{\beta_1}} \dots  \frac{\partial \bar{x}^{\alpha_{D-2}}}{\partial \sigma_{\beta_{D-2}}}  d\sigma_{1} \dots d\sigma_{D-2}
  .
\end{align}
The ideal vortex field $v_\mu(x;V)$ has support only on $V$ and hence it vanishes outside the volume $V$, i.e., $v_\mu(x;V)=0$ for $x \notin V$.  Therefore,  outside the  volume $V$, i.e., $x \in \mathbb{R}^D - V$, the thin vortex field $s_\mu(x;S)$ is the pure gauge due to (\ref{C29-gt}):
\begin{equation}
  s_\mu(x;S_{D-2}) =    iU(x;V_{D-1}) \partial_\mu U^\dagger(x;V_{D-1}) , \quad x \notin V_{D-1}
  ,
\end{equation}
which implies the vanishing field strength $f_{\mu\nu}(x)=0$ outside the volume $V$.  This is reasonable, because the magnetic flux is contained in the vortex sheet $S$.
%The magnetic field computed from the curl of $a_\mu(x;S)$ can be localized on $C^\prime=\partial S$.

The derivation of (\ref{C29-gt}) and (\ref{C29-tv-field}) is as follows.
By using (\ref{C29-solid-angle-D2}),
%the fact that the solid angle is rewritten as
%\begin{equation}
%   \Omega_V(x) := \int_{V_{D-1}} d^{D-1} \tilde{\sigma}_\mu \partial_\mu^{x} D(x-\bar{x}(\sigma))    ,
%\end{equation}
we obtain
\begin{align}
 iU(x;V_{D-1}) \partial_\mu U^\dagger(x;V_{D-1}) 
  =   \Phi \partial_\mu \Omega(x;V_{D-1})
%\nonumber\\
  =  \Phi \int_{V_{D-1}} d^{D-1} \tilde{\sigma}_\nu \partial_\mu^{x} \partial_\nu^{x}  G(x-\bar{x}(\sigma)) 
 .
\end{align}
We have the decomposition:
\begin{align}
  \partial_\mu \partial_\nu
  =   \delta_{\mu\nu} \partial^2 - ( \delta_{\mu\nu} \partial^2 - \partial_\mu \partial_\nu)
%\nonumber\\
%=& \delta_{\mu\nu} \partial^2 - \frac12 \epsilon_{\mu\rho\alpha\beta} \epsilon_{\nu\sigma\alpha\beta} \partial_\rho \partial_\sigma 
=& \delta_{\mu\nu} \partial^2 - \frac{1}{(D-2)!} \epsilon_{\mu\rho\alpha_{1}\cdots\alpha_{D-2}} \epsilon_{\nu\sigma\alpha_{1}\cdots\alpha_{D-2}} \partial_\rho \partial_\sigma 
 ,
\end{align}
following from the identity:
\begin{align}
  \epsilon_{\mu\rho\alpha_{1}\cdots\alpha_{D-2}} \epsilon_{\nu\sigma\alpha_{1}\cdots\alpha_{D-2}} 
  = (D-2)! (\delta_{\mu\nu} \delta_{\rho\sigma} - \delta_{\mu\sigma} \delta_{\rho\nu}) , 
\end{align}
which is consistent with
$ 
\epsilon_{\mu_{1} \cdots\alpha_{D}} \epsilon_{\mu_{1} \cdots\alpha_{D}} = D!
$.
This is used to rewrite the pure gauge form into 
\begin{align}
& iU(x;V_{D-1}) \partial_\mu U^\dagger(x;V_{D-1}) 
%\nonumber\\
%  =&  \Phi \partial_\mu \Omega_V(x)
\nonumber\\
  =& \Phi \int_{V_{D-1}} d^{D-1} \tilde{\sigma}_\mu \partial^2  G(x-\bar{x}(\sigma)) 
%\nonumber\\&
+ \Phi \int_{V_{D-1}} d^{D-1} \tilde{\sigma}_\nu  \epsilon_{\nu\sigma\alpha_{1}\cdots\alpha_{D-2}} \bar\partial_\sigma  \frac{1}{(D-2)!}\epsilon_{\mu\rho\alpha_{1}\cdots\alpha_{D-2}}  \partial_\rho  G(x-\bar{x}(\sigma))  
\nonumber\\
  =& -\Phi \int_{V_{D-1}} d^{D-1} \tilde{\sigma}_\mu  \delta^D(x-\bar{x}(\sigma)) 
%\nonumber\\&
+ \Phi \int_{\partial V_{D-1}} d^{D-2} \tilde{\sigma}_{\alpha_{1}\cdots\alpha_{D-2}}   \frac{1}{(D-2)!} \epsilon_{\mu\rho\alpha_{1}\cdots\alpha_{D-2}}  \partial_\rho  G(x-\bar{x}(\sigma))  
\nonumber\\
  =& - v_\mu(x;V_{D-1}) + s_\mu(x;S_{D-2}=\partial V_{D-1})
%\nonumber\\
%=& -\Phi \int_{V_{D-1}} d^{D-1} \tilde{\sigma}_\mu  \delta^D(x-\bar{x}(\sigma)) 
%\nonumber\\&
%- \Phi \int_{\partial V_{D-1}} d^{D-2} \tilde{\sigma}_{\mu\lambda}  \partial_\lambda   D(x-\bar{x}(\sigma))  
 ,
\end{align}
where we have used the definition of the Green function  $G$ and the Stokes theorem in the second equality. 
In other words, the thin vortex field $s_\mu(x;S)$ is the transverse part of the ideal vortex $v_\mu(x;V)$.

Finally, we find that the surface integral of $f(x;S)$ over $\Sigma$ bounded by the Wilson loop $C$  is equivalent to the line integral of $s(x;S)$ along the closed loop $C$:
\begin{align}
 \int_{\Sigma} f(x;S_{D-2}) 
= \int_{\Sigma} dv(x;V_{D-1}) 
%\nonumber\\
  =&  \oint_{\partial \Sigma=C}  v(x;V_{D-1}) 
\nonumber\\
  =& \oint_{C} [s(x;S_{D-2}) - iU(x;V_{D-1}) d U^\dagger(x;V_{D-1}) ]
\nonumber\\
  =& \oint_{C}  s(x;S_{D-2}) 
%\nonumber\\
  =  \int_{\Sigma}  ds(x;S_{D-2}) 
 ,
\end{align}
since the contribution from the last term $iU(x;V_{D-1}) d U^\dagger(x;V_{D-1})=\Phi d \Omega(x;V_{D-1})$ vanishes for any closed loop $C$.
Then the line integral is cast into 
\begin{align}
&  \oint_{C} dx^\nu s_\nu(x;S_{D-2}) 
 \nonumber\\
  =& \int_{\Sigma} d^2 \sigma_{\mu\nu}(x) \partial_\mu s_\nu(x;S_{D-2}) 
\nonumber\\
  =& \Phi \int_{\Sigma} d^2 \sigma_{\mu\nu}(x)   \int_{S_{D-2}=\partial V_{D-1}} d^{D-2} {\sigma}_{\alpha_{1}\dots\alpha_{D-2}} \  \frac{1}{(D-2)!} \epsilon_{\nu\lambda\alpha_{1}\dots\alpha_{D-2}} \partial_\mu \partial_\lambda G(x-\bar{x}(\sigma))
%\nonumber\\
%  =& \Phi L(C, S_{D-2}=\partial V_{D-1}) 
 .
\label{C29-line-int}
\end{align}

\subsubsection{$D=3$ vortex}
  
%%%%%%%%%%%%%%%%%%%%%%%%%%%%%%%%%%%%%%%%%%%%%%%%%%%%%%%%%%%%%
%%%%%%%%%%%%%%%%%%%%%%%%%%%%%%%%%%%%%%%%%%%%%%%%%%%%%%%%%%%%%
\begin{figure}[ptb]
\begin{center}
\includegraphics[scale=0.40,clip]{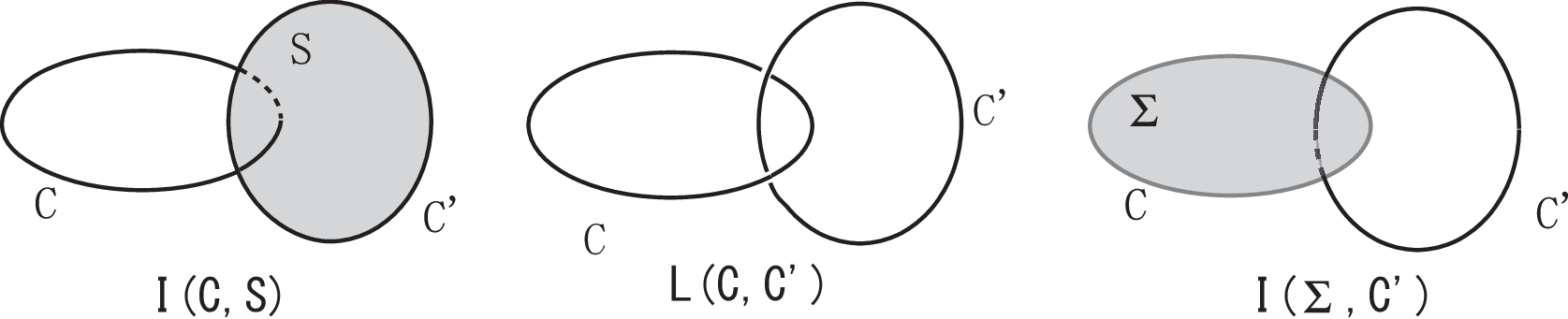}
\end{center} 
\vskip -0.5cm
\caption[]{ \cite{Kondo08b}
  In $D=3$ dimensional space--time,  a non-trivial contribution to the Wilson loop operator comes from    
 (Center panel) 
 the linking $L(C_1,C_1^\prime)$ between the Wilson loop $C_1$ and a  vortex loop $C_1^\prime$ as a boundary of the vortex sheet $S_2$.
(Left panel) intersection $I(C_1=\partial \Sigma_2,S_2)$  between the  Wilson loop $C_1$ and an open vortex sheet $S_2$ bounding the closed loop $C_1^\prime$, 
or 
(Right panel) intersection $I(\Sigma_2,C_1^\prime=\partial S_2)$ between the open surface $\Sigma_2$ bounding the Wilson  loop $C_1$ and a vortex loop $C_1^\prime$. Three are equivalent descriptions: 
$L(C_1,C_1^\prime)=I(C_1=\partial \Sigma_2,S_2)=I(\Sigma_2,C_1^\prime=\partial S_2)$.
}
 \label{C29-fig:linking-3d}
\end{figure}
%%%%%%%%%%%%%%%%%%%%%%%%%%%%%%%%%%%%%%%%%%%%%%%%%%%%%%%%%%%%%
%%%%%%%%%%%%%%%%%%%%%%%%%%%%%%%%%%%%%%%%%%%%%%%%%%%%%%%%%%%%%

%%%%%%%%%%%%%%%%%%%%%%%%%%%%%%%%%%%%%%%%%%%%%%%%%%%%%%%%%%%%%
\begin{figure}[tb]
\begin{center}
\includegraphics[scale=0.30]{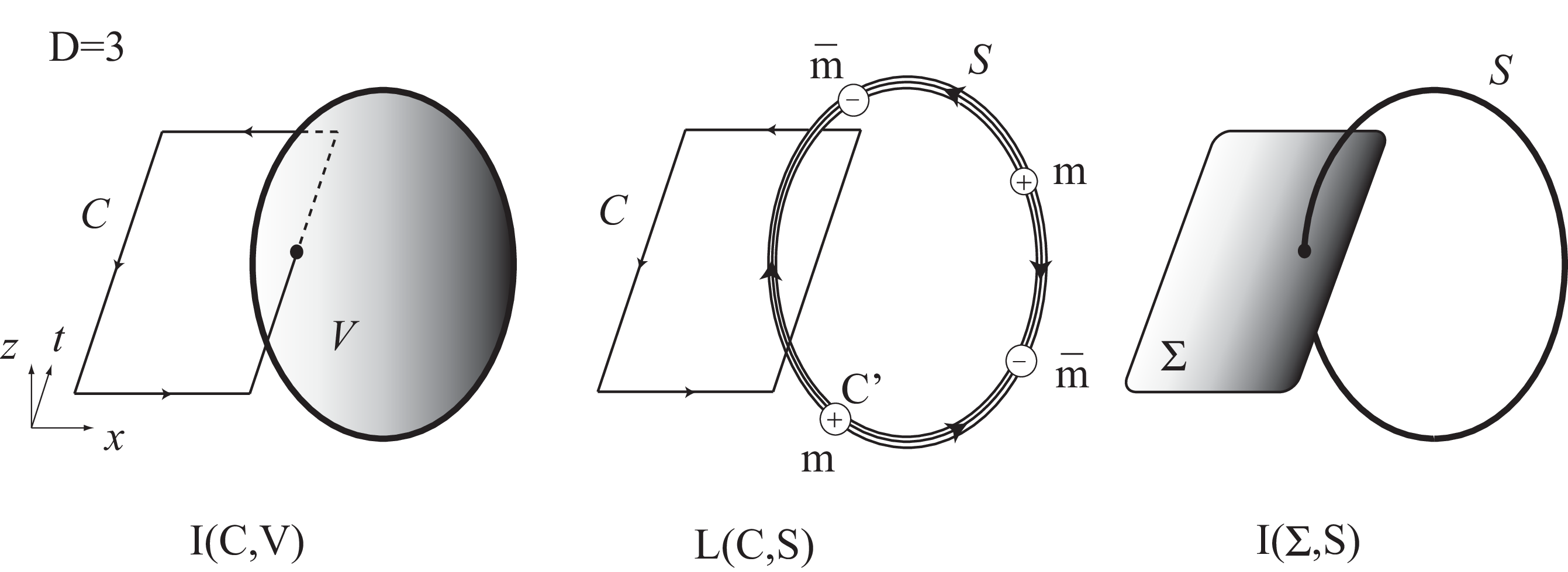}
\end{center} 
\vskip -0.5cm
\caption[]{ %\cite{Kondo08b}
  In $D=3$ dimensional space--time, it is assumed that the magnetic field  emanating from a magnetic monopole $m$ or an antimonopole $\bar m$ is squeezed into the flux tube to form the vortex line $S$.  
A closed vortex line $S$ consists of an even number of connected pieces $S_n'$ $(S=\cup_{n}S_n')$, and each piece $S_n'$ of a vortex line starts at a monopole $m$ with a positive magnetic charge and ends at an antimonopole $\bar m$ with a negative magnetic charge. This is possible if the closed vortex line $S$ is not orientable. The vortex line is closed, but its boundary can be non-trivial due to this non-orientability. 
The magnetic flux going along a vortex is just half of the total magnetic flux emanating from a monopole. 
}
 \label{C29-fig:vortex3d}
\end{figure}
%%%%%%%%%%%%%%%%%%%%%%%%%%%%%%%%%%%%%%%%%%%%%%%%%%%%%%%%%%%%%

For $D=3$, in particular, the magnetic-monopole current is given by
\begin{align}
 k(x) = k(x;C^\prime_{0}) := \Phi  \delta^3(x-\bar{x}(\sigma)) ,
\end{align}
the ideal vortex field $v_\mu$ with the support on the surface $V_2$ is written as
\begin{align}
 v_\mu(x;V_2) =& \Phi   \int_{V_{2}:\partial V_{2}=S_{1}} d^{2} \tilde{\sigma}_{\mu} \ \delta^3(x-\bar{x}(\sigma))  ,
\end{align}
and the thin vortex field $s_\mu$ with the support on the closed loop $S_1=\partial V_2$ is written as
\begin{align}
 s_\mu(x;S_1) 
=& \Phi   \int_{S_{1}=\partial V_{2}} d  \tilde{\sigma}_{\mu\lambda} \ \partial_\lambda G(x-\bar{x}(\sigma))
=  \Phi   \int_{S_{1}=\partial V_{2}} d {\sigma}_{\alpha } \ \epsilon_{\mu\lambda\alpha } \partial_\lambda G(x-\bar{x}(\sigma))
 .
\end{align}
By using (\ref{C29-line-int}), the line integral reads
\begin{align}
   \oint_{C} dx^\nu s_\nu(x;S_1) 
%\nonumber\\
  =& \Phi \int_{\Sigma} d^2 \sigma_{\mu\nu}(x)   \int_{S_{1}=\partial V_{2}} d^{}  {\sigma}_{\alpha }   \epsilon_{\nu\lambda\alpha } \partial_\mu \partial_\lambda G(x-\bar{x}(\sigma))
\nonumber\\
  =& \Phi \int_{\Sigma} d^2 \tilde\sigma_{\rho}(x)   \epsilon_{\mu\nu\rho} \int_{S_{1}=\partial V_{2}} d {\sigma}_{\alpha }   \epsilon_{\nu\lambda\alpha } \partial_\mu \partial_\lambda G(x-\bar{x}(\sigma))
\nonumber\\
  =& \Phi \int_{\Sigma} d^2 \tilde\sigma_{\rho}(x)     \int_{S_{1}=\partial V_{2}} d {\sigma}_{\alpha }   (\delta_{\rho\lambda}\delta_{\mu\alpha} - \delta_{\rho\alpha}\delta_{\mu\lambda}) \partial_\mu \partial_\lambda G(x-\bar{x}(\sigma))
\nonumber\\
  =& \Phi \int_{\Sigma} d^2 \tilde\sigma_{\rho}(x)     \int_{S_{1}=\partial V_{2}} d {\sigma}_{\alpha }    \partial_\alpha \partial_\rho G(x-\bar{x}(\sigma)) 
%\nonumber\\&
-  \Phi \int_{\Sigma} d^2 \tilde\sigma_{\rho}(x)     \int_{S_{1}=\partial V_{2}} d {\sigma}_{\rho}   \partial^2  G(x-\bar{x}(\sigma))
\nonumber\\
  =&   \Phi \int_{\Sigma} d^2 \tilde\sigma_{\rho}(x)     \int_{S_{1}=\partial V_{2}} d {\sigma}_{\rho}      \delta^3(x-\bar{x}(\sigma))
  =    \Phi I(\Sigma,S_1)
,
\end{align}
where we have used the fact that the first term vanishes  in the second to last equality, since it is the line integral on a closed loop $C_1'=\partial S_1=\partial \partial V_2$ due to the Stokes theorem.

Here $I$ is the intersection number between the Wilson surface $\Sigma$ (the string world sheet) and the vortex loop $S_1$. 
\begin{align}
 I(\Sigma_2,S_1) 
= \int_{\Sigma_2} d^2 \tilde\sigma_{\rho}(x) \int_{S_{1}
=\partial V_{2}} d {\sigma}_{\rho}      \delta^3(x-\bar{x}(\sigma)) 
= \int_{\Sigma_2} d^2 \sigma_{\alpha\beta}(x) \frac12 \epsilon_{\alpha\beta\gamma} \int_{S_{1}
=\partial V_{2}} d  {\sigma}_{\gamma}      \delta^3(x-\bar{x}(\sigma))
 .
\end{align}
There are three equivalent descriptions. See Fig.~\ref{C29-fig:linking-3d}. 
The intersection number $I(\Sigma_2,S_1)$ between the Wilson surface $\Sigma$ and the vortex loop $S_1$ is equal to the intersection number $I(C_1,V_2)$ between the Wilson loop $C_1$ and the vortex sheet $V_2$, which is equal to the linking number $L(C_1,S_1)$ between the the Wilson loop $C_1$ and the vortex loop  $S_1$. 
The intersection numbers and the linking number are integer valued: 
\begin{align}
 I(\Sigma_2,S_1)  
=I(C_1=\partial \Sigma_2,V_2)
= L(C_1,S_1)   \in \mathbb{Z} = \{ 0, \pm 1, \pm 2, \cdots \} .
\end{align}

In $D=3$ dimensional space--time, the contribution to the $SU(2)$ Wilson loop operator from the vortices is estimated as the factor: 
\begin{align}
W_C^{\rm vor} := \exp \left[ iJg_{{}_{\rm YM}} \int_{\Sigma} f(x,S) \right] 
=  \exp \left[ iJg_{{}_{\rm YM}} \oint_{S} s(x,S) \right] 
=  \exp \left[ iJg_{{}_{\rm YM}}    \Phi I(\Sigma,S_1) \right] 
. 
\end{align}
If the magnetic flux carries a unit of the magnetic flux according to  the quantization condition:
\begin{align}
\Phi= 2\pi g_{{}_{\rm YM}}^{-1} %n \ (n \in \mathbb{Z})
,
\end{align}
then a vortex as the tube of the magnetic flux contributes the factor (an element of the center group) to the $SU(2)$ Wilson loop operator:
\begin{align}
 W_C^{\rm vor}  
=   e^{ iJ 2\pi  L }
= (-1)^{2J L} \in Z(2), 
\quad 
L = I(\Sigma_2,S_1)  
=I(C_1=\partial \Sigma_2,V_2)
= L(C_1,S_1)
 . 
\end{align}
This becomes non-trivial only when $J$ is a half-integer, $J=\frac12, \frac32, ...$, since the intersection number and the linking number are integers, $L, I \in \mathbb{Z}$. 
This result is consistent with the estimate obtained from the magnetic monopole (\ref{C29-mono-Wcont}). 
This is the most non-trivial contribution, which corresponds to the half of the total magnetic flux emanating from a point magnetic charge.

This is consistent with the claim that a center vortex configuration will appear (after Abelian projection) as a chain of monopoles alternating with antimonopoles, i.e., monopole-antimonopole chain.
If this is the case, then the $4\pi$ monopole flux is not distributed symmetrically on the Abelian-projected lattice, in contrast to the situation one might expect in a Coulomb gas.
Rather, it is collimated in units of $\pm 2\pi$ along the vortex line 
\cite{Greensite03,DFGO97,AGG00}. 
If this picture is correct, then a vortex sheet on $V_2$, a vortex line on $S_1=\partial V_2$, and the magnetic monopoles on $C_0=\partial S_1= \partial\partial V_2$ have the mutual positional relation as shown in Fig.~\ref{C29-fig:vortex3d}.
Here the vortex line is non-orientable. Otherwise, the magnetic monopoles do not exist, since the orientable vortex loop has no boundary $C_0=\partial S_1= \partial\partial V_2= \phi$.

\subsubsection{$D=4$ vortex}

%%%%%%%%%%%%%%%%%%%%%%%%%%%%%%%%%%%%%%%%%%%%%%%%%%%%%%%%%%%%%
%%%%%%%%%%%%%%%%%%%%%%%%%%%%%%%%%%%%%%%%%%%%%%%%%%%%%%%%%%%%%
\begin{figure}[ptb]
\begin{center}
\includegraphics[scale=0.35,clip]{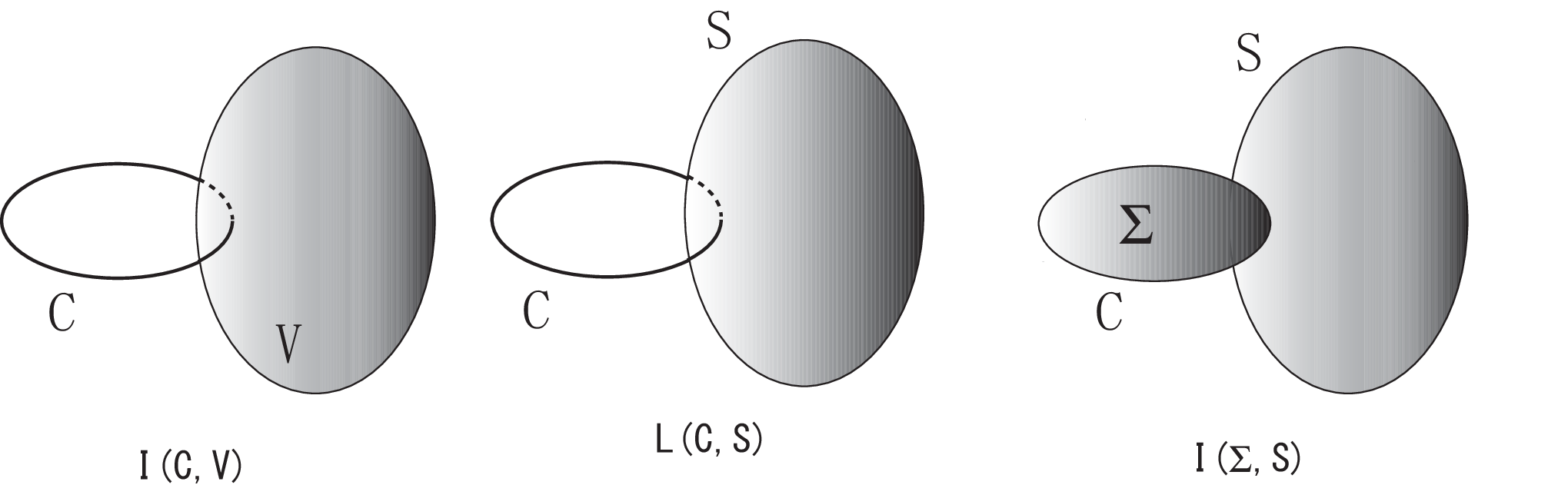}
\end{center} 
\vskip -0.5cm
\caption[]{\cite{Kondo08b}
 In $D=4$ dimensional case,  
(Center panel) Linking $L(C_1,S_2)$ between the Wilson  loop $C_1$ and a closed 2-dimensional vortex surface $S_2$ bounding a 3-dimensional  vortex $V_3$. 
(Left panel) Intersection $I(C_1=\partial \Sigma_2,V_3)$ between the Wilson loop $C_1$ and  the vortex $V_3$.
(Right panel) Intersection $I(\Sigma_2,S_2=\partial V_3)$ between a minimal surface $\Sigma_2$ bounded by the Wilson loop $C_1$ and a closed  vortex surface $S_2$ in 4-dimensional space--time. They are equivalent:
$L(C_1,S_2)=I(C_1=\partial \Sigma_2,V_3)=I(\Sigma_2,S_2=\partial V_3)$.
}
 \label{C29-fig:linking-d4}
\end{figure}
%%%%%%%%%%%%%%%%%%%%%%%%%%%%%%%%%%%%%%%%%%%%%%%%%%%%%%%%%%%%%
%%%%%%%%%%%%%%%%%%%%%%%%%%%%%%%%%%%%%%%%%%%%%%%%%%%%%%%%%%%%%

%%%%%%%%%%%%%%%%%%%%%%%%%%%%%%%%%%%%%%%%%%%%%%%%%%%%%%%%%%%%%
\begin{figure}[tb]
\begin{center}
\includegraphics[scale=0.30]{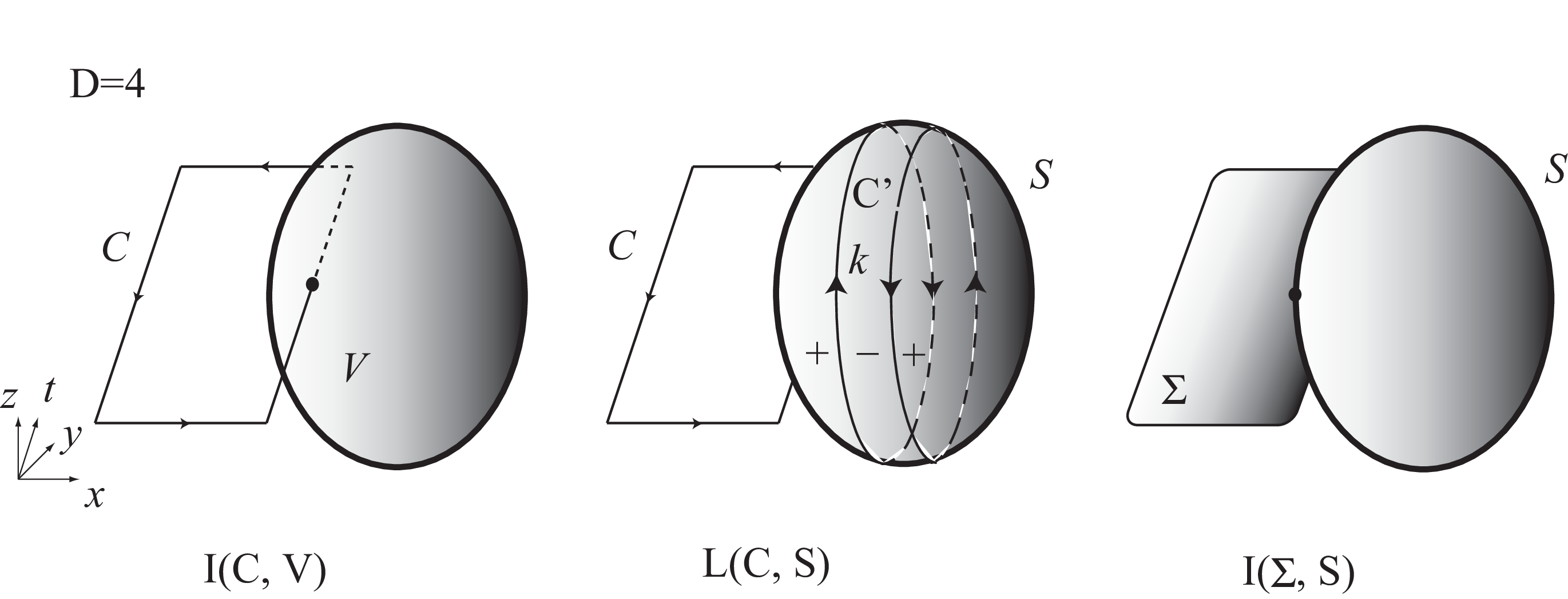}
\end{center} 
\vskip -0.5cm
\caption[]{ %\cite{Kondo08b}
 In $D=4$ dimensional space--time, it is assumed that the magnetic flux emanating from the magnetic-monopole current $k$ is constrained on a surface to form the vortex surface $S$.
A closed vortex surface $S$ consists of a number of connected pieces $S_n$ $(S=\cup_{n}S_n)$, and each piece $S_n$ has the magnetic-monopole current $k$ at its boundary $\partial S_n$.  This is possible if the closed vortex surface is non-orientable, although each piece of the vortex surface is orientable. 
}
 \label{C29-fig:vortex4d}
\end{figure}
%%%%%%%%%%%%%%%%%%%%%%%%%%%%%%%%%%%%%%%%%%%%%%%%%%%%%%%%%%%%%

The above consideration can be extended to higher dimensions.
For $D=4$, $V_3$ is a three-dimensional volume and its boundary $S_2$ is a closed two-dimensional surface $S_2=\partial V_3$. 
If $S_2$ is an oriented closed surface, then its boundary $C_1^\prime=\partial S_2$ is empty, and hence the magnetic current $k^\mu(x;C_1^\prime)$ does not exist in this case, since its support $C_1^\prime=\partial S_2=\partial \partial V_3$ vanishes. 
Therefore, for the non-vanishing magnetic current $k_\mu(x;C_1^\prime)$ to exist in the boundary $C_1^\prime=\partial S_2$ of $S_2$, the vortex surface $S_2=\partial V_3$ must be non-oriented.  
%See \cite{Kondo08b} for more details. 
For $D=4$, we find 
\begin{align}
   \oint_{C} dx^\nu s_\nu(x;S_2) 
%\nonumber\\
%=& \int_{\Sigma} d^2 \sigma_{\mu\nu}(x) \partial_\mu s_\nu(x;S) 
%\nonumber\\
%=& \Phi \int_{\Sigma} d^2 \sigma_{\mu\nu}(x)   \int_{S_{D-2}} d^{D-2} \tilde{\sigma}_{\alpha_{1}\dots\alpha_{D-2}} \  \frac{1}{(D-2)!} \epsilon_{\nu\lambda\alpha_{1}\dots\alpha_{D-2}} \partial_\mu \partial_\lambda D(x-\bar{x}(\sigma))
%\nonumber\\
  =& \Phi \int_{\Sigma} d^2 \sigma_{\mu\nu}(x)   \int_{\partial V_{3}} d^{2} \tilde{\sigma}_{\alpha_{1} \alpha_{2}} \  \frac{1}{2} \epsilon_{\nu\lambda\alpha_{1} \alpha_{2}} \partial_\mu \partial_\lambda G(x-\bar{x}(\sigma))
\nonumber\\
  =& \Phi \frac{1}{2} \int_{\Sigma} d^{2} \tilde{\sigma}_{\rho\sigma}(x) \epsilon_{\mu\nu\rho\sigma} \int_{S_{2}=\partial V_{3}} d^{2}  \sigma_{\alpha\beta}(\bar{x}) \frac{1}{2} \epsilon_{\nu\lambda\alpha\beta} \partial_\mu  \partial_\lambda   G(x-\bar{x}) 
\nonumber\\
  =& \Phi \frac{1}{2} \int_{\Sigma} d^{2} \tilde{\sigma}_{\alpha\beta}(x)  \int_{S_{2}} d^{2}  \sigma_{\alpha\beta}(\bar{x}) \partial^2   G(x-\bar{x}) 
\nonumber\\
  &  + \Phi  \int_{\Sigma} d^{2} \tilde{\sigma}_{\beta\sigma}(x)  \int_{S_{2}} d^{2}  \sigma_{\alpha\beta}(\bar{x}) \partial_\alpha  \partial_\sigma   G(x-\bar{x}) 
\nonumber\\
  =& \Phi \frac{1}{2} \int_{\Sigma} d^{2} \tilde{\sigma}_{\alpha\beta}(x)  \int_{S_{2}} d^{2}  \sigma_{\alpha\beta}(\bar{x}) \delta^4(x-\bar{x}) 
%\nonumber\\
%  =& \Phi L(C, S_{D-2}=\partial V_{D-1}) 
,
\end{align}
where we have used the fact that the second term vanishes due to the Stokes theorem $\partial \partial V_3 =0$ for obtaining the last result. 
Therefore, the line integral is rewritten as
\begin{align}
 \oint_{C} dx^\mu s_\mu(x;S_2) 
 =  \Phi  I(\Sigma_2,S_{2}=\partial V_{3})  
,
\end{align}
in terms of the intersection number $I(\Sigma_2,S_2)$ between the world sheet $\Sigma_2$ of the hadron string and the vortex sheet $S_2$ in $D=4$ dimensions.  
It is known that the intersection number $I(\Sigma_2,S_2)$ is equal to the linking number $L(C_1=\partial \Sigma_2, S_2=\partial V_3)$ between the Wilson loop $C$ and the vortex sheet $S_2$, see Fig.~\ref{C29-fig:linking-d4}:
\begin{align}
I(\Sigma_2,S_{2}=\partial V_{3}) =& 
 L(C_1=\partial \Sigma_2, S_{2}=\partial V_{3}) 
\nonumber\\
  =&  \frac{1}{2} \int_{\Sigma} d^{2} \tilde{\sigma}^{\alpha\beta}(x) \int_{S} d^2 \sigma^{\alpha\beta}(\bar{x}) \delta^4(x-\bar{x})
 .
\end{align}

Thus the thin vortex  contributes to the $SU(2)$ Wilson loop in the fundamental representation:
\begin{align}
\exp \left[ i \frac12 g_{{}_{\rm YM}} \int_{\Sigma} f(x;S) \right]
%  \ni W_C^m 
=  \exp \left[ i \frac12 g_{{}_{\rm YM}}   \Phi L(C, S=\partial V)  \right]
 =  z^{L(C, S=\partial V)}
, \quad
%\end{align}
%where
%\begin{align}
 z := e^{i\frac12 g_{{}_{\rm YM}} \Phi} 
 .
\end{align}
In general, $z$ is a complex number of modulus one, i.e., an element of $U(1)$. 
In the case of $SU(2)$ gauge group, the magnetic charge $q_m$ measured by the Wilson loop is subject to the Dirac quantization condition:  
\begin{equation}
  q_m =  4\pi g_{{}_{\rm YM}} ^{-1} n , \  n \in \mathbb{Z} 
  .
\end{equation}
If the magnetic flux obeys the  quantization condition with an additional factor $f_2$:
\begin{align}
\Phi= 4\pi g_{{}_{\rm YM}} ^{-1} f_2 n \ (n \in \mathbb{Z})
,
\label{C29-qc}
\end{align}
then $z$ reduces to  
\begin{align}
 z = e^{i\frac12 g_{{}_{\rm YM}} \Phi} = e^{i2\pi f_2 n}  
 .
\end{align}
Therefore, $z$ reduces to the center element $\mathbb{Z}_2$ of $SU(2)$:
\begin{align}
  z {\bf 1} \in \mathbb{Z}_2, \quad z = e^{i\pi n} = \pm 1  
 ,
 \label{C29-center}
\end{align}
only when 
\begin{equation}
  f_2 =  \frac12   
  .
\end{equation}
%The quantization condition (\ref{C29-qc}) will be called the  \textit{fractional} quantization condition which happens to agree with the Dirac one for SU(2), which is realized as a special case of the general quantization condition for $SU(N)$ discussed later. 
%If the magnetic charge obeyed the quantization condition
%\begin{align}
%\Phi= 4\pi g^{-1} n \ (n \in \mathbb{Z}),
%\label{C29-qc2}
%\end{align}
%then $z$ would be trivial, i.e., $z=1$.
%Such a vortex can not give a non-trivial contribution to the Wilson loop.
%Therefore, the thin vortex carrying the fractional magnetic charge yields a center element under the fractional magnetic charge quantization (\ref{C29-qc}). 
%Is there any relationship between the non-orientedness of the vortex surface and the fractional magnetic charge (\ref{C29-qc})?  We will try to answer this question in the next section. 

 In the case of $SU(N)$ gauge group, it has been shown \cite{Kondo08} that the magnetic charge $q_m$ measured by the Wilson loop is subject to the quantization condition:  
\begin{equation}
  q_m =  \frac{2\pi}{g_{{}_{\rm YM}} } \sqrt{\frac{2N}{N-1}} n , \  n \in \mathbb{Z} 
  ,
\end{equation}
which is analogous to the Dirac type, but different from it. 
Suppose that the magnetic flux $\Phi$ obeys the fractional quantization condition:
\begin{equation}
  \Phi = f_N \frac{2\pi}{g_{{}_{\rm YM}} } \sqrt{\frac{2N}{N-1}} n , \  n \in \mathbb{Z} 
  .
\end{equation}
Then its  contribution to the Wilson loop is written as 
\begin{align}
\exp \left[ i  \sqrt{\frac{N-1}{2N}} g_{{}_{\rm YM}} \int_{\Sigma} f(x;S) \right]
%  \ni&  W_C^m    
 = \exp \left[ i \sqrt{\frac{N-1}{2N}} g_{{}_{\rm YM}}  \Phi L(C, S=\partial V) \right]
  = z_{N}^{L(C, S=\partial V)}
, 
\end{align}
where
\begin{align}
 z_{N} :=    \exp \left\{ i \sqrt{\frac{N-1}{2N}} g_{{}_{\rm YM}}  \Phi  \right\} 
= e^{2\pi i f_N n} 
 .
\end{align}
The center of $SU(N)$ is $\mathbb{Z}_N$. 
For $z_{N} {\bf 1}$ to belong to the center of $SU(3)$, i.e., $\mathbb{Z}_3 \ni \{  {\bf 1}, e^{2\pi i/3} {\bf 1}, e^{4\pi i/3} {\bf 1}=e^{-2\pi i/3} {\bf 1} \}$, $f_3$ must take the fractional number except for a trivial case $f_N=0$:
\begin{equation}
  f_3 =  \frac13, \quad  \frac23 
  ,
\end{equation}
just as  the aforementioned $SU(2)$ case with the center subgroup  $\mathbb{Z}_2 \ni \{  {\bf 1},   e^{\pi i} {\bf 1}=- {\bf 1} \}$, when 
$
  f_2 =  \frac12   
$.
See Fig.~Fig.~\ref{C29-fig:vortex4d}.
This may be related to the fact that each self-intersection point of a vortex surface contributes $\pm k/N$ ($k= 1, \cdots, N-1$) to the Pontryagin index \cite{ER00}.

Finally, we consider the relationship between center vortices and our gauge-invariant vortices for $SU(N)$ case. 
The  center vortex is obtained by replacing the magnetic flux $\Phi$ with the diagonal matrix $4\pi \mathcal{H}$:
\begin{equation}
   \sqrt{\frac{N-1}{2N}}  g\Phi \rightarrow  4\pi \mathcal{H} 
  ,
\end{equation}
which indeed leads to the non-trivial elements of the center group $\mathbb{Z}_N$:
\begin{align}
z_{N} =&    \exp \left\{ i \sqrt{\frac{N-1}{2N}} g \Phi  \right\} 
\nonumber\\
  \rightarrow & \exp (4\pi i \mathcal{H}) =  e^{2\pi i f_N} {\bf 1}
\in \mathbb{Z}_N , \quad
   f_N = k/N \ (k= 1, \cdots, N-1)
 ,
\end{align}
where $f_N=0$ corresponds to the trivial element ${\bf 1}$.
Thus the center vortices are replaced by our gauge-invariant vortices carrying the fractional magnetic flux $f_N \Phi$. 

If this observation is correct, the magnetic part of the $SU(N)$ Wilson loop with an additional fractional factor $f_N$:
\begin{equation}
 W_C^m    
= \exp \left\{ i g f_N \sqrt{\frac{N-1}{2N}} (k, \omega_{\Sigma})  \right\}
 , \quad
    f_N = k/N \ (k= 1, \cdots, N-1) 
\end{equation}
will reproduce the string tension for any $k$ ($k= 1, \cdots, N-1$), just as confirmed for $SU(2)$ gauge group in \cite{IKKMSS06}.  
This is a conjecture derived in the paper \cite{Kondo08b}. 
This is consistent with the center vortex mechanism for quark confinement. However, it will be rather difficult to identify the vortex structure in $SU(3)$ case in numerical simulations. For the vortex surfaces for $SU(N)$, $N \ge 3$, may branch and the superimposed magnetic fluxes in general also modify the type of vortex flux, i.e., its direction in color space. 
See e.g., \cite{Engelhardt11}.

%%%%%%%%%%%%%%%%%%%%%%%%%%%%%%%%%%%%%%%%%%%%%%%%%%%
%%%%%%%%%%%%%%%%%%%%%%%%%%%%%%%%%%%%%%%%%%%%%%%%%%%
%\subsection{Center vortices}\label{sec:vortex} 
%%%%%%%%%%%%%%%%%%%%%%%%%%%%%%%%%%%%%%%%%%%%%%%%%%%
%%%%%%%%%%%%%%%%%%%%%%%%%%%%%%%%%%%%%%%%%%%%%%%%%%%

The ($Z_N$) vortex condensation theory for quark confinement put forward by 't Hooft \cite{tHooft79}, Mack \cite{Mack80}, Cornwall \cite{Cornwall79} and by Nielsen, Olesen and Ambjorn \cite{NO79} (``Copenhagen vacuum'') is that the QCD vacuum is presumed to be filled with closed magnetic vortices, which carry magnetic flux in the center of the gauge group, and have the topology of tubes for $D=3$ or surfaces for $D=4$ of finite thickness. 

The effect of creating a center vortex linked to a given Wilson loop is as follows: If the area of a Wilson loop is pierced by $n$ such vortices, the value of the Wilson loop in the $SU(N)$ gauge theory is multiplied by a factor of $e^{2i\pi n/N}$, an element of the gauge group center $Z(N)$:
\begin{align}
 W_C \to e^{2i\pi n/N} W_C, \quad (n=1, \dots, N-1).
\end{align}
Quantum fluctuations in the number of $Z_N$ vortices linked to the Wilson loop can be shown to lead to an area law fall-off, assuming that center vortex configurations are condensed in the vacuum. 
The area law is attributed to random fluctuations in the number of $Z_N$ vortices piercing the Wilson loop. 
From the viewpoint of the Aharonov-Bohm effect, see e.g., \cite{CGP98}.

%%%%%%%%%%%%%%%%%%%%%%%%%%%%%%%%%%%%%%%%%%%%%%%%%%%%%%%%%%%%%
%%%%%%%%%%%%%%%%%%%%%%%%%%%%%%%%%%%%%%%%%%%%%%%%%%%%%%%%%%%%%
\subsection{Vortex picture toward the area law}
\label{C29-section:vortex-confinement}%%%%%%%%%%%%%%%%%%%%%%%%%%%%%%%%%%%%%%%%%%%%%%%%%%%%%%%%%%%%%
%%%%%%%%%%%%%%%%%%%%%%%%%%%%%%%%%%%%%%%%%%%%%%%%%%%%%%%%%%%%%

%%%%%%%%%%%%%%%%%%%%% figures %%%%%%%%%%%%%%%%%%%%%%%%%%%
\begin{figure}[thpb]
\begin{center}
%\begin{picture}(0,0)%(0,-5000)
%\put(-900,-1900)
{\includegraphics[scale=0.2]{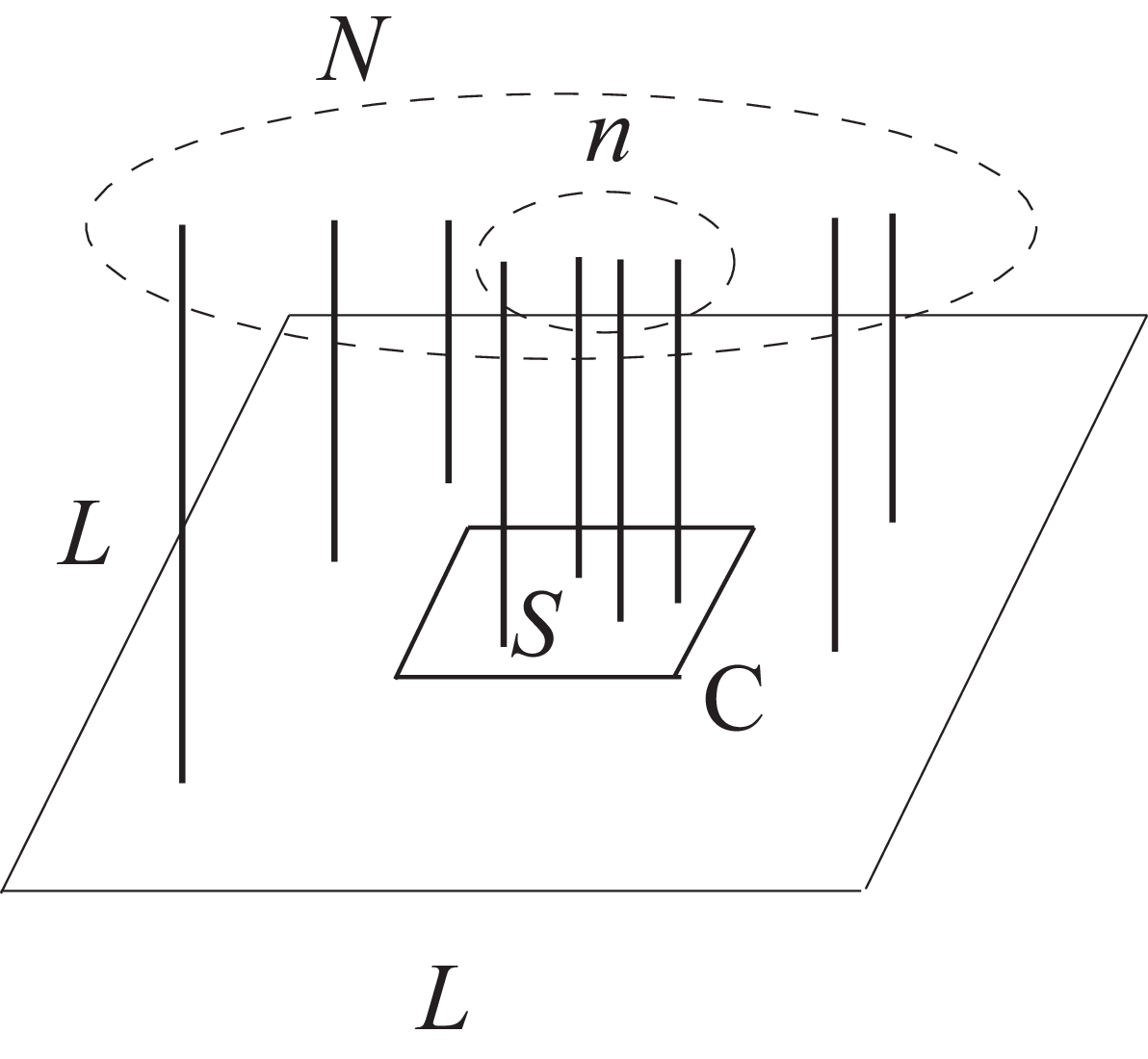}}
%\includegraphics[scale=0.4]{Fig-PR/Wilson-solid-angle.eps}
%\end{picture}
\end{center}
\vskip -0.5cm
\caption[]{
Vortices intersecting with the Wilson surface bounded by the closed loop $C$: $n$ vortices among the total $N$ vortices have the intersection points with the area $S$  where a plane including the area $S$ has the area $L^2$.
}
\label{C29-fig:vortex-picture}
\end{figure}
%%%%%%%%%%%%%%%%%%%%% figures %%%%%%%%%%%%%%%%%%%%%%%%%%%

The vortex condensation picture gives an easy way to understand the area law. 
Let us assume that the vacuum is filled with percolating thin vortices.
Suppose that $N$ random vortices intersect a plane of area $L^2$.  See Fig.~\ref{C29-fig:vortex-picture}.
For simplicity, we consider the $SU(2)$ gauge group in what follows.
Each intersection multiplicatively contributes a factor $(-1)^{2J}$ to the Wilson loop average 
and $n$ intersect within the loop take the value $(-1)^{2Jn}$.  
Then the probability that $n$ of the intersections occur within an area $S$ spanned by a Wilson loop is given by
$(-1)^{2Jn} \left( \frac{S}{L^2} \right)^{n} (+1)^{N-n} \left( 1-\frac{S}{L^2} \right)^{N-n}$. 

Summing over all possibilities with the proper binomial weight yields 
\begin{align}
  W_{C}   
&= \sum_{n=0}^{N} 
\begin{pmatrix}
N \\
n 
\end{pmatrix}
(-1)^{2Jn} \left( \frac{S}{L^2} \right)^{n} (+1)^{N-n} \left( 1-\frac{S}{L^2} \right)^{N-n}
 \nonumber\\
&= \left( 1-\frac{S}{L^2} + (-1)^{2J} \frac{S}{L^2}  \right)^{N}
 \nonumber\\
&= \left( 1- \frac{[1-(-1)^{2J}]\rho S}{N}  \right)^{N}
 \to 
 \exp \left\{ -\sigma_J S \right\} \quad (N \to \infty),
 \quad
 \rho := \frac{N}{L^2}  ,
\end{align}
where
\begin{align}
\sigma_J = [1-(-1)^{2J}]\rho
= \begin{cases}
  \sigma_F = 2\rho & (J=\frac12 , \frac32, \cdots) \cr
 0 & (J=1,2,\cdots) 
 \end{cases}
 ,
\end{align}
where $L$ has been eliminated in favor of the planar vortex density $\rho:=N/L^2$. 
The limit of a large $N \to \infty$ is taken with a constant $\rho$.  Thus one obtains an area law for the Wilson loop average with the string tension $\sigma_J$ determined by the vortex density $\rho$. 

The crucial assumption in this argument is the independence of the intersection points. 
%Let us assume that the vacuum is filled with percolating thin vortices.
%Divide the minimal surface bounded by the loop $C$ into small patches with the area $\epsilon^{2}$. 
%Let $p$ denote the probability that a patch is intersected by a vortex. We assume that intersectionz are random and uncorrelated. 
%Then it is easy to show that 
%\begin{align}
%& W_{C}   = [ (-1)^{2J} p + (+1)(1-p) ]^{\text{Area}(C)/\epsilon^2}
%= \exp [- \sigma_J \text{Area}(C)],
%\\
%& \sigma_J = - \epsilon^{-2} \ln [(-1)^{2J} p  + 1-p ] 
%= \begin{cases}
% - \epsilon^{-2} \ln [1-2p ] \cong 2p \epsilon^{-2} & (J=\frac12 , \frac32, \cdots) \cr
% 0 & (J=1,2,\cdots) 
%\end{cases}
% ,
%\end{align}
The asymptotic string tensions are zero for all integer-$J$ representations (with $N$-ality or ``biality''  being equal to $0$), while they are nonzero and equal for all half-integer $J$  representations (with $N$-ality or ``biality''  being equal to $1$). 
This is the asymptotic string tension. See section \ref{sec:Casimir-scaling} for the string tension in the intermediate distance region.

%We want to construct the vortex such that the vortex ends at the magnetic monopole, that is to say, the non-Abelian orientational zero modes of the vortex endow the endpoint non-Abelian magnetic monopole and antimonopole with same $CP^{N-1}$ zero modes. 

%Such a vortex is obtained in the Higgs phase of a $U(N)$ gauge theory with $SU(N)$ flavor symmetry   where the vortices with non-Abelian $CP^{N-1}$ orientational zero modes exist. This is a candidate of the dual gauge theory for describing the magnetic flux tube as a vortex solution. This is different from the dual Abelian-Higgs model with the magnetic gauge symmetry $U(1)^{N-1}$ suggested from the Abelian projection.  

%%%%%%%%%%%%%%%%%%%%%%%%%%%%%%%%%%%%%%%%%%%%%%%%%%%%%%%%%%%%%
%%%%%%%%%%%%%%%%%%%%%%%%%%%%%%%%%%%%%%%%%%%%%%%%%%%%%%%%%%%%%
\subsection{$SU(2)$ magnetic monopole}
\label{C29-section:SU2-monopole}%%%%%%%%%%%%%%%%%%%%%%%%%%%%%%%%%%%%%%%%%%%%%%%%%%%%%%%%%%%%%
%%%%%%%%%%%%%%%%%%%%%%%%%%%%%%%%%%%%%%%%%%%%%%%%%%%%%%%%%%%%%

For concreteness, we discuss the $SU(2)$ case. 
The gauge transformation of the $SU(2)$ Yang-Mills field $\mathscr{A}_\mu(x)$ by a group element $U(x) \in SU(2)$ is given by  
\begin{equation}
 \mathscr{A}_\mu^\prime(x) := U(x) \mathscr{A}_\mu(x) U(x)^\dagger + ig_{{}_{\rm YM}}^{-1} U(x) \partial_\mu U(x)^\dagger 
. 
% \quad
% \bm{n}^\prime(x) := U(x) \bm{n}(x) U(x)^\dagger = T_3 ,
\end{equation}
We can express a group element $U(x)$ of $SU(2)$ explicitly by the Euler angles:%
%\footnote{Using the formula for the Pauli matrices $\sigma_A$:
%\begin{equation}
%  \exp (i \theta \sigma_A ) = \bm{1} \cos \theta + i \sigma_A \sin \theta \quad (A=1,2,3),
%\end{equation}
%each factor reads
%\begin{equation}
%  e^{i \gamma \sigma_3/2} 
%= \begin{pmatrix}
%  e^{i \frac{\gamma}{2}} & 0 \cr
%  0 & e^{-i \frac{\gamma}{2}} 
% \end{pmatrix}  ,
%\quad 
%   e^{i \beta \sigma_2/2} 
%= \begin{pmatrix}
%  \cos \frac{\beta}{2} & \sin \frac{\beta}{2}  \cr
%  -\sin \frac{\beta}{2}  & \cos \frac{\beta}{2} 
% \end{pmatrix}
%  ,
%\quad 
%   e^{i \alpha \sigma_3/2}  
%= \begin{pmatrix}
%e^{i \frac{\alpha}{2}} & 0 \cr
%  0 & e^{-i \frac{\alpha}{2}} 
% \end{pmatrix}
% .
%\end{equation}
% }
\begin{align}
 U(x) =& e^{i \gamma(x) \sigma_3/2} 
   e^{i \beta(x) \sigma_2/2} 
   e^{i \alpha(x) \sigma_3/2}  
%\nonumber\\
  =  
 \begin{pmatrix}
 e^{{i \over 2}[\alpha(x)+\gamma(x)]} \cos {\beta(x) \over 2} &
 e^{{i \over 2}[-\alpha(x)+\gamma(x)]} \sin {\beta(x) \over 2} \cr
 -e^{{i \over 2}[\alpha(x)-\gamma(x)]} \sin {\beta(x) \over 2} &
 e^{{i \over 2}[-\alpha(x)-\gamma(x)]} \cos {\beta(x) \over 2}   
 \end{pmatrix}
 \in SU(2) 
  ,
  \label{C29-Euler-U}
%\nonumber\\ & 
%\alpha(x) \in [0,2\pi], \beta(x) \in [0,\pi], \gamma(x) \in [0, 2\pi]  ,
\end{align}
where $\sigma_A$ are the Pauli matrices. 
%\begin{equation}
%%T_1=\frac{\sigma_1}{2} = \frac12
%\sigma_1 := \begin{pmatrix}
%  0 & 1 \cr
%  1 & 0 
% \end{pmatrix}   ,
%\quad 
%%T_2=\frac{\sigma_2}{2} = \frac12
%\sigma_2 := \begin{pmatrix}
%  0 & -i \cr
%  i & 0 
% \end{pmatrix}  ,  
%\quad 
%%T_3=\frac{\sigma_3}{2} = \frac12
%\sigma_3 := \begin{pmatrix}
%  1 & 0 \cr 
%  0 & -1 
% \end{pmatrix}  .
%\end{equation}
Then the inhomogeneous term $\Omega_\mu$ is parameterized as  
\begin{align}
  \Omega_\mu(x)  
:=&   ig_{{}_{\rm YM}}^{-1}U(x) \partial_\mu U(x)^\dagger 
 \nonumber\\
 =&  g_{{}_{\rm YM}}^{-1} \frac12
 \begin{pmatrix}
 \cos \beta(x)  \partial_\mu \alpha(x)  + \partial_\mu \gamma(x)  & 
 [-i\partial_\mu \beta(x) -\sin \beta(x)  \partial_\mu \alpha(x) ]e^{i\gamma(x) } \\
  [i\partial_\mu \beta(x) -\sin \beta(x)  \partial_\mu \alpha(x) ]e^{-i\gamma(x) } &
- [\cos \beta(x)  \partial_\mu \alpha(x)  + \partial_\mu \gamma(x) ]  \\
 \end{pmatrix}
  ,
% = \Omega_\mu^A(x) \frac{\sigma_A}{2}
\end{align}
which has the Lie-algebra value:
\begin{align}
  \Omega_\mu(x) =  \Omega_\mu^A(x) \frac{\sigma_A}{2} 
& =  \frac12
\begin{pmatrix}
  \Omega_\mu^3(x) & \Omega_\mu^1(x) -i \Omega_\mu^2(x) \cr
  \Omega_\mu^1(x) + i \Omega_\mu^2(x) & -\Omega_\mu^3(x) 
 \end{pmatrix}
, 
 \nonumber\\
  \Omega_\mu^1(x) =& g_{{}_{\rm YM}}^{-1}[-\sin \beta(x) \cos \gamma(x) \partial_\mu \alpha(x) + \sin \gamma(x) \partial_\mu \beta(x) ] ,
 \nonumber\\
  \Omega_\mu^2(x) =& g_{{}_{\rm YM}}^{-1}[\sin \beta(x) \sin \gamma(x) \partial_\mu \alpha(x) + \cos \gamma(x) \partial_\mu \beta(x) ] ,
 \nonumber\\
  \Omega_\mu^3(x) =& g_{{}_{\rm YM}}^{-1}[\cos \beta(x) \partial_\mu \alpha(x) + \partial_\mu \gamma(x) ] .
 \label{C29-Omega-comp}
\end{align}

First, we examine the Abelian projection picture for magnetic monopoles in Yang-Mills theory.
It is easy to show that an identity holds for $\Omega_\mu$:
\begin{align}
 \partial_\mu \Omega_\nu(x)  - \partial_\nu \Omega_\mu(x)
 = ig_{{}_{\rm YM}} [\Omega_\mu(x), \Omega_\nu(x)] 
+ ig_{{}_{\rm YM}}^{-1} U(x)[\partial_\mu, \partial_\nu]U^\dagger(x) .
\label{C29-ident}
\end{align}
Then the Abelian field strength $f_{\mu\nu}^\Omega(x)$ of the diagonal part $\Omega_\mu^3(x)$ reads
\begin{align}
 f_{\mu\nu}^\Omega(x) := 
 \partial_\mu \Omega_\nu^3(x)  - \partial_\nu \Omega_\mu^3(x)
 =   C_{\mu\nu}^{[\Omega]}(x) 
+ ig_{{}_{\rm YM}}^{-1} (U(x)[\partial_\mu,\partial_\nu]U^\dagger(x))^{(3)} ,
\label{C29-dec}
\end{align}
where we have defined $C_{\mu\nu}^{[\Omega]}(x)$  by
\begin{align}
 C_{\mu\nu}^{[\Omega]} := (ig_{{}_{\rm YM}}[\Omega_\mu, \Omega_\nu])^{(3)}
 = g_{{}_{\rm YM}} \epsilon^{ab3} \Omega_\mu^a \Omega_\nu^b 
 =  g_{{}_{\rm YM}} (\Omega_\mu^1 \Omega_\nu^2 - \Omega_\mu^2 \Omega_\nu^1)
%= ig_{{}_{\rm YM}} (\Omega_\mu^+ \Omega_\nu^- - \Omega_\mu^- \Omega_\nu^+)
 .
\end{align}
Using the above Euler-angle expression (\ref{C29-Omega-comp}) for
$\Omega$, we obtain
\begin{align}
C_{\mu\nu}^{[\Omega]}
%= ig_{{}_{\rm YM}}(\Omega_\mu^+ \Omega_\nu^- - \Omega_\mu^- \Omega_\nu^+)
= g_{{}_{\rm YM}}^{-1} 
 \sin \beta (\partial_\mu \beta \partial_\nu \alpha
 - \partial_\mu \alpha \partial_\nu \beta) .
\label{C29-corr}
\end{align}
Now we show that  $C_{\mu\nu}^{[\Omega]}$ represents the magnetic monopole contribution to the diagonal field strength $f_{\mu\nu}$.
Note that $C_{\mu\nu}^{[\Omega]}$ is generated from the off-diagonal components, $\Omega_\mu^1, \Omega_\mu^2$.

\par
For $D=4$,  the magnetic charge $q_m$ is calculated as the volume integral: 
\begin{align}
 q_m(V^{(3)}) = \int_{V^{(3)}} d^3 \sigma^\mu k_\mu 
\end{align}
from the magnetic current:
\begin{align}
  k_\mu  =   \partial^\nu  {}^{\displaystyle *}f_{\mu\nu}^\Omega ,
\quad
 {}^{\displaystyle *}f_{\mu\nu}^\Omega := {1 \over 2}
\epsilon_{\mu\nu\rho\sigma} f_{\rho\sigma}^\Omega .
\end{align}
By using the Stokes (Gauss) theorem, the volume integral is replaced by the surface integral over the two-dimensional closed surface  $S^{(2)}$ which is the boundary of the three-dimensional volume $V^{(3)}$:
\begin{align}
 q_m(V^{(3)}) %= \int_{V^{(3)}} d^3 \sigma_\mu k_\mu 
 =   \int_{V^{(3)}} d^3 \sigma^\mu
 \partial^\nu  {}^{\displaystyle *}f_{\mu\nu}^\Omega
 =    \oint_{S^{(2)}=\partial V^{(3)}} d^2
\sigma^{\mu\nu} \  {}^{\displaystyle *}f_{\mu\nu}^\Omega .
\end{align}
We can identity the first part $C_{\mu\nu}^{[\Omega]}$  of right-hand-side  of (\ref{C29-dec})  with the  the \textbf{magnetic monopole} part, while the second part $ig_{{}_{\rm YM}}^{-1} (U(x)[\partial_\mu,\partial_\nu]U^\dagger(x))^{(3)}$ with the \textbf{Dirac string} part in the Yang-Mills theory.%
\footnote{
See Appendix of \cite{KondoI} for more details. 
}
This is clearly seen by the explicit calculation, since we can rewrite (\ref{C29-dec}) as
\begin{align}
 f_{\mu\nu}^\Omega 
 = - g_{{}_{\rm YM}}^{-1} \sin \beta (\partial_\mu \beta \partial_\nu \alpha
 - \partial_\mu \alpha \partial_\nu \beta)
 + g_{{}_{\rm YM}}^{-1} ([\partial_\mu, \partial_\nu] \gamma 
 + \cos \beta  [\partial_\mu, \partial_\nu] \alpha) .
\end{align}
For the magnetic charge, the contribution from the magnetic monopole part is given by
\begin{align}
q_{\rm mono}(V^{(3)})  =&  -
{1 \over 2g_{{}_{\rm YM}}}\int_{S^{(2)}} d^2
\sigma_{\mu\nu} \epsilon^{\mu\nu\rho\sigma} 
 \sin \beta (\partial_\rho \beta \partial_\sigma \alpha
 - \partial_\rho \alpha \partial_\sigma \beta) ,
\label{C29-magc1}
\end{align}
while the contribution from the Dirac string part is
\begin{align}
 q_{DS}(V^{(3)})  =&  
{1 \over 2g_{{}_{\rm YM}}}\int_{S^{(2)}} d^2
\sigma_{\mu\nu} \epsilon^{\mu\nu\rho\sigma} 
 ([\partial_\rho, \partial_\sigma] \gamma 
 + \cos \beta  [\partial_\rho, \partial_\sigma] \alpha) .
 \label{C29-defmc}
\end{align}
The first contribution (\ref{C29-magc1}) to the magnetic charge $q_{m}$ leads to the quantized magnetic charge.
This is because the integrand is the Jacobian from $S^2$ to $S^2$.
(See (\ref{C29-mag-charge-calc}) for the details of the calculation.)
Then (\ref{C29-magc1}) gives the magnetic charge $g_m$ satisfying the Dirac quantization condition,
\begin{align}
 q_{\rm mono} ={4\pi n \over g_{{}_{\rm YM}}} , \quad g_{{}_{\rm YM}} q_{\rm mono} = 4\pi n \  (n \in \mathbb{Z}) .
 \label{C29-DQ}
\end{align}
This result is consistent with the non-trivial Homotopy group:
\begin{align}
 \pi_2(SU(2)/U(1)) = \pi_2(S^2) = \mathbb{Z} .
\end{align}
In the second contribution (\ref{C29-defmc}) to the magnetic charge, if we choose $\gamma=-\alpha$ using the residual $U(1)$ gauge invariance, then the Dirac string appears on the negative $Z$ axis, i.e., $\beta=\pi$.  In this case, the surface integral reduces to the line integral along a closed path $S^{(1)}$ around the string using the Stokes theorem:
\begin{align}
 q_{DS}(V^{(3)})  =&   
- {1 \over 2g_{{}_{\rm YM}}} \int_{S^{(2)}} d  \sigma_{\mu\nu} 
\epsilon^{\mu\nu\rho\sigma}
 [\partial_\rho, \partial_\sigma] \alpha(x) 
 \nonumber\\
 =&  - {1 \over 2g_{{}_{\rm YM}}} \int_{S^{(1)}} d  \sigma_{\mu\nu\rho} 
\epsilon^{\mu\nu\rho\sigma} \partial_\sigma \alpha(x) .
\end{align}
This gives the same result (\ref{C29-defmc}) but with the minus sign, which is consistent with the non-trivial Homotopy group:
\begin{align}
  \pi_1(U(1)) =  \mathbb{Z} .
\end{align}
Actually, two description are equivalent, as suggested from the mathematical identity:
\begin{align}
 \pi_2(SU(2)/U(1)) = \pi_1(U(1)) .
\end{align}

\par
If  the contribution from 
$U(x) \mathscr{A}_\mu(x) U^\dagger(x) $ 
is completely neglected, 
%i.e., 
%$\mathscr{A}_\mu^U(x) \equiv \Omega_\mu(x)= i g_{{}_{\rm YM}}^{-1}U(x) \partial_\mu U^\dagger(x)$, 
then (\ref{C29-ident}) implies that the field strength $\mathscr{F}_{\mu\nu}^\Omega(x)$ of $\Omega_\mu(x)$ is given by 
\begin{align}
 \mathscr{F}_{\mu\nu}^\Omega(x) = ig_{{}_{\rm YM}}^{-1}
 U(x) [\partial_\mu, \partial_\nu ]U^\dagger(x) .
\label{C29-fst}
\end{align}
Note that the original Yang-Mills theory does not have the magnetic monopole solution.  However, if we adopt the MA gauge, that is to say, we partially fix the gauge symmetry $G=SU(2)$ to $G/H=SU(2)/U(1)$ and retain the residual $H=U(1)$ gauge symmetry, then the theory can have singular configuration.  This is a reason why the magnetic monopole appears in the Yang-Mills theory which does not have Higgs field.  
The existence of the Dirac string in the right-hand side of (\ref{C29-fst}) reflects the fact that the field strength $\mathscr{F}_{\mu\nu}^\Omega(x)$ does contain the magnetic monopole contribution.  
We have obtained a gauge theory with magnetic monopoles starting from the Yang-Mills theory.  Therefore, the MA gauge  enables us to deduce the magnetic monopole without introducing the scalar field, in contrast to the 't~Hooft-Polyakov monopole.
%See \cite{KondoI} for more details.

%%%%%%%%%%%%%%%%%%%%%%%%%%%%%%%%%%%%%%%%%%%%%%%%%%%%%%%%%%%%%
%%%%%%%%%%%%%%%%%%%%%%%%%%%%%%%%%%%%%%%%%%%%%%%%%%%%%%%%%%%%%
\begin{figure}[ptb]
\begin{center}
\includegraphics[height=1in]{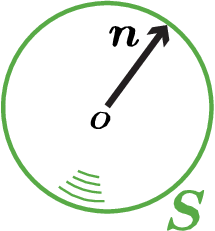}
\end{center} 
 \caption[]{
The color field $\bm{n}$ in $SU(2)$ case.
}
 \label{C29-fig:color-field}
\end{figure}
%%%%%%%%%%%%%%%%%%%%%%%%%%%%%%%%%%%%%%%%%%%%%%%%%%%%%%%%%%%%%
%%%%%%%%%%%%%%%%%%%%%%%%%%%%%%%%%%%%%%%%%%%%%%%%%%%%%%%%%%%%%

Second, we examine magnetic monopoles from the picture of the CDG decomposition. 
The color field has the the target space (See Fig.~\ref{C29-fig:color-field}):
\begin{align}
 \bm{n}(x) 
%=& g_{x} ( \rho -  \mathbf{1}/2 ) g_{x}^\dagger 
=   g(x) \frac{\sigma_3}{2} g(x)^\dagger 
\in SU(2)/U(1) \simeq S^2 \simeq CP^1 .
\end{align}
%with the Pauli matrix $\sigma_3:={\rm diag}(1,-1)$.
%The manifestly gauge--invariant form is
%\\
%\noindent
The existence of magnetic monopole is consistent with a non-trivial Homotopy class of the map $\bm{n}: S^2 \rightarrow SU(2)/U(1)$:  
\begin{equation}
 \pi_2(SU(2)/U(1))=\pi_1(U(1))=\mathbb{Z} .
\end{equation}
The magnetic charge obey the Dirac quantization condition:
\begin{equation}
 q_m := \int d^3x k^0 = 4\pi  g_{{}_{\rm YM}}^{-1} \ell , \ \ell \in \mathbb{Z} .
\end{equation}
%cf. the Abelian magnetic monopole of 't Hooft--Polyakov type:
%\begin{equation}
%  \bm{n}^A \leftrightarrow \hat{\phi}^A(x)/|\hat{\phi}(x)| .
%\end{equation}
We   show that the color field generates the magnetic charge subject to the Dirac quantization condition. 
The color field is expressed in the Lie-algebra form:
\begin{equation}
 \bm{n}(x) := U(x)^\dagger T_3 U(x) = n^A(x) T_A \in su(2)/u(1), %\quad \mathbf{n}(x) = (n_1(x), n_2(x), n_3(x)) \in S^2, 
\end{equation}
or the vector form:
\begin{equation}
 \mathbf{n}(x) = (n^1(x), n^2(x), n^3(x)) \in S^2, 
\end{equation}
By using the expression (\ref{C29-Euler-U}) for $U$, 
%\begin{equation}
%n_1(x) = \sin \beta(x) \cos \alpha(x) , 
%n_2(x) = \sin \beta(x) \sin \alpha(x)  , 
%n_3(x) = \cos \beta(x) . 
%\end{equation}
the color field $\mathbf{n}$ is parameterized by two polar angles $(\alpha,\beta)$ on the target space $S^2 \simeq SU(2)/U(1)$:% 
%each component is given by
\begin{align}
   \mathbf{n}(x) 
  =& \begin{pmatrix} n^1(x) \cr n^2(x) \cr n^3(x) \end{pmatrix}
 = \begin{pmatrix} \sin \beta(x) \cos \alpha(x) \cr
 \sin \beta(x) \sin \alpha(x) \cr
\cos \beta(x)  
 \end{pmatrix}  .
\end{align}
%See Fig.~\ref{C29-fig:color-field}.

%By the gauge transformation by $U$, the color field is diagonalized:
%\begin{equation}
% \bm{n}^\prime(x) := U(x) \bm{n}(x) U(x)^\dagger = %T_3 ,
%\end{equation}

The $SU(2)$ gauge-invariant magnetic current $k=  \delta  {}^{\displaystyle *}f$ is obtained from the $SU(2)$ gauge-invariant field strength:
\begin{align}
%  k^\mu(x) :=& \partial_\nu {}^*G^{\mu\nu}(x) 
%=   (1/2) \epsilon^{\mu\nu\rho\sigma}\partial_{\nu}               G_{\rho\sigma}(x) ,
 f_{\mu\nu} 
 =&  2{\rm tr}( \bm{n}  \mathscr{F}_{\mu\nu}[\mathscr{V}] )
%:=& \bm{n} \cdot (\mathbb{E}_{\mu\nu}+\mathbb{H}_{\mu\nu})
\nonumber\\
=&  \partial_\mu c_\nu  - \partial_\nu c_\mu   -  g_{{}_{\rm YM}}^{-1}  \mathbf{n}  \cdot  (\partial_\mu \mathbf{n} \times \partial_\nu \mathbf{n} ) 
%(=    E_{\mu\nu}+H_{\mu\nu} )
%c_\mu   ={\bm n} \cdot \mathscr{A}_\mu  .
\nonumber\\
=& \partial_\mu [ {n}^A  \mathscr{A}^A_\nu  ] -  \partial_\nu [ {n}^A  \mathscr{A}^A_\mu   ]   
%\nonumber\\ &
  - g_{{}_{\rm YM}}^{-1} \epsilon^{ABC}  {n}^A   \partial_\mu  {n}^B   \partial_\nu  {n}^C  
%  f_{\mu\nu}(x) =& \partial_\mu [\bm{n}(x) \cdot \mathscr{A}_\nu^A(x)] -  \partial_\nu [\bm{n}(x) \cdot \mathscr{A}_\mu(x)]   
%  - g^{-1} \bm{n}(x) \cdot [\partial_\mu \bm{n}(x) \times \partial_\nu \bm{n}(x) ]
\nonumber\\
=&  \partial_\mu 2{\rm tr} \{ \bm{n}  \mathscr{A}_\nu  \} - \partial_\nu 2{\rm tr} \{ \bm{n}  \mathscr{A}_\mu  \} 
%\nonumber\\& 
+ ig_{{}_{\rm YM}}^{-1} 2{\rm tr} \{ \bm{n} [\partial_\mu \bm{n} , \partial_\nu \bm{n}  ] \} .
\label{C29-latCFN-monop}
\end{align}
In fact, the field strength is written in the manifestly gauge-invariant form (\ref{C26-f-inv-form}):
\begin{align}
 f_{\mu\nu}  =& 2 {\rm tr} \left\{  \bm{n}  \mathscr{F}_{\mu\nu}[\mathscr{A}]   
  + ig_{{}_{\rm YM}}^{-1} \bm{n}   [\mathscr{D}_\mu[\mathscr{A}] \bm{n} , \mathscr{D}_\nu[\mathscr{A}] \bm{n}  ] 
 \right\}  
 .
\end{align}

We use the parameterization of the last term:
\begin{align}
\mathbf{n} \cdot (\partial_\mu \mathbf{n}   \times \partial_\nu \mathbf{n}) 
 = \sin \beta (\partial_\mu \beta \partial_\nu \alpha
 - \partial_\mu \alpha \partial_\nu \beta) 
 = \sin \beta {\partial(\beta, \alpha) 
 \over \partial(x^\mu,x^\nu)}  .
\end{align}
Taking into account the fact that 
${\partial(\beta, \alpha) \over \partial(x^\mu,x^\nu)}$ 
is the Jacobian from $(x^\mu,x^\nu) \in S^2_{\rm phy}$ to $(\beta, \alpha) \in S^2_{\rm int} \simeq SU(2)/U(1)$ parameterized
by $(\beta, \alpha)$, thus, we obtain the Dirac quantization condition for the magnetic charge:
\begin{align}
 q_m := \int_{V} d^3x k^0
=& \int_{V} d^3x \frac12 \partial_\ell \epsilon^{\ell jk}     f_{jk} 
 \nonumber\\
=& \oint_{S^2_{\rm phy}} dS_{\ell} \frac12 \epsilon^{\ell jk}  f_{jk} 
 \nonumber\\
=& - \oint_{S^2_{\rm phy}} dS^{jk} 
  g_{{}_{\rm YM}}^{-1} \mathbf{n} \cdot (\partial_j \mathbf{n}   \times \partial_k \mathbf{n}) 
 \nonumber\\
=& -  g_{{}_{\rm YM}}^{-1} \oint_{S^2_{\rm phy}} dS^{jk} 
  {\partial(\beta, \alpha)  \over \partial(x^j,x^k)} \sin \beta
 \nonumber\\
 =& -  g_{{}_{\rm YM}}^{-1}  \oint_{S^2_{\rm int}}  d\beta  d\alpha \sin \beta
 \nonumber\\
 =&  4\pi g_{{}_{\rm YM}}^{-1}   n  \quad (n=0, \pm1, \cdots)
 ,
\label{C29-mag-charge-calc}
\end{align}
since $d\beta  d\alpha \sin \beta$ is the surface element on $S^2_{\rm int}$ and a surface of a unit radius has the area $4\pi$. 
Hence $n$ gives a number of times  $S^2_{\rm int}$ is
wrapped by a mapping from  $S^2_{\rm phys} $ to 
$S^2_{\rm int}$. 
This fact is understood as the Homotopy group:
$\Pi_2(SU(2)/U(1)) = \Pi_2(S^2) = \mathbb{Z}$.

We can introduce the Abelian potential $h_\mu(x)$ for the magnetic monopole:
\begin{equation}
 h_\mu(x)= g_{{}_{\rm YM}}^{-1}[\cos \beta(x) \partial_\mu \alpha(x) + \partial_\mu \gamma(x) ] =\Omega_\mu^3(x) .
\end{equation}
In fact, the field strength of the potential $h_\mu(x)$ agrees with $H_{\mu\nu}$:
\begin{equation}
 \partial_\mu h_\nu  - \partial_\nu h_\mu   
 = -g_{{}_{\rm YM}}^{-1} \sin \beta (\partial_\mu \beta \partial_\nu \alpha
 - \partial_\mu \alpha \partial_\nu \beta) 
= - g_{{}_{\rm YM}}^{-1}  \mathbf{n}  \cdot  (\partial_\mu \mathbf{n} \times \partial_\nu \mathbf{n} ) =H_{\mu\nu}
 .
\end{equation}
Therefore, we call $h_\mu(x)$ the \textbf{magnetic potential}. 
Therefore, non-vanishing magnetic charge is derived only when the magnetic potential $h_\mu(x)$ has singularities. Otherwise, the magnetic current is identically zero due to the Bianchi identity. 
\begin{align}
 q_m 
=& \int_{V} d^3x \partial_\ell \frac12 \epsilon^{\ell jk}     f_{jk} 
 \nonumber\\
=& \int_{S^2_{phy}} dS_{\ell} \frac12 \epsilon^{\ell jk}  (\partial_j h_k  - \partial_k h_j) 
 \nonumber\\
=&  \oint_{S^1_{phy}} dr^{k} h_k
 \nonumber\\
=&  \oint_{S^1_{phy}} dr^{k} g_{{}_{\rm YM}}^{-1} [\cos \beta(x) \partial_k \alpha(x) + \partial_k \gamma(x) ]
 \nonumber\\
=&  g_{{}_{\rm YM}}^{-1} \int_{0}^{2\pi} d\varphi [\cos \beta(x_*) \partial_\varphi \alpha(x_*) + \partial_\varphi \gamma(x_*) ]
 \nonumber\\
=&  g_{{}_{\rm YM}}^{-1} (  \cos \beta(x_*) [ \alpha(x_*)]_{\varphi=0}^{\varphi=2\pi} +  [ \gamma(x_*) ]_{\varphi=0}^{\varphi=2\pi})
 \nonumber\\
 =&  4\pi g_{{}_{\rm YM}}^{-1}   n  \quad (n=0, \pm1, \cdots)
 ,
\end{align}
where we have calculated the line integral around the infinitesimal closed loop around the Dirac string, i.e., line of singularities on the positive or negative $Z$ axis where $\cos \beta(x_*)=\pm 1$ for $\beta(x_*)=0, \pi$.
See Fig.~\ref{C23-fig:Dirac-monopole}.

Finally, we discuss the gauge reducing to the  Abelian projection.
%The Reproduction of Abelian projection
For $SU(2)$, if we choose a special gauge, unitary-like gauge, in which the color field $\bm{n}(x)$ has the uniform color direction everywhere:
\begin{equation}
  \bm{n}(x) =(n_1(x),n_2(x),n_3(x)) \equiv \bm{n}_0, \quad \bm{n}_0:=(0,0,1)  ,
  \label{u-gauge}
\end{equation}
then the gauge-invariant field strength reduces to the Abelian form:
\begin{equation}
  f_{\mu\nu}  =  \partial_\mu  \mathscr{A}^3_\nu    -  \partial_\nu  \mathscr{A}^3_\mu   
  ,
\end{equation}
and the Wilson loop operator reduces to the ``Abelian-projected'' form:
\begin{equation}
 W_C[\mathscr{A}] =  \exp \left[  ig_{{}_{\rm YM}} \int_{\Sigma: \partial \Sigma=C} F \right] 
=  \exp \left[  ig_{{}_{\rm YM}} \frac12 \int_{\Sigma: \partial \Sigma=C} f \right] . \quad
% f_{\mu\nu}  =  \partial_\mu  \mathscr{A}^3_\nu    -  \partial_\nu  \mathscr{A}^3_\mu  .
\end{equation}
%where 
%the two-form $F:=dA=\frac12 F_{\mu\nu}(x) dx^\mu \wedge dx^\nu$ is defined by
%\begin{align}
%  F_{\mu\nu}(x) &=   \frac12 f_{\mu\nu} (x) .
%\end{align}
Therefore, the Abelian projection is reproduced from the gauge-invariant treatment as a special limit of taking the unitary-like gauge  (\ref{u-gauge}).%
%\footnote{
%For reviews of the Abelian projection,  see e.g.,
% Chernodub and Polikarpov  \cite{CP97},
%Greensite \cite{Greensite03}.
%}

%%%%%%%%%%%%%%%%%%%%%%%%%%%%%%%%%%%%%%%%%%%%%%%%%%%%%%%%%%%%%
%%%%%%%%%%%%%%%%%%%%%%%%%%%%%%%%%%%%%%%%%%%%%%%%%%%%%%%%%%%%%
\subsection{$SU(3)$ magnetic monopole}
\label{subsection:SU3mm}%%%%%%%%%%%%%%%%%%%%%%%%%%%%%%%%%%%%%%%%%%%%%%%%%%%%%%%%%%%%%
%%%%%%%%%%%%%%%%%%%%%%%%%%%%%%%%%%%%%%%%%%%%%%%%%%%%%%%%%%%%%

We examine the quantization condition for the magnetic charge.
%First, we consider a special case in which $U(x)$ is rotated to the identity matrix by performing a gauge transformation so that the one-form $V$ takes the form:
%\begin{equation}
% V(x) \rightarrow  {\rm tr}(  \mathcal{H} \mathscr{A}(x)) 
% ,
%\end{equation}
%which reads 
%for $[1,0]$ and $[-1,0]$,  
%\begin{equation}
%  V(x)   \rightarrow  \pm   \left[ \frac{1}{2} \mathscr{A}^3(x) +   \frac{1}{2\sqrt{3}} \mathscr{A}^8(x) \right]
%  =  \pm  (\mathscr{A}^A(x)T_A)_{11}
%  , 
%\end{equation}
%and  $[0,-1]$ and $[0,1]$, 
%\begin{equation}
%  V(x)   \rightarrow  \pm   \left[ -\frac{1}{2} \mathscr{A}^3(x)  +   \frac{1}{2\sqrt{3}} \mathscr{A}^8(x) \right] 
%  =  \pm  (\mathscr{A}^A(x)T_A)_{22}
%  , 
%\end{equation}
%In particular, for $[-1,1]$ and $[1,-1]$,  
%\begin{equation}
% V(x)   \rightarrow   \pm   \left[  \frac{-2}{2\sqrt{3}} \mathscr{A}^8(x) \right]
%  =  \pm   (\mathscr{A}^A(x)T_A)_{33}
%   .
%\end{equation}
The magnetic current  ($(D-3)$-form) $k$ is defined using the gauge-invariant field strength (two-form) 
$
f := 2{\rm tr}\{ \bm{n} \mathscr{F} [\mathscr{V}]\} 
$
 by
\begin{equation}
 k  = \delta {}^{\displaystyle *}
 f = {}^{\displaystyle *}df 
   .
\end{equation}
For $D=4$, the magnetic current reads 
\begin{equation}
 k^\mu = \frac12 \epsilon^{\mu\nu\rho\sigma} \partial_\nu f_{\rho\sigma} 
   .
\end{equation}
Hence, the magnetic charge is defined by  
\begin{equation}
  q_m = \int d^3x k^0 
= \int d^3x \frac12 \epsilon^{jk\ell} \partial_\ell    f_{jk}(x) 
= \int d^2S_\ell  \epsilon^{jk\ell} \frac12   f_{jk}(x) 
%= \int d^2S^{jk}    F_{jk}(x) 
   .
   \label{C29-m-charge}
\end{equation}
%In order to extract the information on the magnetic charge through the non-Abelian Stokes theorem for the Wilson loop operator, we consider the integration of the field strength $F=dV$ (the curvature two-form) over the closed surface $\Sigma$.   

%For the fundamental representation, $F=dV$  agrees with the diagonal components of the non-Abelian field $ \mathscr{A}(x)=\mathscr{A}^A(x)T_A$. 
%This fact is also seen from 
%\begin{equation}
% V(x)   \rightarrow 2{\rm tr}(\bm{m}(x) \mathscr{A}(x))  
%=  m^A(x)  \mathscr{A}^A(x)
%=  \left< \bm{\Lambda} |  G^{\dagger}(x)  \mathscr{A}(x) G(x) |\bm{\Lambda} \right>
% =  \left< U(x), \Lambda |  \mathscr{A} (x)   |U(x), \Lambda \right>
%   ,
%\end{equation}
%where
%$| \bm{\Lambda} >=(1,0,0)$,
%$| \bm{\Lambda} >=(0,1,0)$, and
%$| \bm{\Lambda} >=(0,0,1)$ for three fundamental representations.  

In the $SU(3)$ case, 
 the two kinds of gauge-invariant magnetic-monopole currents
  $k^{(a)}= \delta {}^*f^{(a)}$
%  $k=\frac{1}{\sqrt{3}}\delta {}^*\mathscr{G}$
 are given by the field strength:
% (in the vector notation):
%\begin{align}
% \mathscr{G}_{\mu\nu}  
%  :=   \partial_\mu 2{\rm tr} \{ \bm{n}  \mathscr{A}_\nu  \} - \partial_\nu 2{\rm tr} \{ \bm{n}  \mathscr{A}_\mu  \} 
%\nonumber\\& 
%+ \frac{4}{3} ig_{{}_{\rm YM}}^{-1} 2{\rm tr} \{ \bm{n} [\partial_\mu \bm{n} , \partial_\nu \bm{n}  ] \} 
% .
%\end{align}
The calculations can be done by using the Lie algebra form: 
\begin{align}
 f_{\mu\nu}^{(1)}  
  :=& \bm{n}_3 \cdot \mathscr{F}_{\mu\nu}[\mathscr{V}]
=  \partial_\mu  2{\rm tr}\{ \bm{n}_3 \mathscr{A}_\nu  \} - \partial_\nu 2{\rm tr}\{ \bm{n}_3 \mathscr{A}_\mu  \}
%\nonumber\\& 
-  ig_{{}_{\rm YM}}^{-1} 2{\rm tr}\{  \bm{n}_3  [ \partial_\mu \bm{n}_3  , \partial_\nu \bm{n}_3 ]  + \bm{n}_3 [ \partial_\mu \bm{n}_8  , \partial_\nu \bm{n}_8 ] \} ,
\nonumber\\
 f_{\mu\nu}^{(2)}  
  :=& \bm{n}_8 \cdot \mathscr{F}_{\mu\nu}[\mathscr{V}]
  =   \partial_\mu  2{\rm tr}\{ \bm{n}_8    \mathscr{A}_\nu  \} - \partial_\nu 2{\rm tr}\{ \bm{n}_8  \mathscr{A}_\mu  \}
%\nonumber\\& 
- \frac{4}{3} ig_{{}_{\rm YM}}^{-1} 2{\rm tr}\{   \bm{n}_8 [\partial_\mu \bm{n}_8  , \partial_\nu \bm{n}_8 ]) 
 ,
\end{align}
which is easier than the vector form:
\begin{align}
 f_{\mu\nu}^{(1)}  
 &:= \mathbf{n}_3\cdot\mathbf{F}_{\mu\nu}[\mathbf{V}]
   =\partial_\mu  \{ \mathbf{n}_3  \cdot \mathbf{A}_\nu  \} - \partial_\nu \{ \mathbf{n}_3  \cdot \mathbf{A}_\mu  \}
%\nonumber\\& 
-  g_{{}_{\rm YM}}^{-1}   \mathbf{n}_3  \cdot [ \partial_\mu \mathbf{n}_3  \times \partial_\nu \mathbf{n}_3   +  \partial_\mu \mathbf{n}_8  \times \partial_\nu \mathbf{n}_8  ],
\nonumber\\
 f_{\mu\nu}^{(2)}  
 &:=\mathbf{n}_8\cdot\mathbf{F}_{\mu\nu}[\mathbf{V}]
  =   \partial_\mu  \{ \mathbf{n}_8  \cdot \mathbf{A}_\nu  \} - \partial_\nu \{ \mathbf{n}_8  \cdot \mathbf{A}_\mu  \}
%\nonumber\\& 
- \frac{4}{3} g_{{}_{\rm YM}}^{-1}   \mathbf{n}_8 \cdot (\partial_\mu \mathbf{n}_8  \times \partial_\nu \mathbf{n}_8 ) 
 ,
\end{align}
where $\mathbf{n}_3=(n_3^A)$ and $\mathbf{n}_8=(n_8^A)$ are defined by 
\begin{align}
  \bm{n}_3 = n_3^A T_A := U^\dagger T_3 U, \quad 
  \bm{n}_8 = n_8^A T_A :=  U^\dagger T_8 U, \quad U \in SU(3) 
 .
\end{align}
Here we have used the field strength for the maximal option obtained in  section \ref{subsubsection:field-strength-maximal}:
\begin{align}
\mathscr{F}_{\mu\nu}[\mathscr{V}]
%= \mathscr{E}_{\mu\nu}  + \mathscr{H}_{\mu\nu}  
= \bm{n}_j \{ \partial_\mu (\bm{n}_j \cdot \mathscr{A}_\nu )  - \partial_\nu (\bm{n}_j \cdot \mathscr{A}_\mu ) \} 
+ ig^{-1}   \bm{n}_j(  \bm{n}_k \cdot [\partial_\mu  \bm{n}_k    , \partial_\nu \bm{n}_j]   ) \quad (j,k \in \{ 3, 8 \}) 
 . 
\end{align}

Then it is shown later that one can define two gauge-invariant charges $q_m^{(1)}$ and $q_m^{(2)}$ which obey the different quantization conditions \cite{GNO77}:%
%\footnote{
%See, e.g., % EW(1976). GNO (1977).
%\bibitem{EW76}
%E. Weinberg, 
%\bibitem{GNO77}
%P. Goddard, J. Nuyts, David I. Olive,
%Gauge Theories and Magnetic Charge, 
%Nucl. Phys. B{\bf 125}, 1--28 (1977). 
%\cite{EW76,GNO77}:
%}
\begin{align} 
 q_m^{(1)} 
:=& \int d^3x \frac12 \epsilon^{jk\ell} \partial_\ell f_{jk}^{(1)}(x) 
%({\bm n}_1(x) , \mathscr{F}_{jk}[\mathscr{V}](x))
= \frac{4\pi}{g_{{}_{\rm YM}}} \left( n- \frac12 n' \right) , 
\nonumber\\
 q_m^{(2)} 
:=& \int d^3x \frac12 \epsilon^{jk\ell} \partial_\ell f_{jk}^{(2)}(x) 
%({\bm n}_2(x) , \mathscr{F}_{jk}[\mathscr{V}](x))
= \frac{4\pi}{g_{{}_{\rm YM}}} \frac12 \sqrt{3} n' , \quad n, n' \in \mathbb{Z} 
  ,
\label{C29-qc2}
\end{align}
The existence of magnetic monopole $q_m^{(1)}$ characterized by two integers $n$ and $n'$ is consistent with a fact that 
the map defined by
\begin{equation}
   \bm{n}_3: S^2 \rightarrow SU(3)/[U(1)\times U(1)] \simeq F_2 ,
\end{equation}
 has the nontrivial Homotopy group:  
\begin{equation}
   \pi_2(SU(3)/[U(1) \times U(1)])=\pi_1(U(1) \times U(1)) 
 =\mathbb{Z} + \mathbb{Z} .
\end{equation}
On the other hand, the existence of magnetic monopole $q_m^{(2)}$ characterized by an integer   $n'$ is consistent with a fact that 
the map defined by
\begin{equation}
   \bm{n}_8: S^2 \rightarrow SU(3)/U(2) \simeq \mathbb{C}P^2 = \mathbb{C}P^2 ,
\end{equation}
 has the nontrivial Homotopy group:  
\begin{equation}
   \pi_2(SU(3)/[SU(2) \times U(1)])=\pi_1(SU(2) \times U(1)) 
%\nonumber\\&
=\pi_1(U(1))=\mathbb{Z} .
\end{equation}
%The magnetic charge defined by $Q_m := \int d^3x k^0$ obeys the quantization condition characterized by an integer $\ell$: 
%\begin{equation}
% Q_m := \int d^3x k^0 = 2\pi \sqrt{3} g_{{}_{\rm YM}}^{-1} \ell , \ \ell \in \mathbb{Z} .
%\end{equation}

However, it should be noted that the Wilson loop in the fundamental representation does not distinguish $q_m^{(1)}$ and $q_m^{(2)}$, and can probe only the combinations represented by $Q_m$ as follows.
% through $\bm{n}$ : 
%Thus the magnetic charge in the continuum theory is written as 
%\begin{equation}
%  Q_m = \int d^3x \frac12 \epsilon^{jk\ell} \partial_\ell  
%  (  {\bm{n}}(x) , \mathscr{F}_{jk}[\mathscr{V}](x) )
%   .
%   \label{C29-charge-fs}
%\end{equation}
For $[1,0]$ and $[-1,0]$,  
\begin{equation}
  Q_m 
  = \pm  \left[ q_m^{(1)} +   \frac{1}{\sqrt{3}}  q_m^{(2)} \right] 
  = \pm  \frac{2\pi}{g_{{}_{\rm YM}}} \left[  (2n-n') +     n' \right] 
  = \pm  \frac{4\pi}{g_{{}_{\rm YM}}} n
   ,
\end{equation}
for $[0,-1]$ and $[0,1]$, 
\begin{equation}
  Q_m 
  = \pm  \left[ -q_m^{(1)} +   \frac{1}{\sqrt{3}}  q_m^{(2)} \right] 
  = \pm  \frac{2\pi}{g_{{}_{\rm YM}}} \left[  -(2n-n') +     n' \right] 
  = \pm  \frac{4\pi}{g_{{}_{\rm YM}}} (n'-n)
   ,
\end{equation}
and, in particular, for $[-1,1]$ and $[1,-1]$,  
\begin{equation}
  Q_m 
  = \pm  \left[  \frac{-2}{\sqrt{3}}  q_m^{(2)} \right] 
  = \pm  \frac{2\pi}{g_{{}_{\rm YM}}} \left[  -2  n' \right] 
  =  \mp  \frac{4\pi}{g_{{}_{\rm YM}}}  n'  
   .
\end{equation}
These quantization conditions for $Q_m$ are reasonable because they guarantee that the Wilson loop operator defined originally by the closed loop $C$ should not depend on the choice of the surface $\Sigma$ bounded by the loop $C$ when rewritten in the surface integral form by the non-Abelian Stokes theorem, just as in the $SU(2)$ case.  
\begin{equation}
 1 = \exp \left\{ ig_{{}_{\rm YM}} \frac12 \oint_{S^2: \partial S^2=C}  \frac{1}{\sqrt{3}}f \right\}  = \exp \left\{ ig_{{}_{\rm YM}} \frac12 Q_m \right\} 
  \Longrightarrow 
 Q_m = 4\pi g_{{}_{\rm YM}}^{-1} n 
   .
\end{equation}
%The details is discussed in the next section.
The quantization condition is written as 
\begin{equation}
 1 = \exp \left\{ ig_{{}_{\rm YM}} (q_m^{(1)} T_3 + q_m^{(2)} T_8 ) \right\} 
 = \exp \left\{ ig_{{}_{\rm YM}} \frac12 (q_m^{(1)} \lambda_3 + q_m^{(2)} \lambda_8 ) \right\} 
   .
\end{equation}
The diagonal element leads to the relationship between the magnetic charge and the weight vector $\vec{\Lambda}=(\Lambda_1,\Lambda_2)$ of the fundamental representation:
\begin{equation}
 Q_m = 2 \vec{\Lambda} \cdot \vec{q}_m = 2 (\Lambda_1  q_m^{(1)} + 2 \Lambda_2  q_m^{(2)} )
   .
\end{equation}

%%%%%%%%%%%%%%%%%%%%%%%%%%%%%%%%%%%%%%%%%%%%%%%%%%%%%%%%%%%%%
%%%%%%%%%%%%%%%%%%%%%%%%%%%%%%%%%%%%%%%%%%%%%%%%%%%%%%%%%%%%%
%\section{$SU(3)$ magnetic monopole}
%\label{C29-subsection:SU3mm}%%%%%%%%%%%%%%%%%%%%%%%%%%%%%%%%%%%%%%%%%%%%%%%%%%%%%%%%%%%%%
%%%%%%%%%%%%%%%%%%%%%%%%%%%%%%%%%%%%%%%%%%%%%%%%%%%%%%%%%%%%%

For $SU(3)$ in the minimal option, the color field $\bm{n}(x)$ is defined by the Lie algebra form:
\begin{equation}
 \bm{n}(x) = \bm{n}_8(x) = U(x)^\dagger T_8 U(x) = n_A(x) T_A \in Lie(SU(3)/U(2)), %\quad \mathbf{n}(x) = (n_1(x), n_2(x), n_3(x)) \in S^2, 
\end{equation}
using $T_8=\frac12 \lambda_8$ and an $SU(3)$ group element $U \in SU(3)$. 
The color field is also written in the vector form:
\begin{equation}
 \mathbf{n}(x) = (n_1(x), n_2(x), ..., n_8(x))^t \in \mathbb{R}^8. 
\end{equation}
A group element $U(x)$ of $SU(3)$ is parameterized 
by using  eight angles $\alpha, \beta, \gamma, \alpha^\prime, \beta^\prime, \gamma^\prime, \eta, \xi$  as%
\footnote{
See e.g., \cite{Nelson67,Holland69} for more details. 
%T.J. Nelson,
%A Set of harmonic funcitons for the group SU(3) as specialized matrix elements of a general final transformation,  
%J. Math. Phys. {\bf 8}, 857--863  (1967). 
%D.F. Holland,
%Finite transformations and basis states of su(n), 
%J. Math. Phys. {\bf 10}, 1903--1905  (1969). 
%D.F. Holland,
%Finite transformations of su(3), 
%J. Math. Phys. {\bf 10}, 531--535 (1969).   
} 
\begin{align}
 U(x) &=  V^\prime U_8(\xi) U_7(\eta) V
 \in SU(3) 
  ,
%  \label{C29-Euler-SU3b}
\nonumber\\ & 
U_8(\xi) :=    e^{i  \xi(x) \lambda_8/\sqrt{3}}  , \
U_7(\eta) :=   e^{i  \eta(x) \lambda_7 }  , \
\nonumber\\ & 
V :=   e^{i \gamma(x) \lambda_3/2} 
   e^{i \beta(x) \lambda_2/2} 
   e^{i \alpha(x) \lambda_3/2} , \
%\nonumber\\ & 
V^\prime := e^{i \gamma^\prime(x) \lambda_3/2} 
   e^{i \beta^\prime(x) \lambda_2/2} 
   e^{i \alpha^\prime(x) \lambda_3/2} .
%\\
% U(x) =& e^{i \gamma^\prime(x) \lambda_3/2} 
%   e^{i \beta^\prime(x) \lambda_2/2} 
%   e^{i \alpha^\prime(x) \lambda_3/2}  
%   e^{i  \xi(x) \lambda_8/\sqrt{3}}  
%   e^{i  \eta(x) \lambda_7 }  
%   e^{i \gamma(x) \lambda_3/2} 
%   e^{i \beta(x) \lambda_2/2} 
%   e^{i \alpha(x) \lambda_3/2} 
% \in SU(3)  ,
%\nonumber\\ & 
%\alpha^\prime(x) \in [0,2\pi], 
%\beta^\prime(x) \in [0,\pi], 
%\gamma^\prime(x) \in [0, 2\pi]  ,
%\nonumber\\ & 
%\alpha(x) \in [0,2\pi], 
%\beta(x) \in [0,\pi], 
%\gamma(x) \in [0, 2\pi]  ,
%\nonumber\\ & 
%\xi(x) \in [0, 2\pi]  , 
%\eta(x) \in [0, 2\pi]  .
 \label{C29-Euler-SU3}
\end{align}
By using the expression (\ref{C29-Euler-SU3}) for the $SU(3)$ element $U$, 
%the color field $\bm{n}(x) \in Lie(SU(3)/U(2))$ is represented by four angles $\alpha(x), \beta(x), \gamma(x)$ and $\eta(x)$: 
$\mathbf{n}_3(x)$ and $\mathbf{n}_8(x)$ are represented as follows
\begin{equation}
\mathbf{n}_3(x) 
 =\begin{pmatrix} 
  %\left(
  %\begin{array}{c}
   \{
   \cos\alpha\cos\beta
   \cos(\alpha^\prime+\gamma)
   -\sin\alpha
    \sin(\alpha^\prime+\gamma)
   \}
   \cos\eta
    \sin\beta^\prime
   +\frac14
    \cos\alpha\sin\beta
    (3+\cos2\eta)
    \cos\beta^\prime
  \cr
   \{
   \sin\alpha\cos\beta
   \cos(\alpha^\prime+\gamma)
   +\cos\alpha
    \sin(\alpha^\prime+\gamma)
   \}
   \cos\eta
    \sin\beta^\prime
   +\frac14
    \sin\alpha\sin\beta
    (3+\cos2\eta)
    \cos\beta^\prime
  \cr
   -\sin\beta\cos(\alpha^\prime+\gamma)
    \cos\eta
    \sin\beta^\prime
   +\frac14
    \cos\beta
    (3+\cos2\eta)
    \cos\beta^\prime
  \cr
    \cos(\frac\beta2)
    \cos\{\frac{\alpha+\gamma}2+\alpha^\prime\}
    \sin\eta
    \sin\beta^\prime
    +\frac12
     \sin(\frac\beta2)
     \cos\{\frac{\alpha-\gamma}2\}
     \sin2\eta
     \cos\beta^\prime
  \cr
    \cos(\frac\beta2)
    \sin\{\frac{\alpha+\gamma}2+\alpha^\prime\}
    \sin\eta
    \sin\beta^\prime
    +\frac12
     \sin(\frac\beta2)
     \sin\{\frac{\alpha-\gamma}2\}
     \sin2\eta
     \cos\beta^\prime
  \cr
    \sin(\frac\beta2)
    \cos\{\frac{\alpha-\gamma}2-\alpha^\prime\}
    \sin\eta
    \sin\beta^\prime
    -\frac12
     \cos(\frac\beta2)
     \cos\{\frac{\alpha+\gamma}2\}
     \sin2\eta
     \cos\beta^\prime
  \cr
    -\sin(\frac\beta2)
     \sin\{\frac{\alpha-\gamma}2-\alpha^\prime\}
     \sin\eta
     \sin\beta^\prime
    +\frac12
     \cos(\frac\beta2)
     \sin\{\frac{\alpha+\gamma}2\}
     \sin2\eta
     \cos\beta^\prime
  \cr
   \frac{\sqrt3}4(1-\cos2\eta)
    \cos\beta^\prime
  \cr
  %\end{array}
  %\right),
  \end{pmatrix},
\end{equation}
\begin{align}
   \mathbf{n}_8(x) 
%=& 
%\begin{pmatrix} 
%n_1(x) \cr n_2(x) \cr n_3(x) \cr n_4(x) \cr n_5(x) \cr n_6(x) \cr n_7(x) \cr n_8(x)  
%\end{pmatrix}
=  
\begin{pmatrix} 
 \frac{\sqrt{3}}{4}(1-\cos 2\eta) \sin \beta \cos \alpha  \cr
 \frac{\sqrt{3}}{4}(1-\cos 2\eta) \sin \beta \sin \alpha \cr
 \frac{\sqrt{3}}{4}(1-\cos 2\eta) \cos \beta  \cr
 -\frac{\sqrt{3}}{2}\sin 2\eta  \sin \frac{\beta}{2} \cos \frac{\alpha-\gamma}{2}  \cr
 -\frac{\sqrt{3}}{2}\sin 2\eta   \sin \frac{\beta}{2} \sin \frac{\alpha-\gamma}{2} \cr
 \frac{\sqrt{3}}{2}\sin 2\eta   \cos \frac{\beta}{2}  \cos \frac{\alpha+\gamma}{2}\cr
 -\frac{\sqrt{3}}{2}\sin 2\eta   \cos \frac{\beta}{2} \sin \frac{\alpha+\gamma}{2}  \cr
 \frac{1}{4} +\frac{3}{4}\cos 2\eta 
 \end{pmatrix}  .
\end{align}
In particular, by adopting one of the simplest choice of parameters \cite{Shinohara}
\begin{equation}
\frac{\alpha+\gamma}2+\alpha^\prime=n\varphi,\quad
\frac{\alpha+\gamma}2=-n^\prime\varphi,\quad
\beta=0,\quad
\beta^\prime=\theta,\quad
\eta=\frac\theta2,
\end{equation}
the quantization conditions (\ref{C29-qc2}) are obtained from an explicit expression of the color field written in terms of two angles $\theta, \varphi$ \cite{Cho80c}:
%[Exercise-10] \marginpar{Ex-10}
\begin{align}
   \mathbf{n}_3(x) 
=  
\begin{pmatrix} 
n_1(x) \cr n_2(x) \cr n_3(x) \cr n_4(x) \cr n_5(x) \cr n_6(x) \cr n_7(x) \cr n_8(x)  
\end{pmatrix}
=  
\begin{pmatrix} 
 \sin \theta \cos \frac12 \theta \cos [(n-n') \varphi]  \cr
 \sin \theta \cos \frac12 \theta \sin [(n-n') \varphi]  \cr
 \frac14 \cos \theta (3+\cos \theta)  \cr
 \sin \theta \sin \frac12 \theta \cos (n\varphi)   \cr
 \sin \theta \sin \frac12 \theta \sin (n\varphi)   \cr
 - \frac12 \sin \theta \cos \theta \cos (n' \varphi)  \cr
 - \frac12 \sin \theta \cos \theta \sin (n' \varphi)  \cr
 \frac{1}{4} \sqrt{3} \cos \theta (1- \cos \theta) 
 \end{pmatrix}  .
 \label{C29-color-field-comp}
\end{align}
%These color fields lead  to the quantization condition (\ref{C29-qc}).
In fact, the Lie algebra form of the color field is obtained by substituting the components (\ref{C29-color-field-comp}) into 
\begin{align}
   \bm{n}_3   
=& n^A \frac12 \lambda_A 
=  \frac12
\begin{pmatrix} 
  n_3+\frac{1}{\sqrt{3}}n_8 & n_1-in_2  & n_4-in_5 \cr
  n_1+in_2 & -n_3+\frac{1}{\sqrt{3}}n_8 & n_6-in_7 \cr
  n_4+in_5 & n_6+in_7 & -\frac{2}{\sqrt{3}}n_8 \cr
 \end{pmatrix}  
\nonumber\\
=& \frac12
\begin{pmatrix} 
  \cos \theta & \sin\theta \cos \frac{\theta}{2} e^{-i(n-n')\varphi}  & \sin\theta \sin \frac{\theta}{2} e^{ -i n \varphi}  \cr
  \sin\theta \cos \frac{\theta}{2} e^{i(n-n')\varphi} & - \frac12 \cos \theta(1+\cos \theta) & - \frac12 \sin\theta \cos \theta e^{ -i n' \varphi} \cr
  \sin\theta \sin \frac{\theta}{2} e^{ i n \varphi} & - \frac12 \sin\theta \cos \theta e^{ i n' \varphi} & - \frac12 \cos \theta(1-\cos \theta)  \cr
 \end{pmatrix}  
 .
\end{align}
Then another color field is obtained 
\begin{align}
   \bm{n}_8   
%=& n^A \frac12 \lambda_A 
= \sqrt{3} \{ \bm{n}_3, \bm{n}_3 \} - \frac{1}{\sqrt{3}} \bm{1} 
%\nonumber\\
=    \frac{-1}{4\sqrt{3}}
\begin{pmatrix} 
  -2  & 0  & 0 \cr
  0 & 1-3\cos\theta & -3 \sin\theta e^{-in'\varphi}  \cr
 0 &  -3 \sin\theta e^{in'\varphi} &1+3\cos\theta \cr
 \end{pmatrix}  
 .
\end{align}
Using the unit vectors $\hat{\theta}, \hat{\varphi}$ in the $\theta, \varphi$ directions, we find 
\begin{align}
 -2i {\rm tr}\{   \bm{n}_8 [\partial_j \bm{n}_8  , \partial_k \bm{n}_8 ]) 
 =& \frac{3\sqrt{3}}{8} n' (\hat{\theta}_j \hat{\varphi}_k - \hat{\theta}_k \hat{\varphi}_j) 
 ,
 \nonumber\\
 -2i {\rm tr}\{   \bm{n}_3 [\partial_j \bm{n}_3  , \partial_k \bm{n}_3 ]) 
 =& \frac{1}{8} (8n-4n'-5n' \cos \theta)(\hat{\theta}_j \hat{\varphi}_k - \hat{\theta}_k \hat{\varphi}_j) 
 ,
 \nonumber\\
 -2i {\rm tr}\{   \bm{n}_3 [\partial_j \bm{n}_8  , \partial_k \bm{n}_8 ]) 
 =& -\frac{3}{8} n' \cos\theta (\hat{\theta}_j \hat{\varphi}_k - \hat{\theta}_k \hat{\varphi}_j) 
 ,
\end{align}
which are used to calculate the surface integration over the sphere: 
\begin{equation}
  q_m^{(a)} = \int d^2S_\ell  \epsilon^{jk\ell} \frac12   F_{jk}^{(a)}(x)
= \int d^2 S \ \hat{r}_\ell  \epsilon^{\ell jk} \frac12   F_{jk}^{(a)}(x) 
= \int d^2 S \ \hat{r}_\ell  \epsilon^{\ell jk} \frac12   H_{jk}^{(a)}(x) 
   .
\end{equation}
In other words, the gauge-invariant magnetic fields $H_{\mu\nu}^{(a)}$ are obtained as 
\begin{align}
& H_{\mu\nu}^{(1)} = \mathbf{n}_3 \cdot \mathbf{H}_{\mu\nu}   , \quad
H_{\mu\nu}^{(2)} = \mathbf{n}_8 \cdot \mathbf{H}_{\mu\nu}[V]  ,
\nonumber\\
& \mathbf{H}_{\mu\nu}  = - g_{{}_{\rm YM}}^{-1} \left[ \left( n-\frac12 n' \right) \mathbf{n}_3 + \frac12 \sqrt{3} n' \mathbf{n}_8 \right] \sin\theta (\partial_\mu \theta \partial_\nu \varphi - \partial_\nu \theta \partial_\mu \varphi ) 
 .
\end{align}

%%%%%%%%%%%%%%%%%%%%%%%%%%%%%%%%%%%%%%%%%%%%%%%%%%%%%%%%%%%%
\subsection{non-Abelian magnetic monopole and the magnetic charge quantization condition}\label{sec:nA-monopole} 
%%%%%%%%%%%%%%%%%%%%%%%%%%%%%%%%%%%%%%%%%%%%%%%%%%%%%%%%%%%%

The magnetic charge $g_m$ is defined by the total magnetic flux $\Phi_m$ sourced by the magnetic monopole integrated over a sphere centered on the magnetic monopole, which should be equal to $g_m$. 
On the other hand, the electric coupling $e_0$ of the gauge field with the matter in the fundamental representation of $SU(N)$ is through the minimum coupling constant $g_{{}_{\rm YM}}$. 
In fact, the magnetic flux sourced by a magnetic monopole for $G=SU(2)$ is calculated as
\begin{equation}
  \Phi_m 
%= \int_{S^2} d\mathbf{S} \cdot \mathbf{B}
  = \frac{4\pi}{g_{{}_{\rm YM}}} ,
\end{equation}
\begin{equation}
  \Phi_m = \int_{S^2} d\mathbf{S} \cdot \mathbf{B}
= \int_{S^2} d\mathbf{S} \cdot \frac{g_m}{4\pi r^2} \frac{\bm{r}}{r}
  =  g_m ,
\end{equation}
which leads to
\begin{equation}
  g_m = \frac{4\pi }{g_{{}_{\rm YM}}} .
\end{equation}
Here $g_{{}_{\rm YM}}$ is the electric coupling constant, which enters the Lagrangian as
\begin{align}
  \mathscr{D}_\mu[\mathscr{A}] \Psi 
=& \left( \partial_\mu - i g_{{}_{\rm YM}} \frac{\sigma_A}{2} \mathscr{A}_\mu^A \right) \Psi 
= \left( \partial_\mu - i g_{{}_{\rm YM}} \frac{\sigma_3}{2} \mathscr{A}_\mu^3 + \cdots  \right) \Psi 
\nonumber\\
=& \left( \partial_\mu - i \frac{g_{{}_{\rm YM}}}{2} {\rm diag.}(1,-1) \mathscr{A}_\mu^3 + \cdots  \right) \Psi  ,
\end{align}
where $\Psi$ is an $SU(2)$ doublet matter fields.
This means the minimum electric charge $e_0$:  
\begin{equation}
  e_0 = \frac{g_{{}_{\rm YM}}}{2} . 
\end{equation}
Thus the magnetic charge $g_m$ is given by the minimum electric charge $e_0$:
\begin{equation}
   g_m = \frac{4\pi }{g_{{}_{\rm YM}}} = \frac{4\pi }{2e_0} ,
\end{equation}
which coincides with the Dirac minimum quantum of magnetic charge.

For $G=SU(3)$ in the minimal option corresponding to $SU(3) \rightarrow SU(2) \times U(1)$, the magnetic flux sourced by a magnetic monopole is calculated as
\begin{equation}
    g_m = \Phi_m := \int_{S^2} d\mathbf{S} \cdot \mathbf{B}
  = \frac{2\pi\sqrt{3}}{g_{{}_{\rm YM}}} .
\end{equation}
This means that the minimum magnetic charge is
\begin{equation}
  g_m = \frac{4\pi \sqrt{3}}{2g_{{}_{\rm YM}}} .
\end{equation}
As $\Psi$ enters the Lagrangian in the covariant derivative 
\begin{align}
  \mathscr{D}_\mu[\mathscr{A}] \Psi 
=& \left( \partial_\mu - i g_{{}_{\rm YM}} \frac{\lambda_A}{2} \mathscr{A}_\mu^A \right) \Psi 
= \left( \partial_\mu - i g_{{}_{\rm YM}} \frac{\lambda_8}{2} \mathscr{A}_\mu^8 + ... \right) \Psi 
\nonumber\\
=& \left( \partial_\mu - i \frac{g_{{}_{\rm YM}}}{2\sqrt{3}} {\rm diag}(1,1,-2) \mathscr{A}_\mu^8 + ... \right) \Psi  .
\end{align}
where the factor $2\sqrt{3}$ comes from the normalization of the Gell-Mann matrix $\lambda_8$. 
This means the minimum electric $\mathscr{A}_\mu^8$ charge $e_0$:  
\begin{equation}
  e_0 = \frac{g_{{}_{\rm YM}}}{2\sqrt{3}} . 
\end{equation}
Thus the magnetic charge $g_m$ of the magnetic monopole is given by the minimum electric charge $e_0$:
\begin{equation}
   g_m =  \frac{4\pi }{4e_0} ,
\end{equation}
which is one half of the Dirac minimum quantum of magnetic charge.

Similarly, for $G=SU(N)$ in the minimal option corresponding to $SU(N) \rightarrow SU(N-1) \times U(1)$, the magnetic flux sourced by a magnetic monopole is calculated as
\begin{equation}
   g_m = \Phi_m := \int_{S^2} d\mathbf{S} \cdot \mathbf{B}
  = \frac{2\pi}{g_{{}_{\rm YM}}} \sqrt{\frac{2N}{N-1}} .
\end{equation}
This means that the minimum magnetic charge is
\begin{equation}
  g_m = \frac{4\pi }{g_{{}_{\rm YM}}} \sqrt{\frac{N}{2(N-1)}}.
\end{equation}
As $\Psi$ enters the Lagrangian in the covariant derivative 
\begin{align}
  \mathscr{D}_\mu[\mathscr{A}] \Psi 
=& \left( \partial_\mu - i g_{{}_{\rm YM}} \frac{\lambda_A}{2} \mathscr{A}_\mu^A \right) \Psi 
= \left( \partial_\mu - i g_{{}_{\rm YM}} \frac{\lambda_{N^2-1}}{2} \mathscr{A}_\mu^{N^2-1} + ... \right) \Psi 
\nonumber\\
=& \left( \partial_\mu - i \frac{g_{{}_{\rm YM}}}{\sqrt{2N(N-1)}} {\rm diag}(1,1,...,-(N-1)) \mathscr{A}_\mu^{N^2-1} + ... \right) \Psi  .
\end{align}
where the factor $\sqrt{2N(N-1)}$ comes from the normalization of the Gell-Mann matrix $\lambda_{N^2-1}$. 
This means that the minimum electric $\mathscr{A}_\mu^8$ charge $e_0$ is given by 
\begin{equation}
  e_0 = \frac{g_{{}_{\rm YM}}}{\sqrt{2N(N-1)}} . 
\end{equation}
Thus the magnetic charge $g_m$ of the magnetic monopole is given by by the minimum electric charge $e_0$:
\begin{equation}
   g_m =  \frac{4\pi }{2(N-1)e_0} ,
\end{equation}
which is $1/(N-1)$ of the Dirac minimum quantum of magnetic charge.
This factor of $N-1$ is the degree of the embedding of the fundamental group of the unbroken $U(1)$ into that of the unbroken gauge group.
\footnote{
See e.g., \cite{ABEK03,ABEKM04} for more details.
}

%\newpage

\newpage
%%%%%%%%%%%%%%%%%%%%%%%%%%%%%%%%%%%%%%%%%%%%%%%%%%%%%%%%%%%%
%Chapter :
% 
%%%%%%%%%%%%%%%%%%%%%%%%%%%%%%%%%%%%%%%%%%%%%%%%%%%%%%%%%%%%

%%%%%%%%%%%%%%%%%%%%%%%%%%%%%%%%%%%%%%%%%%%%%%%%%%%%%%%%%%%%
%%%%%%%%%%%%%%%%%%%%%%%%%%%%%%%%%%%%%%%%%%%%%%%%%%%%%%%%%%%%
\section{Infrared dominant field modes and the   high-energy relevant field modes}\label{sec:fandqc} 
%%%%%%%%%%%%%%%%%%%%%%%%%%%%%%%%%%%%%%%%%%%%%%%%%%%%%%%%%%%%
%%%%%%%%%%%%%%%%%%%%%%%%%%%%%%%%%%%%%%%%%%%%%%%%%%%%%%%%%%%%

In order to establish the dual superconductivity  as a gauge-invariant concept, we should cure various shortcomings associated to the MA gauge.

%%%%%%%%%%%%%%%%%%%%%%%%%%%%%%%%%%%%%%%%%%%%%%%%%%
%%%%%%%%%%%%%%%%%%%%%%%%%%%%%%%%%%%%%%%%%%%%%%%%%%
\subsection{Gauge-independent ``Abelian'' dominance for the Wilson loop operator}
%%%%%%%%%%%%%%%%%%%%%%%%%%%%%%%%%%%%%%%%%%%%%%%%%%
%%%%%%%%%%%%%%%%%%%%%%%%%%%%%%%%%%%%%%%%%%%%%%%%%%

%In the light of the reformulation, 
We  pay attention to the Wilson loop operator $W_C[\mathscr{A}]$ for the Yang-Mills gauge connection $\mathscr{A}$:%:= \mathscr{A}_\mu(x)dx^\mu$: 
%in order to reconsider the meaning of the Abelian projection and the resulting Abelian dominance:
\begin{align}
  W_C[\mathscr{A}]  
:=& {\rm tr} \left[ \mathscr{P} \exp \left\{ -ig_{{}_{\rm YM}}  \oint_{C} dx^\mu \mathscr{A}_\mu(x) \right\} \right]/{\rm tr}({\bf 1})  
 .
\end{align}

%In the minimal option of the reformulation, 
We  consider  the decomposition of the  gauge field $\mathscr{A}_\mu(x)$ into two pieces $\mathscr{V}_\mu(x)$ and $\mathscr{X}_\mu(x)$:
\begin{equation}
 \mathscr{A}_\mu(x) = \mathscr{V}_\mu(x) + \mathscr{X}_\mu(x) ,
 \label{C29-decomp}
\end{equation}
so that the resulting new field variables  $\mathscr{X}_\mu(x)$ and $\mathscr{V}_\mu(x)$ have the following remarkable property:
%We define a \textit{gauge-independent ``Abelian'' dominance for the Wilson loop operator (in the operator level)}, which we call the strong Abelian dominance.
For the gauge group $G=SU(2)$,
\begin{enumerate}
\item[(a)]
%The original $SU(2)$ gauge field variable $\mathscr{A}_\mu(x)$ is decomposed into two  pieces $\mathscr{X}_\mu(x)$ and $\mathscr{V}_\mu(x)$:
%\begin{equation}
% \mathscr{A}_\mu(x) = \mathscr{V}_\mu(x) + \mathscr{X}_\mu(x)    ,
%  \label{C29-decomp-0}
%\end{equation} 
%so that 
The Wilson loop operator $W_C[\mathscr{A}]$ defined in terms of the original Yang-Mills field $\mathscr{A}_\mu(x)$ can be reproduced by the \textbf{restricted field variable}  $\mathscr{V}_\mu(x)$ alone and the remaining field variable  $\mathscr{X}_\mu(x)$ is redundant for the Wilson loop operator: 
\begin{equation}
 W_C[\mathscr{A}]= W_C[\mathscr{V}]
 ,
 \label{C29-W-dominant}
\end{equation} 
where $W_C[\mathscr{V}]$ is defined by
\begin{align}
  W_C[\mathscr{V}]  
:=& {\rm tr} \left[ \mathscr{P} \exp \left\{ -ig_{{}_{\rm YM}}  \oint_{C} dx^\mu \mathscr{V}_\mu(x) \right\} \right]/{\rm tr}({\bf 1})  
 .
\end{align}
%Here it is required that  $\mathscr{V}_\mu(x)$ transforms in the same way as $\mathscr{A}_\mu(x)$ under the gauge transformation.
 
\item[(b)]
The non-Abelian field strength $\mathscr{F}_{\mu\nu}[\mathscr{V}](x)$ of the restricted field $\mathscr{V}_\mu(x)$ has the form  
\begin{equation}
\mathscr{F}_{\mu\nu}[\mathscr{V}](x)=F_{\mu\nu}(x)\bm{n}(x)
    .
\end{equation}
If the color field $\bm{n}(x)$  transforms according to the adjoint representation, just like  $\mathscr{F}_{\mu\nu}[\mathscr{V}](x)$ under the gauge transformation, then the magnitude $F_{\mu\nu}(x)=\bm{n}(x)  \cdot \mathscr{F}_{\mu\nu}[\mathscr{V}](x) $ is $SU(2)$ gauge invariant.   

Once the Wilson loop operator $W_C[\mathscr{V}]$ is rewritten in terms of the surface integral over the surface $\Sigma$ bounded by the closed loop $C$,  the the Wilson loop operator is rewritten in terms of the $SU(2)$ invariant ``Abelian'' field strength 
$
F_{\mu\nu}(x)
$
derived from the field strength $\mathscr{F}_{\mu\nu}[\mathscr{V}](x)$ of the restricted field variable $\mathscr{V}_\mu(x)$: 
\begin{equation}
W_C[\mathscr{V}]=W_\Sigma [F], \quad F_{\mu\nu}(x)=\bm{n}(x) \cdot   \mathscr{F}_{\mu\nu}[\mathscr{V}](x) 
    ,
\end{equation}  
where $W_\Sigma[F]$ is defined by
\begin{equation}
W_\Sigma[F]
:= \int [d\mu(g)]_\Sigma \exp \left\{ -iJg_{{}_{\rm YM}} \int_{\Sigma} dS^{\mu\nu} F_{\mu\nu}(x) \right\}
  ,
\end{equation} 
for any representation of $SU(2)$ specified by a half-integer $J=1/2,1,\dots$. 
\end{enumerate}

The set of statements (a) and (b) represents a \textbf{gauge-independent ``Abelian'' dominance} or the \textbf{restricted field dominance} for the Wilson loop operator (in the operator level), which we call  the \textbf{strong Abelian dominance}.
Some remarks are in order.
\begin{itemize}
\item
The strong ``Abelian"  dominance holds for arbitrary closed loop $C$, irrespective of its shape and size.

\item
The strong ``Abelian"  dominance in the above sense does not hold on the lattice. 
It is obtained only in the continuum limit of the lattice spacing going to zero $\epsilon \rightarrow 0$, i.e., after removing the regularization. Therefore, there exists some deviation on a lattice coming from non-zero lattice spacing $\epsilon >0$.  
\footnote{
In \cite{KS08}, the result obtained for $SU(2)$ in the continuum formulation was extended to the $SU(N)$ on the lattice and in the continuum. 
A constructive proof was given by deriving the lattice regularized  version of the non-Abelian Stokes theorem and the lattice versions of the gauge field decomposition which have been constructed and developed  for $SU(2)$ in \cite{IKKMSS06,SKKMSI07} and for $SU(N)$ in \cite{KSSMKI08,SKS10} according to the continuum versions given in \cite{KMS06} and \cite{KSM08}, respectively. 
The constructive approach enables one to compare the result of numerical simulations with the theoretical consideration, or even give a numerical proof of the non-Abelian Stokes theorem. 
It is possible to give a formula estimating a systematic error of the lattice Wilson loop operator from the continuum Wilson loop operator due to the non-zero lattice spacing $\epsilon$. 
In fact, an estimation of this deviation was given up to  $O(\epsilon^2)$. 
Moreover this will shed new light on the role of the MA gauge for the Abelian dominance on a lattice. 
}

\item
The strong ``Abelian"  dominance does NOT immediately imply the dominance of the Wilson loop average for arbitrary closed loop $C$:
\begin{equation}
 \langle W_C[\mathscr{A}] \rangle 
= \langle W_C[\mathscr{V}] \rangle 
    .
\end{equation} 
For the large loop $C$, this equality called the \textbf{infrared Abelian dominance} is expected to hold. 
The reason is explained below (\ref{C29-IAD}). 

%This does not necessarily imply
%$
%\langle W_C[\mathscr{A}] \rangle_{\rm YM}= \langle W_C[\mathscr{V}] \rangle_{\rm YM} 
%$,
%which holds only when the cross term between $\mathscr{V}$ and $\mathscr{X}$ are neglected. 

\end{itemize}

Th restricted field variable  $\mathscr{V}_\mu$ in the new formulation corresponds to the  ``diagonal"  gluon field variable in the Abelian projection scheme, while the remaining field variable  $\mathscr{X}_\mu(x)$ to the ``off-diagonal" gluon field variable. 
The color field $\bm{n}(x)$ represents a space--time-dependent, i.e., \textbf{local embedding of the Abelian direction} into the non-Abelian color space and hence the Abelian direction can vary from point to point of  space--time, while in the Abelian projection the Abelian direction is fixed over the whole space--time.

The general $SU(N)$ case is more involved from the  technical points of view.
For the gauge group $SU(N)$, $N\ge 3$, we define the \textbf{strong Abelian dominance} for the Wilson loop operator $W_C[\mathscr{A}]$ in the fundamental representation, related to the \textbf{gauge-independent ``Abelian''  dominance} or  the \textbf{restricted field dominance} for the Wilson loop average $\langle W_C[\mathscr{A}] \rangle$:
\begin{enumerate}
\item[(a)]
%The original $SU(N)$ gauge field variable $\mathscr{A}_\mu(x)$ is decomposed into two pieces $\mathscr{X}_\mu(x)$ and $\mathscr{V}_\mu(x)$:
%\begin{equation}
% \mathscr{A}_\mu(x)=\mathscr{X}_\mu(x)+\mathscr{V}_\mu(x) ,
%  \label{C29-decomp-SUN}
%\end{equation} 
%so that 
The Wilson loop operator $W_C[\mathscr{A}]$ defined in terms of the original Yang-Mills field $\mathscr{A}_\mu(x)$ can be reproduced by the \textbf{restricted field variable}  $\mathscr{V}_\mu(x)$ alone and the remaining field variable  $\mathscr{X}_\mu(x)$ is redundant for the Wilson loop operator: 
\begin{equation}
W_C[\mathscr{A}]=  W_C[\mathscr{V}]
  ,
\end{equation} 
where $W_C[\mathscr{V}]$ is defined by
\begin{align}
  W_C[\mathscr{V}]  
:=& {\rm tr} \left[ \mathscr{P} \exp \left\{ -ig_{{}_{\rm YM}}  \oint_{C} dx^\mu \mathscr{V}_\mu(x) \right\} \right]/{\rm tr}({\bf 1})  
 .
\end{align}
%Here it is required that $\mathscr{V}_\mu(x)$ transforms in the same way as $\mathscr{A}_\mu(x)$ under the gauge transformation.  

\item[(b)]
Once the Wilson loop operator $W_C[\mathscr{V}]$ is rewritten in terms of the surface integral over the surface $\Sigma$ bounded by the closed loop $C$,  the the Wilson loop operator is rewritten in terms of the $SU(N)$ invariant ``Abelian'' field strength $F_{\mu\nu}(x)$ derived from the field strength  $\mathscr{F}_{\mu\nu}[\mathscr{V}](x)$ of the restricted field variable $\mathscr{V}_\mu(x)$:
%which is proportional to the component of the field strength $\mathscr{F}_{\mu\nu}[\mathscr{V}](x)$ of the restricted variable $\mathscr{V}_\mu(x)$ along the the color direction specified by $\bm{n}(x)$:  
\begin{equation}
W_C[\mathscr{V}]=W_{\Sigma}[F] , \
    F_{\mu\nu}(x):= \sqrt{\frac{2(N-1)}{N}}  {\rm tr}(\bm{n}(x)\mathscr{F}_{\mu\nu}[\mathscr{V}](x)) , 
\end{equation} 
where  $W_\Sigma[F]$ is defined by
\begin{equation}
W_\Sigma[F] := \int [d\mu(g)]_{\Sigma} \exp \left\{ -i g_{{}_{\rm YM}}   \int_{\Sigma} dS^{\mu\nu} F_{\mu\nu}(x) \right\}
    .
\end{equation}

\end{enumerate}

The decomposed fields  $\mathscr V_\mu(x)$ and $\mathscr X_\mu(x)$ are  determined by solving the following defining equations (I) and (II), once the \textbf{color (direction) field} $\bm{n}(x)$ is given: 
\\
\noindent
(I)  $\bm{n}(x)$ is a covariant constant in the background $\mathscr{V}_\mu(x)$:
\begin{align}
  0 = \mathscr{D}_\mu[\mathscr{V}] \bm{n}(x) 
:=\partial_\mu \bm{n}(x) -  ig_{{}_{\rm YM}} [\mathscr{V}_\mu(x), \bm{n}(x)]
 ,
\label{C29-defVL2}
\end{align}
(II)  $\mathscr{X}^\mu(x)$  does not have the $\tilde{H}$-commutative part:
%, i.e., $\mathscr{X}^\mu(x)_{\tilde{H}}=0$:
\begin{align}
 0 =  \mathscr{X}^\mu(x)_{\tilde{H}} :=   \mathscr{X}^\mu(x)  -   2\frac{N-1}{N}  [\bm{n}(x) , [\bm{n}(x) ,  \mathscr{X}^\mu(x) ]]
\label{C29-defXL2}
 ,
\end{align}
where $\tilde{H}$ denotes the \textbf{maximal stability subgroup} of the gauge group $G$ which is determined by specifying the highest-weight state of the representation of the Wilson loop.

A set of conditions:   
\begin{align}
\text{(i)}& \quad 0 = \mathscr{D}_\mu{[\mathscr{V}]}  \bm{n}(x) := \partial_\mu \bm{n}(x)-ig_{{}_{\rm YM}} [\mathscr{V}_\mu(x) , \bm{n}(x)] , %\Longrightarrow (b) ,
\label{C29-DVm}
\\
\text{(ii)}&  \quad 0 = {\rm tr} \left\{ \mathscr{X}_\mu(x) \bm{n}(x) \right\} , %\Longrightarrow (a) ,
\label{C29-def-eq-2}
\end{align}
is a sufficient condition for the gauge-independent ``Abelian"  dominance represented by (a) and (b) for the Wilson loop operator in any representation of $SU(2)$ and in the fundamental representation for $SU(N)$. 
In fact, the condition   (ii) guarantees (a), while the condition  (i) guarantees (b). 
\begin{align}
\text{(i)}&   \Longrightarrow \text{(b)} , \quad
\text{(ii)}    \Longrightarrow \text{(a)} ,
\label{C29-statement1}
\end{align}
This statement will be verified shortly. 
The proofs of (a) and (b) are given in section \ref{sec:NAST}.%of the non-Abelian Stokes theorem. 

For $SU(2)$, the two conditions (\ref{C29-DVm}) and (\ref{C29-def-eq-2}) agree with the first and second defining equations respectively for specifying the conventional Cho-Duan-Ge-Faddeev-Niemi (CDGFN) decomposition.%\cite{DG79,Cho80,FN98,Shabanov99}.
%\footnote{
%See the chapter of Field decomposition and change of variables.
%Cho (1980),
%Duan and Ge (1979). 
%Faddeev and Niemi (1998).
%Shabanov (1999). 
%\cite{Cho80,DG79,FN98,Shabanov99}.
%}
Therefore, the CDGFN decomposition is a sufficient condition for the gauge-independent Abelian dominance for the $SU(2)$ Wilson loop operator.%
\footnote{
This fact that the CDGFN decomposition   leads to the Abelian dominance in the strong sense was pointed out in the continuum formulation through the non-Abelian $SU(2)$ Stokes theorem in 
%Cho (2000). 
\cite{Cho00}.
} 
In other words, the strong Abelian dominance gives a raison d'etre and a physical meaning of gauge covariant decomposition of the gauge field variable \textit{a l\'a}   CDGFN. 
%Therefore, the CDGFNS decomposition is a sufficient condition for the  Abelian dominance in the $SU(2)$ Wilson loop operator.

The general $SU(N)$  ($N \ge 3$) case is more involved than the $SU(2)$ case from the  technical points of view.
%\footnote{
%Kondo and Taira (2000). %\cite{KT00}.
%}
For $SU(N)$, a set of conditions (i) and (ii) follows from a set of the defining equations (I) and (II) which specify the $SU(N)$ gauge field decomposition, including the CDGFNS decomposition for $SU(2)$  as a special case. 
In the minimal option, the gauge field decomposition (\ref{C29-decomp}) has been achieved by way of a single color field $\bm{n}(x)$ for $SU(N)$.
%\footnote{
%Kondo (2008), Kondo-Shinohara-Murakami (2008).%\cite{Kondo08,KSM08}. 
%} 
Therefore, the strong  Abelian dominance in the Wilson loop is an immediate consequence of the gauge field decomposition.

In fact, the defining equation (II) for $SU(N)$:
%, i.e., $\mathscr{X}^\mu(x)_{\tilde{H}}=0$:
\begin{align}
 \mathscr{X}^\mu  =  2\frac{N-1}{N}  [\bm{n} , [\bm{n} ,  \mathscr{X}^\mu  ]] 
 , 
\end{align}
leads to the condition (ii):
\begin{align}
 {\rm tr} \left\{  \bm{n} \mathscr{X}_\mu   \right\} 
  =  2\frac{N-1}{N}  {\rm tr} \left\{  \bm{n} [\bm{n} , [\bm{n} ,  \mathscr{X}^\mu  ]] \right\}
  =  2\frac{N-1}{N}  {\rm tr} \left\{ [ \bm{n} , \bm{n}] [\bm{n} ,  \mathscr{X}^\mu  ]  \right\} = 0
  , 
\end{align}
where we have used an identity:
${\rm tr}(A[B,C])={\rm tr}([A,B]C)$.
The condition (i) agrees with the first  defining equation (I) for the decomposition 
$\mathscr{A}_\mu(x)=\mathscr{V}_\mu(x)+\mathscr{X}_\mu(x)$.
\begin{align}
\text{(I)}&   \Longleftrightarrow \text{(i)} \Longrightarrow  \text{(b)} , \quad
\text{(II)}    \Longrightarrow \text{(ii)} \Longrightarrow  \text{(a)},
\end{align}

Now we proceed to show the relation (\ref{C29-statement1}).
First, we consider the implication of the condition (ii).
The Wilson loop operator has the pre-NAST:
\begin{equation}
 W_C[\mathscr{A}] :=\int [d\mu(g)]_{C} 
\exp \left[ -ig_{{}_{\rm YM}} \oint_{C} A^g  \right]  
 , \ A^g  = A^g_\mu(x)dx^\mu ,
 \label{C29-pre-NAST}
\end{equation}
with the one-form $A^g  = A^g_\mu(x)dx^\mu$ defined by
(see section \ref{subsec:stability-group})
\begin{align}
 A_\mu^g(x) 
:=&  {\rm tr}\{\rho [g(x)^\dagger \mathscr{A}_\mu(x) g(x) + i  g_{{}_{\rm YM}}^{-1} g(x)^\dagger \partial_\mu g(x) ] \}
 \nonumber\\
=&  {\rm tr}\{ g(x) \rho  g(x)^\dagger \mathscr{A}_\mu(x)   \}+ {\rm tr}\{ \rho i  g_{{}_{\rm YM}}^{-1} g(x)^\dagger \partial_\mu g(x)   \} 
 \nonumber\\
=&  {\rm tr}\{ \tilde{\bm{n}}(x) \mathscr{A}_\mu(x)   \}+ {\rm tr}\{ \rho i  g_{{}_{\rm YM}}^{-1} g(x)^\dagger \partial_\mu g(x)   \} 
 ,
\end{align}
where we have defined the traceless field $\tilde{\bm{n}}(x)$  
\begin{equation}
\tilde{\bm{n}}(x) :=  g(x) \left[ \rho - \frac{\bm{1}}{{\rm tr}(\bm{1})} \right]  g(x)^\dagger , 
%= g(x)  \rho g^\dagger(x) - \frac{\bm{1}}{{\rm tr}(\bm{1})}   
%\quad
% \bm{n}(x) := - \sqrt{\frac{N}{2(N-1)}} \tilde{\bm{n}}(x)
\end{equation}
and the normalized and traceless field  $\bm{n}(x)$.
\begin{equation}
 \bm{n}(x) :=   \sqrt{\frac{N}{2(N-1)}} \tilde{\bm{n}}(x)
 = \sqrt{\frac{N}{2(N-1)}} g(x) \left[ \rho - \frac{\bm{1}}{{\rm tr}(\bm{1})} \right]  g^\dagger(x)  
 .
 \label{C29-n-defb}
\end{equation}
For the decomposition $\mathscr{A}_\mu=\mathscr{V}_\mu+\mathscr{X}_\mu$, the one-form $A^g$ is decomposed as
\begin{align}
 A_\mu^g(x) 
=&  {\rm tr}\{ \tilde{\bm{n}}(x) \mathscr{V}_\mu(x)   \} + {\rm tr}\{ \tilde{\bm{n}}(x) \mathscr{X}_\mu(x)   \} + {\rm tr}\{ \rho i  g_{{}_{\rm YM}}^{-1} g^\dagger(x) \partial_\mu g(x)   \} 
  .
\end{align}
Therefore, the condition (ii) means that $A^g$ does not depend on $\mathscr{X}$ and that $\mathscr{A}$ can be replaced by $\mathscr{V}$ in the Wilson loop operator. This is the statement (a).

Next, we consider the implication of the condition (i).
For a while, we use the different notation, $\mathscr{F}_{\mu\nu}^{[\mathscr{V}]}:=\mathscr{F}_{\mu\nu}[\mathscr{V}]$ and $\mathscr{D}_\mu^{[\mathscr{V}]}:=\mathscr{D}_\mu [\mathscr{V}] $, to avoid the confusion in  the commutator $[ \cdot , \cdot ]$.  
The commutator between   $\mathscr{F}_{\mu\nu}^{[\mathscr{V}]}(x)$  and a color field $\bm{n}(x)$ is calculated as  
\begin{equation}
 [ \mathscr{F}_{\mu\nu}^{[\mathscr{V}]}, \bm{n} ]
 =    ig_{{}_{\rm YM}}^{-1} [ \mathscr{D}_\mu^{[\mathscr{V}]},  \mathscr{D}_\nu^{[\mathscr{V}]} ] \bm{n}  
 ,
\end{equation}
using 
\begin{equation}
 \mathscr{F}_{\mu\nu}^{[\mathscr{V}]} 
= ig_{{}_{\rm YM}}^{-1} [ \mathscr{D}_\mu^{[\mathscr{V}]}, \mathscr{D}_\nu^{[\mathscr{V}]} ] ,
\quad
 \mathscr{D}_\mu^{[\mathscr{V}]} := \partial_\mu -ig_{{}_{\rm YM}} [ \mathscr{V}_\mu , \cdot ] .
\end{equation}
Indeed, this is shown as follows.
\begin{align}
 [ \mathscr{F}_{\mu\nu}^{[\mathscr{V}]}, \bm{n} ]
 =& ig_{{}_{\rm YM}}^{-1}  [ [ \mathscr{D}_\mu^{[\mathscr{V}]}, \mathscr{D}_\nu^{[\mathscr{V}]} ], \bm{n}  ] 
 \nonumber\\
 =&   - ig_{{}_{\rm YM}}^{-1} [ [ \mathscr{D}_\nu^{[\mathscr{V}]} , \bm{n} ], \mathscr{D}_\mu^{[\mathscr{V}]}]  - ig_{{}_{\rm YM}}^{-1} [ [ \bm{n}, \mathscr{D}_\mu^{[\mathscr{V}]} ], \mathscr{D}_\nu^{[\mathscr{V}]}]  
 \nonumber\\
 =&   ig_{{}_{\rm YM}}^{-1} [ \mathscr{D}_\mu^{[\mathscr{V}]},  (\mathscr{D}_\nu^{[\mathscr{V}]}  \bm{n}) ]  - ig_{{}_{\rm YM}}^{-1} [  \mathscr{D}_\nu^{[\mathscr{V}]}, (\mathscr{D}_\mu^{[\mathscr{V}]} \bm{n})  ]  
 \nonumber\\
 =&     ig_{{}_{\rm YM}}^{-1} [ \mathscr{D}_\mu^{[\mathscr{V}]},  \mathscr{D}_\nu^{[\mathscr{V}]} ] \bm{n}  
 ,
\end{align}
where we have used the Jacobi identity in the second equality 
and the relation:
$
 [ \mathscr{D}_\mu^{[\mathscr{V}]} , \bm{n} ] f
  = (\mathscr{D}_\mu^{[\mathscr{V}]}  \bm{n}) f 
$
for an arbitrary function $f$ in the third and the last equalities.
Thus, if  $\mathscr{V}_\mu(x)$ and $\bm{n}(x)$ satisfies the condition (i): 
\begin{equation}
 \mathscr{D}_\mu^{[\mathscr{V}]}  \bm{n}(x) = 0
    ,
%    \label{C29-DVm}
\end{equation}
then the field strength $\mathscr{F}_{\mu\nu}^{[\mathscr{V}]}(x)$ of the field $\mathscr{V}_\mu(x)$ and $\bm{n}(x)$ commute:
\begin{equation}
[ \mathscr{F}_{\mu\nu}^{[\mathscr{V}]}(x) , \bm{n}(x) ]
= 0 
 .
 \label{C29-F-m}
\end{equation}

For $SU(2)$, (\ref{C29-F-m}) means that the field strength $\mathscr{F}_{\mu\nu}^{[\mathscr{V}]}(x)$ of the field $\mathscr{V}_\mu(x)$ has only the ``Abelian'' part proportional to $\bm{n}(x)$: 
\begin{equation}
 \mathscr{F}_{\mu\nu}^{[\mathscr{V}]}(x) =   F_{\mu\nu}(x) \bm{n}(x) , \quad 
 F_{\mu\nu}(x) = \bm{n}(x)  \cdot \mathscr{F}_{\mu\nu}^{[\mathscr{V}]}(x)
    ,
    \label{C29-F=Fn}
\end{equation}
since $\mathscr{F}_{\mu\nu}^{[\mathscr{V}]}(x)$ is traceless and cannot have a part proportional to the unit matrix.  
This yields the stronger sense of ``Abelian dominance''.

For $SU(N)$, the above argument is more involved from the technical points of view.  
%This is because the decomposition has some options for SU(N), $N \ge 3$.
%For SU(N), $N \ge 3$, the requirement  for all $\bm{n}_j(x)$  ($j=1, \cdots, N-1$) 
%\begin{equation}
% \mathscr{D}_\mu{[\mathscr{V}]}  \bm{n}_j(x) = 0 \ (j=1, \cdots, N-1) 
%    ,
%    \label{C29-DVn}
%\end{equation}
%leads to 
%\begin{equation}
%[ \mathscr{F}_{\mu\nu}^{[\mathscr{V}]}(x) , \bm{n}_j(x) ]= 0 
% ,
% \label{C29-F-n}
%\end{equation}
%which means 
%\begin{equation}
% \mathscr{F}_{\mu\nu}^{[\mathscr{V}]}(x) =  \sum_{k=1}^{N-1}  F_{\mu\nu}^k(x) \bm{n}_k(x) 
%    .
%\end{equation}
In the $SU(N)$ ($N \ge 3$) case, $\mathscr{F}_{\mu\nu}^{[\mathscr{V}]}(x)$ is not proportional to $\bm{n}(x)$, even if (\ref{C29-F-m}) $[\mathscr{F}_{\mu\nu}^{[\mathscr{V}]}(x) , \bm{n}(x)]=0$ holds:
\begin{equation}
\mathscr{F}_{\mu\nu}^{[\mathscr{V}]}(x) \not= F_{\mu\nu}(x)\bm{n}(x)
    .
\end{equation}
%Note that $\bm{m}(x)=\tilde{\bm{n}}(x)$.
$\mathscr{F}_{\mu\nu}^{[\mathscr{V}]}(x)$ is equal to the $\tilde H$-commutative, i.e., $U(N-1)$-commutative part:
\begin{align}
  \mathscr{F}_{\mu\nu}^{[\mathscr{V}]}(x)
=  \mathscr{F}_{\mu\nu}^{[\mathscr{V}]}(x)_{\tilde{H}}  
%\nonumber\\
=  \bm{n}(x) (\bm{n}(x) \cdot \mathscr{F}_{\mu\nu}^{[\mathscr{V}]}(x)) + \sum_{k=1}^{(N-1)^{2}-1} \bm{u}^{k}(x) \left( \bm{u}^{k}(x) \cdot \mathscr{F}_{\mu\nu}^{[\mathscr{V}]}(x) \right) ,
\end{align}
where $\bm{u}^{k} \in su(N-1)$ are a set of generators of the Lie algebra  $su(N-1)$ of $SU(N-1)$, which commute with $\bm{n}$.

However, the properties (a) and (b) are directly checked by using the explicit expression of decomposed fields which are uniquely determined by solving the defining equation (I) and (II).%
\footnote{ 
See %Kondo (2008) and Kondo, Shinohara and Murakiami (2008)  
 Kondo \cite{Kondo08} and Kondo, Shinohara and Murakami \cite{KSM08}
for the full details. 
}

Therefore, once the Wilson loop operator is rewritten into the surface integral form through the  non-Abelian Stokes theorem using the coherent state, it is represented by the ($SU(N)$ gauge-invariant) ``Abelian'' field strength  $F$ constructed from  the field strength $\mathscr{F}_{\mu\nu}{[\mathscr{V}]}$ of the restricted field $\mathscr{V}$:
\begin{equation}
W_C[\mathscr{V}]
=W_{\Sigma}[F]
:=    \int [d\mu(g)]_{\Sigma} \exp \left( -ig_{{}_{\rm YM}} \  \int_{\Sigma:\partial \Sigma=C} F^g  \right) ,
\ 
F^g :=  {\rm tr}(\tilde{\bm{n}}\mathscr{F}[\mathscr{V}])
    .
    \label{C29-NAST2}
\end{equation} 
Thus we have obtained a gauge-independent  ``Abelian'' dominance in the continuum form: 
\begin{equation}
W_C[\mathscr{A}]=  W_C[\mathscr{V}]=  W_\Sigma[F] ,
\quad 
F^g :=  \sqrt{\frac{2(N-1)}{N}} {\rm tr}(\bm{n}(x)\mathscr{F}[\mathscr{V}](x))
% = {\rm const.} \prod_{x \in S} \int d\mu({g}_{x}) \exp \left( ig \  \int_{S:\partial S=C} F  \right) 
    ,
\end{equation} 
where $\mathscr{V}$ is defined in a gauge-covariant and gauge-independent way and $F$ becomes gauge invariant and hence gauge independent.

Thus we have obtained a sufficient condition for realizing the gauge-independent ``Abelian"  dominance in the $SU(N)$ Wilson loop operator.  
However, it should be remarked that the strong ``Abelian"  dominance holds in the operator level for the Wilson loop operator itself with arbitrary shape and size.
%although the  Abelian projection is obtained as a special gauge fixing from our gauge-invariant framework. 
%It will be interesting to discuss more implications for the infrared Abelian dominance in the usual sense.
Whereas, the ordinary \textbf{infrared Abelian dominance} refers to the expectation value:
\begin{equation}
 \langle W_C[\mathscr{A}] \rangle_{\rm YM} 
\simeq \langle W_C[\mathscr{V}] \rangle_{\rm YM} 
= \langle  W_\Sigma[F]  \rangle_{\rm YM}    
    .
\label{C29-IAD}
\end{equation} 
The strong Abelian dominance for the Wilson loop operator,  
$
 W_C[\mathscr{A}]  =  W_C[\mathscr{V}]   
$, 
does not necessarily imply the dominance of the Wilson loop average 
$
\langle W_C[\mathscr{A}] \rangle_{\rm YM} \simeq \langle W_C[\mathscr{V}] \rangle_{\rm YM} 
$,
%which holds only when the cross term between $\mathscr{V}$ and $\mathscr{X}$ are neglected, since :
which is defined by
\begin{align}
%W(C) %=  \langle W_C[\mathscr{A}] \rangle 
  \langle W_C[\mathscr{A}] \rangle_{\rm YM} 
= Z_{\rm YM}^{-1} \int [d\mathscr{A}] e^{-S_{\rm YM}[\mathscr{A}]  }  W_C[\mathscr{A}] 
= Z_{\rm YM}^{-1} \int [d\mathscr{A}] e^{-S_{\rm YM}[\mathscr{A}]  }  W_C[\mathscr{V}] 
\not= \langle W_C[\mathscr{V}] \rangle_{\rm YM} 
    .
\label{C29-wAD}
\end{align}
Here the original integration measure $[d\mathscr{A}]$ and the action $S_{\rm YM}[\mathscr{A}]$ cannot be written in term of the restricted field $\mathscr{V}$ alone: 
In fact,  the action has the cross term or the interaction term between $\mathscr{V}$ and $\mathscr{X}$, i.e., $S_{\rm YM}[\mathscr{A}] \not= S_{\rm YM}[\mathscr{V}]$, as shown in section \ref{subsubsect:action-minimal}.  
Therefore, the  {infrared Abelian dominance} in the string tension  confirmed numerically from the Wilson loop average does not immediately follows from the  strong ``Abelian"  dominance just defined. 
In the infrared region, however, the field $\mathscr{X}$ decouples from the theory and the effect of the cross term  becomes negligible.
According to the numerical simulations, indeed, there exists  a non-trivial correlation between $\mathscr{V}_\mu(x)$ and $\mathscr{X}_\mu(y)$, but the correlation function $\langle \mathscr{V}_\mu(x) \mathscr{X}_\mu(y) \rangle$ and $\langle \mathscr{X}_\mu(x) \mathscr{X}_\mu(y) \rangle$  fall  off rapidly as the distance between $x$ and $y$ increases.%
\footnote{
For $SU(2)$ case, see Shibata et al. \cite{SKKMSI07}.
For $SU(3)$ in the minimal option,  see Shibata et al. \cite{Shibata-lattice2008}.
%\bibitem{SKKMSI07}
%A. Shibata, S. Kato, K.-I. Kondo, T. Murakami, T. Shinohara and  S. Ito,
%Compact lattice formulation of Cho-Faddeev-Niemi decomposition: Gluon mass generation and infrared Abelian dominance,
%CHIBA-EP-165, KEK-PREPRINT-2007-19,  
%arXiv:0706.2529 [hep-lat],
%Phys.Lett. B{\bf 653}, 101--108  (2007). 
} 
For a sufficiently large loop $C$, finally, the  strong ``Abelian"  dominance for the Wilson loop operator reduces to infrared Abelian dominance for the Wilson loop average.

Finally, we point out that the conventional Abelian projection is reproduced as a special case for the uniform color field:
\begin{equation}
 \bm{n}(x)=T_3=\frac12 \sigma_3 , \ \text{or} \ \mathbf{n}(x)=(0,0,1)
 .
\end{equation} 
the defining equation (II) implies that $\mathscr{X}_\mu(x)$ is the off-diagonal matrix:    
$
\mathscr{X}_\mu(x) = \mathscr{A}_\mu^a(x) \frac{\sigma_a}{2}  \ (a=1,2) 
$, 
while (I) implies that
$\mathscr{V}_\mu(x)$ is the diagonal   matrix:
$
\mathscr{V}_\mu(x) = \mathscr{A}_\mu^3 \frac{\sigma_3}{2}
$.
Thus,  $\mathscr{A}_\mu
= \mathscr{V}_\mu(x)  + \mathscr{X}_\mu(x)$ reduces just to  the Cartan decomposition.
Then the field strength $F$ reduces to the usual definition for the diagonal gauge field $A^3$: $F=dA^3$.
The strong Abelian dominance reads
\begin{equation}
W_C[\mathscr{A}]={\rm const.}W_S[F] \cong {\rm const.}W_C[A^3]
    .
\end{equation} 
%$W_C[\mathscr{A}]={\rm const.}W_S[f]={\rm const.}W_C[A^3]$: Abelian dominance 
% into the diagonal and off-diagonal components:
%$
%\mathscr{A}_\mu(x) = \mathscr{A}_\mu^A(x) \frac{\sigma_A(x)}{2} \ (A=1,2,3) ,
%\ 
%\mathscr{V}_\mu(x) = \mathscr{A}_\mu^3(x) \frac{\sigma_3}{2} ,
%$
Therefore, the color field $\bm{n}(x)$ plays the role of recovering color symmetry which is lost by a global (i.e., space--time independent or uniform) choice of the Abelian direction taken in the conventional approach, e.g., the MA gauge.

%The general SU(N) case is given later for technical reasons.

%%%%%%%%%%%%%%%%%%%%%%%%%%%%%%%%%%%%%%%%%%%%%%%%%%
%%%%%%%%%%%%%%%%%%%%%%%%%%%%%%%%%%%%%%%%%%%%%%%%%%
\subsection{Wilson loop average in the new formulation}
%\setcounter{equation}{0}
%%%%%%%%%%%%%%%%%%%%%%%%%%%%%%%%%%%%%%%%%%%%%%%%%%
%%%%%%%%%%%%%%%%%%%%%%%%%%%%%%%%%%%%%%%%%%%%%%%%%%

The Wilson loop average $W_{C}$ is defined by 
\begin{align}
 W(C)
= \langle W_C[\mathscr{A}] \rangle_{\rm YM}
:=  Z_{{\rm YM}}^{-1} \int \mathcal{D}\mathscr{A}_\mu^A e^{-S_{{\rm YM}}[\mathscr{A}]} W_C[\mathscr{A}] ,
%Z_{{\rm YM}} =  \int \mathcal{D}\mathscr{A}_\mu^A e^{-S_{{\rm YM}}[\mathscr{A}]}  ,
\end{align}
with  the partition function 
$
Z_{{\rm YM}} =  \int \mathcal{D}\mathscr{A}_\mu^A e^{-S_{{\rm YM}}[\mathscr{A}]} 
$
by omitting the gauge fixing procedure to simplify the expression. 
The pre-NAST (\ref{C29-pre-NAST}) for the $SU(3)$ Wilson loop operator tells us that
\begin{equation}
 W(C) =  Z_{{\rm YM}}^{-1} \int   [d\mu(g)]_{C}  \mathcal{D}\mathscr{A}_\mu^A e^{-S_{{\rm YM}}[\mathscr{A}]} 
   e^{  ig_{{}_{\rm YM}} \oint_{C} A^{g} } .
\end{equation}
Inserting the unity:
\begin{align}
1 = \int \mathcal{D}n^\alpha \prod_{x}  \delta \left(\bm{n}(x)-g(x) \frac12 \lambda_8 g^\dagger(x) \right),  \quad g(x) \in G =SU(3)
\end{align}
yields 
\begin{align}
 W(C) =&  Z_{{\rm YM}}^{-1} \int   [d\mu(g)]_{C} \int \mathcal{D}\mathscr{A}_\mu^A  \int \mathcal{D}n^\alpha \delta\left(\bm{n}(x)-g(x) \frac12 \lambda_8 g^\dagger(x) \right)
% \nonumber\\& \times   
 e^{-S_{{\rm YM}}[\mathscr{A}]}  e^{  ig_{{}_{\rm YM}} \oint_{C} A^{g}  } 
.
\end{align}
%where $n^\alpha$  denotes  independent degrees of freedom after solving the constraint $n^A n^A=1$.
Thus, in the reformulated Yang-Mills theory in which $n^\beta(x)$, $\mathscr{C}_\nu^k(x)$, $\mathscr{X}_\nu^b(x)$ are regarded as the {\it independent} field variables,  the Wilson loop average $W(C)$ is written
\begin{align}
 W(C) =&  \tilde{Z}_{{\rm YM}}^{-1} \int   [d\mu(g)]_{\mathbb{R}^D} \int  \mathcal{D}n^\beta \int  \mathcal{D}\mathscr{C}_\nu^k \int  \mathcal{D}\mathscr{X}_\nu^b 
\delta(\tilde{\bm\chi}) \Delta_{\rm FP}^{\rm red}%\delta(\bm{n}(x)-g_{x}(\lambda_8/2)g^\dagger_{x}) 
\tilde{J}
% \nonumber\\&  \times   
e^{-\tilde S_{\rm YM}[\bm n, \mathscr{C},\mathscr{X}]}  e^{  ig_{{}_{\rm YM}} \sqrt{\frac{2(N-1)}{N}} [(j, N_{\Sigma}) + (k, \omega_{\Sigma}) ] }  
 \nonumber\\ 
=& \langle e^{  ig_{{}_{\rm YM}} \sqrt{\frac{2(N-1)}{N}} [(j, N_{\Sigma}) + (k, \omega_{\Sigma}) ] }    \rangle_{\rm YM}
 ,
 \label{Wkj}
\end{align}
where 
%the relation $\delta(\bm{n}(x)-g_{x}(\lambda_8/2)g^\dagger_{x})$ should be understood.
the Yang-Mills action is rewritten in terms of new variables using (\ref{C29-decomp}) and (\ref{C29-NLCV-minimal}), 
\begin{equation}
 S_{\rm YM}[\mathscr A]  = \tilde S_{\rm YM}[\bm n, \mathscr{C},\mathscr{X}] ,
\end{equation}
and the integration measure $\mathcal{D}\mathscr{A}_\mu^A$  is finally transformed to 
\begin{equation}
 \mathcal{D}n^\beta  \mathcal{D}\mathscr{C}_\nu^k \mathcal{D}\mathscr{X}_\nu^b 
\delta(\tilde{\bm\chi}) \Delta_{\rm FP}^{\rm red} \tilde{J} .
%[\bm n,c, \mathscr X]  \tilde{J} 
\end{equation}
The new partition function is introduced:
\begin{equation}
 \tilde{Z}_{{\rm YM}}  
= \int  \mathcal{D}n^\beta \int \mathcal{D}\mathscr{C}_\nu^k \int \mathcal{D}\mathscr{X}_\nu^b  
\delta(\tilde{\bm\chi}) \Delta_{\rm FP}^{\rm red} \tilde{J}
 e^{-\tilde S_{\rm YM}[\bm n, \mathscr{C},\mathscr{X}]} 
.
\label{Z}
\end{equation}
It is shown  that 
\begin{enumerate}
\item[(i)]
the reduction condition $\bm{\chi}[ \mathscr{A},\bm{n}]=0$ is rewritten in terms of new  variables:
\begin{equation}
\tilde{\bm\chi} 
 :=\tilde{\bm\chi} [\bm n, \mathscr{C},\mathscr{X}]
 :=\mathscr{D}^\mu[\mathscr{V}]\mathscr{X}_\mu 
 , 
%\quad
%\mathscr{V}_\mu \equiv c_\mu  \bm{n}  +g^{-1}\partial_\mu \bm{n} \times \bm{n} .
\end{equation} 

\item[(ii)]
 the Jacobian  $\tilde{J}$ 
%is associted with the change of variable:
%$
% (n^\alpha(x), \mathscr{A}_\mu^A(x)) \rightarrow (n^\beta(x), \mathscr{C}_\nu^k(x), \mathscr{X}_\nu^b(x)) ,
%$
%which 
can be simplified, irrespective of the choice of reduction condition (see \cite{KMS06}, section \ref{subsec:path-integral-SU2} for $SU(2)$ and \cite{KSM08}, section \ref{subsec:path-integral-SUN} for $SU(N)$):  
\begin{equation}
  \tilde{J} = 1 ,
\end{equation} 

\item[(iii)]
 the associated Faddeev-Popov determinant  $\Delta_{\rm FP}^{\rm red}$   
%:
%$
% \Delta_{\rm FP}^{\rm red}
%:= \det\left(\frac{\delta\bm\chi}{\delta{\bm\theta}}\right)_{\bm{\chi}=0}
%=   \det\left(\frac{\delta\bm\chi}{\delta\bm n^\theta}\right)_{\bm{\chi}=0} .
%$
is calculable   using the BRST method (see \cite{KMS05}, section \ref{subsec:BRST-SU2} for $SU(2)$ and section \ref{subsec:BRST-SUN} for $SU(N)$).
%\cite{KMS05}
%$
%  \Delta_{\rm FP}^{\rm red}[\bm n,\mathscr{C},\mathscr{X}] 
%= \det \{-D_\mu[\mathscr V+\mathscr X]D_\mu[\mathscr V-\mathscr  X]\}   
%$.
%which reduces to a simple form at one-loop level:
%$
%  \Delta_{\rm FP}^{\rm red}[\bm n,c,\mathscr X] 
%\cong \det \{-D_\mu[\mathscr V]D_\mu[\mathscr V]\}  .
%$

\end{enumerate}

In the previous section, the Wilson loop operator has been exactly rewritten in terms of the gauge-invariant magnetic current $k$ and the electric current $j$.  This shows that the Wilson loop operator can be regarded as a probe of magnetic monopoles.
In this section, moreover, we  have succeeded to connect the Wilson loop  average  with magnetic monopoles which is supposed to be a basic ingredient to cause dual superconductivity as a promising mechanism of quark confinement. 
In fact,  (\ref{Wkj}) tells us what quantity we should examine to see the magnetic monopole contribution $\exp \{  ig_{{}_{\rm YM}} \sqrt{2(N-1)/N} (k, \omega_{\Sigma}) \}$ to the Wilson loop average. This equation is important to give a connection between our formulation and magnetic monopole inherent in the Wilson loop in order to show quark confinement. 
In fact, we give numerical calculations of the potential $V_m(R)$ based on the lattice version of  (\ref{Wkj}) given later. 
The magnetic monopole can be extracted from the restricted field alone.

%%%%%%%%%%%%%%%%%%%%%%%%%%%%%%%%%%%%%%%%%%%%%%%%%%%%%%%%%%%%%
%%%%%%%%%%%%%%%%%%%%%%%%%%%%%%%%%%%%%%%%%%%%%%%%%%%%%%%%%%%%%
\subsection{Extracting the independent field degrees of freedom}
\label{subsubsection:independent-field}%%%%%%%%%%%%%%%%%%%%%%%%%%%%%%%%%%%%%%%%%%%%%%%%%%%%%%%%%%%%%
%%%%%%%%%%%%%%%%%%%%%%%%%%%%%%%%%%%%%%%%%%%%%%%%%%%%%%%%%%%%%

According to the non-Abelian Stokes theorem, the Wilson loop operator is rewritten in terms of $\mathscr{V}_\mu$ and the color field $\bm{n}$, and does not depend on the field variable $\mathscr{X}_\mu$.
Therefore, we can integrate out the field variable $\mathscr{X}_\mu$ in principle, if  the action and the integration measure are rewritten in terms of the new variables. 
In order to perform the integration over $\mathscr{X}_\mu$, the independent field degrees of freedom are to be identified in the integration measure in calculating the Jacobian, while the action can be written in terms of independent degrees of freedom by substituting the decomposition.
 Among the components $\mathscr{X}_\mu^A$ ($A=1, \cdots, {\rm \dim}G$), the component $X_\mu^a$ ($a=1, \cdots, {\rm \dim}(G/\tilde{H})$) in the basis $\bm{e}_a$ are regarded as the independent degrees of freedom.

For $G=SU(2)$, we use $\mathscr{C}_\mu=c_\mu \bm{n}$  and (\ref{C27-basisSU2}) to calculate the covariant derivative of the basis in the background $\mathscr{V}_{\mu}$:
\begin{align}
 \mathscr{D}_\mu[\mathscr{V}]  \bm{e}_a  
 :=& \partial_\mu   \bm{e}_a -ig[\mathscr{V}_\mu,  \bm{e}_a ] 
 \nonumber\\
 =& \mathscr{D}_\mu[\mathscr{B}]  \bm{e}_a -ig[\mathscr{C}_\mu,  \bm{e}_a ] 
 \nonumber\\
 =& g h_\mu \epsilon^{ab}  \bm{e}_b  -ig[c_\mu \bm{n}  ,  \bm{e}_a] 
%\nonumber\\
%=& g h_\mu \epsilon^{ab}  \bm{e}_b + g c_\mu \epsilon^{ab}  \bm{e}_b
 \nonumber\\
 =&  g h_\mu \epsilon^{ab}  \bm{e}_b + g c_\mu \epsilon^{ab}  \bm{e}_b
 = g G_\mu \epsilon^{ab}  \bm{e}_b
  , %\quad
%  G_\rho  := c_\rho  + h_\rho  ,
\end{align}
where we have introduced   
\begin{align}
 G_\mu  := c_\mu  + h_\mu  ,
\end{align}
and used the relation:
\begin{align}
 \mathscr{D}_\mu[\mathscr{B}]  \bm{e}_a   =  g h_\mu \epsilon^{ab}  \bm{e}_b  \quad (f^{ab3}=\epsilon^{ab}) .
  \label{DBeSU2}
\end{align}
Here $h_\mu$ is called the magnetic potential. 
The derivation of the relation (\ref{DBeSU2}) is given in Appendix~\ref{sec:magnetic-potential}.

In the \textbf{minimal option} of $SU(3)$, we find   
\begin{align}
 \mathscr{D}_\mu[\mathscr{V}]  \bm{e}_a  
   =  f^{abj} g G_\mu^j  \bm{e}_b
  , \quad
  G_\mu^j := c_\mu^j + h_\mu^j \quad (j \in \{ 1,2,3,8 \} , \ a,b \in \{ 4,5,6,7 \} ) .
\end{align}
This is derived as follows. We use (\ref{C27-fdec1}) and (\ref{C27-basismin}) to calculate 
\begin{align}
 \mathscr{D}_\mu[\mathscr{V}]  \bm{e}_a  
 :=& \partial_\mu   \bm{e}_a -ig[\mathscr{V}_\mu,  \bm{e}_a ] 
 \nonumber\\
 =& \mathscr{D}_\mu[\mathscr{B}]  \bm{e}_a -ig[\mathscr{C}_\mu ,  \bm{e}_a ] 
 \nonumber\\
 =& g h_\mu^j f^{abj}  \bm{e}_b  -ig[c_\mu^j \bm{u}_j  ,  \bm{e}_a] 
 \nonumber\\
 =&  g h_\mu^j f^{abj}  \bm{e}_b + g c_\mu^j f^{abj}  \bm{e}_b
%\nonumber\\
 =  f^{abj} g G_\mu^j  \bm{e}_b
  , %\quad
%  G_\mu^j := c_\mu^j + h_\mu^j ,
\end{align}
where we have used
\begin{align}
 \mathscr{D}_\mu[\mathscr{B}]  \bm{e}_a  
 =  g h_\mu^j f^{abj}  \bm{e}_b  .
 \label{DBe}
\end{align}
The derivation of (\ref{DBe}) is given in  \ref{sec:magnetic-potential}.

Then the covariant derivative of $\mathscr{X}_\mu$ in the background $\mathscr{V}_{\rho}$  reads
\begin{align}
 \mathscr{D}_\rho[\mathscr{V}] \mathscr{X}_{\mu}  
 =& \mathscr{D}_\rho[\mathscr{V}](X_\mu^a \bm{e}_a) 
%\nonumber\\
 =  \partial_\rho  X_{\mu}^a \bm{e}_a + X_{\mu}^a \mathscr{D}_\rho[\mathscr{V}]  \bm{e}_a 
%\nonumber\\
 =  \partial_\rho  X_{\mu}^a \bm{e}_a + X_{\mu}^a  f^{abj} g G_\rho^j  \bm{e}_b .
 \label{DV-X}
\end{align}
Hence the second covariant derivative reads
\begin{align}
\mathscr{D}^\rho[\mathscr{V}] \mathscr{D}_\rho[\mathscr{V}] \mathscr{X}_{\mu}  
 =& \partial^\rho \partial_\rho  X_{\mu}^a \bm{e}_a + \partial_\rho  X_{\mu}^a \mathscr{D}^\rho[\mathscr{V}]  \bm{e}_a
%\nonumber\\&
+ \partial^\rho (X_{\mu}^a  f^{abj} g G_\rho^j)  \bm{e}_b 
  + X_{\mu}^a  f^{abj} g G_\rho^j  \mathscr{D}^\rho[\mathscr{V}]  \bm{e}_b 
\nonumber\\
 =& \partial^\rho \partial_\rho  X_{\mu}^a \bm{e}_a + \partial^\rho  X_{\mu}^a f^{abj} g G_\rho^j  \bm{e}_b
%\nonumber\\&
+ \partial^\rho (X_{\mu}^a  f^{abj} g G_\rho^j)  \bm{e}_b 
  + X_{\mu}^a  f^{abj} g G_\rho^j  f^{bck} g G^{\rho k}  \bm{e}_c 
  .
\end{align}
Thus the piece quadratic in $\mathscr{X}_\mu$ is rewritten as
\begin{align}
   {\rm tr}\{ \mathscr{X}^\mu (-\mathscr{D}^\rho[\mathscr{V}] \mathscr{D}_\rho[\mathscr{V}] \mathscr{X}_{\mu} ) \}  
% =& \frac12 [ 
%- \partial^\rho \partial_\rho \delta_{ab} +   f^{abj} \partial_\rho  (g G_\rho^j)
 =  \frac12 X^{\mu a} K^{ab} X_\mu^b 
  ,
\end{align}
where we have defined 
\begin{align}
K^{ab} := 
- \partial^\rho \partial_\rho \delta_{ab} 
+  2 f^{abj} g G_\rho^j \partial^\rho 
+   f^{abj} \partial^\rho  (g G_\rho^j)
+    g G_\rho^j (if^{jac})   g G^{\rho k}  (if^{kcb})
  .
\end{align}
Here we have used the normalization convention:
\begin{align}
 {\rm tr}(\bm{e}_a \bm{e}_b) 
= {\rm tr}(e_a^A T_A e_b^B T_B ) 
= \frac12 e_a^A e_b^A  
= \frac12  \bm{e}_a \cdot \bm{e}_b
= \frac12 \delta_{ab} .
\end{align}

In the minimal option of $SU(3)$, the quartic term in $\mathscr{X}$ reads
\begin{align}
  {\rm tr}\{ (i g [ \mathscr{X}_\mu , \mathscr{X}_\nu ])^2 \}
  =&   -g^2{\rm tr}\{  [ \mathscr{X}_\mu , \mathscr{X}_\nu ] [ \mathscr{X}^\mu , \mathscr{X}^\nu ]  \}
  \nonumber\\
  =& -g^2{\rm tr}\{ X_\mu^a  X_\nu^b  [ \bm{e}_a , \bm{e}_b ] X^{\mu c}  X^{\nu d}  [ \bm{e}_c , \bm{e}_d ] \}
  \nonumber\\
  =&  g^2 X_\mu^a  X_\nu^b X^{\mu c}  X^{\nu d}  {\rm tr}\{   f^{abj} \bm{u}_j  f^{cdk} \bm{u}_k  \}
  \nonumber\\
  =&  \frac12 g^2 f^{abj} X_\mu^a  X_\nu^b f^{cdj}  X^{\mu c}  X^{\nu d}       
 ,
  \label{action-minimal-def}
\end{align}
where we have used 
${\rm tr}(\bm{u}_j \bm{u}_k)=\frac12  \delta_{jk}$.

Thus the $SU(3)$ Yang-Mills Lagrangian density in the \textbf{minimal option} (\ref{C27-YM-Lagrangian3}) reads
\begin{align}
   \mathscr{L}_{\rm YM} 
=&  -\frac{1}{2} {\rm tr}( \mathscr{F}_{\mu\nu}[\mathscr{V}] \mathscr{F}^{\mu\nu}[\mathscr{V}] )
-    \frac12   X^{\mu a}  Q_{\mu\nu}^{ab} X^{\nu b} 
-  \frac14 g^2 f^{abj} X_\mu^a  X_\nu^b f^{cdj}  X^{\mu c}  X^{\nu d}
 , 
\nonumber\\
 & (j,k \in \{ 1,2,3,8 \} , \ a,b,c,d \in \{ 4,5,6,7 \} )
\end{align}
where we have defined 
\begin{equation}
Q_{\mu\nu}^{ab}  := K^{ab}  g_{\mu\nu} 
+ 2ig  \mathscr{F}_{\mu\nu}^C [\mathscr{V}] (T_C)^{ab} ,
\label{def:Q}
\end{equation}
with 
\begin{align}
K^{ab} := 
- \partial^\rho \partial_\rho \delta_{ab} 
+    g^2 G_\rho^j (if^{jac})   G^{\rho k}  (if^{kcb})
+    f^{abj}[ 2g G_\rho^j \partial^\rho 
+     \partial^\rho  (g G_\rho^j)]
  .
  \label{def:K}
\end{align}

For $G=SU(2)$, in particular, $j=3$, $a,b \in \{ 1,2 \}$,
\begin{align}
K^{ab} := 
[- \partial^\rho \partial_\rho + g G^{\rho} g G_\rho   ] \delta^{ab} +     [2 g G_\rho \partial^\rho + \partial^\rho  (g G_\rho)
] \epsilon^{ab} 
  ,
\end{align}
where we have used $\epsilon^{ac3} \epsilon^{cb3}=\epsilon^{ac} \epsilon^{cb}=-\delta_{ab}$.

The term quadratic in $X_\mu^a$ is further rewritten using the complex-valued field $X_\mu^\pm$ defined by
$
 X_\mu^\pm := \frac{1}{2} (X_\mu^{1} \pm i X_\mu^{2}) 
$:
\begin{align}
  X_\mu^{a} K^{ab} X_\mu^b
% \mathbb{X}_\mu \cdot [ - D_\rho[\mathbb{V}]D_\rho[\mathbb{V}]] \mathbb{X}_\mu 
%\nonumber\\
%=&    g^2 G_\rho G_\rho \mathbb{X}_\mu  \cdot \mathbb{X}_\mu 
%+   (-X_\mu^1 \partial^2 X_\mu^1 - X_\mu^2 \partial^2 X_\mu^2 ) 
%\nonumber\\
% &- X_\mu^2 gG_\rho \partial_\rho X_\mu^1  +  X_\mu^1 gG_\rho \partial_\rho X_\mu^2 
%- X_\mu^2 \partial_\rho (gG_\rho X_\mu^1) + X_\mu^1 \partial_\rho (gG_\rho X_\mu^2)
%\nonumber\\
=&    
  X_\mu^a [ - \partial^2 \delta^{ab} + g^2 G_\rho G_\rho  \delta^{ab} + 2g\epsilon^{ab}  G_\rho \partial_\rho + g \epsilon^{ab}   \partial_\rho G_\rho ] X_\mu^b 
\nonumber\\
=&    
  X_\mu^+ [ - \partial^2   + g^2 G_\rho G_\rho   + i( 2g G_\rho \partial_\rho + g   \partial_\rho G_\rho) ] X_\mu^-
\nonumber\\&
+ X_\mu^- [ - \partial^2   + g^2 G_\rho G_\rho  - i( 2g G_\rho \partial_\rho + g   \partial_\rho G_\rho) ] X_\mu^+ 
\nonumber\\
=& X_\mu^+ [ -(\partial_\rho-igG_\rho)^2  ] X_\mu^- + X_\mu^- [  -(\partial_\rho+igG_\rho)^2  ] X_\mu^+ .
\end{align}

In the above derivation, we have used the reduction condition. 
In this basis, the reduction condition has the same form as the conventional MA gauge fixing condition: 
\begin{align}
 \mathscr{D}_\mu[\mathscr{V}] \mathscr{X}_{\mu}  = 0 
\Longleftrightarrow 
    \partial_\mu  X_{\nu}^a   - f^{abj} g G_\mu^j  X_{\nu}^b = 0 ,
\end{align}
since (\ref{DV-X}) yields 
\begin{align}
 \mathscr{D}_\mu[\mathscr{V}] \mathscr{X}_{\nu}  
  =  [ \partial_\mu  X_{\nu}^a   - f^{abj} g G_\mu^j  X_{\nu}^b  ] \bm{e}_a .
\end{align}

In the maximal option of $SU(N)$, the diagonal part $a_\mu^j$ and the off-diagonal part $A_\mu^a$ in the conventional Abelian projection are identified with $G_\mu^j$ and $X_\mu^a$ respectively:
\begin{equation}
    G_\mu^j := c_\mu^j+h_\mu^j   \leftrightarrow a_\mu^j , 
\quad 
  X_\mu^a \leftrightarrow  A_\mu^a  .
\end{equation}
For $SU(2)$, in particular, we find
\begin{equation}
 \mathscr{D}_\mu[\mathscr{V}] \mathscr{X}_{\mu}  = 0 
\Longleftrightarrow 
 \partial_\mu X_\mu^a - \epsilon^{ab} g(c_\mu+h_\mu) X_\mu^b=0 .
\end{equation}
%since 
%$
% D_\mu[\mathbb{V}]\mathbb{X}_\nu = \bm{e}_1 [\partial_\mu X_\nu^1 -g(c_\mu+h_\mu)X_\nu^2] + \bm{e}_2 [\partial_\mu X_\nu^2 +g(c_\mu+h_\mu)X_\nu^1]
%$.
But, this does not mean that the reduction condition agrees with the MA gauge. Recall that the reduction condition is introduced to eliminate the extra degrees of freedom in the new reformulation to obtain the same independent degrees of freedom as the original one. Therefore, the above Lagrangian is still  invariant under the local $SU(N)$ gauge transformation, and we can impose any gauge fixing condition, if we want to do so. 
The above formulation reduces to the MA gauge, only when   we set the color field $\bm{n}(x)$ to be uniform in the maximal option, and hence the other bases are also fixed. 
The minimal option for $SU(N)$ ($N \ge 3$) does not agree with the $SU(N)$ MA gauge, even in this limit. 
%The perturbative and loop calculation in this case has been discussed in \cite{Shinohara11}.

%%%%%%%%%%%%%%%%%%%%%%%%%%%%%%%%%%%%%%%%%%%%%%%%%%%%%%%%%%%%%
%%%%%%%%%%%%%%%%%%%%%%%%%%%%%%%%%%%%%%%%%%%%%%%%%%%%%%%%%%%%%
\subsection{Integrating out the infrared non-dominant field   $\mathscr{X}_\mu$}
\label{subsubsection:field-integration}%%%%%%%%%%%%%%%%%%%%%%%%%%%%%%%%%%%%%%%%%%%%%%%%%%%%%%%%%%%%%
%%%%%%%%%%%%%%%%%%%%%%%%%%%%%%%%%%%%%%%%%%%%%%%%%%%%%%%%%%%%%

We proceed to integrate out the field $\mathscr{X}^\mu$ in the functional integration (\ref{Z}). 
For this purpose, we have already rewritten the action (\ref{action-minimal-def}) of the reformulated Yang-Mills theory using the independent variables  $(n^\alpha, C_\mu^j, X_\mu^a)$. 

First, we neglect the term quartic in $\mathscr{X}^\mu$.
The effect of the quartic term will be included later in section \ref{subsection:quartic-X}.

Second, we perform the integration over $\mathscr{X}^\mu$ using the Gaussian (Fresnel) integration, by taking into account the fact that the Jacobian $\tilde J$ for the change of variables is trivial $\tilde J=1$ if the basis adopted in the above is used. 
In the Minkowski space--time, the integration over the field $X_\mu^a$ yields
\begin{align}
 & \int \mathcal{D}X_\mu^a \exp \left\{ -i \frac{1}{2} X^{\mu a}  Q_{\mu\nu}^{ab} X^{\nu b}  \right\} 
 \nonumber\\
 =& (\det Q_{\mu\nu}^{ab})^{-1/2} 
 =  \exp \left[ i \left( \frac{i}{2 } \ln \det Q_{\mu\nu}^{ab} \right) \right] 
 =  \exp \left[i \left( \frac{i}{2 } {\rm Tr} \ln Q_{\mu\nu}^{ab} \right) \right] ,
\end{align} 
while in the Euclidean space, 
\begin{align}
 \int \mathcal{D}X_\mu^a \exp \left\{ -  \frac{1}{2} X_\mu^a Q_{\mu\nu}^{ab} X_\nu^b  \right\} 
%=  (\det Q_{\mu\nu}^{ab})^{-1/2} 
% \nonumber\\
 =  \exp \left[-  \frac{1}{2 } \ln \det Q_{\mu\nu}^{ab} \right] 
 =  \exp \left[-  \frac{1}{2 } {\rm Tr} \ln Q_{\mu\nu}^{ab} \right] .
\end{align}
We have already included the reduction condition $\delta(\tilde{\bm\chi})$. 
In addition, we include  the Faddeev-Popov like term  associated with the reduction condition.  The resulting expression \cite{KMS05,Kondo06} is 
\begin{align}
 \Delta_{\rm FP}^{\rm red}
 = \det K^{ab} 
 =  \exp \left[  i \left( -i \ln \det K^{ab} \right) \right] 
 =  \exp \left[   i \left( -i  {\rm Tr} \ln K^{ab} \right)  \right] .
\end{align}
Thus we obtain the effective $SU(N)$ Yang-Mills Lagrangian density in the  minimal option:
In the Minkowski space, 
\begin{align}
   S_{\rm YM} 
=& - \int  \frac{1}{2} {\rm tr}( \mathscr{F}_{\mu\nu}[\mathscr{V}] \mathscr{F}^{\mu\nu}[\mathscr{V}] )
+   \frac{i}{2 } {\rm Tr} \ln Q_{\mu\nu}^{ab} 
- i    {\rm Tr} \ln K^{ab} ,
\end{align}
while in the Euclidean space, 
\begin{align}
   S_{\rm YM} 
=&  \int  \frac{1}{2} {\rm tr}( \mathscr{F}_{\mu\nu}[\mathscr{V}] \mathscr{F}_{\mu\nu}[\mathscr{V}] )
+   \frac{1}{2} {\rm Tr} \ln Q_{\mu\nu}^{ab} 
-   {\rm Tr} \ln K^{ab} .
\end{align}

For $G=SU(2)$, 
\begin{align}
Q_{\mu\nu}^{ab}  :=& K^{ab}  g_{\mu\nu} 
+ 2ig  \mathscr{F}_{\mu\nu}^C [\mathscr{V}] (T_C)^{ab} ,
\nonumber\\
K^{ab} =& 
[- \partial^\rho \partial_\rho + g G^{\rho} g G_\rho   ] \delta^{ab} +     [2 g G_\rho \partial^\rho + \partial^\rho  (g G_\rho)
] \epsilon^{ab} 
  , \quad ( j=3 ,  a,b \in \{ 1,2 \}) .
\end{align}
Here $K^{ab}$ is written in the matrix form:
\begin{align}
    K^{ab} 
 = \begin{pmatrix}
     - \partial_\rho^2+(gG_\rho)^2 &  2gG^\rho \partial_\rho +\partial^\rho(gG_\rho)          
\\
   -2gG^\rho \partial_\rho -\partial^\rho(gG_\rho)    &  - \partial_\rho^2+(gG_\rho)^2
\\
   \end{pmatrix}
   .
\label{K-def1}
\end{align}
The action for $\mathscr{V}$ is simplified as
\begin{align}
 \frac{1}{2} {\rm tr}( \mathscr{F}_{\mu\nu}[\mathscr{V}]^2 )
=  \frac14 G_{\mu\nu}^2 , 
\quad
%\nonumber\\
 G_{\mu\nu} :=  \bm{n} \cdot \mathscr{F}_{\mu\nu}[\mathscr{V}]
  =  \partial_\mu c_\nu - \partial_\nu c_\mu +i g^{-1} \bm{n} \cdot [\partial_\mu \bm{n} ,  \partial_\nu \bm{n} ] .
\end{align}

This formulation for $SU(2)$ has been already applied to the following problems.
\begin{enumerate} 
\item[i)]
\textbf{Nielsen-Olesen instability} \cite{NO78} of the \textbf{Savvidy vacuum} \cite{Savvidy77}:
 the vacuum (in)\textbf{stability} under the uniform chromomagnetic field or stability of the chromomagnetic condensation. 
See \cite{Kondo04,Kondo06,Kondo14}.

\item[ii)] 
\textbf{dynamical mass generation} for the gluon field through the \textbf{vacuum condensation of mass dimension-two}, i.e., $\langle \mathbf{X}_\mu^2\rangle$ \cite{Kondo06,Kondo14}.

\item[iii)] 
derivation of a low-energy effective field theory in which the propagator of the restricted field has the \textbf{Gribov-Stingl form}  which violates the \textbf{physical positivity}, explaining confinement of the restricted field \cite{Kondo11}.

\item[iv)] 
\textbf{crossover} between chiral-symmetry-breaking/restoration and confinement/deconfinement phase transition at finite temperature \cite{Kondo10}.  

\end{enumerate}

For  $G=SU(3)$ in the \textbf{maximal option}, $K^{ab}$ is given by
\begin{align}
K^{ab} =& 
- \partial^\rho \partial_\rho \delta_{ab} 
-  g G_\rho^j g G^{\rho k}  f^{jac} f^{kcb} 
+  [2 g G_\rho^j    \partial^\rho 
+   \partial^\rho  (g G_\rho^j)] f^{jab} 
 ,
\nonumber\\& 
(j,k \in \{ 3,8 \},  a,b,c  \in \{ 1,2,4,5,6,7 \} )
  .
\end{align}
The structure constants $f^{jab}$ are written in the matrix form ($a,b \in \{ 1,2,4,5,6,7 \}$) which are the off-diagonal matrices:
\begin{align}
 f^{3ab}  %=(T_3)_{ab}
 = \begin{pmatrix}
    0 & 1   & 0 & 0 & 0 & 0   \\
    -1 & 0   & 0 & 0 & 0 & 0   \\
%    \hline
%    0 & 0 & 0 & 0 & 0 & 0 & 0 & 0 \\
%    \hline
    0 & 0   & 0 &  \frac{1}{2}  & 0 & 0   \\
    0 & 0   & -\frac{1}{2} & 0 & 0 & 0   \\
%    \hline
    0 & 0   & 0 & 0 & 0 & -\frac{1}{2}   \\
    0 & 0   & 0 & 0 & \frac{1}{2} & 0   \\
%    \hline
%    0 & 0 & 0 & 0 & 0 & 0 & 0 & 0  
   \end{pmatrix}_{ab}
   , \quad
%\end{align}
%\begin{align}
 f^{8ab}  %=(T_8)_{ab}
 = \begin{pmatrix} 
    0 & 0   & 0 & 0 & 0 & 0   \\
    0 & 0   & 0 & 0 & 0 & 0   \\
%    \hline
%    0 & 0 & 0 & 0 & 0 & 0 & 0 & 0 \\
%    \hline
    0 & 0   & 0 &  \frac{\sqrt{3}}{2} & 0 & 0   \\
    0 & 0   & -\frac{\sqrt{3}}{2} & 0 & 0 & 0   \\
%    \hline
    0 & 0   & 0 & 0 & 0 &  \frac{\sqrt{3}}{2}   \\
    0 & 0   & 0 & 0 & -\frac{\sqrt{3}}{2} & 0   \\
%    \hline
%   0 & 0 & 0 & 0 & 0 & 0 & 0 & 0  
   \end{pmatrix}_{ab}
    ,
    \label{f3abf8ab}
\end{align}
yield the diagonal matrices for the contracted structure constants:
\begin{align}
f^{3ac} f^{3cb}
=& {\rm diag}\left(-1,-1, -\frac14,-\frac14,-\frac14,-\frac14  \right)_{ab}
 ,
\nonumber\\  
f^{3ac} f^{8cb}=& f^{8ac} f^{3cb}
= {\rm diag}\left( 0,0, -\frac{\sqrt{3}}{4}, -\frac{\sqrt{3}}{4}, \frac{\sqrt{3}}{4}, \frac{\sqrt{3}}{4}  \right)_{ab}
 ,
\nonumber\\  
f^{8ac} f^{8cb}
=& {\rm diag}\left(0,0, -\frac34,-\frac34,-\frac34,-\frac34  \right)_{ab}
 .
\end{align}
Thus we find that $K^{ab}$ is decomposed into the diagonal part and the off-diagonal part:
\begin{align}
K^{ab} =& 
- \partial^\rho \partial_\rho \delta_{ab} 
%\nonumber\\&
+    g G_\rho^3 g G^{\rho 3} {\rm diag}\left(1,1, \frac14,\frac14,\frac14,\frac14  \right)_{ab}  
\nonumber\\&
+    g G_\rho^3 g G^{\rho 8}  {\rm diag}\left( 0,0,  \frac{\sqrt{3}}{2},  \frac{\sqrt{3}}{2}, -\frac{\sqrt{3}}{2}, -\frac{\sqrt{3}}{2}  \right)_{ab}
\nonumber\\&
+    g G_\rho^8 g G^{\rho 8} {\rm diag}\left(0,0,  \frac34, \frac34, \frac34, \frac34  \right)_{ab}
\nonumber\\&
+  2 g [G_\rho^3 f^{3ab} + G_\rho^8 f^{8ab}]\partial^\rho 
+   \partial^\rho  (g G_\rho^3) f^{3ab} 
+   \partial^\rho  (g G_\rho^8) f^{8ab} \quad
%\nonumber\\&
 ( a,b   \in \{ 1,2,4,5,6,7 \} )
  ,
\end{align}
which is written in the matrix form (omitting the Lorentz index):
\scriptsize
\begin{align}
  &  K^{ab} 
\nonumber\\
 =& \begin{pmatrix}
     - \partial^2 + G_3^2 & 2G_3 \partial &  0 & 0 & 0 & 0   
\\
    -2G_3 \partial &  - \partial_\rho^2+ G_3^2 &  0 & 0 & 0 & 0   
%\\
%    \hline
%    0 & 0 &  - \partial_0^2 & 0 & 0 & 0 & 0 & 0 
\\
%    \hline
    0 & 0 &    - \partial^2+(\frac12  G_3 +\frac{\sqrt{3}}{2}G_8)^2&   (G_3+\sqrt{3} G_8) \partial_0 & 0 & 0   
\\
    0 & 0 &   - (G_3+\sqrt{3} G_8) \partial &  - \partial_\rho^2+(\frac12  G_3 +\frac{\sqrt{3}}{2}G_8)^2 & 0 & 0   
\\
%    \hline
    0 & 0 &   0 & 0 &  - \partial^2+(\frac12  G_3 -\frac{\sqrt{3}}{2}G_8)^2 & -g(G_3-\sqrt{3} G_8)\partial   
\\
    0 & 0 &   0 & 0 & g(G_3-\sqrt{3} G_8) \partial &  - \partial^2+(\frac12  G_3 -\frac{\sqrt{3}}{2}G_8)^2   
%\\
%    \hline
%    0 & 0 & 0 & 0 & 0 & 0 & 0 &  - \partial_0^2  
   \end{pmatrix}
   .
%\nonumber
\end{align}
\normalsize

For  $G=SU(3)$ in the \textbf{minimal option}, $K^{ab}$ has the form
\begin{align}
K^{ab} =& 
- \partial^\rho \partial_\rho \delta_{ab} 
-    g G_\rho^j  g G^{\rho k}   f^{jac}  f^{kcb} 
+  [2 g G_\rho^j  \partial^\rho 
+   \partial^\rho  (g G_\rho^j)] f^{jab} 
\nonumber\\& 
 (j,k \in \{ 1,2,3,8 \},  a,b,c  \in \{  4,5,6,7 \} )
  ,
\end{align}
which has apparently the same form as that in the maximal option, but the explicit form is different from that, since indices are taken over different values. 
In fact, the structure constants are written in the matrix form:
%\scriptsize
%\footnotesize
\begin{align}
 f^{1ab}  %=(T_1)_{ab}
 =& \begin{pmatrix}
    0 & 0   & 0 &  \frac{1}{2}     \\
    0 & 0   & -\frac{1}{2} & 0    \\
%    \hline
    0 & \frac{1}{2}   & 0 & 0    \\
    -\frac{1}{2} & 0   & 0 & 0     \\
   \end{pmatrix}_{ab}
   , \quad
%\end{align}
%\begin{align}
 f^{2ab}  %=(T_2)_{ab}
 = \begin{pmatrix} 
    0 & 0   &  \frac{1}{2} &   0   \\
    0 & 0   &  0 &  \frac{1}{2}  \\
%    \hline
    - \frac{1}{2} & 0   & 0 & 0     \\
    0 & - \frac{1}{2}   & 0 & 0     \\
   \end{pmatrix}_{ab}
    ,  \quad
%    \label{f1abf2ab}
%\end{align}
\nonumber\\  
%\begin{align}
 f^{3ab}  %=(T_3)_{ab}
 =& \begin{pmatrix}
    0 &  \frac{1}{2}   & 0 &    0   \\
    - \frac{1}{2} & 0    & 0 & 0   \\
%    \hline
    0 & 0  &  0 &  \frac{1}{2}   \\
    0 & 0   & -\frac{1}{2} & 0   \\
   \end{pmatrix}_{ab}
   , \quad
%\end{align}
%\begin{align}
 f^{8ab}  %=(T_8)_{ab}
 = \begin{pmatrix} 
    0 &  \frac{\sqrt{3}}{2} & 0 & 0   \\
     -\frac{\sqrt{3}}{2} & 0 & 0 & 0   \\
%    \hline
     0 & 0 & 0 &  \frac{\sqrt{3}}{2}   \\
     0 & 0 & -\frac{\sqrt{3}}{2} & 0   \\
   \end{pmatrix}_{ab}
    ,
    \label{f3ABf8AB}
\end{align}
\normalsize
yield the diagonal matrices for the contracted structure constants:
\begin{align}
f^{3ac} f^{3cb}
%=&   \begin{pmatrix}
%    0 &  \frac{1}{2}   & 0 &    0   \\
%    - \frac{1}{2} & 0    & 0 & 0   \\
%%    \hline
%    0 & 0  &  0 &  \frac{1}{2}   \\
%    0 & 0   & -\frac{1}{2} & 0   \\
%   \end{pmatrix}
%   \begin{pmatrix}
%    0 &  \frac{1}{2}   & 0 &    0   \\
%    - \frac{1}{2} & 0    & 0 & 0   \\
%%    \hline
%    0 & 0  &  0 &  \frac{1}{2}   \\
%    0 & 0   & -\frac{1}{2} & 0   \\
%   \end{pmatrix}
=&  {\rm diag}\left( -\frac14,-\frac14,-\frac14,-\frac14  \right)_{ab}
 ,
\nonumber\\ 
f^{3ac} f^{8cb}=&  f^{8ac} f^{3cb}
%=  \begin{pmatrix}
%    0 &  \frac{1}{2}   & 0 &    0   \\
%    - \frac{1}{2} & 0    & 0 & 0   \\
%%    \hline
%    0 & 0  &  0 &  \frac{1}{2}   \\
%    0 & 0   & -\frac{1}{2} & 0   \\
%   \end{pmatrix}
%   \begin{pmatrix} 
%    0 &  \frac{\sqrt{3}}{2} & 0 & 0   \\
%     -\frac{\sqrt{3}}{2} & 0 & 0 & 0   \\
%%    \hline
%     0 & 0 & 0 &  \frac{\sqrt{3}}{2}   \\
%     0 & 0 & -\frac{\sqrt{3}}{2} & 0   \\
%   \end{pmatrix}
=  {\rm diag}\left( -\frac{\sqrt{3}}{4}, -\frac{\sqrt{3}}{4}, -\frac{\sqrt{3}}{4}, -\frac{\sqrt{3}}{4}  \right)_{ab}
 ,
\nonumber\\ 
f^{8ac} f^{8cb}
%=& \begin{pmatrix} 
%    0 &  \frac{\sqrt{3}}{2} & 0 & 0   \\
%     -\frac{\sqrt{3}}{2} & 0 & 0 & 0   \\
%%    \hline
%     0 & 0 & 0 &  \frac{\sqrt{3}}{2}   \\
%     0 & 0 & -\frac{\sqrt{3}}{2} & 0   \\
%   \end{pmatrix}
%   \begin{pmatrix} 
%    0 &  \frac{\sqrt{3}}{2} & 0 & 0   \\
%     -\frac{\sqrt{3}}{2} & 0 & 0 & 0   \\
%%    \hline
%     0 & 0 & 0 &  \frac{\sqrt{3}}{2}   \\
%     0 & 0 & -\frac{\sqrt{3}}{2} & 0   \\
%   \end{pmatrix}
=&  {\rm diag}\left( -\frac34,-\frac34,-\frac34,-\frac34  \right)_{ab}
 .
\end{align}
Other terms gives the off-diagonal matrices:
%\footnotesize
\begin{align}
 G^1 f^{1ab} + G^2 f^{2ab}
 = \frac{1}{2}
   \begin{pmatrix}
    0 & 0   & G^2 &    G^1    \\
    0 & 0   & - G^1 &  G^2    \\
%    \hline
    -  G^2 & G^1   & 0 & 0    \\
    - G^1 &  -  G^2   & 0 & 0     \\
   \end{pmatrix}
    ,
\end{align}
and
\begin{align}
  G^3 f^{3ab} + G^8 f^{8ab}
  = \frac{1}{2}
   \begin{pmatrix}
    0 &  G^3 + \sqrt{3}G^8  & 0 &    0   \\
    -  G^3 -  \sqrt{3}   G^8 & 0    & 0 & 0   \\
%    \hline
    0 & 0  &  0 &  G^3 + \sqrt{3}   G^8 \\
    0 & 0   & - G^3 - \sqrt{3}   G^8 & 0   \\
   \end{pmatrix}
    ,
\end{align}
yields 
\begin{align}
 G^1 f^{1ab} + G^2 f^{2ab}+G^3 f^{3ab} + G^8 f^{8ab}
 = \frac{1}{2}
   \begin{pmatrix}
    0 & G^3+\sqrt{3}G^8   & G^2 &    G^1    \\
    -G^3-\sqrt{3}G^8 & 0   & - G^1 &  G^2    \\
%    \hline
    -  G^2 & G^1   & 0 & G^3+\sqrt{3}G^8    \\
    - G^1 &  -  G^2   & -G^3-\sqrt{3}G^8 & 0     \\
   \end{pmatrix}
    ,
\end{align}
Then 
\begin{align}
 (G^1 f^{1ab} + G^2 f^{2ab})^2
 = -\frac{1}{4} ((G^1)^2+(G^2)^2)
   \begin{pmatrix}
    1 & 0   & 0 &    0    \\
    0 & 1   & 0 &  0    \\
%    \hline
    0 & 0   & 1 & 0    \\
    0 &  0   & 0 & 1    \\
   \end{pmatrix}
    ,
\end{align}
%\scriptsize
\begin{align}
 (G^3 f^{3ab} + G^8 f^{8ab})^2
 = -\frac{1}{4}(G^3 + \sqrt{3}G^8)^2
   \begin{pmatrix}
    1 &  0     & 0 &    0   \\
    0   & 1    & 0 & 0   \\
%    \hline
    0 & 0  &  1 &  0 \\
    0 & 0   & 0 & 1    \\
   \end{pmatrix} ,
\end{align}
and
%\scriptsize
%\footnotesize
\begin{align}
& (G^1 f^{1ab} + G^2 f^{2ab}) (G^3 f^{3ab} + G^8 f^{8ab}) 
\nonumber\\
&\quad
 = -\frac{1}{4}
   \begin{pmatrix}
    0 & 0   & G^1(G^3+\sqrt{3}G^8) &    -G^2(G^3+\sqrt{3}G^8)    \\
    0 & 0   & G^2(G^3+\sqrt{3}G^8) &  G^1(G^3+\sqrt{3}G^8)    \\
%    \hline
    G^1(G^3+\sqrt{3}G^8) & G^2(G^3+\sqrt{3}G^8)   & 0 & 0    \\
    -G^2(G^3+\sqrt{3}G^8) &  G^1(G^3+\sqrt{3}G^8)   & 0 & 0     \\
   \end{pmatrix}  
\end{align}
%\normalsize
yields 
%\tiny
\begin{align}
& -(G^1 f^{1ab} + G^2 f^{2ab}+G^3 f^{3ab} + G^8 f^{8ab})^2 
 \nonumber\\
 &\quad
=  \frac{1}{4}
   \begin{pmatrix}
   \scriptstyle (G^1)^2+(G^2)^2+(G^3 + \sqrt{3}G^8)^2
    & 0
    & \scriptstyle 2G^1(G^3+\sqrt{3}G^8) 
    & \scriptstyle -2G^2(G^3+\sqrt{3}G^8)    \\
   0
    & \!\!\!\!\!\!\scriptstyle (G^1)^2+(G^2)^2+(G^3 + \sqrt{3}G^8)^2   
    & \scriptstyle 2G^2(G^3+\sqrt{3}G^8)
    & \scriptstyle 2G^1(G^3+\sqrt{3}G^8)    \\
%    \hline
   \scriptstyle 2G^1(G^3+\sqrt{3}G^8) 
    & \scriptstyle 2G^2(G^3+\sqrt{3}G^8)
    & \!\!\!\!\!\!\scriptstyle (G^1)^2+(G^2)^2+(G^3 + \sqrt{3}G^8)^2
    & 0    \\
   \scriptstyle -2G^2(G^3+\sqrt{3}G^8) 
    & \scriptstyle 2G^1(G^3+\sqrt{3}G^8)   
    & 0 
    & \!\!\!\!\!\!\scriptstyle (G^1)^2+(G^2)^2+(G^3 + \sqrt{3}G^8)^2     \\
   \end{pmatrix}
    .
\end{align}
Finally, we find that $K^{ab}$ has the form:
%\scriptsize
\begin{align}
  K^{ab} 
  =  \begin{pmatrix}
    * 
     & \scriptstyle (G_3+\sqrt{3} G_8) \partial   
     & \scriptstyle G_2 \partial + \frac12 G^1(G^3+\sqrt{3}G^8) 
     & \scriptstyle G_1 \partial - \frac12 G^2(G^3+\sqrt{3}G^8)    \\
    \scriptstyle - (G_3+\sqrt{3} G_8) \partial   
     &  * 
     & \scriptstyle -G_1 \partial +\frac12 G^2(G^3+\sqrt{3}G^8)
     & \scriptstyle G_2 \partial + \frac12  G^1(G^3+\sqrt{3}G^8)    \\
    \scriptstyle -G_2 \partial+\frac12 G^1(G^3+\sqrt{3}G^8) 
     & \scriptstyle G_1 \partial+\frac12 G^2(G^3+\sqrt{3}G^8) 
     &    * 
     & \scriptstyle (G_3+\sqrt{3} G_8) \partial     \\
    \scriptstyle -G_1 \partial-\frac12 G^2(G^3+\sqrt{3}G^8) 
     & \scriptstyle -G_2 \partial+\frac12 G^1(G^3+\sqrt{3}G^8) 
     & \scriptstyle - (G_3+\sqrt{3} G_8) \partial 
     &  *   
   \end{pmatrix}
   ,
\end{align}
%\normalsize
where we have defined 
$
* :=- \partial^2+\frac14 (G_1^2 + G_2^2)+(\frac12  G_3 +\frac{\sqrt{3}}{2}G_8)^2
$.

%%%%%%%%%%%%%%%%%%%%%%%%%%%%%%%%%%%%%%%%%%%%%%%%%%%%%%%%%%%%
%Chapter :
% 
%%%%%%%%%%%%%%%%%%%%%%%%%%%%%%%%%%%%%%%%%%%%%%%%%%%%%%%%%%%%

%%%%%%%%%%%%%%%%%%%%%%%%%%%%%%%%%%%%%%%%%%%%%%%%%%%%%%%%%%%%
%%%%%%%%%%%%%%%%%%%%%%%%%%%%%%%%%%%%%%%%%%%%%%%%%%%%%%%%%%%%
%\subsection{Reformulation of quantum chromodynamics}\label{sec:} 
%%%%%%%%%%%%%%%%%%%%%%%%%%%%%%%%%%%%%%%%%%%%%%%%%%%%%%%%%%%%
%%%%%%%%%%%%%%%%%%%%%%%%%%%%%%%%%%%%%%%%%%%%%%%%%%%%%%%%%%%%

%%%%%%%%%%%%%%%%%%%%%%%%%%%%%%%%%%%%%%%%%%%%%%%%%%%%%%%%%%%%%
%%%%%%%%%%%%%%%%%%%%%%%%%%%%%%%%%%%%%%%%%%%%%%%%%%%%%%%%%%%%%
\subsection{QCD at finite temperature}
\label{subsection:quark-sector}%%%%%%%%%%%%%%%%%%%%%%%%%%%%%%%%%%%%%%%%%%%%%%%%%%%%%%%%%%%%%
%%%%%%%%%%%%%%%%%%%%%%%%%%%%%%%%%%%%%%%%%%%%%%%%%%%%%%%%%%%%%

In this section, we give a theoretical framework to obtain a low-energy effective theory of quantum chromodynamics (QCD) toward a first-principle derivation of confinement/deconfinement and chiral-symmetry breaking/restoration crossover transitions. 
In fact, we demonstrate that an effective theory obtained using simple but non-trivial approximations within this framework enables us to treat both transitions simultaneously on equal footing. 
A resulting effective theory at finite temperature is regarded as a modified and improved form of nonlocal version of the \textbf{Polyakov-loop extended Nambu-Jona-Lasinio (nonlocal PNJL)} models proposed recently by
Hell,  R\"ossner, Cristoforetti and Weise \cite{HRCW08}, Sasaki, Friman and  Redlich \cite{SFR06}, and Blaschke, Buballa, Radzhabov and Volkov \cite{BBRV07}, extending the original (local) PNJL model by Fukushima \cite{Fukushima04}. 
The \textbf{Nambu-Jona-Lasinio (NJL)}  model \cite{NJL61} is well known as a low-energy effective theory of QCD to describe the dynamical breaking of chiral symmetry in QCD  (at least in the confinement phase), see e.g. \cite{Klevansky92,HK94}. 

A novel feature obtained by our results is that the nonlocal NJL coupling depends explicitly on the temperature and Polyakov loop, which affects the entanglement between confinement and chiral symmetry breaking,  together with the cross term introduced through the covariant derivative in the quark sector considered in the conventional PNJL model.  The chiral symmetry breaking/restoration transition is  controlled by the nonlocal NJL interaction, while the confinement/deconfinement transition in the pure gluon sector is specified by the nonperturbative effective potential for the Polyakov loop obtained recently by Marhauser and Pawlowski \cite{MP08}, Fischer,  Maas and  Pawlowski \cite{FMP09} and Braun, Gies and Pawlowski \cite{BGP10}.
The basic ingredients are a reformulation of QCD based on new variables and the flow or functional renormalization group (FRG) equation of the Wetterich type in the Wilsonian renormalization group. 
This framework can be applied to investigate the QCD phase diagram at finite temperature and density.

The action of QCD is written in terms of the gluon field $\mathscr{A}_\mu$ and the quark field $\psi$:
\begin{align}
\label{eq:QCDaction}
 S_{\rm QCD} =& S_{\rm q}  + S_{\rm YM} ,
\nonumber \\
 S_{\rm q}  :=& \int d^Dx \bar{\psi} (i\gamma^\mu \mathcal{D}_\mu[\mathscr{A}] -\hat{m}_q +  \mu_q \gamma^0) \psi ,
\nonumber\\
 S_{\rm YM} :=&  \int d^Dx  \frac{-1}{2} {\rm tr}(\mathscr{F}_{\mu\nu}[\mathscr{A}] \mathscr{F}^{\mu\nu}[\mathscr{A}]) 
 + S_{\rm GF} + S_{\rm FP} 
 ,
\end{align}
where  $\psi$ is the quark field, $\hat{m}_q$ is the quark mass matrix,  $\mu_q$ is the quark chemical potential, $\gamma^\mu$ are the Dirac gamma matrices ($\mu=0, \cdots, D-1$), $\mathcal{D}_\mu[\mathscr{A}]:=\partial_\mu - ig \mathscr{A}_\mu$ is the covariant derivative in the fundamental representation, $g$ is the QCD coupling constant, $\mathscr{A}_\mu=\mathscr{A}_\mu^A T_A$ is the gluon field with $su(N_c)$ generators $T_A$ for the gauge group $G=SU(N_c)$ ($A=1, \cdots, {\rm dim}SU(N_c)=N_c^2-1$) and  $\mathscr{F}_{\mu\nu}[\mathscr{A}]:=\partial_\mu \mathscr{A}_\nu - \partial_\nu \mathscr{A}_\mu -i g [\mathscr{A}_\mu ,  \mathscr{A}_\nu]$ is the field strength.  
In what follows, we suppress the spinor, color and flavor indices. 
Here $S_{\rm GF}$ is the gauge-fixing term and $S_{\rm FP}$ is the associated Faddeev-Popov ghost term. 

By making use of new variables $\mathscr{V}_\mu=\mathscr{V}_\mu^A(x)  T_A$ and $\mathscr{X}_\mu=\mathscr{X}_\mu^A(x)  T_A$ obtained from decomposing the original $SU(N)$ Yang-Mills field:
\begin{equation}
  \mathscr{A}_\mu(x) = \mathscr{V}_\mu(x) + \mathscr{X}_\mu(x)  ,
  \label{decomp}
\end{equation}  
the original QCD action is rewritten into a new form:
\begin{align}
 S_{\rm q}   =& \int d^Dx \Big\{ \bar{\psi} (i\gamma^\mu \mathcal{D}_\mu[\mathscr{V}] -\hat{m}_q +  \mu_q \gamma^0) \psi  
 + gJ^{\mu a} X_\mu^a \Big\} ,
%\nonumber\\&
%+   g\bar{\psi} \gamma^\mu T_A  \psi \mathscr{X}_\mu^A \Big\} ,
\nonumber\\
 S_{\rm YM}  =& \int d^Dx  \Big\{ \frac{-1}{4} (\mathscr{F}_{\mu\nu}^k[\mathscr{V}])^2
%+ \frac{1}{2} \mathscr{F}_{\mu\nu}[\mathscr{V}] \cdot (D_\mu[\mathscr{V}] \mathscr{X}_\nu - D_\nu[\mathscr{V}] \mathscr{X}_\mu)
-     \frac{1}{2} X^{\mu a}  Q_{\mu\nu}^{ab} X^{\nu b} 
%+ \mathcal{O}(\mathscr{X}^3)  ,
%\nonumber\\& 
%+    \frac{1}{2}  (D_\mu[\mathscr{V}] \mathscr{X}_\nu - D_\nu[\mathscr{V}] \mathscr{X}_\mu) \cdot ig[ \mathscr{X}^\mu , \mathscr{X}^\nu ]
%\nonumber\\&
-  \frac14 (\epsilon^{ab}X_\mu^a  X_\nu^b)^2 \Big\}
 + S_{\rm FP},
\label{QCDaction2}
\end{align}
where we have introduced the color current in the basis $e^{A}_{a}$:
\begin{equation}
 J^{\mu a}  
:=  \bar{\psi} \gamma^\mu T_A  \psi e^{A}_{a} ,
\end{equation}
the covariant derivative in the adjoint representation  
$
D_\mu[\mathscr{V}] := \partial_\mu - ig[\mathscr{V}_\mu , \cdot ]
$
and 
%\begin{align} 
%Q_{\mu\nu}^{AB}[\mathscr{V}]  := G^{AB} [\mathscr{V}] g_{\mu\nu} 
%+ 2gf^{ABC} \mathscr{F}_{\mu\nu}^{C}[\mathscr{V}] , 
%\nonumber\\
%G^{AB}[\mathscr{V}]  := - (D_\rho[\mathscr{V}]\mathscr{D}^\rho[\mathscr{V}])^{AB}   .
%\label{Q}
%\end{align}
\begin{align} 
Q_{\mu\nu}^{ab}[\mathscr{V}]  :=& K^{ab} [\mathscr{V}] g_{\mu\nu} 
+ 2gf^{abC} \mathscr{F}_{\mu\nu}^{C}[\mathscr{V}] , 
\quad
%\nonumber\\
%R^{AB}[\mathscr{V}]  := - (D_\rho[\mathscr{V}]\mathscr{D}^\rho[\mathscr{V}])^{AB}   .
%\label{Q2}
%\end{align}
%\begin{align} 
K^{ab}[\mathscr{V}]  :=  - (D_\rho[\mathscr{V}]\mathscr{D}^\rho[\mathscr{V}])^{ab}   
%\nonumber\\
%=& -(\partial_\rho \delta^{AC}+gf^{AEC}\mathscr{V}_\rho^E) (\partial^\rho \delta^{CB}+gf^{CFB}\mathscr{V}^{\rho F})   
%\nonumber\\
%=& - \partial_\rho^2 \delta^{AB} + g^2 f^{AEC}f^{BFC} \mathscr{V}_\rho^E \mathscr{V}^{\rho F} 
%\nonumber\\
%&+ 2g f^{ABE} \mathscr{V}_\rho^E \partial^\rho + gf^{ABE} \partial^\rho \mathscr{V}_\rho^E 
 .
\label{Q}
\end{align}
Here we have explicitly written the independent degrees of freedom for the fields $X_\mu^a$ which are to be integrated out below.

%%%%%%%%%%%%%%%%%%%%%%%%%%%%%%%%%%%%%%%%%%%%%%%%%%%%%%%%%%%%%
%%%%%%%%%%%%%%%%%%%%%%%%%%%%%%%%%%%%%%%%%%%%%%%%%%%%%%%%%%%%%
%\subsection{Effective gauged nonlocal NJL model}
%\label{subsection:quark-sector}%%%%%%%%%%%%%%%%%%%%%%%%%%%%%%%%%%%%%%%%%%%%%%%%%%%%%%%%%%%%%
%%%%%%%%%%%%%%%%%%%%%%%%%%%%%%%%%%%%%%%%%%%%%%%%%%%%%%%%%%%%%

We neglect the quartic term $O(\mathscr{X}^4)$, which will be taken into account later in section \ref{subsection:quartic-X}.
The integration over $\mathscr{X}_\mu^A$ can be achieved by the Gaussian integration:
% according to \cite{Kondo06}. 
\begin{align}
  \int \mathcal{D}X_\mu^a \exp \left\{ -i \frac{1}{2} X_\mu^a Q_{\mu\nu}^{ab} X_\nu^b  + igJ^{\mu c} X_\mu^c \right\} 
 =  (\det Q_{\mu\nu}^{ab})^{-1/2} 
 \exp \left\{ -i \frac{g^2}{2}  J^{\mu a} (Q^{-1})_{\mu\nu}^{ab} J^{\nu b}\right\} 
 .
\end{align} 
Consequently, a nonlocal 4 fermion-interaction is generated:
%an effective 
%interaction between two color currents $J_\mu^A(x)=\bar{\psi}(x) \gamma^\mu   T_A \psi(x)$, i.e., 
%  four-fermion self-interaction of nonlocal type:
\begin{align}
 S_{\rm eff}^{\rm QCD} =&  S_{\rm eff}^{\rm gNJL} + S_{\rm eff}^{\rm glue}  ,
\nonumber\\
S_{\rm eff}^{\rm gNJL} :=&  \int d^Dx \ \bar{\psi} (i\gamma^\mu \mathcal{D}_\mu[\mathscr{V}]  -\hat{m}_q + i  \gamma^0 \mu) \psi 
%\nonumber\\& 
  +  \int d^Dx \int d^Dy  \frac{g^2}{2}  J_{\mu}^{a}(x)  (Q^{-1}[\mathscr{V}])^{\mu\nu}_{ab}(x,y) J_{\nu}^{b}(y)  
\nonumber\\
 S_{\rm eff}^{\rm glue} :=& \int d^Dx  \frac{-1}{4} (\mathscr{F}_{\mu\nu}^k[\mathscr{V}])^2   
%\nonumber\\&
+ \frac{i}{2} \ln \det Q[\mathscr{V}]_{\mu\nu}^{ab} - i \ln \det K[\mathscr{V}]^{ab} 
 ,
%&+ \int d^Dx \int d^Dy  \frac{1}{2}  (\bar{\psi}(x) \gamma^\mu   T_A \psi(x) ) g^2(Q^{-1}[\mathscr{V}])_{\mu\nu}^{AB}(x,y) (\bar{\psi}(y) \gamma^\nu   T_B \psi(y) ) 
%\nonumber\\
%& Q[\mathscr{V}]_{\mu\nu}^{AB}    :=  R[\mathscr{V}]^{AB} g_{\mu\nu} 
%+ 2gf^{ABC} \mathscr{F}_{\mu\nu}^{C}[\mathscr{V}] , 
%\nonumber\\
%& R[\mathscr{V}]^{AB}  := - (D_\rho[\mathscr{V}]\mathscr{D}^\rho[\mathscr{V}])^{AB}   .
\end{align}
%where $Q^{-1}$ is the inverse of $Q$ 
%in the sense that 
%$
%\int d^Dz (Q^{-1})_{\mu\rho}^{AC}(x,z)(Q)_{\rho\nu}^{CB}(z,y)=\delta_{\mu\nu} \delta^{AB} \delta^{(D)}(x-y)
%$
%and 
%$\mathscr{G}_{\mu\nu}^{AB}(x-y)$ is given by (\ref{W}):
%\begin{equation} 
% \mathscr{G}_{\mu\nu}^{AB}(x-y):=g^2(Q^{-1})_{\mu\nu}^{AB}(x,y) ,
%\end{equation}
%and
where the last term $- \ln \det K^{ab}=-{\rm tr
}\ln K^{ab}$ in the action $S_{\rm eff}^{\rm glue}$ comes from the Faddeev-Popov determinant associated with the reduction condition   (see section \ref{subsec:BRST-SU2} and \cite{KMS05} for the precise form). 
% (or the integration of the Faddeev-Popov ghost and antighost field) which is necessary to impose the reduction condition (\ref{reduction-cond}) correctly, see \cite{KMS06,Kondo06} for details. 

In order to estimate the effect of the nonlocal four-fermion interaction, we assume that the inverse  $(Q^{-1})^{\mu\nu}_{ab}[\mathscr{V}]$  is diagonal in the Lorentz indices neglecting the antisymmetric term $2gf^{abC} \mathscr{F}_{\mu\nu}^{C}[\mathscr{V}]$:
\begin{align} 
  (Q^{-1})^{\mu\nu}_{ab}[\mathscr{V}]  
\cong g^{\mu\nu} (K^{-1})^{ab}[\mathscr{V}]  
 .
\end{align}
Then the interaction term reads
\begin{align}
S_{\rm eff}^{\rm int} 
:=& \int d^Dx \int d^Dy  \frac{g^2}{2}  J_{\mu}^{a}(x)  (Q^{-1})^{\mu\nu}_{ab}(x,y) J_{\nu}^{b}(y)  
\nonumber\\
=& \int d^Dx \int d^Dy  \frac{g^2}{2}  J_{\mu}^{a}(x)  (K^{-1})^{ab}(x,y) J_{\mu}^{b}(y)  
 .
\end{align}
Moreover, we consider only the diagonal parts of $(K^{-1})^{ab}[\mathscr{V}]$.  
\footnote{
This approximation is used just for simplifying the Fierz transformation performed below and hence it can be improved by taking into account the off-diagonal parts of $K^{-1}$ if it is necessary to do so.  
}
This is achieved by the procedure
\begin{equation}
  \frac{g^2}{2}   (Q^{-1})_{\mu\nu}^{ab}(x,y) 
\cong \frac{g^2}{2}  g^{\mu\nu} (K^{-1})^{ab}[\mathscr{V}]  
\cong g^{\mu\nu}   \delta^{ab} \mathcal{G}(x-y)
  ,
\end{equation}
which yields
\begin{equation}
 \mathcal{G}(x-y)  = \frac{g^2}{2}  (Q^{-1})_{\mu\nu}^{ab}(x,y)  \frac{g_{\mu\nu}}{D}  \frac{\delta^{ab}}{\delta^{cc}} 
 = \frac{g^2}{2}  (K^{-1})^{ab}[\mathscr{V}]    \frac{\delta^{ab}}{\delta^{cc}} 
 = \frac{g^2}{2}    \frac{ {\rm tr} (K^{-1})[\mathscr{V}]}{\delta^{cc}} 
  .
\end{equation}
where $\delta^{cc}={\rm dim}(G/\tilde H)$.
Then the interaction term reads
\begin{align}
S_{\rm eff}^{\rm int} 
:=& \int d^Dx \int d^Dy  \frac{g^2}{2}  J_{\mu}^{a}(x)  (Q^{-1})^{\mu\nu}_{ab}(x,y) J_{\nu}^{b}(y)  
\nonumber\\
=& \int d^Dx \int d^Dy    J_{\mu}^{a}(x) \mathcal{G}(x-y) J_{\mu}^{a}(y)  
\nonumber\\
=& \int d^Dx \int d^Dy    \bar{\psi}(x) \gamma^\mu T_A  \psi(x) e^{A}_{a}(x)  \mathcal{G}(x-y) \bar{\psi}(y)  \gamma^\mu T_B  \psi(y)  e^{B}_{a}(y)  
 .
\end{align}

For $D=4$,  we use the Fierz identity \cite{Fierz} to rewrite the nonlocal current-current interaction as
\begin{align}
   S_{\rm int}   
   =& \int d^4x \int d^4y      \mathscr{J}^{\mu a}(x) \mathcal{G}(x-y) \mathscr{J}^{\mu a}(y)
  \nonumber\\
=&    \int d^4x  \int d^4y   \mathcal{G}(x-y) \sum_{\alpha} c_\alpha (\bar{\psi}(x) \Gamma_\alpha   \psi(y) ) 
  (\bar{\psi}(y) \Gamma_\alpha  \psi(x) ) 
%  \nonumber\\
%  &=   \int d^4x^\prime \int d^4y^\prime  \sum_{\alpha} c_\alpha (\bar{\psi}(x^\prime) \Gamma_\alpha   \psi(y^\prime) )  \mathcal{G}(x^\prime-y^\prime) (\bar{\psi}(y^\prime) \Gamma_\alpha  \psi(x^\prime) ) 
  \nonumber\\
   =&   \int d^4x   \int d^4z  \mathcal{G}(z) \sum_{\alpha} c_\alpha \{ \bar{\psi}(x+z/2) \Gamma_\alpha   \psi(x-z/2) \}   
   \{ \bar{\psi}(x-z/2) \Gamma_\alpha  \psi(x+z/2) \} ,
\end{align}
where the $\Gamma_\alpha$ are a set of Dirac spinor, color and flavor matrices, resulting from the Fierz transformation, with the property $\gamma_0 \Gamma_\alpha^\dagger \gamma_0=\Gamma_\alpha$. 
Although the Fierz transformation induces mixings and recombinations among operators, the resulting theory must maintain the symmetries of the original QCD Lagrangian. 
A minimal subset of operators satisfying the global chiral symmetry $SU(N_f)_L \times SU(N_f)_R$ which governs low-energy QCD with $N_f$-flavors  is the color-singlet of scalar-isoscalar and pseudoscalar-isovector operators. 
Thus, by restricting $\Gamma_\alpha$ hereafter to  
\begin{equation}
\Gamma_\alpha:=(\mathbf{1}, i\gamma_5 \vec{\tau})
\end{equation}
and ignoring other less relevant operators (vector and axial-vector terms in color singlet and color octet channels), 
 we arrive at a \textbf{nonlocal gauged NJL model}:   
\begin{align}
   S_{\rm eff}^{\rm gNJL} =&  \int d^4x \bar\psi(x) ( i\gamma^\mu \mathcal{D}_\mu[\mathscr{V}] - \hat m_q+i\gamma^0 \mu_q) \psi(x) + S_{\rm int} ,
 \nonumber\\
  S_{\rm int} =&  \int d^4x \int  d^4z \mathcal{G}(z) [\bar \psi(x+z/2) \Gamma_\alpha \psi(x-z/2)  
  \bar \psi(x-z/2)  \Gamma_\alpha \psi(x+z/2)    ] 
 .
 \label{gNJL}
\end{align}
This form is regarded as a gauged version of the nonlocal NJL model proposed in \cite{HRCW08}.    
The function $\mathcal{G}(z)$ is replaced by a coupling constant $G$ times a normalized distribution $\mathcal{C}(z)$:
\begin{equation}
 \mathcal{G}(z)  :=  \frac{G}{2} \mathcal{C}(z) ,
 \quad
 \int d^4z \mathcal{C}(z) = 1 
%\Longrightarrow 
%   \frac{G}{2}  = \int d^4z \mathcal{G}(z) = \tilde{\mathcal{G}}(p=0)
 .
\end{equation}
The standard (local) gauged NJL model follows for the limiting case $\mathcal{C}(z)=\delta^4(z)$ 
with
$\int d^4z \mathcal{C}(z)=1$.

In contrast to \cite{HRCW08}, however, $G$ and $\mathcal{C}$ are determined in conjunction with the behavior of the Polyakov loop $L$  at temperature $T$:
using the Fourier transform $\tilde{\mathcal{G}}(p)$ of $\mathcal{G}$, they are expressed as
\begin{equation}
% \tilde{\mathcal{G}}(p)  :=  \frac{G}{2} \tilde{\mathcal{C}}(p) ,
% \quad
   \frac{G}{2}  
%= \int d^4z \mathcal{G}(z) 
=   \tilde{\mathcal{G}}(p=0) ,
   \quad
   \tilde{\mathcal{C}}(p) = \tilde{\mathcal{G}}(p)/\tilde{\mathcal{G}}(p=0)
%\Longrightarrow 
% \tilde{\mathcal{C}}(p=0)= 1 
  ,
  \label{G}
\end{equation}

The theory given above by $S_{\rm eff}^{\rm QCD}=S_{\rm eff}^{\rm glue}+S_{\rm eff}^{\rm gNJL}$ is able to describe chiral-symmetry breaking/restoration and quark confinement/deconfinement on an equal footing where 
the pure gluon part $S_{\rm eff}^{\rm glue}$  describes confinement/deconfinement transition signaled by the Polyakov loop average. 
We can incorporate the information on confinement/deconfinement transition at finite temperature into the quark sector through the covariant derivative $\mathcal{D}[\mathscr{V}]$ and the nonlocal NJL interaction $\mathcal{G}$ ($G$ and $\mathcal{C}$), in sharp contrast to the conventional PNJL model where the entanglement between chiral-symmetry breaking/restoration and confinement/deconfinement was incorporated  through the covariant derivative $\mathcal{D}[\mathscr{V}]$ alone and the nonlocal NJL interaction $\mathcal{G}$ is fixed to the zero-temperature case. 
In our theory, the nonlocal NJL interaction $\mathcal{G}$ ($G$ and $\mathcal{C}$) is automatically determined through the information of confinement/deconfinement dictated by the Polyakov loop $L$ (non-trivial gluon background), while  in the nonlocal PNJL model \cite{HRCW08} the low-momentum (non-perturbative) behavior of $\mathcal{C}$ was not controlled by first principles and was provided by the instanton model.

%%%%%%%%%%%%%%%%%%%%%%%%%%%%%%%%%%%%%%%%%%%%%%%%%%%%%%%%%%%%%
%%%%%%%%%%%%%%%%%%%%%%%%%%%%%%%%%%%%%%%%%%%%%%%%%%%%%%%%%%%%%
%\subsection{$SU(2)$ gauge group at finite temperature}
%%%%%%%%%%%%%%%%%%%%%%%%%%%%%%%%%%%%%%%%%%%%%%%%%%%%%%%%%%%%%
%%%%%%%%%%%%%%%%%%%%%%%%%%%%%%%%%%%%%%%%%%%%%%%%%%%%%%%%%%%%%

For $SU(2)$ Yang-Mills theory at finite temperature $T$,
  $K^{ab}$ of (\ref{K-def1}) is written in the matrix form 
\begin{align}
    K^{ab} 
 = \begin{pmatrix}
     - \partial^2+(gG)^2 &  2gG\partial          
\\
    -2gG\partial    &  -  \partial^2+(gG)^2      
\\
   \end{pmatrix}
 \cong \begin{pmatrix}
     - \partial_\ell^2-\partial_0^2+T^2 \varphi^2 &  2T\varphi \partial_0            
\\
    -2T\varphi \partial_0   &  - \partial_\ell^2-\partial_0^2+T^2 \varphi^2      
\\
   \end{pmatrix}
   .
\end{align}
where the gauge fixing condition $\partial G=0$ is adopted.
Here we have chosen a trivial background for the spatial component $G_\ell$ ($\ell=1,2,3$) and the uniform background $g^{-1}T \varphi$ for the time component $G_0$ with the angle variable $\varphi$ which specifies the Polyakov loop operator. 
Then this leads to
\begin{align} 
  (Q^{-1})^{\mu\nu}_{ab}[\mathscr{V}]  
= g^{\mu\nu} (K^{-1})^{ab}[\mathscr{V}]  
 =      g^{\mu\nu}
  \begin{pmatrix}
  \frac12 [F_{\varphi}+F_{-\varphi}] & -\frac{1}{2i} [F_{\varphi}-F_{-\varphi}]   \cr
  \frac{1}{2i} [F_{\varphi}-F_{-\varphi}] & \frac12 [F_{\varphi}+F_{-\varphi}]   \cr
   \end{pmatrix} 
 ,
\end{align}
where $F_{\varphi}$ is defined by
\begin{align} 
F_{\varphi}(i\partial) :=&  \frac{1}{(i\partial_\ell)^2+(i\partial_0+T\varphi)^2+R_{k}}
 =   \frac{1}{(i\partial_\mu)^2+(T\varphi)^2+2T\varphi i\partial_0+R_{k}}
 , 
\end{align}
where $R_{k}$ is the infrared cutoff function at the flow parameter $k$ used in the functional renormalization group. 
Then the nonlocal interaction is obtained as
\begin{equation}
 \mathcal{G}(x-y) 
=   \frac{g^2}{2}   \frac{{\rm tr}(K^{-1})}{2} 
=  \frac{g^2}{2}  \frac{F_{\varphi}+F_{-\varphi}}{2}
  ,
  \label{4fGa}
\end{equation}
whose Fourier transform is
\footnote{
In what follows, we correct some errors   in  Kondo \cite{Kondo10}: (82), (88), (103), (104), (105), (107).
These equations should be corrected as given below, which are to be consistent with (\ref{4fGa}) and (\ref{4fGb}).
But qualitative  results obtained in \cite{Kondo10} do not change by these corrections. 
} 
\begin{align}
 \tilde{\mathcal{G}}(p)
=&   \frac{g^2}{2}  \frac{\tilde{F}_{\varphi}(p)+\tilde{F}_{-\varphi}(p) }{2} , 
\quad
\tilde{F}_{\varphi}(p)=  \frac{1}{p^2+(T\varphi)^2+2T\varphi p_0+R_{k}(p)} .
  \label{4fGb}
\end{align}
Note that $\tilde{F}_{\varphi}(p=0)$ and hence $G$ diverge at $T=0$.
This comes from an improper treatment of the $T=0$ part.
To avoid  this IR divergence at $T=0$, we add the $T=0$ contribution $M_0^2 \simeq M_X^2$ and replace $F_{\varphi}(i\partial)$ by  
\begin{align} 
F_{\varphi}(i\partial) 
 = \frac{1}{(i\partial_\ell)^2+(i\partial_0+T\varphi)^2+M_0^2}
 =  \frac{1}{(i\partial_\mu)^2+(T\varphi)^2+2T\varphi i\partial_0+M_0^2}
 , 
\end{align}
and
\begin{align}
\tilde{F}_{\varphi}(p)
 = \frac{1}{\bm{p}^2+(p_0+T\varphi)^2+M_0^2}
 =  \frac{1}{p^2+(T\varphi)^2+2T\varphi p_0+M_0^2} .
\end{align}
In fact, such a contribution $\frac12 M_0^2 $ comes in $G^{AB}$ as an additional term $M_0^2 \delta^{AB}$ from the $O(\mathscr{X}^4)$ terms (Note that $O(\mathscr{X}^3)$ terms are absent for $G=SU(2)$), as already mentioned before. 
%This will be valid for $p < M_X \simeq 1.2$ GeV.
%$M_0 \simeq 0.6 \sim 0.8$ GeV?

At $T=0$ (and at zero chemical potential $\mu_q=0$), QCD must be in the hadron phase where the chiral symmetry is spontaneously broken, which means that the NJL coupling constant $G(0)$ at zero temperature must be greater than the critical NJL coupling constant $G_c$:
\begin{equation}
 G(0) =  g^2  \frac{1}{M_0^2} > G_c .
\end{equation}
The nonlocality function or the form factor $\tilde{\mathcal{C}}(p)$ at $T=0$ behaves as
\begin{equation}
 \tilde{\mathcal{G}}(p)
=   \frac{g^2}{2}     \frac{1}{p^2 +M_0^2 }, \ 
 \tilde{\mathcal{G}}(0)
=   \frac{g^2}{2}    \frac{1}{M_0^2 } ,
\end{equation}
and
\begin{equation}
\tilde{\mathcal{C}}(p) =  \frac{M_0^2}{p^2 +M_0^2 } 
    .
\end{equation}

As an immediate outcome of the resulting effective theory, this determines the temperature-dependence of the coupling constant $G$ of nonlocal NJL model.
Using (\ref{G}), we have
\begin{equation}
 G(T) =    \frac{g^2}{(T\varphi)^2+M_0^2} , \
% G(0) =  g^2  \frac{1}{M_0^2}  ,
\end{equation}
which lead to the NJL coupling constant normalized at $T=0$:
\begin{equation}
 \frac{G(T)}{G(0)} =    \frac{M_0^2}{(T\varphi)^2+M_0^2}  .
\end{equation}

In the presence of the dynamical quark $m_q < \infty$, the Polyakov loop is not an exact order parameter and does not show a sharp change with discontinuous derivatives.  Even in this case, we can introduce the pseudo critical temperature $T_d^*$ as a temperature achieving the peak of the susceptibility. 
Below the deconfinement temperature $T_d^*$, i.e., $T < T_d^*$, therefore, $L \simeq 0$ (or $\varphi \simeq \pi$), the NJL coupling constant $G$ has the temperature-dependence:
\begin{equation}
%T < T_d^* \Longrightarrow L \simeq 0 \Longrightarrow   
G(T)/G(0) \simeq   \frac{M_0^2}{\pi^2 T^2+M_0^2}  \ (T < T_d^*) .
\end{equation}
This naive estimation gives a qualitative understanding for the existence of chiral phase transition.   Since $G(T)$ is (monotonically) decreasing as the temperature $T$ increases, it becomes smaller than the critical NJL coupling constant:  
\begin{equation}
 G(0) =  g^2  \frac{1}{M_0^2} > G_c ,
 \quad 
  T \uparrow \infty \Longrightarrow 
G \downarrow 0 .
\end{equation}
Thus, the chiral transition temperature $T_\chi$ will be determined (if the chiral-symmetry restoration and confinement coexist or the chiral symmetry is restored in the confinement environment before deconfinement takes place, i.e.,  $T_\chi \le T_d^*$ ) by solving
\begin{equation}
 G(T_\chi) \equiv   G(0)  \frac{ M_0^2}{T_\chi^2 \pi^2 +M_0^2}  = G_c 
 .
\end{equation}
Here we have assumed that the nonlocality function $\tilde{\mathcal{C}}(p)$ gives the dominant contribution at $p=0$, namely, $\tilde{\mathcal{C}}(p) \le  \tilde{\mathcal{C}}(0)=\int d^4z \mathcal{C}(z)=1$ and that the occurrence of the chiral transition is determined by the NJL coupling constant alone. 

At finite temperature $T$, the form factor reads
\begin{align}
 \mathcal{C}(\bm{x}-\bm{y}) 
%\nonumber\\ 
 =& T \sum_{n \in \mathbb{Z}} \int \frac{d^3p}{(2\pi)^3} 
  \tilde{\mathcal{C}}(p_0=\omega, \bm{p}) e^{i\bm{p} \cdot (\bm{x}-\bm{y})} 
\nonumber\\ 
 =& T \sum_{n \in \mathbb{Z}} \int \frac{d^3p}{(2\pi)^3} 
\frac{M_0^2}{3} \Big[ \frac{2}{\bm{p}^2+(\omega+T\varphi)^2+M_0^2} 
%\nonumber\\&
+ \frac{1}{\bm{p}^2+\omega^2+M_0^2} \Big] e^{i\bm{p} \cdot (\bm{x}-\bm{y})} 
\nonumber\\ 
 =&   \int \frac{d^3p}{(2\pi)^3} 
\frac{M_0^2}{6\epsilon_p} \Big[ 2\frac{\sinh (\epsilon_p/T)}{\cosh(\epsilon_p/T)-\cos(\varphi)} 
%\nonumber\\&
+  \frac{\sinh (\epsilon_p/T)}{\cosh(\epsilon_p/T)-1}  \Big] e^{i\bm{p} \cdot (\bm{x}-\bm{y})} 
 ,
\end{align}
where we have defined 
$\epsilon_p:=\sqrt{\bm{p}^2+M_0^2}$
and used
\begin{align}
 T \sum_{n \in \mathbb{Z}}  \frac{1}{(\omega+C)^2+\epsilon_p^2}   
 = \frac{1}{2\epsilon_p}   \frac{\sinh (\epsilon_p/T)}{\cosh(\epsilon_p/T)-\cos(C/T)} 
 .
\end{align}
The form factor $\mathcal{C}$ does not change so much around the deconfinement temperature $T \sim T_d^*$ (or $\varphi \sim \pi$). 
This is reasonable since the form factor is nearly equal to the $\mathscr{X}$ correlator $Q^{-1}$, as already  mentioned before. 
The more details can be seen in \cite{Kondo10} .

%%%%%%%%%%%%%%%%%%%%%%%%%%%%%%%%%%%%%%%%%%%%%%%%%%%%%%%%%%%%%%%%%%%%%%%%%%%%%%%%%%%%%%%%%%%%%%%%%%%%%%%%%%
\subsection{Effect of the quartic interaction among $\mathscr{X}$}
\label{subsection:quartic-X}%%%%%%%%%%%%%%%%%%%%%%%%%%%%%%%%%%%%%%%%%%%%%%%%%%%%%
%%%%%%%%%%%%%%%%%%%%%%%%%%%%%%%%%%%%%%%%%%%%%%%%%%%%%

We can introduce the gauge-invariant mass term which is invariant under the local $SU(2)$ gauge transformation II as pointed out in \cite{KMS06}:
\begin{equation}
\mathscr{L}_{\rm m}
= \frac{1}{2} M_X^2 \mathbf{X}_\mu^2 ,
\label{mass1}
\end{equation}
since the mass dimension two operator $\mathbf{X}_\mu^2$ is invariant under the local gauge transformation: 
\begin{align}
 \delta_{\omega'} \mathbf{X}_\mu^2(x) =   0 .
\end{align}
This  gauge-invariant mass term is rewritten in terms of the original variables $\mathscr{A}_\mu$:
\begin{align}
\mathscr{L}_{\rm m}
 =& \frac{1}{2} M_X^2 (\mathbf{A}_\mu - \mathbf{V}_\mu)^2 
 = \frac{1}{2} M_X^2 (\mathbf{A}_\mu -c_\mu \mathbf{n}
  +g^{-1}\partial_\mu \mathbf{n}\times \mathbf{n} )^2 
%\nonumber\\
=  \frac{1}{2g^2} M_X^2 (D_\mu[\mathbf{A}] \mathbf{n})^2 ,
  \label{mass2}
\end{align}
under the understanding that the color field $\mathbf{n}$ is expressed in terms of the original gauge field $\mathbf{A}_\mu$ by solving the reduction condition. 
Therefore, $\mathbf{V}_\mu$ (or $c_\mu$ and $\mathbf{n}$) plays the similar role to the St\"uckelberg field to recover the local gauge symmetry.
Note that   $c_\mu$, $\mathbf{n}$ and $\mathbf{X}_\mu$ are treated as independent variables after the non-linear change of variables and the mass term is a polynomial in the new variable $\mathbf{X}_\mu$, although they might be non-local and non-linear composite operators of the original variables $\mathbf{A}_\mu$.

The proposed mass term (\ref{mass1}) or (\ref{mass2}) for the gluon: 
\begin{align}
\mathscr{L}_{\rm m}
 =&   M_X^2 {\rm tr} \{(\mathscr{A}_\mu - \mathscr{V}_\mu)^2  \} 
 =  M_X^2 {\rm tr} \{ (\mathscr{A}_\mu -c_\mu \bm{n}
  -ig^{-1} [\partial_\mu \bm{n},   \bm{n} ] )^2 \} 
%\nonumber\\
=  \frac{1}{g^2} M_X^2 {\rm tr} \{ (D_\mu[\mathscr{A}] \bm{n})^2 \} ,
  \label{mass4}
\end{align}
should be compared with the conventional gauge-invariant mass term of Kunimasa--Goto type \cite{KG67}:
\begin{equation}
\mathscr{L}_{\rm KG}
 =   M^2 {\rm tr} \{(\mathscr{A}_\mu - ig^{-1}U\partial_\mu U^\dagger)^2 \}
 =   M^2 {\rm tr} \{(UD_\mu[\mathscr{A}]U^\dagger)^2 \} ,
 \quad 
 U(x) = e^{-i\chi(x)/v} .
 \label{mass3}
\end{equation}
This mass term is non-polynomial in the St\"uckelberg field $\chi(x)$.  This fact makes the field theoretical treatment very difficult.

It is possible to  argue that there occurs a novel \textbf{vacuum condensation of mass dimension--two} for the field $\mathbf{X}^\mu$, i.e.,% 
\footnote{
We adopt the Minkowski metric $g_{\mu\nu}={\rm diag}(1,-1,-1,-1)$.  After the Wick rotation to the Euclidean region, the Minkowski metric tensor $g_{\mu\nu}$ is replaced by 
$
 -\delta_{\mu\nu} ={\rm diag}(-1,-1,-1,-1)
$.  Therefore, we have
$
 -\mathbf{X}_\mu^2 \rightarrow (\mathbf{X}^E_\mu)^2 > 0
$.
}
\begin{equation}
 \left< -\mathbf{X}_\mu^2 \right> \ne 0 .
\end{equation}
and that the field $\mathbf{X}^\mu$ acquires the mass dynamically through this condensation.  
A naive way to see this is to apply the mean-field like argument or the Hartree--Fock approximation to the four-gluon interaction, i.e., the quartic self-interaction among $\mathbf{X}^\mu$ which leads to 
the gauge-invariant mass term for  $\mathbf{X}_\mu$ gluons and a gauge-invariant gluon mass $M_X$:
\begin{align}
   -\frac{1}{4}(g \mathbf{X}_\mu \times \mathbf{X}_\nu)^2 
 =&  -\frac{1}{4}(g \mathbf{X}_\mu \times \mathbf{X}_\nu) \cdot (g \mathbf{X}^\mu \times \mathbf{X}^\nu) 
 \nonumber\\
\to & \frac{1}{2}g^2 \mathbf{X}^A_\mu \left[\left\langle -\mathbf{X}^2_\rho \right\rangle \delta^{AB} - \left\langle -\mathbf{X}^A_\rho \mathbf{X}^B_\rho \right\rangle \right] \mathbf{X}^{\mu B} 
=   \frac{1}{2} M_X^2 \mathbf{X}_\mu \cdot \mathbf{X}^\mu ,
\quad 
 M_X^2 = \frac23 g^2 \left\langle -\mathbf{X}^2_\rho \right\rangle .
\label{mass-term}
\end{align} 
Consequently, the gauge-invariant mass for the ``off-diagonal'' gluon $\mathbf{X}_\mu$ is generated. 
Then the $\mathbf{X}_\mu$ gluon modes  decouple in the low-energy (or long-distance) region below the mass scale $M_X$.  
Consequently, the infrared ``Abelian'' dominance for the large Wilson loop average follows immediately from the fact that the Wilson loop operator is written in terms of $\mathbf{V}_\mu$ alone according to the non-Abelian Stokes theorem for the Wilson loop operator and that the Wilson loop average is entirely estimated by the restricted field alone.

The numerical simulations on a lattice \cite{IKKMSS06} have demonstrated the infrared ``Abelian'' dominance and magnetic monopole dominance in the string tension within the compact lattice reformulation, although such phenomena were found for the first time in the MA gauge \cite{SY90,SNW94}.
In fact, the numerical simulations have shown that the string tension calculated from the magnetic part of the Wilson loop average 
%according to (\ref{W2}), (\ref{reducedW}) and (\ref{Z}) 
in the  reformulation of lattice Yang-Mills theory reproduces  90 $\sim$ 95 \% of the full string tension calculated from the original Wilson loop average in the conventional lattice formulation. 
More numerical simulations \cite{SIKKMS06} have shown that the remaining field $\mathbf{X}_\mu$ defined on a lattice acquires the mass  
$M_X \cong 1.2 {\rm GeV}$ which is obtained as the exponential decay rate of the two-point correlation function $\left< \mathbf{X}_\mu(x) \cdot \mathbf{X}_\mu(y) \right>_{\rm YM}$ measured on a lattice: 
$\left< \mathbf{X}_\mu(x) \cdot \mathbf{X}_\mu(y) \right>_{\rm YM} \sim |x-y|^{-\alpha}\exp(-M_X|x-y|)$. 
This value agrees with that of the off-diagonal gluon mass obtained in the MA gauge \cite{AS99}.  
%On the other hand, the same analysis applied to the ``Abelian'' gluon $\mathbf{V}_\mu$   leads to the result $M_V \cong 0.6 {\rm GeV}$. 
This is consistent with the ``Abelian'' dominance. 
See section~\ref{subsubsection:IAD-correlation}.

Moreover, it was pointed out that the existence of dimension--two condensate eliminates a tachyon mode  causing the Nielsen--Olesen instability \cite{NO78} of the  vacuum  with the homogeneous magnetic condensation  $\left< H \right>\ne 0$, which is called  the vacuum of the Savvidy type \cite{Savvidy77}.
Here the magnetic field $H$ is defined by the Lorentz invariant form 
\begin{align}
H:=\sqrt{\frac12 H_{\mu\nu}^2}, \quad 
H_{\mu\nu}=-g\mathbf{n} \cdot (\partial_\mu \mathbf{n} \times \partial_\nu \mathbf{n}) .
\end{align} 
%with $H_{\mu\nu}=-g\mathbf{n} \cdot (\partial_\mu \mathbf{n} \times \partial_\nu \mathbf{n})$.
Therefore, the restoration of the vacuum stability with homogeneous magnetic condensation is obtained as a by-product of the above result.  
The stability of the magnetic vacuum is desirable for the magnetic monopole dominance. 
Thus the non-perturbative Yang-Mills vacuum is characterized by two vacuum condensations, i.e., the condensation $\left< \mathbf{X}_\mu^2 \right>\ne 0$ and the magnetic condensation $\left< H \right>\ne 0$, both of which realize the vacuum energy lower than that of the perturbative vacuum. 
\footnote{
See  Kondo \cite{Kondo06} for more precise treatment of this idea.
}

A first analytical calculation of the effective potential of the composite gluon operator of mass dimension--two $\mathbf{X}_\mu^2$ has been performed for demonstrating the occurrence of the condensation as realized at the minimum located away from the origin.  See \cite{KKMSS05} for numerical simulations for the effective potential on a lattice.  Note that the composite operator $\mathbf{X}_\mu^2$ is gauge-invariant and the resulting mass term for $\mathbf{X}_\mu$ can be induced keeping the original $SU(2)$ gauge invariance intact, contrary to the conventional wisdom.  Consequently, the decoupling of these degrees of freedom is characterized as a gauge-invariant low-energy phenomenon. In other words, this is a dynamical Abelian projection, suggesting the validity of the dual superconductor picture for quark confinement.  Thus the infrared ``Abelian'' dominance  immediately follows in the gauge invariant manner. 
See also \cite{Kondo14} for more recent results along this direction in the framework of the functional renormalization group.

Furthermore, the existence of this condensate enables us to derive the Faddeev-Niemi model describing glueballs as knot solitons, which is regarded as a low-energy effective theory of the Yang-Mills theory, as already pointed out in \cite{Kondo04,KOSSM06}. 
%\noindent
%$\bullet$ glueball mass  
%$\Longrightarrow $ mass gap:
Then the  mass of the knot soliton of the Faddeev-Niemi model could be given by the fractional power of the Hopf topological number $Q_H$ in units of the gluon condensate $\left< \mathbf{X}_\mu^2 \right>$:
\begin{align}
  M_n = C_0 \sqrt{\left< \mathbf{X}_\mu^2 \right>}|Q_H|^{3/4}
\end{align} 
where  $C_0$ is a constant. 
%\noindent
%$\bullet$ string tension (the coefficient of the linear potential)  
%$\Longrightarrow $ confinement:
%\\
The existence of the gluon condensate $\left< \mathbf{X}_\mu^2 \right>$ of dimension two has implications for non-perturbative dynamics of the Yang-Mills theory.
For quark confinement, e.g., there is an argument suggesting that the string tension is proportional to the gluon condensate $\left< \mathbf{X}_\mu^2 \right>$ of dimension two:
\begin{align}
  \sigma = {\rm const.} \left< \mathbf{X}_\mu^2 \right> .
\end{align} 
The gauge-invariant operator $\mathscr{X}_\mu^2$  of mass dimension two corresponds to the \textbf{BRST invariant operator of mass dimension two}, constructed from the gluon and ghost \cite{Kondo01}: 
%The existence of dimension-two condensate $\left< \mathscr{X}_\mu^2 \right> \ne 0$ has been examined by an analytical method \cite{Kondo06}  and numerical one \cite{KKMSS05}. 
for the partial gauge fixing $G \rightarrow H$,  consider the composite operator:
\begin{equation}
\mathcal{O}_{K}(x) :=   {\rm tr}_{\rm G/H} \biggl[ \frac{1}{2} \mathscr{A}_{\mu}(x) \mathscr{A}^{\mu}(x) + \lambda i \mathscr{\bar{C}}(x) \mathscr{C}(x) \biggr] ,
\label{C25-dim2-cond1}
\end{equation}
where $\mathscr{A}$ is the gauge field, $\mathscr{\bar{C}}, \mathscr{C}$ are the ghost fields and $\lambda$ is a gauge fixing parameter.

These are advantages of the reformulation of the Yang-Mills theory based on the non-linear change of variables.

\newpage
%%%%%%%%%%%%%%%%%%%%%%%%%%%%%%%%%%%%%%%%%%%%%%%%%%%%%%%%%%%%
%Chapter :
% 
%%%%%%%%%%%%%%%%%%%%%%%%%%%%%%%%%%%%%%%%%%%%%%%%%%%%%%%%%%%%

%%%%%%%%%%%%%%%%%%%%%%%%%%%%%%%%%%%%%%%%%%%%%%%%%%%%%%%%%%%%
%%%%%%%%%%%%%%%%%%%%%%%%%%%%%%%%%%%%%%%%%%%%%%%%%%%%%%%%%%%%
\section{Reduction condition and magnetic monopole in the topological Yang-Mills background }\label{sec:reduction-topology} 
%%%%%%%%%%%%%%%%%%%%%%%%%%%%%%%%%%%%%%%%%%%%%%%%%%%%%%%%%%%%
%%%%%%%%%%%%%%%%%%%%%%%%%%%%%%%%%%%%%%%%%%%%%%%%%%%%%%%%%%%%

In this section we check that the reduction condition can be used to determine the color direction field, which plays the key role  in the new reformulation of the Yang-Mills theory written in terms of the new variables.
Moreover, we examine whether circular loops of magnetic monopole exist or not  in the four-dimensional Euclidean $SU(2)$ Yang-Mills theory, since they are expected to be dominantly responsible for quark confinement in the dual superconductor picture. 
To achieve these purposes, we solve  in an analytical and numerical ways (the differential equation for) the reduction condition to obtain the color direction field for a given  Yang-Mills field.  
As  the given Yang-Mills fields, we examine  some exact solutions of the classical Yang-Mills field equation of motion characterized by the non-trivial topological invariant, i.e., one-meron, two-merons, one-instanton, and two-instantons.

%%%%%%%%%%%%%%%%%%%%%%%%%%%%%%%%%%%%%%%%%%%%%%%%%

\subsection{Brief history before our works}

%%%%%%%%%%%%%%%%%%%%%%%%%%%%%%%%%%%%%%%%%%%%%%%%%

The \textbf{dual superconductivity} picture \cite{dualsuper} for quark confinement  \cite{Wilson74} was proposed long ago and it is now believed to be a promising mechanism for quark confinement. 
The dual superconductivity is supposed to be realized as an \textbf{electric-magnetic dual} of the ordinary superconductivity.  For the dual superconductivity to be possible, there must exist magnetic monopoles and antimonopoles to be condensed for causing the \textbf{dual Meissner effect}, just as the \textbf{Cooper pairs} exist and they are condensed to cause \textbf{Meissner effect} in the \textbf{ordinary superconductivity}. 
The idea of the dual superconductivity is intuitively easy to understand, but upgrading this idea into a quantitative theory  is not so easy, as can be seen from a fact that we are still involved in this work.

The \textbf{topological soliton} allowed in the $D$-dimensional $SU(2)$ Yang-Mills theory \cite{YM54}  with no matter fields  is only the Yang-Mills \textbf{instanton}
\footnote{
The $D=4$ Euclidean Yang-Mills instanton can be also regarded as a static solution with a finite energy in D=4+1 Minkowski space--time.
Here we use a \textbf{topological soliton} as implying a solution of the Yang-Mills field  equation (of motion) having an invariant which does not change the value under the continuous deformation of the solutions, e.g.,  characterized by a non-trivial Homotopy class. 
}
  \cite{BPST,CF77,tHooft76,Wilczek77}, which gives a finite action integral which is proportional to the integer-valued  \textbf{Pontryagin index} $Q_P$ in $D=4$ dimensional Euclidean space--time $\mathbb{R}^4$ where the continuous map $U$ from the 3-sphere $S^3$ to $G=SU(2) \simeq S^3$:
\begin{align}
 U: S^3 \rightarrow SU(2) \simeq S^3
\end{align}
 is classified by the non-trivial Homotopy group:\begin{align}
\pi_3(S^3)=\mathbb{Z} .
\end{align}%
However, there are some arguments suggesting that the Yang-Mills  instantons do not confine quarks in four dimensions, e.g.,  \cite{Witten79}. 
This is in sharp contrast to the $D=3$ case: in the Georgi-Glashow model, point-like magnetic monopoles exist as instantons in three dimensions, leading to the area law of the Wilson loop average as shown by Polyakov \cite{Polyakov77}.

In view of these, 't Hooft \cite{tHooft81} has proposed an explicit prescription which enables one to extract \textbf{Abelian magnetic monopoles} from the Yang-Mills theory as \textbf{gauge-fixing defects}, which is called the \textbf{Abelian projection method}. 
In this prescription,  the location of an Abelian magnetic monopole in $SU(2)$ Yang-Mills theory is specified by the simultaneous \textbf{zeros} (of first order) in $\mathbb{R}^D$ of a three-component field $\vec{\phi}(x)=\{ \phi^A(x) \}_{A=1,2,3}$, which we call the \textbf{monopole field} hereafter.  
As a result, an Abelian magnetic monopole is a topological object of \textbf{co-dimension} 3, if it exists at all, characterized by a continuous map $\phi$ from the 2-sphere to $SU(2)/U(1)$: 
\begin{align}
 \phi: S^2 \rightarrow SU(2)/U(1) \simeq S^2 ,
\end{align}
 with the non-trivial Homotopy group: 
\begin{align}
\pi_2(S^2)=\mathbb{Z} . 
\end{align}
Therefore, an Abelian magnetic monopole is represented by a point in $D=3$ dimensions, as expected.
In $D=4$ dimensions the world line of a magnetic monopole must draw a closed loop, i.e., \textbf{loop of magnetic monopole}, not merely an open line, due to the topological conservation law $\partial_\mu k^\mu=0$ of the magnetic-monopole current $k^\mu$, which is interpreted as the creation and annihilation of a pair of monopole and antimonopole, since the net magnetic charge must be zero just as in the vacuum. (A magnetic current extending from an infinity to another infinity can be also regarded as a closed loop after the identification of infinities.) 
It is known according to numerical simulations that ``large'' magnetic monopole loops are the most dominant configurations responsible for confinement. See e.g., \cite{CP97} for a review.

The obstacle is the lack of information as to how the magnetic monopole is related to the original Yang-Mills field in a gauge-independent way. 
In the 't Hooft proposal of Abelian projection, the monopole field $\vec{\phi}(x)$ is an arbitrary composite operator constructed from the Yang-Mills field $\mathscr{A}^A_\mu(x)$ as long as $\phi$  transforms according to the adjoint representation under the gauge transformation, e.g., $\phi^A=\mathscr{F}_{12}^A$ (a component of the field strength $\mathscr{F}_{\mu\nu}^A$).  
The \textbf{maximally Abelian gauge (MA gauge, MAG)} \cite{KLSW87} is well-known to be the most effective choice in practical calculations, following from a specific choice for the monopole field in  the Abelian projection method. 
The MA gauge is given by minimizing the gauge-fixing functional written in terms of the off-diagonal gluon fields $A_\mu^a$ ($a=1,2$) by using the gauge degrees of freedom:
\begin{align}
 F_{\rm MAG} = \int d^Dx \frac12 [(A_\mu^1)^2+(A_\mu^2)^2] ,
\end{align}
which transforms the gauge field variable as close as possible to the maximal torus group $U(1)$.

It is important to establish the relationship between the Abelian magnetic monopole in question and an original $SU(2)$ Yang-Mills field.  For this purpose, some works have already been devoted to constructing  explicit configurations of the magnetic monopole loops from  appropriately chosen configurations of the Yang-Mills field.  
In fact, Chernodub and Gubarev \cite{CG95} have pointed out that  within the MA gauge one instanton and a set of instantons arranged along a straight line induce an Abelian  magnetic-monopole current along a straight line going through centers of instantons. 
In this case, the magnetic monopole is given by the standard static \textbf{hedgehog} configuration. 
However, this solution yields a divergent value for the gauge-fixing function 
$
 F_{\rm MAG}% = \int d^4x \frac12 [(A_\mu^1)^2+(A_\mu^2)^2]
$
of the MA gauge.  
Therefore, it must be excluded in four dimensions. 

%\noindent
%$\bullet$ 
%Chernodub and Gubarev,
%Instantons and monopoles in maximal Abelian projection of SU(2) gluodynamics,
%[hep-th/9506026],
%JETP Lett. {\bf 62}, 100 (1995).

%Within  MAG, one instanton or the set of instantons arranged along a straight line induce an Abelian  magnetic monopole current along a straight line going through centers of instantons. 
%The magnetic monopole is given by the standard static hedgehog type. 
%\\
%However, this solution yields a divergent value for the gauge-fixing function of MAG:
%$
% F_{\rm MAG} = \int d^4 \frac12 [(A_\mu^1)^2+(A_\mu^2)^2]
%$. 
%Therefore, it must be excluded in four dimensions.

In a laborious and important work, Brower, Orginos and Tan (BOT) \cite{BOT97} have investigated within the MA gauge whether the magnetic monopole loop represented by a circle with a non-zero and finite radius can exist for some given instanton configurations or not. 
They concluded that  such a stable magnetic monopole loop is absent for one-instanton Yang-Mills background: a circular magnetic monopole loop centered on an instanton is inevitably shrank to the center point, if one imposes the condition of minimizing the MAG functional
$
 F_{\rm MAG} 
$. 
While an instanton-antiinstanton pair seems to support a stable magnetic monopole loop, although such a field configuration is not a solution of the Yang-Mills field equation.  
These results suggest that such instantons are not the topological objects responsible for quark confinement from the viewpoint of the dual superconductivity. 
Incidentally, these conclusions heavily rely on their hard work of numerically solving partial differential equations. 
See e.g., \cite{HT96} for the corresponding result obtained from numerical simulations on a lattice. 

Subsequently, Bruckmann, Heinzl, Vekua and Wipf (BHVW) \cite{BHVW01} have performed a systematic and analytical treatment to this problem within the \textbf{Laplacian Abelian gauge (LA gauge, LAG)} \cite{LAG}, which has succeeded to shed new light on the relationship between an Abelian magnetic monopole and an instanton from a different angle.
In the LA gauge, \textbf{zeros} of an auxiliary Higgs field $\vec{\phi}(x)$ induce \textbf{topological defects} in the gauge potential, upon diagonalization. 
It is important to notice that the nature of the defects depends on the order of the zeros.  For first-order zeros, one obtains \textbf{magnetic monopoles}. The defects obtained from zeros of second order are \textbf{Hopfion} which is characterized by a topological invariant called \textbf{Hopf index} \cite{Hopf31} for the \textbf{Hopf map} $H$ from the 3-sphere to $SU(2)/U(1)$:
\begin{align}
 H: S^3 \rightarrow SU(2)/U(1) \simeq S^2 ,
\end{align}
with non-trivial Homotopy:
\begin{align}
\pi_3(S^2)=\mathbb{Z} .
\end{align}
They have solved the eigenvalue problem of the covariant Laplacian $-D_\mu[\mathbf{A}]D_\mu[\mathbf{A}]$ in the adjoint representation in the Yang-Mills background $\mathbf{A}$ of a single instanton  in the \textbf{singular gauge}:
\begin{align}
 -D_\mu[\mathbf{A}]D_\mu[\mathbf{A}]\vec{\phi}(x)=\lambda   \vec{\phi}(x)
\end{align}
 and have obtained the auxiliary Higgs field $\vec{\phi}(x)$ as the (normalizable) ground state wave function having the lowest eigenvalue $\lambda$.  
Consequently, they have found that the auxiliary Higgs field $\vec{\phi}(x)$ is given by the \textbf{standard Hopf map}: $S^3 \rightarrow S^2$ in the neighborhood of the center of an instanton and by a constant, e.g., $(0,0,1)$ after normalization in the distant region far away from the center where two regions are separated by the scale of the instanton size parameter. 
The zeros of the auxiliary Higgs field $\vec{\phi}(x)$ as the monopole field agree with the origin.  This results enable one to explain the BOT result without numerical calculations.  
In BHVW, however,  $\mathbb{R}^4$ was replaced by a four sphere $S^4$ of a finite radius in order to obtain a finite LAG functional (convergent integral). 
See also \cite{Reinhardt97,Jahn00,TTF00,HY04,BH03} for relationships among various topological objects.

In the course of studying the relationship between Abelian magnetic monopoles and center vortices \cite{Greensite03}, \textbf{merons} \cite{AFF76} are recognized as an important object \cite{ER00,Cornwall98,LNT04}.
Merons \cite{AFF76,AFF77,Actor79} are solutions of the Yang-Mills field equation and are characterized by one half topological charge, i.e., having half-integer Pontryagin index. 
These configurations escaped from the above consideration, since they have  infinite action due to their singular behaviors.  However, once they receive an ultraviolet regularization which does not influence quark confinement as an infrared phenomenon, they can have finite action and contribute to the functional integration over the Yang-Mills field in calculating the Wilson loop average, in the strong coupling region above a critical value based on action and entropy argument. 
In fact, Callan, Dashen and Gross \cite{CDG78} have discussed that merons are the most dominant quark confiner. 
In fact, Reinhardt and Tok \cite{RT01} have investigated the relationship among Abelian magnetic monopoles and center vortices in various Yang-Mills back ground fields: one meron, one instanton, instanton-antiinstanton pair, using both the LA gauge and the \textbf{Laplacian center gauge (LC gauge, LCG)}.
It has been pointed out that an Abelian magnetic monopole and a meron pair are mediated by sheets of center vortices \cite{ER00}.
In fact, Montero and Negele \cite{MN02} have obtained an Abelian magnetic monopole loop and center vortices for two merons (a meron pair) by using numerical simulations on a lattice, see also \cite{meron} for related works.

%Instantons and merons are solutions of the Yang-Mills field equation
%$D_\nu \mathscr{F}_{\mu\nu}=0$.
\footnotesize
%%%%%%%%%%%%%%%%%%%%%%%%%%%%%%%%%%%%%%%%%%%%%%%%%%%%%%%%%%%%%
%%%%%%%%%%%%%%%%%%%%%%%%%%%%%%%%%%%%%%%%%%%%%%%%%%%%%%%%%%%%%
\begin{center}  
\begin{tabular}{l||l|l} \hline\hline
 & instanton & meron  \\  \hline\hline
discovered by & BPST 1975 & DFF 1976    \\ \hline
$D_\nu \mathscr{F}_{\mu\nu}=0$  & YES & YES     \\ \hline
self-duality $ {}^{\displaystyle *}\mathscr{F}=\mathscr{F}$ & YES & NO     \\ \hline
Topological charge $Q_P$ & integer $(0), \pm 1, \pm 2, \cdots$  & half-integer $(0), \pm 1/2, \pm 1, \cdots$    \\ \hline
charge density $D_P$ & $\frac{6\rho^4}{\pi^2} \frac{1}{(x^2+\rho^2)^4}$  & $\frac12 \delta^4(x-a)+\frac12 \delta^4(x-b)$    \\ \hline
solution $ \mathscr{A}_\mu^A(x)$ & $g^{-1}\eta^A_{\mu\nu} \frac{2(x-a)_\nu}{(x-a)^2+\rho^2}$  & $ g^{-1}\left[ \eta^A_{\mu\nu} \frac{(x-a)_\nu}{(x-a)^2} + \eta^A_{\mu\nu}  \frac{(x-b)_\nu}{(x-b)^2} \right] $    \\ \hline
Euclidean  & finite action & (logarithmic) divergent action  \\  
 & $S_{\rm YM}=(8\pi^2/g^2)|Q_P|$ &      \\ \hline
tunneling  & between $Q_P=0$ and $Q_P=\pm 1$   &    $Q_P=0$  and  $Q_P=\pm 1/2$ \\  
  & vacua in the $\mathscr{A}_0=0$ gauge  &  vacua in the Coulomb gauge    \\ \hline
multi-charge solutions & Witten, 't Hooft,  &     \\  
  & Jackiw-Nohl-Rebbi, ADHM  & not known    \\ \hline
Minkowski solution & trivial & everywhere regular \\   
  &   &  finite, non-vanishing action     \\ \hline
\end{tabular}
% \caption[]{After inverting the concentric sphere geometry for a smeared meron about the point $d$ (left panel), we obtain the smeared two meron configuration (right panel). }
\label{table:instanton-meron}
\end{center}
%%%%%%%%%%%%%%%%%%%%%%%%%%%%%%%%%%%%%%%%%%%%%%%%%%%%%%%%%%%%%
%%%%%%%%%%%%%%%%%%%%%%%%%%%%%%%%%%%%%%%%%%%%%%%%%%%%%%%%%%%%%
\normalsize

One more obstacle in these approaches lies in a fact that topological objects such as Abelian magnetic monopoles and center vortices are obtained as gauge fixing defects. 
Therefore, they are not free from criticism of being gauge artifacts. 
%\footnote{
%There is an approach to gauge-invariant Abelian confinement mechanism, see \cite{Suzuki08}. 
%}
Recently, we have given a gauge-invariant (gauge independent) definition of magnetic monopoles  \cite{KMS06,KSM08,Kondo08} and vortices \cite{Kondo08b} in Yang-Mills theory in the framework of a new reformulation of Yang-Mills theory based on change of field variables  
 founded in \cite{Cho80} combined with a non-Abelian Stokes theorem for the Wilson loop operator \cite{DP89,KondoIV,Kondo08}. 
The lattice version has been constructed to support them by numerical simulations \cite{KKMSSI06}. 
% which leads to Abelian magnetic monoples for $G=SU(2)$ and non-Abelian magnetic monopoles for $G=SU(N), N \ge 3$. 

%This paper is organized as follows. 
%In sections 2 and 3, 
We investigate how the gauge-invariant magnetic monopoles are obtained analytically for a given Yang-Mills background field in four dimensions.
%In sections 4 and 5, 
It is shown that the previous results \cite{BOT97,BHVW01,RT01} obtained for one-meron and one-instanton  are easily reproduced within the new  reformulation. 
%In sections 6, we give a new result for  gauge-invariant magnetic monopole loops in Yang-Mills theory:
It is shown in an analytical way that  there exist circular magnetic monopole loops joining two merons. 
This will be the first analytical solution of stable magnetic monopole loops constructed from the Yang-Mills field with non-trivial but finite Pontryagin index.
(Bruckmann and Hansen  \cite{BH03} constructed a ring of magnetic monopole by superposing  infinitely many instantons on a circle.  However, this configuration has infinite action and infinite Pontrayagin index. Therefore, they do not contribute to the path integral.) 
The resulting analytical solution for magnetic monopole loops correspond to the numerical solution found by Montero and Negele \cite{MN02} on a lattice.

The result  has rather interesting implications to  the quark confinement mechanism. As mentioned above, the gauge-invariant magnetic monopole is a complicated object obtained by the non-linear change of variables from the original Yang-Mills field, although they are fundamental objects necessary for the naive dual-superconductivity scenario of quark confinement. 
In other words, the result indicates that a meron pair is one of the most relevant quark confiners if viewed from the original Yang-Mills theory, as Callan, Dashen and Gross \cite{CDG78} suggested long ago.

One instanton is not sufficient to conclude that instantons do not contribute to quark confinement.  We must examine the \textbf{multi-instanton} configurations.  
Although the multi-instanton solution is in general given by the \textbf{Atiyah-Drinfeld-Hitchin-Manin (ADHM)  construction} \cite{ADHM78}, the \textbf{Jackiw-Nohl-Rebbi (JNR)} solution \cite{JNR77} is the most general 2-instanton solution which is explicitly written in the closed form beyond one instanton. 
Therefore, we considered the JNR instanton as the most general 2-instanton solution and a specific 2-instanton of the 't Hooft type which is obtained as a limit of the JNR solution.
The \textbf{'t Hooft multi-instanton} was used in the preceding studies due to its simple structure allowing easy treatment. However, it should be remarked that it is not the most general multi-instanton, since it does not have the sufficient number of collective coordinates which characterize the multi-instanton even for 2-charge. 
We have found that a circular magnetic monopole loop is created around the center of the JNR configuration. 
Thus, the multi-instanton can be a source of magnetic monopoles giving the dominant contribution to quark confinement.

%The result  has rather interesting implications to  the quark confinement mechanism. As mentioned above, the gauge-invariant magnetic monopole is a complicated object obtained by the non-linear change of variables from the original Yang-Mills field, although they are fundamental objects necessary for the naive dual-superconductivity scenario of quark confinement. 
%In other words, the result indicates that a meron pair is one of the most relevant quark confiners if viewed from the original Yang-Mills theory, as Callan, Dashen and Gross \cite{CDG78} suggested long ago. 

%%%%%%%%%%%%%%%%%%%%%%%%%%%%%%%%%%%%%%%%%%%%%%%%%

\subsection{Reduction condition}

%%%%%%%%%%%%%%%%%%%%%%%%%%%%%%%%%%%%%%%%%%%%%%%%%

%In a previous paper \cite{KMS06}, 
In the preceding chapters, we have given a prescription for obtaining the gauge-invariant magnetic monopole from the original Yang-Mills field $\mathscr{A}_\mu(x)$. 
In what follows, we restrict our attention to the $SU(2)$ case. 

(i) For a given $SU(2)$ Yang-Mills field $\mathscr{A}_\mu(x)=\mathscr{A}^A_\mu(x)\frac{\sigma_A}{2}$,  
%$\mathbf{A}_\mu(x)=\mathbf{A}^A_\mu(x)\frac{\sigma_A}{2}$, 
 the color field $\mathbf{n}(x)$ is obtained by solving the differential equation (in the vector notation):
\begin{align}
 \mathbf{n}(x) \times D_\mu[\mathbf{A}]D_\mu[\mathbf{A}] \mathbf{n}(x) = \mathbf{0} 
 ,
 \label{RDE1}
\end{align}
which we call the \textbf{reduction differential equation (RDE)}.
%See Appendix \ref{appendix:rc} for the derivation. 
Here the color field has the unit length
\begin{align}
   \mathbf{n}(x) \cdot \mathbf{n}(x) = 1
   .
   \label{unit}
\end{align}

(ii) Once  the color field $\mathbf{n}(x)$ is known, the gauge-invariant ``magnetic-monopole current''   $k$  is constructed  from the gauge-invariant field strength (curvature two-form) $F$ derived through the non-Abelian Stokes theorem for the Wilson loop operator according to 
\begin{align}
 k:=  \delta  {}^{\displaystyle *}F =  {}^{\displaystyle *}dF 
%\quad
% \nonumber\\
% j:= \delta f 
  ,
  \label{def-k}
\end{align}
where  $d$ is the exterior derivative , $\delta$ is the coderivative (adjoint derivative), $ {}^{\displaystyle *}$ is the Hodge star operation, and 
 the \textit{gauge-invariant} two-form $f$  is defined from the  gauge field (connection one-form) $\mathbf{A}$   by 
\begin{align}
 F_{\alpha\beta}(x) 
%=&  - \sqrt{\frac{N-1}{2N}}
%  {\rm tr}( 2\bm{h}(x) \mathscr{F}_{\mu\nu}[\mathscr{V}](x) )
%  \nonumber\\
  :=&  \partial_\alpha [\mathbf{n}(x) \cdot \mathbf{A}_\beta(x)] - 
  \partial_\beta [\mathbf{n}(x) \cdot \mathbf{A}_\alpha(x)] 
  \nonumber\\&
  +  ig^{-1} \mathbf{n}(x) \cdot [\partial_\alpha \mathbf{n}(x) \times  \partial_\beta \mathbf{n}(x)]  
   .
\end{align}
The  current $k$ is conserved in the sense that 
$\delta k=0$.
In $D=4$ dimensions, especially, we have
\begin{equation}
 k^\mu(x)
 =  \partial_\nu\!\,^\ast F^{\mu\nu}(x)\label{k}
 =  \frac{1}{2} \epsilon^{\mu\nu\alpha\beta} \partial_\nu F_{\alpha\beta}(x) ,
\end{equation}
and the gauge-invariant  magnetic charge $q_m$ is defined in a Lorentz (or Euclidean rotation)  invariant way \cite{Kondo08b} by 
\begin{align}
 q_m :=& \int d^3 \tilde{\sigma}_\mu k^\mu(x)
  ,
  \nonumber\\
 d^3 \tilde{\sigma}_{\mu} 
:=&  \frac{1}{3!}   \epsilon_{\mu\gamma_1\gamma_2\gamma_3} d^3 \sigma^{\gamma_1\gamma_2\gamma_3}
 , \quad
%\nonumber\\
 d^3 \sigma^{\gamma_1\gamma_2\gamma_3}
 :=   \epsilon_{\beta_1\beta_2\beta_3} 
 \frac{\partial \bar{x}^{\gamma_1}}{\partial \sigma_{\beta_1}} \frac{\partial \bar{x}^{\gamma_2}}{\partial \sigma_{\beta_2}}  \frac{\partial \bar{x}^{\gamma_3}}{\partial \sigma_{\beta_3}}  d\sigma_{1} d\sigma_{2} d\sigma_{3} 
  .
\end{align}
where $\bar{x}^\mu$ denotes a parameterization of the 3-dimensional volume $V$ and 
 $d^3 \tilde{\sigma}_{\mu} $ is the dual of the 3-dimensional volume element $d^3 \sigma^{\gamma_1\gamma_2\gamma_3}$. 
The simplest choice is  %$q_m=\int d^3 x k^0$.
\begin{align}
 q_m=\int d^3 x k^0 .
\end{align}
%See \cite{KSM08,Kondo08} for $SU(N)$ ($N \ge 3$) case.
%Once the reduction condition is solved, thus,  
Thus, we can obtain the magnetic-monopole current $k^\mu$ from the original gauge field $\mathscr{A}_\mu$.%
\footnote{
The RDE in the reformulated Yang-Mills theory has the same form as that considered in BOT \cite{BOT97}, but its reasoning behind the RDE is quite different from the previous one, as can be seen from its derivation in  Appendix of \cite{KMS05}.
}

We now give a new form of the RDE (eigenvalue-like equation): 
\begin{align}
 -D_\mu[\mathbf{A}]D_\mu[\mathbf{A}] \mathbf{n}(x) = \lambda(x) \mathbf{n}(x) 
 .
 \label{RDE2}
\end{align}
This implies that solving the RDE (\ref{RDE1}) is equivalent to look for the color field $\mathbf{n}(x)$ such that applying the covariant Laplacian $-D_\mu[\mathbf{A}]D_\mu[\mathbf{A}]$  of a given Yang-Mills field $\mathbf{A}_\mu(x)$  to the color field $\mathbf{n}(x)$ becomes parallel  to itself.  
It should be remarked that $\lambda(x)$ is non-negative, i.e., 
\begin{align}
 \lambda(x) \ge 0
 , 
\end{align}
since $-D_\mu[\mathbf{A}] D_\mu[\mathbf{A}]$ is a non-negative (positive definite) operator. 
The equivalence between (\ref{RDE1}) and (\ref{RDE2}) is given in \cite{KFSS08}.

An advantage of the new form (\ref{RDE2}) of RDE is as follows.
Once  the color field $\mathbf{n}(x)$ satisfying (\ref{RDE2}) is known, the value of the reduction functional $F_{\rm rc}$ is immediately calculable as an integral of the scalar function $\lambda(x)$ over the space--time $\mathbb{R}^D$ as
\begin{align}
 F_{\rm rc} =&  \int d^Dx \frac12 (D_\mu[\mathbf{A}] \mathbf{n}(x)) \cdot (D_\mu[\mathbf{A}] \mathbf{n}(x)) 
 \nonumber\\
 =&   \int d^Dx \frac12  \mathbf{n}(x) \cdot (-D_\mu[\mathbf{A}] D_\mu[\mathbf{A}] \mathbf{n}(x)) 
  \nonumber\\
 =&  \int d^Dx \frac12  \mathbf{n}(x) \cdot \lambda(x) \mathbf{n}(x) 
  \nonumber\\
 =&  \int d^Dx  \frac12  \lambda(x)
 ,
 \label{Rc}
\end{align}
where we have used (\ref{unit}) in the last step.
The solution is not unique. 
Choose the solution giving the smallest value of the reduction functional $F_{\rm rc}$:  integral of the scalar function $\lambda(x)$ over the space--time $\mathbb{R}^D$:

Thus, the problem of solving the RDE has been reduced to another problem: 
 For a given Yang-Mills field $\mathbf{A}_\mu(x)$,  look for the unit vector field $\mathbf{n}(x)$ such that   $-D_\mu[\mathbf{A}] D_\mu[\mathbf{A}]\mathbf{n}(x)$ is proportional to $\mathbf{n}(x)$ 
with the smallest value of the reduction functional $F_{\rm rc}$ which is an integral of the scalar function $\lambda(x)$ over the space--time $\mathbb{R}^D$.
%with the smallest value of 
%$
%  \lambda(x) = \mathbf{n}(x)  \cdot [-D_\mu[\mathbf{A}]D_\mu[\mathbf{A}] \mathbf{n}(x)]  
%$
%at every space--time point $x$.
For the integral (\ref{Rc}) to be convergent, $\lambda(x)$ must decrease rapidly for large $x^2$.

%%%%%%%%%%%%%%%%%%%%%%%%%%%%%%%%%%%%%%%%%%%%%%%%%

\subsection{Simplifying the reduction condition}

%%%%%%%%%%%%%%%%%%%%%%%%%%%%%%%%%%%%%%%%%%%%%%%%%

In what follows, we restrict our considerations to  the four-dimensional ($D=4$) Euclidean Yang-Mills theory.

%%%%%%%%%%%%%%%%%%%%%%%%%%%%%%%%%%%%%%%%%%%%%%%%%
\subsubsection{CFtHW Ansatz for Yang-Mills field}
%%%%%%%%%%%%%%%%%%%%%%%%%%%%%%%%%%%%%%%%%%%%%%%%%

For the $SU(2)$ Yang-Mills field, we adopt the \textbf{Corrigan-Fairlie-'t~Hooft-Wilczek (CFtHW) Ansatz} \cite{CF77,tHooft76,Wilczek77}:
\begin{align}
 g\mathscr{A}_\mu(x) 
=   \frac{\sigma_A}{2} g\mathscr{A}_\mu^A(x) 
=  \frac{\sigma_A}{2} \eta^A_{\mu\nu} f_\nu(x) %\partial_\nu , %\ln \Phi(x)   , 
\quad f_\nu(x) := \partial_\nu \ln \Phi(x) 
 ,
\end{align}
where $\sigma_A$ $(A=1,2,3)$ are the Pauli matrices and $\eta^A_{\mu\nu}=\eta^{(+)}{}^A_{\mu\nu}$ is the symbol defined by 
\begin{align}
 \eta^A_{\mu\nu} \equiv \eta^{(+)}{}^A_{\mu\nu} := \epsilon_{A\mu\nu 4} + \delta_{A\mu}\delta_{\nu 4} - \delta_{\mu 4}\delta_{A \nu}
 = \begin{cases}
    \epsilon_{Ajk} & (\mu=j, \nu=k) \cr
    \delta_{Aj} & (\mu=j, \nu= 4) \cr
    -\delta_{Ak} & (\mu=4, \nu=k)
   \end{cases}
 .
\end{align}
Similarly, we can define
$\bar{\eta}^A_{\mu\nu}=:\eta^{(-)}{}^A_{\mu\nu}$  as
\begin{align}
 \bar{\eta}^A_{\mu\nu} \equiv  \eta^{(-)}{}^A_{\mu\nu} := \epsilon_{A\mu\nu 4} - \delta_{A\mu}\delta_{\nu 4} + \delta_{\mu 4}\delta_{A \nu}
 = \begin{cases}
    \epsilon_{Ajk} & (\mu=j, \nu=k) \cr
    -\delta_{Aj} & (\mu=j, \nu= 4) \cr
    +\delta_{Ak} & (\mu=4, \nu=k)
   \end{cases}
 .
\end{align}
Note that $\eta^A_{\mu\nu}$ is \textbf{self-dual}, 
%i.e., $\eta^A_{\mu\nu}=*\eta^A_{\mu\nu}:=\frac12 \epsilon_{\mu\nu\alpha\beta} \eta^A_{\alpha\beta}$,
while $\bar{\eta}^A_{\mu\nu}$ is \textbf{anti-selfdual}:
% i.e., $-\bar{\eta}^A_{\mu\nu}=*\bar{\eta}^A_{\mu\nu}$.
\begin{align}
   {}^{\displaystyle *}\eta^A_{\mu\nu}:=\frac12 \epsilon_{\mu\nu\alpha\beta} \eta^A_{\alpha\beta} = \eta^A_{\mu\nu} 
 , \quad 
  {}^{\displaystyle *}\bar{\eta}^A_{\mu\nu} = -\bar{\eta}^A_{\mu\nu}  
\Longleftrightarrow 
  {}^{\displaystyle *}\eta^{(\pm)}{}^A_{\mu\nu} = \pm \eta^{(\pm)}{}^A_{\mu\nu} .
\end{align}
Note that $\eta^{(+)}{}^A_{\mu\nu}$ and $\eta^{(-)}{}^A_{\mu\nu}$ are not the Lorentz tensors and therefore they are called the symbols. 
They satisfy the properties:
\begin{align}
  \eta^A_{\mu\nu}=-\eta^A_{\nu\mu}
 , \quad 
 \eta^A_{\mu\alpha} \eta^B_{\mu\beta}
= \delta_{AB}\delta_{\alpha\beta}+\epsilon_{ABC}\eta^C_{\alpha\beta} .
\end{align}

It is interesting that the Yang-Mills field in the  CFtHW  Ansatz satisfies simultaneously the \textbf{Lorenz gauge}:
\begin{align}
 \partial_\mu \mathscr{A}_\mu^A(x)= 0
 ,
\end{align}
and the MA gauge: 
\begin{align}
D_\mu[A^3] A_\mu^{\pm}(x):= (\partial_\mu - ig A_\mu^3) (A_\mu^1(x) \pm i A_\mu^2(x)) = 0
 .
\end{align}
 
Under this Ansatz, it is shown that the RDE is greatly simplified:%
\footnote{
See Appendix B of \cite{KFSS08} for the derivation. 
}
\begin{align}
 \{ [-\partial_\mu \partial_\mu + 2 f_\mu f_\mu ]\delta_{AB} + 2 \epsilon_{ABC} \eta^C_{\mu\nu} f_\nu(x) \partial_\mu \}  {n}_B(x) = \lambda(x)  {n}_A(x)  
 .  
 \label{RDE0}
\end{align}

%%%%%%%%%%%%%%%%%%%%%%%%%%%%%%%%%%%%%%%%%%%%%%%%%
\subsubsection{Euclidean rotation symmetry}
%%%%%%%%%%%%%%%%%%%%%%%%%%%%%%%%%%%%%%%%%%%%%%%%%

In order to further simplify the equation,  we make use of the \textbf{Euclidean rotation group} $SO(4)$.
This symmetry enables one to separate the RDE  into the angular and radial parts. 
We define the generators of four-dimensional Euclidean rotations as 
\begin{align}
  L_{\mu\nu} = -i(x_\mu \partial_\nu - x_\nu \partial_\mu) 
   , \quad \mu, \nu \in \{ 1, 2, 3, 4 \}
  .
   \label{Lmn}
\end{align}
Indeed, it is straightforward to check that $L_{\mu\nu}$ satisfies the Lie algebra of $SO(4)$.  The angular part is expressed in terms of angular momentum derived from the decomposition:
\begin{align}
  so(4) \cong su(2) + su(2)
   .
\end{align}
In analogy with the \textbf{Lorentz group}, we introduce  the angular momentum and boost generators:
\begin{align}
  \mathscr{L}_j := \frac12 \epsilon_{jk\ell} L_{k\ell} , 
  \quad
  \mathscr{K}_j := L_{j4},  \quad  j,k,\ell \in \{ 1,2,3 \} 
   ,
   \label{LjKj}
\end{align}
and their linear combinations:
\begin{align}
 M_A :=& \frac12 (\mathscr{L}_A-\mathscr{K}_A) = - \frac{i}{2} \bar{\eta}^A_{\mu\nu} x_\mu \partial_\nu ,
 \quad  A \in \{ 1,2,3 \} 
 ,
 \nonumber\\
 N_A :=& \frac12 (\mathscr{L}_A+\mathscr{K}_A) = - \frac{i}{2}  \eta^A_{\mu\nu} x_\mu \partial_\nu , 
\quad A \in \{ 1,2,3 \} 
 .
\end{align}
The operators $M_A$ and $N_A$ generate two independent $SU(2)$ subgroups with \textbf{Casimir operators} $\vec{M}^2 :=M_A M_A$ and $\vec{N}^2 :=N_A N_A$ having eigenvalues $M(M+1)$ and $N(N+1)$, respectively:
\begin{align}
 \vec{M}^2 :=& M_A M_A \rightarrow M(M+1), \quad  M \in \{ 0, \frac12, 1, \frac32, \cdots \} 
  ,
 \nonumber\\
 \vec{N}^2 :=& N_A N_A \rightarrow N(N+1), \quad  N \in \{ 0, \frac12, 1, \frac32, \cdots \} 
 .
\end{align}
Here it is important to note that the eigenvalues $M$ and $N$ are half-integers. 

The generators for isospin $S=1$ are given by
\begin{align}
 (S_A)_{BC} := i\epsilon_{ABC} = (S_C)_{AB}  
 .
\end{align}
It is easy to see that $\vec{S}^2$ is a Casimir operator
and
$\vec{S}^2$ has the eigenvalue
\begin{align}
 \vec{S}^2 := S_A S_A \rightarrow S(S+1) = 2
 , 
\end{align}
since   
\begin{align}
 (\vec{S}^2)_{AB} = (S_C)_{AD} (S_C)_{DB} 
= i\epsilon_{DCA} i\epsilon_{BCD} = 2 \delta_{AB} 
 .
\end{align}

Now we introduce the conserved total angular momentum $\vec{J}$ by
\begin{align}
 \vec{J} = \vec{L} + \vec{S}, \quad
 \vec{L}=\vec{M} \ \text{or} \ \vec{L}=\vec{N} 
 ,  
\end{align}
with the eigenvalue 
\begin{align}
 \vec{J}^2 \rightarrow J(J+1), \quad J \in \{ L+1, L, |L-1| \} 
 .
\end{align}
Using the representations (\ref{Lmn}) and (\ref{LjKj}),  we find  that 
\begin{align}
 \vec{N}^2 -\vec{M}^2 = 0 = \vec{\mathscr{L}} \cdot \vec{\mathscr{K}} 
 .
\end{align}
Thus, a complete set of commuting observables is given by the Casimir operators, $\vec{J}^2$, $\vec{L}^2$, $\vec{S}^2$ and their projections, e.g., $J_z, L_z, S_z$.

Since the spherical symmetry allows us to take $\Phi(x)=\tilde{\Phi}(x^2)$ and 
\begin{align}
  f_\nu(x) := \partial_\nu \ln \tilde{\Phi}(x^2) = x_\nu f(x) , \quad f(x) = 2 \frac{d}{dx^2} \ln \tilde{\Phi}(x^2)
 ,
\end{align}
then the RDE is rewritten in the form:
\begin{align}
  \{ -\partial_\mu \partial_\mu \delta_{AB} + 2f(x) (\vec{J}^2-\vec{L}^2-\vec{S}^2)_{AB} + x_\mu x_\mu f^2(x) (\vec{S}^2)_{AB} \}  {n}_B(x) 
%\nonumber\\
 = \lambda(x)  {n}_A(x) 
 ,
\end{align}
where we have used 
\begin{align}
 \vec{S} \cdot \vec{L} = (\vec{J}^2-\vec{L}^2-\vec{S}^2)/2 
 .
\end{align}

The symmetry consideration suggests that $\mathbf{n}(x)$ is separated into the radial and angular part: In the vector (component) notation:
\begin{subequations}
\begin{align}
  {n}_A(x) = \psi(R) Y^A_{(J,L)}(\hat{x}) 
 ,
\end{align}
or in the Lie algebra valued notation:
\begin{align} 
 \bm{n}(x) =&  {n}_A(x) \sigma_A =  \psi(R) Y^A_{(J,L)}(\hat{x}) \sigma_A 
 ,
\nonumber\\
 & R := \sqrt{x_\mu x_\mu} \in \mathbb{R}_{+}, 
 \quad \hat{x}_\mu:=x_\mu/R \in S^3
\end{align}
\end{subequations}
where
$\vec{Y}_{(J,L)}(\hat{x})=\{ Y^A_{(J,L)}(\hat{x}) \}_{A=1,2,3}$ denote the \textbf{vector spherical harmonics} on $S^3$ characterized by 
\begin{align}
 \vec{L}^2 Y^A_{(J,L)}(\hat{x}) =& L(L+1) Y^A_{(J,L)}(\hat{x})
,
 \\
 \vec{J}^2 Y^A_{(J,L)}(\hat{x}) =& J(J+1) Y^A_{(J,L)}(\hat{x})
,
 \\
 \vec{S}^2 Y^A_{(J,L)}(\hat{x}) \sigma_A =& S(S+1) Y^A_{(J,L)}(\hat{x}) \sigma_A 
 ,
\end{align}
with $S=1$.  
The explicit form of the vector spherical harmonics is given later.

In this form, the covariant Laplacian reduces to the diagonal form and RDE reduces to 
\begin{align}
 & [ -\partial_\mu \partial_\mu  + V(x)  ]  {n}_A(x) = \lambda(x) {n}_A(x) 
 ,
\nonumber\\
 & V(x) :=  2f(x) [J(J+1)-L(L+1)-2] + 2x^2 f^2(x)  
 .
 \label{eq1}
\end{align}
This equation does not necessarily mean that the left-hand side of the RDE becomes automatically proportional to $\mathbf{n}(x)$, since $\partial_\mu \partial_\mu \mathbf{n}(x)$ is not guaranteed to be proportional to $\mathbf{n}(x)$. If this is the case, we have
\begin{align}
 \lambda(x) = V(x) + [-\partial_\mu \partial_\mu  {n}_A(x)]/ {n}_A(x) 
 \quad \text{for any A, no sum over A}
 .
\end{align}

Moreover, it is possible to rewrite the Laplacian in terms of the radial coordinate $R$ and the angular coordinates:
\begin{align}
 -\partial_\mu \partial_\mu  
 = - \partial_R \partial_R  - \frac{3}{R} \partial_R + \frac{2(\vec{M}^2+\vec{N}^2)}{R^2} 
 , \quad R := \sqrt{x_\mu x_\mu} 
 ,
\end{align}
which reads
\begin{align}
 -\partial_\mu \partial_\mu  
 = - \partial_R \partial_R  - \frac{3}{R} \partial_R + \frac{4 \vec{L}^2}{R^2} 
  = - \frac{1}{R^3} \frac{\partial}{\partial R} \left( R^3 \frac{\partial}{\partial R}  \right) + \frac{4 \vec{L}^2}{R^2} 
% , \quad R := \sqrt{x_\mu x_\mu} 
 .
\end{align}
Thus, we arrive at another expression of RDE:
\begin{align}
 & \left[ - \partial_R \partial_R  - \frac{3}{R} \partial_R    + \tilde{V}(x)  \right] \mathbf{n}_A(x)  
 =  \lambda(x) \mathbf{n}_A(x) 
 ,
\nonumber\\
 & \tilde{V}(x) 
:=     \frac{4L(L+1)}{x^2}  + V(x) 
%\nonumber\\
% =&   \frac{4L(L+1)}{x^2}  + 2f(x) [J(J+1)-L(L+1)-2] + 2x^2 f^2(x)  
 .
\end{align}
If the left-hand side of the RDE becomes proportional to $\mathbf{n}(x)$, then $\lambda(x)$ is given by
\footnote{
The second term of the right-hand side of (\ref{eq2}) does not contribute to $\lambda(x)$ if and only if 
$
\psi(R)=C_1+C_2/R^2
$.
If $\psi(R) \sim R^{-\gamma}$, then the second term contributes $\gamma(2-\gamma)/R^2$.  This increases $\lambda(x)$ for $0<\gamma<2$, while it decreases $\lambda(x)$ for $\gamma<0$ and $\gamma >2$.
}
\begin{align}
 \lambda(x) = \tilde{V}(x)  - \psi(R)^{-1} \frac{1}{R^3} \frac{\partial}{\partial R} \left( R^3 \frac{\partial}{\partial R} \psi(R) \right)
 .
 \label{eq2}
\end{align}

\subsection{Unit vector condition and angular part}

In rewriting RDE due to $SO(4)$ symmetry,  
 we have not yet used the fact that $\mathbf{n}(x)$ has the unit length: 
%$
%\mathbf{n}(x) \cdot \mathbf{n}(x)=\mathbf{n}_A(x)  \mathbf{n}_A(x)=1
%$:
\begin{align}
 1 = \mathbf{n}(x) \cdot \mathbf{n}(x) =  {n}_A(x)   {n}_A(x) 
= \psi(R) \psi(R)  Y^A_{(J,L)}(\hat{x})  Y^A_{(J,L)}(\hat{x}) 
 .
\end{align}
If the vector spherical harmonics happens to be normalized at every space--time point as
\begin{align}
 1 =  Y^A_{(J,L)}(\hat{x})  Y^A_{(J,L)}(\hat{x}) 
 ,
 \label{YY}
\end{align}
then we can take without loss of generality
\begin{align}
 \psi(R) \equiv 1 
 .
\end{align}
Separating $\mathbb{R}^4$ into the radial and angular parts,  $\mathbf{n}(x)$ is constructed from  the vector spherical harmonics alone: 
\begin{align}
  {n}_A(x)  =  Y^A_{(J,L)}(\hat{x})   ,
\quad
\hat{x}_\mu:=x_\mu/\sqrt{x_\mu x_\mu} \in S^3 
 .
\end{align}
The vector spherical harmonics $Y^A_{(J,L)}(\hat{x})$ is a polynomial in $\hat{x}$ of degree $2L$ with  $(2J+1)(2L+1)$ fold degeneracy.
%The degeneracy of the state $Y^A_{(J,L)}$ is given by $(2J+1)(2L+1)$.  
In this case, the lowest value of $\lambda(x)$ is obtained by minimizing $\tilde{V}(x)$ at every $x$. 
However, (\ref{YY}) is not guaranteed for any set of $(J,L)$ except for some special cases, as we see shortly. 

Usually, the orthonormality of the vector spherical harmonics is given with respect to the integral over $S^3$ with a finite volume:
\begin{align}
 \int_{S^3} d\Omega \ Y^A_{(J,L)}(\hat{x})  Y^A_{(J',L')}(\hat{x}) = \delta_{JJ'} \delta_{LL'} 
 .
\end{align}

%\newpage
%%%%%%%%%%%%%%%%%%%%%%%%%%%%%%%%%%%%%%%%%%%%%%%%%%
%%%%%%%%%%%%%%%%%%%%%%%%%%%%%%%%%%%%%%%%%%%%%%%%%%
\subsection{One-instanton and one-meron case}
%\setcounter{equation}{0}
%%%%%%%%%%%%%%%%%%%%%%%%%%%%%%%%%%%%%%%%%%%%%%%%%%
%%%%%%%%%%%%%%%%%%%%%%%%%%%%%%%%%%%%%%%%%%%%%%%%%%

In order to treat meron and instanton (in the regular gauge) simultaneously, we adopt the form:
\begin{align}
 f_\mu(x) = x_\mu f(x), \quad f(x) = \frac{2\kappa}{x^2+s^2}
 . 
\end{align}
For a given set of $(J,L)$, we have calculated the ``potential'' $\tilde{V}(x)$ and the ``eigenvalue'' $\lambda(x)$, which  are enumerated in the following 
Table. 
 Note that $(J,L)=(0,0)$ is excluded by selection rules for $S=1$.

%%%%%%%%%%%%%%%%%%%%%%%%%%%%%%%%%%%%%%%%%%%%%%%%%%%%%%%%%%%%%
%%%%%%%%%%%%%%%%%%%%%%%%%%%%%%%%%%%%%%%%%%%%%%%%%%%%%%%%%%%%%
\begin{center}
\begin{tabular}{l||l|l||c|c|c} \hline\hline
J & L & S & degeneracy & 1-instanton (zero size) & 1-meron \\  
  &   &   &   & $\tilde{V}(x)$ & $\tilde{V}(x)$ \\ \hline\hline
1 & 0 & 1     & 3  & $8/x^2$ & $2/x^2$ \\ \hline
1/2 & 1/2 & 1 & 4  & $3/x^2$ & $1/x^2$ \\  
3/2 & 1/2 & 1 & 8  & $15/x^2$ & $7/x^2$ \\ \hline
0 & 1 & 1     & 3  & $0$ & $2/x^2$ \\ 
1 & 1 & 1     & 9  & $8/x^2$ & $6/x^2$ \\
2 & 1 & 1     & 15 & $24/x^2$ & $14/x^2$ \\ \hline
\end{tabular}
% \caption[]{After inverting the concentric sphere geometry for a smeared meron about the point $d$ (left panel), we obtain the smeared two meron configuration (right panel). }
\label{table:eigenvalue1}
\end{center}
%%%%%%%%%%%%%%%%%%%%%%%%%%%%%%%%%%%%%%%%%%%%%%%%%%%%%%%%%%%%%
%%%%%%%%%%%%%%%%%%%%%%%%%%%%%%%%%%%%%%%%%%%%%%%%%%%%%%%%%%%%%

\subsubsection{One instanton in the regular (or non-singular) gauge}

The one-instanton configuration in the regular gauge with zero size, i.e., $\kappa=1$, $s=0$, is expressed by
\begin{align}
 f(x) = \frac{2}{x^2} 
 ,
\end{align}
which leads to 
\begin{align}
 V(x) =  \frac{4}{x^2} [J(J+1)-L(L+1)]   
 , \quad
 \tilde{V}(x) =  \frac{4}{x^2} J(J+1)  \ge 0
 .
\end{align}
For one-instanton with zero size in the regular gauge, therefore, $(J,L)=(0,1)$ gives the lowest value of $\tilde{V}(x)$ at every $x$.
Hence the lowest value of $\lambda(x)$ is obtained  $\lambda(x)=\tilde{V}(x)=0$ if we can set $\psi(R) \equiv {\rm const.}$ from (\ref{eq2}).  This is the lowest possible  value, since $\lambda(x) \ge 0$. For this to be satisfied, the corresponding vector harmonics must be orthonormal (\ref{YY}). 
The vector spherical harmonics $Y_{(0,1)}(\hat{x})$ is 3-fold degenerate and is written as a linear combination of three degenerate states:%
 \footnote{
This degeneracy corresponds to the Gribov copies associated with the reduction (partial) gauge fixing from the enlarged gauge symmetry $SU(2)\times SU(2)/U(1)$ to the original gauge symmetry $SU(2)$, see \cite{KMS06}.  
These Gribov copies are true Gribov copies, but are different from those in fixing the original gauge symmetry $SU(2)$.
}
\begin{align}
  \bm{Y}_{(0,1)}(\hat{x})
%\nonumber\\
=& \sum_{B=1}^{3} \hat{a}_B \bm{Y}_{(0,1),(B)}(\hat{x})
 \nonumber\\
=& \hat{a}_1 
 \begin{pmatrix}
 \hat{x}_1^2-\hat{x}_2^2-\hat{x}_3^2+\hat{x}_4^2  \cr
 2(\hat{x}_1\hat{x}_2-\hat{x}_3\hat{x}_4) \cr
 2(\hat{x}_1\hat{x}_3+\hat{x}_2\hat{x}_4)
 \end{pmatrix}
 + \hat{a}_2 
 \begin{pmatrix}
 2(\hat{x}_1\hat{x}_2+\hat{x}_3\hat{x}_4)  \cr
  -\hat{x}_1^2+\hat{x}_2^2-\hat{x}_3^2+\hat{x}_4^2 \cr
 2(\hat{x}_2\hat{x}_3-\hat{x}_1\hat{x}_4)
 \end{pmatrix}
%\nonumber\\&
 + \hat{a}_3 
 \begin{pmatrix}
 2(\hat{x}_1\hat{x}_3-\hat{x}_2\hat{x}_4) \cr
 2(\hat{x}_2\hat{x}_3+\hat{x}_1\hat{x}_4) \cr
 -\hat{x}_1^2-\hat{x}_2^2+\hat{x}_3^2+\hat{x}_4^2
 \end{pmatrix}
 ,
\end{align}
where  $\hat{a}_B$  ($B=1,2,3$)  are coefficients of the linear combination.  Hereafter the vector with the hat symbol denotes a unit vector, e.g., $\hat{a}_B\hat{a}_B=1$.
It is easy to check that $Y_{(0,1)}(\hat{x})$ are orthonormal at every point:
\begin{align}
\bm{Y}_{(0,1),(B)}(\hat{x}) \cdot \bm{Y}_{(0,1),(C)}(\hat{x}) 
:=  Y^A_{(0,1),(B)}(\hat{x}) Y^A_{(0,1),(C)}(\hat{x})
  = \delta_{BC} 
 .
\end{align}
Thus the  solution is given by the linear combination of triplet of vector spherical harmonics $\bm{Y}_{(0,1)}(\hat{x})$, which is written in the manifestly Lorentz covariant Lie-algebra valued form using 
  Pauli matrices $\sigma_A$ and 
\begin{align}
 \bar{e}_\mu := (i\sigma_A, \mathbf{1}), 
 \quad
 e_\mu := (-i\sigma_A, \mathbf{1}) 
 ,
\end{align}
as 
\begin{align}
 \bm{n}(x) := {n}_A(x) \sigma_A 
= \hat{a}_B  Y^A_{(0,1),(B)}(\hat{x}) \sigma_A 
= \hat{a}_B x_\alpha \bar{e}_\alpha \sigma_B x_\beta e_\beta/x^2 
 ,
 \label{Hopf-sol}
\end{align}
or in the vector component 
\begin{align}
  {n}_A(x) 
= \hat{a}_B  Y^A_{(0,1),(B)}(\hat{x}) 
%= \frac12 {\rm tr}[a_B x_\alpha e_\alpha \sigma_B x_\beta \bar{e}_\beta \sigma_A ]/x^2
%= \frac12 a_B x_\alpha x_\beta  {\rm tr}[\sigma_A e_\alpha \sigma_B \bar{e}_\beta  ]/x^2
=  \hat{a}_B x_\alpha x_\beta \bar{\eta}^B_{\alpha \gamma} \eta^A_{\gamma\beta}/x^2
 .
 \label{hedgehog-sol2}
\end{align}
where we have used the formula:
\begin{align}
 {\rm tr}[\sigma_A \bar{e}_\alpha \sigma_B e_\beta  ]
 = -2 \bar{\eta}^B_{\alpha \gamma} \eta^A_{\beta\gamma} 
 .
\end{align}

It is directly checked that (\ref{Hopf-sol}) is indeed the solution of the RDE. 
Explicit calculations show  that (\ref{Hopf-sol}) satisfies 
\begin{align}
  -\partial_\mu \partial_\mu {n}_A(x) = \frac{8}{x^2} {n}_A(x)
 ,
\end{align}
and
\begin{align}
  2 \epsilon_{ABC} \eta^C_{\mu\nu} f_\nu(x) \partial_\mu  {n}_B(x)
  = - 8 f(x) {n}_A(x) 
  = - \frac{16}{x^2} {n}_A(x) 
 .
\end{align}
Then, for $(J,L)=(0,1)$, we arrive at  
\begin{align}
 V(x) =  \frac{-8}{x^2} 
 , \quad
 \tilde{V}(x) = 0 
 ,
\end{align}
and
\begin{align}
 \lambda(x) = V(x) + [-\partial_\mu \partial_\mu {n}_A(x)]/{n}_A(x)
\equiv 0 
 \quad \text{for any A, no sum over A}
 .
\end{align}
Thus this solution is an allowed one, since the solution gives a finite (vanishing) value for the functional $F_{\rm rc}$=0.  
The solution gives a map $\bm{Y}_{(0,1),(B)}$ from $S^3$ to $S^2$, which is known as the \textbf{standard Hopf map}.  
Therefore, the only zeros of $\phi_A(x)$ in the solution
${n}_A(x) =\phi_A(x)/|\phi(x)|=\phi_A(x)/\sqrt{\phi_B(x)\phi_B(x)}$
are the origin and the set of magnetic monopoles consists of the origin only, in other words, the magnetic monopole loop is shrank to a single point. 
Therefore, we have no  monopole loop with a finite and non-zero radius for the given Yang-Mills field of one instanton with zero size in the regular gauge.

For one instanton with  size $\rho$, i.e., $\kappa=1$, $s=\rho$, we must examine
\begin{align}
 f(x^2) = \frac{2}{x^2+\rho^2}
 ,
\end{align}
and
\begin{align}
 V(x) =  \frac{4}{x^2+\rho^2} [J(J+1)-L(L+1) ]  - \frac{8\rho^2}{(x^2+\rho^2)^2} 
 .
\end{align}
The lowest $\lambda(x)$ is realized for distinct set of $(J,L)$  depending on the region of $x$. 
This case is obtained by one-instanton limit of two meron case to be discussed later.

\subsubsection{One instanton in the singular gauge}

For one instanton in the singular gauge, we must take
%\begin{align}
% g\mathscr{A}_\mu(x) 
%=   \frac{\sigma_A}{2} g\mathscr{A}_\mu^A(x) 
%=  \frac{\sigma_A}{2} \bar{\eta}^A_{\mu\nu}  f_\nu(x) , \quad f_\nu(x) := \partial_\nu \ln \tilde{\Phi}(x^2) 
% ,
%\end{align}
\begin{align}
g\mathscr{A}_\mu(x) 
=  \frac{\sigma_A}{2} \bar{\eta}^A_{\mu\nu} x_\nu f(x^2) ,\quad
 f(x^2) = \frac{2\rho^2}{x^2(x^2+\rho^2)}
 .
\end{align}
The results in the previous section hold by replacing $\eta^A_{\mu\nu}$ by $\bar{\eta}^A_{\mu\nu}$.  
In this case, we have 
\begin{align}
 V(x) =  \frac{4\rho^2}{x^2(x^2+\rho^2)} [J(J+1)-L(L+1)-2] +   \frac{8\rho^4}{x^2(x^2+\rho^2)^2}  
 .
\end{align}
Apart from the detailed analysis, we focus on the zero size limit $\rho \rightarrow 0$ (or the distant region $x^2 \rightarrow \infty$): 
\begin{align}
 V(x) \simeq 0
 , \quad
 \tilde{V}(x) \simeq \frac{4L(L+1)}{x^2}   
 .
\end{align}
It is easy to see that the solution is given at $(J,L)=(1,0)$, i.e.,  
$\mathbf{n}(x)=\bm{Y}_{(1,0)}$ (a constant vector) given in (\ref{Y10}), which has 
the lowest value of $\lambda(x)$:  $\lambda(x) \equiv 0$. 
%In order to familiarize the vector spherical harmonics $Y^A_{(J,L)}$, we examine a simple case. 
For  $(J,L)=(1,0)$, the state is 3-fold degenerate:  $\mathbf{n}(x)=\bm{Y}_{(1,0)}$ is written as a linear combination of them:
Writing $\bm{Y}_{(1,0)}$ as a column vector: 
$\bm{Y}_{(1,0)}=(Y^1_{(1,0)},Y^2_{(1,0)},Y^3_{(1,0)})^T$ (T denotes transpose)
\begin{align}
 \bm{Y}_{(1,0)} 
 = \sum_{\alpha=1}^{3} \hat{c}_\alpha \bm{Y}_{(1,0),(\alpha)}
= \hat{c}_1 
 \begin{pmatrix}
 1 \cr
 0 \cr
 0 
 \end{pmatrix}
 + \hat{c}_2 
 \begin{pmatrix}
 0 \cr
 1 \cr
 0 
 \end{pmatrix}
 + \hat{c}_3 
 \begin{pmatrix}
 0 \cr
 0 \cr
 1 
 \end{pmatrix}
 .
 \label{Y10}
\end{align}
It constitutes the orthonormal set:
\begin{align}
\bm{Y}_{(1,0),(\alpha)} \cdot \bm{Y}_{(1,0),(\beta)} 
:=  Y^A_{(1,0),(\alpha)} Y^A_{(1,0),(\beta)}  = \delta_{\alpha \beta} 
 . 
\end{align}
Therefore, 
%if $(J,L)=(1,0)$ state gives the lowest value of $\lambda(x)$, 
the solution is given by a constant:
\begin{align}
 {n}_A(x) 
= \sum_{\alpha=1}^{3} \hat{c}_\alpha Y^A_{(1,0),(\alpha)} 
%= C_\alpha \delta_{A \alpha}/\sqrt{C_\beta C_\beta} 
= \hat{c}_A  
 .
\end{align}
In this limit, 
$
  \partial_\mu {n}_A(x) =0
$,
$
 \partial_\mu \partial_\mu {n}_A(x) =0
$ 
and 
\begin{align}
 \lambda(x) = V(x)  
=   2x^2 f^2(x)  
= \frac{8\rho^4}{x^2(x^2+\rho^2)^2}
 .
\end{align}
One-instanton in the singular gauge yields a finite reduction functional: 
\begin{align}
 F_{\rm rc} = \int d^4x  \lambda(x)  < \infty
 .
\end{align}

%\newpage
%%%%%%%%%%%%%%%%%%%%%%%%%%%%%%%%%%%%%%%%%%%%%%%%%%
%%%%%%%%%%%%%%%%%%%%%%%%%%%%%%%%%%%%%%%%%%%%%%%%%%
\subsubsection{One-meron and magnetic monopole line}
%\setcounter{equation}{0}
%%%%%%%%%%%%%%%%%%%%%%%%%%%%%%%%%%%%%%%%%%%%%%%%%%
%%%%%%%%%%%%%%%%%%%%%%%%%%%%%%%%%%%%%%%%%%%%%%%%%%

In order discuss one-meron configuration, i.e.,  $\kappa=\frac12$, $s=0$, we have
\begin{align}
 f(x^2) = \frac{1}{x^2} 
 ,
\end{align}
which yields 
\begin{align}
 V(x) =  \frac{2}{x^2}  [J(J+1)-L(L+1)-1]   
 , \
 \tilde{V}(x) =  \frac{2}{x^2} [J(J+1)+L(L+1)-1]   > 0
 .
\end{align}
For one meron, we find that $(J,L)=(1/2,1/2)$ gives the lowest $\tilde{V}(x)$. 
This suggests that the solution might be given by 
\begin{align}
 \bm{Y}_{(1/2,1/2)}(\hat{x}) 
 &= \sum_{\mu=1}^{4} \hat{b}_\mu \bm{Y}_{(1/2,1/2),(\mu)}(\hat{x})
\nonumber\\
&= \hat{b}_1 
 \begin{pmatrix}
 -\hat{x}_4 \cr
 \hat{x}_3 \cr
 -\hat{x}_2 
 \end{pmatrix}
 + \hat{b}_2 
 \begin{pmatrix}
 -\hat{x}_3 \cr
 -\hat{x}_4 \cr
 \hat{x}_1 
 \end{pmatrix}
 + \hat{b}_3 
 \begin{pmatrix}
 \hat{x}_2 \cr
 -\hat{x}_1 \cr
 -\hat{x}_4 
 \end{pmatrix}
 + \hat{b}_4 
 \begin{pmatrix}
 \hat{x}_1 \cr
 \hat{x}_2 \cr
 \hat{x}_3 
 \end{pmatrix}
 ,
 \label{Y-hedgehog}
\end{align}
where a unit four-vector $\hat{b}_\mu$ ($\mu=1,2,3,4$) denote four coefficients of the linear combination for 4-fold generate $\bm{Y}_{(1/2,1/2),(\mu)}(\hat{x})$ ($\mu=1,2,3,4$).

However, a subtle point in this case is that $Y^A_{(1/2,1/2),(\mu)}(\hat{x})$ are non-orthonormal sets at every space--time point: 
\begin{align}
\bm{Y}_{(1/2,1/2),(\mu)}(\hat{x}) \cdot \bm{Y}_{(1/2,1/2),(\nu)}(\hat{x}) 
:=  Y^A_{(1/2,1/2),(\mu)}(\hat{x}) Y^A_{(1/2,1/2),(\nu)}(\hat{x})
  \ne \delta_{\mu\nu} 
 .
\end{align}

Nevertheless, we find that the unit vector field:
\begin{align}
  {n}_A(x) 
= b_\nu \eta^A_{\mu\nu} x_\mu /\sqrt{b^2 x^2 -(b \cdot x)^2}
= \hat{b}_\nu \eta^A_{\mu\nu} \hat{x}_\mu /\sqrt{ 1 -(\hat{b} \cdot \hat{x})^2}
 ,
 \label{n-hedgehog-sol}
\end{align}
constructed from 
\begin{align}
 \bm{Y}_{(1/2,1/2),(\mu)}(\hat{x})=\eta^A_{\mu\nu} \hat{x}_\nu  \quad ( \mu=1,2,3,4 )
 ,
\end{align}
can be a solution of RDE. 
In fact, explicit calculations show  that (\ref{n-hedgehog-sol}) satisfies
\begin{align}
  -\partial_\mu \partial_\mu  {n}_A(x) = \frac{2}{ x^2-(\hat{b} \cdot x)^2}  {n}_A(x)
 ,
\end{align}
and
\begin{align}
  2 \epsilon_{ABC} \eta^C_{\mu\nu} f_\nu(x) \partial_\mu   {n}_B(x)
  = - 4 f(x)  {n}_A(x) 
  = - \frac{4}{x^2}  {n}_A(x) 
 .
\end{align}
Then, for $(J,L)=(1/2,1/2)$, we conclude that 
\begin{align}
 V(x) =  \frac{-2}{x^2}  
 , \quad
 \tilde{V}(x) = \frac{1}{x^2} 
 ,
\end{align}
and
\begin{align}
 \lambda(x) =& [-\partial_\mu \partial_\mu  {n}_A(x)]/ {n}_A(x) + V(x) 
%= \frac{2b^2}{b^2 x^2-(b \cdot x)^2}  + \frac{-2}{x^2}
 \quad \text{for any A, no sum over A}
 \nonumber\\
%  =& \frac{2(b \cdot x)^2}{x^2[b^2 x^2-(b \cdot x)^2]}
 =&  \frac{2(\hat{b} \cdot x)^2}{x^2[ x^2-(\hat{b} \cdot x)^2]}
 .
 \label{meron-eigen}
\end{align}
The solution (\ref{n-hedgehog-sol}) is of the \textbf{hedgehog type}.
The magnetic-monopole current is obtained as simultaneous zeros of $\hat{b}_\nu \eta^A_{\mu\nu}x_\mu=0$ for $A=1,2,3$. 
Taking the 4th vector in (\ref{Y-hedgehog}) $\hat{b}_\mu= \delta_{\mu 4}$, the magnetic-monopole current is located at $x_1=x_2=x_3=0$, i.e., on the $x_4$ axis. Whereas, if the 3rd vector in (\ref{Y-hedgehog}) is taken $\hat{b}_\mu= \delta_{\mu 3}$, the magnetic-monopole current flows at $x_1=x_2=x_4=0$, i.e., on $x_3$ axis.

In general, it turns out that the magnetic-monopole current $k_\mu$  is located on the straight line parallel  to $\hat{b}_\mu$ going through the origin. 
Note that the expression (\ref{meron-eigen}) for $\lambda(x)$ is invariant under a subgroup $SO(3)$ of  the Euclidean rotation $SO(4)$. 
In other words, once we select $\hat{b}_\mu$, $SO(4)$ symmetry is broken to $SO(3)$ just as in the spontaneously broken symmetry.  This result is consistent with a fact that the magnetic-monopole current $k_\mu$ flows in the direction of $\hat{b}_\mu$ and the symmetry is reduced to the axial symmetry, the rotation group $SO(3)$, about the axis  in the direction of a four vector $\hat{b}_\mu$.

It is instructive to point out that the Hopf map $Y$ also satisfies the RDE.  
Therefore, it is necessary to compare the value of the reduction functional of $(J,L)=(1/2,1/2)$ with that of $(J,L)=(0,1)$. 
In the $(J,L)=(0,1)$ case, we find
\begin{align}
  \lambda_{(0,1)}(x)
 =   \frac{2}{x^2} = \frac{2}{x_1^2+x_2^2+x_3^2+x_4^2}
 .
\end{align}
For instance, we can choose $\hat{b}_\mu= \delta_{\mu 3}$ without loss of generality: 
\begin{align}
  \lambda_{(1/2,1/2)}(x)
 =   \frac{2x_3^2}{[x_1^2+x_2^2+x_3^2+x_4^2][ x_1^2+x_2^2+x_4^2]}
 .
\end{align}
Note that  the integral of $\lambda_{(1/2,1/2)}(x)$ over the whole space--time $\mathbb{R}^4$ is obviously smaller than that of $\lambda_{(0,1)}(x)$, although 
$\lambda_{(0,1)}(x) < \lambda_{(1/2,1/2)}(x)$ locally 
inside a cone with the symmetric axis $\hat{b}_\mu$, i.e.,
$(\hat{b} \cdot \hat{x})^2 \ge 1/2$.

The reduction functional in $(J,L)=(1/2,1/2)$ case reads 
\begin{align}
 F_{\rm rc} 
 =&  \int d^4x  \frac12 \lambda_{(1/2,1/2)}(x)
 \nonumber\\
 =& \int  dx_3 \int dx_1dx_2dx_4  \frac{x_3^2}{[x_1^2+x_2^2+x_3^2+x_4^2][ x_1^2+x_2^2+x_4^2]}
 \nonumber\\
 =& 4\pi \int_{-L_3}^{L_3} dx_3 x_3^2 \int_{0}^{\infty} dr  \frac{1}{[r^2+x_3^2]}
 \nonumber\\
 =& 4\pi \int_{-L_3}^{L_3} dx_3 x_3^2  \frac{1}{|x_3|} \arctan \frac{r}{|x_3|} \Big|_{r=0}^{r=\infty}
 \nonumber\\
 =& 4\pi \int_{-L_3}^{L_3} dx_3 x_3^2  \frac{1}{|x_3|}  \frac{\pi}{2} 
 \nonumber\\
 =& 4\pi^2 \int_{0}^{L_3} dx_3  x_3  
 ,
\end{align}
where we have defined $r^2:=x_1^2+x_2^2+x_4^2$.
$\lambda_{(1/2,1/2)}(x)$ is zero on the $x_3=0$ hyperplane, i.e.,  three dimensional space $x_1,x_2,x_4$ which is orthogonal to the magnetic current. Therefore, in the three-dimensional space, the magnetic current looks like just a point magnetic charge.

Although $F_{\rm rc}$ remains finite as long as $L_3$ is finite,  it diverges for $L_3 \rightarrow \infty$, i.e., when integrated out literally in the whole space--time $\mathbb{R}^4$.
In the next section, we see that this difficulty is resolved for two meron configuration.

%%%%%%%%%%%%%%%%%%%%%%%%%%%%%%%%%%%%%%%%%%%%%%%%%%
%%%%%%%%%%%%%%%%%%%%%%%%%%%%%%%%%%%%%%%%%%%%%%%%%%
\subsubsection{Summary of one-instanton and one-meron}
%\setcounter{equation}{0}
%%%%%%%%%%%%%%%%%%%%%%%%%%%%%%%%%%%%%%%%%%%%%%%%%%
%%%%%%%%%%%%%%%%%%%%%%%%%%%%%%%%%%%%%%%%%%%%%%%%%%

Summarizing the one-instanton and  one-meron case, 
\begin{enumerate}
\item
 {For one-instanton in the regular gauge with zero size}, the solution is 3-fold ($\hat{a}_{B}$, $B=1,2,3$)  degenerate ($(J,L)=(0,1)$), the solution is a linear combination of the standard Hopf map:
\begin{align} 
  {n}_A(x) 
%= \hat{a}_B  Y^A_{(0,1),(B)}(\hat{x}) 
%= \frac12 {\rm tr}[a_B x_\alpha e_\alpha \sigma_B x_\beta \bar{e}_\beta \sigma_A ]/x^2
%= \frac12 a_B x_\alpha x_\beta  {\rm tr}[\sigma_A e_\alpha \sigma_B \bar{e}_\beta  ]/x^2
=  \sum_{B=1,2,3} \hat{a}_B \sum_{\alpha,\beta,\gamma=1,2,3,4}  \hat{x}_\alpha \hat{x}_\beta \bar{\eta}^B_{\alpha \gamma} \eta^A_{\gamma\beta}   , \
  \hat{x}_\mu :=\frac{x_\mu}{\sqrt{x^2}} ,
\
  \lambda_{(0,1)}(x)=0
. 
\end{align}
The standard Hopf map is singular only at the center of the instanton. Therefore,  {$k_\mu$ is non-zero only at the center of the instanton and  
there is no magnetic monopole loop. 
}
 
\item
 For one-instanton in the singular gauge with zero size, the solution is 3-fold degenerate ($(J,L)=(1,0)$)
\begin{equation} 
  {n}_A(x) = c_A ,
\quad 
  \lambda_{(1,0)}(x)=0 .
\end{equation}

\item
 {For one-meron}, the solution is {4-fold degenerate} ($(J,L)=(1/2,1/2)$)  4d hedgehog given by 
\begin{align} 
   {n}_A(x) 
%= b_\nu \eta^A_{\mu\nu} x_\mu /\sqrt{b^2 x^2 -(b \cdot x)^2}
=&  { \sum_{\nu=1,2,3,4}  b_\nu } \sum_{\mu=1,2,3,4}  \eta^A_{\mu\nu} \hat{x}_\mu /\sqrt{ b^2 -(b \cdot \hat{x})^2} ,
\nonumber\\
 \lambda_{(1/2,1/2)}(x) 
%=& [-\partial_\mu \partial_\mu {n}_A(x)]/{n}_A(x) + V(x) 
%= \frac{2b^2}{b^2 x^2-(b \cdot x)^2}  + \frac{-2}{x^2}
% \quad \text{for any A, no sum over A}
% \nonumber\\
%  =& \frac{2(b \cdot x)^2}{x^2[b^2 x^2-(b \cdot x)^2]}
 =&  \frac{2(b \cdot x)^2}{x^2[ b^2 x^2-(b \cdot x)^2]} .
\end{align}
 {$k_\mu$ denotes a straight magnetic line going  through the center of the meron in the direction $ b_\mu$. 
} 
The solution of RDE is not unique. The Hopf map is also a solution of RDE with 
$\lambda_{(0,1)}(x)
 =   \frac{2}{x^2}
$, but it is excluded, since it 
gives larger $F_{\rm rc}=\int d^4x \lambda(x)$.

\end{enumerate}

%%%%%%%%%%%%%%%%%%%%%%%%%%%%%%%%%%%%%%%%%%%%%%%%%%
%%%%%%%%%%%%%%%%%%%%%%%%%%%%%%%%%%%%%%%%%%%%%%%%%%
\subsection{Two merons and magnetic monopole loop}
%\setcounter{equation}{0}
%%%%%%%%%%%%%%%%%%%%%%%%%%%%%%%%%%%%%%%%%%%%%%%%%%
%%%%%%%%%%%%%%%%%%%%%%%%%%%%%%%%%%%%%%%%%%%%%%%%%%

\subsubsection{Meron pair solution}

The one-meron solution is given by \cite{AFF76,AFF77}
\begin{align}
 g\mathscr{A}^{\rm M}_\mu(x) 
= g\mathscr{A}^A_\mu(x) \frac{\sigma_A}{2}  
= \frac{\sigma_A}{2} \eta^A_{\mu\nu} \frac{x_\nu}{x^2}  
 .
 \label{1-meron}
\end{align} 
The topological charge $Q_P$ is defined using the topological charge density $D_P(x)$: 
\begin{align} 
 Q_P=\int d^4x D_P(x) , \quad 
 D_P(x) := \frac{1}{16\pi^2} {\rm tr}( {}^{\displaystyle *}\mathscr{F}_{\mu\nu}\mathscr{F}_{\mu\nu}) 
  .
 \label{topological-density}
\end{align} 
The meron (\ref{1-meron}) has one half unit of topological charge  concentrated at the origin:
\begin{align} 
 Q_P=\frac12 , \quad D_P(x) = \frac12 \delta^4(x)   
 .
 \label{topological-density-1}
\end{align} 
The one-meron solution can be written in the form:
\begin{align}
 g\mathscr{A}_\mu^{\rm M}(x) = g\mathscr{A}^A_\mu(x) \frac{\sigma_A}{2} = S_{\mu\nu} \frac{x_\nu}{x^2} , \
 S_{\mu\nu} := - \frac{i}{4} (\bar{e}_\mu e_\nu - e_\nu \bar{e}_\mu) = \eta^A_{\mu\nu} \frac{\sigma_A}{2}
 ,
\end{align}
using $\bar{e}_\mu$ and $e_\mu$ defined by the  Pauli matrices $\sigma_A$:
\begin{align}
 \bar{e}_\mu = (i\sigma_A, \mathbf{1}), 
 \quad
 e_\mu := (-i\sigma_A, \mathbf{1}) 
 .
\end{align}
The one-meron solution can be rewritten in another form:
\begin{align}
 g\mathscr{A}_\mu^{\rm M}(x) 
 = \frac12 i U(x) \partial_\mu U^{-1}(x) 
 ,
\end{align}
where
\begin{align}
 U(x) = \frac{\bar{e}_\alpha x_\alpha}{\sqrt{x^2}}, \quad U^{-1}(x) = \frac{e_\alpha x_\alpha}{\sqrt{x^2}}  
 .
\end{align}
These relations are easily checked by using the formulae:
\begin{align}
 \bar{e}_\mu e_\nu = \delta_{\mu\nu} + i \eta^A_{\mu\nu} \sigma_A, \quad 
 e_\mu \bar{e}_\nu = \delta_{\mu\nu} + i \bar{\eta}^A_{\mu\nu} \sigma_A  
 .
\end{align}

While, the one-antimeron solution given by
\begin{align}
 g\mathscr{A}_\mu^{\rm \bar{M}}(x) = g\mathscr{A}^A_\mu(x) \frac{\sigma_A}{2} = \bar{S}_{\mu\nu} \frac{x_\nu}{x^2} , \quad
 \bar{S}_{\mu\nu} := - \frac{i}{4} (e_\mu \bar{e}_\nu - \bar{e}_\nu e_\mu) = \bar{\eta}^A_{\mu\nu} \frac{\sigma_A}{2}
 ,
\end{align}
can be written in the form:
\begin{align}
 g\mathscr{A}_\mu^{\rm \bar{M}}(x) 
 = \frac12 i U^{-1}(x) \partial_\mu U(x) 
 .
\end{align}
Note that the meron and antimeron configurations are not of the pure gauge form, which has an important implications to confinement.

The single meron and antimeron solutions given in the above are singular both at the origin $x^2=0$ and at infinity $x^2=\infty$.  
Using the conformal symmetry of the classical Yang-Mills action, it can be shown that in addition to a meron at the origin, there is a second meron at infinity with another half unit of topological charge.
In fact, the conformal invariance of Yang-Mills theory allows us to displace (map) those singularities to arbitrary points which we define to be the origin  and $d_\mu \in \mathbb{R}^4$.   Explicitly, the conformal transformation
\footnote{
This \textbf{conformal transformation} is obtained by combining
(a) a translation 
$
 x_\mu \rightarrow x_\mu +a_\mu
$,
(b) an \textbf{inversion}
$
 x_\mu \rightarrow - x_\mu/x^2
$,
(c) a \textbf{dilatation} (scale transformation)
$
 x_\mu \rightarrow -2a^2 x_\mu  
$, and
(d) a translation
$
 x_\mu \rightarrow x_\mu -a_\mu
$.
The transformation is constructed so that the origin $x=0$ is transformed to $a$, while $x=\infty$ to $-a$. In addition, $x=a$ is transformed to $0$.
}
\begin{align}
 x_\mu \rightarrow 
z_\mu =  2a^2 \frac{(x+a)_\mu}{(x+a)^2} - a_\mu   
%=  4a^2 \frac{(x+2a)_\mu}{(x+2a)^2} - a_\mu   
 ,
 \label{conformal-transformation}
\end{align}
yields the new solutions
\begin{align}
 g\mathscr{A}_\mu^{\rm M}(x) \rightarrow \frac12 i U(z) \partial_\mu U^{-1}(z) := g\mathscr{A}_\mu^{\rm M\bar{M}}(x)   
 ,
 \nonumber\\
 g\mathscr{A}_\mu^{\rm \bar{M}}(x) \rightarrow  
 \frac12 i U^{-1}(z) \partial_\mu U(z) := g\mathscr{A}_\mu^{\rm \bar{M}M}(x) 
 ,
\end{align}
where we have used the the transformation law:
$
 g\mathscr{A}_\mu^{\rm M}(x) = \partial_\mu z_\nu   g\mathscr{A}_\nu^{\rm M}(z)
$
with  
  $\partial_\mu:=\partial/\partial x_\mu (\not= \partial/\partial z_\mu)$. 

%\footnote{
%This is obtained from the transformation law:
%$
% g\mathscr{A}_\mu^{\rm M}(x) \rightarrow \partial_\mu z_\nu   g\mathscr{A}_\nu^{\rm M}(z)
%$ 
%}
It is shown \cite{AFF77} that $g\mathscr{A}_\mu^{\rm M\bar{M}}$ corresponds to a meron located at $x=-a$ and an antimeron at $x=a$.  Conversely, $g\mathscr{A}_\mu^{\rm \bar{M}M}$ has a meron at $x=a$ and an antimeron at $x=-a$. 
These $g\mathscr{A}_\mu^{\rm M\bar{M}}$ ($g\mathscr{A}_\mu^{\rm \bar{M}M}$) are meron--antimeron (antimeron--meron) solutions. 
The meron-antimeron solution has the explicit expression for $a=(0,0,0,T)$:
\begin{align}
 g\mathscr{A}_\mu^{\rm M\bar{M}}(x) 
=&  \begin{cases}
   \frac{2T}{\tau^2}  x_4 \sigma_\ell x_\ell & (\mu=4) \cr
   \frac{2T}{\tau^2} [\epsilon_{jk\ell} T x_k \sigma_\ell + \frac12 (T^2-x^2) \sigma_j + x_j \sigma_\ell x_\ell ] & (\mu=j) 
   \end{cases}
 ,
 \nonumber\\
  \tau^2 =& (T^2+x^2-2T x_4)(T^2+x^2+2T x_4) = (T^2+x^2)^2-4T^2x_4^2 
 .
\end{align}

It is also shown \cite{AFF77}  that the meron--meron (antimeron--antimeron) solution is given by performing a singular gauge transformation $U(y_{+})$ which changes the antimeron (meron) at $x=-a$ into a meron (antimeron) at the same point, leading from an $\rm M\bar{M}$ ($\rm \bar{M}M$) to an $\rm MM$ ($\rm \bar{M}\bar{M}$) one where
$y_{\pm}:=x\pm a$.
In fact, the singular gauge transformation
\begin{align}
  g\mathscr{A}_\mu^{\rm M\bar{M}}(x) \rightarrow U^{-1}(y_{+})g\mathscr{A}_\mu^{\rm M\bar{M}}(x)U(y_{+})+iU^{-1}(y_{+}) \partial_\mu U(y_{+}) := g\mathscr{A}_\mu^{\rm MM}(x)
 ,
 \label{gauge-transformation}
\end{align}
leads to the dimeron solution
\begin{align}
   g\mathscr{A}_\mu^{\rm MM}(x) = - S_{\mu\nu} \left[ \frac{y_{+}^\nu}{y_{+}^2} + \frac{y_{-}^\nu}{y_{-}^2} \right] 
= - \frac{\sigma_A}{2} \left[ \eta^A_{\mu\nu} \frac{(x+a)_\nu}{(x+a)^2} + \eta^A_{\mu\nu}  \frac{(x-a)_\nu}{(x-a)^2} \right]
 .
 \label{2-meron}
\end{align}
The antidimeron solution $g\mathscr{A}_\mu^{\rm \bar{M}\bar{M}}$ is obtained in the similar way. 
%After the gauge transformation, the gauge field for the two merons take the simple form:
%\begin{align} 
% g\mathscr{A}^{MM}_\mu(x) = \frac{\sigma_A}{2} \left[ \eta^A_{\mu\nu} \frac{x_\nu}{x^2} + \eta^A_{\mu\nu}  \frac{(x-d)_\nu}{(x-d)^2} \right]
% ,
% \label{2-meron}
%\end{align}

The gauge field $g\mathscr{A}_\mu^{\rm MM}(x)$ for a meron pair has infinite action density at $x=\{ 0, d \}$ and the logarithmic singularity of the action integral comes from the delta function concentration of topological charge:
\footnote{
Similar to instantons, a meron pair can be expressed in singular gauge by performing a large gauge transformation about the midpoint of the pair, resulting in a gauge field that falls off faster at large distance $\mathscr{A} \sim x^{-3}$.
} 
\begin{align}
 D_P(x)  = \frac12 \delta^4(x+a) + \frac12 \delta^4(x-a)  
 .
 \label{topological-density-3}
\end{align}

%%%%%%%%%%%%%%%%%%%%%%%%%%%%%%%%%%%%%%%%%%%%%%%%%%%%%%%%%%%%%
\subsubsection{Smeared meron pair}
%%%%%%%%%%%%%%%%%%%%%%%%%%%%%%%%%%%%%%%%%%%%%%%%%%%%%%%%%%%%%

%%%%%%%%%%%%%%%%%%%%%%%%%%%%%%%%%%%%%%%%%%%%%%%%%%%%%%%%%%%%%
%%%%%%%%%%%%%%%%%%%%%%%%%%%%%%%%%%%%%%%%%%%%%%%%%%%%%%%%%%%%%
\begin{figure}[tb]
\begin{center}
\includegraphics[scale=0.35]{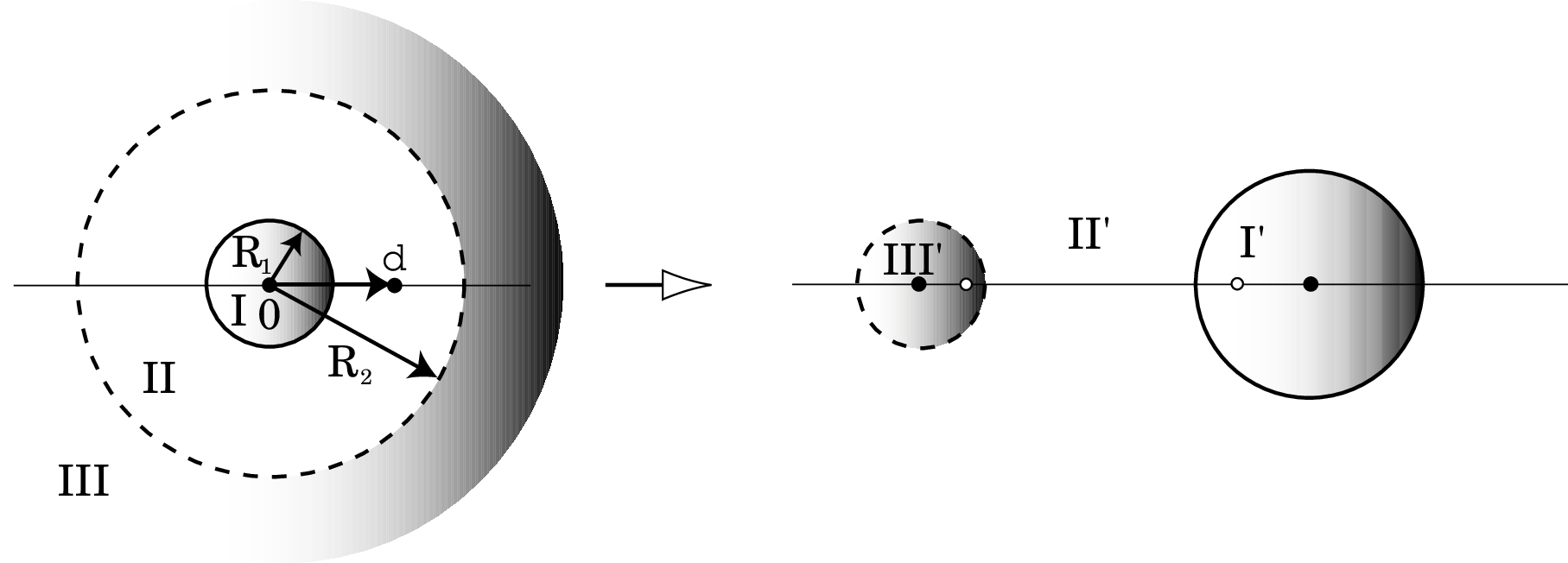}
\end{center} 
 \caption[]{
The concentric sphere geometry for a smeared meron (left panel) is transformed to the smeared two meron configuration (right panel) by the conformal transformation including the inversion about the point $d$. 
}
 \label{C00-fig:smeared-meron}
\end{figure}
%%%%%%%%%%%%%%%%%%%%%%%%%%%%%%%%%%%%%%%%%%%%%%%%%%%%%%%%%%%%%
%%%%%%%%%%%%%%%%%%%%%%%%%%%%%%%%%%%%%%%%%%%%%%%%%%%%%%%%%%%%%

In order to eliminate the singularity in a meron pair configuration, we introduce an  Ansatz of  finite action by replacing the meron pair configuration (\ref{2-meron}) by a  smeared configurations  following Callan, Dashen and Gross \cite{CDG78} (See the left panel of Fig.~\ref{C00-fig:smeared-meron}):
\begin{align}
 \mathscr{A}^{\rm sMM}_\mu(x) = \frac{\sigma_A}{2}  \eta^A_{\mu\nu} x_\nu  \times 
\begin{cases}
 \frac{2}{x^2+R_1^2} & $I$:\sqrt{x^2}<R_1 \ (  Q_P^{\rm I}=\frac12 )  \cr
 \frac{1}{x^2} & $II$: R_1<\sqrt{x^2}<R_2 \ (  Q_P^{\rm II}=0) \cr
 \frac{2}{x^2+R_2^2} & $III$: \sqrt{x^2}>R_2  \ (  Q_P^{\rm III}=\frac12) \cr
\end{cases} 
 ,
 \label{smeared-2-meron}
\end{align}
where  in the region (II) the field is identical to the meron field, while at the inner (outer) radius $R_1(R_2)$ it joins smoothly onto a standard instanton field.  Here the radii $R_1$ and $R_2$ of the inner sphere and the outer sphere  are arbitrary.

The topological charge $Q_P$ is spread out around the origin (I) and infinity (III): The scale size is chosen such that the net topological charge inside I (outside II)  is one-half unit, which agrees with the topological charge carried by each meron. 
This field (\ref{smeared-2-meron}) satisfies the equation of motion everywhere except on the two spheres.%
\footnote{
Although this patching of instanton caps is continuous, the derivatives are not, and therefore the equation of motion are violated at the boundaries of the regions, $\partial \rm I=\partial \rm II$ and $\partial \rm II=\partial \rm III$. 
}
  In fact, it is the solution of the equation of motion under the constraint that there be one-half unit of topological charge both in the inner and outer spheres, i.e., $Q_P^{\rm I}=1/2=Q_P^{\rm III}$. 
In other words, the singular meron fields for I and III are replaced by instanton caps, each containing topological charge 1/2 to agree with (\ref{smeared-2-meron}).

The Yang-Mills action of the new configuration is calculated to be
\begin{align}
 S^{\rm sMM}_{\rm YM} = \frac{8\pi^2}{g^2} +  \frac{3\pi^2}{g^2} \ln \frac{R_2}{R_1} 
 ,
\end{align}
where the first constant term comes from the two half-instantons in (I) and (III) 
\footnote{
There is no angular dependence in this patching, and so the conformal symmetry of the meron pair is retained.  For example, under a dilatation $x_\mu \rightarrow \lambda z_\mu$, both $R_1$ and $R_2$ get multiplied by $1/\lambda$, but the ratio and hence the action remain invariant. 
}
and the second logarithmic term comes from the pure meron region (II) in between.  Furthermore, if we let $|R_1-R_2| \downarrow 0$, this configuration becomes standard instanton. In the one-instanton limit  $|R_1-R_2| \downarrow 0$ ($R_2/R_1 \downarrow 1$),  
$S^{\rm sMM}_{\rm YM}= \frac{8\pi^2}{g^2}$ is finite.
In the one-meron limit  $R_2 \uparrow \infty$ or $R_1 \downarrow 0$ ($R_2/R_1 \uparrow \infty$),
$S^{\rm sMM}_{\rm YM}$ is logarithmic divergent.

We perform the conformal transformation of the configuration about some point $d$ in the region II between $R_1$ and $R_2$:
\begin{align}
 x_\mu \rightarrow d_\mu + \rho^2 \frac{(x-d)_\mu}{(x-d)^2}
 ,
 \label{conformatl-transformation}
\end{align}
with $\rho$ an arbitrary scale factor. 
Because of conformal invariance, this produces an another acceptable solution of the equation of motion.  
The geometry before and after the  conformal transformation  is described in  Fig.~\ref{C00-fig:smeared-meron}.
The conformal transformation maps a sphere into another sphere.
Therefore, the regions I and III, i.e., inner and outer spheres are transformed to two spheres, i.e., regions I$'$ and III$'$ with center coordinates $x_{\rm I'}$, $x_{\rm III'}$ and the scale sizes $R_1'$, $R_2'$, and  the field in region I$'$ and III$'$ is an instanton, since the conformal transformation of an instanton is again an instanton. 
Region II is transformed to region II$'$ and the field in II$'$ is given by
\begin{align}
 \mathscr{A}^{\rm II'}_\mu(x) = \frac{\sigma_A}{2} \left[ \eta^A_{\mu\nu} \frac{(x-x_{\rm I'})_\nu}{(x-x_{\rm I'})^2} + \eta^A_{\mu\nu}  \frac{(x-x_{\rm III'})_\nu}{(x-x_{\rm III'})^2} \right]
 ,
 \label{smeared-2-meron-b}
\end{align}
where  
\begin{align}
 x_{\rm I'} = \frac{R_2d}{R_2-R_1} > d ,
 \quad
 x_{\rm III'} = -\frac{R_1d}{R_2-R_1}<0
 .
\end{align}
The corresponding field strength $\mathscr{F}^{\rm II'}_{\mu\nu}$ falls at infinity as $|x|^{-4}$, leading to a convergent action integral.  
Since the topological charge $Q_p$ is conformal invariant, after transformation we have two spherical regions I$'$, III$'$ of net topological charge one-half surrounded by an infinite region II$'$ of zero topological charge density $D_P(x)=0$.
Therefore, the transformed configuration is a smeared version of two merons at position $x_{\rm I'}$ and $x_{\rm III'}$.

%%%%%%%%%%%%%%%%%%%%%%%%%%%%%%%%%%%%%%%%%%%%%%%%%%%%%%%%%%%%%
%%%%%%%%%%%%%%%%%%%%%%%%%%%%%%%%%%%%%%%%%%%%%%%%%%%%%%%%%%%%%
\begin{figure}%[ptb]
\begin{center}
\includegraphics[scale=0.6]{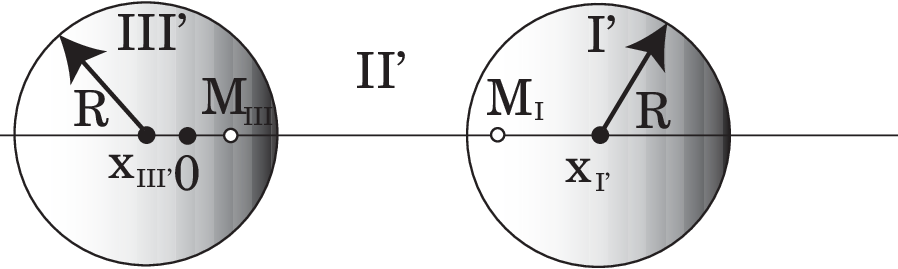}
\end{center} 
 \caption[]{
 Meron pair separated by $d=\sqrt{R_1 R_2}$ regulated with instanton caps. 
The smeared two meron configuration is obtained by the conformal transformation where  $d$ is the scale parameter of the inversion.
The centers of the sphere are 
$
 x_{\rm I'} = \frac{R_2d}{R_2-R_1} ,
$
and
$
 x_{\rm III'} = -\frac{R_1d}{R_2-R_1}
$.
The original positions of the two merons are not the centers of the sphere, nor are they the positions of maximum action density, which occurs with the spheres at 
$
 (M_{\rm I})_\mu = \frac{R_1^2}{R_1^2+d^2} d_\mu,
$
$ 
 (M_{\rm II})_\mu = \frac{R_2^2}{R_2^2+d^2} d_\mu, 
$
with 
$
 S_{\rm max}=\frac{48}{g^2}\frac{(R_1+R_2)^4}{d^8}
$.
The radius of the sphere is 
$
R = \frac{R_1 R_2}{R_2-R_1}
$. 
}
 \label{C00-fig:smeared-meron2}
\end{figure}
%%%%%%%%%%%%%%%%%%%%%%%%%%%%%%%%%%%%%%%%%%%%%%%%%%%%%%%%%%%%%
%%%%%%%%%%%%%%%%%%%%%%%%%%%%%%%%%%%%%%%%%%%%%%%%%%%%%%%%%%%%%

The smoothed meron configuration may be thought of as describing various stages in a sequence of deformations of the instanton, leading from the instanton at one extreme to two widely separated smeared merons at the other. 
In a sense the meron is to be regarded as a constituent of the instanton.
This is realized by holding $R_1$ fixed and increasing $R_2$ from $R_1$ to infinity $\infty$.  For definiteness, we choose $\rho=d:=\sqrt{R_1 R_2}$, see Fig.~\ref{C00-fig:smeared-meron2}.
With these choices, the configuration is two half instantons of scale size $R_1$ and separation $d=$ between the centers of the instanton configuration.
The action is 
\begin{align}
 S^{\rm sMM} = \frac{8\pi^2}{g^2} +  \frac{6\pi^2}{g^2} \ln \frac{d}{R_1} 
 .
\end{align}
As $R_2 \rightarrow R_1$, the regions I$'$ and III$'$ grow without limit in radius and move toward each other, while the centers of the instanton configurations approach each other and the region II$'$ vanishes. 
In the limit the configuration is just given an instanton of scale size $\rho=d=R_1$ split in half through a center.

\subsubsection{Magnetic monopole loop joining the smeared meron pair}

Now we consider the solution of RDE for a smeared (regularized) meron pair configuration based on the finite action Ansatz.  For this purpose, we estimate the value of $\lambda(x)$ in each region. 
For a set of $(J,L)$, $\lambda(x)$ is calculated in each region as follows, see the next Table.
% \ref{table:eigenvalue2}.

\footnotesize
\begin{center}
\begin{tabular}{l|l|l|l|l|l|l} \hline\hline
J & L & S & d* & I:$0<\sqrt{x^2}<R_1$ & II:$R_1<\sqrt{x^2}<R_2$ & III:$\sqrt{x^2}>R_2$ \\  
  &   &   &   & $\kappa=1, s=R_1$ & $\kappa=1/2, s=0$ & $\kappa=1, s=R_2$ \\
\hline\hline
1 & 0 & 1     & 3  & $\frac{8x^2}{(x^2+R_1^2)^2}$ & $\frac{2}{x^2}$ & $\frac{8x^2}{(x^2+R_2^2)^2}$ \\ \hline
1/2 & 1/2 & 1 & 4  & $\frac{2}{x^2-(\hat{b} \cdot x)^2}-\frac{8R_a^2}{(x^2+R_a^2)^2}$ & $\frac{2}{x^2-(\hat{b} \cdot x)^2}-\frac{2}{x^2}$ & $\frac{2}{x^2-(\hat{b} \cdot x)^2}-\frac{8R_a^2}{(x^2+R_2^2)^2}$ \\  
%1/2 & 3/2 & 1 & 8  &   &   &   \\ 
\hline
0 & 1 & 1     & 3  & $\frac{8x^2}{(x^2+R_1^2)^2}$ & $\frac{2}{x^2}$ & $\frac{8x^2}{(x^2+R_2^2)^2}$ \\ 
%1 & 1 & 1     & 9  &   &   &   \\
%1 & 2 & 1     & 15 &   &   &   \\ 
\hline
\end{tabular}
\label{table:eigenvalue2}
\end{center}
\normalsize

For $(J,L)=(1,0)$, 
\begin{align}
 \lambda(x) = V(x)  
=   2x^2 f^2(x)  
= \frac{8\kappa^2 x^2}{(x^2+s^2)^2}
= \begin{cases}
  \frac{8x^2}{(x^2+R_a^2)^2} &  $I$, \ $III$ \cr
  \frac{2}{x^2} & $II$ 
  \end{cases}
 .
\end{align}

For $(J,L)=(1/2,1/2)$, 
\begin{align}
 \lambda(x) 
= \frac{2}{x^2-(\hat{b} \cdot x)^2} + \frac{8\kappa^2 x^2}{(x^2+s^2)^2} - \frac{16\kappa}{x^2+s^2}
= \begin{cases}
  \frac{2}{x^2-(\hat{b} \cdot x)^2}-\frac{8R_a^2}{(x^2+R_a^2)^2} &  $I$, \ $III$ \cr
  \frac{2}{x^2-(\hat{b} \cdot x)^2}-\frac{2}{x^2} & $II$ 
  \end{cases}
 .
\end{align}

For $(J,L)=(0,1)$, 
\begin{align}
 \lambda(x)  
=\frac{8}{x^2} + \frac{8\kappa^2 x^2}{(x^2+s^2)^2} - \frac{8\kappa}{x^2+s^2}
= \begin{cases}
  \frac{8R_a^2}{x^2(x^2+R_a^2)^2} &  $I$, \ $III$ \cr
  \frac{2}{x^2} & $II$ 
  \end{cases}
 .
\end{align}

Comparing the above results, we find,  in order to make the  space--time integral of  $\lambda(x)$ in each region as small as possible, that $(J,L)=(1,0)$ is selected for small $x$, i.e., the region I, $(J,L)=(0,1)$ is for large $x$, i.e., the region III, while in the intermediate region II, $(J,L)=(1/2,1/2)$ can give the smallest value of $\lambda(x)$. 
The result is summarized as
\begin{align}
 \lambda(x) 
= \begin{cases}
  \frac{8x^2}{(x^2+R_1^2)^2}  &  $I$:  (J,L)=(1,0), \  {n}_A(x)=Y^A_{(1,0)}=\text{const.} \cr
  \frac{2(\hat{b} \cdot x)^2}{x^2[x^2-(\hat{b} \cdot x)^2]} & $II$:  (J,L)=(\frac12,\frac12), \  {n}_A(x) \simeq Y^A_{(1/2,1/2)}=\text{hedgehog} \cr
   \frac{8R_2^2}{x^2(x^2+R_2^2)^2} & $III$:  (J,L)=(0,1), \  {n}_A(x)=Y^A_{(0,1)}(x)=\text{Hopf}
  \end{cases}
 .
 \label{two-meron-lambda}
\end{align}

%%%%%%%%%%%%%%%%%%%%%%%%%%%%%%%%%%%%%%%%%%%%%%%%%%%%%%%%%%%%%
%%%%%%%%%%%%%%%%%%%%%%%%%%%%%%%%%%%%%%%%%%%%%%%%%%%%%%%%%%%%%
\begin{figure}[ptb]
\begin{center}
\includegraphics[scale=0.7]{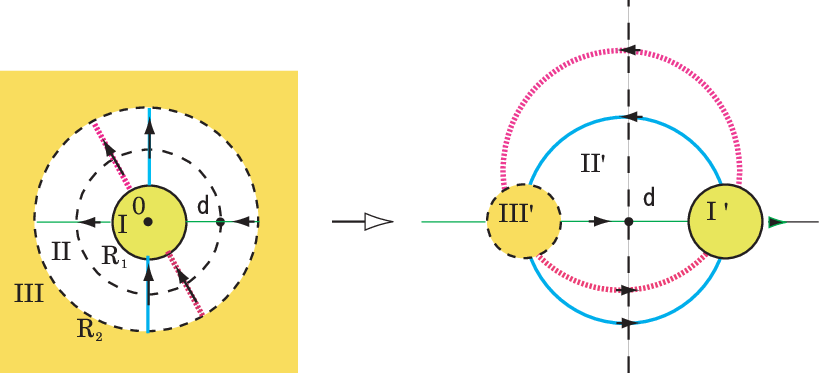}
\end{center} 
 \caption[]{
A magnetic monopole line in II connecting I and III along the direction of $\hat{b}_\mu$ in a smeared meron (left panel) is transformed to a circular magnetic monopole loop in II$'$ connecting I$'$ and III$'$ connecting two merons. 
The magnetic monopole world line going through the center of inversion $d$ in (II) (left panel) is inverted to the  straight line connecting two merons (right panel), which is to be understood as the limit of the circle with infinite radius.  
Here the conformal transformation  maps a sphere to another sphere and preserves the angle between two vectors. 
The angle between two magnetic monopole lines in the left panel is preserved after the transformation in the right panel.  This is also the case for the angles between magnetic lines and concentric spheres in II.
}
\label{C00-fig:meron-monopole}
\end{figure}
%%%%%%%%%%%%%%%%%%%%%%%%%%%%%%%%%%%%%%%%%%%%%%%%%%%%%%%%%%%%%
%%%%%%%%%%%%%%%%%%%%%%%%%%%%%%%%%%%%%%%%%%%%%%%%%%%%%%%%%%%%%

As we have already shown in the previous section, the  magnetic current exists only for $(J,L)=(1/2,1/2)$.  Therefore, in the smeared meron pair configuration, the magnetic current flows only in the region II, while there is no magnetic current in regions I and III.
See the left panel of Fig.~\ref{C00-fig:meron-monopole}. 
For the magnetic current parallel to $d_\mu$, the transformed magnetic current comes from the infinity, goes through two merons and  goes away to infinity, constituting the   straight line, see the right panel of Fig.~\ref{C00-fig:meron-monopole}.
For the magnetic current orthogonal to   $d_\mu$ flowing from $\partial \rm I$ ($\partial \rm III$) to $\partial \rm III$ ($\partial \rm I$), the transformed magnetic current draws a piece of a circle beginning at $\partial \rm I'$ ($\partial \rm III'$) and ending at $\partial \rm III'$ ($\partial \rm I'$).  
Every magnetic current flowing in II is transformed to a circular magnetic monopole loop connecting $\rm I'$ and $\rm III'$.
See Fig.~\ref{C00-fig:meron-monopole}. 
This is easily understood by considering intersections between magnetic lines and concentric spheres in II  from a fact that the conformal transformation  maps a sphere to another sphere and preserves the angle between two vectors.

Note that $\lambda(x)$ obtained in (\ref{two-meron-lambda}) is always finite. 
In addition, due to the rapid decrease of $\lambda(x)$, $\lambda(x) \sim O(x^{-6})$  in region III  
and the asymptotic behavior $\lambda(x) \sim O(x^{2})$  in region I, the reduction functional becomes finite:
\begin{align}
 F_{\rm rc} = \int_{\mathbb{R}^4} d^4x \lambda(x) < \infty
  \quad \text{for $R_1, R_2 >0$} ,
\end{align}
as far as $R_1, R_2 > 0$.
Therefore, this is an allowed solution. 
Thus we have obtained  circular magnetic monopole loops with a non-zero radius $r \ge d/2$  joining a meron pair separated by a distance $d$.

\begin{figure}[tb]
\begin{center}
\includegraphics[scale=0.65]{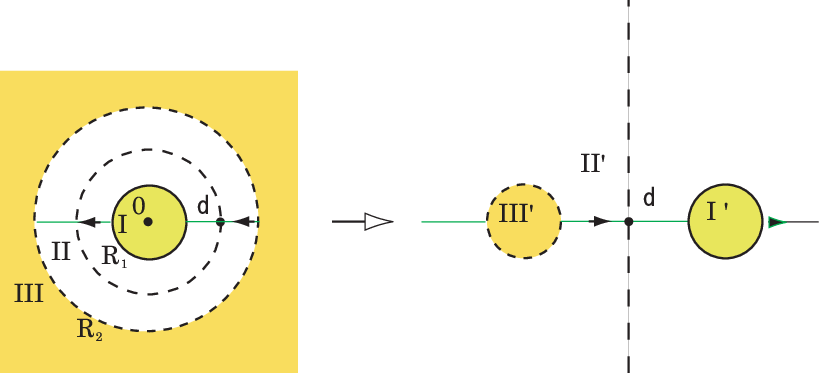}
\quad
\includegraphics[scale=0.65]{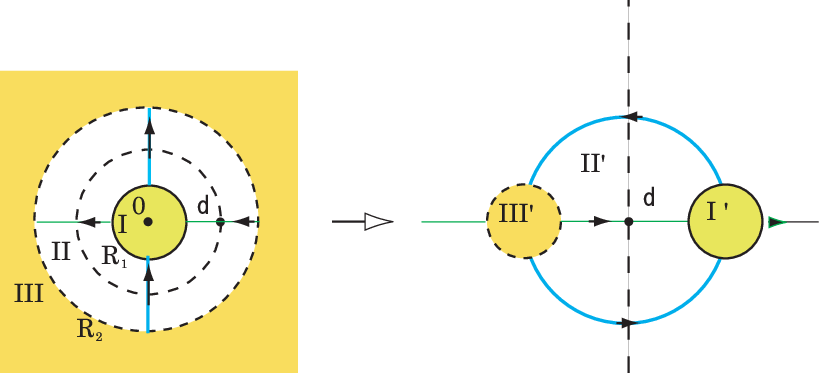}
\quad
\includegraphics[scale=0.65]{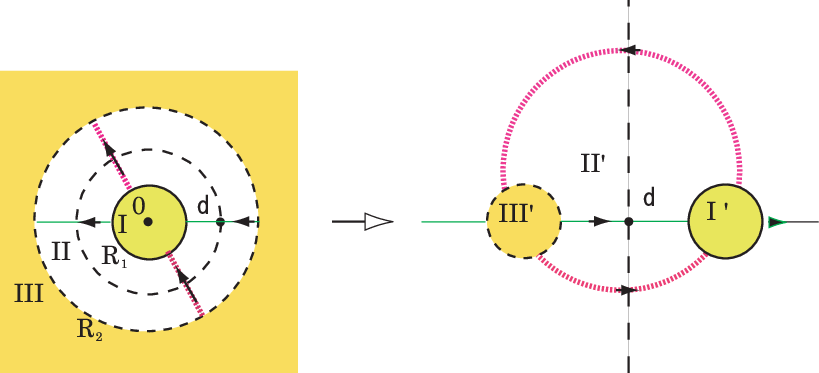}
\end{center} 
 \caption[]{
 Conformal transformations to obtain the closed loops. 
}
\label{C00-fig:meron-monopole2}
\end{figure}

It should be remarked that the reduction functional (\ref{Rc}) is conformal invariant. 
Therefore the color field in the region (II$'$) for the meron pair is obtained  by using  the conformal transformation  (\ref{conformal-transformation}) and a subsequent singular gauge transformation $U(y_{+}=x+a)$ (\ref{gauge-transformation}):
\begin{align}
\bar{\bm{n}}(x)_{\rm II'} 
= \frac{2a^2}{(x+a)^2}
%\propto 
\hat{b}_\nu \eta^A_{\mu\nu} z_\mu U^{-1}(x+a)  \sigma_A U(x+a)/\sqrt{ z^2 -(\hat{b} \cdot z)^2}
 ,
\end{align}
where $z$ is given by (\ref{conformal-transformation}) and $y_+$ is the same as that in (\ref{gauge-transformation}): 
\begin{equation}
z_\mu =  2a^2 \frac{(x+a)_\mu}{(x+a)^2} - a_\mu , 
\quad 
 U(x+a) = \frac{\bar{e}_\alpha (x+a)_\alpha}{\sqrt{(x+a)^2}}
  ,
 \nonumber
\end{equation}
The color field in the region (III$'$) for the meron pair  can be obtained by applying the conformal transformation (\ref{conformal-transformation}) and the gauge transformation (\ref{gauge-transformation}) to the standard Hopf map.
% as
%\begin{align}
% {n}(x)_{III'}  
%= U^{-1}(y_{+}) a_B  Y^A_{(0,1),(B)}(z) \sigma_A U(y_{+})
%= a_B U(y_{+}) z_\alpha e_\alpha \sigma_B z_\beta \bar{e}_\beta U(y_{+})/z^2 
% .
%\end{align}
The color field  in the region (I$'$) for the meron-meron  is trivial. 
%\begin{align}
% {n}(x)_{I'}  
%=  
% .
%\end{align}
See Fig.~\ref{C00-fig:meron-monopole2}.

%%%%%%%%%%%%%%%%%%%%%%%%%%%%%%%%%%%%%%%%%%%%%%%%%%%%%%%%%%%%%
%%%%%%%%%%%%%%%%%%%%%%%%%%%%%%%%%%%%%%%%%%%%%%%%%%%%%%%%%%%%%

The location of the magnetic monopole is dictated by the simultaneous zeros of $\hat{b}_\nu \eta^A_{\mu\nu} z_\mu$ for $A=1,2,3$:
\begin{align}
 0 
%= b_\nu \eta^A_{\mu\nu} z_\mu
 = \hat{b}_\nu \eta^A_{\mu\nu} [2a^2(x_\mu+a_\mu)-(x+a)^2 a_\mu ]
 \quad (A=1,2,3)
 ,
 \label{zeros}
\end{align}
since the gauge transformation $U(y_{+})$ does not change the zeros. 
Without loss of generality, we can fix the direction of connecting two merons as
$a_\mu:=d_\mu/2=\delta_{\mu4}T$. 
For $a_\mu =\delta_{\mu4}T$,  
$
\hat{b}_\nu \eta^A_{\mu\nu}a_\mu
= \epsilon_{Ajk}\hat{b}_k a_j +a_A \hat{b}_4-\hat{b}_A a_4 =-\hat{b}_AT
$ 
and (\ref{zeros}) reads
\begin{align}
 \hat{b}_A x^2  + 2T \hat{b}_k \epsilon_{Ajk}   x_j  + 2T\hat{b}_4  x_A  - \hat{b}_A T^2 = 0 
\quad (A=1,2,3)
 .
 \label{zeros-2}
\end{align}

It is instructive to see two special cases. 
If $\hat{b}_\mu$ is parallel to $a_\mu$, i.e., $\hat{b}_\mu=\delta_{\mu 4}$ (or $\hat{\bm{b}}=\bm{0}$), we find from (\ref{zeros-2}) that the simultaneous zeros are given by 
\begin{align}
 x_A=0 \ ( A=1,2,3 ) ,
\end{align}
i.e., the magnetic current is located on the $x_4$ axis which is parallel to $a_\mu$.   The magnetic-monopole current denotes a straight line going through two merons at $(\mathbf{0}, \pm T)$. 
This straight line can be identified with the maximal circle with infinite radius in the general case discussed below. 
See a horizontal line in the right panel of Fig.~\ref{C00-fig:meron-monopole}.

If $\hat{b}_\mu$ is perpendicular to $a_\mu$ (or $\hat{b}_\mu=\delta_{\mu \ell} \hat{b}_\ell$, $\ell=1,2,3$), i.e., $ \hat{b}_4=0$,  
the simultaneous zeros are obtained on a circle 
\begin{align}
% \bm{x} \parallel \hat{\bm{b}} \quad \& \quad 
 x_\ell^2 + x_4^2 = T^2  
 .
\end{align} 
In this case, the circular magnetic--monopole loop has its center at the origin $0$ in $z$ space and the radius $T=\sqrt{a^2}$ joining two merons at $(\mathbf{0}, \pm T)$ on the plane spanned by $a_\mu$ and $\hat{b}_\ell$ ($\ell=1,2,3$). 
See a minimal circle in the right panel of  Fig.~\ref{C00-fig:meron-monopole}.

In general, it is not difficult to show that the simultaneous zeros are given for $\hat{b}_\mu=(\hat{\bm{b}},\hat{b}_4)$ by   
\begin{align}
 \bm{x} \times \hat{\bm{b}} = \mathbf{0}
 \quad \& \quad 
 \left( \bm{x} + T \frac{\hat{b}_4}{|\hat{\bm{b}}|} \frac{\hat{\bm{b}}}{|\hat{\bm{b}}|} \right)^2 + x_4^2 = T^2 \left( 1 +    \frac{\hat{b}_4^2}{|\hat{\bm{b}}|^2}  \right)
 ,
\end{align}
where $\hat{\bm{b}}$ is the three-dimensional part of unit four vector $\hat{b}_\mu$ ($\hat{b}_\mu \hat{b}_\mu=\hat{b}_4^2+|\hat{\bm{b}}|^2= 1)$.
These equations express circular magnetic monopole loops  joining two merons at $\pm a_\mu$ on the plane  specified by $a_\mu$ and $\hat{\bm{b}}$ where a  circle has the center at 
$
\bm{x}=-T \frac{\hat{b}_4}{|\hat{\bm{b}}|} \frac{\hat{\bm{b}}}{|\hat{\bm{b}}|}
=-\sqrt{a^2} \frac{\hat{b}_4}{|\hat{\bm{b}}|} \frac{\hat{\bm{b}}}{|\hat{\bm{b}}|}
$ and $x_4=0$ with the radius $T\sqrt{1 +     \frac{\hat{b}_4^2}{|\hat{\bm{b}}|^2}}=\frac{\sqrt{a^2}}{|\hat{\bm{b}}|}( \ge T)$. 
See a larger circle in the right panel of Fig.~\ref{C00-fig:meron-monopole}.
The horizontal straight line can be identified with the limit of  infinite radius of the circle.

%We can write the coordinate independent form if we use 
%$a \cdot \hat{b}=Tb_4$, $x_4=a \cdot x/\sqrt{a^2}$  and  $a^2=T^2$.
%\begin{align}
% \bm{x} \parallel \hat{\bm{b}} \quad \& \quad 
% \left( \bm{x} +  \frac{a \cdot \hat{b}}{|\hat{\bm{b}}|} \frac{\hat{\bm{b}}}{|\hat{\bm{b}}|} \right)^2 + \frac{(a \cdot x)^2}{a^2} = a^2 \left( 1 +    \frac{(a \cdot \hat{b})^2}{a^2|\hat{\bm{b}}|^2}  \right)
% .
%\end{align}
These equations express circular magnetic monopole loops  
the center at $\bm{x}=- \frac{a \cdot \hat{b}}{|\hat{\bm{b}}|} \frac{\hat{\bm{b}}}{|\hat{\bm{b}}|}$, $x_4=0$ 
with the radius $\sqrt{a^2 +     \frac{(a \cdot \hat{b})^2}{|\hat{\bm{b}}|^2} } ( \ge \sqrt{a^2})$ 
\begin{align}
 \bm{x} \times \hat{\bm{b}} = \mathbf{0}
 \quad \& \quad 
 \left( \bm{x} +  \frac{a \cdot \hat{b}}{|\hat{\bm{b}}|} \frac{\hat{\bm{b}}}{|\hat{\bm{b}}|} \right)^2 + x_4^2 =  \left( a^2 + \frac{(a \cdot \hat{b})^2}{|\hat{\bm{b}}|^2}  \right)
 ,
\end{align}

%%%%%%%%%%%%%%%%%%%%%%%%%%%%%%%%%%%%%%%%%%%%%%%%%
\begin{figure}[ptb]
\begin{center}
\includegraphics[height=5.5cm]{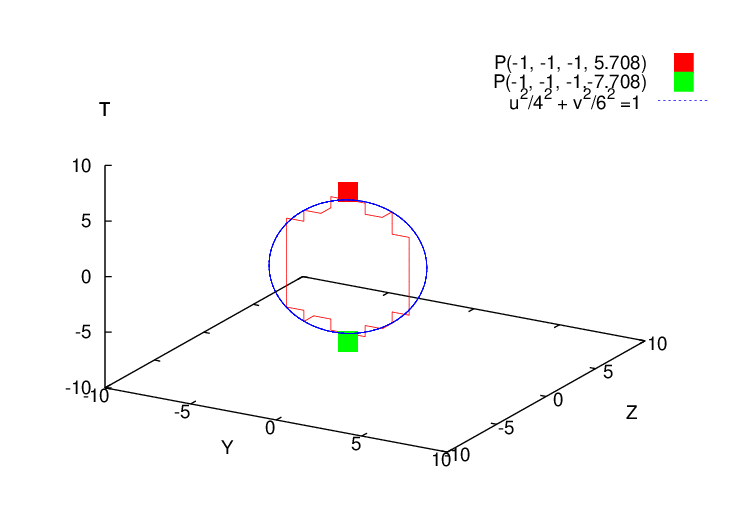}
\quad 
\includegraphics[height=5.5cm]{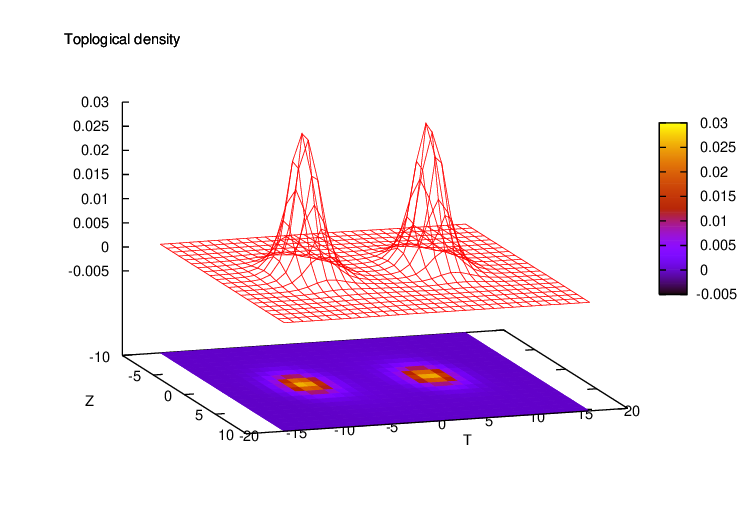}
\end{center}
\vskip -1.0cm
\caption{ Ref.\cite{Shibata-lattice2009}
(Left panel) 
The plot of a magnetic-monopole loop generated by a pair of (smeared) merons in 4-dimensional Euclidean space. The 3-dimensional plot is obtained by projecting the 4-dimensional dual lattice space to the 3-dimensional one, i.e., ($x,y,z,t)$ $\rightarrow$ ($y,z,t$). The positions of two meron sources are described by solid boxes, and the monopole loop by red solid line. 
A circle of blue line is written for guiding eyes. 
%In the lattice of the volume  [-10, 10]$^3 \times$ [-16, 16] with a lattice spacing $\epsilon =1$, the two-merons are located at $( -1, -1, -1, -1 \pm 6.078)$, i.e., equally separated from the origin by the distance  $d = 12.0$  and are smeared with the instanton cap of  size $R = 3.0$ ($d=12$, $R_1=2.833$ and $R_2=50.833$). The monopole loop is confined in the 3-dim. space $x=-1$  and in a 2-dim. plane rotated about $t$-axis by  0.46rad. 
%For guiding the eye, the monopole loop is fitted by an ellipsoid curve (blue  dotted line) with the long radius 6 and the short radius 4.
(Right panel) The plot of the topological charge density for $z-t$ plane  (slice of $x=y=0$). Two peaks of the topological charge density are located at the positions of two merons.
}%
\label{C00-fig:2meron}%
\end{figure}

Finally, we can reproduce the one-instanton case by considering the one-instanton limit  $R_2 \rightarrow R_1$ of the meron pair.
In the one-instanton limit,  the region II$'$ vanishes and the magnetic monopole loop disappears.  This reproduces the previous result \cite{BOT97}  that a circular magnetic monopole loop is shrank to the center of an instanton in one-instanton background field.  
Thus the instanton cannot be the quark confiner which is consistent with the dual superconductivity picture for quark confinement where the magnetic monopole loop must be  the dominant configuration responsible for confinement.

Fig.~\ref{C00-fig:2meron} is a numerical result for a magnetic monopole loop generated by a pair of the smeared merons in 4-dimensional Euclidean lattice \cite{Shibata-lattice2009}. 
For a given  pair of the smeared merons, a circular magnetic-monopole loop joining the two merons is obtained as a numerical solution of the reduction differential equation. 
It is observed that the two sharp peaks of the topological charge density are located at the positions of the two merons. 
The numerical calculation is performed according to the   method explained in the next section. 
This results is consistent with the numerical result of the preceding work by Montero and Negele \cite{MN02}.

%Our analytical solution corresponds to a numerical solution found  on a lattice in \cite{NM02}. 
%See Fig.~\ref{C00-fig:meron-pairs}.
%Montero and J.W. Negele, 
%hep-lat/0202023, 
%Phys. Lett. B533,  322--329 (2002). 

%%%%%%%%%%%%%%%%%%%%%%%%%%%%%%%%%%%%%%%%%%%%%%%%%

\subsection{Reduction condition on a lattice}\label{section:reduction-lattice}

%%%%%%%%%%%%%%%%%%%%%%%%%%%%%%%%%%%%%%%%%%%%%%%%%

We discuss how to carry out the procedures explained in the above in a numerical way. 
For numerical calculations,  we use the lattice regularization where the link variable $U_{x,\mu}$ is related to a gauge field in a continuum theory by
\begin{equation}
 U_{x,\mu}=\text{P}
            \exp
            \left\{ig\int_x^{x+\epsilon\hat{\mu}}dy^\mu \mathscr{A}_\mu(y)\right\} ,
 \label{U}
\end{equation}
where P represents a path-ordered product, $\epsilon$ is a lattice spacing and
$\hat{\mu}$ represents the unit vector in the $\mu$ direction.
The lattice version of the reduction functional for the $SU(2)$ Yang-Mills theory is given by  
\begin{equation}
 F_{\text{red}}[\bm{n},U]
 =\sum_{x,\mu}
  \left\{1-4\,\text{tr}
          \left(U_{x,\mu}\bm{n}_{x+\epsilon\hat{\mu}}U_{x,\mu}^\dagger \bm{n}_x\right)
          /\text{tr}\left({\bf 1}\right)
  \right\},
\label{reduction}
\end{equation}
where ${\bf n}_x$ is a unit color direction field  on a site $x$,
\begin{equation}
 \bm{n}_x=n_x^AT^A,\quad n^A_xn^A_x=1 .
 \label{lattice-n}
\end{equation}
We introduce the Lagrange multiplier $\lambda_x$ to incorporate the constraint of unit length for the color field (\ref{lattice-n}). 
Then the stationary condition for the reduction functional is given by
\begin{equation}
  \frac{\partial }{\partial n_x^A} \left\{ F_{\text{red}}[\bm{n},U] - \frac12  \sum_{x} \lambda_x (n_x^A n_x^A -1) \right\}  =0 .
\end{equation}
When $F_{\text{red}}$ takes a local minimum for a given and fixed configurations $\{U_{x,\mu}\}$, therefore, a Lagrange multiplier $\lambda_x$  satisfies
\begin{equation}
 W_x^A=\lambda_x n_x^A
\label{RDEonLattice1},
\end{equation}
and the color field $n^A_x$ satisfies
\begin{equation}
  n^A_xn^A_x=1 ,
\end{equation}
where
\begin{align}
  W_x^A=4\sum_{\mu=1}^4\text{tr}
         \Big( T^A [U_{x,\mu}\bm{n}_{x+\epsilon\hat{\mu}}U_{x,\mu}^\dagger 
%\notag\\
%&\hspace{2cm}
        +
               U_{x-\epsilon\hat{\mu},\mu}^\dagger \bm{n}_{x-\epsilon\hat{\mu}} U_{x-\epsilon\hat{\mu},\mu} ]
         \Big)/\text{tr}\left({\bf 1}\right)\label{W_x^A}.
\end{align}
Eq.\eqref{RDEonLattice1} is a lattice version of the reduction differential equation (RDE).
We are able to eliminate the Lagrange multiplier to rewrite (\ref{RDEonLattice1}) into
\begin{equation}
 n_x^A=\frac{W^A_x}{\sqrt{W^B_xW^B_x}} .
\label{RDEonLattice2}
\end{equation}
A derivation of this equation is given in Appendix A of \cite{FKSS10}.
The color field configurations $\{ \bm{n}_x\}$  are obtained by solving  \eqref{RDEonLattice2} in a numerical way.

%A magnetic-monopole current on a lattice is constructed from
%$U_{x,\mu}$
%and ${\bf n}_x$ which is a solution of \eqref{RDEonLattice2}
%in the following step.
After obtaining the $\{ \bm{n}_x\}$ configuration for  given configurations $\{U_{x,\mu}\}$ in this way, we introduce a new link variable $V_{x,\mu}$ on a lattice corresponding to the restricted gauge potential $\mathscr{V}_\mu(x)$  by
\begin{align}
 V_{x,\mu}=& \frac{L_{x,\mu}}
                 {\sqrt{\displaystyle{\frac{1}{2}}
                        \ \text{tr}\!
                        \left[L_{x,\mu}L_{x,\mu}^\dagger\right]}} , \quad
%\nonumber\\
 L_{x,\mu}:= U_{x,\mu}+4\bm{n}_xU_{x,\mu}\bm{n}_{x+\epsilon\hat{\mu}} .
\end{align}
Finally, the magnetic-monopole current  $k_{x,\mu}$ on a lattice
is constructed  as 
\begin{equation}
 k_{x,\mu}
 =\sum_{\nu,\rho,\sigma}\frac{\epsilon_{\mu\nu\rho\sigma}}{4\pi}
  \frac{\Theta_{x+\epsilon\hat{\nu},\rho\sigma}[\bm{n},V]
       -\Theta_{x,\rho\sigma}[\bm{n},V]}{\epsilon} ,
 \label{definition_of_k}
\end{equation}
through the angle variable of the plaquette variable:
\begin{align}
  \Theta_{x,\mu\nu}[\bm{n},V]
  =\epsilon^{-2}\arg
  \Big(\text{tr}
        \left\{\left({\bf 1}+2\bm{n}_x\right)
               V_{x,\mu}V_{x+\epsilon\hat{\mu},\nu}
               V_{x+\epsilon\hat{\nu},\mu}^\dagger V_{x,\nu}^\dagger
        \right\}
%\notag\\
%&\hspace{6.2cm}
/\text{tr}\left({\bf 1}\right)
  \Big) .
\end{align}
In this definition, $k_{x,\mu}$ takes an integer value \cite{KKMSSI06,IKKMSS06}.

To obtain the $\{\bm{n}_x\}$ configuration   satisfying \eqref{RDEonLattice2}, we recursively apply \eqref{RDEonLattice2} to $\bm{n}_x$ on each site $x$
and update it keeping $\bm{n}_x$ fixed at a boundary $\partial V$ of a finite lattice $V$
until $F_{\text{red}}$ converges.
Since we calculate the $\{k_{x,\mu}\}$ configuration for the instanton
configuration, we need to decide a boundary condition of the $\{\bm{n}_x\}$ configuration in the instanton case.
We recall that the instanton configuration
approaches  a pure gauge at infinity:
\begin{equation}
 g\mathscr{A}_\mu(x)\rightarrow ih^\dagger(x)\partial_\mu h(x) + O(|x|^{-2}) .
\label{BehaviorOfA}
\end{equation}
%To be more accurate, an instanton turn pure gauge when we neglect $O(1/x^2)$.???
Then,  $\bm{n}(x)$ as a solution of the reduction condition is supposed to behave asymptotically as
\begin{equation}
 \bm{n}(x)\rightarrow h^\dagger(x)T_3 h(x) + O(|x|^{-\alpha}) ,
\label{BehaviorOfn}
\end{equation}
for a certain value of $\alpha>0$.
%at the long distance where we can neglect $O(1/x^2)$.
Under this idea, we adopt a boundary condition:
\begin{equation}
 \bm{n}_x^\text{bound} :=h^\dagger(x)T_3 h(x) , \ x \in \partial V.
\end{equation}
In practice,
we start with an initial state of the $\{\bm{n}_x\}$ configuration:
$\bm{n}_x^\text{init}=h^\dagger(x)T_3 h(x)$ for $x \in V$. Then, we 
repeat updating $\bm{n}_x$ on each site $x$ according to \eqref{RDEonLattice2} except for the configuration $\bm{n}_x^\text{bound}$ on the boundary $\partial V$.

It should be remarked that these asymptotic forms (\ref{BehaviorOfA}) and (\ref{BehaviorOfn}) satisfy the RDE asymptotically in the sense that
\begin{equation}
  D_\mu[\mathscr{A}]{\bf n}(x) \rightarrow 0  \quad (|x| \rightarrow \infty ),
\end{equation}
together with
\begin{equation}
  \lambda(x) \rightarrow 0 \quad (|x| \rightarrow \infty),
\end{equation}
which is necessary to obtain a finite value for the reduction functional \cite{KFSS08}
\begin{equation}
F_{\text{red}} 
= \int d^4x \frac12  \lambda(x) < \infty  .
\end{equation}

%%%%%%%%%%%%%%%%%%%%%%%%%%%%%%%%%%%%%%%%%%%%%%%%%

\subsection{One instantons and magnetic monopole loops on the lattice}

%%%%%%%%%%%%%%%%%%%%%%%%%%%%%%%%%%%%%%%%%%%%%%%%%

The one-instanton solution in the regular (or non-singular) gauge is specified by a constant four-vector representing the center $(b^1,b^2,b^3,b^4) \in \mathbb{R}^4$
and a positive real constant representing the size (width) $\rho \ge 0$:
\begin{equation}
 g\mathscr{A}_\mu(x)= T^A \eta_{\mu\nu}^{A(+)}
                 \frac{2(x^\nu-b^\nu)}{|x-b|^2+\rho^2}, 
\end{equation}
where 
$|x|^2=x_\mu x_\mu$ is the standard Euclidean norm and  $\eta_{\mu\nu}^{A(\pm)}$ is the symbol defined by
\begin{equation}
 \eta_{\mu\nu}^{A(\pm)}
 =\epsilon_{A\mu\nu4}\pm\delta_{A\mu}\delta_{\nu4}\mp\delta_{A\nu}\delta_{\mu4}.
\end{equation}

In this case, we obtain from \eqref{BehaviorOfA} and \eqref{BehaviorOfn}:
\begin{equation}
 h(x)=\frac{x_\mu}{|x|}{e}_\mu,\ 
 \left(e_\mu\equiv(-i\sigma_i,\bm{1})\right) ,
\end{equation}
where $\bm{1}$ is a $2 \times 2$ unit matrix, and
\begin{align}
 &h^\dagger(x)T_3 h(x)=\frac{2\left(x_1x_3-x_2x_4\right)}{x^2}T_1
%\notag\\  &\hspace{3cm}
+\frac{2\left(x_1x_4+x_2x_3\right)}{x^2}T_2
%\notag\\  &\hspace{3cm}
+\frac{-x_1^2-x_2^2+x_3^2+x_4^2}{x^2}T_3.
\end{align}
This exactly agrees with the standard Hopf map \cite{Hopf31}.

The topological charge density is maximal at the point $x=b$ and decreases algebraically with the distance from this point in such a way that the instanton charge $Q_V$ inside the finite lattice $V=[-\epsilon L,\epsilon L]^4$ reproduces  the total instanton charge $Q_P=1$. 
We construct the instanton charge $Q_V$ on a lattice  from the  configuration of link variables $\{U_{x,\mu}\}$ according to 
\begin{align}
 Q_V=& \epsilon^4 \sum_{x\in \{ V-\partial V \} }D_x,
\\
 D_x :=& \frac{1}{2^4}\frac{\epsilon^{-4}}{32\pi^2}
     \sum_{\mu,\nu,\rho,\sigma=\pm1}^{\pm4}
     \hat{\epsilon}^{\mu\nu\rho\sigma} {\rm tr}(\mathbf{1}-U_{x,\mu\nu}U_{x,\rho\sigma}) ,
\\
 U_{x,\mu\nu}=& U_{x,\mu}U_{x+\epsilon\hat{\mu},\nu}
              U_{x+\epsilon\hat{\nu},\mu}^\dagger U_{x,\nu}^\dagger ,
\end{align}
where $D_x$ is a lattice version of the instanton charge density, and
$V-\partial V$ represents the volume without a boundary, and $\hat{\epsilon}$ is related to
the usual $\epsilon$ tensor by
\begin{align}
 &\hat{\epsilon}^{\mu\nu\rho\sigma}
 :=\text{sgn}(\mu)\text{sgn}(\nu)\text{sgn}(\rho)\text{sgn}(\sigma)
  \epsilon^{|\mu||\nu||\rho||\sigma|},
%\notag\\  &\hspace{4cm} 
\quad 
\text{sgn}(\mu):=\frac{\mu}{|\mu|} .
\end{align}

Our interest is the support of $k_{x,\mu}$, namely, a set of links $\{ x,\mu \}$ on which $k_{x,\mu}$ takes non-zero values $k_{x,\mu} \not= 0$.  This expresses the location of the magnetic-monopole current generated for a given instanton configuration. 
By definition \eqref{definition_of_k}, the number of configurations $\{ k_{x,\mu} \}$ are $(2L)^4 \times 4$.
(The number of configurations $\{ k_{x,\mu} \}$ is not equal to $(2L+1)^4\times 4$,
because we cannot calculate $k_{x,\mu}$ at
  positive sides of the boundary $\partial V$ due to the definition of (\ref{definition_of_k}) based on the forward lattice derivative.)

%%%%%%%%%%%%%%%%%%%%%%%%%%%%%%%%%%%%%%%%%%%%%%%%%%%%%%%%%%%%
\begin{table}[htbp]
 \begin{center}
  \begin{tabular}{c||cccc||c}
   $\rho$&$|k_{x,\mu}|>1$&$k_{x,\mu}=-1$&$k_{x,\mu}=0$&$k_{x,\mu}=1$&$Q_V$\\
   \hline
    5&0&4&59105336&4&0.9674\\
   \hline
    10&0&8&59105328&8&0.9804\\
   \hline
    15&0&12&59105320&12&0.9490\\
   \hline
    20&0&12&59105320&12&0.8836
  \end{tabular}
 \end{center}
 \caption{
\cite{FKSS10} The distribution of $k_{x,\mu}$ and the instanton charge $Q_V$ for a given $r$.}
 \label{result_one}
\end{table}
%%%%%%%%%%%%%%%%%%%%%%%%%%%%%%%%%%%%%%%%%%%%%%%%%%%%%%%%%%%%

\begin{figure}[tbp]
%\unitlength=0.001in
%\begin{picture}(7000,4600)(0,0)
%\put(0,2600){\includegraphics[trim=0 0 0 0, width=90mm]%
\includegraphics[scale=0.60]{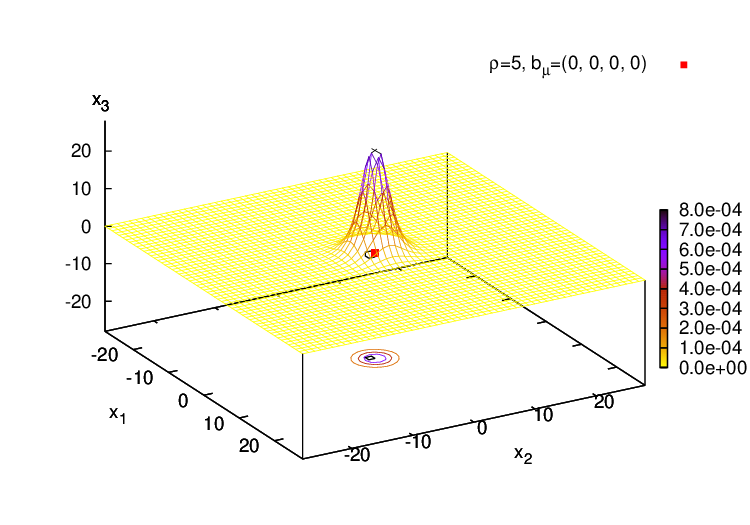}
%}%
%\put(1600,2600){(a)}
%\put(3500,2600){\includegraphics[trim=0 0 0 0, width=90mm]%
\includegraphics[scale=0.60]{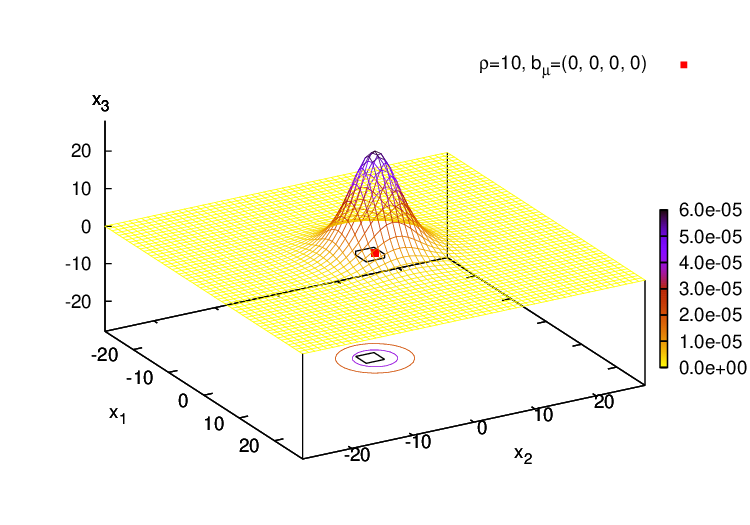}
%}%
%\put(5100,2600){(b)}
%\put(0,150){\includegraphics[trim=0 0 0 0, width=90mm]%
\includegraphics[scale=0.60]{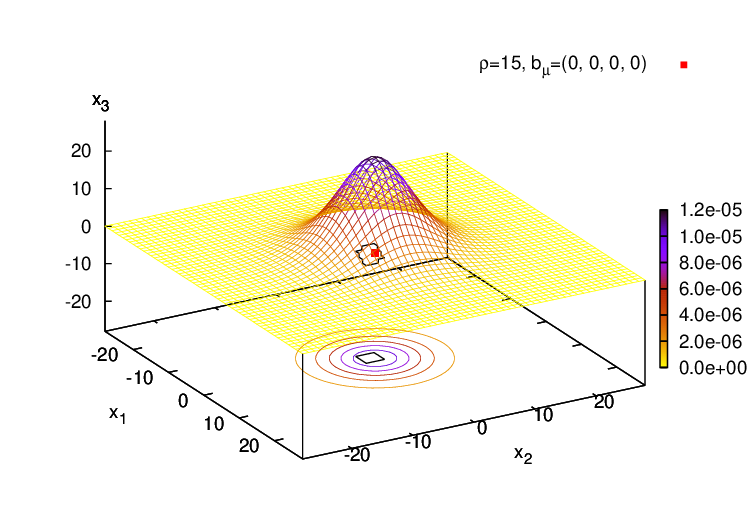}
%}%
%\put(1600,150){(c)}
%\put(3500,150){\includegraphics[trim=0 0 0 0, width=90mm]%
\includegraphics[scale=0.60]{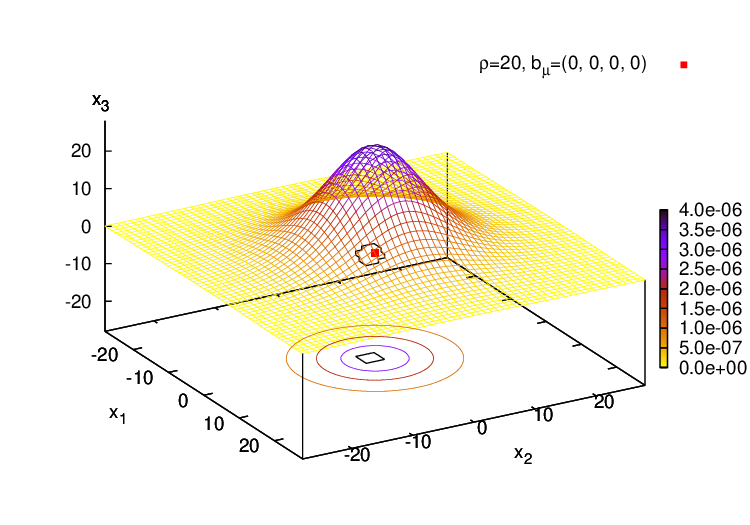}
%}%
%\put(5100,150){(d)}
%\end{picture}
\vskip -0.5cm
 \caption{
Ref.\cite{FKSS10}  One instanton in the regular gauge and the associated magnetic--monopole current $k_{x,\mu}$ for various choice of size parameter $\rho$: (a) $\rho=5\epsilon$, (b) $\rho=10\epsilon$,
          (c) $\rho=15\epsilon$ and (d) $\rho=20\epsilon$.
The grid shows an instanton charge density $D_x$ on $x_1$-$x_2$ ($x_3=x_4=0$) plane.
The black line on the base shows the magnetic monopole loop projected and colored lines shows a contour plot of the instanton charge density.
Figures are drawn in units of the lattice spacing $\epsilon$.
}
 \label{C00-fig:one_loop}
\end{figure}

In our calculations of the magnetic-monopole current configuration, we fix the center on the origin
\begin{equation}
 (b^1,b^2,b^3,b^4)=(0,0,0,0),
\end{equation}
and  change the value of $\rho$.
The results are summarized in Table \ref{result_one} and Fig. \ref{C00-fig:one_loop}.
For a choice of $L=31$, the total number of configurations $\{ k_{x,\mu} \}$ are $62^4\times 4=59105344$.
Although the current $k_{x,\mu}$ is zero on almost all the links $(x,\mu)$, it has  a non-zero value $|k_{x,\mu}|=1$ on a small number of links, e.g., 4+4 links for $\rho=5\epsilon$.  
The number of links with $k_{x,\mu}=+1$ is equal to one with $k_{x,\mu}=-1$, which reflects the fact that the current $k_{x,\mu}$ draws a closed path of links.  
It turns out that there are no configurations such that $|k_{x,\mu}|>1$.

%%%%%%%%%%%%%%%%%%%%%%%%%%%%%%%%%%%%%%%%%%%%%%%%%%%%%%%%%%%%
%%%%%%%%%%%%%%%%%%%%%%%%%%%%%%%%%%%%%%%%%%%%%%%%%%%%%%%%%%%%

We see how $Q_V$ reproduces the total instanton charge $Q_P=1$ for the choice of $\rho$.
The instanton charge density is equal to zero when $\mathscr{A}_\mu(x)$ is a pure gauge, i.e., $\mathscr{F}_{\mu\nu}(x)=0$. 
Therefore, the more rapidly $\mathscr{A}_\mu(x)$ converge to a pure gauge, the more precisely $Q_V$ reproduces the proper value $Q_P$ for a given $L$.
For  one-instanton in the regular gauge, clearly, $\mathscr{A}_\mu(x)$ more rapidly converge to a pure gauge for smaller $\rho$.
Indeed, $Q_V$ for $\rho=10\epsilon$ reproduces $Q_P=1$ more precisely than  $\rho=15\epsilon$ and $\rho=20\epsilon$.
On the contrary, $Q_V$ for $\rho=5\epsilon$  reproduces $Q_P=1$  poorly compared with $\rho=10\epsilon$.  This is because the lattice is too coarse (a lattice spacing is not sufficiently small)  to estimate properly the rapid change of the instanton charge distribution concentrated to the neighborhood of the origin for small $\rho$.

If $\mathscr{A}_\mu(x)$ converge rapidly to a pure gauge $ih^\dagger(x)\partial_\mu h(x)$ so that $Q_V$ gives a good approximation for an expected value $Q_P$, 
then the color field  is expected to behave  as
${\bf n}(x) \rightarrow h^\dagger(x)T_3 h(x)$ asymptotically
and our choice of the boundary condition  
${\bf n}_x^\text{bound} :=h^\dagger(x)T_3 h(x)$ at the boundary $x \in \partial V$  is well motivated.

In Fig. \ref{C00-fig:one_loop},  the support of $k_{x,\mu}$ is drawn by projecting the four-dimensional space on the $x_4=0$ hyperplane (3-dimensional space) for the choice of $\rho=5\epsilon, 10\epsilon, 15\epsilon$  and $20\epsilon$.
This figure shows that the non-zero magnetic-monopole current forms a small loop.
The size of the magnetic monopole loop hardly change while $\rho$ increase.
This is an indication that the magnetic monopole loop  for one-instanton solution disappears in the continuum limit of the lattice spacing $\epsilon$ going to zero. 

Thus, the numerical results are consistent with the analytical ones.

In fact, for $N=1$ the 't Hooft ansatz gives the one-instanton in the singular gauge with the asymptotic behavior: 
\begin{equation}
\mathscr{A}_\mu(x) \sim O(|x|^{-3}) \quad |x| \rightarrow \infty ,
\end{equation}
while the one-instanton in the regular gauge exhibits the asymptotic behavior: 
\begin{equation}
\mathscr{A}_\mu(x) \sim O(|x|^{-1}) \quad |x| \rightarrow \infty .
\end{equation}

%%%%%%%%%%%%%%%%%%%%%%%%%%%%%%%%%%%%%%%%%%%%%%%%%

\subsection{Two instantons and magnetic monopole loops  on the lattice}

%%%%%%%%%%%%%%%%%%%%%%%%%%%%%%%%%%%%%%%%%%%%%%%%%

Finally, we examine the two-instanton solution of Jackiw-Nohl-Rebbi (JNR) \cite{JNR77} from the viewpoint raised above. 
In the conventional studies on  quark confinement, the multi-instanton solution of 't Hooft type \cite{tHooft76} has been used extensively to see the interplay between instantons and magnetic monopoles \cite{BS96,HT96,STSM96,BOT97,RT01,BH03}. 
However, the 't Hooft instanton is not the most general instanton solutions except for the one-instanton case in which the 't Hooft one-instanton agrees with the well-known one-instanton solution in the singular gauge. 
In contrast, the JNR two-instanton solution is the most general two-instanton solution with the full collective coordinates (moduli parameters), while the 't Hooft two-instanton solution   is obtained as a special limit of the JNR solution. 
We demonstrate in a numerical way that a circular loop of magnetic current $k$ is generated for a JNR two-instanton solution.  This is not the case for the 't Hooft two-instanton solution.

In addition, we present the configuration of the color field which plays the crucial role in our formulation.  
The implications of this result for quark confinement will be discussed in the final section.% 
\footnote{
Incidentally, the JNR two instanton was used to study the relationship between dyonic instantons as a supertube connecting two parallel D4-branes and the magnetic monopole string loop as the supertube cross-section  in (4+1) dimensional Yang-Mills-Higgs theory \cite{KL03}, since dyonic instantons of 't Hooft type do not show magnetic string  and D4-branes meet on isolated points, instead of some loop.  
These facts became one of the motivations to study the JNR solution from our point of view. 
}

It is known by the ADHM construction \cite{ADHM78} that the $N$-instanton moduli space has dimension $8N$.  
For $N=1$,   8 moduli parameters are interpreted as 4+1+3 degrees of freedom for the position, size and SU(2) orientation (global gauge rotations), respectively.  

For $N=2$, the  Jackiw-Nohl-Rebbi (JNR) instanton \cite{JNR77} is the most general charge 2 instanton as explained below.   
The explicit form of the JNR two-instanton solution is given by 
\begin{align}
 g\mathscr{A}_\mu^{\rm JNR}(x) 
=& - T_A \eta_{\mu\nu}^{A(-)} \partial_\nu \ln \phi_{\rm JNR} , 
%\\
%=& T^A \eta_{\mu\nu}^{A(-)}
%                  \phi_{\rm JNR}^{-1}                   \sum_{r=0}^2\frac{2%\rho_r^2\left(x^\nu-b^\nu_r\right)}                                    {(|x-b_r|^2)^2} , 
\quad 
\phi_{\rm JNR} := \sum_{r=0}^2\frac{\rho_r^2}{|x-b_r|^2} ,\ 
% |x|^2=x_\mu x^\mu
\end{align}
which is
\begin{align}
 g\mathscr{A}_\mu^{\rm JNR}(x) = T_A \eta_{\mu\nu}^{A(-)}
                  \phi_{\rm JNR}^{-1}
                  \sum_{r=0}^2\frac{2\rho_r^2\left(x^\nu-b^\nu_r\right)}               {(|x-b_r|^2)^2} . 
\end{align}
The JNR two-instanton has $4 \times 3 + 3 +3=18$ parameters, which consist of  three pole positions
$b_0:=(b_0^1,b_0^2,b_0^3,b_0^4)$, $b_1:=(b_1^1,b_1^2,b_1^3,b_1^4)$, $b_2:=(b_2^1,b_2^2,b_2^3,b_2^4)$ 
and three scale parameters  $\rho_0,\rho_1,\rho_2$  including the overall $SU(2)$ orientation.
Note that the number of poles ($r=0,1,2$) is one greater than the number of the instanton charge $Q=2$. 
Although the parameter count of the JNR two-instanton appears to exceed the 16 dimensions of the $N=2$ moduli space,  
the JNR two-instanton has precisely the required number of parameters for the $N=2$ general solution. 
In fact, one parameter is reduced by noting that the multiplication of the scale parameter by a constant does not alter the solution, so only the ratios $\rho_r/\rho_0$ ($r=1,2$) are relevant.  
Moreover, one of the degrees of freedom corresponds to a gauge transformation \cite{JNR77,MS07}.

On the other hand, the 't Hooft two-instanton which is more popular and has been used extensively in the preceding investigations is given by 
\begin{align}
  g\mathscr{A}_\mu^\text{'t Hooft}(x) &= 2T_A
                  \eta_{\mu\nu}^{A(-)}
                  \phi_\text{'t Hooft}^{-1}
                  \sum_{r=1}^2\frac{\rho_r^2\left(x^\nu-b^\nu_r\right)} 
      {|x-b_r|^4} , 
\nonumber\\
 \phi_\text{'t Hooft}&  :=1+\sum_{r=1}^2\frac{\rho_r^2}{|x-b_r|^2} .
\end{align} 
The 't Hooft two-instanton has only $4 \times 2 + 2 +3=13$ parameters, which consist of  two pole positions
$(b_1^1,b_1^2,b_1^3,b_1^4),(b_2^1,b_2^2,b_2^3,b_2^4)$ 
and two scale parameters $\rho_1,\rho_2$ including the overall $SU(2)$ orientation.
The 't Hooft two-instanton solution is reproduced from the JNR  two-instanton solution in the limit  $\rho_0=|b_0| \rightarrow \infty$, namely, the location of the first pole $b_0$ is sent to infinity keeping the relation $\rho_0=|b_0|$.

The crucial difference between the JNR and the 't Hooft solutions is the asymptotic behavior.  
The JNR solution goes to zero slowly, while the gauge potential produced by the 't Hooft ansatz tends rapidly to zero at spatial infinity $|x| \rightarrow \infty$.  
In fact, the JNR two-instanton solution has  the asymptotic behavior at spatial infinity: 
\begin{equation}
\mathscr{A}_\mu^{\rm JNR}(x) \sim O(|x|^{-1}) \quad |x| \rightarrow \infty , 
\end{equation}
while  the 't Hooft multi-instanton has the asymptotic behavior at spatial infinity: 
\begin{equation}
\mathscr{A}_\mu^{\rm 't Hooft}(x) \sim O(|x|^{-3}) \quad |x| \rightarrow \infty .
\end{equation}

In the numerical calculations for the JNR solution \cite{FKSS10}, 
we focus on the special case of equating three size parameters: 
%$\rho_0= \rho_1=\rho_2 \equiv \rho$.  
\begin{align}
 \rho_0=& \rho_1=\rho_2 \equiv \rho .
\end{align}
Then the ratio is uniquely fixed: $\rho_1/\rho_0=\rho_2/\rho_0=\rho/\rho_0 \equiv 1$, and hence $\rho$ can be set to an arbitrary value  without loss of generality in this class. 
In the numerical calculations,  we put $\rho=3\epsilon$ for the lattice spacing $\epsilon$. 

We put three pole positions
$(b_0^1,b_0^2,b_0^3,b_0^4)$, $(b_1^1,b_1^2,b_1^3,b_1^4)$,
$(b_2^1,b_2^2,b_2^3,b_2^4)$
on the $x_3=x_4=0$ plane,  
so that the three poles are located at the vertices of an  equilateral triangle:
\begin{align}
%\rho_0=& \rho_1=\rho_2 \equiv \rho ,
%\\
 b_0=(b_0^1,b_0^2,b_0^3,b_0^4)=& (r,0,0,0)+\Delta ,
\\
 b_1=(b_1^1,b_1^2,b_1^3,b_1^4)
 =& \left(-\frac{r}{2},\frac{\sqrt{3}}{2}r,0,0\right)+\Delta ,
\\
 b_2=(b_2^1,b_2^2,b_2^3,b_2^4)
 =& \left(-\frac{r}{2},-\frac{\sqrt{3}}{2}r,0,0\right)+\Delta ,
\end{align}
where $\Delta$ is a small parameter introduced to avoid the pole singularities at $x=b_r$. 
In the numerical calculation, we choose
\begin{equation}
 \Delta=(0.1\epsilon,\ 0.1\epsilon,\ 0.1\epsilon,\ 0.1\epsilon) .
\end{equation}
Then we have searched for magnetic-monopole currents by changing the size $r$. 
%We have checked that for a given size of the lattice $L$ there is a suitable range of $\rho$, in which the effect of the discretization is sufficiently small.  
We have adopted $L=31$.

\begin{figure}[tbp]
%\unitlength=0.001in
%\begin{picture}(7000,4600)(0,0)
%\put(0,2600){\includegraphics[trim=0 0 0 0, width=90mm]%
\includegraphics[scale=0.65]{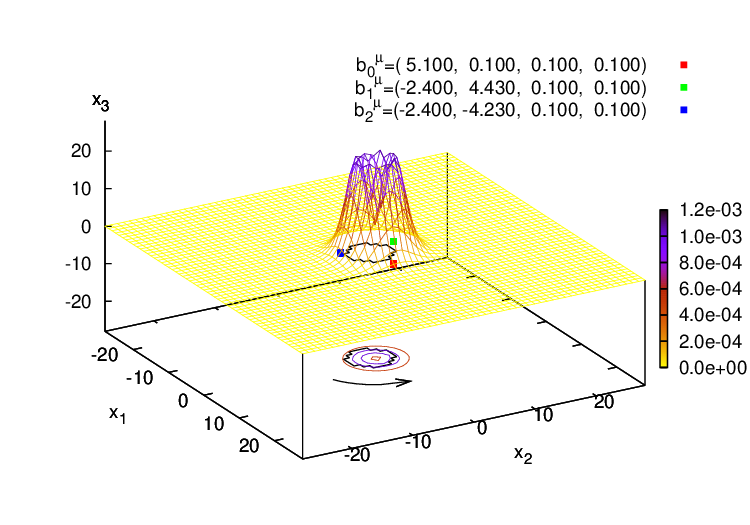}
%}%
%\put(1600,2600){(a)}
%\put(3500,2600){\includegraphics[trim=0 0 0 0, width=90mm]%
\includegraphics[scale=0.65] {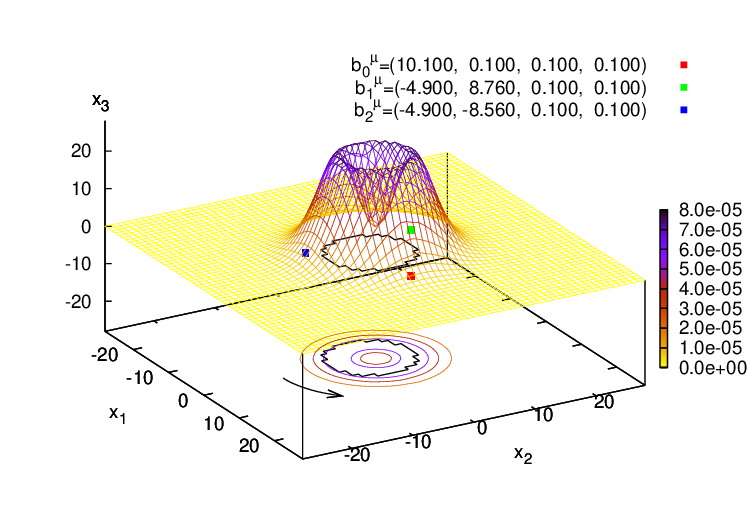}
%}%
%\put(5100,2600){(b)}
%\put(0,150){\includegraphics[trim=0 0 0 0, width=90mm]%
\includegraphics[scale=0.65]{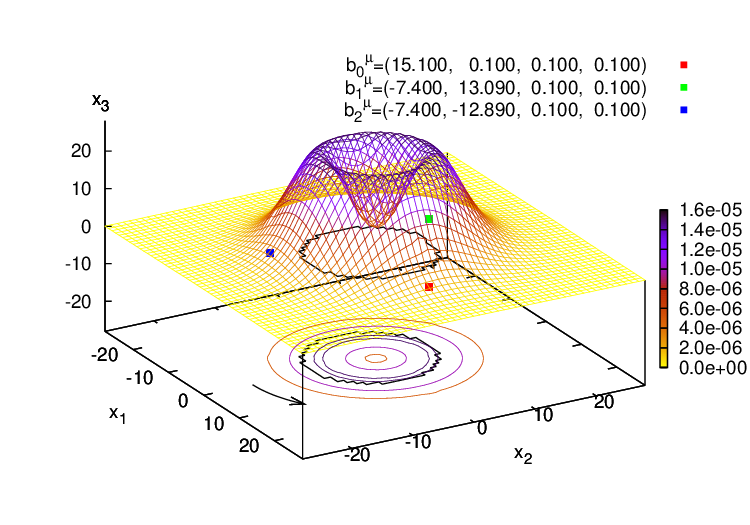}
%}%
%\put(1600,150){(c)}
%\put(3500,150){\includegraphics[trim=0 0 0 0, width=90mm]%
\includegraphics[scale=0.65]{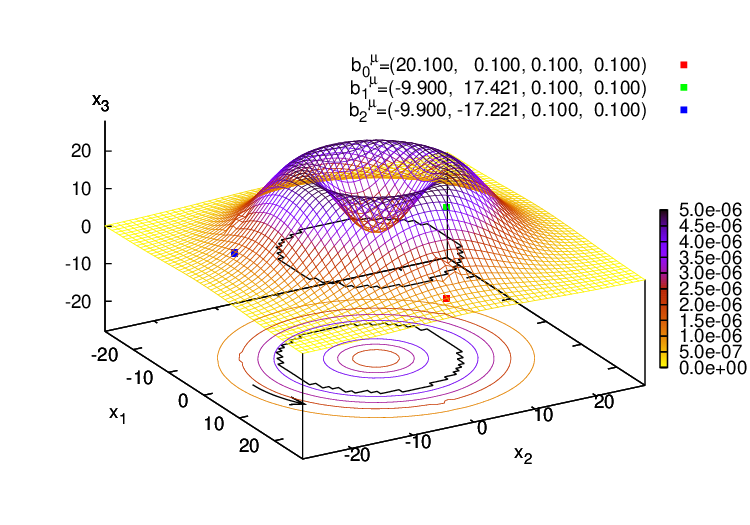}
%}%
%\put(5100,150){(d)}
%\end{picture}
\vskip -0.5cm
\caption{
\cite{FKSS10}  JNR two-instanton and the associated circular loop of the magnetic-monopole current $k_{x,\mu}$.
The JNR two-instanton is defined by fixing three scales   $\rho_0=\rho_1=\rho_2=3\epsilon$ and three pole positions $b_0^\mu,b_1^\mu,b_2^\mu$  which are arranged to be three vertices of an equilateral triangle specified by $r$:   
 (a) $r=5\epsilon$, (b) $r=10\epsilon$, (c) $r=15\epsilon$ and (d) $r=20\epsilon$.
The grid shows an instanton charge density $D_x$ on $x_1$-$x_2$ ($x_3=x_4=0$) plane. %and it is parameterized by the vertical axis.
The associated circular loop of the magnetic-monopole current is located on the same plane as that specified by three poles. 
The black line on the base shows the magnetic monopole loop projected on the $x_1$-$x_2$ plane and the arrow indicates the direction of the magnetic-monopole current, while colored lines on the base show the contour plot for the equi-$D_x$ lines.
Figures are drawn in units of the lattice spacing $\epsilon$.
}
 \label{C00-fig:JNR_loop}
\end{figure}

\begin{figure}[tbp]
%\unitlength=0.001in
%\begin{picture}(7000,3120)(0,0)
%\put(0,50){
%\includegraphics[trim=50 30 0 0, width=110mm]%
\includegraphics[scale=0.75]{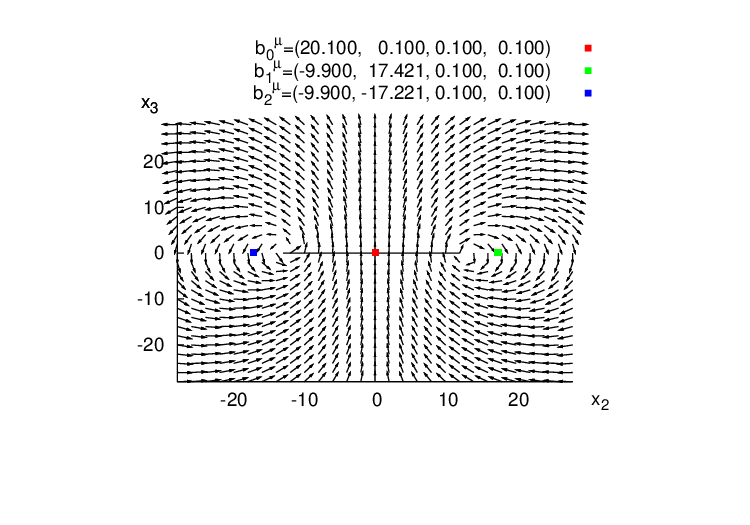}
%}
%\put(1600,10){(a)}
%\put(3500,50){
%\includegraphics[trim=50 30 0 0, width=110mm]%
\includegraphics[scale=0.75]{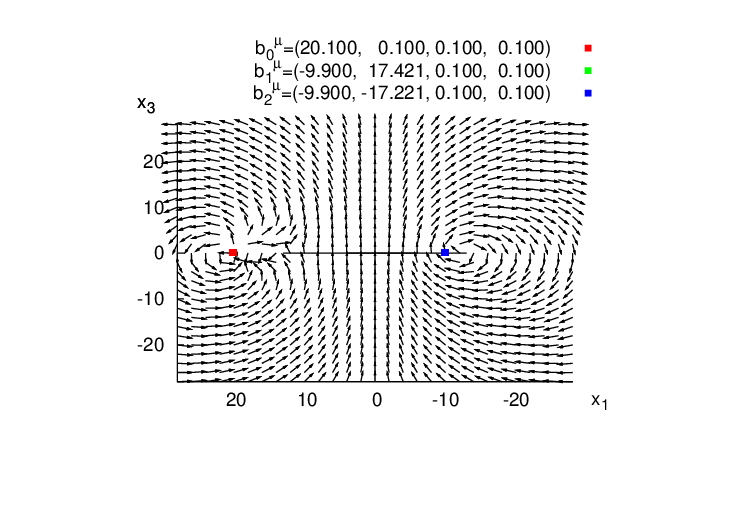}
%}
%\put(5100,10){(b)}
%\end{picture}
\vskip -1.0cm
 \caption{
Ref.\cite{FKSS10}  The configuration of the color field ${\bf n}_x=(n_x^1,n_x^2,n_x^3)$ and a  circular loop of the magnetic-monopole current $k_{x,\mu}$  obtained from the JNR two-instanton solution ($r=20\epsilon$),  viewed in 
(a) the $x_2$-$x_3$ ($x_1=x_4=0$) plane which is off from three poles, and 
(b) the $x_1$-$x_3$ ($x_2=x_4=0$) plane which goes through a pole $b_0^\mu$.
%The JNR solution is specified by $r=20$ and the three pole positions which is projected on the same plane.
The magnetic-monopole current $k_{x,\mu}$  and the three poles of the JNR solution are projected on the same plane.
Here the $SU(2)$ color field $(n_x^1,n_x^2,n_x^3)$ is identified with a unit vector in the three-dimensional space $(x_1,x_2,x_3)$. 
Figures are drawn in units of  the lattice spacing $\epsilon$.
}
 \label{C00-fig:JNR_n}
\end{figure}

The results are summarized in Table \ref{result_JNR}
and Fig. \ref{C00-fig:JNR_loop} and \ref{C00-fig:JNR_n}.
In Fig. \ref{C00-fig:JNR_loop}, as in the case of one-instanton in the regular gauge,
we draw the distribution of the instanton charge density $D_x$ and the support of the magnetic-monopole current $k_{x,\mu}$ projected  on the $x_4=0$ hyperspace for $r=5\epsilon,10\epsilon,15\epsilon,20\epsilon$.

The instanton charge density $D_x$ of the JNR two-instanton takes the maximal value at a circle with radius $R_I$, rather than on the origin,  on the $x_1$-$x_2$ plane.
This is not the case for the 't Hooft two instanton in which the instanton charge distribution concentrates near the two pole positions, as is well known.

We have found that non-vanishing magnetic-monopole currents originating from the JNR two-instanton forms a  circular loop.
The circular loops of the magnetic-monopole current are located on the same plane as that specified by three poles $b_0,b_1,b_2$.
The size of the circular loop, e.g., the radius $R$, increases proportionally as $r$ increases and the circular loops constitute concentric circles with the center at the origin, within the accuracy of our numerical calculations. 
See \cite{FKSS10} for the procedures needed to  reproduce the correct distribution of the instanton charge density $D_x$ in the numerical calculations.

%%%%%%%%%%%%%%%%%%%%%%%%%%%%%%%%%%%%%%%%%%%%%%%%%%%%%%%%%%%%
\begin{table}[tbp]
 \begin{center}
\begin{tabular}{c||cccc||c||c}
  $r$&$|k_{x,\mu}|>1$&$k_{x,\mu}=-1$&$k_{x,\mu}=0$&$k_{x,\mu}=1$&$l/r$&$Q_V$\\
  \hline
   5&0&14&59105316&14&5.6&1.903\\
  \hline
   10&0&26&59105292&26&5.2&1.969\\
  \hline
   15&0&40&59105264&40&5.3&1.950\\
  \hline
   20&0&51&59105242&51&5.1&1.862\\
\end{tabular} 
 \end{center}
 \caption{
Ref.\cite{FKSS10} The distribution of $k_{x,\mu}$ and the instanton charge $Q_V$ for a given size $r$ of the equilateral triangle specifying the JNR solution where  $\ell := \sum_{x,\mu} |k_{x,\mu}|$ is the length of a magnetic-monopole current $k_{x,\mu}$. 
}
 \label{result_JNR}
\end{table}
%%%%%%%%%%%%%%%%%%%%%%%%%%%%%%%%%%%%%%%%%%%%%%%%%%%%%%%%%%%%

%%%%%%%%%%%%%%%%%%%%%%%%%%%%%%%%%%%%%%%%%%%%%%%%%%%%%%%%%%%%
%%%%%%%%%%%%%%%%%%%%%%%%%%%%%%%%%%%%%%%%%%%%%%%%%%%%%%%%%%%%

%%%%%%%%%%%%%%%%%%%%%%%%%%%%%%%%%%%%%%%%%%%%%%%%%%%%%%%%%%%%
%%%%%%%%%%%%%%%%%%%%%%%%%%%%%%%%%%%%%%%%%%%%%%%%%%%%%%%%%%%%

In Table \ref{result_JNR}, we find that the ratio between the length $\ell := \sum_{x,\mu} |k_{x,\mu}|$ of the magnetic-monopole current $k_{x,\mu}$ and the size  $r$ of the equilateral triangle of JNR is nearly constant, 
\begin{equation}
 \ell/r \simeq 5.2  ,
\end{equation}
which  implies  
\begin{equation}
  R_m/r \simeq 0.65  ,
\end{equation}
where we have used a relation $\ell \simeq 8 R_m$ for large $\ell$ since a closed current $k_{x,\mu}$ consists of links on a lattice and the relation $\ell \simeq 2\pi R_m$ in the continuum does not hold. 
Moreover, the magnetic monopole loop passes along the neighborhood of contour giving the absolute maxima of the instanton charge density,  
\begin{equation}
 R_m/R_I \simeq 0.65/0.54 \simeq  1.2 , 
\end{equation}
where $R_I \simeq 0.54r$.

Fig. \ref{C00-fig:JNR_n} shows the relationship between the magnetic monopole loop $k_{x,\mu}$ and the color field ${\bf n}_x$ configuration.
The vector field $\{{\bf n}_x\}$ is winding around the loop,
and it is indeterminate at points where the loop pass.
The configurations of the color field giving the magnetic monopole loop were made available for the first time in this study based on the new reformulation of Yang-Mills theory.

%%%%%%%%%%%%%%%%%%%%%%%%%%%%%%%%%%%%%%%%%%%%%%%%%%%%%%%%%%%%%%%
\begin{figure}[htbp]
%\unitlength=0.001in
%\begin{picture}(7000,9000)(0,0)
%\put(0,6750){\includegraphics[trim=0 0 0 0, width=80mm]%
\includegraphics[scale=0.60]{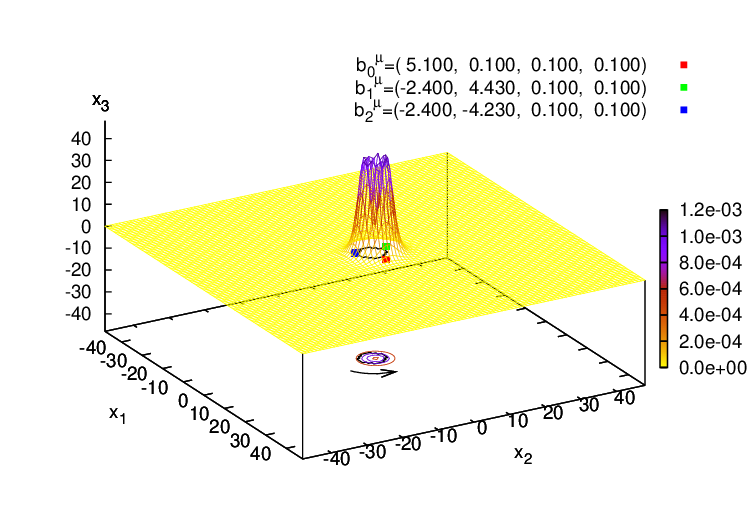}
%}%
%\put(1400,6750){(a)}
%\put(3500,6750){\includegraphics[trim=0 0 0 0, width=80mm]%
\includegraphics[scale=0.60]{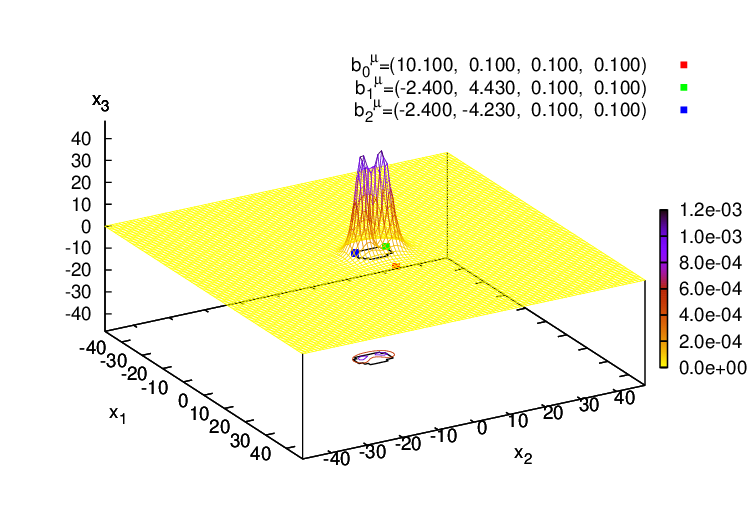}
%}%
%\put(4900,6750){(b)}
%\put(0,4500){\includegraphics[trim=0 0 0 0, width=80mm]%
\includegraphics[scale=0.60]{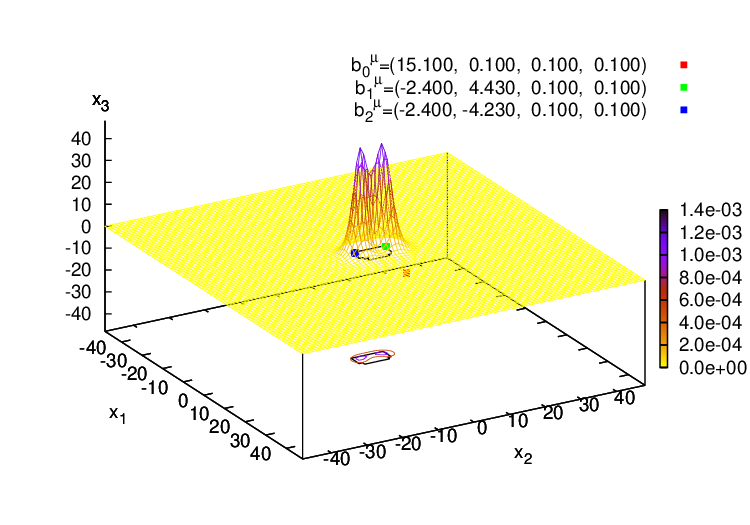}
%}%
%\put(1400,4500){(c)}
%\put(3500,4500){\includegraphics[trim=0 0 0 0, width=80mm]%
\includegraphics[scale=0.60]{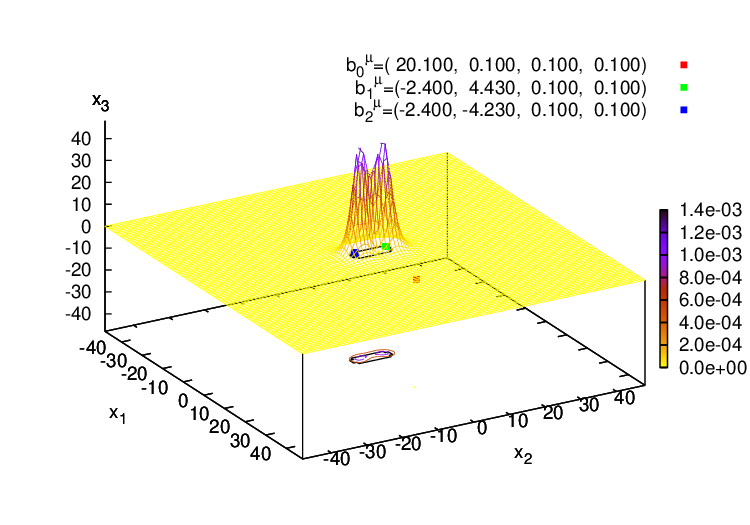}
%}%
%\put(4900,4500){(d)}
%\put(0,2250){\includegraphics[trim=0 0 0 0, width=80mm]%
\includegraphics[scale=0.60]{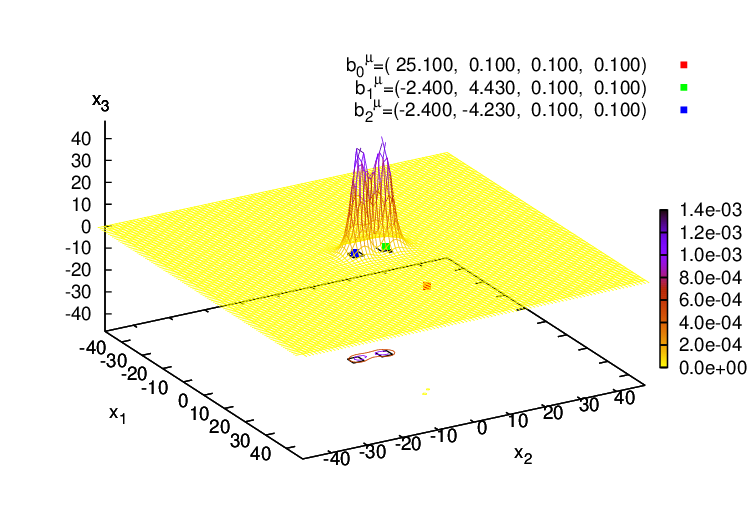}
%}%
%\put(1400,2250){(e)}
%\put(3500,2250){\includegraphics[trim=0 0 0 0, width=80mm]%
\includegraphics[scale=0.60]{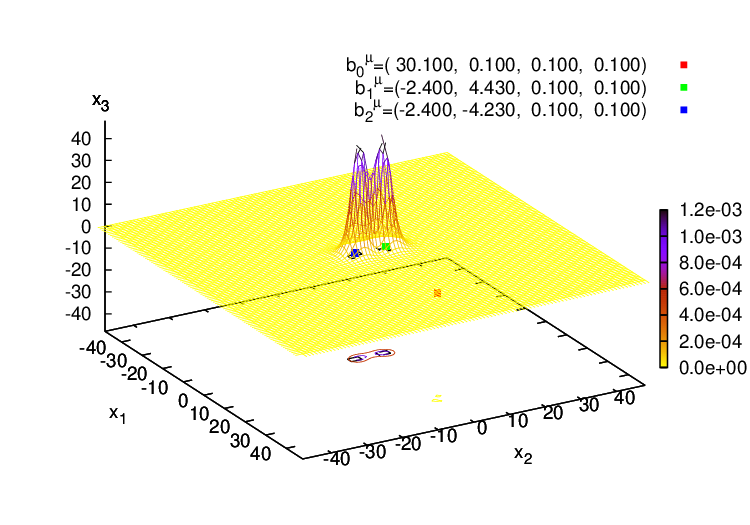}
%}%
%\put(4900,2250){(f)}
%\put(0,0){\includegraphics[trim=0 0 0 0, width=80mm]%
\includegraphics[scale=0.60]{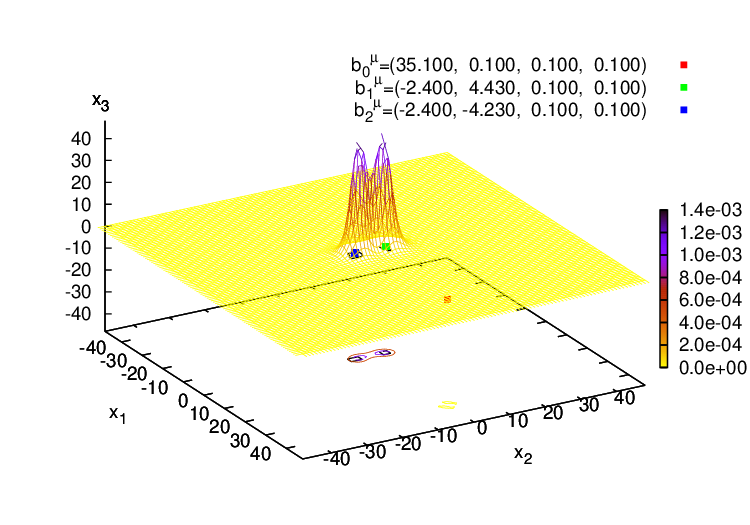}
%}%
%\put(1400,0){(g)}
%\put(3500,0){\includegraphics[trim=0 0 0 0, width=80mm]%
\includegraphics[scale=0.60]{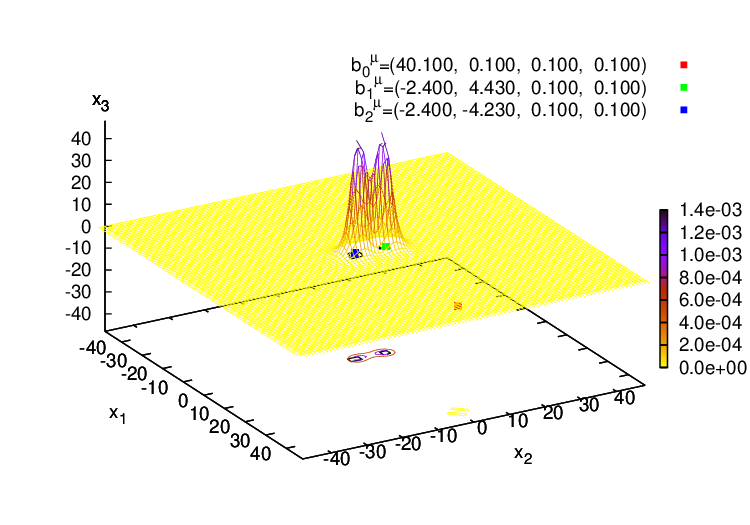}
%}%
%\put(4900,0){(h)}
%\end{picture}
\vskip -0.5cm
\caption{
Ref.\cite{FKSS12}  The magnetic-monopole current $k_{x,\mu}$ generated from the JNR two-instanton
in taking the 't Hooft two-instanton limit.
The parameter  (a) $R=0$, (b) $R=5$, $\dots$, (h)$R=35$ correspond to 
(a) $\rho_0=|b_0|=5$, (b) $10$, $\dots$, (h) $40$.
The grid shows an instanton charge density $D_x$ on $x_1$-$x_2$ ($x_3=x_4=0$) plane.
The black line on the base shows the magnetic monopole loop projected on the $x_1$-$x_2$ plane and the arrow indicates the direction of the magnetic-monopole current, while colored lines on the base show the contour plot for the equi-$D_x$ lines.
Figures are drawn in units of $\epsilon$.
}
 \label{C00-fig:JNRtotHooft_loop}
\end{figure}
%%%%%%%%%%%%%%%%%%%%%%%%%%%%%%%%%%%%%%%%%%%%%%%%%%%%%%%%%%%%%%%

%%%%%%%%%%%%%%%%%%%%%%%%%%%%%%%%%%%%%%%%%%%%%%%%%%%%%%%%%%%%
\begin{figure}[htbp]
 \begin{center}
\includegraphics[scale=0.6]{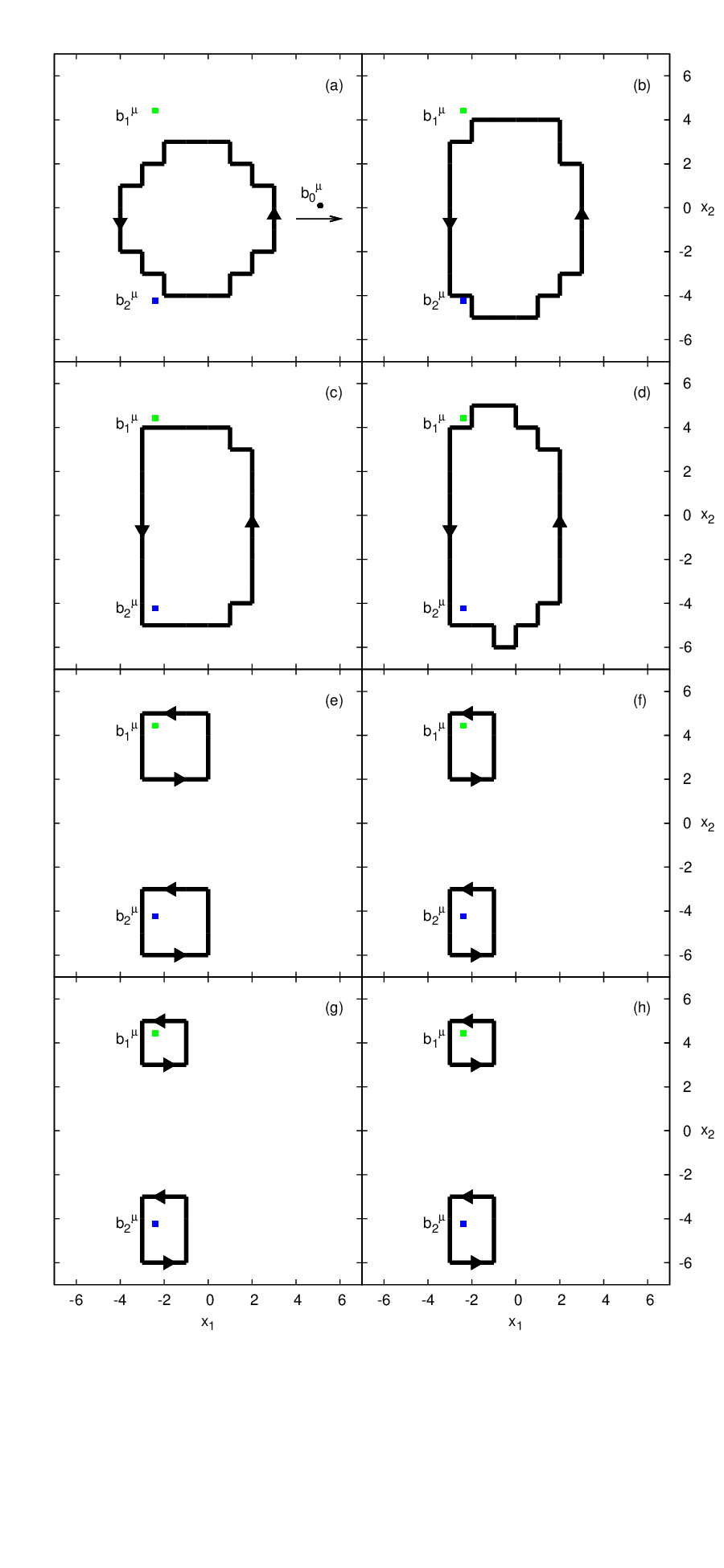}
 \end{center}
% \unitlength=0.001in
% \begin{picture}(7000,9000)(0,0)
%  \put(0,6750){\includegraphics[trim=0 0 0 0, width=80mm]%
%              {figure/JNR_5_above.eps}}%
%  \put(1400,6750){(a)}
%  \put(3500,6750){\includegraphics[trim=0 0 0 0, width=80mm]%
%                 {figure/JNR_10_above.eps}}%
%  \put(4900,6750){(b)}
%  \put(0,4500){\includegraphics[trim=0 0 0 0, width=80mm]%
%           {figure/JNR_15_above.eps}}%
%  \put(1400,4500){(c)}
%  \put(3500,4500){\includegraphics[trim=0 0 0 0, width=80mm]%
%              {figure/JNR_20_above.eps}}%
%  \put(4900,4500){(d)}
%  \put(0,2250){\includegraphics[trim=0 0 0 0, width=80mm]%
%              {figure/JNR_25_above.eps}}%
%  \put(1400,2250){(e)}
%  \put(3500,2250){\includegraphics[trim=0 0 0 0, width=80mm]%
%                 {figure/JNR_30_above.eps}}%
%  \put(4900,2250){(f)}
%  \put(0,0){\includegraphics[trim=0 0 0 0, width=80mm]%
%           {figure/JNR_35_above.eps}}%
%  \put(1400,0){(g)}
%  \put(3500,0){\includegraphics[trim=0 0 0 0, width=80mm]%
%              {figure/JNR_40_above.eps}}%
%  \put(4900,0){(h)}
% \end{picture}
\vskip -3cm
 \caption{
Ref.\cite{FKSS12} Deformation of the magnetic monopole loop viewed from the positive side of $x_3$.
The parameter  (a) $R=0$, (b) $R=5$, $\dots$, (h)$R=35$ correspond to 
(a) $\rho_0=|b_0|=5$, (b) $10$, $\dots$, (h) $40$.
As the pole position $b_0^\mu$ shifts to the positive direction of $x_1$ with fixed $b_1^\mu$,
$b_2^\mu$, the magnetic monopole loop shrinks to fixed poles.
The arrow on a loop shows the direction of the magnetic-monopole current flowing on the loop. 
}
 \label{C00-fig:deformation_loop}
\end{figure}

Finally, we investigate how a magnetic monopole loop generated from the JNR
two-instanton behaves in the course of taking the 't Hooft two-instanton limit.
For this purpose, we set three pole positions and three size parameters to
\begin{align}
 \rho_0=5\epsilon+R&,\quad \rho_1=\rho_2 = 5\epsilon ,
\\
 b_0=(b_0^1,b_0^2,b_0^3,b_0^4)=& (5\epsilon+R,0,0,0)+\Delta ,
\\
 b_1=(b_1^1,b_1^2,b_1^3,b_1^4)
 =& \left(-\frac{5}{2}\epsilon,\frac{5\sqrt{3}}{2}\epsilon,0,0\right)+\Delta ,
\\
 b_2=(b_2^1,b_2^2,b_2^3,b_2^4)
 =& \left(-\frac{5}{2}\epsilon,-\frac{5\sqrt{3}}{2}\epsilon,0,0\right)+\Delta,\\
 \Delta=(0.1\epsilon,&0.1\epsilon,0.1\epsilon,0.1\epsilon)
\end{align}
and calculate  magnetic-monopole current
for various values of $R$: $R=0,5,10,15,20,25,30,35$.
%In the case of the JNR two-instanton, $h(x)$ in Eq.\eqref{BehaviorOfA} is
%\begin{equation}
% h(x)=\frac{x_\mu}{|x|}\bar{e}_\mu,\ 
% \left(\bar{e}_\mu\equiv e_\mu^\dagger=(i\sigma_i,\bm{1})\right)
%\end{equation}
%and a boundary condition for ${\bf n}(x)$ is
%\begin{align}
% &h^\dagger(x)T_3 h(x)=\frac{2\left(x_1x_3+x_2x_4\right)}{x^2}T_1\notag\\
% &\hspace{3cm}+\frac{2\left(-x_1x_4+x_2x_3\right)}{x^2}T_2\notag\\
% &\hspace{3cm}+\frac{-x_1^2-x_2^2+x_3^2+x_4^2}{x^2}T_3.
%\end{align}
Here we have used the same notations as those used in the previous paper \cite{FKSS10}, in which the details on the lattice, instanton discretization and monopole detection are given.

The results are summarized in Fig.~\ref{C00-fig:JNRtotHooft_loop} and
Fig.~\ref{C00-fig:deformation_loop}.
These figures show that a circular monopole loop generated from the JNR two-instanton splits into two smaller loops as $R$ increases.
Eventually, two smaller loops shrink to two fixed poles in the  limit: $\rho_0=|b_0|\rightarrow\infty$.
Such behavior of the magnetic monopole loop is consistent with the result  for the 't Hooft two-instanton obtained in \cite{BOT97}.
Therefore, our result on the JNR two-instanton \cite{FKSS10} does not contradict with the preceding result on the 't Hooft two-instanton \cite{BOT97}.

In Fig.~\ref{C00-fig:deformation_loop}, the separation of a loop into two smaller loops occurs between (d) and (e).
The evolution from (d) to (e) seems to be smooth.  
In order to consider this issue, the direction of magnetic current is indicated by the arrow on the loop in Fig.~\ref{C00-fig:deformation_loop}. 
The directions of magnetic currents of two smaller loops is the same as that of the original loop, which is consistent with the   smoothness of the evolution. 
However, we cannot verify the smoothness by giving the value of the functional to be minimized, since it is difficult to obtain the data at the instance that a loop was just separated into two smaller loops. 
\footnote{
The deformation of the magnetic monopole loop seen in Fig.~\ref{C00-fig:deformation_loop}
agrees very well with the one shown by S. Kim and K. Lee analytically
in $(4+1)$ dimensional supersymmetric Yang-Mills theory \cite{KL03}.
}

%%%%%%%%%%%%%%%%%%%%%%%%%%%%%%%%%%%%%%%%%%%%%%%%%%%%%%%%%%%%
%%%%%%%%%%%%%%%%%%%%%%%%%%%%%%%%%%%%%%%%%%%%%%%%%%%%%%%%%%%%
\subsection{Conclusion and discussion}\label{sec:concldis}
%%%%%%%%%%%%%%%%%%%%%%%%%%%%%%%%%%%%%%%%%%%%%%%%%%%%%%%%%%%%
%%%%%%%%%%%%%%%%%%%%%%%%%%%%%%%%%%%%%%%%%%%%%%%%%%%%%%%%%%%%

For given solutions of the classical  $SU(2)$ Yang-Mills equation in the four-dimensional Euclidean space,  the reduction condition was solved in an analytical and numerical ways to obtain the color field and thereby to define a gauge-invariant magnetic monopole in the new reformulation of the Yang-Mills theory written in terms of the new variables.  
For performing numerical calculations, we have used a lattice regularization \cite{KKMSSI06}.
Then we have constructed the magnetic-monopole current $k_\mu$ on a dual lattice where the resulting magnetic charge is gauge invariant and quantized according to the quantization condition of the Dirac type.

First,  the new reformulation is completed to reproduce some results obtained in the preceding studies based on the conventional methods, i.e.,  specific (partial) gauge fixing procedures called MA gauge,  LA gauge and MC gauge in which topological objects such as Abelian magnetic monopoles and center vortices are regarded as gauge-fixing defects. 
We have obtained the corresponding gauge-invariant results: 
(1) One  instanton configuration cannot support a (gauge-invariant) magnetic monopole loop.
(2) One meron configuration induces a (gauge-invariant)  magnetic-monopole current along a straight line going through the meron. 
(3) However, neither one-instanton nor one-meron supports circular magnetic monopole loops.

Second, it has been shown in an analytical method that there exist circular magnetic monopole loops  supported by a pair of merons smeared  (i.e., ultraviolet regularized) in the sense of Callan, Dashen and Gross \cite{CDG78}, although the corresponding numerical solution has already been found by Montero and Negele \cite{MN02} on a lattice.
Therefore, a meron pair is a first topological object which is found to be consistent with the dual superconductor picture of quark confinement. 
Therefore, the meron pair configurations are candidates for field configurations to be responsible for deriving the area law of the Wilson loop average. 
It will be interesting to see the relationship between this result and \cite{Kondo08,Kondo08b}.
%The detailed analysis will be given in a subsequent paper \cite{Kondo08c} which forms also a relationship  with the recent  papers \cite{Kondo08} and \cite{Kondo08b}.

Third, for the two-instanton solution of the Jackiw-Nohl-Rebbi, we have discovered based on a numerical method that the magnetic-monopole current $k_\mu$ has the support on a circular loop which is located near the maxima of the instanton charge density.  
Thus, we have shown that the two-instanton solution of the Jackiw-Nohl-Rebbi generates the magnetic monopole loop in four-dimensional $SU(2)$ Yang-Mills theory. 
In the same setting, for one instanton the magnetic current has the support only on a plaquette around the center of the one-instanton.  This result confirms that no magnetic monopole loop is generated for one-instanton solution in the continuum limit.

Moreover, it has been investigated numerically how a magnetic monopole loop 
generated from the JNR two-instanton deforms in the course of taking the 't Hooft instanton limit.
In this limit, a circular loop splits into two smaller loops and each loop shrinks to two poles of the 't Hooft instanton. 
This corresponds to the magnetic monopoles demonstrated by  Brower, Orginos and Tan (BOT)  \cite{BOT97} for the 't Hooft instanton. 
As a result, the result that the JNR two-instanton generates a magnetic monopole loop is compatible with the result of BOT \cite{BOT97} that the 't Hooft two-instanton does not generate such a loop of magnetic monopole. 
Thus, there is no contradiction between two results.% 
\footnote{
 A  natural explanation for the difference between the two instantons' monopoles
  will be possible from the fact that in the 't Hooft ansatz the constituents are of same color
  orientation, whereas in the JNR ansatz they are not \cite{JNR77}, and apparently the latter
 is needed to generate a loop of finite size.
 We hope to clarify this issue in future works.  
}

Combining the numerical results with the analytical ones \cite{KFSS08},  we have found that both the JNR two-instanton solution and two-merons solution generate  circular loops of magnetic monopole with the same asymptotic behavior at spatial infinity: 
\begin{equation}
\mathscr{A}_\mu(x) \sim O(|x|^{-1}) \quad |x| \rightarrow \infty , 
\end{equation}
which should be compared with the 't Hooft (multi) instanton with the asymptotic behavior at spatial infinity: 
\begin{equation}
\mathscr{A}_\mu(x) \sim O(|x|^{-3}) \quad |x| \rightarrow \infty .
\end{equation}

It is  expected that these loops of magnetic monopole are responsible for confinement in the dual superconductivity picture. 
This result seems to be consistent with the claim made in \cite{LNT04,Lenz09}.  
However, the correspondence between the instanton and magnetic monopole loop is not one-to-one.  
To draw the final conclusion, we need to collect more data for supporting this claim.
In particular, it is not yet clear which relationship between instanton charge and the magnetic charge holds in our case, as studied in \cite{Reinhardt97,Jahn00,TTF00}.

Moreover, it will be interesting to study the implications of our results to the Faddeev-Niemi model \cite{FNmodel} which is expected to be a low-energy effective theory of Yang-Mills theory, e.g., a relationship between magnetic monopole loop and a Hopfion, i.e., a knot soliton as a topological soliton solution with a non-trivial Hopf index (topological invariant). See e.g., \cite{KOSSM06} for a preliminary result and references therein.

\begin{figure}[ptb]
\begin{center}
%\vspace{-16mm}%
\includegraphics[scale=0.15]{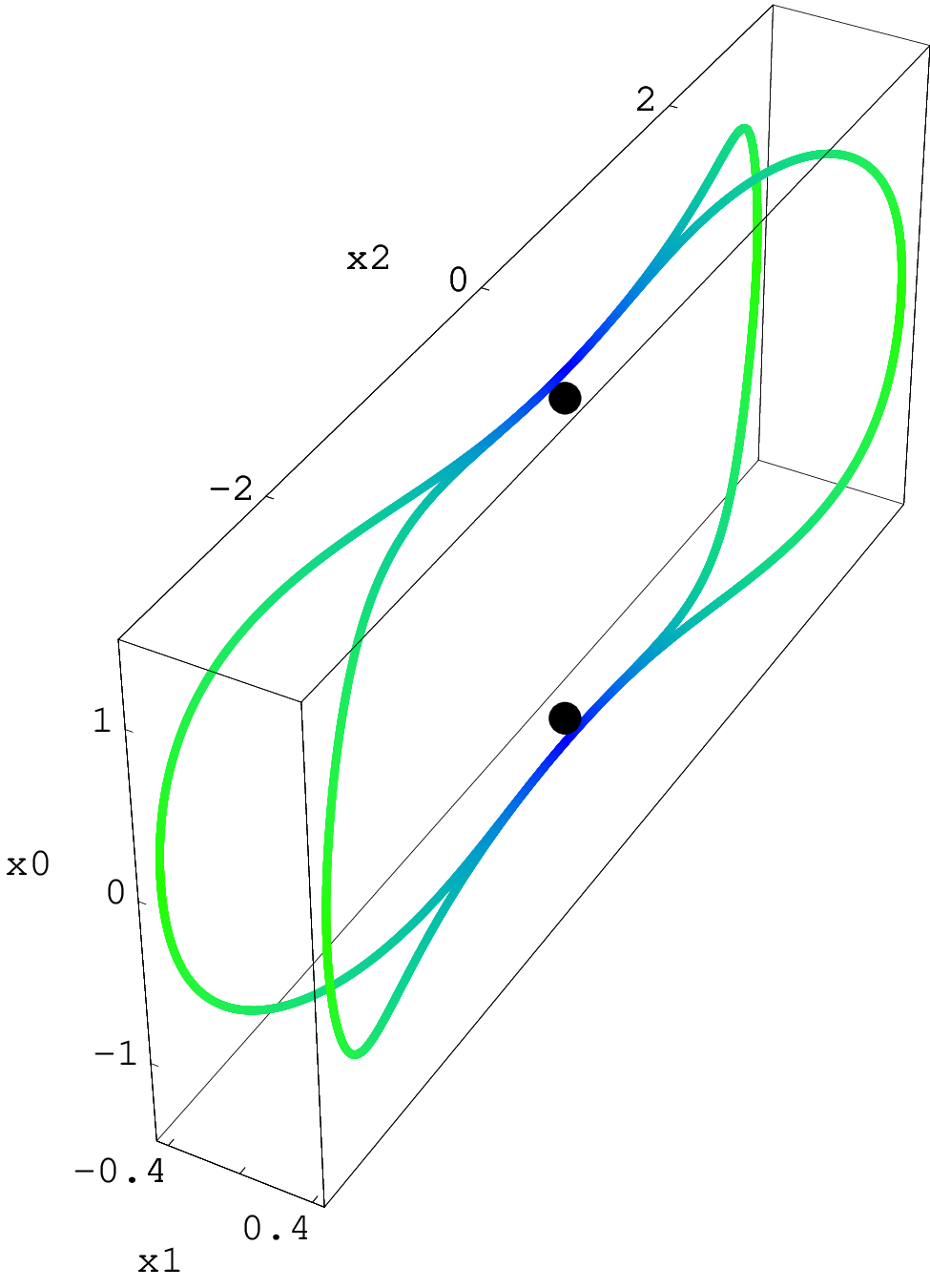}
%\vspace{-8mm}
\end{center}
\vspace{-8mm}
\caption{
Ref.\cite{RT01}
Plot of the two magnetic monopole loops for the gauge potential of 2-instanton %(\ref{inst-gauge-potential}) projected onto the
$x_1-x_2-x_0$-space (dropping the $x_3$-component). 
Rotations with angle $\pi$ around the $x_1$- , $x_2$- and $x_3$-axis interchange the different monopole branches. The thick dots show the positions of the instantons. 
}
\label{C00-fig:2instanton-monopole}%
\end{figure}

\begin{figure}[ptb]
\begin{center}
%\vspace{-16mm}%
%\includegraphics[width=3.0in]{dumpVX.ps}
\includegraphics[scale=0.25]{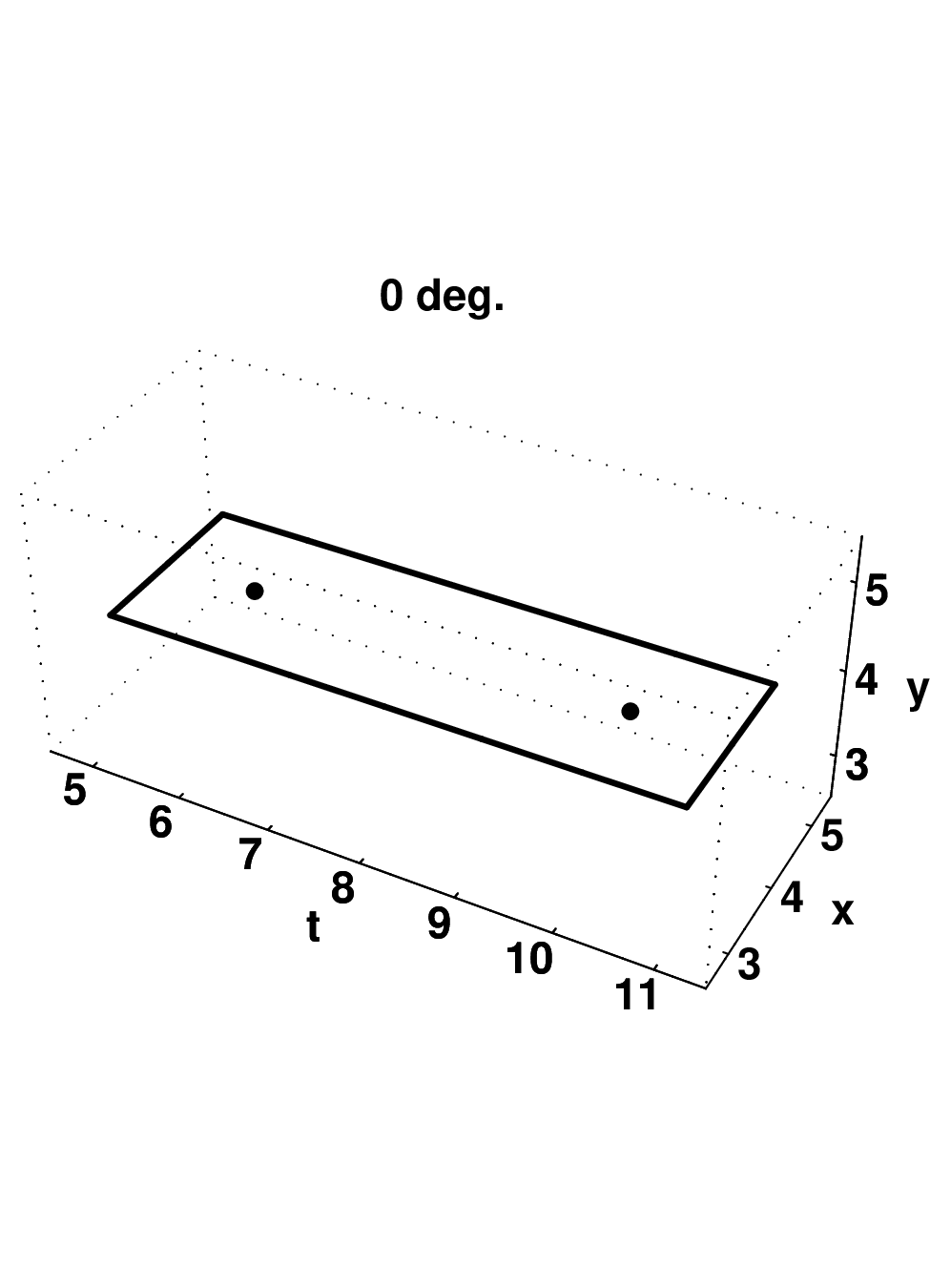}
\includegraphics[scale=0.25]{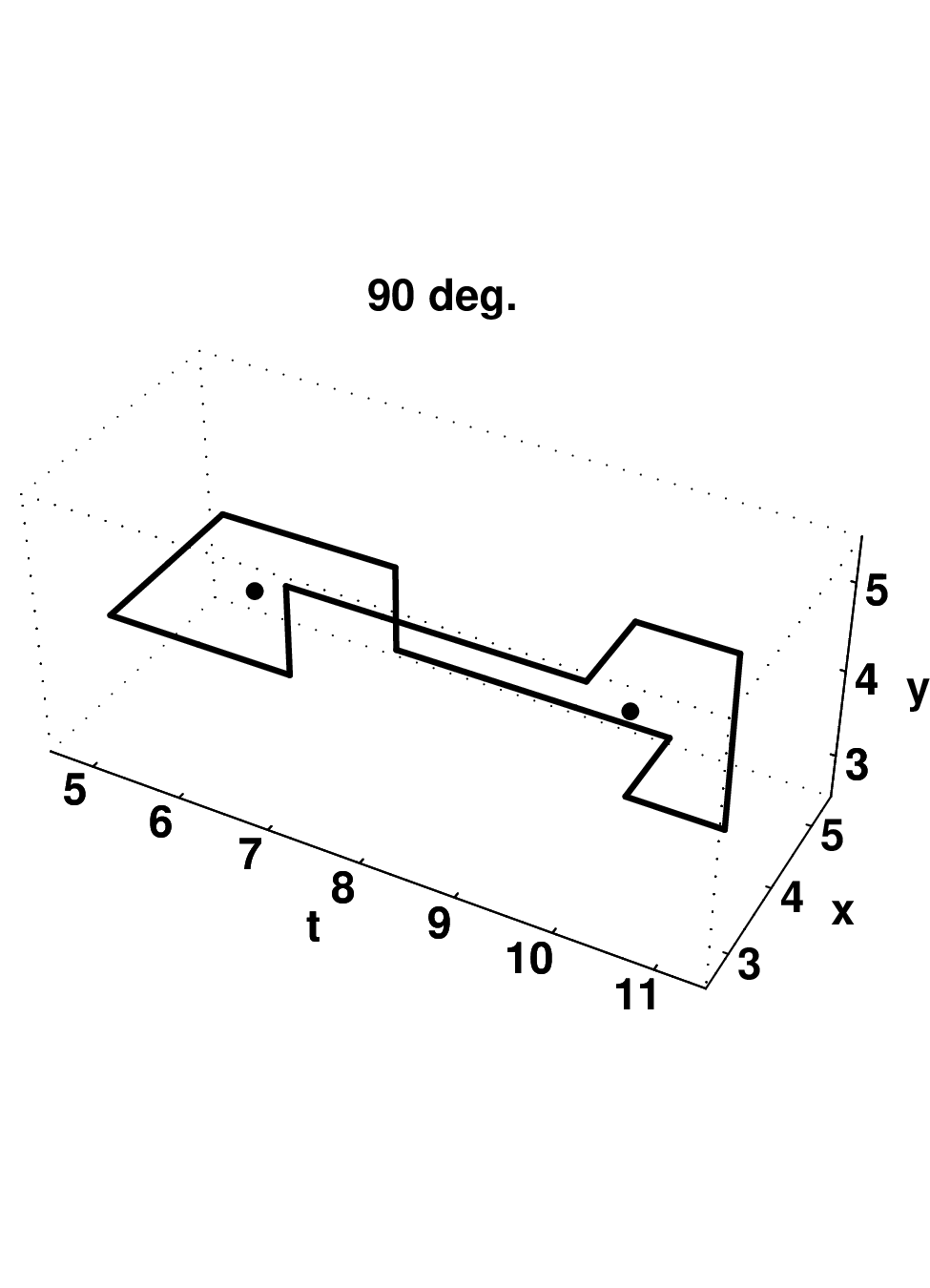}
\includegraphics[scale=0.25]{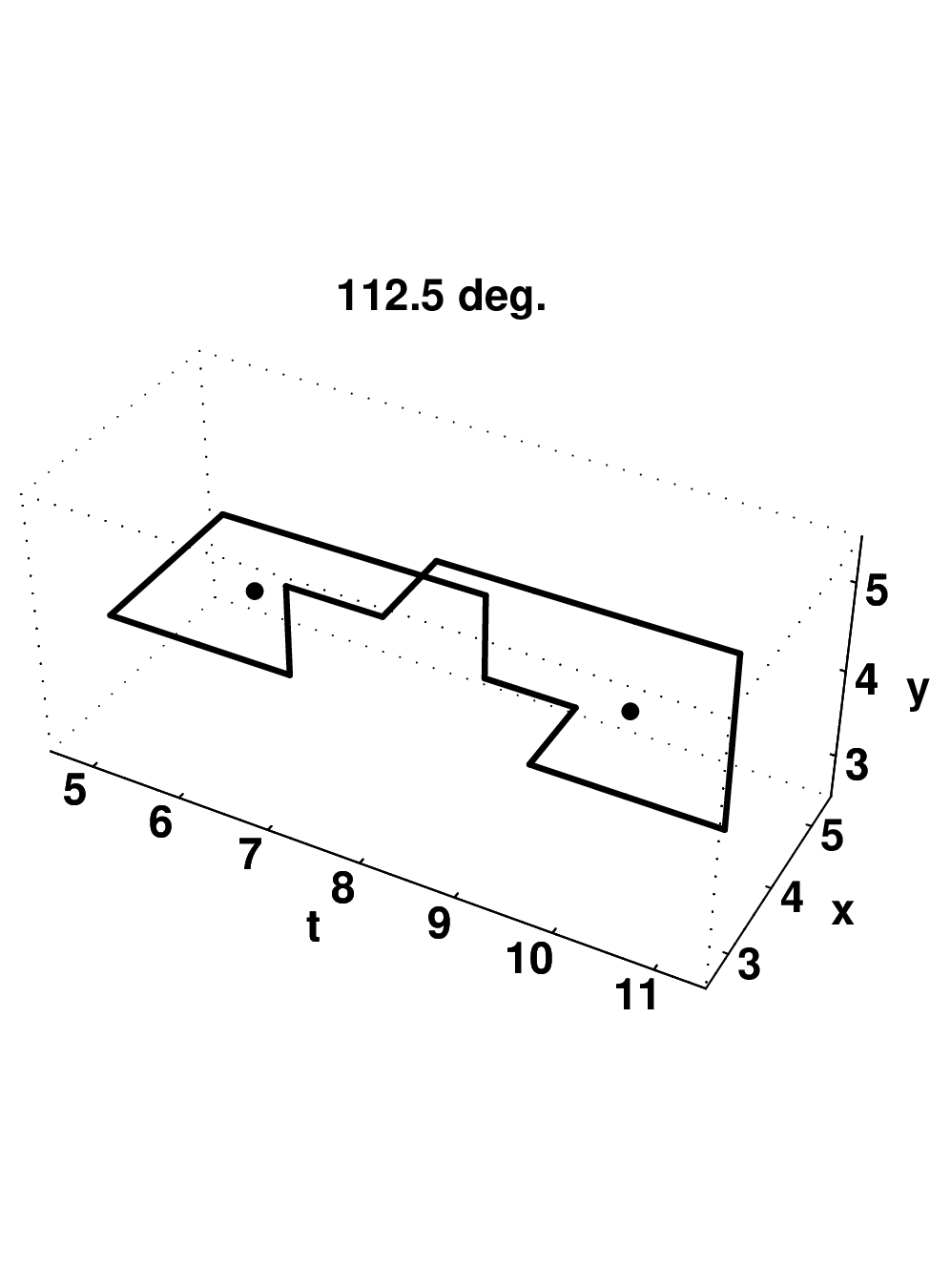}
\includegraphics[scale=0.25]{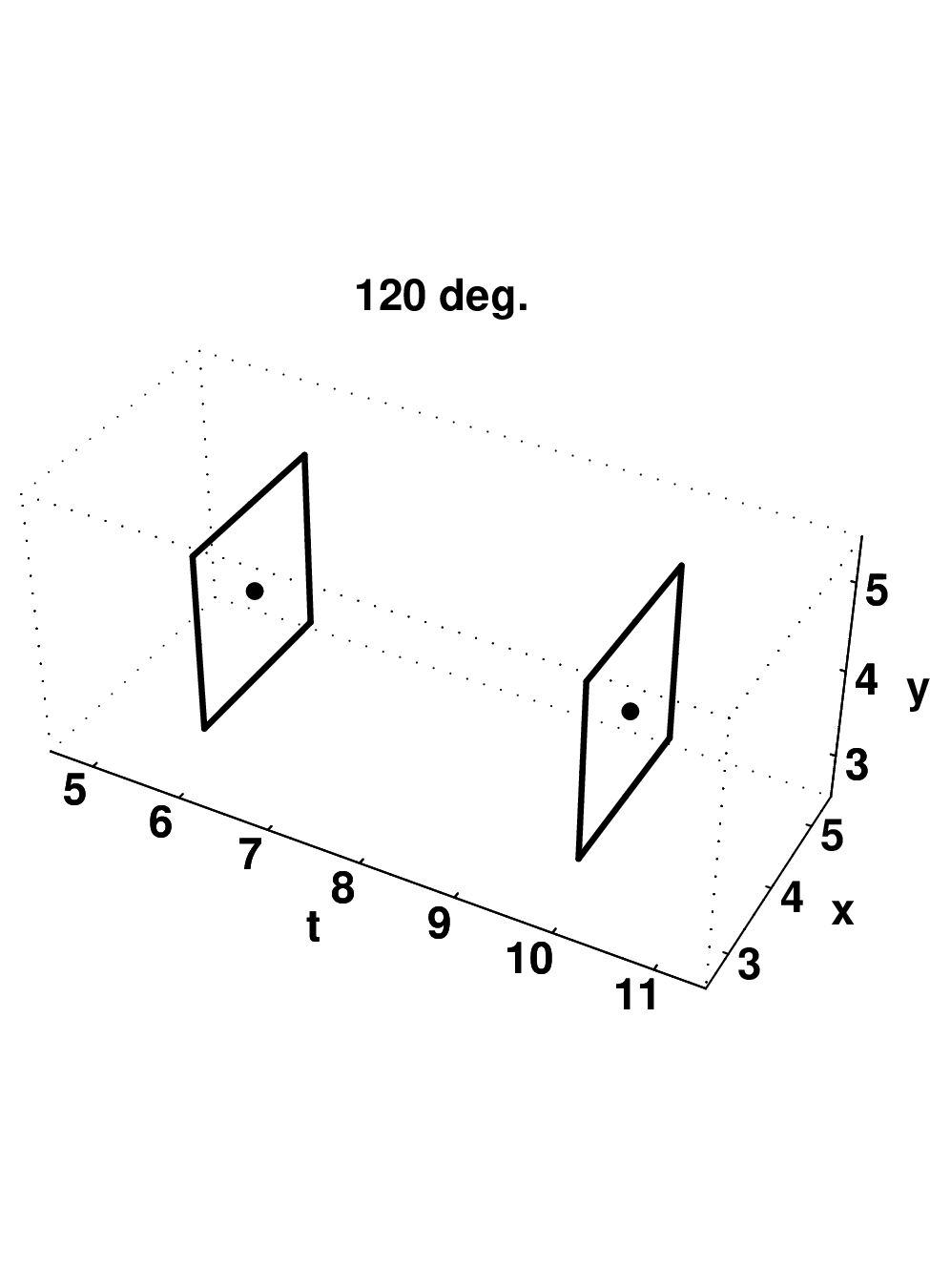}
%\vspace{-8mm}
\end{center}
\vspace{-8mm}
\caption{
{\footnotesize 
Ref.\cite{HT96} Three dimensional projections of the mutual monopole loop
surrounding an instanton--anti-instanton pair (centres marked) of size
$\rho=3$ under increasing rotation angle as detailed in the text. The
loops are flat in the fourth direction.}
}
\label{C00-fig:instanton-pairs}%
\end{figure}

However, we have not yet obtained the analytic solution representing magnetic loops in the 2-instanton background which were found in the numerical way in \cite{RT01} (Fig.~\ref{C00-fig:2instanton-monopole}),
and in the background of an instanton--anti-instanton pair  (see Fig.~\ref{C00-fig:instanton-pairs}).
%\noindent
%\\
%{[Reinhardt \& Tok,
%Merons and instantons in Laplacian Abelian and center gauges in continuum Yang-Mills theory,
%hep-th/0011068, Phys.Lett.B{\bf 505}, 131--140 (2001)]}
%\\
%Reinhardt and Tok,
%Abelian and center gauges in continuum Yang-Mills theory,
%hep-th/0009205.]}

Furthermore, it is known at finite temperature that  there exist self-dual solutions with non-trivial holonomy (calorons) \cite{caloron} which exhibit a non-trivial monopole content by construction.  
It will be interesting to study how  magnetic monopoles to be obtained in our approach from calorons are related to dyons inherent in calorons. 
These issues will be investigated in future works. 
See e.g., \cite{FKSS12} for preliminary work in this direction.

%%%%%%%%%%%%%%%%%%%%%%%%%%%%%%%%%%%%%%%%%%%%%%%%%%%%%%%%%%%%%
%%%%%%%%%%%%%%%%%%%%%%%%%%%%%%%%%%%%%%%%%%%%%%%%%%%%%%%%%%%%%

%A. Hart and M. Teper, 
%e-Print:hep-lat/9511016, 
%Phys.Lett.B371: 261-269, 1996. 
%%%%%%%%%%%%%%%%%%%%%%%%%%%%%%%%%%%%%%%%%%%%%%%%%%%%%%%%%%%%%
%%%%%%%%%%%%%%%%%%%%%%%%%%%%%%%%%%%%%%%%%%%%%%%%%%%%%%%%%%%%%

The extension from $SU(2)$ to $SU(3)$ is an important issue to be investigated in the future works.

\newpage
%%%%%%%%%%%%%%%%%%%%%%%%%%%%%%%%%%%%%%%%%%%%%%%%%%%%%%%%%%%%
%Chapter :
% 
%%%%%%%%%%%%%%%%%%%%%%%%%%%%%%%%%%%%%%%%%%%%%%%%%%%%%%%%%%%%

%%%%%%%%%%%%%%%%%%%%%%%%%%%%%%%%%%%%%%%%%%%%%%%%%%%%%%%%%%%%
%%%%%%%%%%%%%%%%%%%%%%%%%%%%%%%%%%%%%%%%%%%%%%%%%%%%%%%%%%%%
\section{Reformulation of lattice gauge theory and numerical simulations}\label{sec:LGT} 
%%%%%%%%%%%%%%%%%%%%%%%%%%%%%%%%%%%%%%%%%%%%%%%%%%%%%%%%%%%%
%%%%%%%%%%%%%%%%%%%%%%%%%%%%%%%%%%%%%%%%%%%%%%%%%%%%%%%%%%%%

%%%%%%%%%%%%%%%%%%%%%%%%%%%%%%%%%%%%%%%%%%%%%%%%%%
% 
%%%%%%%%%%%%%%%%%%%%%%%%%%%%%%%%%%%%%%%%%%%%%%%%%%

%%%%%%%%%%%%%%%%%%%%%%%%%%%%%%%%%%%%%%%%%%%%%%%%%%
%%%%%%%%%%%%%%%%%%%%%%%%%%%%%%%%%%%%%%%%%%%%%%%%%%
\subsection{Field variable decomposition on the lattice}
%\setcounter{equation}{0}
%%%%%%%%%%%%%%%%%%%%%%%%%%%%%%%%%%%%%%%%%%%%%%%%%%
%%%%%%%%%%%%%%%%%%%%%%%%%%%%%%%%%%%%%%%%%%%%%%%%%%

%%%%%From CHIBA-EP-181 %%%%%

Let $\mathscr{A}_\mu(x)$ be the Yang-Mills field  taking the values in the Lie algebra $\mathscr{G}$ of the Lie group $G$: 
\begin{align}
 \mathscr{A}_\mu(x) = \mathscr{A}_\mu^A(x)T_A \in \mathscr{G}, \quad   \ (A=1,2, \dots, {\rm dim}G) , 
\label{C35-def-A}
\end{align}
where $T_A$ ($A=1,2, \dots, {\rm dim}G$) are the generators of the Lie algebra $\mathscr{G}$. 

We adopt the $D$-dimensional Euclidean lattice $L_\epsilon=(\epsilon \mathbb{Z})^{D}$ with a lattice spacing $\epsilon$. 
In the lattice gauge theory, the gauge field variable taking the values in the gauge group $G$ is represented as the link variable  $U_\ell=U_{x,x+\epsilon \hat\mu}=U_{x,\mu}$ defined on an oriented link $\ell=<x,x+\epsilon \hat\mu> \in L_\epsilon$ running from $x$ to $x+\epsilon \hat\mu$.
We define a unit vector in the $\mu$-th direction by $\hat\mu$ for $\mu=1,\cdots,D$, and we use $\mu:=\epsilon \hat\mu$ to avoid writing the lattice spacing $\epsilon$ explicitly. 
The gauge link variable $U_{x,\mu}$ is related to the gauge potential $\mathscr{A}_\mu(x)$ taking the values in the Lie algebra $\mathscr{G}$ of $G$ as the line integral of a gauge potential $\mathscr{A}_\mu(x)$ along a link $<x,x+ \epsilon \hat\mu>$ from $x$ to $x+\epsilon \hat\mu$:%
\footnote{
See the textbooks of lattice gauge theories, e.g., 
\cite{LGT-text}. % chapter of Lattice gauge theory. 
}
\begin{equation}
 U_{x,\mu} =U_{x,x+\epsilon \hat\mu} =  \mathscr{P} \exp \left\{ -ig \int_{x}^{x+\epsilon \hat\mu} dx^\mu \mathscr{A}_\mu(x) \right\} \in G ,
\end{equation}
where  $g$ is the coupling constant. 
The link variable $U_{x,\mu}$ obeys the well-known lattice gauge transformation:
%The link variable $U_{x,\mu}$ transforms under the original gauge transformation as
\begin{equation}
  U_{x,\mu}  \rightarrow  U_{x,\mu}^\prime = \Omega_{x} U_{x,\mu} \Omega_{x+\mu}^{\dagger} 
  , \quad \Omega_{x} \in G
  \label{C35-U-transf}
 .
\end{equation}

In order to define new variables, we consider the decomposition of the $G$-valued gauge variable $U_\ell=U_{x,\mu} \in G$ into the product of two $G$-valued variables $X_{x,\mu}$ and $V_{x,\mu}$ defined on the same lattice:
\footnote{
Here the lattice variables $V_{x,\mu}$ and $X_{x,\mu}$ are supposed to be related to the Lie-algebra $\mathscr{V}_\mu(x)$ and  $\mathscr{X}_\mu(x)$ in the continuum as  
$ 
  V_{x,\mu} = \exp \{-i\epsilon g\mathscr{V}_\mu(x^\prime) \}  , \ 
  X_{x,\mu} = \exp \{-i\epsilon g\mathscr{X}_\mu(x) \}   
$, 
just as
$
  U_{x,\mu} = \exp \{ -i\epsilon g\mathscr{A}_\mu(x^\prime) \} 
$.
However, the decomposition can be constructed so as to have an intrinsic meaning on a lattice without referring to the naive continuum limit \cite{SKS10}, which includes the result of \cite{KSSMKI08}. 
}
\begin{align}
G \ni U_{x,\mu} = X_{x,\mu}V_{x,\mu}  , \quad X_{x,\mu}  \in G , \ V_{x,\mu} \in G ,
\label{C35-decomp-1}
\end{align}
in such a way that 
  $V_{x,\mu}$ transforms just like a usual link variable under the gauge transformation:
\begin{align}
  V_{x,\mu} \rightarrow V_{x,\mu}' = \Omega_{x} V_{x,\mu} \Omega_{x+\mu}^\dagger    , \quad  \Omega_{x} \in G 
  \label{C35-gt-V}
\end{align}
and thereby $X_{x,\mu}(=  U_{x,\mu}V_{x,\mu}^{-1} )$ transforms like a site  variable under the gauge transformation:
\begin{align}
  X_{x,\mu} \rightarrow  X_{x,\mu}' = \Omega_{x} X_{x,\mu} \Omega_{x}^\dagger   , \quad  \Omega_{x} \in G .
  \label{C35-gt-X}
\end{align}
These features are common to any gauge group $G$.

In order to construct the lattice version of the new reformulation, we need to introduce the \textbf{color (direction) field} $\bm{n}_{x}$ which plays the crucial role in this reformulation defining other new variables.
The color field $\bm{n}_{x}$ on a lattice is regarded as a site variable defined on a site $x$, and has the value in the Lie algebra for the coset space $G/\tilde{H}$ with $\tilde{H}$ being the \textbf{maximal stability group}: 
\begin{equation}
   \bm{n}_{x} = n_x^A T_A\in Lie(G/\tilde{H})=\mathscr{G}-\mathscr{\tilde{H}}  .
\end{equation}
The color field $\bm{n}_{x}$ is used to specify only the color direction in the color space at 
each space--time point and its magnitude is irrelevant ($\bm{n}_{x}^2=\bm{n}_{x} \cdot \bm{n}_{x} =1$).
It should be remarked that the color field $\bm{n}_{x}$ is Hermitian,  $\bm{n}_{x}^\dagger=\bm{n}_{x}$, 
while the gauge field $U_{x,\mu}$ is unitary, $U_{x,\mu}^\dagger=U_{x,\mu}^{-1}$.

The color field transforms in the adjoint way under another (independent) gauge rotation:
\begin{equation}
  \bm{n}_{x}  \rightarrow \Theta_{x} \bm{n}_{x} \Theta_{x}^{\dagger}  
  , \quad \Theta_{x} \in G/\tilde{H}
  .
  \label{C35-n-transf1}
\end{equation}
After applying the  lattice \textbf{reduction condition} which is discussed later in detail, the color field must transform under the gauge transformation in the adjoint way:%
%\footnote{
%This was called the gauge  transformation II in \cite{KMS05}. 
%} 
\begin{equation}
  \bm{n}_{x}  \rightarrow \bm{n}_{x}^\prime = \Omega_{x} \bm{n}_{x} \Omega_{x}^{\dagger}   
  , \quad \Omega_{x} \in G
  .
  \label{C35-n-transf}
\end{equation}
The reduction condition on the lattice has been discussed in \cite{IKKMSS06,SKKMSI07,KKS14,KSSMKI08}.

For $G=SU(2)$, the maximal stability group $\tilde{H}$ is unique and equal to the maximal torus subgroup, i.e.,  a compact $U(1)$ group, $\tilde{H}=H=U(1)$. 
For $G=SU(3)$, there are two possible maximal stability groups, i.e., $\tilde{H}=U(1)\times U(1)$ and $\tilde{H}=U(2)$. 
For $G=SU(N)$ ($N\geq 4)$, there exist more than $N-1$ maximal stability groups. 
Among them,  $\tilde{H}=U(1)^{N-1}$ is called the \textbf{maximal option}, while  $\tilde{H}=U(N-1)$   the \textbf{minimal one}.

%In what follows, we perform the lattice CDGFN decomposition in a constructive way. 

We adopt the correspondence between the Lie group $U$ and the Lie algebra $\mathscr{A}$  by the mid-point definition:
\begin{equation}
 U_{x,\mu} =    \exp \left\{ -i \epsilon g\mathscr{A}_\mu(x^\prime) \right\} ,
\end{equation}
where $x^\prime:=x+\epsilon \hat\mu/2$ is the midpoint of the link $<x,x+\epsilon \hat\mu>$.
The midpoint prescription is useful to suppress as much as possible lattice artifacts coming from a finite (nonzero) lattice spacing, in contrast to the very naive definition  
%$U_{x,\mu} = \exp \left\{ -i \epsilon g\mathscr{A}_\mu(x) \right\}$.
%A naive relationship 
between the gauge link variable $U_{x,\mu}$ and the gauge potential $\mathscr{A}_\mu(x)$ given by 
\begin{align}
U_{x,\mu} = \exp( -i \epsilon g \mathscr{A}_\mu(x)) \in G  . 
\label{C35-def-U}
\end{align}

To understand the meaning of a  lattice version of the \textbf{defining equation}, we  needs  a lattice covariant derivative for an adjoint field.  
We adopt a definition of the covariant lattice derivative for {\it arbitrary} background $V_\mu(x)$:% (which is not necessarily restricted to new variables):
\begin{align}
 D_\mu^{(\epsilon)}[V] \bm{n}_{x} := \epsilon^{-1}[V_{x,\mu} \bm{n}_{x+\mu} - \bm{n}_{x} V_{x,\mu}]  ,  
 \label{C35-cderivative}
\end{align}
by the following reasons. 
\begin{enumerate}
\item[i)]
 When $V_{x,\mu} \equiv {\bf 1}$, the covariant derivative $D_\mu^{(\epsilon)}[V]$ (\ref{C35-cderivative}) reduces to the (forward) lattice derivative $\partial_\mu^{(\epsilon)}$, i.e.,  
$
 \partial_\mu^{(\epsilon)} \bm{n}_{x} := \epsilon^{-1}[\bm{n}_{x+\mu}-\bm{n}_{x}]
$.
\item[ii)]
  The covariant derivative $D_\mu^{(\epsilon)}[V]$ reproduces correctly the continuum covariant derivative for the adjoint field in the naive continuum limit $\epsilon \rightarrow 0$: up to ${\cal O}(\epsilon)$,
\begin{align}
& \epsilon^{-1} [V_{x,\mu} \bm{n}_{x+\mu} - \bm{n}_{x} V_{x,\mu}] 
 \nonumber\\
 =& \epsilon^{-1} [{\bf 1} -i\epsilon g\mathscr{V}_\mu(x) + {\cal O}(\epsilon^2)]\bm{n}_{x+\mu} - \bm{n}_{x} \epsilon^{-1} [{\bf 1} -i\epsilon g\mathscr{V}_\mu(x) + {\cal O}(\epsilon^2)] 
 \nonumber\\
=& \epsilon^{-1}[ \bm{n}_{x+\mu} - \bm{n}_{x}] - i g[\mathscr{V}_\mu(x) \bm{n}_{x+\mu} - \bm{n}_{x} \mathscr{V}_\mu(x)] + {\cal O}(\epsilon) ,
 \nonumber\\
=& \partial_\mu^{(\epsilon)} \bm{n}_{x} - i g[\mathscr{V}_\mu(x), \bm{n}_{x}] + {\cal O}(\epsilon) ,
\label{C35-r1}
\end{align}
where we have used the  ${\cal O}(\epsilon^2)$ ambiguity in the last step.%
\footnote{
%Other equivalent conditions of (\ref{C35-Lcc}) are as follows. 
%\begin{align}
% \bm{n}_{x} V_{x,\mu}\bm{n}_{x+\mu}   = V_{x,\mu}  , \quad 
% \bm{n}_{x} V_{x,\mu}\bm{n}_{x+\mu}V_{x,\mu}^\dagger   =  \bm{1} .
%\end{align}
Adopting another form instead of (\ref{C35-r1}) using the ambiguity of ${\cal O}(\epsilon^2)$, 
$
 0 = \bm{n}_{x+\mu} - \bm{n}_{x}  - i \epsilon g  \mathscr{V}_\mu(x)  \bm{n}_{x}
 +  i \bm{n}_{x+\mu}  \epsilon g  \mathscr{V}_\mu(x)   
   + {\cal O}(\epsilon^2),
$
which is rewritten as
$
  [ \bm{1}   +  i  \epsilon g  \mathscr{V}_\mu(x)  + {\cal O}(\epsilon^2) ]\bm{n}_{x}
 =  \bm{n}_{x+\mu} [ \bm{1} + i  \epsilon g  \mathscr{V}_\mu(x) + {\cal O}(\epsilon^2)] ,
$
we obtain a relation 
$
 V_{x,\mu}^\dagger  \bm{n}_{x}
 = \bm{n}_{x+\mu} V_{x,\mu}^\dagger  .
$
However, this is nothing but the Hermitian conjugate of  (\ref{C35-cderivative}) and does not lead to a new condition. 
The definition of the covariant derivative  could be improved e.g., by using a symmetric difference.
%, as will be discussed in a separate paper. 
By using the mid-point prescription, this is more improved up to ${\cal O}(\epsilon^2)$:
$$
  \epsilon^{-1} [V_{x,\mu} {\bf n}_{x+\mu} - {\bf n}_{x} V_{x,\mu}] 
= \partial_\mu^{(\epsilon)} {\bf n}_{x^\prime} - i g[ \mathbf{V}_\mu(x^\prime), {\bf n}_{x^\prime}] 
-ig \epsilon/2 \{ \mathbf{V}_\mu(x^\prime) , \partial_\mu^{(\epsilon)} {\bf n}_{x^\prime} - i g[ \mathbf{V}_\mu(x^\prime), {\bf n}_{x^\prime}]  \} + {\cal O}(\epsilon^2) 
 .
$$
}

\item[iii)]
  The  covariant derivative $D_\mu^{(\epsilon)}[V]$  obeys the correct transformation property, i.e., the adjoint rotation on a lattice: 
\begin{align}
  D_\mu^{(\epsilon)}[V] \bm{n}_{x} \rightarrow \Omega_{x}(D_\mu^{(\epsilon)}[V] \bm{n}_{x})\Omega_{x+\mu}^\dagger ,
\end{align}
provided that the link variable $V_{x,\mu}$ transforms in the same way as the original link variable $U_{x,\mu}$:
\begin{align}
  V_{x,\mu} \rightarrow \Omega_{x} V_{x,\mu} \Omega_{x+\mu}^\dagger
   = V_{x,\mu}' .
   \label{C35-transf-V}
\end{align}
%Indeed, the solution (\ref{C35-sol}) satisfies this transformation property. 
This is required from the transformation property of the continuum variable $\mathscr{V}_\mu(x)$.
\end{enumerate}
%Therefore, we obtain the desired condition (\ref{C35-Lcc}) between $\bm{n}_{x}$ and $V_{x,\mu}$.
%\begin{align}
% \bm{n}_{x} V_{x,\mu}  = V_{x,\mu} \bm{n}_{x+\mu} .
% \label{C35-Lcc}
%\end{align}

First, we consider the defining equation  which enables us to determine the decomposition uniquely. 
According to the  continuum reformulation  in the minimal option \cite{KSM08}, we introduce just a single color field $\bm{n}_{x}$ for  $G=SU(N)$ ($N \ge 2$).   This unit vector field is an initial or a reference field to construct  possible other color fields which are necessary in the maximal option. 

We have proposed a lattice version of the \textbf{first defining equation} \cite{KSSMKI08,SKS10}:
\\
(I) \textit{
The color field $\bm{n}_{x}$ is covariantly constant in the  background (matrix) field $V_{x,\mu}$:}
\begin{equation}
  0 = \epsilon D_\mu^{(\epsilon)}[V]\bm{n}_{x} := V_{x,\mu} \bm{n}_{x+\mu} - \bm{n}_{x} V_{x,\mu}   ,
  \label{C35-lat-defeq-min1}
\end{equation}
where $D_\mu^{(\epsilon)}[V]$ is the lattice covariant derivative in the adjoint representation (\ref{C35-cderivative}).
%$D_\mu^{(\epsilon)}[V]\bm{n}_{x}:=\epsilon^{-1}(V_{x,\mu} \bm{n}_{x+\mu} - \bm{n}_{x} V_{x,\mu})$. 

This defining equation for an initial color field guarantees that all $N-1$ color fields $\bm{n}_{x}^{(k)}$ ($k=1, \cdots, N-1$) prepared in the maximal option satiety the first defining equation in the maximal option, i.e., covariant constant in the background $V_{x,\mu}$:
\begin{equation}
 0 =  \epsilon D_\mu^{(\epsilon)}[V]\bm{n}_{x}^{(k)} := V_{x,\mu} \bm{n}_{x+\mu}^{(k)} - \bm{n}_{x}^{(k)} V_{x,\mu}   \quad (k=1, \cdots, N-1) .
  \label{C35-lat-defeq-max1}
\end{equation}
The first defining equation  (\ref{C35-lat-defeq-min1}) is just a lattice or group theoretical version of the Lie-algebra valued defining equations given already in the continuum formulation, see Sections 4 and 5.

Next, we give a general consideration to what extent the defining equations determine the decomposition uniquely, before proceeding to solving them explicitly. 
In order to consider the meaning of the second defining equation deeper, we return to the first equation.
It is important to observe that the decomposition (\ref{C35-decomp-1}) is invariant under the simultaneous local transformation of  $V_{x,\mu}$ and $X_{x,\mu}$:
\footnote{
For another decomposition of the form:
$
 U_{x,\mu} = V_{x,\mu} X_{x,\mu} 
  ,
$
it is advantageous to take
$
  V_{x,\mu} \rightarrow V_{x,\mu} g_{x+\mu} , \quad 
  X_{x,\mu} \rightarrow g_{x+\mu}^{-1}  X_{x,\mu} , \quad 
g_{x+\mu} \in G 
  .
%\label{C35-s-rotationb}
$
}
\begin{equation}
 X_{x,\mu} \rightarrow X_{x,\mu} r_{x}^{-1} , \quad
 V_{x,\mu} \rightarrow r_{x} V_{x,\mu} , \quad 
r_{x} \in G 
  .
  \label{C35-s-rotation}
\end{equation}
This is the extra   $G$ degrees of freedom which are absent in the original $G$ lattice theory written in terms of $U_{x,\mu}$.
In order to obtain the unique decomposition (\ref{C35-decomp-1}), we must fix the extra degrees of freedom by imposing suitable conditions. 
%\footnote{
%If $X$ and $V$ are unitary, the group $G$ with such an element $g_{x}$ can be U(N). If we use the speciality of $X$ and $V$, i.e., $\det X=1$ and $\det V=1$, we have $\det G=1$. Therefore, $G$ must be SU(N). 
%$g_{x}=e^{i\theta {\bf 1}}\tilde{G}_{x}$ 
%}
Therefore, we examine to what extent the extra degrees of freedom (\ref{C35-s-rotation}) is fixed by the first defining equation:
\begin{equation}
 \bm{n}_x V_{x,\mu} = V_{x,\mu} \bm{n}_{x+\mu} 
  .
  \label{C35-def-1}
\end{equation}
It is obvious that the diagonal part related to the discrete symmetry of the center $Z(N)$ of $SU(N)$ is undetermined by the first defining equation:
\begin{equation}
  \exp (2\pi i n/N) \mathbf{1} \in Z(N) \quad (n=0,1,  \cdots , N) .
  \label{C35-center}
\end{equation}
Suppose that the first defining equation holds after the local rotations (\ref{C35-s-rotation}):
\begin{equation}
 \bm{n}_x r_{x} V_{x,\mu} = r_{x} V_{x,\mu} \bm{n}_{x+\mu} 
  .
\end{equation}
Combining this with the original equation (\ref{C35-def-1}), we obtain the relationship
\begin{equation}
 \bm{n}_x r_{x} V_{x,\mu} = r_{x} \bm{n}_x V_{x,\mu}
 \Longleftrightarrow [\bm{n}_x , r_{x} ] V_{x,\mu} = 0
  .
\end{equation}
This implies that the degrees of freedom of $r_{x}$ satisfying the following equation cannot be determined by imposing the first defining equation alone. 
%the first defining equation fixes $V_{x,\mu}$ up to $g_{x}$ satisfying
\begin{equation}
  [\bm{n}_x , r_{x} ] = 0
  .
  \label{C35-G-dof}
\end{equation}

For the maximal option, the extra symmetry is $Z(N) \times H$, $H = U(1)^{N-1} \subset SU(N)$:
\begin{equation}
   r_{x} = \exp (2\pi i n/N) \exp \left\{ i \sum_{k=1}^{N-1} \alpha_{x}^{(k)} \bm{n}_{x}^{(k)} \right\} \in Z(N) \times H = Z(N) \times U(1)^{N-1} , 
%\quad \alpha^{(k)} \in \mathbb{R} ,
   \label{C35-extra-max}
\end{equation}
where $\alpha^{(k)} \in \mathbb{R}$ and $\{ \bm{n}_{x}^{(k)} \}$ is a maximal set of mutually commutable Hermitian generators for $U(1)^{N-1}$ with the traceless property ${\rm tr}(\bm{n}_{x}^{(k)})=0$.

For the minimal option, the extra symmetry is $Z(N) \times \tilde{H}$, $\tilde{H} = U(N-1) \subset SU(N)$:
\begin{equation}
   r_{x} = \exp (2\pi i n/N) \exp \left\{ i \alpha_{x} \bm{h}_{x}  \right\}
\exp \left\{ i \sum_{k=1}^{(N-1)^2-1} \beta_{x}^{(k)} \bm{u}_{x}^{(k)} \right\} \in Z(N) \times \tilde{H} = Z(N) \times U(N-1) ,
%= U(1) \times SU(N-1),
   \label{C35-extra-min}
\end{equation}
where $\alpha_{x},  \beta_{x}^{(k)} \in \mathbb{R}$ and $\{ \bm{u}_{x}^{(k)} \}$ is a set of Hermitian generators of $SU(N-1)$ commutable with $\bm{h}_{x}:=\bm{n}_{x}^{(N-1)}$  with the traceless property ${\rm tr}(\bm{u}_{x}^{(k)})=0$.  

Thus, we find that the degrees of freedom corresponding to $H$ or $\tilde{H}$ are left unfixed in the maximal or minimal options, respectively, even after solving the first defining equation.  
%Therefore, we must impose an additional condition to obtain the unique decomposition.  
%For example, for $G=SU(2)$, this is achieved by imposing an extra condition for $X_{x,\mu}$, e.g., 
%\begin{equation}
%   {\rm tr}(X_{x,\mu}\bm{n}_{x} )=0, \quad
%   X_{x,\mu} = U_{x,\mu}V_{x,\mu}^{-1}=U_{x,\mu}V_{x,\mu}^\dagger    ,
%  \label{C35-def-2}
%\end{equation}
In order to obtain the unique decomposition, therefore, we must impose additional conditions to fix  degrees of freedom which remain undetermined by imposing the first defining equation. 
This role is played by the second defining equation.  
In the previous papers \cite{IKKMSS06,SKKMSI07,KSSMKI08}, we used as a second defining equation the condition:
\begin{equation}
  {\rm tr} [ X_{x,\mu} \bm{n}_{x}]  = 0 .
  \label{C35-lat-defeq20}
\end{equation}
This is reasonable from the viewpoint of the naive continuum limit, since it leads to the second defining equation 
\begin{equation}
{\rm tr} [ \mathscr{X}_\mu(x)\bm{n}(x)]=0
  \label{C35-defeq20}
\end{equation}
for the Lie-algebra valued field  $\mathscr{X}_\mu(x)$ in the continuum, as can be seen from 
\begin{align}
  {\rm tr} [ X_{x,\mu} \bm{n}_{x}]  
&= - ig \epsilon {\rm tr} [ \mathscr{X}_\mu(x)\bm{n}(x)] + O(\epsilon^2) 
 ,
\end{align}
using
$
 X_{x,\mu} =    \exp \left\{ -ig \epsilon \mathscr{X}_\mu(x) \right\} 
=  \mathbf{1} -ig \epsilon \mathscr{X}_\mu(x) + O(\epsilon^2) 
$.
Here, however, we are looking for a lattice version of the second defining equation valid for the Lie-group valued field  $X_{x,\mu}$, which is intrinsic for the lattice with arbitrary lattice spacing $\epsilon$. 
In the below, we will observe that (\ref{C35-lat-defeq20}) is valid for $SU(2)$ exceptionally, but it is not valid for $SU(N)$, $N \ge 3$.
We need more care for the second defining equations.
Anyway, imposing simultaneously the first and second defining equations uniquely fix the decomposition (\ref{C35-decomp-1}) by eliminating the extra  gauge degrees of freedom associated to the decomposition.

%%%%%%%%%%%%%%%%%%%%%%%%%%%%%%%%%%%%%%%%%%%%%%%%%%
%%%%%%%%%%%%%%%%%%%%%%%%%%%%%%%%%%%%%%%%%%%%%%%%%%

The explicit forms for the new lattice variables $X_{x,\mu}$ and $V_{x,\mu}$ have been given in terms of the original link variable $U_{x,\mu}$ and the color field $\bm{n}_{x}$  in \cite{IKKMSS06,SKKMSI07} for $G=SU(2)$ and in \cite{KSSMKI08} for $G=SU(3)$. 
They are obtained by solving the defining equations by assuming an ansatz for $V_{x,\mu}$, since it seemed to be difficult to solve the coupled matrix equations.  Quite recently, we have succeeded to solve the defining equations without using any ansatz to obtain the general and exact solutions for $G=SU(N)$ \cite{SKS10} which have remarkably the same form as those obtained previously \cite{IKKMSS06,SKKMSI07,KSSMKI08}  for $G=SU(2)$ and $G=SU(3)$.

In the \textbf{maximal option}, the decomposition is given by %\cite{SKS10}
\begin{align}
   X_{x,\mu}
%   =& U_{x,\mu}  \tilde{V}_{x,\mu}^{-1} 
%   = U_{x,\mu} \tilde{V}_{x,\mu}^{-1} (\sqrt{\tilde{V}_{x,\mu} \tilde{V}_{x,\mu}^\dagger}) (\det(\left(\sqrt{\tilde{V}_{x,\mu} \tilde{V}_{x,\mu}^\dagger}\right)^{-1} \tilde{V}_{x,\mu} ))^{1/N} 
%   \nonumber\\
=  \hat{K}_{x,\mu}^\dagger (\det (\hat{K}_{x,\mu}))^{1/N}  g_{x}^{-1} 
, \quad
V_{x,\mu}
=    g_{x}  \hat{K}_{x,\mu} U_{x,\mu} (\det (\hat{K}_{x,\mu}))^{-1/N}  ,
\label{C35-eq:X}
\end{align}
where
\begin{equation}
  \hat{K}_{x,\mu} := \left(\sqrt{K_{x,\mu} K_{x,\mu}^\dagger} \right)^{-1} K_{x,\mu}, \quad 
  \hat{K}_{x,\mu}^\dagger= K_{x,\mu}^\dagger \left(\sqrt{K_{x,\mu} K_{x,\mu}^\dagger}\right)^{-1} .
\end{equation}
\begin{equation}
 K_{x,\mu} 
:= \mathbf{1} +2N\sum_{k=1}^{N-1}\bm{n}_{x}^{(k)}U_{x,\mu} \bm{n}_{x+\mu}^{(k)}  U_{x,\mu}^{-1} .
\end{equation}
Here  a common factor $g_{x}$  in the above expressions for $X_{x,\mu}$ and $V_{x,\mu}$ is the part undetermined from the first defining equation alone. 
In fact, $g_{x}$ is an element of the extra  symmetry associated with the decomposition: $Z(N) \times {H}$,  ${H} = U(1)^{N-1} \subset SU(N)$, as mentioned above. 
In order to fix it, we must impose further conditions.  
We impose  the \textbf{second defining equation}, e.g., 
\\
(II)~$g_{x}$ is equated with an element $g_{x}^{0}$:
\begin{equation}
  g_{x}=g_{x}^{0} .
\end{equation}
 The simplest one is to take $g_{x}^{0}=\mathbf{1}$, or
\begin{align}
  (\det (\hat{K}_{x,\mu}))^{-1/N}  \hat{K}_{x,\mu} X_{x,\mu}
= g_{x}^{0} =  \mathbf{1}  .
\label{C35-2nd}
\end{align}
Thus the decomposed variables $X_{x,\mu}$ and $V_{x,\mu}$ are completely determined. 
We can check that the naive continuum limit of (\ref{C35-2nd}) reduces to the second defining equation (\ref{C35-defeq20}) in the continuum formulation, see \cite{SKS10} for the detail.
  
Another equivalent form is%\cite{KSSMKI08}%,SKKMSI07b
\begin{align}
   X_{x,\mu}
   =& U_{x,\mu}  \tilde{V}_{x,\mu}^{-1} 
   = U_{x,\mu} \tilde{V}_{x,\mu}^{-1}  \sqrt{\tilde{V}_{x,\mu} \tilde{V}_{x,\mu}^\dagger}  \left\{\det \left[\left(\sqrt{\tilde{V}_{x,\mu} \tilde{V}_{x,\mu}^\dagger}\right)^{-1} \tilde{V}_{x,\mu} \right] \right\}^{1/N} g_{x}^{-1} 
,
\label{C35-eq:X2}
\\
V_{x,\mu} 
%=& X_{x,\mu}^\dagger U_{x,\mu},
=& g_{x} \left(\sqrt{\tilde{V}_{x,\mu} \tilde{V}_{x,\mu}^\dagger} \right)^{-1} \tilde{V}_{x,\mu} \left\{ \det \left[ \left(\sqrt{\tilde{V}_{x,\mu} \tilde{V}_{x,\mu}^\dagger}\right)^{-1} \tilde{V}_{x,\mu} \right] \right\}^{-1/N}   ,
\label{C35-eq:V}
\end{align}where
\begin{equation}
\tilde{V}_{x,\mu} 
:= K_{x,\mu} U_{x,\mu} 
 = U_{x,\mu} +2N\sum_{k=1}^{N-1} \bm{n}_{x}^{(k)}U_{x,\mu} \bm{n}_{x+\mu}^{(k)} .
\end{equation}

In the \textbf{minimal option}, the decomposition is given by  %\cite{SKS10}
\begin{align}
   X_{x,\mu}
=& \hat{L}_{x,\mu}^\dagger (\det (\hat{L}_{x,\mu}))^{1/N} g_{x}^{-1} 
 , \quad
 V_{x,\mu}  
=  g_{x}   \hat{L}_{x,\mu} U_{x,\mu} (\det (\hat{L}_{x,\mu}))^{-1/N}
\label{C35-eq:Xmin}
\end{align}
where
\begin{align}
  \hat{L}_{x,\mu}=& \left(\sqrt{L_{x,\mu} L_{x,\mu}^\dagger} \right)^{-1}  L_{x,\mu}
, \quad
  \hat{L}_{x,\mu}^\dagger  = L_{x,\mu}^\dagger \left( \sqrt{L_{x,\mu} L_{x,\mu}^\dagger}\right)^{-1}  
  \\
L_{x,\mu}  
  :=& \frac{N^{2}-2N+2}{N}\mathbf{1}+\left(  N-2\right)  \sqrt{\frac{2(N-1)}{N}%
}\left(  \bm{n}_{x}+U_{x,\mu}\bm{n}_{x+\mu}U_{x,\mu}^{-1}\right)
\nonumber\\
&
 +4\left(  N-1\right)  \bm{n}_{x}U_{x,\mu}\bm{n}_{x+\mu}U_{x,\mu}%
^{-1} .
\end{align}
We can use the second defining equation which is the same as that used in the maximal option to put $g_{x} =\mathbf{1}$.
This is also rewritten in another equivalent form as in the maximal option. 
For the derivation of these results, see \cite{SKS10}.

In the $SU(2)$ case, there are no differences between maximal and minimal options.
%, and they cannot be distinguished. 
 In fact, $K_{x,\mu}$ and $L_{x,\mu}$ are the same: 
\begin{equation}
K_{x,\mu} =  L_{x,\mu} = \mathbf{1}+ 4 \bm{n}_{x} U_{x,\mu} \bm{n}_{x+\mu} U_{x,\mu}^{-1}  ,
\quad
K_{x,\mu}^\dagger =  L_{x,\mu}^\dagger = \mathbf{1}+ 4  U_{x,\mu}  \bm{n}_{x+\mu}  U_{x,\mu}^\dagger  \bm{n}_{x} .
\end{equation}
The specific feature of the $SU(2)$ case is that $K_{x,\mu} K_{x,\mu} ^\dagger$ is proportional to the unit matrix:%
\footnote{
This was first shown in the footnote 6 of \cite{IKKMSS06} where  $K_{x,\mu} K_{x,\mu} ^\dagger=\tilde{V}_{x,\mu} \tilde{V}_{x,\mu} ^\dagger$. 
}
\begin{equation}
K_{x,\mu} K_{x,\mu} ^\dagger   = \frac12 {\rm tr}(K_{x,\mu} K_{x,\mu} ^\dagger) \mathbf{1} .
\end{equation}
Therefore, $\hat{K}_{x,\mu}$ is proportional to $K_{x,\mu}$, namely, $K_{x,\mu}$ agrees with the unitary $\hat{K}_{x,\mu}$ up to a numerical factor:
\begin{equation}
  \hat{K}_{x,\mu}= \left(\sqrt{K_{x,\mu} K_{x,\mu}^\dagger} \right)^{-1} K_{x,\mu}
=  \left(\sqrt{{\rm tr}(K_{x,\mu} K_{x,\mu} ^\dagger)/2}\right)^{-1} K_{x,\mu} .
\end{equation}
Therefore, we have 
\begin{align}
   X_{x,\mu}
=& \hat{K}_{x,\mu}^\dagger (\det (\hat{K}_{x,\mu}))^{1/2} 
 g_{x}^{-1} 
= \frac{(\det (K_{x,\mu}))^{1/2}}{{\rm tr}(K_{x,\mu} K_{x,\mu} ^\dagger)/2} K_{x,\mu} ^\dagger  g_{x}^{-1} , 
\end{align}
For $SU(2)$, thus, the second defining equation (\ref{C35-lat-defeq20}) is exceptionally satisfied when $g_{x}=\mathbf{1}$, since 
\begin{align}
  {\rm tr} [ X_{x,\mu} \bm{n}_{x}]  
=& \frac{(\det (K_{x,\mu}))^{1/2}}{{\rm tr}(K_{x,\mu} K_{x,\mu} ^\dagger)/2} {\rm tr} [ \bm{n}_{x} K_{x,\mu} ^\dagger  g_{x}^{-1}   ] 
\nonumber\\
=& \frac{(\det (K_{x,\mu}))^{1/2}}{{\rm tr}(K_{x,\mu} K_{x,\mu} ^\dagger)/2} {\rm tr} [ \bm{n}_{x} g_{x}^{-1}   +    U_{x,\mu}  \bm{n}_{x+\mu}  U_{x,\mu}^\dagger    g_{x}^{-1}  ]  
 ,
  \label{C-35lat-defeq20r}
\end{align}
where we take into account ${\rm tr}(\bm{n}_{x})=0$ and the cyclicity of the trace.

%%%%%%%%%%%%%%%%%%%%%%%%%%%%%%%%%%%%%%%%%%%%%%%%%%
%%%%%%%%%%%%%%%%%%%%%%%%%%%%%%%%%%%%%%%%%%%%%%%%%%
\subsection{Lattice reduction condition and color field}
%\setcounter{equation}{0}
%%%%%%%%%%%%%%%%%%%%%%%%%%%%%%%%%%%%%%%%%%%%%%%%%%
%%%%%%%%%%%%%%%%%%%%%%%%%%%%%%%%%%%%%%%%%%%%%%%%%%

First, we generate gauge field configurations of link variables $\{ U_{x,\mu} \}$ on a four-dimensional Euclidean lattice $L_\epsilon=(\epsilon \mathbb{Z})^{4}$ with a lattice spacing $\epsilon$  by using the standard method: 
the Wilson action based on the heat bath method for $G=SU(2)$, 
the Wilson action based on the pseudo heat-bath method  \cite{CM82} for $G=SU(3)$.

Next, we construct the color field variable $\bm{n}_{x}$ according to the following method.  
By introducing the color field $\bm{n}_{x} \in   SU(N)/\tilde{H}$ into the original $SU(N)$  Yang-Mills theory written in terms of 
$U_{x,\mu} \in SU(N)$, we obtain the master $SU(N)$  Yang-Mills theory written in terms of 
$U_{x,\mu} \in SU(N)^{\Omega}$ and $\bm{n}_x \in [SU(N)/\tilde{H}]^{\Theta}$, which has the enlarged local gauge symmetry 
$\tilde{G}^{\Omega,\Theta}_{local}=SU(N)_{local}^{\Omega} \times [SU(N)/\tilde{H}]_{local}^{\Theta}$ 
larger than the local gauge symmetry $SU(N)_{local}^{\Omega}$ in the original Yang-Mills theory \cite{KMS05}. 
  In order to recover the original gauge symmetry $G_{local}=SU(N)_{local}$ by eliminating the extra degrees of freedom in the enlarged local gauge symmetry 
$\tilde{G}^{\Omega,\Theta}_{local}$,  
we must impose sufficient number of constraints, which we call the \textbf{reduction condition}. 
The reduction condition enables us to determine the color field as a functional of the original gauge field. 

For a given set of gauge field configurations $\{ U_{x,\mu} \}$,  a set of color fields $\{ \bm{n}_{x} \}$ is determined by imposing a lattice version of the reduction condition.  
A reduction condition in the minimal option on a lattice is given  by minimizing the reduction functional:
\begin{align}
 F_{\rm red}[\bm{n};U] 
:=& \epsilon^{D} \sum_{x,\mu} {\rm tr}\{ (D_\mu^{(\epsilon)}[U]\bm{n}_{x}) (D_\mu^{(\epsilon)}[U]\bm{n}_{x})^\dagger \}/{\rm tr}(\bf{1}) 
\nonumber\\
=& \epsilon^{D-2} \sum_{x,\mu}[1- 2{\rm tr}( \bm{n}_{x} U_{x,\mu} \bm{n}_{x+\mu} U_{x,\mu}^\dagger ) ]
 ,   
\end{align}
with respect to the color field $\{ \bm{n}_{x} \}$ for  a given set of gauge field configurations  $\{ U_{x,\mu} \}$.
Here $D_\mu^{\epsilon}[U]$ is the lattice covariant derivative in the adjoint representation (\ref{C35-cderivative}).
%defined by 
%$
% D_\mu^{\epsilon}[U] \bm{n}_{x} := \epsilon^{-1}( U_{x,\mu} \bm{n}_{x+\mu} - \bm{n}_{x} U_{x,\mu}) 
%$
%with a lattice spacing $\epsilon$.
Thus, a set of color fields $\{ \bm{n}_{x} \}$  we need is obtained as a set of unit vector fields $\{ \tilde{\bm{n}}_{x} \}$ which realizes the minimum of the reduction functional:
\begin{align}
 F_{\rm red}[\bm{n};U] 
 =&   \min_{\tilde{\bm{n}}} F_{\rm red}[\tilde{\bm{n}};U]
 ,   
\end{align}

After applying the reduction condition,   the color field  transforms under the gauge transformation in the adjoint way: 
\begin{equation}
  \bm{n}_{x}  \rightarrow \Omega_{x} \bm{n}_{x} \Omega_{x}^{-1} = \bm{n}_{x}^\prime
  , \quad \Omega_{x} \in G
  .
  \label{C35-n-transfb}
\end{equation}
The reduction functional $F_{\rm red}$ is invariant under the gauge transformation. Therefore, imposing the reduction condition does not break the original gauge symmetry $G=SU(N)$. 
Therefore, we can impose any gauge fixing afterwards, if necessary. 
%The details for the algorithm of the reduction procedure on a lattice in the SU(3) case will be given in \cite{SKKS11}.

The methods of the minimization are  as follows. 
%A  method of performing the the minimization adopted recently  is as follows  \cite{KKSSI09,KKS14}. 
The two algorithms for solving the reduction equation are available:
\begin{enumerate}
\item[(i)]
%Updating $\{ \bm{n}_x \}$ in terms of the heat bath method.
Updating $\{ \bm{n}_x \}$ via gauge transformation for  solving the reduction condition.
 This method  of the reduction prescription  was adopted in the early studies \cite{IKKMSS06,SKKMSI07}. (This was once called the new MAG.) 

\item[(ii)]
In order to minimize the functional $F_{\rm red}$, we have only to solve the stationary condition:
\begin{align}
\frac{\partial F_{\rm red}[\bm{n};U]}{\partial n^A_x} =0 .
\end{align}
This method of the reduction prescription was adopted in  the recent studies \cite{Kato-lattice2009,KKS14,FKSS10,FKSS12}. 
See section \ref{section:reduction-lattice}.
%This observation has been actually used to find the minimum in SU(2) case \cite{Kato-lattice2009}.
\end{enumerate}
For the second method, it is observed that solving the reduction problem is equivalent to finding the ground state of the spin-glass model, since the reduction functional $F_{\rm red}$ rewritten
\begin{align}
 F_{\rm red}[\bm{n};U] 
 =&  \epsilon^{D-2} \sum_{x,\mu}(1-  J^{AB}_{x,\mu}[U]   {n}_{x}^A   {n}_{x+\mu}^B ) ,
 \ 
%\nonumber\\
 J^{AB}_{x,\mu}[U] :=  2{\rm tr}( T_A U_{x,\mu} T_B U_{x,\mu}^\dagger ) 
 ,
\end{align}
 can be regarded as the energy for a spin-glass model and it is minimized with respect to the color field $\{ \bm{n}_{x} \}$ under the random link interaction $J^{AB}_{x,\mu}[U]$ for given gauge field configurations  $\{ U_{x,\mu} \}$.
In practical application, there exist local minima satisfying this condition.
However, there is no known method for finding a global minimum of the functional. 
Therefore, overrelaxation method/replica exchange (simulated annealing) method should be used in order to approach the global minimum more rapidly. 
%The detail of applying this method to $SU(3)$ will be given in \cite{SKKS11}. 

The first older method goes as follows \cite{IKKMSS06,SKKMSI07}.
We consider  a lattice version of the Landau gauge, i.e., \textbf{lattice Landau gauge (LLG)} playing the role of fixing the original gauge symmetry $G=SU(N)$, which we call the \textbf{overall gauge fixing}.  The LLG is obtained by minimizing the functional $F_{\rm LLG}[U;\Omega]$.
We minimize simultaneously the two functionals $F_{\rm red}$ and $F_{\rm LLG}$ written in terms of gauge (link) variables $U_{x,\mu}$ and  color (site) variables $\bm{n}_{x}$,   
\begin{align}
 F_{\rm red}[\bm{n},U;\Omega,\Theta]
:=&  \sum_{x,\mu}\mathrm{tr}(\mathbf{1}-{}^{\Theta}\bm{n}_{x}{}^{\Omega}U_{x,\mu}{}^{\Theta}\bm{n}_{x+\mu}{}^{\Omega}U_{x,\mu}^{\dagger}),
\\
F_{\rm LLG}[U;\Omega]
:=& \sum_{x,\mu}\mathrm{tr}(\mathbf{1}-{}^{\Omega} U_{x,\mu}) ,
\end{align}
with respect to enlarged lattice gauge transformations: 
\begin{equation} 
 {}^{\Omega}{}U_{x,\mu}:=\Omega_{x}U_{x,\mu}\Omega_{x+\mu}^{\dagger} ,
 \quad
  {}^{\Theta}\bm{n}_{x}:=\Theta_{x} \bm{n}_{x}^{(0)}\Theta_{x}^{\dagger} ,
\end{equation}
for the link variable $U_{x,\mu}$  and  for an initial site variable
 $\bm{n}_{x}^{(0)}$  
where  $\Omega_{x}$ and $\Theta_{x}$
are {\it independent} $SU(N)$ matrices on a site $x$. 
Then we can determine  the configurations  ${}^{\Theta^{*}}\bm{n}_{x}$ and ${}^{\Omega^{*}}U_{x,\mu}$ realizing the minimum of the first functional, up to a common $SU(N)$ transformation $G_{x}$:  
\begin{equation}
 \min_{\Omega,\Theta}F_{\rm red}[{}^{\Theta}\bm{n},{}^{\Omega}U ]
= F_{\rm red}[{}^{\Theta^{*}}\bm{n},{}^{\Omega^{*}}U] .
\end{equation}%$\min_{\Omega,\Theta}F_{nMAG}[U,{n};\Omega,\Theta]=F_{nMAG}[U,{n};\Omega^{*},\Theta^{*}]$, 
This is because the ``common''  gauge transformation $G_{x}$ for $\Theta^{*}$ and $\Omega^{*}$ does not change the value of the functional $F_{\rm red}[\bm{n},U;\Omega,\Theta]$, i.e.,
\begin{equation}
F_{\rm red}[{}^{\Theta^{*}}\bm{n},{}^{\Omega^{*}}U] 
=  F_{\rm red}[{}^{G}(^{\Theta^{*}}\bm{n}),{}^{G}(^{\Omega^{*}}U)] , 
\end{equation}
%$F_{nMAG}[U,\mathbf{n};\Omega^{*},\Theta^{*}]=F_{nMAG}[U,\mathbf{n};G\Omega^{*},G\Theta^{*}]$, 
since 
\begin{align}
& \mathrm{tr}({}^{\Theta}\bm{n}_{x}{}^{\Omega}U_{x,\mu}{}^{\Theta}\bm{n}_{x+\mu}{}^{\Omega}U_{x,\mu}^{\dagger})
\nonumber\\
=& \mathrm{tr}(G_{x} {}^{\Theta}\bm{n}_{x} G_{x+\mu}^\dagger \cdot G_{x+\mu} {}^{\Omega}U_{x,\mu} G_{x+\mu}^\dagger  \cdot G_{x+\mu} {}^{\Theta}\bm{n}_{x+\mu} G_{x+\mu}^\dagger  \cdot G_{x+\mu} {}^{\Omega}U_{x,\mu}^{\dagger} G_{x}^\dagger) .
\end{align}
This degrees of freedom for the $SU(N)$ gauge transformation are fixed by minimizing the second functional  $F_{\rm LLG}[U;\Omega]$ such that the configuration ${}^{\Omega^{**}}U_{x,\mu}$ realizes the minimum of the second functional: 
\begin{equation}
\min_{\Omega}F_{\rm LLG}[{}^{\Omega}U]=F_{\rm LLG}[{}^{\Omega^{**}}U] .
\end{equation}
%$\min_{\Omega}F_{\rm LLG}[U;\Omega]=F_{\rm LLG}[U;\Omega^{**}]$. 
Thus, imposing simultaneously two minimizing conditions removes the $SU(N)$ ambiguity 
\begin{equation}
 \Omega^{**}=G \Omega^{*}  \Longrightarrow G = \Omega^{**}(\Omega^{*})^{-1} ,
\end{equation}
and the color field configuration are decided as 
\begin{equation}
 {\bm{n}}_{x} =\Theta_{x} \bm{n}_{x}^{(0)}\Theta_{x}^{\dagger} ,
\quad
\Theta_{x} = G_{x}\Theta_{x}^{*} , 
\quad 
G_{x}=\Omega_{x}^{**}(\Omega_{x}^{*})^{-1}   .
\end{equation}
% with $\Theta_{x} = G_{x}\Theta_{x}^{*} $ and $G_{x}=\Omega_{x}^{**}(\Omega_{x}^{*})^{-1}$. 
We can choose the initial value $\bm{n}_{x}^{(0)}=\frac12 \sigma_{3}$ for $G=SU(2)$ and $\bm{n}_{x}^{(0)}=\frac12 \lambda_{8}$ for $G=SU(3)$.

It should be remarked that we can choose any overall gauge fixing condition other than the Landau gauge, since our reformulation respecting the original gauge symmetry is gauge independent. 
The Landau gauge is chosen by the reason why it respects the global gauge symmetry, i.e., color symmetry, which is important to discuss color confinement later.

Finally, we compare the reduction condition with the MA gauge. 
For $SU(2)$, the MA gauge is obtained by minimizing the functional:%
\footnote{
This is equivalent to maximizing the functional:
$\sum_{x,\mu}  {\rm tr}( \sigma_3 U_{x,\mu} \sigma_3  U_{x,\mu}^{\dagger})$.
}
\begin{align}
F_{\rm MAG}[U] &=    \sum_{x} \sum_{\mu=1}^{D} {\rm tr}( \bm{1} -\sigma_3 U_{x,\mu} \sigma_3  U_{x,\mu}^{\dagger})  ,
\end{align}
under the gauge transformation:
% $U_{x,\mu} \to U^{\Omega}_{x,\mu} $:
\begin{align}
\min_{\Omega} F_{\rm MAG}[{}^{\Omega}U], \quad
 {}^{\Omega}U_{x,\mu} = \Omega_{x} U_{x,\mu} \Omega_{x+\mu}^{\dagger} .
\end{align}
The resulting Abelian-projected theory has the remnant $U(1)$ symmetry, since this functional is invariant under the residual $U(1)$ gauge transformation: 
\begin{align}
 U_{x,\mu} \to  e^{i\theta_x \sigma_3} U_{x,\mu} e^{-i\theta_{x+\mu} \sigma_3} .
\end{align}
The MA gauge is regarded as an Abelian projection obtained by diagonalizing the operator:
\begin{align}
 \mathcal{X}(x) =  \sum_{\mu=1}^{D} [ U_{x,\mu} \sigma_3  U_{x,\mu}^{\dagger} + U_{x-\mu,\mu}^{\dagger} \sigma_3  U_{x-\mu,\mu} ] .
\end{align}

If we consider the limit in which the color field becomes uniform everywhere: $\bm{n}_{x} \to \frac12 \sigma_3$, then the reduction functional reduces to that of the MA gauge:
\begin{align}
F_{\rm red}[\bm{n};U] \to   F_{\rm MAG}[U]  .
\end{align}
In this limit, there  exist no more color field. 

Note that the gauge-transformed MAG functional:
\begin{align}
F_{\rm MAG}[{}^{\Omega}U]  =&    \sum_{x,\mu} {\rm tr}( \bm{1} -\sigma_3 \Omega_{x} U_{x,\mu} \Omega_{x+\mu}^{\dagger}  \sigma_3  \Omega_{x+\mu} U_{x,\mu}^\dagger \Omega_{x}^{\dagger}  )  
\nonumber\\
=&   \sum_{x,\mu} {\rm tr}[ \bm{1} - (\Omega_{x}^{\dagger}\sigma_3 \Omega_{x}) U_{x,\mu} (\Omega_{x+\mu}^{\dagger}  \sigma_3  \Omega_{x+\mu}) U_{x,\mu}^\dagger   ]  ,
\end{align}
has the same form as the reduction functional under the identification:
\begin{align}
 \bm{n}_x =   \Omega_{x}^{\dagger}\frac12 \sigma_3 \Omega_{x}, \quad \Omega_{x} \in SU(2)  .
\end{align}
Therefore, the color field $\bm{n}_{x}$ plays the role of the Abelian (diagonal) direction embed in the color space, which is allowed to change point to point.
Thus, the introduction of the color field enables us to recover the color symmetry lost by the Abelian projection.

%%%%%%%%%%%%%%%%%%%%%%%%%%%%%%%%%%%%%%%%%%%%%%%%%%%%%
\subsection{Reformulation of lattice $SU(2)$ Yang-Mills theory}
%%%%%%%%%%%%%%%%%%%%%%%%%%%%%%%%%%%%%%%%%%%%%%%%%%%%%

%%%%%%%%%%%%%%%%%%%%%%%%%%%%%%%%%%%%%%%%%%%%%%%%%%%%%
%\subsection{Reduction condition and change of variables}
%%%%%%%%%%%%%%%%%%%%%%%%%%%%%%%%%%%%%%%%%%%%%%%%%%%%%

First, we focus our attentions on the $SU(2)$ case, since the $SU(3)$ case will be studied in the next section.
%Such a decomposition can be performed by introducing another field: 
In the reformulation, we introduce a \textbf{color  direction  field}, or \textbf{color field} in short, as a site variable taking the values in the   Lie algebra  of the gauge group $SU(2)$:
\begin{align}
  \bm{n}_x  :=n_x^A T_A  \in Lie(SU(2)/U(1)) = su(2)-u(1) .
\end{align}
%where $\sigma_A$ $(A=1,2,3)$ are the Pauli matrices.
 Note  that  in the continuum the color field is defined by $\bm{n}(x) :=n^A(x)T_A$

By construction, the site variable $\bm{n}_{x}$ transforms under the gauge transformation  (which was called the gauge transformation II in \cite{KMS05})  as
\begin{align}
%U_{x,\mu} \rightarrow \Omega_{x} U_{x,\mu} \Omega_{x+\mu}^\dagger = U_{x,\mu}' , \quad
  \bm{n}_{x} \rightarrow \Omega_{x} \bm{n}_{x} \Omega_{x}^\dagger =: \bm{n}_{x}', \quad \Omega_{x}  \in SU(2)  .
  \label{C35-gt-n}
\end{align}

It is shown that the decomposition is uniquely determined by imposing the two requirements called the {\bf defining equation}:%
\begin{enumerate}
\item[(i)]
 the color field $\bm{n}_{x}$ is covariantly 
constant in the background (matrix) field $V_{x,\mu}$:
\begin{align}
 \bm{n}_{x} V_{x,\mu}  = V_{x,\mu} \bm{n}_{x+\mu} 
\Longleftrightarrow 
D_\mu^{(\epsilon)}[\mathscr{V}] \bm{n}_{x} = 0 ,
 \label{C35-Lcc}
\end{align}

\item[(ii)]
 the remaining (matrix) field $X_{x,\mu}$ is perpendicular to the color field  $\bm{n}_{x}$:%
\footnote{
This requirement can be replaced by the exact form in the compact formulation, see \cite{SKS10}. 
}
\begin{equation}
 {\rm tr}(\bm{n}_{x} X_{x,\mu} ) \equiv {\rm tr}(\bm{n}_{x} U_{x,\mu} V_{x,\mu}^\dagger) 
  = 0 .
  \label{C35-cond2m}
\end{equation}
\end{enumerate}
 Both conditions (i) and (ii) must be imposed to uniquely determine $V_{x,\mu}$ and $X_{x,\mu}=U_{x,\mu}V_{x,\mu}^{-1}$ for a given set of 
$U_{x,\mu}$ once the color field $\bm{n}_{x}$ is determined. 
They are the naive lattice version of the defining equations in the continuum.
In the naive continuum limit $\epsilon \rightarrow 0$, indeed, these defining equations reduce to the continuum counterparts.
It is important to remark that these defining equations are covariant or form-invariant under the gauge transformation, which is necessary for  the decomposed variables to have the desired transformation property (\ref{C35-gt-V}), (\ref{C35-gt-X}) and (\ref{C35-gt-n}). 
In fact, the  defining equation (\ref{C35-Lcc})   is form-invariant under the gauge transformation (\ref{C35-gt-n}) and (\ref{C35-gt-V}), i.e.,
$ \bm{n}_{x}^\prime V_{x,\mu}^\prime  = V_{x,\mu}^\prime \bm{n}_{x+\mu}^\prime
$. 
This is also the case for the second defining equation (\ref{C35-cond2m}):
$
 {\rm tr}(\bm{n}_{x}^\prime X_{x,\mu}^\prime ) =0 
$.

A lattice version of the orthogonality equation (\ref{C35-cond2m}) is given by 
\begin{equation}
{\rm tr}(\bm{n}_{x} \mathscr{X}_\mu(x))=0 ,
\end{equation}
or
\begin{equation}
 {\rm tr}(\bm{n}_{x} \exp \{-i\epsilon g \mathscr{X}_\mu(x)\})
  =  {\rm tr}(\bm{n}_{x}  \{ {\bf 1}-i\epsilon g  \mathscr{X}_\mu(x) \} ) + {\cal O}(\epsilon^2) = 0 + {\cal O}(\epsilon^2) .
  \label{C35-cond2}
\end{equation}
This implies that the trace vanishes  up to first order of $\epsilon$ apart from the second order term. 
Remembering  the relation $\mathscr{X}_\mu(x)=\mathscr{A}_\mu(x)-\mathscr{V}_\mu(x)$, we can rewrite   (\ref{C35-cond2}) into the second relation (\ref{C35-cond2m}) in terms of $\bm{n}_{x}$ and $U_{x,\mu}$.
Note that the orthogonality condition (\ref{C35-cond2m}) is gauge invariant. 
%\begin{equation}
% {\rm tr}(\bm{n}_{x} U_{x,\mu} V_{x,\mu}^\dagger) 
%  = 0 + {\cal O}(\epsilon^2) .
%  \label{C35-cond2m}
%\end{equation}

We can solve the defining equation (\ref{C35-Lcc}) for the link variable $V_{x,\mu}$ and express it in terms of the site variable $\bm{n}_{x}$ and the original link variable $U_{x,\mu}$, 
just as the continuum variable $\mathscr{V}_\mu(x)$ is expressed in terms of $\bm{n}(x)$ and $\mathscr{A}_\mu(x)$.
By solving the  defining equation (\ref{C35-Lcc}) and (\ref{C35-cond2m}), indeed, 
the link variable $V_{x,\mu}$ is determined  up to an overall normalization constant 
in terms of the site variable $\bm{n}_{x}$ and the original link variable 
$U_{x,\mu}$ \cite{IKKMSS06}:%
\footnote{
This special form has already been invented in a different context in the paper \cite{CGI98} in order to give the gauge-invariant lattice definition of Nambu magnetic monopole with quantized magnetic charge  in the $SU(2)$ Higgs model on a lattice, although we have given more general scheme to find such a form in the paper \cite{IKKMSS06}. 
%\bibitem{CGI98}
%M.N. Chernodub, F.V. Gubarev and E.-M. Ilgenfritz,
%Topological content of the electroweak sphaleron on the lattice,
%Phys. Lett. B {\bf 424}, 106---114 (1998).
%[hep-lat/9710011],
}
%[Exercise-1] \marginpar{Ex-1}
\begin{align}
  \tilde{V}_{x,\mu} = \tilde{V}_{x,\mu}[U,\bm{n}] 
  = U_{x,\mu} +  4\bm{n}_{x} U_{x,\mu} \bm{n}_{x+\mu} .
  \label{C35-sol}
\end{align}
The equation (\ref{C35-Lcc}) is linear in $V_{x,\mu}$. Therefore, the normalization of $V_{x,\mu}$ cannot be determined by this equation alone.
In general,  unitarity is not guaranteed for the general solution of the defining equation and hence a unitarity condition must be imposed afterwards. 
Fortunately, this issue is easily solved at least for $SU(2)$ group,  
since the specialty condition $\det V_{x,\mu}=1$ determines the normalization. 
Then the special unitary link variable $V_{x,\mu}[U,\bm{n}]$ is obtained after the normalization: 
%[Exercise-2] \marginpar{Ex-2}
\begin{align} 
V_{x,\mu} = 
V_{x,\mu}[U,\bm{n}] := 
 \tilde{V}_{x,\mu}/\sqrt{\frac{1}{2}{\rm tr} [\tilde{V}_{x,\mu}^{\dagger}\tilde{V}_{x,\mu}]} .
\label{C35-cfn-mono-4}
\end{align}
It is  shown \cite{IKKMSS06} that the naive continuum limit $\epsilon \rightarrow 0$ 
of the link variable $V_{x,\mu} = \exp (-i\epsilon g \mathscr{V}_\mu(x))$ reduces to the continuum expression:
%[Exercise-3] \marginpar{Ex-3}
\begin{align} 
\mathscr{V}_{\mu}(x) = (n^A(x)A_{\mu}^A(x))\bm{n}(x)
-ig^{-1} [\partial_{\mu}\bm{n}(x) ,  \bm{n}(x) ] ,
\label{C35-cfn-conti-7} 
\end{align}
which  agrees with the expression of the restricted field in the Cho-Duan-Ge decomposition  in the continuum \cite{Cho80,DG79}. 
This is indeed the case for the remaining variable $X_{x,\mu} = \exp (-i\epsilon g  \mathscr{X}_\mu(x))$.

 By including the color field $\bm{n}_x$, the $SU(2)$ Yang-Mills theory written in terms of $U_{x,\mu}$ is extended to a gauge theory written in terms of $U_{x,\mu}$ and $\bm{n}_x$ with the enlarged local gauge symmetry $\tilde{G}_{\omega,\theta}^{local}=SU(2)^{local}_{\omega} \times [SU(2)/U(1)]^{local}_{\theta}$ 
larger than the local gauge symmetry $SU(2)^{local}_{\omega}$ in the original 
Yang-Mills theory \cite{KMS05}. 
%It has been shown that the extended $SU(2)$ Yang-Mills theory written in terms of 
%$U_{x,\mu}$ and ${\bf n}_x$ has the enlarged local gauge symmetry $\tilde{G}^{\omega,\theta}_{local}=SU(2)_{local}^{\omega} \times [SU(2)/U(1)]_{local}^{\theta}$ 
%larger than the local gauge symmetry $SU(2)_{local}^{\omega}$ in the original Yang-Mills theory \cite{KMS05}. 
  In order to eliminate the extra degrees of freedom in the enlarged local gauge symmetry 
$\tilde{G}_{\omega,\theta}^{local}$ for obtaining the Yang-Mills theory which is equipollent to the original Yang-Mills theory, 
we must impose sufficient number of constraints, which we called the \textbf{reduction condition}.

We find that such a  reduction condition is given on a lattice by minimizing the functional:
\begin{align}
F_{\rm red}[\bm{n};U] &=  \sum_{x,\mu} \left[{\rm tr}( \bm{1} - 
                   4\bm{n}_xU_{x,\mu}\bm{n}_{x+\hat{\mu}}U_{x,\mu}^{\dagger})  \right] ,
\end{align}
with respect to the color fields $\{\bm{n}_x\}$ for a given set of link variables $\{U_{x,\mu}\}$.
Thus color field $\bm{n}_x$ is determined by $\bm{n}_x=\bm{n}_x^*$ in such a way that the functional achieves the minimum at $\bm{n}_x=\bm{n}_x^*$:
\begin{align}
{\rm min}_{\bm{n}} F_{\rm red}[\bm{n};U] = F_{\rm red}[\bm{n}^*; U].
\label{C35-reduction_n}
\end{align}
The two algorithms for solving the reduction equation are available, as already mentioned in the previous subsection.
%:
%\begin{enumerate}
%\item[(i)]
%Updating $\{ \bm{n}_x \}$ in terms of the heat bath method.
%Updating $\{ \bm{n}_x \}$ via gauge transformation for  solving the reduction condition.
% This method  of the reduction prescription  was adopted in the early studies \cite{IKKMSS06,SKKMSI07}. (This was called the new MA gauge.) 

%\item[(ii)]
%We solve the stationary condition:
%\begin{align}
%\frac{\partial F_{\rm red}[\bm{n};U]}{\partial n^A_x} =0 ,
%\end{align}
%in order to minimize the functional $F_{\rm red}$.
%This method of the reduction prescription was adopted in the recent studies \cite{KKS14}. 
%\end{enumerate}
%The functional $F_{\rm red}$ can be rewritten in the following way:
%\begin{align}
%F_{\rm red}[\bm{n};U] 
%=&  \sum_{<x,y>} (1-J^{AB}_{x,y}[U]n^A_xn^B_y)    ,   \ 
%\nonumber\\
%J^{AB}_{x,y}[U] =   {\rm tr}(\sigma^A U_{x,\mu}\sigma^B U_{x,\mu}^{\dagger})/{\rm tr}({\bf 1}) .
%\end{align}
%Therefore, the functional $F_{\rm red}$ can be regarded as the energy for the spin-glass system.
%There exist local minima which satisfy the reduction  condition. 
%There is no known method for finding a global minimum of the functional. 
%Therefore, overrelaxation method/replica exchange (simulated annealing) method should be used in order to approach the global minimum more rapidly. 

%%%%%%%%%%%%%%%%%%%%%%%%%%%%%%%%%%%%%%%%%%%%%%%%%%
%%%%%%%%%%%%%%%%%%%%%%%%%%%%%%%%%%%%%%%%%%%%%%%%%%
\subsection{Numerical simulations of lattice $SU(2)$ Yang-Mills theory} 
%%%%%%%%%%%%%%%%%%%%%%%%%%%%%%%%%%%%%%%%%%%%%%%%%%
%%%%%%%%%%%%%%%%%%%%%%%%%%%%%%%%%%%%%%%%%%%%%%%%%%

In what follows, we present the results of numerical simulations. 
First of all, we generate the configurations of $SU(2)$ link variables 
$\{ U_{x,\mu} \}$, 
%$ U_{x,\mu}= \exp [ - ig\epsilon \mathscr{A}_\mu(x) ]$,
using the  (pseudo)  heat bath method for the standard Wilson action. 

Second, we generate the configurations of the color field $\{{\bf n}_x\}$ using the reduction condition (\ref{C35-reduction_n}) for the obtained configurations of $SU(2)$ link variables $\{U_{x,\mu}\}$. 
Then we can construct the restricted field $\{V_{x,\mu}[U,{\bf n}]\}$ according to the change of variables (\ref{C35-cfn-mono-4}).
Moreover, we can construct the magnetic-monopole current $\{k_{x,\mu}\}$ according to (\ref{C35-cfn-conti-20}).

%%%%%%%%%%%%%%%%%%%%%%%%%%%%%%%%%%%%%%%%%%%%%%%%%%
\subsubsection{Color direction field}
%%%%%%%%%%%%%%%%%%%%%%%%%%%%%%%%%%%%%%%%%%%%%%%%%%

The preliminary numerical simulations \cite{KKMSSI06} are performed on the lattice  with the lattice size $8^{4}$ and $16^{4}$ by using the standard Wilson action  for the gauge coupling $\beta=2.2\sim 2.45$ and periodic boundary conditions. 
For $8^4 $ (resp. $16^4$) lattice, we have obtained  50 (resp. 200) configurations (samples)  at intervals of 100 sweeps by starting with cold initial condition and thermalizing 30$\times$100 (resp. 50$\times$100) sweeps.

\begin{figure}[tbp]
%\begin{picture}(0,0)%(0,-3000)
%\put(0,-7500){
\includegraphics[scale=0.4]{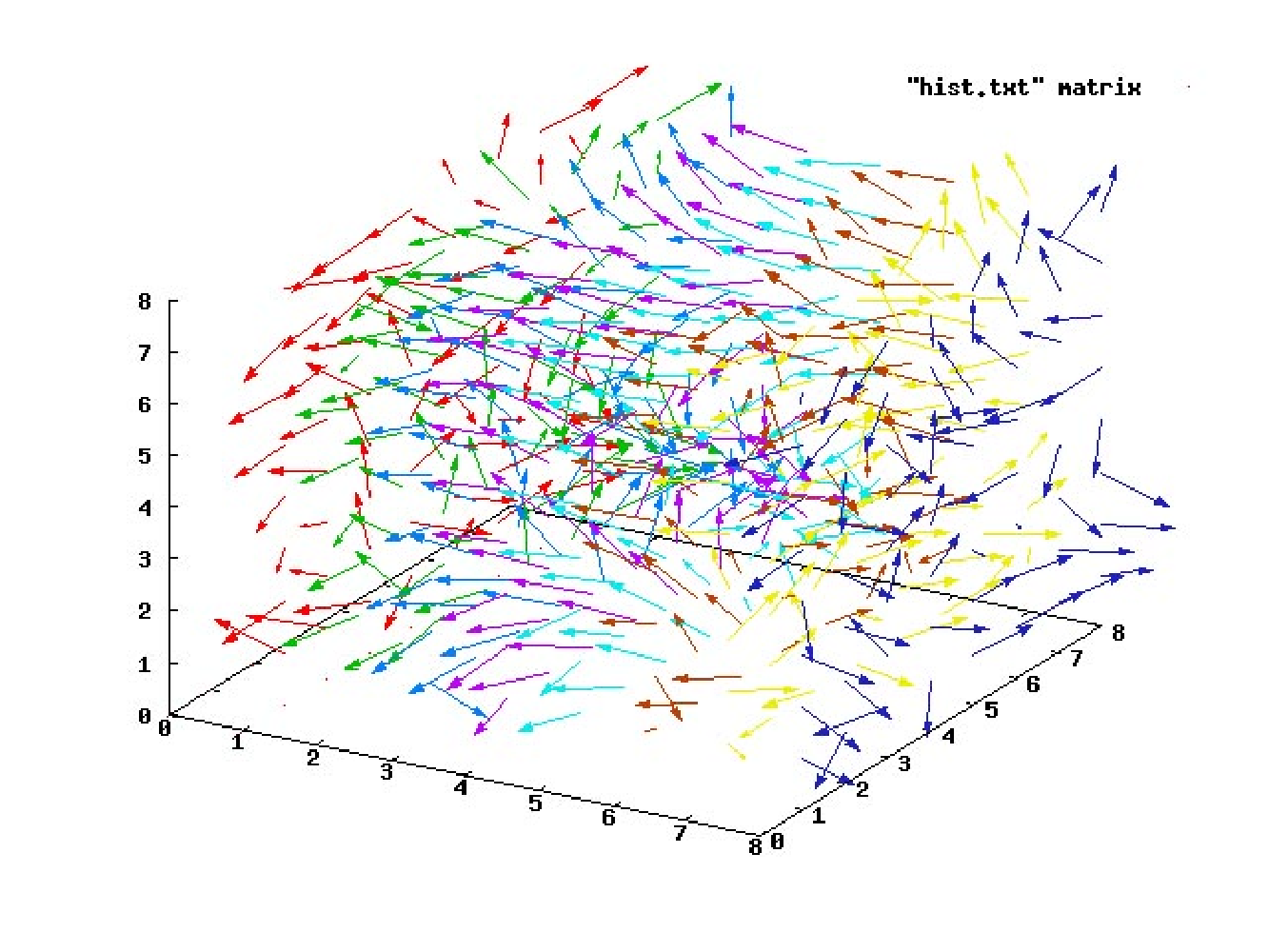}
%}%
\includegraphics[scale=0.4]{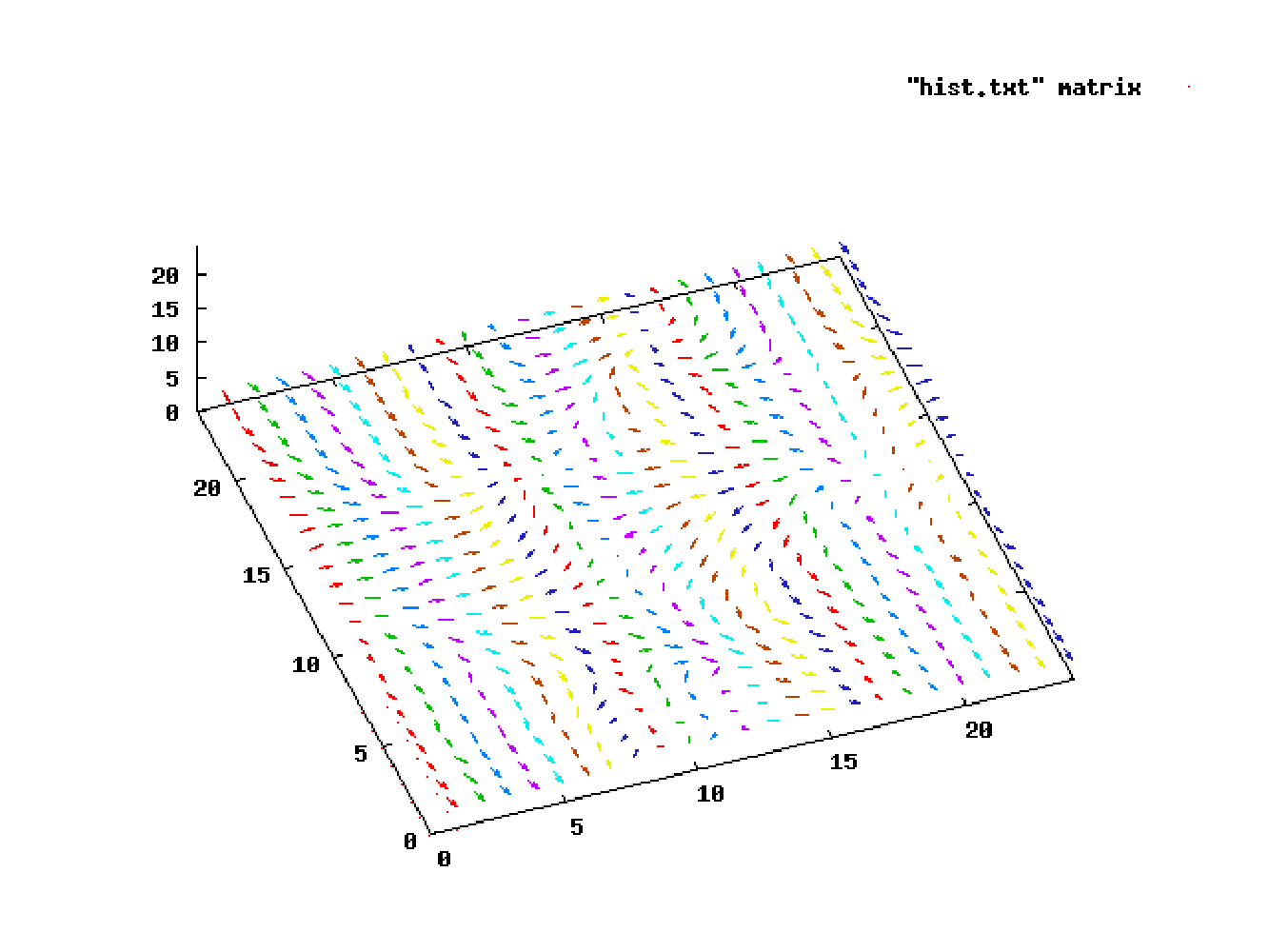}
%\end{picture}
%\vskip 17cm
\caption{%\small 
The color field configurations $\{ \mathbf{n}_{x} \}$ in  $SU(2)$ Yang-Mills theory:
(Left) 
  three-dimensional vectors on a 8$^4$ lattice.
(Right) hedgehog  configurations  on a 16$^4$ lattice.
}
\label{C35-fig:nconfig2}
%\label{C35-fig:gtest}
%\end{center}
\end{figure}

First, we show the results for the color direction field. 
A typical sample of color field configurations generated by numerical simulations are shown in 
Fig.~\ref{C35-fig:nconfig2}.
%Fig.~\ref{C35-fig:nconfig} and Fig.~\ref{C35-fig:hedgehog}.

%%%%%%%%%%%%%%%%%%%%% figures %%%%%%%%%%%%%%%%%%%%%%%
%\begin{figure}[tbp]
%\begin{center}
%%\begin{picture}(0,0)%(0,-3000)
%%\put(0,-7500){\includegraphics[height=19cm]{Fig-PR/Fig-lattice/88.eps}}%
%\includegraphics[height=7cm]{Fig-PR/Fig-lattice/88.eps}
%%\label{C35-fig:nconfig}
%%\end{picture}
%%\vskip 17cm
%\caption{%\small 
%The color field $\{ \mathbf{n}_{x} \}$  as a three-dimensional vector generated in  $SU(2)$ Yang-Mills theory on a 8$^4$ lattice.
%}
%\label{C35-fig:nconfig}
%\end{center}
%\end{figure}
%%%%%%%%%%%%%%%%%%%%% figures %%%%%%%%%%%%%%%%%%%%%%%

%%%%%%%%%%%%%%%%%%%%% figures %%%%%%%%%%%%%%%%%%%%%%%
%\begin{figure}[htbp]
%\begin{center}
%%\begin{picture}(0,0)%(0,-3000)
%%\put(-700,-7500){\includegraphics[height=25cm]{Fig-PR/Fig-lattice/loop358.eps}}%
%\includegraphics[height=7.0cm,width=10cm]{Fig-PR/Fig-lattice/loop358.eps}
%\label{C35-fig:nconfig2}
%\end{picture}
%\caption{%\small 
%hedgehog  configurations of the color field $\{ \mathbf{n}_{x} \}$ in  $SU(2)$ Yang-Mills theory on a 16$^4$ lattice. 
%}
%\label{C35-fig:hedgehog}
%\end{center}
%\end{figure}
%%%%%%%%%%%%%%%%%%%%% figures %%%%%%%%%%%%%%%%%%%%%%%

%%%%%%%%%%%%%%%%%%%%%%%%%%%%%%%%%%%%%%%%%%%%%%%%%%%%%%%%%%%%
%%%%%%%%%%%%%%%%%%%%%%%%%%%%%%%%%%%%%%%%%%%%%%%%%%%%%

%%%%%%%%%%%%%%%%%%%%% Table %%%%%%%%%%%%%%%%%%%%%%%%%%%
\begin{table}[hptb]
\begin{center}
\begin{tabular}{cll}\hline
      & Mean value & Jack knife error(JKbin=2) \\ \hline
$<n^1>$ & -0.0069695  & $\pm$ 0.010294   \\
$<n^2>$ & 0.011511 & $\pm$ 0.015366 \\
$<n^3>$ & 0.0014141 & $\pm$ 0.013791 \\ \hline
\end{tabular}
\caption{\cite{KKMSSI06} The magnetization $<n^A_x>$ on the $16^4$ lattice at $\beta=2.4$.}
\label{C35-table:vevn}
\end{center}
\end{table}
%%%%%%%%%%%%%%%%%%%%% figures %%%%%%%%%%%%%%%%%%%%%%%%%%%

The correlation functions for the color field are as follows. 
The results of the numerical simulations in Table~\ref{C35-table:vevn} show that $n^A_x$ has the vanishing vacuum expectation value:
\begin{align}
  \left< n^A_x \right> = 0  \quad (A=1,2,3) . 
\end{align}

\begin{figure}[ptb]
\begin{center}
\includegraphics[height=5cm]{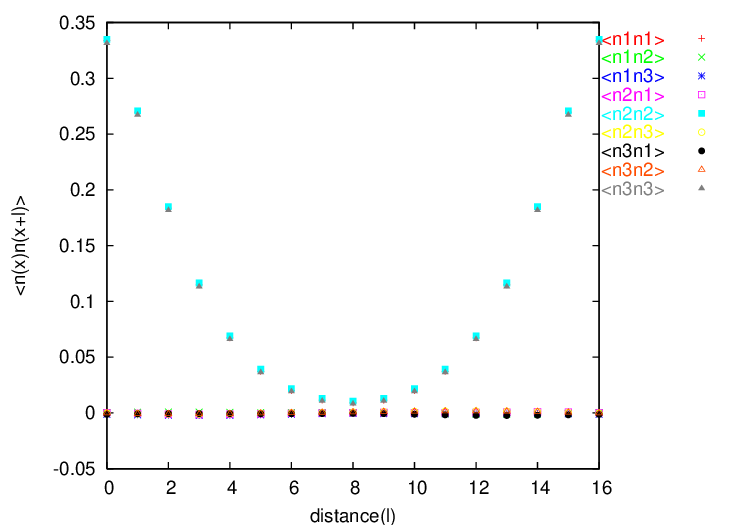}
\caption{\small 
Ref.\cite{KKMSSI06}: The plots of two-point correlation functions 
$\left< n_x^A n_0^B \right>$ for $A,B=1,2,3$
along the lattice axis on the $16^4$ lattice at $\beta=2.4$.
%Staring with cold initial condition and thermalizing 50*100sweeps, we have obtained  200 configurations (samples) at intervals of 100 sweeps.
}
\label{C35-fig:2ptf}
\end{center}
\end{figure}

Moreover, we have measured the two-point correlation functions defined by $\left< n^A_x n^B_0 \right>$, see Fig.~\ref{C35-fig:2ptf}.
The two-point correlation functions $\left< n^A_x n^A_0 \right>$  (no summation over $A$) exhibit almost the same behavior in all the directions ($A=1,2,3$), while $\left< n^A_x n^B_0 \right>$ ($A\not=B$) vanish. Thus, 
we have obtained the correlation function
%$\left< n_x^A n_0^B \right>=\delta^{AB}D(x)$ 
respecting color symmetry:
\begin{align}
  \left< n_x^A n_0^B \right>=\delta^{AB}D(x) \quad (A,B=1,2,3) . 
\end{align}
These results indicate that {\it the global $SU(2)$ symmetry (color symmetry) is   unbroken in our main simulations}.
This property cannot be realized in the MA gauge. 
This is a first remarkable result.

%%%%%%%%%%%%%%%%%%%%% figures %%%%%%%%%%%%%%%%%%%%%%%%%%%
%%%%%%%%%%%%%%%%%%%%% figures %%%%%%%%%%%%%%%%%%%%%%%%%%%

According to our viewpoint \cite{KMS05}, the $\bm{n}_{x}$ field  must respect the global $SU(2)^{\prime}$ symmetry. If this is not the case, the field $\bm{n}_{x}$ cannot be identified with the color field in the CDGFN decomposition of the original gauge potential from our viewpoint. This is a crucial point. 
This is  in sharp contrast to the previous approaches.  
Although the similar technique of constructing the unit vector field $\bm{n}_x$ from a $SU(2)$ matrix $G_{x}$ has already  appeared,  there is a crucial conceptual difference between our approach and others, e.g., \cite{DHW02b,Shabanov01,IS00}.
%\footnote{
%\bibitem{Shabanov01}
%S.V. Shabanov,
%Infrared Yang-Mills theory as a spin system,
%[hep-lat/0110065],
%Phys. Lett. B {\bf 522}, 201%--209
%(2001).
%\bibitem{DHW02b}
%L. Dittmann, T. Heinzl and A. Wipf,
%Effective sigma models and lattice Ward identities,
%[hep-lat/0210021],
%JHEP{\bf 0212}, 014 (2002).
%\bibitem{IS00}
%H. Ichie and H. Suganuma,
%Monopoles and gluon fields in QCD in the maximally abelian gauge,
%[hep-lat/9808054], 
%Nucl. Phys. B {\bf 574}, 70%--106
%(2000).  
%}  

We can perform the global $SU(2)$ rotation at will, since it is not prohibited in our setting.
However, the previous numerical simulations are performed only in a restricted setting where LLG and MA gauge are close to each other by imposing LLG as a preconditioning,
in the sense that the matrices $G$ connecting LLG and MA gauge are on average close to the unit ones.
That is to say, 
$G_x^A \cong 0$ $(A=1,2,3)$, i.e., $G_x \cong g_x^0 I$, 
for the parameterization of $SU(2)$ matrices, 
$
  G_x = G_x^0 I + i G_x^A \sigma^A
$,  
$  
  G_x^0, G_x^A \in \mathbb{R}
$,  
$
   \sum_{\mu=0}^{3} (G_x^\mu)^2 = 1 .
$
Then it has been observed that 
$\bm{n}_x \cong \sigma^3$ or $n_x^A \cong (0,0,1)$, namely,  
 $\bm{n}_x$ are  aligned in the positive 3-direction and hence the non-vanishing vacuum expectation value is observed as 
\begin{align}
  \left< n^A_x \right> = M \delta^{A3} . 
\end{align}
This implies that the global $SU(2)$ symmetry is broken explicitly to a global $U(1)$,
 $SU(2)_{\text{global}} \rightarrow U(1)_{\text{global}}$. 
In the two-point correlation functions, the exponential decay has been observed for 
the parallel propagator  
\begin{align}
  \left< n_x^3 n_0^3 \right> \sim \left< n_0^3 \right>\left< n_0^3 \right> + c e^{-m|x|} = M^2 + c e^{-m|x|} , 
\end{align}
and for the perpendicular propagator
\begin{align}
   \sum_{a=1}^{2} \left< n_x^a n_0^a \right> \sim  c' e^{-m'|x|} ,
\end{align}
with $m$ and $m'$ being different to each other. 
This result was reported in \cite{DHW02b} (and confirmed also by our preliminary simulations \cite{DSB04}). 
%\footnote{
%\bibitem{DSB04}
% K.-I. Kondo, 
%An invited talk given at Workshop on Dynamical Symmetry Breaking, Nagoya University, Dec. 21--22, 2004. 
%The pdf file is available at the homepage:
%http://www.eken.phys.nagoya-u.ac.jp/dsb04/frameset.html?=
%}

In our approach we can identify the lattice field $\bm{n}_x$ as a lattice version of the CDGFN field variable $\bm{n}(x)$ obtained by the CDGFN decomposition of the gauge potential $\mathscr{A}_\mu(x)$ in the original Yang-Mills theory in agreement with the new viewpoint. Moreover, we do not assume any effective theory of Yang-Mills theory written in terms of the unit vector field $\bm{n}_x$, such as the Skyrme-Faddeev model.

%%%%%%%%%%%%%%%%%%%%%%%%%%%%%%%%%%%%%%%%%%%%%%%%%%
%%%%%%%%%%%%%%%%%%%%%%%%%%%%%%%%%%%%%%%%%%%%%%%%%%
\subsubsection{Magnetic monopole and magnetic charge quantization}
%%%%%%%%%%%%%%%%%%%%%%%%%%%%%%%%%%%%%%%%%%%%%%%%%%
%%%%%%%%%%%%%%%%%%%%%%%%%%%%%%%%%%%%%%%%%%%%%%%%%%

By solving the defining equation, 
 the special unitary $SU(2)$ link variable $V_{x,\mu}$ is obtained in terms of the site variable $\bm{n}_{x}$ and the original $SU(2)$ link variable  $U_{x,\mu}$:
\begin{align} 
V_{x,\mu} = 
V_{x,\mu}[U,\bm{n}]  = 
 \tilde{V}_{x,\mu}/\sqrt{\frac{1}{2}{\rm tr} [\tilde{V}_{x,\mu}^{\dagger}\tilde{V}_{x,\mu}]} ,
\ 
  \tilde{V}_{x,\mu}  
  := U_{x,\mu} +  4\bm{n}_{x} U_{x,\mu}\bm{n}_{x+\mu} .
\label{C35-cfn-mono-4b}
\end{align}
%It is easy to show that in the naive continuum limit $\epsilon \rightarrow 0$   the link variable $V_{x,\mu} = \exp (-i\epsilon g {\bf V}_\mu(x))$ reduces to 
%\begin{align} 
%{\bf V}_{\mu}(x) = (n^A(x)A_{\mu}^A(x)){\bf n}(x) +\frac{1}{g}\partial_{\mu}{\bf n}(x)\times {\bf n}(x),
%\label{C35-cfn-conti-7} 
%\end{align}
%which is nothing but the continuum expression of CFN variable. 
Note that the $V_{x,\mu}$ transforms like a usual link variable under the (equipollent) gauge 
transformation as
\begin{equation}
  V_{x,\mu} \rightarrow \Omega_{x} V_{x,\mu} \Omega_{x+\mu}^\dagger = V_{x,\mu}' .
\end{equation}

In the reformulation, therefore, we can define the \textbf{gauge-invariant field strength} $\bar{\Theta}_P[U,{\bf n}]$, i.e., { gauge-invariant flux} 
 as a plaquette variable $\bar{\Theta}_P[U,\bm{n}]$ on a lattice by \cite{IKKMSS06}
\begin{align} 
\bar{\Theta}_{x,\mu\nu}[U,\bm{n}] := \epsilon^{-2}
{\rm arg} [ {\rm tr} \{({\bf 1}+2\bm{n}_x)V_{x,\mu}V_{x+\hat{\mu},\nu}
V_{x+\nu,\mu}^{\dagger}V_{x,\nu}^{\dagger} \}/{\rm tr}({\bf 1})] .
\label{C35-cfn-mono-5}
\end{align}
In fact, taking into account the relation:
\begin{align} 
 V_{P} :=&  V_{x,\mu}V_{x+\hat{\mu},\nu}
V_{x+\nu,\mu}^{\dagger}V_{x,\nu}^{\dagger} 
 = \exp \{ -i\epsilon^2 g \mathscr{F}_{\mu\nu}[\mathscr{V}] \}
,
\nonumber\\
 \mathscr{F}_{\mu\nu}[\mathscr{V}]  :=& \partial_\mu \mathscr{V}_{\nu} - \partial_\nu \mathscr{V}_{\mu} -i g [ \mathscr{V}_{\mu} ,  \mathscr{V}_{\nu} ] ,
\end{align}
and the expansion:
\begin{align} 
& V_{P} =   {\bf 1} -i\epsilon^2 g \mathscr{F}_{\mu\nu}[\mathscr{V}] + O(\epsilon^4) ,
 \quad 
  {\rm tr} ( V_{P} ) =  {\rm tr} ({\bf 1} ) + O(\epsilon^4) ,
\nonumber\\
& {\rm tr} ( \bm{n}_x V_{P} )  
=    -i\epsilon^2 g {\rm tr} ( \mathscr{F}_{\mu\nu}[\mathscr{V}] \bm{n}_x ) + O(\epsilon^4) 
=  - i\epsilon^2 g \frac12  \mathscr{F}_{\mu\nu}[\mathscr{V}] \cdot \bm{n}_x   + O(\epsilon^4)
 ,
\end{align}
it is shown  that the naive continuum limit of (\ref{C35-cfn-mono-5}) reduces to
the gauge-invariant field strength:% (see Appendix A.2 of \cite{IKKMSS06}):
\begin{align} 
\bar{\Theta}_{x,\mu\nu} \simeq&
\partial_{\mu}(n^A(x) \mathscr{A}_{\nu}^A(x))-\partial_{\nu}(n^A(x) \mathscr{A}_{\mu}^A(x))
-i g^{-1} \bm{n}\cdot [\partial_{\mu}\bm{n}, \partial_{\nu}\bm{n} ]
\nonumber\\
=&\frac{-1}{2} \bm{n} \cdot \mathscr{F}_{\mu\nu}[\mathscr{V}]  .
%\quad 
%{\bf F}_{\mu\nu}[{\bf V}]  := \partial_\mu {\bf V}_{\nu} - \partial_\nu {\bf V}_{\mu} + g {\bf V}_{\mu} \times {\bf V}_{\nu} , 
\label{C35-cfn-fs} 
\end{align}
%\begin{equation} 
%\bar{\Theta}_{x,\mu\nu} \simeq
%\partial_{\mu}(n^A(x)\mathscr{A}_{\nu}^A(x))-\partial_{\nu}(n^A(x)\mathscr{A}_{\mu}^A(x))
%+ g^{-1} {\bf n}\cdot (\partial_{\mu}{\bf n}\times\partial_{\nu}{\bf n})
%=\frac{-1}{2}{\rm tr}(2\bm{n}F_{\mu\nu}[\mathscr{V}]),
%\label{C35-cfn-fs} 
%\end{equation}
Here, $\bar{\Theta}_{x,\mu\nu}$  plays the similar role to the 't Hooft  tensor in describing the 
't Hooft-Polyakov magnetic monopole in Georgi-Glashow model.

This definition  for $\bar{\Theta}_{x,\mu\nu}[U,\bm{n}]$ is $SU(2)$ gauge invariant due to   cyclicity of the trace and the transformation property of the link variable and the site variable.
Even if we insert the factor $({\bf 1}+2\bm{n}_x)$ at different corners of a plaquette, $\bar{\Theta}_{x,\mu\nu}[U,\bm{n}]$ does not change  thanks to the (first) defining equation,  e.g., 
$
\bar{\Theta}_{x,\mu\nu}[U,\bm{n}] \equiv \epsilon^{-2}
{\rm arg} [ {\rm tr} \{V_{x,\mu}({\bf 1}+2\bm{n}_{x+\hat{\mu}}) V_{x+\hat{\mu},\nu}
V_{x+\nu,\mu}^{\dagger} V_{x,\nu}^{\dagger} \}]
$.
It should be remarked that the lattice definition which reduces to the continuum form $ \bm{n} \cdot \mathscr{F}_{\mu\nu}[\mathscr{V}]$ in the naive continuum limit is not unique,
 e.g., 
${\rm tr} \{ {\bf 1}+2 \bm{n}_x V_{P} \}$ has the same form as 
${\rm tr} \{({\bf 1}+ 2\bm{n}_x)V_{P} \}$
up to $O(\epsilon^2)$. 
The advantage of using the form ${\rm tr} \{({\bf 1}+ 2\bm{n}_x)V_{P} \}$ is that it guarantees the quantization of the magnetic charge explained below.

Then we can define the \textbf{gauge-invariant magnetic-monopole current}. 
We  use  the gauge-invariant field strength (\ref{C35-cfn-mono-5}) to extract  configurations of 
the  magnetic-monopole current $\{ K_{x,\mu}\}$  defined by
\begin{align}
K_{x,\mu}= 2\pi m_{x,\mu}, \quad 
m_{x,\mu} = 
-\frac{1}{4\pi}{\varepsilon}_{\mu\nu\rho\sigma}
\partial_{\nu}\bar{\Theta}_{x+\mu,\rho\sigma} \in \mathbb{Z} .
\label{C35-cfn-conti-20}
\end{align}
This definition of the magnetic-monopole  current $\{ K_{x,\mu}\}$ agrees with our definition of the magnetic-monopole current in the continuum limit (divided by $2\pi$).
This definition satisfies the quantization of the magnetic charge as will be shown shortly.

%The configurations of the (integer-valued) magnetic-monopole current $\{k_{x,\mu}\}$  are extracted from the gauge-invariant field strength (\ref{C35-cfn-mono-5}) by
%\begin{align}
%k_{x,\mu}= -\frac{1}{4\pi}{\varepsilon}_{\mu\nu\rho\sigma} \partial_{\nu}\bar{\Theta}_{x+\mu,\rho\sigma} .
%\label{C35-cfn-conti-20}
%\end{align}
%This definition agrees with our definition of the magnetic monopole in the continuum (divided by $2\pi$).

In our formulation, on the other hand, we have only the real variable $\bar{\Theta}_P[U,\bm{n}]$ at hand, and we are to calculate the magnetic-monopole current using the final term in (\ref{C35-cfn-conti-20}).  Therefore, it is not so trivial to obtain the integer-valued $K_{x,\mu}$ from the real-valued $\bar{\Theta}_P[U,\bm{n}]$.

We  show  quantization of the magnetic charge  as follows. 
We have constructed a color field  according to the adjoint orbit representation:
\begin{equation}
 \bm{n}_{x} = \Theta_{x} \sigma_3 \Theta_{x}^\dagger , \quad \Theta_{x} \in SU(2) , 
\end{equation}
which yields 
\begin{align}
  \tilde{V}_{x,\mu} 
%= V_{x,\mu}[U,{\bf n}] 
  =& U_{x,\mu} + 4 \bm{n}_{x} U_{x,\mu} \bm{n}_{x+\mu} 
  = \Theta_{x} [  \tilde{U}_{x,\mu} + \sigma_3  \tilde{U}_{x,\mu} \sigma_3] \Theta_{x+\mu}^\dagger , 
  \nonumber\\
   \tilde{U}_{x,\mu} :=&  \Theta_{x}^\dagger U_{x,\mu} \Theta_{x+\mu} .
  \label{C35-sol2}
\end{align}
By representing an $SU(2)$ element $\tilde{U}_{x,\mu}$ in terms of Euler angles and Pauli matrices:
\begin{align}
   \tilde{U}_{x,\mu} :=  \Theta_{x}^\dagger U_{x,\mu} \Theta_{x+\mu} 
   = e^{i\sigma_3 \chi_{\ell}/2} e^{i \sigma_2 \theta_{\ell}/2} e^{i\sigma_3 \varphi_{\ell}/2} , \ 
   \theta_\ell \in [0,\pi), \varphi_\ell, \chi_\ell \in [-\pi,\pi), 
\end{align}
the link variable $\tilde{V}_{x,\mu}$ for a link $\ell=(x,\mu)$ has the representation:
\begin{align}
  \tilde{V}_{x,\mu} =2 \cos \frac{\theta_{\ell}}{2}  V_{x,\mu}, \quad
  V_{x,\mu} = \Theta_{x}
  \begin{pmatrix} e^{i(\varphi_{\ell}+\chi_{\ell})/2} & 0 \\
  0 & e^{-i(\varphi_{\ell}+\chi_{\ell})/2} \\
  \end{pmatrix} \Theta_{x+\mu}^\dagger ,
\label{C35-sol3}
\end{align}
with the normalization factor  
$
 \sqrt{\frac{1}{2} {\rm tr}\tilde{V}_{x,\mu}\tilde{V}_{x,\mu}^\dagger} 
 = \sqrt{4 \cos^2 \frac{\theta_{\ell}}{2}}
 = 2 \cos \frac{\theta_{\ell}}{2}
$.

Thus the gauge-invariant flux  is rewritten in terms of a compact variable $\Phi_\ell:=(\varphi_{\ell}+\chi_{\ell})/2 \in [-\pi,\pi)$:
\begin{align} 
\bar{\Theta}_{x,\mu\nu}[U,\bm{n}] 
 :=& \epsilon^{-2}
{\rm arg} \left[ {\rm tr} \{({\bf 1}+2\bm{n}_x) V_{x,\mu} V_{x+\hat{\mu},\nu}
V_{x+\nu,\mu}^{\dagger} V_{x,\nu}^{\dagger} \}/{\rm tr}({\bf 1}) \right] 
%\nonumber\\
%&=& \epsilon^{-2}
%{\rm arg} ( {\rm tr} \{\Theta_{x}^\dagger({\bf 1}+{\bf n}_x)\Theta_{x} 
%%\hat{V}_{x,\mu}\hat{V}_{x+\hat{\mu},\nu}
%%\hat{V}_{x+\nu,\mu}^{\dagger}\hat{V}_{x,\nu}^{\dagger} 
%\begin{pmatrix} e^{i \sum_{\ell\in P}\Phi_{\ell}} & 0 \\
%  0 & e^{-i\sum_{\ell\in P}\Phi_{\ell}} \\
%  \end{pmatrix} 
%\}/{\rm tr}({\bf 1})) .
\nonumber\\
 =& \epsilon^{-2}
{\rm arg} \left[ {\rm tr} \left\{ ({\bf 1}+\sigma_3) 
%\hat{V}_{x,\mu}\hat{V}_{x+\hat{\mu},\nu}
%\hat{V}_{x+\nu,\mu}^{\dagger}\hat{V}_{x,\nu}^{\dagger} 
\begin{pmatrix} e^{i \sum_{\ell\in P}\Phi_{\ell}} & 0 \\
  0 & e^{-i\sum_{\ell\in P}\Phi_{\ell} } \\
  \end{pmatrix} 
\right\}/{\rm tr}({\bf 1}) \right] 
\nonumber\\
 =&  \epsilon^{-2}
{\rm arg} \exp \{ i \Phi_{P} \} 
= [\Phi_{P}]_{{\rm mod}~2\pi}, 
\end{align}
where
\begin{equation}
 \Phi_{P} :=  (d\Phi)_{P} = \sum_{\ell\in P}\Phi_{\ell} 
 = \Phi_{x,\mu}+\Phi_{x+\mu,\nu}-\Phi_{x+\nu,\mu}-\Phi_{x,\nu} .
\end{equation}
It is important to remark that  $\bar{\Theta}_{x,\mu\nu}$ on a lattice is a compact variable whose range is $[-\pi,\pi)$, although it reduces to the continuum counterpart which is non-compact variable taking the value $(-\infty,\infty )$   in the continuum limit. 
This fact is crucial to  quantization of magnetic charge. 
In the unitary gauge, $\bm{n}_{x} \equiv \sigma_3$, 
which corresponds to $\Theta_{x}\equiv {\bf 1}$ in the above argument, 
$\bar{\Theta}_{x,\mu\nu}[U,\bm{n}]$ agrees with
$\bar{\theta}_{x,\mu\nu}$ in  the DeGrand-Toussaint field strength
where the Abelian tensor $\theta_{\mu\nu}(s) \in [-4\pi,4\pi) \subset \mathbb{R}$ is decomposed into the field strength part $\bar{\theta}_{\mu\nu}(s) \in [-\pi,\pi) \subset \mathbb{R}$
and Dirac string part $n_{\mu\nu}(s)\in  \{-2,-1,0,1,2\} \subset \mathbb{Z}$:
$
 \bar{\theta}_{\mu\nu}(s)=\theta_{\mu\nu}(s)-2\pi n_{\mu\nu}(s) .
$
It is known  that the elementary monopole defined in this way takes an integer-value $[-2,2]$, since the Bianchi identity holds for $\theta_{\mu\nu}(s)$.

The  definition (\ref{C35-cfn-conti-20}) of the magnetic-monopole current should be compared with the conventional magnetic-monopole current on a lattice defined according to DeGrand and Toussaint  through  link variables on the dual lattice \cite{DT80}: 
%\footnote{
%T.A.~DeGrand and D.~Toussaint, 
%Topological excitations and Monte Carlo simulation of Abelian gauge theory, 
%Phys. Rev. D{\bf 22}, 2478 (1980).
%}
\begin{align}
  k_{\mu}(s)=\frac{1}{2}{\varepsilon}_{\mu\nu\rho\sigma}
                  \partial_{\nu}n_{\rho\sigma}(s+\mu)
                   = -\frac{1}{4\pi}{\varepsilon}_{\mu\nu\rho\sigma}
                  \partial_{\nu}\bar{\theta}_{\rho\sigma}(s+\mu) .
\end{align}
The magnetic-monopole current $k_{\mu}(s)$ defined in this way becomes an integer-valued variable, since integer-valued variables $n_{\rho\sigma}$ are used to count the number of Dirac strings going out through a plaquette. 

%First of all, we generate the configurations of SU(2) link variables $\{ U_{x,\mu} \}$ using the standard Wilson action based on the heat bath method. 
%Next, we generate the configurations of the color vector field $\{\bm{n}_x\}$ according to the reduction condition  together with the configurations of SU(2) link variables $\{U_{x,\mu}\}$. 
%Then we can construct $\{V_{x,\mu}[U,\bm{n}]\}$ using (\ref{C35-cfn-mono-4}) and finally $\{k_{x,\mu}\}$ using (\ref{C35-cfn-conti-20}).

%%%%%%%%%%%%%%%%%%%%% figures %%%%%%%%%%%%%%%%%%%%%%%%%%%
\begin{table}[ptb]
\caption{Ref.\cite{IKKMSS06}: Histogram  of the magnetic charge (value of $K(s,\mu)$)   distribution 
for  new and  old monopoles on  $8^4$ lattice at $\beta=2.35$. }
\label{C35-table:hist-of-rmg}
\begin{center}
\begin{footnotesize}
\begin{tabular}{cll}\hline
Charge & Compact \cite{IKKMSS06} & Non-compact \cite{KKMSSI06} \\ \hline
-7.5$\sim$-6.5 & 0   & 0 \\
-6.5$\sim$-5.5 & 299 & 0 \\
-5.5$\sim$-4.5 & 0   & 1 \\
-4.5$\sim$-3.5 & 0   & 19\\
-3.5$\sim$-2.5 & 0   & 52\\
-2.5$\sim$-1.5 & 0   & 149\\
-1.5$\sim$-0.5 & 0   & 1086  \\
-0.5$\sim$0.5  & 15786 & 13801 \\
0.5$\sim$1.5   & 0   & 1035\\
1.5$\sim$2.5   & 0   & 173\\
2.5$\sim$3.5   & 0   & 52\\
3.5$\sim$4.5   & 0   & 16\\
4.5$\sim$5.5   & 0   & 0\\
5.5$\sim$6.5   & 299 & 0 \\
6.5$\sim$7.5   & 0   & 0\\
\hline
\end{tabular}
\end{footnotesize}
\end{center}
\end{table}
%%%%%%%%%%%%%%%%%%%%% figures %%%%%%%%%%%%%%%%%%%%%%%%%%%

%For this purpose, we have performed the numerical simulations on an  $8^4$ lattice at $\beta=2.35$ under thermalization=$3000$.

To check quantization of the magnetic charge, we have made a histogram of
\begin{equation}
 K(s,\mu) %:= 2\pi k_\mu(s) 
= \frac{1}{2}{\varepsilon}_{\mu\nu\rho\sigma}
\partial_{\nu}\bar{\Theta}_{\rho\sigma}(x+\mu) ,
\end{equation}
 i.e., magnetic charge distribution. 
 Note that $K(s,\mu)$ %is a real-valued variable and it
  should become a multiple of $2\pi$ if the magnetic charge is quantized.  
%This is a quite non-trivial problem to be examined. 
Table~\ref{C35-table:hist-of-rmg}  show that $K(s,\mu)$ is completely separated into $0$ or $\pm 2\pi$ within an error of $10^{-10}$.
We have checked that the data in  Table~\ref{C35-table:hist-of-rmg}  exhaust in total all the configurations $N=4\times 8^4=16384$,
because the number $N_l$ of links in the $D$-dimensional lattice with a side length $L$ is given by  $N_l=DL^D$. 
This result clearly shows that the magnetic charge defined anew is quantized as expected from the general argument. 
We have observed that the conservation law of the magnetic-monopole current holds, since the number of $+2\pi$ configurations is the same as that of $-2\pi$ configurations.
In contrast, 
Table~\ref{C35-table:hist-of-rmg} shows that quantization does not occur for the (old) CDGFN monopole constructed in the non-compact formulation \cite{KKMSSI06}.

%%%%%%%%%%%%%%%%%%%%%%%%%%%%%%%%%%%%%%%%%%%%%%%%%%
%%%%%%%%%%%%%%%%%%%%%%%%%%%%%%%%%%%%%%%%%%%%%%%%%%
\subsubsection{Lattice magnetic monopole loops}
%\setcounter{equation}{0}
%%%%%%%%%%%%%%%%%%%%%%%%%%%%%%%%%%%%%%%%%%%%%%%%%%
%%%%%%%%%%%%%%%%%%%%%%%%%%%%%%%%%%%%%%%%%%%%%%%%%%

%%%%%%%%%%%%%%%%%%%%%%%%%%%%%%%%%%%%%%%%%%%%%%%%%%%%%
\begin{figure}[ptb]
\begin{center}
\includegraphics[scale=0.65]{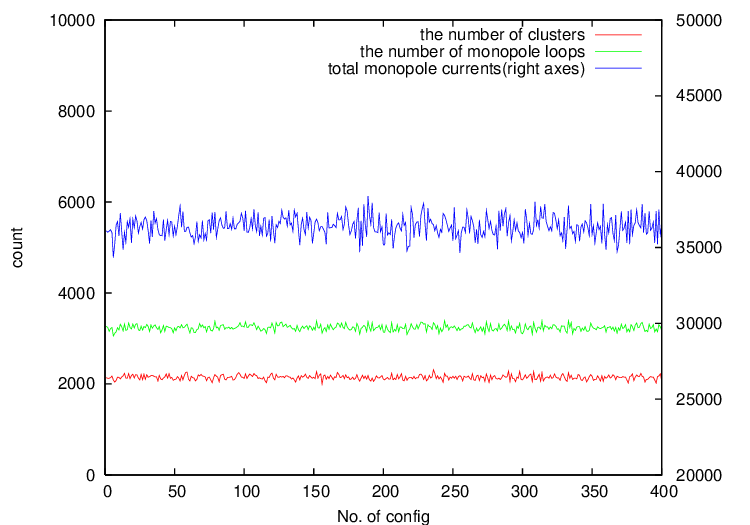}
\includegraphics[scale=0.65]{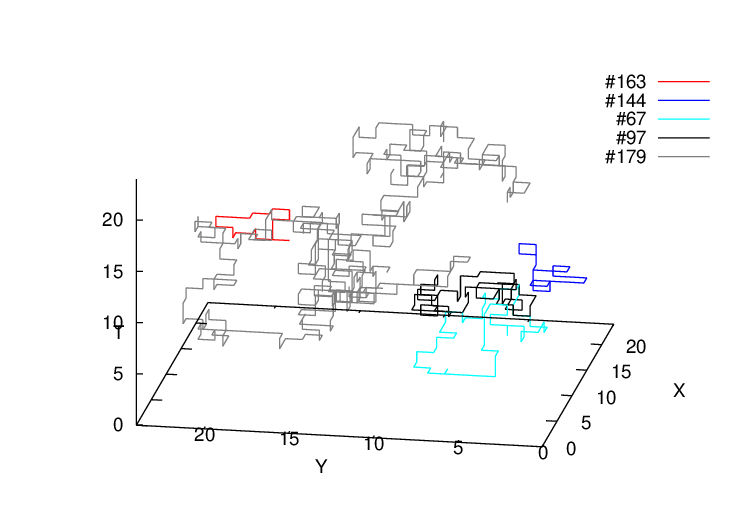}
\end{center}
\vspace{-5mm}
\caption{
Ref.\cite{Shibata-lattice2010}:
The analysis of magnetic monopoles for 400 configurations on $24^{4}$ lattice at $\beta=2.4.$ (Left panel) 
The blue line shows the number of the non-zero magnetic-monopole currents (right vertical axis). The red line and green line show the number of clusters of connected loops and the number of loops for each configuration, respectively (see left vertical axis).
(Right panel) 
A 3-dimensional plot obtained from the detected  magnetic-monopole loops on the 4 dimensional lattice by projecting the 4-dimensional dual lattice space to the 3-dimensional one, i.e., $(x,y,z,t) \rightarrow (x,y,t)$.
%The 3-dimensional plot of detected magnetic monopole loops, where the graph in 4dimensional Euclid space is projected to the 3-dimensional space, i.e., $(x,y,z,t)$ $\rightarrow(x,y,z).$ 
}%
\label{C35-fig:cluster}%
\end{figure}
%%%%%%%%%%%%%%%%%%%%%%%%%%%%%%%%%%%%%%%%%%%%%%%%%%%%%

%%%%%%%%%%%%%%%%%%%%% figures %%%%%%%%%%%%%%%%%%%%%%%
%\begin{figure}[htbp]
%\begin{center}
%\vspace{-5mm}%
%\includegraphics[scale=0.5]{Kondo-book-fig/lattice-nlcv/density.ps}
%\vspace{-5mm}
%\end{center}
%\caption{
%\footnotesize 
%The number vs. length of the magnetic monopole loops
%}
%\label{C35-fig:potential}%
%\end{figure}
%%%%%%%%%%%%%%%%%%%%% figures %%%%%%%%%%%%%%%%%%%%%%%

In the analysis of lattice data for the magnetic currents, it is very hard to manipulate monopole configurations directly, since they contain more than 35000 non-zero magnetic-monopole currents (see the left panel of Figure \ref{C35-fig:cluster}). Therefore, we introduce an algebraic algorithm for topology. The \textbf{CHomP homology software} \cite{KMM04},  provided by the computational homology project,
\footnote{
See 
http://chomp.rutgers.edu/ for computational homology project.
%\bibitem{KMM04}
%Tom Kacynski, Konstantin Mischaikow and Marian Mrazek, 
%Computational Homology, 
%Applied Mathematical Sciences (Book 157)
%(Springer, 2004).
}
 computes a topological invariant called the \textbf{Betti number} of a collection and their generators in the algebraic way. The Betti number is a  part of the information contained in the homology groups of a topological space, which intuitively measure the number of connected components, the number of holes, and the number of enclosed cavities in low dimensions. In our case, the generators of the dimension-one homology group correspond to magnetic monopole loops.
See Shibata et al.~\cite{Shibata-lattice2009} for more details. 

\begin{figure}[ptb]
\begin{center}
\includegraphics[scale=0.65]{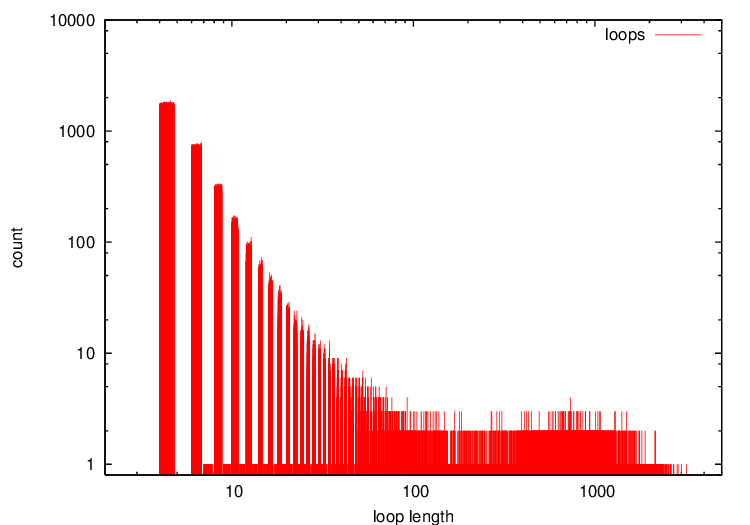}
\includegraphics[scale=0.65]{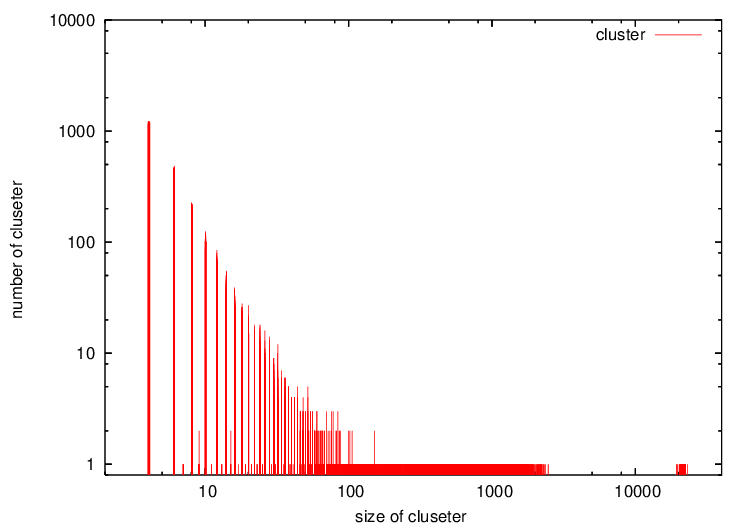}
\end{center}
\caption{
Ref.\cite{Shibata-lattice2010}:
(Left panel) 
The histogram for the length of monopole loops, i.e., for each configuration the number of monopole loops with length $n$ is counted, and is plotted as a truss of poles with the same length$.$ 
(Right panel) 
The histogram of the cluster size. 
}%
\label{C35-fig:anatomy1}%
\end{figure}

We apply the method to the lattice data of $24^{4}$ lattice with periodic boundary condition whose configurations are generated by using the standard Wilson action with the parameter $\beta=2.4.$ 
The left panel of Fig.~\ref{C35-fig:cluster} shows the data of detected magnetic-monopole currents for 400 configurations. The blue line shows the number of non-zero charge currents, i.e., non-zero magnetic-monopole currents occupy about 3\% of the total links. 
The red line and green line show the number of clusters of connected loops (the Betti number of dimension zero) and the number of loops (the Betti number of dimension one) for each configuration, respectively.  
The right panel of Fig.~\ref{C35-fig:cluster} shows an example of detected magnetic monopoles, which are plotted in the 3-dimensional space projected from the 4-dimensional Euclidean space. Fig.~\ref{C35-fig:anatomy1} shows detail of the magnetic monopole configurations. 
%Then, we are ready to investigate the magnetic monopole contribution to the static potential by using extracted monopole loops.
The magnetic monopoles extracted in this way are used to estimate the contribution of the magnetic monopole to the static potential.

%%%%%%%%%%%%%%%%%%%%%%%%%%%%%%%%%%%%%%%%%%%%%%%%%%%%%
%%%%%%%%%%%%%%%%%%%%%%%%%%%%%%%%%%%%%%%%%%%%%%%%%%%%%

%%%%%%%%%%%%%%%%%%%%%%%%%%%%%%%%%%%%%%%%%%%%%%%%%%
%%%%%%%%%%%%%%%%%%%%%%%%%%%%%%%%%%%%%%%%%%%%%%%%%%
\subsubsection{quark potential and the string tension: Abelian dominance and monopole dominance}
%%%%%%%%%%%%%%%%%%%%%%%%%%%%%%%%%%%%%%%%%%%%%%%%%%
%%%%%%%%%%%%%%%%%%%%%%%%%%%%%%%%%%%%%%%%%%%%%%%%%%

The Wilson loop operator $W_{\rm full}[U]$ for a closed loop $C$ on a lattice is defined using the link variable $U_\ell$ in the gauge-invariant way:
\begin{align}
W_{\rm full}[U] := {\rm tr}({\cal P}\prod_{\ell \in C}U_\ell)/{\rm tr}({\bf 1})  .
\end{align}
By replacing the  full $SU(2)$  link variable $U_\ell$ by the restricted variable $V_\ell$, we can define another gauge-invariant quantity  $W_{\rm rest}[V]$ which we call the \textbf{restricted Wilson loop operator}:
\begin{align}
W_{\rm rest}[V] := {\rm tr}({\cal P}\prod_{\ell \in C}V_\ell)/{\rm tr}({\bf 1}) . 
\end{align}
Then we can define the Wilson loop average $W_{\rm full}(C)$ and the restricted Wilson loop average $W_{\rm rest}(C)$ by
\begin{align}
W_{\rm full}(C) := \left< W_{\rm full}[U] \right> , \quad  
W_{\rm rest}(C) := \left< W_{\rm rest}[V] \right> .
\end{align}
Therefore, the respective average  must be independent of the gauge.
Since the restricted field ${\bf V}_{\mu}(x)$ is defined in a gauge-covariant and gauge independent way, 
we have obtained a  gauge-independent definition of the \textbf{``Abelian'' 
dominance} or the \textbf{restricted-field dominance} for the Wilson loop average:
\begin{align}
  W_{\rm full}(C)  \simeq  {\rm const.} W_{\rm rest}(C) .
\end{align}
A gauge-independent definition of Abelian dominance is given in the operator level  $ W_{\rm full}[U]  \simeq   {\rm const.}   W_{\rm rest}[V]$ and a constructive derivation of the Abelian dominance can be discussed  through a non-Abelian Stokes theorem via lattice regularization, see \cite{KS08}.
%It can be shown\cite{KS08} that requirement (I) and (II) are equivalent to defining equation (\ref{C35-Lcc}) and (\ref{C35-cond2m}) 
%which are able to uniquely determine the NLCV. 

In order to study the \textbf{magnetic-monopole dominance}  in the string tension, we proceed to estimate the 
magnetic monopole contribution: 
\begin{align}
W_{\rm mono}(C)=\left< W_{\rm mono}[K] \right> 
\end{align}
to the Wilson loop average $ W_{\rm full}(C)=\left< W_{\rm full}[U] \right> $.  
%We compare the respective coefficient of the area law calculated from the respective average. 
Here we define the magnetic part $W_{\rm mono}[K]$ of the Wilson loop operator $W_{\rm full}[U]$ as the contribution 
from the magnetic-monopole current $K_{x,\mu}$ to the Wilson loop operator:
\begin{align}
 W_{\rm mono}[K]     &:=  \exp \left( i \sum_{x,\mu} K_{x,\mu}\omega_{x,\mu} \right) 
= \exp \left( 2\pi i \sum_{x,\mu}m_{x,\mu}\omega_{x,\mu} \right) ,
% \label{C35-monopole dominance-1}     
\nonumber\\
 \omega_{x,\mu}  &:=  \sum_{x'}\Delta_L^{-1}(x-x')\frac{1}{2}
\epsilon_{\mu\alpha\beta\gamma}\partial_{\alpha}
S^J_{s'+\hat{\mu},\beta\gamma}, 
\quad
\partial'_{\beta}S^J_{x,\beta\gamma} = J_{x,\gamma} ,
\label{C35-monopole dominance-2}
\end{align}
where $\omega_{x,\mu}$  is defined through the external source $J_{x,\mu}$ which is used 
to calculate the static potential, 
$\partial'$ denotes the backward lattice derivative
$\partial_{\mu}^{'}f_x=f_x-f_{x-\mu}$,  $S^J_{x,\beta\gamma}$ denotes a surface bounded by the closed loop $C$ on which the electric source $J_{x,\mu}$ has its support, and $\Delta_L^{-1}(x-x')$ is the inverse Lattice Laplacian. 
Note that $W_{\rm mono}[K]$ is a gauge-invariant operator, since the magnetic-monopole current defined
by (\ref{C35-cfn-conti-20}) is a gauge-invariant variable.
In fact, the form (\ref{C35-monopole dominance-2}) is derived from the non-Abelian Stokes theorem for the Wilson loop operator.
By evaluating the average of $W_{\rm mono}[K]$ from the generated configurations of the monopoles $\{ K_{x,\mu}\}$ we can estimate the contribution to the string tension  from the generated configurations of the magnetic-monopole currents $\{ K_{x,\mu}\}$.

The Wilson loop operator $W_{\rm full}[U]$ is decomposed into the magnetic part $W_{\rm mono}[K] $ and the electric part $W_{\rm elec}[j] $, 
\begin{align}
 W_{\rm full}[U] = W_{\rm mono}[K] W_{\rm elec}[j],
\end{align}
which is derived from the non-Abelian Stokes theorem.
%, see  \cite{Kondo08}.
%  or Appendix B of \cite{Kondo00}.
The magnetic part $W_{\rm mono}[K]$ of the Wilson loop operator $W_{\rm full}[U]$ is used to examine the contribution from the magnetic-monopole current $K_{x,\mu}$ to the Wilson loop operator $W_{\rm full}[U]$, while $W_{\rm elec}[j]$ is expressed by the electric current $j_{\mu}=\partial_{\nu}F_{\mu\nu}$.
In order to establish the monopole dominance in the string tension, we proceed to estimate the magnetic monopole contribution $\left< W_{\rm mono}[K] \right>$ to the Wilson loop average $\left< W_{\rm full}[U] \right>$, i.e., the expectation value of the Wilson loop operator.%  
%\footnote{
It should be remarked that $\left< W_{\rm full}[U] \right> \not= \left< W_{\rm mono}[K] \right> \left< W_{\rm elec}[j] \right>$.
We have not yet calculated the electric contribution $\left< W_{\rm elec}[j]  \right>$ directly  where 
$W_{\rm elec}[j] $ is expressed by the electric current $j_{\mu}=\partial_{\nu}F_{\mu\nu}$. 
%See \cite{Kondo08} for details. 
%}

%%%%%%%%%%%%%%%%%%%%%%%%%%%%%%%%%%%%%%%%%%%%%%%%%%%%%
\subsubsection{Wilson loop average and the quark potential}
%%%%%%%%%%%%%%%%%%%%%%%%%%%%%%%%%%%%%%%%%%%%%%%%%%%%%

For a rectangular Wilson loop $C=(R,T)$ with the spatial length $R$ and the temporal length $T$, we calculate the three kinds of the Wilson loop average $W_{\rm i}(C)$ (\text{i=f(full)}, \text{r(rest)}, \text{m(mono)}).
Then we calculate the static $q\bar q$ potential $V_{\rm i}(R)$ as a function of the interquark distance $R$ using the respective Wilson loop 
average $W_{\rm i}(C)$ according to 
\begin{align}
V_{\rm i}(R) = -\log \left\{ \frac{  W_{\rm i}(R,T) }{ W_{\rm i}(R,T-1) } \right\} 
\quad (\text{i=f(full), r(rest), m(mono)}) .
\label{C35-monopole dominance-3}
\end{align}

The numerical simulations are performed at $\beta=2.4$ on the $16^4$  lattice  and at $\beta=2.5$ on the $24^4$ lattice.%
\footnote{
The lattice spacing in the physical units is given by \cite{SKKMSI07} $\epsilon(\beta=2.4)=0.1201$fm, and 
$\epsilon(\beta=2.5)=0.08320$fm.
}
We thermalize 3000 sweeps, and in particular, we have used 100 configurations for calculating  the full potential $V_{\rm full}$ and restricted potential $V_{\rm rest}$,  while for the monopole potential $V_{\rm mono}$  we have used 50 configurations for the $16^4$ lattice and 500 configurations for the $24^4$ lattice in each case with 100 iterations.%
\footnote{
The results of numerical simulations on the  $16^4$  lattice at $\beta=2.4$ were published in \cite{IKKMSS06,KIKMSS06} only for the full potential $V_{\rm full}$ and the monopole potential $V_{\rm mono}$, while the result on  the restricted potential $V_{\rm rest}$ was separately reported in \cite{Kato-lattice2009}.
} 
 
In order to obtain the full $SU(2)$ and restricted  results, especially, we used the  \textbf{APE smearing method} \cite{Albanese87} as a noise reduction technique.

\begin{figure}[ptb]
\begin{center}
\includegraphics[scale=0.65]{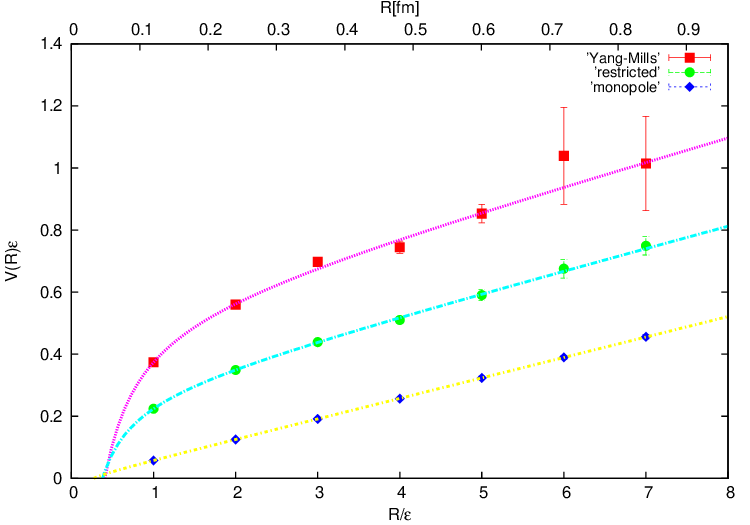} 
\quad
\includegraphics[scale=0.65]{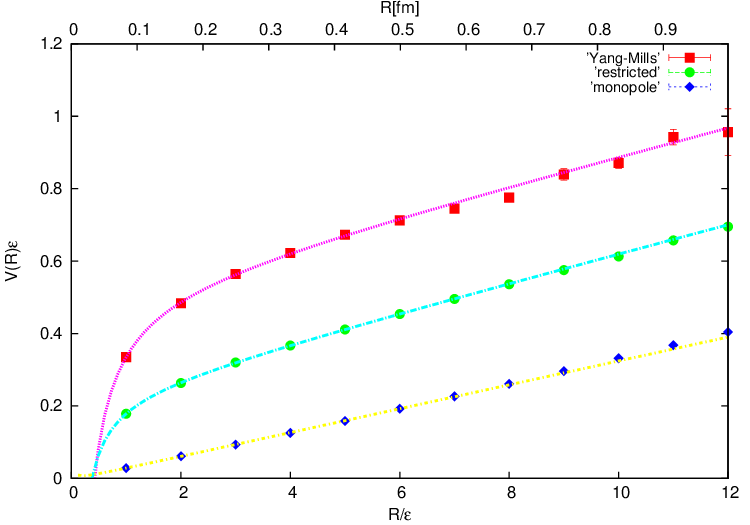}
\end{center}
\vspace{-5mm}
\caption{
Ref.\cite{KKS14}:
The full $SU(2)$ potential $V_f(R)$,  (``Abelian'') restricted potential $V_r(R)$ and magnetic-monopole potential $V_m(R)$ as functions of $R$ 
(Left) on $16^4$ lattice at $\beta=2.4$,
(Right)  on $24^4$ lattice at $\beta=2.5$ 
where the Wilson loop with $T=12$ was used for obtaining  $V_{\rm full}(R)$ and $V_{\rm rest}(R)$, and $T=8$ for  $V_{\rm mono}(R)$. }%
\label{C35-fig:potential}
\end{figure}

Fig.\ref{C35-fig:potential} shows the obtained plot for the respective potential for various values of $R$.
The obtained numerical potential is fitted to the sum of a linear term,  Coulomb term and a constant term: 
\begin{align}
V_{\rm i}(R) = \sigma_{\rm i} R -  \alpha_{\rm i}/R +c_{\rm i} , 
\quad (\text{i=f(full), r(rest), m(mono)}) .
\label{C35-monopole dominance-4}
\end{align}
where $\sigma_{\rm i}$ is the string tension (the coefficient of the area decay), $\alpha_{\rm i}$ is the Coulomb coefficient, 
and $c_{\rm i}$ is the constant which is equal to the coefficient of the perimeter decay:
\begin{equation} 
W_i(R,T)  \sim \exp [-\sigma_i RT -c_i(R+T)+\alpha_i T/R + \cdots] .
\end{equation}
The results are shown in 
Table~\ref{C35-strint-tension-1a} and
 Table~\ref{C35-strint-tension-1}.

%%%%%%%%%%%%%%%%%%%%%%%%%%%%%%%%%%%%%%%%%%%%%%%%%%%%
%%%%%%%%%%%%%%%%%%%%%%%%%%%%%%%%%%%%%%%%%%%%%%%%%%%%

%===============================================
%\begin{figure}[h]
%\vspace*{5mm}
%
%\centerline{
%\epsfxsize=0.8\textwidth
%\epsfxsize=0.7\textwidth
%\epsfbox{figs/potential_nlcv.eps}
%}
%\caption{The full, Abelian and monopole potentials as functions of $R$ at 
%$\beta=2.4$ on $16^4$ lattice.}
%\label{C35-fig:potential}
%\end{figure}
%============================================= 

%%%%%%%%%%%%%%%%%%%%%%%%%%%%%%%%%%%%%%%%%%%%%%%%%%%%%
\begin{table}[t]
\caption{Ref.\cite{KKS14}: String tension and Coulomb coefficient  on $16^4$ lattice at $\beta=2.4$.}
\label{C35-strint-tension-1a}
\begin{center}
\begin{tabular}{lllc}\hline
             & $\sigma$    & $\alpha$ & $\chi^2/N_{dof}$\\ \hline
full     & 0.075(9)    & 0.23(2)  &   1.234 \\
restricted  & 0.070(4)    & 0.11(1) &  0.195 \\
magnetic monopole     & 0.066(2)    & 0.003(7) & 0.198\\\hline
\end{tabular}
\end{center}
\end{table}
%\vspace{-.5cm}
%%%%%%%%%%%%%%%%%%%%%%%%%%%%%%%%%%%%%%%%%%%%%%%%%%%%%

%%%%%%%%%%%%%%%%%%%%%%%%%%%%%%%%%%%%%%%%%%%%%%%%%%%%%%%%%%%%
\begin{table}[h]
\caption{Ref.\cite{KKS14}: String tension and Coulomb coefficient  on $24^4$ lattice at $\beta=2.5$.}
\label{C35-strint-tension-1}
\begin{center}
\begin{tabular}{lllc}\hline
             & $\sigma$    & $\alpha$  & $\chi^2/N_{df}$ \\ \hline
full      & 0.0388(6)    & 0.2245(23)  & 4.73 \\
restricted       & 0.0398(2)    & 0.0912(8)  & 1.82 \\
magnetic monopole    & 0.0330(1)    & -0.0012(4) & 4.81 \\
\hline
\end{tabular}
\end{center}
\end{table}
%\vspace{-.5cm}
%%%%%%%%%%%%%%%%%%%%%%%%%%%%%%%%%%%%%%%%%%%%%%%%%%%%%%%%%%%%

Thus, on the $16^4$ lattice~at $\beta=2.4$  the restricted (``Abelian'') part $\sigma_{\rm rest}$ reproduces 93$\%$ of the full string tension $\sigma_{\rm full}$:
\begin{equation} 
 \frac{\sigma_{\rm rest}}{\sigma_{\rm full}} = (93 \pm 17)\% 
\quad (\text{on} \ 16^4 \ \text{lattice~at} \ \beta=2.4), 
\end{equation}
and the monopole part $\sigma_{\rm mono}$ reproduces 94$\%$ of $\sigma_{\rm rest}$: 
\begin{equation} 
 \frac{\sigma_{\rm mono}}{\sigma_{\rm rest}} =(94 \pm 8)\%  
 \Longrightarrow 
 \frac{\sigma_{\rm mono}}{\sigma_{\rm full}} =  (88 \pm 13)\% 
\quad (\text{on} \ 16^4 \ \text{lattice~at} \ \beta=2.4) . 
\end{equation}

Moreover, on the $24^4$ lattice~at $\beta=2.5$  
the restricted (``Abelian'') part $\sigma_{\rm rest}$ reproduces 100$\%$ of the full string tension $\sigma_{\rm full}$:
\begin{equation} 
 \frac{\sigma_{\rm rest}}{\sigma_{\rm full}} = (102 \pm 2)\% 
\quad (\text{on} \ 24^4 \ \text{lattice~at} \ \beta=2.5), 
\end{equation}
and the monopole part $\sigma_{\rm mono}$ reproduces 83$\%$ of $\sigma_{\rm rest}$: 
\begin{equation} 
 \frac{\sigma_{\rm mono}}{\sigma_{\rm rest}} = (83 \pm 1)\%  
 \Longrightarrow 
 \frac{\sigma_{\rm mono}}{\sigma_{\rm full}} =  (85 \pm 2)\% 
\quad (\text{on} \ 24^4 \ \text{lattice~at} \ \beta=2.5) . 
\end{equation}
In general, the monopole part does not include  the Coulomb term and hence the linear potential 
is obtained to an accuracy better than the full potential. 
Thus, we have confirmed the \textbf{restricted field  dominance} (or \textbf{``Abelian'' dominance}) and the \textbf{magnetic monopole dominance} in the string tension for the $SU(2)$ Yang-Mills theory in our framework.

The  $16^4$ lattice at $\beta=2.4$ and $24^4$ lattice at $\beta=2.5$ have the nearly the same physical size, since the former has the physical size $16 \epsilon(\beta=2.4)=1.92$fm and the latter has the physical size $24 \epsilon(\beta=2.5)=1.99$fm.
Therefore, the data obtained in the above for the different lattice sizes and the gauge coupling constants are consistent to each other within errors.

The results show that the magnetic-monopole potential $V_{\rm mono}(R)$ has a dominant linear term and a negligibly small Coulomb term with small errors.  In general, the monopole part does not include  the Coulomb term and hence the linear potential 
is obtained to an accuracy better than the full potential. 
Consequently, the string tension $\sigma_{\rm mono}$ is obtained within small errors.

%We have proposed a new formulation of the NLCV of Yang-Mills theory, which was once called the CFN decomposition.  

%%%%%%%%%%%%%%%%%%%%% figures %%%%%%%%%%%%%%%%%%%%%%%%%%%
\begin{table}[h]
\caption{String tension and Coulomb coefficient using 
MA gauge and the DeGrand-Toussaint method
reproduced from \cite{SNW94}:% [Stack et al., PRD 50, 3399 (1994)]
}
\label{C35-strint-tension-2}
\begin{center}
\small
\begin{tabular}{cllll}\hline
$\beta$ & $\sigma_f$ & $\alpha_f$ & $\sigma_{DTm}$ & $\alpha_{DTm}$ \\ \hline
{\bf 2.4}($16^4$)  & 0.072(3)   & 0.28(2)  &  {\bf 0.068(2)} & 0.01(1)  \\
2.45($16^4$) & 0.049(1)   & 0.29(1)  &  0.051(1) & 0.02(1)  \\
2.5($16^4$)  & 0.033(2)   & 0.29(1)  &  0.034(1) & 0.01(1) \\
\hline
\end{tabular}
\end{center}
\end{table}
%%%%%%%%%%%%%%%%%%%%% figures %%%%%%%%%%%%%%%%%%%%%%%%%%%

For comparison, we have shown in Table~\ref{C35-strint-tension-2} the data of \cite{SNW94} which has discovered the monopole dominance for the first time in MA gauge on $16^4$ lattice where $\sigma_{DTm}$ reproduces 95$\%$ of $\sigma_f$. 
Here $\sigma_{DTm}$ and $\alpha_{DTm}$ denotes the conventional monopole contribution extracted from the diagonal potential $A_\mu^3$ using  Abelian projection in MA gauge. 
In particular, the comparison of the data on $16^4$ lattice at $\beta=2.4$ between Table~\ref{C35-strint-tension-2} and  Table~\ref{C35-strint-tension-3} reveals that the monopole contributions have exactly the same value between the conventional DeGrand-Toussaint  monopole \cite{DT80} and the gauge-invariant magnetic monopole constructed in the new reformulation.
This is because the monopole part does not include  the Coulomb term and hence the potential is obtained to an accuracy better than the full potential, as pointed above.

%%%%%%%%%%%%%%%%%%%%% figures %%%%%%%%%%%%%%%%%%%%%%%%%%%
\begin{table}[h]
\caption{
Ref.\cite{IKKMSS06}: String tension and Coulomb coefficient in the new reformulation
}
\label{C35-strint-tension-3}
\begin{center}
\begin{tabular}{cllll}\hline
$\beta$ & $\sigma_f$ & $\alpha_f$ & $\sigma_m$ & $\alpha_m$ \\ \hline
2.3($8^4$)  & 0.158(14)   & 0.226(44)  &  0.135(13) & 0.009(36)  \\
2.4($8^4$)  & 0.065(13)   & 0.267(33)  &  0.040(12) & 0.030(34)  \\
{\bf 2.4($16^4$)} & 0.075(9)  & 0.23(2)  &  {\bf 0.068(2)}  & 0.001(5)\\
\hline
\end{tabular}
\end{center}
\end{table}
%%%%%%%%%%%%%%%%%%%%% figures %%%%%%%%%%%%%%%%%%%%%%%%%%%
%

Thus, the Abelian and monopole dominances in the string tension has been shown anew in the gauge invariant way, whereas they have been so far shown only in a special gauge fixing called  MA gauge which breaks the color symmetry explicitly.

%%%%%%%%%%%%%%%%%%%%%%%%%%%%%%%%%%%%%%%%%%%%%%%%%%%%%%%%%%%
\subsubsection{Chromoelectric field and flux tube formation}
%%%%%%%%%%%%%%%%%%%%%%%%%%%%%%%%%%%%%%%%%%%%%%%%%%%%%%%%%%%%

According to the dual superconductor picture for quark confinement, the QCD vacuum must be a dual superconductor so that the chromoelectric field generated by the $q\bar q$ pair is squeezed into the flux tube forming the string structure and hence the energy per unit length of the string gives the string tension of the linear potential. 
In other words, the QCD vacuum exhibits the \textbf{dual Meissner effect}. 
In order to confirm the dual Meissner effect, we measure the chromofield around the $q\bar q$ pair to obtain the information on the distribution or the profile of the chromoelectric field generated by the static $q\bar q$ pair.
These issues are also checked for the restricted field to examine whether or not the restricted field $V$ can reproduce the full results  obtained by the original full field $U$.

\begin{figure}[ptb]
\begin{center}
\includegraphics[height=4cm]{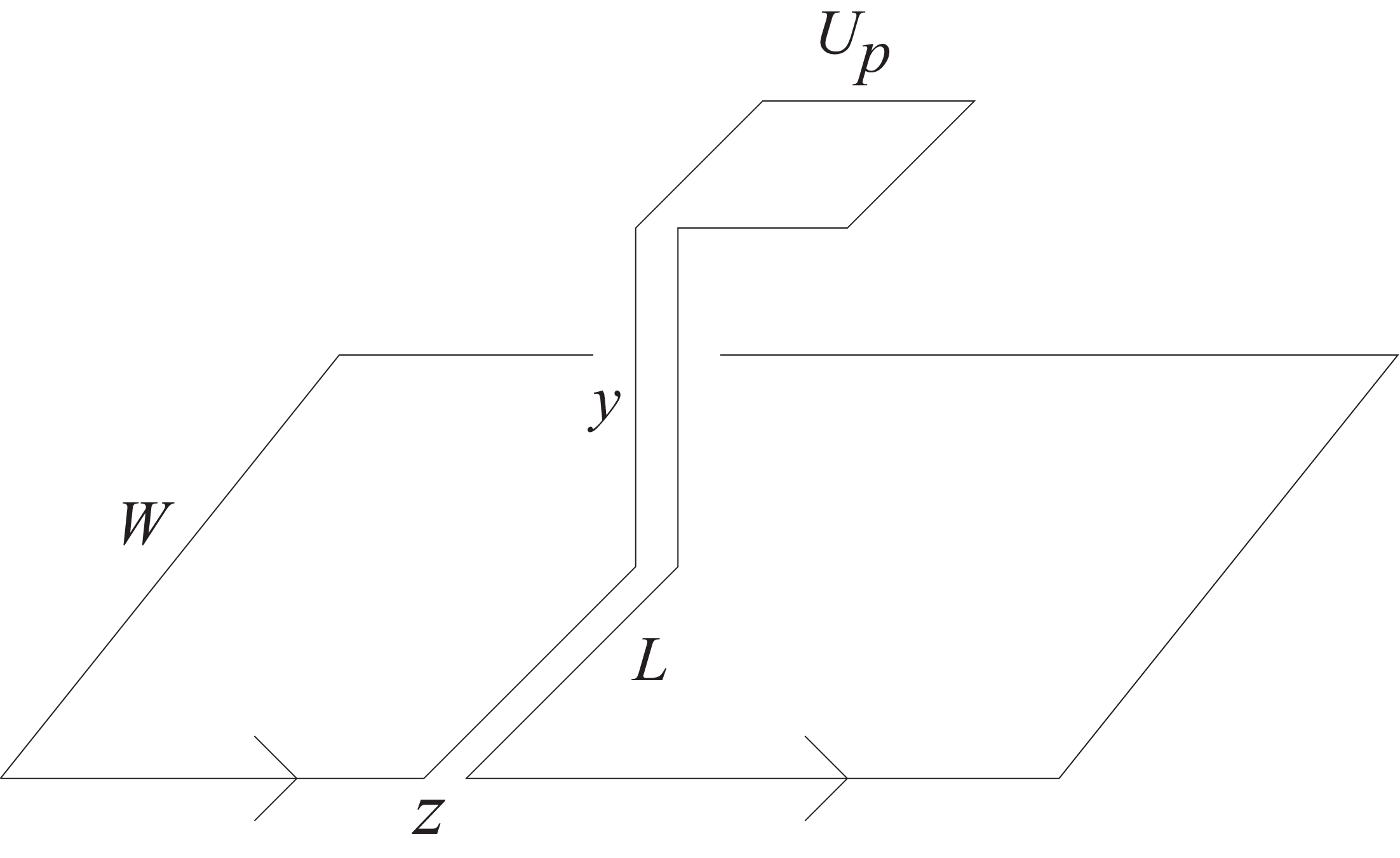} 
\end{center}
\vspace{-5mm}
\caption{
The setup of $WLU_PL^{\dagger}$ in the definition of the operator $\rho_{_{UP}}$: $z$ is the position of the Schwinger line $L$ along the line connecting $\bar{q}$ and $q$ at a fixed Euclidean time $t$, and $y$ is the distance from the plane  spanned by  the Wilson loop $W$ to the plaquette $U_P$. 
}%
\label{C35-cf-fig1}
\end{figure}

For this purpose, we must extract the chromofield in the gauge-invariant way. 
This is a nontrivial issue. 
In order to define the gauge-invariant chromofield strength tensor, we introduce the following  operator representing a gauge-invariant connected correlator   between the Wilson loop operator and a plaquette variable according to Di Giacomo, Maggiore and Olejnik \cite{GMO90,CC95}:
\begin{align} 
 \rho_{_{UP}} :=\frac{\langle {\rm tr}(WLU_PL^{\dagger}) \rangle}{\langle {\rm tr}(W) \rangle}
-\frac{1}{{\rm tr}({\bf 1})}\frac{\langle {\rm tr}(U_P){\rm tr}(W) \rangle}{\langle {\rm tr}(W) \rangle} ,
\label{C35-cf1-1}
\end{align}
where $W$ is the Wilson loop operator representing a pair of quark and antiquark,  $U_P$ is the plaquette variable as the probe for measuring the chromofield strength at the position of the plaquette, and $L$ is the line connecting the plaquette  $U_P$ and the Wilson loop operator $W$, which is called the Schwinger line. See Fig.~\ref{C35-cf-fig1}. 
Here the Schwinger line $L$ is necessary to guarantee the gauge invariance of the correlator $\rho_W$. 
We must pay attention to the orientation between $U_P$ and $W$.
The above definition works for $SU(N)$ gauge group for any $N$ using ${\rm tr}({\bf 1})=N$, and we set ${\rm tr}({\bf 1})=2$ for the gauge group $SU(2)$.

%%%%%%%%%%%%%%%%%%%%%%%%%%%%%%%%%%%%%%%%%%%%%%%%%%%%
%%%%%%%%%%%%%%%%%%%%%%%%%%%%%%%%%%%%%%%%%%%%%%%%%%%%

For
$U_{x,\mu}=\exp (-ig\epsilon \mathscr{A}_\mu(x))$,  the plaquette variable is rewritten as
\begin{equation}
 U_{P}= \exp(-ig\epsilon^2 \mathscr{F}_{\mu\nu} ) 
 = {\bf 1} -ig\epsilon^2 \mathscr{F}_{\mu\nu} + O(\epsilon^4) .
\end{equation}
This leads to the trace: using the cyclicity of the trace and the unitarity $LL^\dagger=L^\dagger L={\bf 1}$, 
\begin{align}
 {\rm tr}(U_PL^{\dagger}WL) =&  {\rm tr}(L^{\dagger}WL)  -ig \epsilon^2 {\rm tr}(\mathscr{F}_{\mu\nu}L^{\dagger}WL) + O(\epsilon^4)
 \nonumber\\
=& {\rm tr}(W)  -ig \epsilon^2 {\rm tr}(\mathscr{F}_{\mu\nu}L^{\dagger}WL) + O(\epsilon^4)  ,
\end{align}
while using the traceless property   ${\rm tr}(\mathscr{F}_{\mu\nu})=0$,
\begin{equation}
 {\rm tr}(U_{P})=  {\rm tr}({\bf 1})   + O(\epsilon^4) .
\end{equation}
Hence the correlator reads
\begin{align} 
 \rho_{_{UP}} 
%:=& \frac{\langle {\rm tr}(WLU_PL^{\dagger}) \rangle}{\langle {\rm tr}(W) \rangle}
%-\frac{1}{{\rm tr}({\bf 1})}\frac{\langle {\rm tr}(U_P){\rm tr}(W) \rangle}{\langle {\rm tr}(W) \rangle}
% \nonumber\\
=  \frac{\left\langle {\rm tr}(W) \right\rangle  -ig \epsilon^2 \left\langle {\rm tr}(\mathscr{F}_{\mu\nu}L^{\dagger}WL) \right\rangle }{\langle {\rm tr}(W) \rangle} 
- \frac{ {\rm tr}({\bf 1}) \langle {\rm tr}(W) \rangle }{{\rm tr}({\bf 1})\langle {\rm tr}(W) \rangle} + O(\epsilon^4) .
\label{C35-cf1-1b}
\end{align}In the naive continuum limit (lattice spacing $\epsilon \to 0$), therefore, the operator $\rho_U$ reduces to the field strength in the presence of the $q\bar{q}$ source: 
\begin{equation}
\rho_{_{UP}}%
\overset{\varepsilon\rightarrow0}{\simeq}g\epsilon^{2}\left\langle
\mathscr{F}_{\mu\nu}\right\rangle _{q\bar{q}}:= - \frac{\left\langle
\mathrm{tr}\left(  ig\epsilon^{2} \mathscr{F}_{\mu\nu}L^{\dag}WL \right)
\right\rangle }{\left\langle \mathrm{tr}\left(  W\right)  \right\rangle
}+O(\epsilon^{4}) .
\label{C35-cf1-2}
\end{equation}
%\begin{align} 
%\rho_{_{UP}} \simeq \epsilon^2g
%\lbra{
%[ \langle  F_{\mu\nu} \rangle_{q\bar{q}}- \langle F_{\mu\nu} \rangle_0] .
%}
%\label{C35-cf1-2}
%\end{align}
Therefore, we can define a gauge-invariant chromofield strength tensor by
\begin{align} 
F_{\mu\nu}[U](x)  :=  \epsilon^{-2} \frac{\sqrt{\beta}}{2} \rho_U(x) , \quad \beta := \frac{2N}{g^2} \quad (\text{for} \ G=SU(N)) . 
\label{C35-cf1-3}
\end{align}

In the definition of the operator $\rho_{_{UP}}$,  $WLU_PL^{\dagger}$ is set up as follows.
Let $z$ be the position of the Schwinger line $L$ along the line connecting $\bar{q}$ and $q$ at a fixed Euclidean time $t$, and $y$ be the distance from the plane spanned by the Wilson loop $W$ to the plaquette $P$. See Fig.~\ref{C35-cf-fig1}.
By changing the distances $y,z$ and the direction of the plaquette $U_P$ relative to the Wilson loop $W$, 
we can scan the chromoelectric and chromomagnetic fields around the $q\bar{q}$ pair.

Similarly, we define the chromofield strength tensor $F_{\mu\nu}[V]$ from  the restricted field $V_{\mu}(x)$ by
\begin{align} 
F_{\mu\nu}[V](x)  &:=  \epsilon^{-2} \frac{\sqrt{\beta}}{2} \rho_{_{V}}(x), \quad 
%\label{C35-cf1-4}
\nonumber\\
\rho_{_{VP}} & :=  \frac{\langle {\rm tr}(W^VL^VV_P{L^V}^{\dagger}) \rangle}{\langle {\rm tr}(W^V) \rangle}
-\frac{1}{{\rm tr}({\bf 1})}\frac{\langle {\rm tr}(V_P){\rm tr}(W^V) \rangle}{\langle {\rm tr}(W^V) \rangle} ,
\label{C35-cf1-5}
\end{align}
where $V_P$ is the plaquette variable for the restricted field (link variable) $V$, $W^V$ and $L^V$ represent respectively the Wilson loop operator and the Schwinger line constructed from the restricted field (link variable) $V$.

%%%%%%%%%%%%%%%%%%%%%%%%%%%%%%%%%%%%%%%%%%%%%%%%%%
%\newpage
%\section{numerical results}

In the numerical simulations,  we have generated the link fields ${U_{x,\mu}}$ using the heat bath method for the standard $SU(2)$ Wilson action.
We have stored 100 configurations for the  
%$L^4=16^4$ at $\beta=2.5$ and 
  $24^4$ lattice at $\beta=2.5$ with 100 iterations. 
%\footnote{
%The lattice spacing in the physical units is given %by $\epsilon(\beta=2.5)=0.08320$fm.
%} 
%\marginpar{$L^4=16^4$}
We take $R=T= 8$ for the size of the Wilson loop operator to calculate the operators (\ref{C35-cf1-3}) and (\ref{C35-cf1-5}). 
%\marginpar{$R=T=L/2-2=6$}
Therefore, the quark and antiquark source is introduced as $R \times T$ Wilson loop $W$ in the $z$-$t$ plane.
The probe $U_{P}$ is set at the center of the Wilson loop and moved along the $y$-direction. 
We have performed the \textbf{hypercubic blocking (HYP)} \cite{Hasenfratz:2001} as a smearing method to obtain  ${U_{x,\mu}}$ for calculating the operators (\ref{C35-cf1-3}) and (\ref{C35-cf1-5}).
See the Appendix of \cite{KKS14} for the details of the HYP.

%%%%%%%%%%%%%%%%%%%%%%%%%%%%%%%%%%%%%%%%%%%%%%%%%%%%
%\begin{figure}[ptb]
%\begin{center}
%\includegraphics[scale=0.5]{figs/u_ebi_vs_y.eps} 
%{figs/u_ei_vs_y.eps} 
%\quad
%\includegraphics[scale=0.7]{figs/u_Ex_3d2.eps} 
%\vspace{-5mm}
%\end{center}
%\caption{
%The chromoelectric and chromomagnetic fields obtained from the full field $U$ on $V=16^4$ lattice at $\beta=2.5$.
%(Left panel) 
%$y$ dependence of the chromoelectric field $E_i(y)=F_{4i}(y)$ ($i=x,y,z$) at fixed $z=3$ %(mid-point of $q\bar q$). 
%(Right panel)
%The distribution of $E_z(y,z)$ obtained for the $6 \times 6$ Wilson loop with $\bar{q}$ at $(y,z)=(0,0)$ and $q$ at $(y,z)=(0,6)$.
%}
%\label{cf-fig2a}
%\end{figure}
%%%%%%%%%%%%%%%%%%%%%%%%%%%%%%%%%%%%%%%%%%%%%%%%%%%%

%%%%%%%%%%%%%%%%%%%%%%%%%%%%%%%%%%%%%%%%%%%%%%%%%%%%
\begin{figure}[ptb]
\begin{center}
\includegraphics[scale=0.65]{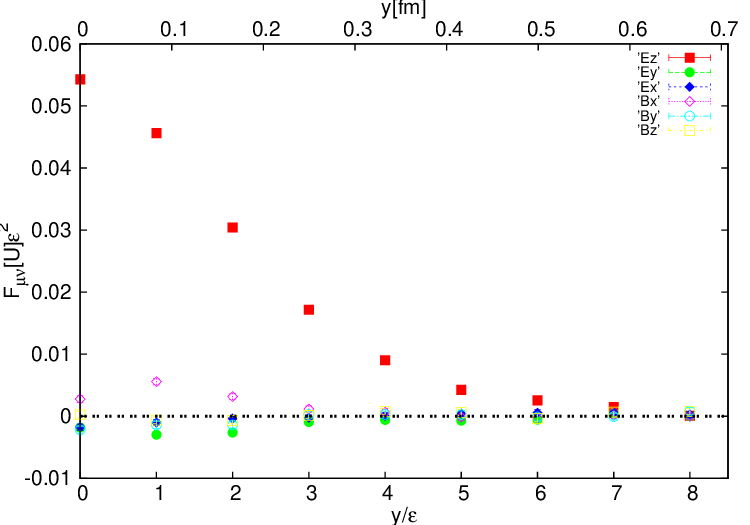}
%{figs/u_ei_vs_y.eps} 
\quad
\includegraphics[scale=0.65]{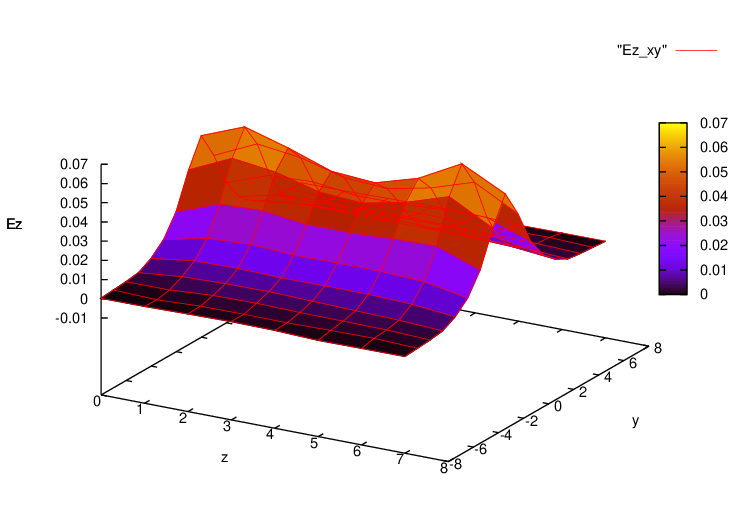} 
\vspace{-5mm}
\end{center}
\caption{Ref.\cite{KKS14}: 
The chromoelectric and chromomagnetic fields obtained from the full field $U$ on $24^4$ lattice at $\beta=2.5$.
(Left panel) 
$y$ dependence of the chromoelectric field $E_i(y)=F_{4i}(y)$ ($i=x,y,z$) at fixed $z=4$ (mid-point of $q\bar q$). 
(Right panel)
The distribution of $E_z(y,z)$ obtained for the $8 \times 8$ Wilson loop with $\bar{q}$ at $(y,z)=(0,0)$ and $q$ at $(y,z)=(0,8)$.
}
\label{C35-cf-fig2}
\end{figure}
%%%%%%%%%%%%%%%%%%%%%%%%%%%%%%%%%%%%%%%%%%%%%%%%%%%%

The results of numerical simulations are shown in Fig.~\ref{C35-cf-fig2}.
In the left panel of Fig.~\ref{C35-cf-fig2}, we find that only the $E_{z}$ component of the chromoelectric field $(E_x,E_y,E_z)=(F_{14},F_{24},F_{34})$ connecting $q$ and $\bar q$ has non-zero value for the original Yang-Mills link field ${U_{x,\mu}}$.
The other components are zero consistently within the numerical errors. 
In other words, the chromoelectric field is directed to the line connecting  quark and antiquark.
%This means that the chromomagnetic field $(B_x,B_y,B_z)=(F_{23},F_{31},F_{12})$ connecting $q$ and $\bar q$ does not exist  
The magnitude of the chromoelectric field $E_{z}$   decreases quickly as the distance $y$ increases in the direction perpendicular to the line.
Thus the obtained profile of the chromoelectric field represents the structure expected for the flux tube.
Therefore, we have confirmed the formation of the chromoelectric flux in $SU(2)$ Yang-Mills theory on a lattice.

To see the profile of the non-vanishing component $E_z$ of the chromoelectric field in detail, we explore the distribution of chromoelectric field on the 2-dimensional plane. 
The right panel of Fig.~\ref{C35-cf-fig2} shows the distribution of $E_{z}$ component of the chromoelectric field, where the quark-antiquark source represented as the $R \times T$ Wilson loop $W$ is placed at $(Y,Z)=(0,R), (0,0)$, and the probe $U_P$ is displaced on the $Y$-$Z$ plane at the midpoint of the $T$-direction. 
The magnitude of $E_{z}$ is shown by the height of the 3D plot.
We find that the magnitude $E_{z}$  is almost uniform for the original part $U$ except for the neighborhoods of the locations of $q$, $\bar q$ source.

\begin{figure}[ptb]
\begin{center}
\includegraphics[scale=0.65]{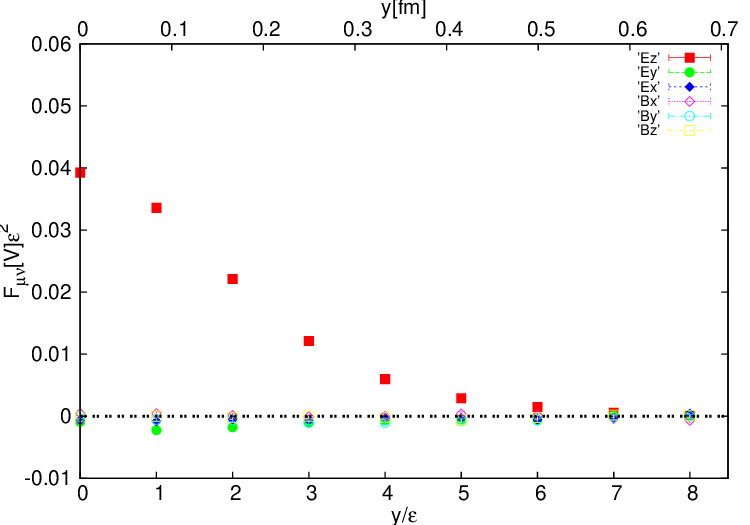} 
%{figs/v_ei_vs_y.eps} 
\quad
\includegraphics[scale=0.65]{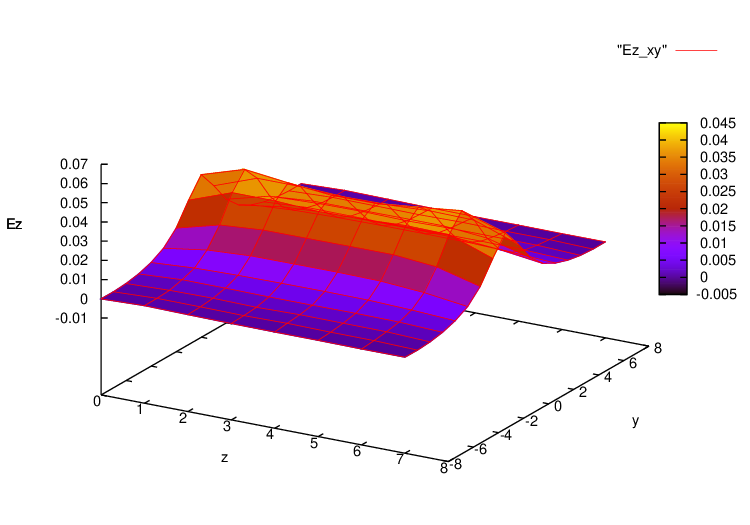} 
\vspace{-5mm}
\end{center}
\caption{Ref.\cite{KKS14}: 
The chromoelectric field obtained from the restricted field $V$ on $24^4$ lattice at $\beta=2.5$.
}
\label{C35-cf-fig3}
\end{figure}

Next, the results for the restricted field $V$ is shown in Fig.~\ref{C35-cf-fig3}.  
From the left panel of Fig.~\ref{C35-cf-fig3}, we find that the strength of the chromoelectric field obtained from the  restricted field becomes smaller than the full one, but the structure of the flux tube survives. 
The ratio of the flux at the origin   $y=0$ is  
%\begin{align}
%E_z^U(0) =9.612\times 10^{-2} , \ E_z^V(0) = 5.832\times 10^{-2}, \ E_z^V(0)/E_z^U(0) = 0.607 
%\quad (\text{for} \ V=16^4 \ \text{and} \ \beta=2.5), 
%\end{align}
\begin{align}
 E^U_z(0)=5.428 \times 10^{-2} , \ E^V_z(0)=3.925\times 10^{-2}, \ E^V_x(0)/E^U_x(0)=0.723 
\quad (\text{on} \ 24^4 \ \text{lattice~at} \ \beta=2.5), 
\end{align}
From the right panel of Fig.~\ref{C35-cf-fig3}, we find that the magnitude $E_{z}$ is quite uniform for the restricted field $V$, compared with the full field. 
This difference is due to  the contributions from the remaining part $X$ which affects only the  short distance, as the correlator of the $X$ field exhibits the exponential fall-off and disappears quickly in the distance as shown in \cite{SKKMSI07}. 
 Thus the restricted field $V$ reproduces the chromoelectric flux tube in the $SU(2)$ Yang-Mills theory on a lattice.

%%%%%%%%%%%%%%%%%%%%%%%%%%%%%%%%%%%%%%%%%%%%%%%%%%%%
%\begin{figure}[ptb]
%\begin{center}
%\includegraphics[scale=0.5]{figs/v_ebi_vs_y.eps} 
%{figs/v_ei_vs_y.eps} 
%\quad
%\includegraphics[scale=0.7]{figs/v_Ex_3d2.eps} 
%\vspace{-5mm}
%\end{center}
%\caption{
%The chromoelectric field obtained from the restricted field $V$ on $V=16^4$ lattice at $\beta=2.5$.
%}
%\label{cf-fig3a}
%\end{figure}
%%%%%%%%%%%%%%%%%%%%%%%%%%%%%%%%%%%%%%%%%%%%%%%%%%%%

%%%%%%%%%%%%%%%%%%%%%%%%%%%%%%%%%%%%%%%%%%%%%%%%%%%%
%%%%%%%%%%%%%%%%%%%%%%%%%%%%%%%%%%%%%%%%%%%%%%%%%%%%

\begin{figure}[ptb]
\begin{center}
%\includegraphics[scale=0.45]{Fig-PR/Fig-lattice/exyz_vs_y_rho2.eps} 
%\quad
\includegraphics[scale=0.65]{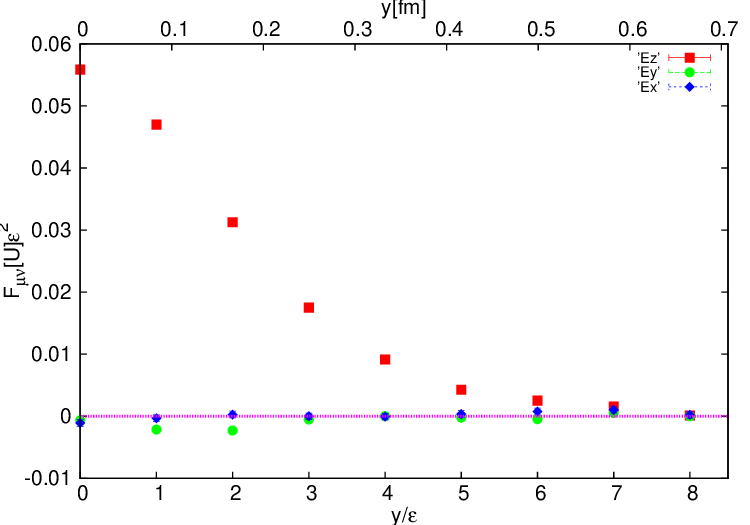} 
\end{center}
\vspace{-5mm}
\caption{Ref.\cite{KKS14}: 
The chromoelectric field $E_j(y)=F_{4j}(y)$ ($j=x,y,z$) obtained from  (\ref{C35-cf2-1}) as a function of the distance $y$, 
%(Left) on $16^4$ lattice at $\beta=2.5$,
%(Right)  
on $24^4$ lattice at $\beta=2.5$. 
}%
\label{C35-cf-fig2-2}
\end{figure}

For comparison, we have calculated also the operator which was estimated by Di Giacomo et al.\cite{GMO90}:
\begin{align} 
 \rho_U' &= \frac{\langle {\rm tr}(U_PL^{\dagger}WL) \rangle}{\langle {\rm tr}(W) \rangle} - \frac{\langle {\rm tr}(U_P) \rangle}{{\rm tr}({\bf 1})}, 
\quad
F'_{\mu\nu}(x)  = \frac{\sqrt{\beta}}{2} \rho_U'(x) 
 .
\label{C35-cf2-1}
\end{align}
It is easy to see that the operator $\rho'$ has the same expression as (\ref{C35-cf1-2}) up to the order $O(\epsilon^2)$ and the difference appears in the order $O(\epsilon^4)$.
The result is shown in Fig.~\ref{C35-cf-fig2-2}.  The comparison of Fig.~\ref{C35-cf-fig2-2} with the left panel of Fig.~\ref{C35-cf-fig2} shows that the value of (\ref{C35-cf2-1}) is consistent with (\ref{C35-cf1-1}).  
The numerical data are given in Table~\ref{C35-cf-tbl2-1}.

%%%%%%%%%%%%%%%%%%%%%%%%%%%%%%%%%%%%%%%%%%%%%%%%%%%%
%%%%%%%%%%%%%%%%%%%%%%%%%%%%%%%%%%%%%%%%%%%%%%%%%%%%

%%%%%%%%%%%%%%%%%%%%%%%%%%%%%%%%%%%%%%%%%%%%%%%%%%%%%
%\begin{table}[hptb]
%\caption{
%The comparison of the chromoelectric field obtained from  (\ref{cf1-1}) and (\ref{cf2-1}) on $16^4$ lattice at $\beta=2.4$. We fix $z$ to be the midpoint, i.e., $z=3$.
%}
%\label{cf-tbl2-1a}
%\begin{center}
%\begin{tabular}{ccc}\hline
%y & $E_z(y)=F_{41}(y)$ &  $E_z'(y)=F'_{41}(y)$ \\ %\hline
%0 &  $9.61173549E-02 (0.16266E-02)$ & $9.91981477E-02 (0.17188E-02)$ \\
%1 &  $7.06650317E-02 (0.14414E-02)$ & $7.26254210E-02 (0.14445E-02)$ \\
%2 &  $3.77369672E-02 (0.13141E-02)$ & $3.81712057E-02 (0.13244E-02)$ \\
%3 &  $1.64689291E-02 (0.11417E-02)$ & $1.67952552E-02 (0.12054E-02)$ \\
%4 &  $7.61921145E-03 (0.94115E-03)$ & $7.88860116E-03 (0.10418E-02)$ \\
%5 &  $2.90196505E-03 (0.11063E-02)$ & $3.01631424E-03 (0.11589E-02)$ \\
%6 &  $2.46271258E-03 (0.10821E-02)$ & $2.14403961E-03 (0.10801E-02)$ \\
%7 &  $1.66375225E-03 (0.13958E-02)$ & $1.19778747E-03 (0.13895E-02)$ \\
%8 &  $1.39909389E-03 (0.11061E-02)$ & $1.60291488E-03 (0.11084E-02)$ \\ \hline
%\end{tabular}
%\end{center} 
%\end{table}
%%%%%%%%%%%%%%%%%%%%%%%%%%%%%%%%%%%%%%%%%%%%%%%%%%%%%

%%%%%%%%%%%%%%%%%%%%%%%%%%%%%%%%%%%%%%%%%%%%%%%%%%%%%
\begin{table}[hptb]
\caption{Ref.\cite{KKS14}: 
The comparison of the chromoelectric field obtained from 
(\ref{C35-cf1-1}) and (\ref{C35-cf2-1}) on $24^4$ lattice at $\beta=2.4$.
We fix $z$ to be the midpoint, i.e., $z=4$.
}
\label{C35-cf-tbl2-1}
\begin{center}
\begin{tabular}{ccc}\hline
y & $E_z(y)=F_{41}(x)$ &  $E_z'(y)=F'_{41}(x)$ \\ \hline
0 &  $5.428(\pm0.062)\times10^{-2} $ & $5.585(\pm0.065)\times10^{-2}$ \\
1 &  $4.560(\pm0.055)\times10^{-2}$ & $4.699(\pm0.059)\times10^{-2}$ \\
2 &  $3.041(\pm0.055)\times10^{-2}$ & $3.127(\pm0.058)\times10^{-2}$ \\
3 &  $1.714(\pm0.050)\times10^{-2}$ & $1.751(\pm0.053)\times10^{-2}$ \\
4 &  $0.901(\pm0.049)\times10^{-2}$ & $0.914(\pm0.054)\times10^{-2}$ \\
5 &  $0.424(\pm0.047)\times10^{-2}$ & $0.426(\pm0.051)\times10^{-2}$ \\
6 &  $0.255(\pm0.048)\times10^{-2}$ & $0.251(\pm0.051)\times10^{-2}$ \\
7 &  $0.149(\pm0.050)\times10^{-2}$ & $0.157(\pm0.054)\times10^{-2}$ \\
8 &  $0.009(\pm0.049)\times10^{-2}$ & $0.011(\pm0.052)\times10^{-2}$ \\\hline
\end{tabular}
\end{center} 
\end{table}
%%%%%%%%%%%%%%%%%%%%%%%%%%%%%%%%%%%%%%%%%%%%%%%%%%%%%

%%%%%%%%%%%%%%%%%%%%%%%%%%%%%%%%%%%%%%%%%%%%%%%%%%%%%
\subsubsection{Magnetic current and  dual Meissner effect}
%%%%%%%%%%%%%%%%%%%%%%%%%%%%%%%%%%%%%%%%%%%%%%%%%%%%%

Although we have confirmed the formation of the chromoelectric flux in $SU(2)$ Yang-Mills theory on a lattice, the existence of the flux tube alone is not sufficient for proving the occurrence of the dual Meissner effect.

Next, we investigate the relation between the chromoelectric flux and the  magnetic current. 
%From the Yang-Mills equation for $\mathbf{V}_{\mu}$ field, 
The magnetic(-monopole) current can be calculated as
\begin{equation}
 k= \delta{}^{\displaystyle *}F[V] ={}^{\displaystyle *}d F[V] , 
\label{C35-def-k}
\end{equation}
where $F[V]$ is the field strength (\ref{C35-cf1-5}) defined from the the restricted field $V$  in the presence of the $q\bar q$ source,
%the field strength (2-form) of the restricted field (1-form) $\mathbf{V}$, 
$d$
the exterior derivative, $\delta$ codifferential, and $^{\ast}$ denotes the Hodge dual operation. 
Note that non-zero magnetic current follows from violation of the Bianchi identity  
(If the field strength was given by the exterior derivative of some  field $A$ (one-form), $ F=dA$, \ we would obtain $k=\delta{}^{\displaystyle *}F={}^{\displaystyle *}d^{2}A=0$).

%%%%%%%%%%%%%%%%%%%%%%%%%%%%%%%%%%%%%%%%%%%%%%%%%%%%%
\begin{figure}[ptb]
\begin{center}
%\vspace{-5mm}
\includegraphics[scale=0.50]{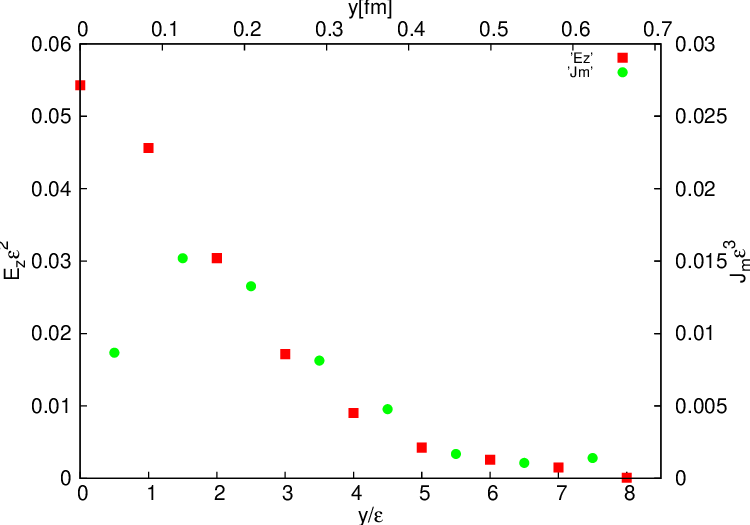}
\includegraphics[scale=0.65]{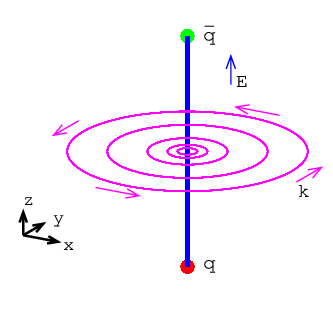} 
\quad
\includegraphics[scale=0.50]{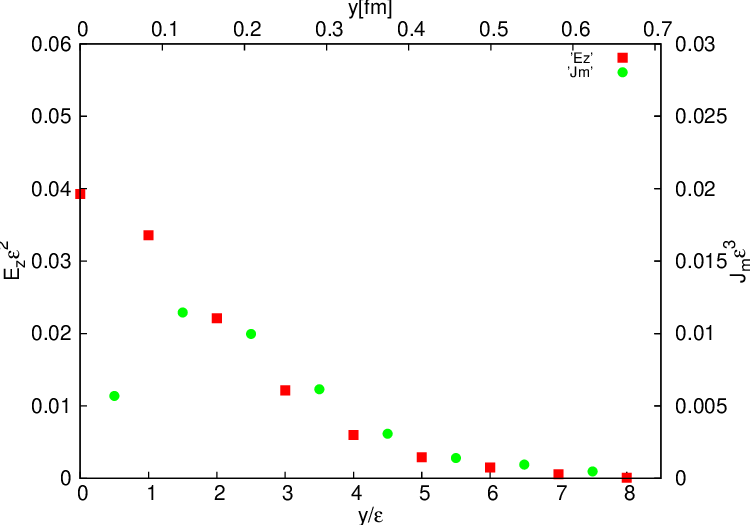}
%{figs/C-flux-M-current.eps} 
\vspace{-5mm}
\end{center}
\caption{Ref.\cite{KKS14}: 
The magnetic-monopole current $\mathbf{k}$ induced around the chromoelectric flux along the $z$ axis connecting a  pair of quark and antiquark.
(Center panel) 
The positional relationship between the chromoelectric field $E_{z}$ and the magnetic current $\mathbf{k}$. 
(Left panel) 
The magnitude of the chromoelectric field $E_{z}$ and the magnetic current  $J_{m}=|\mathbf{k}|$ as functions of the distance $y$ from the $z$ axis calculated from the original full variables. 
(Right panel) 
The counterparts of the left graph calculated from the restricted variables. 
}
\label{C35-fig:Mcurrent}%
\end{figure}
%%%%%%%%%%%%%%%%%%%%%%%%%%%%%%%%%%%%%%%%%%%%%%%%%%%%%

If only the components $E_z=F_{34}=-F_{43}$ are non-vanishing  among $F_{\alpha\beta}$, then $k^{\mu}   =  \frac{1}{2}\epsilon^{\mu\nu\alpha\beta} \partial_{\nu}F_{\alpha\beta}$ reads 
\begin{align}
k^{\mu}  = \epsilon^{\mu\nu34} \partial_{\nu}F_{34}
= \epsilon^{\mu\nu34} \partial_{\nu}E_{z}   ,
\end{align}
and the non-vanishing components of $k^{\mu}$ are given by  $k^1, k^2$ in the $X$-$Y$ plane:
\begin{align}
k^{1} %= \epsilon^{1234} \partial_{y}E_{z} 
= \partial_{y}E_{z}, \quad
k^{2} %= \epsilon^{2134} \partial_{x}E_{z} 
= - \partial_{x}E_{z},  \quad k^3=0,  \quad k^4=0 .
\end{align}

Fig.~\ref{C35-fig:Mcurrent} shows the  magnetic current measured in $X$-$Y$ plane at the midpoint of $q\bar q$ pair in the $Z$-direction. 
The left panel of Fig.~\ref{C35-fig:Mcurrent} shows the positional relationship between chromoelectric flux and  magnetic current.
The right panel of Fig.~\ref{C35-fig:Mcurrent} shows the magnitude of the  chromoelectric field $E_z$ (left
scale) and the magnetic current $k$ (right scale). 
The existence of non-vanishing magnetic current $k$ around the chromoelectric field $E_z$ supports the dual picture of the ordinary superconductor exhibiting the electric current $J$ around the magnetic field $B$.

The above results show the simultaneous formation of the chromoelectric flux tube and the associated magnetic-monopole current  induced around it.    
Thus, we have confirmed the dual Meissner effect in $SU(2)$ Yang-Mills theory on a lattice. 
We have also shown that the restricted field $V$ reproduces the dual Meissner effect in the $SU(2)$ Yang-Mills theory on a lattice.

In our formulation, it is possible to define a gauge-invariant magnetic-monopole current  $K_{\mu}$  by using $V$-field,
\begin{subequations}
\begin{align}
K_{\mu}(x)  &  =2\pi m_{\mu}(x) :=\frac{1}{2}\epsilon_{\mu\nu\alpha\beta} \partial_{\nu} \bar\Theta_{\alpha\beta}(x) ,
 \\
\bar\Theta_{\mu\nu}(x)   &  :=-\arg \left\{ \text{ \textrm{Tr}}\left[  \left(   \mathbf{1}+2\bm{n}_{x}\right)  V_{x,\mu}V_{x+\mu,\mu}V_{x+\nu,\mu}^{\dag}V_{x,\nu}^{\dag} \right]/{\rm tr}(1) \right\} ,
\end{align}
\end{subequations}
which is obtained from the field strength $\mathscr{F}[\mathscr{V}]$ of the restricted field $\mathscr{V}$, as suggested from the non-Abelian Stokes theorem \cite{Kondo08}.
%It should be also noticed that this magnetic-monopole current  is a non-Abelian magnetic monopole extracted from the $V$ field, which corresponds to the maximal stability group $\tilde{H}=U(2)$.
The magnetic-monopole current  $K_{\mu}$ defined in this way can be used to study the magnetic current around the chromoelectric flux tube, instead of the above definition   (\ref{C35-def-k}) of $k$.
The comparison of two magnetic-monopole currents will be a subject in the future work.

%%%%%%%%%%%%%%%%%%%%%%%%%%%%%%%%%%%%%%%%%%%%%%%%%%%
%\newpage
\subsubsection{Ginzburg-Landau parameter and type of dual superconductor}
%%%%%%%%%%%%%%%%%%%%%%%%%%%%%%%%%%%%%%%%%%%%%%%%%%%

Moreover, we investigate the type of the dual superconductor in the QCD vacuum. 
The \textbf{Ginzburg-Landau (GL) parameter} in the superconductor is defined from the penetration depth $\lambda$ and the coherence length $\xi$ by
\begin{align} 
\kappa = \frac{\lambda}{\xi} \ .
\label{C35-cf4-1}
\end{align}
The superconductor is called the type-I when $\kappa<\frac{1}{\sqrt{2}}$, while type-II when $\kappa>\frac{1}{\sqrt{2}}$.
In the type-I superconductor, the attractive force acts between two vortices, while the repulsive force in the type-II superconductor.
There is no interaction at $\kappa=\frac{1}{\sqrt{2}} \simeq 0.707$. 
The preceding studies support that the dual superconductor for the $SU(2)$ lattice Yang-Mills theory is at the border between type-I and type-II, or  weak type-I \cite{Suzuki:1988}.

Usually, in the dual superconductor of the type II, it is justified to use the asymptotic form $K_0(y/\lambda)$ to fit the chromoelectric field in the large $y$ region (as the solution of the Ampere equation in the dual GL theory).  
However, it is clear that this solution cannot be applied to the small $y$ region, as is easily seen from the fact that $K_0(y/\lambda) \to \infty$ as $y \to 0$. 
In order to see the difference between type I and type II, it is crucial to see the relatively small $y$ region.
Therefore, such a simple form cannot be used to detect the type I dual  superconductor. 
However, this important aspect was ignored in the preceding studies except for a work \cite{CCP12}.

We proceed to determine the GL parameter $\kappa$ of the dual superconductor for $SU(2)$ lattice Yang-Mills theory using the numerical data for the chromoelectric field obtained in the previous section.
We can measure the penetration depth $\lambda$ of the chromoelectric field directly from the data obtained in the previous section without any assumption. 
In order to obtain the the coherence length $\xi$, however, we must solve the coupled nonlinear differential equations in the \textbf{GL theory}, i.e., the \textbf{GL equation} and the  Ampere equation. 
In the GL theory, the gauge field $A$ and the scalar field $\phi$ obey simultaneously  the GL equation:
\begin{equation}
 (\partial^\mu -iq A^\mu)(\partial_\mu -iq A_\mu) \phi + \lambda_4 (\phi^* \phi - \eta^2) = 0 ,
\end{equation}
and the Ampere equation:
\begin{equation}
 \partial^\nu F_{\mu\nu} + iq [\phi^* (\partial_\mu \phi -iq A_\mu \phi)  - (\partial_\mu \phi -iq A_\mu \phi)^* \phi] = 0 .
%- iq(\phi^* \partial_\mu \phi - \phi \partial_\mu \phi^*) - 2q^2 A_\mu \phi^* \phi = 0 .
\end{equation}
%In order to examine the type of the dual superconductivity, we apply the formula for the magnetic field  derived by Clem \cite{Clem75} in the ordinary superconductor based on the Ginzburg-Landau (GL) theory to the chromoelectric field in the dual superconductor.
To avoid this, we follow the method given by Clem \cite{Clem75} invented for the ordinary superconductor based on the GL theory, which was recently applied to the dual superconductor for $SU(3)$ lattice Yang-Mills theory by  
\cite{CCP12,SKKS13}.
%Here we do not repeat the advantages of this method. 
The advantage of this method is that it is able to take into account the whole range of $y$ for fitting the data to determine precisely the type of (dual) superconductivity, in sharp contrast to the preceding approach which uses only the asymptotic region at large $y \gg 1$. 
%\footnote{
%See the chapter of Dual superconductivity for details. 
%}
By applying the Clem method to the dual superconductor, the chromoelectric field $E_z(y)$ must obey
\begin{align} 
E_z(y) = \frac{\Phi}{2\pi}\frac{\mu^2}{\alpha}\frac{K_0(\sqrt{\mu^2y^2+\alpha^2})}{K_1(\alpha)} ,
\label{C35-cf4-2}
\end{align}
where $\Phi$ is the external electric flux, $\mu$ and $\alpha$ are defined by 
\begin{align} 
\mu := \frac{1}{\lambda}, \quad {\alpha} := \frac{\zeta}{\lambda} ,
\label{C35-cf4-3}
\end{align}
and $K_0$ and $K_1$ are the modified Bessel functions of zeroth and first order respectively. 
Here $\zeta$ is the variational parameter representing  the core radius. 
The GL parameter $\kappa$ is written in terms of $\alpha$ alone:
\begin{align} 
\kappa = \frac{\sqrt{2}}{\alpha}[1-K_0^2(\alpha)/K_1^2(\alpha)]^{1/2}, \quad {\alpha} := \frac{\zeta}{\lambda} .
\label{C35-cf4-4}
\end{align}

%Here (\ref{C35-cf4-2}) is approximated:
%\begin{align} 
%E_z(y) = \frac{C}{(\sqrt{\mu^2y^2+\alpha^2})^{1/2}}
%e^{-\sqrt{\mu^2y^2+\alpha^2}}
%\left[ 1-\frac{1}{8\sqrt{\mu^2y^2+\alpha^2}}  ?? +\frac{9}{128(\sqrt{\mu^2y^2+\alpha^2})^2} \right] ,
%\label{C35-cf4-6}
%\end{align}
%by taking into account the first order of the asymptotic expansion of the modified Bessel function:
%\begin{align} 
%K_{\nu}[x] = \sqrt{\frac{\pi}{2x}}e^{-x}\sum_{r=1}^{\infty}
% \frac{(4\nu^2-1^2)(4\nu^2-3^2)\cdots\{4\nu^2-(2r-1)^2\}}{r\!(8x)^r} .
%\label{C35-cf4-5}
%\end{align}
Then the obtained value of $\alpha$ is used to determine the GL parameter $\kappa$ according to (\ref{C35-cf4-4}).

%We adopt the following method to determine the GL parameter $\kappa$. 
 
%First, we perform the $\chi^2$-fit of the data for the chromoelectric field $E_z(y)$ given in Fig.~\ref{C35-cf-fig2} and Fig.~\ref{C35-cf-fig3} according to (\ref{C35-cf4-2}) to obtain $\alpha$. 
%Here we have used the Table of the modified Bessel function given in \cite{AS72}.Here (\ref{C35-cf4-2}) is approximated as
%\begin{align} 
 %E_z(y) = \frac{C}{(\sqrt{\mu^2y^2+\alpha^2})^{1/2}} e^{-\sqrt{\mu^2y^2+\alpha^2}}
%\left[ 1-\frac{1}{8\sqrt{\mu^2y^2+\alpha^2}}+\frac{9}{128(\sqrt{\mu^2y^2+\alpha^2})^2} \right] ,
%\label{C35-cf4-6}
%\end{align}
%by taking into account the second order of the asymptotic expansion of the modified Bessel function:
%\begin{align} 
%K_{\nu}[x] = \sqrt{\frac{\pi}{2x}}e^{-x}\sum_{r=1}^{\infty}
%\frac{(4\nu^2-1^2)(4\nu^2-3^2)\cdots\{4\nu^2-(2r-1)^2\}}{r\!(8x)^r} .
%\label{C35-cf4-5}
%\end{align}
%Then the obtained value of $\alpha$ is used to determine the GL parameter $\kappa$ according to (\ref{C35-cf4-4}).

%The graph of the fitting is given in Fig.~\ref{C35-cf-fig5} and the obtained values for the fitted parameters are given in Table~\ref{C35-cf-tbl4-1}. Thus we have obtained the GL parameter for the full field $\kappa_U$ and the restricted field $\kappa_V$:
%\begin{align} 
%\kappa_U = 0.806 \pm 0.248, \quad \kappa_V = 0.446 \pm 0.208.
%\label{C35-cf4-7}
%\end{align}

\begin{figure}[ptb]
\begin{center}
\includegraphics[scale=0.65]{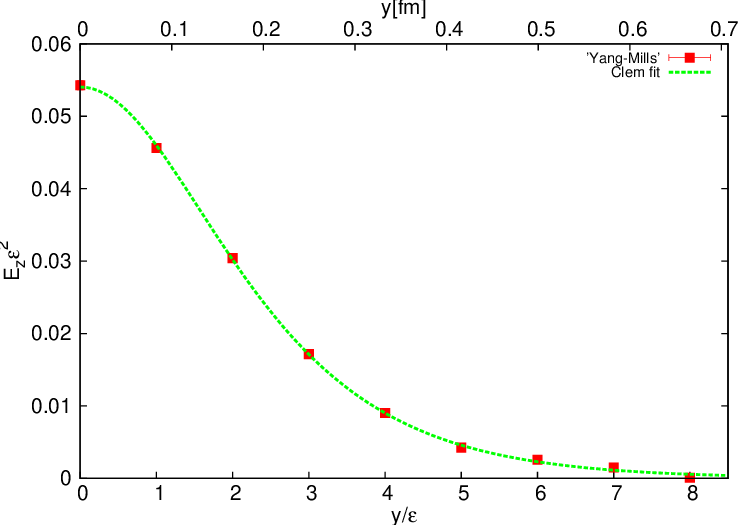} 
\quad
\includegraphics[scale=0.65]{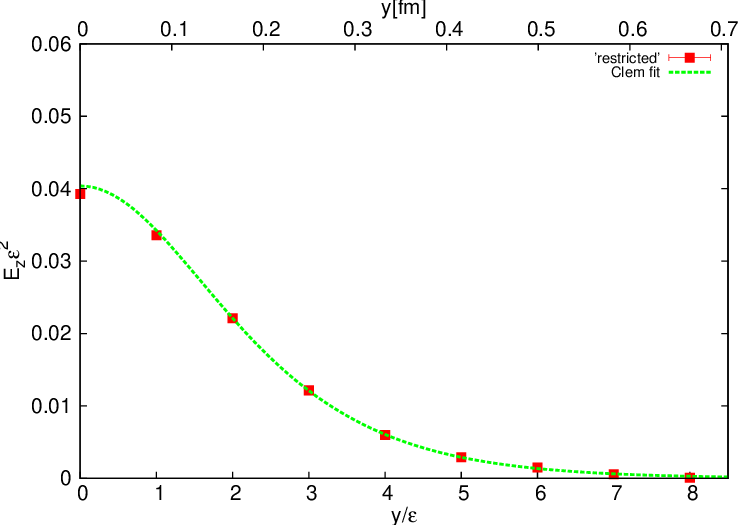} 
\vspace{-5mm}
\end{center}
\caption{Ref.\cite{KKS14}: 
The magnitude of the chromoelectric field $E_z(y)$ as a function of the distance $y$. 
(Left panel) The data of Fig.~\ref{C35-cf-fig2} and the fitted result for the full field. 
(Right panel) The data of Fig.~\ref{C35-cf-fig3} and the fitted result for the restricted field.
The fit range we have used is [0,8] for the full field in the left panel and [2,8] for the restricted field in the right panel. 
}
\label{C35-cf-fig5}
\end{figure}

The graph of the fitting is given in Fig.~\ref{C35-cf-fig5} and the obtained values for the fitted parameters are given in Table~\ref{C35-cf-tbl4-1}
where we have used the fitting function: 
\begin{equation}
E_z(y)=c K_{0}(\sqrt{\mu^{2}y^{2}+\alpha^{2}}) , \quad
c= \frac{\Phi}{2\pi}\frac{\mu^2}{\alpha}\frac{1} {K_{1}(\alpha)}
= \frac{\Phi}{2\pi}\frac{1}{\lambda \zeta}\frac{1} {K_{1}(\zeta/\lambda)} .
%=\left(  \phi/2\pi\right)  (\xi/\lambda)/K_{1}(\xi/\lambda), 
%\quad
%\mu=\frac{1}{\lambda}, 
%\quad
%\alpha= \frac{\zeta}{\lambda}  .
\label{C35-Clem-fit}
\end{equation}
Thus we have obtained the GL parameter for the full field $\kappa_U$ and the restricted field $\kappa_V$:
\begin{align} 
\kappa_U = 0.484 \pm 0.070 \pm 0.026, \quad \kappa_V = 0.377 \pm 0.079 \pm 0.018.
%\kappa_U = 0.404 \pm 0.108, \quad \kappa_V = 0.504 \pm 0.257.
%\kappa_U = 0.717 \pm 0.208, \quad \kappa_V = 0.491 \pm 0.150.
\label{C35-cf4-7}
\end{align}
Here and in what follows, the first error denotes the statistics error and the second one denotes the systematic error or the lattice artifact due to choosing the center or the corner of the plaquette as the representative of $E_z(y)$.

%%%%%%%%%%%%%%%%%%%%%%%%%%%%%%%%%%%%%%%%%%%%%%%%%%%%
%\begin{figure}[hptb]
%\begin{center}
%\includegraphics[scale=0.6]{figs/fit_u_Ex_s.eps} 
%{figs/u_ex_fit.eps} 
%\quad
%\includegraphics[scale=0.6]{figs/fit_v_Ex_s.eps} 
%{figs/v_ex_fit.eps} 
%\vspace{-5mm}
%\end{center}
%\caption{
%The magnitude of the chromoelectric field $E_z(y)$ as a function of the distance $y$. 
%(Left panel) The data of Fig.~\ref{cf-fig2a} and the fitted result for the full field, 
%(Right panel) The data of Fig.~\ref{cf-fig3a} and the fitted result for the restricted field. The fit range we have used are [1,8] for the full field in the left panel and [0,6] for the restricted field in the right panel. 
%}
%\label{cf-fig5a}
%\end{figure}
%%%%%%%%%%%%%%%%%%%%%%%%%%%%%%%%%%%%%%%%%%%%%%%%%%%%

%%%%%%%%%%%%%%%%%%%%%%%%%%%%%%%%%%%%%%%%%%%%%%%%%%%%
%%%%%%%%%%%%%%%%%%%%%%%%%%%%%%%%%%%%%%%%%%%%%%%%%%%%

%%%%%%%%%%%%%%%%%%%%%%%%%%%%%%%%%%%%%%%%%%%%%%%%%%%%%
\begin{table}[htp]
\caption{\cite{KKS14} Summary of the fit values }
\label{C35-cf-tbl4-1}
\begin{center}
\begin{tabular}{lccc}\hline
                                   & $c$    &  $\mu$  &  $\alpha$  \\ \hline
link field $U$           &  $0.355 (0.096)$  & $0.689 (0.039)$  &  $1.767 (0.218)$    \\
restricted field $V$ &  $0.414 (0.187)$  & $0.774 (0.052)$  &  $2.128 (0.400)$    \\ \hline
%link field $U$           &  $0.472 (0.299)$  & $0.666 (0.080)$  &  $2.025 (0.523)$    \\
%restricted field $V$ &  $0.272 (0.369)$  & $0.716 (0.171)$  &  $1.713 (1.188)$    \\ \hline
\end{tabular}
\end{center} 
\end{table}
%%%%%%%%%%%%%%%%%%%%%%%%%%%%%%%%%%%%%%%%%%%%%%%%%%%%%

%%%%%%%%%%%%%%%%%%%%%%%%%%%%%%%%%%%%%%%%%%%%%%%%%%%%%
%\begin{table}[htp]
%\caption{Summary of the fit values }
%\label{C35-cf-tbl4-1}
%\begin{center}
%\begin{tabular}{lccc}\hline
%                                   & $c$    &  $\mu$  &  $\alpha$  \\ \hline
%full link field $U$           &  $0.341 (0.167)$  & $0.781 (0.087)$  &  $1.308 (0.393)$    \\
%restricted field $V$ &  $0.368 (0.249)$  & $0.782 (0.109)$  &  $1.748 (0.548)$    \\ \hline
%\end{tabular}
%\end{center} 
%\end{table}
%%%%%%%%%%%%%%%%%%%%%%%%%%%%%%%%%%%%%%%%%%%%%%%%%%%%%

%%%%%%%%%%%%%%%%%%%%%%%%%%%%%%%%%%%%%%%%%%%%%%%%%%%%%
%\begin{table}[htp]
%\caption{Summary of the fit values }
%\label{C35-cf-tbl4-1}
%\begin{center}
%\begin{tabular}{lccc}\hline
%                                   & $C$    &  $\mu$  &  $\alpha$  \\ \hline
%full field $U$           &  $0.366 (0.115)$  & $0.753 (0.066)$  &  $1.194 (0.227)$    \\
%restricted field $V$ &  $0.538 (0.313)$  & $0.807 (0.092)$  &  $1.879 (0.473)$    \\ \hline
%\end{tabular}
%\end{center} 
%\end{table}
%%%%%%%%%%%%%%%%%%%%%%%%%%%%%%%%%%%%%%%%%%%%%%%%%%%%%

The penetration depth $\lambda$ is obtained  using the first equation of (\ref{C35-cf4-3}) from $\mu$, i.e., $\lambda=1/\mu$, while the coherence length $\xi$ is obtained using (\ref{C35-cf4-1}) from $\lambda$ and the GL parameter $\kappa$ (\ref{C35-cf4-7}), i.e., $\xi=\lambda/\kappa$. 
The full link variable yields
\begin{align} 
\lambda_U = 0.121(7+0) \text{fm}, \quad \xi_U = 0.250(4+1) \text{fm}.
%\lambda_U = 0.125(15) \text{fm}, \quad \xi_U = 0.309(9) \text{fm}.
%\lambda_U = 0.107(12) \text{fm}, \quad \xi_U = 0.149(5) \text{fm} .
\label{C35-cf4-8a}
\end{align}
where we have used the value of scale 
 $\epsilon(\beta=2.5)=0.08320$ fm  of Ref.\cite{SKKMSI07}.%
\footnote{
This corresponds to gauge boson mass $m_A$ and the scalar boson mass $m_\phi$: 
$m_A=1.64{\rm GeV}$, $m_\phi= 1.1 {\rm GeV}$.
} 
While the restricted field gives
\begin{align} 
\lambda_V = 0.107(7+0) \text{fm}, \quad \xi_V = 0.285(7+1) \text{fm}.
%\lambda_V = 0.116(27) \text{fm}, \quad \xi_V = 0.230(14) \text{fm}.
%\lambda_V = 0.106(14) \text{fm}, \quad \xi_U = 0.217(8) \text{fm}.
\label{C35-cf4-8b}
\end{align}
The obtained results are consistent with the result $\lambda=0.1135(27)\text{fm}$ of Ref.\cite{CCP12}.

Our results show  that the dual superconductor for the $SU(2)$ lattice Yang-Mills theory is the weak \textbf{type I}, rather than the border between type-I and type-II.
Our results are to be compared with the preceding works.
First, Cea, Cosmai and Papa \cite{CCP12} have used the same operator and the same fitting function (\ref{C35-cf4-2}) as ours.  In the calculation of the operator, however, they used the cooling method, while we have used HYP. 
In addition, they have confirmed that the GL parameter is shifting toward the type I as the cooling step is forward.
%The result is $\kappa=0.467\pm 0.310$ indicating the weak type I.
They used the constant fit of the data obtained at three points $\beta=2.252,2.55,2.6$ on a lattice $20^4$, and they checked that the value of $\kappa$ does not depend on $\beta$. 
The obtained value of the GL parameter $\kappa_U$ is consistent with the value $\kappa=0.467\pm 0.310$  within errors.
From our point of view, the cooling method cannot be applied to the restricted part, since it amounts to changing the action. 
%The obtained value of the GL parameter $\kappa_U$ is a bit smaller (toward the type I) than the value $\kappa=0.467\pm 0.310$ reported by Cea, Cosmai and Papa \cite{CCP12}, but it is consistent within errors.
%, which is consistent with the preceding results \cite{Suzuki:1988}.
%, since the approximated fitting function (\ref{cf4-6}) tends to yield smaller value of $\alpha$ and hence larger $\kappa$ than using the exact equation (\ref{cf4-2}).

Next, Bali, Schlichter and Schilling \cite{Suzuki:1988} have used the different operator without the Schwinger line, namely, the action distribution proportional to the squared field strength, and applied APE smearing to the Wilson loop alone.
They have used a different fitting function than ours.
%\begin{eqnarray} 
%E_z(x)=\frac{\Phi}{2\pi}\{ \frac{b}{2C_a\lambda^2}\frac{1}{\cosh(x/\lambda)}+\frac{2(1-b)}{\delta^2}\exp(-\frac{x^2}{\delta^2}) \}
%\end{eqnarray}
The result is $\kappa=0.59^{+13}_{-14}$  indicating the border between type I and II.

Third, Suzuki et al. \cite{Suzuki:2009xy} have used the Abelian-projected operator and applied the APE smearing to the Wilson loop alone.
They used the improved Iwasaki action at three values of $\beta=1.10,1.28,1.40$ to calculate  the Wilson loop of small size: $W(R=3,T=5)$, $W(R=5,T=5)$, $W(R=7,T=7)$ where the interquark distance $q-\bar{q}$ is fixed to be 0.32fm.
Moreover, $\langle E_{A_z}(y) \rangle_W$ is fitted to $c_1\exp(-y/\lambda)+c_0$, and $\langle  k_{\mu}^2(y) \rangle_W$ is fitted to $c'_1\exp(-\sqrt{2}y/\xi)+c'_0$ to calculate the GL parameter. The result is $\sqrt{2} \kappa=1.04(7), 1.19(5), 1.09(8)$ indicating the border between type I and II.

%In the end of section 9.4.8 on page 176, add the following paragraph.

Fourth, Koma et al. \cite{KKIS03}
%[Y. Koma, M. Koma, E.-M. Ilgenfritz, and T. Suzuki, 
%Phys. Rev. D{\bf 68}, 114504 (2003)] 
have measured the expectation values of Abelian electric field $E_z$ and monopole current $k_{\phi}$ by using an Abelian Wilson loop, in order to find flux-tube profile in the Maximally Abelian gauge.
The numerical simulations were done at $\beta=2.3 \sim 2.6$ on a lattice $32^4$.
They fit $E_z$ and $k_{\phi}$ with the classical flux-tube solution of the lattice Dual Abelian Higgs model, and found the optimal set of parameters: dual gauge coupling $\beta_g$, dual gauge boson mass $m_B$, and monopole mass (Higgs scalar mass) $m_{\chi}$.
Their result is $\sqrt{2}\kappa= m_{\chi}/m_B=0.87(2)<1$, indicating the weakly type I.
 
%
%\bibitem{KKIS03}
%Y. Koma, M. Koma, E.-M. Ilgenfritz, and T. Suzuki, 
%Phys. Rev. D{\bf 68}, 114504 (2003).

%%%%%%%%%%%%%%%%%%%%%%%%%%%%%%%%%%%%%%%%%%%%%%%%%%%%%
\subsubsection{A new gauge-invariant chromofield strength}
%%%%%%%%%%%%%%%%%%%%%%%%%%%%%%%%%%%%%%%%%%%%%%%%%%%%%

An advantage of the new formulation is that we can give another definition of the gauge-invariant chromofield strength in the presence of the $q\bar q$ source, which does not need the Schwinger line $L$ and $L^\dagger$ to give the gauge-invariant chromofield strength.
We propose a gauge-invariant chromofield strength:
\begin{align} 
 \tilde\rho_V  = \frac{\langle \epsilon^2 \bar{\Theta}_{x,\mu\nu}[V, \bm{n}] {\rm tr}(W) \rangle}{\langle {\rm tr}(W) \rangle}  
 \overset{\varepsilon\rightarrow0}{\simeq}
%g\epsilon^{2}\left\langle \mathscr{F}_{\mu\nu}\right\rangle _{q\bar{q}}:= 
g \frac{\langle  -i \epsilon^2 {\rm tr}(\bm{n}_x \mathscr{F}_{\mu\nu}[\mathscr{V}]) {\rm tr}(W) \rangle}{\langle {\rm tr}({\bf 1}) \rangle \langle {\rm tr}(W) \rangle}  +O(\epsilon^{4}) ,
\quad
\tilde F_{\mu\nu}^{V}(x)  = \frac{\sqrt{\beta}}{2} \tilde\rho_V(x) 
  ,
\label{C35-F-new1}
\end{align}
where 
\begin{align} 
\bar{\Theta}_{x,\mu\nu}[V,\bm{n}] := \epsilon^{-2}
{\rm arg} ( {\rm tr} \{({\bf 1}+ 2\bm{n}_x)V_{P} \}/{\rm tr}({\bf 1})) .
\label{C35-cfn-mono-5b}
\end{align}
Since $\bar{\Theta}_{x,\mu\nu}[V,{\bf n}]$ is   gauge-invariant from the beginning, we do not need the Schwinger line $L$ and $L^\dagger$ to define gauge-invariant chromofield strength.
Note that $\tilde\rho_V$ is equal to 
\begin{align} 
 \tilde\rho_V  &= \frac{\langle {\rm tr} \{({\bf 1}+2\bm{n}_x)V_{P} \}  {\rm tr}(W) \rangle}{\langle {\rm tr}({\bf 1}) \rangle \langle {\rm tr}(W) \rangle} - 1
%\nonumber\\
%&
%\overset{\varepsilon\rightarrow0}{\simeq}g\epsilon^{2}\left\langle
%\mathscr{F}_{\mu\nu}[V]\right\rangle _{q\bar{q}}:= \frac{\langle \epsilon^2 \bar{\Theta}_{x,\mu\nu}[V,\bm{n}] {\rm tr}(W) \rangle}{\langle {\rm tr}(W) \rangle}  +O(\epsilon^{4}) ,
%\quad
%\tilde F_{\mu\nu}^{V}(x)  = \frac{\sqrt{\beta}}{2} \tilde\rho_V(x) 
 .
\label{C35-F[V]-new2}
\end{align}
In view of this, we can also define the gauge-invariant field strength related to the original variable:
\begin{align} 
 \tilde\rho_U  &= \frac{\langle {\rm tr} \{({\bf 1}+ 2\bm{n}_x)U_{P} \}  {\rm tr}(W) \rangle}{\langle {\rm tr}({\bf 1}) \rangle  \langle {\rm tr}(W) \rangle} - 1
%\nonumber\\
%&
\overset{\varepsilon\rightarrow0}{\simeq}
%g\epsilon^{2}\left\langle \mathscr{F}_{\mu\nu}[U]\right\rangle _{q\bar{q}}:=
 \frac{\langle \epsilon^2 \bar{\Theta}_{x,\mu\nu}[U,\bm{n}] {\rm tr}(W) \rangle}{\langle {\rm tr}(W) \rangle}  +O(\epsilon^{4}) ,
\quad
\tilde F_{\mu\nu}^{U}(x)  = \frac{\sqrt{\beta}}{2} \tilde\rho_U(x) 
 .
\label{C35-F[U]-new3}
\end{align}
Although the numerical simulations based on these  operators are in principle possible, the detailed studies  will be postponed to the subsequent works.

%%%%%%%%%%%%%%%%%%%%%%%%%%%%%%%%%%%%%%%%%%%%%%%%%%
%%%%%%%%%%%%%%%%%%%%%%%%%%%%%%%%%%%%%%%%%%%%%%%%%%
\subsubsection{Infrared  Abelian dominance for correlation functions and off-diagonal gluon ``mass''}\label{subsubsection:IAD-correlation}
%\setcounter{equation}{0}
%%%%%%%%%%%%%%%%%%%%%%%%%%%%%%%%%%%%%%%%%%%%%%%%%%
%%%%%%%%%%%%%%%%%%%%%%%%%%%%%%%%%%%%%%%%%%%%%%%%%%

The restricted field $\mathbf{V}_\mu$  can be regarded as the ``Abelian'' part  in the reformulated  Yang-Mills theory by the following reasons.
\begin{enumerate}
\item[i)] 
The restricted part  $\mathbf{V}_\mu^{A}$ corresponds to the diagonal part of the gauge potential $\mathbf{A}_\mu^{A}$ in the context of the conventional MA gauge which is reproduced when the color direction field is aligned in the same direction over the whole space--time, for example,   
\begin{equation}
 \vec{n}(x) \rightarrow \vec{n}_0 :=(0,0,1) .
\label{C35-MAGlimit}
\end{equation}

\item[ii)] 
The Wilson loop operator written in terms of $\mathbf{A}_\mu^{A}$ 
%is rewritten into the reduced Wilson loop average $\tilde{W}(C)$ which 
is entirely rewritten in terms of $\mathbf{V}_\mu^{A}$ in the reformulated Yang-Mills theory, as demonstrated in \cite{Kondo06}.

\item[iii)] 
The mass term for $\mathbf{X}_\mu^{A}$ can be introduced without breaking gauge invariance in this reformulation \cite{KMS05}.  
The correlation function of $\mathbf{X}_\mu^{A}$ falls off quickly in the long distance, while this is not the case for $\mathbf{V}_\mu^{A}$. 
In fact, it has been shown to one-loop order \cite{Kondo06} that such an effective mass term is  generated due to the gauge-invariant dimension two condensate $\left< \mathbf{X}_\mu^A \mathbf{X}_\mu^A \right>$ thanks to the {\it gauge invariant} self-interaction term $\frac14 g^2(\epsilon^{ABC}\mathbf{X}_\mu^B \mathbf{X}_\nu^C)^2$ among $\mathbf{X}_\mu$ gluons, in sharp contrast to the ordinary self-interaction term $\frac14 g^2(\epsilon^{ABC}\mathbf{A}_\mu^B \mathbf{A}_\nu^C)^2$ which is not gauge-invariant.  
\end{enumerate}

According to (iii), in the energy region lower than the mass $M_X$ of the field $\mathbf{X}_\mu$, the remaining components $\mathbf{X}_\mu$ should decouple or negligible and the $\mathbf{V}_\mu$ field could be dominant.
According to (ii), this leads to the infrared restricted field dominance in the string tension  in our reformulation.

Keeping these facts in mind, we proceed to obtain a fitting function of the two--point correlation function.
Suppose that the Yang-Mills theory has the effective mass term:
\begin{equation}
 \frac12 M_X^2 \mathbf{X}_\mu \cdot \mathbf{X}_\mu  =
 \frac12 M_X^2 \mathbf{X}_\mu^A \mathbf{X}_\mu^A  .
\end{equation}
An additional quadratic term in $\mathbf{X}_\mu$ of the following type could be generated from  gauge fixing conditions in the differential form \cite{KMS05}.
\begin{equation}
 - \frac{1}{2\beta} (\partial^\mu \mathbf{X}_\mu^A)^2  .
\end{equation} 
This can be understood as follows. 
Recall that we impose an constraint called the reduction condition to obtain the reformulated Yang-Mills theory with the original gauge symmetry $SU(2)$ even after introducing the color field ${\bf n}(x)$ which apparently increases  gauge degrees of freedom \cite{KMS05}.
Then we introduce a gauge-fixing parameter $\alpha$ for the reduction condition of the form: 
$- \frac{1}{2\alpha} (\mathscr{D}^\mu[\mathbf{V}] \mathbf{X}_\mu)^2$.  
This term does not fix the original $SU(2)$ gauge invariance, since the reduction condition is the gauge fixing condition from the enlarged gauge symmetry to the original gauge symmetry. 

In order to obtain the correlation function that is not gauge invariant, we need to fix the original gauge symmetry.  
Therefore, we adopt the Landau gauge for the overall gauge fixing of $\mathbf{A}_\mu^A$, whose differential form is $\partial_\mu \mathbf{A}_\mu^A=0$.  This gives an additional quadratic term:
$
 - \frac{1}{2\alpha^\prime} (\partial^\mu \mathbf{X}_\mu^A)^2  
$ 
coming from the gauge-fixing term: 
$
 - \frac{1}{2\alpha^\prime} (\partial^\mu \mathbf{A}_\mu^A)^2  
$.  
Therefore, combining two terms yields an additional term quadratic in $\mathbf{X}_\mu^A$: 
$
 - \frac{1}{2\beta} (\partial^\mu \mathbf{X}_\mu^A)^2  
$
with $\beta^{-1}=\alpha^{-1}+\alpha^{\prime}{}^{-1}$.
Thus we assume the effective propagator for $\mathbf{X}$ gluon of the form:
\begin{align}
 D_{\mu\nu}^{XX}(k) =  \frac{-1}{k^2-M_X^2} \left[ \delta_{\mu\nu} - (1-\beta) \frac{k_\mu k_\nu}{k^2-\beta M_X^2} \right] .
%\\
% D_{\mu\mu}^{XX}(k) =  \frac{-1}{k^2-M_X^2} \left[ D - (1-\beta) \frac{k^2}{k^2-\beta M_X^2} \right] .  
\end{align}
In particular, the limit $\beta \rightarrow \infty$ reproduces the Proca case: 
\begin{equation}
 D_{\mu\nu \infty}^{XX}(k) =  \frac{-1}{k^2-M_X^2} \left[  \delta_{\mu\nu} -\frac{k_\mu k_\nu}{M_X^2} \right] . 
%\quad 
% D_{\mu\mu \infty}^{XX}(k) =  \frac{-1}{k^2-M_X^2} \left[ D -\frac{k^2}{M_X^2} \right] . 
\end{equation}
This form was adopted in the study of off-diagonal gluon mass generation in MA gauge \cite{AS99} where the mass term $\frac12 M_{\rm off}^2 A_\mu^a A_\mu^a$ was introduced by hand without preserving the gauge invariance.
Note that both nMA gauge and Landau gauge conditions are exactly satisfied only at $\alpha=0$ and $\alpha^\prime=0$.  
This is realized at $\beta=0$ limit:
\begin{equation}
 D_{\mu\nu 0}^{XX}(k) =  \frac{-1}{k^2-M_X^2} \left[  \delta_{\mu\nu} -  \frac{k_\mu k_\nu}{k^2} \right] ,
 \ 
 D_{\mu\mu 0}^{XX}(k) =  \frac{-(D-1)}{k^2-M_X^2}  
= D_{\mu\mu \infty}^{XX}(k)- \frac{1}{M_X^2} .
\end{equation}
Therefore, the $\beta=0$ limit differs from the previous Proca case used in MA gauge. 
However, it will turn out below that the constant shift of the propagator gives the same decay  rate and hence the same mass $M_X$ of $\mathbf{X}_\mu$ gluon. 
%However, it turns out that both choices yield  the same value for the gauge invariant gluon mass $M_X$ and off-diagonal gluon mass $M_{\rm off}$, as demonstrated in our simulations below.

%%%%%%%%%%%%%%%%%%%%%%%%%%%%%%%%%%%%%%%%%%%%%%%%%%%
%\subsection{Numerical simulations}
%%%%%%%%%%%%%%%%%%%%%%%%%%%%%%%%%%%%%%%%%%%%%%%%%%%

We have generated configurations of link variables $\{U_{x,\mu}\}$ based on the standard heat bath method for the standard
Wilson action. The numerical simulation are performed at $\beta=2.3$, $2.4$ on $24^{4}$ lattice,
at $\beta=2.3$, $2.4$, $2.5$ on $32^{4}$ lattice, 
at $\beta=2.4$, $2.5$, $2.6$ on $36^{4}$ lattice,
and   at $\beta=2.4, 2.5, 2.6$ on $48^4$ lattice by thermalizing 15000 sweeps.  
Here 200 configurations are stored every 300 sweeps. 
[Other settings of numerical simulations are the same as those in the previous paper \cite{IKKMSS06}.]

%We are now ready to study characteristic features of  the reformulated Yang-Mills theory written in terms of new variables ${n}^A(x), c_\mu(x), \mathbf{X}_\mu^A(x)$ defined through NLCV of the original field variable $\mathbf{A}_\mu^A(x)$: infrared ``Abelian'' dominance, magnetic monopole dominance and  non-vanishing gluon mass. 
%Among them, the magnetic monopole dominance in the string tension has already been confirmed in the previous paper \cite{IKKMSS06} using the gauge-invariant magnetic monopole which is guaranteed to have integer-valued magnetic charge subject to the Dirac quantization condition according to our construction of magnetic current based on NLCV.
%\footnote{
%The proposed NLCV enables one to extract the \textquotedblleft Abelian part\textquotedblright\ $\mathbb{V}_{x,\mu}^A$ irrespective of the choice of the gauge fixing preserving the color symmetry.  The Yang-Mills theory in the conventional MA gauge  is reproduced as a very special limit (\ref{C35-MAGlimit}) of our reformulated Yang-Mills theory based on NLCV. 
%} 
%An advantage of our formulation is that we can confirm such characteristic features for any choice of gauge fixing, not restricted to MA gauge, since our formulation allows us to take arbitrary type of gauge fixing for the original variable $\mathbf{A}_\mu^A(x)$. 

To study the \textbf{infrared Abelian dominance} and the \textbf{gluon mass generation} from the viewpoint of the correlation functions in the reformulated Yang-Mills theory, we first define the two-point correlation functions (full propagators) for the independent variables in the new formulation on a lattice, i.e.,  $\bm{n}_{x}^A$, $c_{x,\mu}$ and $\mathbf{X}_{x,\mu}^A$, in addition to the original variable $\mathbf{A}_{x,\mu}^A$. 
For simplicity, we examine just the contracted scalar-type propagator for avoiding the complicated tensor structure: 
\begin{align}
D_{nn}(x-y) =& \left\langle \bm{n}_{x}^A  \ \bm{n}_{y}^A\right\rangle ,
\quad 
D_{cc}(x-y) = \left\langle c_{x',\mu} \  c_{y',\mu}\right\rangle ,
%\chAS{}{}
\nonumber\\
D_{XX}(x-y) =& \left\langle \mathbb{X}_{x,\mu}^A \ \mathbb{X}_{y,\mu}^A
\right\rangle , 
\quad
D_{X'X'}(x-y) = \left\langle \mathbb{X}_{x',\mu}^A \ \mathbb{X}_{y',\mu}^A
\right\rangle ,
%\chAS{}{}
\end{align} 
and
% \chASS{}
\begin{equation}
D_{AA}(x-y)=\left\langle \mathbb{A}_{x',\mu}^A \ \mathbb{A}_{y',\mu}^A \right\rangle ,
%\chAS{}{} 
\end{equation} 
where $x^\prime$ denotes the mid-point between $x$ and $x+\mu$.
Here the Lie-algebra valued gauge potential $\mathbb{A}_{x^{\prime},\mu}$ or $\mathbb{V}_{x^{\prime},\mu}$ on the lattice is defined from the respective link variable by%
\footnote{
The fields $\mathbb{A}_{x^{\prime},\mu},\mathbb{V}_{x^{\prime},\mu},\mathbb{X}_{x^{\prime},\mu}$ on the lattice are counterparts of $\mathbf{A}_{\mu}(x),\mathbf{V}_{\mu}(x),\mathbf{X}_{\mu}(x)$ on the continuum space--time. 
}
\begin{equation}
\mathbb{A}_{x^{\prime},\mu} :=(i/2g\varepsilon) \left[  U_{x,\mu
}-U_{x,\mu}^{\dagger}\right]  , \quad 
\mathbb{V}_{x^{\prime},\mu}= 
(i/2g\varepsilon) \left[  V_{x,\mu}-V_{x,\mu}^{\dagger}\right]  .
\end{equation} 
For the variable $\mathbb{X}_{x,\mu}$, on the other hand, we examined two options:
 one is extracted by decomposing the gauge potential (group-valued): 
\begin{equation}
\mathbb{X}_{x,\mu} :=(i/2g\varepsilon) \left[  X_{x,\mu}-X_{x,\mu}^{\dagger}\right]  ,
\label{C35-Xpro-1}
\end{equation} 
and the other is due to the definition of the decomposition (Lie-algebra-valued):
\begin{equation}
\mathbb{X}_{x^{\prime},\mu} :=\mathbb{A}_{x^{\prime},\mu}-\mathbb{V}%
_{x^{\prime},\mu}.
\label{C35-Xpro-2}
\end{equation} 
The field $c_{x',\mu}$ is defined by 
\begin{equation}
%\chAAS{
 c_{x',\mu} :={\rm tr}( \bm{n}_{x} V_{x,\mu})={\rm tr}( V_{x,\mu}\bm{n}_{x+\mu}) .
%}{
%c_{x,\mu}={\rm tr}( \mathbf{n}_{x^{\prime}} \mathbb{V}_{x^{\prime},\mu})
%}.
\end{equation} 

%%%%%%%%%%%%%%%%%%%%% figures %%%%%%%%%%%%%%%%%%%%%%%%%%%
\begin{figure}[ptb]
\begin{center}
%\vspace{-5mm}%
\includegraphics[scale=0.6]{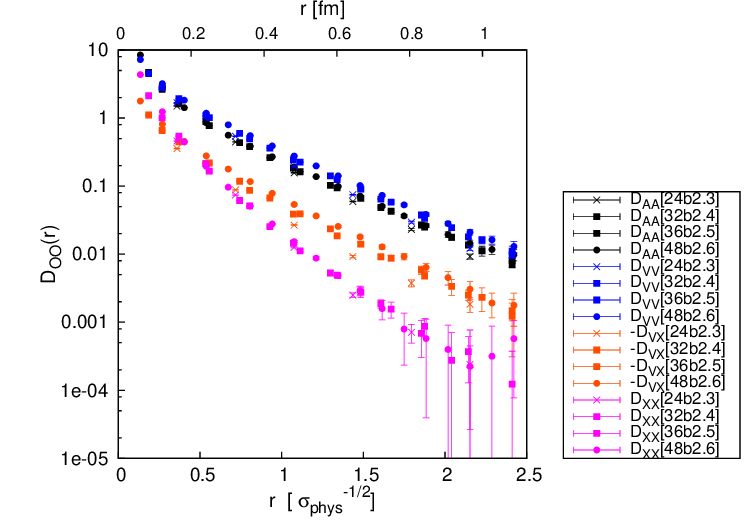}
\includegraphics[scale=0.6]{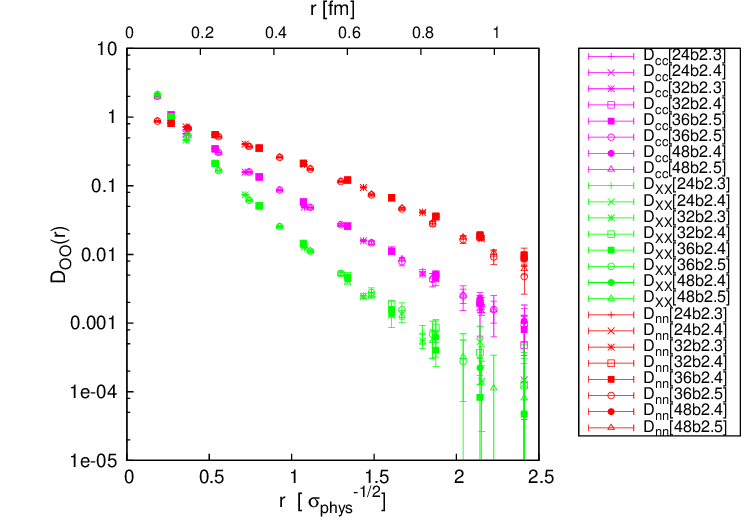}
%\includegraphics[width=3.5in]{prop.ps}
%\includegraphics[width=2.97in]{dumpVX.ps}
%\vspace{-8mm}
\end{center}
\vspace{-0.5cm}
\caption{Ref.\cite{SKKMSI07}: 
%(Left) 
Logarithmic plots of scalar-type two-point correlation functions 
$D_{OO^\prime}(r):=\left\langle \mathcal{O}(x) \mathcal{O}^\prime(y) 
\right\rangle$
as a function of the Euclidean distance $r:=\sqrt{(x-y)^2}$  
for  $\mathcal{O}$ and $\mathcal{O}^\prime$. 
(Left panel)
$\mathcal{O}(x)\mathcal{O}^\prime(y)
= \mathbb{V}_\mu^A(x) \mathbb{V}_\mu^A(y)$, 
$\mathbb{A}_\mu^A(x) \mathbb{A}_\mu^A(y)$,  
$-\mathbb{V}_\mu^A(x) \mathbb{X}_\mu^A(y)$,
   $\mathbb{X}_\mu^A(x) \mathbb{X}_\mu^A(y)$, 
(Right panel)
$\mathcal{O}(x)\mathcal{O}^\prime(y)
= {\bf n}^A(x){\bf n}^A(y)$,  
%$\mathbb{A}_\mu^A(x)\mathbb{A}_\mu^A(y)$,  
$c_\mu(x)c_\mu(y)$,  $\mathbb{X}_\mu^A(x)\mathbb{X}_\mu^A(y)$,  from above to below
% (The vertical axis is on the right side of the graph for the ${\bf n}^A(x){\bf n}^A(y)$ correlation function.), 
using data on the $24^4$ lattice ($\beta=2.3, 2.4$), 
$32^4$  lattice ($\beta=2.3, 2.4$), $36^4$  lattice ($\beta=2.4, 2.5$), 
and    $48^4$  lattice ($\beta=2.4, 2.5, 2.6$).
Here plots are given in the physical unit [fm] or in unit of square root of the string tension $\sqrt{\sigma_{\rm phys}}$. 
%(Right) The logarithmic plot of the rescaled correlation function $r^{3/2}D_{OO}(r)$ as a function of $r$ 
%(in the physical unit [fm] or in unit of square root of the string tension $\sqrt{\sigma_{\rm phys}}$) 
%for $O=V,X$, 
%using data on the lattice $24^4$ at $\beta=2.3, 2.4, 2.5$, and $36^4$ at $\beta=2.5, 2.6, 2.7$,
%where the slope corresponds to the effective mass. 
}
\label{C35-fig:prop}%
\end{figure}
%%%%%%%%%%%%%%%%%%%%% figures %%%%%%%%%%%%%%%%%%%%%%%%%%%

First, we examine the infrared Abelian dominance. 
The numerical results are presented in Fig.~\ref{C35-fig:prop}.
As is quickly observed from the left panel of Fig.~\ref{C35-fig:prop}, 
$D_{VV}(x-y)$ and $D_{AA}(x-y)$ exhibit quite similar behaviors in the measured range of the Euclidean distance $r=|x-y|:=\sqrt{(x-y)^2}$.
In order to determine the physical scale, we have used the relationship between the (inverse) gauge coupling $\beta$ and lattice spacing $\epsilon$ summarized in Table~\ref{C35-table:dictionary} which is  given in \cite{KKNS98}.% 
\footnote{
We use the relationship between the physical units, 
$1 {\rm GeV}^{-1}=0.197327 {\rm fm}$ or $1 {\rm GeV}=5.06773 {\rm fm}^{-1}$.  
This comes from $\hbar c=0.197327 {\rm GeV} \cdot {\rm fm} $.
%A more precise relationship is obtained e.g., from the proton Compton wavelength $\hbar/m_P c=\hbar c/m_Pc^2 = 0.21031  {\rm fm}$ for the proton mass $m_P=0.93827 {\rm GeV}$.
}

From the right panel of Fig.~\ref{C35-fig:prop}, 
$D_{VV}(x-y)$  ($D_{nn}(x-y)$ or $D_{cc}(x-y)$) is dominant compared to $D_{XX}(x-y)$ which decreases more rapidly than other correlation functions in $r$.  
This implies the infrared ``Abelian'' dominance, provided that the components $\mathbf{V}_\mu^A(x)$ composed of ${\bf n}^A_{x}$ and $c_{x,\mu}$ are identified with the ``Abelian'' part of $\mathbf{A}_\mu^A(x)$.  
As is seen from the left  panel of Fig.~\ref{C35-fig:prop},  a non-trivial mixed correlation function
%\chAAS{
$\left\langle \mathbb{V}_\mu^A(x') \mathbb{X}_\mu^A(y') \right\rangle < 0$
%}
%{$\left\langle \mathbb{V}_\mu^A(x) \mathbb{X}_\mu^A(y) \right\rangle < 0$} 
exists, since $\mathbb{V}_\mu^A(x)$ includes a perpendicular component to ${\bf n}^A(x)$. 

Fig.~\ref{C35-fig:prop} demonstrates nice independence of our results against variations of the  ultraviolet cutoff (the lattice spacing $\epsilon$).  The propagators calculated at the lattices with different $\epsilon$ follow the same curve if plotted in the physical units. These accurate plots provide an additional support that the results presented here are definitely not lattice artifacts.

Note that we must impose the gauge fixing condition for the original variable $\mathbb{A}_{x^{\prime},\mu}$  to obtain the correlation function.  In our simulations,  we have chosen the lattice Landau gauge (LLG) for the original field $\mathbb{A}_\mu^A(x)$ for this purpose.  
Thus we have confirmed the infrared ``Abelian'' dominance with color symmetry being kept, since the Landau gauge keeps the color symmetry. 
This is one of our main results.
 
The infrared Abelian dominance was so far obtained only for the MA gauge which breaks the color symmetry explicitly.   
As already mentioned, moreover, we can choose any other gauge and  we can study using this formulation if the infrared ``Abelian'' dominance can be observed in any other gauge. 
%We hope we can report the results in the other gauge in  future investigations. 

%\newpage
%\noindent
%$\bullet$ \underline{Infrared Abelian dominance}

%%%%%%%%%%%%%%%%%%%%% figures %%%%%%%%%%%%%%%%%%%%%%%%%%%
\begin{figure}[tbp]
\begin{center}
%\vspace{-5mm}%
\includegraphics[scale=0.6]{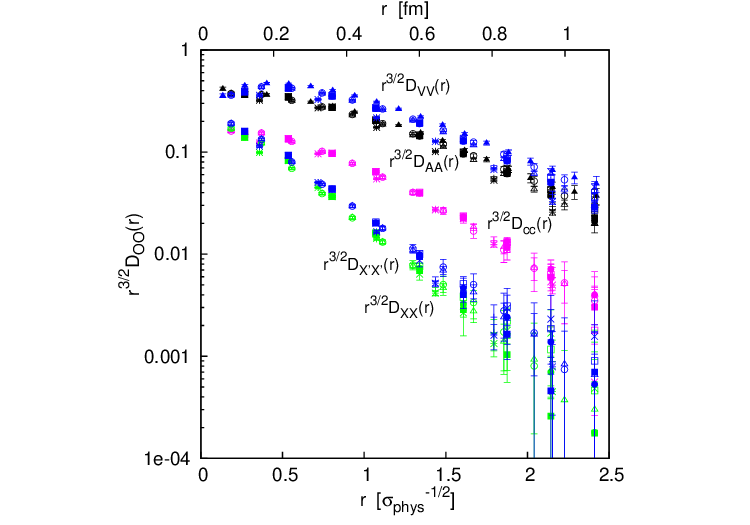}
%\includegraphics[height=5.0cm,width=7cm]{Fig-PR/Fig-lattice/dump-finv2.eps}
%\includegraphics[width=3.5in]{dumpVX.eps}
%\vspace{-8mm}
\end{center}
\vspace{-0.5cm}
\caption{Ref.\cite{SKKMSI07}: 
Logarithmic plots of the rescaled correlation function 
$r^{3/2}D_{OO}(r)$ as a function of $r$ for $O=\mathbb{V}_\mu^A,\mathbb{A}_\mu^A,c_\mu,\mathbb{X}_\mu^A$ (and $\mathbb{X}^\prime{}_\mu^A$)  from above to below, using the same colors and symbols as those in Fig.~\ref{C35-fig:prop}.
Here two sets of data for the correlation function $D_{XX}(x-y)$ are plotted according to the two definitions   of the $\mathbb{X}_\mu^A$ field on a lattice.  
}
\label{C35-fig:rescaled-propagator}%
\end{figure}
%%%%%%%%%%%%%%%%%%%%% figures %%%%%%%%%%%%%%%%%%%%%%%%%%%

%\noindent
%$\bullet$ \underline{Gluon ``mass'' generation}

Next, we determine the gluon mass generated in the non-perturbative way by examining the correlation functions in more detail. 
The gauge boson propagator $D_{\mu\nu}^{XX}(x-y)$ is related to the Fourier transform of the massive propagator $D_{\mu\nu}^{XX}(k)$: 
\begin{equation}
 D_{\mu\nu}^{XX}(r)
=\left\langle \mathbb{X}^A_{\mu}(x)  \mathbb{X}^A_{\nu}%
(y)\right\rangle =\int\frac{d^{4}k}{(2\pi)^{4}}e^{ik(x-y)} 
D_{\mu\nu}^{XX}(k).
\end{equation}
Then the scalar-type propagator $D^{XX}(r):=D^{XX}_{\mu\mu}(x)$  as a function of $r$  should behave for large  $M_X r$ as  
%(See \cite{AS99} for details of the integral calculation.)
\begin{equation}
D_{XX}(r)
=\left\langle \mathbb{X}^A_{\mu}(x)  \mathbb{X}^A_{\mu}(y)\right\rangle 
=\int\frac{d^{4}k}{(2\pi)^{4}}e^{ik(x-y)} \frac{3}{k^{2} +M_X^{2}}
%\left(  4+\frac{k^{2}}{M^{2}}\right)  
\simeq\frac{3\sqrt{M_X}} {2(2\pi)^{3/2}} \frac{e^{-M_X r}}{r^{3/2}} .
\end{equation}
Therefore, the scaled propagator $r^{3/2}D_{XX}(r)$ should be  proportional to $\exp(-M_X r)$ for  $M_X r \gg 1$ with $M$ being the fall-off rate of $r^{3/2}D_{XX}(r)$.  In other words, the ``mass'' $M_X$ of the gauge field $\mathbb{X}_{\mu}$ can be estimated from the slope in the logarithmic plot of the scaled propagator $r^{3/2}D_{XX}(r)$ as a function of $r$.
\footnote{
Here we have assumed that the anomalous dimension is sufficiently small so that the exponent of the power of $r$ is the same as the tree value. 
}

%the linear fitting in the $\ln\left(  r^{3/2}D_{\mu\mu}^{XX}(r;M)\right)$ vs. $r$ plot gives the mass $M$.  

%\newpage
%%%%%%%%%%%%%%%%%%%%% figures %%%%%%%%%%%%%%%%%%%%%%%%%%%
\begin{figure}[ptb]
\begin{center}
%\vspace{-16mm}%
%\includegraphics[width=3.0in]{dumpVX.ps}
\includegraphics[scale=0.6]{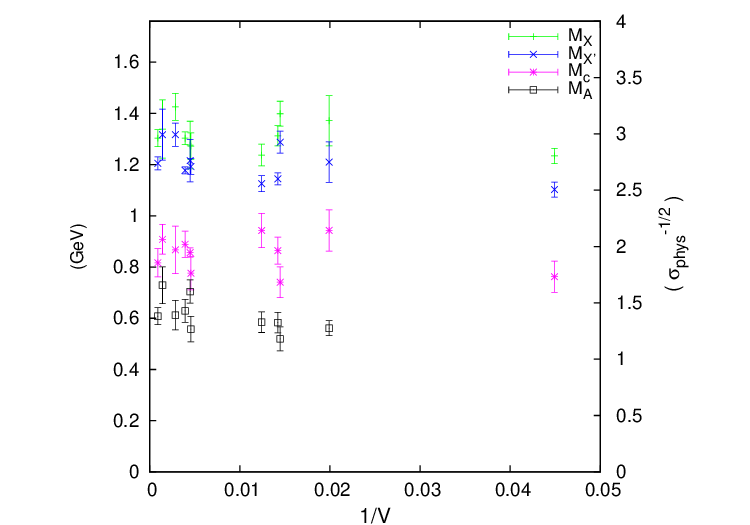}
\includegraphics[scale=0.6]{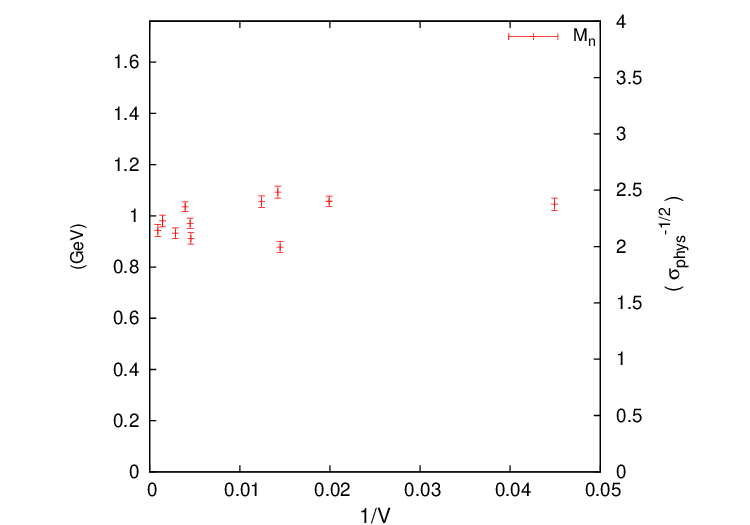}
\vspace{-8mm}
\end{center}
\caption{Ref.\cite{SKKMSI07}: 
Gluon ``mass'' and decay rates (in units of GeV and $\sqrt{\sigma_{\rm phys}}$) as the function of the inverse lattice volume $1/V$ in the physical unit. 
(Left panel) for $\mathcal{O}=\mathbb{X}_\mu^A,  (\mathbb{X}^\prime{}_\mu^A), c_\mu, \mathbb{A}_\mu^A$ from above to below extracted according to the fitting: $\left\langle \mathcal{O}(x) \mathcal{O}(y) \right\rangle \sim r^{-3/2} \exp (-M_{\mathcal{O}}r)$, 
(Right panel) for $\mathbf{n}^A(x)$ extracted according to the fitting: $\left< \mathbf{n}^A(x) \mathbf{n}^A(y) \right> \sim \exp (-M_n r)$. 
}
\label{C35-fig:mass}%
\end{figure}
%%%%%%%%%%%%%%%%%%%%% figures %%%%%%%%%%%%%%%%%%%%%%%%%%%

Fig.~\ref{C35-fig:rescaled-propagator}  shows the logarithmic plots of the scaled scalar-type propagator for 
$\mathbb{A}_{x^{\prime},\mu}$, 
%\chAAS{
$c_{x',\mu}$
%}{$c_{x,\mu}$} 
and 
%\chAAS{
$\mathbb{X}_{x,\mu}$
%}{$\mathbb{X}_{x^{\prime},\mu}$} 
as a function of the distance $r$  measured in the physical unit [fm] and in  unit of square root of the string tension $\sqrt{\sigma_{\rm phys}}=440$ MeV. 
According to Fig.~\ref{C35-fig:rescaled-propagator},  we find just small difference between two types of $D_{XX}(x-y)$ defined by (\ref{C35-Xpro-1}) or (\ref{C35-Xpro-2}) over several choices of lattice spacing  (i.e., several values of $\beta$, $\beta=2.3, 2.4, 2.5, 2.6$).
Therefore, we can use either definition of the lattice variable 
$\mathbb{X}_{x^{\prime},\mu}$ to obtain $D_{X'X'}(x-y)$ in the consistent manner.

In Fig.~\ref{C35-fig:mass}, the measured values for the gluon mass are plotted as the function of the inverse lattice volume $1/V$ in the physical unit, to study the finite-size effect on the mass. 
The finite lattice-size effect seems to be small for the gluon mass $M_X$. Here the error bars originate from the fitting procedure for obtaining the slope, but no systematic errors such as finite-volume are included. 
In this way, we have estimated the mass for the $\mathbb{X}$ gluon:
\begin{align}
  M_{X} &\simeq 2.98  \sqrt{\sigma_{\rm phys}} \simeq   
%\chAS{
1.31 
%}{1.312}
  {\rm GeV} ,
  \nonumber\\
  M_{X'} &\simeq 2.69  \sqrt{\sigma_{\rm phys}} \simeq  
%\chAS{
1.19 
%}{1.186} 
 {\rm GeV} .
\end{align}
%\chASS

Even after the whole gauge fixing, our formulation preserves color symmetry in sharp contrast to the conventional MA gauge. 
In view of the fact that our reformulated Yang-Mills theory reproduces the Yang-Mills theory in MA gauge as a special limit, the remaining part $\mathbb{X}_\mu^A(x)$ could correspond to the  off-diagonal part in this limit. 
From this point of view, our result is consistent with the result obtained for the off-diagonal gluon mass in MA gauge \cite{AS99}.

Moreover, we have simultaneously estimated the decay rate for the new fields 
%\chAAS{
${\bf n}^A(x)$, $c_{\mu}(x)$, $\mathbb{V}_{\mu}^A(x)$
%}{$n_{x}$, $c_{x,\mu}$} 
and the original gauge field 
%\chAAS{
$\mathbb{A}_{\mu}^A(x)$
%}{$\mathbb{A}_{x,\mu}$} 
by imposing the LLG as the overall gauge fixing. 
For $\mathcal{O}=\mathbb{X}_\mu^A, c_\mu, \mathbb{A}_\mu^A, \mathbb{V}_{\mu}^A$,  the decay rate $M_{\mathcal{O}}$ is extracted according to the fitting: $\left\langle \mathcal{O}(x) \mathcal{O}(y) \right\rangle \sim r^{-3/2} \exp (-M_{\mathcal{O}}r)$. 
Fig.~\ref{C35-fig:mass} indicates not so small finite volume effect for data of $1/V > 0.02$.   
Using the data of $1/V<0.02$, therefore, we have estimated the decay rate (or ``mass'') as
\begin{align}
  M_{n} &\simeq 2.24  \sqrt{\sigma_{\rm phys}} 
\simeq 0.986   {\rm GeV} ,
  \nonumber\\
  M_{c} &\simeq 1.94  \sqrt{\sigma_{\rm phys}}
\simeq 0.856  {\rm GeV} ,
  \nonumber\\
  M_{A} &\simeq 1.35  \sqrt{\sigma_{\rm phys}}
\simeq 0.596  {\rm GeV} .
\end{align}
The decay rate $M_c$ obtained from the correlation function of %\chAAS{
$c_\mu(x)$
%}{$c_\mu$}
 field is slightly larger than that expected from the result in MA gauge. 
It should be remarked that the decay rate for the correlation function of $\mathbf{n}^A(x)$ field is extracted according to the fitting function 
$\left< \mathbf{n}^A(x) \mathbf{n}^A(y) \right> \sim \exp (-M_n r)$ which is not yet justified from the theoretical consideration.  This might be an origin of the large value of $M_n$. 
More simulations on the larger lattice are expected to  eliminate finite volume effect for these values. 
However, we have no argument for guaranteeing the gauge invariance of these values or for identifying these values with their ``masses''.  
In fact, the field $c_\mu(x)$ is not gauge invariant.  
These issues should be checked in further investigations based on the new reformulation. 

%\begin{tabular}
%[c]{|c|c|c|c|c|}\hline
%\multicolumn{5}{|c|}{Averaged value of mass dumping}\\\hline
%& \multicolumn{2}{|c|}{$\sqrt{\sigma_{phys}}$} & \multicolumn{2}{|c|}{$GeV$%
%}\\\hline
%$M_{X}$ & $2.98356$ & $0.03683$ & $1.3128$ & $0.0162$\\\hline
%$M_{X^{\prime}}$ & $2.6933$ & $0.03306$ & $1.185$ & $0.01452$\\\hline
%$M_{A}$ & $1.35624$ & $0.03869$ & $0.5960$ & $0.017$\\\hline
%$M_{C}$ & $1.94681$ & $0.03237$ & $0.8566$ & $0.0142$\\\hline
%$M_{nn}$ & $2.24126$ & $0.05084$ & $0.98615$ & $0.0224$\\\hline
%\end{tabular}

%%%%%%%%%%%%%%%%%%%%%%%%%%%%%%%%%%%%%%%%%%%%%%%%%%%%%
\begin{table}
\caption{\rm Ref.\cite{SKKMSI07}: 
The lattice spacing $\epsilon$ and the lattice size $L$ of the lattice volume $L^4$  at various value of  $\beta$ in the physical unit [fm] and the unit given by $\sqrt{\sigma_{phys}}$. 
}
\begin{center}%
\begin{tabular}
[c]{||l|c|c|c|c|c|c||}\hline\hline
& \multicolumn{2}{|c|}{lattice spacing $\epsilon$} &
\multicolumn{4}{||c||}{lattice size $L$ $\mathrm{[fm]}$}\\\hline
\multicolumn{1}{||c|}{$\beta$} & { $[1/\sqrt{\sigma_{phys}}]$ } & {
$\mathrm{[fm]}$ } & \multicolumn{1}{||c|}{$24^{4}$} & $32^{4}$ & $36^{4}$ &
$48^{4}$\\\hline\hline
\multicolumn{1}{||c|}{$2.3$} & $0.35887$ & $0.1609$ & $3.863$ & $5.150$ &
$5.794$ & $7.725$\\\hline
\multicolumn{1}{||c|}{$2.4$} & $0.26784$ & $0.1201$ & $2.883$ & $3.844$ &
$4.324$ & $5.766$\\\hline
\multicolumn{1}{||c|}{$2.5$} & $0.18551$ & $0.08320$ & $1.997$ & $2.662$ &
$2.995$ & $3.993$\\\hline
\multicolumn{1}{||c|}{$2.6$} & $0.13455$ & $0.06034$ & $1.448$ & $1.931$ &
$2.172$ & $2.896$\\\hline\hline
\end{tabular}
\end{center}
\label{C35-table:dictionary}
\end{table}
%\chASS{}
%%%%%%%%%%%%%%%%%%%%%%%%%%%%%%%%%%%%%%%%%%%%%%%%%%%%%

Finally, we comment on the \textquotedblleft Abelian\textquotedblright part 
%\chAAS{
$\mathbb{V}_{\mu}^A(x)$,
%}{$\mathbb{V}_{x,\mu}$
since our treatment of the \textquotedblleft Abelian\textquotedblright part 
%\chAAS{
$\mathbb{V}_{\mu}^A(x)$
%}{$\mathbb{V}_{x,\mu}$} 
is different from the conventional approach based on MA gauge. 
The above result yields the ``mass'' of the \textquotedblleft Abelian\textquotedblright part 
%\chAAS{
$\mathbb{V}_{\mu}^A(x)$
%}{$\mathbb{V}_{x,\mu}$}
:
%\chAAS{
$M_{V} \simeq M_{A} \simeq  0.59 {\rm GeV}$
%}{$M_{V} \simeq M_{A} \simeq  0.58 \sim 0.68$ GeV}
. 
This value is nearly equal to that of the diagonal gluon mass obtained by imposing the Landau gauge in the conventional approach as reported in  \cite{BCGMP03} where the Landau gauge was imposed on the Abelian diagonal part $a_\mu(x)$ in addition to the MA gauge for off-diagonal gluon field $A^a_\mu(x)$ defined by the Cartan decomposition
%\chAAS{
$\mathbf{A}_\mu(x)=A_\mu^a(x) T^a + a_\mu(x) T^3$ ($a=1,2$).
%}{$\mathbb{A}_\mu=A_\mu^a T^a + a_\mu T^3$}
Therefore, the prescription of gauge fixing in \cite{AS99} is different from ours.

%\newpage
%%%%%%%%%%%%%%%%%%%%%%%%%%%%%%%%%%%%%%%%%%%%%%%%%%
\subsection{Reformulation of lattice $SU(3)$ Yang-Mills theory} 
%%%%%%%%%%%%%%%%%%%%%%%%%%%%%%%%%%%%%%%%%%%%%%%%%%

%
%We consider calculating the vacuum expectation value (say, average) of an operator. 
In the path-integral or functional-integral formulation, the basic ingredients are the action and the integration measure, by which the vacuum expectation value, say average of an operator, is to be calculated. We can rewrite the original $SU(3)$ Yang-Mills action and the integration measure using either the \textbf{maximal option} or the \textbf{minimal option} \cite{KSM08}. The resulting two reformulations written in terms of different variables are equivalent to each other, since each formulation corresponds to one of  the choices of the coordinates in the space of gauge field configurations. Therefore, we can use either reformulation (change of variables), instead of the original Yang-Mills theory.

%The difference between the two options, i.e., maximal or minimal, arises when we choose an operator to be calculated.
In what follows, we focus our studies on confinement of quarks in a defining representation, i.e., the \textbf{fundamental representation}. For this purpose, we use the \textbf{Wilson loop   average} for obtaining the \textbf{static quark potential}.
%, magnetic-monopole current, and chromofield strength in a gauge invariant way. 
Remember that the Wilson loop operator is uniquely defined by specifying  a representation $R$, to which the source quark belongs. A remarkable fact is that the Wilson loop operator in the fundamental representation urges us to use the \textbf{minimal option} in the sense that it is exactly rewritten in terms of the field variables (i.e., the \textbf{color field} $n$ and the \textbf{restricted field} $V$) which identified with the field variables used to describe the minimal option.  This was shown in the process of deriving a \textbf{non-Abelian Stokes theorem for the Wilson loop operator} \cite{Kondo08}. Therefore, the set of variables in the minimal option is a  natural and the best choice of coordinate in the space of gauge field configurations to describe the Wilson loop operator in the fundamental representation. At the same time, this fact tells us what is the dominant variable for the Wilson loop average. 
%These are the reasons why we use the minimal option to discuss confinement of quarks in the fundamental representation.
%Thus, the minimal option is superior to the maximal option for discussing confinement of quarks in the fundamental representation.

In view of this, we use the \textbf{reformulation of the $SU(3)$ Yang-Mills theory in the minimal option}  for  discussing confinement of quarks in the fundamental representation. 
Thus, the minimal option is superior to the maximal option for discussing confinement of quarks in the fundamental representation of the $SU(3)$ gauge group.
The reformulation of the lattice $SU(3)$ Yang-Mills theory in the minimal option is quickly reviewed as follows \cite{KSSMKI08,SKS10}.  
% \cite{SCGTKKS08L,exactdecomp}.
%In particular, the minimal option of the formulation enables us to extract the dominant mode for quark confinement, if we focus our attention on quarks belonging to the fundamental representation of the $SU(3)$ gauge group \cite{abeliandomSU(3),lattice2010}. 
For the original $SU(3)$  gauge link variable $U_{x,\mu} \in SU(3)$,
we decompose it into the new variables $V_{x,\mu}$ and $X_{x,\mu}$ which have values in the $SU(3)$ group:%, i.e., $X_{x,\mu}  \in SU(3)$, $V_{x,\mu} \in SU(3)$: 
\begin{equation}
  SU(3) \ni U_{x,\mu}=X_{x,\mu}V_{x,\mu} , \quad X_{x,\mu}, V_{x,\mu} \in SU(3) .
\end{equation} 
Note that $V_{x.\mu}$ could be regarded as the dominant mode for quark confinement, while $X_{x,\mu}$ is the remainder. 
In this decomposition, we require that the restricted field  $V_{x,\mu}$ is transformed in the same way as the original gauge link variable $U_{x,\mu}$ and the remaining field $X_{x,\mu}$ as a site variable under the full $SU(3)$ gauge transformation $\Omega_{x}$:
\begin{subequations}
\begin{align}
%U_{x,\mu} & \longrightarrow U_{x,\mu}^{\prime} = \Omega_{x}U_{x,\mu} \Omega_{x+\mu}^{\dag},
%\\
 V_{x,\mu} & \longrightarrow V_{x,\mu}^{\prime} = \Omega_{x}V_{x,\mu} \Omega_{x+\mu}^{\dagger} , \quad \Omega_{x} \in G=SU(3)
\label{C35-V-transf}
\\
X_{x,\mu} & \longrightarrow X_{x,\mu}^{\prime} = \Omega_{x}X_{x,\mu} \Omega_{x}^{\dagger}  , \quad \Omega_{x} \in G=SU(3)
\label{C35-X-transf}
\end{align}
\label{C35-eq:gaugeTransf} 
\end{subequations}
for
\begin{align}
 U_{x,\mu} & \longrightarrow U_{x,\mu}^{\prime} = \Omega_{x}U_{x,\mu} \Omega_{x+\mu}^{\dagger}  , \quad \Omega_{x} \in G=SU(3) .
\end{align}

First, we introduce the key variable $\bm{h}_{x}$ called the \textbf{color field}. 
In the minimal option of $SU(3)$, a representation of the color field $\bm{h}_{x}$ is given by
\begin{equation}
 \bm{h}_{x}= \Theta_{x} \frac{\lambda^{8}}{2} \Theta_{x}^{\dagger}   \in Lie[SU(3)/U(2)] , \quad \Theta_{x}  \in SU(3), 
\end{equation}
with $\lambda^{8}$ being the Gell-Mann matrix for $SU(3)$ and $g_{x}$ the $SU(3)$ group element.
Once the color field $\bm{h}_{x}$ is introduced, the above decomposition is obtained by solving the (first) \textbf{defining equation}:
%\begin{subequations}
\begin{align}
   D_{\mu}^{(\epsilon)}[V]\bm{h}_{x}:= \epsilon^{-1} \left[  V_{x,\mu}\bm{h}_{x+\mu}-\bm{h}_{x}V_{x,\mu}\right]  =0 .
\label{C35-eq:def1}
%\\
%&  g_{x}:=e^{i2\pi q/3}\exp(-ia_{x}^{0}\mathbf{h}_{x}-i\sum\nolimits_{j=1}^{3}a_{x}^{(j)}\mathbf{u}_{x}^{(i)})=1 . \label{C35-eq:def2}%
\end{align}
%\end{subequations}
%Here the variable
%$g_{x}$ is undetermined parameter from Eq.(\ref{C35-eq:def1}), $\mathbf{u}_{x}^{(j)}$ 's are $su(2)$-Lie algebra values, and $q_{x}$ an integer value $\ 0,1, 2$. 
In fact, this defining equation  can be solved exactly, and the solution is given by%\cite{exactdecomp}
\begin{subequations}
\label{C35-eq:decomp}%
\begin{align}
 X_{x,\mu} &= \widehat{L}_{x,\mu}^{\dag}\det(\widehat{L}_{x,\mu})^{1/3} \hat{g}_{x}^{-1},
\quad
 V_{x,\mu} = X_{x,\mu}^{\dag}U_{x,\mu} = \hat{g}_{x}\widehat{L}_{x,\mu}U_{x,\mu},
\\
 \widehat{L}_{x,\mu} &:= \left(  L_{x,\mu}L_{x,\mu}^{\dag}\right)^{-1/2}L_{x,\mu},
\\
 L_{x,\mu} &:= \frac{5}{3}\mathbf{1} + \sqrt{\frac{4}{3}}(\bm{h}_{x} + U_{x,\mu}\bm{h}_{x+\mu}U_{x,\mu}^{\dag}) + 8\bm{h}_{x}U_{x,\mu}\bm{h}_{x+\mu}U_{x,\mu}^{\dag} .
\end{align}
\label{C35-defeq}
\end{subequations}
Here the variable $\hat{g}_{x}$ is the $U(2)$ part which is undetermined from Eq.(\ref{C35-eq:def1}) alone.
In what follows, therefore, we put the second condition:
\begin{align}
 \hat{g}_{x} =1 ,
%:=e^{i2\pi q/3}\exp(-ia_{x}^{0}\mathbf{h}_{x}-i\sum\nolimits_{j=1}^{3}a_{x}^{(j)}\mathbf{u}_{x}^{(i)})  ,
\label{C35-eq:def2}%
\end{align}
%\end{subequations}
%where  $\mathbf{u}_{x}^{(j)}$ 's are $su(2)$-Lie algebra, and $q_{x}$ an integer value $\ 0,1, 2$, as shown in \cite{exactdecomp}.
so that the above defining equations (\ref{C35-eq:def1}) and (\ref{C35-eq:def2}) correspond  respectively to the continuum version:% \cite{SCGTKKS08}:
\begin{subequations}
\begin{align}
& \mathscr{D}_{\mu}[\mathscr{V}] \bm{h}(x) := \partial_\mu \bm{h}(x) -ig [\mathscr{V}_\mu(x), \bm{h}(x) ] =0, 
\\
& \mathscr{X}_{\mu}(x)-\frac43 [\bm{h}(x),[\bm{h}(x), \mathscr{X}_{\mu}(x)] ] = 0
%\mathrm{tr}(\mathbf{h} (x)\mathcal{X}_{\mu}(x))=0 .
 .
\end{align}
\label{C35-def-eq}
\end{subequations}
%For the physical meaning of the defining equation, see \cite{KondoShibata}. 
In the naive continuum limit, indeed, it is shown directly that (\ref{C35-defeq}) reproduces  the  decomposition in the continuum theory, which is obtained by solving (\ref{C35-def-eq}):% \cite{SCGTKKS08}:
\begin{subequations}
\begin{align}
 \mathscr{A}_{\mu}(x) &= \mathscr{V}_{\mu}(x) + \mathscr{X}_{\mu}(x) ,
\nonumber\\
 \mathscr{V}_{\mu}(x)  &= \mathscr{A}_{\mu}(x)-\frac{4}{3}\left[ \bm{h}(x),\left[  \bm{h}(x),\mathscr{A}_{\mu}(x)\right]
\right]  -ig^{-1}\frac{4}{3}\left[  \partial_{\mu}\bm{h}%
(x), \bm{h}(x)\right]  ,
\\
 \mathscr{X}_{\mu}(x)  &  =\frac{4}{3}\left[ \bm{h}(x),\left[ \bm{h}(x),\mathscr{A}_{\mu}(x)\right]  \right]  +ig^{-1} \frac{4}{3}\left[  \partial_{\mu}\bm{h}(x),\bm{h}(x) \right]  .
\end{align}
\end{subequations}
Thus the decomposition is uniquely determined as   Eqs.(\ref{C35-eq:decomp}) up to the choice of $\hat{g}_{x}$ ($\ref{C35-eq:def2}$), once the color field $\bm{h}_{x}$ is specified.

In order to determine the configuration $\{ \bm{h}_{x}\}$ of color fields, we use the \textbf{reduction condition} which guarantees that  the new theory written in terms of  new variables  is equipollent to the original Yang-Mills  theory. 
Here, we use the reduction condition:
for a given configuration of the original link variables $\{ U_{x,\mu} \}$, a set of color fields $\left\{  \bm{h}_{x}\right\}  $ are obtained by minimizing the functional:%\cite{SCGTKKS08L,exactdecomp}: 
\begin{equation}
F_{\text{red}}[\{ \bm{h}_{x} \}]
= \sum_{x,\mu}\mathrm{tr}\left\{  (D_{\mu}^{(\epsilon)}[U]\bm{h}_{x})^{\dag}(D_{\mu}^{(\epsilon)}[U]\bm{h}_{x})\right\} .
 \label{C35-eq:reduction} 
\end{equation}
Consequently, the color field transforms under the gauge transformation as 
\begin{equation}
  \bm{n}_{x}  \rightarrow  \bm{n}_{x}^\prime = \Omega_{x} \bm{n}_{x} \Omega_{x}^{-1}  , \quad \Omega_{x} \in G=SU(3)
  .
  \label{C35-n-transf-2}
\end{equation}

%%%%%%%%%%%%%%%%%%%%%%%%%%%%%%%%%%%%%%%%%%%%%%%%%%
\subsection{Numerical simulations of lattice $SU(3)$ Yang-Mills theory} 
%{\it Numerical simulations on a lattice}\ ---
%%%%%%%%%%%%%%%%%%%%%%%%%%%%%%%%%%%%%%%%%%%%%%%%%%

%%%%%%%%%%%%%%%%%%%%%%%%%%%%%%%%%%%%%%%%%%%%%%%%%%
\subsubsection{Restricted field dominance and magnetic monopole dominance} 
%%%%%%%%%%%%%%%%%%%%%%%%%%%%%%%%%%%%%%%%%%%%%%%%%%

The lattice version of the Wilson loop operator $W_C[\mathscr{A}]$ is given by
\begin{align}
  W_C[U]  
:=  {\rm tr} \left[ \mathcal{P}  \prod_{<x,x+\mu> \in C}  U_{<x,x+\mu>}  \right]/{\rm tr}({\bf 1})  
 ,
\end{align}
where $\mathcal{P}$ is the path-ordered product. 
In the new formulation, we can define another non-Abelian Wilson loop operator $W_C[\mathscr{V}]$ by replacing the original Yang-Mills field $\mathscr{A}$ by the restricted field $\mathscr{V}$ in the original definition of the Wilson loop operator $W_C[\mathscr{A}]$. 
Similarly, the lattice version of the \textbf{restricted Wilson loop operator} $W_C[\mathscr{V}]$ is easily constructed as
\begin{align}
  W_C[V]  
:=  {\rm tr} \left[ \mathcal{P}  \prod_{<x,x+\mu> \in C}  V_{<x,x+\mu>}  \right]/{\rm tr}({\bf 1})  
 .
\end{align}
This is invariant under the gauge transformation (\ref{C35-V-transf}).

%From the non-Abelian Stokes theorem \cite{KondoNAST,KondoShibata} and the Hodge decomposition of the field strength $\mathscr{F}_{\mu\nu}[\mathscr{V}],$ 
For $G=SU(3)$, the lattice version of the   \textbf{magnetic-monopole current} $K$  is given by using the restricted field $V$  as
\begin{subequations}
\begin{align}
%&  k_{\mu}(x)%=2\pi n_{\mu}(x) 
%= \partial_{\nu} {}^*\Theta_{\mu\nu}(x) 
%=\frac{1}{2}\epsilon_{\mu\nu\alpha\beta}\partial_{\nu}%\Theta_{\alpha\beta}(x) ,
   K_{x,\mu} %=2\pi n_{\mu}(x) 
=& \partial_{\nu}  {}^{\displaystyle *}\Theta_{x,\mu\nu}  
=\frac{1}{2}\epsilon_{\mu\nu\alpha\beta}\partial_{\nu}\Theta_{x,\alpha\beta}  ,
%\\
%& \epsilon^2 \Theta_{\mu\nu}(x) :=-\arg\text{ \textrm{Tr}}\left[  \left(  \frac{1}{3}\mathbf{1}-\frac{2}{\sqrt{3}}\bm{h}_{x}\right)  V_{x,\mu}V_{x+\mu,\nu}V_{x+\nu,\mu}^{\dag}V_{x,\nu}^{\dag}\right]  ,
\\
 \epsilon^2 \Theta_{x,\alpha\beta} 
=&   {\rm arg} \Big[ {\rm tr} \Big\{ \left( \frac13 \bm{1} - \frac{2}{\sqrt{3}} \bm{n}_{x} \right) 
  V_{x,\alpha}V_{x+\alpha,\beta}V_{x+\beta,\alpha}^\dagger V_{x,\beta}^\dagger \Big\} \Big] ,
\\
&  V_{x,\alpha}V_{x+\alpha,\beta}V_{x+\beta,\alpha}^\dagger V_{x,\beta}^\dagger
= \exp\left(-ig \epsilon^2 \mathscr{F}_{\alpha\beta}[\mathscr{V}](x) \right)  .
%=\exp(-ig\Theta_{\mu\nu}^{8}\mathbf{h}(x))
\end{align}
\end{subequations}
%We can define the magnetic-monopole current in the gauge independent way. 
The magnetic-monopole current $K$ just defined in this way is gauge invariant. Indeed, it is easy to observe that $\Theta_{x,\mu\nu}$ is invariant under the gauge transformation (\ref{C35-n-transf-2}) and (\ref{C35-V-transf}),  and hence $K_{x,\mu}$ is also gauge-invariant. 
%Note that the the current $k_{\mu}$ is the non-Abelian magnetic-monopole current, since $V_{x,\mu}$ is the decomposed restricted  field.
Then we can define the magnetic-monopole part of the Wilson loop operator by
\begin{align}
W_C[K] :=&   \exp \left( i  \sum_{x,\mu} K_{x,\mu} \omega^{\Sigma}_{x,\mu} \right)
    ,
\nonumber\\
 \omega^{\Sigma}_{x,\mu} :=& {\displaystyle\sum_{s^{\prime}}}
 \Delta_L^{-1}(s-s^{\prime})\frac{1}{2}\epsilon_{\mu\alpha\beta\gamma} 
\partial_{\alpha}S_{\beta\gamma}^{J}(s^{\prime}+\mu) , \quad
 \partial^\prime_{\alpha} S_{\alpha\beta}^{J}(x)=J_{\beta}(x) ,  
 \label{C35-WK}
\end{align}
where $\omega_{x,\mu}$  is defined through the external source $J_{x,\mu}$ which is used to calculate the static potential,  $S_{\beta\gamma}^{J}(s^{\prime}+\mu)$ is a plaquette variable satisfying $\partial^\prime_{\beta} S_{\beta\gamma}^{J}(x)=J_{\gamma}(x)$ with the external source $J_{x,\mu}$ introduced to calculate the static potential, 
$\partial'$ denotes the backward lattice derivative
$\partial_{\mu}^{'}f_x=f_x-f_{x-\mu}$,  
$S^J_{x,\beta\gamma}$ denotes a surface bounded by the closed loop $C$ on which the electric source $J_{x,\mu}$ has its support, 
and $\Delta_L^{-1}(x-x')$ is the inverse Lattice Laplacian.%, see e.g., \cite{CFKMPS00}. 

The static  \textbf{quark-antiquark potential} $V(R)$ is obtained by taking the limit $T \rightarrow \infty$ from the Wilson loop average $\langle W_C[U]  \rangle$ for a rectangular loop $C=R \times T$.
In order to see the mechanism of quark confinement, we calculate three potentials:
\begin{enumerate}
\item
[(i)] the \textbf{full potential} $V_{\rm full}(R)$  calculated from the standard  $SU(3)$ Wilson loop average 
$\langle W_C[U] \rangle$:
% according to (\ref{C35-WVc}) and (\ref{C35-Vf}),  
\begin{align}
  V_{\rm full}(R) 
=& - \lim_{T \rightarrow \infty} \frac{1}{T} \ln 
\langle W_C[U] \rangle
 , 
 \label{C35-VfR}
\end{align}

\item
[(ii)] the \textbf{restricted potential} $V_{\rm rest}(R)$ calculated from the decomposed variable $\mathscr{V}$ through the restricted Wilson loop average $\langle W_C[V] \rangle$:
% in the same way as (\ref{C35-WVc}) and (\ref{C35-Vf}), 
\begin{align}
  V_{\rm rest}(R) 
=& - \lim_{T \rightarrow \infty} \frac{1}{T} \ln \langle W_C[V] \rangle
 , 
 \label{C35-Vr}
\end{align}

\item
[(iii)] the \textbf{magnetic-monopole potential}  $V_{\rm mono}(R)$ calculated from the lattice counterpart (\ref{C35-WK}) of the continuum quantity   $\langle W_C[K] \rangle=\langle e^{  i (k, \omega_{\Sigma})  }  \rangle$:% according to (\ref{Wkj}):
\begin{align}
V_{\rm mono}(R) 
=& - \lim_{T \rightarrow \infty} \frac{1}{T} \ln  \langle W_C[K] \rangle 
    ,
\end{align}

\end{enumerate}
Three potentials are gauge invariant quantities by construction.

%%%%%%%%%%%%%%%%%%%% Figure %%%%%%%%%%%%%%%%%%%%%%%%%
\begin{figure}[t]
\begin{center}
\includegraphics[scale=0.7]{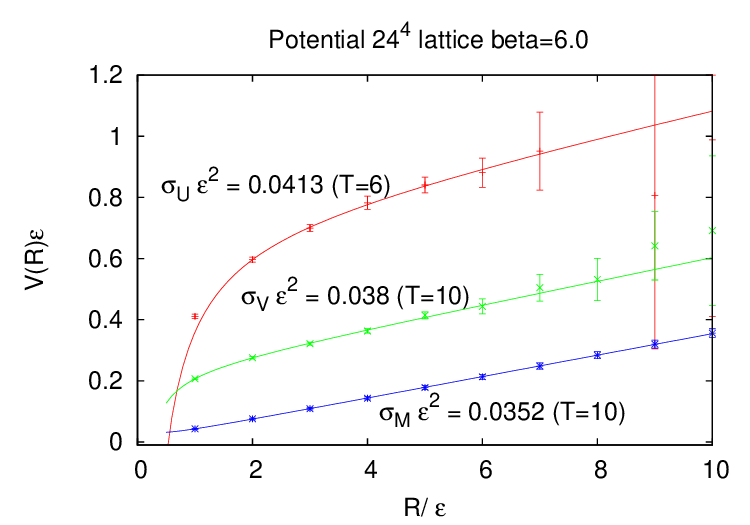}
\end{center}
\vspace{-0.5cm}
\caption{Ref.\cite{KSSK11}: 
$SU(3)$ quark-antiquark potentials as functions of the quark-antiquark distance $R$: (from above to below) 
 (i) full potential $V_{\rm full}(R)$ (red curve),  (ii) restricted part $V_{\rm rest}(R)$ (green curve) and (iii) magnetic--monopole part  $V_{\rm mono}(R)$ (blue curve), measured at $\beta=6.0$ on $24^4$ using 500 configurations where $\epsilon$ is the lattice spacing.
}
%\caption{quark-antiquark potential: (from above to below)  full $SU(3)$ potential $V_f(r)$,  restricted part $V_a(r)$ at $\beta=6.2$ on $24^4$ lattice and  magnetic--monopole part  $V_m(r)$ at $\beta=5.7$ on $16^4$ lattice with $r$ scaled by using the lattice spacing at $\beta=6.2$.}
\label{C35-fig:quark-potential}
\end{figure}
%%%%%%%%%%%%%%%%%%%% Figure %%%%%%%%%%%%%%%%%%%%%%%%%

Numerical simulations are performed for $SU(3)$ Yang-Mills theory on the $24^4$ lattice according to the lattice reformulation explained above. 
%More details of numerical simulations will be given in a subsequent paper \cite{SKKS11}. 
In practice,  we fit  numerical data of $\langle W_C[U] \rangle$ by the two-variable function $W(R,T)$ according to 
%\begin{subequations}
\begin{align}
  \langle W_C[U] \rangle =& \exp (-W(R,T) ),
%\label{C35-WVa}
\nonumber\\
 W(R,T) :=&  V(R)T + (a_1 R+b_1+c_1/R) 
%\nonumber\\&
+ (a_2 R+b_2+c_2/R) T^{-1}   ,
%\label{C35-WVb}
\nonumber\\
 V(R) :=& \sigma R + b + c/R 
 , 
%\label{C35-WVc}
\label{C35-WV}
\end{align}
%\end{subequations}
and determine all coefficients in $W(R,T)$. 
Then we obtain  $V_{\rm full}(R)$  by extrapolating $W(R,T)/T$ to $T \rightarrow \infty$:% from the data obtained for different values of $T$:
\begin{align}
 V_{\rm \cdot }(R) 
=& - \lim_{T \rightarrow \infty} \frac{1}{T} \ln \langle W_C[\cdot ] \rangle
%= \lim_{T \rightarrow \infty}  V(R,T)  
= \lim_{T \rightarrow \infty}  \frac{W(R,T)}{T}   . 
 \label{C35-Vf}
\end{align}
Here the coefficient $\sigma$ of the linear part of the potential $V(R)$ is the string tension which equals   the slope of the curve for large $R$.
In Fig. \ref{C35-fig:quark-potential}, we compare the three quark-antiquark potentials (i), (ii) and (iii).  For each potential, we plot a set of point data for a specified value of $T$ (e.g., $T=6, 10$):
\begin{align}
    - \frac{1}{T} \ln \langle W_{C}[\cdot] \rangle \quad \text{versus} \quad R 
 , 
 \label{C35-VT}
\end{align}
and the curve represented by the function extrapolated to $T \rightarrow \infty$ according to (\ref{C35-WV}) and (\ref{C35-Vf}):
\begin{align}
    V(R) =  \sigma R + b + c/R .
\end{align}

The results of our numerical simulations exhibit the \textbf{infrared restricted variable $\mathscr{V}$ dominance} in the string tension, e.g.,
\begin{equation}
\frac{\sigma_{\rm rest}}{\sigma_{\rm full}} = \frac{0.0380}{0.0413} \simeq 0.92,
\end{equation}
and  the \textbf{non-Abelian  magnetic monopole dominance} in the string tension, e.g.,
\begin{equation}
\frac{\sigma_{\rm mono}}{\sigma_{\rm full}} = \frac{0.0352}{0.0413} \simeq 0.85 .
\end{equation}
However, we know that $\sigma_{\rm full}$ has the largest errors among three string tensions.
Incidentally, if we use the other data for $\epsilon \sqrt{\sigma_{\rm full}*}$  at $\beta=6.0$ given in Table 4 of \cite{EHK98} where 
$\epsilon^2 \sigma_{\rm full}* = (\epsilon \sqrt{\sigma_{\rm full}*})^2 = 0.2154^2 \sim 0.2209^2=0.0464 \sim 0.0488$, 
the ratios of two string tensions $\sigma_{\rm rest}, \sigma_{\rm mono}$ to the total string tension $\sigma_{\rm full}$ are modified  
\begin{align}
\frac{\sigma_{\rm rest}}{\sigma_{\rm full}*} & \cong 0.78  \sim  0.82,
\\
\frac{\sigma_{\rm mono}}{\sigma_{\rm full}*} & \cong 0.72  \sim  0.76 ,
\end{align}
Thus, we have obtained the \textbf{infrared restricted variable $\mathscr{V}$ dominance} in the string tension (78--82\%) and the \textbf{non-Abelian  magnetic monopole dominance} in the string tension  (72--76\%).
Both dominance are obtained in the gauge independent way.% 
\footnote{The method of fitting the data given in this paper is the same as that in \cite{Shibata-lattice2010}, but is different from that used in \cite{Shibata-lattice2008}. 
%This is one of the reasons why the percentages of magnetic monopole dominance in the string tension are so different in two reports. 
}

%%%%%%%%%%%%%%%%%%%%%%%%%%%%%%%%%%%%%%%%%%%%%%%%%%
\subsubsection{Color direction field and color symmetry}
%%%%%%%%%%%%%%%%%%%%%%%%%%%%%%%%%%%%%%%%%%%%%%%%%%

\begin{figure}[t]
\begin{center}
\includegraphics[scale=0.6]{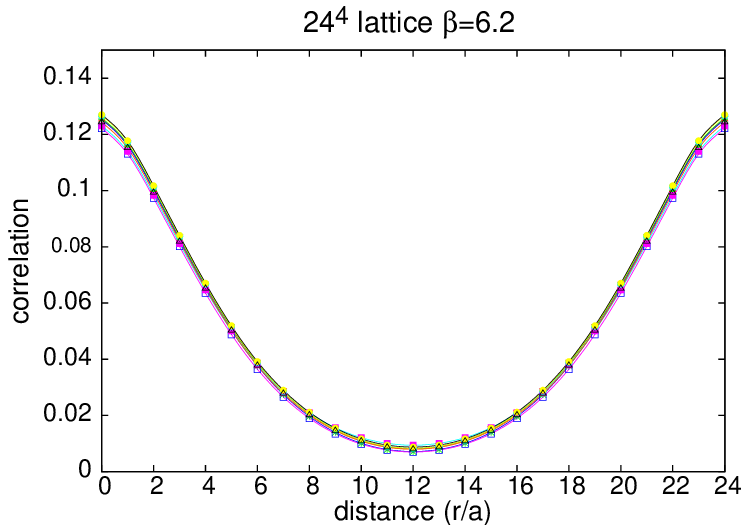}
\includegraphics[scale=0.6]{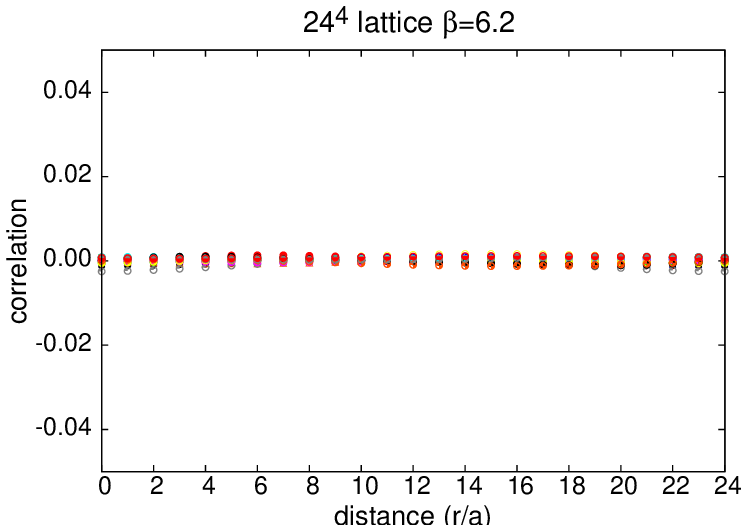}
\end{center}
\vspace{-0.5cm}
\caption{Ref.\cite{KSSK11}: Color field correlators $\langle  n^A(x) n^B(0) \rangle$ ($A,B=1, \cdots, 8$) as functions of the distance $r:=|x|$ measured at  $\beta=6.2$ on   24$^4$ lattice,  using 500 configurations under the Landau gauge. 
(Left panel) $A=B$, 
(Right panel) $A \not= B$.
}
\label{C35-fig:color-field-corr}
\end{figure}

Fig.\ref{C35-fig:color-field-corr} shows  two-point correlation functions of color field $\langle  n^A(x) n^B(0) \rangle$ versus the distance $r:=|x|$. 
All plots of correlators for $A=B=1,2, \cdots, 8$ overlap on top of each other, and hence they can be fitted by a common non-vanishing  function $D(r)$ (left panel), while all correlators for $A \not= B$ are nearly equal to zero (right panel). 
Therefore,  the correlators $\langle  n^A(x) n^B(0) \rangle$ are of the form:
\begin{equation}
\langle  n^A(x) n^B(0) \rangle = \delta^{AB} D(r) \quad (A,B=1,2, \cdots, 8)
 .
\end{equation}
We have also checked that  one-point functions vanish: 
\begin{equation}
\langle  n^A(x)  \rangle = \pm 0.002 \simeq 0 \quad (A =1,2, \cdots, 8)
 .
\end{equation}
These results indicate  that the global $SU(3)$  \textbf{color symmetry} is preserved, that is to say, there is no specific direction in color space.  
This is expected, since the Yang-Mills theory should respect the global gauge symmetry, i.e., color symmetry, even after imposing the Landau gauge.

To obtain correlation functions of field variables, we need to fix the gauge and we have adopted the Landau gauge for the original Yang-Mills field $\mathscr{A}$ so that the global color symmetry is not broken.  This property is desirable to study color confinement, but it is lost in the MA gauge.

%%%%%%%%%%%%%%%%%%%%% figures %%%%%%%%%%%%%%%%%%%%%%%%%%%
%%%%%%%%%%%%%%%%%%%%% figures %%%%%%%%%%%%%%%%%%%%%%%%%%%

%\newpage
%\noindent
%$\bullet$ magnetic-monopole loops for SU(3)

\begin{figure}[ptb]
\begin{center}
\vspace{-5mm}%
\includegraphics[width=4in]{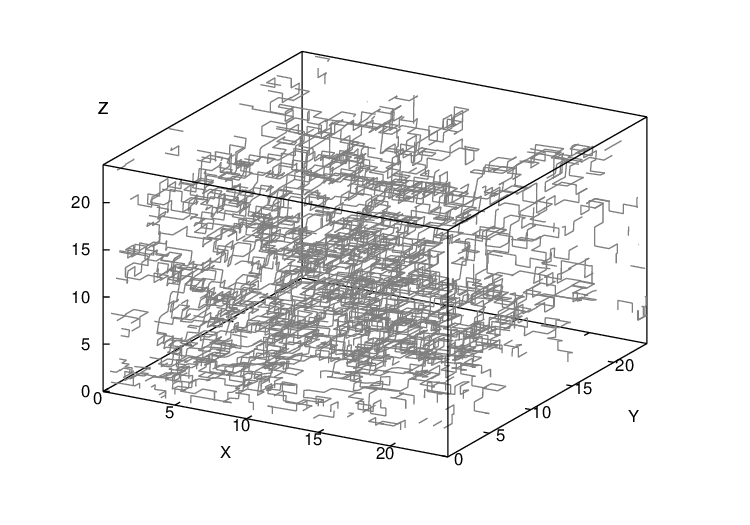}
\vspace{-1.0cm}
\end{center}
\caption{
%\footnotesize 
The  magnetic-monopole loops on the 4-dimensional lattice 
where the 3-dimensional plot is obtained by projecting the 4-dimensional dual lattice space to the 3-dimensional one, i.e., $(x,y,z,t) \rightarrow (x,y,z)$.
}
\label{C35-fig:monopole-loops}%
\end{figure}

Fig.~\ref{C35-fig:monopole-loops} exhibits an example of the magnetic-monopole loops in $SU(3)$ Yang-Mills theory (in the minimal  option) on the 4-dimensional lattice, which is written as the 3-dimensional plot obtained by projecting the 4-dimensional dual lattice space to the 3-dimensional one, i.e., $(x,y,z,t) \rightarrow (x,y,z)$.

\begin{figure}[ptb]
\begin{center}
\includegraphics[
height=1.1in,
]{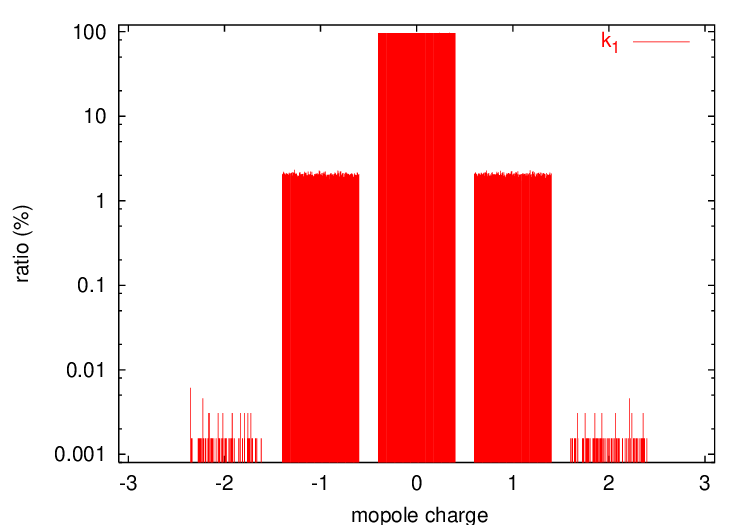}
\includegraphics[
height=1.1in,
]{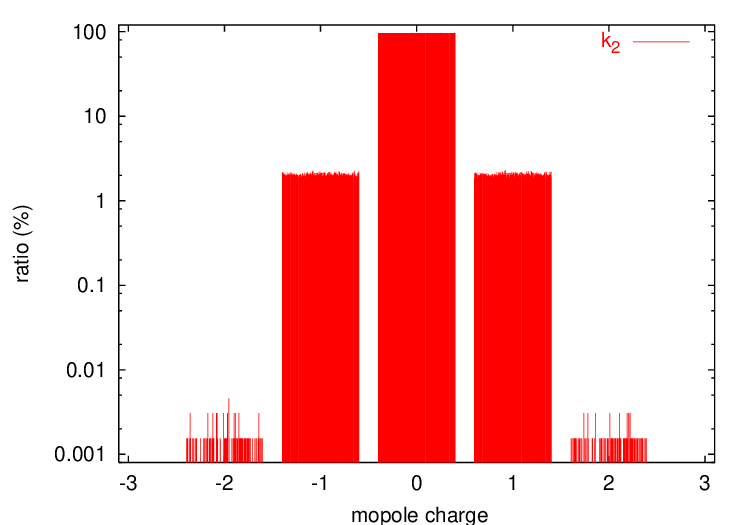}
\includegraphics[
height=1.1in,
]{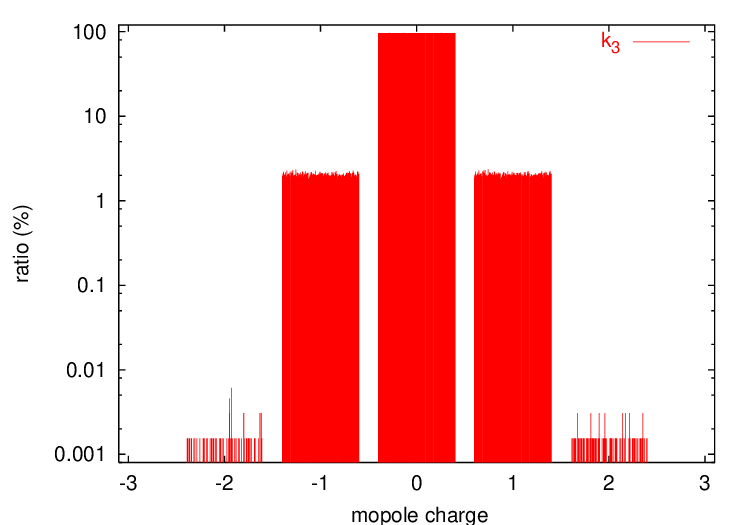}
\includegraphics[
height=1.1in,
]{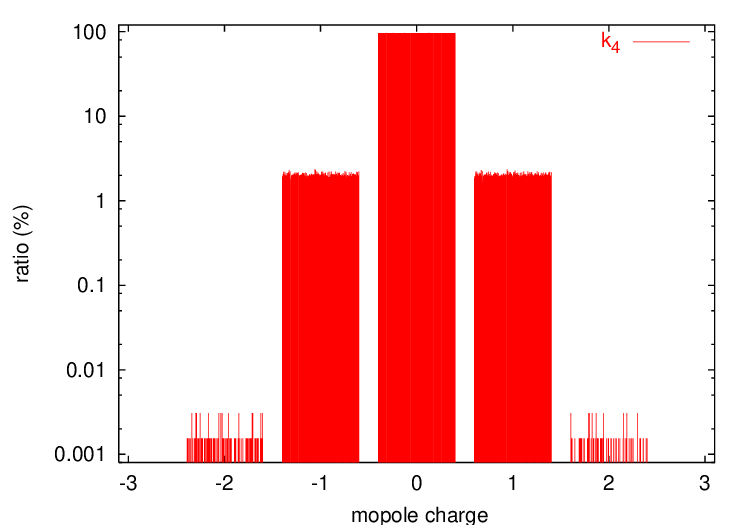}
\end{center}
\caption{Ref.\cite{Shibata-lattice2008}: 
The magnetic monopole charge distribution for minimal option for 
$\beta=5.7$ on the $16^4$ lattice.
The distribution of $k_1$, $k_2$, $k_3$ and $k_4$ are arranged from left to right, respectively. 
%, (Right panel)
%the static interquark potential calculated only from the magnetic monopole part.
}%
\label{C35-fig:monopoleMini}%
\end{figure}

Fig.~\ref{C35-fig:monopoleMini} exhibits the magnetic monopole charge distribution in $SU(3)$ Yang-Mills theory (in the minimal  option) on the $16^4$ lattice  at 
$\beta=5.7$, where the figures are arranged from left to right representing the distribution of $k_1$, $k_2$, $k_3$ and $k_4$, respectively.

%%%%%%%%%%%%%%%%%%%%% figures %%%%%%%%%%%%%%%%%%%%%%%%%%%

%%%%%%%%%%%%%%%%%%%%% figures %%%%%%%%%%%%%%%%%%%%%%%%%%%

%%%%%%%%%%%%%%%%%%%%% figures %%%%%%%%%%%%%%%%%%%%%%%%%%%
%%%%%%%%%%%%%%%%%%%%% figures %%%%%%%%%%%%%%%%%%%%%%%%%%%

%%%%%%%%%%%%%%%%%%%%%%%%%%%%%%%%%%%%%%%%%%%
%%%%%%%%%%%%%%%%%%%%%%%%%%%%%%%%%%%%%%%%%%%
%\subsection{Method and results}
%%%%%%%%%%%%%%%%%%%%%%%%%%%%%%%%%%%%%%%%%%%
%%%%%%%%%%%%%%%%%%%%%%%%%%%%%%%%%%%%%%%%%%%

%%%%%%%%%%%%%%%%%%%%%%%%%%%%%%%%%%%%%%%%%%%%%%%%%%
\subsubsection{Gauge-invariant chromoelectric field and flux tube formation}
%%%%%%%%%%%%%%%%%%%%%%%%%%%%%%%%%%%%%%%%%%%%%%%%%%

We generate configurations of the gauge link variable $\{ U_{x,\mu} \}$ using the standard Wilson action on a $24^{4}$ lattice at $\beta=6.2$.
The gauge link decomposition is obtained according to the framework given in the previous section: the color field configuration  $\{ \bm{h}_{x} \}$ is obtained by solving the reduction condition of minimizing the functional (\ref{C35-eq:reduction}) for each gauge configuration  $\{ U_{x,\mu} \}$, and then the decomposed variables  $\{ V_{x,\mu} \}$,  $\{ X_{x,\mu} \}$ are obtained by using the formula (\ref{C35-defeq}). 
In the measurement of the Wilson loop average, we apply the APE smearing technique to reduce noises \cite{Albanese87}. 

%%%%%%%%%%%%%%%%%%%%% figures %%%%%%%%%%%%%%%%%%%%%%%%%%%
\begin{figure}[ptb]
\begin{center}
\includegraphics[
height=5cm,
]
{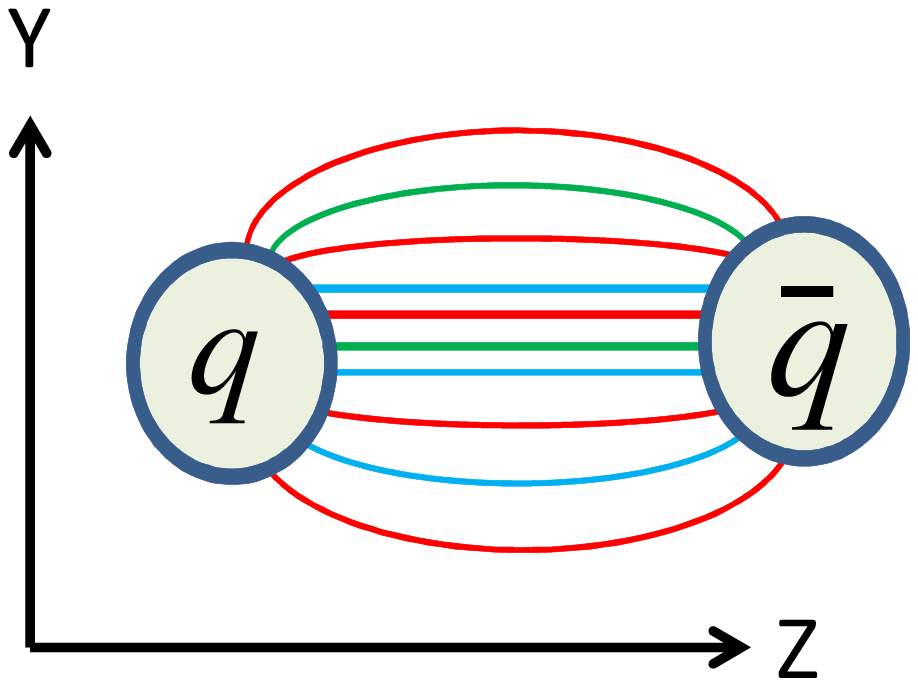} 
\quad\quad
\includegraphics[
origin=c,
height=5.5cm,
%angle=270
]
{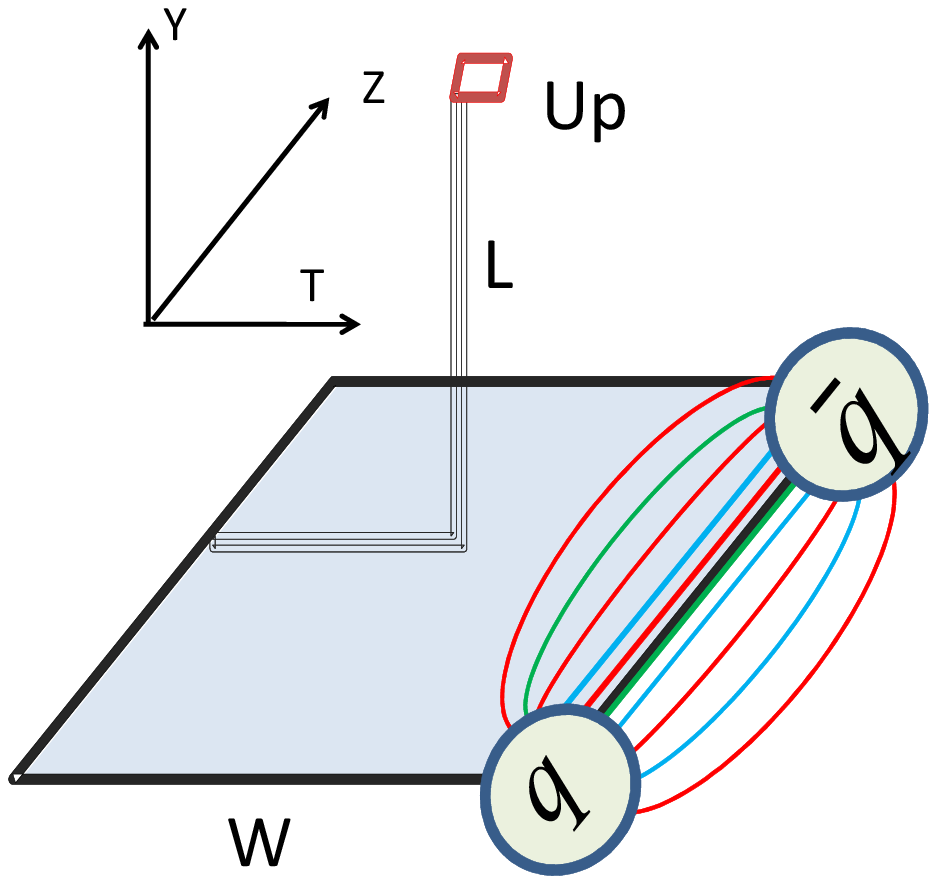}
\end{center}
\vspace{-0.5cm}
\caption{
(Left) 
The setup of measuring the chromo-flux produced by a quark--antiquark pair. 
(Right) 
The gauge-invariant connected correlator ($U_{p}LWL^{\dag})$ between a plaquette $U$ and the Wilson loop $W$.
}%
\label{C35-fig:Operator}%
\end{figure}
%%%%%%%%%%%%%%%%%%%%% figures %%%%%%%%%%%%%%%%%%%%%%%%%%%

We investigate whether or not the non-Abelian dual Meissner effect is the mechanism for quark confinement.
In order to extract the chromo-field, we use a gauge-invariant correlation function proposed by Di Giacomo, Maggiore and Olejnik  \cite{GMO90}.  
The chromo-field created by a quark-antiquark pair in $SU(N)$ Yang-Mills theory is measured by using a gauge-invariant connected correlator between a plaquette and the Wilson loop  
(see Fig.\ref{C35-fig:Operator}):%
\begin{equation}
\rho_{_{U_P}}:=\frac{\left\langle \mathrm{tr}\left(  U_{P}L^{\dag}WL\right)
\right\rangle }{\left\langle \mathrm{tr}\left(  W\right)  \right\rangle
}-\frac{1}{N}\frac{\left\langle \mathrm{tr}\left(  U_{P}\right)
\mathrm{tr}\left(  W\right)  \right\rangle }{\left\langle \mathrm{tr}\left(
W\right)  \right\rangle }, 
\label{C35-eq:Op}%
\end{equation}
where $W$ is the Wilson loop in
$Z$-$T$ plane representing a pair of quark and antiquark, $U_{P}$ a plaquette variable as the probe operator to measure the chromo-field strength at the point $P$, and $L$ the Wilson line connecting the source $W$ and the probe $U_{P}$.
Here $L$ is necessary to guarantee the gauge invariance of the correlator $\rho_{_{U_P}}$ and hence the probe is identified with $LU_PL^\dagger$. 
%, and $N$ the number of color ($N=3$). 
The symbol $\left\langle \mathcal{O}\right\rangle $ denotes the average of the operator $\mathcal{O}$ in the space and the ensemble of the configurations. 
In the naive continuum limit $\epsilon \to 0$, indeed,  $\rho_{_{U_P}}$ reduces to the field strength in the presence of the $q\bar q$ source:
\begin{equation}
\ \rho_{_{U_P}}%
\overset{\varepsilon\rightarrow0}{\simeq}g\epsilon^{2}\left\langle
\mathscr{F}_{\mu\nu}\right\rangle _{q\bar{q}}:=\frac{\left\langle
\mathrm{tr}\left(  ig\epsilon^{2}  \mathscr{F}_{\mu\nu}L^{\dag}WL \right)
\right\rangle }{\left\langle \mathrm{tr}\left(  W\right)  \right\rangle
}+O(\epsilon^{4}) ,
\end{equation}
where we have used $U_{x,\mu}=\exp (-ig\epsilon \mathscr{A}_\mu(x))$ and hence $U_{P}= \exp(-ig\epsilon^2 \mathscr{F}_{\mu\nu} )$.
Thus, the \textbf{gauge-invariant chromo-field strength} $F_{\mu\nu}[U]$ produced by a $q\bar q$ pair is given by 
\begin{equation}
F_{\mu\nu}[U] := \epsilon^{-2} \sqrt{\frac{\beta}{2N}}\rho_{_{U_P}} ,
\end{equation}
where $\beta:=2N/g^2$ is the lattice gauge coupling constant. 
Note that the connected correlator $\rho_{_{U_P}}$ is sensitive to the field strength, while  the disconnected one probes the squared field strength:
\begin{equation}
\rho_{_{U_P}}^{\prime}:=\frac{\left\langle \mathrm{tr}\left(  W\right)
\mathrm{tr}\left(  U_{P}\right)  \right\rangle }{\left\langle \mathrm{tr} 
\left(  W\right)  \right\rangle }-\left\langle \mathrm{tr}\left( U_{P}\right)  \right\rangle \overset{\varepsilon\rightarrow0}{\simeq} 
g\epsilon^{4}\left[  \left\langle \mathscr{F}_{\mu\nu}^{2}\right\rangle_{q\bar{q}} - \left\langle \mathscr{F}_{\mu\nu}^{2}\right\rangle _{0} \right]  .
\end{equation}

%%%%%%%%%%%%%%%%%%%%% figures %%%%%%%%%%%%%%%%%%%%%%%%%%%
\begin{figure}[ptb]
\begin{center}
%\vspace{-7mm}
%\includegraphics[
%width=4.5cm,
%]
%{figs/measure.eps} 
%\quad
\includegraphics[
scale=0.34,
angle=270,
]
{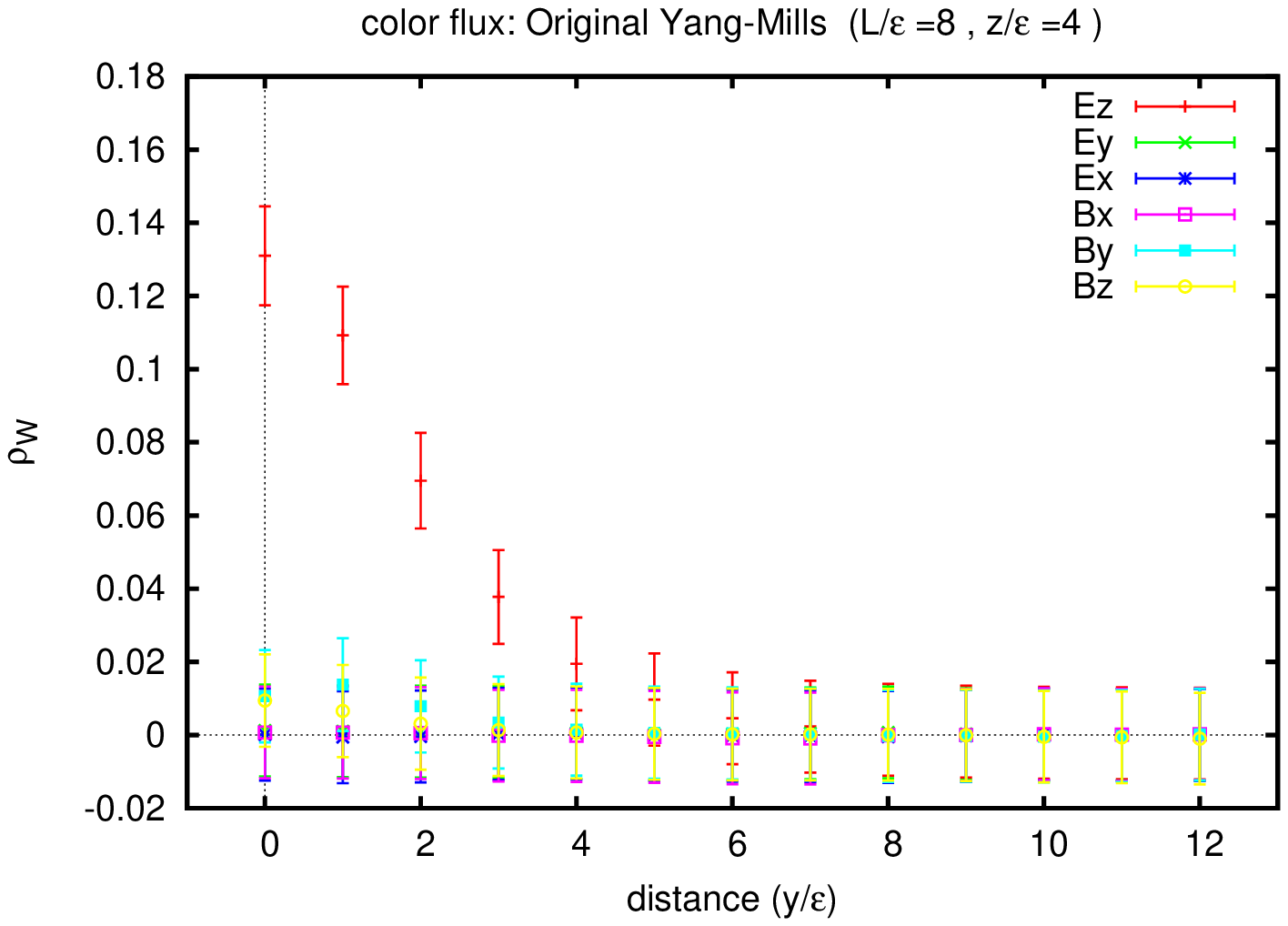}  
\ 
%\begin{minipage}{0.45\textwidth}
\includegraphics[
scale=0.34,
angle=270,
%origin=t
]
{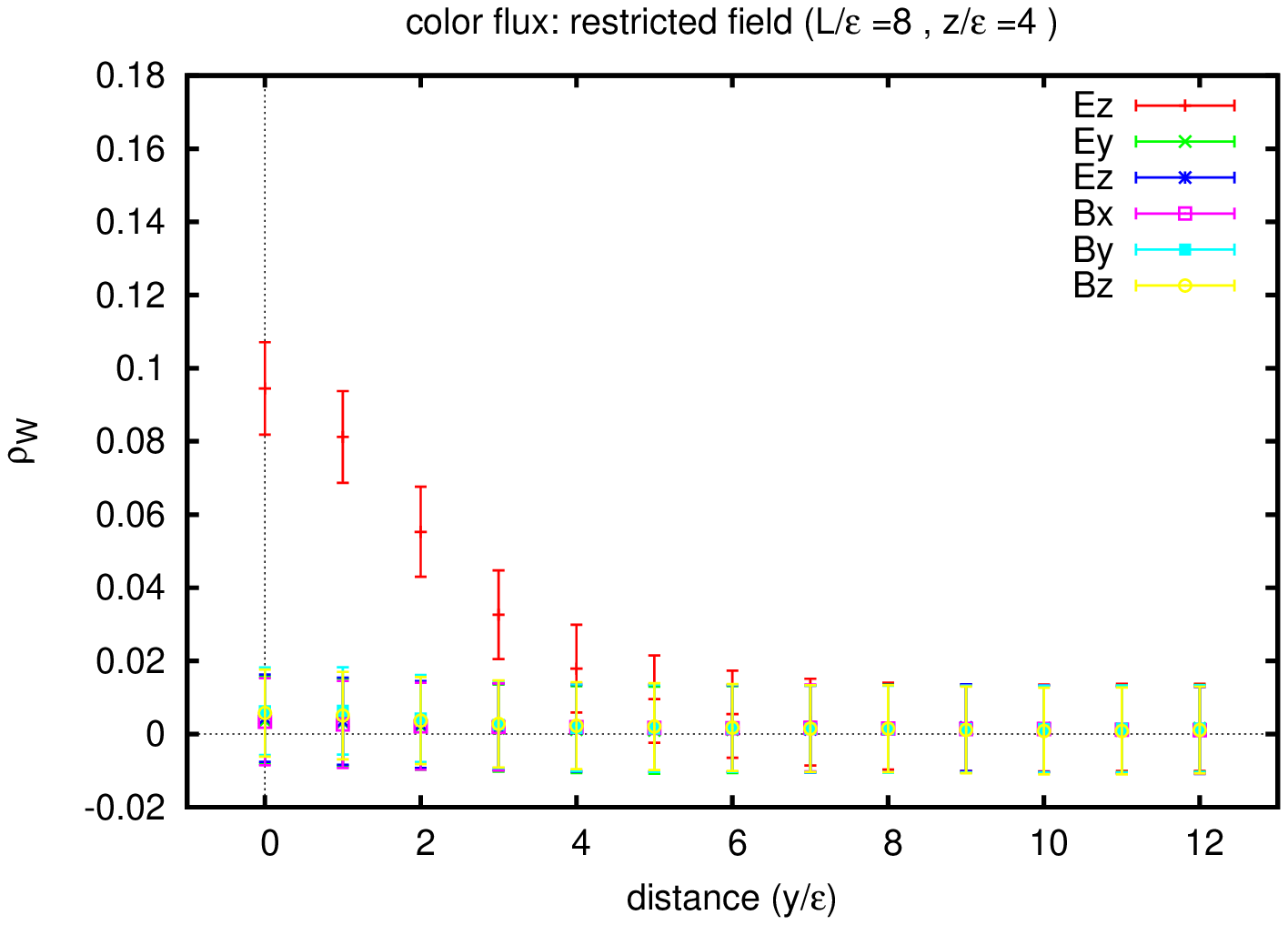}
%\end{minipage}
\vspace{-0.5cm}
\end{center}
\caption{
%The measurement of the chromo-flux connecting a pair of quark and antiquark: 
%(Left) The gauge-invariant connected correlator ($U_{p}LWL^{\dag})$ between a plaquette $U$ and the Wilson loop $W$. 
Ref.\cite{SKKS13}: 
Measurement of components of the chromoelectric field $\bm{E}$ and chromomagnetic field  $\bm{B}$ as functions of the distance $y$ from the $z$ axis. 
(Left panel) the original $SU(3)$ Yang-Mills field, 
(Right panel) the restricted  field. 
}%
\label{C35-fig:measure}%
\end{figure}
%%%%%%%%%%%%%%%%%%%%% figures %%%%%%%%%%%%%%%%%%%%%%%%%%%

We measure correlators between the plaquette $U_P$ and the chromo-field strength of the restricted field $V_{x,\mu}$ as well as the original Yang-Mills field $U_{x,\mu}$.
See the left panel of Fig.~\ref{C35-fig:Operator}.
Here the quark and antiquark source is introduced as $8\times8$ Wilson loop ($W$) in the $Z$-$T$ plane, and the probe $(U_{p})$ is set at the center of the Wilson loop and moved along the $Y$-direction. 
The left and right panel of Fig.~\ref{C35-fig:measure} show respectively the results of measurements for the chromoelectric and chromomagnetic fields $F_{\mu\nu}[U]$ for the original  $SU(3)$ field $U$ and $F_{\mu\nu}[V]$ for the restricted  field $V$, where the field strength $F_{\mu\nu}[V]$  is obtained by using $V_{\,x,\mu}$ in  (\ref{C35-eq:Op}) instead of $U_{x,\mu}$:
\begin{equation}
F_{\mu\nu}[V] := \sqrt{\frac{\beta}{2N}} \tilde\rho_{_{V_P}} , \quad 
\tilde\rho_{_{V_P}} := \frac{\left\langle \mathrm{tr}\left(   V_{P}L^{\dag}WL \right)
\right\rangle }{\left\langle \mathrm{tr}\left(  W\right)  \right\rangle
}-\frac{1}{N}\frac{\left\langle \mathrm{tr}\left(  V_{P}\right)
\mathrm{tr}\left(  W\right)  \right\rangle }{\left\langle \mathrm{tr}\left(
W\right)  \right\rangle } .  
\label{C35-cf1-5-SU3}
\end{equation}
We have checked that even if  $W[U]$ is replaced by $W[V]$, together with replacement of the probe $LU_{P}L^\dagger$ by the corresponding $V$ version, the  change in the magnitude of the field strength $F_{\mu\nu}$ remains within at most a few \%.

From Fig.\ref{C35-fig:measure} we find that  only the $E_{z}$ component of the \textbf{chromoelectric field} $(E_x,E_y,E_z)=(F_{10},F_{20},F_{30})$ connecting $q$ and $\bar q$ has non-zero value for both the restricted field $V$ and the original Yang-Mills field $U$.
The magnitude $E_{z}$ quickly decreases in the distance $y$ away from the Wilson loop.
%, as discussed in more detail in subsection C.

The other components are zero consistently within the numerical errors. 
This means that the chromomagnetic field $(B_x,B_y,B_z)=(F_{23},F_{31},F_{12})$ connecting $q$ and $\bar q$ does not exist  and that the chromoelectric field is parallel to the $z$ axis on which  quark and antiquark are located.

\begin{figure}[ptb]
\begin{center}
%\vspace{-8mm} 
\includegraphics[
height=8.0cm,
angle=270
]
{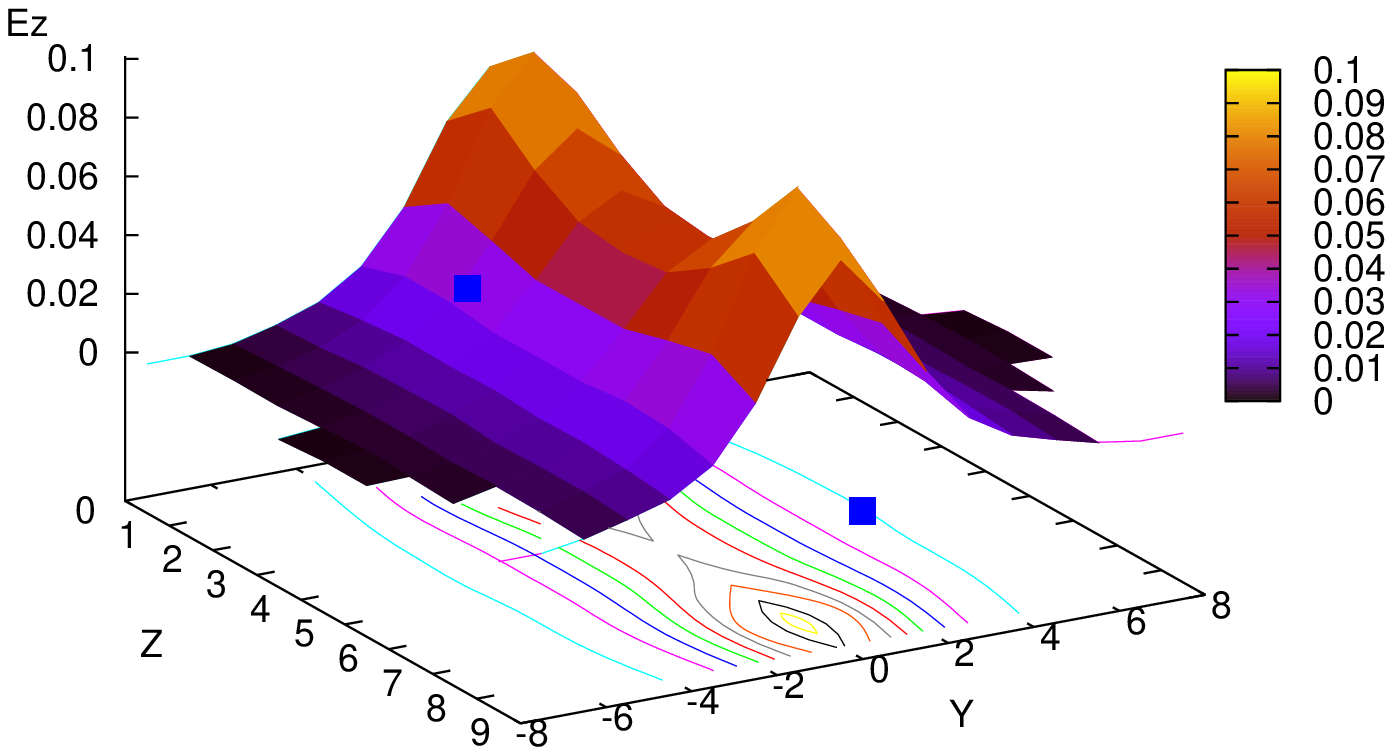} 
\quad
\includegraphics[
height=8.0cm,
angle=270
]
{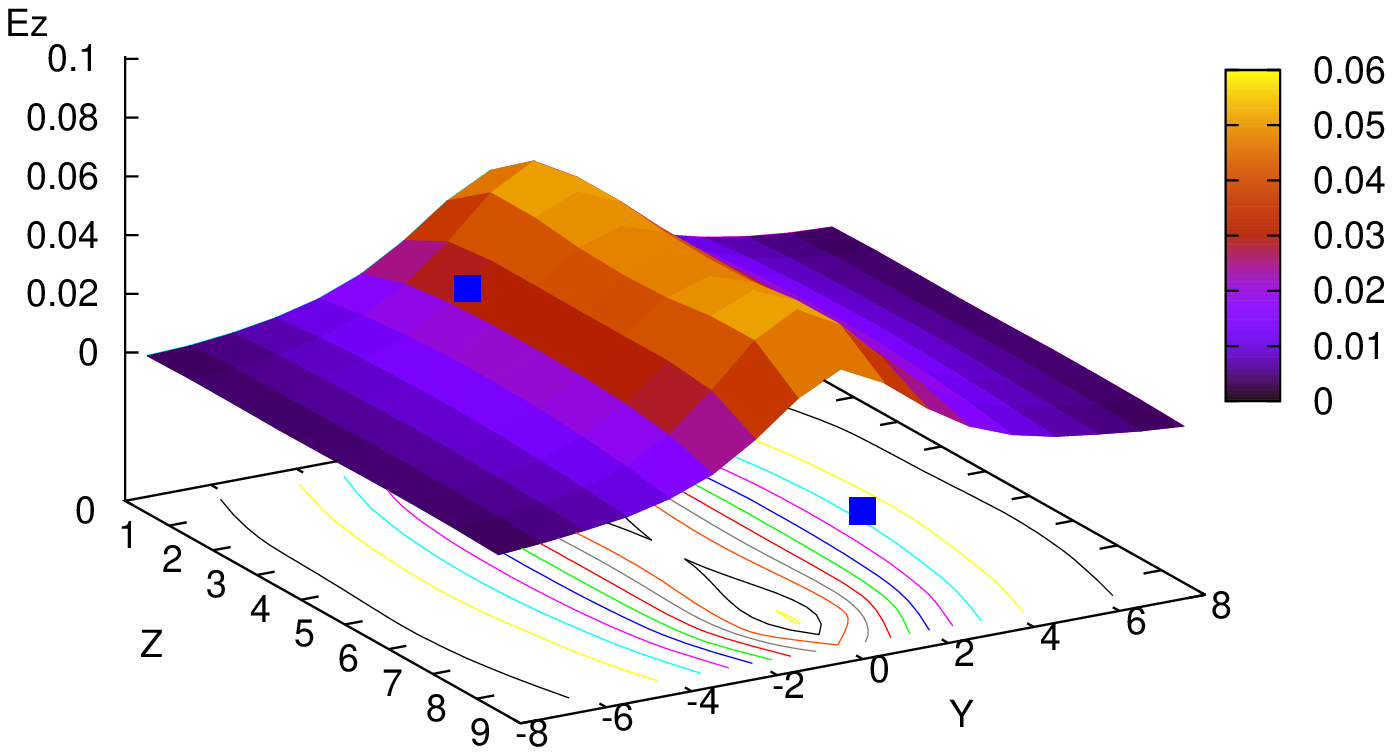} 
\vspace{-0.5cm}
\end{center}
\caption{Ref.\cite{SKKS13}:
The distribution in $Y$-$Z$ plane of the chromoelectric field $E_z$ connecting a pair of quark and antiquark: 
(Left panel) chromoelectric field produced from the original Yang-Mills field, 
(Right panel) chromoelectric field produced from the restricted  field. 
}%
\label{C35-fig:fluxtube}%
\end{figure}

To see the profile of the non-vanishing component $E_z$ of the chromoelectric field in detail, we explore the distribution of chromoelectric field on the 2-dimensional plane. 
 Fig.~\ref{C35-fig:fluxtube} shows the distribution of $E_{z}$ component of the chromoelectric field, where the quark-antiquark source represented as $9\times11$ Wilson loop $W$ is placed at $(Y,Z)=(0,0), (0,9)$, and the probe $U$ is displaced on the $Y$-$Z$ plane at the midpoint of the $T$-direction. 
 The position of a quark and an antiquark is marked by the solid (blue) box. The magnitude of $E_{z}$ is shown by the height of the 3D plot and also the contour plot in the bottom plane.
 The left panel of Fig.~\ref{C35-fig:fluxtube} shows the plot of $E_z$ for the $SU(3)$ Yang-Mills field $U$, and the right panel of Fig.~\ref{C35-fig:fluxtube} for the restricted field $V$. 
We find that the magnitude $E_{z}$ is quite uniform for the restricted part $V$, while it is almost uniform for the original part $U$ except for the neighborhoods of the locations of $q$, $\bar q$ source. 
This difference is due to  the contributions from the remaining part $X$ which affects only the  short distance, as will be discussed later.%in the next section. 

%%%%%%%%%%%%%%%%%%%%% figures %%%%%%%%%%%%%%%%%%%%%%%%%%%
%%%%%%%%%%%%%%%%%%%%% figures %%%%%%%%%%%%%%%%%%%%%%%%%%%

%%%%%%%%%%%%%%%%%%%%%%%%%%%%%%%%%%%%%%%%%%%%%%%%%%
\subsubsection{Magnetic current and dual Meissner effect for $SU(3)$ case}
%%%%%%%%%%%%%%%%%%%%%%%%%%%%%%%%%%%%%%%%%%%%%%%%%%

Next, we investigate the relation between the chromoelectric flux and the  magnetic current. 
%From the Yang-Mills equation for $\mathbf{V}_{\mu}$ field, 
The magnetic(-monopole) current can be calculated as
\begin{equation}
 k= \delta{}^{\displaystyle *}F[V] ={}^{\displaystyle *}d F[V] ,
\label{C35-def-k2}
\end{equation}
where $F[V]$ is the field strength (\ref{C35-cf1-5-SU3}) defined from the restricted field $V$  in the presence of the $q\bar q$ source,
%the field strength (2-form) of the restricted field (1-form) $\mathbf{V}$, 
$d$ the exterior derivative, $\delta$ codifferential, and $^{\ast}$ denotes the Hodge dual operation. 
Note that non-zero magnetic current follows from violation of the Bianchi identity  
(If the field strength was given by the exterior derivative of some  field $A$ (one-form), $ F=dA$, \ we would obtain $k=\delta{}^{\displaystyle *}F={}^{\displaystyle *}d^{2}A=0$).

%%%%%%%%%%%%%%%%%%%%% figures %%%%%%%%%%%%%%%%%%%%%%%%%%%
\begin{figure}[ptb]
\begin{center}
%\vspace{-5mm}
\includegraphics[
scale=1.2
]
{Fig-PR/Fig-lattice/M-current.eps} 
\quad
\includegraphics[
scale=0.8
]
{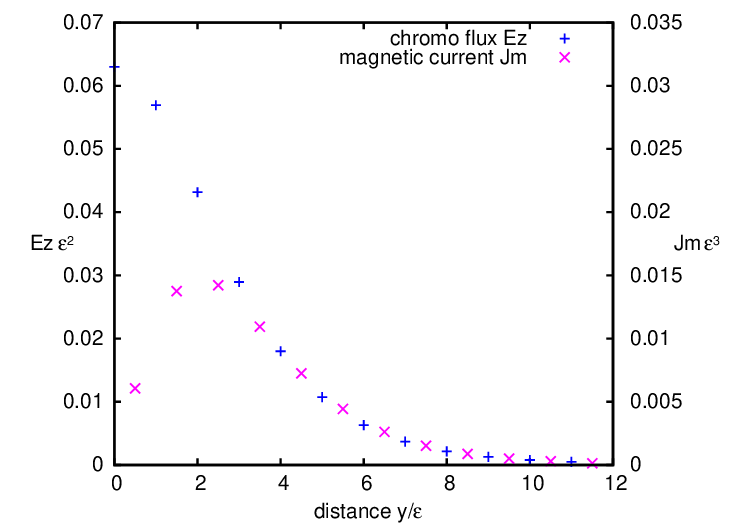} 
\vspace{-0.5cm}
\end{center}
\caption{Ref.\cite{SKKS13}{}:
The magnetic-monopole current $\mathbf{k}$ induced around the flux along the $z$ axis connecting a  quark-antiquark pair.
(Left panel) The positional relationship between the chromoelectric field $E_{z}$ and the magnetic current $\mathbf{k}$. 
(Right panel) The magnitude of the chromo-electronic current $E_{z}$ and the magnetic current  $J_{m}=|\mathbf{k}|$ as functions of the distance $y$ from the $z$ axis. 
}
\label{C35-fig:Mcurrent-SU3}%
\end{figure} 
%%%%%%%%%%%%%%%%%%%%% figures %%%%%%%%%%%%%%%%%%%%%%%%%%%

Fig.~\ref{C35-fig:Mcurrent-SU3} shows the  magnetic current measured in $X$-$Y$ plane at the midpoint of quark and antiquark pair in the $Z$-direction. 
The left panel of Fig.~\ref{C35-fig:Mcurrent-SU3} shows the positional relationship between chromoelectric flux and  magnetic current.
The right panel of Fig.~\ref{C35-fig:Mcurrent-SU3} shows the magnitude of the  chromoelectric field $E_z$ (left
scale) and the magnetic current $k$ (right scale). 
The existence of non-vanishing  \textbf{magnetic current} $k$ around the  \textbf{chromoelectric field} $E_z$ supports the  \textbf{dual superconductivity} which is the dual picture of the ordinary superconductor exhibiting the electric current $J$ around the magnetic field $B$.

In our formulation, it is possible to define a gauge-invariant magnetic-monopole current  $k_{\mu}$  by using $V$-field,
%\begin{subequations}
%\begin{align}
%k_{\mu}  &  =2\pi n_{\mu}:=\frac{1}{2}\epsilon_{\mu\nu\alpha\beta} \partial_{\nu}\Theta_{\alpha\beta} ,
% \\
%\Theta_{\mu\nu}   &  :=-\arg\text{ \textrm{Tr}}\left[  \left(  \frac{1}{3}\mathbf{1}-\frac{2}{\sqrt{3}}\mathbf{h}_{x}\right)  V_{x,\mu}V_{x+\mu,\mu}V_{x+\nu,\mu}^{\dag}V_{x,\nu}^{\dag}\right] ,
%\end{align}
%\end{subequations}
which is obtained from the field strength $\mathscr{F}[\mathscr{V}]$ of the restricted field $\mathscr{V}$, as suggested from the non-Abelian Stokes theorem.
It should be also noticed that this
magnetic-monopole current  is a non-Abelian magnetic monopole extracted from the $V$ field, which corresponds to the maximal stability group $\tilde{H}=U(2)$.
The magnetic-monopole current  $k_{\mu}$ defined in this way can be used to study the magnetic current around the chromoelectric flux tube, instead of the above definition of $k$  (\ref{C35-def-k2}).
The comparison of two magnetic-monopole currents $k$ is to be done in the future works. 

These are numerical evidences supporting   {``non-Abelian'' dual superconductivity due to non-Abelian  magnetic monopoles   as a mechanism for quark confinement in SU(3) Yang-Mills theory}.

%%%%%%%%%%%%%%%%%%%%%%%%%%%%%%%%%%%%%%%%%%%%%%%%%%
\subsubsection{Type of dual superconductivity}
%%%%%%%%%%%%%%%%%%%%%%%%%%%%%%%%%%%%%%%%%%%%%%%%%%

\begin{figure}[ptb]
\begin{center}
\includegraphics[
width=6.0cm,
angle=270
]
{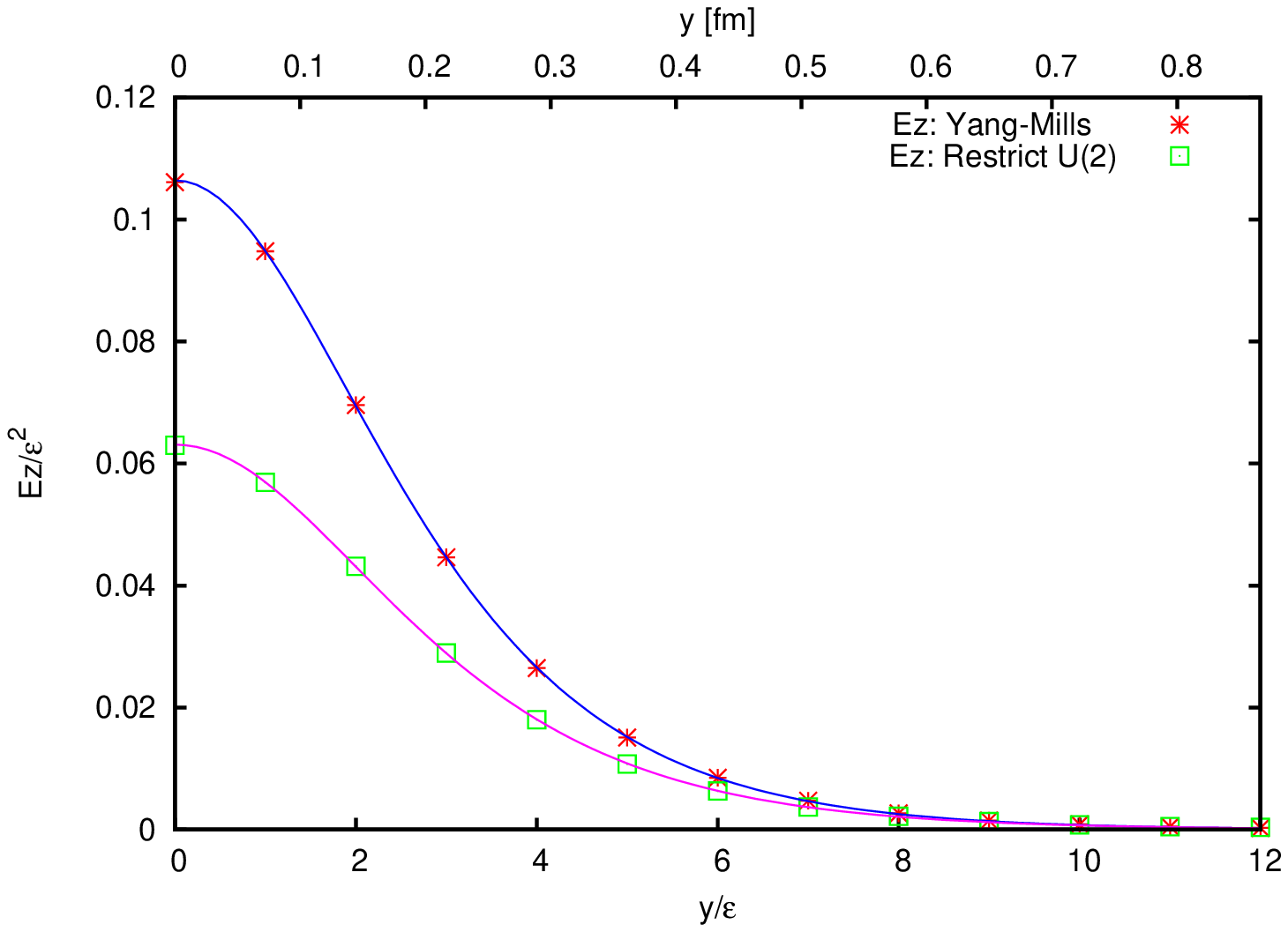} 
\includegraphics[,
width=6.0cm,
angle=270
]
{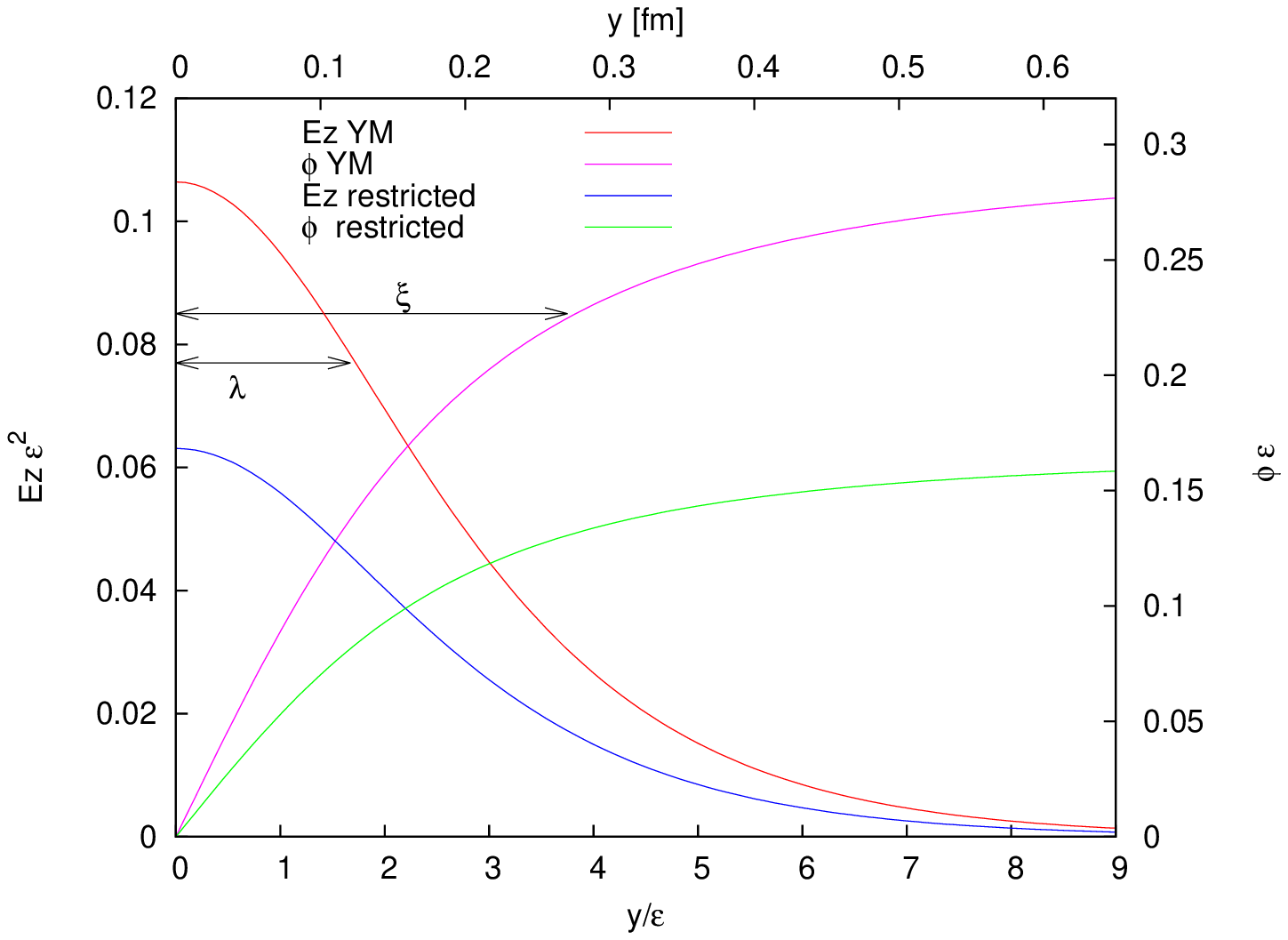}  
\vspace{-0.5cm}
\end{center}
\caption{Ref.\cite{SKKS13}: 
(Left panel)
The plot of the chromoelectric field $E_z$ versus the distance $y$ in units of the lattice spacing $\epsilon$ and the fitting   as a function $E_z(y)$ of  $y$ according to (\ref{C35-Clem-fit2}). 
The red cross for the original $SU(3)$ field and the green square symbol for the restricted field.  
(Right panel) The order parameter $\phi$ reproduced as a function $\phi(y)$ of  $y$ according to (\ref{order-f}), together with the chromoelectric field $E_z(y)$.
}
\label{C35-fig:type}%
\end{figure}

Moreover, we investigate the QCD vacuum, i.e., type of the dual superconductor. 
The left panel of Fig.\ref{C35-fig:type} is the plot for the chromoelectric field $E_z$ as a function of the distance $y$ in units of the lattice spacing $\epsilon$ for the original $SU(3)$ field and for the restricted  field.

In order to examine the \textbf{type of the dual superconductivity}, we apply the formula for the magnetic field  derived by Clem \cite{Clem75} in the ordinary superconductor based on the \textbf{Ginzburg-Landau (GL) theory} to the chromoelectric field in the dual superconductor.
In the GL theory, the gauge field $A$ and the scalar field $\phi$ obey simultaneously  the GL equation:
\begin{equation}
 (\partial^\mu -iq A^\mu)(\partial_\mu -iq A_\mu) \phi + \lambda (\phi^* \phi - \eta^2) = 0 ,
\end{equation}
and the Ampere equation:
\begin{equation}
 \partial^\nu F_{\mu\nu} + iq [\phi^* (\partial_\mu \phi -iq A_\mu \phi)  - (\partial_\mu \phi -iq A_\mu \phi)^* \phi] = 0 .
%- iq(\phi^* \partial_\mu \phi - \phi \partial_\mu \phi^*) - 2q^2 A_\mu \phi^* \phi = 0 .
\end{equation}

Usually, in the dual superconductor of the type II, it is justified to use the asymptotic form $K_0(y/\lambda)$ to fit the chromoelectric field in the large $y$ region (as the solution of the Ampere equation in the dual GL theory).  
However, it is clear that this solution cannot be applied to the small $y$ region, as is easily seen from the fact that $K_0(y/\lambda) \to \infty$ as $y \to 0$. 
In order to see the difference between type I and type II, it is crucial to see the relatively small $y$ region.
Therefore, such a simple form cannot be used to detect the type I dual  superconductor. 
However, this important aspect was ignored in the preceding studies except for a work \cite{Cea:2012qw}.

On the other hand, Clem \cite{Clem75} does not obtain the analytical solution of the GL equation explicitly and use an approximated form for the scalar field  $\phi$ (given below in (\ref{order-f})).
This form is used to solve the Ampere equation exactly to obtain the analytical form for the gauge field $A_\mu$ and the resulting magnetic field $B$.
This method does not change the behavior of the gauge field in the long distance, but it gives a finite value for the gauge field even at the origin. 
Therefore, we can obtain the formula which is valid for any distance (core radius) $y$ from the  axis connecting $q$ and $\bar{q}$: the profile of chromoelectric field in the dual superconductor is obtained:
\begin{equation}
E_{z}(y)=\frac{\Phi}{2\pi}\frac{1}{\zeta\lambda}\frac{K_{0}(R/\lambda)}%
{K_{1}(\zeta/\lambda)},\text{ }R=\sqrt{y^{2}+\zeta^{2}} ,
\label{C35-Clem-fit2}%
\end{equation}
provided that the scalar field is given by (See the right panel of Fig.\ref{C35-fig:type})
\begin{equation}
 \phi(y) = \frac{\Phi}{2\pi} \frac{1}{\sqrt{2}\lambda} \frac{y}{\sqrt{y^2+\zeta^2}} ,
 \label{order-f}
\end{equation}
where $K_{\nu}$ is the modified Bessel function of the $\nu$-th order,
$\lambda$ the parameter corresponding to the London \textbf{penetration length}, $\zeta$
a variational parameter for the core radius, and $\Phi$ external electric flux. 
In the dual superconductor, we define the \textbf{GL parameter} $\kappa$ as the ratio of the London penetration length $\lambda$ and the \textbf{coherence length} $\xi$ which measures the coherence of the magnetic monopole condensate (the dual version of the Cooper pair condensate):
\begin{equation}
\kappa= \frac{\lambda}{\xi}  \ .
\end{equation}  
It is given by \cite{Clem75}
\begin{equation}
\kappa=\sqrt{2} \frac{\lambda}{\zeta} \sqrt{1-K_{0}^{2}(\zeta/\lambda)/K_{1}^{2}(\zeta/\lambda)} .
\end{equation}

%%%%%%%%%%%%%%%%%%%%% figures %%%%%%%%%%%%%%%%%%%%%%%%%%%
%%%%%%%%%%%%%%%%%%%%% figures %%%%%%%%%%%%%%%%%%%%%%%%%%%

%%%%%%%%%%%%%%%%%%%%% Table %%%%%%%%%%%%%%%%%%%%%%%%%%%
%\[%
\begin{table}[t] 
\fontsize{6.5pt}{0pt}\selectfont
%\scriptsize
\begin{tabular}{|l||c|c|c|| c|c|c|c|c|}\hline
& $a\epsilon^2$    & $b\epsilon$    & $c$       & $\lambda/\epsilon$ 
&$ \zeta/\epsilon$ & $\xi/\epsilon$ &  $\Phi$ & $\kappa$ \\
\hline
SU(3) Yang-Mills field 
& $0.804 \pm 0.04$ & $0.598\pm 0.005$ & $1.878\pm 0.04$ & $1.672 \pm 0.014$ 
& $3.14\pm 0.09$ &  $3.75 \pm 0.12$  & $4.36 \pm 0.3$& $0.45 \pm 0.01$ \\
\hline
restricted field 
& $0.435 \pm 0.03 $ & $0.547 \pm 0.007$ & $1.787 \pm 0.05 $ & $1.828 \pm 0.023$ 
& $3.26 \pm 0.13$    & $3.84 \pm 0.19$    &$2.96 \pm 0.3 $     & $0.48 \pm 0.02$ \\
\hline
\end{tabular}
\caption{
The properties of the Yang-Mills vacuum as the dual superconductor obtained by fitting the data of chromoelectric field with the prediction of the dual Ginzburg-Landau theory.  
}
\label{C35-Table:GL-fit}
\end{table}
%\
%\]
%%%%%%%%%%%%%%%%%%%%% figures %%%%%%%%%%%%%%%%%%%%%%%%%%%

According to the formula Eq.(\ref{C35-Clem-fit2}), we estimate the GL parameter $\kappa$ for the dual superconductor of $SU(3)$ Yang-Mills theory, although this formula is obtained for the ordinary superconductor of $U(1)$ gauge field. 
By using the fitting function: 
\begin{equation}
E(y)=aK_{0}(\sqrt{b^{2}y^{2}+c^{2}}) , \quad
a= \frac{\Phi}{2\pi}\frac{1}{\zeta\lambda }\frac{1} {K_{1}(\zeta/\lambda)} ,
%=\left(  \phi/2\pi\right)  (\xi/\lambda)/K_{1}(\xi/\lambda), 
\quad
b=\frac{1}{\lambda}, 
\quad
c= \frac{\zeta}{\lambda}  ,
\label{C35-fitting}
\end{equation}
we obtain the result shown in Table~\ref{C35-Table:GL-fit}.
The superconductor is type I if $\kappa < \kappa_{c}$, while type II if $\kappa > \kappa_{c}$, where the critical value of GL parameter dividing the type of the superconductor is given by $\kappa_{c}=1/\sqrt{2}\simeq0.707$.

See Fig.\ref{C35-fig:type}.
Our data clearly shows that the dual superconductor of $SU(3)$ Yang-Mills theory is \textbf{type I} with 
\begin{equation}
 \kappa=0.45 \pm 0.01 .
\end{equation} 
This result is consistent with a quite recent result obtained independently by  Cea, Cosmai and Papa \cite{Cea:2012qw}. The London penetration length $\lambda=0.1207(17)$fm and the coherence length $\xi=0.2707(86)$fm is obtained in units of the string tension $\sigma_{\text{phys}}=(440\text{MeV})^2$, and data of lattice spacing is taken from the Table I in Ref.\cite{Edward98}.%
\footnote{
This corresponds to gauge boson mass $m_A$ and the scalar boson mass $m_\phi$: 
$m_A=1.64{\rm GeV}$, $m_\phi= 1.0 {\rm GeV}$.
} 
Moreover, our result shows that the restricted  part plays the dominant role in determining the type of the non-Abelian dual superconductivity of the $SU(3)$ Yang-Mills theory, reproducing the same result \cite{ref:confinmentX}\cite{ref:SCGT12}, i.e., type I with  
\begin{equation}
 \kappa=0.48 \pm 0.02 ,
\end{equation} 
$\lambda=0.132(3)$fm and $\xi=0.277(14)$fm.
This is a novel feature overlooked in the preceding studies. 
Thus the \textbf{restricted-field dominance} can be seen also in the determination of the type of dual superconductivity where the discrepancy is just the normalization of the chromoelectric field at the core $y=0$, coming from the difference of the total flux $\Phi$. 
These are  gauge-invariant results. 
Note again that this restricted-field and the non-Abelian magnetic monopole extracted from it reproduce the string tension in the static quark--antiquark potential.% \cite{lattice2010,abeliandomSU(3)}.

Our result should be compared with the result obtained by using the Abelian projection:  Matsubara et. al \cite{Matsubara:1993nq} suggests $\kappa=0.5 \sim 1$(which is $\beta$ dependent), border of type I and type II for both $SU(2)$ and $SU(3)$. 
In $SU(2)$ case, on the other hand, there are other works \cite{Suzuki:2009xy,Chernodub:2005gz} which conclude that the type of vacuum is at the border of type I and type II.
Our results \cite{KKS14} are consistent with the border of type I and type II for the  $SU(2)$ Yang-Mills theory on the lattice, as already shown in the above.

%\begin{figure}[ptb]
%\begin{center}
%%\vspace{-5mm}
%\includegraphics[
%width=5.5cm,
%angle=270,
%]
%{clemfit-cardso-bicdo.ps} 
%\vspace{-5mm}
%\end{center}
%\caption{
%The plot of the chromoelectric field $E_z$ reproduced from the data of \cite{Cardoso:2010kw} and the fitting according  to (\ref{fitting}).
%}
%\label{fig:typeb}%
%\end{figure}

We should mention the work \cite{Cardoso:2010kw} which concludes that  the dual superconductivity of $SU(3)$ Yang-Mills theory is type II with $\kappa=1.2 \sim 1.3$.
This conclusion seems to contradict our result for $SU(3)$. 
If the above formula (\ref{C35-Clem-fit}) is applied to the data of \cite{Cardoso:2010kw}, we have the same conclusion, namely, the type I with $\kappa=0.47 \sim 0.50$. 
%See Fig.~\ref{fig:typeb}. 
Therefore, the data obtained in \cite{Cardoso:2010kw} are consistent with ours. 
The difference between type I and type II is attributed to the way of fitting the data with the formula for the chromo-field.

%%%%%%%%%%%%%%%%%%%%%%%%%%%%%%%%%%%%%%%%%%%%%%%%%%
\subsubsection{Gluon propagators}
%%%%%%%%%%%%%%%%%%%%%%%%%%%%%%%%%%%%%%%%%%%%%%%%%%

%%%%%%%%%%%%%%%%%%%%% figures %%%%%%%%%%%%%%%%%%%%%%%%%%%
\begin{figure}[t]
\begin{center}
\includegraphics[scale=0.65]{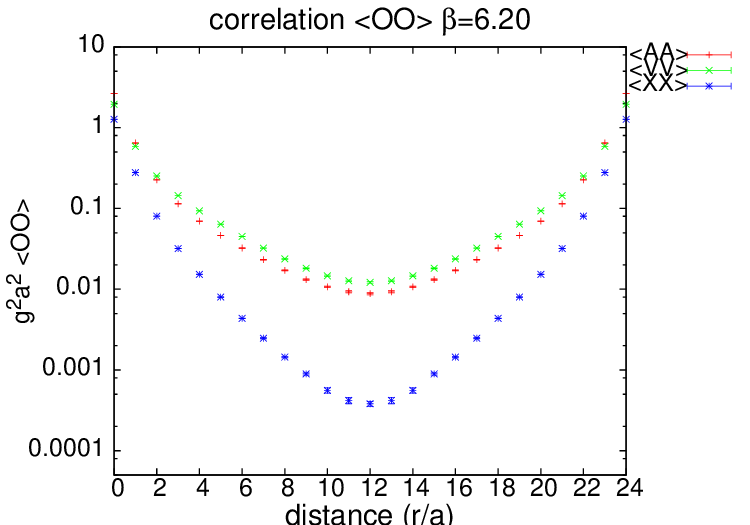}
\end{center}
\vspace{-0.5cm}
\caption{Ref.\cite{KSSK11}: 
Field correlators as functions of the distance $r:=|x|$ (from above to below)
$\langle  \mathscr{V}_\mu^A(x) \mathscr{V}_\mu^A(0) \rangle$, 
$\langle  \mathscr{A}_\mu^A(x) \mathscr{A}_\mu^A(0) \rangle$, 
and
$\langle  \mathscr{X}_\mu^A(x) \mathscr{X}_\mu^A(0) \rangle$.
}
\label{C35-fig:decomp-field-corr}
\end{figure}
%%%%%%%%%%%%%%%%%%%%% figures %%%%%%%%%%%%%%%%%%%%%%%%%%%

%To obtain correlation functions of field variables, we need to fix the gauge and we have adopted the Landau gauge for the original Yang-Mills field $\mathscr{A}$ so that the global color symmetry is not broken.  This property is desirable to study color confinement, but it is lost in the MA gauge. 

We study the 2-point correlation functions (propagators) of the new variables and the original Yang-Mills field variables, which are defined by
\begin{equation}
D_{OO}(x-y):=\left\langle O_{\mu}^{A}(x)O_{\mu}^{A}(y)\right\rangle \text{ for} \  O_{\mu}^{A}(x ) \in \{\mathbb{V}^{A}_{x^{\prime},\mu}, \mathbb{X}^{A}
_{x^{\prime},\mu},\mathbb{A}^{A}_{x^{\prime},\mu}\},
\end{equation}
where an operator $\mathbb{O}_{\mu}(x) =O_{\mu}^{A}(x) T_A$  is defined by the linear type, e.g.,
$\mathbb{A}_{x^{\prime},\mu}:=(U_{x,\mu}-U_{x,\mu}^{\dag})/(2ig\epsilon)$ where $x^{\prime}$ means the mid-point of $x$ and $x+\epsilon \hat\mu$.
In order to calculate the propagators, we must impose a gauge fixing condition, and we have adopted the lattice Landau gauge (LLG).

Fig. \ref{C35-fig:decomp-field-corr} shows the 2-point correlation functions of the new fields $\mathscr{V}$, $\mathscr{X}$, and the original fields $\mathscr{A}$.
This result indicates the \textbf{infrared  dominance  of restricted correlation functions} $\langle  \mathscr{V}_\mu^A(x) \mathscr{V}_\mu^A(0) \rangle$ in the sense that the  correlator of the variable $\mathscr{V}$ behaves just like the correlator  $\langle  \mathscr{A}_\mu^A(x) \mathscr{A}_\mu^A(0) \rangle$ of the original variable $\mathscr{A}$ and dominates  in the long distance, while the correlator $\langle  \mathscr{X}_\mu^A(x) \mathscr{X}_\mu^A(0) \rangle$ of $SU(3)/U(2)$ variable $\mathscr{X}$  decreases quickly in the distance $r$.

%%%%%%%%%%%%%%%%%%%%% figures %%%%%%%%%%%%%%%%%%%%%%%%%%%
\begin{figure}[ptb]
\begin{center}
%\vspace{-8mm}
\includegraphics[
scale=1.0
]
{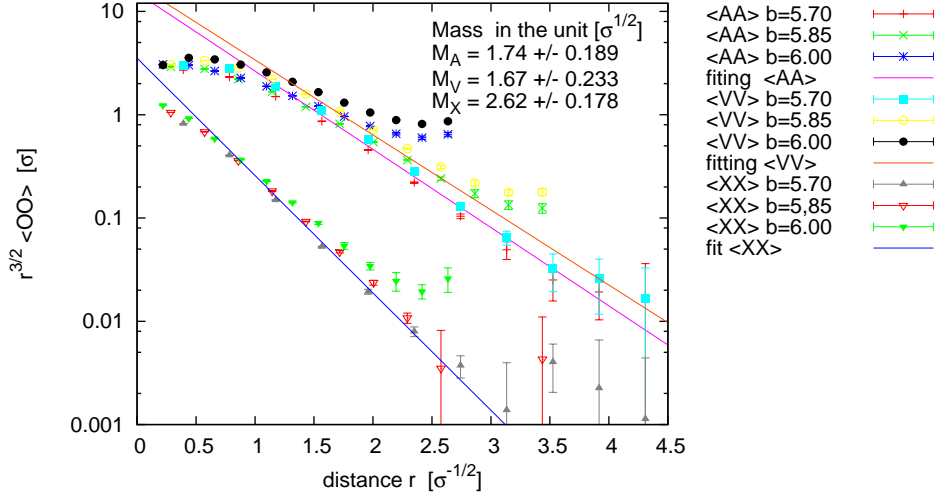} 
\end{center}
\vspace{-0.5cm}
\caption{ Ref.\cite{SKKS13}:
The rescaled correlation functions $r^{3/2}\left\langle
O(r)O(0)\right\rangle$ for $O=\mathbb{A,V,X}$ for $24^{4}$ lattice with $\beta=5.7$, $5.85$, $6.0.$ 
The physical scale is set in units of the string tension $\sigma_{\text{phys}}^{1/2}$. 
The correlation functions have the profile of cosh type because of the periodic boundary condition, and hence we use  data within distance of the half size of lattice.
} 
\label{C35-fig.propagator}%
\end{figure}
%%%%%%%%%%%%%%%%%%%%% figures %%%%%%%%%%%%%%%%%%%%%%%%%%%

For  $\mathscr{X}$,  at least, we can introduce a \textbf{gauge-invariant mass term}: 
\begin{equation}
\frac12 M_X^2 \mathscr{X}_\mu^A \mathscr{X}_\mu^A 
 ,
\end{equation} 
since $\mathscr{X}$ transforms like an adjoint matter field under the gauge transformation. 
In view of this fact, we fit the data of the contracted correlator $\langle  \mathscr{X}_\mu^A(x) \mathscr{X}_\mu^A(0) \rangle$ using the ``massive" propagator for large $r:=|x|$:
\begin{align}
 \langle  \mathscr{X}_\mu^A(x) \mathscr{X}_\mu^A(0) \rangle 
&=   \int \frac{d^4k}{(2\pi)^4} e^{ikx} \frac{3}{k^2+M_X^2} 
 \simeq  {\rm const.} \frac{e^{-M_X r}}{r^{3/2}} 
  .
\end{align}
In the similar way, we estimate the ``mass'' $M_{O}$ (i.e., the rate of exponential fall-off) from the propagator $D_{OO}(r)$ by using the Fourier transformation of the massive
propagator in the Euclidean space, which  behaves for large $M_{O}r$ as
\begin{equation}
D_{OO}(r)
=\left\langle O_{\mu}^{A}(x)O_{\mu}^{A}(y)\right\rangle
=\int\frac{d^{4}k}{(2\pi)^{4}}e^{ik(x-y)}\frac{3}{k^{2}+M_{O}^{2}%
}\simeq\frac{3\sqrt{M_{O}}}{2(2\pi)^{3/2}}\frac{e^{-M_{O}r}}{r^{3/2}}\text{
\ \ (}M_{O}r\gg1\text{)},
\end{equation}
and hence the scaled propagator $r^{3/2}D_{OO}(r)$ should be proportional to $\exp(-M_{O}r)$.

Fig.~\ref{C35-fig.propagator} shows the logarithmic plot of the scaled propagators $r^{3/2}D_{OO}(r)$ as a function of $r=|x-y|$, where the distance $r$ is drawn in units of the string tension $\sigma_{\text{phys}}$, and data of lattice spacing is taken from the Table I in Ref.\cite{Edward98}. 
The propagator $D_{VV}$ fall off slowly and has almost the same fall-off behavior as $D_{AA}$, while the $D_{XX}$ falls off quickly. Thus, from the viewpoint of the propagator, the $V$-field plays the dominant role in the deep infrared region or the long distance, while $X$-field is negligible in the long distance.
The rapid disappearance of $X$ contribution in the long distance is helpful to understand the difference of the profile of the flux tube in Fig.\ref{C35-fig:fluxtube}. 
In order to perform the parameter fitting of $M_{O}$ for $O=\{\mathbf{V}_{x^{\prime},\mu},\mathbf{A}_{x^{\prime},\mu}\}$, 
we use data in the region
$[2.0,4.5]$ and exclude the data near the midpoint of the lattice to eliminate the finite volume effect, while for $O=\mathbf{X}_{x^{\prime},\mu}$ we use the region $[1.0,3.5].$ 

Then the naively estimated ``mass" $M_X$ of $\mathscr{X}$ is 
\begin{equation}
 M_X = 2.409 \sqrt{\sigma_{\rm phys}} = 1.1 {\rm GeV} 
  .
\end{equation} 
%This value should be compared with the result in MA gauge \cite{SAIIMT02}. 
%The monopole potential is linear starting from one lattice spacing without the Coulomb type piece.  
We use $\sigma_{\text{phys}}=(440MeV)^2$ to obtain preliminary result:
\begin{equation}
M_{A}\simeq 0.76  \text{ GeV, \ \ \ }M_{V}\simeq 0.73\text{ GeV , \ \ \ }M_{X}%
\simeq 1.15 \text{ GeV,}
\end{equation}
which should be compared with result of the maximal option \cite{Shibata-lattice2007} in LLG, 
and also result of the Abelian projection in the MA gauge \cite{SAIIMT02,GIS12,GS13}.

For more preliminary results of numerical simulations, see \cite{Shibata-lattice2009} for magnetic monopoles of $SU(2)$,  \cite{Shibata-lattice2007} for the maximal option of $SU(3)$ and \cite{Shibata-lattice2010,Shibata-lattice2008} for the minimal one of $SU(3)$.

%%%%%%%%%%%%%%%%%%%%%%%%%%%%%%%%%%%%%%%%%%%%%%%%%%
\subsubsection{Numerical simulations in the maximal option of $SU(3)$}
%%%%%%%%%%%%%%%%%%%%%%%%%%%%%%%%%%%%%%%%%%%%%%%%%%

%%%%%%%%%%%%%%%%%%%%%%%%%%%%%%%%%%%%%%%%%%%%%%%%%%
\begin{figure}[ptb]
\begin{center}
%\includegraphics[
%bb=116 116 452 721 , height=3in, angle=-90 , origin=br
%]{symmetry.eps}
\includegraphics[
scale=0.6,
]{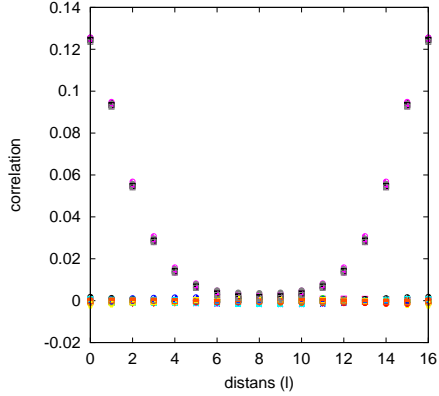}
\end{center}
\vspace{-0.5cm}
\caption{
%(Left panel) The gauge symmetry of the new formulation.
%(Right panel)
Ref.\cite{Shibata-lattice2007}: 
The correlation functions 
$\langle n_{x}^{A}n_{y}^{B} \rangle$  of the color field $\mathbf{n}_{x}$ are plotted for $A<B$.
}%
\label{C35-fig:fig1}%
\end{figure}
%%%%%%%%%%%%%%%%%%%%%%%%%%%%%%%%%%%%%%%%%%%%%%%%%%

For the $SU(3)$ Yang-Mills theory in the maximal option, we have the following results of the numerical simulations \cite{Shibata-lattice2007}. 
Numerical simulations are performed using the standard Wilson action of the $SU(3)$ Yang-Mills theory.
The configurations are generated on a $16^{4}$ lattice at $\beta=5.7$ using the Cabibbo-Marinari heat bath algorithm. After 5000 thermalizing sweeps with the cold start, 120 configurations are stored every 100 sweeps. We choose the lattice Landau gauge (LLG) for the overall gauge fixing of the original Yang-Mills theory. In gauge fixing procedure, we use the over-relaxation algorithm to update link variables by using the gauge transformation of $SU(2)$ sub-groups in the $SU(3)$ gauge transformation. In order to avoid the lattice Gribov copy problem in  both LLG and reduction condition, we try to find out the configuration which absolutely minimizes the gauge fixing functional. In the process of minimizing the gauge fixing functional for $U_{x,\mu}$, we have prepared 16 replicas generated by random gauge transformations from $U_{x,\mu}$, and among them we have selected the configurations which have attained the least value of the functional.

First, we check the color symmetry in our new formulation, which is a global $SU(3)$ symmetry to be preserved in LLG. Under the global gauge transformation, the gauge fixing functional of LLG, 
$F_{\rm LLG}[g]=%
%TCIMACRO{\tsum \nolimits_{x,\mu}}%
%BeginExpansion
{\textstyle\sum\nolimits_{x,\mu}}
%EndExpansion
\mathrm{Tr}({}^{g}U_{x,\mu})$, 
is invariant, while the two color fields
$\bm{n}_{x}$,$\bm{m}_{x}$ change their directions. Therefore, we measure the space--time average of the color   fields 
$n_{x}^{A}=\mathrm{Tr(} \mathbb{\lambda}^{A}\bm{n}_{x}\mathrm{)}$ and $m_{x}^{A}=\mathrm{Tr(}%
\mathbb{\lambda}^{A}\bm{m}_{x}\mathrm{)}$, and their  correlation functions.
Fig.~\ref{C35-fig:fig1} shows, for examples, the correlation functions of $\bm{n}_{x}$. The lattice data show that the color symmetry is preserved: 
\begin{align}
 & \left\langle n^{A}\right\rangle =0 ,  \quad \left\langle m^{A}\right\rangle =0 ,
\nonumber\\
 & \left\langle n_{x}^{A} n_{y}^{B}\right\rangle =\delta^{AB}D_{NN}(l), 
\quad
\left\langle m_{x}^{A} m_{y}^{B}\right\rangle =\delta^{AB}D_{MM}(l), 
\nonumber\\
 & \left\langle n_{x}^{A} m_{y}^{B}\right\rangle =0 
\quad
(y=x+l\hat{\mu}, \mu=4).
\end{align}
The fact that the color symmetry is unbroken is an advantage of our new formulation.

%%%%%%%%%%%%%%%%%%%%% figures %%%%%%%%%%%%%%%%%%%%%%%%%%%
\begin{figure}[ptb]
\begin{center}
\includegraphics[
scale=0.3,
angle=-90
]{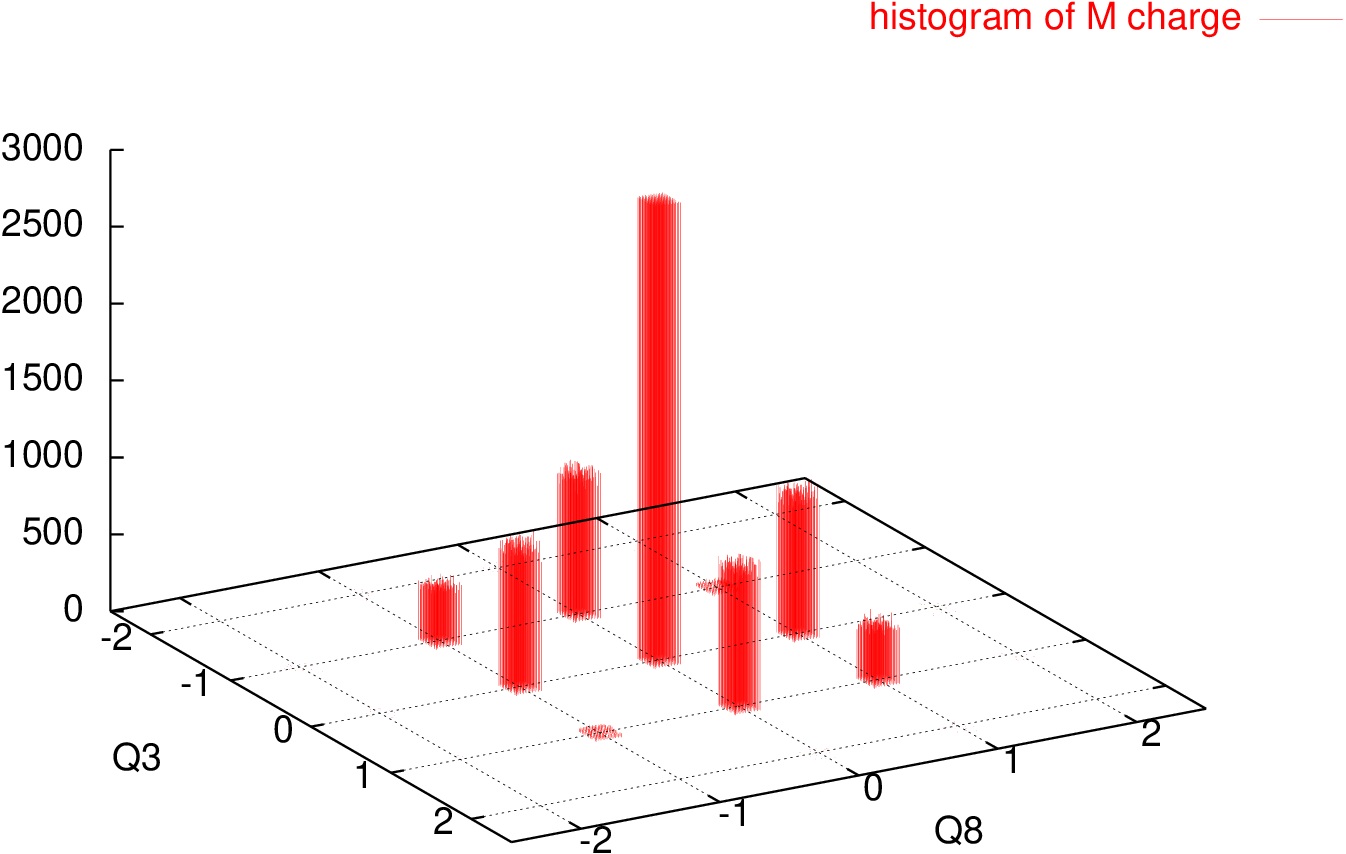}
\includegraphics[
scale=0.3,
angle=-90
]{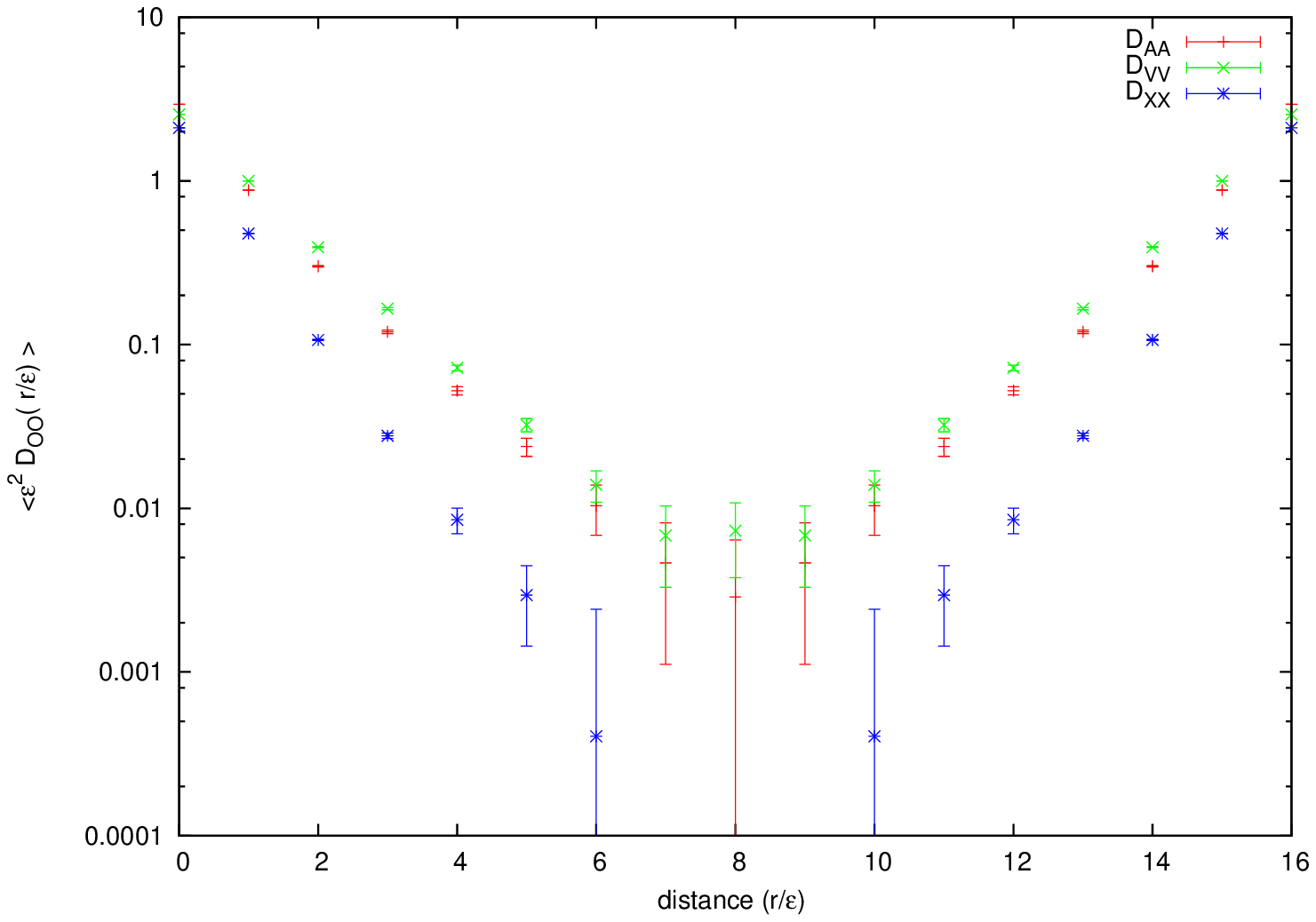}
\end{center}
\vspace{-0.5cm}
\caption{
Ref.\cite{Shibata-lattice2007}:
(Left Panel) 
The histogram of monopole charges $Q^{(1)}$ and $Q^{(2)}$. The number of the monopoles are plotted for 120 configurations. Each distribution with integer valued monopole charges is represented on the grids. 
This is a preliminary result. 
(Right panel) 
The logarithm plot of correlation functions of the new field variables $V,X$ and the original Yang-Mills field $A$:  $D_{VV}$, $D_{XX}$ and $D_{AA}$.
}
\label{C35-fig:fig20}%
\end{figure}
%%%%%%%%%%%%%%%%%%%%% figures %%%%%%%%%%%%%%%%%%%%%%%%%%%

Next, we define a gauge-invariant magnetic monopole using the ``Abelian'' part $V_{x,\mu}$ in the similar way to the $SU(2)$ case. Two kinds of the gauge invariant magnetic-monopole currents  are defined by 
\begin{align}
k_{\mu}^{(a)}  &  :=\frac{1}{2}\epsilon_{\mu\nu\alpha\beta}\partial^{\nu}\Theta_{\alpha\beta}^{(a)} 
\quad ( a=1,2 ),
\nonumber\\
\Theta_{\alpha\beta}^{(1)}  &  :=\arg\mathrm{tr}\left[  \left(  \frac{1}{3}\bm{1}+\bm{n}_{x}+\frac{1}{\sqrt{3}}\bm{m}_{x}\right)  V_{x,\alpha }V_{x+\alpha,\beta}V_{x+\beta,\alpha}^{-1}V_{x,\beta}^{-1}\right]  , 
\nonumber\\
\Theta_{\alpha\beta}^{(2)}  &  :=\arg\mathrm{tr}\left[  \left(  \frac{1}{3}\bm{1}-\frac{2}{\sqrt{3}}\bm{m}_{x}\right)  V_{x,\alpha}V_{x+\alpha,\beta}V_{x+\beta,\mu}^{-1}V_{x,\beta}^{-1}\right]  .
\end{align}
The gauge invariance of $\Theta_{\alpha\beta}^{(a)\text{ }}$ is obvious by definition. 
Note that $\Theta_{\alpha\beta}^{(a)}$ is the $a$-th element of the diagonalized expression of $V_{x,\mu}V_{x+\mu,\nu}V_{x+\nu,\mu}^{-1}V_{x,\nu}^{-1},$ i.e., $\mathrm{diag}( \exp(ig^{2}\epsilon^{2}\Theta_{\mu\nu}^{1}), \exp(ig^{2}\epsilon^{2}\Theta_{\mu\nu}^{2}),
\exp(ig^{2}\epsilon^{2}\Theta_{\mu\nu}^{3}))$. 
The left panel of Fig.~\ref{C35-fig:fig20} shows the histogram of the magnetic monopole charges, indicating that magnetic monopoles with the integer-valued magnetic charge are obtained by this construction.

Finally, we investigate the propagators of the new variables. The correlation functions (propagators) of the original Yang-Mills field and the new variables are  defined by
\begin{equation}
D_{OO}(x-y):=\left\langle O_{\mu}^{A}(x)O_{\mu}%
^{A}(y)\right\rangle \,\text{\ for }O_{\mu}^{A}(x^{\prime})=\mathbb{A}%
_{x^{\prime},\mu}^{A},\mathbb{V}_{x^{\prime},\mu}^{A}\text{, }\mathbb{X}%
_{x^{\prime},\mu}^{A}, \label{eq:corrF}%
\end{equation}
where an operator $\mathbb{O}_{\mu}(x) =O_{\mu}^{A}(x) T_A$ is defined as the linear type, e.g.,
$\mathbb{A}_{x^{\prime},\mu}=\left(  U_{x,\mu} - U_{x,\mu}^{\dag}\right)/\left( 2i\varepsilon g \right)$. 
The right panel of Fig.~\ref{C35-fig:fig20} shows preliminary measurements of correlation functions of $D_{AA},$ $D_{VV}$ and $D_{XX}$. The correlator $D_{VV}$ of the restricted field corresponding to the ``Abelian'' part falls off slowly and has almost the same fall-off rate as $D_{AA}$, while the correlator  $D_{XX}$ of the remaining field corresponding to the ``off-diagonal'' part falls off quickly. This suggests that the restricted part of the gluon propagator is dominated in the infrared region,  and the mass generation of the ``off-diagonal'' gluon.

%%%%%%%%%%%%%%%%%%%%% figures %%%%%%%%%%%%%%%%%%%%%%%%%%%
\begin{figure}[ptb]
\begin{center}
\includegraphics[
scale=0.3, angle=-90
]{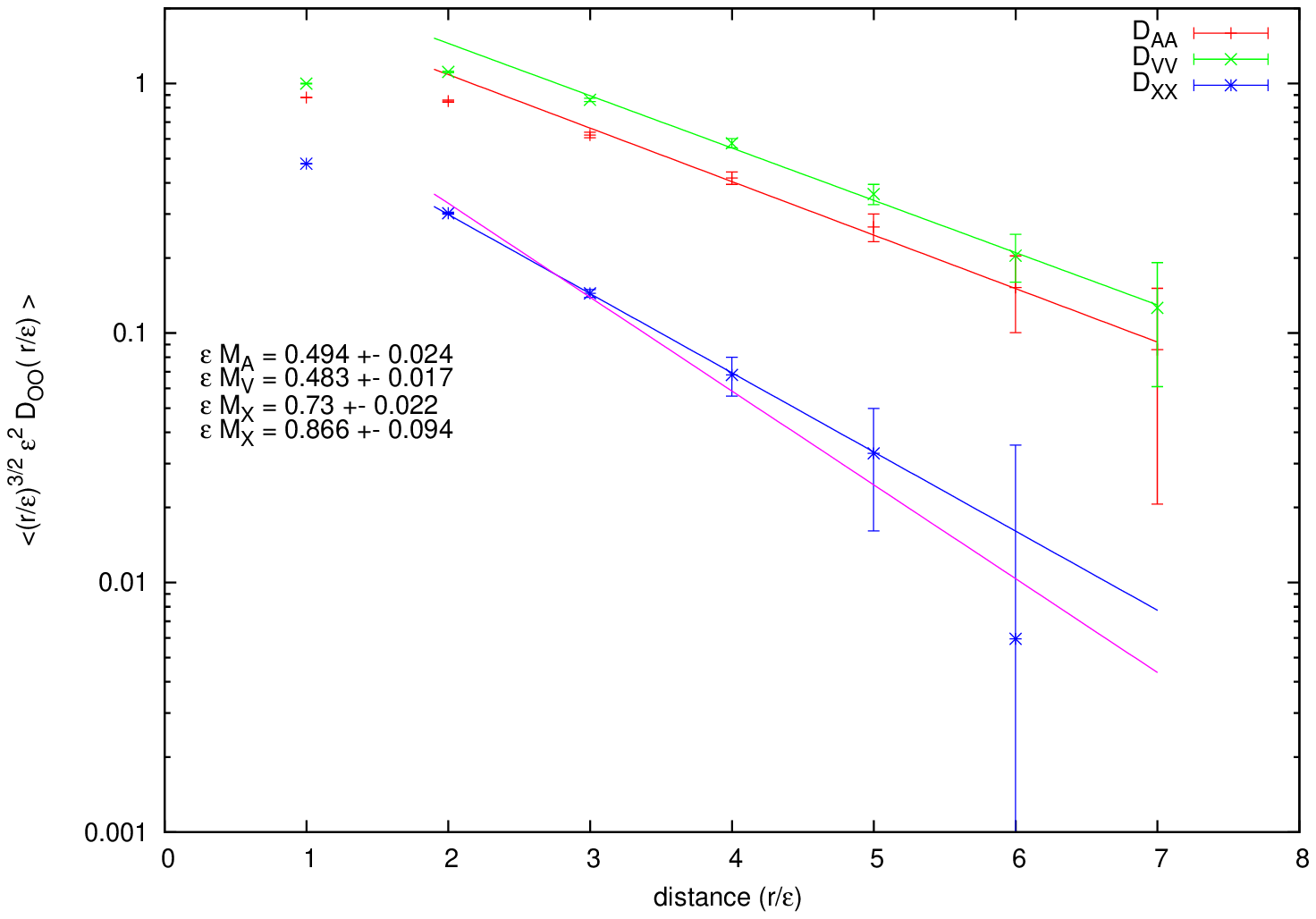}
%\vskip -0.5cm
\caption{ %\cite{Shibata-lattice2007}%
The rescaled correlation functions $r^{3/2}D_{AA}(r)$,  $r^{3/2}D_{VV}(r)$ and $r^{3/2}D_{XX}(r)$. The green line represents the fitted function for $M_A$ and the red line for $M_V$ by using data in the region $[3,7]$.
The blue and magenta lines represent  the fittings for $M_X$ by using the data in the regions $[2,5]$ and $[2,6]$.
}
\label{C35-fig:fig21}%
\end{center}
\end{figure}
%%%%%%%%%%%%%%%%%%%%% figures %%%%%%%%%%%%%%%%%%%%%%%%%%%

Fig.~\ref{C35-fig:fig21} shows the logarithmic plot of the scaled propagators $r^{3/2}D_{OO}(r)$ as a function of $r=|x-y|$.
The propagator $D_{VV}$ falls off slowly and has almost the same fall-off behavior as $D_{AA}$, while the $D_{XX}$ falls off quickly. Thus, from the viewpoint of the propagator, the $V$ field plays the dominant role in the deep infrared region or the long distance, while $X$ field is negligible in the long distance.

In order to perform the parameter fitting of $M_{O}$ for $O=\{\mathbb{V}_{x^{\prime},\mu},\mathbb{A}_{x^{\prime},\mu}\}$, 
we use the data in the fitting range $r/\epsilon \in [3.0,7.0]$  and exclude the data near the midpoint of the lattice to eliminate the finite volume effect, while for $O=\mathbb{X}_{x^{\prime},\mu}$ we use the two regions $[2,5]$ and $[2, 6]$. 
The estimated masses are given by
\begin{align}
M_{A}  & =\sqrt{\sigma_{\rm phys}}\left(  1.2735\pm0.062\right)  =560\pm27\text{\ (MeV),}
\\
M_{V}  & =\sqrt{\sigma_{\rm phys}}(1.245\pm0.043)=547\pm19\text{ \ (MeV)} ,  
\end{align}
and %we obtain  $M_{X}$  as 
\begin{align}
M_{X}  & =\sqrt{\sigma_{phys}}\left(  1.882\pm0.057\right)  =828\pm25\text{
\ (MeV) \ \ for }r/\epsilon\in\lbrack2,5] ,
\nonumber\\
M_{X}  & =\sqrt{\sigma_{phys}}\left(  2.233\pm0.24\right)  =982\pm106\text{
\ (MeV) \ \ for }r/\epsilon\in\lbrack2,6] , 
\end{align}
where we have used 
%the fitting range $\lbrack3,7]$, and 
$\sqrt{\sigma_{\rm phys}}\epsilon
(\beta=5.7)=0.3879$ with $\sigma_{\rm phys}=(440$MeV$)^{2}$. 
Therefore, we obtain
\begin{align}
M_{X}=800\text{\ }\sim\text{ }1000\text{ MeV}. 
\end{align}
In order to determine the physical scale, we have used the relationship between the (inverse) gauge coupling $\beta$ and lattice spacing $\epsilon$ summarized in Table~\ref{C35-table:lattice-size-SU3}. %which is  given in \cite{KKNS98}.% 

%%%%%%%%%%%%%%%%%%%%% Table %%%%%%%%%%%%%%%%%%%%%%%%%%%
\begin{table}[tb]
\vskip -5mm
\caption{ 
Selected world results for the string tension $\sqrt{\sigma}$ and the Sommer scale
$r_{0}$.
}
\begin{center}%
\begin{tabular}
[c]{|c|c|c|c|}\hline
$\beta$ & $\epsilon\sqrt{\sigma}$ & $r_{0}/\epsilon$ & Volume\\\hline\hline
$5.54$ & $0.5727(52)$ & $2.054(13)$ & $12^{4}$\\\hline
$5.6$ & $0.5064(28)$ & $2.344(8)$ & $12^{4}$\\\hline
$5.7$ & $0.3879(39)$ & $2.990(24)$ & $16^{3}.32$\\\hline
$5.85$ & $0.2874(7)$ & $4.103(27)$ & $16^{3}\cdot32$\\\hline
$6.0$ & $0.1289(9)$ & $5.369(9)$ & $16^{3}\cdot32$\\\hline
& $0.2182(16)$ &  & $16^{4}$\\\hline
& $0.21824(19)$ & $5.35(+2)(-3)$ & $16^{4}$\\\hline
& $0.2209(23)$ &  & $32^{4}$\\\hline
& $0.2154(50)$ & $5.47(11)$ & $16^{3}\cdot48$\\\hline
$6.2$ & $0.160(9)$ &  & $32^{4}$\\\hline
& $0.1604(11)$ & $7.73(3)$ & $32^{4}$\\\hline
& $0.1608(23)$ & $7.29(17)$ & $24^{3}\cdot48$\\\hline
$6.4$ & $0.1214(12)$ & $9.89(16)$ & $32^{4}$\\\hline
& $0.1218(28)$ & $9.75(17)$ & $32^{3}\cdot64$\\\hline
$6.5$ & $0.1068(9)$ & $11.23(21)$ & $36^{4}$\\\hline
\end{tabular}
\end{center}
\label{C35-table:lattice-size-SU3}
\end{table}
%\chASS{}
%%%%%%%%%%%%%%%%%%%%%%%%%%%%%%%%%%%%%%%%%%%%%%%%%%%%%

The numerical investigation of the $SU(3)$ Yang-Mills theory in the maximal option is under way.
Quite recently, independent numerical simulations are performed by Cho and his collaborators \cite{CCLL14}.

\newpage
%%%%%%%%%%%%%%%%%%%%%%%%%%%%%%%%%%%%%%%%%%%%%%%%%%%%%%%%%%%%
%%%%%%%%%%%%%%%%%%%%%%%%%%%%%%%%%%%%%%%%%%%%%%%%%%%%%%%%%%%%
\section{Conclusions and remarks}\label{sec:conclrem}
%%%%%%%%%%%%%%%%%%%%%%%%%%%%%%%%%%%%%%%%%%%%%%%%%%%%%%%%%%%%
%%%%%%%%%%%%%%%%%%%%%%%%%%%%%%%%%%%%%%%%%%%%%%%%%%%%%%%%%%%%

%%%%%%%%%%%%%%%%%%%%%%%%%%%%%%%%%%%%%%%%%%%%%%%%%%%%%%%%%%%%
%%%%%%%%%%%%%%%%%%%%%%%%%%%%%%%%%%%%%%%%%%%%%%%%%%%%%%%%%%%%
%\subsection{Conclusions}\label{sec:conclusion} 
%%%%%%%%%%%%%%%%%%%%%%%%%%%%%%%%%%%%%%%%%%%%%%%%%%%%%%%%%%%%
%%%%%%%%%%%%%%%%%%%%%%%%%%%%%%%%%%%%%%%%%%%%%%%%%%%%%%%%%%%%

%%%%%%%%%%%%%%%%%%%%%%%%%%%%%%%%%%%%%%%%%%%%%%%%%%%%%%%%%%%%
%%%%%%%%%%%%%%%%%%%%%%%%%%%%%%%%%%%%%%%%%%%%%%%%%%%%%%%%%%%%
\subsection{Main points}\label{sec:main-results} 
%%%%%%%%%%%%%%%%%%%%%%%%%%%%%%%%%%%%%%%%%%%%%%%%%%%%%%%%%%%%
%%%%%%%%%%%%%%%%%%%%%%%%%%%%%%%%%%%%%%%%%%%%%%%%%%%%%%%%%%%%

The main technical tools presented in this review are the \textbf{novel reformulations of the Yang-Mills theory} (sections 4,5) and the \textbf{non-Abelian Stokes theorem for the Wilson loop operator} (section 6). 
These two subjects can be treated independently to each other. 
In fact, they have been developed independently:  The original field decomposition called the \textbf{CDGFN decomposition} (section 3) which is needed to present the subsequently developed reformulations was invented by Cho \cite{Cho80} and Duan and Ge \cite{DG79} independently, readdressed by Faddeev and Niemi \cite{FN98}, developed by Shabanov \cite{Shabanov99} and furthermore extended by Kondo, Shinohara and Murakami \cite{KMS06,KMS05,Kondo06,KSM08}.
While the version of the non-Abelian Stokes theorem presented in this paper is invented originally by Diakonov and Petrov \cite{DP89} and developed and extended by Kondo and his collaborator \cite{KondoIV,KT00,KT00b,Kondo99Lattice99,Kondo08}.

The combined use of these two tools exhibit their  full power when they are applied to the the problem of quark confinement in the light of the \textbf{Wilson criterion} (section 9), although the reformulation can be applied to other physical problems (section 7). 
In fact, the CHIBA/KEK/Fukui  collaborations with Shibata and Kato have succeeded to  construct the lattice versions  \cite{KKMSSI06,IKKMSS06,KSSMKI08,SKS10} of the reformulated Yang-Mills theory and enabled us to perform the numerical simulations on the lattice \cite{KKMSS05,KKMSSI06,IKKMSS06,SKKMSI07,KSSMKI08,SKS10,KSSK11,SKKS13}, which demonstrates the validity of the new framework equipped with these tools (section 9). 

In this review we stand on the \textbf{dual superconductor picture for quark confinement}, which is supposed to be caused by the condensation of \textbf{(chromo) magnetic monopoles} and is characterized or detected  by the existence of the \textbf{dual Meissner effect}. 
However, for this mechanism to be regarded as the really physical mechanism, we need to establish these phenomena in the gauge-invariant or gauge-independent way. 
This review summarized recent endeavors done mainly by our group toward the goal of understanding quark confinement from the first principle in the gauge-independent way. 

In what follows, we emphasize again the main achievements done in these investigations.

 Even in the pure $SU(N)$ Yang-Mills theory without adjoint scalar fields,  
we have given a definition of a \textbf{gauge-invariant magnetic monopole} by making use of a non-Abelian Stokes theorem for the Wilson loop operator.
The magnetic monopole is guaranteed to be gauge invariant from the beginning by construction. 
This gauge-invariant magnetic monopole is inherent in the Wilson loop operator, and it is detected by the Wilson loop.  
Therefore, we do not need to use the \textbf{Abelian projection} which is used to define the magnetic monopole in the gauge-dependent way (explicit symmetry breaking).  
This suggests that magnetic monopole may be responsible for quark confinement, supporting the dual superconductor picture.

 We have shown that the $SU(N)$ Yang-Mills theory can be reformulated in a number of ways using a different set of new field variables which are obtained by \textbf{change of variables} from the original Yang-Mills field.
Each reformulation is discriminated by the \textbf{maximal stability subgroup} $\tilde H$ of the original gauge group $G=SU(N)$. 
The idea of the field decomposition or  new field variables is originally attributed to Cho, and Faddeev and Niemi. 
In their pioneering works for the $SU(N)$ Yang-Mills theory, $N-1$ color fields $\bm{n}_{j}$ ($j=1,...,N-1$) are introduced in agreement with the number of the Cartan generators, i.e., the rank $N-1$, which corresponds to the maximal stability subgroup $H=U(1)^{N-1}$.
However, our reformulation in the \textbf{minimal option} constructed for the maximal stability subgroup $H=U(N-1)$ is new for $SU(N), N \ge 3$: we introduce  {only a single color field $\bm{n}$ for any $N$}.
We have shown that the minimal option is enough and optimal for reformulating the quantum Yang-Mills theory to describe confinement of  quarks in the fundamental representation.

The new insight obtained by combining two tools is that 
 the optimal description of the gauge-invariant magnetic monopole derived from the Wilson loop operator is described  by the relevant option in the new reformulations. 
Here the relevant maximal stability group to be adopted in this stage is determined by giving the (color group) representation of the source, for which the Wilson loop operator is defined. 
Here the crucial role is played by the \textbf{color direction field} $\bm{n}(x)$ taking the value in the target space $G/\tilde H$ determined by the maximal stability group $\tilde H$:
\begin{align}
 \bm{n}(x) \in G/\tilde H .
\end{align}
In particular, if we restrict the source to \textbf{quarks in the fundamental representation} of $G=SU(N)$, the maximal stability group is given by $\tilde H=U(N-1)$. 
Thus, the magnetic monopole in the $SU(N)$ Yang-Mills theory is described by the color direction field $\bm{n}$ with the target space $G/\tilde H$ which is equal to the complex projective space:
\begin{align}
 \bm{n} \in   SU(N)/U(N-1) \simeq \mathbb{C}P^n  .
\end{align}
This means that  the magnetic monopole in the $SU(2)$ Yang-Mills theory is the \textbf{Abelian magnetic monopole} described by the color field with the target space $SU(2)/U(1)$:
\begin{align}
 G=SU(2) \Longrightarrow \bm{n} \in   SU(2)/U(1) \simeq \mathbb{C}P^1 ,
\end{align}
while the magnetic monopole in the $SU(3)$ Yang-Mills theory is the \textbf{non-Abelian magnetic monopole} described by the color field with the target space $SU(3)/U(2)$:
\begin{align}
G=SU(3)  \Longrightarrow  \bm{n} \in   SU(3)/U(2) \simeq \mathbb{C}P^2 .
\end{align}
For $N \ge 3$, these results based on the gauge invariant magnetic monopole are different from the picture based on the Abelian projection suggesting the \textbf{Abelian magnetic monopoles as the gauge-fixing defect} associated to the partial gauge fixing (explicit breaking of the gauge symmetry) from the original color group $G$ to the maximal torus subgroup $H$:
\begin{align}
  G=SU(N) \to H=U(1)^{N-1}.
\end{align}
which yields the coset space of the Flag space:
\begin{align}
  G/H = SU(N)/U(1)^{N-1} \simeq F_{N-1}.
\end{align}
The reformulation allows a number of options discriminated by the maximal stability group $\tilde{H}$ of the gauge group $G$. 

%\noindent
%{  
% We do not need  the Abelian projection  to define   magnetic monopoles in Yang-Mills theory!
%}

%$\bullet$  We have reformulated the SU(N) Yang-Mills theory in terms of new field variables obtained by change of variables. 

For $G=SU(3)$, in particular, only two options are possible: 
%minimal one with $\tilde{H}$ $=U(2)$ and maximal one with $\tilde{H}=H=   U(1) \times U(1)$.
\begin{itemize}

\item
%\noindent
The \textbf{minimal option} with $\tilde{H}=U(2)$ gives an optimized description of quark confinement through the Wilson loop in the fundamental representation. 

\item
%\noindent
 The \textbf{maximal option} with $\tilde{H}=H=U(1) \times U(1)$ gives  a manifestly gauge-independent reformulation of the conventional Abelian projection in the MA gauge.
\end{itemize}

Moreover, we have constructed the lattice versions of the reformulations of the $SU(N)$ Yang-Mills theory and performed numerical simulations on a lattice for $SU(2)$ and $SU(3)$ cases.
The results are summarized below. 
 
%\noindent
%3) For $SU(3)$, we have confirmed the {infrared dominance of the restricted field} $\mathscr{V}$ and {the non-Abelian magnetic monopole dominance} for quark confinemen in the string tension. This should be compared with the infrared Abelian dominance and Abelian magnetic monopole dominance in the MA gauge. 

%\noindent
%4)  We have presented the numerical evidence of the {dual Meissner effect caused by non-Abelian magnetic monopoles} in $SU(3)$ Yang-Mills theory:  the tube-shaped flux of the chromo-electric field originating from the restricted field and the associated magnetic-monopole current  induced around the flux tube originating from the non-Abelian magnetic monopoles condensations.  To confirm which type of the {non-Abelian dual superconductivity occurs in $SU(3)$ Yang-Mills theory}, we have done further checks, e.g., measurement of the penetrating depth and the determination of the type of dual superconductor. 

In the non-Abelian gauge theory, we can decompose the Yang-Mills field $\mathscr{A}_\mu(x)$ into two pieces:
\begin{equation}
 \mathscr{A}_\mu(x) = \mathscr{V}_\mu(x) + \mathscr{X}_\mu(x) ,
\end{equation}
in the gauge-covariant way, which is characterized by the following properties. 
\begin{itemize}
\item 
{[The infrared dominance of the restricted field $\mathscr{V}_\mu$ (infrared ``Abelian''  dominance)]}

The restricted fields $\mathscr{V}_\mu$ play  the dominant role in the infrared regime which is responsible for quark confinement. 
The Wilson loop average for the restricted field can reproduce the original string tension (\textbf{infrared restricted field dominance}). 
It can be extracted from the original theory by using a \textbf{(nonlinear) change of variables} in the \textit{gauge-independent way}  without breaking the color symmetry. 
Moreover, the restricted field can be used to define the gauge-invariant magnetic monopoles with the magnetic charge which is subject to the  quantization  condition.
%The magnetic monopole dominance in the string tension was confirmed by numerical method. 
%This is interpreted as a gauge-independnet ``Abelian'' projection, by identifying the restricted field with the  the \textit{``Abelian'' part}, i.e., diagonal gluon field in the Abelian projection, 

\item  
{[The infrared suppression of the remaining field $\mathscr{X}_\mu$]}

The remaining fields $\mathscr{X}_\mu$ are suppressed  (or decouple) in the infrared regime. 
The remaining fields $\mathscr{X}_\mu$ decouple in the low-energy regime: The correlation function of the remaining field $\mathscr{X}_\mu$ falls off exponentially in the distance, indicating that the restricted fields  acquire their  mass dynamically without breaking the gauge symmetry (\textbf{dynamical mass generation}).  
The infrared suppression of the remaining field $\mathscr{X}_\mu$ leads to the infrared dominance of the restricted field $\mathscr{V}_\mu$. 
For instance, such a dynamical mass can originate from  the existence of dimension-2 vacuum condensate, e.g., $\left< \mathscr{X}_\mu^2 \right>\ne 0$. 

\item  
{[The remaining field $\mathscr{X}_\mu$ as high-energy modes and the renormalization effect]}

In view of this, the remaining field $\mathscr{X}_\mu$ can be identified with the \textbf{high-energy modes} to be integrated out to obtain the low-energy effective theory which is written in terms of (the low-energy modes of) the restricted field.
Consequently, the remaining field $\mathscr{X}_\mu$ plays the important role of renormalizing the low-energy effective theory.  In fact, it has been shown in \cite{Gies01} that the \textbf{beta function} up to one loop is correctly reproduced by using the Wilson renormalization group:%
\footnote{
In order to obtain the correct beta function, we must simultaneously integrate out the high-energy modes of the restricted field, since the restricted field must have the high-energy modes in addition to the low-energy modes. 
On the other hand, it is supposed that the remaining field has only the high-energy modes without the low-energy modes. 
However, the integration of the high-energy modes of the color direction field (included in the restricted field) is quite hard to be performed, since the color direction field has the fixed length as a non-linear sigma model. 
In the MA gauge, this complication does not occur, since the color direction field disappears. 
Moreover, the off-diagonal gluons can be identified with only the high-energy modes to be integrated out, at least one-loop level. 
}  
\begin{equation}
 \beta(g) :=  \mu \frac{\partial g(\mu)}{\partial \mu} = - \frac{\frac{11N}{3}}{(4\pi)^2} g^3(\mu) .
\end{equation}
This check is a first step toward the quantum equivalence between two theories, i.e., the original Yang-Mills theory and the reformulated Yang-Mills theory. 
In the MA gauge, it is explicitly shown that the integration of the off-diagonal gluons forces the gauge coupling constant of the resulting (low-energy effective)  theory written in terms of the diagonal gluon (in the Abelian-projected theory) to run according to the beta function which agrees with that of the original Yang-Mills theory, as shown in \cite{KondoI}.

\end{itemize}

%\newpage
%%%%%%%%%%%%%%%%%%%%%%%%%%%%%%%%%%%%%%%%%%%%%%%%%%%%%
\subsection{Lattice $SU(2)$ Yang-Mills theory and numerical simulations}
%%%%%%%%%%%%%%%%%%%%%%%%%%%%%%%%%%%%%%%%%%%%%%%%%%%%%

%%%%%%%%%%%%%%%%%%%%%%%%%%%%%%%%%%%%%%%%%%%%%%%%%%
%S. Kato et al., 
%Preprint, CHIBA-EP-155/KEK Preprint 2005-61,
%[hep-lat/0509069], 
%Phys. Lett. B{\bf 632}, 326%--332  (2006).
%%%%%%%%%%%%%%%%%%%%%%%%%%%%%%%%%%%%%%%%%%%%%%%%%%

On a lattice, we have discussed how to implement the CDGFN decomposition (change of variables) of the Yang-Mills field, according to a new viewpoint proposed in \cite{KMS05} for the $SU(2)$ Yang-Mills theory.
% and in \cite{KSM08} for the $SU(N)$ Yang-Mills theory.  
In fact, we have presented the first implementation  \cite{KKMSSI06} of the  CDG or CDGFN decomposition of the $SU(2)$ Yang-Mills field on a lattice. 
A remarkable point is that our approach can retain both the local $SU(2)$ gauge symmetry and the color symmetry (global $SU(2)$ symmetry).
In fact, we have succeeded to perform the numerical simulations in such a way that the color symmetry is unbroken, as explicitly demonstrated by the results of the numerical simulations  \cite{KKMSSI06}. 

%even after imposing a constraint called the reduction condition which is regarded as a new type of gauge fixing imposed to reproduce the original Yang-Mills theory from the enlarged gauge theory with the larger gauge symmetry due to the introduction of the color direction field. 

%We have presented the first implementation of the  CDG or CDGFN decomposition of the $SU(2)$ Yang-Mills field on a lattice. 
%Our construction retains the color symmetry (global $SU(2)$ gauge invariance), %even after a new type of Maximally Abelian gauge, 
%as explicitly demonstrated by numerical simulations. 
%Moreover, we have proposed a gauge-invariant definition of the magnetic monopole (CDGFN monopole) using this formulation and compare the new definition with the conventional  DeGrand and Toussaint (DT) monopole to exhibit its validity. 

%We have given a new definition of the magnetic monopole-current by using the new variables on a lattice. 
%An advantage of our definition of magnetic monopole is that it is  $SU(2)$ gauge invariant. 
We have given a new definition of the magnetic monopole which we call the \textbf{CDGFN monopole}, using the new variables on a lattice.
An advantage of the new magnetic monopole is that it is   gauge invariant for the original gauge group. 
In order to exhibit its validity, we have compared the CDGFN monopole with the conventional  \textbf{DeGrand and Toussaint (DT) monopole} which is originally constructed in compact QED and then used in the Abelian projected Yang-Mills theory on the lattice. 
The numerical simulations show \cite{KKMSSI06} that the CDGFN monopole gives nearly the same value as given by the DT monopole, suggesting the physical equivalence of two definitions on a lattice. 
Therefore, the CDGFN monopole can be used, instead of DT monopole, to study the physics related to quark confinement: the magnetic monopole dominance in the string tension, derivation of the monopole action as a low-energy effective action of the Yang-Mills theory,  the  finite-temperature confinement/deconfinement phase transition, etc..

The new construction of the magnetic current is more similar to the original Abelian projection of 't Hooft rather than that of DT. 
However, the first construction of the CDGFN  monopole  given in \cite{KKMSSI06} does not guarantee the integer-valued magnetic charge which is realized by DT monopole. Indeed, the magnetic charge of the CDGFN  monopole is the real valued, in contrast with the conventional DT monopole. 
This disadvantage can be cured by converting it to the compact variable so as to guarantee the \textbf{quantization of the magnetic charge} from the beginning, as given in a subsequent paper \cite{IKKMSS06}.  
The first construction is called the \textbf{non-compact lattice reformulation}, while the second one is the \textbf{compact lattice reformulation}. The compact reformulation has been used since then. 
The results presented in this review are based on the compact reformulation. 
%It will be interesting to examine the contribution of the CDGFN monopole to the Wilson loop for confirming the monopole dominance in confinement.  

In the compact lattice reformulation, we have shown numerically a \textbf{gauge-independent restricted field (``Abelian'')  dominance} and \textbf{magnetic monopole dominance} in the string tension  extracted from the Wilson loop average in the $SU(2)$ Yang-Mills theory on a lattice.
These results have been obtained in the gauge-independent way  based on a new formulation of the Yang-Mills theory on a lattice, which reduces to the new variables of CDGFN in the continuum limit. 
It should be remarked that the infrared Abelian dominance and magnetic monopole dominance  have been so far shown only in a special Abelian gauge fixing called  MA gauge which breaks the color symmetry explicitly. 
The formulation enables one to reproduce in the gauge-invariant way remarkable  results obtained so far only in the MA gauge.

The formulation enables us to explain the infrared ``Abelian'' dominance, in addition to magnetic monopole dominance,  in the gauge invariant way without relying on the specific gauge fixing called the MA gauge used in the conventional investigations.  
In \cite{SKKMSI07}, we have demonstrated by numerical simulations that gluon degrees of freedom corresponding to the remaining field $\mathbf{X}_\mu(x)$ other than the restricted  field $\mathbf{V}_\mu(x)$ (``Abelian'' part) acquire the mass to be decoupled in the low-energy region leading to the \textbf{infrared ``Abelian'' dominance}.
In order to confirm the \textbf{dynamical mass generation} for the remaining part $\mathbf{X}_\mu(x)$ as a mechanism for the infrared ``Abelian'' dominance, we have measured the two-point correlation function  (the full propagator in real space) in our lattice formulation by imposing LLG for the original gauge field  $\mathbf{A}_\mu(x)$ as the overall gauge fixing (The overall gauge fixing is necessary to determine the propagator which is gauge dependent). 
We have found the infrared ``Abelian'' dominance in the sense that the $\mathbf{X}_\mu(x)$ propagator falls off quite rapidly and is suppressed in the long distance compared to the other new variables ${\bf n}(x)$ and $c_\mu(x)$. 
The estimation of the dynamically generated mass for  $\mathbf{X}_\mu(x)$ is $M_X=1.2 \sim 1.3 ~{\rm GeV}$ according the exponential fall-off in the long distance.  
Consequently, the restricted field becomes dominant in the long distance. 
%  (and $\mathbf{V}_\mu(x)$ propagators as an immediate consequence of dynamically generated mass $M_X=1.2 \sim 1.3 ~{\rm GeV}$ for  $\mathbf{X}_\mu(x)$ (which is larger than the decay rate of other gluon field propagators).

Even after imposing the whole gauge fixing, the new  reformulation can preserve \textbf{color symmetry} by choosing a suitable overall gauge-fixing condition which does not break color symmetry, e.g., Landau gauge.  This fact opens a path to examine  \textbf{color confinement} in the same framework as quark confinement in the dual superconductivity picture, since the gauge-independent criterion for color confinement is not known so far.  This feature is in sharp contrast to the conventional MA gauge which breaks color symmetry explicitly, although the reformulation can reproduce the MA gauge as a special limit (\ref{C35-MAGlimit}). 
It is important to demonstrate explicitly the gauge-fixing independence of our results obtained in a series of papers for  establishing the gauge-invariant mechanism for quark confinement.

%%%%%%%%%%%%%%%%%%%%%%%%%%%%%%%%%%%%%%%%%%%%%%%%%%
% 
%%%%%%%%%%%%%%%%%%%%%%%%%%%%%%%%%%%%%%%%%%%%%%%%%%

%A suitable definition of the link variable $\exp(-i\epsilon g {\bf X}_\mu(x))$ for  ${\bf X}_\mu(x)$ will be given with  relevant numerical results in a separate paper.  
%Extending the promising formulation proposed here for $SU(2)$   to $SU(3)$ will be given in a subsequent paper. 

Moreover, we have investigated the \textbf{dual Meissner effect} and the \textbf{type of the dual superconductor} which is characterized by the \textbf{Ginzburg-Landau parameter} according to the \textbf{Ginzburg-Landau theory}. 
%Our results show  that the dual superconductor for the $SU(2)$ lattice Yang-Mills theory is the border between type-I and type-II, which is consistent with the preceding results \cite{Suzuki:1988,CCP12}.
Our result shows that the dual superconductor for the $SU(2)$ lattice Yang--Mills theory is the border between type-I and type-II, which is consistent with the preceding results \cite{Suzuki:1988,CCP12}, or rather the weakly \textbf{type I}. 
We have confirmed that the same conclusion can be reproduced by the restricted field  on the type of dual superconductor for the $SU(2)$ lattice Yang-Mills theory.
These results establish the existence of the dual Meissner effect in the $SU(2)$ lattice Yang-Mills theory  in the gauge-invariant way, which is responsible for quark confinement.

In our works, the dual Meissner effect is examined by the simultaneous formation of the \textbf{chromoelectric flux tube} and the associated \textbf{magnetic-monopole current} induced around it.    
In a quite recent work \cite{Shibata-lattice2014}, moreover, we have shown that  at finite temperature the chromoelectric flux becomes broader and other components of the flux begin to appear and that the associated magnetic-monopole current vanishes at the critical temperature $T_c$. This indicates the dual Meissner effect disappears above the critical temperature, which is detected by a set of flux tube and the associated magnetic-monopole current.   
It is possible to argue that the condensed monopoles below $T_c$ become thermal monopoles above but close to $T_c$ and dominates the plasma phase there, see e.g., \cite{LS07,CZ09}.

%%%%%%%%%%%%%%%%%%%%%%%%%%%%%%%%%%%%%%%%%%%%%%%%%%

%\newpage
%%%%%%%%%%%%%%%%%%%%%%%%%%%%%%%%%%%%%%%%%%%%%%%%%%
\subsection{Lattice $SU(3)$ Yang-Mills theory and numerical simulations}
%%%%%%%%%%%%%%%%%%%%%%%%%%%%%%%%%%%%%%%%%%%%%%%%%%

We have given new reformulations of the $SU(N)$ Yang-Mills theory on the lattice \cite{KSSMKI08,SKS10} according to a new viewpoint extended to the $SU(N)$ Yang-Mills theory    \cite{KSM08}, which provide  one with an efficient framework to study quark confinement in the gauge-independent manner. 
We have presented the results of numerical simulations  of the lattice $SU(3)$ Yang-Mills theory \cite{KSSK11,SKKS13}, which support  the \textbf{non-Abelian dual superconductivity} for $SU(3)$ Yang-Mills theory proposed in \cite{KSSK11}.

We have shown that the restricted field extracted from the original $SU(3)$ Yang-Mills field plays a dominant role in confinement of quarks in the fundamental representation, i.e., the \textbf{restricted field dominance} in the (fundamental) string tension.
The restricted-field dominance was also confirmed for gluon propagators.

We have given numerical evidences that the non-Abelian magnetic monopoles defined in a gauge-invariant way  are dominant for confinement of fundamental quarks in $SU(3)$ Yang-Mills theory, i.e.,  \textbf{non-Abelian magnetic monopole dominance} in the (fundamental) string tension. 
By using the gauge invariant magnetic current $k$, we have extracted just the $U(1)$ part of  the maximal stability group $U(N-1) \simeq SU(N-1) \times U(1)$ for the non-Abelian magnetic monopole associated with quarks in the fundamental representation, which is consistent with the consideration of the Homotopy group. 
This $U(1)$ part is enough to extract the dominant part of the Wilson loop average.

In order to confirm the existence of the dual Meissner effect in $SU(3)$ Yang-Mills theory,   we have measured the gauge-invariant chromo field strength in the presence of a quark and an antiquark for both the original Yang-Mills field and the restricted  field. 
We have observed the dual Meissner effect in $SU(3)$ Yang-Mills theory: only the chromoelectric field exists in the flux tube connecting a quark and an antiquark and the associated magnetic-monopole current is induced around it. 
Moreover, we have determined the type of the non-Abelian dual superconductivity, i.e., \textbf{type I for the dual superconductivity of  $SU(3)$ Yang-Mills theory}, which should be compared with the border of type I and II for the dual superconductivity of the $SU(2)$ Yang-Mills theory.
These features are reproduced also from the 
restricted part, the restricted field dominance for the dual Meissner effect.
%These results confirm the non-Abelian dual superconductivity picture for quark confinement.
 
In order to draw the definite conclusion on physical quantities in the continuum limit, e.g., the Ginzburg-Landau parameter, however, we must study the scaling of the data obtained in the numerical simulations. For this purpose, we need to accumulate more data at various choices for the gauge coupling on the lattices with different sizes. 
These features will be discussed in the future works. 
In the future, moreover, we hope to study the electric-current contribution to the Wilson loop average and the Abelian dominance and monopole dominance in the \textbf{adjoint Wilson loop} with the  possibilities of their connections to the \textbf{Casimir scaling} in the intermediate region and \textbf{string breaking} as a special case of \textbf{$N$-ality} in the asymptotic region.

%%%%%%%%%%%%%%%%%%%%%%%%%%%%%%%%%%%%%%%%%%%%%%%%%%%%%
\subsection{Some remarks}
%%%%%%%%%%%%%%%%%%%%%%%%%%%%%%%%%%%%%%%%%%%%%%%%%%%%%
 
 The results presented in this report support that the non-Abelian  dual superconductivity could be a mechanism for quark confinement in $SU(3)$ Yang-Mills theory.
However, in order to establish the non-Abelian dual superconductivity caused by condensation of the non-Abelian magnetic monopoles in $SU(3)$ Yang-Mills theory in this sense,  we need to perform more investigations: some of them are enumerated below.

\begin{enumerate}
\item
{[Dual (magnetic) gauge symmetry, the spontaneous symmetry breaking and dual Meissner effect]}
Which symmetry is the \textbf{dual symmetry} to be spontaneously broken due to condensation of the non-Abelian magnetic monopoles.  
%We must confirm that the spontaneous breaking of the dual symmetry is associated with condensation of the non-Abelian magnetic monopoles.  
For this purpose, we need to specify how the relevant low-energy effective theory of the $SU(3)$ Yang-Mills theory looks like, although it is believed to be a dual Ginzburg-Landau model based on the dual superconductor picture for quark confinement. 
For this point, it will be helpful to see e.g., Kondo \cite{KondoI}, and Kato and Kondo \cite{KK05} in the MA gauge for $SU(3)$.

\item
{[Internal degrees of freedom for non-Abelian magnetic monopoles]}
In order to see directly the non-Abelian nature of magnetic monopoles, it will be necessary to study how the interaction among the non-Abelian magnetic monopoles is described in the short distance where the internal non-Abelian $SU(2)$ degrees of freedom in $U(2) \simeq SU(2) \times U(1)$ other than $U(1)$ are expected to become relevant.  
It will be interesting to investigate if  the   interactions between two chromoelectric flux tubes are attractive (expected for the  type I) or repulsive (expected for the type II).
 The interactions should reflect the internal non-Abelian nature which depends on the  distance between two tubes.
The similar and related subjects have been discussed in the references \cite{non-Abelian-vortex}, see e.g., \cite{non-Abelian-vortex-review} for reviews.

\item
{[Gluon confinement]}
It will be interesting to study how the non-Abelian magnetic monopoles contribute to \textbf{gluon confinement}. In the framework of the reformulation using new variables, a first step in this direction was taken recently  for $SU(2)$ gauge group in Kondo \cite{Kondo11}. 
Extending this work to the $SU(3)$ gauge group will be an interesting subject of subsequent works. 

\item
{[Finite temperature deconfinement transition]}
It will be also interesting to extend our results to the finite temperature case to see the fate of the dual superconductivity,  and to see when and how the dual superconductivity disappears above the critical temperature of the \textbf{deconfinement} phase transition.
See Shibata et al.\cite{Shibata-lattice2014} for a preliminary work in this direction.

%\item
%{[Magnetic monopole and other topological configurations, e.g., instantons, merons, knot, calorons.]}

%\item
%{[Fadeev-Niemi model as an low-energy effective model and knot soliton]}

%\item
%{[non-Abelian magnetic monopole and non-Abelian vortex]}

%\item
%{[Nielsen-Olesen instability of the Savvidy vacuum]}

%\item
%{[Chiral symmetry breaking and confinement]}

%\item
%{[Dimension-two condensate]}

%\item
%{[Large $N$ expansion]}

%\item
%{[Casimir scaling in the intermediate region]}

%\item
%{[$N$-ality dependence]}

\end{enumerate}

%%%%%%%%%%%%%%%%%%%%%%%%%%%%%%%%%%%%%%%%%%%%%%%%%%%%%%%%%%%%
%%%%%%%%%%%%%%%%%%%%%%%%%%%%%%%%%%%%%%%%%%%%%%%%%%%%%%%%%%%%
%\subsection{Remarks}\label{sec:remarks} 
%%%%%%%%%%%%%%%%%%%%%%%%%%%%%%%%%%%%%%%%%%%%%%%%%%%%%%%%%%%%
%%%%%%%%%%%%%%%%%%%%%%%%%%%%%%%%%%%%%%%%%%%%%%%%%%%%%%%%%%%%

%%%%%%%%%%%%%%%%%%%%%%%%%%%%%%%%%%%%%%%%%%%%%%%%%%%%%%%%%%%%
%%%%%%%%%%%%%%%%%%%%%%%%%%%%%%%%%%%%%%%%%%%%%%%%%%%%%%%%%%%%
\subsection{Dependence of quark potential on the representation and distance}\label{sec:potential-dependence} 
%%%%%%%%%%%%%%%%%%%%%%%%%%%%%%%%%%%%%%%%%%%%%%%%%%%%%%%%%%%%
%%%%%%%%%%%%%%%%%%%%%%%%%%%%%%%%%%%%%%%%%%%%%%%%%%%%%%%%%%%%

It is known that the static quark potential has the different behaviors depending on both the representation of quark and the distance between a pair of quark and antiquark.
Therefore, it is very important to understand this phenomena to clarify what is the true mechanism of quark confinement. 

%%%%%%%%%%%%%%%%%%%%%%%%%%%%%%%%%%%%%%%%%%%%%%%%%%%
%%%%%%%%%%%%%%%%%%%%%%%%%%%%%%%%%%%%%%%%%%%%%%%%%%%
%\subsubsection{Georgi-Glashow model} 
%%%%%%%%%%%%%%%%%%%%%%%%%%%%%%%%%%%%%%%%%%%%%%%%%%%
%%%%%%%%%%%%%%%%%%%%%%%%%%%%%%%%%%%%%%%%%%%%%%%%%%%

First of all, it is instructive to recall the Georgi-Glashow model in which the magnetic monopole plays the dominant role in confinement. 
In the $D=3$ Georgi-Glashow model, Polyakov has shown the area law of the Wilson loop average in the fundamental representation is derived using an effective Abelian gauge theory, involving only the monopoles and photons associated with the unbroken $U(1)$ subgroup. 
It has been numerically shown that the representation dependence of string tensions is that of the pure Yang-Mills theory in the symmetric phase, but changes abruptly to equal tensions for the $J=1/2, 3/2$ representations, and zero tension for $J=1$, at the transition to the Higgs phase. 
The Abelian gauge field is singled out by a unitary gauge, and for calculating the fundamental string tension it is a reasonable approximation to ignore the contribution of the other color components. 
The infrared dynamics of this model is essentially that of compact QED and it is the Abelian monopole prediction, rather than the Casimir scaling, which agrees with the data. 
This is not the case for the Yang-Mills theory and QCD. 

%%%%%%%%%%%%%%%%%%%%%%%%%%%%%%%%%%%%%%%%%%%%%%%%%%%
%%%%%%%%%%%%%%%%%%%%%%%%%%%%%%%%%%%%%%%%%%%%%%%%%%%
\subsubsection{$N$-ality dependence of the string tension in the asymptotic regime}\label{sec:N-ality} 
%%%%%%%%%%%%%%%%%%%%%%%%%%%%%%%%%%%%%%%%%%%%%%%%%%%
%%%%%%%%%%%%%%%%%%%%%%%%%%%%%%%%%%%%%%%%%%%%%%%%%%%

For $SU(N)$ group, the \textbf{center} elements consist of a set of $N$ group elements proportional to the $N \times N$ unit matrix subject to the condition $\det (g)=1$:
$
g=z_n := \exp (i \frac{2\pi}{N}n) \bm{1} \in Z(N) \ (n=0,1,2,...,N-1)
$.
The center elements of $SU(N)$ form a discrete Abelian subgroup known as $Z(N)$.
The $N$-ality of a representation of $SU(N)$ is equal to the number of boxes in the corresponding \textbf{Young tableau} (mod $N$).
If the $N$-ality of a representation is $k$, then the transformation by $z \in Z(N)$ corresponds to multiplication by a factor $\exp (i \frac{2\pi}{N}nk)$.

The color charge of higher-representation quarks must be eventually screened by gluons.  Hence, the \textbf{asymptotic string tension} $\sigma_R$ for quark in the representation $R$ can only depend on the the \textbf{$N$-ality} $k$ of the representation $R$, i.e., transformation properties of the representation for the quarks under the center of the gauge group: 
\begin{align}
 \sigma_R = \sigma(k) := f(k) \sigma_F .
\end{align}
This could be applied to the asymptotic regime which extends from the color-screening length to infinity.
In particular, the string between quarks in the adjoint representation must break at some distance which presumably depend on the mass of ``gluelumps'', i.e, the energy of a gluon bound to a massive adjoint quark. 
The breaking of the adjoint string is difficult to observe in numerical simulations \cite{Poulis96,CHS04}.

However, the precise form of the $k$-dependence $f(k)$ is not known. 
There are two proposals. 
One is the ``\textbf{Casimir scaling}''\cite{Greensite03}:
\begin{align}
 \sigma(k) = \frac{k(N-k)}{N-1} \sigma_{F} .
\end{align}
Another is known as the ``\textbf{Sine-Law scaling}'' which is suggested by MQCD and softly broken $\mathcal{N}=2$ \cite{DS95,HSZ98}:
\begin{align}
 \sigma(k) = \frac{\sin \frac{\pi k}{N}}{\sin \frac{\pi}{N}} \sigma_{F} .
\end{align}
For the $SU(2)$ gauge group, the asymptotic string tension must satisfy
\begin{align}
 \sigma(k) 
%= k(2-k) \sigma_{F}
= \begin{cases}
  \sigma_{F} & (k=1:\text{J=half-integer}) \\
  0 & (k=0:\text{J=integer}) \\
  \end{cases}  .
\end{align}
%For the $SU(2)$ gauge group, the asymptotic string tension must satisfy
%\begin{align}
% \sigma_J = \begin{cases}
%  \sigma_{1/2} & (\text{J=half-integer}) \\
%  0 & (\text{J=integer}) \\
%  \end{cases}
%\end{align}
For the  $SU(3)$ gauge group, the two proposals give the same prediction:
\begin{align}
 \sigma(k) 
%= \frac{k(3-k)}{2} \sigma_{F}
%=  \frac{\sin \frac{\pi k}{3}}{\sin \frac{\pi}{3}} \sigma_{F}
= \begin{cases}
  \sigma_{F} & (k=1, k=2) \\
  0 & (k=0) \\
  \end{cases}  
 .
\end{align}
The two proposals give different predictions for $N \ge 4$. 
For fixed $k$, the Casimir scaling and the sine-law scaling are identical in the large $N$ limit $N \to \infty$.
\begin{align}
 \sigma(k) =  k  \sigma_{F} .
\end{align}

%%%%%%%%%%%%%%%%%%%%%%%%%%%%%%%%%%%%%%%%%%%%%%%%%%%%%%%%%%%%
%%%%%%%%%%%%%%%%%%%%%%%%%%%%%%%%%%%%%%%%%%%%%%%%%%%%%%%%%%%%
\subsubsection{Casimir scaling of the string tension  in the intermediate regime}\label{sec:Casimir-scaling} 
%%%%%%%%%%%%%%%%%%%%%%%%%%%%%%%%%%%%%%%%%%%%%%%%%%%%%%%%%%%%
%%%%%%%%%%%%%%%%%%%%%%%%%%%%%%%%%%%%%%%%%%%%%%%%%%%%%%%%%%%%

According to numerical simulations of the Yang-Mills theory on the lattice in both $D=3$ and $D=4$ dimensions and for both $SU(2)$ and $SU(3)$ gauge groups \cite{Bali00,Deldar00}, 
%\footnote{
%\bibitem{Bali00}
%G. S. Bali,
%Casimir scaling of SU(3) static potentials, 
%Phys. Rev. D{\bf 62}, 114503  (2000). 
%e-Print: hep-lat/0006022,
%GUTPA-00-06-02 
%DOI: 10.1103/PhysRevD.62.114503 
%
%S. Deldar,
%Static SU(3) potentials for sources in various representations, 
%Phys. Rev. D{\bf 62}, 034509  (2000). 
%DOI: 10.1103/PhysRevD.62.034509 
%e-Print: hep-lat/9911008.
%} 
there is an intermediate range of distance, extending from the onset of the linear potential (confinement) up to the flat potential at some distance (color-screening), which is called the \textbf{Casimir scaling regime}, where flux tubes form and a linear potential is established between heavy quarks in any non-trivial representation of the gauge group. 
The Casimir scaling means that the quark potential $V^{q\bar q}_{R}(r)$ for the quark in the representation $R$ of the color group is proportional to the quadratic Casimir operator of the representation $R$ of the quark \cite{SS00}:
\begin{equation}
V^{q\bar q}_{R}(r) \cong \frac{C_{R}}{C_{F}} V^{q\bar q}_{F}(r) ,\end{equation}
where the subscript $F$ refers to the fundamental representation. 
The total quark potential is well fitted by the sume of two parts: 
$V^{q\bar q}_{R}(r)=V^{\rm ele}_{R}(r)+V^{\rm mag}_{R}(r)$
where $V^{\rm mag}_{R}(r)=\sigma_{R}r$ is the linear potential with the string tension $\sigma_{R}$, and $V^{\rm ele}_{R}(r)=-C_{R}g_{_{\rm YM}}^2 r^{-1}$ is the Coulomb-like potential which is inversely proportional to the inter-quark distance $r$ and is proportional to the quadratic Casimir operator $C_{R}$ of the representation $R$ for the quark as calculated from the one-gluon exchange in the perturbation theory.
Therefore, the Casimir scaling is consistent only when the string tension $\sigma_{R}$ as a coefficient of the linear potential is also proportional to the quadratic Casimir operator.

In this intermediate regime, therefore, the string tension is representation dependent in such a way that the string tension $\sigma_{R}$ of static sources in the representation $R$ of the color group is approximately proportional to the quadratic Casimir invariant $C_R$ of the representation $R$:
\begin{align}
 \sigma_{R} \cong \frac{C_R}{C_F} \sigma_{F} . 
\end{align}
This fact is called the \textbf{Casimir scaling} of the string tension \cite{AOP84,DFGO96}.  
%\footnote{
%\bibitem{AOP84}
%J. Ambjorn, P. Olesen, C. Peterson,
%Stochastic Confinement and Dimensional Reduction. 1. Four-Dimensional SU(2) Lattice Gauge Theory, 
%Nucl. Phys. B{\bf 240},  189--212 (1984). 
%\\
%J. Ambjorn, P. Olesen, C. Peterson,
%Stochastic Confinement and Dimensional Reduction. 2. Three-dimensional SU(2) Lattice Gauge Theory, 
%Nucl. Phys. B{\bf 240},  533--542 (1984).
%\\
%\bibitem{DFGO96}
%L. Del Debbio, M. Faber, J. Greensite, S. Olejnik,
%Casimir scaling versus Abelian dominance in QCD string formation, 
%Phys. Rev. D53, 5891--5897 (1996).  
%hep-lat/9510028,
%SWAT-95-91, LBL-37855 
%DOI: 10.1103/PhysRevD.53.5891 
%}   
This result holds with an accuracy of about 10\%, from the onset of confinement to the onset of color screening. 
%Further numerical evidences are presented in \cite{}. 

For the $SU(2)$ group, any representation is specified by a half integer $J=\frac12, 1, \frac32, 2, \dots$ and the quadratic Casimir invariant $C_R=C_J$ is given by
\begin{align}
 C_J = J(J+1) .   
\end{align}
Hence, the Casimir scaling means that the  string tension $\sigma_{J}$ is related to the fundamental string tension $\sigma_{F}=\sigma_{J=1/2}$ as 
\begin{align}
 \sigma_{J} \cong  \frac{C_J}{C_{1/2}} \sigma_{1/2} = \frac43 J(J+1) \sigma_{1/2} . 
\end{align}
In particular, the adjoint string tension $\sigma_{A}=\sigma_{J=1}$ is related to the fundamental string tension $\sigma_{F}=\sigma_{J=1/2}$ as
\begin{align}
  C_F = C_{J=1/2} = \frac34 , \quad C_A = C_{J=1} = 2 \Longrightarrow \sigma_{A} \cong  \frac83  \sigma_{F} = 2.66... \sigma_{F} .
\end{align}

For the $SU(N)$ group, the quadratic Casimir invariants for the fundamental and adjoint representations are 
\begin{align}
 C_F = \frac{N^2-1}{2N}, \quad C_A = N . 
\end{align}
Hence, the Casimir scaling means that the adjoint string tension $\sigma_{A}$ is related to the fundamental string tension $\sigma_{F}$ as 
\begin{align}
 \sigma_{A} \cong  \frac{2N^2}{N^2-1} \sigma_{F} , 
\end{align}
which is monotonically decreasing in $N$. 
For the $SU(3)$ group, in particular, $\sigma_{A}$ is related to $\sigma_{F}$ as
\begin{align}
 \sigma_{A} \cong  \frac{9}{4} \sigma_{F} = 2.25 \sigma_{F} \quad (N=3) , 
\end{align}
while in the large $N$ limit we have 
\begin{align}
 \sigma_{A} \to  2 \sigma_{F} \quad (N \to \infty).
\label{Casimir-N=infty} 
\end{align}
In the Casimir scaling regime at intermediate distances,  the confining force replaces the Coulombic  behavior at short distances.
In fact, it is only in this regime that the \textbf{QCD string} has been well studied numerically. 
If we understand the Casimir scaling, then we really understand how flux tubes form, leading to the thorough understanding of the quark confinement mechanism.

There are two theoretical arguments supporting the Casimir scaling in $D=4$ $SU(3)$ Yang-Mills theory: \textbf{factorization in the large $N$ limit} \cite{Witten,GH83} and the \textbf{dimensional reduction} to $D=2$ \cite{AOP84}. 
%Note that the string tension satisfies the Casimir scaling exactly for $N=\infty$ colors and for $D=2$ dimensions. 
It is known that the Casimir scaling is exact out to infinite quark separation (i.e., no string breaking) in the \textbf{large $N$ limit $N \to \infty$}. 
Therefore, the approximate Casimir scaling up to some finite range of distances should be observed at finite $N$. 
In the large $N$ limit, in particular, the string tension of the adjoint representation is twice the string tension of the fundamental representation:
\begin{align}
 \sigma_{A}=2\sigma_{F}   \quad (N \to \infty) , 
\end{align}
in agreement with the Casimir scaling (\ref{Casimir-N=infty}). 
This result follows also from the strong coupling expansion of the Wilson loop average $W_{A}(C)$ in the adjoint representation at finite values of $N$ \cite{GH83}: 
\begin{align}
 W_{A}(C) = N^2 e^{-2\sigma_{F} A(C)} + e^{-4\sigma_{F} P(C)}, 
\end{align}
where $A(C)$ is the minimal area of the surface bounded by the loop $C$ and $P(C)$ is the perimeter of the loop $C$. 
%\footnote{
%\bibitem{GH83}
%J. Greensite and M.B. Halpern,
%Suppression of Color Screening at Large N, 
%Phys. Rev. D{\bf 27}, 2545-- (1983).  
%} 
In other words, the string-breaking is a $1/N^2$ suppressed process in the large $N$ case.

In the \textbf{two dimensional case},  the Wilson loop average can be calculated from one-gluon exchange and it is shown that the static potential is linear with the string tension: 
\begin{align}
 \sigma_{R} = \frac12 g_{_{\rm YM}}^2  C_R  \quad (D=2), 
\end{align}
which means that the string tension is exactly proportional to the quadratic Casimir invariants:  
\begin{align}
 \sigma_{R} = \frac{C_R}{C_F} \sigma_{F} \quad (D=2) . 
\end{align}
There is a suggestion known as the \textbf{dimensional reduction} that the string tension of planar Wilson loops in $D=3$ and $D=4$ dimensions could be computed from an effective two-dimensional gauge theory \cite{AOP84}.
Then the ratio of string tensions between quarks in different group representations should equal the ratio of the corresponding quadratic Casimir invariants. 
This was tested numerically, in both $D=3$ and $D=4$ dimensions and found it to be accurate to within 10\%. 
These results have been confirmed by a number of other studies.

%%%%%%%%%%%%%%%%%%%%%%%%%%%%%%%%%%%%%%%%%%%%%%%%%%%%%%%%%%%%
%%%%%%%%%%%%%%%%%%%%%%%%%%%%%%%%%%%%%%%%%%%%%%%%%%%%%%%%%%%%
\subsubsection{Abelian projection for the scaling in the intermediate regime} 
%%%%%%%%%%%%%%%%%%%%%%%%%%%%%%%%%%%%%%%%%%%%%%%%%%%%%%%%%%%%
%%%%%%%%%%%%%%%%%%%%%%%%%%%%%%%%%%%%%%%%%%%%%%%%%%%%%%%%%%%%

The Abelian projection tells us the following result on the scaling in the intermediate region.

For concreteness, we consider the $SU(2)$ case.
The Abelian dominance was checked by the numerical simulations on the lattice. 
For $SU(2)$ Yang-Mills theory in $D=3$ dimensions, the numerical simulations (on the $12^3$ lattice at lattice coupling $\beta=5$ which is just inside the scaling region) show that the hypothesis of Abelian dominance in the MA gauge, which is known to work for Wilson loop averages in the fundamental representation $\sigma_{\rm full}^{J=1/2} \simeq \sigma_{\rm Abel}^{J=1/2}$, fails for Wilson loop averages in higher group representations \cite{DFGO96,DFGO97}:  
In the interval where the Casimir scaling is expected to hold, the string tensions extracted from the Wilson loop averages built from the Abelian projected configurations are the same for the fundamental $(J=1/2)$ and $J=3/2$ representation, and vanish for the adjoint representation $J=1$:
\begin{align}
 \sigma_{\rm Abel}^{J=3/2}=\sigma_{\rm Abel}^{J=1/2} \ne 0 , \quad \sigma_{\rm Abel}^{J=1} =  0 ,
\end{align}
while  the original full string tensions behave according to the Casimir scaling as \cite{Piccioni06}
%\footnote{
%C. Piccioni, 
%}
\begin{align}
 \sigma_{\rm full}^{J=3/2}= 5 \sigma_{\rm full}^{J=1/2}, \quad \sigma_{\rm full}^{J=1} = \frac83 \sigma_{\rm full}^{J=1/2} .
\end{align}
The Abelian monopole prediction for $SU(2)$ Yang-Mills theory is 
\begin{align}
 \sigma_{\rm Abel}^{J}
= \begin{cases}
 \sigma_{\rm Abel}^{J=1/2} & (J=\frac12, \frac32, \frac52, \dots) \\
\\
 \sigma_{\rm Abel}^{J} =  0 & (J=1,2,3, \dots) 
\end{cases}
 ,
\end{align}
This behavior is in fact what one expects in the asymptotic region  due to charge screening.
However, this behavior begins right at the confinement scale has nothing whatever  to do with the physics of charge screening. 
The adjoint sources are unconfined in the Abelian projection theory.  A flux tube between adjoint quarks does not form and then break due to charge screening. 
In this picture, the tube does not form at all.

The \textbf{breakdown of the Abelian dominance and monopole dominance for higher representation} in the $SU(2)$ Yang-Mills theory, not only for the adjoint but also for the $J=3/2$ representation indicates that an effective Abelian gauge theory at the confinement scale involving only degrees of freedom (monopoles and photons) associated with the Cartan subalgebra is inadequate to describe the actual interquark potential in the Yang-Mills theory (unbroken gauge theory).

Later, it was discussed how to cure this pathology in the naive Abelian projection. 
The adjoint Wilson loop operator $W_1(C)$ and the adjoint Polyakov loop operator $P_1(\bm{x})$ contains both the charged term $Q=2$ and the neutral term $Q=0$.
For instance, $W_1$ is decomposed as
\begin{align}
 W_1 = W^{Q=0} +  W^{Q=2} .
 \label{W-charge-decomp}
\end{align}
Then the $Q=2$ charged component $\langle W^{Q=2} \rangle$ of the Wilson loop average show the area law, while the neutral $Q=0$ component $\langle W^{Q=0} \rangle$ is constant. 
Therefore, the straightforward application of the Abelian projection to the adjoint operators leads to vanishing Abelian string tension as far as $\langle W^{Q=0} \rangle \ne 0$, since   
\begin{align}
 \langle W_1  \rangle =  \langle W^{Q=0} \rangle + \langle W^{Q=2} \rangle  
= c + c^\prime e^{-\sigma A(C)} 
\end{align}
leads to
\begin{align}
 \frac{1}{T} \ln \langle W_1  \rangle   
= \frac{1}{T} \ln c + \frac{1}{T} \ln [1+ (c^\prime/c) e^{-\sigma A(C)} ] 
= \frac{1}{T} \ln c + (c^\prime/c) \frac{1}{T} e^{-\sigma A(C)} \to 0 \ (T \to \infty)   .
\end{align}
In order to overcome this difficulty, it has been proposed in \cite{Poulis96} to discard the $Q=0$ component $W^{Q=0}$ of the Wilson loop operator and to consider the $Q=2$ Abelian component $W^{Q=2}$ as the Abelian analogue of the full non-Abelian Wilson loop. Following this recipe, it has been shown in \cite{CHS04} that the string breaking effect can be seen in the $Q=2$ Abelian and monopole components of the potential, and the \textbf{Abelian dominance and monopole dominance for the adjoint string tension} similarly to the case of the fundamental representation. 
The Abelian matter fields, i.e., off-diagonal gluons which become doubly charged Abelian vector fields in the Abelian projection are essential for the \textbf{breaking of the adjoint string} and provide an essential contribution to the total action even in the infrared region. 
However, it should be remarked that the decomposition (\ref{W-charge-decomp}) is projection dependent and each piece $W^{Q=0}$ or $W^{Q=2}$ is not gauge invariant, since only the sum is gauge invariant. 
Therefore, these results must be treated with great care. 

%%%%%%%%%%%%%%%%%%%%%%%%%%%%%%%%%%%%%%%%%%%%%%%%%%%
%%%%%%%%%%%%%%%%%%%%%%%%%%%%%%%%%%%%%%%%%%%%%%%%%%%
\subsubsection{The role of magnetic monopoles and vortices for the Casimir scaling and/or the $N$-ality
} 
%%%%%%%%%%%%%%%%%%%%%%%%%%%%%%%%%%%%%%%%%%%%%%%%%%%
%%%%%%%%%%%%%%%%%%%%%%%%%%%%%%%%%%%%%%%%%%%%%%%%%%%

%This paper studies only the case of the $SU(2)$ group and therefore the Casimir scaling and/or strings of higher $N$-ality in $SU(N)$ gauge theories with $N \ge 3$ are beyond the scope of this paper.
%However, this issue exists already in $SU(2)$ and hence our opinions are briefly described as follows, although the details should be discussed elsewhere.
%In particular, the Casimir scaling is the issue in the intermediate distance and it is not clear to us whether or not it is really explained by magnetic monopoles which could be relevant only to the low-energy or long distance physics.  
%For the $N$-ality for $N=2$, on the other hand, we should say something, since the asymptotic string tension is to be within the reach of this paper. 
%In the Maximal Abelian gauge, we know there are some attempts to explain it. For example, in the paper [40]  

First of all, our claim of the dual superconductor picture (or monopole condensation picture) does not contradict with the vortex condensation picture. 
Indeed, the vortex picture requires the magnetic flux which links with the Wilson loop to give a nontrivial center element, while the dual superconductor picture requires magnetic monopoles to be condensed.   
We imagine that the magnetic flux needed in the vortex   picture is emanating from a magnetic monopole (or antimonopole) specified in the dual superconductor picture. 
In our opinion, a gauge-invariant vortex can be defined through the gauge-invariant magnetic monopole which we have constructed in a series of our works. In this sense, it is very important to study the interplay between magnetic monopoles and vortices. 
A preliminary work along this direction  from the viewpoint of our formulation 
%and some discussions on how to understand the $N$-ality from the magnetic monopole inherent in the Wilson loop operator are given based on the non-Abelian Stokes theorem 
was given by one of the authors \cite{Kondo08b}, 
%\bibitem{Kondo08b}
% K.-I. Kondo, 
%Magnetic monopoles and center vortices as gauge-invariant topological defects simultaneously responsible for confinement,
%Chiba Univ. Preprint, CHIBA-EP-171, 
% arXiv:0802.3829 [hep-th],
%J. Phys. G: Nucl. Part. Phys. {\bf 35}, 085001  (2008).
and a subsequent consideration was done in section 6 of  this review. 
%in the recent review: K.-I. Kondo, et al., 
%Quark confinement: dual superconductor picture
%based on a non-Abelian Stokes theorem and reformulations of  Yang-Mills theory,  
%e-Print: arXiv:1409.1599 [hep-th] (submitted to Physics Report).

 However, in order to deduce the vortex picture from our dual superconductor picture, we need more detailed informations on the profile of the magnetic flux emanating from magnetic monopoles.  As far as we know, such a relation is not yet derived for $SU(3)$ directly from the Yang-Mills theory, although some models for the $SU(3)$ vortex (\textbf{vortex-monopole net}) \cite{Engelhardt11,CZ09,LS07} are proposed and studied, while for $SU(2)$  some works give the numerical evidences for the existence of the \textbf{vortex-monopole chains} \cite{AGG00,CFIS05}.

In order to consider the magnetic monopole contribution to the Casimir scaling and/or N-ality, we remind you of the fact that the Wilson loop operator $W_C[A]$ is separated into the electric part  $W_{\rm elec}[j]$ and the magnetic one  $W_{\rm mono}[K]$ through the non-Abelian Stokes theorem for the Wilson loop operator:
$W_C[A]=W_{\rm elec}[j] W_{\rm mono}[K]$.  Note that each contribution is separately gauge-invariant. 
For the inter-quark potential $V^{q\bar q}_{R}(r)$, the electric part does not yield the linear potential, but it leads to the Coulomb-like potential which is inversely proportional to the inter-quark distance $r$ and is proportional to the quadratic Casimir operator $C_{R}$ of the representation $R$ for the quark as calculated from the one-gluon exchange in the perturbation theory:
$V^{\rm ele}_{R}(r)=-C_{R}g_{_{\rm YM}}^2 r^{-1}$.  Whereas, the magnetic part does not yield the Coulomb-like potential and has the linear potential alone $V^{\rm mag}_{R}(r)=\sigma_{R}r$.  
The Casimir scaling means that the total quark potential $V^{q\bar q}_{R}(r)$ identified with a sum of the electric and magnetic contributions $V^{q\bar q}_{R}(r)=V^{\rm ele}_{R}(r)+V^{\rm mag}_{R}(r)$ is proportional to the quadratic Casimir operator of the representation $R$ of the quark:
$
V^{q\bar q}_{R}(r)= \frac{C_{R}}{C_{F}} V^{q\bar q}_{F}(r) .$
Therefore, the Casimir scaling is consistent only when the string tension $\sigma_{R}$ as a coefficient of the linear potential is also proportional to the quadratic Casimir operator:
$
\sigma_{R}= \frac{C_{R}}{C_{F}} \sigma_{F} 
$,
where $F$ refers to the fundamental representation. 

The magnetic part must dominantly reproduce the string tension irrespective of the representation of the quark, provided that the electric part does not yield the linear potential. 
Therefore, the Casimir scaling can be in principle explained from the magnetic monopole according to the magnetic monopole dominance in the string tension for any representation.

The $N$-ality of the string tension in the asymptotic region is understood based on the vortex picture.  
The $N$-ality is understood for $SU(2)$ relatively easier than the $SU(3)$ case. 
For $SU(2)$, indeed, it is shown that the magnetic flux due to a magnetic monopole is collimated to form the vortex-monopole chain.  
%The origin of the vortex must be the magnetic monopole.  
For $SU(N)$, it is expected that the magnetic flux is grouped into $N$ flux tubes, in other words, a magnetic monopole plays the role of a \textbf{junction} connecting $N$ flux tubes.
If a magnetic flux tube intersects with the surface $\Sigma$ bounded by the Wilson loop $C$, it yields a nontrivial factor corresponding to an element of the center group $Z(N)$ of $SU(N)$. 
At least, the monopole picture is able to explain the $N$-ality in the large size limit of the Wilson loop, i.e., in the large distance limit, since each monopole contributes a half of the total magnetic flux without assuming the details on the profile of the magnetic flux emanating from a monopole, as was explained in \cite{Kondo08b}.
%Kondo (2008) and Kondo et al (2014) cited above. 

For the Casimir scaling, however, even the center vortex approach is not so easy to give the explanation without assuming some structure of the vacuum, e.g., a certain type of \textbf{domain structure} \cite{center-vortex,GLORT07}. 
However, it is known that the Casimir scaling is exact in the large $N$ limit.  This is a nice feature, since our reformulation of $SU(N)$ Yang-Mills theory in the minimal option is compatible with the large $N$ limit. Therefore, we hope to say something about this issue from our point of view in our future works.

\subsection{Other references}

The CDGFN and CFN decompositions have been already used for various purposes, as suggested from the titles of the papers \cite{LMO10,Grigorio11,DM12,MS13,MS14,Upadhyay13,CP06,CPZZ12,CCY13,Xiong13,Xiong14}. 
Recently, a decomposition similar to the minimal option for $SU(N)$ has been studied in \cite{EGKM11}.

%%Acknowledgments
%%%%%%%%%%%%%%%%%%%%%%%%%%%%%%%%%%%%%%%%%%%
%%%%%%%%%%%%%%%%%%%%%%%%%%%%%%%%%%%%%%%%%%%
\section*{Acknowledgments}
%%%%%%%%%%%%%%%%%%%%%%%%%%%%%%%%%%%%%%%%%%%
%%%%%%%%%%%%%%%%%%%%%%%%%%%%%%%%%%%%%%%%%%%
\noindent
This work is financially supported in part by Grant-in-Aid for Scientific Research (C) 24540252 from Japan Society for the Promotion of Science (JSPS).
%The numerical simulations have been done on a supercomputer (NEC SX-8) at Research Center for  Nuclear Physics (RCNP), Osaka University. 
The numerical simulations in the early stage of this project  from 2004 to 2009 have been 
done on a supercomputer (NEC SX-5) at Research Center for Nuclear 
Physics (RCNP), Osaka University.
This work is in part supported by the Large Scale Simulation Program of High Energy Accelerator Research Organization (KEK):
No.~133 (FY2005), No.~22 (FY2006), No.~06--17 (FY2006), No.~07--15 
(FY2007), No.~08--16 (FY2008), No.~09--15 (FY2009), No.~09/10-19 
(FY2009--2010), No.~10--13 (FY2010),No.~T11-15(FY2011), No.~12--13 
(FY2011--2012), No.~12/13--20 (FY2012--2013), No.~13/14--23 
(FY2013--2014), and No.~14/15--20 (FY2014--2015).

%No.08-16 (FY2008), No.09-15 (FY2009), No.T11-15 (FY2011), No.12/13-20 (FY2012-2013) and No.13/14-23 (FY2013-2014) of High Energy Accelerator Research Organization (KEK).
%The numerical calculations are supported by the Large Scale Simulation Program No.T11-15 (FY2011) and No.12-13 (FY2012) of High Energy Accelerator Research Organization (KEK).

%%%%%%%%%%%%%%%%%%%%%%%%%%%%%%%%%%%%%%%%%%%
%%%%%%%%%%%%%%%%%%%%%%%%%%%%%%%%%%%%%%%%%%%
\section*{Notes added in proof}
%%%%%%%%%%%%%%%%%%%%%%%%%%%%%%%%%%%%%%%%%%%
%%%%%%%%%%%%%%%%%%%%%%%%%%%%%%%%%%%%%%%%%%%

A detailed report for the numerical simulations of SU(3) Yang-Mills theory (in the maximal option) has been published,
\cite{CCLL15}.
%N. Cundy, Y.M. Cho, Weonjong Lee, and Jaehoon Leem,
%The Static Quark Potential from the Gauge Independent Abelian Decomposition, 
%Nucl. Phys. B{\bf 895},  64--131 (2015). 
%e-Print: arXiv:1503.07033 [hep-lat]

An application of the reformulation of SU(3) Yang-Mills theory (in the maximal option) to glueballs has appeared,
\cite{CPZXZ15} 
%Y.M. Cho, X.Y. Pham, Pengming Zhang, Ju-Jun Xie, and Li-Ping Zou,
%Glueball Physics in QCD, 
%e-Print: arXiv:1503.08890 [hep-ph]. 

%\newpage 
%%%%%%%%%%%%%%%%%%%%%%%%%%%%%%%%%%%%%%%%%%%%%%%%%%%%%%%%%%%%
%Appendices
%%%%%%%%%%%%%%%%%%%%%%%%%%%%%%%%%%%%%%%%%%%%%%%%%%%%%%%%%%%%

\begin{appendix}

%Miscellanea

%%%%%%%%%%%%%%%%%%%%%%%%%%%%%%%%%%%%%%%%%%%%%%%%%%%%%%%%%%%%%
%%%%%%%%%%%%%%%%%%%%%%%%%%%%%%%%%%%%%%%%%%%%%%%%%%%%%%%%%%%%%
%%%%%%%%%%%%%%%%%%%%%%%%%%%%%%%%%%%%%%%%%%%%%%%%%%%%%%%%%%%%%
\section{Dictionary between two notations}\label{section:dictionary}
%%%%%%%%%%%%%%%%%%%%%%%%%%%%%%%%%%%%%%%%%%%%%%%%%%%%%%%%%%%%%
%%%%%%%%%%%%%%%%%%%%%%%%%%%%%%%%%%%%%%%%%%%%%%%%%%%%%%%%%%%%%
%%%%%%%%%%%%%%%%%%%%%%%%%%%%%%%%%%%%%%%%%%%%%%%%%%%%%%%%%%%%%

In Table~\ref{table:dictionary}, we give a dictionary between two notations used by Cho and ours where (L) means the Lie algebra notion and (V) the vector notation. 

%%%%%%%%%%%%%%%%%%%%%%%%%%%%%%%%%%%%%%%%%%%%%%%%%%%%%%%%%%%%%
%%%%%%%%%%%%%%%%%%%%%%%%%%%%%%%%%%%%%%%%%%%%%%%%%%%%%%%%%%%%%
\begin{table}[htbp]
\begin{center}  
\begin{tabular}{l||l|l} \hline
 & Cho & Ours \\  \hline\hline
 original gauge field (V) & $\vec{A}_\mu=\hat{A}_\mu+\vec{X}_\mu$  &  $\mathbf{A}_\mu=\mathbf{V}_\mu+\mathbf{X}_\mu=(\mathbf{A}_\mu^A)$    \\ 
original gauge field (L) &   &  $\mathscr{A}_\mu=\mathscr{V}_\mu+\mathscr{X}_\mu=\mathscr{A}_\mu^A T_A$    \\  \hline
color field  (V) & $\hat{n}$  (or $\hat{m}$)  &    $\mathbf{n}$    \\
color field (L) &    &    $\bm{n}$    \\  \hline
  restricted field  (V) &  $\hat{A}_\mu=A_\mu \hat{n}+ \vec{C}_\mu$ &    $\mathbf{V}_\mu =\mathbf{C}_\mu+ \mathbf{B}_\mu$    \\ 
 restricted field (L) &    &    $\mathscr{V}_\mu =\mathscr{C}_\mu+\mathscr{B}_\mu$    \\  \hline
 Abelian part  (V) & $\vec{B}_\mu=A_\mu \hat{n}$  &    $\mathbf{C}_\mu=c_\mu \mathbf{n}$     \\ 
(parallel) restricted field (L) &    &    $\mathscr{C}_\mu=\mathscr{C}_\mu^A T_A=c_\mu \bm{n}$  \\ \hline 
 monopole part (V) &  $\vec{C}_\mu= g^{-1}\partial_\mu \hat{n} \times \hat{n}$ &    $\mathbf{B}_\mu=g^{-1}\partial_\mu \mathbf{n} \times \mathbf{n}$    \\
(perpend.) restricted field (L) &   &    $\mathscr{B}_\mu=\mathscr{B}_\mu^A T_A=-i g^{-1} [\partial_\mu\bm{n} ,\bm{n}]$    \\   \hline
Abelian  electric potential  & $A_\mu=\hat{n} \cdot \vec{A}_\mu$  &    $c_\mu=\bm{n} \cdot \mathbf{A}_\mu$    \\  
Abelian  magnetic potential   & $C_\mu$  &    $h_\mu$    \\  
Abelian potential  & $B_\mu=A_\mu+C_\mu$  &    $a_\mu=c_\mu+h_\mu= G_\mu $    \\  \hline
valence gluon field  (V)  & $\vec{X}_\mu$  &    $\mathbf{X}_\mu $    \\ 
remaining field (L) &   &    $\mathscr{X}_\mu=\mathscr{X}_\mu^A T_A$    \\  \hline
covariant derivative  & $\hat{D}_\mu=\partial_\mu+g\hat{A}_\mu $  &    $D_\mu[V]=\partial_\mu+g\mathbf{V}_\mu $    \\  \hline
 original field strength (f. s.) & $\vec{F}_{\mu\nu}=\partial_\mu \vec{A}_\mu - \partial_\nu \vec{A}_\nu  +g\vec{A}_\nu \times \vec{A}_\nu$  &    $\mathbf{F}_{\mu\nu}=\partial_\mu \mathbf{A}_\mu - \partial_\nu \mathbf{A}_\nu  +g\mathbf{A}_\nu \times \mathbf{A}_\nu$    \\   \hline
 restricted f. s. & $\hat{F}_{\mu\nu}=\partial_\mu \hat{A}_\mu - \partial_\nu \hat{A}_\nu  +g\hat{A}_\nu \times \hat{A}_\nu$  &    $\mathbf{F}_{\mu\nu}^{[V]}=\partial_\mu \mathbf{V}_\mu - \partial_\nu \mathbf{V}_\nu  +g\mathbf{V}_\nu \times \mathbf{V}_\nu$   \\   
   & $\hat{F}_{\mu\nu}= (F_{\mu\nu}+H_{\mu\nu})\hat{n}=G_{\mu\nu} \hat{n}$  &    $\mathbf{F}_{\mu\nu}^{[V]}=(F_{\mu\nu}+H_{\mu\nu})\mathbf{n}=G_{\mu\nu}\mathbf{n} $   \\  \hline
 Abelian  f. s.   & $G_{\mu\nu}=\partial_\mu B_\nu - \partial_\nu B_\mu$  &    $G_{\mu\nu}=\partial_\mu a_\nu - \partial_\nu a_\mu=f_{\mu\nu} $    \\  
Abelian electric f. s. & $F_{\mu\nu}=\partial_\mu A_\nu - \partial_\nu A_\mu$  &    $F_{\mu\nu}=\partial_\mu c_\nu - \partial_\nu c_\mu=E_{\mu\nu}$    \\  
Abelian magnetic f. s. & $H_{\mu\nu}=\partial_\mu C_\nu - \partial_\nu C_\mu$  &    $H_{\mu\nu}=\partial_\mu h_\nu - \partial_\nu h_\mu $    \\  \hline
 magnetic f.s. & $\vec{H}_{\mu\nu}=\partial_\mu \vec{C}_\mu - \partial_\nu \vec{C}_\nu  +g\vec{C}_\nu \times \vec{C}_\nu$  &    $\mathbf{H}_{\mu\nu}=\partial_\mu \mathbf{B}_\mu - \partial_\nu \mathbf{B}_\nu  +g\mathbf{B}_\nu \times \mathbf{B}_\nu$    \\  
   & $\vec{H}_{\mu\nu}=H_{\mu\nu}\hat{n} $  &    $\mathbf{H}_{\mu\nu}=H_{\mu\nu}\mathbf{n}$    \\  
 Abelian magnetic f.s.  & $H_{\mu\nu}=-g^{-1}\hat{n} \cdot (\partial_\mu \hat{n}\times \partial_\nu \hat{n})$  &    $H_{\mu\nu}=-g^{-1}\mathbf{n} \cdot (\partial_\mu \mathbf{n} \times \partial_\nu \mathbf{n})$    \\   \hline

\end{tabular}
\end{center}
\caption{
Dictionary between two notations used by Cho and ours.
Here (L) means the Lie algebra notion and (V) the vector notation. 
}
\label{table:dictionary}
\end{table}
%%%%%%%%%%%%%%%%%%%%%%%%%%%%%%%%%%%%%%%%%%%%%%%%%%%%%%%%%%%%%
%%%%%%%%%%%%%%%%%%%%%%%%%%%%%%%%%%%%%%%%%%%%%%%%%%%%%%%%%%%%%

%%%%%%%%%%%%%%%%%%%%%%%%%%%%%%%%%%%%%%%%%%%%%%%%%%%%%%%%%%%%%
%%%%%%%%%%%%%%%%%%%%%%%%%%%%%%%%%%%%%%%%%%%%%%%%%%%%%%%%%%%%%
%%%%%%%%%%%%%%%%%%%%%%%%%%%%%%%%%%%%%%%%%%%%%%%%%%%%%%%%%%%%%
\section{Formulae for Cartan subalgebras}\label{section:Cartan}
%%%%%%%%%%%%%%%%%%%%%%%%%%%%%%%%%%%%%%%%%%%%%%%%%%%%%%%%%%%%%
%%%%%%%%%%%%%%%%%%%%%%%%%%%%%%%%%%%%%%%%%%%%%%%%%%%%%%%%%%%%%
%%%%%%%%%%%%%%%%%%%%%%%%%%%%%%%%%%%%%%%%%%%%%%%%%%%%%%%%%%%%%

The Cartan algebras are written as 
\begin{align}
 H_k =& \frac{1}{\sqrt{2k(k+1)}} {\rm diag}(1, 1, \cdots, 1, -k, 0, \cdots, 0) 
 \nonumber\\
 =&  \frac{1}{\sqrt{2k(k+1)}} \left( \sum_{j=1}^{k} \bm{e}_{jj} - k \bm{e}_{k+1,k+1} \right) 
 ,
\end{align}
where we have defined the matrix $\bm{e}_{AB}$ whose $AB$ element has the value 1 and other elements are zero, i.e.,
$(\bm{e}_{AB})_{ab}=\delta_{Aa} \delta_{Bb}$.  
The the $ab$ element reads
\begin{equation}
 (H_k)_{ab} =  \frac{1}{\sqrt{2k(k+1)}} \left( \sum_{j=1}^{k} \delta_{aj}\delta_{bj} - k \delta_{a,k+1}\delta_{b,k+1}  \right) 
 .
\end{equation}
The unit matrix ${\bf 1}$ with the element $\delta_{ab}$ is written as
\begin{equation}
 {\bf 1} =  {\rm diag}(1, 1, \cdots, 1) 
= \sum_{j=1}^{N} \bm{e}_{jj}
 .
\end{equation}
The product of diagonal generators is decomposed as 
\begin{align}
  H_j H_k = 
  \begin{cases}
  \frac{1}{\sqrt{2j(j+1)}} H_k & (j>k) \cr
  \frac{1}{\sqrt{2k(k+1)}} H_j & (j<k) \cr
  \frac{1}{2N}{\bf 1} + \frac{1-k}{\sqrt{2k(k+1)}} H_k + \sum_{m=k+1}^{N-1}  \frac{1}{\sqrt{2m(m+1)}} H_m & (j=k) \cr
  \end{cases} 
 .
\end{align}
The first and second relations are easily derived from the definition. 
The third relation is derived as follows.
\begin{equation}
 H_k H_k = c_0 {\bf 1} + \sum_{m=1}^{N-1} c_m H_m
 , 
\end{equation}
where
\begin{equation}
 c_m = 2{\rm tr}(H_k H_k H_m) ,  
 \quad 
 c_0 = {\rm tr}(H_k H_k)/{\rm tr}({\bf 1}) 
  ,
\end{equation}
Here the coefficients are calculated as
\begin{align}
  c_m =& 2{\rm tr}\left[  \frac{1}{2k(k+1)} {\rm diag}(1, 1, \cdots, 1,k^2, 0, \cdots, 0)  H_m \right] 
  \nonumber\\
  =& 
  \begin{cases}
  0 & (1 \le m \le k-1) \cr
  \frac{1-k}{\sqrt{2k(k+1)}}  & (m=k) \cr
  \frac{1}{\sqrt{2m(m+1)}}   & (k+1 \le m \le N-1) \cr
  \end{cases} 
 ,
\end{align}
and
\begin{equation}
 c_0 = {\rm tr}(H_k H_k)/{\rm tr}({\bf 1}) 
 = \frac12 \frac1N
 . 
\end{equation}

%%%%%%%%%%%%%%%%%%%%%%%%%%%%%%%%%%%%%%%%%%%%%%%%%%%%%%%%%%%%%
%%%%%%%%%%%%%%%%%%%%%%%%%%%%%%%%%%%%%%%%%%%%%%%%%%%%%%%%%%%%%
%%%%%%%%%%%%%%%%%%%%%%%%%%%%%%%%%%%%%%%%%%%%%%%%%%%%%%%%%%%%%
\section{Decomposition formulae}\label{section:formulae1}
%%%%%%%%%%%%%%%%%%%%%%%%%%%%%%%%%%%%%%%%%%%%%%%%%%%%%%%%%%%%%
%%%%%%%%%%%%%%%%%%%%%%%%%%%%%%%%%%%%%%%%%%%%%%%%%%%%%%%%%%%%%
%%%%%%%%%%%%%%%%%%%%%%%%%%%%%%%%%%%%%%%%%%%%%%%%%%%%%%%%%%%%%

For any $su(N)$ Lie algebra valued function $\mathscr{V}(x)$, the identity holds 
\footnote{
%S.V. Shabanov,
%Yang-Mills theory as an Abelian theory without gauge fixing,
%[hep-th/9907182],
%Phys. Lett. B {\bf 463}, 263--272  (1999).
%S.V. Shabanov,
%On a low energy bound in a class of chiral field theories with solitons,
%[hep-th/0202146],
%J. Math. Phys. {\bf 43}, 4127--4134  (2002).
% \cite{Shabanov99b,Shabanov02}
The SU(2) version of this identity is 
\begin{align}
  \bm{v} =  \bm{n} (\bm{n} \cdot \bm{v}) - \bm{n} \times (\bm{n} \times \bm{v})
=   \bm{v}_\parallel +  \bm{v}_\perp  ,
\end{align}
which follows from a simple identity,
$
 \bm{n} \times (\bm{n} \times \bm{v}) = \bm{n} (\bm{n} \cdot \bm{v}) - (\bm{n} \cdot \bm{n}) \bm{v}.
$
}
\cite{Shabanov02}:%
\begin{align}
  \mathscr{V} =& \sum_{j=1}^{N-1} \bm{n}_j(\bm{n}_j, \mathscr{V}) +  \sum_{j=1}^{N-1} [\bm{n}_j, [\bm{n}_j, \mathscr{V}]]
  \nonumber\\
=& \sum_{j=1}^{N-1} 2{\rm tr}(\mathscr{V} \bm{n}_j)\bm{n}_j +  \sum_{j=1}^{N-1} [\bm{n}_j, [\bm{n}_j, \mathscr{V}]]
% := v_\parallel + v_\perp
 .
\label{C27-idv}
\end{align}
This identity is equivalent to the identity using the structure constants of the Lie algebra: 
%\footnote{
%This is noticed in the reference:
%\cite{FN99a}
%}
\begin{align}
 \delta^{AB}  = n_j^A n_j^B  -f^{ACD} n_j^C f^{DEB} n_j^E  
 .
\end{align}

The identity is proved as follows.
The color field $\bm{n}_j$ can be written without losing  generality in the form:
$
\bm{n}_j = U^\dagger H_j U 
$.
By using the adjoint rotation, $\mathscr{V}'=U\mathscr{V}U^\dagger$, therefore, we have only to prove
\begin{align}
  \mathscr{V}' = \sum_{j=1}^{N-1} H_j (H_j, \mathscr{V}') + \sum_{j=1}^{N-1} [H_j, [H_j, \mathscr{V}']] .
\label{C27-idd}
\end{align}
The Cartan decomposition for $\mathscr{V}=\mathscr{V}'$ reads
\begin{align}
  \mathscr{V} 
=  \sum_{k=1}^{N-1} V^k H_k + \sum_{\alpha=1}^{(N^2-N)/2} (W^*{}^{\alpha} \tilde{E}_{\alpha} + W^{\alpha} \tilde{E}_{-\alpha})  ,
\end{align}
where the Cartan basis is given by
\begin{align}
 \vec{H} =& (H_1, H_2, H_3, \cdots, H_{N-1}) = (T^3, T^8, T^{15}, \cdots, T^{N^2-1}) ,
\nonumber\\
  \tilde{E}_{\pm 1} =& {1 \over \sqrt{2}}(T^1 \pm  i T^2) , \quad
  \tilde{E}_{\pm 2} = {1 \over \sqrt{2}}(T^4 \pm  i T^5) ,  
  \nonumber\\ &
  \cdots,  \quad
  \tilde{E}_{\pm (N^2-N)/2} =  {1 \over \sqrt{2}}(T^{N^2-3} \pm i T^{N^2-2}) ,
\end{align}
and the complex field is defined by
\begin{align}
  W^1 =& {1 \over \sqrt{2}}(V^1+ i V^2) , \quad
  W^2 = {1 \over \sqrt{2}}(V^4+ i V^5) , \quad 
  \nonumber\\ &
  \cdots,  \quad
  W^{(N^2-N)/2} = {1 \over \sqrt{2}}(V^{N^2-3}+ i V^{N^2-2})  .
\end{align}

We now calculate the double commutator as
\begin{align}
&  [H_j, [H_j, \mathscr{V}]]
\nonumber\\
 =& \sum_{j=1}^{N-1} V_k [H_j, [H_j, H_k]] + \sum_{\alpha=1}^{(N^2-N)/2} (W^*{}^{\alpha} [H_j, [H_j,  \tilde{E}_{\alpha}]] + W^{\alpha} [H_j, [H_j,  \tilde{E}_{-\alpha}]]) 
 \nonumber\\
 =& 
  \sum_{\alpha=1}^{(N^2-N)/2} (W^*{}^{\alpha} [H_j, \alpha_j \tilde{E}_{\alpha}] + W^{\alpha} [H_j, -\alpha_j   \tilde{E}_{-\alpha}]) 
 \nonumber\\
 =& 
   \alpha_j  \alpha_j \sum_{\alpha=1}^{(N^2-N)/2} (W^*{}^{\alpha}  \tilde{E}_{\alpha}  + W^{\alpha}  \tilde{E}_{-\alpha} )
 ,
\end{align}
where $\alpha_j$ are root vectors. 
On the other hand, we have 
\begin{align}
& (H_j, \mathscr{V}) 
 \nonumber\\
 =&  \sum_{k=1}^{N-1} V_k (H_j, H_k) + \sum_{\alpha=1}^{(N^2-N)/2} (W^*{}^{\alpha} (H_j, \tilde{E}_{\alpha}) + W^{\alpha} (H_j, \tilde{E}_{-\alpha}))  
 \nonumber\\
 =&   V^j  ,
\end{align}
since 
$(H_j, \tilde{E}_{\alpha})={\rm tr}(H_j \tilde{E}_{\alpha})=0$. 
Thus the RHS of (\ref{C27-idd}) reduces to 
\begin{align}
 \sum_{j=1}^{N-1} V_j H_j  + \sum_{j=1}^{N-1} \alpha_j  \alpha_j \sum_{\alpha=1}^{(N^2-N)/2} (W^*{}^{\alpha}  \tilde{E}_{\alpha}  + W^{\alpha}  \tilde{E}_{-\alpha} ) .
\end{align}
This is equal to the Cartan decomposition of $\mathscr{V}$ itself, since 
\begin{equation}
\sum_{j=1}^{N-1} \alpha_j  \alpha_j=1
 . 
\end{equation}

%%%%%%%%%%%%%%%%%%%%%%%%%%%%%%%%%%%%%%%%%%%%%%%%%%%%%%%%%%%%%
%%%%%%%%%%%%%%%%%%%%%%%%%%%%%%%%%%%%%%%%%%%%%%%%%%%%%%%%%%%%%
%%%%%%%%%%%%%%%%%%%%%%%%%%%%%%%%%%%%%%%%%%%%%%%%%%%%%%%%%%%%%
\section{More decomposition formulae}\label{section:formulae2}
%%%%%%%%%%%%%%%%%%%%%%%%%%%%%%%%%%%%%%%%%%%%%%%%%%%%%%%%%%%%%
%%%%%%%%%%%%%%%%%%%%%%%%%%%%%%%%%%%%%%%%%%%%%%%%%%%%%%%%%%%%%
%%%%%%%%%%%%%%%%%%%%%%%%%%%%%%%%%%%%%%%%%%%%%%%%%%%%%%%%%%%%%

For any $su(N)$ Lie algebra valued function $\mathscr{V}$, the identity holds:
\begin{align}
  \mathscr{V} 
= \sum_{A=1}^{N^2-1} \mathscr{V}^A T_A
=  \tilde{\mathscr{V}} + \bm{h} (\bm{h},\mathscr{V}) 
+  2\frac{N-1}{N}  [\bm{h} , [\bm{h} , \mathscr{V}]]
% := v_\parallel + v_\perp
 ,
\label{C27-idv2}
\end{align}
where we have defined the matrix $\tilde{\mathscr{V}}$  in which all the elements in the last column and the last raw are zero:
\begin{equation}
 \tilde{\mathscr{V}} = \sum_{A=1}^{(N-1)^2-1} \mathscr{V}^A T_A
= \sum_{A=1}^{(N-1)^2-1} (\mathscr{V},T_A) T_A 
= \sum_{A=1}^{(N-1)^2-1} 2{\rm tr}(\mathscr{V}T_A)T_A
 , 
\end{equation}
or
\begin{align}
  \mathscr{V}  =  \tilde{\mathscr{V}} + 2{\rm tr}(\mathscr{V} \bm{h})\bm{h} 
+  2\frac{N-1}{N}  [\bm{h} , [\bm{h} , \mathscr{V}]]
% := v_\parallel + v_\perp
 .
\label{C27-idv3}
\end{align}
Note that $[\tilde{\mathscr{V}},\bm{h}]=0$.

The Cartan decomposition for $\mathscr{V}_N=v' \in su(N)$ reads
\begin{align}
  \mathscr{V}_N 
=&   \sum_{k=1}^{N-1} V_k H_k + \sum_{\alpha=1}^{(N^2-N)/2} (W^*{}^{\alpha} \tilde{E}_{\alpha} + W^{\alpha} \tilde{E}_{-\alpha})  
\nonumber\\
=& \tilde{\mathscr{V}}_{N} + M_{N-1} H_{N-1} 
+ \sum_{\alpha=[(N-1)^2-(N-1)]/2+1}^{(N^2-N)/2} (W^*{}^{\alpha} \tilde{E}_{\alpha} + W^{\alpha} \tilde{E}_{-\alpha})  
 .
\end{align}

Now we calculate the double commutator ($r=N-1$) as
\begin{align}
&  [H_r, [H_r, \mathscr{V}_N]]
\nonumber\\
 =& \sum_{j=1}^{N-1} V_k [H_r, [H_r, H_k]] + \sum_{\alpha=1}^{(N^2-N)/2} (W^*{}^{\alpha} [H_r, [H_r,  \tilde{E}_{\alpha}]] + W^{\alpha} [H_r, [H_r,  \tilde{E}_{-\alpha}]]) 
 \nonumber\\
 =& 
  \sum_{\alpha=[(N-1)^2-(N-1)]/2+1}^{(N^2-N)/2} (W^*{}^{\alpha} [H_r, \alpha_r \tilde{E}_{\alpha}] + W^{\alpha} [H_r, -\alpha_r   \tilde{E}_{-\alpha}]) 
 \nonumber\\
 =& 
   \alpha_r  \alpha_r \sum_{\alpha=[(N-1)^2-(N-1)]/2+1}^{(N^2-N)/2} (W^*{}^{\alpha}  \tilde{E}_{\alpha}  + W^{\alpha}  \tilde{E}_{-\alpha} )
 .
\end{align}
On the other hand, we have 
\begin{align}
& (H_r, \mathscr{V}_N) 
 \nonumber\\
 =&  \sum_{k=1}^{N-1} V_k (H_r, H_k) + \sum_{\alpha=1}^{(N^2-N)/2} (W^*{}^{\alpha} (H_r, \tilde{E}_{\alpha}) + W^{\alpha} (H_r, \tilde{E}_{-\alpha}))  
 \nonumber\\
 =&  V_r  ,
\end{align}
since 
$(H_j, \tilde{E}_{\alpha})={\rm tr}(H_j \tilde{E}_{\alpha})=0$. 

For any Lie algebra valued function $\mathscr{V}_N(x)$, we obtain the identity:
\begin{align}
  \mathscr{V}_N 
=& \sum_{A=1}^{N^2-1} \mathscr{V}_A T_A
\nonumber\\
=&  \tilde{\mathscr{V}_N} + (\mathscr{V}_N,H_r) H_r 
+  \frac{1}{\alpha_r^2}  [H_r , [H_r , \mathscr{V}_N]]
% := v_\parallel + v_\perp
\nonumber\\
=&  \tilde{\mathscr{V}_N} + (\mathscr{V}_N,H_r) H_r 
+  \frac{2(N-1)}{N}  [H_r , [H_r , \mathscr{V}_N]]
 .
\label{C27-idv4}
\end{align}

%%%%%%%%%%%%%%%%%%%%%%%%%%%%%%%%%%%%%%%%%%%%%%%%%%%%%%%%%%%%
%%%%%%%%%%%%%%%%%%%%%%%%%%%%%%%%%%%%%%%%%%%%%%%%%%%%%%%
\section{Reduction condition for $SU(2)$ and the ghost term}
%%%%%%%%%%%%%%%%%%%%%%%%%%%%%%%%%%%%%%%%%%%%%%%%%%%%%%%%%%%%

By identifying the reduction condition $\bm{\chi}=0$ with a gauge-fixing condition for the enlarged gauge symmetry, the gauge-fixing term and associated  ghost term is obtained from
\begin{align}
 {\cal L}_{\rm GF+FP}^\theta
 = -i \bm\delta [\mathbf{\bar C}_\theta \cdot  \bm{\chi}]
=  \mathbf{N}_\theta \cdot \bm{\chi}
+ i \mathbf{\bar C}_\theta \cdot \bm\delta \bm{\chi} , \quad 
\bm{\chi} = D_\mu[\mathbf{V}]\mathbf{X}_\mu .
\end{align}
Here it should be remarked that the antighost field $\mathbf{\bar C}_\theta$ must have the same degrees of freedom of the constraint $\bm{\chi}= D_\mu[\mathbf{V}]\mathbf{X}_\mu$. 
This is achieved by requiring 
$
 \mathbf{n}(x) \cdot  \mathbf{\bar C}_\theta(x)    = 0
$, 
since 
$
  \mathbf{n}(x) \cdot \bm{\chi}(x) =  0
$.
By using
\begin{align}
 \bm{\chi}  = D_\mu[\mathbf{A}]\mathbf{X}_\mu 
= D_\mu[\mathbf{A}] (g^{-1}\mathbf{n} \times D_\mu[\mathbf{A}]\mathbf{n})
= g^{-1} \mathbf{n}\times D_\mu[\mathbf{A}]D_\mu[\mathbf{A}]\mathbf{n}
,
\end{align}
the ghost term is calculated as
\begin{align}
%{\cal L}_{\rm GF+FP}^\theta
%--------------
i \mathbf{\bar C}_\theta \cdot \bm\delta \bm{\chi}
=& g^{-1} i\mathbf{\bar C}_\theta \cdot \bm\delta  \{ \mathbf{n}\times D_\mu[\mathbf{A}]D_\mu[\mathbf{A}]\mathbf{n} \}
  \nonumber\\
  =& g^{-1} i
  \mathbf{\bar C}_\theta \cdot  \{ \bm\delta  \mathbf{n}\times D_\mu[\mathbf{A}]D_\mu[\mathbf{A}]\mathbf{n} 
+ \mathbf{n}\times \bm\delta  ( D_\mu[\mathbf{A}]D_\mu[\mathbf{A}]\mathbf{n} )
\}
  \nonumber\\
  =& g^{-1} i
  \mathbf{\bar C}_\theta \cdot  \{ \bm\delta  \mathbf{n}\times D_\mu[\mathbf{A}]D_\mu[\mathbf{A}]\mathbf{n} 
+ \mathbf{n}\times  D_\mu[\mathbf{A}] \bm\delta  (D_\mu[\mathbf{A}]\mathbf{n} )
%\nonumber\\ &\qquad\qquad\quad
+ \mathbf{n} \times  ( g\bm\delta \mathbf{A}_\mu \times D_\mu[\mathbf{A}]\mathbf{n} )
\}
  \nonumber\\
  =& 
%-i\bm\delta 
%   \mathbf{\bar C}_\theta\cdot(D_\mu[\mathbf{V}]\mathbf{X}_\mu)
%   \nonumber\\
% &\quad
    g^{-1} i \mathbf{\bar C}_\theta \cdot
  \bigl[
  \bm\delta \mathbf{n}\times D_\mu[\mathbf{A}]D_\mu[\mathbf{A}]\mathbf{n}
%\nonumber\\  &\qquad\qquad\quad
  +\mathbf{n}\times D_\mu[\mathbf{A}]D_\mu[\mathbf{A}]\bm\delta \mathbf{n}
%  \nonumber\\
%&\qquad\qquad\quad
  +\mathbf{n}\times D_\mu[\mathbf{A}]
   \left(g\bm\delta \mathbf{A}_\mu\times\mathbf{n}\right)
  \nonumber\\
 &\qquad\qquad\quad
  +\mathbf{n}\times
   \left(g\bm\delta \mathbf{A}_\mu\times D_\mu[\mathbf{A}]\mathbf{n}\right)
  \bigr]
  \nonumber\\
%--------------
=&
%\mathbb N_\theta\cdot(D_\mu[\mathbf{V}]\mathbf{X}_\mu)
%   \nonumber\\
% &\quad
   i\mathbf{\bar C}_\theta\cdot
  \bigl[
  (\mathbf{n}\times\mathbf{C}_\theta)\times D_\mu[\mathbf{A}]D_\mu[\mathbf{A}]\mathbf{n}
%\nonumber\\ &\qquad\qquad\quad
  +\mathbf{n}\times D_\mu[\mathbf{A}]D_\mu[\mathbf{A}](\mathbf{n}\times\mathbf{C}_\theta)
\nonumber\\ &\qquad\qquad\quad
  +\mathbf{n}\times D_\mu[\mathbf{A}]
   \left(D_\mu[\mathbf{A}]\mathbf{C}_\omega\times\mathbf{n}\right)
%\nonumber\\  &\qquad\qquad\quad
  +\mathbf{n}\times
   \left(D_\mu[\mathbf{A}]\mathbf{C}_\omega\times D_\mu[\mathbf{A}]\mathbf{n}\right)
  \bigr] .
%  \nonumber\\
%--------------
\end{align}
The result is simplified as
\begin{align}
i \mathbf{\bar C}_\theta \cdot \bm\delta \bm{\chi}
=&
%\mathbb N_\theta^\perp\cdot(D_\mu[\mathbf{V}]\mathbf{X}_\mu)
%\nonumber\\
%&\quad
  +i\mathbf{\bar C}_\theta\cdot
  \bigl\{
  (\mathbf{n}\cdot D_\mu[\mathbf{A}]D_\mu[\mathbf{A}]\mathbf{n})\mathbf{C}_\theta
  - \mathbf{n} (\mathbf{C}_\theta\cdot D_\mu[\mathbf{A}]D_\mu[\mathbf{A}]\mathbf{n}) 
  \nonumber\\
 &\qquad\qquad\quad
  +\mathbf{n}\times(D_\mu[\mathbf{A}]D_\mu[\mathbf{A}]\mathbf{n}\times\mathbf{C}_\theta)
  \nonumber\\
 &\qquad\qquad\qquad
  +2\mathbf{n}\times(D_\mu[\mathbf{A}]\mathbf{n}\times D_\mu[\mathbf{A}]\mathbf{C}_\theta)
  \nonumber\\
 &\qquad\qquad\qquad
  +\mathbf{n}\times(\mathbf{n}\times D_\mu[\mathbf{A}]D_\mu[\mathbf{A}]\mathbf{C}_\theta)
  \nonumber\\
 &\qquad\qquad\quad
  +\mathbf{n}\times
   \left(D_\mu[\mathbf{A}]D_\mu[\mathbf{A}]\mathbf{C}_\omega\times\mathbf{n}\right)
  \nonumber\\
 &\qquad\qquad\qquad
  +\mathbf{n}\times
   \left(D_\mu[\mathbf{A}]\mathbf{C}_\omega\times D_\mu[\mathbf{A}]\mathbf{n}\right)
  \nonumber\\
 &\qquad\qquad\quad
  + (\mathbf{n}\cdot D_\mu[\mathbf{A}]\mathbf{n})
   D^\mu[\mathbf{A}]\mathbf{C}_\omega
  -(\mathbf{n}\cdot D_\mu[\mathbf{A}]\mathbf{C}_\omega)
   D^\mu[\mathbf{A}]\mathbf{n}
  \bigr\}
  \nonumber\\
%--------------
  =&
%\mathbb N_\theta^\perp\cdot(D_\mu[\mathbb %\nonumber\\
%&\quad
  +i\mathbf{\bar C}_\theta\cdot
  \bigl\{
  (\mathbf{n}\cdot D_\mu[\mathbf{A}]D_\mu[\mathbf{A}]\mathbf{n})\mathbf{C}_\theta
  - \mathbf{n} (\mathbf{C}_\theta\cdot D_\mu[\mathbf{A}]D_\mu[\mathbf{A}]\mathbf{n}) 
  \nonumber\\
 &\qquad\qquad 
  + (\mathbf{n}\cdot \mathbf{C}_\theta)  D_\mu[\mathbf{A}]D_\mu[\mathbf{A}]\mathbf{n} 
  -(\mathbf{n}\cdot D_\mu[\mathbf{A}]D_\mu[\mathbf{A}]\mathbf{n})\mathbf{C}_\theta
  \nonumber\\
 &\qquad\qquad 
  +2(\mathbf{n}\cdot D_\mu[\mathbf{A}]\mathbf{C}_\theta)  D_\mu[\mathbf{A}]\mathbf{n}
  -2(\mathbf{n}\cdot D_\mu[\mathbf{A}]\mathbf{n})  D_\mu[\mathbf{A}]\mathbf{C}_\theta 
  \nonumber\\
 &\qquad\qquad 
  -(\mathbf{n} \cdot \mathbf{n}) D_\mu[\mathbf{A}]D_\mu[\mathbf{A}]\mathbf{C}_\theta
  +\mathbf{n} (\mathbf{n}\cdot D_\mu[\mathbf{A}]D_\mu[\mathbf{A}]\mathbf{C}_\theta)
  \nonumber\\
 &\qquad\qquad 
  + (\mathbf{n} \cdot \mathbf{n}) D_\mu[\mathbf{A}]D_\mu[\mathbf{A}]\mathbf{C}_\omega
  - \mathbf{n} (\mathbf{n}\cdot D_\mu[\mathbf{A}]D_\mu[\mathbf{A}]\mathbf{C}_\omega)
  \nonumber\\
 &\qquad\qquad 
  -(\mathbf{n}\cdot D_\mu[\mathbf{A}]\mathbf{C}_\omega)
   D_\mu[\mathbf{A}]\mathbf{n}
+  (\mathbf{n}\cdot D_\mu[\mathbf{A}]\mathbf{n})  D_\mu[\mathbf{A}]\mathbf{C}_\omega 
  \nonumber\\
 &\qquad\qquad 
  -(\mathbf{n}\cdot D_\mu[\mathbf{A}]\mathbf{C}_\omega)
    D^\mu[\mathbf{A}]\mathbf{n} 
   + (\mathbf{n}\cdot D_\mu[\mathbf{A}]\mathbf{n})  D_\mu[\mathbf{A}]\mathbf{C}_\omega 
  \bigr\}
  \nonumber\\
%--------------
  =&
%\mathbb N_\theta^\perp\cdot(D_\mu[\mathbb %\nonumber\\
%&\quad
  +i\mathbf{\bar C}_\theta\cdot
  \bigl\{
   2(\mathbf{n}\cdot D_\mu[\mathbf{A}]\mathbf{C}_\theta) D_\mu[\mathbf{A}]\mathbf{n} 
  \nonumber\\
 &\qquad\qquad 
  -  D_\mu[\mathbf{A}]D_\mu[\mathbf{A}]\mathbf{C}_\theta
  +   D_\mu[\mathbf{A}]D_\mu[\mathbf{A}]\mathbf{C}_\omega
  \nonumber\\
 &\qquad\qquad 
  -2(\mathbf{n}\cdot D_\mu[\mathbf{A}]\mathbf{C}_\omega)
    D_\mu[\mathbf{A}]\mathbf{n} 
  \bigr\}
  \nonumber\\%--------------
  =&
%\mathbb N_\theta^\perp\cdot(D_\mu[\mathbf{V}]\mathbf{X}_\mu)
%   \nonumber\\
% &\quad
  -i\mathbf{\bar C}_\theta\cdot
    D_\mu[\mathbf{A}]D_\mu[\mathbf{A}](\mathbf{C}_\theta-\mathbf{C}_\omega)
%\nonumber\\  &\quad
  +2 (i\mathbf{\bar C}_\theta\cdot D_\mu[\mathbf{A}]\mathbf{n})
    \{\mathbf{n}\cdot D_\mu[\mathbf{A}](\mathbf{C}_\theta-\mathbf{C}_\omega)\} ,
%  \nonumber\\
%--------------
\label{eq:L_theta}
\end{align}
where we have repeatedly used the Leibniz rule for the covariant derivative and the formula 
(\ref{vector-3product})
%$
%(\mathbf A \times \mathbf B) \times \mathbf C = (\mathbf A \cdot \mathbf C) \mathbf B - \mathbf A (\mathbf B \cdot \mathbf C), 
%$
%$
%\mathbf A \times (\mathbf B \times \mathbf C) = (\mathbf A \cdot \mathbf C) \mathbf B - (\mathbf A \cdot \mathbf B) \mathbf C , 
%$
together with 
%$
% \mathbf{\bar C}_\theta \cdot \mathbf{n} =  \mathbf{n}  \cdot \mathbf{C}_\theta = 0
%$, 
\begin{align}
 \mathbf{n}(x) \cdot  \mathbf{\bar C}_\theta(x) =  0 =  \mathbf{n}(x)  \cdot \mathbf{C}_\theta(x) ,
\end{align}
and
$
  \mathbf{n}\cdot D_\mu[\mathbf{A}]\mathbf{n} =  0
$
following from 
$
\mathbf{n} \cdot \mathbf{n}=1
$.
The result is rewritten in terms of the new variables as
\begin{align}
i \mathbf{\bar C}_\theta \cdot \bm\delta \bm{\chi}
  =&
%\mathbb N_\theta^\perp\cdot(D_\mu[\mathbf{V}]\mathbf{X}_\mu)
%   \nonumber\\
% &\quad
  -i\mathbf{\bar C}_\theta\cdot
    D_\mu[\mathbf{A}]D_\mu[\mathbf{A}](\mathbf{C}_\theta-\mathbf{C}_\omega)
%\nonumber\\&\quad
  +2g\{i\mathbf{\bar C}_\theta\cdot(\mathbf{X}_\mu\times\mathbf{n})\}
    \{\mathbf{n}\cdot D_\mu[\mathbf{A}](\mathbf{C}_\theta-\mathbf{C}_\omega)\}
  \nonumber\\
%--------------
  =&
%\mathbb N_\theta^\perp\cdot(D_\mu[\mathbf{V}]\mathbf{X}_\mu)
%   \nonumber\\
% &\quad
  -i\mathbf{\bar C}_\theta\cdot
    D_\mu[\mathbf{A}]D_\mu[\mathbf{A}](\mathbf{C}_\theta-\mathbf{C}_\omega)
%\nonumber\\&\quad
  +2g\{\mathbf{n}\cdot(i\mathbf{\bar C}_\theta\times\mathbf{X}_\mu)\}
    \{\mathbf{n}\cdot D_\mu[\mathbf{A}](\mathbf{C}_\theta-\mathbf{C}_\omega)\}
  \nonumber\\
%--------------
  =&
%\mathbb N_\theta^\perp\cdot(D_\mu[\mathbf{V}]\mathbf{X}_\mu)
%   \nonumber\\
% &\quad
  -i\mathbf{\bar C}_\theta\cdot
    D_\mu[\mathbf{A}]D_\mu[\mathbf{A}](\mathbf{C}_\theta-\mathbf{C}_\omega)
%\nonumber\\&\quad
  +2g(i\mathbf{\bar C}_\theta\times\mathbf{X}_\mu)
    \cdot D_\mu[\mathbf{A}](\mathbf{C}_\theta-\mathbf{C}_\omega)
   \nonumber\\
%--------------
  =&
%\mathbb N_\theta^\perp\cdot(D_\mu[\mathbf{V}]\mathbf{X}_\mu)
%   \nonumber\\
% &\quad
  -i\mathbf{\bar C}_\theta\cdot
    D_\mu[\mathbf{V}-\mathbf{X}]D_\mu[\mathbf{V}+\mathbf{X}]
    (\mathbf{C}_\theta-\mathbf{C}_\omega) ,
\end{align}
where we have used
$D_\mu[\mathbf{V} ]\mathbf{n}=0$, 
and
$\mathbf{A} \cdot (\mathbf{B} \times \mathbf{C})=\mathbf{B} \cdot (\mathbf{C} \times \mathbf{A})=\mathbf{C} \cdot (\mathbf{A} \times \mathbf{B})=0$ 
.
Thus we obtain
\begin{align}
 {\cal L}_{\rm GF+FP}^\theta
%= -i \bm\delta [\mathbf{\bar C}_\theta \cdot  \bm{\chi}]
=  \mathbf{N}_\theta \cdot D_\mu[\mathbf{V}]\mathbf{X}_\mu
-i\mathbf{\bar C}_\theta^\perp \cdot
    D_\mu[\mathbf{V}-\mathbf{X}]D_\mu[\mathbf{V}+\mathbf{X}]
    (\mathbf{C}_\theta^\perp-\mathbf{C}_\omega) ,
%\bm{\chi} := D_\mu[\mathbf{V}]\mathbf{X}_\mu .
\label{C26-FP1b}
\end{align}
where the conditions imply that $\mathbf{\bar C}_\theta$ and $\mathbf{C}_\theta$ have only the perpendicular components $\mathbf{\bar C}_\theta^\perp$ and $\mathbf{C}_\theta^\perp$ respectively. 

Another way to derive the GF+FP term is as follows.
The BRST transformations of $\mathbf{X}_\mu$ and $\mathbf{V}_\mu$ are given by
\begin{align}
%  \delta_{\omega,\theta} c_\mu(x) 
% =  \mathbf{n}(x) \cdot \partial_\mu \bm{\omega}(x)  + g(\mathbf{n}(x) \times \mathbf{A}_\mu(x)) \cdot (\bm{\omega}_\perp(x) - \bm{\theta}_\perp(x)) 
%  =  \mathbf{n}(x) \cdot \partial_\mu \bm{\omega}(x)  + \mathbf{n}(x) \cdot g(\mathbf{A}_\mu(x) \times    (\bm{\omega}_\perp(x) - \bm{\theta}_\perp(x)))    ,
%\\
  \bm\delta \mathbf{X}_\mu(x) 
= g \mathbf{X}_\mu(x) \times  (\mathbf{C}_\omega^\parallel(x)+\mathbf{C}_\theta^\perp(x)) + D_\mu[\mathbf{V}](\mathbf{C}_\omega^\perp(x)-\mathbf{C}_\theta^\perp(x))  ,
  \nonumber\\
 \bm\delta \mathbf{V}_\mu(x) 
= D_\mu[\mathbf{V}](\mathbf{C}_\omega^\parallel(x)+\mathbf{C}_\theta^\perp(x)) + g \mathbf{X}_\mu(x) \times  (\mathbf{C}_\omega^\perp(x)-\mathbf{C}_\theta^\perp(x))     ,
 \end{align}The ghost term is calculated as
\begin{align}
%{\cal L}_{\rm GF+FP}^\theta
%--------------
i \mathbf{\bar C}_\theta \cdot \bm\delta \bm{\chi}
=&  i\mathbf{\bar C}_\theta \cdot \bm\delta  \{ D_\mu[\mathbf{V}]\mathbf{X}_\mu \}
  \nonumber\\
=&  i\mathbf{\bar C}_\theta \cdot   \{ g \bm\delta\mathbf{V}_\mu \times \mathbf{X}_\mu  + D_\mu[\mathbf{V}]\bm\delta \mathbf{X}_\mu \} 
  \nonumber\\
=&  i\mathbf{\bar C}_\theta \cdot   \{  g\bm\delta\mathbf{A}_\mu \times \mathbf{X}_\mu - g\bm\delta\mathbf{X}_\mu  \times \mathbf{X}_\mu + D_\mu[\mathbf{V}]\bm\delta \mathbf{X}_\mu \} 
  \nonumber\\
=&  i\mathbf{\bar C}_\theta \cdot   \{  g (D_\mu[\mathbf{A}]\mathbf{C}_\omega) \times \mathbf{X}_\mu  
  \nonumber\\
 &\quad\quad -g  [g\mathbf{X}_\mu \times  (\mathbf{C}_\omega^\parallel +\mathbf{C}_\theta^\perp )] \times \mathbf{X}_\mu - gD_\mu[\mathbf{V}](\mathbf{C}_\omega^\perp -\mathbf{C}_\theta^\perp )   \times \mathbf{X}_\mu   
  \nonumber\\
 &\quad\quad +  D_\mu[\mathbf{V}] [ g \mathbf{X}_\mu \times  (\mathbf{C}_\omega^\parallel +\mathbf{C}_\theta^\perp )] + D_\mu[\mathbf{V}] D_\mu[\mathbf{V}](\mathbf{C}_\omega^\perp -\mathbf{C}_\theta^\perp )  \} 
  \nonumber\\
=&  i\mathbf{\bar C}_\theta \cdot   \{  g (D_\mu[\mathbf{A}]\mathbf{C}_\omega) \times \mathbf{X}_\mu  
  \nonumber\\
 &\quad\quad
+   g \mathbf{X}_\mu \times  [  g \mathbf{X}_\mu \times  (\mathbf{C}_\omega^\parallel +\mathbf{C}_\theta^\perp ) ] 
+  D_\mu[\mathbf{V}] [ g \mathbf{X}_\mu \times  (\mathbf{C}_\omega^\parallel +\mathbf{C}_\theta^\perp )]
  \nonumber\\
 &\quad\quad 
+ g\mathbf{X}_\mu \times D_\mu[\mathbf{V}](\mathbf{C}_\omega^\perp -\mathbf{C}_\theta^\perp )  
 + D_\mu[\mathbf{V}] D_\mu[\mathbf{V}](\mathbf{C}_\omega^\perp -\mathbf{C}_\theta^\perp )  \} 
  \nonumber\\
=&  i\mathbf{\bar C}_\theta \cdot   \{  g (D_\mu[\mathbf{A} ]\mathbf{C}_\omega) \times \mathbf{X}_\mu  
%  \nonumber\\ &+ 
 +  D_\mu[\mathbf{V}+\mathbf{X}] [ g \mathbf{X}_\mu \times  (\mathbf{C}_\omega^\parallel +\mathbf{C}_\theta^\perp )]
 \nonumber\\
 &\quad\quad\quad 
+   D_\mu[\mathbf{V}+\mathbf{X}] D_\mu[\mathbf{V}](\mathbf{C}_\omega^\perp -\mathbf{C}_\theta^\perp )  \} .
 %--------------
\end{align}

The result is rewritten in terms of the new variables as
\begin{align}
i \mathbf{\bar C}_\theta \cdot \bm\delta \bm{\chi}
=&  i\mathbf{\bar C}_\theta \cdot   \{  -g \mathbf{X}_\mu \times  D_\mu[\mathbf{V}+\mathbf{X}]\mathbf{C}_\omega   
%  \nonumber\\
% &\quad\quad
+   D_\mu[\mathbf{V}+\mathbf{X}] [ g \mathbf{X}_\mu \times  (\mathbf{C}_\omega - \mathbf{C}_\omega^\perp  +\mathbf{C}_\theta^\perp )]
  \nonumber\\
 &\quad\quad  
+   D_\mu[\mathbf{V}+\mathbf{X}] D_\mu[\mathbf{V}](\mathbf{C}_\omega^\perp -\mathbf{C}_\theta^\perp )  \} 
  \nonumber\\
=&  i\mathbf{\bar C}_\theta \cdot   \{  -g \mathbf{X}_\mu \times  D_\mu[\mathbf{V}+\mathbf{X}]\mathbf{C}_\omega   
%\nonumber\\
%&\quad\quad
+   D_\mu[\mathbf{V}+\mathbf{X}] [ g \mathbf{X}_\mu \times   \mathbf{C}_\omega  ]
  \nonumber\\
 &\quad\quad
-   D_\mu[\mathbf{V}+\mathbf{X}] [ g \mathbf{X}_\mu \times    \mathbf{C}_\omega^\perp ]
+   D_\mu[\mathbf{V}+\mathbf{X}] [ g \mathbf{X}_\mu \times   \mathbf{C}_\theta^\perp ]  \nonumber\\
 &\quad\quad  
+   D_\mu[\mathbf{V}+\mathbf{X}] D_\mu[\mathbf{V}] \mathbf{C}_\omega^\perp - D_\mu[\mathbf{V}+\mathbf{X}] D_\mu[\mathbf{V}] \mathbf{C}_\theta^\perp   \} 
  \nonumber\\
=&  i\mathbf{\bar C}_\theta \cdot   \{  
%-g \mathbf{X}_\mu \times  D_\mu[\mathbf{V}+\mathbf{X}]\mathbf{C}_\omega    
%  \nonumber\\
% &\quad\quad
  g D_\mu[\mathbf{V}]  \mathbf{X}_\mu \times   \mathbf{C}_\omega   
%+  g \mathbf{X}_\mu \times  D_\mu[\mathbf{V}+\mathbf{X}]   \mathbf{C}_\omega 
  \nonumber\\
 &\quad\quad  
+   D_\mu[\mathbf{V}+\mathbf{X}] D_\mu[\mathbf{V}-\mathbf{X}] \mathbf{C}_\omega^\perp - D_\mu[\mathbf{V}+\mathbf{X}] D_\mu[\mathbf{V}-\mathbf{X}] \mathbf{C}_\theta^\perp   \} 
 ,
\end{align}
where we have  used the Leibniz rule for the covariant derivative.

Thus we obtain
\begin{align}
 {\cal L}_{\rm GF+FP}^\theta
%= -i \bm\delta [\mathbf{\bar C}_\theta \cdot  \bm{\chi}]
=&  \mathbf{N}_\theta \cdot D_\mu[\mathbf{V}]\mathbf{X}_\mu
+i\mathbf{\bar C}_\theta^\perp \cdot (g D_\mu[\mathbf{V}]  \mathbf{X}_\mu \times   \mathbf{C}_\omega  )
%\nonumber\\ &
-i\mathbf{\bar C}_\theta^\perp
D_\mu[\mathbf{V}+\mathbf{X}]D_\mu[\mathbf{V}-\mathbf{X}]
    (\mathbf{C}_\theta^\perp-\mathbf{C}_\omega^\perp) .
%\bm{\chi} := D_\mu[\mathbf{V}]\mathbf{X}_\mu .
\label{C26-FP2b}
\end{align}
It is directly shown that (\ref{C26-FP2b}) agrees with (\ref{C26-FP1b}). 
%[Exercise-14] \marginpar{Ex-14}

%%%%%%%%%%%%%%%%%%%%%%%%%%%%%%%%%%%%%%%%%%%%%%%%%%
%%%%%%%%%%%%%%%%%%%%%%%%%%%%%%%%%%%%%%%%%%%%%%%%%%
\section{Field basis}\label{section:field-basis}
%%%%%%%%%%%%%%%%%%%%%%%%%%%%%%%%%%%%%%%%%%%%%%%%%%
%%%%%%%%%%%%%%%%%%%%%%%%%%%%%%%%%%%%%%%%%%%%%%%%%%

In order to define the independent field modes, we introduce the basis for $SU(N)$. 
The minimal option for $SU(3)$ has the maximal stability group  $\tilde H=U(2)=SU(2) \times U(1)$. Therefore, we choose a subset $\{ T_1, T_2, T_3, T_8 \}$ from $\{ T_1, T_2, \cdots,  T_8 \}$ as the generator of $u(2)$.
Taking into account $f_{abc}=0$ and $f_{ajk}=0$ ($j,k, \in \{ 1,2,3,8 \}; a,b,c \in \{ 4,5,6,7 \}$), we set up the basis satisfying 
\begin{align}
  [ \bm{e}_a(x) , \bm{e}_b(x)] =& if_{abj} \bm{u}_j(x) , 
\nonumber\\
 [ \bm{u}_j(x) , \bm{e}_a(x) ] =&  if_{abj} \bm{e}_b(x) ,  
\nonumber\\
 [ \bm{u}_j(x) , \bm{u}_k(x) ] =&  if_{jk\ell} \bm{e}_\ell(x) ,  
\quad 
(j,k,l \in \{ 1,2,3,8 \}; a,b \in \{ 4,5,6,7 \}) .
\label{C27-basismin}
\end{align} 
For $f_{abj}$, non-vanishing structure constants are 
\begin{align}
 & f_{453}=\frac12,  \ f_{458}=\frac{\sqrt{3}}{2},  \  f_{462}=\frac12, \   f_{471}=\frac12,  \  f_{561}=-\frac12,  \  f_{572}=\frac12, 
\nonumber\\&
 f_{673}=-\frac12, \ f_{678}=\frac{\sqrt{3}}{2} .
\end{align}
For $f_{jk\ell}$, non-vanishing structure constants are 
\begin{equation}
f_{123}=1 .
\end{equation}

The maximal option for $SU(3)$ has the maximal stability group $\tilde H=U(1) \times U(1)$.  Therefore, we choose $\{ T_3, T_8 \}$ as the generator of $u(1)+u(1)$.
Since $f_{jk\ell}=0$ and $f_{ajk}=0$ $(j,k,l  \in \{  3,8 \}; a  \in \{ 1,2,4,5,6,7 \})$, we set up the basis satisfying 
\begin{align}
  [ \bm{e}_a(x) , \bm{e}_b(x)] =& if_{abc} \bm{e}_c(x) + if_{abj} \bm{u}_j(x) , 
\nonumber\\
 [ \bm{u}_j(x) , \bm{e}_a(x) ] =&  if_{abj} \bm{e}_b(x) ,  
\nonumber\\
 [ \bm{u}_j(x) , \bm{u}_k(x) ] =&  0 ,  
\quad 
(j,k  \in \{  3,8 \}; a,b,c \in \{ 1,2,4,5,6,7 \}) .
\label{C27-basismax}
\end{align} 
For $f_{abc}$, non-vanishing structure constants are 
\begin{equation}
 f_{147}=\frac12,\ f_{156}=-\frac12, \ f_{246}=-\frac12, \ f_{257}=\frac12  .
\end{equation}
For $f_{abj}$, non-vanishing structure constants are 
\begin{equation}
  f_{123}=1, \ f_{453}=\frac12, \ f_{458}=\frac{\sqrt{3}}{2}, \ f_{673}=-\frac12, \ f_{678}=\frac{\sqrt{3}}{2}  .
\end{equation}

In particular, for $SU(2)$, $\tilde H=H=U(1)$, we set up
\begin{align}
  [ \bm{e}_a(x) , \bm{e}_b(x)] =&   if_{ab3} \bm{u}_3(x) , \quad 
( a,b  \in \{ 1,2  \}) ,
\nonumber\\
 [ \bm{u}_3(x) , \bm{e}_a(x) ] =&  if_{ab3} \bm{e}_b(x) ,  
\nonumber\\
 [ \bm{u}_3(x) , \bm{u}_3(x) ] =&  0 .  
\label{C27-basisSU2}
\end{align}

%%%%%%%%%%%%%%%%%%%%%%%%%%%%%%%%%%%%%%%%%%%%%%%%%%%%%%%%%%%%%
%%%%%%%%%%%%%%%%%%%%%%%%%%%%%%%%%%%%%%%%%%%%%%%%%%%%%%%%%%%%%
\section{Decomposition of the field strength in the minimal option}
\label{section:field-strength-minimal-2}%%%%%%%%%%%%%%%%%%%%%%%%%%%%%%%%%%%%%%%%%%%%%%%%%%%%%%%%%%%%%
%%%%%%%%%%%%%%%%%%%%%%%%%%%%%%%%%%%%%%%%%%%%%%%%%%%%%%%%%%%%%

In the minimal case, the change of variable is given as
\begin{subequations}
\begin{align}
 \mathscr{A}_\mu =  \mathscr{V}_\mu + \mathscr{X}_\mu 
=& \mathscr{C}_\mu + \mathscr{B}_\mu  + \mathscr{X}_\mu ,
\\
 \mathscr{X}_\mu =&   - ig^{-1} \frac{2(N-1)}{N} [  \bm{h}  ,  \mathscr{D}_\mu[\mathscr{A}] \bm{h} ]  
 \in \mathscr{G} - \tilde{\mathscr{H}} 
\\
 \mathscr{C}_\mu =&  \mathscr{A}_\mu - \frac{2(N-1)}{N} [ \bm{h} , [  \bm{h}  ,  \mathscr{A}_\mu ]] 
 \in \tilde{\mathscr{H}} ,
\\
 \mathscr{B}_\mu =&  ig^{-1} \frac{2(N-1)}{N} [  \bm{h}  ,  \partial_\mu \bm{h} ]  
 \in \mathscr{G} - \tilde{\mathscr{H}}  
  .
\end{align}
\end{subequations}
It has already been shown that%
\footnote{
This is shown in (3.68)--(3.70) of \cite{KSM08}.
}
\begin{align}
\mathscr{C}_\mu  \in \tilde{\mathscr{H}}  \Longleftrightarrow 
 [ \bm{h}, \mathscr{C}_\mu ] =& 0 , 
\end{align}
while it is easy to check that 
\begin{align}
 \mathscr{B}_\mu \in \mathscr{G} - \tilde{\mathscr{H}}
\Longleftrightarrow 
  \bm{a} \cdot \mathscr{B}_\mu   :=& 2{\rm tr}[ \bm{a}  \mathscr{B}_\mu ] = 0 \ \text{for arbitrary} \ \bm{a} \in \tilde{\mathscr{H}} , 
\nonumber\\
 \mathscr{X}_\mu \in \mathscr{G} - \tilde{\mathscr{H}}
\Longleftrightarrow 
  \bm{a} \cdot \mathscr{X}_\mu   :=&  2{\rm tr}[ \bm{a}  \mathscr{X}_\mu ] = 0 \ \text{for arbitrary} \ \bm{a} \in \tilde{\mathscr{H}} .  
\end{align}
In fact, for arbitrary $\bm{a} \in \tilde{\mathscr{H}}$ we find 
\begin{align}
  \bm{a} \cdot \mathscr{B}_\mu 
=&      ig^{-1} \frac{2(N-1)}{N} 2{\rm tr}(\bm{a}     [  \bm{h}  ,  \partial_\mu \bm{h} ] )  
%\nonumber\\
=     ig^{-1} \frac{2(N-1)}{N} 2{\rm tr}( [\bm{a}    ,  \bm{h} ]  \partial_\mu \bm{h} ) 
= 0  
 . 
\end{align}
Similarly, it is shown that $\bm{a} \cdot \mathscr{X}_\mu  =0$.%
\footnote{
This is shown in (3.65) of \cite{KSM08}.
}

First, we observe 
\begin{align}
  \mathscr{D}_\mu[\mathscr{V}] \mathscr{X}_\nu, \quad \mathscr{D}_\nu[\mathscr{V}] \mathscr{X}_\mu, \quad \mathscr{D}_\mu[\mathscr{V}] \mathscr{X}_\nu - \mathscr{D}_\nu[\mathscr{V}] \mathscr{X}_\mu  \in \mathscr{G} - \tilde{\mathscr{H}} . 
\end{align}
This is shown  using $\mathscr{D}_\mu[\mathscr{V}]\bm{h}=0$ 
\begin{align}
  \mathscr{D}_\mu[\mathscr{V}] \mathscr{X}_\nu 
=&  - ig^{-1} \frac{2(N-1)}{N} \mathscr{D}_\mu[\mathscr{V}] ([  \bm{h}  ,  \mathscr{D}_\nu[\mathscr{A}] \bm{h} ]  )
%\nonumber\\
=    - ig^{-1} \frac{2(N-1)}{N}  [  \bm{h}  ,  \mathscr{D}_\mu[\mathscr{V}] \mathscr{D}_\nu[\mathscr{A}] \bm{h} ]   
,  
\end{align}
and for arbitrary $\bm{a} \in \tilde{\mathscr{H}}$ 
\begin{align}
  \bm{a} \cdot \mathscr{D}_\mu[\mathscr{V}] \mathscr{X}_\nu 
=&   - ig^{-1} \frac{2(N-1)}{N} {\rm tr}(\bm{a}     [  \bm{h}  ,  \mathscr{D}_\mu[\mathscr{V}] \mathscr{D}_\nu[\mathscr{A}] \bm{h} ] )  
\nonumber\\
=&   - ig^{-1} \frac{2(N-1)}{N} {\rm tr}( [\bm{a}    ,  \bm{h} ]    \mathscr{D}_\mu[\mathscr{V}] \mathscr{D}_\nu[\mathscr{A}] \bm{h}  ) 
= 0  
 .  
\end{align}
Second, we show that 
\begin{align}
  -i g [ \mathscr{X}_\mu , \mathscr{X}_\nu ]   \in  \tilde{\mathscr{H}} 
\Longleftrightarrow
[ [ \mathscr{X}_\mu , \mathscr{X}_\nu ]   , \bm{h}] = 0
. 
\end{align}
This  is shown by taking he commutator with the color field 
\begin{align}
   [ [ \mathscr{X}_\mu , \mathscr{X}_\nu ]   , \bm{h}]
=&  \{ \mathscr{X}_\mu , \{ \mathscr{X}_\nu    , \bm{h} \} \} -  \{ \mathscr{X}_\nu , \{ \mathscr{X}_\mu    , \bm{h} \} \} 
\nonumber\\
=&   \{ \mathscr{X}_\mu , \frac{1}{N} (\mathscr{X}_\nu \cdot \bm{h})  \bm{1} + (\mathscr{X}_\nu * \bm{h})^C T_C  \} 
%\nonumber\\&
-  \{ \mathscr{X}_\nu , \frac{1}{N} (\mathscr{X}_\mu \cdot \bm{h})  \bm{1} + (\mathscr{X}_\mu * \bm{h})^C T_C  \} 
\nonumber\\
=&   \{ \mathscr{X}_\mu , (\mathscr{X}_\nu * \bm{h})^C T_C  \} -  \{ \mathscr{X}_\nu ,  (\mathscr{X}_\mu * \bm{h})^C T_C  \} 
 ,  
\end{align}
where we have used 
$\mathscr{X}_\mu \cdot \bm{h} =0$,
 and the identities:
\begin{align}
% \mathscr{U}  \mathscr{W}=&  [ \mathscr{U}, \mathscr{W} ]  + \{ \mathscr{U}, \mathscr{W} \} ,   
%\nonumber\\ 
 [ \mathscr{U}, \mathscr{W} ] =&  i (\mathscr{U} \times \mathscr{W}), 
\nonumber\\&
\ \mathscr{U} \times \mathscr{W}  := (\mathscr{U} \times \mathscr{W})^C T_C , \
 (\mathscr{U} \times \mathscr{W})^C := f_{ABC} \mathscr{U}^A \mathscr{W}^B ,
\nonumber\\
 \{ \mathscr{U}, \mathscr{W} \} =& \frac{1}{N} (\mathscr{U} \cdot \mathscr{W})  \bm{1} +  \mathscr{U} *\mathscr{W}, 
\nonumber\\&
%\nonumber\\
(\mathscr{U} \cdot \mathscr{W}) :=  \mathscr{U}^A \mathscr{W}^A ,  
\nonumber\\&
  \mathscr{U} *\mathscr{W} :=  (\mathscr{U} *\mathscr{W})^C T_C ,  \     (\mathscr{U} *\mathscr{W})^C := d_{ABC} \mathscr{U}^A \mathscr{W}^B .
\end{align}
Moreover, we have 
\begin{align}
  \bm{h} * \mathscr{X}_\mu   
  =& - ig^{-1} \frac{2(N-1)}{N} 
   \bm{h} * [  \bm{h}  ,  \mathscr{D}_\mu[\mathscr{A}] \bm{h} ]   
\nonumber\\ 
  =&  g^{-1} \frac{2(N-1)}{N} 
   \bm{h} * (  \bm{h} \times  \mathscr{D}_\mu[\mathscr{A}] \bm{h} )  
\nonumber\\ 
  =&  g^{-1} \frac{2(N-1)}{N} 
  (\bm{h} \times (  \bm{h} *  \mathscr{D}_\mu[\mathscr{A}] \bm{h} )    ) 
\nonumber\\ 
  =& -i g^{-1} \frac{2(N-1)}{N} 
  [\bm{h},  (  \bm{h} *  \mathscr{D}_\mu[\mathscr{A}] \bm{h} ) ]
\nonumber\\ 
  =& -i g^{-1} \frac{2(N-1)}{N} 
  [\bm{h},    \frac12 \mathscr{D}_\mu[\mathscr{A}]( \bm{h} * \bm{h} ) ]
\nonumber\\ 
  =& -  \frac12 \kappa i g^{-1} \frac{2(N-1)}{N} 
  [\bm{h},    \mathscr{D}_\mu[\mathscr{A}]  \bm{h}  ]
\nonumber\\ 
  =&  \frac12 \kappa \mathscr{X}_\mu   
 ,  
\end{align}
where we have used in the second equality the identity:%
\begin{equation}
  \bm{h} * (\bm{h} \times \bm{a}) = \bm{h} \times (\bm{h} * \bm{a}) , 
  \label{C27-hha}
\end{equation}
and
\begin{equation}
  \bm{h} * \bm{h}  = \kappa \bm{h} , \quad \kappa := \frac{2(2-N)}{\sqrt{2N(N-1)}} .
\end{equation}
Hence, we have 
\begin{align}
   [ [ \mathscr{X}_\mu , \mathscr{X}_\nu ]   , \bm{h}]
=& \frac12 \kappa \{ \mathscr{X}_\mu ,  \mathscr{X}_\nu  \} - \frac12 \kappa  \{ \mathscr{X}_\nu ,   \mathscr{X}_\mu   \} = 0
 .  
\end{align}
The identity (\ref{C27-hha}) follows from an identity:
\begin{equation}
  \{ \mathscr{A}, [\mathscr{A},C] \} = [\mathscr{A}, \{ \mathscr{A}, C \} ] = \mathscr{A}\mathscr{A}C-C\mathscr{A}\mathscr{A}. 
\end{equation}
In fact, we find 
\begin{align}
  \{ \mathscr{A} , i(\mathscr{A} \times C) \} =&  \frac{1}{N} (\mathscr{A} \cdot i(\mathscr{A} \times C))  \bm{1} + (\mathscr{A} * i(\mathscr{A} \times C))  
\nonumber\\ 
  =& i\mathscr{A} \times   \frac{1}{N} (\mathscr{A} \cdot C)  \bm{1} + i\mathscr{A} \times  (\mathscr{A} * C) ,
\end{align}
leads to 
\begin{equation}
   (\mathscr{A} *  (\mathscr{A} \times C))  
 =    \mathscr{A} \times  (\mathscr{A} * C) ,
\end{equation}
where we have used
$(\mathscr{A} \cdot  i(\mathscr{A} \times C))=2{\rm tr}(\mathscr{A}[\mathscr{A},C])=2{\rm tr}([\mathscr{A},\mathscr{A}]C)=0$ and $i\mathscr{A} \times \bm{1}=[\mathscr{A}, \bm{1}]=0$.

Third, we show that 
\begin{align}
  \mathscr{F}_{\mu\nu}[\mathscr{V}] 
=& \mathscr{F}_{\mu\nu}[\mathscr{B}]  
+ \mathscr{D}_\mu[\mathscr{B}] \mathscr{C}_\nu - \mathscr{D}_\nu[\mathscr{B}] \mathscr{C}_\mu 
%\nonumber\\&
-i g [ \mathscr{C}_\mu , \mathscr{C}_\nu ] \in  \tilde{\mathscr{H}} , 
\end{align}
The fourth term obeys
\begin{align}
 -i g [ \mathscr{C}_\mu , \mathscr{C}_\nu ] \in  \tilde{\mathscr{H}}
\Longleftrightarrow
[ [ \mathscr{C}_\mu , \mathscr{C}_\nu ] , \bm{h}] = 0 ,
\end{align}
which follows from the Jacobi identity:
\begin{align}
 [ [ \mathscr{C}_\mu , \mathscr{C}_\nu ]  , \bm{h} ]
 = -[ [  \bm{h}, \mathscr{C}_\mu ], \mathscr{C}_\nu ] - [ [ \mathscr{C}_\nu ,  \bm{h}] , \mathscr{C}_\mu ]
 ,
\end{align}
and $[  \bm{h}, \mathscr{C}_\mu ]=0$. 
The second and third terms obey
\begin{align}
\mathscr{D}_\mu[\mathscr{B}] \mathscr{C}_\nu  ,
\quad
 \mathscr{D}_\nu[\mathscr{B}] \mathscr{C}_\mu  \in  \tilde{\mathscr{H}} ,
\end{align}
which follows from 
\begin{align}
 [ \mathscr{D}_\mu[\mathscr{B}] \mathscr{C}_\nu , \bm{h} ]
= \mathscr{D}_\mu[\mathscr{B}] [\mathscr{C}_\nu , \bm{h} ] - [ \mathscr{C}_\nu , \mathscr{D}_\mu[\mathscr{B}] \bm{h} ]
= 0  ,
\end{align}
together with $[\mathscr{C}_\nu , \bm{h} ]=0$ and $\mathscr{D}_\mu[\mathscr{B}] \bm{h} = \mathscr{D}_\mu[\mathscr{V}] \bm{h} =0$.

The first term obeys
\begin{align}
 \mathscr{F}_{\mu\nu}[\mathscr{B}]  := \partial_\mu \mathscr{B}_\nu - \partial_\nu \mathscr{B}_\mu 
-i g [ \mathscr{B}_\mu , \mathscr{B}_\nu ]
\in  \tilde{\mathscr{H}} .
\end{align}
This is shown by rewriting the field strength as 
\begin{align}
  \mathscr{F}_{\mu\nu}[\mathscr{B}] 
%=&  \partial_\mu \mathscr{B}_\nu - \partial_\nu \mathscr{B}_\mu -i g [ \mathscr{B}_\mu , \mathscr{B}_\nu ]
%\nonumber\\
=&  ig^{-1} \frac{2(N-1)}{N} ( \partial_\mu [  \bm{h}  ,  \partial_\nu \bm{h} ] 
 -    \partial_\nu [  \bm{h}  ,  \partial_\mu \bm{h} ] )
-i g [ \mathscr{B}_\mu , \mathscr{B}_\nu ]
\nonumber\\
=&  ig^{-1} \frac{2(N-1)}{N} 2 [  \partial_\mu \bm{h}  ,  \partial_\nu \bm{h} ]  
-i g [ \mathscr{B}_\mu , \mathscr{B}_\nu ]
 .
\end{align}
We can observe the fact:
\begin{align}
 [  \partial_\mu \bm{h}  ,  \partial_\nu \bm{h} ]  \in  \tilde{\mathscr{H}} .
\end{align}
This is shown by taking the commutator with the color field  
\begin{align}
   [ [ \partial_\mu \bm{h}  ,  \partial_\nu \bm{h} ]   , \bm{h}]
=&  \{ \partial_\mu \bm{h} , \{ \partial_\nu \bm{h}    , \bm{h} \} \} -  \{ \partial_\nu \bm{h} , \{ \partial_\mu \bm{h}    , \bm{h} \} \} 
\nonumber\\
=&   \{ \partial_\mu \bm{h} , \frac{1}{N} (\partial_\nu \bm{h} \cdot \bm{h})  \bm{1} + (\partial_\nu \bm{h} * \bm{h})^C T_C  \} 
%\nonumber\\&
-  \{ \partial_\nu \bm{h} , \frac{1}{N} (\partial_\mu \bm{h} \cdot \bm{h})  \bm{1} + (\partial_\mu \bm{h} * \bm{h})^C T_C  \} = 0
 .  
\end{align}
where we have used 
\begin{align}
  \partial_\nu \bm{h} * \bm{h}
 = \frac12 \partial_\nu (\bm{h} * \bm{h})
 = \frac12 \kappa \partial_\nu  \bm{h}   
 ,  
\end{align}
and
\begin{align}
 \partial_\nu \bm{h} \cdot \bm{h} 
 = 2{\rm tr}(\partial_\nu \bm{h}   \bm{h})
 = \partial_\nu {\rm tr}(\bm{h}\bm{h}) 
= 0 ,  
\end{align}
which follows from 
\begin{align}
\bm{h}  \bm{h} =& \frac12  \{ \bm{h}, \bm{h} \} + \frac12 [\bm{h} , \bm{h} ] = \frac12  \{ \bm{h}, \bm{h} \} ,
\nonumber\\
&  \{ \bm{h}, \bm{h} \} =  \frac{1}{N} (\bm{h} \cdot \bm{h})  \bm{1} + (\bm{h} *\bm{h})^C T_C 
=  \frac{1}{N}   \bm{1} + \kappa \bm{h}  .
\end{align}
While, we obtain
\begin{align}
 -i g [ \mathscr{B}_\mu , \mathscr{B}_\nu ] \in  \tilde{\mathscr{H}} .
\end{align}
This follows by taking the commutator with the color field reads
\begin{align}
   [ [ \mathscr{B}_\mu , \mathscr{B}_\nu ]   , \bm{h}]
=&  \{ \mathscr{B}_\mu , \{ \mathscr{B}_\nu    , \bm{h} \} \} -  \{ \mathscr{B}_\nu , \{ \mathscr{B}_\mu    , \bm{h} \} \} 
\nonumber\\
=&   \{ \mathscr{B}_\mu , \frac{1}{N} (\mathscr{B}_\nu \cdot \bm{h})  \bm{1} + (\mathscr{B}_\nu * \bm{h})^C T_C  \}
%\nonumber\\& 
 -  \{ \mathscr{B}_\nu , \frac{1}{N} (\mathscr{B}_\mu \cdot \bm{h})  \bm{1} + (\mathscr{B}_\mu * \bm{h})^C T_C  \} 
\nonumber\\
=&   \{ \mathscr{B}_\mu , (\mathscr{B}_\nu * \bm{h})^C T_C  \} -  \{ \mathscr{B}_\nu ,  (\mathscr{B}_\mu * \bm{h})^C T_C  \} 
\nonumber\\
=& \frac12 \kappa \{ \mathscr{B}_\mu ,  \mathscr{B}_\nu  \} - \frac12 \kappa  \{ \mathscr{B}_\nu ,   \mathscr{B}_\mu   \} = 0
 .  \end{align}
where we have used 
$ \mathscr{B}_\mu \cdot \bm{h} =0$
and
\begin{align}
  \bm{h} * \mathscr{B}_\mu   
  =&   ig^{-1} \frac{2(N-1)}{N} 
   \bm{h} * [  \bm{h}  ,  \partial_\mu \bm{h} ]   
\nonumber\\ 
  =&  - g^{-1} \frac{2(N-1)}{N} 
   \bm{h} * (  \bm{h} \times  \partial_\mu \bm{h} )    
\nonumber\\ 
  =& - g^{-1} \frac{2(N-1)}{N} 
  (\bm{h} \times (  \bm{h} *  \partial_\mu \bm{h} )    ) 
\nonumber\\ 
  =&  i g^{-1} \frac{2(N-1)}{N} 
  [\bm{h},  (  \bm{h} *  \partial_\mu \bm{h} ) ]
\nonumber\\ 
  =&  i g^{-1} \frac{2(N-1)}{N} 
  [\bm{h},    \frac12 \partial_\mu ( \bm{h} * \bm{h} ) ]
\nonumber\\ 
  =&    \frac12 \kappa i g^{-1} \frac{2(N-1)}{N} 
  [\bm{h},    \partial_\mu  \bm{h}  ]
\nonumber\\ 
  =&  \frac12 \kappa \mathscr{B}_\mu   
 .  
\end{align}

Thus, the field strength $\mathscr{F}_{\mu\nu}[\mathscr{A}]$ in the minimal case is decomposed into the $\tilde{H}$-commutative part $\mathscr{F}_{\mu\nu}^{\tilde{H}}$ and the remaining $\tilde{H}$-non-commutative part $\mathscr{F}_{\mu\nu}^{G/\tilde{H}}$, which are orthogonal to each other:
\begin{align}
  \mathscr{F}_{\mu\nu}[\mathscr{A}]  
=& \mathscr{F}_{\mu\nu}^{\tilde{H}} + \mathscr{F}_{\mu\nu}^{G/\tilde{H}}  , 
\nonumber\\
 & \mathscr{F}_{\mu\nu}^{\tilde{H}} =  \mathscr{F}_{\mu\nu}[\mathscr{V}] -i g [ \mathscr{X}_\mu , \mathscr{X}_\nu ]  
 \in  \tilde{\mathscr{H}},  
\nonumber\\
 & \mathscr{F}_{\mu\nu}^{G/\tilde{H}} =  \mathscr{D}_\mu[\mathscr{V}] \mathscr{X}_\nu - \mathscr{D}_\nu[\mathscr{V}] \mathscr{X}_\mu 
 \in \mathscr{G} - \tilde{\mathscr{H}} .
\end{align}

%%%%%%%%%%%%%%%%%%%%%%%%%%%%%%%%%%%%%%%%%%%%%%%%%%%%%%%%%%%%
%%%%%%%%%%%%%%%%%%%%%%%%%%%%%%%%%%%%%%%%%%%%%%%%%%%%%%%%%%%%
%%%%%%%%%%%%%%%%%%%%%%%%%%%%%%%%%%%%%%%%%%%%%%%%%%%%%%%%%%%%
%%%%%%%%%%%%%%%%%%%%%%%%%%%%%%%%%%%%%%%%%%%%%%%%%%%%%%%%%%%%
\section{Reduction condition  for $SU(N)$ and the ghost term in the minimal option}\label{sec:FP-ghost-SUN}
%%%%%%%%%%%%%%%%%%%%%%%%%%%%%%%%%%%%%%%%%%%%%%%%%%%%%%%%%%%%
%%%%%%%%%%%%%%%%%%%%%%%%%%%%%%%%%%%%%%%%%%%%%%%%%%%%%%%%%%%%
%%%%%%%%%%%%%%%%%%%%%%%%%%%%%%%%%%%%%%%%%%%%%%%%%%%%%%%%%%%%
%%%%%%%%%%%%%%%%%%%%%%%%%%%%%%%%%%%%%%%%%%%%%%%%%%%%%%%%%%%%

For $SU(N)$ Yang-Mills theory in the minimal option, the Faddeev-Popov ghost term (\ref{C27-FP-ghost-SUN}) associated with the reduction condition is calculated as follows \cite{KSS14}. 
By using
\begin{align}
 \bm{\chi}  = D_\mu[\mathbf{A}]\mathbf{X}_\mu 
= D_\mu[\mathbf{A}] (g^{-1}\mathbf{n} \times D_\mu[\mathbf{A}]\mathbf{n})
= g^{-1} \mathbf{n}\times D_\mu[\mathbf{A}]D_\mu[\mathbf{A}]\mathbf{n}
,
\label{C27-chi-A}
\end{align}
the ghost term is calculated as
\begin{align}
%{\cal L}_{\rm GF+FP}^\theta
%--------------
i \mathbf{\bar C}_\theta \cdot \bm\delta \bm{\chi}
=& g^{-1} i\mathbf{\bar C}_\theta \cdot \bm\delta  \{ \mathbf{n}\times D_\mu[\mathbf{A}]D_\mu[\mathbf{A}]\mathbf{n} \}
  \nonumber\\
  =& g^{-1} i
  \mathbf{\bar C}_\theta \cdot  \{ \bm\delta  \mathbf{n}\times D_\mu[\mathbf{A}]D_\mu[\mathbf{A}]\mathbf{n} 
+ \mathbf{n}\times \bm\delta  ( D_\mu[\mathbf{A}]D_\mu[\mathbf{A}]\mathbf{n} )
\}
  \nonumber\\
  =& g^{-1} i
  \mathbf{\bar C}_\theta \cdot  \{ \bm\delta  \mathbf{n}\times D_\mu[\mathbf{A}]D_\mu[\mathbf{A}]\mathbf{n} 
+ \mathbf{n}\times  D_\mu[\mathbf{A}] \bm\delta  (D_\mu[\mathbf{A}]\mathbf{n} )
  \nonumber\\
   &\qquad\qquad\quad
+ \mathbf{n} \times  ( g\bm\delta \mathbf{A}_\mu \times D_\mu[\mathbf{A}]\mathbf{n} )
\}
  \nonumber\\
  =& 
%-i\bm\delta 
%   \mathbf{\bar C}_\theta\cdot(D_\mu[\mathbf{V}]\mathbf{X}_\mu)
%   \nonumber\\
% &\quad
    g^{-1} i \mathbf{\bar C}_\theta \cdot
  \bigl[
  \bm\delta \mathbf{n}\times D_\mu[\mathbf{A}]D_\mu[\mathbf{A}]\mathbf{n}
  \nonumber\\
 &\qquad\qquad\quad
  +\mathbf{n}\times D_\mu[\mathbf{A}]D_\mu[\mathbf{A}]\bm\delta \mathbf{n}
%  \nonumber\\
%&\qquad\qquad\quad
  +\mathbf{n}\times D_\mu[\mathbf{A}]
   \left(g\bm\delta \mathbf{A}_\mu\times\mathbf{n}\right)
  \nonumber\\
 &\qquad\qquad\quad
  +\mathbf{n}\times
   \left(g\bm\delta \mathbf{A}_\mu\times D_\mu[\mathbf{A}]\mathbf{n}\right)
  \bigr]
  \nonumber\\
%--------------
=&
%\mathbb N_\theta\cdot(D_\mu[\mathbf{V}]\mathbf{X}_\mu)
%   \nonumber\\
% &\quad
   i\mathbf{\bar C}_\theta\cdot
  \bigl[
  (\mathbf{n}\times\mathbf{C}_\theta)\times D_\mu[\mathbf{A}]D_\mu[\mathbf{A}]\mathbf{n}
  \nonumber\\
 &\qquad\qquad\quad
  +\mathbf{n}\times D_\mu[\mathbf{A}]D_\mu[\mathbf{A}](\mathbf{n}\times\mathbf{C}_\theta)
  \nonumber\\
 &\qquad\qquad\quad
  +\mathbf{n}\times D_\mu[\mathbf{A}]
   \left(D_\mu[\mathbf{A}]\mathbf{C}_\omega\times\mathbf{n}\right)
  \nonumber\\
 &\qquad\qquad\quad
  +\mathbf{n}\times
   \left(D_\mu[\mathbf{A}]\mathbf{C}_\omega\times D_\mu[\mathbf{A}]\mathbf{n}\right)
  \bigr]  
  \nonumber\\
%--------------
=&
%\mathbb N_\theta\cdot(D_\mu[\mathbf{V}]\mathbf{X}_\mu)
%   \nonumber\\
% &\quad
   i\mathbf{\bar C}_\theta \cdot
  \bigl[
  (\mathbf{n}\times\mathbf{C}_\theta)\times D_\mu[\mathbf{A}]D_\mu[\mathbf{A}]\mathbf{n}
  \nonumber\\
 &\qquad\qquad\quad
  +\mathbf{n}\times D_\mu[\mathbf{A}]D_\mu[\mathbf{A}](\mathbf{n}\times\mathbf{C}_\theta)
  \nonumber\\
 &\qquad\qquad\quad
  +\mathbf{n}\times 
   \left( D_\mu[\mathbf{A}] D_\mu[\mathbf{A}]\mathbf{C}_\omega\times\mathbf{n}\right)
  \nonumber\\
 &\qquad\qquad\quad
  + 2 \mathbf{n}\times
   \left(D_\mu[\mathbf{A}]\mathbf{C}_\omega\times D_\mu[\mathbf{A}]\mathbf{n}\right)
  \bigr] .
  \label{C27-SUN-ghost2}
\end{align}
First, we consider the third and fourth terms on the right-hand side of (\ref{C27-SUN-ghost2}). 

The third term of (\ref{C27-SUN-ghost2}) is rewritten using the decomposition $\mathscr{F}=\mathscr{F}_{\tilde{H}} + \mathscr{F}_{G/\tilde{H}}$
as 
\begin{align}
 & i\mathbf{\bar C}_\theta \cdot
 [ \mathbf{n}\times 
   \left( D_\mu[\mathbf{A}] D_\mu[\mathbf{A}]\mathbf{C}_\omega\times\mathbf{n}\right) ]
 \nonumber\\
%--------------
=&   i\mathbf{\bar C}_\theta \cdot
 [ - \mathbf{n}\times 
   \left( \mathbf{n} \times  D_\mu[\mathbf{A}] D_\mu[\mathbf{A}]\mathbf{C}_\omega \right) ] 
 \nonumber\\
%--------------
=&  i\mathbf{\bar C}_\theta \cdot 
\frac{N}{2(N-1)}
 [ \left( D_\mu[\mathbf{A}] D_\mu[\mathbf{A}]\mathbf{C}_\omega \right) -( D_\mu[\mathbf{A}] D_\mu[\mathbf{A}]\mathbf{C}_\omega )_{\tilde H}] 
 \nonumber\\
%--------------
=&  \frac{N}{2(N-1)}
 i\mathbf{\bar C}_\theta \cdot 
  \left( D_\mu[\mathbf{A}] D_\mu[\mathbf{A}]\mathbf{C}_\omega \right)  ,
\end{align}
since $\mathbf{\bar C}_\theta$ has no $\tilde H$ commutative part:
\begin{align}
   i\mathbf{\bar C}_\theta \cdot 
( D_\mu[\mathbf{A}] D_\mu[\mathbf{A}]\mathbf{C}_\omega )_{\tilde H} 
= 0 .
\end{align}
The fourth term of (\ref{C27-SUN-ghost2}) is rewritten  as 
  \begin{align}
 & i\mathbf{\bar C}_\theta \cdot
[2 \mathbf{n}\times
   \left(D_\mu[\mathbf{A}]\mathbf{C}_\omega\times D_\mu[\mathbf{A}]\mathbf{n}\right) ]
   \nonumber\\
%--------------
=& - 2 i\mathbf{\bar C}_\theta \cdot [ D_\mu[\mathbf{A}]\mathbf{C}_\omega \times
   \left( D_\mu[\mathbf{A}]\mathbf{n} \times  \mathbf{n}  \right)
    +  D_\mu[\mathbf{A}]\mathbf{n} \times
   \left( \mathbf{n} \times D_\mu[\mathbf{A}]\mathbf{C}_\omega   \right) 
  \bigr] 
 \nonumber\\
%--------------
=& - 2 i\mathbf{\bar C}_\theta \cdot [ D_\mu[\mathbf{A}]\mathbf{C}_\omega \times
   \left( (g\mathbf{X}_\mu \times \mathbf{n}) \times  \mathbf{n}  \right)
    +  (g\mathbf{X}_\mu \times \mathbf{n}) \times
   \left( \mathbf{n} \times D_\mu[\mathbf{A}]\mathbf{C}_\omega   \right) 
  \bigr] 
 \nonumber\\
%--------------
=&   2 i\mathbf{\bar C}_\theta \cdot \left[ \frac{N}{2(N-1)} D_\mu[\mathbf{A}]\mathbf{C}_\omega \times
    g\mathbf{X}_\mu  
    + (\mathbf{n} \times g\mathbf{X}_\mu ) \times
   \left( \mathbf{n} \times D_\mu[\mathbf{A}]\mathbf{C}_\omega   \right) 
  \right] 
 \nonumber\\
%--------------
=&  \frac{N}{2(N-1)}   i\mathbf{\bar C}_\theta \cdot  (D_\mu[\mathbf{A}]\mathbf{C}_\omega \times
    2g\mathbf{X}_\mu )
  ,
\end{align}
where we have used the Jacobi identity in the first equality, the defining equation 
$D_\mu[\mathbf{V}]\mathbf{n}=0$ in the second equality, the expression of $\mathbf{X}_\mu$ in the third equality.
The final equality follows from the fact: 
In the minimal option,  
$(\mathbf{n} \times g\mathbf{X}_\mu ) \times
   \left( \mathbf{n} \times D_\mu[\mathbf{A}]\mathbf{C}_\omega   \right)$
   has only the $\tilde H$ commutative part, which yields 
\begin{align}
  i\mathbf{\bar C}_\theta \cdot [   (\mathbf{n} \times g\mathbf{X}_\mu ) \times
   \left( \mathbf{n} \times D_\mu[\mathbf{A}]\mathbf{C}_\omega   \right) 
  \bigr]  = 0
  ,
\end{align}
since $\mathbf{\bar C}_\theta$ has no $\tilde H$ commutative part.
This is shown as follows.

Note that $\mathbf{n} \times \mathbf{Y}$ has the vanishing $\tilde H$ commutative part, i.e., $(\mathbf{n} \times \mathbf{Y})_{\tilde H}=0$, 
\begin{align}
  \mathbf{n} \times \mathbf{Y} = (\mathbf{n} \times \mathbf{Y})_{G/\tilde H}
  .
\end{align}
When the commutator of two Lie algebras $\mathbf{A}$ and $\mathbf{B}$ in $G/\tilde H$ is written only in terms of $\tilde H$:
\begin{align}
    (\mathbf{A})_{G/\tilde H}  \times (\mathbf{B})_{G/\tilde H} = \mathbf{C}_{\tilde H} 
\Longleftrightarrow 
 [\mathscr{G} - \mathscr{\tilde H}, \mathscr{G} - \mathscr{\tilde H} ] \subset \mathscr{\tilde H} 
  ,
  \label{C27-alge-id}
\end{align}
the coset space $G/H$ is called the \textbf{symmetric space} or the \textbf{homogeneous space}. 
It is known that the coset space $SU(N)/U(N-1)$ is a symmetric space, while this is not the case for the the coset space $SU(N)/U(1)^{N-1}$. 
Hence, we can apply the relation (\ref{C27-alge-id}) to the minimal option, while this is not the case for the maximal option.

Therefore, the third and fourth terms are 
\begin{align}
 &  \frac{N}{2(N-1)}
 i\mathbf{\bar C}_\theta \cdot [
   D_\mu[\mathbf{A}] D_\mu[\mathbf{A}]\mathbf{C}_\omega 
+      D_\mu[\mathbf{A}]\mathbf{C}_\omega \times
    2g\mathbf{X}_\mu   ]
  \nonumber\\
  =& \frac{N}{2(N-1)} 
i\mathbf{\bar C}_\theta \cdot [
   (D_\mu[\mathbf{V}] + g \mathbf{X}_\mu \times ) D_\mu[\mathbf{A}]\mathbf{C}_\omega 
-   2g\mathbf{X}_\mu \times D_\mu[\mathbf{A}]\mathbf{C}_\omega        ] 
  \nonumber\\
  =& \frac{N}{2(N-1)} 
i\mathbf{\bar C}_\theta \cdot [
   (D_\mu[\mathbf{V}] - g \mathbf{X}_\mu \times ) D_\mu[\mathbf{A}]\mathbf{C}_\omega  ]   
  \nonumber\\
  =& \frac{N}{2(N-1)} 
i\mathbf{\bar C}_\theta \cdot 
    D_\mu[\mathbf{V}-\mathbf{X}] D_\mu[\mathbf{A}]\mathbf{C}_\omega 
  .
\label{C27-FP-2}
\end{align}

Next, the first and second terms on the right-hand side of (\ref{C27-SUN-ghost2}) are 
\begin{align}
&   i\mathbf{\bar C}_\theta \cdot
  [  (\mathbf{n}\times\mathbf{C}_\theta)\times D_\mu[\mathbf{A}]D_\mu[\mathbf{A}]\mathbf{n}
%  \nonumber\\
% &\qquad\qquad\quad
  +\mathbf{n}\times D_\mu[\mathbf{A}]D_\mu[\mathbf{A}](\mathbf{n}\times\mathbf{C}_\theta) ]
  \nonumber\\
  =& i\mathbf{\bar C}_\theta \cdot
  \bigl[
  (\mathbf{n}\times\mathbf{C}_\theta)\times D_\mu[\mathbf{A}]D_\mu[\mathbf{A}]\mathbf{n}
%  \nonumber\\
% &\qquad\qquad\quad
  +\mathbf{n}\times ( D_\mu[\mathbf{A}]D_\mu[\mathbf{A}]\mathbf{n}\times\mathbf{C}_\theta)   
\nonumber\\
 &\qquad \quad 
  +\mathbf{n}\times  (D_\mu[\mathbf{A}] \mathbf{n}\times D_\mu[\mathbf{A}] \mathbf{C}_\theta)  
%  \nonumber\\
% &\qquad\qquad\quad
  +\mathbf{n}\times  (\mathbf{n}\times D_\mu[\mathbf{A}] D_\mu[\mathbf{A}] \mathbf{C}_\theta) ] 
.  
\label{C27-ghost-2b}
\end{align}
The first and the second terms of (\ref{C27-ghost-2b}) are summed to give vanishing contribution:
\begin{align}
   & i\mathbf{\bar C}_\theta \cdot
   [   (\mathbf{n}\times\mathbf{C}_\theta)\times D_\mu[\mathbf{A}]D_\mu[\mathbf{A}]\mathbf{n}  
  +\mathbf{n}\times ( D_\mu[\mathbf{A}]D_\mu[\mathbf{A}]\mathbf{n}\times\mathbf{C}_\theta) ] 
  \nonumber\\
=& i\mathbf{\bar C}_\theta \cdot
   [  D_\mu[\mathbf{A}]D_\mu[\mathbf{A}]\mathbf{n} \times  (\mathbf{C}_\theta \times  \mathbf{n})
  +\mathbf{n}\times ( D_\mu[\mathbf{A}]D_\mu[\mathbf{A}]\mathbf{n}\times\mathbf{C}_\theta) ] 
  \nonumber\\
=& i\mathbf{\bar C}_\theta \cdot
   [-  \mathbf{C}_\theta \times  (  \mathbf{n}\times D_\mu[\mathbf{A}]D_\mu[\mathbf{A}]\mathbf{n} ) ] 
  \nonumber\\
=& i\mathbf{\bar C}_\theta \cdot
   [-    \mathbf{C}_\theta \times gD_\mu[\mathbf{A}]\mathbf{X}_\mu ] 
  \nonumber\\
=& i\mathbf{\bar C}_\theta \cdot
   [-    \mathbf{C}_\theta \times gD_\mu[\mathbf{V}]\mathbf{X}_\mu ] 
  \nonumber\\
=& i\mathbf{\bar C}_\theta \cdot
   \left[  \mathbf{C}_\theta \times gD_\mu[\mathbf{V}] \frac{2(N-1)}{N}  (\mathbf{n} \times (\mathbf{n} \times \mathbf{X}_\mu ) ) \right] 
  \nonumber\\
=& g \frac{2(N-1)}{N}  i\mathbf{\bar C}_\theta \cdot
   \left[  \mathbf{C}_\theta \times ( \mathbf{n} \times (\mathbf{n} \times D_\mu[\mathbf{V}] \mathbf{X}_\mu ) ) \right] 
  \nonumber\\
=& g \frac{2(N-1)}{N}  
   \left[  ( \mathbf{n} \times (\mathbf{n} \times D_\mu[\mathbf{V}] \mathbf{X}_\mu ) ) \cdot (i\mathbf{\bar C}_\theta   \times \mathbf{C}_\theta)   \right] 
%\nonumber\\
=  0 ,
\end{align}
where we have used the Jacobi identity in the second equality, (\ref{C27-chi-A}) in the third equality, the defining equation  (\ref{C27-defXL2}) for $\mathbf{X}_\mu$ in the fifth equality, $D_\mu[\mathbf{V}]\mathbf{n} =0$ in the sixth equality, and the fact that
$
( \mathbf{n} \times (\mathbf{n} \times D_\mu[\mathbf{V}] \mathbf{X}_\mu ) )
=(  D_\mu[\mathbf{V}] \mathbf{X}_\mu ) )_{G/\tilde H}
$
and
$
\mathbf{\bar C}_\theta   \times \mathbf{C}_\theta
= (\mathbf{\bar C}_\theta)_{G/\tilde H}   \times (\mathbf{C}_\theta)_{G/\tilde H}
= (\mathbf{\bar C}_\theta   \times \mathbf{C}_\theta)_{\tilde H}
$
due to (\ref{C27-alge-id}) 
yields
$
(  D_\mu[\mathbf{V}] \mathbf{X}_\mu )_{G/\tilde H} \cdot (\mathbf{\bar C}_\theta   \times \mathbf{C}_\theta)_{\tilde H}=0
$
in the last step.

The third and fourth terms of (\ref{C27-ghost-2b}) have the same form as the third and fourth terms of (\ref{C27-SUN-ghost2}).  Similarly, we can show    
\begin{align}
 & i\mathbf{\bar C}_\theta \cdot
   [
    \mathbf{n}\times  (D_\mu[\mathbf{A}] \mathbf{n}\times D_\mu[\mathbf{A}] \mathbf{C}_\theta)   
%  \nonumber\\
% &\qquad\qquad\quad
 + \mathbf{n}\times  (\mathbf{n}\times D_\mu[\mathbf{A}] D_\mu[\mathbf{A}] \mathbf{C}_\theta) ] 
  \nonumber\\
  =& - \frac{N}{2(N-1)} 
i\mathbf{\bar C}_\theta \cdot 
    D_\mu[\mathbf{V}-\mathbf{X}] D_\mu[\mathbf{A}]\mathbf{C}_\theta 
.  
\label{C27-FP-1}
\end{align}
Finally, summing up (\ref{C27-FP-2}) and (\ref{C27-FP-1}) leads to  
\begin{align}
i \mathbf{\bar C}_\theta \cdot \bm\delta \bm{\chi}
=    \frac{N}{2(N-1)} 
i\mathbf{\bar C}_\theta \cdot 
    D_\mu[\mathbf{V}-\mathbf{X}] D_\mu[\mathbf{A}](\mathbf{C}_\omega  - \mathbf{C}_\theta )
.  
\end{align}

%%%%%%%%%%%%%%%%%%%%%%%%%%%%%%%%%%%%%%%%%%%%%%%%%%%%%%%%%%%%
\section{Integration formula}\label{section:integration-formula}
%%%%%%%%%%%%%%%%%%%%%%%%%%%%%%%%%%%%%%%%%%%%%%%%%%%%%%%%%%%%
%%%%%%%%%%%%%%%%%%%%%%%%%%%%%%%%%%%%%%%%%%%%%%%%%%%%%%%%%%%%
%%%%%%%%%%%%%%%%%%%%%%%%%%%%%%%%%%%%%%%%%%%%%%%%%%%%%%%%%%%%
%%%%%%%%%%%%%%%%%%%%%%%%%%%%%%%%%%%%%%%%%%%%%%%%%%%%%%%%%%%%

We can replace the trace of the operator $\mathscr{O}$ with the integral:
\begin{equation}
 \mathcal{N}^{-1} {\rm tr}_{\rm R}(\mathscr{O}) 
 = \int d\mu({g}_0) \left< {g}_0, \Lambda \right| \mathscr{O}\left| {g}_0, \Lambda \right> .
% \label{C29-trace-integral}
\end{equation}
The relation (\ref{C29-trace-integral}) is proved as follows.
%For the proof, see the chapter of Maximal stability subgroup and coherent state. 

First, we show the \textbf{completeness relation}:%
\footnote{
The following relationships hold even if we replace $\xi \in G/\tilde{H}$ by a general element $g \in G$, since they hold on a reference state $\left|  \bm{\Lambda} \right>$.
} 
\begin{equation}
 \int \left| g , \Lambda \right> d\mu(g ) \left< g , \Lambda \right|  = \frac{1}{d_R}{\bf 1}
  ,
  \label{rel1}
\end{equation}
where $d\mu(g )$ is the \textbf{invariant measure} of $G$. 
For this purpose, we derive an important relation: For any operator $\mathcal{O}$ which does not depend on the 
group $g$ explicitly,, the relation holds:
\begin{equation}
 \int  d\mu(g) g \mathcal{O} g^\dagger 
 = \frac{1}{d_R}{\rm tr}(\mathcal{O}) \mathbf{1}  
 ,
 \label{form2}
\end{equation}
which  follows from 
\begin{align}
  \int  d\mu(g) (g \mathcal{O} g^\dagger)_{ab} 
  = \int  d\mu(g) (g)_{ac} \mathcal{O}_{cd} (g^\dagger)_{db} 
  = \frac{1}{d_R} \delta_{ab} \delta_{cd}  \mathcal{O}_{cd} 
  = \frac{1}{d_R} \delta_{ab}  {\rm tr}(\mathcal{O}) 
  .
\end{align}
Here we have used the integration formula for the invariant measure, i.e., \textbf{Haar measure} $dU=d\mu(G)$ of $SU(N)$:
%[Exercise-1] \marginpar{Ex-1}
\begin{equation}
 \int  dU \  U_{ac}(R)   U^\dagger_{db}(R') 
 = \frac{1}{d_R} \delta_{ab} \delta_{cd} \delta_{RR'}
 ,
 \label{Haar-integral}
\end{equation}
where $U(R)$ is the representation matrix belonging to the representation $R$.

Using the relation (\ref{form2}), indeed, the completeness (\ref{rel1}) follows 
\begin{align}
  \int \left| g , \Lambda \right> d\mu(g ) \left< g , \Lambda \right|  
 =&   \int  d\mu(g )  g  \left|  \Lambda \right> \left< \Lambda \right| g^\dagger    
% \nonumber\\
 =   \int  d\mu(g) g  \rho  g^\dagger 
% \nonumber\\
=   \frac{1}{d_R} {\rm tr}(\rho) \mathbf{1} 
 \nonumber\\
=&   \frac{1}{d_R}  \mathbf{1} 
  . 
  \label{comp1}
\end{align}

Second, we show the relation for the trace (\ref{C29-trace-integral}).
If the operator $\mathscr{O}$ does not depend on the group $g$ explicitly, then 
\begin{equation}
 \int d\mu(g) \left< g , \Lambda \right| \mathscr{O}\left| g , \Lambda \right> 
 =  \frac{{\rm tr}_{\rm R}(\mathscr{O})}{{\rm tr}_{\rm R}({\bf 1})} 
 .
 \label{rel2}
\end{equation}
In fact, we have
\begin{align}
   \int d\mu(g) \left< g , \Lambda \right| \mathscr{O}\left| g , \Lambda \right> 
 =& \int d\mu(g) \left<  \Lambda \right| g^\dagger  \mathscr{O} g \left|  \Lambda \right> 
%\nonumber\\
 =  \int d\mu(g) {\rm tr}(\rho g^\dagger  \mathscr{O}g  )
 \nonumber\\
 =& \int d\mu(g) {\rm tr}( \mathscr{O}g \rho g^\dagger  )
% \nonumber\\
 =   {\rm tr}( \mathscr{O} \int d\mu(g)g \rho g^\dagger  )
 \nonumber\\
 =&  {\rm tr}( \mathscr{O} \frac{1}{d_R}{\rm tr}(\rho) {\bf 1}  )
 = \frac{{\rm tr}(\mathscr{O})}{d_R} 
 ,
\end{align}
where we have used 
$
\left<  \Lambda \right|   \mathscr{O}  \left|  \Lambda \right> 
 =  {\rm tr}(\rho  \mathscr{O}  )
$
in the second equality, (\ref{form2}) in the fifth equality and  
the trace property of $\rho$,
$ 
 {\rm tr}(\rho) = \rho_{aa} =  \left|  \Lambda \right>_a \left< \Lambda \right|_a = \lambda_a \lambda_a^* = 1 
$
 in the last equality. 
This completes the proof of  (\ref{C29-trace-integral}).

%In section IV, the resulting expression is further rewritten into another form in terms of a vector field $\mathbf{m}$ or a unit vector field $\mathbf{n}$, which are called color fields. 

%%%%%%%%%%%%%%%%%%%%%%%%%%%%%%%%%%%%%%%%%%%%%%%%%%%%%%%%%%%%
%%%%%%%%%%%%%%%%%%%%%%%%%%%%%%%%%%%%%%%%%%%%%%%%%%%%%%%%%%%%
%%%%%%%%%%%%%%%%%%%%%%%%%%%%%%%%%%%%%%%%%%%%%%%%%%%%%%%%%%%%
%%%%%%%%%%%%%%%%%%%%%%%%%%%%%%%%%%%%%%%%%%%%%%%%%%%%%%%%%%%%
\section{Magnetic potentials}\label{sec:magnetic-potential}
%%%%%%%%%%%%%%%%%%%%%%%%%%%%%%%%%%%%%%%%%%%%%%%%%%%%%%%%%%%%
%%%%%%%%%%%%%%%%%%%%%%%%%%%%%%%%%%%%%%%%%%%%%%%%%%%%%%%%%%%%
%%%%%%%%%%%%%%%%%%%%%%%%%%%%%%%%%%%%%%%%%%%%%%%%%%%%%%%%%%%%
%%%%%%%%%%%%%%%%%%%%%%%%%%%%%%%%%%%%%%%%%%%%%%%%%%%%%%%%%%%%

The relation (\ref{DBeSU2}) is derived as follows.
%(\ref{DBeSU2}) is shown as follows. 
For $\bm{V}_{\mu} = \bm{C}_{\mu} + \bm{B}_{\mu} = c_{\mu} \bm{n} + g^{-1} \partial_{\mu} \bm{n} \times \bm{n}$, we have 
\begin{align}
g\bm{V}_{\mu} \times \bm{e}_{a} 
=& g\bm{C}_{\mu} \times \bm{e}_{a} + g\bm{B}_{\mu} \times \bm{e}_{a}  
\nonumber\\
=& gc_{\mu} \bm{n} \times \bm{e}_{a} +  ( \partial_{\mu} \bm{n} \times \bm{n} ) \times \bm{e}_{a} \nonumber\\
=& gc_{\mu} \epsilon_{ab} \bm{e}_{b} +   [ ( \partial_{\mu} \bm{n} \cdot \bm{e}_{a} ) \bm{n} - \partial_{\mu} \bm{n} ( \bm{n} \cdot \bm{e}_{a} ) ]
\nonumber\\
=& g\epsilon_{ab} c_{\mu} \bm{e}_{b} +  ( \partial_{\mu} \bm{n} \cdot \bm{e}_{a} ) \bm{n} .
\end{align}
Then the covariant derivative of $\bm{e}_{a}$ in the background $\bm{B}_{\mu}$ reads
\begin{align}
D_{\mu} [\bm{B}] \bm{e}_{a} :=& \partial_{\mu} \bm{e}_{a} + g \bm{B}_{\mu} \times \bm{e}_{a} \nonumber\\
=& \partial_{\mu} \bm{e}_{a}   + ( \partial_{\mu} \bm{n} \cdot \bm{e}_{a} ) \bm{n}  
=    \bm{f}_{\mu}^{a} ,
\end{align}
where we have defined
\begin{align}
\bm{f}_{\mu}^{a} :=  \partial_{\mu} \bm{e}_{a} + ( \partial_{\mu} \bm{n} \cdot \bm{e}_{a} ) \bm{n} 
=  \partial_{\mu} \bm{e}_{a} - ( \bm{n} \cdot \partial_{\mu} \bm{e}_{a} ) \bm{n} .
\end{align}
It is easy to show that $\bm{f}_{\mu}^{a}$ is orthogonal to $\bm{e}_{a}$ and $\bm{n}$.
Therefore, $\bm{f}_{\mu}^{a}$ is proportional to $\bm{n} \times \bm{e}_{a} = \epsilon_{ab} \bm{e}_{b}$ and hence $\bm{f}_{\mu}^{a}$ is expressed as
\begin{align}
\bm{f}_{\mu}^{1} =& ( \bm{f}_{\mu}^{1} \cdot \bm{e}_{2} ) \bm{e}_{2} = ( \partial_{\mu} \bm{e}_{1} \cdot \bm{e}_{2} ) \bm{e}_{2} , \nonumber\\
\bm{f}_{\mu}^{2} =& ( \bm{f}_{\mu}^{2} \cdot \bm{e}_{1} ) \bm{e}_{1} = ( \partial_{\mu} \bm{e}_{2} \cdot \bm{e}_{1} ) \bm{e}_{1} = - ( \bm{e}_{2} \cdot \partial_{\mu} \bm{e}_{1} ) \bm{e}_{1} ,
\end{align}
which means
\begin{equation}
\bm{f}_{\mu}^{a} = g h_{\mu} \epsilon_{ab} \bm{e}_{b} ,
\quad 
h_{\mu} := g^{-1} \partial_{\mu} \bm{e}_{1} \cdot \bm{e}_{2} .
\end{equation}
Thus we obtain (\ref{DBeSU2}).

Another derivation of the relation (\ref{DBeSU2}) is as follows. 
The color space is spanned by the basis vectors $\bm{n}$ and $\bm{e}_a$: 
\begin{align}
 D_\mu[\mathscr{B}]  \bm{e}_a   =  C_\mu^{ab}  \bm{e}_b  + d_\mu^a \bm{n} .
  \label{DBeSU2b}
\end{align}
By taking into account $\bm{n} \cdot \bm{n}=1$ and $\bm{n} \cdot \bm{e}_b=0$, the coefficient $d_\mu^a$ is determined from 
\begin{align}
d_\mu^a =&  \bm{n} \cdot D_\mu[\bm{B}]  \bm{e}_a 
\nonumber\\
=& \bm{n} \cdot \partial_{\mu} \bm{e}_{a} + \bm{n} \cdot  (g \bm{B}_{\mu} \times \bm{e}_{a} ) 
\nonumber\\
=& \bm{n} \cdot \partial_{\mu} \bm{e}_{a} + g \bm{B}_{\mu}  \cdot  (\bm{e}_{a}  \times \bm{n}) 
\nonumber\\
=& \bm{n} \cdot \partial_{\mu} \bm{e}_{a} - (\partial_{\mu} \bm{n} \times \bm{n})   \cdot  (\epsilon_{ab} \bm{e}_{b} ) 
\nonumber\\
=& \bm{n} \cdot \partial_{\mu} \bm{e}_{a} - ( \bm{n} \times  \bm{e}_{b} )  \cdot  (\epsilon_{ab} \partial_{\mu} \bm{n}) 
\nonumber\\
=& \bm{n} \cdot \partial_{\mu} \bm{e}_{a} -  \epsilon_{ab}  \epsilon_{bc} \bm{e}_{c}  \cdot  \partial_{\mu} \bm{n}  
\nonumber\\
=& \bm{n} \cdot \partial_{\mu} \bm{e}_{a} +  \bm{e}_{a}  \cdot  \partial_{\mu} \bm{n}  
= \partial_{\mu} (\bm{n} \cdot  \bm{e}_{a}) =  0
 , 
  \label{DBeSU2d}
\end{align}
while using $\bm{e}_a \cdot \bm{e}_b=\delta_{ab}$
 the coefficient $d_\mu^a$ is determined from 
\begin{align}
C_\mu^{ab}  =&  \bm{e}_b \cdot D_\mu[\bm{B}]  \bm{e}_a    .
\nonumber\\
=& \bm{e}_b  \cdot \partial_{\mu} \bm{e}_{a} + \bm{e}_b  \cdot  (g \bm{B}_{\mu} \times \bm{e}_{a} ) 
\nonumber\\
=& \bm{e}_b  \cdot \partial_{\mu} \bm{e}_{a} + g \bm{B}_{\mu}  \cdot  (\bm{e}_{a}  \times \bm{e}_b ) 
\nonumber\\
=& - \partial_{\mu} \bm{e}_b  \cdot  \bm{e}_{a} + g \bm{B}_{\mu}  \cdot  (\epsilon_{ab} \bm{n}) 
\nonumber\\
=& - \partial_{\mu} \bm{e}_b  \cdot  \bm{e}_{a} + (\partial_{\mu} \bm{n} \times \bm{n}) \cdot  (\epsilon_{ab} \bm{n}) 
\nonumber\\
=&  - \bm{e}_{a} \cdot \partial_{\mu} \bm{e}_b   
= \bm{e}_b  \cdot \partial_{\mu} \bm{e}_{a}  
= - C_\mu^{ba}
.
  \label{DBeSU2c}
\end{align}
The shortest proof is given as
\begin{align}
d_\mu^a 
=   \bm{n} \cdot D_\mu[\bm{B}]  \bm{e}_a 
%\nonumber\\
=  - (D_\mu[\bm{B}]  \bm{n}) \cdot  \bm{e}_a  + \partial_{\mu} (\bm{n} \cdot  \bm{e}_{a}) =  0
 , 
  \label{DBeSU2d-2}
\end{align}
where we have used the defining equation 
$0=D_\mu[\bm{V}]  \bm{n}=D_\mu[\bm{B}+\bm{C}]  \bm{n}=D_\mu[\bm{B}]  \bm{n}$
and the orthogonality
$\bm{n} \cdot  \bm{e}_{a}=0$.

Eq.(\ref{DBe}) is derived as follows. 
The color space is spanned by the basis vectors $\bm{u}_j$ and $\bm{e}_a$: 
\begin{align}
 D_\mu[\bm{B}]  \bm{e}_a   =  d_\mu^{aj} \bm{u}_j + C_\mu^{ab}  \bm{e}_b    .
  \label{DBeSU2b2}
\end{align}
First, by taking into account $\bm{u}_j \cdot \bm{u}_k=\delta_{jk}$ and $\bm{u}_j \cdot \bm{e}_a=0$, the coefficient $d_\mu^{aj}$ is determined from 
\begin{align}
d_\mu^{aj} =   \bm{u}_j \cdot D_\mu[\bm{B}]  \bm{e}_a 
%\nonumber\\
%=   - (D_\mu[\bm{B}]  \bm{u}_j) \cdot  \bm{e}_a  +  \partial_{\mu} (\bm{u}_j \cdot  \bm{e}_{a}) 
 . 
\end{align}
We recall the fact that the color field $\bm{n}$ commutes with all the bases $\bm{u}_j$ in the Lie algebra of the maximal stability group:
\begin{align}
 [ \bm{n}, \bm{u}_j]= 0 
 . 
\end{align}
Then the  Leibniz rule for the covariant derivative and the defining equation yields 
\begin{align}
0 
= D_\mu[\bm{B}] [ \bm{n}, \bm{u}_j] 
= [  D_\mu[\bm{B}] \bm{n}, \bm{u}_j] +  [ \bm{n},  D_\mu[\bm{B}] \bm{u}_j] 
= [ \bm{n},  D_\mu[\bm{B}] \bm{u}_j] 
 . 
\end{align}
This means that $D_\mu[\bm{B}] \bm{u}_j$ is represented as the linear combination of the bases $\bm{u}_j$.
By taking into account 
$
 \bm{u}_j \cdot  \bm{e}_{a} = 0 ,
$
we find 
\begin{align}
  (D_\mu[\bm{B}]  \bm{u}_j) \cdot  \bm{e}_a = 0
\end{align}
and therefore 
\begin{align}
d_\mu^{aj} =   - (D_\mu[\bm{B}]  \bm{u}_j) \cdot  \bm{e}_a  +  \partial_{\mu} (\bm{u}_j \cdot  \bm{e}_{a}) 
= 0 .
\end{align}

Next, using $\bm{e}_a \cdot \bm{e}_b=\delta_{ab}$, 
 the coefficient $C_\mu^{ab}$ is determined from 
\begin{align}
C_\mu^{ab}  =&  \bm{e}_b \cdot D_\mu[\bm{B}]  \bm{e}_a    .
\nonumber\\
=& \bm{e}_b  \cdot \partial_{\mu} \bm{e}_{a} + \bm{e}_b  \cdot  (g \bm{B}_{\mu} \times \bm{e}_{a} ) 
\nonumber\\
=& \bm{e}_b  \cdot \partial_{\mu} \bm{e}_{a} + g \bm{B}_{\mu}  \cdot  (\bm{e}_{a}  \times \bm{e}_b )  
\nonumber\\
=& - \partial_{\mu} \bm{e}_b  \cdot  \bm{e}_{a} + g \bm{B}_{\mu}  \cdot  (f^{abj} \bm{u}_j) 
\nonumber\\
=& - \partial_{\mu} \bm{e}_b  \cdot  \bm{e}_{a} + gf^{abj}  {\rm tr}[ \bm{B}_{\mu}  \cdot \bm{u}_j ] 
\nonumber\\
=& - \partial_{\mu} \bm{e}_b  \cdot  \bm{e}_{a} + \frac{2(N-1)}{N}  f^{abj}  {\rm tr}( [\partial_{\mu} \bm{n} ,   \bm{n}]   \bm{u}_j )  
\nonumber\\
=& - \partial_{\mu} \bm{e}_b  \cdot  \bm{e}_{a} + \frac{2(N-1)}{N}  f^{abj}  {\rm tr}(  \partial_{\mu}  \bm{n} [   \bm{n},   \bm{u}_j [)  
\nonumber\\
=&  - \bm{e}_{a} \cdot \partial_{\mu} \bm{e}_b   
= \bm{e}_b  \cdot \partial_{\mu} \bm{e}_{a}  
= - C_\mu^{ba}
.
  \label{DBeSU2c2}
\end{align}
Therefore, we can introduce $h_\mu^j$ so that 
\begin{align}
 C_\mu^{ab} = gf^{abj} h_\mu^j = - C_\mu^{ba} 
  .
\end{align}

Here we have used the relation following from (\ref{C27-fdec2}) and (\ref{C27-basismin}):

\begin{align}
 D_\mu[\mathscr{B}]  \bm{e}_a  
 :=  \partial_\mu   \bm{e}_a -ig[\mathscr{B}_\mu,  \bm{e}_a ] 
% \nonumber\\
% =& \partial_\rho  \bm{e}_a  -ig[B_\rho^b \bm{e}_b ,  \bm{e}_a] 
% \nonumber\\
 =& \partial_\mu  \bm{e}_a  + ig  B_\mu^b  [   \bm{e}_a ,  \bm{e}_b] 
% \nonumber\\
 =  \partial_\mu  \bm{e}_a   -  g  B_\mu^b   f^{abj} \bm{u}_j
  .
\end{align}
The component $B_\rho^b$ reads
\begin{align}
   B_\rho^b =& B_\rho^A e_b^A 
%\nonumber\\
=  2{\rm tr}(\mathscr{B}_\rho \bm{e}_b)
%\nonumber\\
=  ig^{-1} \frac{2(N-1)}{N} 2{\rm tr}(\bm{e}_b [  \bm{n}  ,  \partial_\mu \bm{n} ]  )
%\nonumber\\
=  ig^{-1} \frac{2(N-1)}{N} 2{\rm tr}([\bm{e}_b , \bm{n} ]  \partial_\mu \bm{n}   )
 ,
\end{align}
where we have used
$%\begin{align}
 \mathscr{B}_\mu =   ig^{-1} \frac{2(N-1)}{N} [  \bm{n}  ,  \partial_\mu \bm{n} ]  
 \in \mathscr{G} - \tilde{\mathscr{H}} ,
  .
$%\end{align}

By taking into account $\bm{u}_j \cdot \bm{u}_k=\delta_{jk}$ and $\bm{u}_j \cdot \bm{e}_b=0$, the coefficient $d_\mu^{aj}$ is determined from 
\begin{align}
d_\mu^{aj} =&  \bm{u}_j \cdot D_\mu[\bm{B}]  \bm{e}_a 
\nonumber\\
=& \bm{u}_j \cdot \partial_{\mu} \bm{e}_{a} + \bm{u}_j \cdot  (g \bm{B}_{\mu} \times \bm{e}_{a} ) 
%\nonumber\\
%=& \bm{u}_j \cdot \partial_{\mu} \bm{e}_{a} 
%-  g  B_\mu^b   f^{abk} \bm{u}_j \cdot \bm{u}_k
%\nonumber\\
%=& \bm{u}_j \cdot \partial_{\mu} \bm{e}_{a} 
%-  g  B_\mu^b   f^{abj}  
\nonumber\\
=& \bm{u}_j \cdot \partial_{\mu} \bm{e}_{a} + g \bm{B}_{\mu}  \cdot  (\bm{e}_{a}  \times \bm{u}_j)  
\nonumber\\
=& \bm{u}_j \cdot \partial_{\mu} \bm{e}_{a} + g \bm{B}_{\mu}  \cdot  (f^{abj} \bm{e}_{b}  )  
\nonumber\\
=& \bm{u}_j \cdot \partial_{\mu} \bm{e}_{a} - f^{abj}  \frac{2(N-1)}{N} (\partial_{\mu} \bm{n} \times \bm{n})   \cdot    \bm{e}_{b}  
\nonumber\\
=& \bm{u}_j  \cdot \partial_{\mu} \bm{e}_{a} - f^{abj}   \frac{2(N-1)}{N}  ( \bm{n} \times  \bm{e}_{b} )  \cdot  ( \partial_{\mu} \bm{n}) ***
\nonumber\\
=& \bm{u}_j \cdot \partial_{\mu} \bm{e}_{a} -  f^{abj}   \frac{2(N-1)}{N}  \epsilon_{bc} \bm{e}_{c}  \cdot  \partial_{\mu} \bm{n}  
\nonumber\\
=& \bm{u}_j \cdot \partial_{\mu} \bm{e}_{a} +  \bm{e}_{a}  \cdot  \partial_{\mu} \bm{n}  
= \partial_{\mu} (\bm{n} \cdot  \bm{e}_{a}) =  0
 , 
  \label{DBeSU2d2}
\end{align}

%%%%%%%%%%%%%%%%%%%%%%%%%%%%%%%%%%%%%%%%%%%%%%%%%%%%%%%%%%%%
%%%%%%%%%%%%%%%%%%%%%%%%%%%%%%%%%%%%%%%%%%%%%%%%%%%%%%%%%%%%
%%%%%%%%%%%%%%%%%%%%%%%%%%%%%%%%%%%%%%%%%%%%%%%%%%%%%%%%%%%%
%%%%%%%%%%%%%%%%%%%%%%%%%%%%%%%%%%%%%%%%%%%%%%%%%%%%%%%%%%%%
\section{Weyl symmetry and color direction field}\label{sec:Weyl}
%%%%%%%%%%%%%%%%%%%%%%%%%%%%%%%%%%%%%%%%%%%%%%%%%%%%%%%%%%%%
%%%%%%%%%%%%%%%%%%%%%%%%%%%%%%%%%%%%%%%%%%%%%%%%%%%%%%%%%%%%
%%%%%%%%%%%%%%%%%%%%%%%%%%%%%%%%%%%%%%%%%%%%%%%%%%%%%%%%%%%%
%%%%%%%%%%%%%%%%%%%%%%%%%%%%%%%%%%%%%%%%%%%%%%%%%%%%%%%%%%%%

The \textbf{Weyl group} or Weyl reflection group may be regarded as acting on the real traceless diagonal matrices $M$ by permuting the diagonal elements.  
For $SU(2)$, the Weyl group consists of two elements, $w_1={\bf 1}$ and $w_2=(12)$ acting as follows. 
\begin{equation}
  w_1 = {\bf 1}: 
   w_1 
   \begin{pmatrix} 
   a & 0 \\
   0 & -a 
   \end{pmatrix}
 = \begin{pmatrix} 
   a & 0 \\
   0 & -a 
   \end{pmatrix} ,
   \quad 
w_2 = (12):      w_2 
   \begin{pmatrix} 
   a & 0 \\
   0 & -a 
   \end{pmatrix}
 = \begin{pmatrix} 
   -a & 0 \\
   0 &  a 
   \end{pmatrix}
 .
\end{equation}
These two elements are represented by $SU(2)$ matrices: 
\begin{equation}
 W_1 
 = \begin{pmatrix} 
   1 & 0 \\
   0 & 1 
   \end{pmatrix} ,
   \quad 
W_2 
 = \begin{pmatrix} 
   0 & 1 \\
   -1 & 0 
   \end{pmatrix}
 ,
\end{equation}
acting via the coadjoint action:
\begin{align}
   w_1 
   \begin{pmatrix} 
   a & 0 \\
   0 & -a 
   \end{pmatrix}
 =& W_1^{-1}   \begin{pmatrix} 
   a & 0 \\
   0 & -a 
   \end{pmatrix} 
   W_1
=  \begin{pmatrix} 
   a & 0 \\
   0 &  -a 
   \end{pmatrix}
 ,
\nonumber\\
   w_2 
   \begin{pmatrix} 
   a & 0 \\
   0 & -a 
   \end{pmatrix}
 =& W_2^{-1}   \begin{pmatrix} 
   a & 0 \\
   0 & -a 
   \end{pmatrix} 
   W_2
=  \begin{pmatrix} 
   -a & 0 \\
   0 &  a 
   \end{pmatrix}
= - \begin{pmatrix} 
    a & 0 \\
   0 &  -a 
   \end{pmatrix}
 .
\end{align}
In fact, each $W_a$ is an element of $SU(2)$, since $W_a W_a^\dagger = W_a^\dagger W_a=\bm{1}$ and $\det W_a=1$ for each $a$.
They satisfy
\begin{equation}
   W_1^{-1} 
 = W_1^\dagger =W_1 , 
   \quad 
   W_2^{-1} = W_2^\dagger = -W_2 
 .
\end{equation}

For $SU(3)$, the Weyl group acts on $M$ as the group of permutations of the entries of ${\rm diag}(a,b,c)$, i.e., the diagonal elements of the real traceless diagonal matrices.
Thus the Weyl group of $SU(3)$ consists of six elements, $w_1={\bf 1}, w_2, ..., w_6$ acting as 
\begin{align}
  w_1 = {\bf 1}: 
   w_1 
   {\rm diag}(a,b,c)
= {\rm diag}(a,b,c)
= W_1^{-1} {\rm diag}(a,b,c) W_1 
  ,
\nonumber\\
  w_2 = (12): 
   w_2 
   {\rm diag}(a,b,c)
= {\rm diag}(b,a,c) 
= W_2^{-1} {\rm diag}(a,b,c) W_2 
 ,
\nonumber\\
  w_3 = (23): 
   w_3 
   {\rm diag}(a,b,c)
= {\rm diag}(a,c,b)
= W_3^{-1} {\rm diag}(a,b,c) W_3 
  ,
\nonumber\\
  w_4 = (13): 
   w_4 
   {\rm diag}(a,b,c)
= {\rm diag}(c,b,a)
= W_4^{-1} {\rm diag}(a,b,c) W_4 
  ,
\nonumber\\
  w_5 = (132): 
   w_5 
   {\rm diag}(a,b,c)
= {\rm diag}(c,a,b)
= W_5^{-1} {\rm diag}(a,b,c) W_5 
  ,
\nonumber\\
  w_6 = (123): 
   w_6 
   {\rm diag}(a,b,c)
= {\rm diag}(b,c,a) 
= W_6^{-1} {\rm diag}(a,b,c) W_6 
 ,
 \label{SU3-Weyl-0}
\end{align}
where these six elements $w_a$ ($a=1,\dots,6$) are represented by $SU(3)$ matrices $W_a$  ($a=1,\dots,6$) 
acting via the coadjoint action: for $c=-a-b$,
\begin{align}
& W_1 
 =  \begin{pmatrix} 
   1 & 0 & 0 \\
   0 & 1 & 0 \\
   0 & 0 & 1 \\
   \end{pmatrix} ,
   \quad 
 W_2 
 = \begin{pmatrix} 
   0 & 1 & 0 \\
   -1 & 0 & 0 \\
   0 & 0 & 1 \\
   \end{pmatrix} ,
   \quad 
 W_3 
 =  \begin{pmatrix} 
   1 & 0 & 0 \\
   0 & 0 & 1 \\
   0 & -1 & 0 \\
   \end{pmatrix} ,
   \nonumber\\
& W_4 
 = \begin{pmatrix} 
   0 & 0 & 1 \\
   0 & 1 & 0 \\
   -1 & 0 & 0 \\
   \end{pmatrix} ,
   \quad 
 W_5 
 =  \begin{pmatrix} 
   0 & -1 & 0 \\
   0 & 0 & 1 \\
   -1 & 0 & 0 \\
   \end{pmatrix} ,
   \quad 
 W_6 
 = \begin{pmatrix} 
   0 & 0 & 1 \\
   -1 & 0 & 0 \\
   0 & -1 & 0 \\
   \end{pmatrix} 
    .
\end{align}
In fact, each $W_a$ is an element  of $SU(3)$, since $W_a W_a^\dagger = W_a^\dagger W_a=\bm{1}$ and $\det W_a=1$ for each $a$.

The Wilson loop operator $W_C[\mathscr{A}]$ defined for the original Yang-Mills field $\mathscr{A} (x)=\mathscr{A}_\mu (x)dx^\mu$ is rewritten through the non-Abelian Stokes theorem into 
\begin{equation}
 W_C[\mathscr{A}] :=\int [d\mu(g)]_{\Sigma}
\exp \left[ -ig_{{}_{\rm YM}} \int_{\Sigma: \partial \Sigma=C} F^g  \right] , %\quad
%[d\mu(g)]_{\Sigma} :=\prod_{x \in \Sigma: \partial \Sigma=C}   d\mu(g(x))  ,
\end{equation}
where $g$ is an element of the gauge group $G$, $[d\mu(g)]_{\Sigma}$ is the invariant measure on the surface $\Sigma$ bounding the closed loop $C$, and  the $F^g$ is the two-form  defined by the exterior derivative of the one-form $A^g$ by   
\begin{align}
 F^g  :=& dA^g  ,
\quad A^g=A^g_\mu(x) dx^\mu ,
%= \frac12 F^g_{\mu\nu}(x) dx^\mu \wedge dx^\nu 
%\end{align}
%\begin{align}
 \nonumber\\
   A^g_\mu(x) :=& 
 \langle \Lambda | \mathscr{A}_\mu^g (x)    |\Lambda \rangle  
=  {\rm tr}\{\rho \mathscr{A}^g_\mu(x)   \} 
 \nonumber\\
=& {\rm tr}\{ g(x) \rho  g^\dagger(x)  \mathscr{A}_\mu(x) \} + ig_{{}_{\rm YM}}^{-1} {\rm tr}\{ \rho  g^\dagger(x)  \partial_\mu   g(x)  \} 
 .
 \label{C40-def-Ag}
\end{align}
 Here $|\Lambda \rangle$ is a reference state specifying the representation of the Wilson loop operator and $\rho$ is the matrix defined by 
\begin{align}
 \rho :=|\Lambda \rangle \langle \Lambda | .
\end{align} 
 
Let $w$ be an element  of the Weyl group as a discrete subgroup of $G$. 
Since $gw \in G$ for $g \in G$, we have 
\begin{align}
  {\bf 1}
=&  \int d\mu(g(x) ) \left| g(x) , \Lambda \right>  \left< g(x) , \Lambda \right|  
\nonumber\\
=&     \int d\mu(g(x)w ) \left| g(x)w , \Lambda \right>  \left< g(x)w , \Lambda \right|  
\nonumber\\
=&   \int d\mu(g(x))  g(x)w \left| \Lambda \right>  \left< \Lambda \right|  w^\dagger g^\dagger(x) 
\nonumber\\
=&   \int d\mu(g(x))  g(x)   w\rho  w^\dagger g^\dagger(x) 
   ,
  \label{C40-rel1}
\end{align}
where we have used  the invariance of the integration measure $d\mu(gw )=d\mu(g)$.
Then the Wilson loop operator reads  
\begin{equation}
 W_C[\mathscr{A}] :=\int [d\mu(g)]_{\Sigma}
\exp \left[ -ig_{{}_{\rm YM}} \int_{\Sigma: \partial \Sigma=C} F^{gw}  \right] , %\quad
%[d\mu(g)]_{\Sigma} :=\prod_{x \in \Sigma: \partial \Sigma=C}   d\mu(g(x))  ,
\end{equation}
where
\begin{align}
 F^{gw}  :=& dA^{gw} ,
\quad A^{gw}=A^{gw}_\mu(x) dx^\mu ,
 %= \frac12 F^{gW}_{\mu\nu}(x) dx^\mu \wedge dx^\nu 
%\end{align}
%\begin{align}
 \nonumber\\
   A^{gw}_\mu(x) :=& 
 \langle \Lambda | \mathscr{A}_\mu^{gw} (x)    |\Lambda \rangle  
=  {\rm tr}\{\rho \mathscr{A}^{gw}_\mu(x)   \} 
 \nonumber\\
=& {\rm tr}\{ g(x) w \rho  w^\dagger g^\dagger(x)  \mathscr{A}_\mu(x) \} + ig_{{}_{\rm YM}}^{-1} {\rm tr}\{ w \rho w^\dagger g^\dagger(x)  \partial_\mu   g(x)  \} 
 .
 \label{C40-def-Ag2}
\end{align}
Therefore, $A^{gw}$ and $F^{gw}$ are respectively equal to $A^{g}$ and $F^{g}$ with $\rho$ being replaced by $\rho^w$: 
\begin{equation}
% \rho =:  \left|  \Lambda \right> \left< \Lambda \right|  \Longrightarrow 
 \rho^w := w \rho w^\dagger  =  w \left|  \Lambda \right> \left< \Lambda \right| w^\dagger 
 %\quad 
% \rho_{ab} :=  \left|  \Lambda \right>_a \left< \Lambda \right|_b = \lambda_a \lambda_b^*
 .
\end{equation}
Thus, the traceless color field $\tilde{\bm{n}}(x)$ and the  normalized traceless color field $\bm{n}(x)$  defined through $\rho$:
%and the normalized and traceless field $\bm{n}(x)$
\begin{align}
 \tilde{\bm{n}}(x) :=&  g(x) \left[ \rho - \frac{\bm{1}}{{\rm tr}(\bm{1})} \right]  g^\dagger(x) 
%= g(x)  \rho g^\dagger(x) - \frac{\bm{1}}{{\rm tr}(\bm{1})} 
   ,
\nonumber\\
%\quad
% \bm{n}(x) := - \sqrt{\frac{N}{2(N-1)}} \tilde{\bm{n}}(x)
%\label{C29-tilde-n}
%\end{align}
%\begin{align}
 \bm{n}(x) :=&   \sqrt{\frac{N}{2(N-1)}} \tilde{\bm{n}}(x)
 = \sqrt{\frac{N}{2(N-1)}} g(x) \left[ \rho - \frac{\bm{1}}{{\rm tr}(\bm{1})} \right]  g^\dagger(x)  
 ,
%\label{C29-n-defW}
\end{align}
are replaced by the Weyl-reflected ones:
\begin{align}
 \tilde{\bm{n}}^{w}(x)  
:=&  g(x) w \left[ \rho - \frac{\bm{1}}{{\rm tr}(\bm{1})} \right] w^\dagger  g^\dagger(x) 
=  g(x) \left[ \rho  - \frac{\bm{1}}{{\rm tr}(\bm{1})} \right]^{w}  g^\dagger(x) 
%= g(x)  \rho^{W} g^\dagger(x) - \frac{\bm{1}}{{\rm tr}(\bm{1})}    ,
\nonumber\\
%\quad
% \bm{n}(x) := - \sqrt{\frac{N}{2(N-1)}} \tilde{\bm{n}}(x)
%\label{C29-tilde-n}
%\end{align}
%\begin{align}
 \bm{n}^{w}(x) :=&   \sqrt{\frac{N}{2(N-1)}} \tilde{\bm{n}}(x)
 = \sqrt{\frac{N}{2(N-1)}} g(x)  \left[ \rho - \frac{\bm{1}}{{\rm tr}(\bm{1})} \right]^{w}  g^\dagger(x)  
 ,
 \label{C29-n-defW}
\end{align}
where we have introduced 
\begin{align}
  \left[ \rho - \frac{\bm{1}}{{\rm tr}(\bm{1})} \right]^{w}
:=&   w \left[ \rho - \frac{\bm{1}}{{\rm tr}(\bm{1})} \right] w^\dagger  
=    \rho^{w} - \frac{\bm{1}}{{\rm tr}(\bm{1})}  
 .
\end{align}

For $SU(2)$, the traceless color field $\tilde{\bm{n}}(x)$ and the  normalized traceless color field $\bm{n}(x)$  are constructed as
\begin{equation}
\bm{n}(x) 
=   \tilde{\bm{n}}(x)
%= \sqrt{3} \bm{m}(x) 
=   g(x) \left( \rho - \frac12 \mathbf{1} \right) g(x)^\dagger , \   g(x) \in SU(2) .
%= \sqrt{3} g(x) \mathcal{H} g(x)^\dagger \  (g(x) \in SU(3))
\end{equation}
Then the Weyl-reflection color field is obtained as
\begin{equation}
\bm{n}^{w}(x) 
%= \frac{\sqrt{3}}{2}  \tilde{\bm{n}}(x)
%= \sqrt{3} \bm{m}(x) 
=  g(x) \left( \rho - \frac12 \mathbf{1} \right)^{w} g(x)^\dagger , \   g(x) \in SU(2) .
%= \sqrt{3} g(x) \mathcal{H} g(x)^\dagger \  (g(x) \in SU(3))
\end{equation}
If we choose the \textbf{highest-weight state} as a reference state of the fundamental representation of $SU(2)$:
\begin{equation}
| \Lambda \rangle 
=  
\begin{pmatrix}
 1 \\
 0 
\end{pmatrix} 
 ,
\end{equation}
then the traceless version of $\rho$ is transformed under the Weyl transformation as  
\begin{align}
& \rho :=  | \Lambda \rangle \langle \Lambda |
= 
\begin{pmatrix}
 1 \\
 0 
\end{pmatrix}
 (1,0)
= 
\begin{pmatrix}
 1 & 0 \\
 0 & 0 
\end{pmatrix} 
\Longrightarrow 
 \rho - \frac12 \mathbf{1} 
= \frac12 \left(
  \begin{array}{cc}
   1 & 0 \\
   0 & -1 \\
  \end{array}
 \right) 
= \frac{\sigma_3}{2} 
 , 
\nonumber\\
& \Longrightarrow  
  \left( \rho - \frac12 \mathbf{1}  \right)^{w_2}
= -\frac12 \left(
  \begin{array}{cc}
   1 & 0 \\
   0 & -1 \\
  \end{array}
 \right) 
= -\frac{\sigma_3}{2} 
  .
\end{align}
For the normalized traceless color field:
\begin{align}
  \bm{n}(x) 
%=  \frac{\sqrt{3}}{2} g_{x} \left(  \rho - \frac13 \mathbf{1} \right) g_{x}^\dagger 
=   {g(x)}
\frac{\sigma_3}{2}
   {g^\dagger(x)} 
\in SU(2)/U(1) \simeq CP^1 ,
\end{align}
therefore, we find that the Weyl-reflected color field reads 
\begin{align}
  {\bm{n}}^{w_1}(x) =   {\bm{n}}(x), \quad 
 \bm{n}^{w_2}(x) = -\bm{n} (x) .
\end{align}
In other words, $\bm{n}^{w_2}(x)$ is equal to the color field $\bm{n}(x)$ constructed from the lowest-weight state:
\begin{equation}
| \Lambda \rangle 
=  
\begin{pmatrix}
 0 \\
 1 
\end{pmatrix} 
 ,
\end{equation}
namely, the Weyl transformation $w_2$ exchanges the highest and lowest states.

For  $SU(3)$, the traceless color field $\tilde{\bm{n}}(x)$ and the  normalized traceless color field $\bm{n}(x)$   are constructed as
\begin{equation}
\bm{n}(x) 
= \frac{\sqrt{3}}{2}  \tilde{\bm{n}}(x)
%= \sqrt{3} \bm{m}(x) 
= \frac{\sqrt{3}}{2} g(x) \left( \rho - \frac13 \mathbf{1} \right) g(x)^\dagger , \   g(x) \in SU(3) .
%= \sqrt{3} g(x) \mathcal{H} g(x)^\dagger \  (g(x) \in SU(3))
\end{equation}
Then the Weyl-reflection color field is obtained as
\begin{equation}
\bm{n}^{w}(x) 
%= \frac{\sqrt{3}}{2}  \tilde{\bm{n}}(x)
%= \sqrt{3} \bm{m}(x) 
= \frac{\sqrt{3}}{2} g(x) \left( \rho - \frac13 \mathbf{1} \right)^{w} g(x)^\dagger , \   g(x) \in SU(3) .
%= \sqrt{3} g(x) \mathcal{H} g(x)^\dagger \  (g(x) \in SU(3))
\end{equation}
If we choose the {highest-weight state} as a reference state of the fundamental representation:
\begin{equation}
| \Lambda \rangle 
=  \small
\begin{pmatrix}
 1 \\
 0 \\
 0 
\end{pmatrix} ,
\end{equation}
we obtain the  operator $\rho$ and its traceless version:
\begin{align}
  \rho :=  | \Lambda \rangle \langle \Lambda |
%= 
%\begin{pmatrix}
% 1 \\
% 0 \\
% 0 
%\end{pmatrix}
% (1,0,0)
= 
\begin{pmatrix}
 1 & 0 & 0 \\
 0 & 0 & 0 \\
 0 & 0 & 0 \\
\end{pmatrix} 
%\nonumber\\
  \Longrightarrow \quad
%\mathcal{H}=  \frac12 \left(  \rho - \frac13 \mathbf{1} \right) 
\rho - \frac13 \mathbf{1}  = \frac{-1}{3} 
\begin{pmatrix}
 -2 & 0 & 0 \\
 0 & 1 & 0 \\
 0 & 0 & 1 \\
\end{pmatrix} 
= \frac{-1}{3} {\rm diag}(-2,1,1) .
\end{align}
Hence, the the traceless color field $\tilde{\bm{n}}(x)$ and the  normalized traceless color field $\bm{n}(x)$  are written as 
\begin{align}
  \bm{n}(x) 
%=  \frac{\sqrt{3}}{2} g_{x} \left(  \rho - \frac13 \mathbf{1} \right) g_{x}^\dagger 
=   {g(x)}
\frac{-1}{2\sqrt{3}} 
{\rm diag}(-2,1,1)
   {g(x)^\dagger} 
\in SU(3)/U(2) \simeq CP^2 .
\end{align}
%with the Pauli matrix $\sigma_3:={\rm diag.}(1,-1)$.
By taking into account the Weyl-reflection (\ref{SU3-Weyl-0}):
\begin{align} 
\left( \rho - \frac13 \mathbf{1}  \right)^{w_1} 
=& \left( \rho - \frac13 \mathbf{1}  \right)^{w_3} 
= \frac{-1}{3} {\rm diag}(-2,1,1)
%= \frac{-1}{3} (-3 H_1 - \sqrt{3} H_2 )  
=   2 \left( \frac12 H_1 + \frac{1}{2\sqrt{3}} H_2 \right)   ,
\nonumber\\
\left( \rho - \frac13 \mathbf{1}  \right)^{w_2} 
=& \left( \rho - \frac13 \mathbf{1}  \right)^{w_5} 
= \frac{-1}{3} {\rm diag}(1,-2,1)
%= \frac{-1}{3} (3 H_1 - \sqrt{3} H_2 ) 
=  2 \left( \frac{-1}{2} H_1 + \frac{1}{2\sqrt{3}} H_2 \right)   ,
\nonumber\\
\left( \rho - \frac13 \mathbf{1}  \right)^{w_4} 
=& \left( \rho - \frac13 \mathbf{1}  \right)^{w_6} 
= \frac{-1}{3} {\rm diag}(1,1,-2) 
%= \frac{-1}{3}  2\sqrt{3} H_2 
= 2 \frac{-1}{\sqrt{3}}   H_2,
\end{align}
 the Weyl-reflected color fields satisfy
\begin{align}
& \bm{n}^{w_1}(x) =\bm{n}^{w_3}(x)  = \bm{n}(x) 
=   {g(x)} \frac{-1}{2\sqrt{3}}{\rm diag}(-2,1,1) {g(x)^\dagger} \in SU(3)/U(2) \simeq CP^2, 
\nonumber\\
& 
 \bm{n}^{w_2}(x) =\bm{n}^{w_5}(x) 
=   {g(x)} \frac{-1}{2\sqrt{3}} {\rm diag}(1,-2,1) {g(x)^\dagger} \in SU(3)/U(2) \simeq CP^2 , 
\nonumber\\
& 
 \bm{n}^{w_4}(x) =\bm{n}^{w_6}(x) 
=   {g(x)} \frac{-1}{2\sqrt{3}} {\rm diag}(1,1,-2) {g(x)^\dagger} \in SU(3)/U(2) \simeq CP^2.
\end{align}
The Weyl-reflected color fields are  constructed from the other reference state vectors of the fundamental representations: 
\begin{align}
& | \Lambda \rangle 
=  \small
\begin{pmatrix}
 1 \\
 0 \\
 0 
\end{pmatrix} \Longrightarrow \bm{n}^{w_1}(x) =\bm{n}^{w_3}(x) = \bm{n}(x)  ,  
\nonumber\\
& | \Lambda \rangle 
=  \small
\begin{pmatrix}
 0 \\
 1 \\
 0 
\end{pmatrix} \Longrightarrow \bm{n}^{w_2}(x) =\bm{n}^{w_5}(x),  
\nonumber\\
& | \Lambda \rangle 
=  \small
\begin{pmatrix}
 0 \\
 0 \\
 1 
\end{pmatrix} \Longrightarrow \bm{n}^{w_4}(x) =\bm{n}^{w_6}(x) .
\end{align}

On the other hand, by introducing the weight vector $\bm{\Lambda}=(\Lambda_1,\Lambda_2)$ of the fundamental representations, the traceless version of the operator $\rho$ is rewritten using the weight vector $\bm{\Lambda}=(\Lambda_1,\Lambda_2)$ and the Cartan generators $H_1, H_2$:
\begin{align}
 \mathcal{H} 
=  \frac{1}{2}  \left( \rho - \frac13 \mathbf{1}  \right) 
=  \Lambda_1 H_1  + \Lambda_2 H_2  
 \Longrightarrow 
 \mathcal{H}^{w} 
=  (\Lambda_1 H_1  + \Lambda_2 H_2)^{w}  
%\frac{1}{2}  \left( \rho - \frac13 \mathbf{1}  \right)^{w} 
%=  \Lambda_1 H_1^{W}  + \Lambda_2 H_2^{W}  
 ,
\end{align}
and the  color field is constructed as
\begin{equation}
\bm{n}(x) 
%= \frac{\sqrt{3}}{2}  \tilde{\bm{n}}(x)
%= \sqrt{3} \bm{m}(x) 
%= \frac{\sqrt{3}}{2} g(x) \left( \rho - \frac13 \mathbf{1} \right) g(x)^\dagger , \   g(x) \in SU(3) ,
= \sqrt{3} g(x) \mathcal{H} g(x)^\dagger \quad   g(x) \in SU(3) .
\end{equation}
Then  the Weyl-reflected color field reads 
\begin{equation}
\bm{n}^{w}(x) 
%= \frac{\sqrt{3}}{2}  \tilde{\bm{n}}(x)
%= \sqrt{3} \bm{m}(x) 
%= \frac{\sqrt{3}}{2} g(x) \left( \rho - \frac13 \mathbf{1} \right) g(x)^\dagger , \   g(x) \in SU(3) ,
= \sqrt{3} g(x) \mathcal{H}^{w} g(x)^\dagger, 
%\quad \mathcal{H}^{w} =  \frac{1}{2}  \left( \rho - \frac13 \mathbf{1}  \right)^{w} ,
 \quad   g(x) \in SU(3) .
\end{equation}

The Dynkin indices and weight vectors of the fundamental representations of $SU(3)$ are given by 
\begin{subequations}
\begin{align}
%\begin{enumerate}
%\item[[1,0]]:
%\begin{equation}
[1,0]:&
 \bm{\Lambda} 
= \left( \frac{1}{2}, \frac{1}{2\sqrt{3}} \right) := \vec{\nu}_1 
 , \quad
  \mathcal{H} = -\frac{1}{6} {\rm diag} \left( -2,  1,  1 \right) 
   ,
   \label{C40-fr1}
%\end{equation}
\\
%\item[[-1,1]]:
%\begin{equation}
[-1,1]:&
 \bm{\Lambda} 
= \left( \frac{-1}{2}, \frac{1}{2\sqrt{3}} \right) := \vec{\nu}_2 
 , \quad
  \mathcal{H} = -\frac{1}{6} {\rm diag} \left(  1,-2,  1 \right)
   ,
   \label{C40-fr2}
%\end{equation}
\\
%\item[[0,-1]]:
%\begin{equation}
[0,-1]:&
 \bm{\Lambda} 
= \left( 0, \frac{-1}{\sqrt{3}} \right) := \vec{\nu}_3 
 , \quad
  \mathcal{H} = -\frac{1}{6} {\rm diag} \left(  1,  1, -2 \right)
  = \frac{-1}{\sqrt{3}} \frac{\lambda_8}{2} 
   ,
   \label{C40-fr3}
%\end{equation}
%\end{enumerate}
\end{align}
\end{subequations}
while their conjugates $\textbf{3$^*$}$ are given by 
\begin{subequations}
\begin{align}
%\begin{enumerate}
%\item[[0,1]]:
%\begin{equation}
[0,1]:&
 \bm{\Lambda} 
= \left( 0, \frac{1}{\sqrt{3}} \right) = -\vec{\nu}_3 
 , \ 
  \mathcal{H} = - \frac{1}{6} {\rm diag} \left( 1,  1 ,-2  \right)
  = \frac{-1}{\sqrt{3}} \frac{\lambda_8}{2} 
   ,
%\end{equation}
\\
%\item[[1,-1]]:
%\begin{equation}
[1,-1]:&
\bm{\Lambda} 
= \left( \frac{1}{2}, \frac{-1}{2\sqrt{3}} \right) = -\vec{\nu}_2 
 , \ 
  \mathcal{H} = - \frac{1}{6} {\rm diag} \left( 1, -2, 1 \right)
   ,
%\end{equation}
\\
%\item[[-1,0]]:
%\begin{equation}
[-1,0]:&
 \bm{\Lambda} 
= \left( \frac{-1}{2}, \frac{-1}{2\sqrt{3}} \right) = -\vec{\nu}_1 
 , \ 
  \mathcal{H} = - \frac{1}{6} {\rm diag} \left(  -2, 1, 1 \right)
   .
%\end{equation}
%\end{enumerate}
\end{align}
\end{subequations}

By using  the weight vectors $\vec{\nu}_1,\vec{\nu}_2,\vec{\nu}_3$ of the fundamental representation and the Cartan generators $\vec{H}=(H_1, H_2)$, 
the Weyl-reflected color field are written as  
\begin{align}
& 
 \bm{n}^{w_1}(x) =\bm{n}^{w_3}(x) = \bm{n}(x) = \sqrt{3} g(x) \vec{\nu_1} \cdot \vec{H} g(x)^\dagger ,  
\nonumber\\
& 
 \bm{n}^{w_2}(x) =\bm{n}^{w_5}(x)= \sqrt{3} g(x) \vec{\nu_2} \cdot \vec{H} g(x)^\dagger ,   
\nonumber\\
& 
 \bm{n}^{w_4}(x) =\bm{n}^{w_6}(x) = \sqrt{3} g(x) \vec{\nu_3} \cdot \vec{H} g(x)^\dagger ,  
\end{align}
where we have used the fact:
\begin{align} 
\mathcal{H}^{w_1}
=& \mathcal{H}^{w_3} 
=    \frac12 H_1 + \frac{1}{2\sqrt{3}} H_2 
=   \vec{\nu_1} \cdot \vec{H}   ,
\nonumber\\
\mathcal{H}^{w_2} 
=& \mathcal{H}^{w_5}
=    \frac{-1}{2} H_1 + \frac{1}{2\sqrt{3}} H_2   
=   \vec{\nu_2} \cdot \vec{H}   ,
\nonumber\\
\mathcal{H}^{w_4} 
=& \mathcal{H}^{w_6}
=  \frac{-1}{\sqrt{3}}   H_2 
=   \vec{\nu_3} \cdot \vec{H}   .
\end{align}

The six root vectors of $SU(3)$ are given by 
\begin{align}
%\vec{\alpha} =& 
\pm \vec{\alpha}^{(1)} := \pm ( 1,  0 ), \ 
%\rightarrow 
%v_\alpha  = \frac{1}{\sqrt{6}} (1, \pm i, 0, 0, 0, 0, 0, 0)^T ,
%\nonumber\\
%\vec{\alpha} =&  
\pm \vec{\alpha}^{(2)} := \pm \left(  \frac12 ,   \frac{\sqrt{3}}{2} \right), \
%\rightarrow 
%v_\alpha  =   \frac{1}{\sqrt{6}} (0, 0, 0, 1, \pm i, 0, 0, 0, 0)^T ,
%\nonumber\\
%\vec{\alpha} =& 
\pm \vec{\alpha}^{(3)} := \pm \left(- \frac12,  \frac{\sqrt{3}}{2} \right) .
%\rightarrow 
%v_\alpha  = \frac{1}{\sqrt{6}} (0, 0, 0, 0, 0, 1, \pm i, 0)^T ,
  \label{C40-eigen-SU3}
\end{align}
It should be remarked that each weight vector of the fundamental representation is orthogonal to the two of the six root vectors:
\begin{equation}
     \pm \vec{\alpha}^{(3)} \cdot  \vec{\nu_1}  = 0 , \
     \pm \vec{\alpha}^{(2)} \cdot  \vec{\nu_2}  = 0 , \
     \pm \vec{\alpha}^{(1)} \cdot  \vec{\nu_3}  = 0 . 
\end{equation}
Therefore, the Weyl-reflected color fields satisfy
\begin{align}
& 
 \bm{n}^{w_1}(x) =\bm{n}^{w_3}(x) = \bm{n}(x) = \sqrt{3} g(x) \vec{\nu_1} \cdot \vec{H} g(x)^\dagger \in SU(3)/U(2) \simeq CP^2,   
\nonumber\\
& 
 \bm{n}^{w_2}(x) =\bm{n}^{w_5}(x)= \sqrt{3} g(x) \vec{\nu_2} \cdot \vec{H} g(x)^\dagger \in SU(3)/U(2) \simeq CP^2,    
\nonumber\\
& 
 \bm{n}^{w_4}(x) =\bm{n}^{w_6}(x) = \sqrt{3} g(x) \vec{\nu_3} \cdot \vec{H} g(x)^\dagger \in SU(3)/U(2) \simeq CP^2 .   
\end{align}

On the other hand, if we adopt the color field $\bm{n}_3(x)$ with the flag space $F^2$ as the target space:
\begin{align}
  \bm{n}_3(x) 
%=  \frac{\sqrt{3}}{2} g_{x} \left(  \rho - \frac13 \mathbf{1} \right) g_{x}^\dagger 
=   {g(x)}
\frac{ 1}{2} 
{\rm diag}(1,-1,0)
   {g(x)^\dagger} 
\in SU(3)/(U(1)\times U(1)) \simeq F^2 ,
\end{align}
then the Weyl-reflected color fields are obtained as
\begin{align}
& 
 \bm{n}_3^{w_1}(x) = - \bm{n}_3^{w_2}(x) = \bm{n}_3(x) 
=   {g(x)} \frac{1}{2} {\rm diag}(1,-1,0) {g(x)^\dagger} \in SU(3)/(U(1)\times U(1))  ,   
\nonumber\\
& 
 \bm{n}_3^{w_3}(x) =- \bm{n}_3^{w_6}(x) 
=   {g(x)} \frac{1}{2} {\rm diag}(1,0,-1) {g(x)^\dagger} \in SU(3)/(U(1)\times U(1))   ,    
\nonumber\\
& 
 \bm{n}_3^{w_5}(x) =- \bm{n}_3^{w_4}(x) 
=   {g(x)} \frac{1}{2} {\rm diag}(0,1,-1) {g(x)^\dagger}  \in SU(3)/(U(1)\times U(1))   .   
\end{align}
We find that they are rewritten using the root vectors $\vec{\alpha}^{(1)},\vec{\alpha}^{(2)},\vec{\alpha}^{(3)}$ (weight vectors of the adjoint representation) and the Cartan generators $\vec{H}=(H_1, H_2)$ as  
\begin{align}
& 
 \bm{n}_3^{w_1}(x) = - \bm{n}_3^{w_2}(x) = \bm{n}_3(x) =   g(x) \vec{\alpha}^{(1)}  \cdot \vec{H} g(x)^\dagger 
%\in SU(3)/(U(1)\times U(1))
  ,   
\nonumber\\
& 
 \bm{n}_3^{w_3}(x) =- \bm{n}_3^{w_6}(x)=   g(x) \vec{\alpha}^{(2)} \cdot \vec{H} g(x)^\dagger 
%\in SU(3)/(U(1)\times U(1)) 
  ,    
\nonumber\\
& 
 \bm{n}_3^{w_4}(x) =- \bm{n}_3^{w_5}(x) =   g(x) \vec{\alpha}^{(3)} \cdot \vec{H} g(x)^\dagger 
%\in SU(3)/(U(1)\times U(1)) 
  .   
\end{align}

The \textbf{ color reflection group} introduced in \cite{Cho80} is neither the center group nor the Weyl reflection group. It is a generalization of the Weyl group.%  
\footnote{
This is according to Cho \cite{Cho80}. 
Cho call this symmetry the color reflection symmetry. 
}
For $SU(2)$, the reflection group consists of four elements written in the matrix form:
\begin{align}
&  R_1
 = W_1 
 = \begin{pmatrix} 
   1 & 0 \\
   0 & 1 
   \end{pmatrix} = \bm{1},
   \quad 
 R_2
 =  W_2 
 = \begin{pmatrix} 
   0 & 1 \\
   -1 & 0 
   \end{pmatrix} ,
   \quad 
\nonumber\\&
   R_3 
   = - W_1 
 = \begin{pmatrix} 
   -1 & 0 \\
   0 & -1 
   \end{pmatrix} =-\bm{1},
   \quad 
   R_4 
   = - W_2
 = \begin{pmatrix} 
   0 & -1 \\
    1 & 0 
   \end{pmatrix}
 .
\end{align}
Here $W_1$ and $W_2$ are elements of the Weyl group.  Indeed, $W_a$ are elements of $SU(2)$, since $W_a W_a^\dagger = W_a^\dagger W_a=\bm{1}$ and $\det W_a=1$. 
Note that $W_2^\dagger=-W_2$ and $W_2W_2=-\bm{1}$, which means that $-\bm{1}$ and $-W_2$ are also elements of $SU(2)$. Under the group multiplication, the four elements $R_a$ of the reflection group form a closed set. 

%%%%%%%%%%%%%%%%%%%%%%%%%%%%%%%%%%%%%%%%%%%%%%%%%%%%%%%%%%%%%
%%%%%%%%%%%%%%%%%%%%%%%%%%%%%%%%%%%%%%%%%%%%%%%%%%%%%%%%%%%%%
\begin{center}  
\begin{tabular}{c||c|c}  
 & $W_1$ & $W_2$   \\  \hline\hline
$W_1$ & $W_1$ & $W_2$   \\  \hline 
$W_2$ & $W_2$ & $-\bm{1}$      \\   
\end{tabular}
% \caption[]{After inverting the concentric sphere geometry for a smeared meron about the point $d$ (left panel), we obtain the smeared two meron configuration (right panel). }
\label{table:Weyl-ref1}
\end{center}
%%%%%%%%%%%%%%%%%%%%%%%%%%%%%%%%%%%%%%%%%%%%%%%%%%%%%%%%%%%%%
%%%%%%%%%%%%%%%%%%%%%%%%%%%%%%%%%%%%%%%%%%%%%%%%%%%%%%%%%%%%%

\noindent
Then the color-reflected color field reads
\begin{align}
  {\bm{n}}^{w_1}(x) =  {\bm{n}}^{w_3}(x) =   {\bm{n}}(x), \quad 
 \bm{n}^{w_2}(x) = {\bm{n}}^{w_4}(x) =  -\bm{n} (x) .
\end{align}
Among the $q\bar q$ mesons and the $qq$ baryons one may easily find that only the color-singlet states satisfy the reflection invariance. Therefore, the color reflection invariance of the magnetic condensation excludes any colored states from the physical spectrum of the theory \cite{Cho80}. 

For $SU(3)$, the reflection group consists of 24(=4*6) elements written in the matrix form:
\begin{align}
 &
 R_c=\omega_a W_b , \quad (a=1,2,3,4; b=1,...,6)
\nonumber\\&
   \omega_1 
 = \begin{pmatrix} 
   1 & 0 & 0 \\
   0 & 1 & 0 \\
   0 & 0 & 1 
   \end{pmatrix} ,
   \quad 
   \omega_2 
 = \begin{pmatrix} 
   1 & 0 & 0 \\
   0 & -1 & 0 \\
   0 & 0 & -1 
   \end{pmatrix} ,
   \quad 
%\nonumber\\&
   \omega_3 
 = \begin{pmatrix} 
   -1 & 0 & 0 \\
   0 & 1 & 0 \\
   0 & 0 & -1 
   \end{pmatrix} ,
   \quad 
   \omega_4 
 = \begin{pmatrix} 
   -1 & 0 & 0 \\
   0 & -1 & 0 \\
   0 & 0 & 1 
   \end{pmatrix}
 .
\end{align}
Here each $W_b$ $(b=1,\dots,6)$ is a matrix element forming the Weyl group and  each $\omega_a$ $(a=1,2,3,4)$ is an element of $SU(3)$.
Note that $\omega_a \omega_a=\bm{1}$ for each $a$. 
It is shown that the 24 elements $R_c$ of the reflection group form a closed set under the group multiplication. 
%[Ex-1] Check that . 

%%%%%%%%%%%%%%%%%%%%%%%%%%%%%%%%%%%%%%%%%%%%%%%%%%%%%%%%%%%%%
%%%%%%%%%%%%%%%%%%%%%%%%%%%%%%%%%%%%%%%%%%%%%%%%%%%%%%%%%%%%%
%\begin{center}  
%\begin{tabular}{c||c|c|c|c|c|c}   
% & $W_1$ & $W_2$ & $W_3$ & $W_4$ & $W_5$ & $W_6$  \\  \hline\hline
%$W_1$ & $W_1$ & $W_2$ & $W_3$ & $W_4$ & $W_5$ & $W_6$  \\  \hline 
%$W_2$ & $W_2$ & $\omega_4$ & $W_6$ &  $W_5\omega_2$ & $W_4$ & $$     \\  \hline 
%$W_3$ & $W_3$ & $$ & $\omega_2$ & $W_6$ & $$ & $$  \\  \hline 
%$W_4$ & $W_4$ & $$ & $$ & $\omega_3$ & $$ & $$  \\  \hline 
%$W_5$ & $W_5$ & $$ & $$ & $$ & $$ & $\omega_2$  \\  \hline 
%$W_6$ & $W_6$ & $$ & $$ & $$ & $\omega_3$ & $W_5\omega_3$  \\    
%\end{tabular}
% \caption[]{After inverting the concentric sphere geometry for a smeared meron about the point $d$ (left panel), we obtain the smeared two meron configuration (right panel). }
%\label{table:eigenvalue1}
%\end{center}
%%%%%%%%%%%%%%%%%%%%%%%%%%%%%%%%%%%%%%%%%%%%%%%%%%%%%%%%%%%%%
%%%%%%%%%%%%%%%%%%%%%%%%%%%%%%%%%%%%%%%%%%%%%%%%%%%%%%%%%%%%%

The \textbf{center group} of $G$ is a discrete subgroup of $G$, such that each element of the center group commutes with all the elements of $G$.  
Therefore, the center element $g$ is proportional to the unit matrix, i.e., $g=z\bm{1}$ ($z \in \mathbb{C}$). 
The center of $SU(N)$ is $Z(N)$, since    
the properties $g^\dagger g=gg^\dagger=\bm{1}$ and $\det g=1$ lead to $zz^*=1$ and $z^N=1$ respectively:  
\begin{align}
 \text{Center}(SU(N))= Z(N) = \{  e^{2\pi i\frac{n}{N} } \bm{1} ; n=0,1, \dots, N-1  \} .
\end{align}
For $SU(2)$, we find
\begin{align}
  \text{Center}(SU(2))=Z(2) = \{ \bm{1}, -  \bm{1} \}   .
\end{align}
For $SU(3)$, we find
\begin{align}
  \text{Center}(SU(2))=Z(3) = \{  \bm{1}, e^{ i\frac{2\pi}{3}} \bm{1}, e^{ -i\frac{2\pi}{3}} \bm{1} \}   .
\end{align}

Some remarks are in order.
\begin{itemize}

\item
We have discussed three kinds of discrete symmetries: Weyl symmetry, center symmetry and color reflection symmetry.
It should be remarked that the Weyl symmetry and the center symmetry are the symmetries existing in the original theory. 
%while the (color) reflection symmetry for the color field exists only in the reformulated theory (e.g.,  $\bm{n}(x) \rightarrow -\bm{n}(x)$ for $SU(2)$). 
It should be kept in mind that the Weyl symmetry and the color reflection symmetry are  global symmetries, while the center symmetry can be made a local symmetry as a subgroup of the original local gauge symmetry. Therefore, the Weyl symmetry is a subgroup of color symmetry in the original theory. 

\item
If the target space of the color field $\bm{n}(x)$ is given by $G/\tilde H$, then the target space of the Weyl-reflected color field $\bm{n}^{w}(x)$ has the same target space $G/\tilde H$: 
\begin{align}
  \bm{n}(x)  \in G/\tilde H    
 \Longrightarrow \bm{n}^{w}(x)  \in G/\tilde H  .
\end{align}
For $SU(3)$, e.g., two options are possible:
\begin{align}
 & \bm{n}(x)  \in SU(3)/U(2)    
 \Longrightarrow \bm{n}^{w}(x)  \in SU(3)/U(2)  ,
 \nonumber\\
 & \bm{n}(x)  \in SU(3)/U(1)^2    
 \Longrightarrow \bm{n}^{w}(x)  \in SU(3)/U(1)^2  .
\end{align}

\item
For $G=SU(2)$, the Weyl transformation can be identified  with the reflection for the color field: $\bm{n}^{w}(x)=-\bm{n}(x)$. 
This is not the case for $G=SU(3)$.
For $G=SU(3)$, the Wely symmetry cannot be identified  with the simple reflection: $\bm{n}^{w}(x) \not= -\bm{n}(x)$. This is also the case for $\bm{n}_3^{w}(x) \not= -\bm{n}_3(x)$, as well as $\bm{n}_8^{w}(x) \not= -\bm{n}_8(x)$.

\end{itemize}

In the following, we discuss the role of the discrete symmetries in constructing the new reformulation.

 For $SU(2)$, the reflection symmetry $\bm{n}(x) \rightarrow -\bm{n}(x)$ is required in the new reformulation by the following reasons.
The ordinary superconductivity is caused by the condensation of the Cooper pairs, i.e., electron pairs with  charges of the same signature. 
In  ordinary superconductors the magnetic flux   is accompanied by the surrounding supercurrent which is exclusively made of the electron pairs alone.  
There is no supercurrent made of the positron pairs in the ordinary superconductor in the real world. 
(We can imagine a virtual world constructed from the antimatters.  In such a world, the positron pairs will cause the similar phenomenon to the superconductivity where the matters are replaced by the corresponding antimatters.)

The invariance of the theory under the Weyl reflection means that  there is no way for us to tell whether the magnetic condensation of the dual superconductor vacuum is made of monopoles or antimonopoles.  

The theory supporting the dual superconductivity has the reflection symmetry, since the reflection $\bm{n}(x) \rightarrow -\bm{n}(x)$ leads to the opposite magnetic charge $-q_m$, if the magnetic charge $q_m$ is derived from the color field $\bm{n}(x)$.

Consequently,  the magnetic ``supercurrent'  that confines the color electric flux in the dual superconductor of QCD has to be made of the symmetric combination of the two oppositely charged monopoles. 
The dual superconductivity is supposed to be caused by condensation of monopole-antimonopole pairs due to the attractive force, which should be distinguished from the monopole-monopole condensation or antimonopole-antimonopole condensation. 
In this respect we find that the two confinement mechanisms are not exactly dual to each other.

 The color field defined in the original way is not Weyl symmetric.
In arriving at the master Yang-Mills theory by introducing the color field, therefore, we can impose the extra Weyl symmetry on the color field, which is however different from the original Weyl symmetry existing in the original gauge symmetry $G$ respected by the original Yang-Mills theory, if we wish to do so.  
In order to obtain the equipollent theory, however, the extra Weyl symmetry should be removed by imposing the appropriate reduction condition. 
Then the equipollent theory has only the original Weyl symmetry.

 For $SU(2)$, the reduction condition respects the reflection symmetry. 
For a given original field $\mathscr{A}_\mu(x)$, if the color field $\bm{n}(x)$ is a solution, the reflected color field $-\bm{n}(x)$ is also a solution. 
This is seen from the explicit form of the differential reduction condition:
$\bm{n} \times D_\mu[\mathscr{A}] D_\mu[\mathscr{A}] \bm{n}=0$.

For $SU(2)$, the defining equations for the new  field variable are reflection covariant. Consequently, the decomposed  variables are reflection invariant.

%\item
%The non-Abelian Stokes theorem must be used in the reformulated theory (YM'). 

%The defining equation for obtaining the Weyl-reflected new variables is as follows. 

On the lattice, the color field respecting the discrete symmetries is constructed as follows.
As an initial configuration of the color field, we prepare the initial color field: 
\begin{align}
 \bm{n}^{(0)}(x) := g^{(0)}(x) T_0 g^{(0)}{}^\dagger(x) \in Lie(G/\tilde H) , \quad g^{(0)}(x) \in G=SU(3)  . 
\end{align}
For instance, in the minimal option of $SU(3)$, we choose 
\begin{align}
 T_0 = \frac12 \lambda_8  . 
\end{align}
The color field is locally updated according to the procedure:
\begin{align}
 \bm{n}^{(0)}(x) \to  \bm{n}^{(1)}(x) := g^{(1)}(x) \bm{n}^{(0)}(x) g^{(1)}{}^\dagger(x)  , \quad g^{(1)}(x) \in G=SU(3) .
\end{align}
This procedure is repeated 
\begin{align}
 \bm{n}^{(n-1)}(x) \to \bm{n}^{(n)}(x) := g^{(n)}(x) \bm{n}^{(n-1)}(x) g^{(n)}{}^\dagger(x) , \quad g^{(n)}(x) \in G=SU(3),
\end{align}
until the minimum of the reduction functional $F_{\rm red}[\bm{n}^{(n)}]$ is reached.

The Weyl group is a discrete subgroup of $G$ and therefore an element of the Weyl group is contained in the element of $G$. 
Using an element $w$ of the Weyl group, we can replace the initial group element $g^{(0)}(x)$ by
\begin{align}
 g^{(0)}(x) \to  g^{(0)}(x) w  . 
\end{align}
Then the color field is modified to the Weyl-reflected field:
\begin{align}
 \bm{n}^{(0)}(x) \to  \bm{n}^{(0)}{}^{w}(x) := g^{(0)}(x) w T_0 w^\dagger g^{(0)}{}^\dagger(x) . 
\end{align}
By repeating the above step sufficiently many times, therefore, the Weyl symmetry is respected in the simulations.

Moreover, it is easy to see that the Center symmetry is respected by the simulation.  
The color field is invariant under the center symmetry transformation: 
\begin{align}
 \bm{n}^{(0)}(x) \to  \bm{n}^{(0)}{}^{z}(x) := g^{(0)}(x) z \bm{1} T_0 \bm{1} z^* g^{(0)}{}^\dagger(x) = g^{(0)}(x) T_0 g^{(0)}{}^\dagger(x) = \bm{n}^{(0)} (x) . 
\end{align}

\end{appendix}

\newpage
%%%%%%%%%%%%%%%%%%%%%%
%Bibliography
%%%%%%%%%%%%%%%%%%%%%%

\end{document}